\documentclass[12pt]{iopart}

\pdfoutput=1

\usepackage[perpage,marginal]{footmisc}
\makeatletter
\renewcommand\footnoterule{%
  \kern-3\p@
  \hrule\@width2.5cm
  \kern2.6\p@}
\makeatother

\usepackage{color, xcolor}
\usepackage[hypertexnames=false, unicode, colorlinks, allcolors=blue!70!black, linktocpage, pdfusetitle]{hyperref}
\usepackage[all]{hypcap}
\usepackage{amssymb,amsbsy}
\usepackage{indentfirst}
\usepackage[pdftex]{graphicx}
\usepackage[numbers,square,comma,sort&compress]{natbib}
\usepackage{url}

\usepackage[T1]{fontenc}
\usepackage[utf8]{inputenc}
\usepackage{lmodern}

\newcommand{\be}{\begin{equation}}
\newcommand{\ee}{\end{equation}}
\newcommand{\ba}{\begin{eqnarray}}
\newcommand{\ea}{\end{eqnarray}}
\def\bea{\begin{eqnarray}}
\def\eea{\end{eqnarray}}
\newcommand{\bes}{\begin{subequations}}
\newcommand{\ees}{\end{subequations}}

\usepackage{soul}
\usepackage{bm}
\definecolor{linkcolor}{rgb}{0.0,0.3,0.5}

\def\simgreat{\lower2pt\hbox{$\buildrel {\scriptstyle >}
   \over {\scriptstyle\sim}$}}
\def\simless{\lower2pt\hbox{$\buildrel {\scriptstyle <}
   \over {\scriptstyle\sim}$}}
\def\msun{\,{\rm M_\odot}}

\def\kms{\,{\rm km\,s^{-1}}}

\def\msun{{\rm ~M}_{\odot}}
\def\rsun{{\rm ~R}_{\odot}}
\def\zsun{{\rm ~Z}_{\odot}}
\def\gpy{{\rm ~Gpc}^{-3} {\rm ~yr}^{-1}}
\def\kms{{\rm ~km} {\rm ~s}^{-1}}

\usepackage{bm}
\newcommand{\Msun}{\ensuremath \mathrm{M}_{\odot}}

\newcommand{\yr}{\ensuremath{\rm yr}}
\newcommand{\Gpc}{\ensuremath{\rm Gpc}}
\newcommand{\Mpc}{\ensuremath{\rm Mpc}}
\newcommand{\pc}{\ensuremath{\rm pc}}
\DeclareMathAccent{\dot}    {\mathalpha}{operators}{'137} 
\DeclareMathAccent{\ddot}    {\mathalpha}{operators}{'177} 

\newcommand{\araa}{ARA\&A}
\newcommand{\apj}{ApJ}
\newcommand{\apss}{Ap\&SS}
\newcommand{\aap}{A\&A}
\newcommand{\aapr}{A\&AR}
\newcommand{\mnras}{MNRAS}
\newcommand{\na}{New Astron.}
\newcommand{\nar}{New A Rev.}
\newcommand{\pasj}{PASJ}
\newcommand{\ssr}{Space~Sci.~Rev.}

\begin{document}

\title[Black holes, gravitational waves and fundamental physics: a roadmap]{Black holes, gravitational waves and fundamental physics: a roadmap}

\hypersetup{pdfauthor={Barack, Cardoso, Nissanke, Sotiriou, et al.}}

\author{
Leor Barack$^{1}$, Vitor Cardoso$^{2,3}$, Samaya Nissanke$^{4,5,6}$, Thomas P.~Sotiriou$^{7,8}$ (editors)
}

\author{
Abbas Askar$^{9,10}$,
Krzysztof Belczynski$^{9}$,
Gianfranco Bertone$^5$, 
Edi Bon$^{11,12}$, 
Diego Blas$^{13}$, 
Richard Brito$^{14}$, 
Tomasz Bulik$^{15}$,
Clare Burrage$^{8}$, 
Christian T.~Byrnes$^{16}$,
Chiara Caprini$^{17}$,
Masha Chernyakova$^{18,19}$, 
Piotr Chru\'sciel$^{20,21}$,
Monica Colpi$^{22,23}$,
Valeria Ferrari$^{24}$,
Daniele Gaggero$^{5}$,
Jonathan Gair$^{25}$, 
Juan Garc\'ia-Bellido$^{26}$, 
S. F. Hassan$^{27}$, 
Lavinia Heisenberg$^{28}$,
Martin Hendry$^{29}$,
Ik Siong Heng$^{29}$, 
Carlos Herdeiro$^{30}$, 
Tanja Hinderer$^{4,14}$, 
Assaf Horesh$^{31}$, 
Bradley J.~Kavanagh$^{5}$,
Bence Kocsis$^{32}$,
Michael Kramer$^{33,34}$,
Alexandre Le Tiec$^{35}$,
Chiara Mingarelli$^{36}$,
Germano Nardini$^{37a,37b}$, 
Gijs Nelemans$^{4,6}$
Carlos Palenzuela$^{38}$, 
Paolo Pani$^{24}$, 
Albino Perego$^{39,40}$,
Edward K. Porter$^{17}$,
Elena M. Rossi$^{41}$,
Patricia Schmidt$^{4}$, 
Alberto Sesana$^{42}$,
Ulrich Sperhake$^{43,44}$, 
Antonio Stamerra$^{45,46}$,
Leo C. Stein$^{43}$,
Nicola Tamanini$^{14}$,
Thomas M. Tauris$^{33,47}$,
L. Arturo Urena-L\'opez$^{48}$, 
Frederic Vincent$^{49}$, 
Marta Volonteri$^{50}$, 
Barry Wardell$^{51}$,
Norbert Wex$^{33}$,
Kent Yagi$^{52}$ 
(Section coordinators)
}

\author{
Tiziano Abdelsalhin$^{24}$,
Miguel \'Angel Aloy$^{53}$,
Pau Amaro-Seoane$^{54,55,56}$,
Lorenzo Annulli$^{2}$,
Manuel Arca-Sedda$^{57}$,
Ibrahima Bah$^{58}$,
Enrico Barausse$^{50}$,
Elvis Barakovic$^{59}$,
Robert Benkel$^{7}$,
Charles L. Bennett$^{58}$
% ...
}

\newpage
\vskip 2cm

\begin{abstract} 
The grand challenges of contemporary fundamental physics---dark matter, dark energy, vacuum energy, inflation and early universe cosmology, singularities and the hierarchy problem---all involve {\em gravity} as a key component. And of all gravitational phenomena, {\em black holes} stand out in their elegant simplicity, while harbouring some of the most remarkable predictions of General Relativity: event horizons, singularities and ergoregions.

The hitherto invisible landscape of the gravitational Universe is being unveiled before our eyes: the historical direct detection of gravitational waves by the LIGO-Virgo collaboration marks the dawn of a new era of scientific exploration. Gravitational-wave astronomy will allow us to test models of black hole formation, growth and evolution, as well as models of gravitational-wave generation and propagation. It will provide evidence for event horizons and ergoregions, test the theory of General Relativity itself, and may reveal the existence of new fundamental fields. The synthesis of these results has the potential to radically reshape our understanding of the cosmos and of the laws of Nature.

The purpose of this work is to present a concise, yet comprehensive overview of the state of the art in the relevant fields of research, summarize important open problems, and lay out a roadmap for future progress. This write-up is an initiative taken within the framework of the European Action on ``Black holes, Gravitational waves and Fundamental Physics''.
\end{abstract}


\newpage

\section*{Glossary}
Here we provide an overview of the acronyms used throughout this
paper and also in common use in the literature. \\

\begin{tabular}{ll}
	BBH   &  Binary black hole\\
  BH    &  Black hole \\
	BNS   &  Binary neutron star\\
  BSSN  &  Baumgarte-Shapiro-Shibata-Nakamura \\
	CBM   &  Compact binary mergers \\
	CMB   &  Cosmic microwave background \\
	DM    &  Dark matter \\
	ECO   &  Exotic Compact Object\\
  EFT   &  Effective Field theory \\
	EMRI  &  Extreme-mass-ratio inspiral\\
	EOB   &  Effective One Body model\\
	EOS   &  Equation of state\\
	eV    &  electron Volt\\
  GR    &  General Relativity \\
	GSF   &  Gravitational self-force \\
	GRB   &  Gamma-ray burst\\
  GW    &  Gravitational Wave \\
	HMNS  &  Hypermassive neutron star\\
	IMBH  &  Intermediate-mass black hole\\
	IVP   &  Initial Value Problem\\
	LVC   &  LIGO Scientific and Virgo Collaborations\\
	MBH   &  Massive black hole\\
	NK    &  Numerical kludge model\\
	NSB   &  Neutron star binary\\
	NS    &  Neutron star\\
	NR    &  Numerical Relativity\\
	PBH   &  Primordial black hole\\
  PN    &  Post-Newtonian \\
	PM    &  Post-Minkowskian\\
	QNM   &  Quasinormal modes\\
	sBH   &  Black hole of stellar origin\\
	SGWB  &  Stochastic GW background \\
	SM    &  Standard Model \\
	SMBBH &  Supermassive binary black hole\\
	SOBBH &  Stellar-origin binary black hole\\
	SNR   &  Signal-to-noise ratio \\
	ST    &  Scalar-tensor
\end{tabular}

\newpage

\maketitle

\tableofcontents

\markboth{Black holes, gravitational waves and fundamental physics: a roadmap}{Black holes, gravitational waves and fundamental physics: a roadmap}
%
%
%
%
%
%

\newpage
\section*{Preface}

The long-held promise of {\it gravitational-wave astronomy} as a new window onto the universe has finally materialized with the dramatic discoveries of the LIGO-Virgo collaboration in the past few years. We have taken but the first steps along a new, exciting avenue of exploration that has now opened before us. The questions we will tackle in the process are cross-cutting and multidisciplinary, and the answers we will get will no doubt reshape our understanding of black-hole-powered phenomena, of structure formation in the universe, and of gravity itself, at all scales.

The harvesting of useful information from gravitational-wave (GW) signals and the understanding of its broader implications demand a cross-disciplinary effort. What exactly will GWs tell us about how, when and in which environment black holes were formed? How fast do black holes spin and how have some of them grown to become supermassive? GWs from merging black holes probe the environment in which they reside, potentially revealing the effect of dark matter or new fundamental degrees of freedom. The analysis of GWs will allow for precise tests of General Relativity, and of the black hole paradigm itself. However, to be able to collect and interpret the information encoded in the GWs, one has to be equipped with faithful and accurate theoretical models of the predicted waveforms. To accomplish the far-reaching goals of gravitational-wave science it is of paramount importance to bring together expertise over a very broad range of topics, from astrophysics and cosmology, through general-relativistic source modelling to particle physics and other areas of fundamental science.

In 2016, a short time before the announcement of the first gravitational-wave detection, a cross-disciplinary initiative in Europe led to the establishment of the new COST networking Action on ``Black holes, gravitational waves and fundamental physics'' (``GWverse''). GWverse aims to maintain and consolidate leadership in black-hole physics and gravitational-wave science, linking three scientific communities that are currently largely disjoint: one specializing in gravitational-wave detection and analysis, another in black-hole modelling (in both astrophysical and general-relativistic contexts), and a third in strong-gravity tests of fundamental physics. The idea is to form a single, interdisciplinary exchange network, facilitating a common language and a framework for discussion, interaction and learning.  The Action will support the training of the next generation of leaders in the field, and the very first ``native'' GW/multi-messenger astronomers, ready to tackle the challenges of high-precision GW astronomy with ground and space-based detectors.

\vskip 1cm

\noindent Leor Barack

\noindent Vitor Cardoso

\noindent Samaya Nissanke

\noindent Thomas Sotiriou

\newpage

\phantomsection
\addcontentsline{toc}{part}{\bf Chapter I: The astrophysics of compact object mergers: prospects and challenges}
\begin{center}
{\large \bf Chapter I: The astrophysics of compact object mergers: prospects and challenges}
\end{center}
\begin{center}
Editor: Samaya Nissanke
\end{center}

\vskip 1cm
\section{Introduction} \label{Sec:introduction1}

In the last two years, strong-field gravity astrophysics research has been undergoing a momentous transformation thanks to the recent discoveries of five binary black hole (BBH) mergers that were observed in gravitational waves (GWs) by the LIGO and Virgo detectors. This was compounded last year by the multi-messenger discovery of a binary neutron star (BNS) merger measured in both GWs and detected in every part of the electromagnetic (EM) spectrum, allowing us to place compact object mergers in their full astrophysical context. These measurements have opened up an entirely new window onto the Universe, and given rise to a new rapidly growing and observationally-driven field of GW astrophysics. 

Despite the multiple scientific breakthroughs and ``firsts'' that these discoveries signify, the measured properties of the BBH and BNS mergers have immediately bought up accompanying challenges and pertinent questions to the wider astrophysics community as a whole. Here, we aim to provide an up-to-date and encompassing review of the astrophysics of compact object mergers and future prospects and challenges. Section~\ref{Sec:LVC} first introduces and briefly details the LIGO and Virgo observations of BBH and BNS mergers. Section~\ref{Sec:BHgenesis} then discusses the astrophysics of BHs, in particular BBHs, from their genesis to archeology, for BHs that span more than ten decades in their mass range. In the case of stellar-mass BBH mergers, Section~\ref{Sec:binaryevolution} details the formation of compact binary mergers, in particular BBHs, through isolated stellar binary evolution. Section~\ref{Sec:nbody} then reviews how one could dynamically form such events in order to explain the observed merger rate and distribution of masses, mass ratios, and spins. Section~\ref{Sec:PBHandDM} explores the intriguing possibility that at least a fraction of dark matter (DM) in the Universe is in the form of primordial BHs (PBHs), an area that has recently been invigorated by the recent LIGO and Virgo observations of BBH mergers. Section~\ref{Sec:FormSMBH} presents an overview on the formation of supermassive BBHs through galaxy mergers, and Section~\ref{Sec:PTA} introduces efforts underway to probe the astrophysics of such supermassive BBHs with pulsar timing arrays. Turning our attention to the mergers themselves as multi-messenger sources, Section~\ref{Sec:NumericalRelativity} reviews the state-of-the-art numerical modelling of compact object mergers, in particular, systems with NSs in which we have already observed accompanying EM radiation. For detailed and exhaustive discussion of the modelling of BBH mergers, we refer the reader to Chapter II. Section~\ref{Sec:EMfollowup} provides a summary of the observational efforts by a wide range of facilities and instruments in following up GW mergers in light of the first BNS merger discovery measured in both GWs and EM. Focusing entirely on EM observations, Section~\ref{Sec:BHBandAGN} reviews observations of active galactic nuclei as probes of BBH systems and Section~\ref{Sec:XrayGammaRay} concludes by summarising recent advances in high-energy observations of X-ray binaries. Finally, Section~\ref{Sec:cosmography} provides an extensive review on how observations of GWs can impact the field of cosmology, that is, in our understanding of the origins, evolution and fate of the Universe. 

\section{LIGO and Virgo Observations of Binary Black Hole Mergers and a Binary Neutron Star } \label{Sec:LVC}
\vspace{-3mm}
{\it Contributors:} E.~Porter, M.~Hendry, I.~S.~Heng
\vspace{3mm}

On September 15th 2014 the discovery of GWs from the merger of two BHs during the first advanced detectors era run, commonly called O1, by the two LIGO observatories heralded the dawn of GW astronomy~\cite{Abbott:2016blz}.  This event was quickly followed up by two other BBH mergers: one of lower significance on October 12th, 2015, and another on December 26th, 2015~\cite{Abbott:2016nmj,TheLIGOScientific:2016pea}; see Table~\ref{tab:definition_sources} for the source properties of the published GW mergers.  These detections, as exemplified by this white paper, have had a major impact on the fields of astrophysics and fundamental physics~\cite{TheLIGOScientific:2016wfe,TheLIGOScientific:2016htt,TheLIGOScientific:2016pea,Abbott:2016ymx,TheLIGOScientific:2016xzw,Abbott:2016cjt}.

The detection of GWs from only BBH mergers from all O1 detections has had significant ramifications on our understanding of astrophysical populations~\cite{Abbott:2016nhf,Abbott:2016drs,TheLIGOScientific:2016pea,Abbott:2016ymx}.  The detected BHs were more massive than any BHs that had been previously detected in low mass X-ray binaries, requiring a re-evaluation of the models of stellar evolution in binary systems~\cite{TheLIGOScientific:2016htt}.  From just these three events, the LIGO Scientific and Virgo collaborations (LVC) constrained the rate of BBH mergers to between 9-240 Gpc$^{-3}$ yr$^{-1}$~\cite{TheLIGOScientific:2016pea} (see \cite{Abbott:2017vtc} for an updated BBH merger rate of 12-213 Gpc$^{-3}$ yr$^{-1}$).  The non-detection of BNSs and NS-BH binaries allowed constraints of $< 12,600$ Gpc$^{-3}$ yr$^{-1}$ and $< 3,600$ Gpc$^{-3}$ yr$^{-1}$ respectively~\cite{Abbott:2016ymx}. At the time of this run, the LVC had over 60 MOUs signed with external telescopes, satellites and neutrino detectors.  No EM counterparts were found relating to the BBH mergers~\cite{Abbott:2016gcq,Abbott:2016iqz,Aasi:2013wya}.

To detect and extract astrophysical information, GW astronomy uses the method of matched filtering~\cite{1991PhRvD..44.3819S}.  This method is the optimal linear filter for signals buried in noise, and is very much dependent on the phase modelling of a GW template.  Within the LVC, the GW templates are constructed using both analytical and numerical relativity~\cite{TheLIGOScientific:2016qqj}.  In this case, the phase evolution of the template is a function of a number of frequency dependent coefficients.  Alternative theories of gravity predict that these coefficients should be individually modified if general relativity (GR) is not the correct theory of gravity.  While GR predicts specific values for these coefficients, one can treat each coefficient as a free variable and use Bayesian inference to test for deviations in the values of the parameters from the nominal GR value.  All tests conducted by the LVC displayed no deviations from GR~\cite{TheLIGOScientific:2016src,TheLIGOScientific:2016pea,Abbott:2017vtc,Abbott:2017oio}  

In addition, searches for generic GW transients, or GW-bursts, typically do not require a well-known or accurate waveform model and are robust against uncertainties in the GW signature. GW-burst searches are designed to detect transients with durations between $10^{-3} - 10$ seconds with minimal assumptions about the expected signal waveform. Such searches are, therefore, sensitive to GW transients from a wide range of progenitors, ranging from known sources such as BBH mergers to poorly-modeled signals such as core-collapse supernovae as well as transients that have yet to be discvered. An overview of GW-burst searches performed by LVC can be found here~\cite{TheLIGOScientific:2016uux}. Both GW-burst and compact binary coalescences (CBC) searches detected the first GW signal from BBH mergers, GW150914. 

In November 2016, the {\it second} Advanced Era Observation run, O2, began.  Once again, in January and June 2017, two BBH mergers were observed by the two LIGO detectors~\cite{Abbott:2017vtc,Abbott:2017oio}.  At the end of July 2017, the Advanced Virgo detector joined the global network of detectors.   On August 14th, all three detectors observed the merger of a BBH system. In previous detections, using only the two LIGO detectors, the sources were located to 1000s of square degrees in the sky.  In this case, due to the addition of Advanced Virgo, this system was localised to within 60 square degrees.  While not greatly advancing our understanding of the formation mechanisms of such systems, this detection did have a major effect in the field of fundamental physics.  Due to the misalignment of the three detectors, for the first time we were able to test the tensorial nature of GWs.  This event allowed the LVC to conclude that the GW signals were tensorial in nature, as is predicted by GR~\cite{Abbott:2017oio}.

Burst searches were also used as an independent analysis to complement matched filtering analyses for the detection of GW170104~\cite{Abbott:2017vtc}. Burst searches further identified a coherent signal, corresponding to GW170608, with a false-alarm rate of 1 in $\sim 30$ years~\cite{Abbott:2017gyy} and validated the detection of GW170814 with a false-alarm rate $< 1$ in 5900 years~\cite{Abbott:2017oio}. Note that, given the ``unmodelled'' nature of burst searches, the estimated event significances from burst searches tend to be lower than matched-filtered searches for the same event, especially for lower-mass compact binary signals.

On August 17th, the first BNS merger was observed by the LIGO and Virgo detectors~\cite{TheLIGOScientific:2017qsa}. This event was very quickly associated with a short gamma-ray burst (sGRB) detected by both the Fermi and Integral satellites~\cite{GBM:2017lvd}.  Within 10 hours, the host galaxy had been optically identified.  Within 16 days, the source had been identified across all bands of the  EM spectrum.  This single event heralded the true beginning of multi-messenger astronomy, and raised as many questions as it answered.

While confirming the hypothetical link between BNS mergers and sGRBs, the delay between the gamma and X-ray signals (9 days) suggested that not all sGRBs are the same~\cite{Monitor:2017mdv}.  This fact generated a number of studies regarding equation of state models, and the possible remnant of such mergers.  This one event also allowed the LVC to update the BNS event rate from $< 12,600$ Gpc$^{-3}$ yr$^{-1}$ in O1, to 320-4740 Gpc$^{-3}$ yr$^{-1}$ in O2~\cite{TheLIGOScientific:2017qsa}.

Perhaps, the most interesting results from this event concern fundamental physics. The delay between the detection of GWs and gamma-rays was 1.74 seconds.  This places a bound on the difference between the speed of light and the speed of GWs of $3\times10^{-15}\leq|\Delta c/c|\leq 7\times 10^{-16}$~\cite{Monitor:2017mdv}.  This single result has implications for certain alternative theories of gravity. For instance, the fact that GWs seem to travel at the same speed as that of strongly constrains the family of alternative theories of gravity that require $v_+, v_{\mathrm{\times}} \neq v_{\rm{light}}$ (e.g., beyond Horndeski, quartic/quintic Galileon, Gauss-Bonnet, if they are supposed to explain cosmology), as well as theories that predict a massive graviton. Furthermore, by investigating the Shapiro delay, the GW170817 detection also rules out MOND and DM emulator MOND-like theories (e.g., TeVeS), as according to these theories, the GWs would have arrived 1000 days after the gamma-ray detection.

The detection of GWs by the Advanced LIGO and Advanced Virgo detectors have had a major effect on our understanding of the Universe, sparking the fields of GW and multi-messenger astronomy and cosmology~\cite{GBM:2017lvd,Abbott:2017xzu}.  It is becoming increasingly clear that combining EM and GW information will be the only way to better explain observed phenomena in our Universe.  The third Advanced Detector Observation run (O3) will begin in early 2019, and will run for a year~\cite{Aasi:2013wya}.  We expect the detected events to be dominated by BBH mergers at a rate of one per week.  However, we also expect on the order of ten BNS events during this time, and possibly a NS-BH discovery (and potentially more than one such system).  Given the effects of one GW detection on both astrophysics and fundamental physics, we expect O3 to fundamentally change our view of the Universe.

\begin{table*}[t]
\begin{center}
\scriptsize
\begin{tabular}{|c|c|c|c|c|c|c|c|}
\hline
 & GW150914 & GW151226  & LVT151012  & GW170104 & GW170608 & GW170814 & GW170817   \\
\hline
$m_1/\msun$ & $36.2^{+5.2}_{-3.8}$ & $14.2^{+8.3}_{-3.7}$ & $23^{+18}_{-6}$ & $31.2^{+8.4}_{-6.0}$ & $12^{+7}_{-2}$ & $30.5^{+5.7}_{-3.0}$ & $(1.36,1.60)$  \\
$m_2/\msun$ & $29.1^{+3.7}_{-4.4}$ & $7.5^{+2.3}_{-2.3}$ & $13^{+4}_{-5}$ & $19.4^{+5.3}_{-5.9}$ & $7^{+2}_{-2}$ & $25.3^{+2.8}_{-4.2}$ & $(1.16,1.36)$  \\
${\mathcal M}/\msun$ & $28.1^{+1.8}_{-1.5}$ & $8.88^{+0.33}_{-0.28}$ & $15.1^{+1.4}_{-1.1}$ & $21.1^{+2.4}_{-2.7}$ & $7.9^{+0.2}_{-0.2}$ & $24.1^{+1.4}_{-1.1}$  & $1.186^{+0.001}_{-0.001}$\\
$q$ & $0.81^{+0.17}_{-0.20}$ & $0.52^{+0.40}_{-0.29}$ & $0.57^{+0.38}_{-0.37}$ & $0.62^{+}_{-}$ & $0.6^{+0.3}_{-0.4}$ & $0.83^{+}_{-}$ & $(0.73,1)$  \\
$M_f/\msun$ & $62.3^{+3.7}_{-3.1}$ & $20.8^{+6.1}_{-1.7}$ & $35^{+14}_{-4}$ & $48.7^{+5.7}_{-4.6}$ & $18.0^{+4.8}_{-0.9}$ & $53.2^{+3.2}_{-2.5}$ & $---$  \\
$\chi_{eff}$ & $-0.06^{+0.14}_{-0.29}$ & $0.21^{+0.20}_{-0.10}$ & $0.03^{+0.31}_{-0.20}$ & $-0.12^{+0.21}_{-0.30}$ & $0.07^{+0.23}_{-0.09}$ & $0.06^{+0.12}_{-0.12}$ & $0.00^{+0.02}_{-0.01}$  \\
$a_f$ & $0.68^{+0.05}_{-0.06}$ & $0.74^{+0.06}_{-0.06}$ & $0.66^{+0.09}_{-0.10}$ & $0.64^{+0.09}_{-0.20}$ & $0.69^{+0.04}_{-0.05}$ & $0.70^{+0.07}_{-0.05}$ & $---$  \\
$D_L$ / Mpc & $420^{+150}_{-180}$ & $440^{+180}_{-190}$ & $1020^{+500}_{-490}$ & $880^{+450}_{-390}$ & $340^{+140}_{-140}$ & $540^{+130}_{-210}$ & $40$  \\
$z$ & $0.090^{+0.029}_{-0.036}$ & $0.094^{+0.035}_{-0.039}$ & $0.201^{+0.086}_{-0.091}$ & $0.18^{+0.08}_{-0.07}$ & $0.07^{+0.03}_{-0.03}$ & $0.11^{+0.03}_{-0.04}$ & $0.0099$ \\

\hline 
\end{tabular}
\end{center}
\caption{Source properties of the published BBH and BNS discoveries (June 2018) by the LIGO and Virgo detectors}
\label{tab:definition_sources}
\end{table*}

\section{Black hole genesis and archaeology } \label{Sec:BHgenesis}
\vspace{-3mm}
{\it Contributors:} M.~Colpi and M.~Volonteri 
\vspace{3mm}

\subsection{Black Hole Genesis}

Gravity around BHs is so extreme that gravitational energy is converted into EM and kinetic energy with high efficiency, when gas and/or stars skim the event horizon of astrophysical BHs. Black holes of stellar origin (sBHs) with masses close to those of known stars power galactic X-ray sources in binaries, while supermassive black holes (SMBHs) with masses up to billions of solar masses power luminous quasars and active nuclei at the centre of galaxies. BHs are key sources in EM in our cosmic landscape.

According to General Relativity (GR), Kerr BHs, described by their mass $M_{\rm BH}$ and spin vector ${\bf S}=\chi_{\rm spin} \, G \, M_{\rm BH}/c$ (with $-1\leq \chi_{\rm spin}\leq 1$) are the unique endstate of unhalted gravitational collapse.
Thus understanding astrophysical BHs implies understanding the conditions under which gravitational equilibria 
lose their stability irreversibly. The chief and only example we know is the case of NSs which can exist 
up to a maximum mass $M_{\rm max}^{\rm NS}$, around $2.2\msun-2.6\msun$.  No baryonic microphysical state emerges
in nuclear matter, described by the standard model, capable to reverse the collapse to a BH state, during the contraction of the 
iron core of a supernova progenitor.  The
existence of  $M_{\rm max}^{\rm NS}$ is due to the non linearity of gravity which is sourced not only by the mass ``charge'' but also by pressure/energy density, according to the Oppenheimer-Volkoff equation. Thus, sBHs carry a mass exceeding  $M_{\rm max}^{\rm NS}$. Discovering sBHs lighter than this value (not known yet to high precision) would provide direct evidence of the existence of PBHs arising from phase transitions in the early universe.

As of today, we know formation scenarios in the mass range between  $5\msun -40\msun$, resulting from the core-collapse of
 very massive stars.  The high masses of the sBHs
revealed by the LVC, up to $36\msun$, hint formation sites of low-metallicity, $Z$\footnote{In astrophysics ``metallicity'' refers to the global content of heavy elements above those produced by primordial nucleosynthesis}, below $0.5\%$ of the solar value $Z_\odot=0.02$
\cite{Belczynski:2010tb,Spera15,Belczynski:2016obo}.
Theory extends this range up to about $40-60\msun$~\cite{Giacobbo18} and predicts the existence of a gap, between about $60 \, \simless \, M_{\rm BH}/\msun \, \simless \, 150,$ 
since in this window pair instabilities during oxygen burning lead either to substantial mass losses or (in higher mass stellar progenitors) the complete disruption of the star~\cite{Woosley17,Belczynski:2016jno,Spera:2017fyx}.  sBHs heavier than $150\msun$ can form at $Z \, < \, 1\% \,  Z_\odot$, if the initial mass function
of stars extends further out, up to hundreds of solar masses.

The majestic discovery of BBHs, detected by LVC interferometers~\cite{Abbott:2016blz,TheLIGOScientific:2016htt,Abbott:2016nmj,Abbott:2017vtc,Abbott:2017gyy,Abbott:2017oio}, at the time of their coalescence further indicates, from an astrophysical standpoint, that in nature sBHs have the capability of pairing to form binary systems, contracted to such an extent that GW
emission drives their slow inspiral and final merger, on observable cosmic timescales.  As GWs carry exquisite information on the individual masses and spins of the BHs, and on the luminosity distance of the source, detecting a population of coalescing sBHs with LVC in their advanced configurations, and with the next-generation of ground-based detectors~\cite{Evans:2016mbw,2012CQGra..29l4013S}, will let us reconstruct the mass spectrum and evolution of sBHs out to very large redshifts.

Observations teach us that astrophysical BHs interact with their environment, and that there are two ways to increase the mass: either through accretion, or through a merger, or both. These are the two fundamental processes that drive BH mass and spin evolution.   Accreting gas or stars onto BHs carry angular momentum, either positive or negative, depending on the orientation of the disk angular momentum relative to the BH spin. As a consequence the spin changes in magnitude and direction~\cite{King05,Perego09,Berti:2008af}. In a merger, the spin of the new BH is the sum of the individual and orbital angular momenta of the two BHs, prior to merging~\cite{Barausse:2012qz,Rezzolla16}. An outstanding and unanswered question is can sequences of multiple accretion-coalescence events let sBHs grow, in some (rare) cases, up to the realm of SMBHs? If this were true, the ``only'' collapse to a BH occurring in nature would be driven by the concept of instability of NSs at $M_{\rm max}^{\rm NS}$.

\begin{figure}[!t]
\begin{center}
\includegraphics[width=1.00\textwidth]{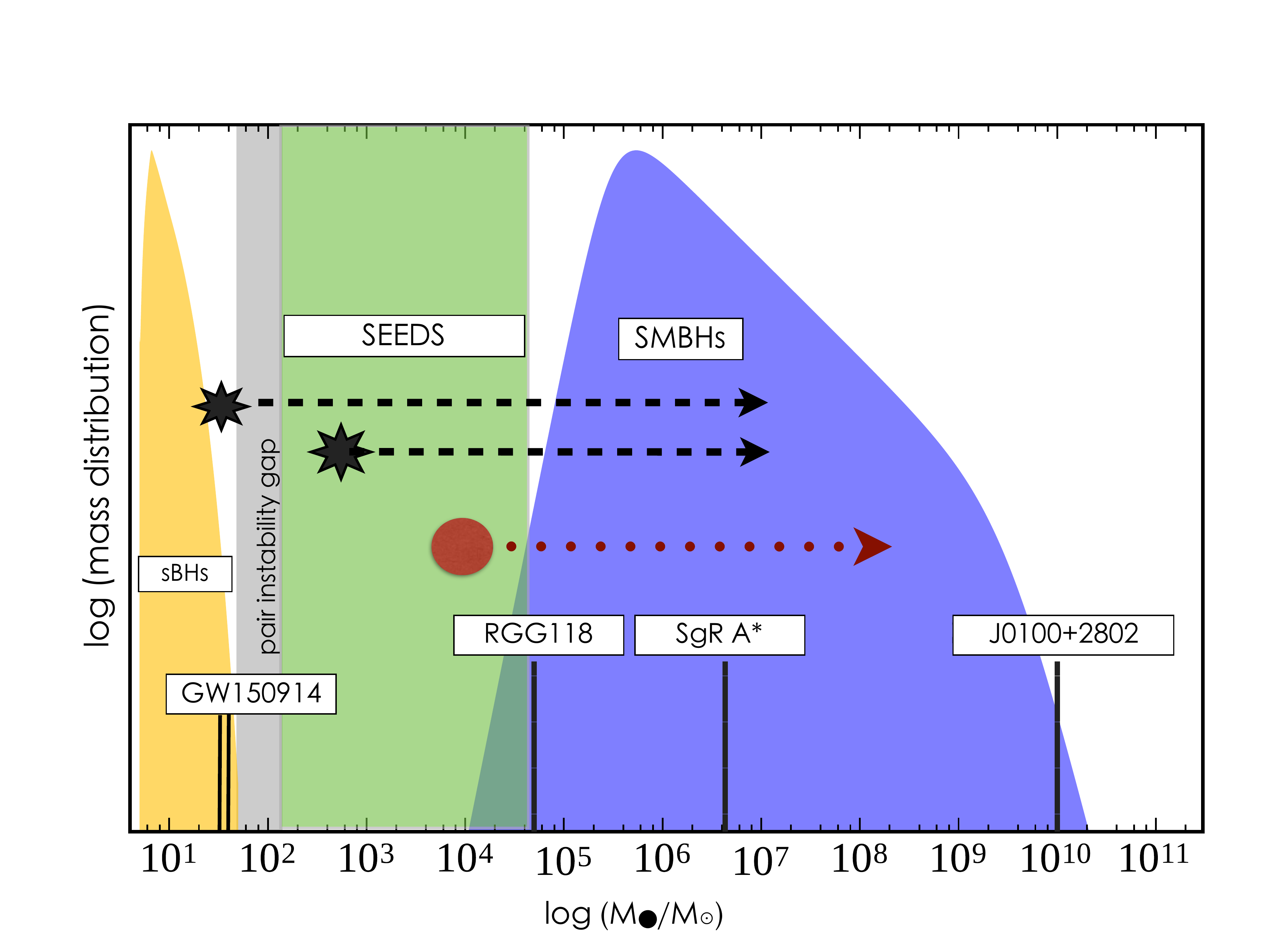}
\caption{Cartoon illustrating the BH mass spectrum encompassing the whole astrophysical relevant range, from sBHs to SMBHs, through the unexplored (light-green) zone where BH seeds are expected to form and grow. 
 Vertical black-lines denote the two sBH masses in GW150914, the mass $M_\bullet$ of RGG118 (the lightest SMBH known as of today in the dwarf galaxy RG118), 
 of SgrA* in the Milky Way, and of J0100+2802 (the heaviest SMBH ever recorded).
 The mass distribution of sBHs, drawn from the observations of the Galactic sBH candidates, has been extended to account for the high-mass tail
 following the discovery of GW150914. The minimum (maximum) sBHs is set equal to  $3\msun$ ($60\msun$), and the theoretically predicted pair-instability gap is depicted as a narrow darker-grey strip.  The SMBH distribution has been drawn scaling their mass according to the local galaxy mass function and $M_\bullet$-$\sigma$ correlation. The decline below $\sim 10^5\msun$ is set arbitrarily: BH of $\sim 10^{4-5}\msun$ may not be ubiquitous in low-mass galaxies as often a nuclear star cluster is in place in these galaxies, which may or may not host a central IMBH~\cite{Graham09}.  The black stars and dashed tracks illustrate the possibility that a SMBH at high redshift forms as sBH-only  (born on the left side of the sBH gap) or as light seed (on the right of the gap) which then grows through phases of super-Eddington accretion~\cite{Lupi16}. The red circle and dotted track illustrates the possibility of a {\it genetic} divide between
sBHs and SMBHs, and that a heavy seed forms through the direct collapse of a supermassive protostar in a metal free,
atomic-hydrogen cooling, DM halo~\cite{Latif13,Schleicher13}. The seed later grows via gas accretion and mergers with SMBHs in other black halos.
}
\label{seeds-BHmass-function} 
\end{center} 

\end{figure}

SMBHs are observed as luminous quasars and active galactic nuclei, fed by accretion of gas~\cite{Merloni16}, or as massive dark objects at the centre of quiescent galaxies which perturb the stellar and/or gas dynamics in the nuclear regions~\cite{Kormendy13}.  The SMBH mass spectrum currently observed extends from about $5\times 10^4\msun$ (the SMBH in the galaxy RGG118~\cite{Baldassare15}) up to  about $1.2\times 10^{10}\msun$ (SDSS J0100+2802~\cite{Wu15}), as illustrated in Figure \ref{seeds-BHmass-function}.  The bulk of active and  quiescent SMBHs are nested at the centre of their host galaxies, where the potential well is the deepest.  The correlation between the SMBH mass  $M_\bullet$ and the stellar velocity dispersion $\sigma$ in nearby spheroids, and 
even in disk/dwarf galaxies~\cite{2018ApJ...858..118N} hints towards a concordant evolution which establishes in the centre-most region controlled by powerful AGN outflows.
Extrapolated to lower mass disk or dwarf galaxies, this correlation predicts BH masses of $M_\bullet\sim 10^3\msun$ at $\sigma$ as low as $10\kms$, typical of nuclear star clusters (globular clusters) \cite{vandenbosch16}.  We remark that only BHs of mass in excess of $10^3\msun$ can grow a stellar cusp. The lighter BHs would random walk, and thus would have a gravitational sphere of 
influence smaller than the mean stellar separation and of the random walk mean pathlength.

Observations suggest that SMBHs have grown in mass through repeated episodes  of gas accretion and (to a minor extent)  through mergers with other BHs. This complex process initiates with the formation of a {\it seed}  BH of yet unknown origin~\cite{Volonteri10}. The concept of seed has emerged to explain the appearance of a large number of  SMBHs of billion suns at $z\sim 6,$  shining when the universe was only 1 Gyr old~\cite{Jiang16}. 
Furthermore, the comparison between the local SMBH mass density, as inferred from the $M_\bullet-\sigma$ relation,
with limits imposed by the cosmic X-ray background light, resulting from unresolved AGN powered by SMBHs in the mass interval between $10^{8-9}\msun$, indicates that radiatively efficient accretion played a large part in the building of SMBHs below $z\sim 3,$ and that information is lost upon their initial mass spectrum~\cite{Marconi04}.
Thus, SMBHs are believed to emerge from a population of seeds of 
yet unconstrained initial mass, in a mass range {\it intermediate} between those of sBHs and SMBHs, about $10^2\msun$ to $10^5\msun$, and therefore they are sometimes dubbed 
Intermediate-mass BHs (IMBHs).

Seeds are IMBHs that form ``early'' in cosmic history (at redshift $z\sim 20$, when the universe was only 180 Myr old). They form
in extreme environments, and grow over cosmic time by accretion and mergers. 
Different formation channels have been proposed for the seeds~\cite{Volonteri10,Schleicher13,Latif:2016qau}.  {\it Light seeds}  refer to IMBHs of about $100\msun$ that form from the relativistic collapse of massive Pop III stars, but the concept extends to higher masses, up to $\sim 10^3\msun$. These seeds likely arise from runaway collisions of massive stars in dense star clusters of low metallicity \cite{Mapelli16,Devecchi12}, or from mergers of sBHs in star clusters subjected to gas-driven evolution~\cite{Lupi14}.  The progenitors of light seeds are massive stars. However, they could be also the end result of repeated mergers among BHs~\cite{Gerosa:2017kvu}.  Finding merging sBHs with the LVC detectors with masses in the pair instability gap would be a clear hint of a second generation of mergers resulting from close dynamical interactions.

Accretion on sBHs occurs in X-ray binaries, and there is no evidence of accretion from the interstellar medium onto isolated sBHs in the Milky Way.  But, in gas-rich, dense environments characteristic of  galaxy halos at high redshifts, single sBHs might accrete to grow sizably, despite their initial small gravitational sphere of influence, if specific dynamical conditions are met.  For instance, in rare cases they may be captured in dense gas clouds within the galaxy~\cite{Lupi16}. Another possibility is that a sBH forms at the very center of the galaxy, where large inflows may temporarily deepen the potential well and allow it to grow significantly. This ``winning sBH'' must be significantly more massive than all other sBHs in the vicinity to avoid being ejected by scatterings and to be retained at the center of the potential well by dynamical friction. Similar conditions can also be present in nuclear star clusters characterized by high escape velocities. After ejection of the bulk of the sBHs, the only (few) remaining isolated BH can grow by tidally disrupting stars and by gas accretion~\cite{Stone17}  sparking their growth to become an IMBH.
 
{\it Heavy seeds} refer instead to IMBHs of about  $10^{4-5}\msun $ resulting from the monolithic collapse of massive gas clouds, forming in metal-free halos with virial temperatures  $T_{\rm vir} \, \simgreat \, 10^4$ K, which happen to be exposed to an intense H$_2$ photodissociating ultraviolet flux~\cite{Latif13,Dijkstra14,Habouzit16,Regan17,Latif:2016qau}. These gas clouds do not fragment and condense in a single massive proto-star which is constantly fueled by an influx of gas that lets the proto-star grow large and massive. Then, the star contracts sizably and may form a quasi-star \cite{Begelman10}, or it may encounter the GR instability that leads the whole star to collapse directly into a BH. Heavy seeds might also form in major gas-rich galaxy mergers over a wider range of redshifts, as mergers trigger massive nuclear inflows \cite{Mayer15}. Figure \ref{seeds-BHmass-function} is a cartoon summarising the current knowledge of BHs in our Universe, and the link that may exist between sBHs and SMBHs, which is established by seed BHs along the course of cosmic evolution.
 
The seeds of the first SMBHs are still elusive to most instruments that exist today, preventing us to set constraints on their nature. Seed  BHs are necessarily a {\it transient} population of objects and inferring their initial mass function and spin distribution from observations is possible only 
if they can be detected either through EM or GW observations at very high $z$, as high as $\sim 20$ (even $z\sim 40$ as discussed recently).
Since, according to GR,  BHs of any flavour captured in binaries are loud sources of GWs at the time of their merging, 
unveiling seeds and MBHs through cosmic ages via their GW emission at coalescence would provide unique and invaluable
information on the {\sl BH  genesis and evolution}.  The {\sl Gravitational Wave Universe} is the universe we can sense using GWs as messengers~\cite{Colpi17,Audley:2017drz}. In this universe, BBHs are key sources carrying invaluable information on their masses, spins and luminosity distance that are encoded in the  GW signal.  There is one key condition that needs to be fulfilled: that the BHs we aim at detecting
pair and form a binary with GW coalescence time smaller that the Hubble time, possibly close to the redshift of their formation. 
This condition, enabling the detection of seeds at very high redshifts, is extremely challenging to be fulfilled. BHs in binaries form at ``large'' separation. Thus, nature has to provide additional dissipative processes leading to the contraction of the BBH down to the scale where GWs drive the inspiral. This requires a strong coupling of the two BHs with the environment, before and after forming a binary system. As we now discuss, understanding this coupling is a current challenge in contemporary astrophysics, cosmology and computational physics~\cite{Colpi14}.

\subsection{Black Hole Binaries: the difficulty of pairing}

Due to the weakness of gravity, BBH inspirals driven by GW emission occur on a timescale:
 
\begin{eqnarray}
\nonumber
 t_{\rm coal} & =& {\frac{5 c^5}{256G^3}} {\cal G}(e)(1-e^2)^{7/2} {\frac{{a}^4}{\nu\, M_{\rm BBH}^3}} \\ 
& = & {\frac{5\cdot 2^4}{256}} {\cal G}(e)(1-e^2)^{7/2}{\frac{GM_{\rm BBH}}{ \nu c^3}{\tilde a}^4},
 \label{tcoal}
 \end{eqnarray}
where $M_{\rm BBH}$ is the total mass of the BBH, $a$ and $e$ the semi-major axis and eccentricity respectively (${\cal G}(e)$ a weak function of $e$, and ${\cal G}(0)=1$) and $\nu=\mu/M_{\rm BBH}$ the symmetric mass ratio ($\nu=1/4$ for equal mass binaries), with $\mu$ the reduced mass of the binary. The values of $a$ and $e$
 at the time of formation of the binary determine $ t_{\rm coal}$, and this is the longest timescale. A (circular) binary hosting two equal-mass seed BHs of  $10^3\msun$  ($M_{\rm BBH}=10^5\msun$)  would reach coalescence in $0.27$ Gyrs, corresponding to the cosmic time at redshift $z\sim 15,$ if the two BHs
 are at an initial separation of  $ a\sim \nu^{1/4} 4.84\times 10^4 R_{\rm G}$ ( $ \nu^{1/4} \,1.5\times 10^4 R_{\rm G}$) corresponding to 
 $a\sim \nu^{1/4}\,0.1$ AU, ($\nu^{1/4}\,30$ AU).  For the case of two equal-mass MBHs of $10^6\msun$  coalescing at $z\sim 3$
 (close to the peak of the star-formation rate and AGN rate of activity in the universe) $ a\sim \nu^{1/4} 4.84\times 10^6 R_{\rm G}$ 
 corresponding to about one milli-parsec.   These are {\it tiny} scales, and to reach these separations the binary needs to harden
 under a variety of dissipative processes. The quest for efficient mechanisms of binary shrinking, on 
 AU-scales for sBHs and  BH seeds, and sub-galactic scales for MBHs, make merger rate predictions extremely challenging, as
 Nature has to set rather fine-tuned conditions for a BBH to reach these critical separations. Only below these
 critical distances the binary contracts driven by GW emission. The merger occurs when 
  the GW frequency (to leading order equal to twice the Keplerian frequency) reaches a maximum value, 
\begin{equation}
f_{\rm GW}^{\rm max}\sim {\frac{c^3} {\pi 6^{3/2} GM_{\rm BBH}}}=4.4\times 10^3 \left  ({\frac{\msun}{ M_{\rm BBH}}}\right )\,\rm Hz. 
\label{fmax}
\end{equation}
This frequency $f_{\rm GW}^{\rm max}$ scales with the inverse of the total mass of the binary, $M_{\rm BBH}$ as it is determined by the size of the horizon of the two BHs, at the time of coalescence.

 Coalescing sBHs in binaries occur in galactic fields~\cite{Belczynski:2010tb,Dominik:2013tma}, or/and in stellar systems 
 such as globular clusters or/and nuclear star clusters \cite{Portegies00,Benacquista13,Rodriguez:2016avt,Antonini16,Haster16}. Thus, sBHs  describe  phenomena {\it inside galaxies}. Since there is a time delay  between formation of the binary and its coalescence, dictated by the efficiency of the hardening processes, sBHs can merge in galaxies of all types, as in this lapse time that can be of the order of Gyrs, host galaxies undergo strong evolution.
Instead, coalescing IMBHs, seeds of the SMBHs, formed in DM halos at high redshifts, and thus track
 a different environment.  Forming in pristine gas clouds their pairing either requires in-situ formation, e.g., from
 fissioning of rotating super-massive stars \cite{Reisswig13}, or via halo-halo mergers 
 on cosmological scales subjected to rapid evolution and embedded in a cosmic web that feed baryons through gaseous stream~\cite{Valiante17}. Coalescing MBHs refer exclusively to galaxy-galaxy mergers of different morphological types \cite{Colpi14} occurring during the clustering of cosmic structures, which encompass a wide range of redshifts, from $z\sim 9$ to $z\sim 0$ passing through the era of cosmic reionization, and of cosmic high noon when the averaged star formation rate has its peak.
  
In the following Sections we describe in detail the different channels proposed for the formation and pairing of BBHs at all scales. For each physical scenario we review the state of the art, challenges and unanswered questions and the most promising lines of research for the future. Ample space is devoted to stellar mass objects (BNS and BH-NS binaries and BBHs, with a particular focus on the latter), for which we discuss separately the three main formation channels: pairing of isolated binaries in the field, the various flavors of dynamical formation processes, relics from the early universe. We then move onto discuss the state of the art of our understanding of MBH binary pairing and evolution, the current theoretical and observational challenges, and the role of future surveys and pulsar timing arrays (PTAs) in unveiling the cosmic population of these elusive systems.

\section{The formation of compact object mergers through classical binary stellar evolution}\label{Sec:binaryevolution}
\vspace{-3mm}
{\it Contributors:} K.~Belczynski, T.~Bulik, T.~M. Tauris, G.~Nelemans
\vspace{3mm}


\subsection{Stellar-origin black holes}

The LIGO/Virgo detections of BBH mergers can be explained with stellar-origin BHs~\cite{Belczynski:2016obo} or by primordial BHs
that have formed from density fluctuations right after Big Bang \cite{Garcia-Bellido:2017imq},
Stars of different ages and
chemical compositions can form BHs and subsequently BBH mergers. In particular, 
the first metal-free (population~III) stars could have produced BBH mergers in the early 
Universe ($z\approx10$), while the local ($z\approx 0-2$) Universe is most likely 
dominated by mergers formed by subsequent generation of more metal-rich population~II 
and I stars~\cite{Belczynski:2016ieo}. The majority of population~I/II stars (hereafter: stars) are found 
in galactic fields ($\sim 99\%$) where they do not experience frequent or strong dynamical interactions
with other stars. In contrast, some small fraction of stars 
($\sim 1\%$) are found in dense structures like globular or nuclear clusters, in which stellar densities are high enough that stars 
interact dynamically with other stars. Here, we briefly summarize basic concepts of  
isolated (galactic fields) stellar and binary evolution that leads to the formation 
of BBH mergers.

\subsubsection{Single star evolution}
Detailed 
evolutionary calculations with numerical stellar codes that include rotation (like BEC, MESA or 
the Geneva code, e.g., \cite{ywl10,Paxton2015,Eggenberger2012,Georgy2013}) allow us to calculate the evolution 
of massive stars. Note that these are not (detailed) hydrodynamic nor multi-dimensional calculations (as such computations are 
well beyond current computing capabilities), but they solve the basic equations of 
stellar structure/radiation, energy transport and element diffusion with corrections for
effects of rotation. These calculations are burdened with uncertainties in 
treatment of various physical processes (nuclear reaction rates, convection and 
mixing, transport of angular momentum within a star, wind mass loss, pulsations 
and eruptions), yet progress is being made to improve on stellar modeling.  
Stellar models are used to predict the structure and physical properties of
massive stars at the time of core-collapse, after nuclear fusion stops 
generating energy and (radiation) pressure that supports the star. This is also a point 
in which transition (at latest) to hydrodynamical calculations is being made to assess the 
fate of a collapsing star \cite{Oconnor2011,Fryer2012,Ertl2016}. 

For a star to form a BH, it is required that either the explosion engine is weak or delayed (so energy 
can leak from the center of the collapsing star) or that the infalling stellar layers 
are dense and massive enough to choke the explosion engine adopted in a given 
hydrodynamical simulation. In consequence, BHs form either in  weak supernovae 
explosions (with some material that is initially ejected falling back and accreting 
onto the BH or without a supernova explosion at all (in a so-called direct collapse). 
Note that signatures of BH formation may already have been detected. For example, in SN~1987A there is no sign of a pulsar \cite{Alp:2018oek}, although the pulsar may still appear when
dust obscuration decreases or it simply beams in another direction. Further evidence is the disappearance with no sign of a supernova of a 
$25\msun$ supergiant star \cite{Adams2017}, although this can be a potentially 
long period pulsating Mira variable star that will re-emerge after a deep decline 
in luminosity.

Stellar evolution and core-collapse simulations favor the formation of BHs 
with masses $M_{\rm BH} \sim 5-50 \msun$ and possibly with very high masses 
$M_{\rm BH} \gtrsim 135 \msun$. The low-mass limit is set by the so-called 
``first mass gap'', coined after the scarcity of compact objects in mass range $2-5\msun$ 
\cite{Bailyn1998,Ozel2010}. However, this mass gap may be narrower than previously 
thought as potential objects that fill the gap are discovered \cite{frb+08,vbk11,Linares:2018ppq}.
The second gap arises from the occurrence of pair-instability SNe (PISN) as discussed below.

The first mass gap may be explained either by an observational bias in
determination of BH masses in Galactic X-ray binaries
\cite{Kreidberg2012} or in terms of a timescale of development of the
supernova explosion engine: for short timescales ($\sim\!100\;{\rm ms}$) a
mass gap is expected, while for longer timescales ($\sim\!1\;{\rm s}$) a mass
gap does not appear and NSs and BHs should be present in the $2-5\msun$
mass range \cite{Belczynski:2011bn}. The mass threshold between NSs and
BHs is not yet established, but realistic equations-of-state indicate
that this threshold lies somewhere in range $2.0-2.6 \msun$. The
second limit at $M_{\rm BH} \sim 50 \msun$ is caused by (pulsational) 
PISNe
\cite{Heger2002,Woosley2016}. Massive stars with He-cores in the mass
range $45 \lesssim M_{\rm He} \lesssim 65 \msun$ are subject to pulsational PISNe
before they undergo a core-collapse. These pulsations are predicted to remove the
outer layers of a massive star (above the inner $40-50\msun$) and
therefore this process limits the BH mass to $\sim 50\msun$.  BHs
within these two limits ($M_{\rm BH} \sim 5-50 \msun$) are the result
of the evolution of stars with an initial mass $M_{\rm
  ZAMS} \approx 20-150\msun$. For high-metallicity stars (typical of
stars in the Milky Way disk, $Z=\zsun=0.01-0.02$;
\cite{Asplund2009,Villante2014}) BHs form up to $\sim 15\msun$, for
medium-metallicity stars ($Z=10\% \zsun$) BHs form up to $\sim
30\msun$, while for low-metallicity stars ($Z=1\% \zsun$) BHs form
up to $\sim 50\msun$~\cite{Belczynski:2009xy,Kruckow:2018slo}.

The remaining question is whether stars can form BHs above 
$\sim 50\msun$. Stars with He-cores in mass range: $65\lesssim M_{\rm He} \lesssim 
135 \msun$ are subject to PISNe \cite{Bond1984a,Fryer2001,Heger2002}  
that totally disrupts the star and therefore does not produce a BH. 
However, it is expected that stars with He cores as massive as $M_{\rm He}
\gtrsim 135 \msun$, although subject to pair instability, are too massive to be 
disrupted and they could possibly form massive BHs ($M_{\rm BH} \gtrsim 135 \msun$). 
If these massive BHs exist, then second mass gap will emerge with no BHs in the mass
range $M_{\rm BH} \simeq 50-135 \msun$ \cite{Heger2003,Yusof2013,Spera2015,Marchant2016}.
If these massive BHs exist, and if they find their way into merging BBH binaries 
then GW observatories will eventually find them \cite{Belczynski:2014iua,Marchant2016}.
The existence of very massive BHs will constrain the extend of the stellar initial 
mass function (IMF) and wind mass-loss rates for the most massive stars
($M_{\rm ZAMS}>300\msun$) that can produce these BHs. So far, there are no
physical limitations for the existence of such massive stars  
\cite{Massey2011,Crowther2012}. Note that the most massive stars known today 
are found in the LMC with current masses of $\sim 200\msun$ \cite{Crowther2010}. 

BH formation may be accompanied by a natal kick. Natal kicks are
observed for Galactic radio pulsars, that move significantly faster (with average 3-dimensional speeds of
$\sim 400\kms$, e.g., \cite{Hobbs2005})
than their typical progenitor star ($10-20\kms$). These high velocities are argued to be
associated with some supernova asymmetry: 
either asymmetric mass ejection \cite{Janka1994,Tamborra2014,jan17}
or asymmetric neutrino
emission \cite{Kusenko1996,Fryer2006b,Socrates2005}. Note that
neutrino kick models all require (possibly unrealistic) strong magnetic fields, 
and simulations of core collapse without magnetic fields are unable to
produce significant neutrino kicks. Naturally, in these simulations
the authors find the need for asymmetric mass ejection to explain
natal kicks (e.g., \cite{Tamborra2014}). Although BH natal kicks as
high as observed for NSs cannot yet be observationally excluded, it is
unlikely for BHs to receive such large natal kicks \cite{Mandel2016b,Repetto2017}. It appears that some of the BHs may form without a
natal kick \cite{ntv99,Mirabel2017}, while some may form with a kick of the
order of $\sim 100 \kms$ \cite{Repetto2017}. 

The BH natal spin may simply depend on the angular momentum content
of the progenitor star at the time of core collapse. Massive stars are known to
rotate; with the majority of massive stars spinning at moderate
surface velocities (about 90\% at $\sim 100 \kms$) and with some stars
spinning rather rapidly (10\% at $\sim 400 \kms$). During its
evolution, a star may transport angular momentum from its interior to
its atmosphere. Then angular momentum is lost from the star when the
outer star layers are removed. The envelope removal in massive stars
that are progenitors of BBH mergers is easily accomplished either by
stellar winds or by mass transfer to a close companion star. However,
the efficiency of angular momentum transport is unknown. Two
competitive models are currently considered in the literature: very
effective angular momentum transport by a magnetic dynamo
\cite{Spruit1999,Spruit2002} included in the {\tt MESA} stellar
evolution code that leads to solid body rotation of the entire star, and
mild angular momentum transport through meridional currents
\cite{Eggenberger2012,Ekstrom2012} included in the {\tt Geneva} code
that leads to differential rotation of the star. Asteroseismology that
probes the internal rotation of stars has not yet provided any data on
massive stars (i.e. progenitors of BHs). The available measurements
for intermediate-mass stars (B~type main-sequence stars) show that
some stars are well described by solid body rotation and some by
differential rotation \cite{Aerts2008}. Depending on an the adopted
model, the angular momentum content of a star at core-collapse could
be very different. During BH formation some angular momentum may be
lost affecting the natal BH spin if material is ejected in a supernova
explosion. Whether BH formation is accompanied by mass loss is not at
all clear and estimates that use different assumptions on mass
ejection in the core-collapse process are underway~\cite{Belczynski:2017gds,Schroder:2018hxk}. At the moment, from the modeling
perspective, the BH natal spin is mostly unconstrained.

\subsubsection{Binary star evolution}
The majority ($\gtrsim 70\%$) of massive O/B stars, the potential progenitors of 
NSs and BHs, are found in close binary systems \cite{Sana2012}.
The evolution of massive stars in binaries deviates significantly from that of single stars \cite{pjh92,wl99,bhl+01,lan12}.
The main uncertainties affecting the calculation of BH merger rates are the metallicity, the common-envelope phase and the natal kick a BH receives at birth. These factors also determine the two main BH properties: mass and spin. 

Two main scenarios were proposed for BBH merger formation from
stars that evolve in galactic
fields: classical isolated binary evolution similar to that developed
for double neutron stars
(e.g., \cite{Tutukov1993,Lipunov1997,Belczynski:2001uc,Voss2003,Mennekens2014,Eldridge2016,Belczynski:2016obo,Stevenson2017}) and chemically homogeneous
evolution (e.g., \cite{Maeder1987,
  Yoon2005,Marchant2016,Mandel2016a,deMink2016,Woosley2016}). Classical
binary evolution starts with two massive stars in a wide orbit ($a
\gtrsim 50-1000 \rsun$), and then binary components interact with each
other through mass transfers decreasing the orbit below $\sim 50\rsun$
in common envelope (CE) evolution \cite{Webbink1984,Ivanova2013}.
Depending on their mass, both stars collapse to BHs, either with or
without supernova explosion, forming a compact BBH binary. The orbital
separation of two BHs which merge within a Hubble time is
below $\sim 50\rsun$ (for a circular orbit and two $30\msun$ BHs \cite{Peters:1964zz}). \cite{Heuvel2017} highlight that for the massive
stars that are expected to form BHs, the mass ratio in the second
mass-transfer phase is much less extreme, which means a
CE phase may be avoided. 

In the chemically homogeneous evolution scenario, two
massive stars in a low-metallicity environment form in a very close binary ($\lesssim 50\rsun$) and
interact strongly through tides \cite{Zahn1992,Kushnir2016}. Tidal
interactions lock the stars in rapid rotation and allow for the very
effective mixing of elements in their stellar interior that inhibits radial
expansion of the stars. Hence, these stars remain compact throughout their evolution and
collapse to BHs without experiencing a CE phase \cite{Marchant2016}. This
evolutionary scheme may well explain the most massive
LIGO/Virgo BBH mergers, as the enhanced tidal mixing required in
this channel only works for most massive stars ($\gtrsim 30\msun$). It
also predicts that both binary components evolve while rotating fairly rapidly and this may
produce rapidly spinning BHs, unless angular momentum is lost
very efficiently in the last phases of stellar evolution 
or during BH formation.

\subsubsection{Reconciling observations and theory}

There seems to be some confusion in the community as to what was expected and predicted
by stellar/binary evolution models prior to LIGO/Virgo detections of the first sources.
In particular, it is striking that often it is claimed that LIGO/Virgo detections of 
BBH mergers with very massive BHs were surprising or unexpected. Before 2010, in fact 
most models were indicating that BNS are dominant GW sources for 
ground-based detectors (however, see also Ref.~\cite{Voss2003}, predicting LIGO detection rates strongly dominated by BBH binaries), and that stellar-origin BHs are formed with small masses of  
$\sim 10 \msun$ \cite{Abadie:2010cf}. The models before 2010 were limited to calculations 
for stars with high metallicity (typical of the current Milky Way population) and this has 
introduced a dramatic bias in predictions.  
However, already around 2010 it was shown that stars at low metallicities 
can produce much more massive ($30-80 \msun$) BHs than observed in the Milky Way 
\cite{Zampieri2009,Mapelli2009,Belczynski:2009xy}. Additionally, it was demonstrated that 
binaries at low metallicities are much more likely to 
produce BBH mergers than high metallicity stars by one or two ordes of magnitude~\cite{Belczynski:2010tb}. 
This led directly to the pre-detection predictions that {\em (i)} the first LIGO/Virgo  
detection was expected when the detector (BNS) sensitivity range reached about $50-100$ Mpc 
(the first detection was made at $70$ Mpc), that {\em (ii)} BBH mergers will be the 
first detected sources, and that {\em (iii)} the BBH merger chirp-mass distribution may 
reach $30\msun$~\cite{Belczynski:2010tb,Dominik:2012kk,Dominik:2013tma,Dominik:2014yma}. 
Additionally studies of the future evolution of X-ray binaries like IC10 X-1 and NGC300 X-1
\cite{Bulik2011} suggested that there exists a large population of merging BH binaries with 
masses in excess of $20 \msun$.

\begin{figure}
\begin{center}
   \hspace*{-0.2cm}
   \includegraphics[width=9.2cm]{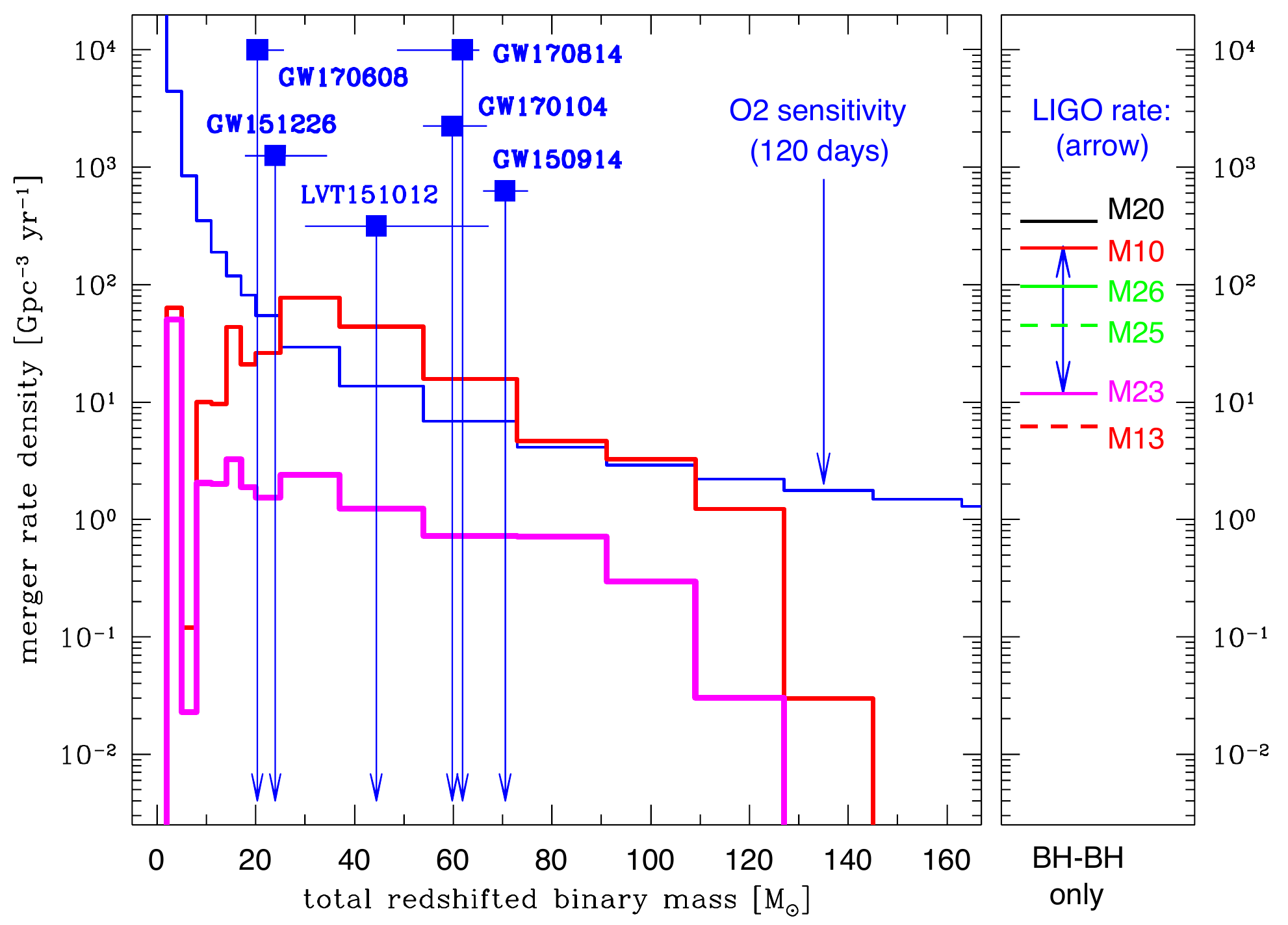}
   \begin{center}\caption{
    \emph{Left:}
    Redshifted total merger mass distribution for two population synthesis models
    \cite{Belczynski:2017gds}: M10 (low BH natal kicks) and M23 (high BH natal kicks).
    The O2 LIGO sensitivity is marked; the most likely detections
    are expected when models are closest to the sensitivity curve. We also
    mark LIGO/Virgo BBH merger detections (vertical positions have no
    meaning), all of which fall within the most likely detection region       between $20-100\msun$.
    \emph{Right:}
    Source frame BBH merger-rate density of several population synthesis models 
    for the local Universe ($z=0$). The current LIGO O1/O2 BBH merger rate is
    $12$--$213\gpy$ (blue double-headed arrow). Note that the models with
    fallback-attenuated BH natal kicks (M10, M20) are at the LIGO upper limit, 
    while models with high BH natal kicks are at the LIGO lower limit (M13, M23). 
    Models with small (M26) and intermediate (M25) BH kicks fall near the middle 
    of the LIGO estimate.
   } \end{center}
   \label{fig.nature}
\end{center}
\end{figure}

\begin{figure}
\begin{center}
  \includegraphics[width=\columnwidth]{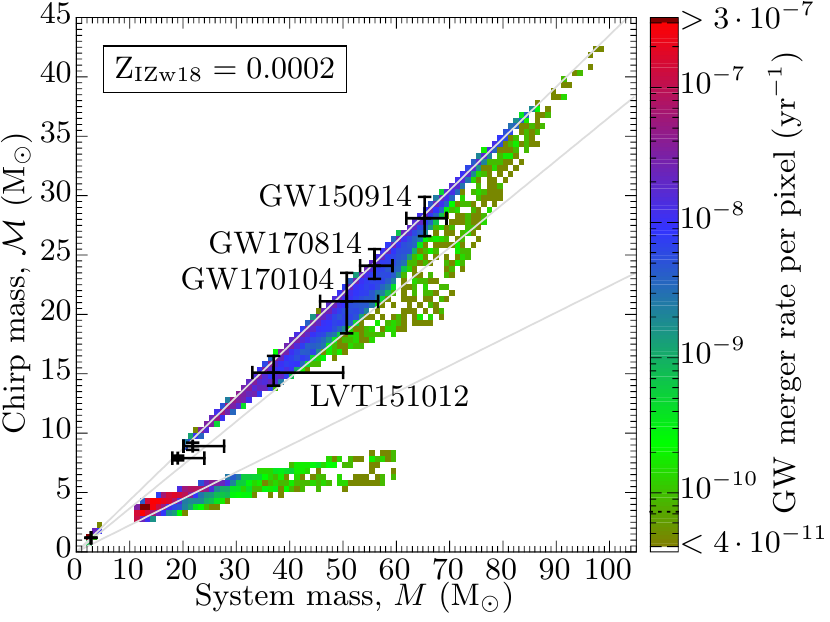}
  \caption{\label{fig:kruckow}Distribution of simulated double compact object binaries in the total mass–
chirp mass plane for a metallicity of Z = 0.0002. Three islands of data are visible,
corresponding to BBH, mixed BH-NS and BNS systems. The colour code indicates
the merger rate per pixel for a Milky~Way equivalent galaxy. The three solid grey lines indicate a constant mass ratio
of 1, 3 and 10 (from top to bottom). Observed LIGO/Virgo sources are shown with
black crosses and event names are given for the four most massive cases. The lowest
mass BBH mergers can only be reproduced with a higher metallicity. Figure taken from Ref.~\cite{Kruckow:2018slo}.}
\end{center}
\end{figure}

Post-detection binary evolution studies expanded on earlier work to show agreement of 
calculated BBH merger rates and BBH masses with LIGO/Virgo observations~\cite{Eldridge2016,Belczynski:2016obo,Stevenson2017,Kruckow:2018slo}. The range of calculated merger
rates ($10-300 \gpy$) comfortably coincides with the observed rate estimates 
($12-213 \gpy$ for the LIGO/Virgo $90\%$ credible interval). Note that these 
classical binary evolution rates are typically much higher than rates predicted for 
dynamical BBH formation channels ($5-10 \gpy$, \cite{Rodriguez2016b,Askaretal2017}).
The most likely detection mass range that is predicted from classical isolated binary
evolution is found in the total BBH merger mass range of $20-100 \msun$ (e.g.\cite{Kruckow:2018slo}).
Examples of merger rate and mass predictions for BBH mergers are given in Figures~\ref{fig.nature} and \ref{fig:kruckow}. 
A similar match between observed LIGO/Virgo BH masses and model predictions is obtained from the dynamical formation channel \cite{Rodriguez2016b,Askaretal2017}. Note that this
makes these two channels indistinguishable at the moment, although the merger rates are likely to be much smaller for the dynamical channel.

A caveat of concern for the prospects of LIGO/Virgo detecting BBH mergers with masses above the PISN gap is related to the relatively low GW frequencies of the such massive BBH binaries with chirp masses above $100\;\msun$. During the in-spiral, the emitted frequencies are expected to peak approximately at the innermost stable circular orbit (ISCO), before the plunge-in phase and the actual merging. Hence, the emitted frequencies are most likely less than 100~Hz, and with redshift corrections the frequencies to be detected are easily lower by a factor of two or more. A frequency this low is close to the (seismic noise) edge of the detection window of LIGO/Virgo and may not be resolved.

The current LIGO/Virgo broad range of an empirically determined local BBH merger-rate density
($12-213 \gpy$) can easily be explained by uncertainties in key input physics parameters of population synthesis modelling, such as the slope of the IMF or the efficiency of the CE ejection, see e.g. Table~5 in \cite{Kruckow:2018slo}. Alternatively, it may be explained by altering BH natal kicks~\cite{Belczynski:2016obo} from full
NS natal kicks corresponding to a low rate estimate) to almost no BH
kicks (high rate estimate). Once LIGO/Virgo narrows its empirical estimate it
may be possible to use the merge-rate density to constrain the input physics applied in modelling, although it should be cautioned that there is a large degree of degeneracy \cite{Voss2003,Kruckow:2018slo}.

LIGO/Virgo provides an estimate of the effective spin parameter that measures the projected BH spin 
components ($a_1, a_2$) parallel to binary angular momentum, weighted by BH masses 
($M_1, M_2$):  
\begin{equation}
\chi_{\rm eff}\equiv \frac{M_1 a_1 \cos\Theta_1 + M_2 a_2 \cos\Theta_2}{M_1+M_2}\;,
\label{eq.xeff}
\end{equation}
where $\Theta_{1,2}$ are the angles between the BH spins and the orbital angular 
momentum vector. So far, for the six LIGO/Virgo BBH detections the effective spins cluster 
around $|\chi_{\rm eff}| < 0.35$ (e.g., \cite{Abbott:2017vtc}).  
This defies the expectations for the main BBH formation channels. If spin magnitudes 
of stellar-origin BHs are significant, as estimated for several Galactic and
extra-galactic X-ray binaries \cite{Fragos2015}, then the dynamical formation channel 
(random BH capture) predicts an isotropic $\chi_{\rm eff}$ distribution, while the 
classical binary evolution channel mostly predicts aligned BH spins (aligned stellar 
spins that are only moderately misaligned by BH natal kicks). Hence, in the latter case one expects a distribution 
peaked at high values of $\chi_{\rm eff}$. On the one hand this tension is rather 
unfortunate, as it does not allow to distinguish between these two very different 
scenarios of BBH merger formation. On the other hand, this is a great opportunity  
to learn something new about stars and BHs that was so far not expected and 
is not easily understood in the framework of current knowledge. 

There are five potential explanations of this apparent tension. First,
there could be a mechanism that puts both BH spins in the plane of the binary orbit, producing $\chi_{\rm eff}=0$ independent of BH masses
and spin magnitudes. Such a mechanism is proposed to operate in
triple stars \cite{Antonini:2017tgo}. Note that triple stars are a minority
of stars ($10-20\%$ of all field stars) and that the proposed
mechanism requires very specific tuning to operate, so it is not clear
how likely it is that it worked for all LIGO/Virgo sources. Second, there could be a mechanism that forces the BH spins to be in
opposite directions so that they cancel out. For approximately equal
mass BBH binaries (typical of LIGO/Virgo sources) this would imply
180 degree flip of spins. No mechanism is known to produce such
configurations. Third, both BH spin magnitudes may be very small
reducing effective spin parameter to $\chi_{\rm eff}=0$, independent of
other merger parameters. This was already proposed and is used in
studies of angular momentum transport in stars
\cite{Belczynski:2017gds}. Fourth, LIGO/Virgo BHs may not have been
produced by stars, but for example they come from a primordial
population for which small spins are naturally expected
\cite{Garcia-Bellido:2017imq,Clesse:2017bsw}. Fifth, it may be the case that the
spin of a BH (at least its direction) is not mainly determined by the angular momentum of the progenitor star, but a result of the
physics of the collapse (spin tossing, e.g., \cite{Tauris2017}). In that case, there is no reason to assume the
spins in mergers formed from isolated binary evolution are aligned and
they may be isotropic.
Note that with these five options, that need to be tested and developed further, one cannot determine the main formation channel of BBH mergers, as it was proposed in several 
recent studies \cite{Kushnir2016,Zaldarriagaetal2018,Hotokezaka2017,Hotokezaka:2017dun}.

The issue of spins is rather fundamental as the effective spin parameter
most likely contains information on natal BH spin magnitudes and therefore
information on stellar astrophysics regarding angular momentum transport in massive stars, 
which is still unconstrained by electromagnetic observations. Possibly the second-formed 
BH spin could have been increased in binary evolution by accretion of mass from 
a companion star. However, it is argued that BHs cannot accrete a significant 
amount of mass in the binary evolution leading to the BBH formation~\cite{Belczynski:2007xg,Belczynski:2016obo,Belczynski:2017gds}. This is partly due to very low 
accretion rates during a CE of $1-10\%$ of the Bondi-Hoyle accretion rate 
\cite{Ricker2008,MacLeod2017,Murguia-Berthier:2017bdf,Holgado:2017vut}) and fast timescale Roche-lobe overflow (RLO)
in massive progenitor binaries that leads to ejection of most of the exchanged matter  
from the binary system (due to super-Eddington mass-transfer rates). The amount of
mass accreted by BHs in binary systems ($\lesssim 1-3 \msun$) cannot
significantly spin up massive BHs ($10-30\msun$) that are detected by 
LIGO/Virgo.  

It is important to note the most challenging parts of the evolutionary predictions 
in the context of BBH formation. In the classical binary evolutionary 
channel, the two most uncertain aspects of input physics are related to the CE evolution and the 
natal BH kicks. Although some observational constraints on both processes exist,
they are rather weak. Systems entering CE evolution have recently been 
reported. However, they are not as massive as stars that could produce NSs or BHs 
\cite{Tylenda2016}. The search for CE traces as IR outburts has so far yielded no 
clear detection of emerging or receding X-ray binaries as expected in this scenario
\cite{Oskinova:2018sfn}.

 BH natal kicks are only measured indirectly by positions
and motions of X-ray binaries hosting BHs, and usually only lower limits on
natal kicks are derived \cite{ntv99,Mandel2016b,Repetto2017,Belczynski:2015tba}. 
On theoretical grounds, reliable models for CE \cite{Ivanova2013} and
supernovae \cite{Janka2016} are missing. In the chemically homogeneous
evolution channel the largest uncertainty is connected with the efficiency of the mixing, the number of
massive binaries that can form in very close orbits, and the strength of
tidal interactions in close binaries. Since initial orbital period
distributions are measured only in the very local Universe \cite{Sana2012}, it
is not clear whether they apply to the majority of all stars and thus it
is not fully understood how many stars are potentially subject to this kind
of evolution. Even a deeper problem exists with our understanding of tides and
their effectiveness in close binaries \cite{Zahn1992,Kushnir2016}, and
effective tides are the main component of input physics in chemically  homogeneous evolution.  

Astrophysical inferences from GW observations are currently limited. First, it is not known which formation channel (or what
mixture of them) produces the known LIGO/Virgo BBH mergers. Since each
channel is connected to specific set of conclusions (for example, the isolated
binary channel informs about CE evolution and natal kicks; while the dynamical
channel informs predominantly about stellar interactions in dense clusters) 
it is not clear which physics we are testing with GW observations. Second, within each channel there is degeneracy such that multiple model parameters 
are only weakly constrained by observations. As nobody so far was able to deliver a
comprehensive study of the large multi-dimensional parameter space, the
inferences on various model parameters (e.g. the strength of BH natal kicks or the 
CE efficiency) are hindered by various and untested model parameter
degeneracies. However, it is already possible to test several aspects of 
stellar evolution as some processes leave unambiguous signatures in GW observations. For example, the existence of the first and the 
second mass gap, if confirmed by LIGO/Virgo, will constrain core-collapse 
supernovae and PISNe, respectively. Careful studies 
with detailed exposure of caveats are needed to transform  future observations 
into astrophysical inferences. 

It is expected that GW events resulting from the merger of stellar-mass BHs are unlikely to produce electromagnetic counterparts. Nevertheless, a (marginal) transient signal detected by the Fermi gamma-ray burst monitor, 0.4~seconds after GW150914, was reported \cite{cbg+16}. This claim encouraged several theoretical speculations for a possible origin. It has been suggested \cite{dk17}  that a tiny fraction of a solar mass of baryonic material could be retained in a circumbinary disk around the progenitor binary which probably shed a total mass of $>10\;\msun$ during its prior evolution. The sudden mass loss and recoil of the merged BH may then shock and heat this material, causing a transient electromagnetic signal. 
It will be interesting see if any further electromagnetic signals will be associated with BBH mergers in the near future.

\subsection{BNS mergers}
The formation of double NSs has been studied in detail following the
discovery of the Hulse-Taylor pulsar \cite{ht75} and currently more
than 20 such BNS systems are known -- see \cite{Tauris2017} for a details and a review
on their formation and evolution. Whereas no BBH binaries had been detected prior to GW150914, detailed knowledge on BNS systems was known for many years from Galactic radio pulsar observations \cite{lk12}.

LIGO/Virgo has currently only detected one BNS merger event (GW170817), located in the lenticular (S0) galaxy NGC~4993, and thus the local empirical BNS merger rate density still remains rather uncertain: $1540^{+3200}_{-1220}
\gpy$ ($90\%$ credible limits~\cite{TheLIGOScientific:2017qsa}). The study of double
NSs is relevant for the study of BHs because it gives independent
constraints on the evolution of similar massive binary populations
from which binary BHs are formed. In particular the question which stars for NSs and which form BHs and if and how this depends on previous binary interactions is a question that likely only can be answered observationally by significant statistics on the relative abundance of double BHs, BNS and NS-BH binaries.

There are two major sites to produce BNS mergers: isolated
binaries in galactic fields (the main contributor), and dense environments in globular and nuclear clusters. None of these sites (nor any combination of them) can easily
reproduce the preliminary estimated LIGO/Virgo event rate, even if all elliptical host
galaxies are included within the current LIGO/Virgo horizon~\cite{Belczynski:2017mqx}. 
The local supernova rate can be estimated to
be about $10^5 \gpy$, so the current empirical BNS merger rate from LIGO/Virgo would imply a
very high efficiency of BNS binary formation.

This apparent tension may be solved if BNS mergers are allowed to originate 
from a wide range of host galaxies and if the low-end of the LIGO/Virgo
merger-rate estimate is used ($320 \gpy$). Population synthesis studies seem
to agree that rates as high as $200-600 \gpy$ can possibly be reached if favorable
conditions are assumed for classical binary evolution 
\cite{Chruslinska2018,Kruckow:2018slo,Vigna-Gomez:2018dza}. In the coming years, the statistics of the empirical BNS merger rate will improve significantly and reveal whether current theoretical BNS merger rates need a revision.
It is interesting to notice, however, that calibrations with the rates of observed short gamma-ray bursts and the rate of mergers required to reproduce the abundances of heavy r-process elements favor a merger-rate density significantly smaller than the current empirical rate announced by LIGO/Virgo \cite{Kruckow:2018slo}.

The main uncertainties of the theoretically predicted merger rate of BNS binaries are also related to CE evolution and SNe (similar to the case of BBH mergers). A CE evolution is needed to efficiently remove orbital angular momentum to tighten the binary orbit and allow a merger event within a Hubble time. However, the onset criterion and the efficiency of the in-spiral in a CE remain uncertain \cite{Ivanova2013,ktl+16}.
The kick velocities imparted on newborn NSs span a wide range from a few ${\rm km\,s}^{-1}$ (almost symmetric SNe) to more than 1000~${\rm km\,s}^{-1}$ and are sometimes difficult to determine \cite{Verbunt18}. The kick magnitude seems to be related to the mass of the collapsing core, its density structure and the amount of surrounding envelope material \cite{jan17}. Additional important factors for the predicted merger rates include the slope of the initial-mass function and the efficiency of mass accretion during RLO \cite{Kruckow:2018slo}.

All taken together, the predicted merger rate of double NSs in a Milky~Way equivalent galaxy vary by more than two orders of magnitude~\cite{Abadie:2010cf}. The empirical merger rate that LIGO/Virgo will detect at design sensitivity in a few years is of uttermost importance for constraining the input physics behind the rate predictions. 
Besides the detection rates, the mass spectrum and the spin rates of the double NS mergers will reveal important information about their origin. Although their precise values cannot be determined due to degeneracy, the overall distribution of estimated NS masses will reveal information on their formation process (electron capture vs iron-core collapse SNe), as well as constraining the nuclear-matter equation-of-state. The latter will also be constrained from tidal deformations of the NSs in their last few orbits~\cite{TheLIGOScientific:2017qsa}.

An important observational signature of the merger event of BNS binaries is the detection of the ring-down signal of either a meta-stable highly massive NS or a BH remnant. Such information would set constraints on the fundamental equation-of-state of nuclear matter at high densities. 
Whereas LIGO/Virgo is not sensitive enough to detect a ring-down signal, it is the hope that third-generation GW detectors might be able to do so.
Another important observational input is the distribution of mass ratios in BNS merger events. This distribution could provide important information about the formation of NSs and the nature of the supernovae  (e.g. electron-capture vs iron core-collapse supernovae).

Optical follow-up will in many cases reveal the location of a double NS merger (e.g., \cite{GBM:2017lvd}). This will provide information on their formation environments \cite{sha+17,Coulter:2017wya} and kinematics \cite{fb13}, besides crucial information on heavy r-process nucleosynthesis \cite{2010MNRAS.406.2650M}.

\subsection{Mixed BH-NS mergers}

The formation of mixed BH/NS mergers is expected to follow similar scenarios as double NS or double BH \cite{Tutukov1993,Voss2003} with all the associated uncertainties. 

It is perhaps somewhat surprising that LIGO/Virgo detected a double NS merger (GW170817) before a mixed BH/NS merger, since (at least some) population synthesis codes predict a detection rate of mixed BH/NS systems which is an order of magnitude larger that the expected detection rate of double NS systems (with large uncertainties, e.g.,\cite{Kruckow:2018slo}). Hence, if these predictions are correct, GW170817 is a statistical rare event and detections of mixed BH/NS systems are expected already in the upcoming O3/O4 LIGO runs.
The detection of mixed BH-NS mergers is interesting for two reasons: (i) a key question is whether BH-NS and NS-BH binaries may be distinguished from one another (i.e. the formation order of the two compact objects, which leads to a (mildly) recycled pulsar in the latter case), and (ii)  the detected in-spiral of mixed BH-NS mergers may reveal interesting deviations from GR and pure quadrupole radiation given the difference in compactness between BHs and NSs \cite{wil94}.

\section{Dynamical Formation of Stellar-mass Binary Black Holes}\label{Sec:nbody}
\vspace{-3mm}
{\it Contributors:} B.~Kocsis and A.~Askar
\vspace{3mm}


\subsection{Introduction}\label{sec:dynamicsintroduction}
The recent GW observations from six BBH mergers (GW150914, LVT151012, GW151226, GW170104, GW170608, GW170814) and a BNS merger (GW170817) opened ways to test the astrophysical theories explaining the origin of these sources
 \cite{Abbott:2016nmj,Abbott:2016blz,Abbott:2017vtc,Abbott:2017gyy,Abbott:2017oio,TheLIGOScientific:2017qsa}
. As discussed earlier, the component masses of these merging sources span a range between 8--$35\Msun$~\cite{Abbott:2017vtc}, which is different from the distribution of BHs seen in X-ray binaries, $5$ -- $17\Msun$ \cite{Farretal2011} with two possible exceptions (NGC300X-1 and IC10X-1). The event rates of BBH mergers are estimated to be between $40$--$213\,\Gpc^{-3}\yr^{-1}$ for a power-law BH mass function and between $12$--$65\,\Gpc^{-3}\yr^{-1}$ for a uniform-in-log BH mass function~\cite{Abbott:2017vtc}, which is higher than previous theoretical expectations of dynamically formed mergers, for instance see \cite{Abadie:2010cf}. The event rates of BNS mergers is currently based on a single measurement which suggests a very high value of $1540^{+3200}_{-1220}\,\Gpc^{-3}\yr^{-1}$~\cite{TheLIGOScientific:2017qsa} (c.f.~\cite{Belczynski:2017mqx}). How do we explain the observed event rates and the distribution of masses, mass ratios, and spins?

Several astrophysical merger channels have been proposed to explain observations. Here we review some of the recent findings related to {\bf dynamics}, their limitations and directions for future development. These ideas represent alternatives to the classical binary evolution picture, in which the stars undergo poorly understood processes, such as common envelope evolution. In all of these models the separation between the compact objects is reduced dynamically to less than an AU, so that GWs may drive the objects to merge within a Hubble time, $t_{\rm Hubble}=10^{10}\,\yr$.

\subsection{Merger rate estimates in dynamical channels}

\paragraph{Dynamical formation and mergers in globular clusters} Although about 0.25\% of the stellar mass is currently locked in globular clusters (GCs) \cite{Rosenblatt1988,Boker2008,Harris2015}, dynamical encounters greatly catalyze the probability of mergers compared to that in the field. Within the first few million years of GC evolution, BHs become the most massive objects. Due to dynamical friction, they will efficiently segregate to the cluster center \cite{Spitzer1969} where they can dynamically interact and form binaries with other BHs  \cite{Sigurdsson1993,Ivanova2005}. The dense environments of GCs can also lead to binary-single and binary-binary encounters involving BHs that could result in their merger. Collisional systems like GCs can also undergo core collapse, during which central densities can become very large leading to many strong dynamical interactions. The encounter rate density is proportional to $\mathcal{R}\sim \int dV \,\langle n_{*}^2 \rangle \, \sigma_{cs} v$, where $n_*$ is the stellar number density, $\sigma_{\rm cs}\sim GMb/v^2$ is the capture cross section, $M$ is the total mass, $b$ is the impact parameter, $v$ is the typical velocity dispersion. Note the scaling with $\langle n_*^2\rangle$, where $\langle n_{*}^2\rangle^{1/2} \sim 10^5\pc^{-3}$ in GCs and $\sim 1\pc^{-3}$ in the field. 

Estimates using Monte Carlo method to simulate realistic GCs yield merger rates of at least $\mathcal{R}_{\rm GC}\sim 5 \, \Gpc^{-3}\yr^{-1}$ \cite{Rodriguez2016b,Askaretal2017}, falling below the current limits on the observed rates. Rate estimates from results of direct $N$--body simulations also yield a similar value of $\mathcal{R}_{\rm GC}\sim 6.5 \, \Gpc^{-3}\yr^{-1}$ \cite{Park2017}. In particular, these papers have shown that the low-mass GCs below $10^5\mathrm{M}_{\odot}$ have a negligible contribution to the rates. However, they also show that initially more massive GCs (more massive than $10^6\mathrm{M}_{\odot}$) contribute significantly to the rates. \cite{Askaretal2017} argue that actual merger rates from BHs originating in GCs could be 3 to 5 times larger than their estimated value of $\sim 5 \, \Gpc^{-3}\yr^{-1}$ due to uncertainties in initial GC mass function, initial mass function of stars in GCs, maximum initial stellar mass and evolution of BH progenitors.
Furthermore, BBH merger rates can be significant in young clusters with masses $\sim 10^4\,\Msun$ \cite{2016MNRAS.459.3432M,2017MNRAS.467..524B}.

A simple robust upper limit may be derived by assuming that all BHs merge once in each GC in a Hubble time: 
\begin{eqnarray}
\mathcal{R} & \leq & \frac12f_{\rm BH} N_{*}\frac{ n_{\rm GC}}{t_{\rm Hubble}} \nonumber \\
& < & \frac12 
\frac{ \int_{20\Msun}^{150\Msun} f_{\rm IMF}(m) dm}{ 
    \int_{0.08\Msun}^{150\Msun} m f_{\rm IMF}(m) dm}
    \times 10^{5.5}\Msun
    \times\frac{0.8\Mpc^{-3}}{10^{10}\yr} \nonumber \\ & = & 38\,\Gpc^{-3}\yr^{-1} \label{eq:GCrate}
\end{eqnarray}
where $f_{\rm BH}$ is the fraction of stars that turn into BHs from a given stellar initial mass function $f_{\rm IMF}$, $N_{*}$ is the initial number of stars in a GC, and $n_{\rm GC}$ is the cosmic number density of GCs, and $f_{\rm IMF}(m) \propto (m/0.5\Msun)^{-2.3}$ for $m>0.5\Msun$ and $(m/0.5\Msun)^{-1.3}$ otherwise \cite{2001MNRAS.322..231K}. The result is not sensitive to the assumed upper bound on mass of the BH progenitor, which is set by the pair instability supernova. Recent estimates find 110--$130\,\Msun$ for GC metallicities \cite{Spera:2017fyx}. However, note that the mass in GCs may have been much higher than currently by a factor $\sim 5$, since many GCs evaporated or got tidally disrupted \cite{Gnedinetal2014,2014MNRAS.444.3738A}. This effect increases the rates by at most a factor 2 at $z<0.3$, but more than a factor 10 at $z>2.5$~\cite{Fragione:2018vty}.

\paragraph{Dynamical and relativistic mergers in galactic nuclei} The densest regions of stellar BHs in the Universe are expected to be in the centers of nuclear stars clusters (NSC), whose mass-segregated inner regions around the SMBH exceed $n_*\sim 10^{10}\pc^{-3}$ \cite{OLearyetal2009}. In contrast to GCs, the escape velocity of the SMBH in the central regions of NSCs is so high that compact objects are not expected to be ejected by dynamical encounters or supernova birth kicks. In these regions, close BH binaries may form due to GW emission in close single-single encounters \cite{OLearyetal2009}. Binary mergers may also be induced by the secular Kozai-Lidov effect of the SMBH \cite{Antonini_Perets2012,Hoang:2017fvh,PetrovichAntonini2017,2018MNRAS.477.4423A,Arca-Sedda:2017wea,Hamers:2018hxv} and tidal effects \cite{Fernandez:2018uiy}. Detailed estimates give from the $\mathcal{R}_{\rm NSC}\sim 5 - 15\,\Gpc^{-3}\yr$ \cite{OLearyetal2009,Hoang:2017fvh,Antonini_Perets2012,PetrovichAntonini2017,Belczynski:2017mqx}, below the observed value. Higher values may be possible for top heavy BH mass functions and if black holes are distributed in disk configurations \cite{OLearyetal2009,Szolgyen:2018zra}.

These numbers are sensitive to the uncertain assumptions on the total supply of BHs in the NSCs, either formed in situ or delivered by infalling GCs \cite{2012ApJ...750..111A,2013ApJ...763...62A,Gnedinetal2014,ArcaSedda2015,Arca-Sedda:2017qcq}. If all BHs merged in galactic nuclei once, an upper limit similar to Equation~(\ref{eq:GCrate}) is $\mathcal{R}_{\rm NSC}< 30\,\Gpc^{-3}\yr^{-1}$. Due to the high escape velocity from NSCs and a rate of infalling objects, this bound may be in principle exceeded.

\paragraph{Mergers facilitated by ambient gas} Dynamical friction facilitates mergers in regions where a significant amount of gas embeds the binary, which may carry away angular momentum efficiently. Particularly, this may happen in a star forming regions \cite{2016MNRAS.462.3812T,Belczynski:2016ieo,2018ApJ...856...47T}, in accretion disks around SMBHs in active galactic nuclei (AGN) \cite{Barausse:2007dy,Kocsisetal2011} or if the stellar envelope is ejected in a stellar binary \cite{Tagawa:2018vls}. In AGN, the accretion disk may serve to capture stellar mass binaries from the NSC \cite{Bartosetal2017}, or help to form them in its outskirts by fragmentation \cite{Stoneetal2017}. The rate estimates are at the order of $1\,\Gpc^{-3}\yr^{-1}$, below the observed value. Nevertheless, mergers in this channel deserve attention as they have electromagnetic counterparts, the population of AGNs, which may be used in a statistical sense \cite{Bartosetal2017b}.

\paragraph{Isolated triples} 
The stellar progenitors of BHs are massive stars, which mostly reside not only in binaries, but in many cases in triples. The gravity of the triple companion drives eccentricity oscillations through the Kozai-Lidov effect, which may lead to GW mergers after close encounters. However, the rate estimates are around or below $6 \, \Gpc^{-3}\yr^{-1}$, below the current observational range \cite{Silsbee_Tremaine2017,2017ApJ...841...77A}. The rates may be higher $2$--$25 \, \Gpc^{-3}\yr^{-1}$ for low metallicity triples \cite{Rodriguez:2018jqu} and further increased by non-hierarchical configurations \cite{Arca-Sedda:2018qgq,Banerjee:2018pmh} and by quadrupole systems \cite{Fang_Thompson_Hirata2018}.

\paragraph{Mergers in dark matter halos}

The first metal-free stars (Pop III) may form binaries dynamically and merge in DM halos \cite{2014MNRAS.442.2963K}, but the expected rates are below the observed rates~\cite{Belczynski:2016ieo}.
Futhermore, two dynamical channels have been suggested leading to a high number of BH mergers in DM halos. If PBHs constitute a significant fraction of DM, the merger rates following GW captures in single-single encounters match the observed rates \cite{Bird:2016dcv}. However, this would also lead to the dispersion of weakly bound GCs in ultrafaint dwarf galaxies, contradicting an observed system \cite{Brandt:2016aco}. More recent estimates show that the LIGO rates are matched by this channel even if only $1\%$ of the DM is in PBHs \cite{Ali-Haimoud:2017rtz}. The second PBH channel requires only a $0.1\%$-- fraction of DM to be PBHs, given that they form binaries dynamically in the very early universe \cite{Sasaki:2016jop}. While these sources can match the observed rates, as discussed in the following Sec.~\ref{Sec:PBHandDM}, we await further strong theoretical arguments or observational evidence for the existence of these PBHs \cite{Sasaki:2018dmp}.

\subsection{Advances in numerical methods in dynamical modeling}
Recent years have brought significant advances in modelling the dynamical environments leading to GW events. 
\subsubsection{Direct $N$-body integration}
State of the art direct $N$--body simulations have been used to model the dynamics of GCs to interpret GW observations. These methods now reach $N=10^6$, so that stars are represented close to 1:1 in the cluster \cite{Wangetal2016}. Comparisons between the largest direct $N$--body and Monte Carlo simulations show an agreement \cite{Wangetal2016,Rodriguezetal2016a}. However, due to their high numerical costs, only a very low number of initial conditions have been examined to date. Further development is needed to account for a higher number of primordial binaries, larger initial densities, a realistic mass spectrum, and a nonzero level of rotation and anisotropy.

Large direct $N$-body simulation have also been used to study the dynamics in nuclear star clusters (NSC) in galactic nuclei with an SMBH and the formation of NSCs from the infall of GCs \cite{ArcaSedda2015}. Recent simulations have reached $N=10^6$ in regions embedding NSCs \cite{Panamarev:2018bwq}. Here, the number of simulated stars to real stars is 1:100. To interpret GW mergers in these systems, a 1:1 simulation of the innermost region of the NSC would be most valuable, even if the total simulated number is of the order $10^5$--$10^6$. Including a mass distribution and primordial binaries would also be useful.

Direct $N$-body methods were also recently used to simulate the NSC in AGN 
with stellar captures by the accretion disk, to predict the rate of tidal disruption events and the formation of a nuclear stellar disk \cite{Kennedyetal2016,Panamarev:2018who}. The most important place for development is to relax the assumption of a rigid fixed disk in these simulations. Indeed, the disk may be expected to be significantly warped by the star cluster \cite{Kocsis_Tremaine2011,Peretsetal2018}, and the stars and BHs captured in the disk may grow into IMBHs and open gaps in the disk \cite{Kocsisetal2011}. Furthermore, an initially nonzero number of binaries, the binary-disk interaction, and a mass spectrum would be important to incorporate to make GW predictions.
\subsubsection{Monte-Carlo methods}
State of the art Monte-Carlo methods have also improved significantly during the past years, providing a numerically efficient way to model the evolution of GCs accurately. Recent developments showed that the BH subclusters do not necessarily decouple and evaporate at short time scales and that GCs with longer relaxation times can retain BHs up to a Hubble time \cite{Morscheretal2015,Arcasedda2018,Askar2018}. These methods have been used to predict the evolutionary pathways to some of the observed mergers~\cite{Rodriguez:2016avt,Chatterjeeetal2017} and to interpret the distributions of masses and spins \cite{Rodriguez2016b,Askaretal2017}. These methods have been used to study the formation of IMBHs~\cite{Gierszetal2015}. Most recently, 2.5-order post-Newtonian dissipative effects were incorporated in order to re-simulate binary-single interactions involving three BHs from results of these Monte Carlo codes which increased the rate of eccentric mergers by a factor of 100~\cite{Rodriguez:2017pec,Samsingetal2017,Samsing:2018isx}. The implementation of post-Newtonian terms and tidal dissipation \cite{Samsing:2018uoo} for computing results of strong binary-single and binary-binary encounters involving at least two BHs could further increase merger rates for BBHs expected to originate from GCs. Moreover, implementation of two body gravitational and tidal capture within these codes could provide more insights into the role of dynamics in forming potentially merging BBHs.

Further development is needed to include resonant multibody effects. Moreover, simulations of galactic nuclei with Monte Carlo methods would be valuable to  tracking the evolution of binaries, and accounting for long term global effects such as resonant relaxation.
\subsubsection{Secular Symplectic $N$-body integration}
Systems which are described by a spherical gravitational potential such as a galactic nucleus or a GC are affected by strong global resonances, in which the anisotropies of the system drive a rapid secular change in the angular momentum vectors of the objects, called resonant relaxation \cite{Rauch_Tremaine1996}. Vector and scalar resonant relaxation, which affect the distribution of orbital planes and the eccentricity, respectively, are expected to reach statistical equilibrium within a few Myr to a few 100 Myr, respectively. A secular symplectic $N$-body integration method was recently developed \cite{Kocsis_Tremaine2015}. Preliminary results show that objects such as BHs, which are heavier than the average star in a GC or a nuclear star cluster tend to be distributed in a disk \cite{Szolgyen:2018zra}. Since LVC mergers happen more easily in BH disks, if they exist, future studies are necessary to explore the formation, evolution, and the expected properties of such configurations.

The statistical equilibrium phase space distribution of resonant relaxation is known only for a limited number of idealized configurations \cite{Touma_Tremaine2014,Roupasetal2017,Takacs:2017wnn,Fouvry2018}. Interestingly resonant relaxation has strong similarities to other systems in condensed matter physics such as point vortices and liquid crystals \cite{Touma_Tremaine2014,Kocsis_Tremaine2015,Roupasetal2017}. This multidisciplinary connection may be used  to study these probable sites of BH mergers \cite{Roupasetal2017}.

\subsubsection{Semianalytical methods}
Semianalytical methods are developed to include the highest number of physical effects in an approximate way. The formation of the Galactic bulge from the infall of GCs was examined in great detail with this technique \cite{Gnedinetal2014,2014MNRAS.444.3738A}. It was shown, that during this process, more than 95\% of the initial GC population is destroyed. Thus, GCs were much more common at cosmologically early times. Since one way to form IMBHs in GCs is by runaway collisions of stars or BHs \cite{PortegiesZwart_McMillan2002}, their numbers may also expected to be higher than previously thought (however, this conclusion may be different in other IMBH formation channels \cite{Gierszetal2015}). GCs generate a high rate of mergers between stellar mass and $10^2$--$10^3\Msun$ IMBHs detectable in the near future with advanced LIGO and Virgo detectors at design sensitivity at $z>0.6$ \cite{Fragione:2017blf,2018MNRAS.477.4423A,Arca-Sedda:2017wea}.  BH mergers with IMBHs with $10^3$--$10^4\Msun$ may also be detected with LISA from the local Universe.

\subsection{Astrophysical interpretation of dynamical sources}
How can dynamical models test the astrophysical interpretation of GW sources? GW detectors can measure the binary component masses, the spin magnitudes and direction, the binary orbital plane orientation, the eccentricity, the distance to the source and the sky location. The observed distributions of these parameters may be compared statistically to the predicted distributions for each channel. Some smoking gun signatures of the astrophysical environment are also known for individual sources, discussed below.

\subsubsection{Mass distribution}
The mass distribution of mergers depends on the theoretically poorly known BH mass function times the mass-dependent likelihood of mergers. 
Particularly, the mergers of massive objects are favored in GCs, due to the mass dependence of binary formation in triple encounters, binary exchange interactions, dynamical friction, and the Spitzer instability. Monte Carlo and $N$-body simulations show that the likelihood of merger is proportional to $M^4$ in GCs \cite{OLearyetal2016}. The 2-dimensional total mass and mass ratio distribution of mergers may be used to test this prediction \cite{Zevinetal2017} and to determine the underlying BH mass function in these environments. The mass function of mergers may also vary with redshift as the most massive BHs are ejected earlier \cite{Rodriguez2016b}.

Recently, Ref.~\cite{Kocsisetal2018} introduced a parameter to discriminate among different astrophysical channels:
\begin{equation}
 \alpha = -(m_1+m_2)^2 \frac{\partial^2 \mathcal{R}}{\partial m_1\partial m_2}\,.
\end{equation}
This dimensionless number is $1 \pm 0.03$ for PBHs formed dynamically in the early universe (if the PBH mass
function is assumed to be flat. For arbitrary PBH mass function the results can be substantially different~\cite{Chen:2018czv}), $10/7$ for BHs which form by GW emission in collisionless systems such as DM halos. For BHs which form by GW emission in collisional systems which exhibit mass segregation, this $\alpha$ varies for different component masses. In galactic nuclei it ranges between $10/7$ for the low mass components in the population to $-5$ for the highest mass component. It would be very useful to make predictions for $\alpha$ for all other merger channels. 

\subsubsection{Spin distribution}
Using the empirically measured rotation rate of Wolf-Rayet (WR) stars, the birth spin of the massive BHs is expected to be small~\cite{Amaro-Seoane:2015umi}. If BHs have undergone previous mergers, their spin is distributed around 0.7 \cite{Gerosa:2017kvu,Fishbachetal2017}.
Dynamical effects may also spin up the WR-descendent BH \cite{Zaldarriagaetal2018}. 
If the BH acquires a significant amount of mass due to accretion, it is expected to be highly spinning. Thus, second generation mergers are distinguishable in mass and spin from first generation mergers. Monte Carlo simulations predict that $10\%$ are second generation in GCs
\cite{OLearyetal2016,Rodriguez:2017pec}. Results from Monte Carlo simulations have also been utilized to investigate the role of spin in gravitational recoil kicks on merged BHs \cite{Morawski:2018kfs}. This has important implications for repeated mergers in dense environments like GCs, results from Ref.~\cite{Morawski:2018kfs} suggest that that about 30\% of merging BHs could be retained in GCs. According to a recent X-ray observing campaign, 7 out of the 22 Active Galactic Nuclei analyzed are candidates for being high spin SMBHs (with spin $>98\%$), see Table 1 of Ref.~\cite{Brenneman:2013oba}. Some of the X-ray binaries show evidence of highly spinning stellar mass BHs \cite{McClintocketal2014}. However, with the exception of a single source, current LVC sources are consistent with zero effective spin \cite{Farretal2017}. If spinning, the relative direction of spins is expected to be uncorrelated (isotropic) for spherical dynamical systems. This is different from the standard common envelope channel, where spin alignment is generally expected with the angular momentum vector and counteralignment is not likely, in case the BHs are spinning \cite{Farretal2017,Farretal2018}.

\subsubsection{Eccentricity distribution}
Since GW emission circularizes binaries \cite{Peters:1964zz}, they are expected to be nearly circular close to merger unless they form with a very small pericenter separation. Indeed, GW sources in GCs are expected to have a relatively small eccentricity in the LIGO band close to merger \cite{OLearyetal2006}. Moderate eccentricity of $e=0.1$ is expected from $10\%$ of GC sources at the low-frequency edge of the Advanced LIGO design sensitivity \cite{Rodriguez:2017pec,Samsingetal2017}. However, they may have a high eccentricity in the LISA band \cite{OLearyetal2006,Breiviketal2016,Samsing:2018isx,DOrazio:2018jnv,Samsing:2018ykz} or the DeciHz band~\cite{Chen:2017gfm}. Field triples may also yield some eccentric LIGO mergers \cite{Silsbee_Tremaine2017,2017ApJ...841...77A,Rodriguez:2018jqu,Arca-Sedda:2018qgq}. Eccentricity may be much higher for sources forming by GW emission in close encounters \cite{OLearyetal2009}.

The eccentricity distribution of merging BH binaries in GCs is expected to have three distinct peaks corresponding to binaries which merge outside of the cluster after ejection following a binary-single hardening interaction, binary mergers which happen within a cluster in between two binary-single interactions, and mergers which happen during a binary-single interactions \cite{Samsing:2018isx,DOrazio:2018jnv}.

For GW capture binaries, the typical eccentricity at the last stable orbit (LSO, extrapolating \cite{Peters:1964zz}) is set by the velocity dispersion of the source population $\sigma$ as 
$e_{\rm LSO} \sim 0.03 (\sigma/1000{\rm km}{\rm s}^{-1})^{35/32}\,(4\eta)^{-35/16}$, where $\eta=m_1m_2/(m_1+m_2)^2$ is the symmetric mass ratio \cite{Gondan:2017wzd}.
The heavier stellar BHs and IMBHs are expected to merge with a higher eccentricity at around $0.1$, while the lower mass BHs will have an eccentricity around $10^{-3}$ \cite{Gondan:2017wzd}. At design sensitivity Advanced LIGO and VIRGO may be more sensitive to eccentricity well before merger \cite{Gondan:2017hbp}.

\subsubsection{Sky location distribution}
Since GW sources are observed from beyond $100\,\Mpc$, they are expected to be isotropically distributed. The sky location measurement accuracy is typically insufficient to identify a unique host galaxy counterpart for individual mergers. However, the power spectrum of sky location of all mergers may be measured. This may be useful to determine the typical galaxy types of mergers, particularly, whether the sources are in active galactic nuclei \cite{Bartosetal2017b}.

\subsubsection{Smoking gun signatures}
Which other features may help to conclusively identify the host environment of individual GW sources? 

\paragraph{Resolved clusters with electromagnetic counterparts} Recently it was shown that several inspiraling stellar mass black hole binaries of Milky Way GCs are expected to be in the LISA band \cite{Kremer:2018tzm}. Therefore if LISA identifies these GW sources on the sky, it will constrain the event rates corresponding to the GC channel.

\paragraph{Modulation of the GW phase due to environmental effects} In the case of LISA sources, extreme mass ratio inspirals may be significantly affected by an a SMBH companion or gas-driven migration \cite{Yunesetal2011,Kocsisetal2011}. For stellar-mass BH mergers, the most important effect of a perturbing object is a Doppler phase shift, which accumulates mostly at low GW frequencies \cite{Meironetal2017}. LIGO and LISA together will be able to measure the SMBH provided that the orbital period around the SMBH is less than $\mathcal{O}(\yr)$ \cite{Meironetal2017,Inayoshi:2017hgw}.

\paragraph{GW astrophysical echos} The identification of a secondary lensed signal by the SMBH using LVC may confirm that a LVC merger takes place in a galactic nucleus. This manifests as a GW echo with a nearly identical chirp waveform as the primary signal but with a typically fainter amplitude depending on direction and distance from the SMBH \cite{Kocsis2013}. If the distance between the source and the SMBH is less than $10^3 M_{\rm SMBH}$, the relative amplitude of the echo is of the order of $10\%$, and the time-delay is less than a few hours. Studies of the expected source parameters of sources with GW echos are underway.

\paragraph{Double mergers and mergers with electromagnetic counterparts}
Finally, a case is conceivable in which in which a binary-single or binary-binary encounter results in a double merger, i.e. two mergers from the same direction within a short timespan \cite{Samsing:2017xod}. Such an observation would indicate a dense dynamical host environment. Furthermore, binary-single or binary-binary encounters involving at least 2 BHs and one or more stellar object may also lead to BBH mergers with with transient electromagnetic counterparts associated with the disruption or accretion of stellar
matter on the BHs. Observation of such a counterpart would also indicate that the merging BHs originated in a dense dynamical environment. The inclusion of tidal and gravitational dissipation effects during the computation of such strong encounters \cite{Samsing:2018uoo} within simulations of GCs could help to constrain rates for BBH mergers in which an electromagnetic counterpart could be expected.

\section {Primordial Black Holes and Dark Matter}\label{Sec:PBHandDM}
\vspace{-3mm}
{\it Contributors:} G.~Bertone, C.~T.~Byrnes, D.~Gaggero, J.~Garc\'ia-Bellido, B.~J.~Kavanagh
\vspace{3mm}

\subsection{Motivation and Formation Scenarios.}

The nature of Dark Matter (DM) is one of the most pressing open problems of Modern Cosmology. The evidence for this mysterious form of matter started to grow in the early 20th century~\cite{Bertone:2016nfn}, and it is today firmly established thanks to a wide array of independent observations~\cite{Bertone:2010zza}. Its nature is still unknown, and the candidates that have been proposed to explain its nature span over 90 orders of magnitude in mass, from ultra-light bosons, to massive BHs~\cite{Bergstrom:2000pn,Bertone:2004pz,Feng:2010gw}.  
An intriguing possibility is that at least a fraction of DM is in the form of PBHs~\cite{Carr:2016drx}. The recent LVC detections~\cite{Bird:2016dcv,Clesse:2016vqa}, have revived the interest in these objects, and prompted a reanalysis of the existing bounds~\cite{Carr:2016drx,Garcia-Bellido:2017xvr} and new prospects for phenomenological signatures~\cite{Garcia-Bellido:2017fdg,Clesse:2017bsw}.

PBHs might be produced from early universe phase transitions, topological defects (e.g., cosmic strings, domain walls), condensate fragmentation, bubble nucleation and large amplitude small scale fluctuations produced during inflation (see the reviews~\cite{Green:2014faa,Sasaki:2018dmp} and references therein). Excluding the possibility of BH relics, non-evaporating PBHs could range in mass from $10^{-18}$ to $10^6$ solar masses today, although the most interesting range today is that associated with the observed LIGO BBHs, around a few to tens of solar masses.
For those PBHs, amongst the most promising production mechanisms was the one first proposed in
\cite{GarciaBellido:1996qt} from high peaks in the matter power spectrum generated during inflation. When those fluctuations re-enter the horizon during the radiation era, their gradients induce a gravitational collapse that cannot be overcome even by the radiation pressure of the expanding plasma, producing BHs with a mass of order the horizon mass~\cite{Clesse:2015wea}. Most of these PBH survive until today, and may dominate the Universe expansion at matter-radiation equality.

One important question is what fraction $f_{\rm PBH}= \Omega_\mathrm{PBH}/\Omega_\mathrm{DM}$ of DM should be made of O(10M) PBHs in order to match the BBH merger rate inferred by LIGO/Virgo ($\mathcal{R} \simeq$ 10 - 200 Gpc$^{-3}$ yr$^{-1}$ \cite{Abbott:2017vtc}).
If one considers PBH binaries that form within virialized structures, the corresponding rate is compatible with $f_{\rm PBH} = 1$ \cite{Bird:2016dcv}.  However, PBH binary systems can form in the early Universe (deep in the radiation era) as well \cite{Nakamura:1997sm,Ioka:1998nz}, if the orbital parameters of the pair allow the gravitational pull to overcome the Hubble flow and decouple from it. A recent calculation of the associated merger rate today \cite{Sasaki:2016jop} -- significantly extended  in \cite{Ali-Haimoud:2017rtz} -- provides a much larger estimate (compared to the former scenario): This result can be translated into a bound on $f_{\rm PBH}$, which is potentially stronger than any other astrophysical and cosmological constraint in the same range. The constraint has been recently put on more solid grounds in \cite{Kavanagh:2018ggo} by taking into account the impact of the DM mini-halos around PBHs, which have a dramatic effect on the orbits of PBH binaries, but surprisingly subtle effect on their merger rate today. Several aspects of this calculation are still under debate, including the PBH mass distribution, the role of a circumbinary accretion disk of baryons, the impact of initial clustering of PBHs (see \cite{Ali-Haimoud:2018dau}) and the survival until present time of the binary systems that decoupled in the radiation era.

The critical collapse threshold to form a PBH is typically taken to be $\delta_c\equiv {\delta\rho_c/\rho}\sim0.5$, with an exponential sensitivity on the background equation of state, the initial density profile and angular momentum. Because any initial angular momentum suppresses PBH formation, PBHs are expected to spin slowly, a potential way to discriminate between them and (rapidly rotating) astrophyical BHs \cite{Chiba:2017rvs}.

Inflationary models which generate PBHs are required to generate a much larger amplitude of perturbations on small scales compared to those observed on CMB scales. This can be achieved either through multiple-field effects or an inflection point in single-field inflation~\cite{Ezquiaga:2017fvi,Garcia-Bellido:2017mdw}. These typically generate a reasonably broad mass fraction of PBHs, in contrast to the monochromatic mass spectrum usually assumed when interpreting observational constraints. In addition, the softening of the equation of state during the QCD phase transition greatly enhances the formation probability of solar mass PBHs compared to the more massive merging BHs LIGO detected \cite{Byrnes:2018clq}. BHs below the Chandrasekhar mass limit would be a smoking gun of a primordial origin. 

At non-linear order, scalar and tensor perturbations couple, implying that the large amplitude perturbations required to generate PBHs will also generate a stochastic background of GWs. For the LIGO mass range the corresponding GW frequency is constrained by pulsar timing arrays unless the scalar perturbations are non-Gaussian~\cite{Nakama:2016gzw,Garcia-Bellido:2017aan}.

\subsection{Astrophysical probes}

\begin{figure}
\centering
\includegraphics[width=0.79\textwidth]{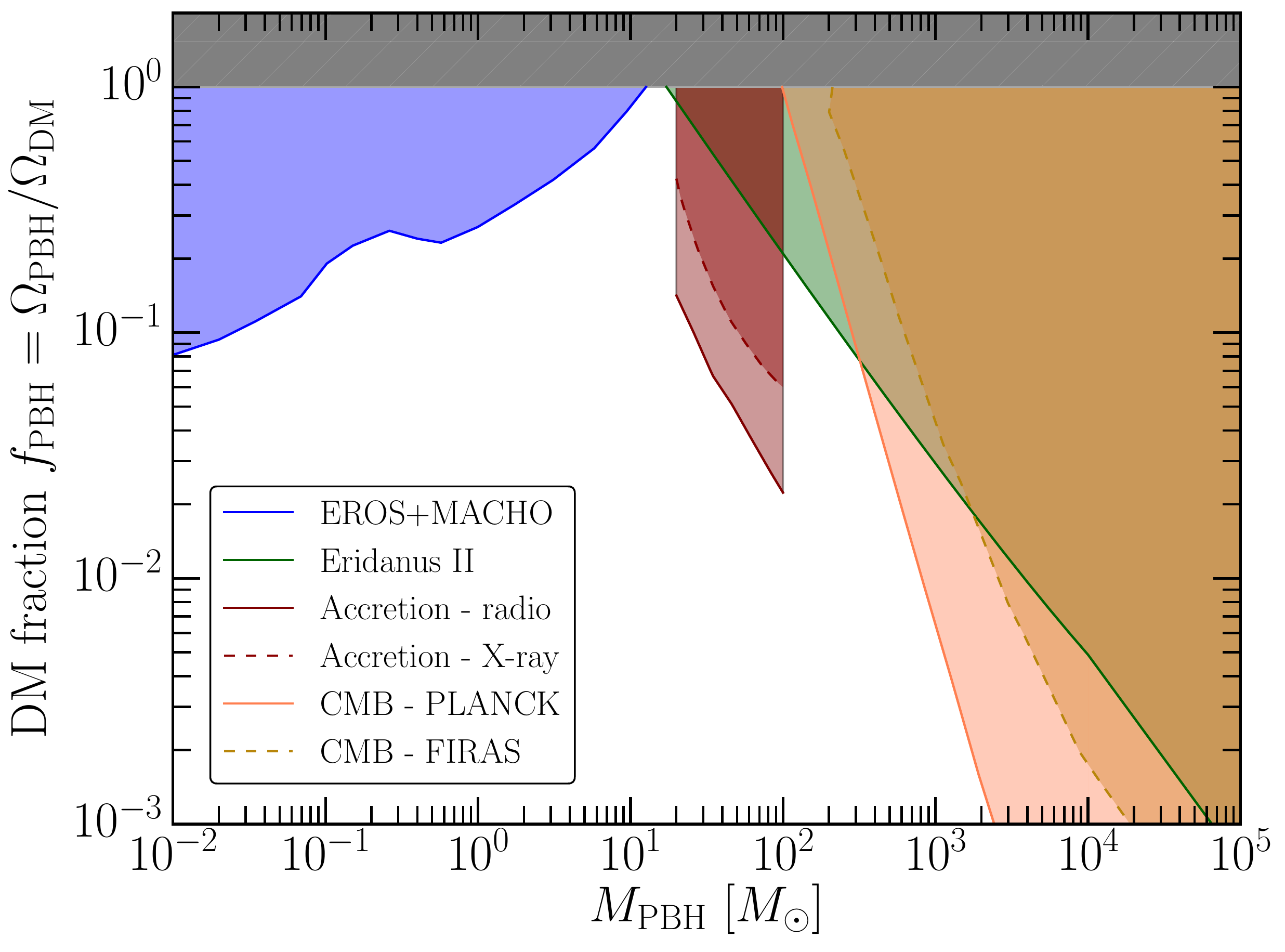}
\caption{\textbf{Summary of astrophysical constraints on PBHs in the mass range $M \in [10^{-2},\,10^5]\,M_\odot$.} Details of the constraints are given in the main text and we plot here the most conservative. We emphasize that astrophysical constraints may have substantial systematic uncertainties and that the constraints shown here apply only for monochromatic mass functions.}
\label{fig:constraints}
\end{figure}
Constraints on the PBH abundance are typically phrased in terms of $f_\mathrm{PBH} = \Omega_\mathrm{PBH}/\Omega_\mathrm{DM}$, the fraction of the total DM density which can be contributed by PBHs. 
Even in the case where $f_\mathrm{PBH} < 1$, a future detection via astrophysical probes or GW searches remains hopeful \cite{Sasaki:2016jop}. We outline selected astrophysical constraints below, summarizing these in Fig.~\ref{fig:constraints}, where we focus on PBHs around the Solar mass range, most relevant for GW signals.

{\it Micro-lensing:}  The MACHO \cite{Allsman:2000kg} and EROS \cite{Tisserand:2006zx} collaborations searched for micro-lensing events in the Magellanic Clouds in order to constrain the presence of Massive Compact Halo Objects (MACHOs) in the Milky Way Halo. Considering events on timescales of $\mathcal{O}$(1-800) days constrains $f_\mathrm{PBH} < 1$ for masses in the range $M_\mathrm{PBH} \in [10^{-7}, 30]\,M_\odot$ 
(though these constraints come with a number of caveats, see e.g. \cite{Hawkins:2015uja,Garcia-Bellido:2017xvr}).
Another promising target is M31: an important fraction of the Andromeda and Milky Way DM halo can be probed by micro-lensing surveys, and an interesting hint comes from the observation of $56$ events in the $\mathcal{O}$(1-100) days range from this region of interest, which may suggest a MACHO population with $f \sim 0.5$ and mass $1 \,M_\odot$ or lighter \cite{CalchiNovati:2005cd,Lee:2015boa}.
A recent search for lensing of Type Ia Supernovae obtained constraints on all PBH masses larger than around $10^{-2}\,M_\odot$, although substantial PBH fractions $f_\mathrm{PBH} \lesssim 0.6$ are still compatible with the data \cite{Zumalacarregui:2017qqd}. For wide mass distributions the SN lensing constraints go away \cite{Garcia-Bellido:2017imq} and PBH could still constitute all of the DM. 

{\it Early Universe constraints:} PBHs are expected to accrete gas in the Early Universe, emitting radiation and injecting energy into the primordial plasma. This in turn can lead to observable distortions of the CMB spectrum and can affect CMB anisotropies \cite{Carr1981}. Early calculations over-estimated the effect \cite{Ricotti:2007au,Chen:2016pud}, with more recent studies finding weaker constraints \cite{Blum:2016cjs,Ali-Haimoud:2016mbv}. In spite of this, data from COBE/FIRAS \cite{Blum:2016cjs,Clesse:2016ajp} and PLANCK \cite{Ali-Haimoud:2016mbv} can still rule out PBHs with masses $M_\mathrm{PBH} \gtrsim 100 \,M_\odot$ as the dominant DM component. It should be noted that these constraints depend sensitively on the details of accretion onto PBHs in the Early Universe (see e.g.~Ref.~\cite{Poulin:2017bwe}).

{\it Dynamical constraints:} The presence of PBHs is also expected to disrupt the dynamics of stars. Wide halo binaries may be perturbed by PBHs and the actual observation of such systems constrains the PBHs abundance above  $\simeq 20 \,M_\odot$ \cite{Monroy-Rodriguez:2014ula} (although significant fractions $f_\mathrm{PBH} \lesssim 0.2$ are still allowed). PBHs are also expected to dynamically heat and thereby deplete stars in the centre of dwarf galaxies. Observations of stellar clusters in Eridanus II \cite{Brandt:2016aco} and Segue I \cite{Koushiappas:2017chw} have been used to constrain PBHs heavier than $\mathcal{O}(1)\,M_\odot$ to be sub-dominant DM components, unless they have an IMBH at their center~\cite{Li:2016utv}.

{\it Radio and X-ray constraints:}
If PBHs exist in the inner part of the Galaxy, which contains high
gas densities, a significant fraction of them would inevitably form an accretion disk and emit a broad-band spectrum of radiation. A comparison with existing catalogs of radio and X-ray sources in the Galactic Ridge region already rules out the possibility that PBHs constitute all of the DM in the Galaxy, even under conservative assumptions on the physics of accretion~\cite{Gaggero:2016dpq}. 
During the next decade, the SKA experiment will provide an unprecedented increase in sensitivity in the radio band; in particular, SKA1-MID will have the unique opportunity to probe the presence of a subdominant population of PBHs in the Galaxy in the $10 \div 100 \, M_\odot$ mass range, even if it amounts to a fraction as low as $\simeq 1$\% of the DM.

In Fig.~\ref{fig:constraints}, we show only the most conservative  of these constraints, which suggest that PBHs may still constitute all of the DM for masses close to $10\,M_\odot$, or a smaller fraction ($f_\mathrm{PBH} \gtrsim 0.1$) for masses up to $\mathcal{O}(100\,M_\odot)$. We highlight that such astrophysical constraints may have large systematic uncertainties and that these constraints apply only to PBHs with a mono-chromatic mass function. Recent studies have begun re-evaluating these constraints for extended mass functions \cite{Green:2016xgy,Bellomo:2017zsr,Kuhnel:2017pwq,Garcia-Bellido:2017xvr}. For physically motivated mass functions, it may be possible to achieve up to $f_\mathrm{PBH} \sim 0.1$ for PBHs in the mass range $25-100 \,M_\odot$ \cite{Carr:2017jsz}. Relaxing certain dynamical constraints (which typically have large uncertainties) or considering more general mass functions may accommodate an even larger PBH fraction \cite{Carr:2017jsz,Lehmann:2018ejc}.

\subsection{Discriminating PBHs from astrophysical BHs} 

A number of observations may help discriminating PBHs from ordinary astrophysical BHs. The detection of GWs from a binary system including a compact object lighter than standard stellar BHs, say below 1 solar mass, would point towards the existence of PBHs. This can be deduced from the highest frequency reached in the GW chirp signal, $f_{\rm ISCO} = 4400\,{\rm Hz}\, (M_\odot/M)$, and it is in principle possible already with Advanced LVC in the next run O3 ~\cite{Garcia-Bellido:2017fdg}. The detection of GWs 
at redshift $z \gtrsim 40$ would imply a non-standard cosmology, or the existence of PBHs~\cite{Koushiappas:2017kqm}.
Further insights on the origin of BHs might be obtained through the analysis of `environmental' effects, which are discussed in Sec.~\ref{Sec:DM}, and through the analysis of the spatial distribution and mass function of X-ray and radio sources powered by BHs.

\section{Formation of supermassive black hole binaries in galaxy mergers}\label{Sec:FormSMBH}
\vspace{-3mm}
{\it Contributors:} M.~Colpi, M.~Volonteri, A.~Sesana 
\vspace{3mm}

\begin{figure}
\begin{center}
\includegraphics[width=1.00\textwidth]{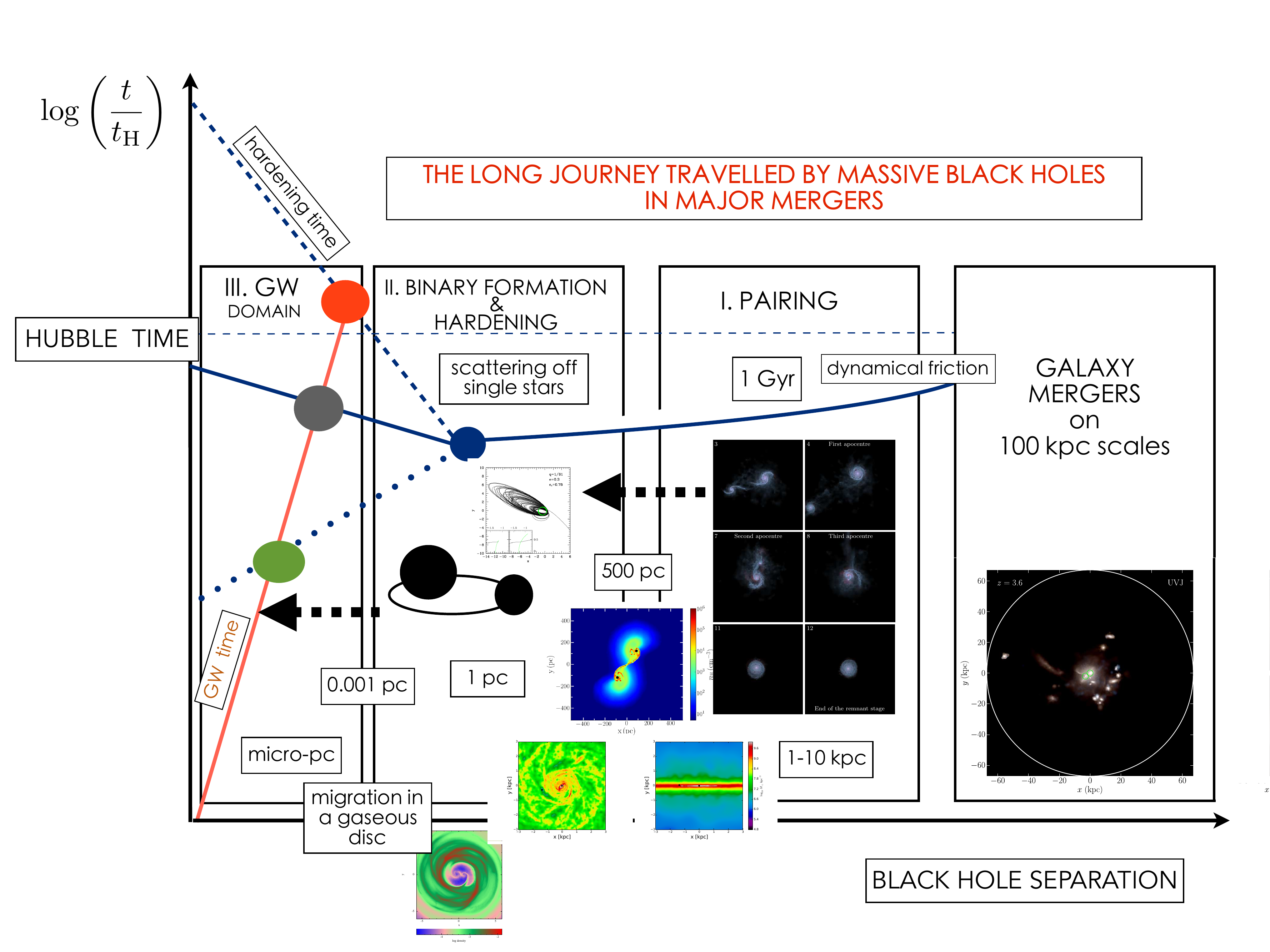}
\caption{Cartoon illustrating the journey travelled by SMBHs of masses in the range $10^{6-8}\msun$ during major galaxy-galaxy collisions. The $x-$ axis informs on the SMBH separation given in various panels, while the $y-$ informs on the timescale.
The journey starts when two galaxies (embedded in their DM halos) collide on kpc scales (right-most plot).
The inset shows a selected group of galaxies from the cosmological simulation described in Ref.~\cite{Khan:2016vln}.
The inset in the second panel (from right to left), from Ref.~\cite{2015MNRAS.447.2123C}, depicts the merger of two disc galaxies and their embedded SMBHs. Pairing occurs when the two SMBHs are in the midst of the new galaxy that has formed, at separations of a few kpc. The SMBHs sink under the action of star-gas-dynamical friction.
In this phase, SMBHs may find themselves embedded in star forming nuclear discs, so that their dynamics can be altered by the presence of massive gas-clouds. Scattering off the clouds makes the SMBH orbit stochastic, potentially broadening the distribution of sinking times during the pairing phase~\cite{Fiacconi13,2015MNRAS.446.1765L,2017MNRAS.464.2952T,2015MNRAS.453.3437L}. Furthermore, feedback from supernovae and AGN triggering by one or both the SMBHs affect the dynamics as these processes alter the thermodynamics of the gas and its density distribution, which in turn affect the process of gas-dynamical friction on the massive BHs. It is expected that eventually  the SMBHs form a Keplerian binary, on pc scales. Then, individual scattering off stars harden the binary down to the GW-driven domain. This is an efficient mechanism (and not a bottleneck) if the relic galaxy displays some degree of triaxiality and/or rotation. In this case, there exists a large enough number  of stars on low-angular momentum orbits capable to interact with the binary and extract orbital energy~\cite{2015ApJ...810...49V}. The binary in this phase can also be surrounded by a (massive) circum-binary disc \cite{2012A&A...545A.127R}. In a process reminiscent to type II planet migration, the two SMBHs can continue to decrease their separation and they eventually cross the GW boundary. Then, GW radiation controls the orbital decay down to coalescence.} 
\label{journey} 
\end{center} 
\end{figure}

When two galaxies, each hosting a central SMBH merge, the SMBHs start a long journey that should bring them from separations of tens of kpc (1 kpc = $3.086\times 10^{21}$ cm) down to milli-parsec scales, below which they merge through GWs. The initial {\it pairing} of BHs in merging galaxies, their binding into a {\it binary} on parsec scales, and their crossing to enter the GW-driven inspiral are the three main stages of this dynamical journey, sketched in Figure \ref{journey} \cite{Colpi14}. In some cases, the two SMBHs become bound in the core of the merger remnant, cross to the gravitational-driven regime and eventually coalesce. However, there are cases in which the two SMBHs never form a binary: one of the SMBHs may remain stranded and unable to pair and bind \cite{1994MNRAS.271..317G}.

The cosmic context, the growth of DM halos through mergers with other halos gives us the backdrop upon which all the subsequent evolution develops. This is the first clock: the halo merger rate evolves over cosmic time and peaks at different redshift for different halo masses. Furthermore, the cosmic merger rate predicts that not all mergers occur with the same frequency: {\it major} mergers, of halos of comparable masses (from 1:1 down to $\sim$ 1:4) are rare, while  minor mergers (mass ratio $<$ 1 : 10) are more common. However, the dynamical evolution is much faster in major mergers than in minor mergers: if the mass ratio of the merging halos and galaxies is too small, the satellite halo is tidally disrupted early on in its dynamical  evolution and its central SMBH is left on a wide orbit, too far from the center of the larger galaxy to ever merge with its SMBH \cite{Callegari09,2018MNRAS.tmp..160T}. A population of wandering SMBHs is therefore predicted to exist in galaxies \cite{2005MNRAS.358..913V,2010ApJ...721L.148B,2016MNRAS.460.2979V}.

For mergers where the SMBHs can pair, i.e., find themselves in the newly formed core of the merger remnant, the second hurdle is to get close enough to form a gravitationally bound binary \cite{Yu2002,2007Sci...316.1874M,2015MNRAS.449..494R,2017MNRAS.471.3646P}.  For SMBHs embedded in a singular isothermal sphere, a binary forms at a separation of $a = GM/2\sigma^2 \simeq 0.2 \, {\rm pc} \, (M_{\rm BBH}/10^6{\msun}) \, \left(\sigma/100 \kms\right)^{-2}$, where the mass of the SMBHs exceeds the enclosed mass in stars, gas, and DM. The journey from the beginning of the merger, at tens of kpc,  to the formation of the binary on pc to subpc scales, takes between a few tens of Myr for compact galaxies at very high redshift, $z>3-4$ to several Gyr for larger, less dense galaxies at low redshift \cite{Yu2002,Volonteri:2002vz,2017ApJ...840...31D}.  

In the cases where the SMBHs form a bound binary within the Hubble time, the final crossing into the GW regime hinges on exchanges of energy through scattering with low angular momentum stars in the nucleus of the galaxy \cite{Begelman:1980vb} or on extraction of angular momentum through gravitational torques coming from a gas disc that may result on the shrinking of its separation \cite{1980ApJ...241..425G}, or on a combination of the two processes. Recent results of direct N-body simulations, Monte Carlo methods and scattering experiments are giving an optimistic view of what has been considered to be the main bottleneck of the binary evolution for almost 40 years: the ``final parsec problem,'' i.e.,  running out of low-angular momentum stars \cite{Begelman:1980vb}.  The evolution of SMBH binaries through stellar scattering seems to continue at nearly a constant rate leading to merger in less than $\sim$1 Gyr \cite{2006ApJ...642L..21B,2015ApJ...810...49V,2015MNRAS.454L..66S} once rotation, triaxiality and the granularity of the stellar distribution are taken into account.

\begin{figure}
\begin{center}
\includegraphics[width=1.00\textwidth]{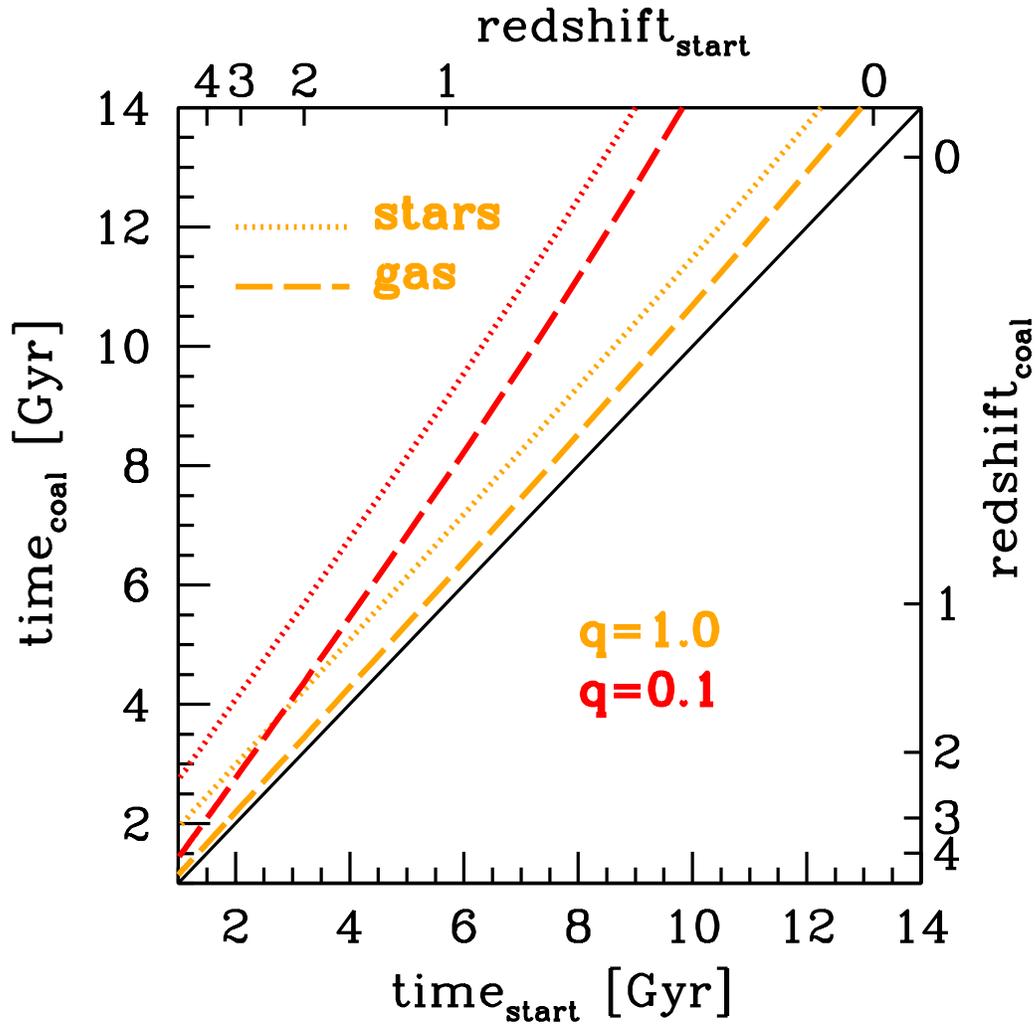}
\caption{Relation between ``time$_{\rm start}$,'' corresponding to the cosmic time
of onset of a galaxy-galaxy collision  hosting SMBHs, and `` time$_{\rm coal}$'' at which the SMBHs merge for circular binaries with the primary BH of $10^8 \msun$ and mass ratios $q=1$, and 0.1. Redshifts at start and coalescence are given in the same plot.
Dotted and dashed lines refer to sinking times associated to the total of the merger time for the host halos and galaxies, the pairing of SMBH, and then the shrinking of the binary subject to either stellar and gaseous processes. The black solid line represents ``time$_{\rm coal}={\rm time}_{\rm start}$  The Figure shows that galaxies colliding at $z\sim 1$ can display  time delays as long as 3 Gyr. 
\label{delay}}
\end{center} 
\end{figure}

If the binary environment is dominated by gas, rather than stars, the binary is expected to evolve through interaction with the accretion disk(s) surrounding the SMBHs, through the so called circumbinary disk phase. The disks are formed by the gas that inflows towards the SMBHs with an even small amount of angular momentum: in the single SMBH case these are referred to as ``accretion discs'' \cite{1973A&A....24..337S}. Depending on the mass ratio $q$, the binary may or may not clear a gap within the circumbinary disk. In the former case (valid for $q \ll 1$) the less massive hole behaves as a fluid element within the accretion disk of the primary, thus experiencing what is known as Type I migration. In the latter case (generally valid for $q > 0.01$), the formation of a central cavity slows down the evolution of the binary (Type II migration). The energy and angular momentum transfer between the binary and the circumbinary disk is mediated by streams leaking from the inner rim into the cavity. The pace of the supermassive binary BH (SMBBH) evolution depends on the detailed dynamics of the streams crossing the gap, which are partially accreted by the two SMBHs and partially flung back to the disk, extracting the binary energy.
Simulations generally concur that accretion into the central cavity is efficient enough to bring the SMBBH into the GW-dominated regime ($\sim 0.01$ pc) within $\sim$1-100 Myr \cite{ArmitageNarajan2005,2008ApJ...672...83M,Haiman09,rds+11,2012A&A...545A.127R,2012ApJ...749..118S,2013MNRAS.436.2997D, 2015ApJ...807..131S}. The role of AGN feedback on this scales, however, is still largely unexplored and may bring surprises.

In summary, SMBHs in merging galaxies have a long journey, that takes between 1 and 10 Gyr depending on the cosmic time,  mass and mass ratios of the host galaxies, galaxy morphologies, stellar and gas content, and on the masses of the SMBH themselves. Figure \ref{delay} is an illustration of the delay timescales in the dynamics of SMBH mergers, during the pairing process in major mergers. This is a simplified example for circular binaries with the primary SMBH of $10^8 \msun$ and mass ratios $q=1$, and 0.1. Here we included the time for halos to merge \cite{Boylan2008}, and then added an additional timescale for BHs to pair, based on the formalism of \cite{2014ApJ...789..156M}, motivated by the recent results of \cite{2018MNRAS.tmp..160T} who find that the halo merger timescale is insufficient to estimate the pairing timescale. We assumed 100 Myr from binary formation to coalescence for the gas-driven case and the fit proposed by \cite{2015MNRAS.454L..66S} for the stellar-driven case.

 Theoretical studies and predictions, however, are still far from being complete: this is a severely multi-scale problem, intimately connected  with the processes of galaxy clustering on cosmological scales, which involves a rich physics. Determining the distribution of merging times in SMBH mergers is critical, as only through the detailed knowledge of this distribution and associated processes, we are able
to bracket the uncertainties in the estimates of the SMBH merger rates, relevant for the LISA mission, and to provide reliable estimate of the expected GW signal in the nHz band, relevant to PTAs.

\section{Probing supermassive black hole binaries with pulsar timing arrays} \label{Sec:PTA}
\vspace{-3mm}
{\it Contributors:} C.~Mingarelli, M.~Kramer, A.~Sesana
\vspace{3mm}

Millisecond pulsars are excellent clocks over long timespans, making them ideal tools to search for GWs~\cite{saz78,det79, Backer:1982, hd83, r89, Burke-Spolaor:2015xpf,l15, l17}. Indeed, an array of millisecond pulsars forms a galactic-scale nanohertz GW detector called a Pulsar Timing Array (PTA, \cite{fb90}). The GWs change the proper distance between the Earth and the pulsars, which induces a timing delay or advance in the pulsar pulses. The difference between the expected pulse arrival time and the actual arrival time, called the timing residual, is used to search for signatures of low-frequency GWs. The frequency band where PTAs operate is set by the length of pulsar observations, and the cadence of the observations. Briefly, the lower limit is set by $1/T_\mathrm{obs}$, where $T_\mathrm{obs}$ is the total observation time, and the high-frequency limit is set by $1/\Delta t$, where $\Delta t$ is the cadence of the pulsar observations. This sets the sensitivity band between 1 nHz and 100 nHz.

The most promising signals in the PTA frequencies are due to the expected cosmic population of SMBBHs. The systems of interest here have masses $>10^8~M_\odot$, and can spend tens of millions of years emitting GWs in the relevant frequency band. The incoherent superposition of their signals gives rise to a stochastic GW background (GWB, see, e.g. \cite{rm95, jb03, wl03, shm+04, svc08}). On top of it, particularly nearby and/or very massive SMBBHs might be individually resolved ~\cite{svv09, mls+2017}, providing interesting targets for multimessenger observations. Moreover, the memory effect following the merger of those systems may also give rise to detectable bursts of GW radiation ~\cite{vHL10, mcc17}. Gravitational radiation may also originate from cosmic strings~\cite{Siemens:2006vk, lss09, sbs13}, and primordial GWs from e.g. inflation~\cite{Lasky:2015lej}.

Here we focus on the stochastic GWB and continuous GW sources, but it is worth mentioning that the correlation function present in GWB searches depends on both the underlying theory of gravity, the distribution of GW power on the sky, and the intra-pulsar distances. Indeed, additional GWB polarizations such as breathing modes can in principal be detected with PTA experiments, e.g. \cite{ljp08, ss12}, as can departures from GWB isotropy~\cite{msmv13, tg13, ms14, Cornish:2014rva, grt+14,TaylorEtAl:2015}. Clustering of large-scale structure, resulting in an overdensity of merging SMBBHs, can lead to GWB anisotropy, as can nearby continuous GW sources which are individually unresolvable, but can contribute to GWB anisotropy at level of $\sim 20\%$ of the isotropic component, see \cite{mls+2017}. Moreover, pulsars separated by less than $\sim 3 ^{\circ}$ violate the short-wavelength approximation~{ms14, mm18}, used to write down the Hellings and Downs curve, and exhibit an enhanced response to GWs which may help in their detection.

World-wide, PTA experiments have been taking data for over a decade, with efforts in North American governed by the North American Nanohertz Observatory for GWs (NANOGrav, see e.g. \cite{Arzoumanian:2017puf}), in Europe by the European PTA (EPTA)~\cite{desvignes+:2016}, and in Australia by the Parkes PTA (PPTA)~\cite{ShannonEtAl:2015}. The union of these PTAs forms the International Pulsar Timing Array (IPTA)~\cite{VerbiestEtAl:2016}, where the PTAs share data in the effort to accelerate the first detection of nHz gravitational radiation, and to learn more about the individual millisecond pulsars. For a more comprehensive overview of PTA experiments, see e.g.~\cite{Burke-Spolaor:2015xpf,l15,l17}

\subsection{The Gravitational-Wave Background}
The cosmic merger history of SMBBHs is expected to form a low-frequency GWB, which may be detected by PTAs in the next few years~\cite{Siemens:2013zla,rsg15,tve+16}. The time to detection depends strongly on the number of pulsars in the array, the total length of the dataset, and the underlying astrophysics which affects the SMBBH mergers, such as stellar hardening process, interactions with accretion disks, binary eccentricity, and potentially SMBH stalling. Searches for the GWB have resulted in evermore stringent limits on the amplitude $A$ of the GWB reported at a reference frequency of 1/yr:
\begin{equation}
h_c = A \left( \frac{f}{\mathrm{yr^{-1}}}\right)^{-2/3} \, ,
\end{equation}
where $h_c$ is the characteristic strain of the GWB~\cite{Phinney:2001di}. This simple power-law scaling assumes the the SMBBH systems are circular when emitting GWs, and that they are fully decoupled from their environment. Binary eccentricity and interactions with gas and stars around the binary can deplete the GW signal at very low frequencies (equivalently at wide binary separations), causing the GW strain spectrum to turn over~\cite{ks11, rws+14, scm15, ArzoumanianEtAl:2016}. On the other hand, the amplitude of the GWB is affected by the abundance and mass range of the cosmic population of SMBHs. Therefore, future detections of a stochastic GWB will allow to constrain both the overall population of SMBBHs and the physics driving their local dynamics~\cite{Sesana13,2017PhRvL.118r1102T, 2017MNRAS.468..404C, rph+18, lbh+17,rhh+16, iih18}. Current non-detections have been used in \cite{Middleton:2015oda} to challenge the popular $M_\mathrm{BH}-M_\mathrm{bulge}$ from \cite{kh13}. However, a full analysis taking into account uncertainties in the merger rate, SMBBH dynamics and the possibility of stalling is needed to draw firm conclusions \cite{2018NatCo...9..573M}.

The current upper limit on $A$ from the various PTA experiments are similar and are improving with time. From the EPTA is $A<3\times 10^{-15}$~\cite{Lentati:2015qwp}, from the PPTA this is $A<1\times 10^{-15}$~\cite{ShannonEtAl:2015}, from NANOGrav this is $A <1.45\times 10^{-15}$~\cite{Arzoumanian:2018saf}, and the IPTA limit is $1.7\times 10^{-15}$~\cite{VerbiestEtAl:2016}. In order to improve those, further more sensitive observations are needed, which can and will be provided by improved instrumentation and new telescopes like FAST, MeerKAT and eventually the SKA \cite{Janssen:2014dka}. Additionally, systematics needs to be addressed. For instance, solar system ephemeris errors can mimic a GWB signal, if the underlying data are sufficiently sensitive~\cite{thk+16}. Here, mitigation techniques can be applied, as already done in the recent analysis of the NANOGrav 11-year data \cite{Arzoumanian:2018saf}, while other effects, such as the interstellar weather may be best addressed with multi-frequency observations. Future IPTA results will take those and other effects into account.

\begin{figure*}
		\centering
		\includegraphics[width=4in]{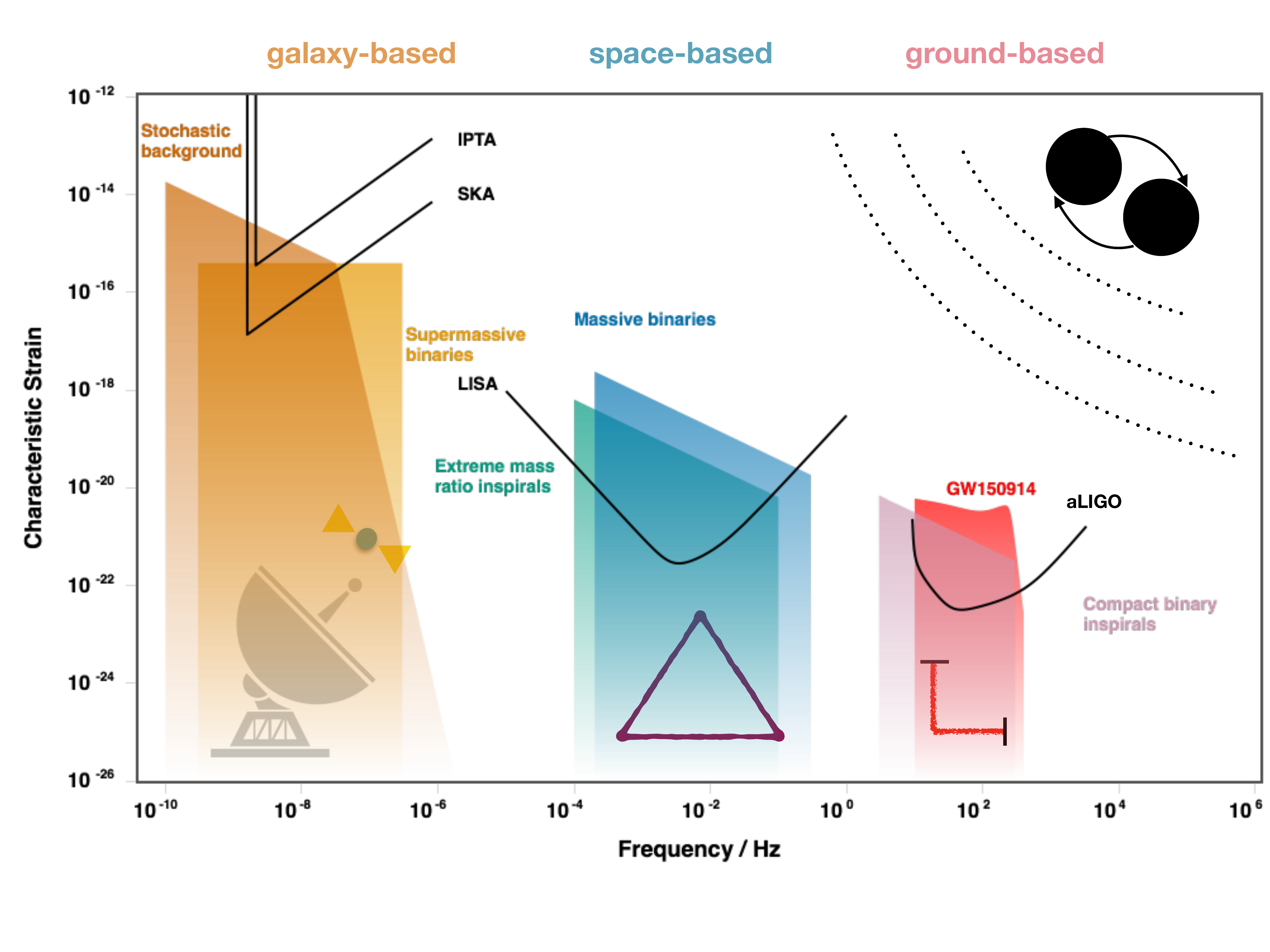}
		\caption[The Gravitational-Wave Landscape]{The spectrum of gravitational radiation from low-frequency (PTA) to high-frequency (LIGO). At very low frequencies pulsar timing arrays can detect both the GWB from supermassive black hole binaries, in the $10^8-10^{10}~M_\odot$ range, as well as radiation from individual binary sources which are sufficiently strong. We assume 20 pulsars with 100 ns timing precision with a 15 year dataset for IPTA, and 100 pulsars timed for 20 years with 30~ns timing precision for SKA. Both estimates assume 14-day observation cadence. This Figure was reproduced with minor changes from~\cite{mm18}, who in turn also used free software from~\cite{Moore:2014lga}.} 
		\label{fig:gwspectrum}
\end{figure*}

\subsection{Continuous Gravitational Waves}
Individual nearby and very massive SMBBHs emitting nHz GWs can also be detected with PTA experiments~\cite{lwk+11,2012PhRvD..85d4034B,abb+14,Babak:2015lua}. These SMBBHs are likely in giant elliptical galaxies, though the timescale between the galaxy merger and the subsequent SMBH merger is poorly understood. Unlike LIGO or LISA sources, these SMBBHs are in the PTA band for many millions of years, and will likely merge outside both the PTA and LISA band. The sky and polarization averaged strain of a continuous GW source is
\begin{equation}
h = \sqrt{\frac{32}{5}}\frac{\mathcal{M}_c^{5/3}[\pi f (1+z)]^{2/3}}{D_L} \, ,
\end{equation}
where $\mathcal{M}_c^{5/3} = [q/(1+q)^2] M^{5/3}$ is the binary chirp mass, $q<1$ is the binary mass ratio, $M$ is the total binary mass, $f$ is the GW frequency and $D_L$ is the luminosity distance to the binary.

The time to detection of continuous GW sources has been estimated in various ways: namely by simulating a GW source population based on cosmological simulations~\cite{svv09,sv10,rsg15}, or by using an underlying galaxy catalog to estimate which nearby galaxies can host SMBBH systems which are PTA targets~\cite{mls+2017,spl+14}. Both these approaches largely agree that at least one SMBBH system will be detected in the next decade or so, with the details depending on the amount of red noise in the pulsars.
Detecting individual SMBBH systems is expected to shed light on the so-called {\it final parsec problem}, providing valuable insights into how SMBBHs merge. In fact, pulsar distances are well-measured, it may be possible to measure the SMBH spins using the pulsar terms~\cite{Mingarelli2012}. 

Importantly, resolving an individual SMBBH with PTAs will open new avenues in multimessenger astronomy \cite{srr+12, tmh12, Burke-Spolaor:2013aba, m17}. The combined GW and electromagnetic signals will allow to pin down the properties of the binary and its environment, to study the dynamics of the SMBBH pairing and the coupling with the surrounding gaseous disc, if present~\cite{Sesana13,Burke-Spolaor:2015xpf}. Even a single joint detection will allow to nail down the distinctive electromagnetic properties of accreting SMBBHs. Analog systems can then be searched in the archival data of large surveys, to quantify their overall cosmic population. A promising way forward is to check candidates mined systematically in time domain surveys for peculiar spectral signatures, hence producing credible targets for PTAs.

\section{Numerical Simulations of Stellar-mass Compact Object Mergers}\label{Sec:NumericalRelativity}
\vspace{-3mm}
{\it Contributor:} A.~Perego
\vspace{3mm}

\subsection{Motivations}

Compact binary mergers (CBM) comprising at least one NS (i.e. BNS or BH-NS mergers) 
are unique cosmic laboratories for fundamental physics at the extreme. These powerful stellar collisions 
involve all fundamental interactions in a highly dynamical and intense regime (see, e.g., \cite{Rosswog:2015nja,Shibata:2011jka,Baiotti:2016qnr,Paschalidis:2016agf} and 
references therein for more comprehensive overviews). Their study is extremely challenging and requires multidimensional, multiphysics and multiscale models. General relativistic hydrodynamical (GRHD) simulations, including a detailed, microphysical description of matter and radiation, are necessary to model the merger and to produce robust predictions for a large variety of observables. 
Their comparison with observations will provide an unprecedented test for our understanding of the fundamental laws of Nature, in regimes that will never be accessible in terrestrial laboratories.

CBMs are events of primary relevance in astrophysics and fundamental physics. They are intense sources of neutrinos \cite{1989Natur.340..126E} and GWs \cite{Peters:1964zz}, and primary targets for the present generation of ground-based GW detectors. Both theoretical and observational arguments support CBMs as progenitors of sGRB \cite{1986ApJ...308L..43P,Berger:2013jza}. At the same time, they are the places where the heaviest elements in the Universe (including Gold and Uranium) are synthesized and ejected into space, via the so-called $r$-process nucleosynthesis \cite{1974ApJ...192L.145L,1989ApJ...343L..37M,1999ApJ...525L.121F}. The radioactive decay of these freshly synthesized, neutron-rich elements in the ejecta powers a peculiar EM transient, called kilonova (or macronova), hours to days after the merger \cite{1998ApJ...494L..45P,Kulkarni:2005jw}. Despite happening far away from the merger remnant, the $\gamma$-ray emission, the $r$-process nucleosynthesis, and the kilonova transient are extremely sensitive to physical processes happening where the gravitational curvature is stronger. In particular, the equation of state (EOS) of NS matter is thought to play a central role in all the emission processes, since it determines the compactness of the merging NSs, their tidal deformations, the lifetime and spin frequency of the remnant, the amount and the properties of the ejected mass.

On August, 17th 2017, the first detection of GWs from a CBM event compatible with a BNS merger marked the beginning of the multimessenger astronomy era \cite{TheLIGOScientific:2017qsa,GBM:2017lvd} and remarkably confirmed several years of intense and productive lines of research. The GW signal, GW170817, provided the link to associate a sGRB, GRB170817A \cite{Monitor:2017mdv}, and the kilonova AT2017gfo transient \cite{Pian:2017gtc,Tanvir:2017pws,Coulter:2017wya} to a CBM localized in the NGC4993 galaxy. Moreover, this single event set constraints on the EOS of NS matter and gave an independent measure of the Hubble constant \cite{Abbott:2017xzu}.

Many relevant aspects of the merger and of the emission processes are, however, not yet fully understood and relate to open questions in fundamental physics and astrophysics. The interpretation of the observational data strongly relies on the theoretical modeling of compact binary sources in GR. During the last years the unprecedented growth of computing power has allowed the development of increasingly sophisticated numerical models. The most advanced simulations in Numerical Relativity (NR) are set up using a first-principles and 
an ab-initio approach.
Neutrino radiation and magnetic fields are thought to play a key role during the merger and its aftermath. Their inclusion in detailed NR simulations, in association with microphysical, finite temperature EOS for the description of matter at and above nuclear saturation density is one of the present challenges in the field.
Reliable predictions do not only require the inclusion of all the relevant physics, but also accurate numerical schemes and numerically converging results. Robust and computationally stable discretizations are also essential to perform long-term and high-resolution simulations.

\subsection{Recent results for binary neutron star mergers}

\subsubsection{Gravitational waves and remnant properties}
A primary outcome of BNS simulations are accurate gravitational waveforms. Numerical models provide consistent and continuous GW signals for all three major phases: 
inspiral, merger, and post-merger/ring-down. As discussed earlier, key parameters encoded in the waveform are the masses and the spins of the coalescing objects, the quadrupolar tidal polarizability (depending on the NS EOS), and possibly the residual binary eccentricity. 
For the inspiral phase, state-of-the-art simulations cover at least 10
to 20 orbits before coalescence~\cite{Bernuzzi:2014owa,2017PhRvD..96h4060K}. Due to the cold
character of this phase, zero-temperature EOSs for the NS matter are
often used. High-order finite difference operators have been shown to
be key to reach high-confidence numerical results
\cite{Radice:2013hxh}. Error control on the numerical results is essential for using NR waveforms for the analysis
of data streams coming from GW detectors. State-of-art analysis of the error budget include the study of the numerical convergence, requiring multiple resolutions (4 or 5), and an estimate of the error due to the finite distance extraction of the GW signal within the computational domain. Only recently, initial conditions for rotating NSs have become available \cite{Tichy:2012rp}. They were used to follow consistently, for the first time, the inspiral phase in numerical relativity, including spin precession as well as spin-spin and spin-orbit couplings \cite{Bernuzzi:2013rza,2015PhRvD..92l4007D,2017PhRvD..95d4045D}.
The high computational costs have limited the exploration of the BNS wide parameter space. However, thanks to the large set of presently available waveforms, detailed comparisons with Analytical Relativity results (including post-Newtonian and Effective One Body approaches) are nowadays possible~\cite{Bernuzzi:2014owa,2016PhRvD..93f4082H,Hinderer:2016eia}.
Moreover, large databases of NR waveforms led to the production of purely NR-based waveform models \cite{2017CQGra..34j5014D}. These results are critical to improve the quality of semi-analytic waveforms employed in current GW data analysis. 

The merger and its aftermath are highly dynamical and non-linear phases. A systematic study of the remnant properties and of its GW emission is presently made possible by large and extensive sets of simulations in NR (e.g., \cite{2013PhRvD..88d4026H,Bauswein:2013jpa,Zappa:2017xba}), sometimes extending for tens of ms after the merger. Detailed analysis of the post-merger gravitational waveforms revealed the presence of characteristic high-frequency peaks, which will be possibly accessible by third generation GW detectors. These features in the GW spectrum are associated with properties of the nuclear EOS \cite{2013PhRvD..88d4026H,Bauswein:2015vxa,Rezzolla:2016nxn}, with the presence of a magnetised long-lived massive NS \cite{2017PhRvD..95f3016C}, with the development of one-arm spiral instabilities \cite{Paschalidis:2015mla,Radice:2016gym,Lehner:2016wjg} or convective excitation of inertial modes inside the remnant \cite{DePietri:2018tpx}. However, a firm understanding of these structures in the GW frequency spectrum is still in progress.
The remnant fate depends primarily on the masses of the colliding NS and on the nuclear EOS. The collapse timescales of metastable remnants to BHs crucially relies on physical processes that are still not fully understood, including angular momentum redistribution (possibly of magnetic origin) and neutrino cooling. A prompt collapse to a BH, expected in the case of massive enough NSs and soft EOS, results in the most luminous GW emissions at merger. For stiffer EOS or lower NS masses, differential rotation and thermal support can temporarily prevent the collapse of an object whose mass is larger than the mass of a maximally rotating NS, forming a so-called hypermassive NS (HMNS). 
Mergers producing a HMNS emit the largest amount of energy in GWs, since an intense luminosity is sustained for several dynamical timescales. If the remnant mass is below the maximum mass of a maximally rotating NS or even of a non-rotating NS, a supramassive or massive NS can form, respectively. Gravitational waveforms for a supramassive or massive NS are similar to those of a HMNS case, but weaker. The detection of this part of the spectrum is beyond the capability of the present generation of GW detectors \cite{Abbott:2017dke}. 
NR simulations take consistently into account the angular momentum emitted in GWs and in ejected mass, and predict the angular momentum of the final remnant to be $0.6 \lesssim J/M^2 \lesssim 0.85$. In the HMNS and supramassive NS cases, the super-Keplerian value of $J$ provides a large reservoir to power subsequent angular momentum ejections. Within the first dynamical timescale, this leads to the formation of a massive, thick disk around the central remnant.

\subsubsection{Matter ejection and electromagnetic counterparts}

Numerical simulations are necessary to quantitatively model matter ejection from BNS. Different ejection mechanisms result in different ejecta properties. In addition to GW emission, BNS simulations predict the ejection of both tidal (cold) and shocked (hot) \textit{dynamic ejecta}, a few ms after the GW peak \cite{2012MNRAS.426.1940K,Bauswein:2013jpa,2013PhRvD..88d4026H,Sekiguchi:2015dma,Foucart:2015gaa,Radice:2016dwd,2017PhRvD..96l4005B}. The most recent NR simulations employing microphysical, finite temperature EOS predict between $10^{-4}$ and a few $10^{-3}$ M$_{\odot}$ of dynamic ejecta, moving at $v \sim 0.3\,c$. This range covers both intrinsic variability in the binary (e.g., NS masses) and uncertainties in the nuclear EOS. However, results for the same systems from different groups agree only within a factor of a few. Differences in the treatments of the NS surface and of the floor atmosphere, as well as in the microphysical content and in the extraction of matter properties at finite radii possibly account for these discrepancies. Differently from what was previously thought, state-of-the-art simulations including weak reactions show that neutrino emission and absorption decrease the neutron richness of the equatorial ejecta and, more significantly, of the high latitude ejecta. The former can still synthesize the heaviest $r$-process elements \cite{Wanajo:2014wha}, while for the latter the formation of lanthanides can be inhibited. These results have been recently confirmed by parametric studies based on NR models \cite{Goriely:2015,2018CQGra..35c4001M}. 

Long term (100s of ms) simulations of the merger aftermath show \textit{wind ejecta} coming from the remnant and the accretion disk. These winds are powered by neutrino absorption \cite{Perego:2014fma,2017ApJ...846..114F} or, on the longer disk lifetime, by viscous processes of magnetic origin and nuclear recombination inside the disk \cite{2013MNRAS.435..502F,Just:2015ypa,Foucart:2015gaa,Siegel:2017jug,Fujibayashi:2017puw}. If a HMNS has formed, the neutrino emission is expected to be more intense and the differential rotation can also power winds of magnetic origin \cite{Siegel:2014ita}. The unbound mass can sum up to several $10^{-2}$ M$_\odot$, depending on the disk mass, and moves at $\sim 0.1~c$. Weak processes, including neutrino irradiation, decrease the neutron richness, particularly at high latitudes and for very long-lived remnants \cite{2017MNRAS.472..904L}. This implies the production of the first {\it r}-process peak in $\nu$-driven winds and viscous ejecta. In the former case, the nucleosynthesis of  lanthanides is highly suppressed, while in the latter the wider distribution of neutron richness leads to the production of possibly all r-process elements, from the first to the third peak (e.g.,\cite{Siegel:2017jug, 2017MNRAS.472..904L,2017PhRvD..96l3015W,Just:2015ypa}. Only a few simulations of the merger aftermath are presently available and only few of them were performed in a NR framework. While the study of the viscosity-driven ejecta led to a partial agreement between different groups, simulations of $\nu$- and magnetically-driven winds are still very few and additional work is required.

The different compositions and kinematic properties of the different ejecta channels have  implications  for  the  magnitude,  color,  and  duration  of  the  kilonova (see \cite{2016ARNPS..66...23F} and \cite{Metzger:2016pju} for recent reviews).  
The presence of lanthanides in the ejecta is expected to significantly increase the photon opacity due to the presence of millions of mostly unknowns absorption lines \cite{Barnes:2013wka}. Detailed radiative transfer codes provide the most accurate models for the EM emission over timescales ranging from a few hours up to a few weeks \cite{Tanaka:2013ana,2015MNRAS.450.1777K,Wollaeger:2017ahm}. Due to the uncertainties in the opacity treatment and in the geometry of the ejecta, large differences could still arise between different models. More phenomenological approaches, characterized by significantly lower computational costs and accuracy, allow a more systematic and extensive exploration of the different parameters in the ejecta properties \cite{2014MNRAS.439..757G,Rosswog:2016dhy}. Based on these models, it was suggested that, due to the absorption of neutrinos at high latitudes, the presence of lanthanide-free material (characterized by lower opacity) could result in a bluer and earlier ($\sim$ a few hours) emission, compared with the redder and dimmer emission powered by material enriched in lanthanides \cite{2014MNRAS.441.3444M,Perego:2014fma,Martin:2015hxa}.     

\subsubsection{GW170817 and its counterparts}
The detection of GW170817 and of its counterparts represents an unprecedented chance to test our understanding of BNS. The analysis of the gravitational waveform and the subsequent parameter estimation made use of large sets of approximated waveform templates. The most recent results obtained in NR could potentially improve this analysis and set more stringent constraints on the intrinsic binary parameters and on the EOS of NS matter. The interpretation of AT2017gfo as a kilonova transient revealed the need to consider at least two different components to explain both the early blue and the subsequent red emission \cite{Tanvir:2017pws,Villar:2017wcc}. More recently, the observed light curves have been traced back to the properties and to the geometry of the ejecta predicted by the most recent numerical models. This analysis has confirmed the multi-component and anisotropic character of the BNS ejecta, as well as the central role of weak interaction in setting the ejecta composition \cite{Tanaka:2017qxj,Perego:2017wtu,2017ApJ...848L..34M}. Finally, the combination of the information extracted from the GW and from the EM signal enabled the possibility to set more stringent constraints on the nuclear EOS, in a genuine multimessenger approach \cite{2017ApJ...850L..34B,Radice:2017lry}. 
Both very stiff and very soft EoSs seem to be disfavored by this kind of analysis. A similar approach has been used to show that the formation of a HMNS is the most probable outcome of this event \cite{2017ApJ...850L..19M}.

\subsection{Recent results for black hole-neutron star mergers}
Modelling of BH-NS mergers in GR shares many similarities with BNS merger modelling, but reveals also differences and peculiarities (see \cite{Shibata:2011jka} for a review). The larger ranges in masses and spins expected for stellar mass BHs significantly increase the parameter space. The large mass ratio between the colliding objects makes NR simulations more expensive, since the inspiral requires larger initial distances to cover a sufficient number of orbits, and the dissimilar lengthscales require higher resolutions. These numerical challenges have limited the number of models and waveforms that are presently available in comparison with the BNS case. If the final remnant is always represented by a spinning BH, the presence of a disk around it depends on the mass ratio and BH spin \cite{2013PhRvD..88d1503K,2013ApJ...776...47D,Foucart:2014}. Larger BHs, with moderate spins, swallow the NS during the plunge dynamical phase, while the decrease of the last stable orbit in the case of aligned, fast spinning BHs leads more easily to the tidal disruption of the NS (in the form of mass shedding) and to a massive accretion disk formation (with masses possibly in excess of 0.1 M$_\odot$). Misalignment between the orbital and the BH angular momentum induces non-trivial precessions, encoded both in the GW signal and in the amount of mass outside the BH horizon. The GW spectrum shows a characteristic cutoff frequency directly related with the orbital frequency at the NS tidal disruption or at the last stable orbit. Since this frequency decreases with increasing BH spin and with decreasing compactness, its detection could provide direct constraints on the NS matter EOS. Dynamical mass ejection from BH-NS merging system is expected to be caused by tidal torque when the NS is disrupted, and to happen predominantly along the equatorial plane. The ejected mass can be significantly larger than in the BNS case and neutrino irradiation is expected to have only a minor effect in changing the ejecta composition~\cite{Roberts:2016igt}. Similarly to the BNS case, the viscous evolution of the disk drives significant matter outflows on the disk lifetime, while the absence of a central NS reduces the emission of neutrinos and the ejection of $\nu$- or magnetically-driven winds \cite{2015PhRvD..92f4034K}. The weaker effect of neutrino processes on the dynamic ejecta results always in robust $r$-process nucleosynthesis between the second and the third $r$-process peaks, while viscous disk winds can still synthesize all {\it r}-process nuclei. The high lanthanide content in the ejecta is expected to produce a redder kilonova transient a few days after the merger \cite{2017CQGra..34o4001F}. 

\subsection{Perspectives and future developments}

A complete and dense exploration of the wide parameter space, including a consistent treatment of NS and BH spins and of their evolution, is one of the major challenges in the study of CBMs in NR. The extraction of waveforms from long inspiral simulations at high resolution, employing accurate high-order numerical schemes, is necessary to build a robust NR waveform database. These templates, in combination with more analytical approaches, can guide the construction of complete and coherent semi-analytical waveforms for the GW data analysis, spanning many orbits from the inspiral to the actual coalescence. The large variety of properties in the colliding system potentially translates to a broad distribution of properties for the remnant and for the ejecta. Long term simulations of the merger and of its aftermath are the necessary tool to provide a complete and accurate description of the ejecta. 
The inclusion of the physics necessary to model the remnant and the matter ejection is still at its outset. In particular, the consistent inclusion of neutrino radiation and magnetic field evolution in NR model is extremely challenging. Leakage scheme for neutrino radiation are presently implemented in many CBM models (see, e.g., the introduction section of \cite{2016ApJS..223...22P} for a detailed discussion). They provide a robust and physically motivated treatment for neutrino cooling. However, they are too inaccurate to model long term evolution and neutrino absorption in optically thin conditions. State-of-the-art simulations include gray moment scheme \cite{Radice:2016dwd,2016PhRvD..94l3016F}. Nevertheless, the significant dependence of neutrino cross-sections on particle energy requires spectral schemes. Large velocity gradients inside the computational domain make the accurate transformation of the neutrino energy spectrum between different observers and its transport highly non-trivial. Moreover, the application of moment schemes for colliding rays in free streaming conditions leads to closure artifacts in the funnel above the remnant. Monte Carlo radiation schemes represent an appealing alternative, but their computational cost is still far beyond our present capabilities. The usage of Monte Carlo techniques to provide more physical closures in moment scheme seems a more viable approach \cite{Foucart:2018}. 
Neutrino masses have a potentially high impact on the propagation and flavor evolution of neutrinos from CBMs. New classes of neutrino oscillations, including the so-called matter neutrino resonances, can appear above the remnant of BNS and BH-NS mergers and influence the ejecta nucleosynthesis \cite{2016PhRvD..93d5021M,2017PhRvD..96d3001T}. However, numerical modelling of these phenomena is still at its dawn \cite{2016PhRvD..94j5006Z,2017PhRvD..95b3011F,2017PhRvD..96l3015W}.

General relativistic magneto-hydrodynamics codes are presently available and they have started to access the spatial scales necessary to consistently resolve the magneto-rotational instabilities (MRIs, \cite{1960PNAS...46..253C}). The latter appear in the presence of differential rotation and locally amplify the magnetic field. However, global simulations obtained with unprecedented spatial resolution (of the order of 12.5 m) do not show convergence yet, proving that we are still not able to resolve the most relevant scales for magnetic field amplification in a self-consistent way \cite{Kiuchi:2017zzg}. In the past, sub-grid models have been used as alternative approaches to ab-initio treatments \cite{2015PhRvD..92d4045P,2016CQGra..33p4001E,2017PhRvD..95f3016C}. More recently, two different formulations of effective MHD-turbulent viscosity in General Relativity have been proposed and used in both global BNS and long term aftermath simulations \cite{Radice:2017zta,Shibata:2017jyf}. Moreover, the role of bulk viscosity in the post-merger phase
has started being investigated \cite{Alford:2017rxf}.

Accurate modelling of the GW and EM signals for CBMs are key to set tight constraints on the EOS of NS matter, which still represents one of the major sources of uncertainty in both fundamental physics and in numerical models. If the connection between CBMs and short GRBs will be confirmed by future detections, the knowledge of the remnant fate, as well as of the environment around it, will be crucial to address the problem of short GRB engines, including jet formation, collimation, and break-out. The accurate modelling of small-scale magnetic field amplification, as well as of heat redistribution due to neutrino transport, is key to predict the lifetime of the remnant for cases of astrophysical interest. In fact, the presence of highly rotating and magnetized massive NSs \cite{Perego:2017fho,2017PhRvD..95f3016C}, or of fast spinning BHs, is anticipated to play an essential role in solving the puzzle of the short GRB central engine.

The high computational costs required by long-term, high-resolution, numerically accurate and multiphysics models of CBMs point to the need of developing a new generation of numerical schemes and codes for the new generations of large supercomputers. These codes will need, for example, to improve scalability and to employ more heavily vectorization in the hybrid (shared \& distributed) parallelization paradigm. This is perhaps the greatest challenge in the NR field for the years to come.

\section{Electromagnetic Follow-up of Gravitational Wave Mergers }\label{Sec:EMfollowup}
\vspace{3mm}
{\it Contributor:} A.~Horesh
\vspace{3mm}

The year 2017 will be remembered as the year in which extraordinary achievements in observational astrophysics have been made. On 2017, August 17, the LIGO and Virgo detectors detected for the first time GWs from the merger of two NSs, dubbed GW\,170817. Adding to the excitement was the detection of gamma-ray emission only two seconds after the merger event, by the {\it Fermi} satellite. The sensational discovery of a GW signal with a coincident EM emission led to one of the most comprehensive observational campaigns worldwide. A few hours after the GW detection, the LIGO and Virgo detectors managed to pinpoint the position of the GW event to an error circle of $34$ sq. degrees in size (see Fig.~\ref{fig:gw17} below). This area was small enough so that an international teams of astronomers encompassing more than a hundred instruments around the world on ground and in space could conduct an efficient search for EM counterparts. 

Roughly $11$ hours after the GW event, an optical counterpart was announced by the 1M2M (`Swope') project team \cite{Coulter:2017wya,Kilpatrick:2017mhz}. Many other teams around the world independently identified and observed the counterpart in parallel to the SWOPE team discovery. The optical detection of the counterpart pinpointed the source to a precise location in the galaxy NGC4993 at a distance of $40$\,Mpc, (which is also consistent with the distance estimate from the GW measurements; see Sec.~\ref{Sec:LVC}). Overall, the counterpart (for which we will use the official IAU name AT\,2017gfo) was detected across the spectrum with detections in the ultraviolet, optical, infrared (e.g., \cite{Coulter:2017wya,Kilpatrick:2017mhz,Drout:2017ijr,Smartt:2017fuw,Tanvir:2017pws,Chornock:2017sdf,Arcavi:2017xiz,Shappee:2017zly,Kasliwal:2017ngb,Andreoni:2017ppd,McCully:2017lgx,Pian:2017gtc,Cowperthwaite:2017dyu}) and later on also in the X-ray~(e.g., \cite{Troja:2017nqp,Haggard:2017qne,Margutti:2017cjl}) and in the radio~(e.g., \cite{Hallinan:2017woc,Mooley:2017enz}).

Below we briefly summarize the observational picture with respect to theoretical predictions.

\subsection{The High-energy Counterpart}

For many years it was hypothesized that short gamma-ray bursts (GRBs) that are by now routinely observed, are originating from NS mergers (e.g., \cite{Narayan:1992iy}). The theoretical model includes the launching of a relativistic jet during the short period of accretion onto the merger remnant. This energetic jet is believed to be responsible for the prompt gamma-ray emission. In addition, the interaction of the jet with the interstellar medium will produce an afterglow emission in the optical, X-ray and also radio wavelengths.

The short GRB ($T_{90} = 2.0 \pm 0.5$\,sec) that has been discovered about $1.7$ second after the GW170817 merger event \cite{Goldstein:2017mmi} was irregular compared to other short GRBs observed so far. Most notable is the low equivalent isotropic energy $E_{iso} \approx 5 \times 10^{46}$\,erg (in the $10 - 1000$\,kev band), which is $\sim 4$ orders of magnitude lower than a typical short GRB energy. In addition to the initial detection by {\it Fermi-GBM}, there were observations made by the Integral satellite. Integral also detected the short GRB with a flux of $\sim 1.4 \times 10^{-7}$\,erg\,cm$^{-2}$ (at $75 - 2000$\,kev~\cite{Savchenko:2017ffs}).

Following the initial detection and once the candidate counterpart had been localized, X-ray observations were made within less than a day of the merger event. The initial X-ray observations by both the Neil Gehrels Swift Observatory and by the Nuclear Spectroscopic Telescope Array ({\it NuSTAR}) resulted in null-detections. However, later on, on day $\sim 9$\, after the merger, the {\it Chandra} telescope detected X-ray emission from the AT\,2017gfo for the first time \cite{Troja:2017nqp} with an initial isotropic luminosity of $\approx 9 \times 10^{38}$\,erg\,s$^{-1}$. 
In the following days, the X-ray luminosity appeared to continue to slowly rise \cite{Troja:2017nqp,Haggard:2017qne}. Late-time observations ($109$\,days after the merger) by {\it Chandra} still show that the X-ray emission continues to slowly rise \cite{Margutti:2018xqd}. 

In order to reconcile the relatively faint prompt gamma-ray emission and the late onset of the X-ray emission, it was suggested that AT\,2017gfo was a regular short GRB but one that is observed slightly off the main axis by $\sim 10\,$degrees of the jet~\cite{Troja:2017nqp}. However, this interpretation has yet to be tested against observations in other wavelengths, such as the radio (see below).

\subsection{The Optical and Infrared Counterpart}

Over four decades ago, a prediction was made by \cite{1974ApJ...192L.145L,Eichler:1989} that neutron-rich material can be tidally ejected in a NS merger. About a decade later numerical simulations also showed that NS mergers will exhibit such mass ejections, where the mass ejected is expected to be in the mass range $10^{-5} - 10^{-2}$\,M$_{\odot}$ with velocities of $0.1 - 0.3$\,c (e.g., \cite{Rosswog:1998hy}). It was also predicted that heavy elements would form in the neutron rich ejected material via r-processes. As discussed in detail in the previous Section~\ref{Sec:NumericalRelativity}, additional material is also expected to be ejected by accretion disk winds and from the interface of the two merging stars (the latter material will be ejected mostly in the polar direction). In general, each ejected mass component may have a different neutron fraction leading to a different r-process element composition.

As the ejecta from the NS merger is radioactive, it can power transient emission, as proposed in Ref.~\cite{Li:1998bw}. The initial prediction was that the emission will be supernova-like and will be blue and peaking on a one-day timescale. Later on, more detailed calculations by \cite{2010MNRAS.406.2650M} showed that the peak of the emission will be weaker at a level of $10^{41}$\,erg\,s$^{-1}$. The next theoretical developments were made by \cite{Barnes:2013wka,Tanaka:2013ana,2014MNRAS.439..757G} who for the first time calculated the effect of the heavy r-process element opacities on the emission. They found that due to very high opacities, the emission is expected to be even weaker, with its spectral peak now in the infrared (instead of the optical) and with the peak timescale being delayed. Since the various ejecta mass components may have different compositions, they therefore may have different opacities. Thus in principle, one component may form a blue short-lived emission (the so-called `blue component') while another may emit in the infrared with week-long timescales (the so-called `red component').

The optical emission from the AT\,2017gfo peaked at a faint absolute magnitude of $M_{V} \approx -16$ in one day and began to rapidly decline, while the infrared emisson had a somewhat fainter and later (at approximately 2 days) peak compared to the optical emission. The overall observed brightness of the source in the optical and infrared bands and its evolution over time have shown in general an astonishing agreement with the predicted properties of such an event. 

The multiple photometric and spectroscopic measurements sets obtained by the various groups paints the following picture: At early times, at about $0.5$ days after the merger, the ejecta temperature was high at $T\sim 11,000$\,K (e.g. \cite{Shappee:2017zly}). A day later the spectral peak was at $\approx 6000$\,Angstrom, and the temperature decreased to $T\approx 5000$\,K~~\cite{Pian:2017gtc}. From this point onwards the spectral peak quickly moved into the infrared band. The combined optical and IR emission and its evolution were found to be consistent with an energy source powered by the radioactive decay of r-process elements (e.g. \cite{Drout:2017ijr,Smartt:2017fuw,Tanvir:2017pws}). Based on both the photometric and spectroscopic analysis an estimate of the ejecta mass and its velocity were found to be ${\rm M}_{ej} \approx 0.02 - 0.05$\,M$_{\odot}$ with a velocity in the range $0.1 - 0.3$\,c~\cite{Chornock:2017sdf,Arcavi:2017xiz,Smartt:2017fuw,Shappee:2017zly,Andreoni:2017ppd,Drout:2017ijr,McCully:2017lgx,Pian:2017gtc,Cowperthwaite:2017dyu}. 

One of the main claims with regards to this event is that the observations provide evidence for the formation of r-process elements. This conclusion is mainly driven based on the evolution of the infrared emission~\cite{Drout:2017ijr,Kasliwal:2017ngb}. \cite{Tanvir:2017pws}, for example, obtained late-time IR measurements using the Hubble Space Telescope and argue that the slower evolution of the IR emission compared to the optical require high opacity r-process heavy elements with atomic number $> 195$. The infrared spectrum shows broad features which are presumably comprised of blended r-process elements \cite{Tanvir:2017pws}. These broad spectral features were compared to existing model predictions and show a general agreement, albeit there are still inconsistencies that need to be explained (e.g., \cite{Kasen:2017sxr,Metzger:2016pju,Kasliwal:2017ngb}). There is still an ongoing debate about the composition of the heavier elements. Ref.~\cite{Smartt:2017fuw} claims that the observed infrared spectral features can be matched with CS and Te lines. Others estimate the overall fraction of the lanthanides in the ejecta and find it to be in the range ${\rm X}_{lan} \approx 10^{-4}  - 10^{-2}$. It seems that at early times (up to 3-5 days), the ejecta that dominates the emission has very low lanthanide fraction and that there are discrepancies between the predicted and observed optical light curves (e.g., \cite{Arcavi:2017xiz,McCully:2017lgx}. Several works (e.g., \cite{Kasen:2017sxr,Metzger:2016pju,Cowperthwaite:2017dyu,Nicholl:2017ahq,Chornock:2017sdf,Kasliwal:2017ngb,Pian:2017gtc,Smartt:2017fuw}) explain the early {\it vs.} late-time behaviour of the emission by having an ejecta with two components (as also predicted in the literature). The first component is the ``blue'' kilonova, with a high electron fraction and thus a low fraction of heavy elements, which dominates the optical emission early only. A ``red'' kilonova is the second component that is comprised by heavy r-process elements and that produces the slow IR evolution and the broad spectral features observed at late-times. Still there are some claims that even at late times, the emission can be originating from an ejecta with only low-mass elements~\cite{Waxman:2017sqv}. 

\subsection{The Radio Counterpart}

In addition to the radio afterglow emission on the short time scale (days), radio emission is also expected on longer time scales (month to years). The latter is not a result of the relativistic jet but rather originating from the interaction of the slower dynamical ejecta with the instersteIlar medium (ISM)~\cite{Nakar:2011cw}. This emission is expected to be rather weak where the strength of the peak emission depends on the velocity of the dynamical ejecta, the ISM density and on some microphysical parameters. While in the past, radio afterglows of short GRBs (not accompanied by GWs) were detected (e.g., \cite{Fong:2015oha}), long-term radio emission was never observed until GW\,170817 (including in the previous cases where kilonova candidates were discovered~\cite{Horesh:2016dah,Fong:2016orv}). 

Similar to any other wavelength, radio observations were undertaken within the day of the GW170817 merger discovery and the following days. The early time observations performed by The Jansky Very Large Array (VLA), The Australian Compact Array (ATCA), the Giant Meterwave Radio Telescope (GMRT) and the Atacama Large Millimeter Array (ALMA), all resulted in null-detections \cite{Hallinan:2017woc,Alexander:2017aly,Kim:2017skw}. Late-time VLA observations, however, finally revealed a radio counterpart $\approx 16$\,days after the merger \cite{Hallinan:2017woc}. The initial radio emission was weak at a level of a couple of tens of $\mu$Jy only (at both $3$ and $6$\,GHz). Follow-up ATCA observations confirmed the radio detection.  Upon detection, a long-term radio monitoring campaign was initiated, and the results are reported in Ref.~~\cite{Mooley:2017enz}. They report that the radio emission still continued to rise at $> 100$\,days after the merger. The multiple frequency observations also show that the radio emission is optically thin with a spectral index of $\alpha = -0.6$.  

\cite{Hallinan:2017woc} compared the radio observations to several predictions including an on-axis jet, a slightly off-axis jet, and being completely off-axis. In addition, they included in their comparison a model in which the jet forms a hot wide-angled mildly relativistic cocoon. This cocoon formed as the jet is working its way out of the dynamical ejecta, may also lead to gamma-ray, X-ray, and radio emission, but with different characteristics than the emission formed by either the highly relativistic jet or the slower dynamical ejecta. In fact, \cite{Hallinan:2017woc} find that both an on-axis jet model, or even a slightly off-axis jet one, are expected to produce bright radio emission in the first few days after the merger which by day 16 should start fading, a prediction which does not match the rising observed radio source. The model that is the most consistent with current data is the cocoon model with either a choked or successful or `structured' jet (see e.g., \cite{Lamb:2017ych,Margutti:2018xqd}). In fact, the choked-jet cocoon model may also explain the relatively low energy of the gamma ray emission \cite{Gottlieb:2017pju}.

If both the late-time radio and X-ray emission originate from interaction of the material (whether it is the dynamical ejecta or a cocoon) with the ISM, one can test whether the observed emission in both wavelengths is indeed connected as expected. For now, it seems that the observed X-ray emission fits the prediction which is based on extrapolating the observed radio emission into higher energies. Both the X-rays and the radio emission have also roughly the same spectral slope. This suggests that there is no additional power source in play at this time. 

\subsection{Many Open Questions}

While vast amounts of data have been collected for this amazing merger event, and while a flood of papers report many types of analysis and conclusions (by no means are we attempting to cover all of them here), there are still many open questions remaining. For example, is this event connected to short GRBs or do we still have to prove this connection? Which r-process elements form and at what time? Do the observations really require producing heavy r-process elements or can they all be explained by lighter-element components? How many components does the ejecta have and which one dominates the emission and when? Are there any other emission power sources in play such as a magnetar (even if at short time) ? Are we really seeing a cocoon with a choked or successful jet or is there some other scenario that can explain the radio emission combined with all the other evidence? 

As scientists around the world are still working on analyzing all the data in hand for this event and are also still collecting new data, they are also gearing up for the future, and preparing for the next year when the LIGO and Virgo detectors switch back on. At this time, hopefully more events with EM signatures will be discovered and more answers (and surely more new questions) will present themselves.

\section{X-ray and gamma-ray binaries } \label{Sec:XrayGammaRay}
\vspace{-3mm}
{\it Contributor:} M. ~Chernyakova
\vspace{3mm}

The population of Galactic X-ray sources  above 2 keV is dominated by the X-ray binaries, see e.g. \cite{Grimm2002}.
A typical X-ray binary contains either a NS or a BH accreting material from a companion star. Due to angular momentum in the system, accreted material does not flow directly onto the compact object,  forming a differentially rotating disk around the BH known as an accretion disk .
X-ray binaries can be further divided into two different classes, regardless  the  nature  of  the  compact  object,  according  to  the
mass of the companion star: high-mass X-ray binaries and low-mass X-ray binaries. The secondary of low-mass X-ray binary systems is a low-mass star, which transfers matter by Roche-lobe overflow. High-mass X-ray binaries comprise a compact object orbiting a massive OB class star. High-mass X-ray binaries systems are a strong X-ray emitter via the accretion of matter from the OB companion. 
At the moment 114 high-mass X-ray binaries~\cite{Liu2006} and 187 low-mass X-ray binaries~\cite{Liu2007} are known. 

Black hole X-ray binaries are interacting binary systems where X-rays are produced by material accreting from a secondary companion star onto a BH primary \cite{1973A&A....24..337S}. While some material accretes onto the BH, a portion of this inward falling material may also be removed from the system via an outflow in the form of a relativistic plasma jet or an accretion disk  wind, see e.g. \cite{BHrev2006} for a review. Currently, the known Galactic BH X-ray population is made up of 19 dynamically confirmed BHs, and 60 BH candidates \cite{BHcat2016}. The vast majority of these Galactic BH X-ray objects are low-mass X-ray binaries. Most of these systems are transient, cycling between periods of quiescence and outburst. This behaviour is associated with changing geometries of mass inflow and outflow, e.g. \cite{BHrev2006}.

At higher energies, however the situation is drastically different. While current Cherenkov telescopes have detected around  80 Galactic sources (see the TeVCat catalogue~\cite{TeVCat}), only 7 binary systems are regularly observed at TeV energies. Properties of PSR B1259-63, LS 5039, LSI +61 303, HESS J0632+057 and 1FGL J1018.6-5856 are reviewed in \cite{dubus_review13}. Since 2013 two more Galactic binaries have been discovered at TeV sky,  PSR J2032+4127 \cite{PSRJ2032_ATel_TeV} and HESS J1832-093 \cite{2016MNRAS.457.1753E}, but still the number of binaries observed at TeV sky  is extremely small, and the reason why these systems are able to accelerate particles so efficiently is not known yet. These systems are called gamma-ray-loud binaries (GRLB), as the peak of their spectral energy distribution lies at GeV - TeV energy range.

All GRLB systems host compact objects orbiting around massive young star of O or Be spectral type. This allows
to suggest, that the observed $\gamma$-ray emission is produced in the result of interaction of the relativistic outflow from
the compact object with the non-relativistic wind and/or radiation field of the companion massive star. However,
neither the nature of the compact object (BH or NS?) nor the geometry (isotropic or anisotropic?)
of relativistic wind from the compact object are known in the most cases. Only in PSR B1259-63 and PSR J2032+4127 systems the compact object is known to be a young rotation powered pulsar which produces relativistic pulsar wind. Interaction of the pulsar wind with the wind of the Be star leads to the huge GeV flare, during which up to 80\% of the spin-down luminosity is released \cite{Abdo2011,Chernyakova2015}.

In all other cases the source of the high-energy activity of GRLBs is uncertain. It can be either accretion onto or dissipation of rotation energy of the compact object. In these systems the orbital period is much shorter than in PSR B1259-63 and PSR J2032+4127, and the compact object spend most of the time in the dense wind of the companion star. The optical depth of the wind to free-free absorption is big enough to suppress most of the radio emission within the orbit, including the pulsed signal of the rotating NS~\cite{zdz10}, making impossible direct detection of the possible pulsar.

In Ref.~\cite{Massi2017} authors tried to deduce the nature of the compact source in LSI +61 303 studying the relation between X-ray luminosity
and the photon index of its X-ray spectrum. It turned out that existing X-ray observations of the system follows the same anti-correlation trend as BH X-ray binaries \cite{2015PKAS...30..565Y}. The hypothesis on microquasar nature of LSI +61 303 allowed to explain the observed radio morphology \cite{BoschRamon:2004se} and explains the observed superorbital period as a beat frequency between the orbital and jet-precession periods~\cite{Massi2016}. At the same time it was shown that the model in which the compact source is a pulsar allowed
naturally explain the keV-TeV spectrum of LSI +61 303~\cite{zdz10}. Authors argued, that the radio source has a complex, varying
morphology, and the jet emission is unlikely to dominate the spectrum through the whole orbit. Within this model
the superorbital period of the source is explained as timescale of the gradual build-up and decay of the disk of the
Be star. This hypothesis is also supported by the optical observations confirming the superorbital variability of the Be-star disk \cite{fortuny15}.
A number of multi-wavelength campaigns are currently ongoing aiming to resolve the nature of these peculiar systems.

GeV observations revealed a few more binaries visible up to few GeV. Among them are Cyg X-1 \cite{Zanin2016,Zdz_CygX1_2017} and Cyg X-3 \cite{2016ATel.9502....1C}. However, contrary to the GRLBs
described above these systems are transients and seen only during the
ares, or, in the case of Cyg X-1, during the hard state. In addition to this the peak of the spectral
energy distribution of these systems happens at much lower energies than in the case of binaries visible at TeV energies.

However contrary to the GRLBs described above these systems are transients and seen only during the flares, or ,in the case of  Cyg X-1, during the hard state. In addition to this the peak of the spectral energy distribution of these system  happens at much lower energies than in the case of binaries visible at TeV energies. From these observations it seems that wind collision can accelerate particles more efficient that the accretion, but more sensitive observations are needed to prove it and understand the reason. Hopefully CTA \cite{CTA} observations will be able to shed light on the details of the physical processes taking place in these systems.

\section{Supermassive black hole binaries in the cores of AGN} \label{Sec:BHBandAGN}
\vspace{-3mm}
{\it Contributors:} E.~Bon, E.~M.~Rossi, A.~Sesana, A.~Stamerra
\vspace{3mm}

Following mergers it is expected that the galaxy cores should eventually end up close to each other. In this process, the term dual SMBHs refers to the stage where the two embedded SMBHs are still widely separated (gravitationally-bound to the surrounding gas and stars and not to one another), while SMBBHs denotes the evolutionary stage where they are gravitationally bounded in the close-orbiting system of SMBHs. 

Bound SMBBHs on centi-pc scales are the most relevant to GW emission (and therefore to this roadmap). In the approximation of circular orbits, these systems emit GWs at  twice their orbital frequency, i.e. $f_{\rm GW}=2/P_{\rm orb} \gtrsim 1$ nHz. As we discuss in the next Section, this is the frequency at which PTAs are most sensitive \cite{Lentati:2015qwp,ShannonEtAl:2015,VerbiestEtAl:2016,Arzoumanian:2017puf}, having a concrete chance to make a direct detection of these systems within the next decade \cite{Siemens:2013zla,rsg15,tve+16}. The presence of a SMBBH with an orbital period of several years introduces a natural timescale in the system. In fact, numerical simulations of SMBBHs embedded in circumbinary accretion disks display consistent periodic behaviors of the gas leaking into the cavity \cite{1994ApJ...421..651A,2008ApJ...672...83M,2012A&A...545A.127R,2012ApJ...749..118S,2013MNRAS.436.2997D,Farris:2014iga}. This led to the notion that SMBBHs might be detected via periodicity of their lightcurve. However, the diffusion time of the gas within the mini-disks surrounding the two holes is generally longer than the binary period, and it is not at all clear that the periodic supply of gas through the cavity will in turn result in periodicities in the binary lightcurves or their spectra \cite{srr+12}. Even though the cross-gap accretion rate is generally periodic, only light generated at the accretion stream/minidisk or outgoing stream/circumbinary disk shocks is guaranteed to follow this modulation \cite{2014ApJ...785..115R,Farris:2014iga}. Since some periodicity seems inevitable, we focus on this signature in the following discussion. We notice however, that several spectral signatures of close SMBBHs have also been proposed, including a dimming at UV wavelengths \cite{tmh12}, double K$\alpha$ lines \cite{srr+12}, notches in the spectral continuum \cite{2014ApJ...785..115R}, steepening of the thermal spectrum compared to the standard thin disk model \cite{Ryan:2016vcm}.

Among many mechanisms proposed to explain the emission variability of active galactic nuclei (AGN) besides outflows, jet precession, disk precession, disk warping, spiral arms, flares, and other kinds of accretion disk instabilities, one of the most intriguing possibilities involves the existence of a SMBBH system in their cores~\cite{Komossa:2003wz,Bogdanovic:2007bu,Bogdanovic:2014cua,Graham:2015gma,Nguyen:2016qnk}, and the tidal disruption event (see Ref.~\cite{2015JHEAp...7..148K} and references therein). 

The light variability emitted from AGN was tracked much before they were recognized as active galaxies. There are light curves showing AGN variability of over 100 years of observations, see for example \cite{Guo:2014ora,Oknyanskij2016}, with variability timescale of over decades. In fact, many AGN show variability of different time scales depending on time scales of processes that drive the variability, such as speeds at which variations propagate, for example the speed of light c $\sim$ 3 $\cdot$ 10$^5$ km/s or the speed of sound $v_s\sim(kT/m)^{1/2}$, where $c > v_{orb} \gg v_s$, or the time scale of the orbital motion $v_{orb}\sim(GM/R)^{1/2}$. The shortest timescale corresponds to the light crossing timescale, on which the reverberation mapping campaigns  are based on. Orbital timescales are longer~\cite{Hagai2013},  and are dominant in the case of SMBBH systems. Recently, possible connection of AGN variability time scales and orbital radius is presented in~\cite{Bon:2018ibp}, indicating that variability time scales may not be random, and that they might correspond to the orbiting time scales.

To identify possible candidates, we search for periodic variations in their light and radial velocity curves. We expect that periodic variability should correspond to orbital motion exclusively, while the other processes could produce only quasi periodic signals. Unfortunately, AGN were identified only about 70 years ago, so observing records are long for few decades only, which is of order of orbiting time scales, and therefore not long enough to trace many orbits in historical light curves, not to mention the radial velocity curves \cite{Begelman:1980vb,Popovic:2011uy,Bon:2012,Bon:2016-NGC5548,Li:2016hcm}, which are harder to obtain because of their faintness, and even shorter records. Therefore, for AGN it is very hard to prove that the signal is actually periodic, especially if they are compared to the red noise like variability curves, that in fact, AGN light curves are very similar to \cite{Simm:2015ova,Vaughan:2016pyf}. Therefore, standard methods like Fourier and Lomb-Scargle \cite{Scargle:1982bw} may show peaks of high looking significance but the derived p-value may not be valid \cite{Vaughan:2016pyf, Bon:2016-NGC5548, Bon:2017tgh}.  

Keeping in mind the aforementioned difficulties, a number of AGN have been proposed to display a significant periodical variability in their light curves~\cite{Sillanpaa:1988zz,Graham:2015gma,Graham:2015tba,Charisi:2016fqw,Liu:2016msr,Bhatta:2016dsn,Kovacevic:2017sjl}. Notable examples include the  blazar OJ287 (11.5 yr period, \cite{Sillanpaa:1988zz, Bhatta:2016dsn}, the quasar PG1302-102 (6 yr period, \cite{Graham:2015gma}), the blazar PG 1553+113 (2 yr period, \cite{Ackermann:2015wda}). Among those candidates, there are only few that could indicate periodic light and radial velocity curves in the same time \cite{Bon:2016-NGC5548,Bon:2012,Li:2016hcm,Li:2017eqf}, which therefore could be recognized as SMBBH candidates, like NGC4151 with a 15.9 year periodicity \cite{Bon:2012}, NGC5548 with a $\approx$15 year periodicity \cite{Li:2016hcm,Bon:2016-NGC5548}, and Akn120 with a $\approx$20 year periodicity \cite{Li:2017eqf}. We note that simulating the emission from such systems is very complex \cite{Cuadra:2008xn,Farris:2014iga,Kimitake:2007fs,Tang:2017eiz,Ryan:2016vcm}, especially for the eccentric high-mass ratio systems \cite{Bogdanovic:2007bu,2017MNRAS.466.1170M,2013PASJ...65...86H}.  

Among AGN, the class of blazars is dominated by the emission from the jet due to beaming effects caused by the small angle of sight. Blazars show high variability in all wavebands from radio to gamma-rays. High energy emission is likely originated in small jet scales and therefore can be modulated by the orbital motion of the SMBH binary system. The modulation can be funneled through variations on the accretion rate induced by the perturbation on the disk by the companion SMBH, as suggested for OJ287, or in helical paths induced by precession \cite{Sobacchi:2016yez} as suggested to explain the clear signature of a periodic modulation on the gamma-ray blazar PG 1553+113 \cite{Ackermann:2015wda}. More complex interplay among the different components in the jet, emitting at different wavelength is possible in the framework of a binary SMBH system (\cite{Rieger:2004ay}). We mention that quasi-periodicities in the jet emission can be induced by intrinsic oscillatory disk instabilities that can mimic periodical behaviour. The continuous gamma-ray monitoring of blazars by the {\em Fermi}-LAT satellite is providing new possible candidates showing periodic or quasi-periodic emission (see e.g., \cite{Prokhorov:2017amk}, \cite{Sandrinelli:2015ijk}). Similarly, a 14 year periodicity is found in the X-ray and optical variations of 3C 273, while in OJ 287, the optical variability may not always be consistent with radio. Even, a detection of periodic variations of spinning jet could indicate presence of SMBBH \cite{Kun:2014tva}.

The time domain window has only been opened in the past decade with dedicated surveys such as CRTS, PTF and Pan-STARRS, and already produced several SMBBH candidates. Many of them, however, have been already severely constrained by PTA upper limits on the stochastic GW background they would imply \cite{Sesana:2017lnk}. This confirms our poor understanding of SMBBH appearance. More sophisticated numerical simulations, including 3-D grids, radiative transport schemes, feedback from the accreting sources, etc., are needed to better understand the emission properties and peculiar signatures of SMBBHs. Under their guidance, future candidates should be then proposed based on systematic cross check of variability coupled with peculiar spectral features.

\subsection{Modeling electromagnetic signatures of merging SMBBHs}
Although the identification of compact SMBBHs might have an important impact on GW observations with PTAs in the near future, looking further ahead, LISA is expected to detect tens to hundreds of coalescing SMBBHs throughout the Universe. Electromagnetic observations related to the merger of SMBBHs are important both for cosmology and the astrophysics of galactic nuclei. Pinning down the host galaxy of the merger will allow us to combine redshift and distance from gravitational waves to constrain cosmological parameters (\cite{Tamanini:2016zlh}, see next Section) and to study the large scale galactic environment of merging SMBBHs, adding to our understanding to the process galaxy formation. On the other hand, electromagnetic observations will give us access to the properties of matter in the relative close environment of a merger and to the gas an stars (hydro)-dynamics as they adjust in response to the merger. 

Given the importance of such identification, there has been an extensive effort to predict observable electromagnetic signatures that can occur in nearly coincidence  with the event (``prompt signals'') or afterwards (``afterglows''). In the following few examples will be given; notably, some of them also inspired recent models for possible electromagnetic counterparts to SOBBH mergers, of the kind detected by LIGO and VIRGO. As mentioned previously, SMBBH mergers, especially at high redshifts, can happen in a gaseous environment that provides each SMBH with a ``minidisc,'' fed through streams leaking from a circumbinary disc. Those minidiscs are likely to be retained even after the orbital decay due to gravitational radiation dominates, providing distinctive modulation of the emerging luminosity as the binary spirals in \cite{2018MNRAS.476.2249T,2018ApJ...853L..17B}. During the final orbits, the surviving gas between the black holes gets squeezed possibly producing super-Eddington outflows as discussed in \cite{armitage,2016MNRAS.457..939C}, but see \cite{2017MNRAS.468L..50F}. Full GR simulations have also been employed to study the possible formation of precessing jets during the inspiral and merger, in the attempt of identifying distinctive signatures \cite{Palenzuela:2010nf,2012ApJ...749L..32M,2014PhRvD..90j4030G}. General relativity predicts that a newly formed black hole suffers a recoil because GWs carry away a non-zero linear momentum (e.g., \cite{1983MNRAS.203.1049F,Campanelli:2007ew}). This recoil affects the circumbinary disc, bound to the black hole: a kick is imparted that shocks the gas producing a slowly rising, $\sim 10$ yr lasting afterglows \cite{2008ApJ...676L...5L,2008ApJ...684..835S,2008ApJ...682..758S,2009PhRvD..80b4012M,rossi}.  In stellar mass black hole merger, similar phenomena but on much shorter timescales can occur \cite{perna,dk17}. Contrary to the SMBH case, however, providing a gas rich environment for the merger is a challenge. A possible venue involves cold relic discs, formed as a result of weak supernovae, where accretion is suppressed until either the ``squeezing'' or the ``kicking'' heat them up again in the same configuration envisaged for SMBHs.

As already mentioned, the next theoretical challenge for these dynamical models is to predict realistic lightcurves and spectra, which will require non-trivial radiative transfer calculations (see, e.g., \cite{2016ApJ...819...48S}). With solid predictions in hand, appropriate strategies can be devised to coordinate electromagnetic follow-ups, to take full advantage of multimessenger astronomy in the LISA era.

\section{Cosmology and cosmography with gravitational waves}\label{Sec:cosmography}
\vspace{-3mm}
{\it Contributors:} C.~Caprini, G.~Nardini, N.~Tamanini
\vspace{3mm}

The recent direct measurement of GWs by the Earth-based interferometers LIGO and Virgo opened up a new observational window onto the universe and, right from the first detection, led to the discovery of a new, unexpected source: fairly massive stellar-origin BBHs. This demonstrates the great potential of GW observations to improve our knowledge of the universe. Concerning cosmology, it is beyond doubt that the possible detection of a stochastic GW background (SGWB) from the early universe would be revolutionary from this point of view: similar to the discovery of the CMB, which constitutes a milestone in our understanding of the universe, rich of consequences that we are still investigating. Furthermore, the plethora of new GW detections expected in the next decades by both Earth and space-based interferometers will not only deliver fundamental information on the emitting astrophysical sources, but it will also bring complementary and independent data, with respect to standard EM observations, that can be used for cosmological purposes. In particular, by means of GW detections, we can probe the history of the universe both at early and late times, shedding new light on some of the most elusive cosmological mysteries, such as dark energy, DM and the origin of cosmic inhomogeneities. In this Section, we overview how the observation of GWs can enhance our knowledge of the history of the universe.

\subsection{Standard sirens as a probe of the late universe}
Within the theory of GR, a binary system composed by two compact astrophysical objects orbiting around each other, emits a GW signal with the two polarizations~\cite{Maggiore:1900zz}
\begin{eqnarray}
h_\times(t) & = &\frac{4}{d_L(z)} \left( \frac{G \mathcal{M}_c(z)}{c^2} \right)^{5/3} \left( \frac{\pi f(t)}{c} \right)^{2/3}  \sin[\Phi(t)] \cos\iota\,, \nonumber \\
& & \label{eq:hx}\\
 h_+(t) & = & h_\times(t) \frac{1+\cos^2 \iota}{2 \cos\iota} \cot[\Phi(t)] ~,
 \label{eq:h+}
\end{eqnarray}
where $h_{\times,+}(t)$ are the GW strains in the transverse-traceless gauge (we neglect here post-Newtonian contributions).
In these expressions $\iota$ is the orientation of the orbital plane with respect to the detector, $z$ is the redshift of the source, $d_L(z)$ is the luminosity distance, $f$ is the GW frequency at the observer, $\Phi(t)$ is the phase of the GW, and $\mathcal{M}_c(z)=(1+z)(m_1 m_2)^{3/5}/(m_1+m_2)^{1/5}$ is the redshifted chirp mass, with $m_{1,2}$ being the masses of the two binary bodies.
An accurate detection of the GW signal allows to reconstruct all the parameters in Eqs.~(\ref{eq:hx}) and (\ref{eq:h+}) within some (correlated) uncertainties~\cite{Schutz:1986gp}.
In particular, thanks to the reconstruction of $d_L$, binaries can be employed as reliable cosmological distance indicators.

Eqs.~(\ref{eq:hx}) and (\ref{eq:h+}) highlight three key aspects of using inspiralling binaries as cosmological distance indicators:
\textit{i)} The measurement of $d_L$ from GW signals is not affected by any systematic uncertainties in the modelling of the source, since the dynamics of compact binary systems is directly determined by GR. This is in contrast with supernovae type-Ia (SNIa), which require the cosmic distance ladder, i.e.~cross calibration with local measurements of sources of known distance, to overcome unknown systematics in the determination of their luminosity distance.
\textit{ii)} Due to the scaling $\propto d_L^{-1}$, GW cosmological indicators are suitable even at large distances where EM sources, whose intensity scales as $d_L^{-2}$, are too faint. This implies that given the same amount of emitted energy, a source producing GWs can be observed at higher distances with respect to a source emitting EM waves.
\textit{iii)} The measurement of the quoted waveform does not allow to determine the redshift of the source: in fact Eqs.~(\ref{eq:h+}) and (\ref{eq:hx}) are invariant under the transformation $m_i\to m_i (1+z)$ plus $d_L \to d_L (1+z)$. In other words the waveform detected from any system with masses $m_i$ at a distance $d_L$ will be equivalent to a waveform produced by a system with masses $m_i (1+z)$ at a distance $d_L (1+z)$.

The luminosity distance $d_L$ is tightly linked to the redshift $z$ in a given cosmological setup. For a homogeneous and isotropic universe (at large scale), the luminosity distance is given by
\begin{equation}
	d_L(z) = \frac{c}{H_0}\frac{1+z}{\sqrt{\Omega_k}} \sinh \left[ \sqrt{\Omega_k} \int_0^z \frac{H_0}{H(z')} dz' \right] \,,
	\label{eq:dL_z}
\end{equation}
with $\Omega_k$ being the present value of the density parameter of the spatial curvature, $H(z)$ being the Hubble rate as a function of the redshift, and $H_0=H(z=0)$.
Therefore, if besides $d_L$ (reconstructed from the waveform) one can independently establish the redshift $z$ of a GW source, then one obtains a data point useful to constrain the relation~(\ref{eq:dL_z}). The parameters of any given cosmological model, which are implicitly included in $H(z)$ since its dynamics is determined by the Einstein equations, can then be statistically constrained by fitting Eq.~(\ref{eq:dL_z}) against a number of $(z,d_L(z))$ data points. This can be done using observations from GW inspiralling binaries, if a determination of their redshift is available by some means. For this reason, well detectable binaries, for which the redshift is also known (or at least estimated), are dubbed {\it standard sirens}, in analogy with SNIa used as standard candles in cosmography~\cite{Holz:2005df}.

Of course, to obtain a robust bound, it is crucial to fit Eq.~(\ref{eq:dL_z}) with as many standard sirens as possible at the most diverse redshifts, especially if the cosmological model at hand contains many parameters.
The outcome of such an analysis can be remarkable.  It constitutes the first robust cosmological test not using EM radiation as the only messenger of astronomical information, and it allows to probe the validity of the cosmic distance ladder up to far distances. As demonstrated by the recent analysis of the LVC detection GW170817, discussed in section \ref{sub:status} below, it also provides a measurement of $H_0$ that is independent of the calibrations necessary to establish constraints using SNIa. 
In particular, if the current tension between the $H_0$ measurement from SNIa and CMB (assuming $\Lambda$CDM) will persist, standard sirens will be a very useful observable to decipher the origin of this tension.

\subsubsection{Redshift information}
\label{sub:method}
$ $\\[1mm]
\noindent 
The cosmography procedure just described assumes that the redshift of the GW sources can be acquired. This is however not straightforward. Because of their intrinsic nature, coalescing BHs do not guarantee any signature besides GWs. Nevertheless, EM counterparts can be envisaged for BBHs surrounded by matter, or for compact binaries where at least one of the two bodies is not a BH.
For this reason, an EM counterpart is expected for merging massive BHs at the centre of galaxies, which may be surrounded by an accretion disk, and for BNSs, whose merger produces distinctive EM emissions, including gamma ray bursts and kilonovae.
On the other hand, extreme mass ratio inspiral (EMRI) systems and stellar-origin BBHs (SOBBHs) are not expected to produce significant EM radiation at merger, although their merging environment is still unclear.  
Of course, in order to fit Eq.~(\ref{eq:dL_z}) with as many data points ($z$, $d_L$) as possible, it would help substantially to determine the redshift of all detectable standard sirens, independently of whether they do or do not exhibit an EM counterpart.
There are mainly two different ways to obtain redshift information for a standard siren, depending on the observation or not of an EM counterpart:

\begin{description}
\item[\it Method with EM counterpart:]
This method relies on EM telescopes to recognize the galaxy hosting the GW source \cite{Schutz:1986gp}.
Reaching a good sky localization ($\mathcal O(10\,$deg$^2)$ or below) as soon as possible after the detection of the GW signal, is essential to alert EM telescopes and point them towards the solid angle determining the direction of the GW event to look for EM transients.
Once such a transient is detected, the GW event can be associated with the nearest galaxy whose redshift can be measured either spectroscopically or photometrically.
It is important for this method to have GW detectors able to rapidly reach a well-beamed sky localization of the GW source.
A network of GW interferometers not only improves the sensitivity to a standard siren signal (improving the reconstruction of $d_L$) but also the identification of the sky solid angle containing the source thanks to spatial triangulation.

\item[\it Method without EM counterpart:] 
This method allows to determine the sky localization of the standard siren much after the GW event, with clear practical advantages, in particular, the presence of the source does not need to be recognized in real time, and moreover the SNR required for a good sky localization does not need to be reached before the stage at which the EM signal might be triggered).
It adopts a statistical approach \cite{Schutz:1986gp,MacLeod:2007jd}.
Indeed, given a galaxy catalogue, the galaxy hosting the standard siren has to be one of those contained in the box given by the identified solid angle (with its experimental error) times a properly-guessed redshift range.
This range is obtained by applying a reasonable prior to the redshift obtained inverting Eq.~(\ref{eq:dL_z}).
In this procedure one must of course take into account the dependency upon the cosmological parameters of $H(z)$, which will affect the final posterior on the parameters themselves.
The redshift of the standard siren can be estimated as the weighted average of the redshifts of all the galaxies within the error box (an additional prior on the conformation of each galaxy may be also included).
Due to the large uncertainty of this statistical approach, this method is effective only when a limited amount of galaxies can be identified within the volume error box and only if a sufficiently large amount of GW events is observed.
\end{description}

These procedures suffer from some major uncertainties. The redshift appearing in Eq.~(\ref{eq:dL_z}) is the one due to the Hubble flow, and consequently the contribution due to the peculiar velocity of the host galaxy to the measured redshift should be subtracted.
Moreover, the real geodesic followed by the GW is not the one resulting from the homogeneous and isotropic metric assumed in Eq.~(\ref{eq:dL_z}), and in fact the lensing contribution to $d_L$ due to cosmic inhomogeneities should be removed.
Although in principle very precise lensing maps and galaxy catalogues are helpful to estimate such source of uncertainty \cite{Jonsson:2006vc,Shapiro:2009sr,Hirata:2010ba,Hilbert:2010am}, still the lensing and peculiar velocity effects have to be treated as a (large) systematic error that can be reduced only by means of numerous detections (the uncertainty due to peculiar velocities dominating at low redshift, and the lensing uncertainty dominating at large redshift).

\subsubsection{Standard sirens with current GW data: GW170817} 
\label{sub:status}
\begin{figure}[t]
\begin{minipage}[h]{.45\linewidth}
\centering
\includegraphics[width=\linewidth]{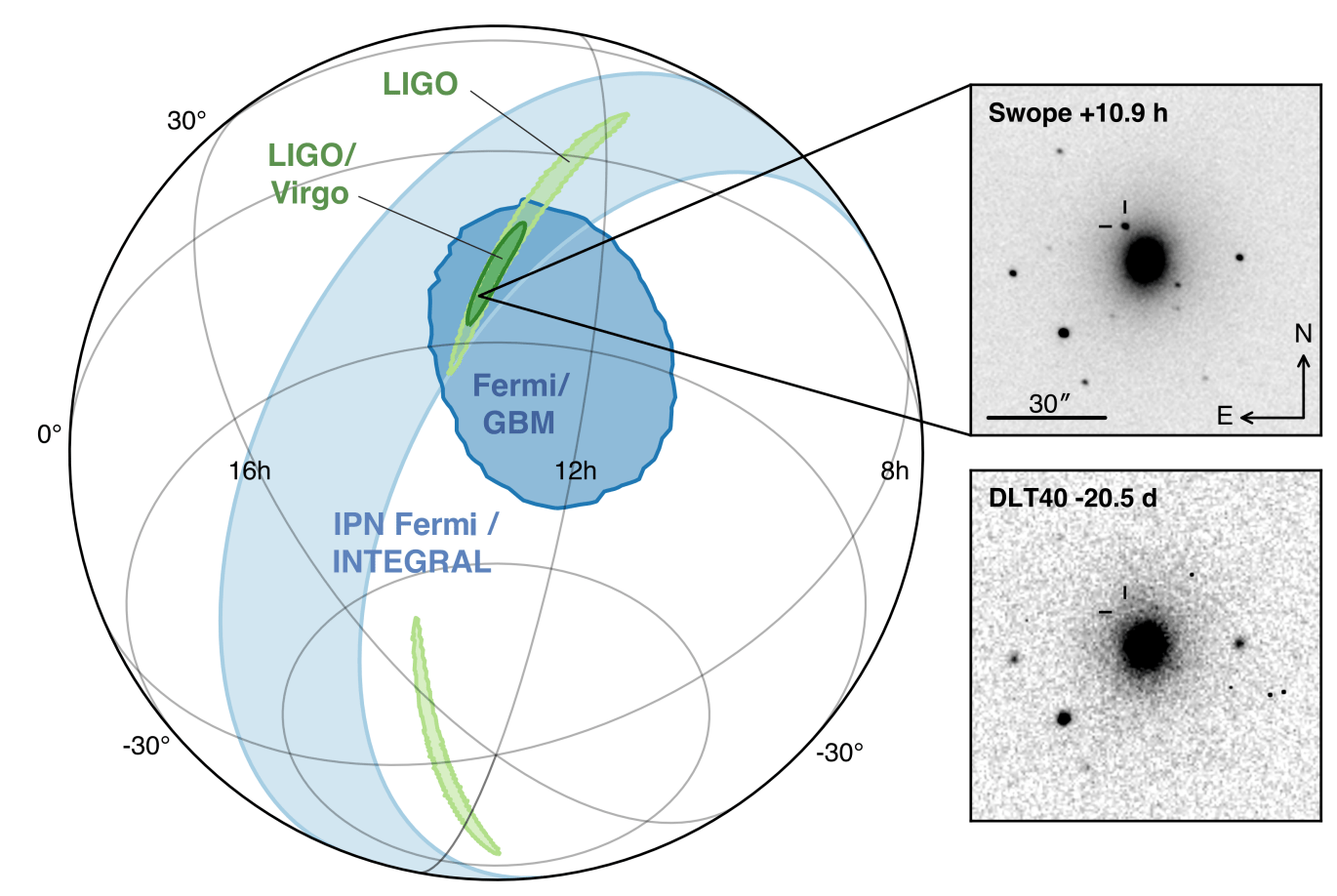}
\end{minipage}\hfill
\begin{minipage}[h]{0.49\linewidth}
            \vspace*{5mm}\centering
            \includegraphics[width=\linewidth]{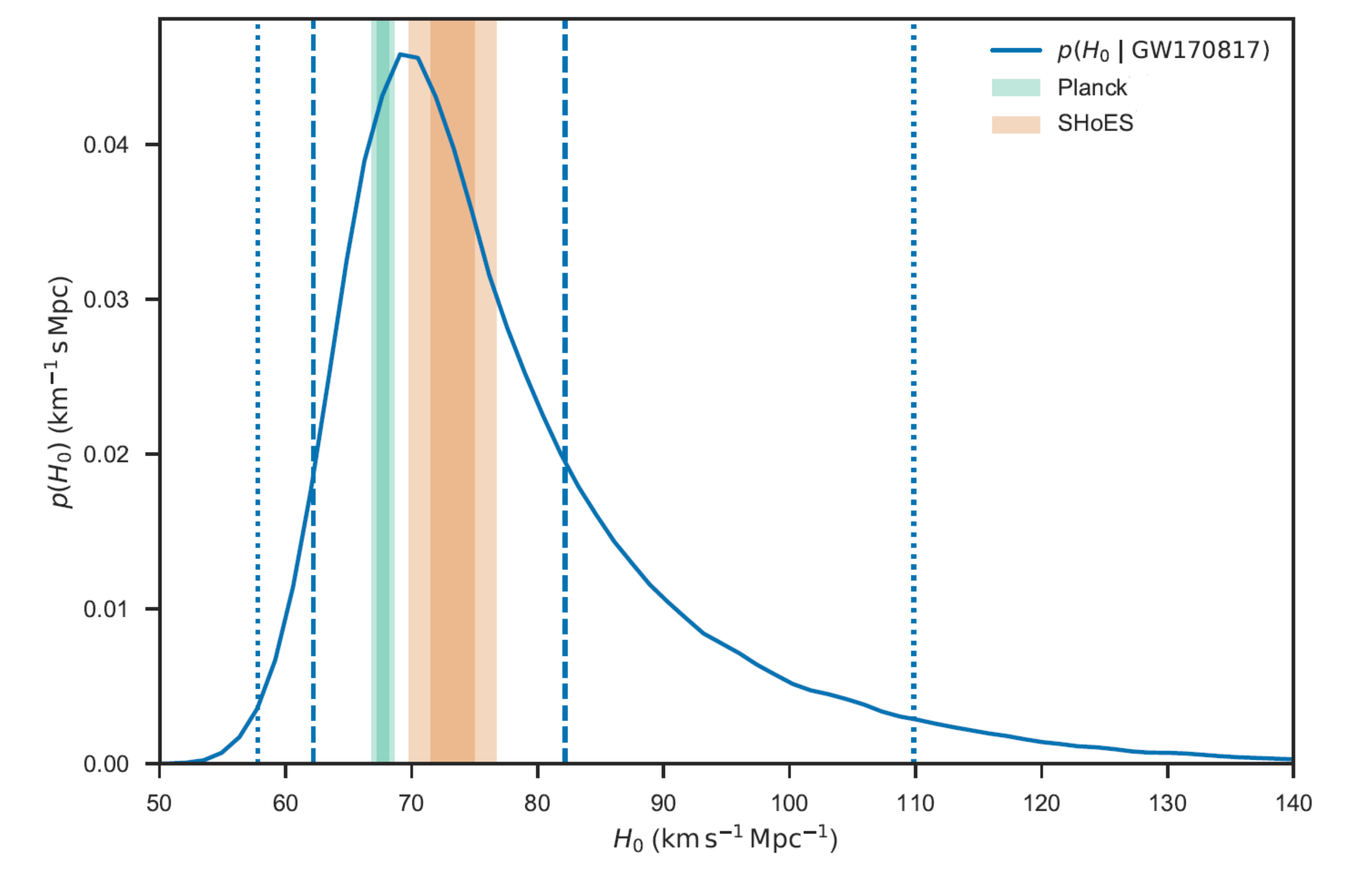}
        \end{minipage}
        \caption{\it{Left panel: The multi-messanger sky localization of GW170817 and the identification of the host galaxy. Right panel: The posterior distribution of $H_0$ compared to the recent CMB and SNIa constraints~\cite{Ade:2015xua, Riess:2016jrr}. Figures taken from  Refs.~\cite{GBM:2017lvd,Abbott:2017xzu}}.
    }
    \label{fig:gw17}
    \end{figure}

$ $\\[1mm]
\noindent 
The GW170817 event, corresponding to the coalescence of a BNS, is the exquisite progenitor of cosmography via standard sirens with EM counterpart.
For that event, the LVC interferometer network recovered the luminosity distance $d_L= 43.8^{+2.9}_{-6.9}$ Mpc at 68\% C.L., and a sky localization of 31 deg$^2$ \cite{TheLIGOScientific:2017qsa,GBM:2017lvd}, corresponding to the solid angle shown in the left panel of Fig.~\ref{fig:gw17} (green area).
The optical telescopes exploring this portion of the sky identified an EM transient in association to the galaxy NGC4993, which is known to be departing from us at the speed of $3327\pm 72\,$\,km s$^{-1}$.
Although part of the peculiar velocity of the galaxy NGC4993 could be subtracted \cite{GBM:2017lvd}, remaining uncertainties on the peculiar motion eventually yield the estimate $v_H=3017\pm 166$\,km s$^{-1}$ for its Hubble flow velocity, corresponding to $z\simeq 10^{-2}$ (notice that at this redshift the lensing uncertainty is negligible).
Using the Hubble law $d_L=z/H_0$, which is the leading order term in the expansion of Eq.~(\ref{eq:dL_z}) at small $z$, one obtains the posterior distribution for $H_0$ presented in Fig.~\ref{fig:gw17} (right panel), providing the measurement $H_0 = 70.0^{+12.0}_{-8.0}\,\rm{km\, s^{-1} Mpc^{-1}}$ at 68\% C.L.~\cite{Abbott:2017xzu}.
The uncertainty on $H_0$ is too large to make the measurement competitive with CMB \cite{Ade:2015xua} and SNIa constraints \cite{Riess:2016jrr}. Nevertheless, this represents a local estimate of $H_0$ that is not dependent on the cosmic distance ladder and the first cosmological measurement not relying only on EM radiation.
Moreover, the large number of events similar to GW170817 expected to be observed in the future will eventually yield constraints on $H_0$ 
at the level of both local and CMB measurements.

\subsubsection{Cosmological forecasts with standard sirens} $ $\\[1mm]
\noindent 
The potential of current and forthcoming GW detectors for cosmology have been widely studied in the literature. To appreciate the capability of standard sirens to probe the late universe, here we report on the most recent forecasts (see e.g.~Refs.~\cite{Dalal:2006qt,Nissanke:2009kt,DelPozzo:2011yh,Nissanke:2013fka} (LVC) and \cite{Menou:2008dv,Stavridis:2009ys,VanDenBroeck:2010fp,Shang:2010ta,Sereno:2011ty,Petiteau:2011we} (LISA) for previous studies). Such analyses are expected to continuously improve over the upcoming years of data taking.

\begin{description}
\item[\it Forecasts with ground-based interferometers:]
Ground-based interferometers can detect binary systems composed of NSs and/or stellar-origin BHs, with masses ranging from few solar masses up to tens of solar masses. 
By means of the planned LVC-KAGRA-LIGO India network, BNSs with counterpart will allow to put a tight constraint on $H_0$.
For example, assuming the $\Lambda$CDM model, $H_0$ is expected to be measured roughly with a $\sim$1\% error after $\sim$100 detections with counterpart~\cite{Chen:2017rfc, Seto:2017swx}.
The result would be even tighter if, for some SOBBHs, the host galaxy could be identified, or if BNSs with counterpart are considered instead, as only $\sim$10 detections would yield the same level of accuracy~\cite{Chen:2017rfc,Nishizawa:2016ood}.
The forecasts drastically improve for third generation ground-based detectors, such as the Einstein Telescope (ET) \cite{Sathyaprakash:2009xt,Taylor:2012db,Cai:2016sby}.
In addition, besides the two approaches mentioned above for the redshift measurement, a further method, feasible only with an ET-like device, will become available~\cite{Messenger:2011gi} (see also \cite{Taylor:2011fs} for a similar idea).
Thanks to the tidal effects of a BNS waveform, the redshift-mass degeneracy can be broken.
Consequently, by measuring the tidal effect in the waveform and assuming a prior knowledge on the NS equation of state, the redshift $z$ entering the redshifted chirp mass can be reconstructed from the waveform itself.
In this way, with more than 1000 detected BNS events, the ET is expected to constrain the parameters $H_0$ and $\Omega_m$ of the $\Lambda$CDM model roughly with an uncertainty of $\sim 8\%$ and $\sim 65\%$, respectively~\cite{DelPozzo:2015bna}.
However, if the rate of BNS mergers in the universe will result to be in the higher limit of the currently allowed range, the ET will be able to see up to $\sim\!\!10^7$ BNS mergers, ameliorating these constraints by two orders of magnitude.

\item[\it Forecasts with space-based interferometers:]
Space-based GW interferometry will open the low-frequency window (mHz to Hz) in the GW landscape, which is complementary to Earth-based detectors (Hz to kHz) and PTA experiments (nHz).
The Laser Interferometer Space Antenna (LISA) is currently the only planned space mission designed to detect GWs, as it has been selected by ESA~\cite{Audley:2017drz}.
Several new GW astrophysical sources will be observed by LISA, including SOBBHs, EMRIs and massive BBHs from $10^4$ to $10^7$ solar masses.
These sources can not only be conveniently employed as standard sirens, but they will be detected at different redshift ranges, making LISA a unique cosmological probe, able to measure the expansion rate of the universe from local ($z\sim 0.01$) to very high ($z\sim 10$) redshift.
The current forecasts, produced taking into account only massive BBHs \cite{Tamanini:2016zlh} (for which an EM counterpart is expected) or SOBBHs \cite{Kyutoku:2016zxn,DelPozzo:2017kme} (for which no EM counterpart is expected), estimate constraints on $H_0$ down to a few percent.
However, joining all possible GW sources that can be used as standard sirens with LISA in the same analysis, should not only provide better results for $H_0$, which will likely be constrained to the sub-percent level, but it will open up the possibility to constrain other cosmological parameters.
The massive BBH data points at high redshifts will, moreover, be useful to test alternative cosmological models, predicting deviations from the $\Lambda$CDM expansion history at relatively early times \cite{Caprini:2016qxs,Cai:2017yww}.
Finally, more advanced futuristic missions, such as DECIGO or BBO, which at the moment have only been proposed on paper, may be able to probe the cosmological parameters, including the equation of state of dark energy, with ultra-high precision~\cite{Cutler:2009qv,Nishizawa:2010xx,Kawamura:2011zz,Arabsalmani:2013bj}. 
They might also be able to detect the effect of the expansion of the universe directly on the phase of the binary GW waveform \cite{Seto:2001qf,Nishizawa:2011eq}, although the contribution due to peculiar accelerations would complicate such a measurement \cite{Bonvin:2016qxr,Inayoshi:2017hgw}.

\end{description}

\subsubsection{Future prospects}$ $\\[1mm]
\noindent
The recent GW170817 event has triggered numerous studies about the use of standard sirens for cosmography, and the forthcoming experimental and theoretical developments might change the priorities in the field. As mentioned above, the network of Earth-based GW interferometers is expected to detect an increasing number of GW sources employable as standard sirens, with or without EM counterparts.
This will eventually yield a measurement of the Hubble constant competitive with CMB and SNIa constraints, which might be used to alleviate the tension between these two datasets.
On the other hand, higher redshift ($z \gtrsim 1$) standard sirens data will probably be obtained only with third generation detectors or space-based interferometers.
With these high redshift data we will start probing the expansion of the universe at large distances, implying that a clean, and not exclusively EM-based, measurement of other cosmological parameters will become a reality. 
Furthermore, the precision of cosmography via standard sirens might be boosted by improvements in other astronomical observations.
This is the case if e.g.~galactic catalogues and lensing maps improve substantially in the future, as expected.    

Concerning present forecasts, there is room for improvements, especially regarding third generation Earth-based interferometers and space-borne GW detectors.
The full cosmological potential of future GW experiments such as ET and LISA has still to be assessed.
For ET we still do not have a full cosmological analysis taking into account all possible GW sources that will be used as standard sirens, using both the method with counterpart (BNSs) and the method without counterpart (SOBBHs and BNSs).
Moreover, we still need to fully understand up to what extent the information on the NS equation of state can be used to infer the redshift of BNSs and NS-BH binaries, with the latterpotentially representing a new interesting future source of standard sirens \cite{Nissanke:2009kt,Nissanke:2013fka,Vitale:2018wlg}. Regarding LISA, there are still several issues to be addressed in order to produce reliable forecasts.
For instance, the EMRI detection rate is still largely unknown \cite{Babak:2017tow}, although these sources might turn out to be excellent standard siren candidates at redshift $0.1\lesssim z \lesssim 1$~\cite{Tamanini:2016uin}. In addition, the prospects for massive BBH mergers as standard sirens would be more robust if an up-to-date model of the EM counterpart were implemented in the investigations. Performing these analyses and combining the results for all the different types of standard sirens observable by LISA will allow for a full assessment of the cosmological potential of the mission.

Most of the literature on standard sirens assumes GR and an homogeneous and isotropic universe at large scales. However, breakthroughs may occur by generalizing the aforementioned analyses to theories beyond GR, and in fact forecasts for scenarios not fulfilling these assumptions are flourishing topics with potentially revolutionary results.
For example, some models of modified gravity, formulated to provide cosmic acceleration, have been strongly constrained by the measurement of the speed of GW (compared to the speed of light) with GW170817 \cite{Creminelli:2017sry,Ezquiaga:2017ekz,Baker:2017hug,Sakstein:2017xjx}. In general, in theories beyond GR or admitting extra dimensions, the cosmological propagation of GWs changes and a comparison between the values of $d_L$ measured from GWs and inferred from EM observations can provide a strong test of the validity of these theories (see \cite{Belgacem:2017ihm,Amendola:2017ovw,Linder:2018jil,Pardo:2018ipy} for recent works).

Finally we mention that the large number of GW sources expected to be observed by third generation interferometers such as ET or by future space-borne detectors such as DECIGO/BBO, might even be used to test the cosmological principle \cite{Yagi:2011bt,Namikawa:2015prh,Cai:2017aea} and to extract cosmological information by cross-correlating their spatial distribution with galaxy catalogues or lensing maps \cite{Camera:2013xfa,Oguri:2016dgk,Raccanelli:2016fmc}.

\subsection{Interplay between GWs from binaries and from early-universe sources}

The gravitational interaction is so weak that GWs propagate practically unperturbed along their path from the source to us.  GWs produced in the early universe thus can carry a unique imprint of the pre-CMB era, in which the universe was not transparent to photons. In this sense, GW detection can provide for the first time direct, clean access to epochs that are very hard to probe by any other observational means. GWs of cosmological origin appear (pretty much as the CMB) to our detectors as a SGWB \cite{Romano:2016dpx,Caprini:2018mtu}. 

Several pre-CMB phenomena sourcing GWs might have occurred along the cosmological history, from inflation to the epoch of the QCD phase transition. In such a case, the SGWB would be constituted by the sum of all single contributions, each of them potentially differing from the others in its spectral shape as a function of frequency, or because of other properties such as chirality and/or gaussianity. It is hence crucial to have precise predictions of all proposed cosmological sources to possibly isolate all components in the detected SGWB, with the aim of reconstructing the early time history of the universe.

On the other hand, binary systems can also be detected as a SGWB. This happens for those (independent) astrophysical events that are overall too weak to be individually resolved. The level of their ``contamination'' to the cosmological SGWB thus depends on the sensitivity and on the resolution of the available detector. Crucially, at LIGO-like and LISA-like experiments the astrophysical component might be stronger than the cosmological one, so that the information hidden in the pre-CMB signal would be impossible to recover, unless the astrophysical contribution is known in great detail and methods are found to subtract it.

\subsubsection{Status and future prospects}
$ $\\[-1mm]

\noindent No measurement of the SGWB has been done yet, but upper bounds have been inferred from the observations. These are usually expressed in terms of the SGWB energy density, which is given by $\Omega_{\rm GW}(f)=(f/\rho_c) ~\partial \rho_{\rm GW}/\partial f$, with $\rho_c$ and $\rho_{\rm GW}$ being the critical and the SGWB energy densities, respectively. Specifically, by assuming the frequency shape $\Omega_{\rm GW}(f)=\Omega_\alpha (f/25$Hz)$^\alpha$, the LVC found the 95\% C.L.~limit $\Omega_{\alpha=0,2/3,3} < 17\times 10^{-7},13\times 10^{-7}$ and $1.7\times 10^{-8}$ in the $\mathcal O (10)$ -- $\mathcal O (100)\,$Hz frequency band~\cite{TheLIGOScientific:2016dpb}. The latest Pulsar Timing Array analysis, done with the data of the NANOGrav collaboration, yields  $h^2\Omega_{\alpha=0} < 3.4\times 10^{-10}$ at 95\% C.L.~at $3.17\times 10^{-8}$\,Hz~\cite{Arzoumanian:2018saf}. Since the astrophysical sources are not expected to produce a SGWB with higher amplitude than these bounds, the latter are particularly relevant only for the most powerful cosmological signals~\cite{Lasky:2015lej, Abbott:2017mem}. 

However, it is probably only a matter of some years to achieve the first detection of the astrophysical SGWB by the LVC.  Based on the current detection rates, the SOBBHs and BNSs lead to the power-law SGWBs $\Omega_{\rm BBH}(f)=1.2^{+1.9}_{-0.9}\times 10^{-9}\left(f/25\,{\rm Hz}\right)^{2/3}$ and $\Omega_{\rm BBH+BNS}(f)=1.8^{+2.7}_{-1.3}\times 10^{-9}\left(f/25\,{\rm Hz}\right)^{2/3}$ in the frequency band of both LVC and LISA~\cite{Abbott:2017xzg}. These signals are expected to be measurable by the LIGO and Virgo detectors in around 40 months of observation time~\cite{Abbott:2017xzg}, while in LISA they will reach a Signal to Noise Ratio (SNR) of $\mathcal O(10)$ in around one month of data taking~\cite{CosWGpreparation}. On the other hand, given the large uncertainties on EMRIs~\cite{Babak:2017tow}, it is not clear whether they can also give rise to an observable SGWB component in the LISA band.

Focusing exclusively on the SGWB contribution from astrophysical binaries, two general aspects about it are particularly worth investigating. The first regards the detailed prediction and characterization of this component. For the time being, the literature does not suggest any technique to disentangle the cosmological SGWB component from a generic astrophysical one. At present, it is believed that component separation will be (partially) feasible in the LISA data only for what concerns the SGWB signal due to galactic binaries. In this case, indeed, the anisotropy of the spatial distribution of the source (confined to the galactic plane), together with the motion of the LISA constellation, generates a SGWB signal modulated on a yearly basis, which allows to distinguish and subtract this component from the total SGWB~\cite{Adams:2013qma}. Although a similar approach is not possible for the extra-galactic SGWB component, this example well illustrates that detailed predictions of the astrophysical signatures might highlight subtraction techniques allowing to perform component separation and hopefully isolate the information coming from the early universe (similarly to what done for foreground subtraction in CMB analyses). 

The second aspect concerns the possibility of exploiting third generation detectors, such as the ET and Cosmic Explorer, to subtract the astrophysical SGWB component \cite{Regimbau:2016ike}. Their exquisite sensitivity may allow to resolve many of the SOBBHs and BNSs that give rise to the SGWB in present detectors, including the full LVC-KAGRA-LIGO India network. This would clean the access to the cosmological SGWB down to a level of $\Omega_{\rm GW}\simeq 10^{-13}$ after five years of observation \cite{Regimbau:2016ike}. This technique not only applies in the frequency bandwidth of the Earth-based devices, but it can also be used to clean the LISA data and reach a potential SGWB of cosmological origin in the LISA band. Note that LISA might as well help in beating down the level of the SGWB from SOBBHs by exploiting possible multi-band detections of the same source \cite{Sesana:2016ljz}. The SOBBH would be detected first by LISA, during its inspiral phase; some years later, when the binary has arrived to the merger stage, it would reappear in Earth-based interferometers. These latter can therefore be alerted in advance, possibly leading to an increase in the number of detected BBHs (depending on their LISA SNR). 


\newpage

\phantomsection
\addcontentsline{toc}{part}{\bf Chapter II: Modelling black-hole sources of gravitational waves: prospects and challenges}
\begin{center}
{\large \bf Chapter II: Modelling black-hole sources of gravitational waves: prospects and challenges}
\end{center}
\begin{center}
Editor: Leor Barack
\end{center}

\vskip 1cm
\setcounter{section}{0}
\section{Introduction} \label{Sec:introduction2}

The detection and characterization of gravitational-wave (GW) sources rely heavily on accurate models of the expected waveforms. This is particularly true for black hole binaries (BBHs) and other compact objects, for which accurate models are both necessary and hard to obtain. To appreciate the state of affairs, consider the following three examples. (i) While GW150914, the first BH merger event detected by LIGO, had initially been identified using a template-free search algorithm, some of the subsequent events, which were not as bright, would likely have been altogether missed if template-based searches had not been performed. (ii) While the error bars placed on the extracted physical parameters of detected BH mergers have so far come primarily from instrumental noise statistics, systematic errors from the finite accuracy of available signal models are only marginally smaller and would actually dominate the total error budget for sources of some other spin configurations or greater mass disparity; in the case of the binary neutron star (BNS) GW170817, deficiencies in available models of tidal effects already restrict the quality of science extractable from the signal. (iii) Even with a perfectly accurate model at hand, analysis of GW170817 would not have been possible within the timescale of weeks in which it was carried out, without the availability of a suitable reduced-order representation of the model, necessary to make such an analysis computationally manageable.

Indeed, accurate and computationally efficient models underpin all of GW data analysis. They do so now, and will increasingly do so even more in the future as more sensitive and broader-band instruments go online. The response of detectors like ET, and especially LISA, will be source-dominated, with some binary GW signals occurring at high signal-to-noise ratio (SNR) and remaining visible in band through many more wave cycles. The accuracy standard of models needs to increase commensurably with detector sensitivity, or else modelling error would restrict our ability to fully exploit the detected signals. As a stark example, consider that, in a scenario that is not unlikely, LISA's output will be dominated by a bright massive BH (MBH) merger signal visible with SNR of several 100s. This signal would have to be carefully ``cleaned out'' of the data in order to enable the extraction and analysis of any other sources buried underneath; any model inaccuracies would form a systematic noise residual, potentially hiding dimmer sources. In the case of Extreme Mass Ratio
Inspirals (EMRIs), where $O(10^5)$ wave cycles are expected in the LISA band at a low SNR, a precise model is a crucial prerequisite for both detection and parameter extraction.

This chapter reviews the current situation with regard to the modelling of GW sources within GR, identifying the major remaining challenges and drawing a roadmap for future progress. 
To make our task manageable, we focus mainly on sources involving a pair of BHs in a vacuum environment in GR, but,
especially in Sec.~\ref{Sec:HE}, we also touch upon various extensions beyond vacuum, GR and the standard
model (SM) of particle physics.

Isolated vacuum BHs in GR are remarkably simple objects, described in exact form
by the Kerr family of solutions to Einstein’s field equations (see, however, Sec.~\ref{Sec:MR} of this chapter). But let two such BHs interact with each other, and the resulting system displays a remarkably complicated dynamics, with no known exact solutions. Even numerical solutions have for decades proven elusive, and despite much progress following the breakthrough of 2005 they remain computationally very expensive---prohibitively so for mass ratios smaller than $\sim 1:10$---and problematic for certain astrophysically relevant BH spin configurations. Systematic analytical approximations are possible and have been developed based around expansions of the field equations in the weak-field or extreme mass-ratio regimes, and these may be combined with fully numerical solutions to inform waveform models across broader areas of the parameter space. To facilitate the fast production of such waveforms, suitable for GW search pipelines, effective and phenomenological models have been developed, which package together and interpolate results from systematic numerical simulations and analytical approximations. With the rapid progress in GW experiments, there is now, more than ever, a need for a concentrated community effort to improve existing models in fidelity, accuracy and parameter-space reach, as well as in computational efficiency. 

The structure of this chapter is as follows. In Secs.\ \ref{Sec:perturbations} and \ref{Sec:PN} we review the two main systematic approximations to the BBH problem: perturbation theory including the gravitational self-force (GSF) and post-Newtonian (PN) approaches, respectively.
Section \ref{Sec:NR} surveys progress and prospects in the Numerical Relativity (NR) modelling of inspiralling and merging BHs in astrophysical settings, and Sec.\ \ref{Sec:HE} similarly reviews the role of NR in studying the dynamics of compact objects in the context of alternative theories of gravity and beyond the SM. Section \ref{Sec:EOB} then reviews the Effective One Body (EOB) approach to the BBH problem, and the various phenomenological models that have been developed to facilitate fast production of waveform templates. 
Section \ref{Sec:DA} reviews the unique and highly involved challenge of data-analysis in GW astronomy, with particular emphasis on the role of source models; this data-analysis challenge sets the requirements and accuracy standards for such models. Finally, Sec.\ \ref{Sec:MR} gives a mathematical relativist's point of view, commenting on a variety of (often overlooked) foundational questions that are yet to be resolved in order to enable a mathematically rigorous and unambiguous interpretation of GW observations.

\section{Perturbation Methods} \label{Sec:perturbations}
\vspace{-3mm}
{\it Contributor:} B. Wardell
\vspace{3mm}

Exact models for GWs from BBHs can only be obtained by exactly
solving the full Einstein field equations. However, there is an important regime in which a
perturbative treatment yields a highly accurate approximation. For BBH systems in
which one of the BHs
is much less massive than the other, one may treat the mass ratio as a small
perturbation parameter. Then, the Einstein equations are amenable to a perturbative expansion in
powers of this parameter. Such an expansion is particularly suitable for EMRIs, systems in which the mass ratio may be as small as $10^{-6}$, or even smaller
\cite{Magorrian:1997hw,Gair:2004iv,AmaroSeoane:2007aw,Gair:2008bx,AmaroSeoane:2012km}. In such cases, it has been
established \cite{Hinderer:2008dm,Isoyama:2012bx,Burko:2013cca} that it will be necessary to incorporate information at
second-from-leading perturbative order to achieve the accuracy that will be required for optimal
parameter estimation by the planned LISA mission
\cite{LISA,Barack:2003fp,Babak:2010ej,Gair:2010bx,Babak:2017tow}. Aside from EMRIs, a
perturbative expansion is likely to also be useful as a model for Intermediate Mass Ratio Inspirals
(IMRIs): systems where the mass ratio may be as large as $\sim10^{-2}$. Such systems, if they exist, are detectable
in principle by Advanced LIGO and Virgo, and are indeed being looked for in the data of these experiments\cite{Abbott:2017iws,Haster:2015cnn}.

The perturbative approach (often called the \emph{self-force}\footnote{Strictly speaking, the term
\emph{self-force} refers to the case where local information about the perturbation in the vicinity
of the smaller object is used; other calculations of, e.g., the flux of GW energy far
from the binary also rely on a perturbative expansion
\cite{Glampedakis:2002ya,Mino:2003yg,Hughes:2005qb,Sago:2005fn,Drasco:2005is,Drasco:2005kz,Sundararajan:2007jg,Fujita:2009us}, but are not referred to as GSF calculations.} approach) yields a set of equations for the motion of the smaller object about the larger one.
For a detailed technical review of self-force physics, see \cite{Poisson:2011nh}, and for a most recent, pedagogical review, see \cite{Barack:2018yvs}. 
At zeroth order in the
expansion, one recovers the standard geodesic equations for a test particle in orbit around
(i.e.~moving in the background spacetime of) the larger BH. At first order, we obtain
coupled equations for an accelerated worldline forced off a geodesic by the GSF, which itself arises from the metric perturbation to the background spacetime sourced
by the stress-energy of the smaller BH. In the GSF approach, it can be convenient to
treat the smaller BH as a ``point particle'', with the GSF being computed from an
effective \emph{regularized} metric perturbation
\cite{DeWitt:1960fc,Hobbs:1968a,Mino:1996nk,Quinn:1996am,Detweiler:2002mi}. Such an assumption is
not strictly necessary, but has been validated by more careful treatments whereby both BHs
are allowed to be extended bodies, and the point-particle-plus-regularization prescription is
recovered by appropriately allowing the smaller BH to shrink down to zero size
\cite{Gralla:2008fg,Pound:2009sm,Pound:2010pj,Pound:2012dk}. These more careful treatments have also allowed the GSF
prescription to be extended to second perturbative order
\cite{Rosenthal:2005it,Rosenthal:2006iy,Detweiler:2011tt,Gralla:2012db,Pound:2012nt,Pound:2012dk,Pound:2014xva}
and to the fully non-perturbative case \cite{Harte:2008xq,Harte:2009yr,Harte:2011ku}.

The goal of the GSF approach is to develop efficient methods for computing the motion of the smaller object and 
the emitted GWs in astrophysically-relevant scenarios. These involve, most generally, a spinning (Kerr)
large BH, and a small (possibly spinning) compact object in a generic (possibly inclined and eccentric) 
inspiral orbit around it. The data analysis goals of the LISA mission (which
demand that the phase of the extracted waveform be accurate to within a fraction of a radian over the entire inspiral) require all
contributions to the metric perturbation at first order, along with the dissipative contributions at
second order \cite{Hinderer:2008dm}.

\subsection{Recent developments}

Focused work on the GSF problem has been ongoing for at least the last two
decades, during which time there has been substantial progress. Early work developed much of the
mathematical formalism, particularly in how one constructs a well-motivated and unambiguous
regularized first-order metric perturbation
\cite{DeWitt:1960fc,Hobbs:1968a,1982JPhA...15.3737G,Mino:1996nk,Quinn:1996am,Detweiler:2002mi,Gralla:2008fg,Pound:2009sm}.
More recent work has addressed the conceptual challenges around how these initial results can be
extended to second perturbative order
\cite{Rosenthal:2005it,Rosenthal:2006iy,Detweiler:2011tt,Gralla:2012db,Pound:2012nt,Pound:2012dk,Pound:2014xva},
and on turning the formal mathematical prescriptions into practical numerical schemes \cite{Poisson:Wiseman:1998,Barack:1999wf,Barack:2001gx,Anderson:2005gb,Barack:2007jh,Vega:2007mc}. As a result,
we are now at the point where first-order GSF calculations are possible for
almost any orbital configuration in a Kerr background spacetime \cite{vandeMeent:2017bcc}. Indeed,
we now have not one, but three practical schemes for computing the regularized first-order
GSF \cite{Wardell:2015kea}. Some of the most recent highlights from the
substantial body of work created by the GSF community are discussed below.

\subsubsection{First-order gravitational self-force for generic orbits in Kerr spacetime}

Progress in developing tools for GSF calculations has been incremental, starting out with
the simplest toy model of a particle with scalar charge in a circular orbit about a
Schwarzschild BH, and then extending to eccentric, inclined, and generic orbits about a spinning
Kerr BH \cite{Barack:2000zq,Burko:2000xx,Burko:2001kr,Detweiler:2002gi,DiazRivera:2004ik,Haas:2007kz,Ottewill:2007mz,Ottewill:2008uu,Lousto:2008mb,Vega:2009qb,Canizares:2009ay,Dolan:2010mt,Thornburg:2010tq,Canizares:2010yx,Warburton:2010eq,Warburton:2011hp,Diener:2011cc,Dolan:2011dx,Casals:2012qq,Ottewill:2012aj,Casals:2013mpa,Warburton:2013lea,Vega:2013wxa,Wardell:2014kea,Warburton:2014bya,Gralla:2015rpa,Thornburg:2016msc,Heffernan:2017cad}.
Progress has also been made towards developing tools for the conceptually
similar, but computationally more challenging GSF problem; again starting out
with simple circular orbits in Schwarzschild spacetime before extending to eccentric equatorial orbits and,
most recently, fully generic orbits in Kerr spacetime \cite{Barack:2002ku,Barack:2005nr,Anderson:2005gb,Barack:2007tm,Sago:2009zz,Shah:2010bi,Akcay:2010dx,Barack:2010tm,Keidl:2010pm,Warburton:2011fk,Dolan:2012jg,Shah:2012gu,Dolan:Capra16,Akcay:2013wfa,Pound:2013faa,Isoyama:2014mja,Merlin:2014qda,Osburn:2015duj,Gralla:2016qfw}. Along the way, there have been many
necessary detours in order to establish the most appropriate choice of gauge \cite{Keidl:2010pm,Hopper:2012ty,Pound:2013faa,Osburn:2014hoa,Merlin:2014qda,Pound:2015fma,Shah:2016juc,Chen:2016plo}, reformulations of the regularization procedure \cite{Haas:2006ne,Barack:2007we,Wardell:2011gb,Vega:2011wf,Dolan:2012jg,Heffernan:2012su,Heffernan:2012vj,Pound:2013faa,Warburton:2013lea,Linz:2014pka,Wardell:2015ada}, and various numerical
methods and computational optimisations \cite{Barack:2008ms,Vega:2009qb,Canizares:2009ay,Field:2009kk,Field:2010xn,Canizares:2010yx,Hopper:2010uv,Akcay:2010dx,Hopper:2012ty,Dolan:2012jg,Akcay:2013wfa,Osburn:2014hoa,Hopper:2015jxa,Hopper:2017iyq,Barack:2017oir}.

\subsubsection{Extraction of gauge-invariant information}

Both the regularized metric perturbation and the GSF associated with it are themselves gauge-dependent \cite{Barack:2001ph,Gralla:2011zr,Pound:2015fma}, but their combination encapsulates gauge-invariant information.  
A series of works have derived a set of gauge-invariant quantities accessible from the regularized metric and GSF, which quantify {\em conservative} aspects of the dynamics in EMRI systems beyond the geodesic approximation. (Strictly speaking, they are only ``gauge invariant'' within a particular, physically motivated class
of gauge transformations. Nevertheless, even with this restriction the gauge invariance is very
useful for comparisons with other results.) 
Examples include Detweiler's \emph{red-shift} invariant \cite{Detweiler:2008ft}, 
the frequency of the innermost stable circular orbit (ISCO) in Schwarzschild \cite{Barack:2009ey} and Kerr 
\cite{Isoyama:2014mja}, the periastron advance of slightly eccentric orbits in Schwarzschild \cite{Barack:2010ny} and Kerr \cite{vandeMeent:2016hel}, spin (geodetic) precession \cite{Dolan:2013roa,Shah:2015nva,Akcay:2016dku,Akcay:2017azq}, and, most recently, quadrupolar and octupolar tidal invariants \cite{Dolan:2014pja,Nolan:2015vpa}. The most important outcome from the development of these gauge invariants is the synergy it has enabled, both within the GSF programme \cite{Sago:2008id} (e.g., by allowing for direct
comparisons between results computed in different gauges) and, as described next, with other approaches to the two-body problem.

\subsubsection{Synergy with PN approximations, EOB theory, and NR}

One of the most fruitful outcomes arising from the development of GSF gauge-invariants is
the synergy it has enabled between the GSF and PN and EOB theories. With a
gauge-invariant description of the physical problem available, it is possible to make direct
connections between GSF, PN and EOB approximations. This synergy has worked in
a bidirectional way: GSF calculations have been used to determine previously-unknown
coefficients in both PN and EOB expansions
\cite{Detweiler:2008ft,Blanchet:2010zd,Barack:2010ny,Blanchet:2009sd,Akcay:2012ea,Bini:2014zxa,Blanchet:2014bza,Bini:2015xua,Bini:2015bfb,Kavanagh:2015lva,Akcay:2015pjz,Akcay:2015pza,Hopper:2015icj,Kavanagh:2016idg,Bini:2016qtx,Bini:2016dvs,Kavanagh:2017wot}; 
and EOB and PN calculations have been used to validate GSF results \cite{Bini:2014zxa},
and even to assess the region of validity of the perturbative approximation
\cite{LeTiec:2011bk,LeTiec:2011dp,Tiec:2013twa}. More on this in Sec.~\ref{Sec:PN}.

Mirroring the synergy between GSF and PN/EOB, there have emerged methods for making
comparisons between GSF and NR. This started out with direct comparisons of the periastron advance of slightly eccentric orbits \cite{LeTiec:2011bk,Tiec:2013twa}. More recently, a similar comparison was made possible for Detweiler's redshift \cite{Zimmerman:2016ajr,LeTiec:2017ebm},
facilitated by an emerging understating of the relation between Detweiler's red-shift and the horizon surface gravity of the small BH.

\subsubsection{New and efficient calculational approaches}

Despite the significant progress in developing numerical tools for computing the GSF, it is still a computationally challenging problem, particularly in cases where high
accuracy is required. This challenge has prompted the development of new and efficient
calculational approaches to the problem.

Initial GSF results were obtained in the Lorenz gauge \cite{Barack:2005nr,Barack:2007tm,Barack:2009ey,Barack:2010tm,Akcay:2010dx,Barack:2011ed,Warburton:2011fk,Akcay:2013wfa}, where the
regularization procedure is best understood. Unfortunately, the details of a Lorenz-gauge
calculation---in which one must solve coupled equations for the $10$ components of the metric
perturbation---are tedious and cumbersome, making it difficult to implement and even more
difficult to achieve good accuracy. In the Schwarzschild case, other calculations based on variations of Regge-Wheeler gauge \cite{Detweiler:2005kq,Field:2009kk,Field:2010xn,Hopper:2010uv,Hopper:2012ty,Osburn:2015duj,Hopper:2017qus,Hopper:2017iyq}
were found to be much easier to implement and yielded much more accurate results. However, with
regularization in Regge-Wheeler gauge less well understood, those calculations were restricted to
the computation of gauge-invariant quantities. Perhaps more
importantly, they are restricted to Schwarzschild spacetime, meaning they can not be used in
astrophysically realistic cases where the larger BH is spinning. The radiation gauge---in
which one solves the Teukolsky equation \cite{Teukolsky:1972my,Teukolsky:1973ha} for a single complex pseudo-scalar $\psi_4$ (or,
equivalently, $\psi_0$)---retains much of the simplicity of the Regge-Wheeler gauge, but has the
significant benefit of being capable of describing perturbations of a Kerr BH. Furthermore,
recent progress has clarified subtle issues related to regularization \cite{Pound:2013faa} (including metric completion \cite{Merlin:2016boc,vandeMeent:2017fqk}) in radiation gauge, paving
the way for high-accuracy calculations of both gauge-invariant quantities and the GSF~\cite{vandeMeent:2015lxa,vandeMeent:2016pee,vandeMeent:2016hel,vandeMeent:2017bcc}.

Within these last two approaches (using Regge-Wheeler and radiation gauges) \emph{functional
methods} \cite{Leaver:1986JMP,Mano:1996vt,Sasaki:2003xr} have emerged as a particularly efficient means of achieving high accuracy when computing
the metric perturbation. Fundamentally, these methods rely on the fact that solutions of the
Teukolsky (or Regge-Wheeler) equation can be written as a convergent series of hypergeometric
functions. This essentially reduces the problem of computing the metric perturbation to the problem
of evaluating hypergeometric functions. The approach has proved very successful, enabling both highly accurate numerical calculations \cite{Shah:2013uya,Shah:2014tka,Johnson-McDaniel:2015vva,vandeMeent:2017bcc} and even exact results in the low-frequency--large-radius
(i.e. PN) regime \cite{Bini:2014zxa,Bini:2015bla,Bini:2015mza,Bini:2015xua,Bini:2015bfb,Kavanagh:2015lva,Hopper:2015icj,Kavanagh:2016idg,Bini:2016qtx,Bini:2016dvs,Kavanagh:2017wot}.
A different type of analytic treatment is possible for modelling the radiation from the last stage of inspiral into a nearly extremal BH, thanks to the enhanced conformal symmetry in this scenario \cite{Porfyriadis:2014fja,Hadar:2014dpa,Hadar:2015xpa,Hadar:2016vmk}. This method, based on a scheme of matched asymptotic expansions, has so far been applied to equatorial orbits \cite{Gralla:2015rpa,Compere:2017hsi}, with the GSF neglected.

\subsubsection{Cosmic censorship}

Independently of the goal of producing accurate waveforms for LISA data analysis, the GSF programme has also yielded several other important results. One particular area of interest has
been in the relevance of the GSF to answering questions about cosmic censorship.
Calculations based on test-particle motion made the surprising discovery that a test particle
falling into a Kerr BH had the potential to increase the BH spin past the extremal
limit, thus yielding a naked singularity \cite{Jacobson:2009kt}. (Analogous cases exist
where an electric charge falling into a charged (Reissner-Nordstr\"om) BH may cause the
charge on the BH to increase past the extremal limit
\cite{Hubeny:1998ga,Saa:2011wq,Gao:2012ca}.) The intuitive expectation is that this is an artifact
of the test-particle approximation, and that by including higher-order terms in this approximation
the GSF may in effect act as a ``cosmic censor'' by preventing over-charging and restoring
cosmic censorship \cite{Barausse:2010ka}. Several works have explored this issue in detail, studying
the self-force on electric charges falling into a Reissner-N\"ordstrom BH
\cite{Zimmerman:2012zu,Revelar:2017sem} and on a massive particle falling into a Kerr BH
\cite{Colleoni:2015afa,Colleoni:2015ena}. These works demonstrated with explicit calculations how 
the overspinning or overcharging scenarios are averted once the full effect of the self-force is 
taken into account (a result later rigorously proven in a more general context \cite{Sorce:2017dst}).

\subsection{Remaining challenges and prospects}

While the perturbative (self-force) approach has proven highly effective to date, there remain
several important and challenging areas for further development. Here, we list a number of the most
important future challenges and prospects for the GSF programme.

\subsubsection{Efficient incorporation of self-force information into waveform models}

There are at least two key aspects to producing an EMRI model: (i) computing the GSF; and (ii) using the GSF to actually drive an inspiral. Unfortunately, despite the
substantial advances in calculational approaches, GSF calculations are still much too slow
to be useful on their own as a means for producing LISA gravitational waveforms. Existing work has
been able to produce first-order GSF driven inspirals for a small number of cases \cite{Diener:2011cc,Warburton:2011fk,Osburn:2015duj,Warburton:2017sxk} but when
one takes into account the large parameter space of EMRI systems it is clear that these existing
methods are inadequate. It is therefore important to develop efficient methods for incorporating
GSF information into EMRI models and waveforms. There has been some promising recent
progress in this direction, with fast ``kludge'' codes producing approximate (but not sufficiently
accurate) inspirals \cite{Glampedakis:2002ya,Hughes:2005qb,Drasco:2005kz,Sundararajan:2007jg,Sundararajan:2008zm,Chua:2017ujo,Berry:2012im}, and with the emergence of mathematical frameworks based on near-identity
transformations \cite{vandeMeent:2018rms}, renormalization group methods~\cite{Galley:2016zee} and
two-timescale expansions \cite{Moxon:2017ozd}.

To complicate matters, 
inspirals generically go through a number of transient resonances, when the momentary radial and polar frequencies of the orbit occur in a small rational ratio. During such resonances, approximations based on adiabaticity break down \cite{Flanagan:2010cd,Flanagan:2012kg,vandeMeent:2013sza,vandeMeent:2014raa,Brink:2013nna}. Works so far have mapped the locations of resonances in the inspiral parameter space, studied how the orbital parameters (including energy and angular momentum) experience a ``jump'' upon resonant crossing, and illustrated how the magnitude of the jump depends sensitively on the precise resonant phase (the relative phase between the radial and polar motions at resonance). The impact of resonances on the detectability of EMRIs with LISA was studied in Refs.~\cite{Berry:2016bit,Berry:2017cty} using an approximate model of the resonant crossing. But so far there has been no actual calculation of the orbital evolution through a resonance. Now that GSF codes for generic orbits are finally at hand, such calculations become possible, in principle. There is a vital need to perform such calculations, in order to allow orbital evolution methods to safely pass through resonances without a significant loss in the accuracy of the inspiral model.

\subsubsection{Producing accurate waveform models: self-consistent evolution and second-order gravitational self-force}

Possibly the most challenging outstanding obstacle to reaching the sub-radian phase accuracy
required for LISA data analysis is the fact that the first-order GSF on its
own is insufficient and one must also incorporate information at second perturbative order \cite{Hinderer:2008dm}. There are ongoing efforts to develop tools for computing the second order GSF \cite{Pound:2009sm,Pound:2012nt,Warburton:2013lea,Pound:2014xva,Pound:2014koa,Pound:2015wva,Wardell:2015ada,Miller:2016hjv,Pound:2017psq,Moxon:2017ozd}, but, despite significant
progress, a full calculation of the second-order metric perturbation has yet to be completed.

One of the challenges of the GSF problem when considered through second order is that it is
naturally formulated as a self-consistent problem, whereby the coupled equations for the metric
perturbation and for the particle worldline are evolved simultaneously. Indeed, this
self-consistent evolution has yet to be completed even for the first-order GSF (it has,
however, been done for the toy-model scalar charge case \cite{Diener:2011cc}). Even when methods
are developed for computing the second order GSF, it will remain a further challenge to
incorporate this information into a self-consistent evolution scheme.

\subsubsection{Gravitational Green Function}

One of the first proposals for a practical method for computing the GSF was based on writing
the regularized metric perturbation in terms of a convolution, integrating the Green function for
the wave equation along the worldline of the particle \cite{Poisson:Wiseman:1998,Anderson:2005gb}. It took several years for this idea to
be turned into a complete calculation of the GSF, and even then the results were restricted
to the toy-model problem of a scalar charge moving in Schwarzschild spacetime \cite{Casals:2013mpa,Wardell:2014kea}. Despite this
deficiency, the Green function approach has produced a novel perspective on the GSF problem.

In principle, the methods used for a scalar field in Schwarzschild spacetime should be applicable to 
the GSF problem in Kerr. The challenge is two-fold: (i) to actually adapt the methods to the Kerr problem and to the relevant wave equation, which is the linearized Einstein equation in the Lorenz gauge;
and (ii) to explore whether and how one can instead work with the much simpler Teukolsky wave equation.  

\subsubsection{Internal-structure effects}

The vast majority of GSF calculations to date have been based on the assumption that the smaller object is spherically symmetric and non-spinning. 
This idealization ignores the possibility that the smaller BH may be spinning, or more generally that other internal-structure effects may be relevant. Unfortunately, this picture is inadequate; certain internal-structure effects can make important contributions to the equations of motion. For example, the coupling of the
small body's spin to the larger BH's spacetime curvature (commonly referred to as the
Mathisson-Papapetrou force) is expected to contribute to the phase evolution of a typical EMRI 
at the same order as the conservative piece of the first-order GSF
\cite{Ruangsri:2015cvg}. Furthermore, finite internal-structure is likely to be even more important in
the case that the smaller body is a NS.

While there has been some progress in assessing the contribution from the smaller body's spin to
the motion
\cite{Faye:2006gx,Han:2010tp,Ruangsri:2015cvg,Harms:2016ctx,Warburton:2017sxk,Maia:2017gxn,Maia:2017yok,Lukes-Gerakopoulos:2017vkj},
existing work has focused on flux-based calculations or on the PN regime. It remains an outstanding challenge to determine the influence of internal-structure effects on the GSF, especially at second order.

\subsubsection{EMRIs in alternative theories of gravity}

In all of the discussion so far, we have made one overarching assumption: that BHs behave
as described by General Relativity (GR). However, with EMRIs we have the exciting prospect of not simply
assuming this fact, but of testing its validity with exquisite precision.  
There is initial work on the self-force in the context of scalar-tensor gravity \cite{Zimmerman:2015hua}, but much more remains to be done to
establish exactly what EMRIs can do to test the validity of GR (and how) when
pushed to its most extreme limits. Much more on this in Chapter III (see, in particular, Sec.~\ref{sec:ringdown} therein)

\subsubsection{Open tools and datasets}

While there has been significant progress in developing tools for computing the GSF, much of it has been ad-hoc, with individual groups developing their own private tools
and codes. Now that a clear picture has emerged of exactly which are the most useful methods and
tools, the community has begun to combine their efforts. This has lead to the development of a
number of initiatives, including (i) tabulated results for Kerr quasinormal modes and their excitation
factors~\cite{Cardosoweb,Bertiweb}; (ii) open source ``kludge'' codes for generating an approximate waveform for EMRIS
\cite{EKS,Chua:2017ujo}; and (ii) online repositories of self-force results \cite{BHPC}. It
is important for such efforts to continue, so that the results of the many years of development of GSF tools and methods are available to the widest possible user base. One promising
initiative in this direction is the ongoing development of the \emph{Black Hole Perturbation
Toolkit} \cite{BHPT}, a free and open source set of codes and results produced by the GSF community.

\section{Post-Newtonian and Post-Minkowskian Methods} \label{Sec:PN}
\vspace{-3mm}
{\it Contributor:} A. Le Tiec
\vspace{3mm}

\subsection{Background}

The PN formalism is an approximation method in GR that is well suited to describe the orbital motion and the GW emission from binary systems of compact objects, in a regime where the orbital velocity is small compared to the speed of light and the gravitational fields are weak. This approximation method has played a key role in the recent detections, by the LIGO and Virgo observatories, of GWs generated by inspiralling and merging BH and NS binaries \cite{Abbott:2016blz,Abbott:2016nmj,Abbott:2017vtc,Abbott:2017oio,TheLIGOScientific:2017qsa}, by providing accurate template waveforms to search for those signals and to interpret them. Here we give a brief overview of the application of the PN approximation to binary systems of compact objects, focusing on recent developments and future prospects. See the review articles \cite{FuIt.07,Blanchet:2011wga,Schafer:2009dq,Foffa:2013qca,Rothstein:2014sra,Blanchet:2013haa,Porto:2016pyg,Schafer:2018kuf} and the textbooks \cite{Maggiore:1900zz,PoissonWill} for more information.

In PN theory, relativistic corrections to the Newtonian solution are incorporated in a systematic manner into the equations of motion (EOM) and the radiation field, order by order in the small parameter $v^2/c^2 \sim Gm/(c^2r)$, where $v$ and $r$ are the typical relative orbital velocity and binary separation, $m$ is the sum of the component masses, and we used the fact that $v^2 \sim Gm/r$ for bound motion. (The most promising sources for current and future GW detectors are bound systems of compact objects.)

Another important approximation method is the post-Minkowskian (PM) approximation, or non-linearity expansion in Newton's gravitational constant $G$, which assumes weak fields ($Gm/c^2 r \ll 1$) but unrestricted speeds ($v^2 / c^2 \lesssim 1$), and perturbs about the limit of special relativity. In fact, the construction of accurate gravitational waveforms for inspiralling compact binaries requires a combination of PN and PM techniques in order to solve two coupled problems, namely the problem of motion and that of wave generation.

\hspace{-0.07cm}The two-body EOM have been derived in a PN framework using three  well-developed sets of techniques in classical GR: (i) the PN iteration of the Einstein field equations in harmonic coordinates \cite{Blanchet:1998vx,Blanchet:2000ub,deAndrade:2000gf,Blanchet:2003gy,Mitchell:2007ea}, (ii) the Arnowitt-Deser-Misner (ADM) canonical Hamiltonian formalism \cite{Jaranowski:1997ky,Damour:2000kk,Damour:2001bu}, and (iii) a surface integral approach pioneered by Einstein, Infeld and Hoffmann \cite{Itoh:2001np,Itoh:2003fy,Itoh:2003fz}. By the early 2000s, each of these approaches has independently produced a computation of the EOM for binary systems of non-spinning compact objects through the 3rd PN order (3PN).\footnote{By convention, ``$n$PN'' refers to EOM terms that are $O(1/c^{2n})$ smaller than the Newtonian acceleration, or, in the radiation field, smaller by that factor relative to the standard quadrupolar field.} More recently, the application of effective field theory (EFT) methods \cite{Goldberger:2004jt}, inspired from quantum field theory, has provided an additional independent derivation of the 3PN EOM \cite{Foffa:2011ub}. All of those results were shown to be in perfect agreement. Moreover, the 3.5PN terms---which constitute a 1PN relative correction to the leading radiation-reaction force---are also known \cite{Iyer:1993xi,Iyer:1995rn,Jaranowski:1996nv,Pati:2002ux,Konigsdorffer:2003ue,Nissanke:2004er}.

At the same time, the problem of radiation (i.e., computing the field in the far/wave zone) has been extensively investigated within the multipolar PM wave generation formalism of Blanchet and Damour \cite{Blanchet:1985sp,Blanchet:1986dk,Blanchet:1998in}, using the ``direct integration of the relaxed Einstein equation'' approach of Will, Wiseman and Pati \cite{Will:1996zj,Pati:2000vt}, and more recently with EFT techniques \cite{Goldberger:2009qd,Ross:2012fc}. The application of these formalisms to non-spinning compact binaries has, so far, resulted in the computation of the GW phase up to the relative 3.5PN (resp. 3PN) order for quasi-circular (resp.~quasi-eccentric) orbits \cite{Blanchet:2001aw,Blanchet:2001ax,Blanchet:2004ek,Blanchet:2005tk,Arun:2007rg,Arun:2007sg,Arun:2009mc,Moore:2016qxz}, while amplitude corrections in the GW polarizations are known to 3PN order, and even to 3.5PN order for the quadrupolar mode \cite{Arun:2004ff,Kidder:2007gz,Kidder:2007rt,Blanchet:2008je,Faye:2012we,Mishra:2015bqa}.
\subsection{Recent developments}
Over the last five years or so, significant progress on PN modelling of compact binary systems has been achieved on multiple fronts, including (i) the extension of the EOM to 4PN order for non-spinning bodies (with partial results also obtained for aspects of the two-body dynamics at the 5PN order \cite{Blanchet:2010zd,Barack:2010ny,Foffa:2013gja}), (ii) the inclusion of spin effects in the binary dynamics and waveform, (iii) the comparison of several PN predictions to those from GSF theory, and (iv) the derivation of general laws controlling the mechanics of compact binaries.

\subsubsection{4PN equations of motion for non-spinning compact-object binaries}

Recently, the computation of the two-body EOM has been extended to 4PN order, by using both the canonical Hamiltonian framework in ADM-TT coordinates \cite{Jaranowski:2012eb,Jaranowski:2013lca,Damour:2014jta,Jaranowski:2015lha,Damour:2016abl,Damour:2017ced} and a Fokker Lagrangian approach in harmonic coordinates \cite{Bernard:2015njp,Bernard:2016wrg,Bernard:2017bvn,Marchand:2017pir,Bernard:2017ktp}. Partial results at 4PN order have also been obtained using EFT techniques \cite{Foffa:2012rn,Foffa:2016rgu,Porto:2017dgs}. All of those high-order PN calculations resort to a point-particle model for the (non-spinning) compact objects, and rely on dimensional regularization to treat the local ultraviolet (UV) divergences that are associated with the use of point particles. The new 4PN results have been used to inform the EOB framework \cite{Damour:2015isa}, a semi-analytic model of the binary dynamics and wave emission (see Sec.\ \ref{Sec:EOB} below).

The occurrence at the 4PN order of infrared (IR) divergences of spatial integrals led to the introduction of several \textit{ambiguity parameters}; one in the ADM Hamiltonian approach and two in the Fokker Lagrangian approach. One of those IR ambiguity parameters was initially fixed by requiring agreement with an analytical GSF calculation \cite{Bini:2013zaa} of the so-called Detweiler redshift along circular orbits \cite{Detweiler:2008ft,Sago:2008id}. Recently, however, Marchand et al. \cite{Marchand:2017pir} gave the first complete (i.e., ambiguity-free) derivation of the 4PN EOM. The last remaining ambiguity parameter was determined from first principles, by resorting to a matching between the near-zone and far-zone fields, together with a computation of the conservative 4PN tail effect in $d$ dimensions, allowing to treat both UV and IR divergences using dimensional regularization.

Another interesting (and related) feature of the binary dynamics at the 4PN order is that it becomes \textit{non-local} in time \cite{Damour:2014jta,Bernard:2015njp}, because of the occurence of a GW tail effect at that order: gravitational radiation that gets scattered off the background spacetime curvature backreacts on the orbital motion at later times, such that the binary's dynamics at a given moment in time depends on its entire past history \cite{Blanchet:1987wq,Foffa:2011np,Galley:2015kus}.

\subsubsection{Spin effects in the binary dynamics and gravitational waveform}

\begin{figure}[h!]
	\includegraphics[width=0.7\linewidth]{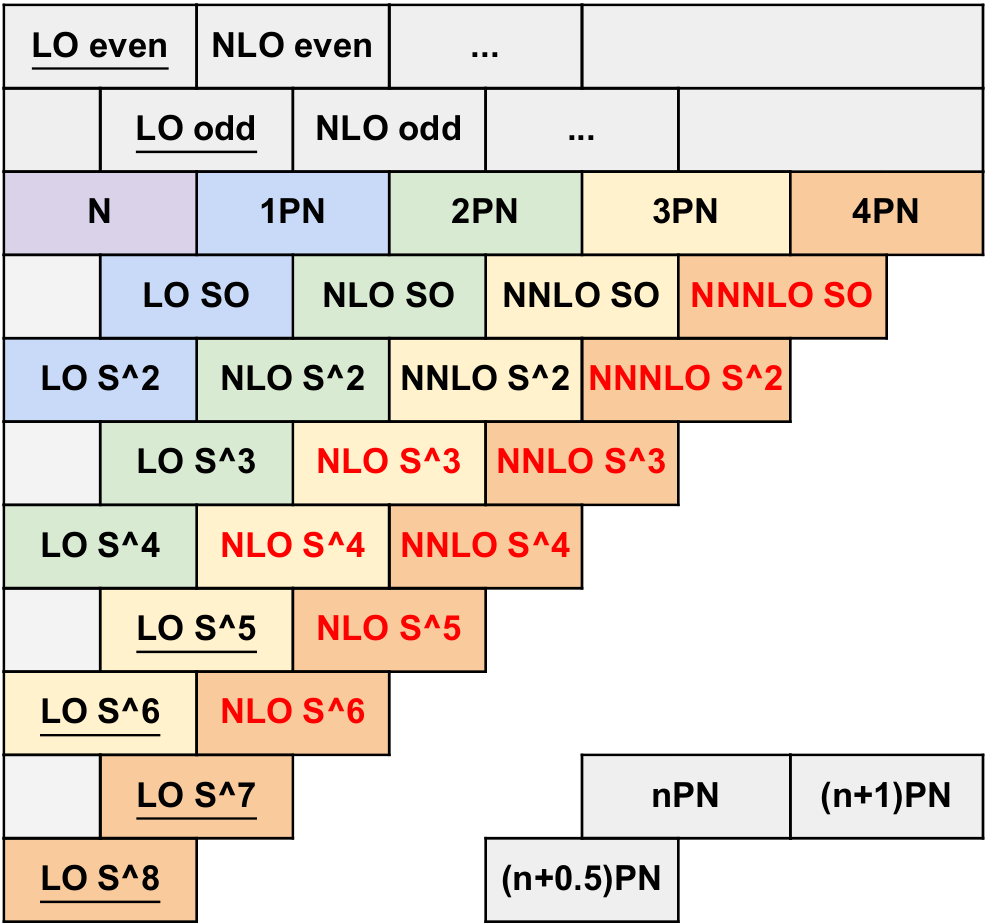}
	\caption{Contributions to the two-body Hamiltonian in the PN spin expansion, for arbitrary-mass-ratio binaries with spin induced multipole moments. Contributions in red are yet to be calculated. LO stands for ``leading order'', NLO for ``next-to-leading order'', and so on. SO stands for ``spin-orbit''. Figure from Ref.~\cite{Vines:2016qwa}.}
	\label{fig:PNtable}
\end{figure}

\hspace{-0.05cm}Since stellar-mass and/or supermassive BHs may carry significant spins \cite{Reynolds:2013rva,Miller:2014aaa}, much effort has recently been devoted to include spin effects in PN template waveforms. In particular, spin-orbit coupling terms linear in either of the two spins have been computed up to the next-to-next-to leading order, corresponding to 3.5PN order in the EOM, using the ADM Hamiltonian framework \cite{Hartung:2011te,Hartung:2013dza}, the PN iteration of the Einstein field equations in harmonic coordinates \cite{Marsat:2012fn,Bohe:2012mr}, and EFT techniques \cite{Levi:2015uxa}. Spin-spin coupling terms proportional to the product of the two spins have also been computed to the next-to-next-to leading order, corresponding to 4PN order in the EOM, using the ADM Hamiltonian and EFT formalisms \cite{Hartung:2011ea,Levi:2011eq,Levi:2014sba,Levi:2015ixa,Levi:2016ofk}. The leading order 3.5PN cubic-in-spin and 4PN quartic-in-spin contributions to the binary dynamics are also known for generic compact bodies \cite{Hergt:2007ha,Hergt:2008jn,Marsat:2014xea,Levi:2014gsa,Vaidya:2014kza}, as well as all higher-order-in-spin contributions for BBHs (to leading PN order) \cite{Vines:2016qwa}. All these results are summarized in Fig.~\ref{fig:PNtable}.
2PN BH binary spin precession was recently revisited using multi-timescale methods~\cite{Gerosa:2015tea,Kesden:2014sla},
uncovering new phenomenology such as precessional instabilities~\cite{Gerosa:2015hba} and nutational resonances~\cite{Zhao:2017tro}.

Spin-related effects on the far-zone field have also been computed to high orders, for compact binaries on quasi-circular orbits. To linear order in the spins, those effects are known up to the relative 4PN order in the GW energy flux and phasing \cite{Bohe:2013cla,Marsat:2013caa}, and to 2PN in the wave polarizations~\cite{Buonanno:2012rv,Mishra:2016whh}. At quadratic order in the spins, the contributions to the GW energy flux and phasing have been computed to 3PN order \cite{Porto:2010zg,Bohe:2015ana}, and partial results were derived for amplitude corrections to 2.5PN order \cite{Porto:2012as}. The leading 3.5PN cubic-in-spin effects in the GW energy flux and phasing are known as well \cite{Marsat:2014xea}.

Finally, some recent works have uncovered remarkable relationships between the PN \cite{Vines:2016qwa} and PM \cite{Vines:2017hyw} dynamics of a binary system of spinning BHs with an arbritrary mass ratio on the one hand, and that of a test BH in a Kerr background spacetime on the other hand. Those results are especially relevant for the ongoing development of EOB models for spinning BH binaries (see Sec.\ \ref{Sec:EOB}), and in fact give new insight into the energy map at the core of such models.

\subsubsection{Comparisons to perturbative gravitational self-force calculations}

The GWs generated by a coalescing compact binary system are not the only observable of interest. As we have described in Sec.~\ref{Sec:perturbations}, over recent years, several \textit{conservative} effects on the orbital dynamics of compact-object binaries moving along quasi-circular orbits have been used to compare the predictions of the PN approximation to those of the GSF framework, by making use of gauge-invariant quantities such as (i) the Detweiler redshift \cite{Detweiler:2008ft,Blanchet:2009sd,Blanchet:2010zd,Damour:2009sm,Blanchet:2013txa,Blanchet:2014bza}, (ii) the relativistic periastron advance \cite{Barack:2010ny,LeTiec:2011bk,Tiec:2013twa,vandeMeent:2016hel}, (iii) the geodetic spin precession frequency \cite{Dolan:2013roa}, and (iv) various tidal invariants \cite{Dolan:2014pja,Nolan:2015vpa}, all computed as functions of the circular-orbit frequency of the binary. Some of these comparisons were extended to generic bound (eccentric) orbits~\cite{Barack:2011ed,Akcay:2015pza,Akcay:2016dku}. All of those comparisons showed perfect agreement in the common domain of validity of the two approximation schemes, thus providing crucial tests for both methods. Building on recent progress on the second-order GSF problem \cite{Pound:2012nt,Pound:2012dk,Pound:2014xva,Pound:2015fma,Miller:2016hjv,Pound:2017psq}, we expect such comparisons to be extended to second order in the mass ratio, e.g. by using the redshift variable \cite{Pound:2014koa}.

Independently, the BH perturbative techniques of Mano, Suzuki and Takasugi \cite{Mano:1996vt,Mano:1996gn} have been applied to compute analytically, up to very high orders, the PN expansions of the GSF contributions to the redshift for circular \cite{Bini:2013rfa,Bini:2014nfa,Bini:2015bla,Kavanagh:2016idg} and eccentric \cite{Bini:2015xua,Bini:2015bfb,Hopper:2015icj,Bini:2016qtx,Bini:2016dvs} orbits, the geodetic spin precession frequency \cite{Bini:2014ica,Bini:2015mza,Kavanagh:2017wot}, and various tidal invariants \cite{Bini:2014zxa,Shah:2015nva,Kavanagh:2015lva}. Additionally, using similar techniques, some of those quantities have been computed numerically, with very high accuracy, allowing the extraction of the exact, analytical values of many PN coefficients \cite{Shah:2013uya,Johnson-McDaniel:2015vva,vandeMeent:2015lxa}.

\subsubsection{First law of compact binary mechanics}

The conservative dynamics of a binary system of compact objects has a fundamental property now known as the \textit{first law of binary mechanics} \cite{LeTiec:2011ab}. Remarkably, this variational formula can be used to relate local physical quantities that characterize each body (e.g. the redshift) to global quantites that characterize the binary system (e.g. the binding energy). For point-particle binaries moving along circular orbits, this law is a particular case of a more general result, valid for systems of BHs and extended matter sources \cite{Friedman:2001pf}.

Using the ADM Hamiltonian formalism, the first law of \cite{LeTiec:2011ab} was generalized to spinning point particles, for spins (anti-)aligned with the orbital angular momentum~\cite{Blanchet:2012at}, and to non-spinning binaries moving along generic bound (eccentric) orbits \cite{Tiec:2015cxa}. The derivation of the first law for eccentric motion was then extended to account for the non-locality in time of the orbital dynamics due to the occurence at the 4PN order of a GW tail effect~\cite{Blanchet:2017rcn}. These various laws were derived on general grounds, assuming only that the conservative dynamics of the binary derives from an autonomous canonical Hamiltonian. (First-law-type relationships have also been derived in the context of linear BH perturbation theory and the GSF framework \cite{Gralla:2012dm,Tiec:2013kua,Fujita:2016igj}.) Moreover, they have been checked to hold true up to 3PN order, and even up to 5PN order for some logarithmic terms.

So far the first laws have been applied to (i) determine the numerical value of the aforementioned ambiguity parameter appearing in derivations of the 4PN two-body EOM \cite{Bini:2013zaa}, (ii) calculate the exact linear-in-the-mass-ratio contributions to the binary's binding energy and angular momentum for circular motion \cite{LeTiec:2011dp}, (iii) compute the shift in the frequencies of the Schwarzschild and Kerr innermost stable circular orbits induced by the (conservative) GSF~\cite{Barack:2009ey,Damour:2009sm,Barack:2010tm,LeTiec:2011dp,Akcay:2012ea,Isoyama:2014mja,vandeMeent:2016hel}, (iv) test the weak cosmic censorship conjecture in a scenario where a massive particle subject to the GSF falls into a nonrotating BH along unbound orbits \cite{Colleoni:2015afa,Colleoni:2015ena}, (v) calibrate the effective potentials that enter the EOB model for circular~\cite{Barausse:2011dq,Akcay:2012ea} and mildly eccentric orbits~\cite{Akcay:2015pjz,Bini:2015bfb,Bini:2016qtx}, and spin-orbit couplings for spinning binaries \cite{Bini:2015xua}, and (vi) define the analogue of the redshift of a particle for BHs in NR simulations, thus allowing further comparisons to PN and GSF calculations \cite{Zimmerman:2016ajr,LeTiec:2017ebm}.

\subsection{Prospects}

On the theoretical side, an important goal is to extend the knowledge of the GW phase to the relative 4.5PN accuracy, at least for non-spinning binaries on quasi-circular orbits. (Partial results for some specific tail-related effects were recently derived \cite{Marchand:2016vox}.) This is essential in order to keep model systematics as a sub-dominant source of error when processing observed GW signals \cite{Abbott:2016wiq}. Accomplishing that will require, in particular, the calculation of the mass-type quadrupole moment of the binary to 4PN order and the current-type quadrupole and mass-type octupole moments to 3PN order. Moreover, some spin contributions to the waveform---both in the phasing and amplitude corrections---still have to be computed to reach the 4PN level, especially at quadratic order in the spins.

Most of the PN results reviewed here have been established for circularized binaries, and often for spins aligned or anti-aligned with the orbital angular momentum. It is important to extend this large body of work to generic, eccentric, precessing systems. Progress on the two-body scattering problem would also be desirable \cite{Damour:2014afa,Bini:2018ywr,Bini:2017wfr}. Additionally, much of what has been achieved in the context of GR could be done as well for well-motivated alternative theories of gravitation, such as for scalar-tensor gravity (e.g. \cite{Mirshekari:2013vb,Lang:2013fna,Lang:2014osa,Sennett:2016klh,Bernard:2018hta}) or in quadratic gravity \cite{Yagi:2011xp,Yagi:2013mbt}.

The first law of binary mechanics reviewed above could be extended to generic, precessing spinning systems, and the effects of higher-order spin-induced multipoles should be investigated. Recent work on the PM approximation applied to the gravitational dynamics of compact binaries has given new insight into the EOB model \cite{Damour:2016gwp,Bini:2017xzy,Vines:2017hyw,Damour:2017zjx}, and in particular into the energy map therein. These two lines of research may improve our physical understanding of the general relativistic two-body problem.

\begin{figure}[h!]
	\includegraphics[width=\linewidth]{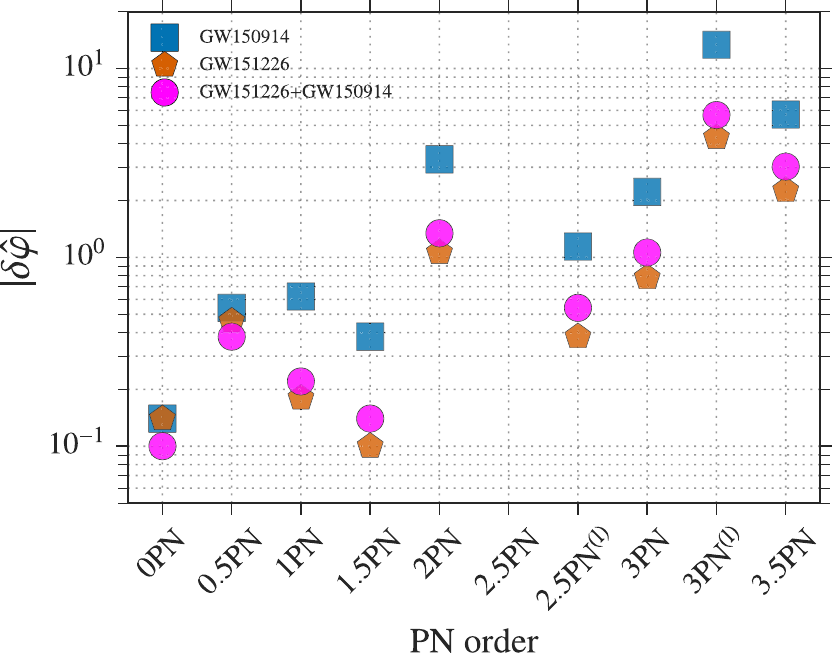}
	\caption{Two GW detections of inspiralling BH binaries were used to set bounds on possible deviations from their GR values of various PN coefficients that appear in the expression for the GW phase. Figure from Ref.~\cite{TheLIGOScientific:2016pea}.}
	\label{fig:test_GR}
\end{figure}

On the observational side, future GW detections from inspiralling compact-object binaries will allow testing GR in the strong-field/radiative regime, by constraining possible deviations from their GR values of the various PN coefficients that appear in the expression of the phase. This, in particular, can be used to test some important nonlinear features of GR such as the GW tail, tail-of-tail and nonlinear memory effects. Indeed, the first detections of GWs from inspiralling BH binaries have already been used to set bounds on these PN coefficients, including an $O(10\%)$ constraint on the leading tail effect at the 1.5PN order~\cite{TheLIGOScientific:2016pea}; see Fig.~\ref{fig:test_GR} (or Ref.~\cite{Abbott:2017oio} for a more up-to-date version thereof). More detections with a wider network of increasingly senstive interferometric GW detectors will of course improve those bounds.

Finally, following the official selection of the LISA mission by the European Space Agency (ESA), with a launch planned for 2034, we foresee an increased level of activity in source modelling of binary systems of MBHs and EMRIs, two promising classes of sources for a mHz GW antenna in space. This will motivate more work at the interface between the PN approximation and GSF theory.

\section{Numerical Relativity and the Astrophysics of Black Hole Binaries }\label{Sec:NR}
\vspace{-3mm}
{\it Contributor:} P. Schmidt
\vspace{3mm}

The year of $2005$ marked a remarkable breakthrough: 
the first successful numerical simulation of---and the extraction of the GWs from---an inspiraling pair of BHs through their merger and final ringdown~\cite{Pretorius:2005gq, Campanelli:2005dd, Baker:2005vv} (see e.g. \cite{Sperhake:2014wpa} for a review). 

NR provides us with accurate gravitational waveforms as predicted by GR. BBHs cover an eight-dimensional parameter space spanned by the mass ratio $q=m_1/m_2$, the spin angular momenta $\mathbf{S}_i$ and the eccentricity $e$.  Simulations are computationally extremely expensive, thus the large BBH parameter space is still sparsely sampled. Nevertheless, NR waveforms already play a crucial part in the construction and verification of semi-analytic waveform models used in GW searches, which facilitated the first observations of GWs from BBH mergers~\cite{Abbott:2016blz,Abbott:2016nmj,TheLIGOScientific:2016pea,Abbott:2017vtc,Abbott:2017oio}. Furthermore, they play a key role in the estimation of source properties and in facilitating important tests of GR in its most extreme dynamical regime. 

Since the initial breakthrough, NR has made significant progress: from the first simulations of equal-mass non-spinning BBHs spanning only the last few orbits~\cite{Pretorius:2005gq,Campanelli:2005dd, Baker:2005vv}, to a realm of simulations exploring aligned-spin~\cite{Campanelli:2006uy, Hannam:2007wf, Dain:2008ck} as well as precessing quasi-circular binaries~\cite{Campanelli:2006fy, Campanelli:2008nk, Schmidt:2010it}, eccentric-orbit binaries~\cite{Sperhake:2007gu, Hinder:2007qu}, and evolutions long enough to reach into the early-inspiral regime where they can be matched onto PN models~\cite{Szilagyi:2015rwa}. 

Today, several codes are capable of stably evolving BBHs and extracting their GW signal. They can roughly be divided into two categories: finite-differencing codes including BAM~\cite{Bruegmann:2006at}, the Einstein Toolkit~\cite{Loffler:2011ay}, LazEv~\cite{Campanelli:2005dd, Campanelli:2006gf, Campanelli:2006uy, Campanelli:2007ew}, MAYA~\cite{Vaishnav:2007nm}, LEAN~\cite{Sperhake:2006cy} as well as the codes described in Refs.~\cite{Baker:2006yw, Pretorius:2004jg, Pretorius:2005gq};
and (pseudo-)spectral codes such as the Spectral Einstein Code (\texttt{SpEC})~\cite{Scheel:2006gg}. Other evolution codes currently under development include~\cite{Hilditch:2015aba} and~\cite{Clough:2015sqa}.
To date, these codes have together produced several thousands of BBH simulations~\cite{Mroue:2013xna, Chu:2015kft, Jani:2016wkt, Healy:2017psd}. 

\subsection{Current status}
\label{s:NR}

{\it Excision and puncture.} In BBH simulations one numerically solves the vacuum Einstein equations with initial conditions that approximate a pair of separate BHs at some initial moment.  An obvious complication is the presence of spacetime singularities inside the BHs, where the solution diverges. There are two approaches to this problem. The first is {\it excision}~\cite{0264-9381-4-5-013}, whereby a region around the singularity is excised (removed) from the numerical domain. As no information can propagate outwards from the interior of the BH, the physical content of the numerical solution outside the BHs is unaffected. Since the excision boundary is spacelike (not timelike), one cannot and does not specify boundary conditions on it.
Instead, the main technical challenge lies in ensuring that the excision boundary remains spacelike. The risk, for example, is that part of the boundary may become timelike if numerical noise isn't properly controlled~\cite{Cook:2004kt,Szilagyi:2014fna,Kidder:2000yq}. It must also be ensured that non-physical gauge modes do not lead to numerical instabilities. Excision is used in \texttt{SpEC}, for example.

The second common way to deal with the BH singularities is to choose singularity-avoiding coordinates. This is achieved by representing BHs as compactified topological wormholes~\cite{Brandt:1997tf} or infinitely long cylinders (``trumpets'') \cite{Hannam:2008sg}, known as {\it puncture} initial data. Specific gauge conditions allow the punctures to move across the numerical grid, giving this approach the name moving punctures~\cite{Campanelli:2005dd, Baker:2005vv, vanMeter:2006vi, Gundlach:2006tw, Hannam:2008sg}.  

{\it Initial Data.}
No exact solutions are known, in general, for the BBH metric on the initial spatial surface, so one resorts to approximate initial conditions. 
Two types of initial data are commonly used: conformally flat and conformally curved. Most simulations that incorporate moving punctures use conformally flat initial data. Under this assumption, three of the four constraint equations (themselves a subset of the full Einstein field equations) are given analytically in terms of the Bowen-York solutions~\cite{Bowen:1980yu}. The maximal possible angular momentum in this approach is $a/m=0.93$, known as the Bowen-York limit~\cite{Dain:2002ee}. In order to go beyond this limit, conformally curved initial data have to be constructed. 
For codes that use excision, these can be obtained by solving the extended conformal thin sandwich (CTS) equations~\cite{York:1998hy} with quasi-elliptical boundary conditions~\cite{Gourgoulhon:2001ec, Grandclement:2001ed, Cook:2001wi,Pfeiffer:2002iy}. The initial spatial metric is proportional to a superposition of the metrics of two boosted Kerr-Schild BHs~\cite{Lovelace:2008tw}.
More recently, the first non-conformally flat initial data within the moving punctures framework have been constructed~\cite{Ruchlin:2014zva, Zlochower:2017bbg} by superposing the metrics and extrinsic curvatures of two Lorentz-boosted, conformally Kerr BHs. \\

{\it Evolution Systems.}
The successful evolution of a BBH spacetime further requires a numerically stable formulation of the Einstein field equations and appropriate gauge choices. Long-term stable evolutions today are most commonly performed with either a variant of the generalised harmonic~\cite{Pretorius:2005gq, Friedrich:2000qv, Lindblom:2005qh} or the Baumgarte-Shapiro-Shibata-Nakamura (BSSN) formulation~\cite{PhysRevD.52.5428, Baumgarte:1998te}. Another formulation, Z4, combines constraint preserving boundary conditions with an evolution system very close to BSSN~\cite{Bona:2003fj, Bona:2004yp, Gundlach:2005eh,Hilditch:2012fp}. More advanced Z4-type formulations, which are conformal and traceless, were developed in Refs.~\cite{Bernuzzi:2009ex,Alic:2011gg,Dumbser:2017okk}\\

{\it Gravitational Wave Extraction.}
The GW signal emitted through the inspiral, merger and ringdown is usually extracted from the Newman-Penrose curvature scalar $\Psi_4$~\cite{Newman:1961qr, Stewart:1990aa} associated with the computed metric. 
The extraction of the signal is typically performed on spheres of constant coordinate radius some distance from the binary, followed by extrapolation to infinity. The method of Cauchy-Characteristic Extraction (CCE)~\cite{Bishop:1996gt, Winicour:1999ba} allows us to extract the observable GW signal directly at future null infinity by matching the Cauchy evolution onto a characteristic evolution that extends the simulation to null infinity. In the Cauchy-Characteristic Extraction, the gravitational waveform is most naturally extracted from the Bondi news function~\cite{Bondi21, Sachs103}.

\subsection{Challenges}
\label{s:challenges}

{\it High spins.} Numerical simulations of close to maximally spinning BHs are still challenging to carry out due to difficulties in the construction of initial data as well as increasingly demanding accuracy requirements during the evolutions. This particularly affects binary configurations of unequal masses and arbitrary spin orientation, despite significant developments made in the past five years~\cite{Mroue:2013xna, Chu:2015kft, Jani:2016wkt, Healy:2017psd, Husa:2015iqa}. While spins up to 0.994 have been evolved stably~\cite{Scheel:2014ina}, extensive work is still underway to reach the proposed Novikov-Thorne limit of 0.998~\cite{Thorne:1974ve} and beyond.  \\

{\it High mass ratios.} The highest mass ratio BBH fully general-relativistic numerical simulation available to date is of $q =100$~\cite{Lousto:2010ut}. (However, this simulation follows a relatively small number of orbital cycles.)  For spinning BHs, simulations of mass ratio $q=18$ have been performed~\cite{Husa:2015iqa}. Both of these have been obtained in the moving punctures approach. For spectral methods, mass ratios higher than $q\sim 10$ are numerically very challenging. Generally, higher mass ratios are more computationally expensive as the disparate lengthscales demand higher spatial resolution, and because the number of orbital cycles increases in proportion to $q$. 
The choice of a Courant factor poses another problem in explicit evolution schemes, as it forces a small time step, which becomes increasingly challenging for high mass ratios. Implicit schemes could provide a potential solution to this~\cite{Lau:2011we}, but have not yet been applied to the fully relativistic binary problem.
Improvements in numerical technique, and perhaps a synergy with the perturbative methods reviewed in Sec.\ \ref{Sec:perturbations}, will be crucial for overcoming this problem in the hope of extending the reach of NR towards the extreme-mass-ratio regime.  \\

{\it Long simulations.} It is crucial for simulations to track the binary evolution from the early inspiral, where it can be matched to a PN model, all the way down to the final merger and ringdown. However, such long evolutions are computationally very expensive, requiring many months of CPU time with existing codes. Note that the run time is not simply linear in the evolution time: a longer run would usually also require a larger spatial domain in order to keep spatial boundaries out of causal contact.     
To date, only one complete numerical simulation reaching well into the PN regime has been produced, using a modified version of \texttt{SpEC}~\cite{Szilagyi:2015rwa}. Significant modifications to existing codes would need to be made in order to be able to generate long simulations on a production scale. \\

{\it Waveform accuracy.} As the sensitivity of GW detectors increases, further work will be required to  further improve models and assess their systematic errors \cite{Abbott:2016wiq}. 
Already, for large precessing spins, or large mass ratios, or non-negligible eccentricities, systemic modelling errors limit the accuracy with which LIGO and Virgo can extract source parameters
(see~\cite{Hannam:2009rd} for a more detailed discussion). This is due, in part, to relatively large errors in current models of high multipole-mode contributions, which we expect in the coming years to become resolvable in detected signals. 
\\

{\it Beyond General Relativity.} GW observations from merging compact binaries allow us to probe the strong-field regime and test GR (cf.\ Chapter III). While theory-agnostic ways to test deviations from GR are commonly used, testing a selection of well-motivated alternative theories through direct waveform comparison is desirable. First steps have been taken to simulate BBHs in alternative theories of gravity that admit BH solutions~\cite{Berti:2013gfa, Okounkova:2017yby}. However, many issues remain to be addressed, not least of which the fact that some of these theories may not even possess a well-posed initial value formulation (see~\cite{Papallo:2017qvl, Delsate:2014hba} for examples).

\subsection{Numerical Relativity and GW observations}
\label{s:obs}
NR plays an important role in GW astrophysics and data analysis. The semi-analytical waveform models employed to search for BBHs (see Sec.\ \ref{Sec:EOB} below) model the complete radiative evolution from the early inspiral, through the merger and to the final ringdown stage, with the later stages calibrated to BBH simulations~\cite{Khan:2015jqa, Bohe:2016gbl}. These models and more sophisticated ones incorporating more physical effects, for example precession~\cite{Apostolatos:1994mx, Kidder:1995zr}, are used to determine the fundamental properties of the GW source, i.e. its masses and spin angular momenta as well as extrinsic quantities such as the orbital orientation relative to the observatories~\cite{Hannam:2013oca, Pan:2013rra}. 

Alternatively, pure NR waveforms may be used directly~\cite{Abbott:2016apu,Lange:2017wki} or by the means of surrogate models~\cite{Blackman:2015pia, Blackman:2017dfb, Blackman:2017pcm}. But due to the computational cost of simulations, these have only been attempted so far on a very restricted portion of the parameter space. Despite this restriction, pure NR surrogate models have the advantage of incorporating more physical effects that may be limited or entirely neglected in currently available semi-analytic models.

The ringdown phase after the merger may be described by perturbation theory (see e.g.~\cite{Kokkotas:1999bd} for a review), but the amplitudes of the excited quasi-normal modes can only be obtained from numerical simulations. These are of particular interest as measuring the amplitudes of individual quasinormal modes would allow to map the final state of the merger to the properties of the progenitor BHs~\cite{Kamaretsos:2011um, Kamaretsos:2012bs} and to test the BH nature of the source~\cite{Berti:2005ys} (see also Sections~\ref{sec:nohair} and \ref{sec:hairyBHs}). 

In order to estimate the mass and spin angular momentum of the remnant BH, fits to NR simulations are essential~\cite{Barausse:2009uz, Hemberger:2013hsa, Hofmann:2016yih, Healy:2016lce, Jimenez-Forteza:2016oae,Buonanno:2007sv,Rezzolla:2007rz,Lousto:2009mf,Bernuzzi:2009ex,Alic:2011gg,Dumbser:2017okk}. Independent measurements of the binary properties from the inspiral portion of the GW signal and from the later stages of the binary evolution using fits from NR allow, when combined, to test the predictions from GR~\cite{Ghosh:2016qgn, TheLIGOScientific:2016src}.
While the phenomenological fit formulae have seen much improvement in recent years, modelling the final state of precessing BH binaries from numerical simulations still remains an open challenge.

Generically, when BHs merge, anisotropic emission of GWs leads to the build up of a net linear momentum. Due to the conservation of momentum, once the GW emission subsides, the remnant BH recoils (``kick''). While the recoil builds up during the entire binary evolution, it is largest during the non-linear merger phase. Kick velocities can be as high as several thousands ${\rm km}/{\rm s}$, with astrophysical consequences: a BH whose recoil velocity is larger than the escape velocity of its galaxy may leave its host. Numerical simulations are necessary to predict the recoil velocities~\cite{Gonzalez:2006md, Gonzalez:2007hi, Brugmann:2007zj, Lousto:2011kp} (a convenient surrogate model for these velocities was constructed recently in \cite{Gerosa:2018qay}). 
It has been suggested that it may be possible to directly measure the recoil speed from the GW signal by observing the induced differential Doppler shift throughout the inspiral, merger and ringdown~\cite{Gerosa:2016vip}.

Numerical simulations of the strong field regime are also crucial for exploring other spin-related phenomena such as ``spin-flips''~\cite{Lousto:2015uwa} caused by spin-spin coupling, whose signatures may be observable. Understanding the spin evolution and correctly modelling spin effects is crucial for mapping out the spin distribution of astrophysical BHs from GW observations.

Waveform models as well as fitting formulae for remnant properties are prone to systematic modelling errors. For GW observations with low SNR, the statistical uncertainty dominates over the systematic modelling error in the parameter measurement accuracy. Improved sensitivities for current and future GW observatories (including, especially, LISA \cite{Berti:2006ew,Cutler:2007mi}) will allow for high SNR observations reaching into the regime where systematic errors are accuracy limiting. This includes strongly inclined systems, higher-order modes, eccentricity, precession and kicks, all of which can be modelled more accurately through the inclusion of results from NR (see~\cite{Abbott:2016wiq} for a detailed discussion in the context of the first BBH observation GW150914).

\section{Numerical relativity in fundamental physics }\label{Sec:HE}
\vspace{-3mm}
{\it Contributor:} U. Sperhake
\vspace{3mm}

The standard model of particle physics and Einstein's theory
of GR provide us with an exquisite theoretical
framework to understand much of what we observe in the Universe.
From high-energy collisions at particle colliders to planetary motion,
the GW symphony of NS and BH
binaries and the cosmological evolution of the Universe at large,
the theoretical models give us remarkably accurate descriptions.
And yet, there are gaps in this picture that prompt us
to believe that something is wrong or incomplete. Galactic rotation
curves, strong gravitational lensing effects, X-ray observations
of galactic halos and the cosmic microwave background cannot
be explained in terms of the expected gravitational effects of the
visible matter \cite{Bertone:2004pz}. Either we are prepared to
accept the need to modify the laws of gravity, or there exists a
form of {\em dark matter} (DM) that at present we can not explain
satisfactorily with the SM (or both). Different DM candidates and their status are reviewed in Section~\ref{sec:BH}.
Likewise, the accelerated expansion of
the Universe \cite{Kowalski:2008ez} calls for an exotic form of
matter dubbed {\em dark energy} or (mathematically equivalently)
the introduction of a cosmological constant with a value many
orders of magnitude below the zero-point energy estimated
by quantum field theory---the {\em cosmological constant problem}.
Further chinks in the
armor of the SM+GR model of the universe include the
{\em hierarchy problem}, i.e.~the extreme weakness of gravity
relative to the other forces, and the seeming irreconcilability
of GR with quantum theory.

Clearly, gravity is at the center of some of the most profound contemporary
puzzles in physics. But it has now also given us a new observational handle 
on these puzzles, in the form of GWs \cite{Abbott:2016blz}.
Furthermore, 
the aforementioned 2005 breakthrough in NR
\cite{Pretorius:2005gq,Baker:2005vv,Campanelli:2005dd} has given
us the tools needed to systematically explore the non-linear
strong-field regime of gravity.
For much of its history, NR was motivated
by the modeling of astrophysical sources of GWs, as we have 
described in Sec.~\ref{Sec:NR}. As early as 1992, however, Choptuik's
milestone discovery of critical behaviour in gravitational collapse
\cite{Choptuik:1992jv} has demonstrated the enormous potential of
NR as a tool for exploring a much wider range of gravitational phenomena.
In this section we review the discoveries made in this field
and highlight the key challenges and goals for future work.

\subsection{Particle laboratories in outer space}
DM, by its very definition, generates little if any electromagnetic
radiation; rather, it interacts with its environment through gravity.
Many DM candidates have been suggested (see also Section~\ref{Sec:DM}),
ranging from primordial BH clouds to
weakly interacting massive particles (WIMPs)
and ultralight bosonic fields
\cite{Bertone:2004pz,Feng:2010gw,Khlopov:2017vcj}.
The latter are a particularly attractive candidate
in the context of BH and GW physics due to their
specific interaction with BHs. Ultralight fields,
also referred to as weakly interacting slim particles (WISPs), typically
have mass parameters orders of magnitude below an
electron volt and arise in extensions of the SM of particle physics
or extra dimensions in string theory. These include {\em axions}
or axion-like particles, dark photons, non-topological solitons
(so-called {\em Q-balls}) and condensations of bosonic states
\cite{Bertone:2004pz,Arvanitaki:2010sy,Ackerman:mha,Klasen:2015uma}.
For reference, we note that a mass $10^{-10}\,{\rm eV}$ ($10^{-20}\,{\rm eV}$)
corresponds to a Compton wavelength of $\mathcal{O}(1)\,{\rm km}$
($\mathcal{O}(10^2)\,{\rm AU}$) which covers the range of
Schwarzschild radii of astrophysical BHs.

At first glance, one might think the interaction between such fundamental
fields and BHs is simple; stationary states are constrained by the
no-hair theorems and the field either falls into the BH or disperses away.
In practice, however, the situation is more complex---and more exciting.
NR simulations have illustrated the existence of
long-lived, nearly periodic
states of single, massive scalar or Proca (i.e.~vector)
fields around BHs
\cite{Witek:2012tr,Okawa:2014nda,Zilhao:2015tya}. Intriguingly,
these configurations are able to extract rotational energy from spinning
BHs through {\em superradiance}
\cite{Zeldovich:1971,Zeldovich1972,Misner:1972kx,Brito:2015oca}, an effect
akin to the Penrose process \cite{Penrose:1971uk,Brito:2015oca}. Further details on this process are provided in Section~\ref{sec:superradiance}. In the presence
of a confining mechanism that prevents the field from escaping to infinity,
this may even lead to a runaway instability dubbed the {\em BH bomb}
\cite{Press:1972zz,Cardoso:2004nk}. Another peculiar consequence arising in the same context is the possibility of floating
orbits where dissipation of energy through GW emission is compensated
by energy gain through superradiance~\cite{Cardoso:2011xi,Fujita:2016yav}.

Naturally, non-linear effects
will limit the growth of the field amplitude or the lifetime of floating
orbits, and recent years have seen the first numerical studies
to explore the role of non-linearities in superradiance. These simulations
have shown that massive, real scalar fields around BHs can become trapped
inside a potential barrier outside the horizon, form a bound state
and may grow due to superradiance \cite{Witek:2012tr}. Furthermore,
beating phenomena result in a more complex structure in
the evolution of the scalar field. These findings were confirmed
in \cite{Okawa:2014nda}, which also demonstrated that the scalar clouds
can source GW emission over long timescales. The amplification of GWs through
superradiance around a BH spinning close to extremality has been modeled in Ref.~\cite{East:2013mfa}
and found to be maximal if the
peak frequency of the wave is above the superradiant threshold
but close to the dominant quasi-normal mode frequency of the BH.
The generation of GW templates for the inspiral and merger of hairy BHs
and the identification of possible smoking gun effects
distinguishing them from their vacuum counterparts remains a key
challenge for future NR simulations.

Numerical studies of the non-linear saturation of superradiance are
very challenging due to the long time scales involved. A particularly convenient
example of superradiance in spherical symmetry
arises through the interaction of
charged, massless scalar fields with Reissner-Nordstr{\"o}m BHs in asymptotically anti-de Sitter
spacetime or in a cavity; here energy is extracted from the BH charge
rather than its rotation.
The scalar field initially grows in accordance
with superradiance, but eventually saturates, leaving behind a stable
hairy BH \cite{Bosch:2016vcp,Sanchis-Gual:2016tcm}. Recently the instability of spinning BHs in AdS backgrounds was studied in full generality~\cite{Chesler:2018txn}.
The system displays extremely rich dynamics and it is unclear if there is a stationary final state.

The superradiant growth of a complex, massive vector field
around a near-extremal Kerr BH has recently been modeled
in axisymmetry \cite{East:2017ovw}.
Over $9\,\%$ of the BH mass can be extracted
until the process gradually saturates. Massive scalar fields
can also pile up at the center of ``normal'' stars. Due to
gravitational cooling, this pile-up does not lead to BH formation
but stable configurations composed of the star with a ``breathing''
scalar field \cite{Brito:2015yga}.

The hairy BH configurations considered so far are long-lived but not
strictly stationary. A class of genuinely stationary hairy BH spacetimes
with scalar or vector fields has been identified in a series of papers
\cite{Herdeiro:2014goa,Brihaye:2014nba,Benone:2014ssa,Herdeiro:2016tmi,Chodosh:2015oma}.
The main characteristic of these systems is that the scalar field
is {\em not stationary}---thus bypassing the no-hair theorems---but 
the spacetime metric and energy-momentum are.
A subclass of these solutions smoothly interpolates between
the Kerr metric and boson stars. A major challenge for numerical explorations
is to evaluate whether these solutions are non-linearly stable and might thus be
astrophysically viable. A recent study of linearized perturbations around the hairy BH
solution of \cite{Herdeiro:2014goa} has found unstable modes with
characteristic growth rates similar to or larger than
those of a massive scalar field on a fixed Kerr background. However, such solution may still be of some astrophysical relevance~\cite{Degollado:2018ypf}. See also Sec.~\ref{sec:circumvent} of Chapter III for a related discussion.

\subsection{Boson stars\label{sec:boson_stars}}
The idea of stationary localized, soliton-like configurations made up of the
type of fundamental fields discussed in the previous section
goes back to Wheeler's ``gravitational-electromagnetic entities'' or {\em geons} of the 1950s \cite{Wheeler:1955zz}.
While Wheeler's solutions turn out to be unstable,
replacing the electromagnetic field with a complex scalar field leads
to ``Klein-Gordon geons'' first discovered by Kaup \cite{Kaup:1968zz} and now
more commonly referred to as {\em boson stars}. The simplest type of
boson stars, i.e.~stationary solutions to the Einstein-complex-Klein-Gordon
system, is obtained for a non-self-interacting field
with harmonic time dependence $\phi \propto e^{i\omega t}$ where the potential
contains only a mass term $V(\phi)= m^2 |\phi|^2$. The resulting
one-parameter family of these so-called {\em mini boson star}
solutions is characterized by the central
amplitude of the scalar field and leads to a mass-radius diagram
qualitatively similar to that of static NSs; a maximum mass value of
$M_{\rm max}=0.633\,M_{\rm Planck}^2/m$ separates the stable und
unstable branches~\cite{Jetzer:1991jr,Mundim:2010hi,Liebling:2012fv}.
For a particle mass $m=30\,{\rm GeV}$, for instance,
one obtains $M_{\rm max}\sim 10^{10}\,{\rm kg}$ with radius $R\sim 6\times
10^{-18}\,m$ and a density $10^{48}$ times that of a NS
\cite{Mielke:1997re}.

A wider range of boson star models is obtained by adding self-interaction
terms to the potential $V(\phi)$ which result in more astrophysically
relevant bulk properties of the stars. For a quartic term
$\lambda |\phi|^4/4$, for example, the maximal mass is given by
$M_{\rm Planck} (0.1\,{\rm GeV}^2)\,M_{\odot}\,\lambda^{1/2}/m^2$
\cite{Colpi:1986ye}; for further types of boson stars with different
potential terms see the reviews
\cite{Mielke:1997re,Liebling:2012fv,Helfer:2016ljl}
and references therein.
A particularly intriguing feature of rotating boson stars exemplifies
their macroscopic quantum-like nature: the ratio of angular momentum
to the conserved particle number must be of integer value which
prevents a continuous transition from rotating to non-rotating
configurations \cite{Kobayashi:1994qi,Mielke:1997re}.
More recently, stationary, soliton-like configurations have also
been found for complex, massive Proca fields
\cite{Brito:2015pxa}. For real scalar fields, in contrast, stationary
solutions do not exist, but localized, periodically oscillating solutions
dubbed {\em oscillatons} have been identified in Ref.~\cite{Seidel:1991zh}.

Boson stars are natural candidates for DM~\cite{Liebling:2012fv}, but may also act as BH mimickers
\cite{Guzman:2009zz,Cardoso:2016olt}. In the new era of
GW observations, it is vital to understand the GW generation in
boson-star binaries and search for specific signatures that may
enable us to distinguish them from BH or NS systems.
Recent perturbative calculations of the tidal deformation of boson
stars demonstrate that the inspiral part of the waveform may
allow us to discriminate boson stars,
at least with third-generation detectors
\cite{Sennett:2017etc}.
Numerical studies of dynamic boson stars have so far mostly focused
on the stability properties of single stars and confirmed the
stable and unstable nature of the branches either side of the maximum
mass configuration; see
e.g.~\cite{Balakrishna:1997ej,ValdezAlvarado:2012xc,Collodel:2017biu,Sanchis-Gual:2017bhw}.
The modeling of head-on collisions of boson stars
\cite{Palenzuela:2006wp,Bezares:2017mzk} reveals rich structure
in the scalar radiation and that the merger leads to the formation
of another boson star.
Head-on collisions have also served as a testbed for confirming the
validity of the hoop conjecture in high-energy collisions
\cite{Choptuik:2009ww}.
Inspiralling configurations result
either in BH formation, dispersion of the scalar field to infinity
or non-rotating stars
\cite{Palenzuela:2007dm,Palenzuela:2017kcg},
possibly a consequence of the quantized
nature of the angular momentum that makes it difficult to form spinning
boson stars instead.

Binary boson star systems thus remain largely uncharted territory,
especially regarding the calculation of waveform templates
for use in GW data analysis and the quantized nature of spinning
boson stars and their potential constraints on forming rotating stars
through mergers.

\subsection{Compact objects in modified theories of gravity}
New physical phenomena and the signature of new ``modified'' theories
are typically encountered when probing extreme regimes not accessible
to previous experiments and observations. Quantum effects, for instance,
become prominent on microscopic scales and their observation
led to the formulation of quantum theory
while classical physics still provides an accurate description of
macroscopic systems. Likewise, Galilean invariance and Newtonian theory
accurately describe slow motion and weakly gravitating
systems but break down at velocities comparable to the speed of light
or in the regime of strong gravity. We therefore expect modifications
of GR, if present, to reveal themselves in the study
of extreme scales such as the large-scale dynamics of the universe
or the strong curvature regime near compact objects.

The dawn of GW observations provides us with unprecedented
opportunities to probe such effects. The modeling of compact
objects in alternative theories of gravity represents one of the
most important challenges for present and future NR studies,
so that theoretical predictions can be confronted with observations.
This challenge faces additional mathematical, numerical and
conceptual challenges as compared to the GR case; a more
detailed discussion of these is given in Sec.~\ref{IVPandNumerics} below.
A convenient way to classify the numerous modified
theories of gravity is provided by the assumptions underlying Lovelock's
theorem and, more specifically, which of these assumptions are dropped
\cite{Berti:2015itd}. Unfortunately, for most of these candidate
theories, well-posed formulations are not known (cf.~Table 1 in
\cite{Berti:2015itd}) or presently available only in the form
of a continuous limit to GR or linearization around some background
\cite{Delsate:2014hba,Okounkova:2017yby}. Prominent exceptions
are (single- or multi-) scalar-tensor (ST) theories of gravity
\cite{Fujii:2003pa}
which includes Brans-Dicke theory \cite{Brans:1961sx} and,
through mathematical equivalence, a subset of $f(R)$ theories
\cite{DeFelice:2010aj}.
ST theories inherit the well-posedness of GR through the Einstein
frame formulation \cite{Damour:1992we}; see also
\cite{Salgado:2005hx,Salgado:2008xh,KijowskiJakubiecUniversality}. Furthermore, ST physics would present
the most conspicuous strong-field deviation from GR discovered
so far, the {\em spontaneous scalarization} of NSs
\cite{Damour:1993hw} (for a similar effect in vector-tensor
theory see \cite{Ramazanoglu:2017xbl}).
For these reasons, almost all NR studies have
focused on this class of theories, even though its parameter
space is significantly constrained by solar system tests
and binary pulsar observations \cite{Bertotti:2003rm}.

The structure of equilibrium models of NSs in ST theories
has been studied extensively in the literature
(e.g.~\cite{Damour:1993hw,Doneva:2013qva,Mendes:2014ufa,Silva:2014fca,
Horbatsch:2015bua,Palenzuela:2015ima,Ramazanoglu:2016kul,Morisaki:2017nit})
and leads to a mass-radius diagram that contains one branch of
GR or GR-like models plus possible additional branches of strongly
scalarized stars. The presence or absence of these additional branches
depends on the coupling between the scalar and tensor sector of the
theory.

Early numerical studies considered the collapse of dust in
spherical symmetry
\cite{Shibata:1994qd,Scheel:1994yr,Scheel:1994yn,Harada:1996wt}
which leads to a hairless BH in agreement with the no-hair theorems,
even though departures from GR are possible during the dynamic
stages of the collapse. In a sequence of papers, Novak {\em et al}
studied the collapse of NSs into a BH
\cite{Novak:1997hw}, the transition of NSs between
stable branches \cite{Novak:1998rk} and the formation of
NSs through gravitational collapse \cite{Novak:1999jg}.
In all cases, strong scalar radiation is generated for
that part of the ST theory's parameter space that admits
spontaneously scalarized equilibrium models. In \cite{Gerosa:2016fri},
the collapse of stellar cores to BHs was found to be the most promising
scenario to generate detectable scalar radiation for parameters
allowed by the Cassini and pulsar observations; galactic sources
at a distance $D=10\,{\rm kpc}$ may be detected with present and
future GW observatories or used to further constrain the theory's
parameters. All these simulations, however, consider {\em massless
ST theory}. For massive fields, low frequency interactions
decay exponentially with distance, so that the pulsar and Cassini constraints may no longer apply \cite{Alsing:2011er}. In consequence, massive ST theory
still allows for very strongly scalarized equilibrium stars if
$m\gtrsim 10^{-15}\,{\rm eV}$
\cite{Ramazanoglu:2016kul,Morisaki:2017nit}.
This has dramatic consequences for the GW signals that can be
generated in massive ST theory as compared with its massless counterpart:
GW amplitudes can be orders of magnitude larger
and the waves are spread out into a nearly monochromatic signal
over years or even centuries due to the dispersion
of the massive field. GW searches may therefore be directed at historic
supernovae such as SN1987A and either observe a signal or constrain the
theory's parameter space \cite{Sperhake:2017itk}.

Numerical studies of binary systems in ST theory are rather scarce.
The no-hair theorems strongly constrain possible deviations of pure BH
spacetimes from GR. They can be bypassed, however,
through non-trivial potentials \cite{Healy:2011ef} or boundary
conditions \cite{Berti:2013gfa} which leads to scalar wave emission.
Nevertheless, NS systems appear to be the more natural candidate
to search for imprints of ST theory. Dynamical scalarization has indeed
been observed in simulations of the merger of two NSs with initially vanishing
scalar charge \cite{Barausse:2012da,Palenzuela:2013hsa}.
Beyond GR and ST theory, we are only aware of one numerical study
\cite{Okounkova:2017yby}, which simulated the evolution of BH binaries in the dynamical
Chern-Simons (dCS) theory linearized around GR. LIGO observations
may then measure or constrain the dCS length scale to $\lesssim
\mathcal{O}(10)\,{\rm km}$.

With the dawn of GW astronomy \cite{Abbott:2016blz}, the topics
discussed so far in this section are becoming important subjects
of observational studies with LIGO, Virgo and future GW detectors.
These studies are still in an early stage and the generation of
precision waveforms for scenarios involving modifications of gravity,
fundamental fields or more exotic compact objects will be a key
requirement for fully exploiting the scientific potential of this
new channel to observing the universe.

The analysis of GW events has so far concentrated on testing the
consistency of the observed signals with GR predictions, establishing
bounds on phenomenological parametrizations of deviations from
GR and obtaining constraints from the propagation of the GW signal.
An extended study of GW150914 demonstrated consistency between
the merger remnant's mass and spins obtained separately from the
low-frequency inspiral and the high-frequency postinspiral signal
\cite{TheLIGOScientific:2016src}. The Compton wavelength of the
graviton was constrained with a 90\,\% lower bound of
$10^{13}$\,km (corresponding to an upper bound for the
mass of $\sim 10^{-22}\,{\rm eV}$)
and parametrizations of violations
of GR using high-order PN terms have been constrained;
see also \cite{Abbott:2017gyy, Abbott:2017vtc}.
The first NSB observation \cite{TheLIGOScientific:2017qsa},
combined with electromagnetic observations, limited the difference
between the speed of propagation of GWs and that of light to within
$-3\times 10^{-15}$ and $7\times 10^{-16}$ times the speed of light
\cite{Monitor:2017mdv}. Analysis of the polarization of the first
triple coincidence detection GW170814 found Bayes' factors of 200
(1000) favoring a purely tensor polarization against
purely vector (purely scalar) radiation.
In summary, the GW observations have as yet not identified
an inconsistency with the predictions of vacuum GR for BBH
signals or GR predictions for NS systems.

Chapter III contains detailed discussions on many of the topics raised here, including motivation and brief description of certain broad classes of alternative to GR (Sec.~\ref{sec:theories}), numerics beyond GR (Sec.~\ref{sec:NRbeyond}), and the nature of compact objects beyond GR (Sec.~\ref{sec:compactbeyond}).

\subsection{High-energy collisions of black holes}
The hierarchy problem of physics consists in the vast discrepancy
between the weak coupling scale $(\approx 246\,{\rm GeV})$ and the
Planck scale $1.31\times 10^{19}\,{\rm GeV}$ or, equivalently,
the relative weakness of gravity compared with the other interactions.
A possible explanation has been suggested in the form of
``large'' extra spatial dimensions
\cite{Antoniadis:1998ig,ArkaniHamed:1998rs}
or extra dimensions with a warp factor
\cite{Randall:1999ee,Randall:1999vf}.
On short lengthscales $\lesssim 10^{-4}\,{\rm m}$, gravity is at present poorly constrained by experiment and would,
according to these models, be diluted
due to the steeper fall off in higher dimensions. All other interactions,
on the other hand, would be constrained to a 3+1 dimensional brane and,
hence, be unaffected. In these {\em braneworld} scenarios, the
fundamental Planck scale would be much smaller than the four-dimensional
value quoted above, possibly as low as $\mathcal{O}({\rm TeV})$ which
inspired the name {\em TeV gravity}. This fundamental Planck scale
determines the energy regime where gravity starts dominating the
interaction, leading to the exciting possibility that BHs may
be formed in particle collisions at the LHC or in high-energy cosmic rays
hitting the atmosphere \cite{Banks:1999gd,Dimopoulos:2001hw,Giddings:2001bu}.

The analysis of experimental data employs so-called
Monte-Carlo event generators \cite{Frost:2009cf} which require
as input the cross section for BH formation and the loss of energy
and momentum through GW emission. In the ultrarelativistic limit,
the particles may be modeled as pointlike or, in GR, as BHs.

In $D=4$ spacetime dimensions, high-energy collisions of BHs are
by now well understood. Head-on collisions near the speed of light
radiate about $14\,\%$ of the center-of-mass energy $M$ in GWs
\cite{Sperhake:2008ga,Healy:2015mla}. The impact parameter separating
merging from scattering collisions with boost velocity $v$
is $b/M=(2.50\pm0.05)/v$ \cite{Shibata:2008rq}. Grazing collisions
exhibit zoom-whirl behaviour \cite{Pretorius:2007jn,Sperhake:2009jz}
and can radiate up to $\sim 50\,\%$ of the total energy in GWs
\cite{Sperhake:2012me}. The collision dynamics furthermore
become insensitive to the structure of the colliding
objects -- BHs or matter balls -- at high velocities
\cite{Choptuik:2009ww,East:2012mb,Rezzolla:2012nr,Sperhake:2012me,
Sperhake:2015siy} as expected when kinetic energy dominates the budget.
Finally, NR simulations of hyperbolic BH encounters are in good
agreement with predictions by the EOB method
\cite{Damour:2014afa}.

The key challenges are to generalize these results to the higher $D$
scenarios relevant for TeV gravity (there are partial, perturbative results~\cite{Galtsov:2010vtu,Berti:2010gx,Galtsov:2012pcw}). Considering the symmetries
of the collision experiments, it appears plausible to employ
rotational symmetry in higher $D$ numerics which is vital to
keep the computational costs under control and underlies all NR studies
performed so far
\cite{Pretorius:2004jg,Zilhao:2010sr,Yoshino:2011zz,Yoshino:2011zza,
Cook:2016soy}.
Nonetheless, NR in higher $D$ faces new challenges. The extraction
of GWs is more complex, but now tractable either with perturbative
methods \cite{Kodama:2003jz,Kodama:2003kk,Witek:2010xi},
using the Landau-Lifshitz pseudotensor
\cite{Yoshino:2009xp} or projections of the Weyl tensor
\cite{Godazgar:2012zq,Cook:2016qnt}. Likewise, initial data
can be obtained by generalizing four-dimensional techniques
\cite{Zilhao:2011yc}. Studies performed so far, however, indicate that,
for reasons not yet fully understood,
obtaining numerically stable evolutions is harder than in $D=4$
\cite{Okawa:2011fv,Cook:2017fec}. Results for the scattering threshold
are limited to $v\lesssim 0.5\,c$ \cite{Okawa:2011fv} and
the emission of GWs has only been computed for non-boosted collisions
of equal and unequal mass BH binaries
\cite{Witek:2010az,Witek:2014mha,Cook:2017fec}. These simulations show a
strong suppression of the radiated energy with $D$ beyond its peak value
$E_{\rm rad}/M \approx 9\times 10^{-4}$ at $D=5$ for equal-mass systems,
but reveal more complex behaviour for low mass ratios.
A remarkable outcome of BH grazing collisions in $D=5$ is the possibility of
super-Planckian curvature in a region outside the BH horizons
\cite{Okawa:2011fv}.

\subsection{Fundamental properties of black holes and non-asymptotically flat spacetimes}
Recent years have seen a surge of NR applications to non-asymptotically
flat spacetimes in the context of the gauge-gravity duality,
cosmological settings and for the exploration of fundamental properties
of BH spacetimes. We list here a brief selection of some results and
open questions; more details can be found in the reviews
\cite{Sperhake:2013qa,Cardoso:2014uka}.

Cosmic censorship has for a long time been a topic of interest
in NR, but to date no generic, regular and asymptotically flat class of
initial data are known to result in the formation of naked singularities
in four spacetime dimensions. Higher-dimensional BHs, however, have a
much richer phenomenology \cite{Emparan:2008eg}, including in particular
black rings which may be subject to the Gregory-Laflamme instability
\cite{Gregory:1993vy}. Thin black rings have indeed been found to
cascade to a chain of nearly circular BHs connected by ever thinner
segments in finite time \cite{Figueras:2015hkb} in the same way
as infinite black strings \cite{Lehner:2010pn}.
Similarly, ultraspinning topologically spherical BHs in $D\ge 6$
dimensions are unstable \cite{Shibata:2010wz} and ultimately form
ever thinner rings in violation of the weak cosmic censorship
conjecture \cite{Figueras:2017zwa}.

We have already mentioned Choptuik's discovery of critical phenomena
in the collapse of spherical scalar fields \cite{Choptuik:1992jv}.
In asymptotically Anti-de Sitter (AdS) spacetimes,
the dynamics change through the confining mechanism of the AdS boundary,
allowing the scalar field to recollapse again and again until a BH forms
\cite{Bizon:2011gg,Jalmuzna:2011qw,Buchel:2012uh,Bizon:2013xha,Olivan:2015fmy};
see also \cite{Bantilan:2017kok} for non-spherically symmetric
configurations.
NR simulations of asymptotically AdS spacetimes are very challenging
due to the complex outer boundary conditions, in particular away
from spherical symmetry, but recent years have seen the first
simulations of BH collisions in AdS \cite{Bantilan:2012vu,Bantilan:2014sra}
which, assuming gauge-gravity duality, would imply a far-from hydrodynamic
behaviour in heavy-ion collisions during the early time after merger.

Using a cosmological constant with an opposite sign leads to asymptotically
de Sitter (dS) spacetimes widely believed to describe
the Universe we live in. NR studies in dS
have explored the possible impact of local structures on the
cosmological expansion \cite{Bentivegna:2012ei,Yoo:2013yea,Yoo:2014boa,
Bentivegna:2015flc,Mertens:2015ttp}
and found such inhomogeneities to not significantly affect the global expansion.
Further work has explored the robustness of inflation under
inhomogeneities \cite{Clough:2016ymm}, the propagation of light in
an expanding universe \cite{Bentivegna:2016fls} and the impact of
extreme values of the cosmological constant on the physics of BH collisions
\cite{Zilhao:2012bb}.

\section{Effective-One-Body and Phenomenological models}\label{Sec:EOB}
\vspace{-3mm}
{\it Contributor:} T. Hinderer
\vspace{3mm}


This section surveys the status of two main classes of models that are currently used in GW data analysis: (i) the EOB approach, which describes both the dynamics and waveforms in the time domain for generic spinning binaries, and (ii) the phenomenological approach (Phenom), which provides a closed-form description of the waveform in the frequency domain and includes the dominant spin effects. The discussion below reviews mainly the current state-of-the-art models; a more comprehensive overview of prior work can be found in review articles such as Refs.~\cite{Buonanno:2014aza,Hannam:2013pra,Damour:2008yg}.


\subsection{Effective-one-body models}
The EOB approach was introduced in ~\cite{Buonanno:1998gg,Buonanno:2000ef} as a method to combine information from the test-particle limit with PN results (for early ideas in this spirit, see~Ref.\cite{Maheshwari:2016edp}). The model comprises a Hamiltonian for the inspiral dynamics, a prescription for computing the GWs and corresponding radiation reaction forces, and a smooth transition to the ringdown signals from a perturbed final BH. The idea is to map the conservative dynamics of the relative motion of two bodies, with masses $m_{1,2}$, spins ${\bm S}_{1,2}$, orbital separation $\bm x$ and relative momentum $\bm p$, onto an auxiliary Hamiltonian description of an effective particle of mass $\mu =m_1\,m_2/(m_1+m_2)$ and effective spin $\bm{S}_*(\bm{S}_1,\bm{S}_2,\bm{x},\bm{p})$ 
moving in an effective spacetime $g_{\mu \nu}^{\rm eff}(M,\bm{S}_{\rm Kerr};\nu)$ that is characterized by the total mass $M =m_1+m_2$, symmetric mass ratio $\nu= \mu/M$, and total spin  
$\bm{S}_{\rm Kerr}(\bm{S}_1,\bm{S}_2)$. The basic requirements on this mapping are that (i) the test-particle limit reduces to a particle in Kerr spacetime, and (ii) in the weak-field, slow-motion limit the EOB Hamiltonian reduces to the PN Hamiltonian, up to canonical transformations. These considerations, together with insights from results in quantum-electrodynamics~\cite{Brezin:1970zr}, scattering theory~\cite{Damour:2016gwp,Vines:2017hyw}, and explicit comparisons of the action variables~\cite{Buonanno:1998gg} all lead to the ``energy map'' given by
\be
H=M \sqrt{1+2\nu \left(\frac{H_{\rm eff}}{\mu}-1\right)},
\ee
where $H$ and $H_{\rm eff}$ are the Hamiltonians describing the physical binary system and the effective particle respectively.

The details of the effective Hamiltonian $H_{\rm eff}$ are less constrained by theoretical considerations, and different descriptions have been proposed that differ in the structure of the Hamiltonian, the form of the potentials characterizing the effective spacetime, and the spin mapping between the physical and effective systems. The differences between these choices reflect current limitations of theoretical knowledge; they all agree in the PN and nonspinning test-particle limit. For the structure of $H_{\rm eff}$, the incarnation of the EOB model of Refs.~\cite{Barausse:2009aa, Barausse:2009xi, Barausse:2011ys,Taracchini:2013rva,Bohe:2016gbl} imposes that the limiting case of a spinning test-particle in Kerr spacetime must be recovered, to linear order in the test-spin. The version of the model of Refs.~\cite{Damour:2001tu,Damour:2008qf,Nagar:2011fx,Balmelli:2015lva,Damour:2014sva} does not include test-particle spin effects, which enables a more compact description. These different choices also result in different spin mappings, for both the Kerr parameter and the spin of the effective particle. Finally, these two branches of EOB models also employ different ways to re-write the potentials that are calculated as a Taylor series  in a PN expansion in a ``re-summed'' form. This means that an empirically motivated non-analytic representation is used that consists either of a Pade-resummation ~\cite{Damour:2001tu,Damour:2008qf,Nagar:2011fx} or has a logarithmic form~\cite{Barausse:2009xi, Barausse:2011ys}. In addition to the above choices for describing the strong-field, comparable-mass regime, the models also include parameterized terms whose coefficients are functions of the mass and spin parameters that are fixed by comparisons to NR results. In the model of Ref.~\cite{Barausse:2009xi, Barausse:2011ys,Taracchini:2013rva,Bohe:2016gbl}, these calibration parameters are constrained by the requirement that the model must reproduce the GSF results for the ISCO shift in Schwarzschild~\cite{Barack:2009ey}.

The radiative sector in the EOB model is described by so-called factorized waveforms (instead of a PN Taylor series expansion) that are motivated from the structure of waveforms in the test-particle limit and have the form~\cite{Damour:2007xr,Damour:2008gu}
\begin{equation}
\label{hlm}
h_{\ell m}^{\rm insp-plunge}(t)=h_{\ell m}^{(N,\epsilon)}\,\hat{S}_{\rm eff}^{(\epsilon)}\, T_{\ell m}\, e^{i\delta_{\ell m}}\,
f_{\ell m}\,N_{\ell m}\,.
\end{equation}
Here, $h_{\ell m}^{(N,\epsilon)}$ is the Newtonian contribution, and  $\hat{S}_{\rm eff}^{(\epsilon)}$ is a certain effective ``source term'' that, depending on the parity $\epsilon$ of the mode, is related to either the energy or the angular momentum. The factor $T_{\ell m}$ 
contains the leading order logarithms arising from tail effects, the term $e^{i\delta_{\ell m}}$ 
is a phase correction due to sub-leading order logarithms in hereditary contributions, while the function 
$f_{\ell m}$ collects the remaining PN terms. Finally, the factor $N_{\ell m}$ 
is a phenomenological correction that accounts for deviations from 
quasi-circular motion~\cite{Damour:2002vi} and is calibrated to NR results. The modes from Eq.~(\ref{hlm}) are used to construct dissipative forces $\bm{\mathcal{F}}$, in terms of which the equations of motion are given by
\begin{eqnarray}
&&\frac{d\bm{x}}{d t}=\frac{\partial H}{\partial \bm{p}}\,, \quad \quad 
\frac{d\bm{p}}{d t}=-\frac{\partial H}{\partial \bm{x}} 
    +\bm{\mathcal{F}}\,, \qquad \qquad\\
 &&   \frac{d\bm{S}_{1,2}}{d t}=\frac{\partial H}{\partial \bm{S}_{1,2}}\times \bm{S}_{1,2}.
\end{eqnarray}

This EOB description of the inspiral-plunge dynamics is smoothly connected to the merger-ringdown signal in the vicinity of the peak in the amplitude $|h_{22}|$. To perform this matching, initial nonspinning models and the aligned-spin models SEOBNRv1~\cite{Barausse:2011ys} and SEOBNRv2~\cite{Taracchini:2013rva,Taracchini:2012ig} (as well as in the models of the IHES group~\cite{Damour:2012ky,Damour:2001tu}) used a superposition of damped sinusoids similar to quasi-normal modes~\cite{Buonanno:2000ef}. This method is inspired by results for the infall of a test-particle into a BH and the close-limit approximation~\cite{Price:1994pm}. More recently, a simpler phenomenological fit of the amplitude and phase inspired by the rotating source approximation \cite{Baker:2008mj}, which was adapted to the EOB context in \cite{Damour:2014yha}, has become standard in SEOBNRv4~\cite{ Bohe:2016gbl}. This method provides a more stable and controlled way to connect the inspiral-plunge to the ringdown. A further key input into the merger-ringdown model is the frequency of the least-damped quadrupolar quasinormal mode $\sigma_{220}$ of the remnant based on a fitting formula from NR for the mass and spin of the final object given the initial parameters, with the currently most up-to-date fit from ~\cite{ Bohe:2016gbl}.

Generic spin precession effects are also included in the model, by starting from a calibrated spin-aligned model and transforming to the precessing frame as dictated by the precession equations derived from the EOB Hamiltonian \cite{Pan:2010hz,Babak:2016tgq}, without further calibrations. The most recent refinement for generic precessing binaries from Ref.~\cite{Babak:2016tgq} is known as SEOBNRv3. 

For non-vacuum binaries involving e.g. neutron stars, exotic compact objects, or condensates of fundamental fields around BHs, several effects of matter change the GWs from the inspiral relative to those from a BH binary, as described in Chapter III, Sec. 4.4, where also the signatures from such systems generated during the merger and ringdown are covered.  Effects during the inspiral include spin-induced deformations of the objects, tidal effects, the presence of a surface instead of an event horizon, tidal excitation of internal oscillation modes of the objects, and more complex spin-tidal couplings. Two classes of EOB models for such systems are currently available, corresponding to the two different baseline EOB models for BHs. One is  known as TEOBRESUMS \cite{Nagar:2018zoe} and incorporates the effects of rotational deformation and adiabatic tidal effects \cite{Damour:2009wj,Bini:2012gu} in re-summed form that was inspired by \cite{Bini:2014zxa} and augmented as described in \cite{Bernuzzi:2014owa}. The other model is known as SEOBNRv4T and includes the spin-induced quadrupole as well as dynamical tides from the objects' fundamental oscillation modes \cite{Steinhoff:2016rfi,Hinderer:2016eia}.
\subsection{Phenomenological (Phenom) models}
The aim of the Phenom models is to provide a simple, efficient, closed-form expression for the GW signal in the frequency domain~\cite{Ajith:2007qp} by assuming the schematic form
\be
\tilde{h}_{\rm phen}(f;\vec{\alpha}; \vec{\beta}) := A(f;\vec{\alpha})e^{i\Psi(f;\vec{\beta})},
\ee
where $\vec{\alpha}$ and $\vec{\beta}$ are amplitude and phase parameters in the model.
Phenomenological (``Phenom'') models were first developed for nonspinning binaries in Refs.~\cite{Ajith:2007qp,Ajith:2007kx} and subsequently refined to include aligned spins~\cite{Ajith:2009bn} known as ``PhenomB''. This model employed only a single weighted combination of the individual BH spins characterizing the dominant spin effect in the GWs, and was further refined in ~\cite{Santamaria:2010yb} (``PhenomC''), and in ~\cite{Khan:2015jqa,Husa:2015iqa} (``PhenomD''). The latter has been calibrated using NR data for mass ratios up to 1:18 and dimensionless spins up to $0.85$ (with a larger spin range for equal masses). An effective description of the dominant precession effects has also been developed~\cite{Hannam:2013oca} (``PhenomP'')~\cite{Schmidt:2012rh, Schmidt:2014iyl}. The PhenomP model provides an approximate mapping for obtaining a precessing waveform from any non-precessing model, with PhenomD being currently used as the baseline.

To construct these state-of-the-art Phenom models, the GW signal is divided into three main regimes: an inspiral, intermediate region, and merger-ringdown. The ansatz for the inspiral model is the PN result for the frequency-domain phasing obtained from energy balance in the stationary phase approximation (``TaylorF2''), accurate to 3.5PN in the nonspinning and linear-in-spin sector, and to 2PN in spin-spin terms. This has the form
\be
\Psi_{\rm TF2}=2\pi f t_c-\phi_c-\pi/4+\frac{3}{128\nu}(\pi M f)^{-5/3}\sum_{i=0}^7 c_i (\pi M f)^{i/3}, \label{eq:TF2phase}
\ee
where $c_i$ are PN  coefficients. The Phenom models add to this auxiliary terms so that the inspiral phase becomes
\be
\Psi=\Psi_{\rm TF2}+\frac{1}{\nu}\sum_{i=0}^6 \sigma_i f^{i/3},
\ee
where $\sigma_i$ are phenomenological coefficient with $\sigma_{1}=\sigma_2=0$.

A different ansatz is made for the late inspiral and merger-ringdown signals that are likewise closed-form expressions involving the frequency and phenomenological parameters; in total the model involves 17 phenomenological parameters. For the late inspiral portion, these are mapped to the set of two physical parameters
$(\nu, \chi_{\rm PN})$, where $\chi_{\rm PN}$ is defined by
\begin{eqnarray}
\chi_{\rm PN}&=& \chi_{\rm eff}-\frac{38\nu}{113}\left(\chi_1+\chi_2\right)\,,\\
\chi_{\rm eff}&=&\frac{m_1}{M} \vec{\chi}_1\cdot \hat{L}_N+\frac{m_2}{M} \vec{\chi}_2\cdot \hat{L}_N\,,\label{chi}
\end{eqnarray}
where $\hat L_N$ is the direction of the Newtonian angular momentum and $\chi_i=S_i/m_i^2$. $\chi_{\rm PN}$ describes the dominant spin effect (the dependence on $M$ is only a rescaling). The mapping is done by assuming a polynomial dependence $\sigma_i=\sum b_{ijk}\nu^j \chi_{\rm eff}^k$ to quadratic order in $\nu$ and third order in $\chi_{\rm PN}$ so that the coefficients vary smoothly across the parameter space. Finally, the coefficients are calibrated to a large set of hybrid waveforms, which themselves are formed using the uncalibrated SEOBNRv2 model.

Spin effects in the Phenom models are described by several different combinations of parameters. In the aligned-spin baseline model,  the description of the early inspiral depends on both spin parameters $\chi_1,\chi_2$ from PN. In the later inspiral regime, spin effects are described by the effective combination $\chi_{\rm PN}$ described above, while the merger-ringdown model is expressed in terms of the total spin. For generic spin orientations, an additional parameter $\chi_p$ that characterized the most readily measurable effects of precession is included in the model. The defnition of the precessional parameter $\chi_p$ is motivated by the observation that waveforms are simpler in the coprecessing frame that is aligned with the direction of the dominant radiation emission~\cite{Schmidt:2010it}; it is given by~\cite{Schmidt:2014iyl}
\be
\chi_p=\frac{1}{B_1 m_1^2}{\rm max}(B_1 S_{1\perp},B_2 S_{2\perp}),
\ee
where $B_1=2+3 m_2/(2 m_1)$ and $B_2=2+3 m_1/(2 m_2)$ assuming the convention $m_1>m_2$. Starting from an aligned-spin frequency-domain waveform model, an approximate precessing waveform is constructed by ``twisting up'' the non-precessing waveform with the precessional motion
based on a single-spin PN description for the precession angles~\cite{Schmidt:2012rh,Hannam:2013oca}. While there are some broad similarities of PhenomP with the way the precessing EOB model is constructed, the two approaches differ in several details, as explained in Sec.\ IV of Ref.~\cite{Babak:2016tgq}.

For non-BH objects, the effects of rotational and tidal deformations are included in the phasing using either PN information \cite{Krishnendu:2017shb,Vines:2011ud,Damour:2012yf} or, for the case of binary neutron stars, using a model calibrated to NR \cite{Dietrich:2017aum,Dietrich:2018uni}.
\subsection{Remaining challenges} 
Important advances that all EOB and Phenom models aim to address in the near-future are higher modes, parameter space coverage (especially in the mass ratio) and inclusion of all available theoretical knowledge, reduction of systematic errors, and going beyond circular orbits. 

To date, most of the effort in calibrating the EOB and Phenom models has focused on the $(2,2)$ mode, although for the special case of nonspinning binaries an EOB multipolar approximant is available~\cite{Pan:2011gk}. An accurate model of higher modes is important for robust data analysis, especially for spinning binaries. Work is ongoing to address this within the EOB model~\cite{Cotesta:2018fcv} and with studies that will inform the Phenom approach~\cite{Bustillo:2015ova}. 
The parameter space over which current models have been calibrated and tested is limited by available NR simulations of sufficient accuracy and length (see Sec.\ \ref{Sec:NR}), and systematic uncertainties remain a concern. 

Extending the range in parameter space ties into the pressing issue that many available results from GSF calculations are not currently incorporated in these models. One of the obstacles is that GSF data for circular orbits only make sense above the ``light ring'' radius and cannot directly inform the EOB model across and below that radius.  
This issue is a further subject of ongoing work.
 Efforts are also underway to include effects of eccentricity~\cite{Huerta:2016rwp,Hinder:2017sxy,Hinderer:2017jcs}. Effects such as the motion of the center-of-mass, BH radiation absorption, and radiation reaction for spins are also not yet included in current models. 

For EOB models, efficiency for data analysis is a further concern, because computing EOB waveforms is relatively time-consuming. To overcome this issue, reduced-order models for EOB waveforms in the frequency domain have been developed for aligned-spin binaries~\cite{Purrer:2015tud, Bohe:2016gbl}, with work being underway on the larger parameter space of precessing binaries. Also ongoing is work on theoretical aspects of spin effects in the EOB model, aimed at devising an improved description that is more robust and computationally efficient. 

\subsection{Tests of GR with EOB and Phenom models}
The effective models described above have been used for tests of GR and exotic objects, e.g. in \cite{TheLIGOScientific:2016src}. One of the parameterized tests makes use of the general framework ``TIGER'' \cite{Agathos:2013upa} that has been implemented for data analysis, where parameterized fractional deviations from the GR coefficients are included in the frequency-domain phase evolution from Eq.~(\ref{eq:TF2phase}).
In recent studies, e.g.~\cite{TheLIGOScientific:2016src,Abbott:2017oio}, this framework was used on top of the PhenomD inspiral model \cite{Khan:2015jqa,Husa:2015iqa} to obtain bounds on non-GR effects, where the tests also included parameterized deviations in the intermediate and merger-ringdown regime. Work is ongoing to implement tests of GR within the EOB approach~\cite{Brito:2018rfr}. Finally, work is also underway to include in these models the option to test for exotic physics manifest in spin-quadrupole and tidal effects during the inspiral. Despite this recent progress on frameworks to test GR and the nature of BHs, substantial further work will be necessary to refine the level of sophistication and physical realism of such tests.

\section {Source modelling and the data-analysis challenge}\label{Sec:DA}
\vspace{-3mm}
{\it Contributor:} J. R. Gair
\vspace{3mm}

Models of GW sources form an essential component of data analysis for GW detectors. Indeed, many of the scientific objectives of current and future GW detectors cannot be realised without having readily available accurate waveform models. In this section we first provide a brief overview of the primary techniques used for GW data analysis, and follow this with a discussion of the additional challenges that data analysis for future GW detectors will pose and the corresponding requirements that this will place on waveform models.

\subsection{Overview of current data analysis methods}
The LIGO-Virgo Collaboration (LVC) uses a variety of techniques to first identify and then characterise GW transients identified in their data. The majority of these make use of signal models in some way. We briefly describe these methods in the following.

\subsubsection{Unmodelled searches}
A number of LVC analysis pipelines are targeted towards ``burst'' sources for which the waveforms are not well modelled. These included Coherent Wave Burst~\cite{2008CQGra..25k4029K}, X-pipeline~\cite{2010NJPh...12e3034S,2012PhRvD..86b2003W} and BayesWave~\cite{2015CQGra..32m5012C}. All three algorithms make use of the fact that there are multiple detectors in the LVC network, which allows signals (common between different detectors) to be distinguished from noise (generally uncorrelated between detectors). The algorithms differ in their implementation. Coherent Wave Burst and X-pipeline operate on spectrograms (time-frequency) maps of the data, computed using wavelet transforms in the first case and a Fourier transform in the second. Both algorithms then identify ``hot'' pixels that have particularly large power, carry out clustering to identify candidate events, i.e., continuous tracks of excess power, in individual detectors, and then impose consistency in the properties of the tracks identified in the different detectors. BayesWave takes a slightly different approach, using Bayesian techniques to construct a model for the noise in each detector and any signal present. The signal is constructed as a superposition of wavelets, and reversible jump Markov Chain Monte Carlo is used to add or remove components from the signal and noise models. 

These model-free searches are powerful tools for source identification. Indeed, the first algorithm to find GW150914, the first GW event detected by LIGO, was the Coherent Wave Burst~\cite{Abbott:2016blz,TheLIGOScientific:2016uux}, because it was the only online search pipeline running at the time of the event. However, these algorithms are not as sensitive as searches based on models and the second clear GW event, GW151226, was only found with high significance by the template-based searches~\cite{Abbott:2016nmj}. In addition, parameter estimation can only be done using models. Moreover, of the three algorithms, only BayesWave is truly independent of waveform models. Both Coherent Wave Burst and X-pipeline uses signal injections in order to determine the optimal choice of the various tunable thresholds in the algorithms that maximizes distinction between signal and noise. The injections use realistic GW signal waveforms, for which models are needed. BayesWave does not do tuning in this way, although the development of that algorithm and demonstration of its performance was done using signal injections.

\subsubsection{Template based searches}
The primary search pipelines within the LVC Compact Binary Coalescence (CBC) group use matched filtering. This involves using a precomputed bank of templates of possible GW signals and computing their overlap, i.e., inner product, with the data. The template bank needs to fully cover the parameter space so that if a signal is present, at least one of the templates will recover it confidently. There are two primary matched filtering based searches used by the LVC --- pyCBC~\cite{Canton:2014ena,Usman:2015kfa} and gstLAL~\cite{2012ApJ...748..136C,2014PhRvD..89b4003P}. The two searches differ in a number of ways, such as the choice of template bank and placement, and in the various consistency checks that are used to compare the signals identified in the  different detectors and to compare the post-signal extraction residual to the expected noise distribution. We refer the interested reader to the previous references for more details. All the BBH systems identified by LIGO to date were found by both of these pipelines with very high significance, and several of them would not have been identified using template-free methods only. Matched filtering is only possible if models of GW signals are available that have sufficient fidelity to true signals.

\subsubsection{Parameter estimation}
The other primary area where GW signal models are essential is in parameter estimation. Once a potential signal has been identified by one of the search pipelines described above, the signal is characterized using the separate LALInference pipeline~\cite{2015PhRvD..91d2003V} (though other parameter inference methods are also used~\cite{Pankow:2015cra,Lange:2018pyp}). LALInference constructs a posterior probability distribution for the source parameters using Bayes theorem. This relies on a model for the likelihood which is taken to be the likelihood of the noise (assumed Gaussian on the short stretches of data around each signal). The noise is computed as observed data minus signal, and is therefore a function of the signal parameters and requires an accurate model of potential signals. LALInference includes two different algorithms --- LALInferenceNest~\cite{2010PhRvD..81f2003V} and LALInferenceMCMC~\cite{2008ApJ...688L..61V}. The first uses nested sampling to determine the posterior and associated model evidence, while the latter uses Markov Chain Monte Carlo techniques based on the Metropolis-Hastings algorithm.

Parameter estimation is essential to extract physical information from identified GW events, and the resulting posterior distributions summarise all our information about the properties of the source. Any physical effect that we wish to probe using GW observations must be included in the signal model used in parameter estimation. The recent NSB event observed by LIGO, GW170817~\cite{TheLIGOScientific:2017qsa},  showed some evidence for tidal effects in the signal, which highlighted deficiencies in signal models with tides and the need for further work in that area~\cite{Abbott:2018wiz}. 

\subsection{Challenges posed by future detectors}
The next LIGO-Virgo observing run, O3, is scheduled to start in early 2019, and is expected to have a factor of $\sim2$ improvement in sensitivity over the O2 run that finished in August 2017. LIGO/Virgo are expected to complete their first science runs at design sensitivity in the early 2020s~\cite{Aasi:2013wya}. Based on the earlier science runs, several tens of BBH systems and a small number of NSB events~\cite{Abbott:2016nhf,TheLIGOScientific:2017qsa} are likely to be detected in O3. These events will be similar to sources previously identified and so the primary challenge will be computational --- the LVC will need the capability to process multiple events simultaneously, which implies a need for accurate waveform models that are as fast to generate as possible. Further in the future there are plans for third generation ground based detectors, such as the Einstein Telescope~\cite{2012CQGra..29l4013S} and Cosmic Explorer~\cite{Evans:2016mbw}, and a space-based GW detector, the Laser Interferometer Space Antenna (LISA)~\cite{Audley:2017drz}. These new detectors will observe new types of source which will pose new modelling challenges.

\subsubsection{Third-generation ground-based detectors}
Planned third-generation ground-based detectors, like the Einstein Telescope (ET)~\cite{2012CQGra..29l4013S} or the Cosmic Explorer~\cite{Evans:2016mbw}, aim to improve on advanced detectors in two ways. Firstly, they aim to have an increase in sensitivity by an order of magnitude. Secondly, they aim to improve low-frequency sensitivity, with the ultimate goal of sensitivity as low as $1$Hz, compared to $\sim30$Hz for the LIGO/Virgo detectors in O2, and $10$Hz at design sensitivity. For the ET, these aims will be achieved by increasing the arm-length to $10$km, sitting the detector underground and using cryogenic cooling of the mirrors. ET will also have three arms in a triangular configuration, so that it is equivalent to having two detectors at the same site. 

The increase in sensitivity will increase the number of sources detected by about a factor of a thousand. Data analysis must be able to keep pace with such a high source volume, which means algorithms must run quickly. This places constraints on the computational cost of waveform models, which is discussed further below. The improvement in low frequency sensitivity significantly increases the duration of any given signal in band. At leading order, the time to coalescence of a binary scales like $f^{-8/3}$~\cite{Peters:1964zz}. The NSB event GW170817 was in the LIGO band for ~$40s$, starting from $30$Hz, and generated 3000 waveform cycles~\cite{TheLIGOScientific:2017qsa}. For a detector with low frequency cut-off at $3$Hz, the same source would be in band for $\sim5$h and generate $\sim10^5$ cycles. This longer duration places more stringent requirements on waveform models, since fractional errors in the waveforms will need to be small enough that the templates are accurate to within 1 cycle of the $10^5$ in band. This is mitigated partially by the fact that most of the additional cycles are generated in the weak field where analytic models are well understood.

Third-generation detectors also offer the prospect of detection of new classes of sources. These include higher-mass BH systems, made possible by the improved low-frequency sensitivity~\cite{2012CQGra..29l4013S}, and possibly intermediate mass ratio inspirals (the inspirals of stellar origin BHs into intermediate mass BHs with mass $\sim 100M_\odot$)~\cite{brown:2007,Gair:2010dx}. The former do not pose additional modelling challenges, as these signals will be well represented by rescaling templates for lower mass systems. The latter, however, lie in a regime where both finite mass-ratio effects and higher-order PN effects become important. The latter require numerical techniques, but these are limited in terms of the number of cycles that can be modelled, while the former requires perturbative techniques, but are then limited by the size of mass ratio corrections. Any IMRIs observed are likely to be at very low SNR and so will only be identifiable in the data if accurate templates are available. Initial attempts to construct hybrid IMRI models have been made~\cite{Huerta:2011a,Huerta:2011b,Huerta:2012a,Huerta:2012b}, but considerable work is still required.

It is also hoped that it will be possible to test GR to high precision with third-generation ground-based detectors~\cite{2012CQGra..29l4013S}. This requires development of models in alternative theories. This challenge is common to space-based detectors and is discussed further in the next section.

\subsubsection{The Laser Interferometer Space Antenna}
LISA is a space-based GW detector that has been selected as the third large mission that ESA will launch in its current programme, with a provisional launch date of 2034. LISA will comprise three satellites, arranged in an approximately equilateral triangular formation with 2.5Mkm long arms and with laser links passing in both directions along each arm. By precisely measuring the phase of the outgoing and incoming laser light in each arm LISA can do interferometry and detect GWs. It will operate at lower frequency than the LIGO/Virgo detectors, in the range $0.1$mHz--$0.1$Hz, with peak sensitivity around a few mHz. The lower frequency sensitivity means that the typical systems that LISA will observe have higher mass, $M\sim 10^4$--$10^7M_\odot$. Such massive BHs (MBHs) are observed to reside in the centres of lower mass galaxies. LISA is expected to observe mergers of binaries comprised of two such MBHs, which are expected to occur following mergers between the BH host galaxies. MBHs are typically surrounded by clusters of stars, which include BHs similar to those observed by LIGO that were formed as the endpoint of stellar evolution. LISA is also expected to observe the EMRIs of such stellar origin BHs into MBHs. In addition to these MBH sources, LISA will also observe stellar compact binaries in the Milky Way, it could detect some of the stellar origin BH binaries that LIGO will observe and may detect sources of cosmological origin~\cite{Audley:2017drz}. The latter sources do not pose particular modelling challenges, but the BH sources do.

In contrast to the LIGO-Virgo network, there will be only one LISA constellation. While the three LISA arms allow, in principle, the construction of two independent data streams, there will inevitably be some correlation between noise in these channels. In addition, LISA sources are long-lived, lasting months or years in the data set, and so there will be hundreds of sources overlapping in the data. These properties make it much more difficult to construct unmodelled source pipelines like those used in LIGO and so LISA will rely even more heavily on having models of potential signals in order to identify them in the data. Typical MBH binary signals will have SNR in the tens to hundreds, with a few as high as a thousand. This allows much more precise estimates of parameters, but places much higher demands on the fidelity of waveform models. A template accurate to a few percent is fine for characterising a source that has SNR of a few tens as is typical for LIGO/Virgo, but for an SNR of one thousand, the residual after extracting that source will have SNR in the several tens, biasing parameter estimation and contaminating the extraction of subsequent sources. Templates need to be two orders of magnitude more accurate for use in LISA. This accuracy comes coupled to the need for longer duration templates, as for the third-generation ground based detectors, since the signals are present in the data stream for months. MBH binaries detected by LISA are additionally expected to have high spins~\cite{Barausse:2012fy}, in contrast to the observed LIGO/Virgo sources which are all consistent with low or zero spin~\cite{TheLIGOScientific:2016pea,Abbott:2017vtc,Abbott:2017oio,Abbott:2017gyy,Wysocki:2017isg,Wysocki:2018mpo}. Finally, MBH binaries are more likely to show precession. The likely presence of these physical effects in observed signals, coupled with the necessity of model-based searches for LISA places strong requirements on the MBH binary waveform models that must be available by the time LISA flies.

For EMRIs, expected SNRs are in the tens~\cite{Babak:2017tow,Berry:2016bit}, but this SNR is accumulated over $\sim10^5$ cycles. This makes unmodelled EMRI searches impossible and imposes the requirement on modelled searches that the EMRI templates match the true signals to better than one cycle over $10^5$. In addition, all of these cycles are generated in the strong-field where accurate modelling of the signals is more challenging. This drives the requirement for GSF EMRI models accurate to second order in the mass ratio described above. These models must include the effect of high spin in the central MBH, and eccentricity and inclination of the orbit of the smaller BH~\cite{AmaroSeoane:2007aw,AmaroSeoane:2012tx}. EMRI models will also have to be computationally efficient. Naively, to match $10^5$ cycles in a parameter space with $8$ intrinsic (and $6$ extrinsic) parameters, requires $(10^5)^8 \sim 10^{40}$ templates. This is a crude overestimate, but illustrates the complexity of a template-based EMRI search, and the need to be able to generate large numbers of templates in a small computational time. Semi-coherent methods have been proposed~\cite{Gair:2004iv} that have less stringent requirements, but still need templates accurate for $10^3$ or more cycles.

Finally, one of LISA's primary science objectives is to carry out high precision tests of GR, including both tests of strong-field gravity and tests of the nature of compact objects by using EMRIs to probe the gravitational field structure in the vicinity of BHs. Many different tests have been proposed and we refer the reader to~\cite{Gair:2012nm} for a comprehensive review. Several methods exist for phenomenological tests, which assess consistency of the observed signal with the predictions of GR. However, understanding the significance of any constraints that LISA places, and interpreting any deviations that are identified requires models for deviations from the predictions of GR in alternate theories. We require strong-field predictions, most likely relying on numerical simulations, to compare to the merger signals from MBH binaries, which will be observable with high significance. We also require predictions for the sorts of deviations that might be present in the inspiral phase of EMRIs. The latter need to be accurate to a part in $10^5$ and must allow for confusion between gravitational effects and effects of astrophysical origin, e.g., the presence of matter in the system, perturbations from distant objects, etc.

\subsection{Computationally efficient models}
As described above, a significant obstacle to data analysis for future detectors is computational. In order to analyse a large number of potential signals and search large parameter spaces, we require models that capture all the main physics reliably but can be evaluated rapidly. We describe the current status and outstanding challenges here.

\subsubsection{Reduced-order models}
In the context of LIGO/Virgo, interest in developing computationally efficient representations of waveform models arose due to the high cost of parameter estimation algorithms~\cite{2015PhRvD..91d2003V}. This led to the development of reduced-order or surrogate models. The gstLAL search algorithm~\cite{2012ApJ...748..136C,2014PhRvD..89b4003P} uses a singular-value decomposition (SVD) of the signal and noise parameter spaces to efficiently identify candidate signals. However, for parameter estimation we also need to quickly map such a representation onto physical parameters. One approach is to take Fourier representations of gravitational waveforms, construct an SVD of both of these and finally construct a fit to the dependence of the SVD coefficients on source parameters~\cite{2014CQGra..31s5010P,Purrer:2015tud}. An alternative approach is to find a reduced-order representation of the waveform parameter space. A set of templates covering the whole parameter space is constructed and then waveforms are added sequentially to the reduced-basis set using a greedy algorithm. At each stage the template that is least well represented by the current reduced basis is added to the set~\cite{2011PhRvL.106v1102F,2012PhRvD..86h4046F}. This procedure is terminated once the representation error of the basis is below a specified threshold, and typically the number of templates in the final basis is one or more orders of magnitude less than the size of the original set. Representing an arbitrary template with the basis is then achieved by requiring the basis representation to match the waveform at a specific set of points, i.e., interpolation rather than projection. These points are chosen by another greedy algorithm --- picking the points at which the difference between the next basis waveform and its interpolation on the current basis is biggest. This method of representation then allows the likelihood of the data to be written as a quadrature rule sum over the values of the desired waveform at the interpolation points~\cite{2013PhRvD..87l4005C}. This is typically a far smaller number of points than the original time series and so is much cheaper to evaluate than the full waveform. For numerical waveforms an additional step is needed in which the values of the waveform at the desired points is interpolated across parameter space~\cite{2014PhRvX...4c1006F}.

The first GW application of this reduced-order quadrature method used a simple sine-Gaussian burst model~\cite{2013PhRvD..87l4005C} and the first implementation within the LIGO/Virgo analysis infrastructure was for a NS waveform model~\cite{2015PhRvL.114g1104C}. Reduced-order models and associated quadrature rules have subsequently been developed for a number of other waveform models, including NR simulations for higher mass BH binaries~\cite{Blackman:2015pia,Blackman:2017dfb,Blackman:2017pcm} and phenomenological inspiral-merger-ringdown waveform models~\cite{2016PhRvD..94d4031S}. Parameter estimation for GW170817~\cite{TheLIGOScientific:2017qsa} would arguably not have been possible within the timescale on which that paper was written if the reduced order model waveform had not been available.

Some of the existing reduced-order models and reduced-order quadratures will be useful for analysis of data from future detectors. However, the increased duration of signals will typically increase the size of the reduced basis and so work will be needed to optimize the models. In addition, new types of waveform models including higher spins, lower mass ratios or eccentricities do not yet have available reduced-order models, and so these will have to be developed for application to LISA searches for massive BHs and EMRIs. It is possible that some eventual advanced phenomenological models may on their own provide a competitive alternative to Reduced-order models.
\subsubsection{Kludges}
To model EMRIs, ``kludge'' models have been developed that capture the main qualitative features of true inspiral signals. There are two main kludge approaches. The analytic kludge (AK~\cite{Barack:2003fp}) starts from GW emission from Keplerian orbits, as described by~\cite{1963PhRv..131..435P,Peters:1964zz}, and then adds various strong-field effects on top---evolution of the orbit through GW emission, perihelion precession and orbital-plane precession. The model is semi-analytic and hence cheap to generate, which means it has been used extensively to scope out LISA data analysis~\cite{Gair:2004iv,2007CQGra..24S.551A,Babak:2007zd}. However, it rapidly dephases from true EMRI signals. Versions of the AK that include deviations from GR arising from an excess quadrupole moment~\cite{Barack:2006pq} and generic changes in the spacetime structure~\cite{Gair:2011ym} have also been developed. The numerical kludge model (NK) is built around Kerr geodesic trajectories, with the parameters of the geodesic evolved as the object inspirals, using a combination of PN results and fits to perturbative computations~\cite{Gair:2005ih}. A waveform is generated from the trajectory by identifying the Boyer-Lindquist coordinates in the Kerr spacetime with spherical polar coordinates and using a flat-space GW emission formula~\cite{Babak:2006uv}. The NK model is quite accurate when compared to trajectories computed using BH perturbation theory~\cite{Babak:2006uv}. Versions of the NK model including conservative GSF corrections have also been developed~\cite{2009PhRvD..79h4021H}.

Current kludge models are fast but not sufficiently accurate to be used in LISA data analysis. The NK model can be improved using fits to improved perturbative results as these become available. An osculating element formalism has been developed~\cite{Pound:2007th,Gair:2010iv} that can be used to compute the contribution appropriate corrections to the phase and orbital parameter evolution for an arbitrary perturbing force (see~\cite{Warburton:2011fk} for an example using the GSF). This model is likely to be sufficiently reliable for use in preliminary data analysis, but will have to be continually improved as the GSF calculations described in Sec.\ \ref{Sec:perturbations} are further developed. Although much faster than GSF models, the NK model is likely to be too expensive computationally for the first stages of data analysis, in which source candidates are identified using a large number of templates. The AK model in its current form is potentially fast enough, but is not sufficiently accurate. However, recently a hybrid model called the ``Augmented Analytic Kludge'' was proposed~\cite{2015CQGra..32w2002C,Chua:2017ujo} that corrects the AK phase and orbital evolution equations so that they match numerical kludge trajectories. The ``Augmented Analytic Kludge'' model is almost as fast as the AK model and is almost as accurate as the NK model, so it is a promising approach to data analysis, though much work is needed to develop the preliminary model to include the full range of physical effects.

Computationally efficient models will be essential for identifying sources in data from current and future GW detectors. However, they can only be constructed if accurate physical waveform models have been developed, and the final stage of any parameter estimation pipeline must use the most accurate physical model available in order to best extract the source parameters. Accurate source modelling underpins all GW data analysis.

\section {The view from Mathematical GR}\label{Sec:MR}
\vspace{-3mm}
{\it Contributor:} Piotr T.~Chru\'sciel
\vspace{3mm}

The detection of gravitational waves presents formidable challenges to the mathematical relativist: can one invoke mathematical GR to provide an unambiguous interpretation of observations, and a rigorous underpinning for some of the interpretations already made? Here we briefly discuss some of the issues arising, focusing on questions that are tractable using available methods in mathematical GR. We shall sidestep a host of other important problems, such as that of the global well-posedness of the Cauchy problem for binary BHs, or the related problem of \emph{cosmic censorship},  which are currently completely out of reach to mathematical GR.

\subsection{Quasi-Kerr remnants?}
 \label{ss23V18.1}
A working axiom in the GW community is that \emph{Kerr BHs exhaust the collection of BH end states}. There is strong evidence for this, based on the ``no-hair'' theorem, which asserts that \emph{suitably regular stationary, analytic, non-degenerate, connected, asymptotically flat vacuum BHs belong to the Kerr family}; see Ref.~\cite{Chrusciel:2012jk} for precise definitions and  a list of many contributors (see also Secs.~\ref{sec:nohair} and \ref{sec:hairyBHs} of Chapter III for a discussion on no-hair theorems in the context of beyond-GR phenomenology). The analyticity assumption is highly undesirable, being rather curious: it implies that knowledge of a metric in a small space-time region determines the metric throughout the universe. This assumption was relaxed  in~\cite{Alexakis:2009ch} for near-Kerr configurations and in~\cite{AIK3} for slowly rotating horizons, but the general case remains open. Next, the non-degeneracy assumption is essentially that of non-vanishing surface gravity: the maximally spinning Kerr BHs are degenerate in this sense. One often discards the extremal-Kerr case by declaring that it is unstable. This might well be the case, but it is perplexing that the spin range of observed SMBH candidates is clearly biased towards high spin, with a possible peak at extremality~\cite{Brenneman:2013oba,Reynolds:2013qqa}. (Note, however, that these studies have an a priori assumption built-in, that the BHs are Kerr.)
Finally, while the connectedness assumption is irrelevant when talking about an isolated BH, it is quite unsatisfactory to have it around: what mechanism could keep a many-BH vacuum configuration in a stationary state?
 All this makes it clear that removing the undesirable assumptions of \emph{analyticity, non-degeneracy, and connectedness} in the uniqueness theorems should be a high priority to mathematically-minded relativists.
\subsection{Quasi-normal ringing?}
 \label{ss23V18.2}
The current paradigm is that, after a merger, the final BH settles to a stationary one by emitting radiation which, at the initial stage, has a characteristic profile determined by  \emph{quasi-normal modes}. This characteristic radiation field provides a very useful tool for determining the final mass and angular momentum of the solution.  (Note that the extraction of these parameters from the ringdown profile assumes the final state to be Kerr, taking us back to the issues raised in the previous paragraph...) All this seems to be well supported by numerics. But we are still far behind by way of mathematical proof. Work by Dyatlov~\cite{Dyatlov,Dyatlov2} has rigorously established that  quasi-normal modes  are part of an asymptotic expansion of the time-behaviour of linear waves on the \emph{Kerr-de Sitter} background with a nonzero cosmological constant. The asymptotically flat Kerr case turns out to be much more difficult to tackle,  and so far no rigorous mathematical statements are available for it in the literature. The whole problem becomes even more difficult when non-linearities are introduced, with no results available so far. The recently announced proofs of nonlinear stability of the Schwarzschild solution within specific families of initial data~\cite{DHRT,KlainermanSzeftel} might serve as a starting point for further studies of this important problem.
\subsection{Quasi-local masses, momenta and angular momenta?}
It was mind-blowing, and still is, when LIGO detected two BHs of about 35 and 30 solar masses merging into one, of about 60 solar masses. But the question arises, what do these numbers mean? Assuming  the end state to be Kerr, the last part of the statement is clear. But how can the pre-merger  masses be determined, or even defined, given the non-stationarity of the BBH configuration? 
Each NR group uses its own method for assigning mass and spin parameters to the initial datasets used in their simulations, and it is not {\it a priori} clear how these numbers relate to each other. EOB models likewise employ their own definitions of mass and spin.  Any differences are most likely irrelevant at the current level of detection accuracy. But one hopes that it will become observationally important at some stage, and is a fundamental issue in our formulation of the problem in any case. 

To address the issue requires choosing benchmark definitions of quasi-local mass, momentum, and angular momentum that can be blindly calculated on numerical initial datasets, without knowing whether the  data were obtained, e.g., by solving the spacelike constraints with an Ansatz, or by evolving some other Ansatz, or by matching spacelike and characteristic data.
One can imagine a strategy whereby one first locates the apparent horizons within a unique preferred time-slicing (e.g., maximal, keeping in mind that apparent horizons are slicing-dependent), and then calculates the chosen quasi-local quantities for those. There exists a long list of candidates for quasi-local mass, with various degrees of computational difficulty, which could be used after extensive testing, all to be found in~\cite{Szabados:2009eka}; alphabetically: Bartnik's, Brown-York's, Hawking's (directly readable from the area), Kijowski's, Penrose's, with the currently-most-sophisticated one due to Chen, Wang and Yau (see~\cite{ChenWangYau1804} and references therein).
The issue is not whether there is an \emph{optimal} definition of quasi-local quantities, which is an interesting theoretical question on its own, unlikely to find a universally agreed answer, but whether there is a \emph{well-defined and computationally convenient one} that can be used for scientific communication. Such a definition would need to have an unambiguous Newtonian limit, necessary for making contact with non-GR astrophysical observations. Incidentally, in  recent work~\cite{Chen:2019obg}, the Wang and Yau mass has been calculated for balls of fixed radius receding to infinity along null geodesics. The key new observation is that the leading order volume integrand is the square of the Bondi news function. This is accompanied by an integral over the surface of the bounding spheres which deserves further investigation, numerically or otherwise.

\subsection{Quasi-mathematical numerical relativity?}

To an outsider, NR looks like a heroic struggle with a quasi-impossible task, which, after years of inspired attempts, resulted in a maze of incredibly complicated codes that manage to perform the calculations related to the problem at hand. It is conceivable that there is no way to control the whole construction in a coherent way. However, some mathematically minded outsiders would like to get convinced that the numerical calculations are doing what they are supposed to do; compare~\cite{KGO,Sarbach:2012pr}. In other words, is it possible to show that, at least  some, if not most, if not all of the current  numerical approximations to Einstein equations would converge to a real solution of the problem at hand if the numerical accuracy could be increased without limit?  Standard convergence tests, or checks that the constraints are preserved, are of course very important, but they cannot on their own settle the point. A more rigorous proof of convergence is desirable.


\newpage

\phantomsection
\addcontentsline{toc}{part}{\bf Chapter III: Gravitational waves and fundamental physics}
\begin{center}
{\large \bf Chapter III: Gravitational waves and fundamental physics}
\end{center}
\begin{center}
Editor: Thomas P.~Sotiriou
\end{center}

\vskip 1cm
\newcommand{\pp}[1]{{\textcolor{red}{\sf{[PP: #1]}} }}
\setcounter{section}{0}
\section{Introduction} \label{Sec:introduction3}

Gravity is arguably the most poorly understood fundamental interaction. It is clear that General Relativity (GR) is not sufficient for describing the final stages of gravitational collapse or the very early universe. Additionally, a deeper understanding of the role of gravitation seems to be a necessary ingredient for solving almost any other major challenge in fundamental physics, cosmology, and astrophysics, such as the hierarchy problem, or the dark matter (DM) and dark energy problems. Gravitational waves (GWs) promise to turn gravity research into a data-driven field and potentially provide much needed insights. However, one needs to overcome a major obstacle first: that of extracting useful information about fundamental physics from GW observations. The reason this is a challenge should have become apparent in the previous chapter. The  signal is buried inside noise and extracting it requires precise modelling. Doing so in GR is already a major feat and it only gets harder when one tries to add new ingredients to the problem. Nonetheless there is very strong motivation. Black hole binary (BBH) mergers are among the most violent events in the universe and some of the most interesting and exotic phenomena are expected to take place in the vicinity of BHs.

This chapter focusses on how BHs can be used to probe fundamental physics through GWs. The next section sets the background by discussing some examples of beyond-GR scenarios. Sec.~\ref{Sec:detection} summarises the techniques that are being used or developed in order to probe new fundamental physics through GWs. Sec.~\ref{Sec:nonKerr} is devoted to the efforts to probe the nature and structure of the compact objects that are involved in binary mergers. Finally, Sec.~\ref{Sec:DM} provide a thorough discussion on what GW observation might reveal on the nature of DM. 

\section{Beyond GR and the standard model} \label{Sec:beyondGR}

\subsection{Alternative theories as an interface between new physics and observations}

\label{sec:theories}

The most straightforward way to test new fundamental physics with GWs from BBH coalescences would be the following: select one's favourite scenario and model the system in full detail, extract a waveform (or better a template bank), and then look for the prediction in the data. The technical difficulties of doing so will become apparent below, where it will also be apparent that the required tools are at best incomplete and require further development. However, there is a clear non-technical drawback in this approach as well: it assumes that one knows what new fundamental physics to expect and how to model it to the required precision. One should contrast this with the fact that most quantum gravity candidates, for example, are not developed enough to give unique and precise predictions for the dynamics of a binary system. Moreover, ideally one would want to obtain the maximal amount of information from the data, instead of just looking for very specific predictions, in order not to miss unexpected new physics. Hence, the question one needs to ask is what is the optimal way to extract new physics from the data?

GW observations test gravity itself and the way matter interacts through gravity. Hence, at the theoretical level, they can test GR and the way GR couples to the Standard Model (SM) and its extensions. This suggests clearly that the new fundamental physics that can leave an imprint on GW observations can most likely be effectively modelled using an alternative theory of gravity. Recall that alternative theories of gravity generically contain extra degrees of freedom that can be nonminimally coupled to gravity. 

The advantages of this approach are: (i) Alternative theories of gravity can act as {\em effective field theories} for describing certain effects and phenomena ({\em e.g.}~violations of Lorentz symmetry or parity) or be eventually linked to a specific more fundamental theory; (ii) they can provide a complete framework for obtaining predictions for binary evolutions and waveforms, as they come with fully nonlinear field equations; (iii) Their range of validity is broad, so they allow one to combine constraints coming from the strong gravity regime with many other bounds from {e.g.}~the weak field regime, cosmology, astrophysics, laboratory tests, etc. The major drawback of this approach is that it requires theory-dependent modelling, which can be tedious and requires one to focus on specific alternative theories of gravity. 

In this Section we will give a brief overview of some alternative theories that can be used to model new physics in GW observations. This is not meant to be a comprehensive list nor is it our intent to pinpoint interesting candidates. We simply focus on theories that have received significant attention in the literature in what regards their properties in the strong gravity regime and to which we plan to refer to in the coming Sections. 

It is worth mentioning that a complementary approach is to use theory-agnostic {\em strong field parametrisations}. Their clear advantage is that they simplify the modelling drastically and they render it theory-independent. However, they fall short in points (i)-(ii) above. In specific, it is no always straightforward to physical interpret them, they typically describe only part of the waveform, and they do not allow one to combine constraints with other observations unless interpreted in the framework of a theory.  Strong field parametrisations and their advantanges and disadvantages will be discussed in Secs.~\ref{sec:propagation}, \ref{sec:inspiral} and \ref{sec:ringdown}.

\subsection{Scalar-tensor theories}
\vspace{-3mm}
{\it Contributors:}  T.~P.~Sotiriou, K.~Yagi
\vspace{3mm}

\label{sec:sttheories}

One of the simplest ways to modify GR is to introduce a scalar field that is nonminimally coupled to gravity. In many cases, the term scalar-tensor theories refers to theories described by the action
\begin{eqnarray}
\label{staction}
S_{\rm st}&=&\frac{1}{16\pi }\int d^4x \sqrt{-g} \Big(\varphi 
R-\frac{\omega(\varphi)}{\varphi} \nabla^\mu \varphi 
\nabla_\mu \varphi-V(\varphi) \Big)
+S_m(g_{\mu\nu},\psi)\,,
\end{eqnarray}
where $g$ is the determinant of the spacetime metric $g_{\mu\nu}$, $R$ is the Ricci scalar of the metric, and $S_m$ is the matter action. $\psi$ is used to collectively denote the matter fields, which are  coupled minimally to $g_{\mu\nu}$. The functions $\omega(\varphi)$ and $V(\varphi)$ need to be specified to identify a specific theory within the class. Brans-Dicke theory is a special case with $\omega=\omega_0=$constant and $V=0$~\cite{Brans:1961sx}.

It is common in the literature to rewrite the action in terms of a different set of variables. In particular, the 
conformal  transformation $\hat{g}_{\mu\nu}=G \varphi 
\, g_{\mu\nu}$, where $G$ is Newton's constant, and the scalar field redefinition
$2\varphi d\phi=\sqrt{2\omega(\varphi)+3} \, d\varphi$,
can be employed to bring the action~(\ref{staction}) to the form
\begin{equation}
\label{stactionein}
S_{\rm st}=\frac{1}{16\pi G}\int d^4x \sqrt{-\hat{g}} \Big(\hat{R}-2 \hat{g}^{\nu\mu}\partial_\nu \phi \partial_\mu \phi-U(\phi)\Big)+S_m(g_{\mu\nu},\psi)\,,
\end{equation}
where  $U(\phi)=V(\varphi)/(G\varphi)^2$. The set of variables $(\hat{g}_{\mu\nu}, \phi)$ is known as the {\em Einstein frame}, whereas the original set of variables $(g_{\mu\nu},\varphi)$ is known as the {\em Jordan frame}. Quantities with a hat are defined with $\hat{g}_{\mu\nu}$. In the Einstein frame, scalar-tensor theories seemingly take the form of Einstein's theory with a minimally coupled scalar field (hence the name). However, this comes at a price: $\phi$ now couples to the matter field $\psi$, as can be seen if $S_m$ is written in terms of $\hat{g}_{\mu\nu}$, $\phi$, and $\psi$. Hence, if one sticks to the Einstein frame variables, one would infer that matter experiences an interaction with $\phi$ (fifth force), and hence it cannot follow geodesics of $\hat{g}_{\mu\nu}$. In the Jordan frame instead, there is no interaction between $\varphi$ and $\psi$ (by definition), and this implies that matter follows geodesics of $g_{\mu\nu}$. This line of thinking allows one to ascribe a specific physical meaning to the Jordan frame metric, but it does not change the fact that the two frames are equivalent descriptions of the same theory (see Ref.~\cite{Sotiriou:2007zu} and references therein for a more detailed discussion). 

Scalar-tensor theories are well-studied but generically disfavoured by weak field observations, {\em e.g.}~\cite{Bertotti:2003rm,Perivolaropoulos:2009ak}. In brief, weak field tests dictate that the scalar field, if it exists, should be in a trivial configuration around the Sun and around weakly gravitating objects, such as test masses in laboratory experiments. These precision  tests translate to very strong constraints on scalar-tensor theories; strong enough to make it unlikely that there can be any interesting phenomenology in the strong gravity regime, which is our focus here. There is one notable known exception~\cite{Damour:1993hw}. To introduce these models it is best to first set $V=U=0$ and then consider the equation of motion for $\phi$, as obtained from varying action (\ref{stactionein}). It reads
\begin{equation}
\hat{\Box} \phi=-4\pi G \alpha(\phi) T \,,
\end{equation}
 where $T\equiv T^\mu_{\phantom{a}\mu}$, $T_{\mu\nu}\equiv -2 (-g)^{-1/2} \delta S_m/\delta g^{\mu\nu}$ is the  Einstein frame stress energy tensor, and 
\begin{equation}
\alpha(\phi)=- [2\omega(\varphi)+3]^{-1/2}\,.
\end{equation}
Solution with $\phi=\phi_0=$ constant are admissible only if $\alpha(\phi_0)=0$. This condition identifies a special class of theories and translates to $\omega(\varphi_0)\to \infty$ in the Jordan frame, where $\varphi_0$ is the corresponding value of $\varphi$.  For $\phi=\phi_0$ solutions the corresponding metric is identical to the GR solution for the same system. However, these solutions are not unique. It turns out that they are preferable for objects that are less compact than a certain threshold, controlled by the value of $\alpha'(\phi_0)$. When compactness exceeds the threshold, a nontrivial scalar configuration is preferable, leading to deviations from GR~\cite{Damour:1992we,Damour:1993hw,Damour:1996ke}. The phenomenon is called {\em spontaneous scalarization} and it is known to happen for neutron stars (NSs).

A NSB endowed with non-trivial scalar monopolar configurations would lose energy through dipole emission~\cite{Will:2014kxa} and this would affect the orbital dynamics. More specifically, this would be an additional, lower-order, contribution to the usual quadrupolar GW emission that changes the period of binary pulsars. Hence, binary pulsar constraints are extremely efficient in constraining spontaneous scalarization and the original scenario is almost ruled out \cite{Freire:2012mg,Shao:2017gwu}, with the exception of rapidly rotating stars. However, dipolar emission can be avoided if the scalar-scalar interaction between the stars is suppressed. This is the case is the scalar is massive \cite{Alsing:2011er,Ramazanoglu:2016kul}. The mass defines a characteristic distance, beyond which the fall-off of the scalar profile is exponential (as opposed to $1/r$), so if the value of the mass is in the right range, the stars can be scalarized and yet the binary will not exhibit any appreciable dipolar emission above a given separation. One can then avoid binary pulsar constraints and still hope to see some effect in GW emission from the late inspiral, merger, and ringdown.  There is also the possibility that NSs will \emph{dynamically} scalarize once they come close enough to each other~\cite{Barausse:2012da,Palenzuela:2013hsa,Shibata:2013pra,Taniguchi:2014fqa,Sennett:2016rwa}.  

It should be noted that the known scalarization scenario faces the following issue. Usually, calculations assume flat asymptotics with a vanishing gradient for the scalar. However, realistically the value of the scalar far away from the star changes at cosmological timescales and it turns out that this change is sufficient to create conflict with observations~\cite{Sampson:2014qqa,Anderson:2017phb}. This problem can be pushed into the future by tuning initial data, so it can technically be easily avoided. Nonetheless, one would eventually need to do away with the tuning, presumably by improving the model.

Black holes in scalar-tensor theories have received considerable attention, most notably in the context of no-hair theorems ({\em e.g.}~\cite{Hawking:1972qk,Bekenstein:1995un,Sotiriou:2011dz}) and their evasions (see also Ref.~\cite{Sotiriou:2015pka} for a discussion). Sections ~\ref{sec:nohair} and \ref{sec:hairyBHs} are entirely devoted to this topic, so we will not discuss it further here. 

The action of eq.~(\ref{staction}) is the most general one can write that is quadratic in the first derivatives of the scalar, and hence it presents itself as a sensible effective field theory for a scalar field nonminimally coupled to the metric. However, this action can be generalized further if one is willing to include more derivatives. Imposing the requirement that variation leads to field equations that are second order in derivatives of both the metric and the scalar leads to the action 
\begin{eqnarray}
\label{hdaction}
S_H&=&\int d^4 x \sqrt{-g}\left(L_2+L_3+L_4+L_5\right)\,,
\end{eqnarray}
where
\begin{eqnarray}
L_2 &=& K(\phi,X)     ,
\\
L_3 &=& -G_3(\phi,X) \Box \phi     ,
\\
L_4 &=& G_4(\phi,X) R + G_{4X} \left[ (\Box \phi)^2 
-(\nabla_\mu\nabla_\nu\phi)^2 \right]     ,
\\
L_5 &=& G_5(\phi,X) G_{\mu\nu}\nabla^\mu \nabla^\nu \phi 
\nonumber\\
&&\qquad- 
 \frac{G_{5X}}{6} \left[ (\Box \phi)^3 - 3\Box 
\phi(\nabla_\mu\nabla_\nu\phi)^2 + 2(\nabla_\mu\nabla_\nu\phi)^3 \right] \,,
\end{eqnarray}
where $X\equiv-\frac{1}{2} \nabla^\mu \phi \nabla_\mu \phi$, $G_i$ need to be specified to pin down a model within the class, and $G_{iX}\equiv \partial G_i/\partial X$. Horndeski was the first to write down this action in an equivalent form~\cite{Horndeski:1974wa} and it was rediscovered fairly recently in Ref.~\cite{Deffayet:2009mn}. Theories described by this action are referred to as {\em generalized scalar-tensor theories}, Horndeski theories, {\em generalized galileons}, or simply scalar-tensor theories. We will reserve the last term here for the action~(\ref{staction}) in order to avoid confusion. The term generalized galileons comes from the fact that these theories were (re)discovered as curved space generalization of a class of flat space scalar theories that are symmetric under  $\phi \to \phi + c_\mu x^\mu +c$, where $c_\mu$ is a constant one-form and $c$ is a constant (Galilean symmetry) \cite{Nicolis:2008in}. In fact, the numbering of the $L_i$ terms in the Lagrangian is a remnant of the original Galileons, where the $i$ index indicates the number of copies of the field. This is no longer true in action (\ref{hdaction}),  but now the $L_i$ term contains $i-2$ second derivatives of the scalar.

In what regards strong gravity phenomenology, a lot of attention has been given to the shift-symmetric version of action (\ref{hdaction}). If $G_i=G_i(X)$, {\em i.e.}~$\partial G_i/\partial \phi=0$, then (\ref{hdaction}) is invariant under $\phi \to \phi+$constant. This symmetry protects the scalar from acquiring a mass from quantum corrections. It has been shown in Ref.~\cite{Hui:2012qt}  that asymptotically flat, static, spherically symmetric BHs in shift-symmetric Horndeski theories have trivial (constant) scalar configuration and are hence described by the Schwarzschild solution. This proof can  be extended to slowly-rotating BHs, in which case the solution is the slowly rotating limit of the Kerr spacetime~\cite{Sotiriou:2013qea}. However, there is a specific, unique term in (\ref{hdaction}) that circumvents the no-hair theorem and leads to BHs that differ from those of GR and have nontrivial scalar configurations (hairy BHs) \cite{Sotiriou:2013qea}: $\phi {\cal G}$, where ${\cal G}\equiv R^2 - 4 R_{\mu\nu} R^{\mu\nu} + R_{\mu\nu\rho\sigma}R^{\mu\nu\rho\sigma}$ is the the Gauss-Bonnet invariant.\footnote{Though this term does not seem to be part of action (\ref{hdaction}), it actually corresponds to the choice $G_5\propto \ln|X|$~\cite{Kobayashi:2011nu}.} This singles out
\begin{equation}
\label{phiG}
S_{\phi{\cal G}}=\frac{1}{16\pi G}\int d^4x \sqrt{-g} \Big(R-\partial^\nu \phi \partial_\mu \phi+\alpha\phi {\cal G}\Big)+S_m(g_{\mu\nu},\psi)\,,
\end{equation}
as the simplest action within the shift-symmetric Horndeski class that has hairy BHs.  $\alpha$ is a coupling constant with dimension of a length squared. Hairy BHs in this theory \cite{Yunes:2011we,Sotiriou:2014pfa} are very similar to those found earlier in the (non-shift-symmetric) theory with exponential coupling \cite{Campbell:1991kz,Kanti:1995vq}, {\em i.e.}~$e^\phi {\cal G}$. The theory with exponential coupling is known as Einstein-dilaton Gauss-Bonnet theory and it arises as a low-energy effective theory of heterotic strings~\cite{Metsaev:1987zx,Maeda:2009uy}.

The subclass of theories described by the action
\begin{equation}
\label{fG}
S_{f(\phi){\cal G}}=\frac{1}{16\pi G}\int d^4x \sqrt{-g} \Big(R-\frac{1}{2}\partial^\nu \phi \partial_\mu \phi+f(\phi){\cal G}\Big)+S_m(g_{\mu\nu},\psi)\,,
\end{equation}
seems to have interesting BH phenomenology in general. This theory admits GR solutions if $f'(\phi_0)=0$ for some constant $\phi_0$. Assuming this is the case, it has recently been proven in Ref.~\cite{Silva:2017uqg} that stationary, asymptotically flat BH solutions will be identical to those of GR, provided $f''(\phi_0){\cal G}<0$. A similar proof, but restricted to spherical symmetry,  can be found in Ref.~\cite{Antoniou:2017acq}. Theories that do not satisfy the $f''(\phi_0){\cal G}<0$ exhibit an interesting phenomenon \cite{Doneva:2017bvd,Silva:2017uqg}: BH scalarization, similar to the NS scalarization discussed above. See also Ref.~\cite{Antoniou:2017hxj} for a  study of hairy BH solution in this class of theories and Ref.~\cite{Blazquez-Salcedo:2018jnn} for a very recent exploration of linear stability.

The same class of theories can exhibit spontaneous scalarization in NSs~\cite{Silva:2017uqg}. On the other hand, in shift-symmetric Horndeski theories it is known that for NSs the scalar configuration will be trivial~\cite{Barausse:2015wia,Barausse:2017gip,Lehebel:2017fag}, provided that the $\phi {\cal G}$ terms is absent. Even if it is there, there will be a faster than $1/r$ fall-off \cite{Yagi:2011xp,Yagi:2015oca}. Hence, for shift-symmetric theories BH binaries are probably the prime strong gravity system of interest. 

It should be noted that the speed of GWs in generalized scalar tensor theories can differ from the speed of light under certain circumstances and this has been used to obtain constraint recently~\cite{Lombriser:2016yzn,Lombriser:2015sxa,Sakstein:2017xjx,Ezquiaga:2017ekz,Creminelli:2017sry,Baker:2017hug}. These will be discussed in Sec.~\ref{sec:propagation}. 

As mentioned above, action (\ref{hdaction}) is the most general one that leads to second order equations upon variation. Nonetheless, it has also been shown that there exist theories outside this class that contain no other degrees of freedom than the metric and the scalar field \cite{Zumalacarregui:2013pma,Gleyzes:2014dya,Gleyzes:2014qga,Langlois:2015cwa,Crisostomi:2016czh,Achour:2016rkg}. These models are often referred to as {\em beyond-Horndeski} theories.

Last but not least, dynamical Chern-Simons (dCS) gravity~\cite{Jackiw:2003pm,Alexander:2009tp} is a scalar-tensor theory that introduces gravitational parity violation at the level of the field equations. It draws motivation from the standard model~\cite{Weinberg:1996kr}, heterotic superstring theory~\cite{Green:1987mn}, loop quantum gravity~\cite{Ashtekar:1988sw,Alexander:2004xd,Taveras:2008yf,Calcagni:2009xz,Gates:2009pt} and effective field theories for inflation~\cite{Weinberg:2008hq}. dCS gravity is not a member of the Horndeski class. The scalar is actually a pseudo-scalar, {\em i.e.}~it changes sign under a parity transformation. Moreover, it couples to the Pontryagin density $*R^{\alpha\phantom{a}\gamma\delta}_{\phantom{a}\beta}  R_{\phantom{a}\alpha\gamma\delta}^{\beta}$, where $R_{\phantom{a}\beta\gamma\delta}^{\alpha}$ is the Riemann tensor, $*R^{\alpha\phantom{a}\gamma\delta}_{\phantom{a}\beta} \equiv {1\over2}  \epsilon^{\gamma\delta\mu\nu}R_{\phantom{a}\beta\mu\nu}^{\alpha}$ and $\epsilon^{\gamma\delta\mu\nu}$ is the Levi-Civita tensor. This coupling term gives rise to third-order derivatives in the field equations. The fact that the theory has higher-order equations implies that it is unlikely it is actually predictive as it stands~\cite{Delsate:2014hba} (see also \cite{Crisostomi:2017ugk}). Hence, most work has used a specific approach to circumvent that problem that is inspired by effective field theory and relies on allowing the coupling term to introduce only perturbative correction. This issue will be discussed in some detail in Sec.~\ref{IVPandNumerics}.

 dCS gravity has received a lot of attention in the strong gravity regime. Non-rotating BH and NS solutions are the same as in GR, as such solutions do not break parity. On the other hand, slowly-rotating~\cite{Yunes:2009hc,Konno:2009kg,Yagi:2012ya}, rapidly-rotating~\cite{Stein:2014xba} and extremal~\cite{McNees:2015srl,Chen:2018jed} BHs and slowly-rotating NSs~\cite{Yagi:2013mbt} all differ from the corresponding GR  solutions. In particular, such rotating compact objects acquire a scalar \emph{dipole} hair and a correction to the quadrupole moment, which modify gravitational waveforms from compact binary inspirals~\cite{Sopuerta:2009iy,Pani:2011xj,Yagi:2012vf,Canizares:2012is} and mergers~\cite{Okounkova:2017yby}. Future GW observations will allow us to place bounds on the theory that are orders of magnitude stronger than the current bounds from Solar System~\cite{AliHaimoud:2011fw} and table-top experiments~\cite{Yagi:2012ya}. Another interesting phenomenon in dCS gravity is gravitational amplitude birefringence, where the right-handed (left-handed) circularly-polarized GWs are either enhanced or suppressed (suppressed or enhanced) during their propagation from a source to Earth, which can be used to probe the evolution of the pseudo-scalar field~\cite{Alexander:2007kv,Yunes:2008bu,Yunes:2010yf}. Such GW observations are complementary~\cite{Yagi:2017zhb} to Solar System~\cite{Smith:2007jm} and binary pulsar~\cite{Yunes:2008ua,AliHaimoud:2011bk} probes. 

For a general introduction to standard scalar-tensor theories see Refs.~\cite{Fujii:2003pa,2004cstg.book.....F} and for generalized scalar-tensor theories see Ref.~\cite{Deffayet:2013lga}, and for a recent brief review on gravity and scalar fields see Ref.~\cite{Sotiriou:2015lxa}.

\subsection{Lorentz violations}
\vspace{-3mm}
{\it Contributor:} T.~P.~Sotiriou
\vspace{3mm}
\label{sec:LVtheories}

Lorentz symmetry is a fundamental symmetry for the standard model of particle physics, hence it is important to question whether it plays an equally important role in gravity. To address this question one needs to study Lorentz-violating (LV) theories of gravity for two distinct reasons: (i) to quantify how much one is allowed to deviated from GR without contradicting {\em combined} observations that are compatible with Lorentz symmetry one needs to model the deviations in the framework of a consistent theory; (ii) studying the properties of such theories can provide theoretical insights. We will briefly review two examples of LV theories here.

Einstein-aether theory (\ae-theory) \cite{Jacobson:2000xp} is described by the action
\begin{equation} \label{Saetheory}
S_{ae}= \frac{1}{16\pi G}\int d^{4}x\sqrt{-g} (-R -M^{\alpha\beta}{}_{\mu\nu} \nabla_\alpha u^\mu \nabla_\beta u^\nu)
 \end{equation}
where 
\begin{equation} M^{\alpha\beta}{}_{\mu\nu} = c_1 g^{\alpha\beta}g_{\mu\nu}+c_2\delta^{\alpha}_{\mu}\delta^{\beta}_{\nu}
+c_3 \delta^{\alpha}_{\nu}\delta^{\beta}_{\mu}+c_4 u^\alpha u^\beta g_{\mu\nu}\,,
\end{equation}
$c_i$ are dimensionless coupling constants and the  aether $u_\mu$ is forced to satisfy the constraint $u^2\equiv u^\mu u_\mu=1$. This constraint can be enforced by adding to the action the Lagrange multiplier term $\lambda (u^2-1)$ or by restricting the variation of the aether so as to respect the constraint. Due to the constraint, the aether has to be timelike, thereby defining a {\em preferred threading} of spacetime by timelike curves, exactly like an observer would. It cannot vanish in any configuration, including flat space. Hence, local Lorentz symmetry, and more precisely boost invariance, is violated. (\ref{Saetheory}) is the most general action that is quadratic in the first derivatives of a vector field that couples to GR and satisfies the unit constraint. Indeed, \ae-theory is a good effective field theory for Lorentz-symmetry breaking with this field content \cite{Withers:2009qg}. It should be emphasised that the aether brings 3 extra degrees of freedom into the theory, 2 vector modes and a scalar (longitudinal) mode. The speed of any mode, including the usual spin-2 mode that corresponds to GWs, can differ from the speed of light and is controlled by a certain combination of the $c_i$ parameters.

One can straightforwardly obtain an LV theory with a smaller field content from action (\ref{Saetheory}) by restricting the aether to be hypersurface orthogonal \cite{Jacobson:2010mx}. This amounts to imposing
\begin{equation}
\label{ho}
u_\mu=\frac{\partial_\mu T}{\sqrt{g^{\lambda\nu}\partial_\lambda T \partial_\nu T}}\,
\end{equation}
where $T$ is a scalar and the condition is imposed before the variation ({\em i.e.~}one varies with respect to $T$). This condition incorporates already the unit constraint $u^2=1$ and removes the vector modes from the theory. It can be used to rewrite the action in terms of the metric and $T$ only. $T$ cannot vanish in any regular configuration that solves the field equations and its gradient will always be timelike. Hence, the aether is always orthogonal to some spacelike hypersurfaces over which $T$ is constant. That is, $T$ defines a preferred foliation and acts as a preferred time coordinate. Note that it appears in the action only through $u_\mu$ and the latter is invariant under $T\to f(T)$, so this is a symmetry of the theory and the preferred foliation can be relabelled freely. 

One can actually use the freedom to choose the time coordinate and write this theory directly in the preferred foliation \cite{Jacobson:2010mx}. Indeed, if $T$ is selected as the time coordinate, the $T=$ constant surfaces define a foliation by spacelike hypersurfaces with extrinsic curvature $K_{ij}$ and $N$ and $N^i$ are the lapse function and shift vector of this foliation. Then $u_\mu=N \delta^0_\mu$,  and action (\ref{Saetheory}) takes the form 
\begin{equation}
\label{SBPSHIR}
S_T= \frac{1}{16\pi G'}\int dT d^3x \, N\sqrt{h}\left(K_{ij}K^{ij} - \lambda K^2 + \xi {}^{(3)}\!R + \eta a_ia^i \right)\,,
\end{equation}
where $a_i\equiv \partial_i \ln N$ and 
\begin{equation}
\label{HLpar}
\frac{G'}{G}=\xi=\frac{1}{1-(c_{1}+c_3)}, \quad \lambda=\frac{1+c_2}{1-(c_{1}+c_{3})},\quad \eta=\frac{c_1+c_4}{1-(c_1+c_3)}\,.
\end{equation}
Since $T$ was chosen as a time coordinate, action (\ref{SBPSHIR}) is no longer invariant under general diffeomorphisms. It is, however, invariant under the subset of diffeomorphisms that preserve the folation, or else the transformations $T\to T'=f(T)$ and $x^i\to x'^i=x'^i(T,x^i)$, which are now the defining symmetry of the theory from a EFT perspective. In fact, this action can be thought of as the infrared (low-energy) limit of a larger theory with the same symmetry, called Ho\v rava gravity \cite{Horava:2009uw}. The most general action of Ho\v rava gravity is \cite{Blas:2009qj} 
\begin{equation}
\label{SBPSHfull}
S_{H}= \frac{1}{16\pi G_{H}}\int dT d^3x \, N\sqrt{h}\left(L_2+\frac{1}{M_\star^2}L_4+\frac{1}{M_\star^4}L_6\right)\,,
\end{equation}
where $L_2$ is precisely the Lagrangian of action (\ref{SBPSHIR}). $L_4$ and $L_6$ collectively denote all the operators that are compatible with this symmetry and contain 4 and 6 spatial derivatives respectively (in the $T$ foliation). The existence of higher order spatial derivatives implies that perturbation will satisfy higher order dispersion relations and $M_\star$ is a characteristic energy scale that suppresses the corresponding terms. 
Ho\v rava gravity has been argued to be power-counting renormalizable and hence a quantum gravity candidate \cite{Horava:2009uw}. The basic principle is that the higher-order spatial derivative operators modify the behaviour of the propagator in the ultraviolet and remove/regulate divergences (see Refs.~\cite{Visser:2009fg,Visser:2009ys} for a discussion in the context of a simple scalar model). The {\em projectable} version of the theory \cite{Horava:2009uw,Sotiriou:2009gy}, in which the lapse $N$ is restricted to have vanishing spatial derivatives, has actually been shown to be renormalizable (beyond power-counting) in 4 dimensions \cite{Barvinsky:2015kil}, whereas the 3-dimensional version \cite{Sotiriou:2011dr} is asymptotically \cite{Barvinsky:2017kob}. This restricted version does however suffer from serious viability issues at low energies \cite{Horava:2009uw,Sotiriou:2009bx,Charmousis:2009tc,Blas:2009yd,Koyama:2009hc}.

We will not discuss the precise form of $L_4$ and $L_6$ any further, as it not important for low-energy physics. It should be stressed though that such terms lead to higher order corrections to the dispersion relation of GWs and that of the extra scalar excitation Ho\v rava gravity propagates. This implies that the speeds of GW polarizations can differ from the speed of light and acquire a frequency dependence. The relevant constraints from GW observations are discussed in Sec.~\ref{sec:propagation} (see also \cite{Sotiriou:2017obf} for a critical discussion). At large wavenumber $k$, the dispersion relations scale as $\omega^2 \propto k^6$ (since by construction the leading order ultraviolet operators in the action contain 2 temporal and 6 spatial derivatives) and this implies that an excitation can reach any speed, provided it has short enough wavelength. More surprisingly, it has been shown that Ho\v rava gravity exhibits instantaneous propagation even in the infrared limit, described by action (\ref{SBPSHIR}). In this limit, dispersion relations are linear but, both at perturbative \cite{Blas:2011ni} and nonperturbative level \cite{Bhattacharyya:2015uxt}, there exists an extra degree of freedom that satisfies an elliptic equation on constant preferred time hypersurfaces. This equation is not preserved by time evolution and hence it transmits information instantaneously within these hypersurfaces. For a detailed discussion see Ref.~\cite{Bhattacharyya:2015gwa}. 

It should be clear from the discussion above that both \ae-theory and Ho\v rava gravity can exhibit superluminal propagation and this makes BHs particularly interesting in these theories. For both theories BHs  differ from their GR counterparts and have very interesting feature that will be discussed further in Sec.~\ref{sec:hairyBHs}. For a review see Ref.~\cite{Barausse:2013nwa}.

For early reviews on \ae-theory and Ho\v rava gravity see Refs.~\cite{Jacobson:2008aj} and \cite{Sotiriou:2010wn}, respectively. For the most recent {\em combined} constraints see Refs.~\cite{Gumrukcuoglu:2017ijh,Oost:2018tcv} and references therein.

\subsection{Massive gravity and tensor-tensor
  theories}
\vspace{-3mm}
{\it Contributor:} S.~F.~Hassan
\vspace{3mm}
\label{sec:massivegrav}

In the appropriate limit GR reproduces Newton's famous inverse square law of gravitation. At the same time, linear excitations in GR (GWs) satisfy a linear dispersion relation without a mass term. In particle theory terminology, both of these statements imply that the graviton --- the boson that mediates the gravitational interaction --- is massless. Extensions of GR that attempt to give the graviton a mass are commonly referred to
as massive gravity, bimetric gravity, and multimetric gravity
theories. One of the original motivations for the study of these
theories was the hope that they could provide a resolution to the
cosmological constant problem by changing the gravitational interaction beyond some length scale (controlled by the mass of the graviton). However, a more fundamental motivation
emerges from comparing the structure of GR with that
of the standard model of particle physics that successfully accounts
for all known particles interactions. This is elaborated below.

The standard model describes particle physics in terms of fields of
spin 0 (the Higgs field), spin 1/2 (leptons and quarks) and spin 1
(gauge bosons). At the most basic level, these fields are governed by
the respective field equation, namely, the Klein-Gordon equation, the
Dirac equation and the Maxwell/Yang-Mills/Proca equations. The
structures of these equations are uniquely fixed by the spin and mass
of the field along with basic symmetry and consistency requirements,
in particular, the absence of ghost instabilities, for example, in
spin-1 theories. From this point of view, GR is the
unique ghost-free theory of a massless spin-2 field; the metric
$g_{\mu\nu}$. Then the Einstein equation, $R_{\mu\nu} - \frac{1}{2}
g_{\mu\nu} R=0$, with all its intricacies is simply the counterpart
for a spin 2 field of the massless Klein-Gordon equation $\Box
\phi=0$, or of the Maxwell equation $\partial_\mu
F^{\mu\nu}=0$. However, to describe physics, the standard model
contains a second level of structure: At the fundamental level, all
fields appear in multiplets allowing them to transform under
symmetries, in particular under gauge symmetries which are in turn
responsible for the renormalizability and hence, for the quantum
consistency of the theory. The observed fields are related to the
fundamental ones through spontaneous symmetry breaking and through
mixings. For example, to obtain a single Higgs field, one needs to
start with four scalar fields arranged in a complex $SU(2)$ doublet,
rather than the one scalar field of the final theory. Similarly,
although electrodynamics is very well described by Maxwell's equations
as a standalone theory, its origin in the standard model is much more
intricate, requiring the spontaneous breaking of a larger gauge
symmetry and a specific mixing of the original gauge fields into the
vector potential of electromagnetism. Now, one may contrast these
intricate features of the standard model with GR which is
the simplest possible theory of a single massless spin 2 field with no
room for any further structure. The comparison suggests embedding GR
in setups with more than one spin 2 field and investigating the
resulting models. The aim would be to see if the new structures could
address some of the issues raised above, either in analogy with the
standard model, or through new mechanisms. This will shed light on a
corner of the theory space that has remained unexplored so far.

However, implementing this program is not as straightforward as it may
seem. It was known since the early 70's that adding other spin 2 fields to
GR generically leads to ghost instabilities. The instability was first
noted in the context of massive gravity, which contains an interaction
potential between the gravitational metric $g_{\mu\nu}$ and a
nondynamical predetermined metric $f_{\mu\nu}$. However it also
plagues dynamical theories of two or more spin 2 fields.  Recent
developments since 2010 have shown that one can indeed construct ghost
free theories of massive gravity (with a fixed auxiliary metric)
\cite{deRham:2010kj,Hassan:2011hr}, or fully dynamical ghost free
bimetric theories \cite{Hassan:2011zd}. In particular, the bimetric
theory is the theory of gravitational metric that interacts with an
extra spin-2 field in specific ways and contains several
parameters. It propagates one massless and one massive spin-2 mode
which are combinations of the two metrics. Consequently, it contains
massive gravity and GR as two different limits. For a
review of bimetric theory see \cite{Schmidt-May:2015vnx}.

In the massive gravity limit, as the name suggests, the theory has a
single massive spin-2 field that mediates gravitational interactions
in the background of a fixed auxiliary metric. The most relevant
parameter then is the graviton mass which is constrained by different
observations (for a review see, \cite{deRham:2016nuf}). In particular,
consistency with the detection of GWs by LIGO
restricts the graviton mass to $m < 7.7\times 10^{-23}{\rm eV}$
\cite{TheLIGOScientific:2016src,Abbott:2017vtc}. However, in this limit, it becomes
difficult to describe both large scale and small scale phenomena
consistently with the same fixed auxiliary metric.

Of course, the bimetric theory is more relevant phenomenologically
close to the GR limit, where gravity is described by a
mostly massless metric, interacting with a massive spin-2 field. Now
the two most relevant parameters are the spin-2 mass $m$ and the
Planck mass $M$ of the second spin-2 field, $M\ll M_P$, where $M_P$ is
the usual Planck mass of the gravitational metric. The presence of the
second parameter $M$ significantly relaxes the bounds on the spin-2 mass
$m$. In the most commonly assumed scenarios, the stability of
cosmological perturbations in the very early Universe requires $M
\lesssim 100 {\rm GeV}$ \cite{Akrami:2015qga}. In this case, GWs will contain a very small admixture of the massive mode, hence,
$m$ is not strongly constrained by LIGO and can be large. A new
feature is that now the massive spin-2 field can be a DM candidate. Depending on the specific assumptions, the relevant allowed
mass ranges are estimated as $10^{-4} {\rm eV}\lesssim m \lesssim 10^7 {\rm eV}$
\cite{Aoki:2017cnz}, or the higher range, $1 {\rm TeV} \lesssim m \lesssim 
100 {\rm TeV}$ \cite{Babichev:2016bxi}.

While some studies of cosmology in bimetric theory have been carried
out, many important features of the theory are yet to be fully
explored. One of these is the problem of causality and the
well-posedness of the initial value problem. In particular, bigravity
provides a framework for investigating models where the causal metric
need not always correspond to the gravitational metric \cite{Hassan:2017ugh,Kocic:2018yvr}. Another
interesting question is if the bimetric interaction potential can be
generated through some kind of Higgs mechanism, in analogy with the
Proca mass term for spin 1 fields. From a phenomenological point of
view, it is interesting to investigate the massive spin 2 field
present in the theory as a DM candidate. Strong field effects
in bimetric gravity are expected to show larger deviations from GR as
compared to weak field effects.

While ghost free bimetric models have interesting properties, they do
not admit enlarged symmetries, since the two spin 2 fields have only
nonderivative interactions. Constructing ghost free theories of spin 2
multiplets with gauge or global symmetries is a more difficult problem
which requires constructing ghost free derivative interactions. So
far, this is an unresolved problem which needs to be further
investigated. If such theories exist, they will provide the closest
analogues of the standard model in the spin 2 sector.

\section{Detecting new  fields in the strong gravity regime}\label{Sec:detection}

As has been discussed in some detail in Section \ref{Sec:beyondGR}, if new physics makes an appearance in the strong gravity regime, it can most likely be modelled as a new field that couples to gravity. In order to detect this field with GW observations, one has to model its effects on the dynamics of the binary system and produce waveforms that can be compared with observations. The 3 distinct phases of a binary coalescence --- inspiral, merger, and ringdown --- require different modelling techniques. The inspiral and the ringdown are susceptible to different types of perturbative treatment. 

During most of the inspiral, the two members of the binary are well separated and, even though they are strongly gravitating objects individually, they  interact weakly with each other. Hence, one can model this interaction perturbatively and this modeling can be used to produce a partial waveform for the inspiral phase. It should be stressed that this gives rise to a perturbation expansion whose parameters depend on the nature and structure of the members of the binary. Computing these parameters, known as {\em sensitivities}, generally requires modelling the objects composing the binary in a nonperturbative fashion. Moreover, if the objects are endowed with extra field configurations, then one needs to model them as translating in an ambient field (in principle created by the companion), in order to faithfully capture the extra interaction. This is the same calculation one performs to obtain constraints from binary pulsars~\cite{Yagi:2013qpa,Sampson:2014qqa}. An alternative is to try to prepare theory-agnostic, parametrised templates. Three different ways to achieve this will be discussed in detail in Sec.~\ref{sec:inspiral}.

The ringdown can be modelled with linear perturbation theory around a known configuration for a compact object, but there are significant challenges. The most obvious ones are: (i) one needs to first have fully determined the structure of the quiescent object that will be used as background; (ii) the equations describing linear perturbations around such a background are not easy to solve ({\em e.g.}~separability is an issue); (iii) both the background and the perturbation equations depend crucially on the scenario one is considering. An in-depth discussion in perturbation theory around BHs, and its use to describe the ringdown phase, will be given in Sec.~\ref{sec:ringdown}.

The merger phase is highly nonlinear and modeling it requires full-blown numerical simulations. A prerequisite for performing such simulations is to have an initial value formulation and to show that the {\em initial value problem} (IVP) is well-posed (in an appropriate sense, as will be discussed in more detail below). This has not been addressed for the vast majority of gravity theories that are currently being considered. Sec.~\ref{IVPandNumerics} focuses on the current state of the art in modeling the merger phase. It contains a critical discussion of the IVP in alternative theories of gravity and a summary of known numerical results.

Last but not least, one could attempt to detect new fields (polarisations) by the effect they have on GW propagation, as opposed to generation. This amounts to parametrizing the dispersion relation of the GWs and constraining the various parameters that enter this parameterisation, such as the mass, the speed, and the parameters of terms that lead to dispersive effects. The next section is devoted to propagation constraints. 

\subsection{Gravitational wave propagation}
\vspace{-3mm}
{\it Contributor:} T. P. Sotiriou
\vspace{3mm}
\label{sec:propagation}

According to GR, GWs are massless and propagate at the speed of light, $c$. Hence, they satisfy the dispersion relation  $E^2=c^2 p^2$, where $E$ is  energy and $p$ is  momentum. Deviations from GR can introduce modifications to this dispersion relation. LIGO uses the following parametrisation to model such deviations: $E^2=c^2 p^2+A c^\alpha p^\alpha$, $\alpha>0$ \cite{Mirshekari:2011yq}\footnote{A more general parametrisation \cite{Kostelecky:2016kfm} has been used in Ref.~\cite{Monitor:2017mdv}.}. When $\alpha=0$, $A$ represents a mass term in the dispersion relation. The current upper bound on the the mass from GW observations is $7.7\times 10^{-23}{\rm eV}/c^2$ \cite{Abbott:2017vtc}.  When $\alpha=2$, $A$ represents a correction to the speed of GWs. The fact that GWs from the NSB merger GW170817 almost coincided with the gamma ray burst GRB 170817A has given a spectacular double-sided constraint on the speed of GWs, which should agree with the speed of light to less than a part in $10^{15}$~\cite{Monitor:2017mdv,TheLIGOScientific:2017qsa}.

All other corrections introduce dispersive effects which accumulate with the distance the wave had to travel to reach the detector. This enhances the constraints in principle. However, it should be noted that $A$ has dimensions $[A]=[E]^{2-\alpha}$, which implies that corrections that have $\alpha>2$ are suppressed by positive powers of $c p/M_\star$. $M_\star$ denotes some characteristic energy scale and it is useful to think of the constraints on $A$ as constraints on $M_\star$. GWs have long wavelengths, so one expects $\alpha>2$ corrections to be heavily suppressed unless the aforementioned energy scale is unexpectedly small. For instance, for $\alpha=3$ the current upper bound on $M_\star$ is in the $10^9\,{\rm GeV}$ range, whereas for $\alpha=4$ it drops well below the eV range \cite{Abbott:2017vtc,Mirshekari:2011yq,Yunes:2016jcc,Sotiriou:2017obf}.

It is important to stress that all of the aforementioned constraints need to be interpreted within the framework of a theory (as is the case for any parametrisation). The mass bound can be turned to a constraint on the mass of the graviton in massive gravity theories, discussed in Sec.~\ref{sec:massivegrav}. Interestingly, it is not the strongest constraint one can obtain \cite{deRham:2016nuf,Will:2018gku}. All other bounds can in principle be interpreted as constraints on Lorentz symmetry. However, terms with odd powers of $\alpha$ require spatial parity violations as well and the bounds on terms with $\alpha>2$ are probably too weak to give interesting constraints on Lorentz-violating gravity theories \cite{Sotiriou:2017obf}. The double-sided bound on the speed of GWs certainly yields a very strong constraint on Lorentz-violating theories of gravity. However, one has to take into account that such models generically have a multidimensional parameter space even at low energies and generically exhibit extra polarisations, the existence/absence of which can potentially give  stronger or additional constraints \cite{Sotiriou:2017obf}. For details on the implications of this bound for Ho\v rava gravity and for Einstein-\ae ther theory, discussed in Sec.~\ref{sec:LVtheories}, see Refs.~\cite{Gumrukcuoglu:2017ijh} and \cite{Oost:2018tcv} respectively.

Lorentz symmetry breaking is not the only scenario in which GWs propagate at a different speed than light. Lorentz invariant theories can exhibit such behaviour as well if they satisfy two conditions: (i) they have extra fields that are in a nontrivial configuration between the emission and the detection locations; (ii) these fields couple to metric perturbations in a way that modifies their propagation. The second conditions generally requires specific types of nonminimal coupling between the extra fields and gravity. It is satisfied by certain generalised scalar tensor theories \cite{Lombriser:2015sxa}, discussed in Sec.~\ref{sec:sttheories}. Indeed, the bound on the speed has been used to obtain strong constraints on such theories as models of dark energy~\cite{Lombriser:2015sxa,Lombriser:2016yzn,Sakstein:2017xjx,Ezquiaga:2017ekz,Creminelli:2017sry,Baker:2017hug}. It should be stressed, however, that, due to condition (i),  such constraints are inapplicable if one does not require that the scalar field be the dominant cosmological contribution that drives the cosmic speed-up. Hence, they should not be interpreted as viability constraints on the theories themselves, but only on the ability of such models to account for dark energy. 

In summary, modifications in the dispersion relation of GWs can be turned into constraints on deviations from GR. How strong these constraints are depends on the theory or scenario one is considering. The clear advantage of propagation constraints is that they are straightforward to obtain and they do not require modeling of the sources in some alternative scenario. This reveals a hidden caveat: they are intrinsically conservative, as they assume that there is no deviation from the GR waveform at the source. Moreover, as should be clear from the discussion above, propagation constraints have other intrinsic limitations and can only provide significant constraints for models that lead to deviations in the dispersion relation of GWs for relatively long wavelengths. Hence, one generally expects to get significantly stronger or additional constraints by considering the deviation from the GR waveform at the source.

\subsection{Inspiral and parametrized templates}
\vspace{-3mm}
{\it Contributor:} K. Yagi
\vspace{3mm}
\label{sec:inspiral}

\subsubsection{Post-Newtonian parametrisation}

Rather than comparing GW data to template waveforms in non-GR theories to probe each theory one at a time, sometimes it is more efficient to prepare parameterized templates that capture deviations away from GR in a generic way so that one can carry out model-independent tests of gravity with GWs. We will mainly focus on the parameterized modification in the inspiral part of the waveform and discuss what needs to be done to construct parameterized waveforms in the merger-ringdown part of the waveform. 

One way to prepare such parameterized templates is to treat coefficients of each post-Newtonian (PN) term in the waveform phase as an independent coefficient~\cite{Arun:2006yw,Arun:2006hn,Mishra:2010tp}. For gravitational waveforms of non-spinning BBH mergers in GR, such coefficients only depend on the masses of individual BHs. Thus, if GR is correct, any two measurements of these coefficients give us the masses, and the measurement of a third coefficient allows us to carry out a consistency test of GR. This idea is similar to what has been done with binary pulsar observations with measurements of post-Keplerian parameters~\cite{Stairs:2003eg,Perrodin:2017bxr}. Though, it is nontrivial how to perform such tests for spinning BBHs. Moreover, this formalism does not allow us to probe non-GR corrections entering at negative PN orders due to e.g. scalar dipole emission. 

\begin{figure}[t]
  \includegraphics[width=0.7\textwidth]{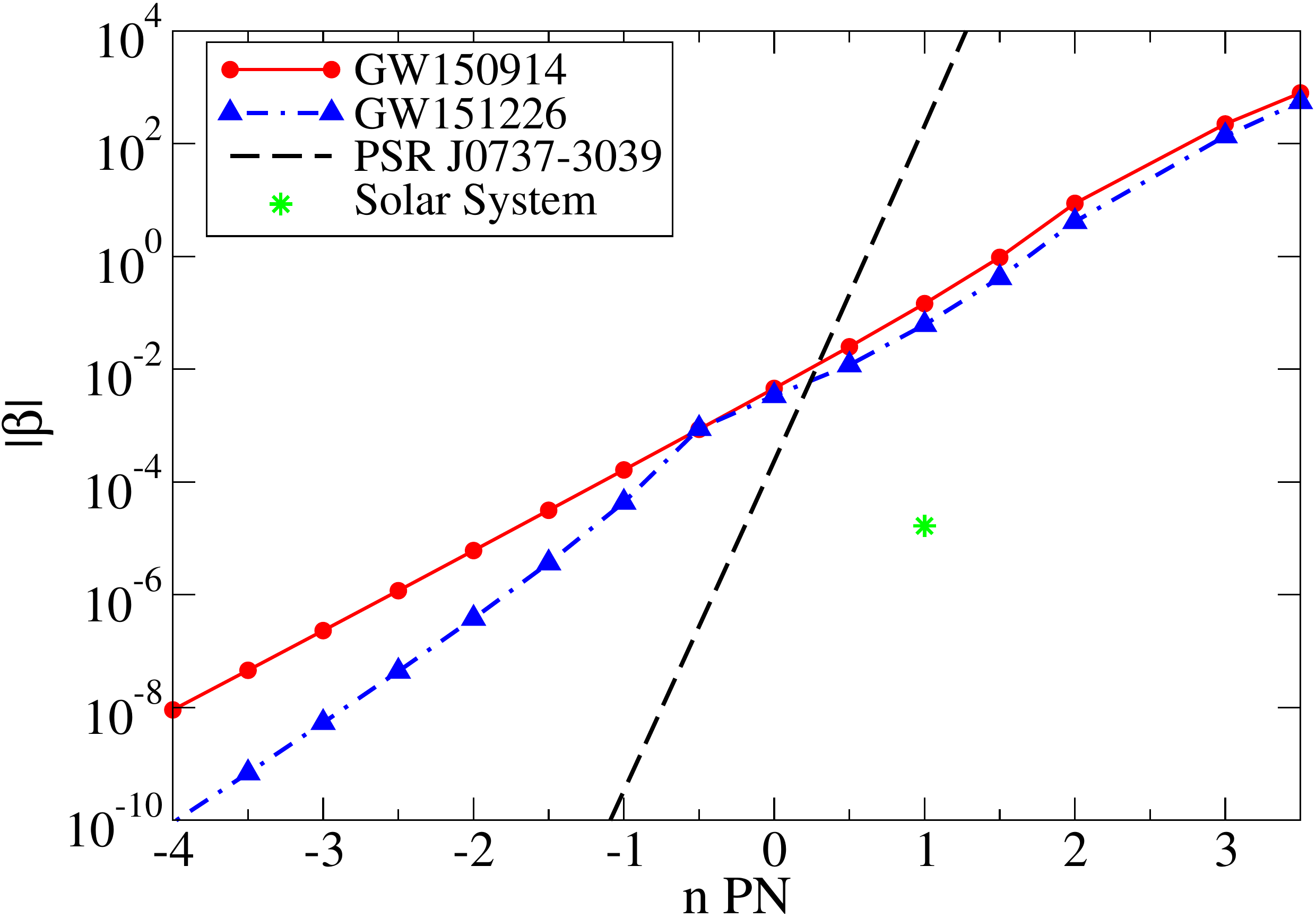}
\caption{\label{beta-bound} Upper bounds on the ppE parameter $\beta$ with GW150914 and GW151226 entering at different PN orders. For comparison, we also present bounds from binary pulsar observations and Solar System experiments.
 This figure is taken and edited from~\cite{Yunes:2016jcc}.
}
\end{figure}

\subsubsection{Parameterized post-Einsteinian (ppE) formalism}

In this  formalism the modified waveform in the Fourier domain is given by $\tilde h (f) = \tilde h_{\mbox{\tiny GR}}(f)\, e^{i \beta v^{2n-5}}$~\cite{Yunes:2009ke}. Here $\tilde h_{\mbox{\tiny GR}}$ is the GR waveform, $\beta$ is the ppE parameter representing the magnitude of the non-GR deviation while $v$ is the relative velocity of the binary components. The above correction enters at $n$th PN order. Since $n$ is arbitrary, this formalism can capture deviations from GR that enter at  negative PN orders. It has been used in~\cite{Yunes:2016jcc} to probe strong and dynamical field gravity with GW150914~\cite{Abbott:2016blz} and GW151226~\cite{Abbott:2016nmj}. Bounds on $\beta$ at each PN order using a Fisher analysis are shown in Fig.~\ref{beta-bound}. Here, the authors included non-GR corrections only in the inspiral part of the waveform. Having these bounds on generic parameters at hand, one can map them to bounds on violations in fundamental pillars of GR. For example, a -4PN deviation maps to time variation of the gravitational constant $G$, that allows us to check the strong equivalence principle. 0PN (2PN) corrections can be used to probe violations in Lorentz (parity) invariance. 

\subsubsection{Phenomenological waveforms}
The LIGO/Virgo Collaboration (LVC) used a generalized IMRPhenom waveform model~\cite{TheLIGOScientific:2016src} by promoting phenomenological parameters $\vec p$ in the GR waveform to $\vec p(1 + \delta p)$, where $\delta p$ corresponds to the fractional deviation from GR. For the inspiral part of the waveform, one can show that these waveforms are equivalent to the ppE waveform. LVC's bounds on non-GR parameters using a Bayesian analysis on the real data set are consistent with the Fisher bound in Fig.~\ref{beta-bound}.

One important future direction to pursue is to come up with a meaningful parameterization of the waveform in the merger and ringdown phases. The generalized IMRPhenom waveform model used by LVC does contain non-GR parameters in the merger-ringdown part of the waveform, though it is not clear what they mean physically or how one can map such parameters to e.g. coupling constants in each non-GR theory.
To achieve the above goal, one needs to carry out as many BBH merger simulations as possible in theories beyond GR~\cite{Healy:2011ef,Berti:2013gfa,Okounkova:2017yby,Jai-akson:2017ldo,Hirschmann:2017psw} to give us some insight on how we should modify the waveform from GR in the merger-ringdown phase.

\subsection{Ringdown and black hole perturbations beyond General Relativity}
\vspace{-3mm}
{\it Contributor:} P. Pani
\vspace{3mm}
\label{sec:ringdown}

Depending on the physics being tested, the post-merger ``ringdown'' phase of a compact-binary coalescence might be better suited than the inspiral or the merger. For instance, in the inspiral phase the nature of the binary components shows up only at high PN order (though corrections due to extra polarizations can show up at low PN order) whereas the merger phase requires time-consuming and theory-dependent numerical simulations. On the other hand, the post-merger phase --- during which the remnant BH settles down to a stationary configuration --- can be appropriately described by perturbation theory and it is therefore relatively simple to model beyond GR. If the remnant is a BH in GR, the ringdown phase is well-described by a superposition of exponentially damped sinusoids, 
\begin{equation}
h_+ +i h_\times \sim \sum_{lmn} A_{lmn}(r) e^{-t/\tau_{lmn}}\sin (\omega_{lmn} t+\phi_{lmn})\,, \label{ringdown}
\end{equation}
called quasinormal modes (QNMs)~\cite{Kokkotas:1999bd,Ferrari:2007dd,Berti:2009kk,Konoplya:2003ii}. Here, $\omega_{lmn}$ is the characteristic oscillation frequency of the final BH, $\tau_{lmn}$ is the damping time, $A_{lmn}$ is the amplitude at a distance $r$, $\phi_{lmn}$ is the phase, $l$ and $m$ ($|m|\leq l$) are angular indices describing
how radiation is distributed on the final BH's sky, and $n$ is an overtone index labeling the (infinite, countable) number of modes.  

Equation~(\ref{ringdown}) sets the ground for \emph{GW spectroscopy}: QNMs are the fingerprints of the final BH and extracting their frequency and damping time allows for a consistency check of the GW templates~\cite{TheLIGOScientific:2016qqj}, tests of gravity~\cite{Berti:2015itd,TheLIGOScientific:2016src,Yunes:2016jcc}, and of the no-hair properties of the Kerr solution~\cite{Cardoso:2016ryw}.
As a corollary of the GR uniqueness theorem~\cite{Carter:1971zc,Robinson},
the entire QNM spectrum of a Kerr BH eventually depends only on its mass and spin. Thus, detecting several modes allows for multiple null-hypothesis tests of the Kerr metric~\cite{Gossan:2011ha,Berti:2016lat,Meidam:2014jpa} and, in turn, of one of the pillars of GR.

\subsubsection{Background}
BH perturbation theory beyond GR is well established in the case of nonspinning background solutions. Extending seminal work done in GR~\cite{Vishveshwara,Regge:1957td,Zerilli:1970se,Chandra}, the BH metric is approximated as $g_{\mu\nu} = g_{\mu\nu}^{(0)}+h_{\mu\nu}$,
where $g_{\mu\nu}^{(0)}$ is the background solution, namely the static BH metric that solves the modified Einstein equations in a given theory, whereas $h_{\mu\nu}\ll1$ are the metric perturbations. Likewise, all possible extra fields (e.g.\ the scalar field in a scalar-tensor theory) are linearized around their own background value. Perturbations of spherically symmetric objects can be conveniently decomposed in a basis of spherical harmonics, and modes with different harmonic indices $(l,m)$ and different parity decouple from each other. 

At linear order, the dynamics is described by two systems of ordinary differential equations: the axial (or odd parity) sector and the polar (or even parity) sector. When supplemented by physical boundary conditions at the horizon and at infinity~\cite{Kokkotas:1999bd,Berti:2009kk}, each resulting linear system defines an eigenvalue problem whose solutions are the (complex) QNMs, $\omega_{lmn}-i/\tau_{lmn}$. 
Remarkably, for a Schwarzschild BH within GR the odd and even parity sectors are isospectral~\cite{Chandra}, whereas this property is generically broken in other gravity theories.

Perturbations of spinning BHs beyond GR are much more involved. In GR, the gravitational perturbations of a Kerr BH can be separated using the Newman-Penrose formalism~\cite{Teukolsky:1973ha,Chandra} and the corresponding QNMs are described by a single master equation. This property does not generically hold beyond GR or for other classes of modes. To treat more generic cases, one could perform a perturbative analysis in the spin (the so-called slow-rotation approximation~\cite{Pani:2012bp,Pani:2013pma}) or instead solve the corresponding set of partial differential equations with spectral methods and other elliptic solvers~\cite{Dias:2015nua}.
In general, the spectrum of spinning BHs beyond GR is richer and more difficult to study. This fact has limited the development of parametrized approaches to ringdown tests~\cite{Barausse:2014tra}, which clearly require taking into account the spin of the final object.

Remarkably, there is a tight relation between the BH QNMs in the eikonal limit ($l\gg1$) and some geodesic properties associated to the spherical photon orbit (the photon sphere)~\cite{Ferrari:1984zz,Cardoso:2008bp,Yang:2012he}. 
This ``null geodesic correspondence'' is useful as it requires only manipulation of background quantities which are easy to obtain. Furthermore, it provides a clear physical insight into the BH QNMs in terms of waves trapped within the photon sphere, slowly leaking out on a timescale given by the geodesic instability time scale.
Within this approximation, the QNMs of BHs beyond GR have been recently studied in Refs.~\cite{Blazquez-Salcedo:2016enn} and a parametrized approach has been proposed in Ref.~\cite{Glampedakis:2017dvb}. There are however two important limitations. First, the correspondence is strictly valid only in the eikonal limit, $l\gg1$, whereas the GW signal is typically dominated by the lowest-$l$ modes. More importantly, the geodesic properties are ``kinematical'' and do not take into account dynamical aspects, e.g.\ those related to extra degrees of freedom~\cite{Blazquez-Salcedo:2016enn} or nonminimal couplings~\cite{Konoplya:2017wot}. For example, Schwarzschild BHs in two different theories of gravity will share the same geodesics but will have in general different QNM spectra~\cite{Barausse:2008xv,Blazquez-Salcedo:2016enn}.

\subsubsection{Signatures}
There are several distinctive features of linearized BH dynamics that can be used to perform tests of gravity in the strong-field regime.

\paragraph{Ringdown tests \& the no-hair theorem}
For BBH mergers within GR, the QNMs excited to larger amplitudes are usually the $(2,2,0)$ and $(3,3,0)$ gravitational modes~\cite{Buonanno:2006ui,Berti:2007fi,Barausse:2011kb,Baibhav:2017jhs}.
Because the Kerr BH depends only on two parameters, extracting $\omega_{220}$ and $\tau_{220}$ allows to estimate the mass and spin of the final object, whereas extracting further subleading modes provides multiple independent consistency checks of the Kerr metric, since the QNMs are generically different in extensions of GR. 
Since QNMs probe a highly dynamical aspect of the theory, the latter statement 
is true even for those theories in which the Kerr metric is the only stationary 
vacuum solution~\cite{Barausse:2008xv}. 
These tests requires high SNR in the ringdown phase and will be best performed 
with next generation detectors, especially with 
LISA~\cite{Berti:2004bd,Berti:2005ys,Berti:2007zu,Berti:2016lat}, although 
coherent mode-stacking methods may be feasible also with advanced ground-based 
detectors~\cite{Yang:2017zxs}.
Furthermore, the ability to perform accurate tests will rely on understanding theoretical issues such the starting time of the ringdown~\cite{Baibhav:2017jhs,Bhagwat:2017tkm} and on the modelling of higher harmonics~\cite{Brito:2018rfr}.
Such tests are based on the prompt ringdown response and implicitly assume that the remnant object has an horizon. If an ultracompact exotic compact object rather than a BH forms as a results of beyond-GR corrections, the ringdown signal depends strongly on the final object's compactness. If the compactness is comparable or slightly larger than that of a neutron star, the prompt ringdown will show distinctive signatures~\cite{Pani:2009ss,Macedo:2013jja,Chirenti:2016hzd}. On the other hand, for objects as compact as BHs, the prompt ringdown is identical to that of a Kerr BH, but the signal is characterized by ``echoes'' at late times~\cite{Cardoso:2016rao,Cardoso:2016oxy,Cardoso:2017cqb} (see also Sec.~\ref{sec:ECOs}).
\paragraph{Extra ringdown modes}
In addition to the shift of gravitational modes discussed above, virtually any 
extension of GR predicts extra degrees of freedom~\cite{Berti:2015itd} which may 
be excited during the 
merger~\cite{Blazquez-Salcedo:2016enn,Okounkova:2017yby,Tattersall:2017erk,
Tattersall:2018nve}. Although the amplitude of such modes is still poorly 
estimated, this type of tests calls for novel ringdown searches including two 
modes of different nature~\cite{Brustein:2017koc}.

\paragraph{Isospectrality}
Isospectrality of Schwarzschild BHs in GR has a bearing also on Kerr BHs. Since this property is generically broken in alternative theories, a clear signature of beyond-GR physics would be the appearance of a ``mode doublet'' (i.e., two modes with very similar frequency and damping time) in the waveform~\cite{Barausse:2014tra,Blazquez-Salcedo:2016enn}.

\paragraph{Instabilities}

BHs in Einstein's theory are (linearly mode) stable. A notable exception relates to the superradiant instability triggered by minimally-coupled light bosonic fields~\cite{Brito:2015oca} (see also Section~\ref{Sec:DM} below). For astrophysical BHs, this instability is ineffective except for ultralight bosons with masses below $10^{-10}\,{\rm eV}$, in which case it has important phenomenological consequences (see Sec.~\ref{sec:superradiance}).
The situation is different in modified gravity, for example certain BH solutions 
are known to be unstable in scalar-tensor theories in the presence of 
matter~\cite{Cardoso:2013fwa,Cardoso:2013opa}, in theories with higher-order 
curvature couplings~\cite{Silva:2017uqg,Blazquez-Salcedo:2018jnn}, and in 
massive (bi)-gravity~\cite{Babichev:2013una,Brito:2013wya}. These instabilities 
are not necessarily pathological, but simply indicate the existence of a 
different, stable, ``ground state'' in the spectrum of BH solutions to a given 
theory.
Finally, as discussed in Sec.~\ref{sec:ECOs}, extensions of GR might predict 
exotic compact objects without an event horizon. These objects are typically 
unstable and their instability can lead to peculiar signatures also in the GW 
signal~\cite{Cardoso:2017cqb,Cunha:2017qtt}.

\subsubsection{A parametrised ringdown approach?}

It should be clear from the previous discussion that the Teukolsky formalism is 
insufficient to study BH perturbation theory in alternative theories of gravity. 
Extending it is hence a major open problem. One might be even more ambitious and 
attempt to develop a theory-agnostic parametrisation for the ringdown, similar 
in spirit to that discussed in Sec.~\ref{sec:inspiral} for the inspiral phase. 
This would generally be a two-faceted problem, since the quiescent BH that acts 
as the endpoint of evolution is not a Kerr BH, in general. Hence, one first 
needs to parametrise deviations from the Kerr spacetime and then propose a 
meaningful and accurate parametrisation of the perturbations around such 
non-Kerr background. 

There have been several attempts to parametrise stationary deviations from the Kerr spacetime. \emph{Bumpy} metrics were constructed in Refs.~\cite{Collins:2004ex,Vigeland:2009pr,Vigeland:2010xe} within GR, starting from a perturbation around the Schwarzschild metric and applying the Newman-Janis transformation~\cite{Newman:1965tw} to construct a rotating configuration. 
The \emph{quasi-Kerr} metric uses the Hartle-Thorne metric~\cite{Hartle:1968si}, valid for slowly-rotating configurations, and modifies the quadrupole moment~\cite{Glampedakis:2005cf}. 
Non-Kerr metrics can also be found, with the extra requirement that a Carter-like, second-order Killing tensor exists~\cite{Vigeland:2011ji}. This approach and metric were later simplified in Ref.~\cite{Johannsen:2015pca}. Note however that rotating BH solutions in alternative theories of gravity  do not possess such a Killing tensor in general. Johannsen and Psaltis~\cite{Johannsen:2011dh} proposed the \emph{modified Kerr} metric by starting from a spherically symmetric and static metric (that does not necessarily satisfy the Einstein equations) and applying the Newman-Janis transformation. This approach was later extended in Ref.~\cite{Cardoso:2014rha}. Finally, modified axisymmetric spacetime geometries can be found by adopting a continued-fraction expansion to obtain a rapid convergence in the series~\cite{Rezzolla:2014mua,Konoplya:2016jvv}.  
See e.g. Refs.~\cite{Lin:2015oan,Ghasemi-Nodehi:2016wao} for other parameterizations for deviations from Kerr. Properties of various parameterized Kerr metrics have been studied in~\cite{Johannsen:2013rqa}. See also Ref.~\cite{Yagi:2016jml} for more details on such generic parameterization of the Kerr metric. 

It should be stressed that a shortfall of all of the aforementioned 
parametrisation is that they do not have a clear physical interpretation and 
this makes them rather {\em ad hoc}. Moreover, they only address half of the 
problem of parametrizing the ringdown. So far their use in modelling waveforms 
has been limited, but see Refs.~\cite{Gair:2011ym,Moore:2017lxy} and 
Ref.~\cite{Glampedakis:2017dvb} for applications to extreme-mass ratio inspirals 
and BH ringdown respectively. At this stage there has been very little progress 
on how one can take the second step, {\em i.e.}~parametrise linear perturbations 
around such spacetimes in a theory-agnostic way. Along these lines, a general 
theory of gravitational perturbations of a 
Schwarzschild metric has been developed in Ref.~\cite{Tattersall:2017erk} and 
applied to Horndeski theory in Ref.~\cite{Tattersall:2018nve}.

\subsection{Merger, numerics, and complete waveforms beyond General Relativity\label{IVPandNumerics}}
\vspace{-3mm}
{\it Contributors:} C. Palenzuela,  T. P. Sotiriou
\vspace{3mm}
\label{sec:NRbeyond}

The merger of binary compact objects will test the highly dynamical and strongly non-linear regime of gravity, that can only be modelled by using numerical simulations. 
As already discussed above, the recent direct detection of GWs from BBHs \cite{Abbott:2016blz,Abbott:2016nmj,Abbott:2017vtc,Abbott:2017oio,Abbott:2017gyy}, and of the NSB merger~\cite{TheLIGOScientific:2017qsa} with an associated short GRB \cite{Monitor:2017mdv} and a plethora
of concurrent electromagnetic signals from the same source \cite{GBM:2017lvd} has already provided constraints on deviations from GR in various forms and contexts ({\em e.g.~}\cite{Yunes:2016jcc,TheLIGOScientific:2016src,Sakstein:2017xjx,Ezquiaga:2017ekz,Creminelli:2017sry,Baker:2017hug,Gumrukcuoglu:2017ijh,Oost:2018tcv} respectively. However, most of these constraints use either partial, potentially parametrised, waveforms or rely on propagation effects. In fact, the true potential of testing GR is currently limited by the lack of knowledge of GW emission during the merger phase in  alternatives to GR~\cite{Yunes:2016jcc}. This problem is particularly acute for heavier BH mergers, where only a short part of the inspiral is detected. 

This suggests that constraints could be strengthened significantly in most cases if one had complete, theory-specific, waveforms. However, performing stable and accurate numerical simulations that would produce such waveforms requires an understanding of several complex issues. Probably the first among them is the well-posedness of the system of equations that describe evolution of a given alternative scenario to GR. In the next section we briefly describe the issue of well-posedness and discuss some techniques that have been used so far to obtain a well-posed evolution system in alternative theories. In Sec.~\ref{sec:numerics} we overview some recent numerical studies of nonlinear evolution beyond GR.

\subsubsection{Initial value formulation and predictivity beyond GR}
\label{sec:ivp}
Modelling the evolution of a binary system for given initial data is a type of initial value problem (IVP). An IVP is well-posed if the solution exists, is unique and does not change abruptly for small changes in the data. A theory with an ill-posed IVP cannot make predictions. The IVP is well-posed in GR \cite{1952AcM....88..141F} but it is not clear if this is true for most of its contenders. This is a vastly overlooked issue and a systematic exploration of the IVP in many interesting alternative theories, such as those discussed in Sec.~\ref{sec:theories}, is pending. 

A class of alternative theories of gravity that are known to be well-posed is scalar-tensor theories described by action (\ref{staction}). As discussed in Sec.~\ref{sec:sttheories}, after suitable field redefinitions, these theories take the form of GR plus a scalar field with a canonical kinetic term and potential non-minimal couplings between the scalar and standard model fields, see action (\ref{stactionein}). Since these couplings do not contain more than two derivatives, one can argue for well-posedness using the known results for GR and the fact that lower-order derivative terms are not relevant for this discussion. Interestingly, most alternative theories actually include modifications to the leading order derivative terms in the modified Einstein's equations, so a theory-specific study is necessary. This has been attempted  in very few cases and results are mostly very preliminary. In particular, there is evidence that dynamical Chern-Simons gravity is ill-posed~\cite{Delsate:2014hba}. Certain generalised scalar-tensor theories appear to be ill-posed in a generalised harmonic gauge when linearised over a generic, weak field background \cite{Papallo:2017ddx}; however, note that this result is gauge-dependent, and hence not conclusive. Finally, in certain Lorentz-violating theories that exhibit instantaneous propagation, describing evolution might require solving a mixture of hyperbolic and elliptic equations (where the latter are not constraints as in GR) \cite{Bhattacharyya:2015uxt,Bhattacharyya:2015gwa}.

\subsubsection{Well-posedness and effective field theories}

One is tempted to use well-posedness as a selection criterion for alternative theories of gravity, as a physical theory certainly needs to be predictive (in an appropriate sense). However, alternatives to GR can be thought of as effective field theories --- truncations of a larger theory and hence inherently limited in their range of validity. This complicates the question of what one should do when a given theory turns out to be ill-posed. In fact, it even affects how one should view its field equations and the dynamics they describe in the first place. Effective field theories (EFTs) can often contain spurious degrees of freedom ({\em e.g.}~ghosts) that lead to pathological dynamics. In linearised theories it is easy to remove such degrees of freedom and the corresponding pathologies, but there is no unique prescription to doing so in general. Hence, instead of setting aside theories that appear to be ill-posed when taken at face value, perhaps one should be looking for a way to `cure' them and render them predictive at nonlinear level. 

A very well-known EFT, viscous relativistic hydrodynamics, requires such `curing' to control undesirable effects of higher order derivatives and a prescription for doing so has been given long ago~\cite{Israel:1976tn,1976PhLA...58..213I,Israel:1979wp}. A similar prescription applicable to gravity theories has been given recently~\cite{2017PhRvD..96h4043C}. Roughly speaking, this approach treats the theory as a gradient expansion and, hence, it considers as the cause of the pathologies runaway energy transfer to the ultraviolet, that in turn renders the gradient expansion inapplicable. As a results, it attempts to modify the equations so as to prevent such transfer and to ensure that the system remains within the regime of validity of the effective descriptions throughout the evolution. 

Another approach is to consider the theory as arising from a perturbative expansion is a certain parameter, say $\lambda$. If that were the case, then higher order corrections in $\lambda$ have been neglected and, consequently, the solutions can only be trusted only up to a certain order in $\lambda$. Moreover, they have to be perturbatively close (in $\lambda$) to solutions with $\lambda=0$, which are solution of GR. Hence, one can iteratively generate fully dynamical solutions order-by-order in $\lambda$. This process is effectively an {\em order-reduction} algorithm and yields a well-posed system of equations. It has a long history in gravity theories in different contexts ({\em e.g.~}\cite{Bel:1985zz,Simon:1990ic,Sotiriou:2008ya}), but it has only recently been used for nonlinear, dynamical evolution in alternative theories \cite{Benkel:2016kcq,Benkel:2016rlz,Okounkova:2017yby}.

These two `cures' do not necessarily give the same results. Moreover, there is no reason to believe that there cannot be other ways to address this problem, so this remains a crucial open question.  In principle, the way that an EFT is obtained from a more fundamental theory should strongly suggest which is the way forward when considering nonlinear evolution. However, in practice one often starts with an EFT and hopes to eventually relate it to some fundamental theory. Hence, it seems wise to consider all possible approaches. In principle different theories might require different approaches. 

\subsubsection{Numerical simulations in alternative theories}
\label{sec:numerics}

As mentioned above, one can straightforwardly argue that the IVP is well-posed in standard scalar-tensor theories described by action (\ref{staction}). However, as will be discussed in detail in Sec.~\ref{sec:nohair}, BHs in this class of theories are generically identical to GR and carry no scalar configuration. Hence, even though gravitational radiation in these theories can in principle contain a longitudinal component, it is highly unlikely it will get excited in a BBH. In Sec.~\ref{sec:hairyBHs} we outline the conditions under which BHs can differ from their GR counterparts in standard scalar-tensor theories.\footnote{It might hence be preferable for readers that are not familiar with the literature on hairy BHs to return to this section after going through Secs.~\ref{sec:nohair} and \ref{sec:hairyBHs}.} When these conditions are satisfied there should be an imprint in GWs from BH binaries.  It is worth highlighting here a couple of cases where numerical simulations have to be used to address this question. The first has to do with the role of asymptotics. It has been shown that time-dependent asymptotic for the scalar could lead to scalar radiation during the coalescence~\cite{Berti:2013gfa}, though this emission would probably be undetectable for realistic asymptotic values of the scalar field gradient. The second case has to do with whether matter in the vicinity of a BH can force it to develop a non-trivial configuration. This has been shown in idealised setups only in Refs. \cite{Cardoso:2013fwa,Cardoso:2013opa}. However, numerical simulations for the same phenomenon in NSs have been performed in~\cite{Barausse:2012da,Shibata:2013pra,Palenzuela:2013hsa}.

$f(R)$ theories of gravity (see Ref.~\cite{Sotiriou:2008rp} for a review), which are dynamically equivalent to a specific subclass of scalar-tensor theories, are also well-posed \cite{LanahanTremblay:2007sg,Sotiriou:2008rp}.  A comparative study of NSB mergers in GR with those of a one-parameter model of $f(R)=R+aR^2$ gravity is performed in~\cite{Sagunski:2017nzb}.

A well-posed extension of scalar-tensor theories is Einstein-Maxwell-Dilaton gravity. It has its origin in low energy approximations of string theory and it includes, apart from the metric and a scalar, a U(1) gauge field. The scalar field couples exponentially to the gauge field, allowing for deviations from GR even for BHs with asymptotically constant scalar field. BBH simulations have shown that these deviations are rather small for reasonable values of the hidden charge~\cite{Hirschmann:2017psw}, leading to weak constraints on the free parameters of the theory.

Finally, some first numerical results have recently appeared  in scalar-Gauss--Bonnet gravity, and specifically the theory described by action (\ref{phiG}), and in Chern-Simons gravity (see Sec.~\ref{sec:sttheories} for more details on the theories). Refs.~\cite{Benkel:2016rlz,Benkel:2016kcq} studied scalar evolution in scalar-Gauss--Bonnet gravity and Ref.~\cite{Okounkova:2017yby} performed the first binary evolution in Chern-Simons gravity. These results are notable for using a perturbative expansion in the free parameter of the theory, described in the previous section, in order to circumvent potential issues with well-posedness (as discussed above, if the field equation are taken at face value, well-posedness is known to be an issue for Chern-Simons gravity~\cite{Delsate:2014hba} and it might be an issue for scalar-Gauss--Bonnet gravity~\cite{Papallo:2017ddx}).

\section{The nature of Compact Objects}\label{Sec:nonKerr}

\label{sec:compactbeyond}

\subsection{No-hair theorems}
\vspace{-3mm}
{\it Contributors:}  C.~Herdeiro,  T.~P.~Sotiriou
\vspace{3mm}
\label{sec:nohair}

The discovery of the Schwarzschild solution in 1915-16~\cite{Schwarzschild:1916uq}, shortly after Einstein's presentation of GR~\cite{Einstein:1915ca}, triggered a half-a-century long quest for its rotating counterpart, which ended with the discovery of the Kerr solution in 1963~\cite{Kerr:1963ud} - see~\cite{Kerr:2007dk} for a historical account of this discovery. At this time, unlike in Schwarzschild's epoch, it was already considered plausible that these metrics could represent the endpoint of the gravitational collapse of stars, even though the name \textit{black hole} to describe them only became widespread after being used by Wheeler in 1967-68~\cite{Wheeler:1998vs}.

The extraordinary significance of the Kerr metric became clear with the establishment of the \textit{uniqueness theorems} for BHs in GR - see~\cite{Chrusciel:2012jk} for a review. The first such theorem was that of Israel's for the static case~\cite{Israel:1967wq}. Then, Carter~\cite{Carter:1971zc} and Robinson~\cite{Robinson:1975bv} constructed a theorem stating: \textit{An asymptotically flat stationary and axisymmetric vacuum spacetime that is non-singular on and outside a connected event horizon is a member of the two-parameter Kerr family}. The axial symmetry assumption was subsequently shown to be unnecessary (assuming analyticity, nondegenerary, and causality \cite{Carter:1997im}); for BHs, stationarity actually  implies axial symmetry, via the ``rigidity theorem,'' which relates the teleologically defined ``event horizon'' to the locally defined ``Killing horizon''~\cite{Hawking:1971vc}. The upshot of these theorems is that a stationary vacuum BH, despite possessing an infinite number of non-trivial, appropriately defined multipolar moments~\cite{Hansen:1974zz}, has only two degrees of freedom: its mass and angular momentum. This simplicity was written in stone by Wheeler's dictum ``\textit{BHs have no hair}''~\cite{Ruffini:1971bza}.

In parallel to these developments the astrophysical evidence for BHs piled up over the last half century - see $e.g.$~\cite{Narayan:2013gca}. In fact, the very conference where the Kerr solution was first presented (the first \textit{Texas symposium on relativistic astrophysics}) held in December 1963, was largely motivated by the discovery of \textit{quasars} in the 1950s and the growing belief that relativistic phenomena related to highly compact objects should be at the source of the extraordinary energies emitted by such very distant objects~\cite{Schucking:1989}.  Gravitational collapse was envisaged to play a key role in the formation of these highly compact objects and such collapse starts from  matter distributions, rather than vacuum. Thus, one may ask if BHs in the presence of matter, rather than in vacuum, still have \textit{no-hair}. In this respect, Wheeler's dictum was a conjecture, rather than a synthetic expression of the uniqueness theorems. It hypothesised a much more general and ambitious conclusion than that allowed solely by these theorems. The conjecture stated that: \textit{gravitational collapse leads to equilibrium BHs uniquely determined by mass, angular momentum and electric charge -asymptotically measured quantities subject to a Gauss law and no other independent characteristics (hair)}~\cite{Misner:1974qy}. Here, the Gauss law plays a key role to single out the physical properties that survive gravitational collapse, because they are anchored to a fundamental (gauged) symmetry, whose conserved charge cannot be cloaked by the event horizon. The uniqueness theorems (which can be generalised to electro-vacuum~\cite{Chrusciel:2012jk}), together with the no-hair conjecture, lay down the foundation for the \textit{Kerr hypothesis}, that the myriad of  BHs that exist in the Cosmos are well described, when near equilibrium, by the Kerr metric. This astonishing realisation was elegantly summarised by S. Chandrasekhar~\cite{Chandrasekhar:1989}: \textit{In my entire scientific life, extending over forty-five years, the most shattering experience has been the realisation that an exact solution of Einstein's field equations of GR, discovered by the New Zealand mathematician, Roy Kerr, provides the absolutely exact representation of untold numbers of massive BHs that populate the Universe.}

The Kerr hypothesis can be tested both theoretically and by confrontation with observations. On the theory front, the key question is whether BHs that are not described by the Kerr metric can constitute solutions of reasonable field equations that describe gravity and potentially other fields. Additionally, one can ask whether such BH solutions can form dynamically and are (sufficiently) stable as to be astrophysically relevant. On the observational front, the question is to which extent the BHs we observe are compatible with the Kerr hypothesis and what constraints can one extract from that. 

A set of results on the non-existence of non-Kerr BHs in specific models, either including different matter terms in Einstein's GR or by considering theories of gravity beyond Einstein's GR (or both) have been established since the 1970s and are known as \textit{no-hair theorems} - see $e.g.$ the reviews~\cite{Bekenstein:1996pn,Herdeiro:2015waa,Sotiriou:2015pka,Volkov:2016ehx}.\footnote{Often, the uniqueness theorems are also called no-hair theorems, under the rationale that they show that in vacuum (or electro-vacuum), BHs have no independent multipolar hair, and only two (or three, ignoring magnetic charge) degrees of freedom. Here we use no-hair theorems to refer to no-go results for non-Kerr BHs in models beyond electro-vacuum GR.} No-hair theorems apply to specific models or classes of models and rely on assumptions regarding symmetries, asymptotics, etc.  As will become evident in the next section, dropping one or more of these assumptions in order to circumvent the corresponding no-hair theorem can be a good tactic for finding non-Kerr BHs (see also Ref.~\cite{Sotiriou:2015pka} for a discussion). More generally, understanding the various sort of theorems, their assumptions and limitations is instructive. 

A large body of work was dedicated to the case of scalar hair, partly due to the conceptual and technical simplicity of scalar fields. Indeed, one of the earliest examples of a no-hair theorem was provided by Chase~\cite{Chase:1970}, following~\cite{Penney:1968zz}. Chase inquired if a static, regular BH spacetime could be found in GR minimally coupled to a real, massless scalar field. Without assuming spherical symmetry, he could show that the scalar field necessarily becomes singular at a simply-connected event horizon. Thus, a static BH cannot support scalar hair, in this model. The 1970s witnessed the formulation of influential no-hair theorems, notably, by Bekenstein, who developed a different approach for establishing the non-existence of scalar hair, applicable to higher spin fields as well~\cite{Bekenstein:1971hc,Bekenstein:1972ky,Bekenstein:1972ny}. An elegant no-hair theorem was given by Hawking~\cite{Hawking:1972qk} for minimally coupled scalar fields without self interactions and for Brans-Dicke theory \cite{Brans:1961sx}. Hawking's theorem assumes stationarity (as opposed to staticity) and no other symmetry and its proof is particularly succinct. This theorem has been later generalised by Sotiriou and Faraoni~\cite{Sotiriou:2011dz} to scalars that exhibit self interactions and to the class of scalar-tensor theories described by action (\ref{staction}). Hui and Nicolis~\cite{Hui:2012qt} have proven a no-hair theorem for shift-symmetric (and hence massless) scalar fields that belong to the Horndeski class discussed in Sec.~\ref{sec:sttheories}. This proof assumes staticity and spherical symmetry, though it can be easily generalized to slowly-rotating BHs~\cite{Sotiriou:2013qea}. It is also subject to a notable exception, as pointed out in Ref.~\cite{Sotiriou:2013qea}: a linear coupling between the scalar and the Gauss-Bonnet invariant is shift symmetric and yet circumvents the theorem [see Sec.~\ref{sec:sttheories} and action (\ref{phiG})]. Finally, a no-hair theorem  has been given in Ref.~\cite{Silva:2017uqg} for stationary BH solutions for the theory described by action (\ref{fG}), provided $f''(\phi_0){\cal G}<0$;  a similar proof, but restricted to spherical symmetry,  can be found in Ref.~\cite{Antoniou:2017acq}.

\subsection{Non-Kerr black holes}
\vspace{-3mm}
{\it Contributors:} C.~Herdeiro, T.~P.~Sotiriou
\vspace{3mm}
\label{sec:hairyBHs}

Despite the many no-hair theorems discussed above, hairy BH solutions are known to exist in many different contexts. In GR, the 1980s and 1990s saw the construction of a variety of hairy BH solutions, typically in theories with non-linear matter sources - see $e.g.$ the reviews~\cite{Bizon:1994dh,Bekenstein:1996pn,Volkov:1998cc}. The paradigmatic example are the coloured BHs, found in Einstein-Yang-Mills theory~\cite{Volkov:1989fi,Volkov:1990sva,1990JMP....31..928K,Bizon:1990sr}. However, these BHs are unstable and have no known formation mechanism. Hence, they should not be considered as counter examples to the weak no-hair conjecture, and it is unlikely that they play a role in the astrophysical scene. 

Below we will review some more recent attempt to find hairy BH solutions that are potentially astrophysically relevant. We first focus on cases that are covered by no-hair theorems and discuss how dropping certain assumptions can lead to hairy BHs. We then move on to various cases of hairy BHs in theories and models that are simply not covered by no-hair theorems.  This is not meant to be an exhaustive review but merely a discussion of some illuminating examples in order to highlight the interesting BH physics that GWs could potentially probe.

\subsubsection{Circumventing no-hair theorems} 
\label{sec:circumvent}

No-hair theorems rely on assumptions, such as: (i) the asymptotic flatness of the metric and other fields; (ii) stationarity (or more restrictive symmetries); (iii) the absence of matter; (iv) stability, energy conditions and other theorem-specific conditions. There is no doubt that removing any one of these assumptions leads to BH hair. For example, standard scalar-tensor theories described by the action (\ref{staction}) are covered by  the no-hair theorems  discussed in the previous section \cite{Chase:1970,Bekenstein:1971hc,Bekenstein:1972ky,Bekenstein:1972ny,Hawking:1972qk,Sotiriou:2011dz}. Nonetheless, BHs with scalar hair exist in these theories if one allows for anti-de Sitter (AdS) asymptotics ({\em e.g.}~Ref.~\cite{Torii:2001pg}), or attempts to ``embed'' the BHs in an evolving universe \cite{Jacobson:1999vr}, or allows for matter to be in their vicinity \cite{Cardoso:2013fwa}. The case of AdS asymptotics, though interesting theoretically, is not relevant for astrophysics. The fact that cosmic evolution could potentially endow  astrophysical BHs with scalar hair is certainly enticing but the effect should be very small  \cite{Horbatsch:2011ye}. Finally, it is not yet clear if the presence of matter in the vicinity of a BH could induce a  scalar charge that is detectable with the current observations.  See Ref.~\cite{Sotiriou:2015pka} for a more detailed discussion. Circumventing the theorems by violating stability arguments will be discussed in Sec.~\ref{sec:BHscalarization}.

A perhaps less obvious way to obtain hairy solutions is to strictly impose the symmetry assumptions on the metric only and relax them for other fields. Even though the field equations relate the fields their symmetries do not need to match exactly in order to be compatible (see, for instance, Ref.~\cite{Smolic:2016dmh} for a recent discussion). A well-studied case is that of a complex, massive scalar field minimally coupled to gravity \cite{Herdeiro:2014goa,Herdeiro:2015gia}. The scalar has a time-dependent phase, but  its stress-energy tensor remains time-independent (thanks to some tuning) and the metric can remain stationary. See also~\cite{Kleihaus:2015iea,Herdeiro:2015tia} for generalisations and~\cite{Chodosh:2015oma} for an existence proof of the solutions. BHs with Proca (massive vector) hair have also been found using the same approach \cite{Herdeiro:2016tmi}.


Whether or not these hairy BH are relevant for astrophysical phenomena depends on three main factors: (1) the existence of (ultra-light) massive bosonic fields in Nature; (2) the existence of a formation mechanism and (3) their stability properties. 
The first factor is an open issue. Ultra-light bosonic fields of the sort necessary for the existence of these BHs with astrophysical masses do not occur in the standard model of particle physics. Some beyond the standard model scenarios, however, motivate the existence of this type of ``matter,'' notably the axiverse scenario in string theory~\cite{Arvanitaki:2009fg}. The second point has been positively answered by fully non-linear numerical simulations~\cite{East:2017ovw,Herdeiro:2017phl}. There is at least one formation channel for these BHs, via the superradiant instability of Kerr BHs, in the presence of ultralight bosonic fields - see~\cite{Brito:2015oca} for a review of this instability. The hairy BHs that result from this dynamical process, however, are never very hairy, being always rather Kerr-like. Concerning the third point, it has been recognised since their discovery, that these BHs could be afflicted by the superradiant instability themselves~\cite{Herdeiro:2015gia,Herdeiro:2014jaa}. Recent work~\cite{Ganchev:2017uuo} succeeded in computing the corresponding timescales, and reported that the timescale of the strongest superradiant instability that can afflict hairy BH is roughly 1000 times longer than that of the strongest instability that can afflict the Kerr BH. This study, albeit based on the analysis of a very small sample of solutions, leaves room for the possibility that some of these BHs can form dynamically and be sufficiently stable to play a role in astrophysical processes \cite{Degollado:2018ypf}. However, a more detailed analysis needs to be performed before definite conclusions can be drawn.

Hairy BHs have also been found in certain shift-symmetric scalar tensor theories by allowing the scalar to depend linearly on time while requiring that the metric be static \cite{Babichev:2013cya,Babichev:2017guv}. This is possible because shift symmetry implies that the scalar only appears in the equations through its gradient and the linear time dependence renders the gradient time independent. The linear stability on such solutions has been explored in Refs.~\cite{Ogawa:2015pea,Babichev:2017lmw,Babichev:2018uiw}.  It remains largely unexplored whether these BHs can form dynamically and hence whether they are relevant for astrophysics.

\subsubsection{Black holes with scalar hair}

The most straightforward way to find hairy BH solutions is to consider theories that are not covered by no-hair theorems. In the case of scalar-tensor theories, it is known that couplings between the scalar (or pseudoescalar) and quadratic curvature invariants, such as the Gauss-Bonnet invariant ${\cal G}\equiv R^2 - 4 R_{\mu\nu} R^{\mu\nu} + R_{\mu\nu\rho\sigma}R^{\mu\nu\rho\sigma}$ or the Pontryagin density $*R^{\alpha\phantom{a}\gamma\delta}_{\phantom{a}\beta}  R_{\phantom{a}\alpha\gamma\delta}^{\beta}$, can lead to scalar hair, see {\em e.g.}~\cite{Campbell:1991kz,Kanti:1995vq,Yunes:2011we,Sotiriou:2013qea,Sotiriou:2014pfa}. Indeed,  the existence of hairy BHs singles out specific class of theories, such as dynamical Chern-Simons (dCS) gravity~\cite{Jackiw:2003pm,Alexander:2009tp} and the scalar-Gauss-Bonnet theories described by the action (\ref{fG}) of Sec.~\ref{sec:sttheories} (which have hairy BH solutions provided that $f'(\phi_0)\neq 0$ for any constant $\phi_0$ \cite{Sotiriou:2013qea,Silva:2017uqg}).
Sec.~\ref{sec:sttheories} already contains a discussion about these theories and their BH phenomenology, so we refer the reader there for details.  See also Sec.~\ref{sec:numerics} for a discussion on the first attempts to simulate dynamical evolution of hairy BHs and to produce waveforms for binaries that contain them. 
%
%
%


\subsubsection{Black hole scalarization}
\label{sec:BHscalarization}

In  the previous section the focus was on theories that are not covered by no hair theorems and have hairy BH for all masses. An interesting alternative was recently pointed out in Refs.~\cite{Doneva:2017bvd,Silva:2017uqg}: that certain theories that marginally manage to escape no-hair theorems \cite{Silva:2017uqg,Antoniou:2017acq} can have hairy BHs only in certain mass ranges, outside which BHs are identical to those of GR. The theories belong in the class describe by action (\ref{fG}) and the phenomenon resembles the``spontaneous scalarization'' of stars that happens in certain models of the standard scalar-tensor theories of action (\ref{staction}); see Sec.~\ref{sec:sttheories} for more details. In the mass ranges where hairy BHs exist the GR BHs of the same mass are expected to be unstable and this instability is supposed to gives rise to the scalar hair. However, depending on the model, the hairy solutions can also be unstable \cite{Blazquez-Salcedo:2018jnn} and this issue requires further investigation. It is also not clear if other theories can also exhibit BH scalarization and the astrophysical significance of the effect has not yet be explored.

\subsubsection{Black holes in theories with vector fields and massive/bimetric theories}

In analogy to the generalised scalar-tensor theories discussed in Sec.~\ref{sec:sttheories} and described by action (\ref{hdaction}), one can construct the most general vector-tensor theories with second-order equations of motion. These theories can be considered as generalised Proca theories and they contain new vector interactions \cite{Heisenberg:2014rta,Jimenez:2016isa}. 
Their phenomenology can be distinct from that of Horndeski scalars and BH solutions with vector hair are known to exist \cite{Chagoya:2016aar,Minamitsuji:2016ydr,Heisenberg:2017xda,Heisenberg:2017hwb,Fan:2016jnz,Cisterna:2016nwq,Babichev:2017rti}.
Depending on the order of derivative interactions, these BH solutions can have primary ({\em i.e.}~the new charge is independent) or secondary ({\em i.e.}~the new charge is not independent) Proca hair. The presence of a temporal vector component
significantly enlarges the possibility for the existence of hairy BH solutions without the need of tuning the models \cite{Heisenberg:2017xda,Heisenberg:2017hwb}. 

Horndeski scalar-tensor and the generalised Proca theories can be unified 
into scalar-vector-tensor theories  for both the gauge invariant and the gauge broken cases \cite{Heisenberg:2018acv}. In the $U(1)$ gauge invariant
and shift symmetric theories the presence of cubic scalar-vector-tensor interactions is crucial for obtaining  scalar hair, which manifests itself
around the event horizon. The inclusion of the quartic order scalar-vector-tensor interactions enables the presence of regular BH solutions endowed with 
scalar and vector hairs \cite{Heisenberg:2018vti}. It is worth mentioning that this new type of hairy BH solutions are stable against odd-parity
perturbations \cite{Heisenberg:2018mgr}.

In massive and bimetric gravity theories the properties of the BH solutions depend heavily on choices one makes for the fiducial or dynamical second metric \cite{Volkov:2012wp,Volkov:2013roa,Brito:2013xaa,Volkov:2014ooa,Babichev:2015xha,Torsello:2017cmz}. If the two metrics are forced to be diagonal in the same coordinate system (bidiagonal solutions) their horizons need to coincide if they are not singular \cite{Deffayet:2011rh}. It should be stressed that generically there is not enough coordinate freedom to enforce the metrics to be bidiagonal. The analysis of linear perturbations reveals also an unstable mode with a Gregory-Laflamme instability \cite{Babichev:2013una,Brito:2013wya}, which remains in bi-Kerr geometry.  Perturbations of the non-bidiagonal geometry is better behaved \cite{Babichev:2014oua}. Non-singular and time dependent numerical solutions were recently found in \cite{Rosen:2017dvn}.

\subsubsection{Black holes in Lorentz-violating theories}

 Lorentz symmetry  is central to the definition of a BH thanks to the assumption that the speed of light is the maximum attainable speed. Superluminal propagation is typical of Lorentz violating theories\footnote{It might be worth stressing that superluminal propagation does not necessarily lead to causal conundrums in Lorentz-violating theories.} 
 and GW observations have already given  a spectacular constraint: the  detection of a NSB merger (GW170817) with coincident gamma ray emission has constrained the speed of GWs to a part in $10^{15}$ \cite{Monitor:2017mdv}. However, Lorentz-violating theories generically exhibit extra polarizations \cite{Sotiriou:2017obf}, whose speed remains virtually unconstrained \cite{ Gumrukcuoglu:2017ijh}. 
 See also Sec.~\ref{sec:propagation}. Moreover, BHs in such theories are expected to always be hairy. The reason has essentially been explained in the discussion of Sec.~\ref{sec:LVtheories}: Lorentz-violating theories can generally be written in a covariant way (potentially after restoring diffeomorphism invariance through the introduction of a Stueckelberg field) and in this setup Lorentz symmetry breaking can be attributed to some extra field that has the property of being nontrivial in any solution to the field equation. That same property will then endow BHs with hair. Indeed, this is known to be true for the theories reviewed in Sec.~\ref{sec:LVtheories}. The  structure of such BHs depends on the causal structure of the corresponding theory. 
 
 In particular, in Einstein-aether theory (\ae-theory), though speeds of linear perturbations can exceed the speed of light, they satisfy an upper bound set by the choice of the $c_i$ parameters. Indeed, massless excitations travel along null cones of effective metrics composed by the metric that defines light propagation and the aether \cite{Eling:2006ec}. This makes the causal structure quasi-relativistic, when one adopts the perspective of the effective metric with the widest null cone \cite{Bhattacharyya:2015gwa}. Stationary BHs turn out to have multiple nested horizons, each of which is a Killing horizon of one of the effective metrics and acts as a causal boundary of a specific excitation \cite{Eling:2006ec,Barausse:2011pu}. 
 
 In Ho\v rava gravity there is no maximum speed, as already explained is Sec.~\ref{sec:LVtheories} . There is a preferred foliation and all propagating modes have dispersion relations that scale as $\omega^2\propto k^6$ for large wave numbers $k$ and there is arbitrarily fast propagation. Even at low momenta, where the dispersion relations are truncated to be linear, there is also an instantaneous (elliptic) mode, both at  perturbative \cite{Blas:2011ni} and  nonperturbative level \cite{Bhattacharyya:2015uxt}. It would be tempting to conclude that BHs cannot exist in Ho\v rava gravity. Nonetheless, there is a new type  of causal horizon that allows one to properly define a BH, dubbed the universal horizon \cite{Barausse:2011pu,Blas:2011ni}. It is not a null surface of any metric, but a spacelike leaf of the preferred foliation that cloaks the singularity. Future-directed signals can only cross it in one direction and this shields the exterior from receiving any signal, even an instantaneous one, from the interior.  Universal horizons persist in rotating BHs in Ho\v rava gravity \cite{Barausse:2012qh,Sotiriou:2014gna,Bhattacharyya:2015gwa,Barausse:2013nwa,Barausse:2015frm} and have been rigorously defined without resorting to symmetries in Ref.~\cite{Bhattacharyya:2015gwa}.

Even though BHs in Lorentz-violating theories and their GR counterparts have very different causal structure, their exteriors can be very similar and hard to tell apart with electromagnetic observations  \cite{Barausse:2011pu,Barausse:2013nwa,Barausse:2015frm}. Nevertheless, the fact that  Lorentz-violating theories have additional polarisations and their BHs are have hair suggests very strongly that binary systems will emit differently than in GR. Hence, confronting model-dependent waveforms with observations is likely to yields strong constraints for Lorentz violation in gravity.

\subsection{Horizonless exotic compact objects\label{sec:ECOs}}
\vspace{-3mm}
{\it Contributors:} V.~Cardoso, V.~Ferrari,  P.~Pani,  F.~H.~Vincent
\vspace{3mm}

BHs are the most dramatic prediction of GR. 
They are defined by an event horizon, a null-like surface that causally 
disconnects the interior from our outside world. While initially considered only 
as curious mathematical solutions to GR, BHs have by now acquired a central role 
in astrophysics: they are commonly accepted to be the outcome of gravitational 
collapse, to power active galactic nuclei, and to grow hierarchically through 
mergers during galaxy evolution. The BH paradigm explains a variety of 
astrophysical phenomena and is so far the simplest explanation for all 
observations.
It is, however, useful to remember that the only evidence for BHs in our universe is the existence of dark, compact and massive objects. 
In addition, BHs come hand in hand with singularities and unresolved quantum effects.
Hence, it is useful to propose and to study exotic compact objects (ECOs) -- dark objects more compact than NSs but without an event horizon.

Perhaps the strongest theoretical motivation to study ECO's follows from quantum gravity. Quantum modifications to GR are expected to shed light on theoretical issues such as the pathological inner structure of Kerr BHs and information loss in BH evaporation. While there is a host of recent research on the quantum structure of BHs, it is not clear whether these quantum effects are of no consequence on physics outside the horizon or will rather lead to new physics that resolves singularities and does away with horizons altogether. The consequence of any such detection, however small the chance, would no less than shake the foundations of physics.  As there is no complete quantum gravity approach available yet, the study of ECOs, in various degrees of phenomenology, is absolutely crucial to connect to current GW observations.

In summary, ECOs should be used both as a testing ground to {\it quantify} the observational evidence for BHs, but also to understand the signatures of alternative proposals and to search for them in GW and electromagnetic data~\cite{Cardoso:2017cqb}.

{\em Classification:} Compact objects can be conveniently classified according to their compactness, $M/R$, where $M$ and $R$ are the mass and (effective) radius of the object, respectively. NSs have $M/R\approx 0.1-0.2$, a Schwarzschild BH has $M/R=1/2$, whereas a nearly extremal Kerr BH has $M/R\approx1$. The study of the dynamics of light rays and of GWs shows
that the ability of ECOs to mimic BHs is tied to the existence of an unstable light ring in the geometry~~\cite{Abramowicz:2002vt,Cardoso:2017cqb,Cardoso:2017njb}.
Accordingly, two important categories of this classification are~\cite{Cardoso:2017cqb,Cardoso:2017njb}:
\begin{enumerate}
 \item \emph{Ultracompact objects (UCOs)}, whose exterior spacetime has a photon sphere. In the static case, this requires $M/R>1/3$. UCOs are expected to be very similar to BHs in terms of geodesic motion. NSs are not in this category, unless they are unrealistically compact and rapidly spinning.
 \item \emph{Clean-photon-sphere objects (ClePhOs)}, so compact that the light travel time from the photon sphere to the surface and back is longer than the characteristic time scale, $\tau_{\rm geo}\sim M$, associated with null geodesics along the photon sphere. These objects are therefore expected to be very similar to BHs at least over dynamical time scales $\sim M$. In the static case, this requires $M/R>0.492$~\cite{Cardoso:2017cqb,Cardoso:2017njb}. 
\end{enumerate}

Some models of ECOs have been proposed in attempts to challenge the BH paradigm. Others are motivated by (semiclassical) quantum gravity scenarios that suggest that
BHs would either not form at all~\cite{Mazur:2004fk,Mathur:2005zp,Mathur:2008nj,Barcelo:2015noa,Danielsson:2017riq,Danielsson:2017pvl,Berthiere:2017tms,Cardoso:2017njb}, or that new physics should drastically modify the structure of the horizon~\cite{Giddings:2013kcj,Giddings:2014nla,Giddings:2017mym}. 

In the presence of exotic matter fields (e.g. ultralight scalars), even classical GR can lead to ECOs. The most characteristic case is perhaps that of \emph{boson stars}~\cite{feinblum68,Kaup:1968zz,ruffini69}: self-gravitating solutions of the Einstein-Klein-Gordon theory. They have no event horizon, no hard surface, and are regular in their interior.
Depending on the scalar self-interactions, they can be as massive as supermassive compact objects and --~when rapidly spinning~-- as compact as UCOs.
Further details about these objects can be found elsewhere~\cite{liebling17}. Though boson stars are solutions of a well defined theory, most ECO-candidates currently arise from  phenomenological considerations and employ some level of speculation. For instance, certain attempts of including quantum gravity effects in the physics of gravitational collapse postulate corrections in the geometry at a Planck distance away from the horizon, regardless of the mass (and, hence, of the curvature scale) of the object. Since $l_{\rm Planck}\ll M$, these objects naturally classify as ClePhOs. 
One example are fuzzballs, ensembles of horizonless microstates that emerge as rather generic solutions to several string theories~\cite{Mathur:2005zp,Mathur:2008nj}. In these solutions the classical horizon arises as a coarse-grained description of the (horizonless) microstate geometries. Another example are so-called gravitational-vacuum stars, or gravastars, ultracompact objects supported by a negative-pressure fluid~\cite{Mazur:2001fv,Mazur:2004fk,Cattoen:2005he,Chirenti:2007mk}, which might arise from one-loop effects in curved spacetime in the hydrodynamical limit~\cite{Mottola:2006ew}.
Other examples of ClePhOs inspired by quantum corrections include black stars~\cite{Barcelo:2009tpa}, superspinars~\cite{Gimon:2007ur}, collapsed polymers~\cite{Brustein:2016msz,Brustein:2017kcj}, $2-2$-holes~\cite{Holdom:2016nek}, etc~\cite{Cardoso:2017njb}. 
We should highlight that for most of these models, $1-2M/R\sim 10^{-39}-10^{-46}$ for stellar to supermassive dark objects. Thus, their phenomenology is very different from that of boson stars, for which $1-2M/R\sim {\cal O}(0.1)$ at most.

The speculative nature of most ECOs implies that their formation process has not been consistently explored (with notable exceptions~\cite{liebling17}). If they exist they are likely subject to the so-called ergoregion instability. This was first found in Ref.~\cite{Friedman:1978wla} for scalar and electromagnetic perturbations (see also~\cite{CominsSchutz,YoshidaEriguchi}), then shown to affect  gravitational perturbations as well in Ref.~\cite{KokkotasRuoffAndersson}, and recently proved rigorously in Ref.~\cite{Moschidis:2016zjy}). The instability affects any spinning compact object with an ergoregion but without a horizon and its time scale may be shorter than the age of the Universe~\cite{Cardoso:2007az,Chirenti:2008pf}. The endpoint of the instability is unknown but a possibility is that the instability removes angular momentum~\cite{Cardoso:2014sna}, leading to a slowly-spinning ECO~\cite{Brito:2015oca}, or perhaps it is quenched by dissipation within the object~\cite{Maggio:2017ivp} (although the effects of viscosity in ECOs are practically unknown). An interesting question is also the outcome of a potential ECO coalescence. All known equilibrium configurations of ECOs have a maximum mass above which the object is unstable against radial perturbations and is expected to collapse. The configuration with maximum mass is also the most compact (stable) one. Therefore, the more an ECO mimics a BH, the closer it is to its own maximum mass. Thus, if two ECOs are compact enough as to reproduce the inspiral of a BH coalescence (small enough deviations in the multipole moments, tidal heating, and tidal Love numbers relative to the BH case; see below), their merger would likely produce an ECO with mass above the critical mass of the model. Hence the final state could be expected to be a BH. 
In other words, one might face a ``short blanket'' problem: ECOs can mimic well \emph{either} the post-merger ringdown phase of a BH \emph{or} the pre-merger inspiral of two BHs, but they may find it difficult to mimic the entire inspiral-merger-ringdown signal of a BH coalescence with an \emph{ECO+ECO $\to$ ECO} process.
The only way out of this problem is to have a mass-radius diagram that resembles that of BHs for a wide range of masses. No classical model is known that behaves this way, yet.

It is worth noting that, even if the BH paradigm is correct, ECOs might be lurking in the universe along with BHs and NSs. Modelling the EM and GW signatures of these exotic sources is necessary to detect them, and may even provide clues for other outstanding puzzles in physics, such as DM (see also Section~\ref{Sec:DM} below).

\subsection{Testing the nature of compact objects}
\vspace{-3mm}
{\it Contributors:} V.~Cardoso,  V.~Ferrari, M.~Kramer, P.~Pani,  F.~H.~Vincent, N.~Wex
\vspace{3mm}

\subsubsection{EM diagnostics}

EM radiation emitted from the vicinity of BH candidates
can probe the properties of spacetime in the strong-field region
and may lead to constraining the nature of the compact object.
We focus our discussion on three popular EM probes, namely \textit{shadows},
\textit{X-ray spectra}, and \textit{quasi-periodic oscillations}.
We do not discuss here polarization, nor effects on stellar trajectories.
Because all EM tests can be eventually traced back to geodesic motion, EM probes may distinguish between BHs and ECOs with $M/R<0.492$, whereas it is much more challenging to tell a ClePhO from a BH through EM measurements.

\paragraph{Shadows}

BHs appear on a bright background as a dark patch on the sky, due to photons captured by the horizon.
This feature is known as the BH shadow~\cite{Bardeen1973,FalckeMeliaAgol2000}. 
The Event Horizon Telescope~\cite{Doeleman:2009te},
aiming at obtaining sub-millimeter images of the shadow of the supermassive
object Sgr~A* at the center of the Milky Way and of the supermassive object at the center of the elliptical galaxy M87, is a strong
motivation for these studies.

A rigorous
and more technical definition of the BH shadow is the following~\cite{Cunha:2018acu}: 
it is the set of directions on an observer's local sky that asymptotically
approach the event horizon when photons are ray traced
backwards in time along them. Thus, by this definition, shadows are intrinsically linked with
the existence of a horizon and, strictly speaking, an ECO cannot have a shadow.
However, in practice ultracompact horizonless objects (in particular UCOs and ClePhOs) might be very efficient in mimicking the exterior spacetime of a Kerr BH. It is therefore interesting to study photon trajectories in such spacetimes and to contrast them with those occurring in the Kerr case.

In the analysis of shadows, one generally either considers
parametrized spacetimes~\cite{Johannsen:2011dh,grenzebach14,Ghasemi-Nodehi:2015raa,wang17} (that allow to tune the departure from
Kerr but might not map to known solutions of some theory), 
or takes into account a specific 
alternative theoretical framework~\cite{amarilla12,amarilla13,wei13,vincent14,tinchev14,moffat15,schee16,cunha17} 
or a particular compact object within GR~\cite{hioki09,nedkova13,cunha15,vincent16}.
Most of these studies obtain differences with respect to the standard
Kerr spacetime that are smaller than the current instrumental resolution ($< 10\,\mu$as).
Some studies report more noticeable differences in the case of naked singularities~\cite{hioki09}, exotic matter that violates some energy condition~\cite{nedkova13}, or
in models that allow for large values of the non-Kerrness parameters~\cite{tinchev14,moffat15}. 
Moreover, demonstrating a clear difference in the shadow is not enough
to infer that a test can be made, since such difference might be degenerate with the mass, spin, distance, and inclination of the source.
Finally, the fact that horizonless objects lead to shadow-like regions
that can share some clear resemblance with Kerr~\cite{vincent16} shows the extreme difficulty
of an unambiguous test based on such observables (for perspective tests with current observations see, e.g., Refs.~\cite{2015ApJ...814..115P,Mizuno:2018lxz}).

\paragraph{$K\alpha$ iron line and continuum-fitting}
The X-ray spectra of BHs in X-ray binaries and AGNs are routinely
used to constrain the spin parameter of BHs, assuming a Kerr
metric~\cite{reynolds03,mcclintock11}. In particular, the iron K$\alpha$ emission line, that is
the strongest observed and is strongly affected by relativistic effects,
is an interesting probe.

Many authors have recently investigated the X-ray spectral observables associated
to non-Kerr spacetimes~\cite{Bambi:2015kza}. The same distinction presented above for the tests of shadows can be made between
parametric studies~\cite{johannsen13b,moore15,ni16}, specific alternative theoretical
frameworks~\cite{harko09,harko09b,vincent14,moore15,schee09}, and
alternative objects within classical GR~\cite{cao16,ni16b}.

Although differences with respect to Kerr are commonly found in various
frameworks, these are generally degenerate with other parameters, such as mass, spin and inclination of the source. Moreover, the precise
shape of the iron line depends on the subtle radiative transfer in the accretion
disk, which is ignored in theoretical studies that generally favor simple
analytic models. This is a great advantage of the shadow method, which is
probably the EM probe least affected by astrophysical systematics.

\paragraph{QPOs}

Quasi-periodic oscillations (QPOs) are narrow peaks in the power spectra, routinely observed in the X-ray light curves of binaries. QPOs are
of the order of Keplerian frequencies in the innermost regions of the
accretion flow~\cite{remillard06}.

A series of recent works were devoted to studying the QPO observables 
associated to alternative compact objects, be it in the context
of parametric spacetimes~\cite{johannsen11,bambi16}, alternative theoretical frameworks~\cite{vincent14,maselli15,chen16},
or alternative objects within GR~\cite{franchini17}.
Although the QPO frequencies can be measured with great accuracy, QPO diagnostics suffer from the same limitations as the X-ray spectrum:
degeneracies and astrophysical uncertainties. The non-Kerr parameters
are often degenerate with the object's spin. Moreover, it is currently
not even clear what the correct model for QPOs is.

\paragraph{Pulsar-BH systems} 

High-precision timing observations of radio pulsars provide very sensitive probes of the spacetime in the vicinity of compact objects. Indeed, the first evidence for the existence of gravitational waves were obtained from observations of binary pulsars \cite{tay94}; light propagation in strong gravitational effects can be precisely tested with Shapiro delay experiments \cite{ksm+06}; relativistic spin-precession can be studied by examining pulse shape and polarisation properties of pulsars \cite{kra98,bkk+08}. The very same techniques can be applied for pulsars to be found around BHs \cite{wk99}. 

While shadows, accretion-disk spectra, and QPOs are probing the near field of the BH, i.e.\ on a scale of a few Schwarzschild radii, it is not expected to find a pulsar that close to a BH. The lifetime of such a pulsar-BH system due to gravitational wave damping is very small, which makes a discovery of such a system extremely unlikely. This is also true for a pulsar around Sgr A$^*$, although observational evidence stills point towards an observable pulsar population in the Galactic Centre\cite{Wharton:2011dv}. Consequently, a pulsar-BH system, once discovered, is expected to provide a far field test, i.e. a test  of only the leading multipole moments of the BH spacetime, in particular mass $M_\bullet$, spin $S_\bullet$ and — for IMBHs and Sgr~A$^*$ — the quadrupole moment $Q_\bullet$ \cite{Liu:2011ae,Liu:2014uka,Zhang:2017qbb}. On the one hand, the measurement of $M_\bullet$, $S_\bullet$ and $Q_\bullet$ can be used to test the {\em Kerr hypothesis}. On the other hand, a pulsar can provide a complementary test to the near field test, and break potential degeneracies of, for instance, a test based on the shadow of the BH\cite{Ghasemi-Nodehi:2015raa,Psaltis:2015uza,Mizuno:2018lxz}. 

A pulsar in a sufficiently tight orbit (period $\lesssim$ few days) about a stellar-mass BH is expected to show a measurable amount of orbital shrinkage due to the emission of gravitational waves. Alternatives to GR generally show the existence of  ``gravitational charges'' associated with additional long-range fields, like e.g.\ the scalar field in scalar-tensor theories. Any asymmetry in such ``gravitational charge'' generally leads to the emission of dipolar radiation, which is a particular strong source of energy loss, as it appears at the 1.5 post-Newtonian level in the equations of motion. Of particular interest here are theories that give rise to extra gravitational charges only for BHs  and therefore evade any present binary pulsar tests. Certain shift-symmetric Horndeski theories known to have such properties \cite{Yunes:2011we,Yagi:2015oca,Sotiriou:2013qea,Sotiriou:2014pfa}, where a star that collapses to a BH suddenly acquires a scalar charge in a nontrivial manner \cite{Benkel:2016rlz,Benkel:2016kcq} (cf.~Sections~\ref{sec:sttheories} and \ref{sec:hairyBHs}). Based on mock timing data, Liu et al.\ \cite{Liu:2014uka} have demonstrated the capability of utilizing a suitable pulsar-BH system to put tight constraints on any form of dipolar radiation.

Finally, there are interesting considerations how a pulsar-BH system could be used to constrain quantum effects related to BHs. For instance, there could be a change in the orbital period caused by the mass loss of an enhanced evaporation of the BH, for instance due to an extra spatial dimension. An absence of any such change in the timing data of the pulsar would lead to constraints on the effective length scale of the extra spatial dimension \cite{Simonetti:2010mk}. If quantum fluctuations of the spacetime geometry outside a BH do occur on macroscopic scales, the radio signals of a pulsar viewed in an (nearly) edge-on orbit around a BH could be used to look for such metric fluctuations. Such fluctuations are expected to modify the propagation of the radio signals and therefore lead to characteristic residuals in the timing data \cite{Estes:2016wgv}.

Given the prospects and scientific rewards promised by PSR-BH systems, searches are on-going to discover these elusive objects. Pulsars orbiting stellar-mass BHs are expected to be found in or near the Galactic plane. Since binary evolution requires such system to survive two supernova explosions, this implies a low systemic velocity, placing it close to its original birth place. On-going deep Galactic plane surveys, like those as part in the High Resolution Universe Survey \cite{cck+18} or upcoming surveys using the MeerKAT or FAST telescopes, clearly have the potential to uncover such systems. Looking at regions of high stellar density, one can expect even to find millisecond pulsars around BHs due to binary exchange interactions, making globular clusters \cite{Hessels:2014yja} and the Galactic Centre \cite{Eatough:2015jka} prime targets for current and future surveys. 
As discussed, finding a pulsar orbiting Sgr A* would be particularly rewarding. Past surveys are likely to have been limited by sensitivity and scattering effects due to the turbulent interstellar medium although the discovery of a radio magnetar in the Galactic Center \cite{efk+13} indicates that the situation may be more complicated than anticipated \cite{ddb+17}. Searches with sensitive high-frequency telescopes will ultimate provide the answer \cite{Goddi:2017pfy}. Meanwhile, sensitive timing observations of pulsars in Globular cluster can also probe the proposed existence of Intermediate Mass Black Holes (IMBHs) in the cluster centres by sensing how the pulsars ``fall'' in the cluster potential.  In some clusters, prominent claims \cite{kbl17} can be safely refuted  \cite{frk+17}, while in other clusters IMBHs may still exist in the centre \cite{psl+17a,psl+17b} or their potential mass can at least be constrained meaningfully \cite{brf+17}.

In summary, current and future radio pulsars observations have the potential to study BH properties over a wide mass range, from stellar-mass to super-massive BHs providing important complementary observations presented in this chapter;

\subsubsection{GW diagnostics}
In contrast to EM probes, GW tests are able to probe also the interior of compact objects and are much less affected by the astrophysical environment~\cite{Barausse:2014tra}. 
Thus, they are best suited to constraining all classes of ECOs.

\paragraph{GW tests based on the ringdown phase}
The remnant of a binary merger is a highly distorted object that approaches a 
stationary configuration by emitting GWs during the ``ringdown'' phase. If the 
remnant is a BH in GR, the ringdown can be modeled as a superposition of damped 
sinusoids described by the quasinormal modes (QNMs) of the 
BH~\cite{Kokkotas:1999bd,Ferrari:2007dd,Berti:2009kk} (see 
Sec.~\ref{sec:ringdown}). 
If the remnant is an ECO, the ringdown signal is different:
\begin{itemize}
 \item For UCOs, the ringdown signal can be qualitatively similar to that of a BH, but the QNMs are different from their Kerr counterpart~\cite{Pani:2009ss,Macedo:2013jja,Chirenti:2016hzd}. The rates of binary mergers that allow for QNM spectroscopic tests depends on the astrophysical models of BH population~\cite{Berti:2016lat}. The estimated rates are lower than $1/{\rm yr}$ for current detectors even at design sensitivity. On the other hand, rates are higher for third-generation, Earth-based detectors and range between a few to $100$ events per year for LISA, depending on the astrophysical model~\cite{Berti:2016lat}. Even if the ringdown frequencies of a single source are hard to measure with current detectors, coherent mode stacking methods using multiple sources may enhance the ability of aLIGO/aVirgo to perform BH spectroscopy~\cite{Yang:2017zxs}. Such procedure is dependent upon careful control of the dependence of ringdown in alternative theories on the parameters of the system (mass, spin, etc). 
 \item For ClePhOs, the prompt post-merger ringdown signal is identical to that of a BH, because it excited at the light ring~\cite{Cardoso:2016rao,Cardoso:2017cqb}. 
However, ClePhOs generically support quasi-bound trapped modes~\cite{1991RSPSA.434..449C,Chandrasekhar:1992ey} which produce a modulated train of pulses at late times. 
The frequency and damping time of this sequence of pulses is described by the QNMs of the ClePhO (which are usually completely different from those of the corresponding BH)~\cite{Correia:2018apm,Vicente:2018mxl}. These modes were dubbed ``GW echoes'' and appear after a delay time that, in many models, scales as $\tau_{\rm echo}\sim M\log(1-2M/R)$~\cite{Cardoso:2016oxy}. Such logarithmic dependence is key to allow for tests of Planckian corrections at the horizon scale~\cite{Cardoso:2016oxy,Abedi:2016hgu,Cardoso:2017cqb}.
Models of ultracompact stars provide GW echoes with a different scaling~\cite{Cardoso:2017njb,Pani:2018flj}, the latter being a possible smoking gun of exotic state of matter in the merger remnant.

\item In addition to gravitational modes, matter modes can be excited in ECOs~\cite{Yunes:2016jcc}. So far, this problem has been studied only for boson stars~\cite{Bezares:2017mzk,Palenzuela:2017kcg} and it is unclear whether matter QNMs would be highly redshifted for more compact ECOs~\cite{Cardoso:2017njb}.
\end{itemize}

\paragraph{GW tests based on the inspiral phase}
The nature of the binary components has a bearing also on the GW inspiral phase, chiefly through three effects:
\begin{itemize}
\item \emph{Multipolar structure.} As a by-product of the BH uniqueness and no-hair
theorems~\cite{Robinson}, the mass and current multipole moments $(M_{\ell}, {S}_\ell)$ of any stationary, isolated BH can be written
only in terms of mass $M\equiv M_0$ and spin $\chi\equiv {S_1}/{M^2}$.
The quadrupole moment of the binary components enters the GW phase at $2$PN relative order, whereas higher multipoles enter at higher PN order~\cite{Blanchet:2013haa}. 
The multipole moments of an ECO are generically different, e.g. ${M}_2^{\rm ECO}={M}_2^{\rm Kerr}+\delta q(\chi, M/R)M^3$,
and it is therefore possible to constrain the dimensionless deviation $\delta q$ by measuring the $2$PN coefficient of the inspiral waveform. This was recently used to constrain ${\cal O}(\chi^2)$ parametrized deviations in $\delta q$~\cite{Krishnendu:2017shb}. It should be mentioned that ECOs will generically display higher-order spin corrections in $\delta q$ and that --~at least for the known models of rotating ultracompact objects~\cite{Pani:2015tga,Uchikata:2016qku,Yagi:2015hda}~-- the multipole moments approach those of a Kerr BH in the high-compactness limit. Moreover, the quadrupole PN correction is degenerate with the spin-spin coupling. Such degeneracy can be broken using the I-Love-Q relations~\cite{Yagi:2013bca,Yagi:2016bkt} for ECOs, as computed for instance in the case of gravastars~\cite{Pani:2015tga,Uchikata:2016qku}.

\item \emph{Tidal heating.} In a binary coalescence the inspiral proceeds through energy and angular momentum loss in the form of GWs emitted to infinity. If the binary components are BHs, a fraction of the energy loss is due to absorption at the horizon~\cite{Alvi:2001mx,Hughes:2001jr,Taylor:2008xy,Poisson:2009di,Chatziioannou:2012gq,Cardoso:2012zn}. This effect introduces a $2.5$PN $\times\log v$ correction to the GW phase of spinning binaries, where $v$ is the orbital velocity. The sign of this correction depends on the spin~\cite{Hughes:2001jr,Taylor:2008xy,Poisson:2009di}, since highly spinning BHs can amplify radiation due to superradiance~\cite{Brito:2015oca}. In the absence of a horizon, GWs are not expected to be absorbed significantly, and tidal heating is negligible~\cite{Maselli:2017cmm,Maggio:2017ivp}. Highly spinning supermassive binaries detectable by LISA will provide a golden test of this effect~\cite{Maselli:2017cmm}.

\item \emph{Tidal deformability.} The gravitational field of the companion induces tidal deformations, which produce an effect at $5$PN relative order, proportional to the so-called tidal Love numbers of the object~\cite{PoissonWill}. Remarkably, the tidal Love numbers of a BH are identically zero for static BHs~\cite{Binnington:2009bb,Damour:2009vw,Fang:2005qq,Gurlebeck:2015xpa}, and for spinning BHs to first order in the spin~\cite{Poisson:2014gka,Pani:2015hfa,Pani:2015nua}, and to second order in the axisymmetric perturbations~\cite{Pani:2015hfa}. On the other hand, the tidal Love numbers of ECOs are small but finite~\cite{Pani:2015tga,Uchikata:2016qku,Porto:2016zng,Cardoso:2017cfl}. Thus, the nature of ECOs can be probed by measuring the tidal deformability, similarly to what is done to infer the nuclear equation of state in NSBs~\cite{Flanagan:2007ix,Hinderer:2007mb,TheLIGOScientific:2017qsa}. Analysis of the LIGO data shows that interesting bounds on the tidal deformability can be imposed already, to the level that some boson star models (approximated through a polytropic fluid) can be excluded~\cite{Johnson-McDaniel:2018uvs}. The tidal Love numbers of a ClePhO vanish logarithmically in the BH limit~\cite{Cardoso:2017cfl}, providing a way to probe horizon scales. For Planckian corrections near the horizon, the tidal Love numbers are about $4$ orders of magnitude smaller than those of a NS. It is therefore expected that current and future ground-based detectors will not be sensitive enough to detect such small effect, while LISA might be able to measure the tidal deformability of highly-spinning supermassive binaries~\cite{Maselli:2017cmm}.

\end{itemize}

Finally, it is possible that ECOs' low-frequency modes are excited during the inspiral, leading to resonances in the emitted GW flux~\cite{Pani:2010em,Macedo:2013qea,Macedo:2013jja}. Low-frequency modes certainly arise in the gravitational sector, as we discussed already. In addition, fluid modes at low frequency can also be excited, although this issue is poorly studied.

\paragraph{Challenges in modeling ECO coalescence waveforms}

With the notable exception of boson stars~\cite{liebling17}, very little is known about the dynamical formation of isolated ECOs or about their coalescence. 
While the early inspiral and post-merger phases can be modelled within a PN expansion and perturbation theory, respectively, searches for ECO coalescence require a full inspiral-merger-ringdown waveform. Some combination of numerical and semianalytical techniques --~analog to what is done to model precisely the waveform of BH binaries~\cite{Taracchini:2013rva,Husa:2015iqa,Khan:2015jqa}~-- is missing.

Concerning the post-merger ringdown part alone, it is important to develop accurate templates of GW echoes. While considerable progress has been recently done~\cite{Nakano:2017fvh,Mark:2017dnq,Volkel:2017kfj,Bueno:2017hyj,Maselli:2017tfq,Wang:2018mlp,Correia:2018apm,Wang:2018gin}, a complete template which is both accurate and practical is missing. This is crucial to improve current searches for echoes in LIGO/Virgo data~\cite{Abedi:2016hgu,Ashton:2016xff,Abedi:2017isz,Conklin:2017lwb,Westerweck:2017hus,Abedi:2018pst,Nielsen:2018lkf,Lo:2018sep}. 
Model-independent burst searches have recently been reported, and will be instrumental in the absence of compelling models~\cite{Tsang:2018uie}.

\section{The dark matter connection\label{Sec:DM}} 
\vspace{-3mm}
{\it Contributors:} G.~ Bertone, D.~Blas, R.~ Brito, C.~Burrage, V.~Cardoso, L. Urena-L\'opez
\vspace{3mm}

The nature and properties of DM and dark energy in the Universe are among the outstanding open issues of modern Cosmology. They seem to be the responsible agents for the formation of large scale structure and the present accelerated phase of the cosmic expansion. Quite surprisingly, there is a concordance model that fits all available set of observations at hand. This model is dubbed $\Lambda$CDM because it assumes that the main matter components at late times are in the form of a cosmological constant for the dark energy~\cite{Carroll:2000fy,Copeland:2006wr} and a pressureless component known as cold DM~\cite{Peter:2012rz}. These two assumptions, together with the theoretical basis for GR, make up a consistent physical view of the Universe~\cite{Ade:2015xua}.

 The nature of the missing mass in the Universe has proven difficult to determine, because it 
interacts very feebly with ordinary matter. Very little is known about the fundamental nature of DM, and models range from ultralight bosons with masses $\sim 10^{-22} \mbox{ \rm eV}$ to 
BHs with masses of order $10\, M_{\odot}$.
Looking for matter with unknown properties is extremely challenging, and explains to a good extent why DM 
has never been directly detected in any experiment so far. However, the equivalence principle upon which GR
stands --  tested to remarkable accuracy with known matter -- offers a solid starting point. All forms of energy gravitate and fall
similarly in an external gravitational field. Thus, gravitational physics can help us unlock the mystery of DM: even if blind to other interactions, it should still ``see'' gravity. The feebleness with which DM interacts with baryonic matter, along with its small density in local astrophysical environments poses the question of how to look for DM signals with GWs.

\subsection{BHs as DM} \label{sec:BH}
In light of the LIGO discoveries there has been a revival of interest in the possibility that DM could be composed of BHs with masses in the range $1-100 M_{\odot}$ \cite{Kashlinsky:2016sdv,Clesse:2016vqa,Bird:2016dcv,Sasaki:2016jop,Wang:2016ana}. To generate enough such BHs to be DM, they would need to be produced from the collapse of large primordial density fluctuations \cite{Carr:1974nx,Carr:1975qj,Carr:2016drx}.  The distribution of the BH masses that form depends on the model of inflation \cite{Byrnes:2012yx}. Such BHs can be produced with sufficiently large masses that they would not have evaporated by the current epoch. Alternatively, DM could be composed of ultracompact horizonless objects for which Hawking radiation is suppressed~\cite{Raidal:2018eoo}. Different formation scenarios and constraints on such objects were reviewed in Chapter I, Section~\ref{Sec:PBHandDM}.

If all of the DM is composed of such heavy compact objects a key signature is the frequency of microlensing events \cite{Paczynski:1985jf}. Microlensing is the amplification, for a short period of time, of the light from a star when a compact object passes close to the line of sight between us and the star. How frequent and how strong these events will be, depends on the distribution of BH masses \cite{Griest:1990vu, DeRujula:1990wq,Alcock:1996yv}. It has been claimed that DM composed entirely of primordial BHs in this mass range is excluded entirely by microlensing \cite{Carr:2016drx}; however, such study assumed that the BH mass distribution was a delta function. If the mass distribution is broadened the tension with the microlensing data weakens~\cite{Green:2017qoa,Green:2016xgy,Bellomo:2017zsr}, although whether realistic models can be compatible with the data remains a subject of debate~\cite{Clesse:2017bsw}. 
Further observational signatures include the dynamical heating of dwarf galaxies, through two body interactions between BHs and stars~\cite{Koushiappas:2017chw,Brandt:2016aco}, electromagnetic signatures if regions of high BH density also have high gas densities (such as the center of the galaxy) \cite{Gaggero:2016dpq,Inoue:2017csr}, constraints from the CMB due to energy injection into the plasma in the early universe \cite{Ali-Haimoud:2016mbv,Poulin:2017bwe} and from the (absence of) a stochastic background of GWs~\cite{Wang:2016ana}. 

At the very least primordial BHs in the LIGO mass range can be a component of the DM in our universe. Future GW observations determining the mass and spatial distribution of BHs in our galaxy will be key to testing this hypothesis.

\subsection{The DM Zoo and GWs}
Despite the accumulation of evidence for the existence of DM, the determination of its fundamental properties remains elusive. The observations related to GWs may change this.  GWs and their progenitors `live' in a DM environment to which they are sensitive (at least gravitationally though other interactions may occur). Furthermore, the DM sector may include new particles and forces that can change the nature and dynamics of the sources.

Before making any classification of DM candidates, we remind that the preferred models are those that: {\it i)} explain {\it all}  the DM, {\it ii)} include a {\it natural} 
generation mechanism, {\it iii)} can be added to the SM to solve other (astrophysical or fundamental) puzzles, {\it iv)} can be tested beyond the common DM gravitational effects.  None of these properties is really necessary, but their advantages are clear.

A first distinction one can make when classifying DM models is whether the state of DM is a distribution of fundamental particles or of compact objects. The masses of compact objects are determined by astrophysical evolution from initial overdensities in a matter field. As a consequence, the range and distribution of masses is quite broad. These models and their bounds are discussed in Sec.~\ref{sec:BH} (see also Sec.~\ref{Sec:BHgenesis} for more details about astrophysical BHs).

Fundamental particles have fixed mass, which constrains the mass to be  $m_\chi \lesssim M_P$ to treat gravity within an EFT formalism  (see \cite{Garny:2015sjg} for a recent study of the top range). Furthermore, there is no candidate within the SM for these fundamental particles. Particles with typical energy scales in the electroweak range (WIMPs) have been preferred in the past because they satisfy the four points above \cite{Bertone:2004pz}. Other popular candidates include axions and sterile neutrinos (see e.g. Ref. \cite{Feng:2010gw}).

 The first classification of particle models is based on the DM spin. Fermionic DM can only exist with masses above\footnote{All the limits of this section apply to candidates that represent {\it all} the DM. The bounds are less stringent for fractional components.} $m_\chi \gtrsim\,$keV. This is because the phase-space available for fermions in virial equilibrium in dwarf spheroidals is limited \cite{Tremaine:1979we}. The limit $m_\chi \gtrsim\,$keV always applies for models where the DM was generated thermally.
In this case, the distribution of DM particles is too `hot' for  lighter masses and there is a suppression of the growth of cosmological structures at small scales, in tension with observations  \cite{Baur:2015jsy}. Other `out-of-equilibrium' production mechanisms allow for 'cold' distributions for all possible masses, see e.g. \cite{DAgnolo:2017dbv} and references therein. The extreme situation are models where a bosonic candidate is generated `at rest' as happens in the cases of misalignment or inflationary generation, see \cite{Marsh:2015xka}. The fundamental limitation is then  $m_\chi \gtrsim 10^{-22}\,$eV, which comes from the wiping of structure at scales smaller than the de Broglie wavelength of the candidate. This scale should allow for the existence of dwarf spheroidals. Ultra-light candidates incorporate the existence of oscillating modes with coherence times long enough to generate new phenomena as compared to standard candidates (see below).
 
The complexity of the DM sector is also unknown. Models that incorporate natural candidates for DM (e.g. supersymmetric models or axion models) provide concrete proposals, but one should keep an open mind when thinking about DM phenomenology. The existence of a coupling DM-SM beyond the gravitational one may mean an extra interaction of gravitating bodies  among themselves or with the DM background. This can modify the orbital properties and open new emission channels, if the DM candidate is light enough. Besides, different DM models may admit different DM distributions, depending on their production mechanism and type of interactions. The common feature is the existence of spherical DM halos, but more complex structures (mini-halos, DM disks, breathing modes,....) may occur.

Finally, the distribution of DM close to the galactic center is also quite uncertain~\cite{Pato:2015dua}. This is because baryonic effects are more important. Still, one may be able to distinguish certain features, as the existence of huge overdensities, or solitons, characteristic of some DM models~\cite{Schive:2014dra}. For instance,  the prediction  of solitons in center of galaxies for masses $m\sim (10^{-22}\div 10^{-21})\,$eV
 has been recently used in \cite{Bar:2018acw} to claim that this mass range is in tension with data.

\subsection{The cosmological perspective on DM: scalar field dark matter}
Requiring a consistent cosmology limits some of the DM candidates, {\it if} they are to solve both the dark energy and DM puzzles. 
 An intense experimental search in the last years has successfully imposed stringent limits on the interactions between DM and ``ordinary'' matter~\cite{Aprile:2017iyp,Cui:2017nnn,Hoferichter:2017olk}. The Weakly Interacting Massive Particle (WIMP) hypothesis~\cite{Queiroz:2017kxt}, while appealing, is under extreme pressure. Thus, alternative models need to be seriously considered~\cite{Hooper:2017eje}. Among the vast number of possibilities, scalar fields have become increasingly tempting~\cite{Copeland:2006wr,Joyce:2016vqv}.
Scalar field DM is a model where the properties of the DM can be represented by a relativistic scalar field $\phi$ endowed with an appropriate scalar potential $V(\phi)$~\cite{Matos:2000ss,Hu:2000ke,Hui:2016ltb,Suarez:2013iw,Marsh:2015xka}. In the relativistic regime, the equation of motion for the scalar is the Klein-Gordon equation $\partial_\mu \partial^\mu \phi - \partial_\phi V=0$, whereas the non-relativistic regime leads to a Schr\"odinger-type equation for the wave function. At the quantum level, the scalar represents the mean field value of a coherent state of boson particles with a large occupation number.

Although many possibilities exist for the potential $V(\phi)$ in DM models, it should possess a minimum at some critical value $\phi_c$ around which we can define a mass scale $m$ for the boson particle, $m\equiv \partial^2_\phi V(\phi_c)$. The simplest possibility is the parabolic potential $V(\phi) = (1/2) m^2 \phi^2$ (with $\phi_c =0$), but higher order terms  could play a role at higher energies. One representative example of the scalar field hypothesis is the axion from the Peccei-Quinn mechanism to solve the strong CP problem, for which the potential has the trigonometric form $V(\phi) = m^2 f^2 [1-\cos(\phi/f)]$, where $f$ is the so-called decay constant of the axion particle~\cite{Marsh:2015xka}. In contrast to the parabolic potential, the axion potential is periodic, has an upper energy limit $V(\phi) \leq m^2 f^2$ and includes contributions from higher-order terms $\phi^4,\phi^6,\ldots$~\cite{Zhang:2017dpp,Cedeno:2017sou}. For simplicity, we present below a summary of results obtained for the parabolic potential and their constraints arising from cosmological and astrophysical observations.

\subsubsection{Cosmological background dynamics}
Because of the resemblance of the KG equation to that of the harmonic oscillator, there are two main stages in the evolution of the scalar field: a damped phase with the Hubble constant $H \gg m$ during which $\phi \simeq {\rm const.}$, and a stage of rapid field oscillations around the minimum of the potential that corresponds to $H \ll m$. A piece-wise solution for the two regimes can be envisaged from semi-analytical studies, but the real challenge arises in numerical simulations for which one desires a continuum solution for all the field variables. 
The two stages can be put together smoothly in numerical simulations~\cite{Urena-Lopez:2015gur}. The evolution of the energy density $\rho_\phi = (1/2) \dot{\phi}^2 + (1/2) m^2\phi^2$ should transit from $\rho_\phi = {\rm const}.$ to $\rho_\phi \propto a^{-3}$ (the expected one for a DM component) before the time of radiation-matter equality to obtain the correct amount of DM at the present time. A lower bound on the boson mass $m \geq 10^{-26} {\rm eV}$ arises from such a requirement~\cite{Urena-Lopez:2015gur}.

\subsubsection{Cosmological linear perturbations}
For this to be a successful model, the scalar field DM must allow the formation of structure. Generically, it is found that small density fluctuations
$\delta_\rho/\rho$ grow (decay) for wavelengths $k^2 < 2 a^2 m H$ ($k^2 > 2 a^2 m H$), which indicates the existence of an effective time-dependent Jeans wavenumber defined by $k^2_J = 2 a^2 m H$,
where $a$ is the scale factor. The most noticeable aspect about the growth of linear perturbations is that the power spectrum has the same amplitude as that of CDM for large enough scales (ie $k < k_J$), and that a sharp-cut-off appears for small enough scales (ie $k > k_J$). The straightforward interpretation is that the SFDM models naturally predicts a dearth of small scale structure in the Universe, and that such dearth is directly related to the boson mass. This time the lower bound on the boson mass is somehow improved to $m \geq 10^{-24} {\rm eV}$~\cite{Urena-Lopez:2015gur,Hlozek:2014lca}, but one is limited by the non-linear effects on smaller scales to impose a better constraint.

\subsubsection{Cosmological non-linear perturbations}
Although $N$-body simulations have been adapted to the field description required for SFDM, they cannot capture the full field dynamics and the best option remains to solve directly the so-called Schr\"odinger-Poisson system. Some studies suggest that the non-linear process of structure formation proceeds as in CDM for large enough scales~\cite{Schive:2014dra,Mocz:2017wlg,Schwabe:2016rze}. 

The gravitationally bound objects that one could identify with DM galaxy halos all have a common structure: a central soliton surrounded by a NFW-like envelope created by the interference of the Schr\"odinger wave functions~\cite{Schive:2014dra,Schive:2014hza}. In terms of the standard nomenclature, the scalar-field DM model belongs to the so-called non-cusp, or cored, types of DM models (towards which evidence is marginally pointing). Simulations suggest a close relationship between the soliton $M_s$ and the total halo mass $M_h$ of the form $M_s = \gamma(a) M^{1/3}_h$, so that massive halos also have massive central solitons~\cite{Schive:2014dra,Schive:2014hza,Mocz:2017wlg}. Here $\gamma(a)$ is a time-dependent coefficient of proportionality.

 These results then predict a minimum halo mass $M_{h,{\rm min}} = \gamma^{3/2}(1) \simeq (m/10^{-22}\, {\rm eV})^{-3/2} \, 10^7 M_\odot$ when the galaxy halo is just the central soliton. 
But the soliton scale radius $r_s$ is related to its mass via  $M_s r_ s \simeq 2 (m_{\rm Pl}/m)^2$, where $m_{\rm Pl} = G^{-1/2}$ is the Planck mass. For the aforementioned values of the soliton mass, $r_s = 300 \, {\rm pc}$ if $M_s=10^7 \, M_\odot$, in remarkable agreement with observations~\cite{Urena-Lopez:2017tob} (see, however, \cite{Deng:2018jjz}). Furthermore, the previous relation between the soliton mass and the host halo mass can be used to 
predict the presence of a second peak in the velocity of rotation curves, closer to the galactic center \cite{Bar:2018acw}. 
This prediction is in tension with data for masses in the range  $m\sim (10^{-22}\div 10^{-21})\,$eV \cite{Bar:2018acw}. 

Finally, non-linear results also constrain the scalar mass through comparison with observations inferred from the Lyman-$\alpha$ forest~\cite{Palanque-Delabrouille:2013gaa}. These constraints require dedicated $N$-body simulations with gas and stars, so that the flux of quasars can be calculated. The constraints are obtained indirectly and subjected to uncertainties arising from our limited knowledge on the formation of realistic galaxies. In a first approximation we can estimate the power spectrum of the transmitted flux of the Lyman-$\alpha$ forest at linear order, and obtain a lower limit on the boson mass $m > 10^{-21} {\rm eV}$ if SFDM is to fulfill the DM budget completely~\cite{Kobayashi:2017jcf}.

In conclusion, scalar-field models of dark energy can account for the whole of DM as well, provided the scalar has a mass of order  $10^{-21}\, {\rm eV}$. 
This model explains also the seemingly cored density profile in the central parts of the galaxies.

\subsection{The DM environment}
In the following we provide a (hopefully agnostic) view on how different DM models may affect GW emission by compact objects.
The environment in which compact objects live -- be it the interstellar medium plasma, accretion disks or DM --
can influence the GW signal in different ways:
\begin{itemize}

\item {\it By modifying the structure of the compact objects themselves}. Accretion disks, for example, alter the spacetime multipolar structure relative to the standard Kerr geometry. Accretion disks are also known to limit the parameter space of Kerr BHs, preventing them from spinning beyond $cJ/(GM^2)\sim 0.998$~\cite{Thorne:1974ve}. Likewise, if DM behaves as heavy particles, its effects parallel that of baryonic matter either in accretion disks or in the  interstellar medium.

In models where a particle description is insufficient (for example, if DM assumes the form of fundamental light bosons),
Kerr BHs can either be unstable, or not be an equilibrium solution of the field equations at all~\cite{Cardoso:2016ryw,Sotiriou:2015pka,Herdeiro:2015waa} (see also discussion in Section~\ref{Sec:nonKerr}). Some of the simplest forms of axionic DM have a more mild effect instead, contributing to the spinning down of astrophysical BHs~\cite{Arvanitaki:2010sy,Brito:2014wla,Brito:2017wnc}.

Independently of the nature of DM, compact stars evolving in DM-rich environments may accrete a significant amount of DM:
DM is captured by the star due to gravitational deflection and a non-vanishing
cross-section for collision with the star material~\cite{Press:1985ug,Gould:1989gw,Goldman:1989nd,Bertone:2007ae,Brito:2015yga}. 
The DM material eventually thermalizes with the star, and accumulates inside a finite-size core~\cite{Brito:2015yfh,Brito:2015yga,Gould:1989gw,Goldman:1989nd}.

\item {\it By altering the way that GWs are generated.} The different equilibrium configuration of the compact objects, together with the possible self-gravity effects of DM and different structure of the field equations gives rise, generically,
to a different GW output from that predicted in vacuum GR. 

\item The environment -- interstellar dust, a cosmological constant or DM -- must of course affect also the way that GWs {\it and} electromagnetic waves propagate. It is a common-day experience that light can be significantly affected by even low-density matter. 
For example, low-frequency light in a plasma travels at the group velocity $v_g=c\sqrt{1-\omega_p^2/\omega^2}$, with $\omega_p$ the plasma frequency $\omega_p^2=\frac{n_0 q^2}{4\pi\epsilon_0 \,m}$. Here, $n_0$ is the particle number density, $q$ the electron charge and $m$ its mass. 
%
%
%
Thus, light is delayed with respect to GWs, by an amount $\delta t$ directly proportional to the total distance they both traversed 
\begin{eqnarray}
\delta t&=&\frac{d \rho_0 q^2}{8\pi\epsilon_0\,c\,m^2 \omega^2}\nonumber\\
&=& 6.7 \frac{\rho_0}{\rm 1\, {\rm GeV/cm}^3} \left(\frac{\rm 6 GHz}{\omega}\right)^2\,{\rm days}\,,
\end{eqnarray}
where in the last equality we substituted numbers for the very latest observation of GWs with an electromagnetic counterpart~\cite{GBM:2017lvd}. These numbers could be interesting: given the observed time delay of several days for radio waves, one may get constraints for models where DM consists of mili-charged matter (i.e., models where $q$ and $m$ may be a fraction of the electron charge and mass).

Cold DM affects the propagation of GWs by inducing a small frequency dependent modification of the propagation speed \cite{Flauger:2017ged}. Furthermore, the effect of viscous fluids or oscillators on the passage of GWs are discussed in Refs.~\cite{Thorne_notes,Annulli:2018quj}. These calculations indicate that the effect may be too small to be measurable in the near future.

\end{itemize}

These different effects have, so far, been explored and quantified in the context of a few possible mechanisms where DM plays a role.

\subsection{Accretion and gravitational drag}
%
\begin{figure}[th]
\begin{tabular}{c}
\includegraphics[scale=1,clip]{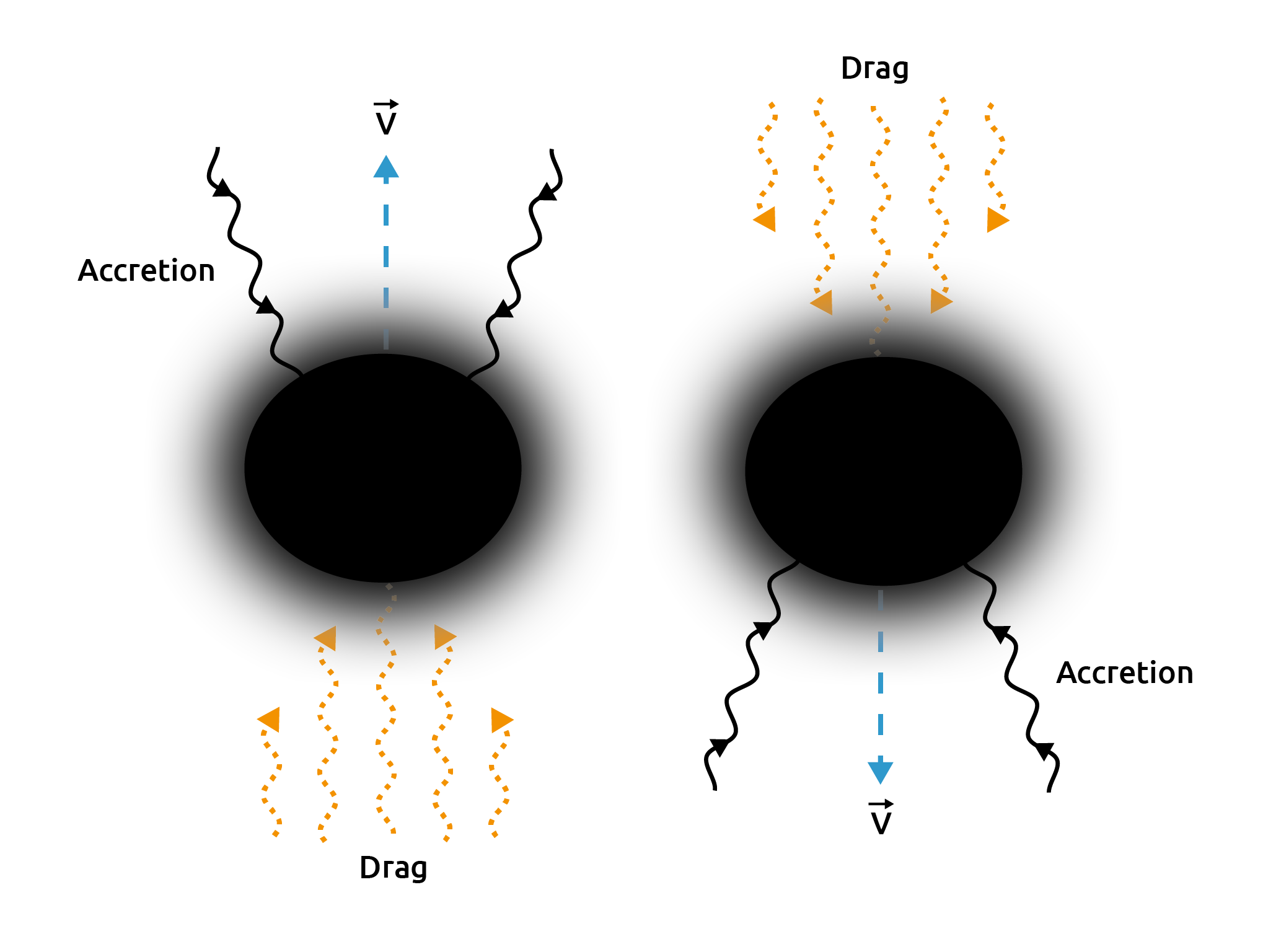}
\end{tabular}
\caption{\label{Fig_accretion_drag} Depiction of a BBH evolving in a possible interstellar or DM environment. Each individual BH accretes and exherts a gravitational pull on the surrounding matter. Both effects contribute to decelerate the BHs, leading to a faster inspiral. Credit: Ana Sousa.}
\end{figure}
The formation and growth of compact objects may lead to large overdensities of DM around them \cite{Gondolo:1999ef,Bertone:2005xz,Zhao:2005zr}. Whatever the nature and amount of DM surrounding them, a binary of two BHs or compact stars evolving in a DM-rich environment will be subjected to
at least two effects, depicted in Fig.~\ref{Fig_accretion_drag}. As it moves through the medium it accretes
the surrounding material, but also exherts a gravitational pull (known as ``gravitational drag'') on all the medium, which affects the inspiral dynamics.
To quantify these effects, it is important to know how DM behaves. Collisionless DM causes, generically, a gravitational drag different from that of normal fluids~\cite{Macedo:2013qea,Macedo:2013jja}. The gravitational drag caused by DM which is coherent on large scales may be suppressed~\cite{Hui:2016ltb}, but further work is necessary to understand this quantitatively.

The phase evolution of a pair of BHs, taking gravitational radiation, accretion and drag as well as the DM self-gravity was studied recently~\cite{Barausse:2014tra,Macedo:2013qea,Macedo:2013jja,Eda:2014kra,Yue:2017iwc,Hannuksela:2018izj}. These estimates are all Newtonian, and the results depend crucially on the local DM density. Perhaps more importantly, back-reaction on the DM distribution was not taken into account in a systematic manner. The only studies available including backreaction, look only at the very last orbits of the binary, and are unlikely to capture well the full physics of DM dispersal~\cite{Berti:2013gfa}.

\subsection{The merger of two dark stars}
Gravitational drag and accretion have a cumulative effect: even though their magnitude is small,
the effect piles up over a large number of orbits. By contrast, drag and accretion have a tiny effect during merger.
The merger signal carries, mostly, an imprint of the configuration of the colliding objects. For BHs,
the no-hair results strongly suggest that the colliding objects belong to the Kerr family, and it seems hopeless 
to find any DM imprints here. However, the no-hair results can be by-passed, as they assume stationarity of the spacetime
or certain matter properties, as explained in Sec.~\ref{Sec:nonKerr}~\cite{Cardoso:2016ryw,Sotiriou:2015pka,Herdeiro:2015waa}.
Detailed investigations of BHs other than Kerr are still to be done. Smoking-gun effects arising from the merger of such objects are, likewise, unknown.

As we discussed, compact stars can grow DM cores at their center. It is conceivable that the DM core might imprint a signature on different phases of the coalescence of two of these objects. The effects of tidal deformations within the gravitational wave signal produced by binary DM stars have been investigated in \cite{Maselli:2017vfi}. Moreover, the imprint on the post-merger spectrum, so far only for analog-mechanical models, has been studied in~\cite{Ellis:2017jgp}.

\subsection{Non-perturbative effects: superradiance and spin-down.}
\label{sec:superradiance}
Spinning BHs are huge energy reservoirs from which all  rotational energy can ultimately be extracted. In particular, due to the absence of a Pauli exclusion principle for bosons, bosonic waves can effectively extract energy from spinning BHs through a process known as rotational superradiance~\cite{Brito:2015oca}. The superradiant energy extraction might be used to rule out (or detect) ultralight bosons that might be a significant component of DM, such as the QCD axion or axion-like particles predicted by the string axiverse scenario~\cite{Arvanitaki:2010sy}, dark photons~\cite{Pani:2012vp,Pani:2012bp} and even massive spin-2 fields~\cite{Brito:2013wya}. For boson masses in the range $10^{-21}$~--~$10^{-11}$ eV, their Compton wavelength is of the order of the horizon of typical astrophysical BHs, the gravitational coupling of these two objects is strongest, and long-lived quasi-bound states of bosonic particles around BHs can form. For rotating BHs these states are typically superradiant, thus becoming an effective tap of rotational energy. The extracted energy condenses as a bosonic cloud around the BH producing ``gravitational atoms''~\cite{Arvanitaki:2010sy,Brito:2014wla,East:2017ovw}. The evolution of these systems leads to several observational channels. For complex bosonic fields, GW emission is suppressed and the end-state of the superradiant instability might be a Kerr BH with bosonic hair~\cite{Herdeiro:2014goa,Herdeiro:2016tmi}. These solutions are themselves unstable~\cite{Ganchev:2017uuo} which, over the age of the Universe, would likely lead to a slowly-rotating BH surrounded by a bosonic cloud populating the most unstable modes. On the other hand for real bosonic fields, this cloud would disperse over long timescales though the emission of nearly-monochromatic GWs, either observable individually or as a very strong stochastic GW background~\cite{Arvanitaki:2014wva,Arvanitaki:2016qwi,Yoshino:2014wwa,Baryakhtar:2017ngi,Brito:2017zvb,Brito:2017wnc}. These are appealing GW emitters with appropriate amplitude and frequency range for Earth- and space-based detectors. Although estimates from Ref.~\cite{Brito:2017zvb,Brito:2017wnc} suggest that current LIGO data can already be used constrain some range of scalar field masses, an analysis on real data has not yet been performed.

These models also predict that BHs affected by the instability would spin-down in the process, and therefore the mere observation of rotating BHs can place severe limits on these particles, thereby strongly constraining the theory. In fact, observations of spinning BHs can already be used to bound the photon~\cite{Pani:2012vp,Pani:2012bp} and graviton masses~\cite{Brito:2013wya}. These studies demonstrate that BHs have an enormous potential as theoretical laboratories, where particle physics can be put to the test using current observations. 

All these studies neglect the possible decay of ultralight bosonic fields into Standard Model particles which is justified given the current constraints on ultralight bosons in this mass range (see e.g.~\cite{Essig:2013lka,Marsh:2015xka}). However, the superradiant instability leads to the generation of boson condensates around BHs with a very large particle number, which could enhance their decay through channels other than gravitational. For the coupling typical of the QCD axion with photons, stimulated decay into photon pairs becomes significant for axion masses above $\gtrsim 10^{-8}$ eV, and consequently for BHs with masses $\lesssim 0.01 M_{\odot}$~\cite{Rosa:2017ury}. If such systems exist, they would lead to extremely bright lasers with frequencies in the radio band. For the expected mass of the QCD axion $\sim 10^{-5}$ eV, such lasing effect would occur around primordial BHs with masses around $\sim 10^{24}\,{\rm Kg}$ and lead to millisecond radio-bursts, which could suggest a link to the observed fast radio bursts~\cite{Rosa:2017ury}.

In addition to these effects, the coupling of light DM bosons with Standard Model particles may also turn NSs unstable against superradiant amplification. Rotational superradiance is in fact a generic process and can happen in any rotating object, as long as some sort of dissipation mechanism is at work~\cite{Brito:2015oca}. Thus, even NSs can be superradiantly unstable against some models of DM, such as dark photons coupled to charged matter, which can already be used to put constraints on these models~\cite{Cardoso:2017kgn}.

An important question that remains to be fully investigated is: if ultralight bosons do exist and we detect them through one or several of these channels, will we be able to start using compact objects to do precision particle physics, in particular can one infer fundamental properties such as their mass, spin and fundamental interactions? This will require a better understanding of higher-spin fields around rotating BHs. 

Although some progress has been made to deal with massive vector fields using time-domain simulations, either in full GR~\cite{Zilhao:2015tya,East:2017ovw} or by fixing the background geometry to be a Kerr BH~\cite{Witek:2012tr,East:2017mrj}, frequency-domain computations, which are more appropriate to span the parameter space, have mainly been dealt with by  employing a small-spin approximation~\cite{Pani:2012vp,Pani:2012bp,Brito:2013wya},  using an analytical matched asymptotics approach~\cite{Baryakhtar:2017ngi} or  using a field theory approach~\cite{Endlich:2016jgc}, these last two being only valid in the Newtonian limit. Computations of the instability rates without employing any of these approximations have only recently started to be available~\cite{Cardoso:2018tly,Frolov:2018ezx}. An estimate of the stochastic background coming from vector or tensor fields has also not been computed. Finally, besides some some estimates on the GWs from a bosenova collapse due to axion self-interations~\cite{Yoshino:2012kn}, a detailed analysis of how self-interactions or interactions with other fields could affect the GW emission and its detectability is still missing.

\subsection{Pulsar timing}

The gravitational  drag from particle DM candidates in the orbital motion of binary pulsars has been studied in \cite{Pani:2015qhr,Caputo:2017zqh}. 
This may be important to constrain DM models that generate a thick DM disk \cite{Caputo:2017zqh} or to study the DM density closer to the galactic center, if a pulsar binary is discovered there and can be timed to high precision. 

DM, specially if under the form of light bosons, can also produce other peculiar effects, accessible through
EM observations. In models where DM comes in the form of an ultralight boson, the DM configuration is able to support itself through pressure, which is in turn caused by a time-dependence of the field. This time-periodicity gives rise to an oscillating gravitational potential which is able to produce tiny, but potentially measurable, differences in the time-of-arrival signal from precise clocks, such as pulsars~\cite{Khmelnitsky:2013lxt,Blas:2016ddr}.

\subsection{Mini-galaxies around BHs}
In addition, in these theories, BHs can grow (permanent or long-lived) time-oscillating, non-axisymmetric bosonic structures~\cite{Brito:2015oca}. Planets or stars in the vicinity of these BH are subjected to a periodic forcing, leading to Lindblad and co-rotation resonances. This phenomenon is akin to the forcing that spiral-armed galaxies exhert on its planets and stars, and it is conceivable that the same type of pattern appears, on smaller-scales, around supermassive hairy BHs~\cite{Ferreira:2017pth,Hannuksela:2018izj}.

\subsection{GW detectors as DM probes}
Finally, there are interesting proposals to use the GW detectors themselves as probes of DM, when used as extremely sensitive accelerometers~\cite{Bekenstein:2006fi,Hall:2016usm,Englert:2017det}.


\newpage

\phantomsection
\addcontentsline{toc}{part}{\bf Postscript}
\begin{center}
{\large \bf Postscript}
\end{center}

Gravity sits at the heart of some of the most important open problems in astrophysics, cosmology and fundamental physics, making it a subject of strong interdisciplinarity. Black holes are, in some ways, the ``atoms'' of gravity, the ``simplest'' astrophysical objects, yet they harbour some of the most remarkable predictions of GR, including that of its own ultimate failure. Gravitational-wave astronomy will allow us to test models of BH formation, growth and evolution, as well as models of GW generation and propagation. It will provide evidence for event horizons and ergoregions, test GR itself and may reveal the existence of new fundamental fields. The synthesis of these results has the potential to shed light on some of the most enigmatic issues in contemporary physics. This write-up summarized our present knowledge and key open challenges. We hope that it can serve as a guide for the exciting road ahead.

\section*{Acknowledgements}
This article is based upon work from COST Action CA16104 ``GWverse'', supported by COST (European Cooperation in Science and Technology).
We would like to thank Walter del Pozzo for useful comments.
A.A. acknowledges partial support from the Polish National Science Center (NCN) through the grant UMO-2016/23/B/ST9/02732 and is currently supported by the Carl Tryggers Foundation through the grant CTS 17:113.
L.B. acknowledges support from STFC through Grant No. ST/R00045X/1.
V.C. acknowledges financial support provided under the European Union's H2020 ERC Consolidator Grant ``Matter and strong-field gravity: New frontiers in Einstein's theory'' grant agreement no. MaGRaTh--646597.
T.P.S. acknowledges partial support from Science and Technology Facilities Council Consolidated Grant ST/P000703/1.
K. B. acknowledges support from the Polish National Science Center
(NCN) grant: Sonata Bis 2 (DEC-2012/07/E/ST9/01360).
E. B. acknowledges financial support from projects 176003 and 176001 by the Ministry of Education and Science of the Republic of Serbia. 
T. B. was supported by the TEAM/2016-3/19 grant from FNP.
P. C. acknowledges support from the Austrian Research Fund (FWF), Project P 29517-N16, and
by the Polish National Center of Science (NCN) under grant 2016/21/B/ST1/00940.
B. K. acknowledges support from the European Research Council (ERC) under the European Union's Horizon 2020 research and innovation programme ERC-2014-STG under grant agreement No 638435 (GalNUC) and from the Hungarian National Research, Development, and Innovation Office grant NKFIH KH-125675.
G. N. is supported by the Swiss National Science Foundation (SNF) under grant 200020-168988.
P. P. acknowledges financial support provided under the European
Union's H2020 ERC, Starting Grant agreement no.~DarkGRA--757480.
U. S. acknowledges H2020-MSCA-RISE-2015 Grant No.~690904, NSF
Grant No.~PHY-090003 and PRACE Tier-0 Grant No.~PPFPWG.
The Flatiron Institute is supported by the Simons Foundation.
P. A. S. acknowledges support from the Ram{\'o}n y Cajal Programme of the Ministry
of Economy, Industry and Competitiveness of Spain.
E. B. supported by NSF Grants No. PHY-1607130 and AST-1716715.
K. B., A. C., A. G., N. I., T. P. and G. Z. acknowledge financial support from the Slovenian Research Agency.
S.~B. acknowledges support by the EU H2020 under ERC Starting Grant, no.~BinGraSp-714626. 
P. C-D. is supported by the Spanish MINECO (grants AYA2015-66899- C2-1-P and RYC-2015-19074) and the Generalitat Valenciana (PROMETEOII-2014-069).
D. G. is supported by NASA through Einstein Postdoctoral Fellowship Grant No. PF6–170152 by the Chandra X-ray Center, 
operated by the Smithsonian Astrophysical Observatory for NASA under Contract NAS8–03060.
R. E. acknowledges financial support provided by the Scientific and Technical Research Council of Turkey (TUBITAK) under the grant no. 117F296.
J. A. F. is supported by the Spanish MINECO (AYA2015-66899-C2-1-P), by the
Generalitat Valenciana (PROMETEOII-2014-069), and by the H2020-MSCA-RISE-2017 Grant No.~FunFiCO-777740.
F. M. R. is supported by the Scientific and Technological Research Council of Turkey (T\"{U}B\.{I}TAK) project 117F295.
I. R. was supported by the POLONEZ programme of the National Science Centre of Poland which has received funding from the European Union`s Horizon 2020 research and innovation programme under the Marie Sk{\l}odowska-Curie grant agreement No.~665778.
A. S. thanks the Spanish Ministry of Economy, Industry and Competitiveness,
the Spanish Agencia Estatal de Investigaci\'on,
the Vicepresidència i Conselleria d'Innovaci\'o, Recerca i Turisme del Govern de les Illes Balears,
the European Commission, the European Regional Development Funds (ERDF).
N. S. acknowledges support from DAAD Germany-Greece Grant (ID 57340132) and GWAVES (pr002022) of ARIS-GRNET(Athens).
H. W. acknowledges financial support by the Royal Society UK under the University Research Fellowship UF160547-BHbeyGR and the Research Grant $RGF-R1-180073$.
%
K. Y. is supported by T\"UB\.ITAK-117F188.
\vskip 2cm
\address{$^{1}$ Mathematical Sciences, University of Southampton, Southampton SO17 1BJ, United Kingdowm}
\address{${^2}$ CENTRA, Departamento de F\'{\i}sica, Instituto Superior T\'ecnico -- IST, Universidade de Lisboa -- UL,
Avenida Rovisco Pais 1, 1049 Lisboa, Portugal}
\address{${^3}$ Perimeter Institute for Theoretical Physics, 31 Caroline Street North Waterloo, Ontario N2L 2Y5, Canada}
\address{$^{4}$ Department of Astrophysics/IMAPP, Radboud University Nijmegen, P.O. Box 9010, 6500 GL Nijmegen, The Netherlands}
\address{$^{5}$ GRAPPA, University of Amsterdam, Science Park 904, 1098 XH Amsterdam, The Netherlands}
\address{$^{6}$ Nikhef, Science Park, 1098 XG Amsterdam, The Netherlands}
\address{$^{7}$ School of Mathematical Sciences, University of Nottingham, University Park, Nottingham, NG7 2RD, UK}
\address{$^{8}$ School of Physics and Astronomy, University of Nottingham, University Park, Nottingham, NG7 2RD, UK}
\address{$^{9}$ Nicolaus Copernicus Astronomical Center, Polish Academy of Sciences,
ul. Bartycka 18, 00-716 Warsaw, Poland}
\address{$^{10}$ Lund Observatory, Department of Astronomy, and Theoretical Physics, Lund University, Box 43, SE-221 00 Lund, Sweden}
\address{$^{11}$ Astronomical Observatory, Volgina 7, 11060 Belgrade, Serbia}
\address{$^{12}$ Isaac Newton Institute of Chile, Yugoslavia branch, Serbia}
\address{$^{13}$ Theoretical Particle Physics and Cosmology Group, Physics Department, King's College London, University of London, Strand, London WC2R 2LS, UK}
\address{$^{14}$ Max Planck Institute for Gravitational Physics (Albert Einstein Institute), Am Muhlenberg 1, Potsdam-Golm, 14476, Germany}
\address{$^{15}$ Astronomical Observatory, University of Warsaw, Aleje Ujazdowskie 4, 00478 Warsaw, Poland}
\address{$^{16}$ Department of Physics \& Astronomy, University of Sussex, Brighton BN1 9QH, United Kingdom}
\address{$^{17}$ APC, AstroParticule et Cosmologie, Universit\'e Paris Diderot, CNRS/IN2P3, CEA/Irfu,
Observatoire de Paris,  Sorbonne Paris Cit\'e,  F-75205 Paris Cedex 13,  France}
\address{$^{18}$ School of Physical Sciences and Centre for Astrophysics \& Relativity, Dublin City University, Glasnevin, Dublin 9, Ireland}
\address{$^{19}$ Dublin Institute of Advanced Studies, 31 Fitzwilliam Place, Dublin 2, Ireland}
\address{$^{20}$ University of Vienna, Boltzmanngasse 5, A 1090 Wien, Austria}
\address{$^{21}$ Institut des Hautes \'Etudes Scientifiques, Le Bois-Marie, 35 route de Chartres, 91440 Bures-sur-Yvette, France}

\address{$^{22}$ Universit\`a degli Studi di Milano-Bicocca, Dip. di Fisica “G. Occhialini”, Piazza della Scienza 3, I-20126 Milano, Italy}
\address{$^{23}$ INFN – Sezione di Milano-Bicocca, Piazza della Scienza 3, I-20126 Milano, Italy}
\address{$^{24}$ Dipartimento di Fisica, Universit\`a di Roma ``Sapienza'' \& Sezione INFN Roma1, P.A. Moro 5, 00185, Roma, Italy}
\address{$^{25}$ School of Mathematics, University of Edinburgh, James Clerk Maxwell Building, Peter Guthrie Tait Road, Edinburgh EH9 3FD, UK}
\address{$^{26}$ Instituto de F\'isica Te\'orica UAM/CSIC, Universidad Aut\'onoma de Madrid, Cantoblanco 28049 Madrid, Spain}
\address{$^{27}$ Department of Physics \& The Oskar Klein Centre, Stockholm University, AlbaNova University Cetre, SE-106 91 Stockholm}
\address{$^{28}$ Institute for Theoretical Studies, ETH Zurich, Clausiusstrasse 47, 8092 Zurich, Switzerland}
\address{$^{29}$ SUPA, University of Glasgow, Glasgow G12 8QQ, United Kingdom}
\address{$^{30}$ Departamento de F\'{\i}sica da Universidade de Aveiro and CIDMA, Campus de Santiago,
3810-183 Aveiro, Portugal}
\address{$^{31}$ Racah Institute of Physics, The Hebrew University of Jerusalem, Jerusalem, 91904, Israel}
\address{$^{32}$ Institute of Physics, E\"otv\"os University, P\'azm\'any P. s. 1/A, Budapest, 1117, Hungary}
\address{$^{33}$ Max-Planck-Institut f\"ur Radioastronomie, Auf dem H\"{u}gel 69, D-53121 Bonn, Germany}
\address{$^{34}$ Jodrell Bank Centre for Astrophysics, The University of Manchester, M13 9PL, United Kingdom}
\address{$^{35}$ LUTH, Observatoire de Paris, PSL Research University, CNRS,
Universit\'e Paris Diderot, Sorbonne Paris Cit\'e, 92190 Meudon, France}
\address{$^{36}$ Center for Computational Astrophysics, Flatiron Institute, 162 5th Ave, New York, NY 10010 US}
\address{$^{37a}$ Universit\"at Bern, Institute for Theoretical Physics, Sidlerstrasse 5, CH-3012 Bern, Switzerland}
\address{$^{37b}$ Faculty of Science and Technology, University of Stavanger, 4036 Stavanger, Norway}
\address{$^{38}$ Departament de F\'{\i}sica \& IAC3, Universitat de les Illes Balears and Institut d'Estudis Espacials de Catalunya, Palma de Mallorca, Baleares E-07122, Spain} 
\address{$^{39}$ INFN, Sezione di Milano-Bicocca, gruppo collegato di Parma, Parco Area delle Scienze 7/A, IT-43124 Parma, Italy} 
\address{$^{40}$ Dipartimento di Fisica G. Occhialini, Università degli Studi di Milano-Bicocca, Piazza della Scienza 3, IT-20126 Milano, Italy} 
\address{$^{41}$ Leiden Observatory, Leiden University, PO Box 9513 2300 RA Leiden, the Netherlands}
\address{$^{42}$ School of Physics and Astronomy and Institute of Gravitational Wave Astronomy, University of Birmingham, Edgbaston B15 2TT, United Kingdom}
\address{$^{43}$ Theoretical Astrophysics, Walter Burke Institute for Theoretical Physics,
California Institute of Technology, Pasadena, California 91125, USA}
\address{$^{44}$ Department of Applied Mathematics and Theoretical Physics, Centre for Mathematical Sciences,
University of Cambridge, Wilberforce Road, Cambridge CB3 0WA, UK}
\address{$^{45}$ INAF, Osservatorio Astronomico di Roma,  via Frascati 33, Monte Porzio Catone, 00078, Rome, Italy}

\address{$^{46}$ Scuola Normale Superiore, Piazza dei Cavalieri, 7, 56126 Pisa, Italy}
\address{$^{47}$ Argelander-Institut f\"ur Astronomie, Universit{\"a}t Bonn, Auf dem H\"{u}gel 71, D-53121 Bonn, Germany}
\address{$^{48}$ Departamento de F\'{\i}sica, DCI, Campus Le\'on, Universidad de
Guanajuato, 37150, Le\'on, Guanajuato, M\'exico}
\address{$^{49}$ LESIA, Observatoire de Paris, PSL Research University, CNRS, Sorbonne Universit \'es, UPMC Univ.  Paris 06, Univ.  Paris Diderot,
Sorbonne Paris Cit\'e, 5 place Jules Janssen, 92195 Meudon, France}
\address{$^{50}$ Institut d'Astrophysique de Paris, CNRS \& Sorbonne Universit\'es, UMR 7095, 98 bis bd Arago, 75014 Paris, France}
\address{$^{51}$ School of Mathematics and Statistics, University College Dublin, Belfield, Dublin 4, Ireland}
\address{$^{52}$ Department of Physics, University of Virginia, Charlottesville, Virginia 22904, USA}


\address{$^{53}$ Departamento de Astronom\'{\i}a y Astrof\'{\i}sica, Universitat de Val\`encia,
C/ Dr.~Moliner 50, 46100, Burjassot (Valencia), Spain}

\address{$^{54}$ Institute of Space Sciences (ICE, CSIC) \& Institut d'Estudis
Espacials de Catalunya (IEEC)\\
at Campus UAB, Carrer de Can Magrans s/n 08193 Barcelona, Spain}

\address{$^{55}$ Institute of Applied Mathematics, Academy of Mathematics and Systems
Science, CAS, Beijing 100190, China}

\address{$^{56}$ Kavli Institute for Astronomy and Astrophysics, Beijing 100871, China}

\address{$^{57}$ Zentrum f\"ur Astronomie der Universit\"at Heidelberg, Astronomisches Rechen-Institut, M\"onchhofstr. 12-14, D-69120, Heidelberg Germany}

\address{$^{58}$ Department of Physics and Astronomy, Johns Hopkins University, 3400 N. Charles Street, Baltimore, MD 21218, USA}

\address{$^{59}$ Department of Mathematics, Faculty of Natural Sciences and Mathematics, University of Tuzla, Univerzitetska 4, 75000 Tuzla, Bosnia and Herzegovina}

\address{$^{60}$ Theoretisch-Physikalisches Institut, Friedrich-Schiller-Universit{\"a}t Jena, 07743, Jena, Germany}

\address{$^{61}$ Department of Physics and Astronomy, The University of Mississippi, University, MS 38677, USA}

\address{$^{62}$ Ikerbasque and Department of Theoretical Physics, University of the Basque Country, UPV/EHU 48080, Bilbao
Spain}

\address{$^{63}$ Institut f\"{u}r Physik, Universit\"{a}t Oldenburg, Postfach 2503, D-26111 Oldenburg, Germany} 

\address{$^{64}$ DiSAT, Universit\`a dell'Insubria, Via Valleggio 11, 22100 Como, Italy}

\address{$^{65}$ Department of Astronomy, Petnica Science Center,
P. O. B. 6, 14104 Valjevo, Serbia}

\address{$^{66}$ Faculty of Electrical Engineering and Computing,
University of Zagreb, 10000 Zagreb, Croatia}

\address{$^{67}$ Center for Astrophysics and Cosmology, University of Nova Gorica, Vipavska 13, 5000 Nova Gorica, Slovenia}

\address{$^{68}$ Center for Theoretical Astrophysics and Cosmology, Institute for Computational Science, University of Zurich, Winterthurerstrasse 190, CH-8057 Z{\"u}rich, Switzerland}

\address{$^{69}$ Laboratoire de Physique Th\'eorique, CNRS, Univ. Paris-Sud, Universit\'e Paris-Saclay, 
F-91405 Orsay, France}

\address{$^{70}$ AIM, UMR 7158, CEA/DRF, CNRS, Universit\'e Paris Diderot, Irfu/Service d'Astrophysique, Centre de Saclay, FR-91191 Gif-sur-Yvette, France}

\address{$^{71}$ Eberhard Karls University of
Tuebingen, Tuebingen 72076, Germany}

\address{$^{72}$ Universit\'e Libre de Bruxelles, Campus Plaine CP231, 1050 Brussels, Belgium} 

\address{$^{73}$ Departamento de Matem\'aticas, Universitat de Val\`encia, C/ Dr. Moliner 50, E-46100 Burjassot, Val\`encia, Spain} 

\address{$^{74}$ Department of Mathematics and Applied Mathematics, University of Cape Town, Private Bag, Rondebosch, South Africa, 7700}

\address{$^{75}$ HAS Wigner Research Centre for Physics, H-1525 Budapest, Hungary}

\address{$^{76}$ Department of Physics, University of Naples "Federico II", Via Cinthia,
I-80126, Naples, Italy}

\address{$^{77}$ Istituto Nazionale di Fisica Nucleare (INFN), Sez. di Napoli, Via Cinthia
9, I-80126, Napoli, Italy}

\address{$^{78}$ INRNE - Bulgarian Academy of Sciences, 1784 Sofia, Bulgaria}

\address{$^{79}$ Physics Department, University of Lancaster, UK}

\address{$^{80}$ Instituci\'o Catalana de Recerca i Estudis Avan\c cats (ICREA), Passeig Llu\'{\i}s Companys 23, E-08010 Barcelona, Spain}

\address{$^{81}$ Departament de F{\'\i}sica Qu\`antica i Astrof\'{\i}sica, Institut de Ci\`encies del Cosmos, Universitat de Barcelona, Mart\'{\i} i Franqu\`es 1, E-08028 Barcelona, Spain}

\address{$^{82}$ Department of Physics, Izmir Institute of Technology, Izmir, Turkey}

\address{$^{83}$ Astrophysics,  University of Oxford,  DWB, Keble Road,  Oxford OX1 3RH, UK}

\address{$^{84}$ Institute of Space Sciences and Astronomy of the University of Malta}

\address{$^{85}$ Observatori Astron\`omic, Universitat de Val\`encia, C/ Catedr\'atico
 Jos\'e Beltr\'an 2, 46980, Paterna (Val\`encia), Spain}

\address{$^{86}$ Institute for Theoretical Physics, KULeuven, Celestijnenlaan 200D, 3001 Leuven, Belgium}

\address{$^{87}$ Faculty of Physics, Moscow State University, 119899, Moscow, Russia}

\address{$^{88}$ Kazan Federal University, 420008 Kazan, Russia}

\address{$^{89}$ Department of Mathematics, Rhodes University, Grahamstown 6140, South Africa}

\address{$^{90}$ Departamento de F\'isica, Universidad de Murcia, Murcia E-30100, Spain}

\address{$^{91}$ Physics Department, Ben Gurion University, Beer Sheva, Israel}

\address{$^{92}$ DiSAT, Universit\`a dell'Insubria, Como, Italy}

\address{$^{93}$ Niels Bohr Institute, Copenhagen University,
Blegdamsvej 17, DK-2100 Copenhagen, Denmark}

\address{$^{94}$ Department of Physics and Astronomy, Earlham College, Richmond, IN 47374 US}

\address{$^{95}$ ASTRON, the Netherlands Institute for Radio Astronomy, Postbus 2,
7990 AA, Dwingeloo, The Netherlands}

\address{$^{96}$ Anton Pannekoek Institute for Astronomy, University of Amsterdam,
Science Park 904, 1098 XH, Amsterdam, The Netherlands}

\address{$^{97}$ Department of Physics, University of Zurich, Winterthurerstrasse 190, CH-8057 Zurich}

\address{$^{98}$ Section of Astrophysics, Astronomy and Mechanics, Department of Physics,
National and Kapodistrian University of Athens, 15784 Zografos, Athens, Greece}

\address{$^{99}$  Astrophysics Research Institute, LJMU, IC2, Liverpool Science Park, 146 Brownlow Hill, Liverpool L3 5RF, UK}

\address{$^{100}$ Center for Applied Space Technology and Microgravity (ZARM), University of Bremen, Am Fallturm, 28159 Bremen, Germany}

\address{$^{101}$ Monash Centre for Astrophysics, School of Physics and Astronomy, Monash University, VIC 3800, Australia}

\address{$^{102}$ OzGrav: The ARC Centre of Excellence for Gravitational-wave Discovery, Clayton, VIC 3800, Australia}

\address{$^{103}$ SISSA, International School for Advanced Studies, Via Bonomea 265, 34136 Trieste, Italy }

\address{$^{104}$ INFN Sezione di Trieste, Via Valerio 2, 34127 Trieste, Italy}

\address{$^{105}$ School of Chemistry and Physics, University of KwaZulu-Natal, Westville Campus, Private Bag X54001, Durban, 4000, South Africa}

\address{$^{106}$ NAOC-UKZN Computational Astrophysics Centre (NUCAC), University of KwaZulu-Natal, Durban, 4000, South Africa} 

\address{$^{107}$ Purple Mountain Observatory, Chinese Academy of Sciences, Nanjing 210008, China} 

\address{$^{108}$ Consortium for Fundamental Physics, School of Mathematics and Statistics, University of Sheffield, Hicks Building, Hounsfield Road, Sheffield, Sheffield S3 7RH, United Kingdom} 

\address{$^{109}$ Van Swenderen Institute, University of Groningen, 9747 AG, Groningen, The Netherlands}

\address{$^{110}$ Institute of Theoretical Astrophysics, University of Oslo, 0315 Oslo, Norway}

\address{$^{111}$ Institute for Nuclear Research and Nuclear Energy,
Bulgarian Academy of Sciences, Boul. Tsarigradsko Chausse 72, BG-1784 Sofia, Bulgaria}

\address{$^{112}$ Service de Physique Th\'eorique et Math\'ematique,
Universit\'e Libre de Bruxelles and International Solvay Institutes,
Campus de la Plaine, CP 231, B-1050 Brussels, Belgium}

\address{$^{113}$ Department of Mathematics, Faculty of Natural Sciences and Mathematics, University of Tuzla, Univerzitetska 4, 75000 Tuzla, Bosnia and Herzegovina}

\address{$^{114}$ Faculty of Natural Sciences and Mathematics, University of Banjaluka, Mladena Stojanovića 2, 78000 Banjaluka, Republic of Srpska, Bosnia and Herzegovina}

\address{$^{115}$ Department of Astronomy, Faculty of Mathematics, University of Belgrade, Studentski Trg 16, 11000 Belgrade, Serbia}

\address{$^{116}$ Faculty of Physics, University of Warsaw, Ludwika Pasteura 5, 02-093 Warsaw, Poland}

\address{$^{117}$ Wigner RCP, H-1121 Budapest, Konkoly Thege Mikl\'{o}s \'{u}t 29-33, Hungary}

\address{$^{118}$ Department of Physics, Ko\c{c} University, Rumelifeneri Yolu, 34450 Sariyer, Istanbul, Turkey}

\address{$^{119}$ Institute of Physics, Maria Curie-Skłodowska University,
Pl. Marii Curie-Skłodowskiej 1, Lublin, 20-031, Poland}

\address{$^{120}$ Janusz Gil Institute of Astronomy, University of Zielona Gora, Licealna 9, 65-417, Zielona Gora, Poland}

\address{$^{121}$ Departamento de F\'{\i}sica-ETSIDI, Universidad Polit\'ecnica de Madrid, 28012 Madrid, Spain}

\address{$^{122}$ INAF -- Osservatorio Astronomico di Brera - Merate, via Emilio Bianchi 46, I-23807 Merate (LC), Italy}

\address{$^{123}$ Astronomical Observatory, Volgina 7, 11060 Belgrade, Serbia}

\address{$^{124}$ Institut de Ci\`encies de l'Espai (ICE, CSIC), Campus UAB, Carrer de Can Magrans s/n, 08193 Cerdanyola del Vall\`es, Spain}

\address{$^{125}$ Institut d'Estudis Espacials de Catalunya (IEEC), Carrer del Gran Capit\`a 2-4, desp. 201, 08034 Barcelona, Spain}

\address{$^{126}$ Department of Physics, Aristotle University of Thessaloniki, 54124 Thessaloniki, Greece}

\address{$^{127}$ Department of Physics and Astronomy, FI-20014 University of Turku, Finland}

\address{$^{128}$ INAF-Observatory or Rome, via Frascati 33, Monte Porzio Catone, 00078, Rome, Italy}

\address{$^{129}$ Faculdade de Ci\^encias, Universidade da Beira Interior, R. Marqu\^es d'\'Avila e Bolama, 6201-001 Covilh\~a, Portugal}

\address{$^{130}$ Department of Natural Sciences, The Open University of Israel, Raanana, Israel}

\address{$^{131}$ Research and Education Center for Natural Sciences, Keio University, 4-1-1 Hiyoshi, Kanagawa 223-8521, Japan}

\address{$^{132}$ Department of Physics, University of Florida, Gainesville, FL 32611, USA}

\address{$^{133}$ Department of Astronomy and Space Sciences, University of Ege, 35100, {\.I}zmir, Turkey}

\address{$^{134}$ Department of Theoretical Physics, Faculty of Physics, Sofia University, Sofia 1164, Bulgaria}


\newcommand{\newblock}{}
\raggedright{}
\nocite{apsrev41Control}
\bibliographystyle{apsrev4-1}
\bibliography{apsrev-controls,References}

\begin{thebibliography}{1888}%
\makeatletter
\providecommand \@ifxundefined [1]{%
 \@ifx{#1\undefined}
}%
\providecommand \@ifnum [1]{%
 \ifnum #1\expandafter \@firstoftwo
 \else \expandafter \@secondoftwo
 \fi
}%
\providecommand \@ifx [1]{%
 \ifx #1\expandafter \@firstoftwo
 \else \expandafter \@secondoftwo
 \fi
}%
\providecommand \natexlab [1]{#1}%
\providecommand \enquote  [1]{``#1''}%
\providecommand \bibnamefont  [1]{#1}%
\providecommand \bibfnamefont [1]{#1}%
\providecommand \citenamefont [1]{#1}%
\providecommand \href@noop [0]{\@secondoftwo}%
\providecommand \href [0]{\begingroup \@sanitize@url \@href}%
\providecommand \@href[1]{\@@startlink{#1}\@@href}%
\providecommand \@@href[1]{\endgroup#1\@@endlink}%
\providecommand \@sanitize@url [0]{\catcode `\\12\catcode `\$12\catcode
  `\&12\catcode `\#12\catcode `\^12\catcode `\_12\catcode `\%12\relax}%
\providecommand \@@startlink[1]{}%
\providecommand \@@endlink[0]{}%
\providecommand \url  [0]{\begingroup\@sanitize@url \@url }%
\providecommand \@url [1]{\endgroup\@href {#1}{\urlprefix }}%
\providecommand \urlprefix  [0]{URL }%
\providecommand \Eprint [0]{\href }%
\providecommand \doibase [0]{http://dx.doi.org/}%
\providecommand \selectlanguage [0]{\@gobble}%
\providecommand \bibinfo  [0]{\@secondoftwo}%
\providecommand \bibfield  [0]{\@secondoftwo}%
\providecommand \translation [1]{[#1]}%
\providecommand \BibitemOpen [0]{}%
\providecommand \bibitemStop [0]{}%
\providecommand \bibitemNoStop [0]{.\EOS\space}%
\providecommand \EOS [0]{\spacefactor3000\relax}%
\providecommand \BibitemShut  [1]{\csname bibitem#1\endcsname}%
\let\auto@bib@innerbib\@empty
\bibitem [{\citenamefont {Abbott}\ \emph
  {et~al.}(2016{\natexlab{a}})\citenamefont {Abbott} \emph
  {et~al.}}]{Abbott:2016blz}%
  \BibitemOpen
  \bibfield  {author} {\bibinfo {author} {\bibfnamefont {B.~P.}\ \bibnamefont
  {Abbott}} \emph {et~al.} (\bibinfo {collaboration} {Virgo, LIGO
  Scientific}),\ }\href {\doibase10.1103/PhysRevLett.116.061102} {\bibfield
  {journal} {\bibinfo  {journal} {Phys. Rev. Lett.}\ }\textbf {\bibinfo
  {volume} {116}},\ \bibinfo {pages} {061102} (\bibinfo {year}
  {2016}{\natexlab{a}})},\ \Eprint {http://arxiv.org/abs/1602.03837}
  {arXiv:1602.03837 [gr-qc]}\BibitemShut {NoStop}%
\bibitem [{\citenamefont {Abbott}\ \emph
  {et~al.}(2016{\natexlab{b}})\citenamefont {Abbott} \emph
  {et~al.}}]{Abbott:2016nmj}%
  \BibitemOpen
  \bibfield  {author} {\bibinfo {author} {\bibfnamefont {B.~P.}\ \bibnamefont
  {Abbott}} \emph {et~al.} (\bibinfo {collaboration} {Virgo, LIGO
  Scientific}),\ }\href {\doibase10.1103/PhysRevLett.116.241103} {\bibfield
  {journal} {\bibinfo  {journal} {Phys. Rev. Lett.}\ }\textbf {\bibinfo
  {volume} {116}},\ \bibinfo {pages} {241103} (\bibinfo {year}
  {2016}{\natexlab{b}})},\ \Eprint {http://arxiv.org/abs/1606.04855}
  {arXiv:1606.04855 [gr-qc]}\BibitemShut {NoStop}%
\bibitem [{\citenamefont {Abbott}\ \emph
  {et~al.}(2016{\natexlab{c}})\citenamefont {Abbott} \emph
  {et~al.}}]{TheLIGOScientific:2016pea}%
  \BibitemOpen
  \bibfield  {author} {\bibinfo {author} {\bibfnamefont {B.~P.}\ \bibnamefont
  {Abbott}} \emph {et~al.} (\bibinfo {collaboration} {Virgo, LIGO
  Scientific}),\ }\href {\doibase10.1103/PhysRevX.6.041015} {\bibfield
  {journal} {\bibinfo  {journal} {Phys. Rev.}\ }\textbf {\bibinfo {volume}
  {X6}},\ \bibinfo {pages} {041015} (\bibinfo {year} {2016}{\natexlab{c}})},\
  \Eprint {http://arxiv.org/abs/1606.04856} {arXiv:1606.04856
  [gr-qc]}\BibitemShut {NoStop}%
\bibitem [{\citenamefont {Abbott}\ \emph
  {et~al.}(2016{\natexlab{d}})\citenamefont {Abbott} \emph
  {et~al.}}]{TheLIGOScientific:2016wfe}%
  \BibitemOpen
  \bibfield  {author} {\bibinfo {author} {\bibfnamefont {B.~P.}\ \bibnamefont
  {Abbott}} \emph {et~al.} (\bibinfo {collaboration} {Virgo, LIGO
  Scientific}),\ }\href {\doibase10.1103/PhysRevLett.116.241102} {\bibfield
  {journal} {\bibinfo  {journal} {Phys. Rev. Lett.}\ }\textbf {\bibinfo
  {volume} {116}},\ \bibinfo {pages} {241102} (\bibinfo {year}
  {2016}{\natexlab{d}})},\ \Eprint {http://arxiv.org/abs/1602.03840}
  {arXiv:1602.03840 [gr-qc]}\BibitemShut {NoStop}%
\bibitem [{\citenamefont {Abbott}\ \emph
  {et~al.}(2016{\natexlab{e}})\citenamefont {Abbott} \emph
  {et~al.}}]{TheLIGOScientific:2016htt}%
  \BibitemOpen
  \bibfield  {author} {\bibinfo {author} {\bibfnamefont {B.~P.}\ \bibnamefont
  {Abbott}} \emph {et~al.} (\bibinfo {collaboration} {Virgo, LIGO
  Scientific}),\ }\href {\doibase10.3847/2041-8205/818/2/L22} {\bibfield
  {journal} {\bibinfo  {journal} {Astrophys. J.}\ }\textbf {\bibinfo {volume}
  {818}},\ \bibinfo {pages} {L22} (\bibinfo {year} {2016}{\natexlab{e}})},\
  \Eprint {http://arxiv.org/abs/1602.03846} {arXiv:1602.03846
  [astro-ph.HE]}\BibitemShut {NoStop}%
\bibitem [{\citenamefont {Abbott}\ \emph
  {et~al.}(2016{\natexlab{f}})\citenamefont {Abbott} \emph
  {et~al.}}]{Abbott:2016ymx}%
  \BibitemOpen
  \bibfield  {author} {\bibinfo {author} {\bibfnamefont {Benjamin~P.}\
  \bibnamefont {Abbott}} \emph {et~al.} (\bibinfo {collaboration} {Virgo, LIGO
  Scientific}),\ }\href {\doibase10.3847/2041-8205/832/2/L21} {\bibfield
  {journal} {\bibinfo  {journal} {Astrophys. J.}\ }\textbf {\bibinfo {volume}
  {832}},\ \bibinfo {pages} {L21} (\bibinfo {year} {2016}{\natexlab{f}})},\
  \Eprint {http://arxiv.org/abs/1607.07456} {arXiv:1607.07456
  [astro-ph.HE]}\BibitemShut {NoStop}%
\bibitem [{\citenamefont {Abbott}\ \emph
  {et~al.}(2017{\natexlab{a}})\citenamefont {Abbott} \emph
  {et~al.}}]{TheLIGOScientific:2016xzw}%
  \BibitemOpen
  \bibfield  {author} {\bibinfo {author} {\bibfnamefont {Benjamin~P.}\
  \bibnamefont {Abbott}} \emph {et~al.} (\bibinfo {collaboration} {Virgo, LIGO
  Scientific}),\ }\href {\doibase10.1103/PhysRevLett.118.121102} {\bibfield
  {journal} {\bibinfo  {journal} {Phys. Rev. Lett.}\ }\textbf {\bibinfo
  {volume} {118}},\ \bibinfo {pages} {121102} (\bibinfo {year}
  {2017}{\natexlab{a}})},\ \Eprint {http://arxiv.org/abs/1612.02030}
  {arXiv:1612.02030 [gr-qc]}\BibitemShut {NoStop}%
\bibitem [{\citenamefont {Abbott}\ \emph
  {et~al.}(2017{\natexlab{b}})\citenamefont {Abbott} \emph
  {et~al.}}]{Abbott:2016cjt}%
  \BibitemOpen
  \bibfield  {author} {\bibinfo {author} {\bibfnamefont {B.~P.}\ \bibnamefont
  {Abbott}} \emph {et~al.} (\bibinfo {collaboration} {Virgo, IPN, LIGO
  Scientific}),\ }\href {\doibase10.3847/1538-4357/aa6c47} {\bibfield
  {journal} {\bibinfo  {journal} {Astrophys. J.}\ }\textbf {\bibinfo {volume}
  {841}},\ \bibinfo {pages} {89} (\bibinfo {year} {2017}{\natexlab{b}})},\
  \Eprint {http://arxiv.org/abs/1611.07947} {arXiv:1611.07947
  [astro-ph.HE]}\BibitemShut {NoStop}%
\bibitem [{\citenamefont {Abbott}\ \emph
  {et~al.}(2016{\natexlab{g}})\citenamefont {Abbott} \emph
  {et~al.}}]{Abbott:2016nhf}%
  \BibitemOpen
  \bibfield  {author} {\bibinfo {author} {\bibfnamefont {B.~P.}\ \bibnamefont
  {Abbott}} \emph {et~al.} (\bibinfo {collaboration} {Virgo, LIGO
  Scientific}),\ }\href {\doibase10.3847/2041-8205/833/1/L1} {\bibfield
  {journal} {\bibinfo  {journal} {Astrophys. J.}\ }\textbf {\bibinfo {volume}
  {833}},\ \bibinfo {pages} {L1} (\bibinfo {year} {2016}{\natexlab{g}})},\
  \Eprint {http://arxiv.org/abs/1602.03842} {arXiv:1602.03842
  [astro-ph.HE]}\BibitemShut {NoStop}%
\bibitem [{\citenamefont {Abbott}\ \emph
  {et~al.}(2016{\natexlab{h}})\citenamefont {Abbott} \emph
  {et~al.}}]{Abbott:2016drs}%
  \BibitemOpen
  \bibfield  {author} {\bibinfo {author} {\bibfnamefont {B.~P.}\ \bibnamefont
  {Abbott}} \emph {et~al.} (\bibinfo {collaboration} {Virgo, LIGO
  Scientific}),\ }\href {\doibase10.3847/0067-0049/227/2/14} {\bibfield
  {journal} {\bibinfo  {journal} {Astrophys. J. Suppl.}\ }\textbf {\bibinfo
  {volume} {227}},\ \bibinfo {pages} {14} (\bibinfo {year}
  {2016}{\natexlab{h}})},\ \Eprint {http://arxiv.org/abs/1606.03939}
  {arXiv:1606.03939 [astro-ph.HE]}\BibitemShut {NoStop}%
\bibitem [{\citenamefont {Abbott}\ \emph
  {et~al.}(2017{\natexlab{c}})\citenamefont {Abbott} \emph
  {et~al.}}]{Abbott:2017vtc}%
  \BibitemOpen
  \bibfield  {author} {\bibinfo {author} {\bibfnamefont {Benjamin~P.}\
  \bibnamefont {Abbott}} \emph {et~al.} (\bibinfo {collaboration} {VIRGO, LIGO
  Scientific}),\ }\href {\doibase10.1103/PhysRevLett.118.221101} {\bibfield
  {journal} {\bibinfo  {journal} {Phys. Rev. Lett.}\ }\textbf {\bibinfo
  {volume} {118}},\ \bibinfo {pages} {221101} (\bibinfo {year}
  {2017}{\natexlab{c}})},\ \Eprint {http://arxiv.org/abs/1706.01812}
  {arXiv:1706.01812 [gr-qc]}\BibitemShut {NoStop}%
\bibitem [{\citenamefont {Abbott}\ \emph
  {et~al.}(2016{\natexlab{i}})\citenamefont {Abbott} \emph
  {et~al.}}]{Abbott:2016gcq}%
  \BibitemOpen
  \bibfield  {author} {\bibinfo {author} {\bibfnamefont {B.~P.}\ \bibnamefont
  {Abbott}} \emph {et~al.} (\bibinfo {collaboration} {InterPlanetary Network,
  DES, INTEGRAL, La Silla-QUEST Survey, MWA, Fermi-LAT, J-GEM, GRAWITA, Pi of
  the Sky, Fermi GBM, MASTER, Swift, iPTF, VISTA, ASKAP, SkyMapper, PESSTO,
  TOROS, Pan-STARRS, Virgo, Liverpool Telescope, BOOTES, LIGO Scientific,
  LOFAR, C2PU, MAXI}),\ }\href {\doibase10.3847/2041-8205/826/1/L13} {\bibfield
   {journal} {\bibinfo  {journal} {Astrophys. J.}\ }\textbf {\bibinfo {volume}
  {826}},\ \bibinfo {pages} {L13} (\bibinfo {year} {2016}{\natexlab{i}})},\
  \Eprint {http://arxiv.org/abs/1602.08492} {arXiv:1602.08492
  [astro-ph.HE]}\BibitemShut {NoStop}%
\bibitem [{\citenamefont {Abbott}\ \emph
  {et~al.}(2016{\natexlab{j}})\citenamefont {Abbott} \emph
  {et~al.}}]{Abbott:2016iqz}%
  \BibitemOpen
  \bibfield  {author} {\bibinfo {author} {\bibfnamefont {B.~P.}\ \bibnamefont
  {Abbott}} \emph {et~al.} (\bibinfo {collaboration} {InterPlanetary Network,
  DES, INTEGRAL, La Silla-QUEST Survey, MWA, Fermi-LAT, J-GEM, Zadko, GRAWITA,
  Pi of the Sky, MASTER, Swift, iPTF, VISTA, ASKAP, SkyMapper, PESSTO, TOROS,
  Pan-STARRS, Virgo, Algerian National Observatory, Liverpool Telescope,
  BOOTES, LIGO Scientific, LOFAR, TAROT, C2PU, MAXI, Fermi-GBM}),\ }\href
  {\doibase10.3847/0067-0049/225/1/8} {\bibfield  {journal} {\bibinfo
  {journal} {Astrophys. J. Suppl.}\ }\textbf {\bibinfo {volume} {225}},\
  \bibinfo {pages} {8} (\bibinfo {year} {2016}{\natexlab{j}})},\ \Eprint
  {http://arxiv.org/abs/1604.07864} {arXiv:1604.07864
  [astro-ph.HE]}\BibitemShut {NoStop}%
\bibitem [{\citenamefont {Abbott}\ \emph
  {et~al.}(2018{\natexlab{a}})\citenamefont {Abbott} \emph
  {et~al.}}]{Aasi:2013wya}%
  \BibitemOpen
  \bibfield  {author} {\bibinfo {author} {\bibfnamefont {Benjamin~P.}\
  \bibnamefont {Abbott}} \emph {et~al.} (\bibinfo {collaboration} {VIRGO,
  KAGRA, LIGO Scientific}),\ }\href {\doibase10.1007/s41114-018-0012-9}
  {\bibfield  {journal} {\bibinfo  {journal} {Living Rev. Rel.}\ }\textbf
  {\bibinfo {volume} {21}},\ \bibinfo {pages} {3} (\bibinfo {year}
  {2018}{\natexlab{a}})},\ \Eprint {http://arxiv.org/abs/1304.0670}
  {arXiv:1304.0670 [gr-qc]}\BibitemShut {NoStop}%
\bibitem [{\citenamefont {{Sathyaprakash}}\ and\ \citenamefont
  {{Dhurandhar}}(1991)}]{1991PhRvD..44.3819S}%
  \BibitemOpen
  \bibfield  {author} {\bibinfo {author} {\bibfnamefont {B.~S.}\ \bibnamefont
  {{Sathyaprakash}}}\ and\ \bibinfo {author} {\bibfnamefont {S.~V.}\
  \bibnamefont {{Dhurandhar}}},\ }\href {\doibase10.1103/PhysRevD.44.3819}
  {\bibfield  {journal} {\bibinfo  {journal} {Physical Review D}\ }\textbf
  {\bibinfo {volume} {44}},\ \bibinfo {pages} {3819--3834} (\bibinfo {year}
  {1991})}\BibitemShut {NoStop}%
\bibitem [{\citenamefont {Abbott}\ \emph
  {et~al.}(2016{\natexlab{k}})\citenamefont {Abbott} \emph
  {et~al.}}]{TheLIGOScientific:2016qqj}%
  \BibitemOpen
  \bibfield  {author} {\bibinfo {author} {\bibfnamefont {B.~P.}\ \bibnamefont
  {Abbott}} \emph {et~al.} (\bibinfo {collaboration} {Virgo, LIGO
  Scientific}),\ }\href {\doibase10.1103/PhysRevD.93.122003} {\bibfield
  {journal} {\bibinfo  {journal} {Phys. Rev.}\ }\textbf {\bibinfo {volume}
  {D93}},\ \bibinfo {pages} {122003} (\bibinfo {year} {2016}{\natexlab{k}})},\
  \Eprint {http://arxiv.org/abs/1602.03839} {arXiv:1602.03839
  [gr-qc]}\BibitemShut {NoStop}%
\bibitem [{\citenamefont {Abbott}\ \emph
  {et~al.}(2016{\natexlab{l}})\citenamefont {Abbott} \emph
  {et~al.}}]{TheLIGOScientific:2016src}%
  \BibitemOpen
  \bibfield  {author} {\bibinfo {author} {\bibfnamefont {B.~P.}\ \bibnamefont
  {Abbott}} \emph {et~al.} (\bibinfo {collaboration} {Virgo, LIGO
  Scientific}),\ }\href {\doibase10.1103/PhysRevLett.116.221101} {\bibfield
  {journal} {\bibinfo  {journal} {Phys. Rev. Lett.}\ }\textbf {\bibinfo
  {volume} {116}},\ \bibinfo {pages} {221101} (\bibinfo {year}
  {2016}{\natexlab{l}})},\ \Eprint {http://arxiv.org/abs/1602.03841}
  {arXiv:1602.03841 [gr-qc]}\BibitemShut {NoStop}%
\bibitem [{\citenamefont {Abbott}\ \emph
  {et~al.}(2017{\natexlab{d}})\citenamefont {Abbott} \emph
  {et~al.}}]{Abbott:2017oio}%
  \BibitemOpen
  \bibfield  {author} {\bibinfo {author} {\bibfnamefont {B.~P.}\ \bibnamefont
  {Abbott}} \emph {et~al.} (\bibinfo {collaboration} {Virgo, LIGO
  Scientific}),\ }\href {\doibase10.1103/PhysRevLett.119.141101} {\bibfield
  {journal} {\bibinfo  {journal} {Phys. Rev. Lett.}\ }\textbf {\bibinfo
  {volume} {119}},\ \bibinfo {pages} {141101} (\bibinfo {year}
  {2017}{\natexlab{d}})},\ \Eprint {http://arxiv.org/abs/1709.09660}
  {arXiv:1709.09660 [gr-qc]}\BibitemShut {NoStop}%
\bibitem [{\citenamefont {Abbott}\ \emph
  {et~al.}(2016{\natexlab{m}})\citenamefont {Abbott} \emph
  {et~al.}}]{TheLIGOScientific:2016uux}%
  \BibitemOpen
  \bibfield  {author} {\bibinfo {author} {\bibfnamefont {B.~P.}\ \bibnamefont
  {Abbott}} \emph {et~al.} (\bibinfo {collaboration} {Virgo, LIGO
  Scientific}),\ }\href {\doibase10.1103/PhysRevD.93.122004} {\bibfield
  {journal} {\bibinfo  {journal} {Phys. Rev.}\ }\textbf {\bibinfo {volume}
  {D93}},\ \bibinfo {pages} {122004} (\bibinfo {year} {2016}{\natexlab{m}})},\
  \bibinfo {note} {[Addendum: Phys. Rev.D94,no.6,069903(2016)]},\ \Eprint
  {http://arxiv.org/abs/1602.03843} {arXiv:1602.03843 [gr-qc]}\BibitemShut
  {NoStop}%
\bibitem [{\citenamefont {Abbott}\ \emph
  {et~al.}(2017{\natexlab{e}})\citenamefont {Abbott} \emph
  {et~al.}}]{Abbott:2017gyy}%
  \BibitemOpen
  \bibfield  {author} {\bibinfo {author} {\bibfnamefont {B.~P.}\ \bibnamefont
  {Abbott}} \emph {et~al.} (\bibinfo {collaboration} {Virgo, LIGO
  Scientific}),\ }\href {\doibase10.3847/2041-8213/aa9f0c} {\bibfield
  {journal} {\bibinfo  {journal} {Astrophys. J.}\ }\textbf {\bibinfo {volume}
  {851}},\ \bibinfo {pages} {L35} (\bibinfo {year} {2017}{\natexlab{e}})},\
  \Eprint {http://arxiv.org/abs/1711.05578} {arXiv:1711.05578
  [astro-ph.HE]}\BibitemShut {NoStop}%
\bibitem [{\citenamefont {Abbott}\ \emph
  {et~al.}(2017{\natexlab{f}})\citenamefont {Abbott} \emph
  {et~al.}}]{TheLIGOScientific:2017qsa}%
  \BibitemOpen
  \bibfield  {author} {\bibinfo {author} {\bibfnamefont {B.~P.}\ \bibnamefont
  {Abbott}} \emph {et~al.} (\bibinfo {collaboration} {Virgo, LIGO
  Scientific}),\ }\href {\doibase10.1103/PhysRevLett.119.161101} {\bibfield
  {journal} {\bibinfo  {journal} {Phys. Rev. Lett.}\ }\textbf {\bibinfo
  {volume} {119}},\ \bibinfo {pages} {161101} (\bibinfo {year}
  {2017}{\natexlab{f}})},\ \Eprint {http://arxiv.org/abs/1710.05832}
  {arXiv:1710.05832 [gr-qc]}\BibitemShut {NoStop}%
\bibitem [{\citenamefont {Abbott}\ \emph
  {et~al.}(2017{\natexlab{g}})\citenamefont {Abbott} \emph
  {et~al.}}]{GBM:2017lvd}%
  \BibitemOpen
  \bibfield  {author} {\bibinfo {author} {\bibfnamefont {B.~P.}\ \bibnamefont
  {Abbott}} \emph {et~al.} (\bibinfo {collaboration} {GROND, SALT Group,
  OzGrav, DFN, INTEGRAL, Virgo, Insight-Hxmt, MAXI Team, Fermi-LAT, J-GEM,
  RATIR, IceCube, CAASTRO, LWA, ePESSTO, GRAWITA, RIMAS, SKA South
  Africa/MeerKAT, H.E.S.S., 1M2H Team, IKI-GW Follow-up, Fermi GBM, Pi of Sky,
  DWF (Deeper Wider Faster Program), Dark Energy Survey, MASTER, AstroSat
  Cadmium Zinc Telluride Imager Team, Swift, Pierre Auger, ASKAP, VINROUGE,
  JAGWAR, Chandra Team at McGill University, TTU-NRAO, GROWTH, AGILE Team, MWA,
  ATCA, AST3, TOROS, Pan-STARRS, NuSTAR, ATLAS Telescopes, BOOTES, CaltechNRAO,
  LIGO Scientific, High Time Resolution Universe Survey, Nordic Optical
  Telescope, Las Cumbres Observatory Group, TZAC Consortium, LOFAR, IPN, DLT40,
  Texas Tech University, HAWC, ANTARES, KU, Dark Energy Camera GW-EM, CALET,
  Euro VLBI Team, ALMA}),\ }\href {\doibase10.3847/2041-8213/aa91c9} {\bibfield
   {journal} {\bibinfo  {journal} {Astrophys. J.}\ }\textbf {\bibinfo {volume}
  {848}},\ \bibinfo {pages} {L12} (\bibinfo {year} {2017}{\natexlab{g}})},\
  \Eprint {http://arxiv.org/abs/1710.05833} {arXiv:1710.05833
  [astro-ph.HE]}\BibitemShut {NoStop}%
\bibitem [{\citenamefont {Abbott}\ \emph
  {et~al.}(2017{\natexlab{h}})\citenamefont {Abbott} \emph
  {et~al.}}]{Monitor:2017mdv}%
  \BibitemOpen
  \bibfield  {author} {\bibinfo {author} {\bibfnamefont {B.~P.}\ \bibnamefont
  {Abbott}} \emph {et~al.} (\bibinfo {collaboration} {Virgo, Fermi-GBM,
  INTEGRAL, LIGO Scientific}),\ }\href {\doibase10.3847/2041-8213/aa920c}
  {\bibfield  {journal} {\bibinfo  {journal} {Astrophys. J.}\ }\textbf
  {\bibinfo {volume} {848}},\ \bibinfo {pages} {L13} (\bibinfo {year}
  {2017}{\natexlab{h}})},\ \Eprint {http://arxiv.org/abs/1710.05834}
  {arXiv:1710.05834 [astro-ph.HE]}\BibitemShut {NoStop}%
\bibitem [{\citenamefont {Abbott}\ \emph
  {et~al.}(2017{\natexlab{i}})\citenamefont {Abbott} \emph
  {et~al.}}]{Abbott:2017xzu}%
  \BibitemOpen
  \bibfield  {author} {\bibinfo {author} {\bibfnamefont {B.~P.}\ \bibnamefont
  {Abbott}} \emph {et~al.} (\bibinfo {collaboration} {LIGO Scientific,
  VINROUGE, Las Cumbres Observatory, DES, DLT40, Virgo, 1M2H, Dark Energy
  Camera GW-E, MASTER}),\ }\href {\doibase10.1038/nature24471} {\bibfield
  {journal} {\bibinfo  {journal} {Nature}\ }\textbf {\bibinfo {volume} {551}},\
  \bibinfo {pages} {85--88} (\bibinfo {year} {2017}{\natexlab{i}})},\ \Eprint
  {http://arxiv.org/abs/1710.05835} {arXiv:1710.05835
  [astro-ph.CO]}\BibitemShut {NoStop}%
\bibitem [{\citenamefont {Belczynski}\ \emph
  {et~al.}(2010{\natexlab{a}})\citenamefont {Belczynski}, \citenamefont
  {Dominik}, \citenamefont {Bulik}, \citenamefont {O'Shaughnessy},
  \citenamefont {Fryer},\ and\ \citenamefont {Holz}}]{Belczynski:2010tb}%
  \BibitemOpen
  \bibfield  {author} {\bibinfo {author} {\bibfnamefont {K.}~\bibnamefont
  {Belczynski}}, \bibinfo {author} {\bibfnamefont {M.}~\bibnamefont {Dominik}},
  \bibinfo {author} {\bibfnamefont {T.}~\bibnamefont {Bulik}}, \bibinfo
  {author} {\bibfnamefont {R.}~\bibnamefont {O'Shaughnessy}}, \bibinfo {author}
  {\bibfnamefont {C.~L.}\ \bibnamefont {Fryer}}, \ and\ \bibinfo {author}
  {\bibfnamefont {D.~E.}\ \bibnamefont {Holz}},\ }\href
  {\doibase10.1088/2041-8205/715/2/L138} {\bibfield  {journal} {\bibinfo
  {journal} {Astrophys. J.}\ }\textbf {\bibinfo {volume} {715}},\ \bibinfo
  {pages} {L138} (\bibinfo {year} {2010}{\natexlab{a}})},\ \Eprint
  {http://arxiv.org/abs/1004.0386} {arXiv:1004.0386 [astro-ph.HE]}\BibitemShut
  {NoStop}%
\bibitem [{\citenamefont {{Spera}}\ \emph
  {et~al.}(2015{\natexlab{a}})\citenamefont {{Spera}}, \citenamefont
  {{Mapelli}},\ and\ \citenamefont {{Bressan}}}]{Spera15}%
  \BibitemOpen
  \bibfield  {author} {\bibinfo {author} {\bibfnamefont {M.}~\bibnamefont
  {{Spera}}}, \bibinfo {author} {\bibfnamefont {M.}~\bibnamefont {{Mapelli}}},
  \ and\ \bibinfo {author} {\bibfnamefont {A.}~\bibnamefont {{Bressan}}},\
  }\href {\doibase10.1093/mnras/stv1161} {\bibfield  {journal} {\bibinfo
  {journal} {MNRAS}\ }\textbf {\bibinfo {volume} {451}},\ \bibinfo {pages}
  {4086--4103} (\bibinfo {year} {2015}{\natexlab{a}})},\ \Eprint
  {http://arxiv.org/abs/1505.05201} {arXiv:1505.05201
  [astro-ph.SR]}\BibitemShut {NoStop}%
\bibitem [{\citenamefont {Belczynski}\ \emph
  {et~al.}(2016{\natexlab{a}})\citenamefont {Belczynski}, \citenamefont {Holz},
  \citenamefont {Bulik},\ and\ \citenamefont
  {O'Shaughnessy}}]{Belczynski:2016obo}%
  \BibitemOpen
  \bibfield  {author} {\bibinfo {author} {\bibfnamefont {Krzysztof}\
  \bibnamefont {Belczynski}}, \bibinfo {author} {\bibfnamefont {Daniel~E.}\
  \bibnamefont {Holz}}, \bibinfo {author} {\bibfnamefont {Tomasz}\ \bibnamefont
  {Bulik}}, \ and\ \bibinfo {author} {\bibfnamefont {Richard}\ \bibnamefont
  {O'Shaughnessy}},\ }\href {\doibase10.1038/nature18322} {\bibfield  {journal}
  {\bibinfo  {journal} {Nature}\ }\textbf {\bibinfo {volume} {534}},\ \bibinfo
  {pages} {512} (\bibinfo {year} {2016}{\natexlab{a}})},\ \Eprint
  {http://arxiv.org/abs/1602.04531} {arXiv:1602.04531
  [astro-ph.HE]}\BibitemShut {NoStop}%
\bibitem [{\citenamefont {{Giacobbo}}\ \emph {et~al.}(2018)\citenamefont
  {{Giacobbo}}, \citenamefont {{Mapelli}},\ and\ \citenamefont
  {{Spera}}}]{Giacobbo18}%
  \BibitemOpen
  \bibfield  {author} {\bibinfo {author} {\bibfnamefont {N.}~\bibnamefont
  {{Giacobbo}}}, \bibinfo {author} {\bibfnamefont {M.}~\bibnamefont
  {{Mapelli}}}, \ and\ \bibinfo {author} {\bibfnamefont {M.}~\bibnamefont
  {{Spera}}},\ }\href {\doibase10.1093/mnras/stx2933} {\bibfield  {journal}
  {\bibinfo  {journal} {MNRAS}\ }\textbf {\bibinfo {volume} {474}},\ \bibinfo
  {pages} {2959--2974} (\bibinfo {year} {2018})},\ \Eprint
  {http://arxiv.org/abs/1711.03556} {arXiv:1711.03556
  [astro-ph.SR]}\BibitemShut {NoStop}%
\bibitem [{\citenamefont {{Woosley}}(2017)}]{Woosley17}%
  \BibitemOpen
  \bibfield  {author} {\bibinfo {author} {\bibfnamefont {S.~E.}\ \bibnamefont
  {{Woosley}}},\ }\href {\doibase10.3847/1538-4357/836/2/244} {\bibfield
  {journal} {\bibinfo  {journal} {Astrophysical Journal}\ }\textbf {\bibinfo
  {volume} {836}},\ \bibinfo {eid} {244} (\bibinfo {year} {2017})},\ \Eprint
  {http://arxiv.org/abs/1608.08939} {arXiv:1608.08939
  [astro-ph.HE]}\BibitemShut {NoStop}%
\bibitem [{\citenamefont {Belczynski}\ \emph
  {et~al.}(2016{\natexlab{b}})\citenamefont {Belczynski} \emph
  {et~al.}}]{Belczynski:2016jno}%
  \BibitemOpen
  \bibfield  {author} {\bibinfo {author} {\bibfnamefont {K.}~\bibnamefont
  {Belczynski}} \emph {et~al.},\ }\href {\doibase10.1051/0004-6361/201628980}
  {\bibfield  {journal} {\bibinfo  {journal} {Astron. Astrophys.}\ }\textbf
  {\bibinfo {volume} {594}},\ \bibinfo {pages} {A97} (\bibinfo {year}
  {2016}{\natexlab{b}})},\ \Eprint {http://arxiv.org/abs/1607.03116}
  {arXiv:1607.03116 [astro-ph.HE]}\BibitemShut {NoStop}%
\bibitem [{\citenamefont {Spera}\ and\ \citenamefont
  {Mapelli}(2017)}]{Spera:2017fyx}%
  \BibitemOpen
  \bibfield  {author} {\bibinfo {author} {\bibfnamefont {Mario}\ \bibnamefont
  {Spera}}\ and\ \bibinfo {author} {\bibfnamefont {Michela}\ \bibnamefont
  {Mapelli}},\ }\href {\doibase10.1093/mnras/stx1576} {\bibfield  {journal}
  {\bibinfo  {journal} {Mon. Not. Roy. Astron. Soc.}\ }\textbf {\bibinfo
  {volume} {470}},\ \bibinfo {pages} {4739--4749} (\bibinfo {year} {2017})},\
  \Eprint {http://arxiv.org/abs/1706.06109} {arXiv:1706.06109
  [astro-ph.SR]}\BibitemShut {NoStop}%
\bibitem [{\citenamefont {Abbott}\ \emph
  {et~al.}(2017{\natexlab{j}})\citenamefont {Abbott} \emph
  {et~al.}}]{Evans:2016mbw}%
  \BibitemOpen
  \bibfield  {author} {\bibinfo {author} {\bibfnamefont {Benjamin~P.}\
  \bibnamefont {Abbott}} \emph {et~al.} (\bibinfo {collaboration} {LIGO
  Scientific}),\ }\href {\doibase10.1088/1361-6382/aa51f4} {\bibfield
  {journal} {\bibinfo  {journal} {Class. Quant. Grav.}\ }\textbf {\bibinfo
  {volume} {34}},\ \bibinfo {pages} {044001} (\bibinfo {year}
  {2017}{\natexlab{j}})},\ \Eprint {http://arxiv.org/abs/1607.08697}
  {arXiv:1607.08697 [astro-ph.IM]}\BibitemShut {NoStop}%
\bibitem [{\citenamefont {{Sathyaprakash}}\ \emph {et~al.}(2012)\citenamefont
  {{Sathyaprakash}}, \citenamefont {{Abernathy}}, \citenamefont {{Acernese}},
  \citenamefont {{Ajith}}, \citenamefont {{Allen}}, \citenamefont
  {{Amaro-Seoane}}, \citenamefont {{Andersson}}, \citenamefont {{Aoudia}},
  \citenamefont {{Arun}}, \citenamefont {{Astone}},\ and\ \citenamefont
  {et~al.}}]{2012CQGra..29l4013S}%
  \BibitemOpen
  \bibfield  {author} {\bibinfo {author} {\bibfnamefont {B.}~\bibnamefont
  {{Sathyaprakash}}}, \bibinfo {author} {\bibfnamefont {M.}~\bibnamefont
  {{Abernathy}}}, \bibinfo {author} {\bibfnamefont {F.}~\bibnamefont
  {{Acernese}}}, \bibinfo {author} {\bibfnamefont {P.}~\bibnamefont {{Ajith}}},
  \bibinfo {author} {\bibfnamefont {B.}~\bibnamefont {{Allen}}}, \bibinfo
  {author} {\bibfnamefont {P.}~\bibnamefont {{Amaro-Seoane}}}, \bibinfo
  {author} {\bibfnamefont {N.}~\bibnamefont {{Andersson}}}, \bibinfo {author}
  {\bibfnamefont {S.}~\bibnamefont {{Aoudia}}}, \bibinfo {author}
  {\bibfnamefont {K.}~\bibnamefont {{Arun}}}, \bibinfo {author} {\bibfnamefont
  {P.}~\bibnamefont {{Astone}}}, \ and\ \bibinfo {author} {\bibnamefont
  {et~al.}},\ }\href {\doibase10.1088/0264-9381/29/12/124013} {\bibfield
  {journal} {\bibinfo  {journal} {Classical and Quantum Gravity}\ }\textbf
  {\bibinfo {volume} {29}},\ \bibinfo {eid} {124013} (\bibinfo {year}
  {2012})},\ \Eprint {http://arxiv.org/abs/1206.0331} {arXiv:1206.0331
  [gr-qc]}\BibitemShut {NoStop}%
\bibitem [{\citenamefont {{King}}\ \emph {et~al.}(2005)\citenamefont {{King}},
  \citenamefont {{Lubow}}, \citenamefont {{Ogilvie}},\ and\ \citenamefont
  {{Pringle}}}]{King05}%
  \BibitemOpen
  \bibfield  {author} {\bibinfo {author} {\bibfnamefont {A.~R.}\ \bibnamefont
  {{King}}}, \bibinfo {author} {\bibfnamefont {S.~H.}\ \bibnamefont {{Lubow}}},
  \bibinfo {author} {\bibfnamefont {G.~I.}\ \bibnamefont {{Ogilvie}}}, \ and\
  \bibinfo {author} {\bibfnamefont {J.~E.}\ \bibnamefont {{Pringle}}},\ }\href
  {\doibase10.1111/j.1365-2966.2005.09378.x} {\bibfield  {journal} {\bibinfo
  {journal} {MNRAS}\ }\textbf {\bibinfo {volume} {363}},\ \bibinfo {pages}
  {49--56} (\bibinfo {year} {2005})},\ \Eprint
  {http://arxiv.org/abs/astro-ph/0507098} {astro-ph/0507098}\BibitemShut
  {NoStop}%
\bibitem [{\citenamefont {{Perego}}\ \emph {et~al.}(2009)\citenamefont
  {{Perego}}, \citenamefont {{Dotti}}, \citenamefont {{Colpi}},\ and\
  \citenamefont {{Volonteri}}}]{Perego09}%
  \BibitemOpen
  \bibfield  {author} {\bibinfo {author} {\bibfnamefont {A.}~\bibnamefont
  {{Perego}}}, \bibinfo {author} {\bibfnamefont {M.}~\bibnamefont {{Dotti}}},
  \bibinfo {author} {\bibfnamefont {M.}~\bibnamefont {{Colpi}}}, \ and\
  \bibinfo {author} {\bibfnamefont {M.}~\bibnamefont {{Volonteri}}},\ }\href
  {\doibase10.1111/j.1365-2966.2009.15427.x} {\bibfield  {journal} {\bibinfo
  {journal} {MNRAS}\ }\textbf {\bibinfo {volume} {399}},\ \bibinfo {pages}
  {2249--2263} (\bibinfo {year} {2009})},\ \Eprint
  {http://arxiv.org/abs/0907.3742} {arXiv:0907.3742}\BibitemShut {NoStop}%
\bibitem [{\citenamefont {Berti}\ and\ \citenamefont
  {Volonteri}(2008)}]{Berti:2008af}%
  \BibitemOpen
  \bibfield  {author} {\bibinfo {author} {\bibfnamefont {Emanuele}\
  \bibnamefont {Berti}}\ and\ \bibinfo {author} {\bibfnamefont {Marta}\
  \bibnamefont {Volonteri}},\ }\href {\doibase10.1086/590379} {\bibfield
  {journal} {\bibinfo  {journal} {Astrophys. J.}\ }\textbf {\bibinfo {volume}
  {684}},\ \bibinfo {pages} {822--828} (\bibinfo {year} {2008})},\ \Eprint
  {http://arxiv.org/abs/0802.0025} {arXiv:0802.0025 [astro-ph]}\BibitemShut
  {NoStop}%
\bibitem [{\citenamefont {Barausse}\ \emph
  {et~al.}(2012{\natexlab{a}})\citenamefont {Barausse}, \citenamefont
  {Morozova},\ and\ \citenamefont {Rezzolla}}]{Barausse:2012qz}%
  \BibitemOpen
  \bibfield  {author} {\bibinfo {author} {\bibfnamefont {Enrico}\ \bibnamefont
  {Barausse}}, \bibinfo {author} {\bibfnamefont {Viktoriya}\ \bibnamefont
  {Morozova}}, \ and\ \bibinfo {author} {\bibfnamefont {Luciano}\ \bibnamefont
  {Rezzolla}},\ }\href {\doibase10.1088/0004-637X/758/1/63} {\bibfield
  {journal} {\bibinfo  {journal} {Astrophys. J.}\ }\textbf {\bibinfo {volume}
  {758}},\ \bibinfo {pages} {63} (\bibinfo {year} {2012}{\natexlab{a}})},\
  \bibinfo {note} {[Erratum: Astrophys. J.786,76(2014)]},\ \Eprint
  {http://arxiv.org/abs/1206.3803} {arXiv:1206.3803 [gr-qc]}\BibitemShut
  {NoStop}%
\bibitem [{\citenamefont {{Rezzolla}}(2016)}]{Rezzolla16}%
  \BibitemOpen
  \bibfield  {author} {\bibinfo {author} {\bibfnamefont {L.}~\bibnamefont
  {{Rezzolla}}},\ }in\ \href {\doibase10.1007/978-3-319-19416-5_1} {\emph
  {\bibinfo {booktitle} {Lecture Notes in Physics, Berlin Springer Verlag}}},\
  \bibinfo {series} {Lecture Notes in Physics, Berlin Springer Verlag}, Vol.\
  \bibinfo {volume} {905},\ \bibinfo {editor} {edited by\ \bibinfo {editor}
  {\bibfnamefont {F.}~\bibnamefont {{Haardt}}}, \bibinfo {editor}
  {\bibfnamefont {V.}~\bibnamefont {{Gorini}}}, \bibinfo {editor}
  {\bibfnamefont {U.}~\bibnamefont {{Moschella}}}, \bibinfo {editor}
  {\bibfnamefont {A.}~\bibnamefont {{Treves}}}, \ and\ \bibinfo {editor}
  {\bibfnamefont {M.}~\bibnamefont {{Colpi}}}}\ (\bibinfo {year} {2016})\
  p.~\bibinfo {pages} {1}\BibitemShut {NoStop}%
\bibitem [{\citenamefont {{Merloni}}(2016)}]{Merloni16}%
  \BibitemOpen
  \bibfield  {author} {\bibinfo {author} {\bibfnamefont {A.}~\bibnamefont
  {{Merloni}}},\ }in\ \href {\doibase10.1007/978-3-319-19416-5_4} {\emph
  {\bibinfo {booktitle} {Lecture Notes in Physics, Berlin Springer Verlag}}},\
  \bibinfo {series} {Lecture Notes in Physics, Berlin Springer Verlag}, Vol.\
  \bibinfo {volume} {905},\ \bibinfo {editor} {edited by\ \bibinfo {editor}
  {\bibfnamefont {F.}~\bibnamefont {{Haardt}}}, \bibinfo {editor}
  {\bibfnamefont {V.}~\bibnamefont {{Gorini}}}, \bibinfo {editor}
  {\bibfnamefont {U.}~\bibnamefont {{Moschella}}}, \bibinfo {editor}
  {\bibfnamefont {A.}~\bibnamefont {{Treves}}}, \ and\ \bibinfo {editor}
  {\bibfnamefont {M.}~\bibnamefont {{Colpi}}}}\ (\bibinfo {year} {2016})\ p.\
  \bibinfo {pages} {101},\ \Eprint {http://arxiv.org/abs/1505.04940}
  {arXiv:1505.04940 [astro-ph.HE]}\BibitemShut {NoStop}%
\bibitem [{\citenamefont {{Kormendy}}\ and\ \citenamefont
  {{Ho}}(2013{\natexlab{a}})}]{Kormendy13}%
  \BibitemOpen
  \bibfield  {author} {\bibinfo {author} {\bibfnamefont {J.}~\bibnamefont
  {{Kormendy}}}\ and\ \bibinfo {author} {\bibfnamefont {L.~C.}\ \bibnamefont
  {{Ho}}},\ }\href {\doibase10.1146/annurev-astro-082708-101811} {\bibfield
  {journal} {\bibinfo  {journal} {\araa}\ }\textbf {\bibinfo {volume} {51}},\
  \bibinfo {pages} {511--653} (\bibinfo {year} {2013}{\natexlab{a}})},\ \Eprint
  {http://arxiv.org/abs/1304.7762} {arXiv:1304.7762}\BibitemShut {NoStop}%
\bibitem [{\citenamefont {{Baldassare}}\ \emph {et~al.}(2015)\citenamefont
  {{Baldassare}}, \citenamefont {{Reines}}, \citenamefont {{Gallo}},\ and\
  \citenamefont {{Greene}}}]{Baldassare15}%
  \BibitemOpen
  \bibfield  {author} {\bibinfo {author} {\bibfnamefont {V.~F.}\ \bibnamefont
  {{Baldassare}}}, \bibinfo {author} {\bibfnamefont {A.~E.}\ \bibnamefont
  {{Reines}}}, \bibinfo {author} {\bibfnamefont {E.}~\bibnamefont {{Gallo}}}, \
  and\ \bibinfo {author} {\bibfnamefont {J.~E.}\ \bibnamefont {{Greene}}},\
  }\href {\doibase10.1088/2041-8205/809/1/L14} {\bibfield  {journal} {\bibinfo
  {journal} {Astrophysical Journal}\ }\textbf {\bibinfo {volume} {809}},\
  \bibinfo {eid} {L14} (\bibinfo {year} {2015})},\ \Eprint
  {http://arxiv.org/abs/1506.07531} {arXiv:1506.07531}\BibitemShut {NoStop}%
\bibitem [{\citenamefont {{Wu}}\ \emph {et~al.}(2015)\citenamefont {{Wu}},
  \citenamefont {{Wang}}, \citenamefont {{Fan}}, \citenamefont {{Yi}},
  \citenamefont {{Zuo}}, \citenamefont {{Bian}}, \citenamefont {{Jiang}},
  \citenamefont {{McGreer}}, \citenamefont {{Wang}}, \citenamefont {{Yang}},
  \citenamefont {{Yang}}, \citenamefont {{Thompson}},\ and\ \citenamefont
  {{Beletsky}}}]{Wu15}%
  \BibitemOpen
  \bibfield  {author} {\bibinfo {author} {\bibfnamefont {X.-B.}\ \bibnamefont
  {{Wu}}}, \bibinfo {author} {\bibfnamefont {F.}~\bibnamefont {{Wang}}},
  \bibinfo {author} {\bibfnamefont {X.}~\bibnamefont {{Fan}}}, \bibinfo
  {author} {\bibfnamefont {W.}~\bibnamefont {{Yi}}}, \bibinfo {author}
  {\bibfnamefont {W.}~\bibnamefont {{Zuo}}}, \bibinfo {author} {\bibfnamefont
  {F.}~\bibnamefont {{Bian}}}, \bibinfo {author} {\bibfnamefont
  {L.}~\bibnamefont {{Jiang}}}, \bibinfo {author} {\bibfnamefont {I.~D.}\
  \bibnamefont {{McGreer}}}, \bibinfo {author} {\bibfnamefont {R.}~\bibnamefont
  {{Wang}}}, \bibinfo {author} {\bibfnamefont {J.}~\bibnamefont {{Yang}}},
  \bibinfo {author} {\bibfnamefont {Q.}~\bibnamefont {{Yang}}}, \bibinfo
  {author} {\bibfnamefont {D.}~\bibnamefont {{Thompson}}}, \ and\ \bibinfo
  {author} {\bibfnamefont {Y.}~\bibnamefont {{Beletsky}}},\ }\href
  {\doibase10.1038/nature14241} {\bibfield  {journal} {\bibinfo  {journal}
  {Nature}\ }\textbf {\bibinfo {volume} {518}},\ \bibinfo {pages} {512--515}
  (\bibinfo {year} {2015})},\ \Eprint {http://arxiv.org/abs/1502.07418}
  {arXiv:1502.07418}\BibitemShut {NoStop}%
\bibitem [{\citenamefont {{Nguyen}}\ \emph {et~al.}(2018)\citenamefont
  {{Nguyen}}, \citenamefont {{Seth}}, \citenamefont {{Neumayer}}, \citenamefont
  {{Kamann}}, \citenamefont {{Voggel}}, \citenamefont {{Cappellari}},
  \citenamefont {{Picotti}}, \citenamefont {{Nguyen}}, \citenamefont
  {{B{\"o}ker}}, \citenamefont {{Debattista}}, \citenamefont {{Caldwell}},
  \citenamefont {{McDermid}}, \citenamefont {{Bastian}}, \citenamefont
  {{Ahn}},\ and\ \citenamefont {{Pechetti}}}]{2018ApJ...858..118N}%
  \BibitemOpen
  \bibfield  {author} {\bibinfo {author} {\bibfnamefont {D.~D.}\ \bibnamefont
  {{Nguyen}}}, \bibinfo {author} {\bibfnamefont {A.~C.}\ \bibnamefont
  {{Seth}}}, \bibinfo {author} {\bibfnamefont {N.}~\bibnamefont {{Neumayer}}},
  \bibinfo {author} {\bibfnamefont {S.}~\bibnamefont {{Kamann}}}, \bibinfo
  {author} {\bibfnamefont {K.~T.}\ \bibnamefont {{Voggel}}}, \bibinfo {author}
  {\bibfnamefont {M.}~\bibnamefont {{Cappellari}}}, \bibinfo {author}
  {\bibfnamefont {A.}~\bibnamefont {{Picotti}}}, \bibinfo {author}
  {\bibfnamefont {P.~M.}\ \bibnamefont {{Nguyen}}}, \bibinfo {author}
  {\bibfnamefont {T.}~\bibnamefont {{B{\"o}ker}}}, \bibinfo {author}
  {\bibfnamefont {V.}~\bibnamefont {{Debattista}}}, \bibinfo {author}
  {\bibfnamefont {N.}~\bibnamefont {{Caldwell}}}, \bibinfo {author}
  {\bibfnamefont {R.}~\bibnamefont {{McDermid}}}, \bibinfo {author}
  {\bibfnamefont {N.}~\bibnamefont {{Bastian}}}, \bibinfo {author}
  {\bibfnamefont {C.~C.}\ \bibnamefont {{Ahn}}}, \ and\ \bibinfo {author}
  {\bibfnamefont {R.}~\bibnamefont {{Pechetti}}},\ }\href
  {\doibase10.3847/1538-4357/aabe28} {\bibfield  {journal} {\bibinfo  {journal}
  {\apj}\ }\textbf {\bibinfo {volume} {858}},\ \bibinfo {eid} {118} (\bibinfo
  {year} {2018})},\ \Eprint {http://arxiv.org/abs/1711.04314}
  {arXiv:1711.04314}\BibitemShut {NoStop}%
\bibitem [{\citenamefont {{van den Bosch}}(2016)}]{vandenbosch16}%
  \BibitemOpen
  \bibfield  {author} {\bibinfo {author} {\bibfnamefont {R.~C.~E.}\
  \bibnamefont {{van den Bosch}}},\ }\href
  {\doibase10.3847/0004-637X/831/2/134} {\bibfield  {journal} {\bibinfo
  {journal} {Astrophysical Journal}\ }\textbf {\bibinfo {volume} {831}},\
  \bibinfo {eid} {134} (\bibinfo {year} {2016})},\ \Eprint
  {http://arxiv.org/abs/1606.01246} {arXiv:1606.01246}\BibitemShut {NoStop}%
\bibitem [{\citenamefont {{Volonteri}}(2010)}]{Volonteri10}%
  \BibitemOpen
  \bibfield  {author} {\bibinfo {author} {\bibfnamefont {M.}~\bibnamefont
  {{Volonteri}}},\ }\href {\doibase10.1007/s00159-010-0029-x} {\bibfield
  {journal} {\bibinfo  {journal} {\aapr}\ }\textbf {\bibinfo {volume} {18}},\
  \bibinfo {pages} {279--315} (\bibinfo {year} {2010})},\ \Eprint
  {http://arxiv.org/abs/1003.4404} {arXiv:1003.4404}\BibitemShut {NoStop}%
\bibitem [{\citenamefont {{Jiang}}\ \emph {et~al.}(2016)\citenamefont
  {{Jiang}}, \citenamefont {{McGreer}}, \citenamefont {{Fan}}, \citenamefont
  {{Strauss}}, \citenamefont {{Ba{\~n}ados}}, \citenamefont {{Becker}},
  \citenamefont {{Bian}}, \citenamefont {{Farnsworth}}, \citenamefont {{Shen}},
  \citenamefont {{Wang}}, \citenamefont {{Wang}}, \citenamefont {{Wang}},
  \citenamefont {{White}}, \citenamefont {{Wu}}, \citenamefont {{Wu}},
  \citenamefont {{Yang}},\ and\ \citenamefont {{Yang}}}]{Jiang16}%
  \BibitemOpen
  \bibfield  {author} {\bibinfo {author} {\bibfnamefont {L.}~\bibnamefont
  {{Jiang}}}, \bibinfo {author} {\bibfnamefont {I.~D.}\ \bibnamefont
  {{McGreer}}}, \bibinfo {author} {\bibfnamefont {X.}~\bibnamefont {{Fan}}},
  \bibinfo {author} {\bibfnamefont {M.~A.}\ \bibnamefont {{Strauss}}}, \bibinfo
  {author} {\bibfnamefont {E.}~\bibnamefont {{Ba{\~n}ados}}}, \bibinfo {author}
  {\bibfnamefont {R.~H.}\ \bibnamefont {{Becker}}}, \bibinfo {author}
  {\bibfnamefont {F.}~\bibnamefont {{Bian}}}, \bibinfo {author} {\bibfnamefont
  {K.}~\bibnamefont {{Farnsworth}}}, \bibinfo {author} {\bibfnamefont
  {Y.}~\bibnamefont {{Shen}}}, \bibinfo {author} {\bibfnamefont
  {F.}~\bibnamefont {{Wang}}}, \bibinfo {author} {\bibfnamefont
  {R.}~\bibnamefont {{Wang}}}, \bibinfo {author} {\bibfnamefont
  {S.}~\bibnamefont {{Wang}}}, \bibinfo {author} {\bibfnamefont {R.~L.}\
  \bibnamefont {{White}}}, \bibinfo {author} {\bibfnamefont {J.}~\bibnamefont
  {{Wu}}}, \bibinfo {author} {\bibfnamefont {X.-B.}\ \bibnamefont {{Wu}}},
  \bibinfo {author} {\bibfnamefont {J.}~\bibnamefont {{Yang}}}, \ and\ \bibinfo
  {author} {\bibfnamefont {Q.}~\bibnamefont {{Yang}}},\ }\href
  {\doibase10.3847/1538-4357/833/2/222} {\bibfield  {journal} {\bibinfo
  {journal} {Astrophysical Journal}\ }\textbf {\bibinfo {volume} {833}},\
  \bibinfo {eid} {222} (\bibinfo {year} {2016})},\ \Eprint
  {http://arxiv.org/abs/1610.05369} {arXiv:1610.05369}\BibitemShut {NoStop}%
\bibitem [{\citenamefont {{Marconi}}\ \emph {et~al.}(2004)\citenamefont
  {{Marconi}}, \citenamefont {{Risaliti}}, \citenamefont {{Gilli}},
  \citenamefont {{Hunt}}, \citenamefont {{Maiolino}},\ and\ \citenamefont
  {{Salvati}}}]{Marconi04}%
  \BibitemOpen
  \bibfield  {author} {\bibinfo {author} {\bibfnamefont {A.}~\bibnamefont
  {{Marconi}}}, \bibinfo {author} {\bibfnamefont {G.}~\bibnamefont
  {{Risaliti}}}, \bibinfo {author} {\bibfnamefont {R.}~\bibnamefont {{Gilli}}},
  \bibinfo {author} {\bibfnamefont {L.~K.}\ \bibnamefont {{Hunt}}}, \bibinfo
  {author} {\bibfnamefont {R.}~\bibnamefont {{Maiolino}}}, \ and\ \bibinfo
  {author} {\bibfnamefont {M.}~\bibnamefont {{Salvati}}},\ }\href
  {\doibase10.1111/j.1365-2966.2004.07765.x} {\bibfield  {journal} {\bibinfo
  {journal} {MNRAS}\ }\textbf {\bibinfo {volume} {351}},\ \bibinfo {pages}
  {169--185} (\bibinfo {year} {2004})},\ \Eprint
  {http://arxiv.org/abs/astro-ph/0311619} {astro-ph/0311619}\BibitemShut
  {NoStop}%
\bibitem [{\citenamefont {{Schleicher}}\ \emph {et~al.}(2013)\citenamefont
  {{Schleicher}}, \citenamefont {{Palla}}, \citenamefont {{Ferrara}},
  \citenamefont {{Galli}},\ and\ \citenamefont {{Latif}}}]{Schleicher13}%
  \BibitemOpen
  \bibfield  {author} {\bibinfo {author} {\bibfnamefont {D.~R.~G.}\
  \bibnamefont {{Schleicher}}}, \bibinfo {author} {\bibfnamefont
  {F.}~\bibnamefont {{Palla}}}, \bibinfo {author} {\bibfnamefont
  {A.}~\bibnamefont {{Ferrara}}}, \bibinfo {author} {\bibfnamefont
  {D.}~\bibnamefont {{Galli}}}, \ and\ \bibinfo {author} {\bibfnamefont
  {M.}~\bibnamefont {{Latif}}},\ }\href {\doibase10.1051/0004-6361/201321949}
  {\bibfield  {journal} {\bibinfo  {journal} {\aap}\ }\textbf {\bibinfo
  {volume} {558}},\ \bibinfo {eid} {A59} (\bibinfo {year} {2013})},\ \Eprint
  {http://arxiv.org/abs/1305.5923} {arXiv:1305.5923}\BibitemShut {NoStop}%
\bibitem [{\citenamefont {Latif}\ and\ \citenamefont
  {Ferrara}(2016)}]{Latif:2016qau}%
  \BibitemOpen
  \bibfield  {author} {\bibinfo {author} {\bibfnamefont {Muhammad~A.}\
  \bibnamefont {Latif}}\ and\ \bibinfo {author} {\bibfnamefont {Andrea}\
  \bibnamefont {Ferrara}},\ }\href {\doibase10.1017/pasa.2016.41} {\bibfield
  {journal} {\bibinfo  {journal} {Publ. Astron. Soc. Austral.}\ }\textbf
  {\bibinfo {volume} {33}},\ \bibinfo {pages} {e051} (\bibinfo {year}
  {2016})},\ \Eprint {http://arxiv.org/abs/1605.07391} {arXiv:1605.07391
  [astro-ph.GA]}\BibitemShut {NoStop}%
\bibitem [{\citenamefont {{Mapelli}}(2016{\natexlab{a}})}]{Mapelli16}%
  \BibitemOpen
  \bibfield  {author} {\bibinfo {author} {\bibfnamefont {M.}~\bibnamefont
  {{Mapelli}}},\ }\href {\doibase10.1093/mnras/stw869} {\bibfield  {journal}
  {\bibinfo  {journal} {MNRAS}\ }\textbf {\bibinfo {volume} {459}},\ \bibinfo
  {pages} {3432--3446} (\bibinfo {year} {2016}{\natexlab{a}})},\ \Eprint
  {http://arxiv.org/abs/1604.03559} {arXiv:1604.03559}\BibitemShut {NoStop}%
\bibitem [{\citenamefont {{Devecchi}}\ \emph {et~al.}(2012)\citenamefont
  {{Devecchi}}, \citenamefont {{Volonteri}}, \citenamefont {{Rossi}},
  \citenamefont {{Colpi}},\ and\ \citenamefont {{Portegies
  Zwart}}}]{Devecchi12}%
  \BibitemOpen
  \bibfield  {author} {\bibinfo {author} {\bibfnamefont {B.}~\bibnamefont
  {{Devecchi}}}, \bibinfo {author} {\bibfnamefont {M.}~\bibnamefont
  {{Volonteri}}}, \bibinfo {author} {\bibfnamefont {E.~M.}\ \bibnamefont
  {{Rossi}}}, \bibinfo {author} {\bibfnamefont {M.}~\bibnamefont {{Colpi}}}, \
  and\ \bibinfo {author} {\bibfnamefont {S.}~\bibnamefont {{Portegies
  Zwart}}},\ }\href {\doibase10.1111/j.1365-2966.2012.20406.x} {\bibfield
  {journal} {\bibinfo  {journal} {MNRAS}\ }\textbf {\bibinfo {volume} {421}},\
  \bibinfo {pages} {1465--1475} (\bibinfo {year} {2012})},\ \Eprint
  {http://arxiv.org/abs/1201.3761} {arXiv:1201.3761}\BibitemShut {NoStop}%
\bibitem [{\citenamefont {{Lupi}}\ \emph {et~al.}(2014)\citenamefont {{Lupi}},
  \citenamefont {{Colpi}}, \citenamefont {{Devecchi}}, \citenamefont
  {{Galanti}},\ and\ \citenamefont {{Volonteri}}}]{Lupi14}%
  \BibitemOpen
  \bibfield  {author} {\bibinfo {author} {\bibfnamefont {A.}~\bibnamefont
  {{Lupi}}}, \bibinfo {author} {\bibfnamefont {M.}~\bibnamefont {{Colpi}}},
  \bibinfo {author} {\bibfnamefont {B.}~\bibnamefont {{Devecchi}}}, \bibinfo
  {author} {\bibfnamefont {G.}~\bibnamefont {{Galanti}}}, \ and\ \bibinfo
  {author} {\bibfnamefont {M.}~\bibnamefont {{Volonteri}}},\ }\href
  {\doibase10.1093/mnras/stu1120} {\bibfield  {journal} {\bibinfo  {journal}
  {MNRAS}\ }\textbf {\bibinfo {volume} {442}},\ \bibinfo {pages} {3616--3626}
  (\bibinfo {year} {2014})},\ \Eprint {http://arxiv.org/abs/1406.2325}
  {arXiv:1406.2325}\BibitemShut {NoStop}%
\bibitem [{\citenamefont {Gerosa}\ and\ \citenamefont
  {Berti}(2017)}]{Gerosa:2017kvu}%
  \BibitemOpen
  \bibfield  {author} {\bibinfo {author} {\bibfnamefont {Davide}\ \bibnamefont
  {Gerosa}}\ and\ \bibinfo {author} {\bibfnamefont {Emanuele}\ \bibnamefont
  {Berti}},\ }\href {\doibase10.1103/PhysRevD.95.124046} {\bibfield  {journal}
  {\bibinfo  {journal} {Phys. Rev.}\ }\textbf {\bibinfo {volume} {D95}},\
  \bibinfo {pages} {124046} (\bibinfo {year} {2017})},\ \Eprint
  {http://arxiv.org/abs/1703.06223} {arXiv:1703.06223 [gr-qc]}\BibitemShut
  {NoStop}%
\bibitem [{\citenamefont {{Graham}}\ and\ \citenamefont
  {{Spitler}}(2009)}]{Graham09}%
  \BibitemOpen
  \bibfield  {author} {\bibinfo {author} {\bibfnamefont {A.~W.}\ \bibnamefont
  {{Graham}}}\ and\ \bibinfo {author} {\bibfnamefont {L.~R.}\ \bibnamefont
  {{Spitler}}},\ }\href {\doibase10.1111/j.1365-2966.2009.15118.x} {\bibfield
  {journal} {\bibinfo  {journal} {MNRAS}\ }\textbf {\bibinfo {volume} {397}},\
  \bibinfo {pages} {2148--2162} (\bibinfo {year} {2009})},\ \Eprint
  {http://arxiv.org/abs/0907.5250} {arXiv:0907.5250}\BibitemShut {NoStop}%
\bibitem [{\citenamefont {{Lupi}}\ \emph {et~al.}(2016)\citenamefont {{Lupi}},
  \citenamefont {{Haardt}}, \citenamefont {{Dotti}}, \citenamefont
  {{Fiacconi}}, \citenamefont {{Mayer}},\ and\ \citenamefont
  {{Madau}}}]{Lupi16}%
  \BibitemOpen
  \bibfield  {author} {\bibinfo {author} {\bibfnamefont {A.}~\bibnamefont
  {{Lupi}}}, \bibinfo {author} {\bibfnamefont {F.}~\bibnamefont {{Haardt}}},
  \bibinfo {author} {\bibfnamefont {M.}~\bibnamefont {{Dotti}}}, \bibinfo
  {author} {\bibfnamefont {D.}~\bibnamefont {{Fiacconi}}}, \bibinfo {author}
  {\bibfnamefont {L.}~\bibnamefont {{Mayer}}}, \ and\ \bibinfo {author}
  {\bibfnamefont {P.}~\bibnamefont {{Madau}}},\ }\href
  {\doibase10.1093/mnras/stv2877} {\bibfield  {journal} {\bibinfo  {journal}
  {MNRAS}\ }\textbf {\bibinfo {volume} {456}},\ \bibinfo {pages} {2993--3003}
  (\bibinfo {year} {2016})},\ \Eprint {http://arxiv.org/abs/1512.02651}
  {arXiv:1512.02651}\BibitemShut {NoStop}%
\bibitem [{\citenamefont {{Latif}}\ \emph {et~al.}(2013)\citenamefont
  {{Latif}}, \citenamefont {{Schleicher}}, \citenamefont {{Schmidt}},\ and\
  \citenamefont {{Niemeyer}}}]{Latif13}%
  \BibitemOpen
  \bibfield  {author} {\bibinfo {author} {\bibfnamefont {M.~A.}\ \bibnamefont
  {{Latif}}}, \bibinfo {author} {\bibfnamefont {D.~R.~G.}\ \bibnamefont
  {{Schleicher}}}, \bibinfo {author} {\bibfnamefont {W.}~\bibnamefont
  {{Schmidt}}}, \ and\ \bibinfo {author} {\bibfnamefont {J.~C.}\ \bibnamefont
  {{Niemeyer}}},\ }\href {\doibase10.1093/mnras/stt1786} {\bibfield  {journal}
  {\bibinfo  {journal} {MNRAS}\ }\textbf {\bibinfo {volume} {436}},\ \bibinfo
  {pages} {2989--2996} (\bibinfo {year} {2013})},\ \Eprint
  {http://arxiv.org/abs/1309.1097} {arXiv:1309.1097}\BibitemShut {NoStop}%
\bibitem [{\citenamefont {{Stone}}\ \emph
  {et~al.}(2017{\natexlab{a}})\citenamefont {{Stone}}, \citenamefont
  {{K{\"u}pper}},\ and\ \citenamefont {{Ostriker}}}]{Stone17}%
  \BibitemOpen
  \bibfield  {author} {\bibinfo {author} {\bibfnamefont {N.~C.}\ \bibnamefont
  {{Stone}}}, \bibinfo {author} {\bibfnamefont {A.~H.~W.}\ \bibnamefont
  {{K{\"u}pper}}}, \ and\ \bibinfo {author} {\bibfnamefont {J.~P.}\
  \bibnamefont {{Ostriker}}},\ }\href {\doibase10.1093/mnras/stx097} {\bibfield
   {journal} {\bibinfo  {journal} {MNRAS}\ }\textbf {\bibinfo {volume} {467}},\
  \bibinfo {pages} {4180--4199} (\bibinfo {year} {2017}{\natexlab{a}})},\
  \Eprint {http://arxiv.org/abs/1606.01909} {arXiv:1606.01909}\BibitemShut
  {NoStop}%
\bibitem [{\citenamefont {{Dijkstra}}\ \emph {et~al.}(2014)\citenamefont
  {{Dijkstra}}, \citenamefont {{Ferrara}},\ and\ \citenamefont
  {{Mesinger}}}]{Dijkstra14}%
  \BibitemOpen
  \bibfield  {author} {\bibinfo {author} {\bibfnamefont {M.}~\bibnamefont
  {{Dijkstra}}}, \bibinfo {author} {\bibfnamefont {A.}~\bibnamefont
  {{Ferrara}}}, \ and\ \bibinfo {author} {\bibfnamefont {A.}~\bibnamefont
  {{Mesinger}}},\ }\href {\doibase10.1093/mnras/stu1007} {\bibfield  {journal}
  {\bibinfo  {journal} {MNRAS}\ }\textbf {\bibinfo {volume} {442}},\ \bibinfo
  {pages} {2036--2047} (\bibinfo {year} {2014})},\ \Eprint
  {http://arxiv.org/abs/1405.6743} {arXiv:1405.6743}\BibitemShut {NoStop}%
\bibitem [{\citenamefont {{Habouzit}}\ \emph {et~al.}(2016)\citenamefont
  {{Habouzit}}, \citenamefont {{Volonteri}}, \citenamefont {{Latif}},
  \citenamefont {{Dubois}},\ and\ \citenamefont {{Peirani}}}]{Habouzit16}%
  \BibitemOpen
  \bibfield  {author} {\bibinfo {author} {\bibfnamefont {M.}~\bibnamefont
  {{Habouzit}}}, \bibinfo {author} {\bibfnamefont {M.}~\bibnamefont
  {{Volonteri}}}, \bibinfo {author} {\bibfnamefont {M.}~\bibnamefont
  {{Latif}}}, \bibinfo {author} {\bibfnamefont {Y.}~\bibnamefont {{Dubois}}}, \
  and\ \bibinfo {author} {\bibfnamefont {S.}~\bibnamefont {{Peirani}}},\ }\href
  {\doibase10.1093/mnras/stw1924} {\bibfield  {journal} {\bibinfo  {journal}
  {MNRAS}\ }\textbf {\bibinfo {volume} {463}},\ \bibinfo {pages} {529--540}
  (\bibinfo {year} {2016})},\ \Eprint {http://arxiv.org/abs/1601.00557}
  {arXiv:1601.00557}\BibitemShut {NoStop}%
\bibitem [{\citenamefont {{Regan}}\ \emph {et~al.}(2017)\citenamefont
  {{Regan}}, \citenamefont {{Visbal}}, \citenamefont {{Wise}}, \citenamefont
  {{Haiman}}, \citenamefont {{Johansson}},\ and\ \citenamefont
  {{Bryan}}}]{Regan17}%
  \BibitemOpen
  \bibfield  {author} {\bibinfo {author} {\bibfnamefont {J.~A.}\ \bibnamefont
  {{Regan}}}, \bibinfo {author} {\bibfnamefont {E.}~\bibnamefont {{Visbal}}},
  \bibinfo {author} {\bibfnamefont {J.~H.}\ \bibnamefont {{Wise}}}, \bibinfo
  {author} {\bibfnamefont {Z.}~\bibnamefont {{Haiman}}}, \bibinfo {author}
  {\bibfnamefont {P.~H.}\ \bibnamefont {{Johansson}}}, \ and\ \bibinfo {author}
  {\bibfnamefont {G.~L.}\ \bibnamefont {{Bryan}}},\ }\href
  {\doibase10.1038/s41550-017-0075} {\bibfield  {journal} {\bibinfo  {journal}
  {Nature Astronomy}\ }\textbf {\bibinfo {volume} {1}},\ \bibinfo {eid} {0075}
  (\bibinfo {year} {2017})},\ \Eprint {http://arxiv.org/abs/1703.03805}
  {arXiv:1703.03805}\BibitemShut {NoStop}%
\bibitem [{\citenamefont {{Begelman}}(2010)}]{Begelman10}%
  \BibitemOpen
  \bibfield  {author} {\bibinfo {author} {\bibfnamefont {M.~C.}\ \bibnamefont
  {{Begelman}}},\ }\href {\doibase10.1111/j.1365-2966.2009.15916.x} {\bibfield
  {journal} {\bibinfo  {journal} {MNRAS}\ }\textbf {\bibinfo {volume} {402}},\
  \bibinfo {pages} {673--681} (\bibinfo {year} {2010})},\ \Eprint
  {http://arxiv.org/abs/0910.4398} {arXiv:0910.4398}\BibitemShut {NoStop}%
\bibitem [{\citenamefont {{Mayer}}\ \emph {et~al.}(2015)\citenamefont
  {{Mayer}}, \citenamefont {{Fiacconi}}, \citenamefont {{Bonoli}},
  \citenamefont {{Quinn}}, \citenamefont {{Ro{\v s}kar}}, \citenamefont
  {{Shen}},\ and\ \citenamefont {{Wadsley}}}]{Mayer15}%
  \BibitemOpen
  \bibfield  {author} {\bibinfo {author} {\bibfnamefont {L.}~\bibnamefont
  {{Mayer}}}, \bibinfo {author} {\bibfnamefont {D.}~\bibnamefont {{Fiacconi}}},
  \bibinfo {author} {\bibfnamefont {S.}~\bibnamefont {{Bonoli}}}, \bibinfo
  {author} {\bibfnamefont {T.}~\bibnamefont {{Quinn}}}, \bibinfo {author}
  {\bibfnamefont {R.}~\bibnamefont {{Ro{\v s}kar}}}, \bibinfo {author}
  {\bibfnamefont {S.}~\bibnamefont {{Shen}}}, \ and\ \bibinfo {author}
  {\bibfnamefont {J.}~\bibnamefont {{Wadsley}}},\ }\href
  {\doibase10.1088/0004-637X/810/1/51} {\bibfield  {journal} {\bibinfo
  {journal} {Astrophysical Journal}\ }\textbf {\bibinfo {volume} {810}},\
  \bibinfo {eid} {51} (\bibinfo {year} {2015})},\ \Eprint
  {http://arxiv.org/abs/1411.5683} {arXiv:1411.5683}\BibitemShut {NoStop}%
\bibitem [{\citenamefont {{Colpi}}\ and\ \citenamefont
  {{Sesana}}(2017)}]{Colpi17}%
  \BibitemOpen
  \bibfield  {author} {\bibinfo {author} {\bibfnamefont {M.}~\bibnamefont
  {{Colpi}}}\ and\ \bibinfo {author} {\bibfnamefont {A.}~\bibnamefont
  {{Sesana}}},\ }\enquote {\bibinfo {title} {{Gravitational Wave Sources in the
  Era of Multi-Band Gravitational Wave Astronomy}},}\ in\ \href
  {\doibase10.1142/9789813141766_0002} {\emph {\bibinfo {booktitle} {An
  Overview of Gravitational Waves: Theory, Sources and Detection}}},\ \bibinfo
  {editor} {edited by\ \bibinfo {editor} {\bibfnamefont {G.}~\bibnamefont
  {{Augar}}}\ and\ \bibinfo {editor} {\bibfnamefont {E.}~\bibnamefont
  {{Plagnol}}}}\ (\bibinfo  {publisher} {World Scientific Publishing Co},\
  \bibinfo {year} {2017})\ pp.\ \bibinfo {pages} {43--140}\BibitemShut
  {NoStop}%
\bibitem [{\citenamefont {Amaro-Seoane}\ \emph {et~al.}(2017)\citenamefont
  {Amaro-Seoane} \emph {et~al.}}]{Audley:2017drz}%
  \BibitemOpen
  \bibfield  {author} {\bibinfo {author} {\bibfnamefont {Pau}\ \bibnamefont
  {Amaro-Seoane}} \emph {et~al.} (\bibinfo {collaboration} {LISA}),\
  }\href@noop {} {\  (\bibinfo {year} {2017})},\ \Eprint
  {http://arxiv.org/abs/1702.00786} {arXiv:1702.00786
  [astro-ph.IM]}\BibitemShut {NoStop}%
\bibitem [{\citenamefont {{Colpi}}(2014)}]{Colpi14}%
  \BibitemOpen
  \bibfield  {author} {\bibinfo {author} {\bibfnamefont {M.}~\bibnamefont
  {{Colpi}}},\ }\href {\doibase10.1007/s11214-014-0067-1} {\bibfield  {journal}
  {\bibinfo  {journal} {\ssr}\ }\textbf {\bibinfo {volume} {183}},\ \bibinfo
  {pages} {189--221} (\bibinfo {year} {2014})},\ \Eprint
  {http://arxiv.org/abs/1407.3102} {arXiv:1407.3102}\BibitemShut {NoStop}%
\bibitem [{\citenamefont {Dominik}\ \emph {et~al.}(2013)\citenamefont
  {Dominik}, \citenamefont {Belczynski}, \citenamefont {Fryer}, \citenamefont
  {Holz}, \citenamefont {Berti}, \citenamefont {Bulik}, \citenamefont
  {Mandel},\ and\ \citenamefont {O'Shaughnessy}}]{Dominik:2013tma}%
  \BibitemOpen
  \bibfield  {author} {\bibinfo {author} {\bibfnamefont {Michal}\ \bibnamefont
  {Dominik}}, \bibinfo {author} {\bibfnamefont {Krzysztof}\ \bibnamefont
  {Belczynski}}, \bibinfo {author} {\bibfnamefont {Christopher}\ \bibnamefont
  {Fryer}}, \bibinfo {author} {\bibfnamefont {Daniel~E.}\ \bibnamefont {Holz}},
  \bibinfo {author} {\bibfnamefont {Emanuele}\ \bibnamefont {Berti}}, \bibinfo
  {author} {\bibfnamefont {Tomasz}\ \bibnamefont {Bulik}}, \bibinfo {author}
  {\bibfnamefont {Ilya}\ \bibnamefont {Mandel}}, \ and\ \bibinfo {author}
  {\bibfnamefont {Richard}\ \bibnamefont {O'Shaughnessy}},\ }\href
  {\doibase10.1088/0004-637X/779/1/72} {\bibfield  {journal} {\bibinfo
  {journal} {Astrophys. J.}\ }\textbf {\bibinfo {volume} {779}},\ \bibinfo
  {pages} {72} (\bibinfo {year} {2013})},\ \Eprint
  {http://arxiv.org/abs/1308.1546} {arXiv:1308.1546 [astro-ph.HE]}\BibitemShut
  {NoStop}%
\bibitem [{\citenamefont {{Portegies Zwart}}\ and\ \citenamefont
  {{McMillan}}(2000)}]{Portegies00}%
  \BibitemOpen
  \bibfield  {author} {\bibinfo {author} {\bibfnamefont {S.~F.}\ \bibnamefont
  {{Portegies Zwart}}}\ and\ \bibinfo {author} {\bibfnamefont {S.~L.~W.}\
  \bibnamefont {{McMillan}}},\ }\href {\doibase10.1086/312422} {\bibfield
  {journal} {\bibinfo  {journal} {Astrophysical Journal}\ }\textbf {\bibinfo
  {volume} {528}},\ \bibinfo {pages} {L17--L20} (\bibinfo {year} {2000})},\
  \Eprint {http://arxiv.org/abs/astro-ph/9910061}
  {astro-ph/9910061}\BibitemShut {NoStop}%
\bibitem [{\citenamefont {{Benacquista}}\ and\ \citenamefont
  {{Downing}}(2013)}]{Benacquista13}%
  \BibitemOpen
  \bibfield  {author} {\bibinfo {author} {\bibfnamefont {M.~J.}\ \bibnamefont
  {{Benacquista}}}\ and\ \bibinfo {author} {\bibfnamefont {J.~M.~B.}\
  \bibnamefont {{Downing}}},\ }\href {\doibase10.12942/lrr-2013-4} {\bibfield
  {journal} {\bibinfo  {journal} {Living Reviews in Relativity}\ }\textbf
  {\bibinfo {volume} {16}} (\bibinfo {year} {2013}),\ 10.12942/lrr-2013-4},\
  \Eprint {http://arxiv.org/abs/1110.4423} {arXiv:1110.4423
  [astro-ph.SR]}\BibitemShut {NoStop}%
\bibitem [{\citenamefont {Rodriguez}\ \emph {et~al.}(2016)\citenamefont
  {Rodriguez}, \citenamefont {Haster}, \citenamefont {Chatterjee},
  \citenamefont {Kalogera},\ and\ \citenamefont {Rasio}}]{Rodriguez:2016avt}%
  \BibitemOpen
  \bibfield  {author} {\bibinfo {author} {\bibfnamefont {Carl~L.}\ \bibnamefont
  {Rodriguez}}, \bibinfo {author} {\bibfnamefont {Carl-Johan}\ \bibnamefont
  {Haster}}, \bibinfo {author} {\bibfnamefont {Sourav}\ \bibnamefont
  {Chatterjee}}, \bibinfo {author} {\bibfnamefont {Vicky}\ \bibnamefont
  {Kalogera}}, \ and\ \bibinfo {author} {\bibfnamefont {Frederic~A.}\
  \bibnamefont {Rasio}},\ }\href {\doibase10.3847/2041-8205/824/1/L8}
  {\bibfield  {journal} {\bibinfo  {journal} {Astrophys. J.}\ }\textbf
  {\bibinfo {volume} {824}},\ \bibinfo {pages} {L8} (\bibinfo {year} {2016})},\
  \Eprint {http://arxiv.org/abs/1604.04254} {arXiv:1604.04254
  [astro-ph.HE]}\BibitemShut {NoStop}%
\bibitem [{\citenamefont {{Antonini}}\ and\ \citenamefont
  {{Rasio}}(2016)}]{Antonini16}%
  \BibitemOpen
  \bibfield  {author} {\bibinfo {author} {\bibfnamefont {F.}~\bibnamefont
  {{Antonini}}}\ and\ \bibinfo {author} {\bibfnamefont {F.~A.}\ \bibnamefont
  {{Rasio}}},\ }\href {\doibase10.3847/0004-637X/831/2/187} {\bibfield
  {journal} {\bibinfo  {journal} {Astrophysical Journal}\ }\textbf {\bibinfo
  {volume} {831}},\ \bibinfo {eid} {187} (\bibinfo {year} {2016})},\ \Eprint
  {http://arxiv.org/abs/1606.04889} {arXiv:1606.04889
  [astro-ph.HE]}\BibitemShut {NoStop}%
\bibitem [{\citenamefont {{Haster}}\ \emph {et~al.}(2016)\citenamefont
  {{Haster}}, \citenamefont {{Antonini}}, \citenamefont {{Kalogera}},\ and\
  \citenamefont {{Mandel}}}]{Haster16}%
  \BibitemOpen
  \bibfield  {author} {\bibinfo {author} {\bibfnamefont {C.-J.}\ \bibnamefont
  {{Haster}}}, \bibinfo {author} {\bibfnamefont {F.}~\bibnamefont
  {{Antonini}}}, \bibinfo {author} {\bibfnamefont {V.}~\bibnamefont
  {{Kalogera}}}, \ and\ \bibinfo {author} {\bibfnamefont {I.}~\bibnamefont
  {{Mandel}}},\ }\href {\doibase10.3847/0004-637X/832/2/192} {\bibfield
  {journal} {\bibinfo  {journal} {Astrophysical Journal}\ }\textbf {\bibinfo
  {volume} {832}},\ \bibinfo {eid} {192} (\bibinfo {year} {2016})},\ \Eprint
  {http://arxiv.org/abs/1606.07097} {arXiv:1606.07097
  [astro-ph.HE]}\BibitemShut {NoStop}%
\bibitem [{\citenamefont {{Reisswig}}\ \emph {et~al.}(2013)\citenamefont
  {{Reisswig}}, \citenamefont {{Ott}}, \citenamefont {{Abdikamalov}},
  \citenamefont {{Haas}}, \citenamefont {{M{\"o}sta}},\ and\ \citenamefont
  {{Schnetter}}}]{Reisswig13}%
  \BibitemOpen
  \bibfield  {author} {\bibinfo {author} {\bibfnamefont {C.}~\bibnamefont
  {{Reisswig}}}, \bibinfo {author} {\bibfnamefont {C.~D.}\ \bibnamefont
  {{Ott}}}, \bibinfo {author} {\bibfnamefont {E.}~\bibnamefont
  {{Abdikamalov}}}, \bibinfo {author} {\bibfnamefont {R.}~\bibnamefont
  {{Haas}}}, \bibinfo {author} {\bibfnamefont {P.}~\bibnamefont {{M{\"o}sta}}},
  \ and\ \bibinfo {author} {\bibfnamefont {E.}~\bibnamefont {{Schnetter}}},\
  }\href {\doibase10.1103/PhysRevLett.111.151101} {\bibfield  {journal}
  {\bibinfo  {journal} {Physical Review Letters}\ }\textbf {\bibinfo {volume}
  {111}},\ \bibinfo {eid} {151101} (\bibinfo {year} {2013})},\ \Eprint
  {http://arxiv.org/abs/1304.7787} {arXiv:1304.7787}\BibitemShut {NoStop}%
\bibitem [{\citenamefont {{Valiante}}\ \emph {et~al.}(2018)\citenamefont
  {{Valiante}}, \citenamefont {{Schneider}}, \citenamefont {{Graziani}},\ and\
  \citenamefont {{Zappacosta}}}]{Valiante17}%
  \BibitemOpen
  \bibfield  {author} {\bibinfo {author} {\bibfnamefont {R.}~\bibnamefont
  {{Valiante}}}, \bibinfo {author} {\bibfnamefont {R.}~\bibnamefont
  {{Schneider}}}, \bibinfo {author} {\bibfnamefont {L.}~\bibnamefont
  {{Graziani}}}, \ and\ \bibinfo {author} {\bibfnamefont {L.}~\bibnamefont
  {{Zappacosta}}},\ }\href {\doibase10.1093/mnras/stx3028} {\bibfield
  {journal} {\bibinfo  {journal} {\mnras}\ }\textbf {\bibinfo {volume} {474}},\
  \bibinfo {pages} {3825--3834} (\bibinfo {year} {2018})},\ \Eprint
  {http://arxiv.org/abs/1711.11033} {arXiv:1711.11033}\BibitemShut {NoStop}%
\bibitem [{\citenamefont {Garcia-Bellido}\ \emph {et~al.}(2018)\citenamefont
  {Garcia-Bellido}, \citenamefont {Clesse},\ and\ \citenamefont
  {Fleury}}]{Garcia-Bellido:2017imq}%
  \BibitemOpen
  \bibfield  {author} {\bibinfo {author} {\bibfnamefont {Juan}\ \bibnamefont
  {Garcia-Bellido}}, \bibinfo {author} {\bibfnamefont {Sebastien}\ \bibnamefont
  {Clesse}}, \ and\ \bibinfo {author} {\bibfnamefont {Pierre}\ \bibnamefont
  {Fleury}},\ }\href {\doibase10.1016/j.dark.2018.04.005} {\bibfield  {journal}
  {\bibinfo  {journal} {Phys. Dark Univ.}\ }\textbf {\bibinfo {volume} {20}},\
  \bibinfo {pages} {95--100} (\bibinfo {year} {2018})},\ \Eprint
  {http://arxiv.org/abs/1712.06574} {arXiv:1712.06574
  [astro-ph.CO]}\BibitemShut {NoStop}%
\bibitem [{\citenamefont {Belczynski}\ \emph
  {et~al.}(2017{\natexlab{a}})\citenamefont {Belczynski}, \citenamefont {Ryu},
  \citenamefont {Perna}, \citenamefont {Berti}, \citenamefont {Tanaka},\ and\
  \citenamefont {Bulik}}]{Belczynski:2016ieo}%
  \BibitemOpen
  \bibfield  {author} {\bibinfo {author} {\bibfnamefont {K.}~\bibnamefont
  {Belczynski}}, \bibinfo {author} {\bibfnamefont {T.}~\bibnamefont {Ryu}},
  \bibinfo {author} {\bibfnamefont {R.}~\bibnamefont {Perna}}, \bibinfo
  {author} {\bibfnamefont {E.}~\bibnamefont {Berti}}, \bibinfo {author}
  {\bibfnamefont {T.~L.}\ \bibnamefont {Tanaka}}, \ and\ \bibinfo {author}
  {\bibfnamefont {T.}~\bibnamefont {Bulik}},\ }\href
  {\doibase10.1093/mnras/stx1759} {\bibfield  {journal} {\bibinfo  {journal}
  {Mon. Not. Roy. Astron. Soc.}\ }\textbf {\bibinfo {volume} {471}},\ \bibinfo
  {pages} {4702--4721} (\bibinfo {year} {2017}{\natexlab{a}})},\ \Eprint
  {http://arxiv.org/abs/1612.01524} {arXiv:1612.01524
  [astro-ph.HE]}\BibitemShut {NoStop}%
\bibitem [{\citenamefont {{Yoon}}\ \emph {et~al.}(2010)\citenamefont {{Yoon}},
  \citenamefont {{Woosley}},\ and\ \citenamefont {{Langer}}}]{ywl10}%
  \BibitemOpen
  \bibfield  {author} {\bibinfo {author} {\bibfnamefont {S.-C.}\ \bibnamefont
  {{Yoon}}}, \bibinfo {author} {\bibfnamefont {S.~E.}\ \bibnamefont
  {{Woosley}}}, \ and\ \bibinfo {author} {\bibfnamefont {N.}~\bibnamefont
  {{Langer}}},\ }\href {\doibase10.1088/0004-637X/725/1/940} {\bibfield
  {journal} {\bibinfo  {journal} {The Astrophysical Journal}\ }\textbf
  {\bibinfo {volume} {725}},\ \bibinfo {pages} {940--954} (\bibinfo {year}
  {2010})},\ \Eprint {http://arxiv.org/abs/1004.0843} {arXiv:1004.0843
  [astro-ph.SR]}\BibitemShut {NoStop}%
\bibitem [{\citenamefont {{Paxton}}\ \emph {et~al.}(2015)\citenamefont
  {{Paxton}}, \citenamefont {{Marchant}}, \citenamefont {{Schwab}},
  \citenamefont {{Bauer}}, \citenamefont {{Bildsten}}, \citenamefont
  {{Cantiello}}, \citenamefont {{Dessart}}, \citenamefont {{Farmer}},
  \citenamefont {{Hu}}, \citenamefont {{Langer}}, \citenamefont {{Townsend}},
  \citenamefont {{Townsley}},\ and\ \citenamefont {{Timmes}}}]{Paxton2015}%
  \BibitemOpen
  \bibfield  {author} {\bibinfo {author} {\bibfnamefont {B.}~\bibnamefont
  {{Paxton}}}, \bibinfo {author} {\bibfnamefont {P.}~\bibnamefont
  {{Marchant}}}, \bibinfo {author} {\bibfnamefont {J.}~\bibnamefont
  {{Schwab}}}, \bibinfo {author} {\bibfnamefont {E.~B.}\ \bibnamefont
  {{Bauer}}}, \bibinfo {author} {\bibfnamefont {L.}~\bibnamefont {{Bildsten}}},
  \bibinfo {author} {\bibfnamefont {M.}~\bibnamefont {{Cantiello}}}, \bibinfo
  {author} {\bibfnamefont {L.}~\bibnamefont {{Dessart}}}, \bibinfo {author}
  {\bibfnamefont {R.}~\bibnamefont {{Farmer}}}, \bibinfo {author}
  {\bibfnamefont {H.}~\bibnamefont {{Hu}}}, \bibinfo {author} {\bibfnamefont
  {N.}~\bibnamefont {{Langer}}}, \bibinfo {author} {\bibfnamefont {R.~H.~D.}\
  \bibnamefont {{Townsend}}}, \bibinfo {author} {\bibfnamefont {D.~M.}\
  \bibnamefont {{Townsley}}}, \ and\ \bibinfo {author} {\bibfnamefont {F.~X.}\
  \bibnamefont {{Timmes}}},\ }\href {\doibase10.1088/0067-0049/220/1/15}
  {\bibfield  {journal} {\bibinfo  {journal} {The Astrophysical Journal
  Supplement}\ }\textbf {\bibinfo {volume} {220}},\ \bibinfo {eid} {15}
  (\bibinfo {year} {2015})},\ \Eprint {http://arxiv.org/abs/1506.03146}
  {arXiv:1506.03146 [astro-ph.SR]}\BibitemShut {NoStop}%
\bibitem [{\citenamefont {{Eggenberger}}\ \emph {et~al.}(2012)\citenamefont
  {{Eggenberger}}, \citenamefont {{Montalb{\'a}n}},\ and\ \citenamefont
  {{Miglio}}}]{Eggenberger2012}%
  \BibitemOpen
  \bibfield  {author} {\bibinfo {author} {\bibfnamefont {P.}~\bibnamefont
  {{Eggenberger}}}, \bibinfo {author} {\bibfnamefont {J.}~\bibnamefont
  {{Montalb{\'a}n}}}, \ and\ \bibinfo {author} {\bibfnamefont {A.}~\bibnamefont
  {{Miglio}}},\ }\href {\doibase10.1051/0004-6361/201219729} {\bibfield
  {journal} {\bibinfo  {journal} {\aap}\ }\textbf {\bibinfo {volume} {544}},\
  \bibinfo {eid} {L4} (\bibinfo {year} {2012})},\ \Eprint
  {http://arxiv.org/abs/1207.1023} {arXiv:1207.1023 [astro-ph.SR]}\BibitemShut
  {NoStop}%
\bibitem [{\citenamefont {{Georgy}}\ \emph {et~al.}(2013)\citenamefont
  {{Georgy}}, \citenamefont {{Ekstr{\"o}m}}, \citenamefont {{Eggenberger}},
  \citenamefont {{Meynet}}, \citenamefont {{Haemmerl{\'e}}}, \citenamefont
  {{Maeder}}, \citenamefont {{Granada}}, \citenamefont {{Groh}}, \citenamefont
  {{Hirschi}}, \citenamefont {{Mowlavi}}, \citenamefont {{Yusof}},
  \citenamefont {{Charbonnel}}, \citenamefont {{Decressin}},\ and\
  \citenamefont {{Barblan}}}]{Georgy2013}%
  \BibitemOpen
  \bibfield  {author} {\bibinfo {author} {\bibfnamefont {C.}~\bibnamefont
  {{Georgy}}}, \bibinfo {author} {\bibfnamefont {S.}~\bibnamefont
  {{Ekstr{\"o}m}}}, \bibinfo {author} {\bibfnamefont {P.}~\bibnamefont
  {{Eggenberger}}}, \bibinfo {author} {\bibfnamefont {G.}~\bibnamefont
  {{Meynet}}}, \bibinfo {author} {\bibfnamefont {L.}~\bibnamefont
  {{Haemmerl{\'e}}}}, \bibinfo {author} {\bibfnamefont {A.}~\bibnamefont
  {{Maeder}}}, \bibinfo {author} {\bibfnamefont {A.}~\bibnamefont {{Granada}}},
  \bibinfo {author} {\bibfnamefont {J.~H.}\ \bibnamefont {{Groh}}}, \bibinfo
  {author} {\bibfnamefont {R.}~\bibnamefont {{Hirschi}}}, \bibinfo {author}
  {\bibfnamefont {N.}~\bibnamefont {{Mowlavi}}}, \bibinfo {author}
  {\bibfnamefont {N.}~\bibnamefont {{Yusof}}}, \bibinfo {author} {\bibfnamefont
  {C.}~\bibnamefont {{Charbonnel}}}, \bibinfo {author} {\bibfnamefont
  {T.}~\bibnamefont {{Decressin}}}, \ and\ \bibinfo {author} {\bibfnamefont
  {F.}~\bibnamefont {{Barblan}}},\ }\href {\doibase10.1051/0004-6361/201322178}
  {\bibfield  {journal} {\bibinfo  {journal} {\aap}\ }\textbf {\bibinfo
  {volume} {558}},\ \bibinfo {eid} {A103} (\bibinfo {year} {2013})},\ \Eprint
  {http://arxiv.org/abs/1308.2914} {arXiv:1308.2914 [astro-ph.SR]}\BibitemShut
  {NoStop}%
\bibitem [{\citenamefont {{O'Connor}}\ and\ \citenamefont
  {{Ott}}(2011)}]{Oconnor2011}%
  \BibitemOpen
  \bibfield  {author} {\bibinfo {author} {\bibfnamefont {E.}~\bibnamefont
  {{O'Connor}}}\ and\ \bibinfo {author} {\bibfnamefont {C.~D.}\ \bibnamefont
  {{Ott}}},\ }\href {\doibase10.1088/0004-637X/730/2/70} {\bibfield  {journal}
  {\bibinfo  {journal} {The Astrophysical Journal}\ }\textbf {\bibinfo {volume}
  {730}},\ \bibinfo {eid} {70} (\bibinfo {year} {2011})},\ \Eprint
  {http://arxiv.org/abs/1010.5550} {arXiv:1010.5550 [astro-ph.HE]}\BibitemShut
  {NoStop}%
\bibitem [{\citenamefont {{Fryer}}\ \emph {et~al.}(2012)\citenamefont
  {{Fryer}}, \citenamefont {{Belczynski}}, \citenamefont {{Wiktorowicz}},
  \citenamefont {{Dominik}}, \citenamefont {{Kalogera}},\ and\ \citenamefont
  {{Holz}}}]{Fryer2012}%
  \BibitemOpen
  \bibfield  {author} {\bibinfo {author} {\bibfnamefont {C.~L.}\ \bibnamefont
  {{Fryer}}}, \bibinfo {author} {\bibfnamefont {K.}~\bibnamefont
  {{Belczynski}}}, \bibinfo {author} {\bibfnamefont {G.}~\bibnamefont
  {{Wiktorowicz}}}, \bibinfo {author} {\bibfnamefont {M.}~\bibnamefont
  {{Dominik}}}, \bibinfo {author} {\bibfnamefont {V.}~\bibnamefont
  {{Kalogera}}}, \ and\ \bibinfo {author} {\bibfnamefont {D.~E.}\ \bibnamefont
  {{Holz}}},\ }\href {\doibase10.1088/0004-637X/749/1/91} {\bibfield  {journal}
  {\bibinfo  {journal} {The Astrophysical Journal}\ }\textbf {\bibinfo {volume}
  {749}},\ \bibinfo {eid} {91} (\bibinfo {year} {2012})}\BibitemShut {NoStop}%
\bibitem [{\citenamefont {{Ertl}}\ \emph {et~al.}(2016)\citenamefont {{Ertl}},
  \citenamefont {{Janka}}, \citenamefont {{Woosley}}, \citenamefont
  {{Sukhbold}},\ and\ \citenamefont {{Ugliano}}}]{Ertl2016}%
  \BibitemOpen
  \bibfield  {author} {\bibinfo {author} {\bibfnamefont {T.}~\bibnamefont
  {{Ertl}}}, \bibinfo {author} {\bibfnamefont {H.-T.}\ \bibnamefont {{Janka}}},
  \bibinfo {author} {\bibfnamefont {S.~E.}\ \bibnamefont {{Woosley}}}, \bibinfo
  {author} {\bibfnamefont {T.}~\bibnamefont {{Sukhbold}}}, \ and\ \bibinfo
  {author} {\bibfnamefont {M.}~\bibnamefont {{Ugliano}}},\ }\href
  {\doibase10.3847/0004-637X/818/2/124} {\bibfield  {journal} {\bibinfo
  {journal} {The Astrophysical Journal}\ }\textbf {\bibinfo {volume} {818}},\
  \bibinfo {eid} {124} (\bibinfo {year} {2016})},\ \Eprint
  {http://arxiv.org/abs/1503.07522} {arXiv:1503.07522
  [astro-ph.SR]}\BibitemShut {NoStop}%
\bibitem [{\citenamefont {Alp}\ \emph {et~al.}(2018)\citenamefont {Alp} \emph
  {et~al.}}]{Alp:2018oek}%
  \BibitemOpen
  \bibfield  {author} {\bibinfo {author} {\bibfnamefont {Dennis}\ \bibnamefont
  {Alp}} \emph {et~al.},\ }\href@noop {} {\  (\bibinfo {year} {2018})},\
  \Eprint {http://arxiv.org/abs/1805.04526} {arXiv:1805.04526
  [astro-ph.HE]}\BibitemShut {NoStop}%
\bibitem [{\citenamefont {{Adams}}\ \emph {et~al.}(2017)\citenamefont
  {{Adams}}, \citenamefont {{Kochanek}}, \citenamefont {{Gerke}}, \citenamefont
  {{Stanek}},\ and\ \citenamefont {{Dai}}}]{Adams2017}%
  \BibitemOpen
  \bibfield  {author} {\bibinfo {author} {\bibfnamefont {S.~M.}\ \bibnamefont
  {{Adams}}}, \bibinfo {author} {\bibfnamefont {C.~S.}\ \bibnamefont
  {{Kochanek}}}, \bibinfo {author} {\bibfnamefont {J.~R.}\ \bibnamefont
  {{Gerke}}}, \bibinfo {author} {\bibfnamefont {K.~Z.}\ \bibnamefont
  {{Stanek}}}, \ and\ \bibinfo {author} {\bibfnamefont {X.}~\bibnamefont
  {{Dai}}},\ }\href {\doibase10.1093/mnras/stx816} {\bibfield  {journal}
  {\bibinfo  {journal} {Monthly Notices of The Royal Astronomical Society}\
  }\textbf {\bibinfo {volume} {468}},\ \bibinfo {pages} {4968--4981} (\bibinfo
  {year} {2017})},\ \Eprint {http://arxiv.org/abs/1609.01283} {arXiv:1609.01283
  [astro-ph.SR]}\BibitemShut {NoStop}%
\bibitem [{\citenamefont {{Bailyn}}\ \emph {et~al.}(1998)\citenamefont
  {{Bailyn}}, \citenamefont {{Jain}}, \citenamefont {{Coppi}},\ and\
  \citenamefont {{Orosz}}}]{Bailyn1998}%
  \BibitemOpen
  \bibfield  {author} {\bibinfo {author} {\bibfnamefont {C.~D.}\ \bibnamefont
  {{Bailyn}}}, \bibinfo {author} {\bibfnamefont {R.~K.}\ \bibnamefont
  {{Jain}}}, \bibinfo {author} {\bibfnamefont {P.}~\bibnamefont {{Coppi}}}, \
  and\ \bibinfo {author} {\bibfnamefont {J.~A.}\ \bibnamefont {{Orosz}}},\
  }\href {\doibase10.1086/305614} {\bibfield  {journal} {\bibinfo  {journal}
  {The Astrophysical Journal}\ }\textbf {\bibinfo {volume} {499}},\ \bibinfo
  {pages} {367--374} (\bibinfo {year} {1998})},\ \Eprint
  {http://arxiv.org/abs/astro-ph/9708032} {astro-ph/9708032}\BibitemShut
  {NoStop}%
\bibitem [{\citenamefont {{{\"O}zel}}\ \emph {et~al.}(2010)\citenamefont
  {{{\"O}zel}}, \citenamefont {{Psaltis}}, \citenamefont {{Narayan}},\ and\
  \citenamefont {{McClintock}}}]{Ozel2010}%
  \BibitemOpen
  \bibfield  {author} {\bibinfo {author} {\bibfnamefont {F.}~\bibnamefont
  {{{\"O}zel}}}, \bibinfo {author} {\bibfnamefont {D.}~\bibnamefont
  {{Psaltis}}}, \bibinfo {author} {\bibfnamefont {R.}~\bibnamefont
  {{Narayan}}}, \ and\ \bibinfo {author} {\bibfnamefont {J.~E.}\ \bibnamefont
  {{McClintock}}},\ }\href {\doibase10.1088/0004-637X/725/2/1918} {\bibfield
  {journal} {\bibinfo  {journal} {The Astrophysical Journal}\ }\textbf
  {\bibinfo {volume} {725}},\ \bibinfo {pages} {1918--1927} (\bibinfo {year}
  {2010})},\ \Eprint {http://arxiv.org/abs/1006.2834}
  {arXiv:1006.2834}\BibitemShut {NoStop}%
\bibitem [{\citenamefont {{Freire}}\ \emph {et~al.}(2008)\citenamefont
  {{Freire}}, \citenamefont {{Ransom}}, \citenamefont {{B{\'e}gin}},
  \citenamefont {{Stairs}}, \citenamefont {{Hessels}}, \citenamefont {{Frey}},\
  and\ \citenamefont {{Camilo}}}]{frb+08}%
  \BibitemOpen
  \bibfield  {author} {\bibinfo {author} {\bibfnamefont {P.~C.~C.}\
  \bibnamefont {{Freire}}}, \bibinfo {author} {\bibfnamefont {S.~M.}\
  \bibnamefont {{Ransom}}}, \bibinfo {author} {\bibfnamefont {S.}~\bibnamefont
  {{B{\'e}gin}}}, \bibinfo {author} {\bibfnamefont {I.~H.}\ \bibnamefont
  {{Stairs}}}, \bibinfo {author} {\bibfnamefont {J.~W.~T.}\ \bibnamefont
  {{Hessels}}}, \bibinfo {author} {\bibfnamefont {L.~H.}\ \bibnamefont
  {{Frey}}}, \ and\ \bibinfo {author} {\bibfnamefont {F.}~\bibnamefont
  {{Camilo}}},\ }\href {\doibase10.1086/526338} {\bibfield  {journal} {\bibinfo
   {journal} {The Astrophysical Journal}\ }\textbf {\bibinfo {volume} {675}},\
  \bibinfo {pages} {670--682} (\bibinfo {year} {2008})},\ \Eprint
  {http://arxiv.org/abs/0711.0925} {arXiv:0711.0925}\BibitemShut {NoStop}%
\bibitem [{\citenamefont {{van Kerkwijk}}\ \emph {et~al.}(2011)\citenamefont
  {{van Kerkwijk}}, \citenamefont {{Breton}},\ and\ \citenamefont
  {{Kulkarni}}}]{vbk11}%
  \BibitemOpen
  \bibfield  {author} {\bibinfo {author} {\bibfnamefont {M.~H.}\ \bibnamefont
  {{van Kerkwijk}}}, \bibinfo {author} {\bibfnamefont {R.~P.}\ \bibnamefont
  {{Breton}}}, \ and\ \bibinfo {author} {\bibfnamefont {S.~R.}\ \bibnamefont
  {{Kulkarni}}},\ }\href {\doibase10.1088/0004-637X/728/2/95} {\bibfield
  {journal} {\bibinfo  {journal} {The Astrophysical Journal}\ }\textbf
  {\bibinfo {volume} {728}},\ \bibinfo {eid} {95} (\bibinfo {year} {2011})},\
  \Eprint {http://arxiv.org/abs/1009.5427} {arXiv:1009.5427
  [astro-ph.HE]}\BibitemShut {NoStop}%
\bibitem [{\citenamefont {Linares}\ \emph {et~al.}(2018)\citenamefont
  {Linares}, \citenamefont {Shahbaz},\ and\ \citenamefont
  {Casares}}]{Linares:2018ppq}%
  \BibitemOpen
  \bibfield  {author} {\bibinfo {author} {\bibfnamefont {Manuel}\ \bibnamefont
  {Linares}}, \bibinfo {author} {\bibfnamefont {Tariq}\ \bibnamefont
  {Shahbaz}}, \ and\ \bibinfo {author} {\bibfnamefont {Jorge}\ \bibnamefont
  {Casares}},\ }\href {\doibase10.3847/1538-4357/aabde6} {\bibfield  {journal}
  {\bibinfo  {journal} {Astrophys. J.}\ }\textbf {\bibinfo {volume} {859}},\
  \bibinfo {pages} {54} (\bibinfo {year} {2018})},\ \Eprint
  {http://arxiv.org/abs/1805.08799} {arXiv:1805.08799
  [astro-ph.HE]}\BibitemShut {NoStop}%
\bibitem [{\citenamefont {{Kreidberg}}\ \emph {et~al.}(2012)\citenamefont
  {{Kreidberg}}, \citenamefont {{Bailyn}}, \citenamefont {{Farr}},\ and\
  \citenamefont {{Kalogera}}}]{Kreidberg2012}%
  \BibitemOpen
  \bibfield  {author} {\bibinfo {author} {\bibfnamefont {L.}~\bibnamefont
  {{Kreidberg}}}, \bibinfo {author} {\bibfnamefont {C.~D.}\ \bibnamefont
  {{Bailyn}}}, \bibinfo {author} {\bibfnamefont {W.~M.}\ \bibnamefont
  {{Farr}}}, \ and\ \bibinfo {author} {\bibfnamefont {V.}~\bibnamefont
  {{Kalogera}}},\ }\href {\doibase10.1088/0004-637X/757/1/36} {\bibfield
  {journal} {\bibinfo  {journal} {The Astrophysical Journal}\ }\textbf
  {\bibinfo {volume} {757}},\ \bibinfo {eid} {36} (\bibinfo {year} {2012})},\
  \Eprint {http://arxiv.org/abs/1205.1805} {arXiv:1205.1805
  [astro-ph.HE]}\BibitemShut {NoStop}%
\bibitem [{\citenamefont {Belczynski}\ \emph {et~al.}(2012)\citenamefont
  {Belczynski}, \citenamefont {Wiktorowicz}, \citenamefont {Fryer},
  \citenamefont {Holz},\ and\ \citenamefont {Kalogera}}]{Belczynski:2011bn}%
  \BibitemOpen
  \bibfield  {author} {\bibinfo {author} {\bibfnamefont {K.}~\bibnamefont
  {Belczynski}}, \bibinfo {author} {\bibfnamefont {G.}~\bibnamefont
  {Wiktorowicz}}, \bibinfo {author} {\bibfnamefont {C.}~\bibnamefont {Fryer}},
  \bibinfo {author} {\bibfnamefont {D.}~\bibnamefont {Holz}}, \ and\ \bibinfo
  {author} {\bibfnamefont {V.}~\bibnamefont {Kalogera}},\ }\href
  {\doibase10.1088/0004-637X/757/1/91} {\bibfield  {journal} {\bibinfo
  {journal} {Astrophys. J.}\ }\textbf {\bibinfo {volume} {757}},\ \bibinfo
  {pages} {91} (\bibinfo {year} {2012})},\ \Eprint
  {http://arxiv.org/abs/1110.1635} {arXiv:1110.1635 [astro-ph.GA]}\BibitemShut
  {NoStop}%
\bibitem [{\citenamefont {{Heger}}\ and\ \citenamefont
  {{Woosley}}(2002)}]{Heger2002}%
  \BibitemOpen
  \bibfield  {author} {\bibinfo {author} {\bibfnamefont {A.}~\bibnamefont
  {{Heger}}}\ and\ \bibinfo {author} {\bibfnamefont {S.~E.}\ \bibnamefont
  {{Woosley}}},\ }\href {\doibase10.1086/338487} {\bibfield  {journal}
  {\bibinfo  {journal} {The Astrophysical Journal}\ }\textbf {\bibinfo {volume}
  {567}},\ \bibinfo {pages} {532--543} (\bibinfo {year} {2002})},\ \Eprint
  {http://arxiv.org/abs/astro-ph/0107037} {astro-ph/0107037}\BibitemShut
  {NoStop}%
\bibitem [{\citenamefont {{Woosley}}(2016)}]{Woosley2016}%
  \BibitemOpen
  \bibfield  {author} {\bibinfo {author} {\bibfnamefont {S.~E.}\ \bibnamefont
  {{Woosley}}},\ }\href {\doibase10.3847/2041-8205/824/1/L10} {\bibfield
  {journal} {\bibinfo  {journal} {The Astrophysical Journal Letters}\ }\textbf
  {\bibinfo {volume} {824}},\ \bibinfo {eid} {L10} (\bibinfo {year} {2016})},\
  \Eprint {http://arxiv.org/abs/1603.00511} {arXiv:1603.00511
  [astro-ph.HE]}\BibitemShut {NoStop}%
\bibitem [{\citenamefont {{Asplund}}\ \emph {et~al.}(2009)\citenamefont
  {{Asplund}}, \citenamefont {{Grevesse}}, \citenamefont {{Sauval}},\ and\
  \citenamefont {{Scott}}}]{Asplund2009}%
  \BibitemOpen
  \bibfield  {author} {\bibinfo {author} {\bibfnamefont {M.}~\bibnamefont
  {{Asplund}}}, \bibinfo {author} {\bibfnamefont {N.}~\bibnamefont
  {{Grevesse}}}, \bibinfo {author} {\bibfnamefont {A.~J.}\ \bibnamefont
  {{Sauval}}}, \ and\ \bibinfo {author} {\bibfnamefont {P.}~\bibnamefont
  {{Scott}}},\ }\href {\doibase10.1146/annurev.astro.46.060407.145222}
  {\bibfield  {journal} {\bibinfo  {journal} {\araa}\ }\textbf {\bibinfo
  {volume} {47}},\ \bibinfo {pages} {481--522} (\bibinfo {year} {2009})},\
  \Eprint {http://arxiv.org/abs/0909.0948} {arXiv:0909.0948
  [astro-ph.SR]}\BibitemShut {NoStop}%
\bibitem [{\citenamefont {{Villante}}\ \emph {et~al.}(2014)\citenamefont
  {{Villante}}, \citenamefont {{Serenelli}}, \citenamefont {{Delahaye}},\ and\
  \citenamefont {{Pinsonneault}}}]{Villante2014}%
  \BibitemOpen
  \bibfield  {author} {\bibinfo {author} {\bibfnamefont {F.~L.}\ \bibnamefont
  {{Villante}}}, \bibinfo {author} {\bibfnamefont {A.~M.}\ \bibnamefont
  {{Serenelli}}}, \bibinfo {author} {\bibfnamefont {F.}~\bibnamefont
  {{Delahaye}}}, \ and\ \bibinfo {author} {\bibfnamefont {M.~H.}\ \bibnamefont
  {{Pinsonneault}}},\ }\href {\doibase10.1088/0004-637X/787/1/13} {\bibfield
  {journal} {\bibinfo  {journal} {The Astrophysical Journal}\ }\textbf
  {\bibinfo {volume} {787}},\ \bibinfo {eid} {13} (\bibinfo {year}
  {2014})}\BibitemShut {NoStop}%
\bibitem [{\citenamefont {Belczynski}\ \emph
  {et~al.}(2010{\natexlab{b}})\citenamefont {Belczynski}, \citenamefont
  {Bulik}, \citenamefont {Fryer}, \citenamefont {Ruiter}, \citenamefont
  {Vink},\ and\ \citenamefont {Hurley}}]{Belczynski:2009xy}%
  \BibitemOpen
  \bibfield  {author} {\bibinfo {author} {\bibfnamefont {Krzysztof}\
  \bibnamefont {Belczynski}}, \bibinfo {author} {\bibfnamefont {Tomasz}\
  \bibnamefont {Bulik}}, \bibinfo {author} {\bibfnamefont {Chris~L.}\
  \bibnamefont {Fryer}}, \bibinfo {author} {\bibfnamefont {Ashley}\
  \bibnamefont {Ruiter}}, \bibinfo {author} {\bibfnamefont {Jorick~S.}\
  \bibnamefont {Vink}}, \ and\ \bibinfo {author} {\bibfnamefont {Jarrod~R.}\
  \bibnamefont {Hurley}},\ }\href {\doibase10.1088/0004-637X/714/2/1217}
  {\bibfield  {journal} {\bibinfo  {journal} {Astrophys. J.}\ }\textbf
  {\bibinfo {volume} {714}},\ \bibinfo {pages} {1217--1226} (\bibinfo {year}
  {2010}{\natexlab{b}})},\ \Eprint {http://arxiv.org/abs/0904.2784}
  {arXiv:0904.2784 [astro-ph.SR]}\BibitemShut {NoStop}%
\bibitem [{\citenamefont {Kruckow}\ \emph {et~al.}(2018)\citenamefont
  {Kruckow}, \citenamefont {Tauris}, \citenamefont {Langer}, \citenamefont
  {Kramer},\ and\ \citenamefont {Izzard}}]{Kruckow:2018slo}%
  \BibitemOpen
  \bibfield  {author} {\bibinfo {author} {\bibfnamefont {Matthias~U.}\
  \bibnamefont {Kruckow}}, \bibinfo {author} {\bibfnamefont {Thomas~M.}\
  \bibnamefont {Tauris}}, \bibinfo {author} {\bibfnamefont {Norbert}\
  \bibnamefont {Langer}}, \bibinfo {author} {\bibfnamefont {Michael}\
  \bibnamefont {Kramer}}, \ and\ \bibinfo {author} {\bibfnamefont {Robert~G.}\
  \bibnamefont {Izzard}},\ }\href@noop {} {\  (\bibinfo {year} {2018})},\
  \Eprint {http://arxiv.org/abs/1801.05433} {arXiv:1801.05433
  [astro-ph.SR]}\BibitemShut {NoStop}%
\bibitem [{\citenamefont {{Bond}}\ \emph {et~al.}(1984)\citenamefont {{Bond}},
  \citenamefont {{Arnett}},\ and\ \citenamefont {{Carr}}}]{Bond1984a}%
  \BibitemOpen
  \bibfield  {author} {\bibinfo {author} {\bibfnamefont {J.~R.}\ \bibnamefont
  {{Bond}}}, \bibinfo {author} {\bibfnamefont {W.~D.}\ \bibnamefont
  {{Arnett}}}, \ and\ \bibinfo {author} {\bibfnamefont {B.~J.}\ \bibnamefont
  {{Carr}}},\ }\href {\doibase10.1086/162057} {\bibfield  {journal} {\bibinfo
  {journal} {The Astrophysical Journal}\ }\textbf {\bibinfo {volume} {280}},\
  \bibinfo {pages} {825--847} (\bibinfo {year} {1984})}\BibitemShut {NoStop}%
\bibitem [{\citenamefont {{Fryer}}\ and\ \citenamefont
  {{Kalogera}}(2001)}]{Fryer2001}%
  \BibitemOpen
  \bibfield  {author} {\bibinfo {author} {\bibfnamefont {C.~L.}\ \bibnamefont
  {{Fryer}}}\ and\ \bibinfo {author} {\bibfnamefont {V.}~\bibnamefont
  {{Kalogera}}},\ }\href {\doibase10.1086/321359} {\bibfield  {journal}
  {\bibinfo  {journal} {The Astrophysical Journal}\ }\textbf {\bibinfo {volume}
  {554}},\ \bibinfo {pages} {548--560} (\bibinfo {year} {2001})},\ \Eprint
  {http://arxiv.org/abs/astro-ph/9911312} {astro-ph/9911312}\BibitemShut
  {NoStop}%
\bibitem [{\citenamefont {{Heger}}\ \emph {et~al.}(2003)\citenamefont
  {{Heger}}, \citenamefont {{Fryer}}, \citenamefont {{Woosley}}, \citenamefont
  {{Langer}},\ and\ \citenamefont {{Hartmann}}}]{Heger2003}%
  \BibitemOpen
  \bibfield  {author} {\bibinfo {author} {\bibfnamefont {A.}~\bibnamefont
  {{Heger}}}, \bibinfo {author} {\bibfnamefont {C.~L.}\ \bibnamefont
  {{Fryer}}}, \bibinfo {author} {\bibfnamefont {S.~E.}\ \bibnamefont
  {{Woosley}}}, \bibinfo {author} {\bibfnamefont {N.}~\bibnamefont {{Langer}}},
  \ and\ \bibinfo {author} {\bibfnamefont {D.~H.}\ \bibnamefont {{Hartmann}}},\
  }\href {\doibase10.1086/375341} {\bibfield  {journal} {\bibinfo  {journal}
  {The Astrophysical Journal}\ }\textbf {\bibinfo {volume} {591}},\ \bibinfo
  {pages} {288--300} (\bibinfo {year} {2003})},\ \Eprint
  {http://arxiv.org/abs/astro-ph/0212469} {astro-ph/0212469}\BibitemShut
  {NoStop}%
\bibitem [{\citenamefont {{Yusof}}\ \emph {et~al.}(2013)\citenamefont
  {{Yusof}}, \citenamefont {{Hirschi}}, \citenamefont {{Meynet}}, \citenamefont
  {{Crowther}}, \citenamefont {{Ekstr{\"o}m}}, \citenamefont {{Frischknecht}},
  \citenamefont {{Georgy}}, \citenamefont {{Abu Kassim}},\ and\ \citenamefont
  {{Schnurr}}}]{Yusof2013}%
  \BibitemOpen
  \bibfield  {author} {\bibinfo {author} {\bibfnamefont {N.}~\bibnamefont
  {{Yusof}}}, \bibinfo {author} {\bibfnamefont {R.}~\bibnamefont {{Hirschi}}},
  \bibinfo {author} {\bibfnamefont {G.}~\bibnamefont {{Meynet}}}, \bibinfo
  {author} {\bibfnamefont {P.~A.}\ \bibnamefont {{Crowther}}}, \bibinfo
  {author} {\bibfnamefont {S.}~\bibnamefont {{Ekstr{\"o}m}}}, \bibinfo {author}
  {\bibfnamefont {U.}~\bibnamefont {{Frischknecht}}}, \bibinfo {author}
  {\bibfnamefont {C.}~\bibnamefont {{Georgy}}}, \bibinfo {author}
  {\bibfnamefont {H.}~\bibnamefont {{Abu Kassim}}}, \ and\ \bibinfo {author}
  {\bibfnamefont {O.}~\bibnamefont {{Schnurr}}},\ }\href
  {\doibase10.1093/mnras/stt794} {\bibfield  {journal} {\bibinfo  {journal}
  {Monthly Notices of The Royal Astronomical Society}\ }\textbf {\bibinfo
  {volume} {433}},\ \bibinfo {pages} {1114--1132} (\bibinfo {year}
  {2013})}\BibitemShut {NoStop}%
\bibitem [{\citenamefont {{Spera}}\ \emph
  {et~al.}(2015{\natexlab{b}})\citenamefont {{Spera}}, \citenamefont
  {{Mapelli}},\ and\ \citenamefont {{Bressan}}}]{Spera2015}%
  \BibitemOpen
  \bibfield  {author} {\bibinfo {author} {\bibfnamefont {M.}~\bibnamefont
  {{Spera}}}, \bibinfo {author} {\bibfnamefont {M.}~\bibnamefont {{Mapelli}}},
  \ and\ \bibinfo {author} {\bibfnamefont {A.}~\bibnamefont {{Bressan}}},\
  }\href {\doibase10.1093/mnras/stv1161} {\bibfield  {journal} {\bibinfo
  {journal} {Monthly Notices of The Royal Astronomical Society}\ }\textbf
  {\bibinfo {volume} {451}},\ \bibinfo {pages} {4086--4103} (\bibinfo {year}
  {2015}{\natexlab{b}})}\BibitemShut {NoStop}%
\bibitem [{\citenamefont {{Marchant}}\ \emph {et~al.}(2016)\citenamefont
  {{Marchant}}, \citenamefont {{Langer}}, \citenamefont {{Podsiadlowski}},
  \citenamefont {{Tauris}},\ and\ \citenamefont {{Moriya}}}]{Marchant2016}%
  \BibitemOpen
  \bibfield  {author} {\bibinfo {author} {\bibfnamefont {P.}~\bibnamefont
  {{Marchant}}}, \bibinfo {author} {\bibfnamefont {N.}~\bibnamefont
  {{Langer}}}, \bibinfo {author} {\bibfnamefont {P.}~\bibnamefont
  {{Podsiadlowski}}}, \bibinfo {author} {\bibfnamefont {T.~M.}\ \bibnamefont
  {{Tauris}}}, \ and\ \bibinfo {author} {\bibfnamefont {T.~J.}\ \bibnamefont
  {{Moriya}}},\ }\href {\doibase10.1051/0004-6361/201628133} {\bibfield
  {journal} {\bibinfo  {journal} {\aap}\ }\textbf {\bibinfo {volume} {588}},\
  \bibinfo {eid} {A50} (\bibinfo {year} {2016})},\ \Eprint
  {http://arxiv.org/abs/1601.03718} {arXiv:1601.03718
  [astro-ph.SR]}\BibitemShut {NoStop}%
\bibitem [{\citenamefont {Belczynski}\ \emph {et~al.}(2014)\citenamefont
  {Belczynski}, \citenamefont {Buonanno}, \citenamefont {Cantiello},
  \citenamefont {Fryer}, \citenamefont {Holz}, \citenamefont {Mandel},
  \citenamefont {Miller},\ and\ \citenamefont {Walczak}}]{Belczynski:2014iua}%
  \BibitemOpen
  \bibfield  {author} {\bibinfo {author} {\bibfnamefont {Krzysztof}\
  \bibnamefont {Belczynski}}, \bibinfo {author} {\bibfnamefont {Alessandra}\
  \bibnamefont {Buonanno}}, \bibinfo {author} {\bibfnamefont {Matteo}\
  \bibnamefont {Cantiello}}, \bibinfo {author} {\bibfnamefont {Chris~L.}\
  \bibnamefont {Fryer}}, \bibinfo {author} {\bibfnamefont {Daniel~E.}\
  \bibnamefont {Holz}}, \bibinfo {author} {\bibfnamefont {Ilya}\ \bibnamefont
  {Mandel}}, \bibinfo {author} {\bibfnamefont {M.~Coleman}\ \bibnamefont
  {Miller}}, \ and\ \bibinfo {author} {\bibfnamefont {Marek}\ \bibnamefont
  {Walczak}},\ }\href {\doibase10.1088/0004-637X/789/2/120} {\bibfield
  {journal} {\bibinfo  {journal} {Astrophys. J.}\ }\textbf {\bibinfo {volume}
  {789}},\ \bibinfo {pages} {120} (\bibinfo {year} {2014})},\ \Eprint
  {http://arxiv.org/abs/1403.0677} {arXiv:1403.0677 [astro-ph.HE]}\BibitemShut
  {NoStop}%
\bibitem [{\citenamefont {{Massey}}(2011)}]{Massey2011}%
  \BibitemOpen
  \bibfield  {author} {\bibinfo {author} {\bibfnamefont {P.}~\bibnamefont
  {{Massey}}},\ }in\ \href@noop {} {\emph {\bibinfo {booktitle} {UP2010: Have
  Observations Revealed a Variable Upper End of the Initial Mass Function?}}},\
  \bibinfo {series} {Astronomical Society of the Pacific Conference Series},
  Vol.\ \bibinfo {volume} {440},\ \bibinfo {editor} {edited by\ \bibinfo
  {editor} {\bibfnamefont {M.}~\bibnamefont {{Treyer}}}, \bibinfo {editor}
  {\bibfnamefont {T.}~\bibnamefont {{Wyder}}}, \bibinfo {editor} {\bibfnamefont
  {J.}~\bibnamefont {{Neill}}}, \bibinfo {editor} {\bibfnamefont
  {M.}~\bibnamefont {{Seibert}}}, \ and\ \bibinfo {editor} {\bibfnamefont
  {J.}~\bibnamefont {{Lee}}}}\ (\bibinfo {year} {2011})\ p.~\bibinfo {pages}
  {29},\ \Eprint {http://arxiv.org/abs/1008.1014} {arXiv:1008.1014
  [astro-ph.SR]}\BibitemShut {NoStop}%
\bibitem [{\citenamefont {{Crowther}}\ \emph {et~al.}(2012)\citenamefont
  {{Crowther}}, \citenamefont {{Hirschi}}, \citenamefont {{Walborn}},\ and\
  \citenamefont {{Yusof}}}]{Crowther2012}%
  \BibitemOpen
  \bibfield  {author} {\bibinfo {author} {\bibfnamefont {P.~A.}\ \bibnamefont
  {{Crowther}}}, \bibinfo {author} {\bibfnamefont {R.}~\bibnamefont
  {{Hirschi}}}, \bibinfo {author} {\bibfnamefont {N.~R.}\ \bibnamefont
  {{Walborn}}}, \ and\ \bibinfo {author} {\bibfnamefont {N.}~\bibnamefont
  {{Yusof}}},\ }in\ \href@noop {} {\emph {\bibinfo {booktitle} {Proceedings of
  a Scientific Meeting in Honor of Anthony F. J. Moffat}}},\ \bibinfo {series}
  {Astronomical Society of the Pacific Conference Series}, Vol.\ \bibinfo
  {volume} {465},\ \bibinfo {editor} {edited by\ \bibinfo {editor}
  {\bibfnamefont {L.}~\bibnamefont {{Drissen}}}, \bibinfo {editor}
  {\bibfnamefont {C.}~\bibnamefont {{Robert}}}, \bibinfo {editor}
  {\bibfnamefont {N.}~\bibnamefont {{St-Louis}}}, \ and\ \bibinfo {editor}
  {\bibfnamefont {A.~F.~J.}\ \bibnamefont {{Moffat}}}}\ (\bibinfo {year}
  {2012})\ p.\ \bibinfo {pages} {196},\ \Eprint
  {http://arxiv.org/abs/1209.6157} {arXiv:1209.6157 [astro-ph.SR]}\BibitemShut
  {NoStop}%
\bibitem [{\citenamefont {{Crowther}}\ \emph {et~al.}(2010)\citenamefont
  {{Crowther}}, \citenamefont {{Schnurr}}, \citenamefont {{Hirschi}},
  \citenamefont {{Yusof}}, \citenamefont {{Parker}}, \citenamefont
  {{Goodwin}},\ and\ \citenamefont {{Kassim}}}]{Crowther2010}%
  \BibitemOpen
  \bibfield  {author} {\bibinfo {author} {\bibfnamefont {P.~A.}\ \bibnamefont
  {{Crowther}}}, \bibinfo {author} {\bibfnamefont {O.}~\bibnamefont
  {{Schnurr}}}, \bibinfo {author} {\bibfnamefont {R.}~\bibnamefont
  {{Hirschi}}}, \bibinfo {author} {\bibfnamefont {N.}~\bibnamefont {{Yusof}}},
  \bibinfo {author} {\bibfnamefont {R.~J.}\ \bibnamefont {{Parker}}}, \bibinfo
  {author} {\bibfnamefont {S.~P.}\ \bibnamefont {{Goodwin}}}, \ and\ \bibinfo
  {author} {\bibfnamefont {H.~A.}\ \bibnamefont {{Kassim}}},\ }\href
  {\doibase10.1111/j.1365-2966.2010.17167.x} {\bibfield  {journal} {\bibinfo
  {journal} {Monthly Notices of The Royal Astronomical Society}\ }\textbf
  {\bibinfo {volume} {408}},\ \bibinfo {pages} {731--751} (\bibinfo {year}
  {2010})},\ \Eprint {http://arxiv.org/abs/1007.3284} {arXiv:1007.3284
  [astro-ph.SR]}\BibitemShut {NoStop}%
\bibitem [{\citenamefont {{Hobbs}}\ \emph {et~al.}(2005)\citenamefont
  {{Hobbs}}, \citenamefont {{Lorimer}}, \citenamefont {{Lyne}},\ and\
  \citenamefont {{Kramer}}}]{Hobbs2005}%
  \BibitemOpen
  \bibfield  {author} {\bibinfo {author} {\bibfnamefont {G.}~\bibnamefont
  {{Hobbs}}}, \bibinfo {author} {\bibfnamefont {D.~R.}\ \bibnamefont
  {{Lorimer}}}, \bibinfo {author} {\bibfnamefont {A.~G.}\ \bibnamefont
  {{Lyne}}}, \ and\ \bibinfo {author} {\bibfnamefont {M.}~\bibnamefont
  {{Kramer}}},\ }\href {\doibase10.1111/j.1365-2966.2005.09087.x} {\bibfield
  {journal} {\bibinfo  {journal} {Monthly Notices of The Royal Astronomical
  Society}\ }\textbf {\bibinfo {volume} {360}},\ \bibinfo {pages} {974--992}
  (\bibinfo {year} {2005})}\BibitemShut {NoStop}%
\bibitem [{\citenamefont {{Janka}}\ and\ \citenamefont
  {{Mueller}}(1994)}]{Janka1994}%
  \BibitemOpen
  \bibfield  {author} {\bibinfo {author} {\bibfnamefont {H.-T.}\ \bibnamefont
  {{Janka}}}\ and\ \bibinfo {author} {\bibfnamefont {E.}~\bibnamefont
  {{Mueller}}},\ }\href@noop {} {\bibfield  {journal} {\bibinfo  {journal}
  {\aap}\ }\textbf {\bibinfo {volume} {290}},\ \bibinfo {pages} {496--502}
  (\bibinfo {year} {1994})}\BibitemShut {NoStop}%
\bibitem [{\citenamefont {Tamborra}\ \emph {et~al.}(2014)\citenamefont
  {Tamborra}, \citenamefont {Hanke}, \citenamefont {Janka}, \citenamefont
  {Müller}, \citenamefont {Raffelt},\ and\ \citenamefont
  {Marek}}]{Tamborra2014}%
  \BibitemOpen
  \bibfield  {author} {\bibinfo {author} {\bibfnamefont {Irene}\ \bibnamefont
  {Tamborra}}, \bibinfo {author} {\bibfnamefont {Florian}\ \bibnamefont
  {Hanke}}, \bibinfo {author} {\bibfnamefont {Hans-Thomas}\ \bibnamefont
  {Janka}}, \bibinfo {author} {\bibfnamefont {Bernhard}\ \bibnamefont
  {Müller}}, \bibinfo {author} {\bibfnamefont {Georg~G.}\ \bibnamefont
  {Raffelt}}, \ and\ \bibinfo {author} {\bibfnamefont {Andreas}\ \bibnamefont
  {Marek}},\ }\href {\doibase10.1088/0004-637X/792/2/96} {\bibfield  {journal}
  {\bibinfo  {journal} {Astrophys. J.}\ }\textbf {\bibinfo {volume} {792}},\
  \bibinfo {pages} {96} (\bibinfo {year} {2014})},\ \Eprint
  {http://arxiv.org/abs/1402.5418} {arXiv:1402.5418 [astro-ph.SR]}\BibitemShut
  {NoStop}%
\bibitem [{\citenamefont {{Janka}}(2017)}]{jan17}%
  \BibitemOpen
  \bibfield  {author} {\bibinfo {author} {\bibfnamefont {H.-T.}\ \bibnamefont
  {{Janka}}},\ }\href {\doibase10.3847/1538-4357/aa618e} {\bibfield  {journal}
  {\bibinfo  {journal} {The Astrophysical Journal}\ }\textbf {\bibinfo {volume}
  {837}},\ \bibinfo {eid} {84} (\bibinfo {year} {2017})},\ \Eprint
  {http://arxiv.org/abs/1611.07562} {arXiv:1611.07562
  [astro-ph.HE]}\BibitemShut {NoStop}%
\bibitem [{\citenamefont {{Kusenko}}\ and\ \citenamefont
  {{Segr{\`e}}}(1996)}]{Kusenko1996}%
  \BibitemOpen
  \bibfield  {author} {\bibinfo {author} {\bibfnamefont {A.}~\bibnamefont
  {{Kusenko}}}\ and\ \bibinfo {author} {\bibfnamefont {G.}~\bibnamefont
  {{Segr{\`e}}}},\ }\href {\doibase10.1103/PhysRevLett.77.4872} {\bibfield
  {journal} {\bibinfo  {journal} {Physical Review Letters}\ }\textbf {\bibinfo
  {volume} {77}},\ \bibinfo {pages} {4872--4875} (\bibinfo {year} {1996})},\
  \Eprint {http://arxiv.org/abs/hep-ph/9606428} {hep-ph/9606428}\BibitemShut
  {NoStop}%
\bibitem [{\citenamefont {{Fryer}}\ and\ \citenamefont
  {{Kusenko}}(2006)}]{Fryer2006b}%
  \BibitemOpen
  \bibfield  {author} {\bibinfo {author} {\bibfnamefont {C.~L.}\ \bibnamefont
  {{Fryer}}}\ and\ \bibinfo {author} {\bibfnamefont {A.}~\bibnamefont
  {{Kusenko}}},\ }\href {\doibase10.1086/500933} {\bibfield  {journal}
  {\bibinfo  {journal} {The Astrophysical Journal Supplement}\ }\textbf
  {\bibinfo {volume} {163}},\ \bibinfo {pages} {335--343} (\bibinfo {year}
  {2006})},\ \Eprint {http://arxiv.org/abs/astro-ph/0512033}
  {astro-ph/0512033}\BibitemShut {NoStop}%
\bibitem [{\citenamefont {{Socrates}}\ \emph {et~al.}(2005)\citenamefont
  {{Socrates}}, \citenamefont {{Blaes}}, \citenamefont {{Hungerford}},\ and\
  \citenamefont {{Fryer}}}]{Socrates2005}%
  \BibitemOpen
  \bibfield  {author} {\bibinfo {author} {\bibfnamefont {A.}~\bibnamefont
  {{Socrates}}}, \bibinfo {author} {\bibfnamefont {O.}~\bibnamefont {{Blaes}}},
  \bibinfo {author} {\bibfnamefont {A.}~\bibnamefont {{Hungerford}}}, \ and\
  \bibinfo {author} {\bibfnamefont {C.~L.}\ \bibnamefont {{Fryer}}},\ }\href
  {\doibase10.1086/431786} {\bibfield  {journal} {\bibinfo  {journal} {The
  Astrophysical Journal}\ }\textbf {\bibinfo {volume} {632}},\ \bibinfo {pages}
  {531--562} (\bibinfo {year} {2005})},\ \Eprint
  {http://arxiv.org/abs/astro-ph/0412144} {astro-ph/0412144}\BibitemShut
  {NoStop}%
\bibitem [{\citenamefont {{Mandel}}(2016)}]{Mandel2016b}%
  \BibitemOpen
  \bibfield  {author} {\bibinfo {author} {\bibfnamefont {I.}~\bibnamefont
  {{Mandel}}},\ }\href {\doibase10.1093/mnras/stv2733} {\bibfield  {journal}
  {\bibinfo  {journal} {Monthly Notices of The Royal Astronomical Society}\
  }\textbf {\bibinfo {volume} {456}},\ \bibinfo {pages} {578--581} (\bibinfo
  {year} {2016})},\ \Eprint {http://arxiv.org/abs/1510.03871} {arXiv:1510.03871
  [astro-ph.HE]}\BibitemShut {NoStop}%
\bibitem [{\citenamefont {{Repetto}}\ \emph {et~al.}(2017)\citenamefont
  {{Repetto}}, \citenamefont {{Igoshev}},\ and\ \citenamefont
  {{Nelemans}}}]{Repetto2017}%
  \BibitemOpen
  \bibfield  {author} {\bibinfo {author} {\bibfnamefont {S.}~\bibnamefont
  {{Repetto}}}, \bibinfo {author} {\bibfnamefont {A.~P.}\ \bibnamefont
  {{Igoshev}}}, \ and\ \bibinfo {author} {\bibfnamefont {G.}~\bibnamefont
  {{Nelemans}}},\ }\href {\doibase10.1093/mnras/stx027} {\bibfield  {journal}
  {\bibinfo  {journal} {Monthly Notices of The Royal Astronomical Society}\
  }\textbf {\bibinfo {volume} {467}},\ \bibinfo {pages} {298--310} (\bibinfo
  {year} {2017})},\ \Eprint {http://arxiv.org/abs/1701.01347} {arXiv:1701.01347
  [astro-ph.HE]}\BibitemShut {NoStop}%
\bibitem [{\citenamefont {{Nelemans}}\ \emph {et~al.}(1999)\citenamefont
  {{Nelemans}}, \citenamefont {{Tauris}},\ and\ \citenamefont {{van den
  Heuvel}}}]{ntv99}%
  \BibitemOpen
  \bibfield  {author} {\bibinfo {author} {\bibfnamefont {G.}~\bibnamefont
  {{Nelemans}}}, \bibinfo {author} {\bibfnamefont {T.~M.}\ \bibnamefont
  {{Tauris}}}, \ and\ \bibinfo {author} {\bibfnamefont {E.~P.~J.}\ \bibnamefont
  {{van den Heuvel}}},\ }\href@noop {} {\bibfield  {journal} {\bibinfo
  {journal} {\aap}\ }\textbf {\bibinfo {volume} {352}},\ \bibinfo {pages}
  {L87--L90} (\bibinfo {year} {1999})},\ \Eprint
  {http://arxiv.org/abs/astro-ph/9911054} {astro-ph/9911054}\BibitemShut
  {NoStop}%
\bibitem [{\citenamefont {{Mirabel}}(2017)}]{Mirabel2017}%
  \BibitemOpen
  \bibfield  {author} {\bibinfo {author} {\bibfnamefont {F.}~\bibnamefont
  {{Mirabel}}},\ }\href {\doibase10.1016/j.newar.2017.04.002} {\bibfield
  {journal} {\bibinfo  {journal} {\nar}\ }\textbf {\bibinfo {volume} {78}},\
  \bibinfo {pages} {1--15} (\bibinfo {year} {2017})}\BibitemShut {NoStop}%
\bibitem [{\citenamefont {{Spruit}}(1999)}]{Spruit1999}%
  \BibitemOpen
  \bibfield  {author} {\bibinfo {author} {\bibfnamefont {H.~C.}\ \bibnamefont
  {{Spruit}}},\ }\href@noop {} {\bibfield  {journal} {\bibinfo  {journal}
  {\aap}\ }\textbf {\bibinfo {volume} {349}},\ \bibinfo {pages} {189--202}
  (\bibinfo {year} {1999})},\ \Eprint {http://arxiv.org/abs/astro-ph/9907138}
  {astro-ph/9907138}\BibitemShut {NoStop}%
\bibitem [{\citenamefont {{Spruit}}(2002)}]{Spruit2002}%
  \BibitemOpen
  \bibfield  {author} {\bibinfo {author} {\bibfnamefont {H.~C.}\ \bibnamefont
  {{Spruit}}},\ }\href {\doibase10.1051/0004-6361:20011465} {\bibfield
  {journal} {\bibinfo  {journal} {\aap}\ }\textbf {\bibinfo {volume} {381}},\
  \bibinfo {pages} {923--932} (\bibinfo {year} {2002})},\ \Eprint
  {http://arxiv.org/abs/astro-ph/0108207} {astro-ph/0108207}\BibitemShut
  {NoStop}%
\bibitem [{\citenamefont {{Ekstr{\"o}m}}\ \emph {et~al.}(2012)\citenamefont
  {{Ekstr{\"o}m}}, \citenamefont {{Georgy}}, \citenamefont {{Eggenberger}},
  \citenamefont {{Meynet}}, \citenamefont {{Mowlavi}}, \citenamefont
  {{Wyttenbach}}, \citenamefont {{Granada}}, \citenamefont {{Decressin}},
  \citenamefont {{Hirschi}}, \citenamefont {{Frischknecht}}, \citenamefont
  {{Charbonnel}},\ and\ \citenamefont {{Maeder}}}]{Ekstrom2012}%
  \BibitemOpen
  \bibfield  {author} {\bibinfo {author} {\bibfnamefont {S.}~\bibnamefont
  {{Ekstr{\"o}m}}}, \bibinfo {author} {\bibfnamefont {C.}~\bibnamefont
  {{Georgy}}}, \bibinfo {author} {\bibfnamefont {P.}~\bibnamefont
  {{Eggenberger}}}, \bibinfo {author} {\bibfnamefont {G.}~\bibnamefont
  {{Meynet}}}, \bibinfo {author} {\bibfnamefont {N.}~\bibnamefont {{Mowlavi}}},
  \bibinfo {author} {\bibfnamefont {A.}~\bibnamefont {{Wyttenbach}}}, \bibinfo
  {author} {\bibfnamefont {A.}~\bibnamefont {{Granada}}}, \bibinfo {author}
  {\bibfnamefont {T.}~\bibnamefont {{Decressin}}}, \bibinfo {author}
  {\bibfnamefont {R.}~\bibnamefont {{Hirschi}}}, \bibinfo {author}
  {\bibfnamefont {U.}~\bibnamefont {{Frischknecht}}}, \bibinfo {author}
  {\bibfnamefont {C.}~\bibnamefont {{Charbonnel}}}, \ and\ \bibinfo {author}
  {\bibfnamefont {A.}~\bibnamefont {{Maeder}}},\ }\href
  {\doibase10.1051/0004-6361/201117751} {\bibfield  {journal} {\bibinfo
  {journal} {\aap}\ }\textbf {\bibinfo {volume} {537}},\ \bibinfo {eid} {A146}
  (\bibinfo {year} {2012})},\ \Eprint {http://arxiv.org/abs/1110.5049}
  {arXiv:1110.5049 [astro-ph.SR]}\BibitemShut {NoStop}%
\bibitem [{\citenamefont {{Aerts}}(2008)}]{Aerts2008}%
  \BibitemOpen
  \bibfield  {author} {\bibinfo {author} {\bibfnamefont {C.}~\bibnamefont
  {{Aerts}}},\ }in\ \href {\doibase10.1017/S1743921308020541} {\emph {\bibinfo
  {booktitle} {Massive Stars as Cosmic Engines}}},\ \bibinfo {series} {IAU
  Symposium}, Vol.\ \bibinfo {volume} {250},\ \bibinfo {editor} {edited by\
  \bibinfo {editor} {\bibfnamefont {F.}~\bibnamefont {{Bresolin}}}, \bibinfo
  {editor} {\bibfnamefont {P.~A.}\ \bibnamefont {{Crowther}}}, \ and\ \bibinfo
  {editor} {\bibfnamefont {J.}~\bibnamefont {{Puls}}}}\ (\bibinfo {year}
  {2008})\ pp.\ \bibinfo {pages} {237--244}\BibitemShut {NoStop}%
\bibitem [{\citenamefont {Belczynski}\ \emph
  {et~al.}(2017{\natexlab{b}})\citenamefont {Belczynski} \emph
  {et~al.}}]{Belczynski:2017gds}%
  \BibitemOpen
  \bibfield  {author} {\bibinfo {author} {\bibfnamefont {K.}~\bibnamefont
  {Belczynski}} \emph {et~al.},\ }\href@noop {} {\  (\bibinfo {year}
  {2017}{\natexlab{b}})},\ \Eprint {http://arxiv.org/abs/1706.07053}
  {arXiv:1706.07053 [astro-ph.HE]}\BibitemShut {NoStop}%
\bibitem [{\citenamefont {Schrøder}\ \emph {et~al.}(2018)\citenamefont
  {Schrøder}, \citenamefont {Batta},\ and\ \citenamefont
  {Ramirez-Ruiz}}]{Schroder:2018hxk}%
  \BibitemOpen
  \bibfield  {author} {\bibinfo {author} {\bibfnamefont {Sophie~L.}\
  \bibnamefont {Schrøder}}, \bibinfo {author} {\bibfnamefont {Aldo}\
  \bibnamefont {Batta}}, \ and\ \bibinfo {author} {\bibfnamefont {Enrico}\
  \bibnamefont {Ramirez-Ruiz}},\ }\href@noop {} {\  (\bibinfo {year} {2018})},\
  \Eprint {http://arxiv.org/abs/1805.01269} {arXiv:1805.01269
  [astro-ph.HE]}\BibitemShut {NoStop}%
\bibitem [{\citenamefont {{Sana}}\ \emph {et~al.}(2012)\citenamefont {{Sana}},
  \citenamefont {{de Mink}}, \citenamefont {{de Koter}}, \citenamefont
  {{Langer}}, \citenamefont {{Evans}}, \citenamefont {{Gieles}}, \citenamefont
  {{Gosset}}, \citenamefont {{Izzard}}, \citenamefont {{Le Bouquin}},\ and\
  \citenamefont {{Schneider}}}]{Sana2012}%
  \BibitemOpen
  \bibfield  {author} {\bibinfo {author} {\bibfnamefont {H.}~\bibnamefont
  {{Sana}}}, \bibinfo {author} {\bibfnamefont {S.~E.}\ \bibnamefont {{de
  Mink}}}, \bibinfo {author} {\bibfnamefont {A.}~\bibnamefont {{de Koter}}},
  \bibinfo {author} {\bibfnamefont {N.}~\bibnamefont {{Langer}}}, \bibinfo
  {author} {\bibfnamefont {C.~J.}\ \bibnamefont {{Evans}}}, \bibinfo {author}
  {\bibfnamefont {M.}~\bibnamefont {{Gieles}}}, \bibinfo {author}
  {\bibfnamefont {E.}~\bibnamefont {{Gosset}}}, \bibinfo {author}
  {\bibfnamefont {R.~G.}\ \bibnamefont {{Izzard}}}, \bibinfo {author}
  {\bibfnamefont {J.-B.}\ \bibnamefont {{Le Bouquin}}}, \ and\ \bibinfo
  {author} {\bibfnamefont {F.~R.~N.}\ \bibnamefont {{Schneider}}},\ }\href
  {\doibase10.1126/science.1223344} {\bibfield  {journal} {\bibinfo  {journal}
  {Science}\ }\textbf {\bibinfo {volume} {337}},\ \bibinfo {pages} {444}
  (\bibinfo {year} {2012})}\BibitemShut {NoStop}%
\bibitem [{\citenamefont {{Podsiadlowski}}\ \emph {et~al.}(1992)\citenamefont
  {{Podsiadlowski}}, \citenamefont {{Joss}},\ and\ \citenamefont
  {{Hsu}}}]{pjh92}%
  \BibitemOpen
  \bibfield  {author} {\bibinfo {author} {\bibfnamefont {P.}~\bibnamefont
  {{Podsiadlowski}}}, \bibinfo {author} {\bibfnamefont {P.~C.}\ \bibnamefont
  {{Joss}}}, \ and\ \bibinfo {author} {\bibfnamefont {J.~J.~L.}\ \bibnamefont
  {{Hsu}}},\ }\href {\doibase10.1086/171341} {\bibfield  {journal} {\bibinfo
  {journal} {The Astrophysical Journal}\ }\textbf {\bibinfo {volume} {391}},\
  \bibinfo {pages} {246--264} (\bibinfo {year} {1992})}\BibitemShut {NoStop}%
\bibitem [{\citenamefont {{Wellstein}}\ and\ \citenamefont
  {{Langer}}(1999)}]{wl99}%
  \BibitemOpen
  \bibfield  {author} {\bibinfo {author} {\bibfnamefont {S.}~\bibnamefont
  {{Wellstein}}}\ and\ \bibinfo {author} {\bibfnamefont {N.}~\bibnamefont
  {{Langer}}},\ }\href@noop {} {\bibfield  {journal} {\bibinfo  {journal}
  {\aap}\ }\textbf {\bibinfo {volume} {350}},\ \bibinfo {pages} {148--162}
  (\bibinfo {year} {1999})},\ \Eprint {http://arxiv.org/abs/astro-ph/9904256}
  {astro-ph/9904256}\BibitemShut {NoStop}%
\bibitem [{\citenamefont {{Brown}}\ \emph {et~al.}(2001)\citenamefont
  {{Brown}}, \citenamefont {{Heger}}, \citenamefont {{Langer}}, \citenamefont
  {{Lee}}, \citenamefont {{Wellstein}},\ and\ \citenamefont
  {{Bethe}}}]{bhl+01}%
  \BibitemOpen
  \bibfield  {author} {\bibinfo {author} {\bibfnamefont {G.~E.}\ \bibnamefont
  {{Brown}}}, \bibinfo {author} {\bibfnamefont {A.}~\bibnamefont {{Heger}}},
  \bibinfo {author} {\bibfnamefont {N.}~\bibnamefont {{Langer}}}, \bibinfo
  {author} {\bibfnamefont {C.-H.}\ \bibnamefont {{Lee}}}, \bibinfo {author}
  {\bibfnamefont {S.}~\bibnamefont {{Wellstein}}}, \ and\ \bibinfo {author}
  {\bibfnamefont {H.~A.}\ \bibnamefont {{Bethe}}},\ }\href
  {\doibase10.1016/S1384-1076(01)00077-X} {\bibfield  {journal} {\bibinfo
  {journal} {\na}\ }\textbf {\bibinfo {volume} {6}},\ \bibinfo {pages}
  {457--470} (\bibinfo {year} {2001})},\ \Eprint
  {http://arxiv.org/abs/astro-ph/0102379} {astro-ph/0102379}\BibitemShut
  {NoStop}%
\bibitem [{\citenamefont {{Langer}}(2012)}]{lan12}%
  \BibitemOpen
  \bibfield  {author} {\bibinfo {author} {\bibfnamefont {N.}~\bibnamefont
  {{Langer}}},\ }\href {\doibase10.1146/annurev-astro-081811-125534} {\bibfield
   {journal} {\bibinfo  {journal} {\araa}\ }\textbf {\bibinfo {volume} {50}},\
  \bibinfo {pages} {107--164} (\bibinfo {year} {2012})},\ \Eprint
  {http://arxiv.org/abs/1206.5443} {arXiv:1206.5443 [astro-ph.SR]}\BibitemShut
  {NoStop}%
\bibitem [{\citenamefont {{Tutukov}}\ and\ \citenamefont
  {{Yungelson}}(1993)}]{Tutukov1993}%
  \BibitemOpen
  \bibfield  {author} {\bibinfo {author} {\bibfnamefont {A.~V.}\ \bibnamefont
  {{Tutukov}}}\ and\ \bibinfo {author} {\bibfnamefont {L.~R.}\ \bibnamefont
  {{Yungelson}}},\ }\href {\doibase10.1093/mnras/260.3.675} {\bibfield
  {journal} {\bibinfo  {journal} {Monthly Notices of The Royal Astronomical
  Society}\ }\textbf {\bibinfo {volume} {260}},\ \bibinfo {pages} {675--678}
  (\bibinfo {year} {1993})}\BibitemShut {NoStop}%
\bibitem [{\citenamefont {{Lipunov}}\ \emph {et~al.}(1997)\citenamefont
  {{Lipunov}}, \citenamefont {{Postnov}},\ and\ \citenamefont
  {{Prokhorov}}}]{Lipunov1997}%
  \BibitemOpen
  \bibfield  {author} {\bibinfo {author} {\bibfnamefont {V.~M.}\ \bibnamefont
  {{Lipunov}}}, \bibinfo {author} {\bibfnamefont {K.~A.}\ \bibnamefont
  {{Postnov}}}, \ and\ \bibinfo {author} {\bibfnamefont {M.~E.}\ \bibnamefont
  {{Prokhorov}}},\ }\href@noop {} {\bibfield  {journal} {\bibinfo  {journal}
  {Astronomy Letters}\ }\textbf {\bibinfo {volume} {23}},\ \bibinfo {pages}
  {492--497} (\bibinfo {year} {1997})}\BibitemShut {NoStop}%
\bibitem [{\citenamefont {Belczynski}\ \emph {et~al.}(2001)\citenamefont
  {Belczynski}, \citenamefont {Kalogera},\ and\ \citenamefont
  {Bulik}}]{Belczynski:2001uc}%
  \BibitemOpen
  \bibfield  {author} {\bibinfo {author} {\bibfnamefont {Krzysztof}\
  \bibnamefont {Belczynski}}, \bibinfo {author} {\bibfnamefont {Vassiliki}\
  \bibnamefont {Kalogera}}, \ and\ \bibinfo {author} {\bibfnamefont {Tomasz}\
  \bibnamefont {Bulik}},\ }\href {\doibase10.1086/340304} {\bibfield  {journal}
  {\bibinfo  {journal} {Astrophys. J.}\ }\textbf {\bibinfo {volume} {572}},\
  \bibinfo {pages} {407--431} (\bibinfo {year} {2001})},\ \Eprint
  {http://arxiv.org/abs/astro-ph/0111452} {arXiv:astro-ph/0111452
  [astro-ph]}\BibitemShut {NoStop}%
\bibitem [{\citenamefont {{Voss}}\ and\ \citenamefont
  {{Tauris}}(2003)}]{Voss2003}%
  \BibitemOpen
  \bibfield  {author} {\bibinfo {author} {\bibfnamefont {R.}~\bibnamefont
  {{Voss}}}\ and\ \bibinfo {author} {\bibfnamefont {T.~M.}\ \bibnamefont
  {{Tauris}}},\ }\href {\doibase10.1046/j.1365-8711.2003.06616.x} {\bibfield
  {journal} {\bibinfo  {journal} {Monthly Notices of The Royal Astronomical
  Society}\ }\textbf {\bibinfo {volume} {342}},\ \bibinfo {pages} {1169--1184}
  (\bibinfo {year} {2003})},\ \Eprint {http://arxiv.org/abs/astro-ph/0303227}
  {astro-ph/0303227}\BibitemShut {NoStop}%
\bibitem [{\citenamefont {{Mennekens}}\ and\ \citenamefont
  {{Vanbeveren}}(2014)}]{Mennekens2014}%
  \BibitemOpen
  \bibfield  {author} {\bibinfo {author} {\bibfnamefont {N.}~\bibnamefont
  {{Mennekens}}}\ and\ \bibinfo {author} {\bibfnamefont {D.}~\bibnamefont
  {{Vanbeveren}}},\ }\href {\doibase10.1051/0004-6361/201322198} {\bibfield
  {journal} {\bibinfo  {journal} {\aap}\ }\textbf {\bibinfo {volume} {564}},\
  \bibinfo {eid} {A134} (\bibinfo {year} {2014})},\ \Eprint
  {http://arxiv.org/abs/1307.0959} {arXiv:1307.0959 [astro-ph.SR]}\BibitemShut
  {NoStop}%
\bibitem [{\citenamefont {{Eldridge}}\ and\ \citenamefont
  {{Stanway}}(2016)}]{Eldridge2016}%
  \BibitemOpen
  \bibfield  {author} {\bibinfo {author} {\bibfnamefont {J.~J.}\ \bibnamefont
  {{Eldridge}}}\ and\ \bibinfo {author} {\bibfnamefont {E.~R.}\ \bibnamefont
  {{Stanway}}},\ }\href {\doibase10.1093/mnras/stw1772} {\bibfield  {journal}
  {\bibinfo  {journal} {Monthly Notices of The Royal Astronomical Society}\
  }\textbf {\bibinfo {volume} {462}},\ \bibinfo {pages} {3302--3313} (\bibinfo
  {year} {2016})},\ \Eprint {http://arxiv.org/abs/1602.03790} {arXiv:1602.03790
  [astro-ph.HE]}\BibitemShut {NoStop}%
\bibitem [{\citenamefont {{Stevenson}}\ \emph {et~al.}(2017)\citenamefont
  {{Stevenson}}, \citenamefont {{Vigna-G{\'o}mez}}, \citenamefont {{Mandel}},
  \citenamefont {{Barrett}}, \citenamefont {{Neijssel}}, \citenamefont
  {{Perkins}},\ and\ \citenamefont {{de Mink}}}]{Stevenson2017}%
  \BibitemOpen
  \bibfield  {author} {\bibinfo {author} {\bibfnamefont {S.}~\bibnamefont
  {{Stevenson}}}, \bibinfo {author} {\bibfnamefont {A.}~\bibnamefont
  {{Vigna-G{\'o}mez}}}, \bibinfo {author} {\bibfnamefont {I.}~\bibnamefont
  {{Mandel}}}, \bibinfo {author} {\bibfnamefont {J.~W.}\ \bibnamefont
  {{Barrett}}}, \bibinfo {author} {\bibfnamefont {C.~J.}\ \bibnamefont
  {{Neijssel}}}, \bibinfo {author} {\bibfnamefont {D.}~\bibnamefont
  {{Perkins}}}, \ and\ \bibinfo {author} {\bibfnamefont {S.~E.}\ \bibnamefont
  {{de Mink}}},\ }\href {\doibase10.1038/ncomms14906} {\bibfield  {journal}
  {\bibinfo  {journal} {Nature Communications}\ }\textbf {\bibinfo {volume}
  {8}},\ \bibinfo {eid} {14906} (\bibinfo {year} {2017})},\ \Eprint
  {http://arxiv.org/abs/1704.01352} {arXiv:1704.01352
  [astro-ph.HE]}\BibitemShut {NoStop}%
\bibitem [{\citenamefont {{Maeder}}(1987)}]{Maeder1987}%
  \BibitemOpen
  \bibfield  {author} {\bibinfo {author} {\bibfnamefont {A.}~\bibnamefont
  {{Maeder}}},\ }\href@noop {} {\bibfield  {journal} {\bibinfo  {journal}
  {\aap}\ }\textbf {\bibinfo {volume} {178}},\ \bibinfo {pages} {159--169}
  (\bibinfo {year} {1987})}\BibitemShut {NoStop}%
\bibitem [{\citenamefont {{Yoon}}\ and\ \citenamefont
  {{Langer}}(2005)}]{Yoon2005}%
  \BibitemOpen
  \bibfield  {author} {\bibinfo {author} {\bibfnamefont {S.-C.}\ \bibnamefont
  {{Yoon}}}\ and\ \bibinfo {author} {\bibfnamefont {N.}~\bibnamefont
  {{Langer}}},\ }\href {\doibase10.1051/0004-6361:20054030} {\bibfield
  {journal} {\bibinfo  {journal} {\aap}\ }\textbf {\bibinfo {volume} {443}},\
  \bibinfo {pages} {643--648} (\bibinfo {year} {2005})},\ \Eprint
  {http://arxiv.org/abs/astro-ph/0508242} {astro-ph/0508242}\BibitemShut
  {NoStop}%
\bibitem [{\citenamefont {{Mandel}}\ and\ \citenamefont {{de
  Mink}}(2016)}]{Mandel2016a}%
  \BibitemOpen
  \bibfield  {author} {\bibinfo {author} {\bibfnamefont {I.}~\bibnamefont
  {{Mandel}}}\ and\ \bibinfo {author} {\bibfnamefont {S.~E.}\ \bibnamefont {{de
  Mink}}},\ }\href {\doibase10.1093/mnras/stw379} {\bibfield  {journal}
  {\bibinfo  {journal} {Monthly Notices of The Royal Astronomical Society}\
  }\textbf {\bibinfo {volume} {458}},\ \bibinfo {pages} {2634--2647} (\bibinfo
  {year} {2016})},\ \Eprint {http://arxiv.org/abs/1601.00007} {arXiv:1601.00007
  [astro-ph.HE]}\BibitemShut {NoStop}%
\bibitem [{\citenamefont {{de Mink}}\ and\ \citenamefont
  {{Mandel}}(2016)}]{deMink2016}%
  \BibitemOpen
  \bibfield  {author} {\bibinfo {author} {\bibfnamefont {S.~E.}\ \bibnamefont
  {{de Mink}}}\ and\ \bibinfo {author} {\bibfnamefont {I.}~\bibnamefont
  {{Mandel}}},\ }\href {\doibase10.1093/mnras/stw1219} {\bibfield  {journal}
  {\bibinfo  {journal} {Monthly Notices of The Royal Astronomical Society}\
  }\textbf {\bibinfo {volume} {460}},\ \bibinfo {pages} {3545--3553} (\bibinfo
  {year} {2016})},\ \Eprint {http://arxiv.org/abs/1603.02291} {arXiv:1603.02291
  [astro-ph.HE]}\BibitemShut {NoStop}%
\bibitem [{\citenamefont {{Webbink}}(1984)}]{Webbink1984}%
  \BibitemOpen
  \bibfield  {author} {\bibinfo {author} {\bibfnamefont {R.~F.}\ \bibnamefont
  {{Webbink}}},\ }\href {\doibase10.1086/161701} {\bibfield  {journal}
  {\bibinfo  {journal} {The Astrophysical Journal}\ }\textbf {\bibinfo {volume}
  {277}},\ \bibinfo {pages} {355--360} (\bibinfo {year} {1984})}\BibitemShut
  {NoStop}%
\bibitem [{\citenamefont {{Ivanova}}\ \emph {et~al.}(2013)\citenamefont
  {{Ivanova}}, \citenamefont {{Justham}}, \citenamefont {{Chen}}, \citenamefont
  {{De Marco}}, \citenamefont {{Fryer}}, \citenamefont {{Gaburov}},
  \citenamefont {{Ge}}, \citenamefont {{Glebbeek}}, \citenamefont {{Han}},
  \citenamefont {{Li}}, \citenamefont {{Lu}}, \citenamefont {{Marsh}},
  \citenamefont {{Podsiadlowski}}, \citenamefont {{Potter}}, \citenamefont
  {{Soker}}, \citenamefont {{Taam}}, \citenamefont {{Tauris}}, \citenamefont
  {{van den Heuvel}},\ and\ \citenamefont {{Webbink}}}]{Ivanova2013}%
  \BibitemOpen
  \bibfield  {author} {\bibinfo {author} {\bibfnamefont {N.}~\bibnamefont
  {{Ivanova}}}, \bibinfo {author} {\bibfnamefont {S.}~\bibnamefont
  {{Justham}}}, \bibinfo {author} {\bibfnamefont {X.}~\bibnamefont {{Chen}}},
  \bibinfo {author} {\bibfnamefont {O.}~\bibnamefont {{De Marco}}}, \bibinfo
  {author} {\bibfnamefont {C.~L.}\ \bibnamefont {{Fryer}}}, \bibinfo {author}
  {\bibfnamefont {E.}~\bibnamefont {{Gaburov}}}, \bibinfo {author}
  {\bibfnamefont {H.}~\bibnamefont {{Ge}}}, \bibinfo {author} {\bibfnamefont
  {E.}~\bibnamefont {{Glebbeek}}}, \bibinfo {author} {\bibfnamefont
  {Z.}~\bibnamefont {{Han}}}, \bibinfo {author} {\bibfnamefont {X.-D.}\
  \bibnamefont {{Li}}}, \bibinfo {author} {\bibfnamefont {G.}~\bibnamefont
  {{Lu}}}, \bibinfo {author} {\bibfnamefont {T.}~\bibnamefont {{Marsh}}},
  \bibinfo {author} {\bibfnamefont {P.}~\bibnamefont {{Podsiadlowski}}},
  \bibinfo {author} {\bibfnamefont {A.}~\bibnamefont {{Potter}}}, \bibinfo
  {author} {\bibfnamefont {N.}~\bibnamefont {{Soker}}}, \bibinfo {author}
  {\bibfnamefont {R.}~\bibnamefont {{Taam}}}, \bibinfo {author} {\bibfnamefont
  {T.~M.}\ \bibnamefont {{Tauris}}}, \bibinfo {author} {\bibfnamefont
  {E.~P.~J.}\ \bibnamefont {{van den Heuvel}}}, \ and\ \bibinfo {author}
  {\bibfnamefont {R.~F.}\ \bibnamefont {{Webbink}}},\ }\href
  {\doibase10.1007/s00159-013-0059-2} {\bibfield  {journal} {\bibinfo
  {journal} {\aapr}\ }\textbf {\bibinfo {volume} {21}},\ \bibinfo {eid} {59}
  (\bibinfo {year} {2013})}\BibitemShut {NoStop}%
\bibitem [{\citenamefont {Peters}(1964)}]{Peters:1964zz}%
  \BibitemOpen
  \bibfield  {author} {\bibinfo {author} {\bibfnamefont {P.~C.}\ \bibnamefont
  {Peters}},\ }\href {\doibase10.1103/PhysRev.136.B1224} {\bibfield  {journal}
  {\bibinfo  {journal} {Phys. Rev.}\ }\textbf {\bibinfo {volume} {136}},\
  \bibinfo {pages} {B1224--B1232} (\bibinfo {year} {1964})}\BibitemShut
  {NoStop}%
\bibitem [{\citenamefont {{van den Heuvel}}\ \emph {et~al.}(2017)\citenamefont
  {{van den Heuvel}}, \citenamefont {{Portegies Zwart}},\ and\ \citenamefont
  {{de Mink}}}]{Heuvel2017}%
  \BibitemOpen
  \bibfield  {author} {\bibinfo {author} {\bibfnamefont {E.~P.~J.}\
  \bibnamefont {{van den Heuvel}}}, \bibinfo {author} {\bibfnamefont {S.~F.}\
  \bibnamefont {{Portegies Zwart}}}, \ and\ \bibinfo {author} {\bibfnamefont
  {S.~E.}\ \bibnamefont {{de Mink}}},\ }\href {\doibase10.1093/mnras/stx1430}
  {\bibfield  {journal} {\bibinfo  {journal} {Monthly Notices of The Royal
  Astronomical Society}\ }\textbf {\bibinfo {volume} {471}},\ \bibinfo {pages}
  {4256--4264} (\bibinfo {year} {2017})},\ \Eprint
  {http://arxiv.org/abs/1701.02355} {arXiv:1701.02355
  [astro-ph.SR]}\BibitemShut {NoStop}%
\bibitem [{\citenamefont {{Zahn}}(1992)}]{Zahn1992}%
  \BibitemOpen
  \bibfield  {author} {\bibinfo {author} {\bibfnamefont {J.-P.}\ \bibnamefont
  {{Zahn}}},\ }\href@noop {} {\bibfield  {journal} {\bibinfo  {journal} {\aap}\
  }\textbf {\bibinfo {volume} {265}},\ \bibinfo {pages} {115--132} (\bibinfo
  {year} {1992})}\BibitemShut {NoStop}%
\bibitem [{\citenamefont {{Kushnir}}\ \emph {et~al.}(2016)\citenamefont
  {{Kushnir}}, \citenamefont {{Zaldarriaga}}, \citenamefont {{Kollmeier}},\
  and\ \citenamefont {{Waldman}}}]{Kushnir2016}%
  \BibitemOpen
  \bibfield  {author} {\bibinfo {author} {\bibfnamefont {D.}~\bibnamefont
  {{Kushnir}}}, \bibinfo {author} {\bibfnamefont {M.}~\bibnamefont
  {{Zaldarriaga}}}, \bibinfo {author} {\bibfnamefont {J.~A.}\ \bibnamefont
  {{Kollmeier}}}, \ and\ \bibinfo {author} {\bibfnamefont {R.}~\bibnamefont
  {{Waldman}}},\ }\href {\doibase10.1093/mnras/stw1684} {\bibfield  {journal}
  {\bibinfo  {journal} {Monthly Notices of The Royal Astronomical Society}\
  }\textbf {\bibinfo {volume} {462}},\ \bibinfo {pages} {844--849} (\bibinfo
  {year} {2016})},\ \Eprint {http://arxiv.org/abs/1605.03839} {arXiv:1605.03839
  [astro-ph.HE]}\BibitemShut {NoStop}%
\bibitem [{\citenamefont {Abadie}\ \emph {et~al.}(2010)\citenamefont {Abadie}
  \emph {et~al.}}]{Abadie:2010cf}%
  \BibitemOpen
  \bibfield  {author} {\bibinfo {author} {\bibfnamefont {J.}~\bibnamefont
  {Abadie}} \emph {et~al.} (\bibinfo {collaboration} {VIRGO, LIGO
  Scientific}),\ }\href {\doibase10.1088/0264-9381/27/17/173001} {\bibfield
  {journal} {\bibinfo  {journal} {Class. Quant. Grav.}\ }\textbf {\bibinfo
  {volume} {27}},\ \bibinfo {pages} {173001} (\bibinfo {year} {2010})},\
  \Eprint {http://arxiv.org/abs/1003.2480} {arXiv:1003.2480
  [astro-ph.HE]}\BibitemShut {NoStop}%
\bibitem [{\citenamefont {{Zampieri}}\ and\ \citenamefont
  {{Roberts}}(2009)}]{Zampieri2009}%
  \BibitemOpen
  \bibfield  {author} {\bibinfo {author} {\bibfnamefont {L.}~\bibnamefont
  {{Zampieri}}}\ and\ \bibinfo {author} {\bibfnamefont {T.~P.}\ \bibnamefont
  {{Roberts}}},\ }\href {\doibase10.1111/j.1365-2966.2009.15509.x} {\bibfield
  {journal} {\bibinfo  {journal} {Monthly Notices of The Royal Astronomical
  Society}\ }\textbf {\bibinfo {volume} {400}},\ \bibinfo {pages} {677--686}
  (\bibinfo {year} {2009})},\ \Eprint {http://arxiv.org/abs/0909.1017}
  {arXiv:0909.1017 [astro-ph.HE]}\BibitemShut {NoStop}%
\bibitem [{\citenamefont {{Mapelli}}\ \emph {et~al.}(2009)\citenamefont
  {{Mapelli}}, \citenamefont {{Colpi}},\ and\ \citenamefont
  {{Zampieri}}}]{Mapelli2009}%
  \BibitemOpen
  \bibfield  {author} {\bibinfo {author} {\bibfnamefont {M.}~\bibnamefont
  {{Mapelli}}}, \bibinfo {author} {\bibfnamefont {M.}~\bibnamefont {{Colpi}}},
  \ and\ \bibinfo {author} {\bibfnamefont {L.}~\bibnamefont {{Zampieri}}},\
  }\href {\doibase10.1111/j.1745-3933.2009.00645.x} {\bibfield  {journal}
  {\bibinfo  {journal} {Monthly Notices of The Royal Astronomical Society}\
  }\textbf {\bibinfo {volume} {395}},\ \bibinfo {pages} {L71--L75} (\bibinfo
  {year} {2009})},\ \Eprint {http://arxiv.org/abs/0902.3540} {arXiv:0902.3540
  [astro-ph.HE]}\BibitemShut {NoStop}%
\bibitem [{\citenamefont {Dominik}\ \emph {et~al.}(2012)\citenamefont
  {Dominik}, \citenamefont {Belczynski}, \citenamefont {Fryer}, \citenamefont
  {Holz}, \citenamefont {Berti}, \citenamefont {Bulik}, \citenamefont
  {Mandel},\ and\ \citenamefont {O'Shaughnessy}}]{Dominik:2012kk}%
  \BibitemOpen
  \bibfield  {author} {\bibinfo {author} {\bibfnamefont {Michal}\ \bibnamefont
  {Dominik}}, \bibinfo {author} {\bibfnamefont {Krzysztof}\ \bibnamefont
  {Belczynski}}, \bibinfo {author} {\bibfnamefont {Christopher}\ \bibnamefont
  {Fryer}}, \bibinfo {author} {\bibfnamefont {Daniel}\ \bibnamefont {Holz}},
  \bibinfo {author} {\bibfnamefont {Emanuele}\ \bibnamefont {Berti}}, \bibinfo
  {author} {\bibfnamefont {Tomasz}\ \bibnamefont {Bulik}}, \bibinfo {author}
  {\bibfnamefont {Ilya}\ \bibnamefont {Mandel}}, \ and\ \bibinfo {author}
  {\bibfnamefont {Richard}\ \bibnamefont {O'Shaughnessy}},\ }\href
  {\doibase10.1088/0004-637X/759/1/52} {\bibfield  {journal} {\bibinfo
  {journal} {Astrophys. J.}\ }\textbf {\bibinfo {volume} {759}},\ \bibinfo
  {pages} {52} (\bibinfo {year} {2012})},\ \Eprint
  {http://arxiv.org/abs/1202.4901} {arXiv:1202.4901 [astro-ph.HE]}\BibitemShut
  {NoStop}%
\bibitem [{\citenamefont {Dominik}\ \emph {et~al.}(2015)\citenamefont
  {Dominik}, \citenamefont {Berti}, \citenamefont {O'Shaughnessy},
  \citenamefont {Mandel}, \citenamefont {Belczynski}, \citenamefont {Fryer},
  \citenamefont {Holz}, \citenamefont {Bulik},\ and\ \citenamefont
  {Pannarale}}]{Dominik:2014yma}%
  \BibitemOpen
  \bibfield  {author} {\bibinfo {author} {\bibfnamefont {Michal}\ \bibnamefont
  {Dominik}}, \bibinfo {author} {\bibfnamefont {Emanuele}\ \bibnamefont
  {Berti}}, \bibinfo {author} {\bibfnamefont {Richard}\ \bibnamefont
  {O'Shaughnessy}}, \bibinfo {author} {\bibfnamefont {Ilya}\ \bibnamefont
  {Mandel}}, \bibinfo {author} {\bibfnamefont {Krzysztof}\ \bibnamefont
  {Belczynski}}, \bibinfo {author} {\bibfnamefont {Christopher}\ \bibnamefont
  {Fryer}}, \bibinfo {author} {\bibfnamefont {Daniel~E.}\ \bibnamefont {Holz}},
  \bibinfo {author} {\bibfnamefont {Tomasz}\ \bibnamefont {Bulik}}, \ and\
  \bibinfo {author} {\bibfnamefont {Francesco}\ \bibnamefont {Pannarale}},\
  }\href {\doibase10.1088/0004-637X/806/2/263} {\bibfield  {journal} {\bibinfo
  {journal} {Astrophys. J.}\ }\textbf {\bibinfo {volume} {806}},\ \bibinfo
  {pages} {263} (\bibinfo {year} {2015})},\ \Eprint
  {http://arxiv.org/abs/1405.7016} {arXiv:1405.7016 [astro-ph.HE]}\BibitemShut
  {NoStop}%
\bibitem [{\citenamefont {{Bulik}}\ \emph {et~al.}(2011)\citenamefont
  {{Bulik}}, \citenamefont {{Belczynski}},\ and\ \citenamefont
  {{Prestwich}}}]{Bulik2011}%
  \BibitemOpen
  \bibfield  {author} {\bibinfo {author} {\bibfnamefont {T.}~\bibnamefont
  {{Bulik}}}, \bibinfo {author} {\bibfnamefont {K.}~\bibnamefont
  {{Belczynski}}}, \ and\ \bibinfo {author} {\bibfnamefont {A.}~\bibnamefont
  {{Prestwich}}},\ }\href {\doibase10.1088/0004-637X/730/2/140} {\bibfield
  {journal} {\bibinfo  {journal} {The Astrophysical Journal}\ }\textbf
  {\bibinfo {volume} {730}},\ \bibinfo {eid} {140} (\bibinfo {year} {2011})},\
  \Eprint {http://arxiv.org/abs/0803.3516} {arXiv:0803.3516}\BibitemShut
  {NoStop}%
\bibitem [{\citenamefont {{Rodriguez}}\ \emph
  {et~al.}(2016{\natexlab{a}})\citenamefont {{Rodriguez}}, \citenamefont
  {{Chatterjee}},\ and\ \citenamefont {{Rasio}}}]{Rodriguez2016b}%
  \BibitemOpen
  \bibfield  {author} {\bibinfo {author} {\bibfnamefont {C.~L.}\ \bibnamefont
  {{Rodriguez}}}, \bibinfo {author} {\bibfnamefont {S.}~\bibnamefont
  {{Chatterjee}}}, \ and\ \bibinfo {author} {\bibfnamefont {F.~A.}\
  \bibnamefont {{Rasio}}},\ }\href {\doibase10.1103/PhysRevD.93.084029}
  {\bibfield  {journal} {\bibinfo  {journal} {Physical Review D}\ }\textbf
  {\bibinfo {volume} {93}},\ \bibinfo {eid} {084029} (\bibinfo {year}
  {2016}{\natexlab{a}})},\ \Eprint {http://arxiv.org/abs/1602.02444}
  {arXiv:1602.02444 [astro-ph.HE]}\BibitemShut {NoStop}%
\bibitem [{\citenamefont {{Askar}}\ \emph {et~al.}(2017)\citenamefont
  {{Askar}}, \citenamefont {{Szkudlarek}}, \citenamefont
  {{Gondek-Rosi{\'n}ska}}, \citenamefont {{Giersz}},\ and\ \citenamefont
  {{Bulik}}}]{Askaretal2017}%
  \BibitemOpen
  \bibfield  {author} {\bibinfo {author} {\bibfnamefont {A.}~\bibnamefont
  {{Askar}}}, \bibinfo {author} {\bibfnamefont {M.}~\bibnamefont
  {{Szkudlarek}}}, \bibinfo {author} {\bibfnamefont {D.}~\bibnamefont
  {{Gondek-Rosi{\'n}ska}}}, \bibinfo {author} {\bibfnamefont {M.}~\bibnamefont
  {{Giersz}}}, \ and\ \bibinfo {author} {\bibfnamefont {T.}~\bibnamefont
  {{Bulik}}},\ }\href {\doibase10.1093/mnrasl/slw177} {\bibfield  {journal}
  {\bibinfo  {journal} {MNRAS}\ }\textbf {\bibinfo {volume} {464}},\ \bibinfo
  {pages} {L36--L40} (\bibinfo {year} {2017})},\ \Eprint
  {http://arxiv.org/abs/1608.02520} {arXiv:1608.02520
  [astro-ph.HE]}\BibitemShut {NoStop}%
\bibitem [{\citenamefont {{Fragos}}\ and\ \citenamefont
  {{McClintock}}(2015)}]{Fragos2015}%
  \BibitemOpen
  \bibfield  {author} {\bibinfo {author} {\bibfnamefont {T.}~\bibnamefont
  {{Fragos}}}\ and\ \bibinfo {author} {\bibfnamefont {J.~E.}\ \bibnamefont
  {{McClintock}}},\ }\href {\doibase10.1088/0004-637X/800/1/17} {\bibfield
  {journal} {\bibinfo  {journal} {The Astrophysical Journal}\ }\textbf
  {\bibinfo {volume} {800}},\ \bibinfo {eid} {17} (\bibinfo {year} {2015})},\
  \Eprint {http://arxiv.org/abs/1408.2661} {arXiv:1408.2661
  [astro-ph.HE]}\BibitemShut {NoStop}%
\bibitem [{\citenamefont {Antonini}\ \emph {et~al.}(2017)\citenamefont
  {Antonini}, \citenamefont {Rodriguez}, \citenamefont {Petrovich},\ and\
  \citenamefont {Fischer}}]{Antonini:2017tgo}%
  \BibitemOpen
  \bibfield  {author} {\bibinfo {author} {\bibfnamefont {Fabio}\ \bibnamefont
  {Antonini}}, \bibinfo {author} {\bibfnamefont {Carl~L.}\ \bibnamefont
  {Rodriguez}}, \bibinfo {author} {\bibfnamefont {Cristobal}\ \bibnamefont
  {Petrovich}}, \ and\ \bibinfo {author} {\bibfnamefont {Caitlin~L.}\
  \bibnamefont {Fischer}},\ }\href@noop {} {\  (\bibinfo {year} {2017})},\
  \Eprint {http://arxiv.org/abs/1711.07142} {arXiv:1711.07142
  [astro-ph.HE]}\BibitemShut {NoStop}%
\bibitem [{\citenamefont {Clesse}\ and\ \citenamefont
  {Garc\'ia-Bellido}(2017{\natexlab{a}})}]{Clesse:2017bsw}%
  \BibitemOpen
  \bibfield  {author} {\bibinfo {author} {\bibfnamefont {Sebastien}\
  \bibnamefont {Clesse}}\ and\ \bibinfo {author} {\bibfnamefont {Juan}\
  \bibnamefont {Garc\'ia-Bellido}},\ }\href@noop {} {\  (\bibinfo {year}
  {2017}{\natexlab{a}})},\ \Eprint {http://arxiv.org/abs/1711.10458}
  {arXiv:1711.10458 [astro-ph.CO]}\BibitemShut {NoStop}%
\bibitem [{\citenamefont {{Tauris}}\ \emph {et~al.}(2017)\citenamefont
  {{Tauris}}, \citenamefont {{Kramer}}, \citenamefont {{Freire}}, \citenamefont
  {{Wex}}, \citenamefont {{Janka}}, \citenamefont {{Langer}}, \citenamefont
  {{Podsiadlowski}}, \citenamefont {{Bozzo}}, \citenamefont {{Chaty}},
  \citenamefont {{Kruckow}}, \citenamefont {{van den Heuvel}}, \citenamefont
  {{Antoniadis}}, \citenamefont {{Breton}},\ and\ \citenamefont
  {{Champion}}}]{Tauris2017}%
  \BibitemOpen
  \bibfield  {author} {\bibinfo {author} {\bibfnamefont {T.~M.}\ \bibnamefont
  {{Tauris}}}, \bibinfo {author} {\bibfnamefont {M.}~\bibnamefont {{Kramer}}},
  \bibinfo {author} {\bibfnamefont {P.~C.~C.}\ \bibnamefont {{Freire}}},
  \bibinfo {author} {\bibfnamefont {N.}~\bibnamefont {{Wex}}}, \bibinfo
  {author} {\bibfnamefont {H.-T.}\ \bibnamefont {{Janka}}}, \bibinfo {author}
  {\bibfnamefont {N.}~\bibnamefont {{Langer}}}, \bibinfo {author}
  {\bibfnamefont {P.}~\bibnamefont {{Podsiadlowski}}}, \bibinfo {author}
  {\bibfnamefont {E.}~\bibnamefont {{Bozzo}}}, \bibinfo {author} {\bibfnamefont
  {S.}~\bibnamefont {{Chaty}}}, \bibinfo {author} {\bibfnamefont {M.~U.}\
  \bibnamefont {{Kruckow}}}, \bibinfo {author} {\bibfnamefont {E.~P.~J.}\
  \bibnamefont {{van den Heuvel}}}, \bibinfo {author} {\bibfnamefont
  {J.}~\bibnamefont {{Antoniadis}}}, \bibinfo {author} {\bibfnamefont {R.~P.}\
  \bibnamefont {{Breton}}}, \ and\ \bibinfo {author} {\bibfnamefont {D.~J.}\
  \bibnamefont {{Champion}}},\ }\href {\doibase10.3847/1538-4357/aa7e89}
  {\bibfield  {journal} {\bibinfo  {journal} {The Astrophysical Journal}\
  }\textbf {\bibinfo {volume} {846}},\ \bibinfo {eid} {170} (\bibinfo {year}
  {2017})},\ \Eprint {http://arxiv.org/abs/1706.09438} {arXiv:1706.09438
  [astro-ph.HE]}\BibitemShut {NoStop}%
\bibitem [{\citenamefont {{Zaldarriaga}}\ \emph {et~al.}(2018)\citenamefont
  {{Zaldarriaga}}, \citenamefont {{Kushnir}},\ and\ \citenamefont
  {{Kollmeier}}}]{Zaldarriagaetal2018}%
  \BibitemOpen
  \bibfield  {author} {\bibinfo {author} {\bibfnamefont {M.}~\bibnamefont
  {{Zaldarriaga}}}, \bibinfo {author} {\bibfnamefont {D.}~\bibnamefont
  {{Kushnir}}}, \ and\ \bibinfo {author} {\bibfnamefont {J.~A.}\ \bibnamefont
  {{Kollmeier}}},\ }\href {\doibase10.1093/mnras/stx2577} {\bibfield  {journal}
  {\bibinfo  {journal} {MNRAS}\ }\textbf {\bibinfo {volume} {473}},\ \bibinfo
  {pages} {4174--4178} (\bibinfo {year} {2018})},\ \Eprint
  {http://arxiv.org/abs/1702.00885} {arXiv:1702.00885
  [astro-ph.HE]}\BibitemShut {NoStop}%
\bibitem [{\citenamefont {{Hotokezaka}}\ and\ \citenamefont
  {{Piran}}(2017)}]{Hotokezaka2017}%
  \BibitemOpen
  \bibfield  {author} {\bibinfo {author} {\bibfnamefont {K.}~\bibnamefont
  {{Hotokezaka}}}\ and\ \bibinfo {author} {\bibfnamefont {T.}~\bibnamefont
  {{Piran}}},\ }\href {\doibase10.3847/1538-4357/aa6f61} {\bibfield  {journal}
  {\bibinfo  {journal} {The Astrophysical Journal}\ }\textbf {\bibinfo {volume}
  {842}},\ \bibinfo {eid} {111} (\bibinfo {year} {2017})},\ \Eprint
  {http://arxiv.org/abs/1702.03952} {arXiv:1702.03952
  [astro-ph.HE]}\BibitemShut {NoStop}%
\bibitem [{\citenamefont {Hotekezaka}\ and\ \citenamefont
  {Piran}(2017)}]{Hotokezaka:2017dun}%
  \BibitemOpen
  \bibfield  {author} {\bibinfo {author} {\bibfnamefont {Kenta}\ \bibnamefont
  {Hotekezaka}}\ and\ \bibinfo {author} {\bibfnamefont {Tsvi}\ \bibnamefont
  {Piran}},\ }\href@noop {} {\  (\bibinfo {year} {2017})},\ \Eprint
  {http://arxiv.org/abs/1707.08978} {arXiv:1707.08978
  [astro-ph.HE]}\BibitemShut {NoStop}%
\bibitem [{\citenamefont {Belczynski}\ \emph {et~al.}(2008)\citenamefont
  {Belczynski}, \citenamefont {Taam}, \citenamefont {Rantsiou},\ and\
  \citenamefont {van~der Sluys}}]{Belczynski:2007xg}%
  \BibitemOpen
  \bibfield  {author} {\bibinfo {author} {\bibfnamefont {Krzysztof}\
  \bibnamefont {Belczynski}}, \bibinfo {author} {\bibfnamefont {Ronald~E.}\
  \bibnamefont {Taam}}, \bibinfo {author} {\bibfnamefont {Emmanouela}\
  \bibnamefont {Rantsiou}}, \ and\ \bibinfo {author} {\bibfnamefont {Marc}\
  \bibnamefont {van~der Sluys}},\ }\href {\doibase10.1086/589609} {\bibfield
  {journal} {\bibinfo  {journal} {Astrophys. J.}\ }\textbf {\bibinfo {volume}
  {682}},\ \bibinfo {pages} {474} (\bibinfo {year} {2008})},\ \Eprint
  {http://arxiv.org/abs/astro-ph/0703131} {arXiv:astro-ph/0703131
  [ASTRO-PH]}\BibitemShut {NoStop}%
\bibitem [{\citenamefont {{Ricker}}\ and\ \citenamefont
  {{Taam}}(2008)}]{Ricker2008}%
  \BibitemOpen
  \bibfield  {author} {\bibinfo {author} {\bibfnamefont {P.~M.}\ \bibnamefont
  {{Ricker}}}\ and\ \bibinfo {author} {\bibfnamefont {R.~E.}\ \bibnamefont
  {{Taam}}},\ }\href {\doibase10.1086/526343} {\bibfield  {journal} {\bibinfo
  {journal} {The Astrophysical Journal Letters}\ }\textbf {\bibinfo {volume}
  {672}},\ \bibinfo {pages} {L41--L44} (\bibinfo {year} {2008})}\BibitemShut
  {NoStop}%
\bibitem [{\citenamefont {{MacLeod}}\ \emph {et~al.}(2017)\citenamefont
  {{MacLeod}}, \citenamefont {{Antoni}}, \citenamefont {{Murguia-Berthier}},
  \citenamefont {{Macias}},\ and\ \citenamefont
  {{Ramirez-Ruiz}}}]{MacLeod2017}%
  \BibitemOpen
  \bibfield  {author} {\bibinfo {author} {\bibfnamefont {M.}~\bibnamefont
  {{MacLeod}}}, \bibinfo {author} {\bibfnamefont {A.}~\bibnamefont {{Antoni}}},
  \bibinfo {author} {\bibfnamefont {A.}~\bibnamefont {{Murguia-Berthier}}},
  \bibinfo {author} {\bibfnamefont {P.}~\bibnamefont {{Macias}}}, \ and\
  \bibinfo {author} {\bibfnamefont {E.}~\bibnamefont {{Ramirez-Ruiz}}},\ }\href
  {\doibase10.3847/1538-4357/aa6117} {\bibfield  {journal} {\bibinfo  {journal}
  {The Astrophysical Journal}\ }\textbf {\bibinfo {volume} {838}},\ \bibinfo
  {eid} {56} (\bibinfo {year} {2017})},\ \Eprint
  {http://arxiv.org/abs/1704.02372} {arXiv:1704.02372
  [astro-ph.SR]}\BibitemShut {NoStop}%
\bibitem [{\citenamefont {Murguia-Berthier}\ \emph {et~al.}(2017)\citenamefont
  {Murguia-Berthier}, \citenamefont {MacLeod}, \citenamefont {Ramirez-Ruiz},
  \citenamefont {Antoni},\ and\ \citenamefont
  {Macias}}]{Murguia-Berthier:2017bdf}%
  \BibitemOpen
  \bibfield  {author} {\bibinfo {author} {\bibfnamefont {Ariadna}\ \bibnamefont
  {Murguia-Berthier}}, \bibinfo {author} {\bibfnamefont {Morgan}\ \bibnamefont
  {MacLeod}}, \bibinfo {author} {\bibfnamefont {Enrico}\ \bibnamefont
  {Ramirez-Ruiz}}, \bibinfo {author} {\bibfnamefont {Andrea}\ \bibnamefont
  {Antoni}}, \ and\ \bibinfo {author} {\bibfnamefont {Phillip}\ \bibnamefont
  {Macias}},\ }\href {\doibase10.3847/1538-4357/aa8140} {\bibfield  {journal}
  {\bibinfo  {journal} {Astrophys. J.}\ }\textbf {\bibinfo {volume} {845}},\
  \bibinfo {pages} {173} (\bibinfo {year} {2017})},\ \Eprint
  {http://arxiv.org/abs/1705.04698} {arXiv:1705.04698
  [astro-ph.SR]}\BibitemShut {NoStop}%
\bibitem [{\citenamefont {Holgado}\ \emph {et~al.}(2018)\citenamefont
  {Holgado}, \citenamefont {Ricker},\ and\ \citenamefont
  {Huerta}}]{Holgado:2017vut}%
  \BibitemOpen
  \bibfield  {author} {\bibinfo {author} {\bibfnamefont {A.~Miguel}\
  \bibnamefont {Holgado}}, \bibinfo {author} {\bibfnamefont {Paul~M.}\
  \bibnamefont {Ricker}}, \ and\ \bibinfo {author} {\bibfnamefont {E.~A.}\
  \bibnamefont {Huerta}},\ }\href {\doibase10.3847/1538-4357/aab6a9} {\bibfield
   {journal} {\bibinfo  {journal} {Astrophys. J.}\ }\textbf {\bibinfo {volume}
  {857}},\ \bibinfo {pages} {38} (\bibinfo {year} {2018})},\ \Eprint
  {http://arxiv.org/abs/1706.09413} {arXiv:1706.09413
  [astro-ph.HE]}\BibitemShut {NoStop}%
\bibitem [{\citenamefont {{Tylenda}}\ and\ \citenamefont
  {{Kami{\'n}ski}}(2016)}]{Tylenda2016}%
  \BibitemOpen
  \bibfield  {author} {\bibinfo {author} {\bibfnamefont {R.}~\bibnamefont
  {{Tylenda}}}\ and\ \bibinfo {author} {\bibfnamefont {T.}~\bibnamefont
  {{Kami{\'n}ski}}},\ }\href {\doibase10.1051/0004-6361/201527700} {\bibfield
  {journal} {\bibinfo  {journal} {\aap}\ }\textbf {\bibinfo {volume} {592}},\
  \bibinfo {eid} {A134} (\bibinfo {year} {2016})},\ \Eprint
  {http://arxiv.org/abs/1606.09426} {arXiv:1606.09426
  [astro-ph.SR]}\BibitemShut {NoStop}%
\bibitem [{\citenamefont {Oskinova}\ \emph {et~al.}(2018)\citenamefont
  {Oskinova}, \citenamefont {Bulik},\ and\ \citenamefont
  {Gómez-Morán}}]{Oskinova:2018sfn}%
  \BibitemOpen
  \bibfield  {author} {\bibinfo {author} {\bibfnamefont {Lidia~M.}\
  \bibnamefont {Oskinova}}, \bibinfo {author} {\bibfnamefont {Tomasz}\
  \bibnamefont {Bulik}}, \ and\ \bibinfo {author} {\bibfnamefont {Ada~Nebot}\
  \bibnamefont {Gómez-Morán}},\ }\href {\doibase10.1051/0004-6361/201832925}
  {\bibfield  {journal} {\bibinfo  {journal} {Astron. Astrophys.}\ }\textbf
  {\bibinfo {volume} {613}},\ \bibinfo {pages} {L10} (\bibinfo {year}
  {2018})},\ \Eprint {http://arxiv.org/abs/1805.08080} {arXiv:1805.08080
  [astro-ph.HE]}\BibitemShut {NoStop}%
\bibitem [{\citenamefont {Belczynski}\ \emph
  {et~al.}(2016{\natexlab{c}})\citenamefont {Belczynski}, \citenamefont
  {Repetto}, \citenamefont {Holz}, \citenamefont {O'Shaughnessy}, \citenamefont
  {Bulik}, \citenamefont {Berti}, \citenamefont {Fryer},\ and\ \citenamefont
  {Dominik}}]{Belczynski:2015tba}%
  \BibitemOpen
  \bibfield  {author} {\bibinfo {author} {\bibfnamefont {Krzysztof}\
  \bibnamefont {Belczynski}}, \bibinfo {author} {\bibfnamefont {Serena}\
  \bibnamefont {Repetto}}, \bibinfo {author} {\bibfnamefont {Daniel~E.}\
  \bibnamefont {Holz}}, \bibinfo {author} {\bibfnamefont {Richard}\
  \bibnamefont {O'Shaughnessy}}, \bibinfo {author} {\bibfnamefont {Tomasz}\
  \bibnamefont {Bulik}}, \bibinfo {author} {\bibfnamefont {Emanuele}\
  \bibnamefont {Berti}}, \bibinfo {author} {\bibfnamefont {Christopher}\
  \bibnamefont {Fryer}}, \ and\ \bibinfo {author} {\bibfnamefont {Michal}\
  \bibnamefont {Dominik}},\ }\href {\doibase10.3847/0004-637X/819/2/108}
  {\bibfield  {journal} {\bibinfo  {journal} {Astrophys. J.}\ }\textbf
  {\bibinfo {volume} {819}},\ \bibinfo {pages} {108} (\bibinfo {year}
  {2016}{\natexlab{c}})},\ \Eprint {http://arxiv.org/abs/1510.04615}
  {arXiv:1510.04615 [astro-ph.HE]}\BibitemShut {NoStop}%
\bibitem [{\citenamefont {{Janka}}\ \emph {et~al.}(2016)\citenamefont
  {{Janka}}, \citenamefont {{Melson}},\ and\ \citenamefont
  {{Summa}}}]{Janka2016}%
  \BibitemOpen
  \bibfield  {author} {\bibinfo {author} {\bibfnamefont {H.-T.}\ \bibnamefont
  {{Janka}}}, \bibinfo {author} {\bibfnamefont {T.}~\bibnamefont {{Melson}}}, \
  and\ \bibinfo {author} {\bibfnamefont {A.}~\bibnamefont {{Summa}}},\ }\href
  {\doibase10.1146/annurev-nucl-102115-044747} {\bibfield  {journal} {\bibinfo
  {journal} {Annual Review of Nuclear and Particle Science}\ }\textbf {\bibinfo
  {volume} {66}},\ \bibinfo {pages} {341--375} (\bibinfo {year} {2016})},\
  \Eprint {http://arxiv.org/abs/1602.05576} {arXiv:1602.05576
  [astro-ph.SR]}\BibitemShut {NoStop}%
\bibitem [{\citenamefont {{Connaughton}}\ \emph {et~al.}(2016)\citenamefont
  {{Connaughton}}, \citenamefont {{Burns}}, \citenamefont {{Goldstein}},\ and\
  \citenamefont {{et~al}}}]{cbg+16}%
  \BibitemOpen
  \bibfield  {author} {\bibinfo {author} {\bibfnamefont {V.}~\bibnamefont
  {{Connaughton}}}, \bibinfo {author} {\bibfnamefont {E.}~\bibnamefont
  {{Burns}}}, \bibinfo {author} {\bibfnamefont {A.}~\bibnamefont
  {{Goldstein}}}, \ and\ \bibinfo {author} {\bibnamefont {{et~al}}},\ }\href
  {\doibase10.3847/2041-8205/826/1/L6} {\bibfield  {journal} {\bibinfo
  {journal} {The Astrophysical Journal Letters}\ }\textbf {\bibinfo {volume}
  {826}},\ \bibinfo {eid} {L6} (\bibinfo {year} {2016})},\ \Eprint
  {http://arxiv.org/abs/1602.03920} {arXiv:1602.03920
  [astro-ph.HE]}\BibitemShut {NoStop}%
\bibitem [{\citenamefont {{de Mink}}\ and\ \citenamefont
  {{King}}(2017)}]{dk17}%
  \BibitemOpen
  \bibfield  {author} {\bibinfo {author} {\bibfnamefont {S.~E.}\ \bibnamefont
  {{de Mink}}}\ and\ \bibinfo {author} {\bibfnamefont {A.}~\bibnamefont
  {{King}}},\ }\href {\doibase10.3847/2041-8213/aa67f3} {\bibfield  {journal}
  {\bibinfo  {journal} {The Astrophysical Journal Letters}\ }\textbf {\bibinfo
  {volume} {839}},\ \bibinfo {eid} {L7} (\bibinfo {year} {2017})},\ \Eprint
  {http://arxiv.org/abs/1703.07794} {arXiv:1703.07794
  [astro-ph.HE]}\BibitemShut {NoStop}%
\bibitem [{\citenamefont {{Hulse}}\ and\ \citenamefont
  {{Taylor}}(1975)}]{ht75}%
  \BibitemOpen
  \bibfield  {author} {\bibinfo {author} {\bibfnamefont {R.~A.}\ \bibnamefont
  {{Hulse}}}\ and\ \bibinfo {author} {\bibfnamefont {J.~H.}\ \bibnamefont
  {{Taylor}}},\ }\href {\doibase10.1086/181708} {\bibfield  {journal} {\bibinfo
   {journal} {The Astrophysical Journal}\ }\textbf {\bibinfo {volume} {195}},\
  \bibinfo {pages} {L51--L53} (\bibinfo {year} {1975})}\BibitemShut {NoStop}%
\bibitem [{\citenamefont {{Lorimer}}\ and\ \citenamefont
  {{Kramer}}(2012)}]{lk12}%
  \BibitemOpen
  \bibfield  {author} {\bibinfo {author} {\bibfnamefont {D.~R.}\ \bibnamefont
  {{Lorimer}}}\ and\ \bibinfo {author} {\bibfnamefont {M.}~\bibnamefont
  {{Kramer}}},\ }\href@noop {} {\emph {\bibinfo {title} {{Handbook of Pulsar
  Astronomy}}}}\ (\bibinfo  {publisher} {Cambridge University Press},\ \bibinfo
  {year} {2012})\BibitemShut {NoStop}%
\bibitem [{\citenamefont {Belczynski}\ \emph
  {et~al.}(2017{\natexlab{c}})\citenamefont {Belczynski} \emph
  {et~al.}}]{Belczynski:2017mqx}%
  \BibitemOpen
  \bibfield  {author} {\bibinfo {author} {\bibfnamefont {K.}~\bibnamefont
  {Belczynski}} \emph {et~al.},\ }\href@noop {} {\  (\bibinfo {year}
  {2017}{\natexlab{c}})},\ \Eprint {http://arxiv.org/abs/1712.00632}
  {arXiv:1712.00632 [astro-ph.HE]}\BibitemShut {NoStop}%
\bibitem [{\citenamefont {{Chruslinska}}\ \emph {et~al.}(2018)\citenamefont
  {{Chruslinska}}, \citenamefont {{Belczynski}}, \citenamefont {{Klencki}},\
  and\ \citenamefont {{Benacquista}}}]{Chruslinska2018}%
  \BibitemOpen
  \bibfield  {author} {\bibinfo {author} {\bibfnamefont {M.}~\bibnamefont
  {{Chruslinska}}}, \bibinfo {author} {\bibfnamefont {K.}~\bibnamefont
  {{Belczynski}}}, \bibinfo {author} {\bibfnamefont {J.}~\bibnamefont
  {{Klencki}}}, \ and\ \bibinfo {author} {\bibfnamefont {M.}~\bibnamefont
  {{Benacquista}}},\ }\href {\doibase10.1093/mnras/stx2923} {\bibfield
  {journal} {\bibinfo  {journal} {Monthly Notices of The Royal Astronomical
  Society}\ }\textbf {\bibinfo {volume} {474}},\ \bibinfo {pages} {2937--2958}
  (\bibinfo {year} {2018})},\ \Eprint {http://arxiv.org/abs/1708.07885}
  {arXiv:1708.07885 [astro-ph.HE]}\BibitemShut {NoStop}%
\bibitem [{\citenamefont {Vigna-Gómez}\ \emph {et~al.}(2018)\citenamefont
  {Vigna-Gómez} \emph {et~al.}}]{Vigna-Gomez:2018dza}%
  \BibitemOpen
  \bibfield  {author} {\bibinfo {author} {\bibfnamefont {Alejandro}\
  \bibnamefont {Vigna-Gómez}} \emph {et~al.},\ }\href@noop {} {\  (\bibinfo
  {year} {2018})},\ \Eprint {http://arxiv.org/abs/1805.07974} {arXiv:1805.07974
  [astro-ph.SR]}\BibitemShut {NoStop}%
\bibitem [{\citenamefont {{Kruckow}}\ \emph {et~al.}(2016)\citenamefont
  {{Kruckow}}, \citenamefont {{Tauris}}, \citenamefont {{Langer}},
  \citenamefont {{Sz{\'e}csi}}, \citenamefont {{Marchant}},\ and\ \citenamefont
  {{Podsiadlowski}}}]{ktl+16}%
  \BibitemOpen
  \bibfield  {author} {\bibinfo {author} {\bibfnamefont {M.~U.}\ \bibnamefont
  {{Kruckow}}}, \bibinfo {author} {\bibfnamefont {T.~M.}\ \bibnamefont
  {{Tauris}}}, \bibinfo {author} {\bibfnamefont {N.}~\bibnamefont {{Langer}}},
  \bibinfo {author} {\bibfnamefont {D.}~\bibnamefont {{Sz{\'e}csi}}}, \bibinfo
  {author} {\bibfnamefont {P.}~\bibnamefont {{Marchant}}}, \ and\ \bibinfo
  {author} {\bibfnamefont {P.}~\bibnamefont {{Podsiadlowski}}},\ }\href
  {\doibase10.1051/0004-6361/201629420} {\bibfield  {journal} {\bibinfo
  {journal} {\aap}\ }\textbf {\bibinfo {volume} {596}},\ \bibinfo {eid} {A58}
  (\bibinfo {year} {2016})},\ \Eprint {http://arxiv.org/abs/1610.04417}
  {arXiv:1610.04417 [astro-ph.SR]}\BibitemShut {NoStop}%
\bibitem [{\citenamefont {{Verbunt}}\ \emph {et~al.}(2017)\citenamefont
  {{Verbunt}}, \citenamefont {{Igoshev}},\ and\ \citenamefont
  {{Cator}}}]{Verbunt18}%
  \BibitemOpen
  \bibfield  {author} {\bibinfo {author} {\bibfnamefont {F.}~\bibnamefont
  {{Verbunt}}}, \bibinfo {author} {\bibfnamefont {A.}~\bibnamefont
  {{Igoshev}}}, \ and\ \bibinfo {author} {\bibfnamefont {E.}~\bibnamefont
  {{Cator}}},\ }\href {\doibase10.1051/0004-6361/201731518} {\bibfield
  {journal} {\bibinfo  {journal} {\aap}\ }\textbf {\bibinfo {volume} {608}},\
  \bibinfo {eid} {A57} (\bibinfo {year} {2017})},\ \Eprint
  {http://arxiv.org/abs/1708.08281} {arXiv:1708.08281
  [astro-ph.HE]}\BibitemShut {NoStop}%
\bibitem [{\citenamefont {{Soares-Santos}}\ \emph {et~al.}(2017)\citenamefont
  {{Soares-Santos}}, \citenamefont {{Holz}}, \citenamefont {{Annis}},\ and\
  \citenamefont {{et~al.}}}]{sha+17}%
  \BibitemOpen
  \bibfield  {author} {\bibinfo {author} {\bibfnamefont {M.}~\bibnamefont
  {{Soares-Santos}}}, \bibinfo {author} {\bibfnamefont {D.~E.}\ \bibnamefont
  {{Holz}}}, \bibinfo {author} {\bibfnamefont {J.}~\bibnamefont {{Annis}}}, \
  and\ \bibinfo {author} {\bibnamefont {{et~al.}}},\ }\href
  {\doibase10.3847/2041-8213/aa9059} {\bibfield  {journal} {\bibinfo  {journal}
  {The Astrophysical Journal Letters}\ }\textbf {\bibinfo {volume} {848}},\
  \bibinfo {eid} {L16} (\bibinfo {year} {2017})},\ \Eprint
  {http://arxiv.org/abs/1710.05459} {arXiv:1710.05459
  [astro-ph.HE]}\BibitemShut {NoStop}%
\bibitem [{\citenamefont {Coulter}\ \emph {et~al.}(2017)\citenamefont {Coulter}
  \emph {et~al.}}]{Coulter:2017wya}%
  \BibitemOpen
  \bibfield  {author} {\bibinfo {author} {\bibfnamefont {D.~A.}\ \bibnamefont
  {Coulter}} \emph {et~al.},\ }\href {\doibase10.1126/science.aap9811}
  {\bibfield  {journal} {\bibinfo  {journal} {Science}\ } (\bibinfo {year}
  {2017}),\ 10.1126/science.aap9811},\ \bibinfo {note}
  {[Science358,1556(2017)]},\ \Eprint {http://arxiv.org/abs/1710.05452}
  {arXiv:1710.05452 [astro-ph.HE]}\BibitemShut {NoStop}%
\bibitem [{\citenamefont {{Fong}}\ and\ \citenamefont {{Berger}}(2013)}]{fb13}%
  \BibitemOpen
  \bibfield  {author} {\bibinfo {author} {\bibfnamefont {W.}~\bibnamefont
  {{Fong}}}\ and\ \bibinfo {author} {\bibfnamefont {E.}~\bibnamefont
  {{Berger}}},\ }\href {\doibase10.1088/0004-637X/776/1/18} {\bibfield
  {journal} {\bibinfo  {journal} {The Astrophysical Journal}\ }\textbf
  {\bibinfo {volume} {776}},\ \bibinfo {eid} {18} (\bibinfo {year} {2013})},\
  \Eprint {http://arxiv.org/abs/1307.0819} {arXiv:1307.0819
  [astro-ph.HE]}\BibitemShut {NoStop}%
\bibitem [{\citenamefont {{Metzger}}\ \emph {et~al.}(2010)\citenamefont
  {{Metzger}}, \citenamefont {{Mart{\'{\i}}nez-Pinedo}}, \citenamefont
  {{Darbha}}, \citenamefont {{Quataert}}, \citenamefont {{Arcones}},
  \citenamefont {{Kasen}}, \citenamefont {{Thomas}}, \citenamefont {{Nugent}},
  \citenamefont {{Panov}},\ and\ \citenamefont
  {{Zinner}}}]{2010MNRAS.406.2650M}%
  \BibitemOpen
  \bibfield  {author} {\bibinfo {author} {\bibfnamefont {B.~D.}\ \bibnamefont
  {{Metzger}}}, \bibinfo {author} {\bibfnamefont {G.}~\bibnamefont
  {{Mart{\'{\i}}nez-Pinedo}}}, \bibinfo {author} {\bibfnamefont
  {S.}~\bibnamefont {{Darbha}}}, \bibinfo {author} {\bibfnamefont
  {E.}~\bibnamefont {{Quataert}}}, \bibinfo {author} {\bibfnamefont
  {A.}~\bibnamefont {{Arcones}}}, \bibinfo {author} {\bibfnamefont
  {D.}~\bibnamefont {{Kasen}}}, \bibinfo {author} {\bibfnamefont
  {R.}~\bibnamefont {{Thomas}}}, \bibinfo {author} {\bibfnamefont
  {P.}~\bibnamefont {{Nugent}}}, \bibinfo {author} {\bibfnamefont {I.~V.}\
  \bibnamefont {{Panov}}}, \ and\ \bibinfo {author} {\bibfnamefont {N.~T.}\
  \bibnamefont {{Zinner}}},\ }\href {\doibase10.1111/j.1365-2966.2010.16864.x}
  {\bibfield  {journal} {\bibinfo  {journal} {MNRAS}\ }\textbf {\bibinfo
  {volume} {406}},\ \bibinfo {pages} {2650--2662} (\bibinfo {year} {2010})},\
  \Eprint {http://arxiv.org/abs/1001.5029} {arXiv:1001.5029
  [astro-ph.HE]}\BibitemShut {NoStop}%
\bibitem [{\citenamefont {{Will}}(1994)}]{wil94}%
  \BibitemOpen
  \bibfield  {author} {\bibinfo {author} {\bibfnamefont {C.~M.}\ \bibnamefont
  {{Will}}},\ }\href {\doibase10.1103/PhysRevD.50.6058} {\bibfield  {journal}
  {\bibinfo  {journal} {Physical Review D}\ }\textbf {\bibinfo {volume} {50}},\
  \bibinfo {pages} {6058--6067} (\bibinfo {year} {1994})},\ \Eprint
  {http://arxiv.org/abs/gr-qc/9406022} {gr-qc/9406022}\BibitemShut {NoStop}%
\bibitem [{\citenamefont {{Farr}}\ \emph {et~al.}(2011)\citenamefont {{Farr}},
  \citenamefont {{Sravan}}, \citenamefont {{Cantrell}}, \citenamefont
  {{Kreidberg}}, \citenamefont {{Bailyn}}, \citenamefont {{Mandel}},\ and\
  \citenamefont {{Kalogera}}}]{Farretal2011}%
  \BibitemOpen
  \bibfield  {author} {\bibinfo {author} {\bibfnamefont {W.~M.}\ \bibnamefont
  {{Farr}}}, \bibinfo {author} {\bibfnamefont {N.}~\bibnamefont {{Sravan}}},
  \bibinfo {author} {\bibfnamefont {A.}~\bibnamefont {{Cantrell}}}, \bibinfo
  {author} {\bibfnamefont {L.}~\bibnamefont {{Kreidberg}}}, \bibinfo {author}
  {\bibfnamefont {C.~D.}\ \bibnamefont {{Bailyn}}}, \bibinfo {author}
  {\bibfnamefont {I.}~\bibnamefont {{Mandel}}}, \ and\ \bibinfo {author}
  {\bibfnamefont {V.}~\bibnamefont {{Kalogera}}},\ }\href
  {\doibase10.1088/0004-637X/741/2/103} {\bibfield  {journal} {\bibinfo
  {journal} {Astrophysical Journal}\ }\textbf {\bibinfo {volume} {741}},\
  \bibinfo {eid} {103} (\bibinfo {year} {2011})},\ \Eprint
  {http://arxiv.org/abs/1011.1459} {arXiv:1011.1459}\BibitemShut {NoStop}%
\bibitem [{\citenamefont {{Rosenblatt}}\ \emph {et~al.}(1988)\citenamefont
  {{Rosenblatt}}, \citenamefont {{Faber}},\ and\ \citenamefont
  {{Blumenthal}}}]{Rosenblatt1988}%
  \BibitemOpen
  \bibfield  {author} {\bibinfo {author} {\bibfnamefont {E.~I.}\ \bibnamefont
  {{Rosenblatt}}}, \bibinfo {author} {\bibfnamefont {S.~M.}\ \bibnamefont
  {{Faber}}}, \ and\ \bibinfo {author} {\bibfnamefont {G.~R.}\ \bibnamefont
  {{Blumenthal}}},\ }\href {\doibase10.1086/166466} {\bibfield  {journal}
  {\bibinfo  {journal} {The Astrophysical Journal}\ }\textbf {\bibinfo {volume}
  {330}},\ \bibinfo {pages} {191--200} (\bibinfo {year} {1988})}\BibitemShut
  {NoStop}%
\bibitem [{\citenamefont {{B{\"o}ker}}(2008)}]{Boker2008}%
  \BibitemOpen
  \bibfield  {author} {\bibinfo {author} {\bibfnamefont {T.}~\bibnamefont
  {{B{\"o}ker}}},\ }\href {\doibase10.1086/527033} {\bibfield  {journal}
  {\bibinfo  {journal} {The Astrophysical Journal Letters}\ }\textbf {\bibinfo
  {volume} {672}},\ \bibinfo {pages} {L111} (\bibinfo {year} {2008})},\ \Eprint
  {http://arxiv.org/abs/0711.4542} {arXiv:0711.4542}\BibitemShut {NoStop}%
\bibitem [{\citenamefont {{Harris}}\ \emph {et~al.}(2015)\citenamefont
  {{Harris}}, \citenamefont {{Harris}},\ and\ \citenamefont
  {{Hudson}}}]{Harris2015}%
  \BibitemOpen
  \bibfield  {author} {\bibinfo {author} {\bibfnamefont {W.~E.}\ \bibnamefont
  {{Harris}}}, \bibinfo {author} {\bibfnamefont {G.~L.}\ \bibnamefont
  {{Harris}}}, \ and\ \bibinfo {author} {\bibfnamefont {M.~J.}\ \bibnamefont
  {{Hudson}}},\ }\href {\doibase10.1088/0004-637X/806/1/36} {\bibfield
  {journal} {\bibinfo  {journal} {The Astrophysical Journal}\ }\textbf
  {\bibinfo {volume} {806}},\ \bibinfo {eid} {36} (\bibinfo {year} {2015})},\
  \Eprint {http://arxiv.org/abs/1504.03199} {arXiv:1504.03199}\BibitemShut
  {NoStop}%
\bibitem [{\citenamefont {{Spitzer}}(1969)}]{Spitzer1969}%
  \BibitemOpen
  \bibfield  {author} {\bibinfo {author} {\bibfnamefont {L.}~\bibnamefont
  {{Spitzer}}, \bibfnamefont {Jr.}},\ }\href {\doibase10.1086/180451}
  {\bibfield  {journal} {\bibinfo  {journal} {The Astrophysical Journal
  Letters}\ }\textbf {\bibinfo {volume} {158}},\ \bibinfo {pages} {L139}
  (\bibinfo {year} {1969})}\BibitemShut {NoStop}%
\bibitem [{\citenamefont {{Sigurdsson}}\ and\ \citenamefont
  {{Hernquist}}(1993)}]{Sigurdsson1993}%
  \BibitemOpen
  \bibfield  {author} {\bibinfo {author} {\bibfnamefont {S.}~\bibnamefont
  {{Sigurdsson}}}\ and\ \bibinfo {author} {\bibfnamefont {L.}~\bibnamefont
  {{Hernquist}}},\ }\href {\doibase10.1038/364423a0} {\bibfield  {journal}
  {\bibinfo  {journal} {Nature}\ }\textbf {\bibinfo {volume} {364}},\ \bibinfo
  {pages} {423--425} (\bibinfo {year} {1993})}\BibitemShut {NoStop}%
\bibitem [{\citenamefont {{Ivanova}}\ \emph {et~al.}(2005)\citenamefont
  {{Ivanova}}, \citenamefont {{Belczynski}}, \citenamefont {{Fregeau}},\ and\
  \citenamefont {{Rasio}}}]{Ivanova2005}%
  \BibitemOpen
  \bibfield  {author} {\bibinfo {author} {\bibfnamefont {N.}~\bibnamefont
  {{Ivanova}}}, \bibinfo {author} {\bibfnamefont {K.}~\bibnamefont
  {{Belczynski}}}, \bibinfo {author} {\bibfnamefont {J.~M.}\ \bibnamefont
  {{Fregeau}}}, \ and\ \bibinfo {author} {\bibfnamefont {F.~A.}\ \bibnamefont
  {{Rasio}}},\ }\href {\doibase10.1111/j.1365-2966.2005.08804.x} {\bibfield
  {journal} {\bibinfo  {journal} {Monthly Notices of The Royal Astronomical
  Society}\ }\textbf {\bibinfo {volume} {358}},\ \bibinfo {pages} {572--584}
  (\bibinfo {year} {2005})},\ \Eprint {http://arxiv.org/abs/astro-ph/0501131}
  {astro-ph/0501131}\BibitemShut {NoStop}%
\bibitem [{\citenamefont {{Park}}\ \emph {et~al.}(2017)\citenamefont {{Park}},
  \citenamefont {{Kim}}, \citenamefont {{Lee}}, \citenamefont {{Bae}},\ and\
  \citenamefont {{Belczynski}}}]{Park2017}%
  \BibitemOpen
  \bibfield  {author} {\bibinfo {author} {\bibfnamefont {D.}~\bibnamefont
  {{Park}}}, \bibinfo {author} {\bibfnamefont {C.}~\bibnamefont {{Kim}}},
  \bibinfo {author} {\bibfnamefont {H.~M.}\ \bibnamefont {{Lee}}}, \bibinfo
  {author} {\bibfnamefont {Y.-B.}\ \bibnamefont {{Bae}}}, \ and\ \bibinfo
  {author} {\bibfnamefont {K.}~\bibnamefont {{Belczynski}}},\ }\href
  {\doibase10.1093/mnras/stx1015} {\bibfield  {journal} {\bibinfo  {journal}
  {Monthly Notices of The Royal Astronomical Society}\ }\textbf {\bibinfo
  {volume} {469}},\ \bibinfo {pages} {4665--4674} (\bibinfo {year} {2017})},\
  \Eprint {http://arxiv.org/abs/1703.01568} {arXiv:1703.01568
  [astro-ph.HE]}\BibitemShut {NoStop}%
\bibitem [{\citenamefont
  {{Mapelli}}(2016{\natexlab{b}})}]{2016MNRAS.459.3432M}%
  \BibitemOpen
  \bibfield  {author} {\bibinfo {author} {\bibfnamefont {M.}~\bibnamefont
  {{Mapelli}}},\ }\href {\doibase10.1093/mnras/stw869} {\bibfield  {journal}
  {\bibinfo  {journal} {Monthly Notices of The Royal Astronomical Society}\
  }\textbf {\bibinfo {volume} {459}},\ \bibinfo {pages} {3432--3446} (\bibinfo
  {year} {2016}{\natexlab{b}})},\ \Eprint {http://arxiv.org/abs/1604.03559}
  {arXiv:1604.03559}\BibitemShut {NoStop}%
\bibitem [{\citenamefont {{Banerjee}}(2017)}]{2017MNRAS.467..524B}%
  \BibitemOpen
  \bibfield  {author} {\bibinfo {author} {\bibfnamefont {S.}~\bibnamefont
  {{Banerjee}}},\ }\href {\doibase10.1093/mnras/stw3392} {\bibfield  {journal}
  {\bibinfo  {journal} {Monthly Notices of The Royal Astronomical Society}\
  }\textbf {\bibinfo {volume} {467}},\ \bibinfo {pages} {524--539} (\bibinfo
  {year} {2017})},\ \Eprint {http://arxiv.org/abs/1611.09357} {arXiv:1611.09357
  [astro-ph.HE]}\BibitemShut {NoStop}%
\bibitem [{\citenamefont {{Kroupa}}(2001)}]{2001MNRAS.322..231K}%
  \BibitemOpen
  \bibfield  {author} {\bibinfo {author} {\bibfnamefont {P.}~\bibnamefont
  {{Kroupa}}},\ }\href {\doibase10.1046/j.1365-8711.2001.04022.x} {\bibfield
  {journal} {\bibinfo  {journal} {Monthly Notices of The Royal Astronomical
  Society}\ }\textbf {\bibinfo {volume} {322}},\ \bibinfo {pages} {231--246}
  (\bibinfo {year} {2001})},\ \Eprint {http://arxiv.org/abs/astro-ph/0009005}
  {astro-ph/0009005}\BibitemShut {NoStop}%
\bibitem [{\citenamefont {{Gnedin}}\ \emph {et~al.}(2014)\citenamefont
  {{Gnedin}}, \citenamefont {{Ostriker}},\ and\ \citenamefont
  {{Tremaine}}}]{Gnedinetal2014}%
  \BibitemOpen
  \bibfield  {author} {\bibinfo {author} {\bibfnamefont {O.~Y.}\ \bibnamefont
  {{Gnedin}}}, \bibinfo {author} {\bibfnamefont {J.~P.}\ \bibnamefont
  {{Ostriker}}}, \ and\ \bibinfo {author} {\bibfnamefont {S.}~\bibnamefont
  {{Tremaine}}},\ }\href {\doibase10.1088/0004-637X/785/1/71} {\bibfield
  {journal} {\bibinfo  {journal} {Astrophysical Journal}\ }\textbf {\bibinfo
  {volume} {785}},\ \bibinfo {eid} {71} (\bibinfo {year} {2014})},\ \Eprint
  {http://arxiv.org/abs/1308.0021} {arXiv:1308.0021}\BibitemShut {NoStop}%
\bibitem [{\citenamefont {{Arca-Sedda}}\ and\ \citenamefont
  {{Capuzzo-Dolcetta}}(2014)}]{2014MNRAS.444.3738A}%
  \BibitemOpen
  \bibfield  {author} {\bibinfo {author} {\bibfnamefont {M.}~\bibnamefont
  {{Arca-Sedda}}}\ and\ \bibinfo {author} {\bibfnamefont {R.}~\bibnamefont
  {{Capuzzo-Dolcetta}}},\ }\href {\doibase10.1093/mnras/stu1683} {\bibfield
  {journal} {\bibinfo  {journal} {Monthly Notices of The Royal Astronomical
  Society}\ }\textbf {\bibinfo {volume} {444}},\ \bibinfo {pages} {3738--3755}
  (\bibinfo {year} {2014})},\ \Eprint {http://arxiv.org/abs/1405.7593}
  {arXiv:1405.7593}\BibitemShut {NoStop}%
\bibitem [{\citenamefont {Fragione}\ and\ \citenamefont
  {Kocsis}(2018)}]{Fragione:2018vty}%
  \BibitemOpen
  \bibfield  {author} {\bibinfo {author} {\bibfnamefont {Giacomo}\ \bibnamefont
  {Fragione}}\ and\ \bibinfo {author} {\bibfnamefont {Bence}\ \bibnamefont
  {Kocsis}},\ }\href@noop {} {\  (\bibinfo {year} {2018})},\ \Eprint
  {http://arxiv.org/abs/1806.02351} {arXiv:1806.02351
  [astro-ph.GA]}\BibitemShut {NoStop}%
\bibitem [{\citenamefont {{O'Leary}}\ \emph {et~al.}(2009)\citenamefont
  {{O'Leary}}, \citenamefont {{Kocsis}},\ and\ \citenamefont
  {{Loeb}}}]{OLearyetal2009}%
  \BibitemOpen
  \bibfield  {author} {\bibinfo {author} {\bibfnamefont {R.~M.}\ \bibnamefont
  {{O'Leary}}}, \bibinfo {author} {\bibfnamefont {B.}~\bibnamefont {{Kocsis}}},
  \ and\ \bibinfo {author} {\bibfnamefont {A.}~\bibnamefont {{Loeb}}},\ }\href
  {\doibase10.1111/j.1365-2966.2009.14653.x} {\bibfield  {journal} {\bibinfo
  {journal} {MNRAS}\ }\textbf {\bibinfo {volume} {395}},\ \bibinfo {pages}
  {2127--2146} (\bibinfo {year} {2009})},\ \Eprint
  {http://arxiv.org/abs/0807.2638} {arXiv:0807.2638}\BibitemShut {NoStop}%
\bibitem [{\citenamefont {{Antonini}}\ and\ \citenamefont
  {{Perets}}(2012)}]{Antonini_Perets2012}%
  \BibitemOpen
  \bibfield  {author} {\bibinfo {author} {\bibfnamefont {F.}~\bibnamefont
  {{Antonini}}}\ and\ \bibinfo {author} {\bibfnamefont {H.~B.}\ \bibnamefont
  {{Perets}}},\ }\href {\doibase10.1088/0004-637X/757/1/27} {\bibfield
  {journal} {\bibinfo  {journal} {Astrophysical Journal}\ }\textbf {\bibinfo
  {volume} {757}},\ \bibinfo {eid} {27} (\bibinfo {year} {2012})},\ \Eprint
  {http://arxiv.org/abs/1203.2938} {arXiv:1203.2938}\BibitemShut {NoStop}%
\bibitem [{\citenamefont {Hoang}\ \emph {et~al.}(2018)\citenamefont {Hoang},
  \citenamefont {Naoz}, \citenamefont {Kocsis}, \citenamefont {Rasio},\ and\
  \citenamefont {Dosopoulou}}]{Hoang:2017fvh}%
  \BibitemOpen
  \bibfield  {author} {\bibinfo {author} {\bibfnamefont {Bao-Minh}\
  \bibnamefont {Hoang}}, \bibinfo {author} {\bibfnamefont {Smadar}\
  \bibnamefont {Naoz}}, \bibinfo {author} {\bibfnamefont {Bence}\ \bibnamefont
  {Kocsis}}, \bibinfo {author} {\bibfnamefont {Frederic~A.}\ \bibnamefont
  {Rasio}}, \ and\ \bibinfo {author} {\bibfnamefont {Fani}\ \bibnamefont
  {Dosopoulou}},\ }\href {\doibase10.3847/1538-4357/aaafce} {\bibfield
  {journal} {\bibinfo  {journal} {Astrophys. J.}\ }\textbf {\bibinfo {volume}
  {856}},\ \bibinfo {pages} {140} (\bibinfo {year} {2018})},\ \Eprint
  {http://arxiv.org/abs/1706.09896} {arXiv:1706.09896
  [astro-ph.HE]}\BibitemShut {NoStop}%
\bibitem [{\citenamefont {{Petrovich}}\ and\ \citenamefont
  {{Antonini}}(2017)}]{PetrovichAntonini2017}%
  \BibitemOpen
  \bibfield  {author} {\bibinfo {author} {\bibfnamefont {C.}~\bibnamefont
  {{Petrovich}}}\ and\ \bibinfo {author} {\bibfnamefont {F.}~\bibnamefont
  {{Antonini}}},\ }\href {\doibase10.3847/1538-4357/aa8628} {\bibfield
  {journal} {\bibinfo  {journal} {The Astrophysical Journal}\ }\textbf
  {\bibinfo {volume} {846}},\ \bibinfo {eid} {146} (\bibinfo {year} {2017})},\
  \Eprint {http://arxiv.org/abs/1705.05848} {arXiv:1705.05848
  [astro-ph.HE]}\BibitemShut {NoStop}%
\bibitem [{\citenamefont {{Arca-Sedda}}\ and\ \citenamefont
  {{Gualandris}}(2018)}]{2018MNRAS.477.4423A}%
  \BibitemOpen
  \bibfield  {author} {\bibinfo {author} {\bibfnamefont {M.}~\bibnamefont
  {{Arca-Sedda}}}\ and\ \bibinfo {author} {\bibfnamefont {A.}~\bibnamefont
  {{Gualandris}}},\ }\href {\doibase10.1093/mnras/sty922} {\bibfield  {journal}
  {\bibinfo  {journal} {Monthly Notices of The Royal Astronomical Society}\
  }\textbf {\bibinfo {volume} {477}},\ \bibinfo {pages} {4423--4442} (\bibinfo
  {year} {2018})},\ \Eprint {http://arxiv.org/abs/1804.06116}
  {arXiv:1804.06116}\BibitemShut {NoStop}%
\bibitem [{\citenamefont {Arca-Sedda}\ and\ \citenamefont
  {Capuzzo-Dolcetta}(2017)}]{Arca-Sedda:2017wea}%
  \BibitemOpen
  \bibfield  {author} {\bibinfo {author} {\bibfnamefont {M.}~\bibnamefont
  {Arca-Sedda}}\ and\ \bibinfo {author} {\bibfnamefont {Roberto}\ \bibnamefont
  {Capuzzo-Dolcetta}},\ }\href@noop {} {\  (\bibinfo {year} {2017})},\ \Eprint
  {http://arxiv.org/abs/1709.05567} {arXiv:1709.05567
  [astro-ph.GA]}\BibitemShut {NoStop}%
\bibitem [{\citenamefont {Hamers}\ \emph {et~al.}(2018)\citenamefont {Hamers},
  \citenamefont {Bar-Or}, \citenamefont {Petrovich},\ and\ \citenamefont
  {Antonini}}]{Hamers:2018hxv}%
  \BibitemOpen
  \bibfield  {author} {\bibinfo {author} {\bibfnamefont {Adrian~S.}\
  \bibnamefont {Hamers}}, \bibinfo {author} {\bibfnamefont {Ben}\ \bibnamefont
  {Bar-Or}}, \bibinfo {author} {\bibfnamefont {Cristobal}\ \bibnamefont
  {Petrovich}}, \ and\ \bibinfo {author} {\bibfnamefont {Fabio}\ \bibnamefont
  {Antonini}},\ }\href@noop {} {\  (\bibinfo {year} {2018})},\ \Eprint
  {http://arxiv.org/abs/1805.10313} {arXiv:1805.10313
  [astro-ph.HE]}\BibitemShut {NoStop}%
\bibitem [{\citenamefont {Fernández}\ and\ \citenamefont
  {Kobayashi}(2018)}]{Fernandez:2018uiy}%
  \BibitemOpen
  \bibfield  {author} {\bibinfo {author} {\bibfnamefont {Joseph~John}\
  \bibnamefont {Fernández}}\ and\ \bibinfo {author} {\bibfnamefont {Shiho}\
  \bibnamefont {Kobayashi}},\ }\href@noop {} {\  (\bibinfo {year} {2018})},\
  \Eprint {http://arxiv.org/abs/1805.09593} {arXiv:1805.09593
  [astro-ph.HE]}\BibitemShut {NoStop}%
\bibitem [{\citenamefont {Szölgyén}\ and\ \citenamefont
  {Kocsis}(2018)}]{Szolgyen:2018zra}%
  \BibitemOpen
  \bibfield  {author} {\bibinfo {author} {\bibfnamefont {Ákos}\ \bibnamefont
  {Szölgyén}}\ and\ \bibinfo {author} {\bibfnamefont {Bence}\ \bibnamefont
  {Kocsis}},\ }\href@noop {} {\  (\bibinfo {year} {2018})},\ \Eprint
  {http://arxiv.org/abs/1803.07090} {arXiv:1803.07090
  [astro-ph.GA]}\BibitemShut {NoStop}%
\bibitem [{\citenamefont {{Antonini}}\ \emph {et~al.}(2012)\citenamefont
  {{Antonini}}, \citenamefont {{Capuzzo-Dolcetta}}, \citenamefont
  {{Mastrobuono-Battisti}},\ and\ \citenamefont
  {{Merritt}}}]{2012ApJ...750..111A}%
  \BibitemOpen
  \bibfield  {author} {\bibinfo {author} {\bibfnamefont {F.}~\bibnamefont
  {{Antonini}}}, \bibinfo {author} {\bibfnamefont {R.}~\bibnamefont
  {{Capuzzo-Dolcetta}}}, \bibinfo {author} {\bibfnamefont {A.}~\bibnamefont
  {{Mastrobuono-Battisti}}}, \ and\ \bibinfo {author} {\bibfnamefont
  {D.}~\bibnamefont {{Merritt}}},\ }\href {\doibase10.1088/0004-637X/750/2/111}
  {\bibfield  {journal} {\bibinfo  {journal} {The Astrophysical Journal}\
  }\textbf {\bibinfo {volume} {750}},\ \bibinfo {eid} {111} (\bibinfo {year}
  {2012})},\ \Eprint {http://arxiv.org/abs/1110.5937}
  {arXiv:1110.5937}\BibitemShut {NoStop}%
\bibitem [{\citenamefont {{Antonini}}(2013)}]{2013ApJ...763...62A}%
  \BibitemOpen
  \bibfield  {author} {\bibinfo {author} {\bibfnamefont {F.}~\bibnamefont
  {{Antonini}}},\ }\href {\doibase10.1088/0004-637X/763/1/62} {\bibfield
  {journal} {\bibinfo  {journal} {The Astrophysical Journal}\ }\textbf
  {\bibinfo {volume} {763}},\ \bibinfo {eid} {62} (\bibinfo {year} {2013})},\
  \Eprint {http://arxiv.org/abs/1207.6589} {arXiv:1207.6589}\BibitemShut
  {NoStop}%
\bibitem [{\citenamefont {{Arca-Sedda}}\ \emph {et~al.}(2015)\citenamefont
  {{Arca-Sedda}}, \citenamefont {{Capuzzo-Dolcetta}}, \citenamefont
  {{Antonini}},\ and\ \citenamefont {{Seth}}}]{ArcaSedda2015}%
  \BibitemOpen
  \bibfield  {author} {\bibinfo {author} {\bibfnamefont {M.}~\bibnamefont
  {{Arca-Sedda}}}, \bibinfo {author} {\bibfnamefont {R.}~\bibnamefont
  {{Capuzzo-Dolcetta}}}, \bibinfo {author} {\bibfnamefont {F.}~\bibnamefont
  {{Antonini}}}, \ and\ \bibinfo {author} {\bibfnamefont {A.}~\bibnamefont
  {{Seth}}},\ }\href {\doibase10.1088/0004-637X/806/2/220} {\bibfield
  {journal} {\bibinfo  {journal} {Astrophysical Journal}\ }\textbf {\bibinfo
  {volume} {806}},\ \bibinfo {eid} {220} (\bibinfo {year} {2015})},\ \Eprint
  {http://arxiv.org/abs/1501.04567} {arXiv:1501.04567}\BibitemShut {NoStop}%
\bibitem [{\citenamefont {Arca-Sedda}\ \emph {et~al.}(2017)\citenamefont
  {Arca-Sedda}, \citenamefont {Kocsis},\ and\ \citenamefont
  {Brandt}}]{Arca-Sedda:2017qcq}%
  \BibitemOpen
  \bibfield  {author} {\bibinfo {author} {\bibfnamefont {Manuel}\ \bibnamefont
  {Arca-Sedda}}, \bibinfo {author} {\bibfnamefont {Bence}\ \bibnamefont
  {Kocsis}}, \ and\ \bibinfo {author} {\bibfnamefont {Timothy}\ \bibnamefont
  {Brandt}},\ }\href@noop {} {\  (\bibinfo {year} {2017})},\ \Eprint
  {http://arxiv.org/abs/1709.03119} {arXiv:1709.03119
  [astro-ph.GA]}\BibitemShut {NoStop}%
\bibitem [{\citenamefont {{Tagawa}}\ \emph {et~al.}(2016)\citenamefont
  {{Tagawa}}, \citenamefont {{Umemura}},\ and\ \citenamefont
  {{Gouda}}}]{2016MNRAS.462.3812T}%
  \BibitemOpen
  \bibfield  {author} {\bibinfo {author} {\bibfnamefont {H.}~\bibnamefont
  {{Tagawa}}}, \bibinfo {author} {\bibfnamefont {M.}~\bibnamefont {{Umemura}}},
  \ and\ \bibinfo {author} {\bibfnamefont {N.}~\bibnamefont {{Gouda}}},\ }\href
  {\doibase10.1093/mnras/stw1877} {\bibfield  {journal} {\bibinfo  {journal}
  {Monthly Notices of The Royal Astronomical Society}\ }\textbf {\bibinfo
  {volume} {462}},\ \bibinfo {pages} {3812--3822} (\bibinfo {year} {2016})},\
  \Eprint {http://arxiv.org/abs/1602.08767} {arXiv:1602.08767}\BibitemShut
  {NoStop}%
\bibitem [{\citenamefont {{Tagawa}}\ and\ \citenamefont
  {{Umemura}}(2018)}]{2018ApJ...856...47T}%
  \BibitemOpen
  \bibfield  {author} {\bibinfo {author} {\bibfnamefont {H.}~\bibnamefont
  {{Tagawa}}}\ and\ \bibinfo {author} {\bibfnamefont {M.}~\bibnamefont
  {{Umemura}}},\ }\href {\doibase10.3847/1538-4357/aab0a4} {\bibfield
  {journal} {\bibinfo  {journal} {The Astrophysical Journal}\ }\textbf
  {\bibinfo {volume} {856}},\ \bibinfo {eid} {47} (\bibinfo {year} {2018})},\
  \Eprint {http://arxiv.org/abs/1802.07473} {arXiv:1802.07473}\BibitemShut
  {NoStop}%
\bibitem [{\citenamefont {Barausse}\ and\ \citenamefont
  {Rezzolla}(2008)}]{Barausse:2007dy}%
  \BibitemOpen
  \bibfield  {author} {\bibinfo {author} {\bibfnamefont {Enrico}\ \bibnamefont
  {Barausse}}\ and\ \bibinfo {author} {\bibfnamefont {Luciano}\ \bibnamefont
  {Rezzolla}},\ }\href {\doibase10.1103/PhysRevD.77.104027} {\bibfield
  {journal} {\bibinfo  {journal} {Phys. Rev.}\ }\textbf {\bibinfo {volume}
  {D77}},\ \bibinfo {pages} {104027} (\bibinfo {year} {2008})},\ \Eprint
  {http://arxiv.org/abs/0711.4558} {arXiv:0711.4558 [gr-qc]}\BibitemShut
  {NoStop}%
\bibitem [{\citenamefont {{Kocsis}}\ \emph {et~al.}(2011)\citenamefont
  {{Kocsis}}, \citenamefont {{Yunes}},\ and\ \citenamefont
  {{Loeb}}}]{Kocsisetal2011}%
  \BibitemOpen
  \bibfield  {author} {\bibinfo {author} {\bibfnamefont {B.}~\bibnamefont
  {{Kocsis}}}, \bibinfo {author} {\bibfnamefont {N.}~\bibnamefont {{Yunes}}}, \
  and\ \bibinfo {author} {\bibfnamefont {A.}~\bibnamefont {{Loeb}}},\ }\href
  {\doibase10.1103/PhysRevD.84.024032} {\bibfield  {journal} {\bibinfo
  {journal} {Physical Review D}\ }\textbf {\bibinfo {volume} {84}},\ \bibinfo
  {eid} {024032} (\bibinfo {year} {2011})},\ \Eprint
  {http://arxiv.org/abs/1104.2322} {arXiv:1104.2322}\BibitemShut {NoStop}%
\bibitem [{\citenamefont {Tagawa}\ \emph {et~al.}(2018)\citenamefont {Tagawa},
  \citenamefont {Kocsis},\ and\ \citenamefont {Saitoh}}]{Tagawa:2018vls}%
  \BibitemOpen
  \bibfield  {author} {\bibinfo {author} {\bibfnamefont {Hiromichi}\
  \bibnamefont {Tagawa}}, \bibinfo {author} {\bibfnamefont {Bence}\
  \bibnamefont {Kocsis}}, \ and\ \bibinfo {author} {\bibfnamefont
  {Takayuki~R.}\ \bibnamefont {Saitoh}},\ }\href@noop {} {\  (\bibinfo {year}
  {2018})},\ \Eprint {http://arxiv.org/abs/1802.00441} {arXiv:1802.00441
  [astro-ph.HE]}\BibitemShut {NoStop}%
\bibitem [{\citenamefont {{Bartos}}\ \emph
  {et~al.}(2017{\natexlab{a}})\citenamefont {{Bartos}}, \citenamefont
  {{Kocsis}}, \citenamefont {{Haiman}},\ and\ \citenamefont
  {{M{\'a}rka}}}]{Bartosetal2017}%
  \BibitemOpen
  \bibfield  {author} {\bibinfo {author} {\bibfnamefont {I.}~\bibnamefont
  {{Bartos}}}, \bibinfo {author} {\bibfnamefont {B.}~\bibnamefont {{Kocsis}}},
  \bibinfo {author} {\bibfnamefont {Z.}~\bibnamefont {{Haiman}}}, \ and\
  \bibinfo {author} {\bibfnamefont {S.}~\bibnamefont {{M{\'a}rka}}},\ }\href
  {\doibase10.3847/1538-4357/835/2/165} {\bibfield  {journal} {\bibinfo
  {journal} {Astrophysical Journal}\ }\textbf {\bibinfo {volume} {835}},\
  \bibinfo {eid} {165} (\bibinfo {year} {2017}{\natexlab{a}})},\ \Eprint
  {http://arxiv.org/abs/1602.03831} {arXiv:1602.03831
  [astro-ph.HE]}\BibitemShut {NoStop}%
\bibitem [{\citenamefont {{Stone}}\ \emph
  {et~al.}(2017{\natexlab{b}})\citenamefont {{Stone}}, \citenamefont
  {{Metzger}},\ and\ \citenamefont {{Haiman}}}]{Stoneetal2017}%
  \BibitemOpen
  \bibfield  {author} {\bibinfo {author} {\bibfnamefont {N.~C.}\ \bibnamefont
  {{Stone}}}, \bibinfo {author} {\bibfnamefont {B.~D.}\ \bibnamefont
  {{Metzger}}}, \ and\ \bibinfo {author} {\bibfnamefont {Z.}~\bibnamefont
  {{Haiman}}},\ }\href {\doibase10.1093/mnras/stw2260} {\bibfield  {journal}
  {\bibinfo  {journal} {MNRAS}\ }\textbf {\bibinfo {volume} {464}},\ \bibinfo
  {pages} {946--954} (\bibinfo {year} {2017}{\natexlab{b}})},\ \Eprint
  {http://arxiv.org/abs/1602.04226} {arXiv:1602.04226}\BibitemShut {NoStop}%
\bibitem [{\citenamefont {{Bartos}}\ \emph
  {et~al.}(2017{\natexlab{b}})\citenamefont {{Bartos}}, \citenamefont
  {{Haiman}}, \citenamefont {{Marka}}, \citenamefont {{Metzger}}, \citenamefont
  {{Stone}},\ and\ \citenamefont {{Marka}}}]{Bartosetal2017b}%
  \BibitemOpen
  \bibfield  {author} {\bibinfo {author} {\bibfnamefont {I.}~\bibnamefont
  {{Bartos}}}, \bibinfo {author} {\bibfnamefont {Z.}~\bibnamefont {{Haiman}}},
  \bibinfo {author} {\bibfnamefont {Z.}~\bibnamefont {{Marka}}}, \bibinfo
  {author} {\bibfnamefont {B.~D.}\ \bibnamefont {{Metzger}}}, \bibinfo {author}
  {\bibfnamefont {N.~C.}\ \bibnamefont {{Stone}}}, \ and\ \bibinfo {author}
  {\bibfnamefont {S.}~\bibnamefont {{Marka}}},\ }\href
  {\doibase10.1038/s41467-017-00851-7} {\bibfield  {journal} {\bibinfo
  {journal} {Nature Communications}\ }\textbf {\bibinfo {volume} {8}},\
  \bibinfo {eid} {831} (\bibinfo {year} {2017}{\natexlab{b}})},\ \Eprint
  {http://arxiv.org/abs/1701.02328} {arXiv:1701.02328
  [astro-ph.HE]}\BibitemShut {NoStop}%
\bibitem [{\citenamefont {{Silsbee}}\ and\ \citenamefont
  {{Tremaine}}(2017)}]{Silsbee_Tremaine2017}%
  \BibitemOpen
  \bibfield  {author} {\bibinfo {author} {\bibfnamefont {K.}~\bibnamefont
  {{Silsbee}}}\ and\ \bibinfo {author} {\bibfnamefont {S.}~\bibnamefont
  {{Tremaine}}},\ }\href {\doibase10.3847/1538-4357/aa5729} {\bibfield
  {journal} {\bibinfo  {journal} {Astrophysical Journal}\ }\textbf {\bibinfo
  {volume} {836}},\ \bibinfo {eid} {39} (\bibinfo {year} {2017})},\ \Eprint
  {http://arxiv.org/abs/1608.07642} {arXiv:1608.07642
  [astro-ph.HE]}\BibitemShut {NoStop}%
\bibitem [{\citenamefont {{Antonini}}\ \emph {et~al.}(2017)\citenamefont
  {{Antonini}}, \citenamefont {{Toonen}},\ and\ \citenamefont
  {{Hamers}}}]{2017ApJ...841...77A}%
  \BibitemOpen
  \bibfield  {author} {\bibinfo {author} {\bibfnamefont {F.}~\bibnamefont
  {{Antonini}}}, \bibinfo {author} {\bibfnamefont {S.}~\bibnamefont
  {{Toonen}}}, \ and\ \bibinfo {author} {\bibfnamefont {A.~S.}\ \bibnamefont
  {{Hamers}}},\ }\href {\doibase10.3847/1538-4357/aa6f5e} {\bibfield  {journal}
  {\bibinfo  {journal} {The Astrophysical Journal}\ }\textbf {\bibinfo {volume}
  {841}},\ \bibinfo {eid} {77} (\bibinfo {year} {2017})},\ \Eprint
  {http://arxiv.org/abs/1703.06614} {arXiv:1703.06614}\BibitemShut {NoStop}%
\bibitem [{\citenamefont {Rodriguez}\ and\ \citenamefont
  {Antonini}(2018)}]{Rodriguez:2018jqu}%
  \BibitemOpen
  \bibfield  {author} {\bibinfo {author} {\bibfnamefont {Carl~L.}\ \bibnamefont
  {Rodriguez}}\ and\ \bibinfo {author} {\bibfnamefont {Fabio}\ \bibnamefont
  {Antonini}},\ }\href@noop {} {\  (\bibinfo {year} {2018})},\ \Eprint
  {http://arxiv.org/abs/1805.08212} {arXiv:1805.08212
  [astro-ph.HE]}\BibitemShut {NoStop}%
\bibitem [{\citenamefont {Arca-Sedda}\ \emph {et~al.}(2018)\citenamefont
  {Arca-Sedda}, \citenamefont {Li},\ and\ \citenamefont
  {Kocsis}}]{Arca-Sedda:2018qgq}%
  \BibitemOpen
  \bibfield  {author} {\bibinfo {author} {\bibfnamefont {Manuel}\ \bibnamefont
  {Arca-Sedda}}, \bibinfo {author} {\bibfnamefont {Gongjie}\ \bibnamefont
  {Li}}, \ and\ \bibinfo {author} {\bibfnamefont {Bence}\ \bibnamefont
  {Kocsis}},\ }\href@noop {} {\  (\bibinfo {year} {2018})},\ \Eprint
  {http://arxiv.org/abs/1805.06458} {arXiv:1805.06458
  [astro-ph.HE]}\BibitemShut {NoStop}%
\bibitem [{\citenamefont {Banerjee}(2018)}]{Banerjee:2018pmh}%
  \BibitemOpen
  \bibfield  {author} {\bibinfo {author} {\bibfnamefont {Sambaran}\
  \bibnamefont {Banerjee}},\ }\href@noop {} {\  (\bibinfo {year} {2018})},\
  \Eprint {http://arxiv.org/abs/1805.06466} {arXiv:1805.06466
  [astro-ph.HE]}\BibitemShut {NoStop}%
\bibitem [{\citenamefont {{Fang}}\ \emph {et~al.}(2018)\citenamefont {{Fang}},
  \citenamefont {{Thompson}},\ and\ \citenamefont
  {{Hirata}}}]{Fang_Thompson_Hirata2018}%
  \BibitemOpen
  \bibfield  {author} {\bibinfo {author} {\bibfnamefont {X.}~\bibnamefont
  {{Fang}}}, \bibinfo {author} {\bibfnamefont {T.~A.}\ \bibnamefont
  {{Thompson}}}, \ and\ \bibinfo {author} {\bibfnamefont {C.~M.}\ \bibnamefont
  {{Hirata}}},\ }\href {\doibase10.1093/mnras/sty472} {\bibfield  {journal}
  {\bibinfo  {journal} {MNRAS}\ }\textbf {\bibinfo {volume} {476}},\ \bibinfo
  {pages} {4234--4262} (\bibinfo {year} {2018})},\ \Eprint
  {http://arxiv.org/abs/1709.08682} {arXiv:1709.08682
  [astro-ph.HE]}\BibitemShut {NoStop}%
\bibitem [{\citenamefont {{Kinugawa}}\ \emph {et~al.}(2014)\citenamefont
  {{Kinugawa}}, \citenamefont {{Inayoshi}}, \citenamefont {{Hotokezaka}},
  \citenamefont {{Nakauchi}},\ and\ \citenamefont
  {{Nakamura}}}]{2014MNRAS.442.2963K}%
  \BibitemOpen
  \bibfield  {author} {\bibinfo {author} {\bibfnamefont {T.}~\bibnamefont
  {{Kinugawa}}}, \bibinfo {author} {\bibfnamefont {K.}~\bibnamefont
  {{Inayoshi}}}, \bibinfo {author} {\bibfnamefont {K.}~\bibnamefont
  {{Hotokezaka}}}, \bibinfo {author} {\bibfnamefont {D.}~\bibnamefont
  {{Nakauchi}}}, \ and\ \bibinfo {author} {\bibfnamefont {T.}~\bibnamefont
  {{Nakamura}}},\ }\href {\doibase10.1093/mnras/stu1022} {\bibfield  {journal}
  {\bibinfo  {journal} {Monthly Notices of The Royal Astronomical Society}\
  }\textbf {\bibinfo {volume} {442}},\ \bibinfo {pages} {2963--2992} (\bibinfo
  {year} {2014})},\ \Eprint {http://arxiv.org/abs/1402.6672} {arXiv:1402.6672
  [astro-ph.HE]}\BibitemShut {NoStop}%
\bibitem [{\citenamefont {Bird}\ \emph {et~al.}(2016)\citenamefont {Bird},
  \citenamefont {Cholis}, \citenamefont {Mu\~noz}, \citenamefont {Ali-Haimoud},
  \citenamefont {Kamionkowski}, \citenamefont {Kovetz}, \citenamefont
  {Raccanelli},\ and\ \citenamefont {Riess}}]{Bird:2016dcv}%
  \BibitemOpen
  \bibfield  {author} {\bibinfo {author} {\bibfnamefont {Simeon}\ \bibnamefont
  {Bird}}, \bibinfo {author} {\bibfnamefont {Ilias}\ \bibnamefont {Cholis}},
  \bibinfo {author} {\bibfnamefont {Julian~B.}\ \bibnamefont {Mu\~noz}},
  \bibinfo {author} {\bibfnamefont {Yacine}\ \bibnamefont {Ali-Haimoud}},
  \bibinfo {author} {\bibfnamefont {Marc}\ \bibnamefont {Kamionkowski}},
  \bibinfo {author} {\bibfnamefont {Ely~D.}\ \bibnamefont {Kovetz}}, \bibinfo
  {author} {\bibfnamefont {Alvise}\ \bibnamefont {Raccanelli}}, \ and\ \bibinfo
  {author} {\bibfnamefont {Adam~G.}\ \bibnamefont {Riess}},\ }\href
  {\doibase10.1103/PhysRevLett.116.201301} {\bibfield  {journal} {\bibinfo
  {journal} {Phys. Rev. Lett.}\ }\textbf {\bibinfo {volume} {116}},\ \bibinfo
  {pages} {201301} (\bibinfo {year} {2016})},\ \Eprint
  {http://arxiv.org/abs/1603.00464} {arXiv:1603.00464
  [astro-ph.CO]}\BibitemShut {NoStop}%
\bibitem [{\citenamefont {Brandt}(2016)}]{Brandt:2016aco}%
  \BibitemOpen
  \bibfield  {author} {\bibinfo {author} {\bibfnamefont {Timothy~D.}\
  \bibnamefont {Brandt}},\ }\href {\doibase10.3847/2041-8205/824/2/L31}
  {\bibfield  {journal} {\bibinfo  {journal} {Astrophys. J.}\ }\textbf
  {\bibinfo {volume} {824}},\ \bibinfo {pages} {L31} (\bibinfo {year}
  {2016})},\ \Eprint {http://arxiv.org/abs/1605.03665} {arXiv:1605.03665
  [astro-ph.GA]}\BibitemShut {NoStop}%
\bibitem [{\citenamefont {Ali-Haïmoud}\ \emph {et~al.}(2017)\citenamefont
  {Ali-Haïmoud}, \citenamefont {Kovetz},\ and\ \citenamefont
  {Kamionkowski}}]{Ali-Haimoud:2017rtz}%
  \BibitemOpen
  \bibfield  {author} {\bibinfo {author} {\bibfnamefont {Yacine}\ \bibnamefont
  {Ali-Haïmoud}}, \bibinfo {author} {\bibfnamefont {Ely~D.}\ \bibnamefont
  {Kovetz}}, \ and\ \bibinfo {author} {\bibfnamefont {Marc}\ \bibnamefont
  {Kamionkowski}},\ }\href {\doibase10.1103/PhysRevD.96.123523} {\bibfield
  {journal} {\bibinfo  {journal} {Phys. Rev.}\ }\textbf {\bibinfo {volume}
  {D96}},\ \bibinfo {pages} {123523} (\bibinfo {year} {2017})},\ \Eprint
  {http://arxiv.org/abs/1709.06576} {arXiv:1709.06576
  [astro-ph.CO]}\BibitemShut {NoStop}%
\bibitem [{\citenamefont {Sasaki}\ \emph {et~al.}(2016)\citenamefont {Sasaki},
  \citenamefont {Suyama}, \citenamefont {Tanaka},\ and\ \citenamefont
  {Yokoyama}}]{Sasaki:2016jop}%
  \BibitemOpen
  \bibfield  {author} {\bibinfo {author} {\bibfnamefont {Misao}\ \bibnamefont
  {Sasaki}}, \bibinfo {author} {\bibfnamefont {Teruaki}\ \bibnamefont
  {Suyama}}, \bibinfo {author} {\bibfnamefont {Takahiro}\ \bibnamefont
  {Tanaka}}, \ and\ \bibinfo {author} {\bibfnamefont {Shuichiro}\ \bibnamefont
  {Yokoyama}},\ }\href {\doibase10.1103/PhysRevLett.117.061101} {\bibfield
  {journal} {\bibinfo  {journal} {Phys. Rev. Lett.}\ }\textbf {\bibinfo
  {volume} {117}},\ \bibinfo {pages} {061101} (\bibinfo {year} {2016})},\
  \Eprint {http://arxiv.org/abs/1603.08338} {arXiv:1603.08338
  [astro-ph.CO]}\BibitemShut {NoStop}%
\bibitem [{\citenamefont {Sasaki}\ \emph {et~al.}(2018)\citenamefont {Sasaki},
  \citenamefont {Suyama}, \citenamefont {Tanaka},\ and\ \citenamefont
  {Yokoyama}}]{Sasaki:2018dmp}%
  \BibitemOpen
  \bibfield  {author} {\bibinfo {author} {\bibfnamefont {Misao}\ \bibnamefont
  {Sasaki}}, \bibinfo {author} {\bibfnamefont {Teruaki}\ \bibnamefont
  {Suyama}}, \bibinfo {author} {\bibfnamefont {Takahiro}\ \bibnamefont
  {Tanaka}}, \ and\ \bibinfo {author} {\bibfnamefont {Shuichiro}\ \bibnamefont
  {Yokoyama}},\ }\href {\doibase10.1088/1361-6382/aaa7b4} {\bibfield  {journal}
  {\bibinfo  {journal} {Class. Quant. Grav.}\ }\textbf {\bibinfo {volume}
  {35}},\ \bibinfo {pages} {063001} (\bibinfo {year} {2018})},\ \Eprint
  {http://arxiv.org/abs/1801.05235} {arXiv:1801.05235
  [astro-ph.CO]}\BibitemShut {NoStop}%
\bibitem [{\citenamefont {{Wang}}\ \emph {et~al.}(2016)\citenamefont {{Wang}},
  \citenamefont {{Spurzem}}, \citenamefont {{Aarseth}}, \citenamefont
  {{Giersz}}, \citenamefont {{Askar}}, \citenamefont {{Berczik}}, \citenamefont
  {{Naab}}, \citenamefont {{Schadow}},\ and\ \citenamefont
  {{Kouwenhoven}}}]{Wangetal2016}%
  \BibitemOpen
  \bibfield  {author} {\bibinfo {author} {\bibfnamefont {L.}~\bibnamefont
  {{Wang}}}, \bibinfo {author} {\bibfnamefont {R.}~\bibnamefont {{Spurzem}}},
  \bibinfo {author} {\bibfnamefont {S.}~\bibnamefont {{Aarseth}}}, \bibinfo
  {author} {\bibfnamefont {M.}~\bibnamefont {{Giersz}}}, \bibinfo {author}
  {\bibfnamefont {A.}~\bibnamefont {{Askar}}}, \bibinfo {author} {\bibfnamefont
  {P.}~\bibnamefont {{Berczik}}}, \bibinfo {author} {\bibfnamefont
  {T.}~\bibnamefont {{Naab}}}, \bibinfo {author} {\bibfnamefont
  {R.}~\bibnamefont {{Schadow}}}, \ and\ \bibinfo {author} {\bibfnamefont
  {M.~B.~N.}\ \bibnamefont {{Kouwenhoven}}},\ }\href
  {\doibase10.1093/mnras/stw274} {\bibfield  {journal} {\bibinfo  {journal}
  {MNRAS}\ }\textbf {\bibinfo {volume} {458}},\ \bibinfo {pages} {1450--1465}
  (\bibinfo {year} {2016})},\ \Eprint {http://arxiv.org/abs/1602.00759}
  {arXiv:1602.00759 [astro-ph.SR]}\BibitemShut {NoStop}%
\bibitem [{\citenamefont {{Rodriguez}}\ \emph
  {et~al.}(2016{\natexlab{b}})\citenamefont {{Rodriguez}}, \citenamefont
  {{Morscher}}, \citenamefont {{Wang}}, \citenamefont {{Chatterjee}},
  \citenamefont {{Rasio}},\ and\ \citenamefont
  {{Spurzem}}}]{Rodriguezetal2016a}%
  \BibitemOpen
  \bibfield  {author} {\bibinfo {author} {\bibfnamefont {C.~L.}\ \bibnamefont
  {{Rodriguez}}}, \bibinfo {author} {\bibfnamefont {M.}~\bibnamefont
  {{Morscher}}}, \bibinfo {author} {\bibfnamefont {L.}~\bibnamefont {{Wang}}},
  \bibinfo {author} {\bibfnamefont {S.}~\bibnamefont {{Chatterjee}}}, \bibinfo
  {author} {\bibfnamefont {F.~A.}\ \bibnamefont {{Rasio}}}, \ and\ \bibinfo
  {author} {\bibfnamefont {R.}~\bibnamefont {{Spurzem}}},\ }\href
  {\doibase10.1093/mnras/stw2121} {\bibfield  {journal} {\bibinfo  {journal}
  {MNRAS}\ }\textbf {\bibinfo {volume} {463}},\ \bibinfo {pages} {2109--2118}
  (\bibinfo {year} {2016}{\natexlab{b}})},\ \Eprint
  {http://arxiv.org/abs/1601.04227} {arXiv:1601.04227
  [astro-ph.IM]}\BibitemShut {NoStop}%
\bibitem [{\citenamefont {Panamarev}\ \emph
  {et~al.}(2018{\natexlab{a}})\citenamefont {Panamarev}, \citenamefont {Just},
  \citenamefont {Spurzem}, \citenamefont {Berczik}, \citenamefont {Wang},\ and\
  \citenamefont {Sedda}}]{Panamarev:2018bwq}%
  \BibitemOpen
  \bibfield  {author} {\bibinfo {author} {\bibfnamefont {Taras}\ \bibnamefont
  {Panamarev}}, \bibinfo {author} {\bibfnamefont {Andreas}\ \bibnamefont
  {Just}}, \bibinfo {author} {\bibfnamefont {Rainer}\ \bibnamefont {Spurzem}},
  \bibinfo {author} {\bibfnamefont {Peter}\ \bibnamefont {Berczik}}, \bibinfo
  {author} {\bibfnamefont {Long}\ \bibnamefont {Wang}}, \ and\ \bibinfo
  {author} {\bibfnamefont {Manuel~Arca}\ \bibnamefont {Sedda}},\ }\href@noop {}
  {\  (\bibinfo {year} {2018}{\natexlab{a}})},\ \Eprint
  {http://arxiv.org/abs/1805.02153} {arXiv:1805.02153
  [astro-ph.GA]}\BibitemShut {NoStop}%
\bibitem [{\citenamefont {{Kennedy}}\ \emph {et~al.}(2016)\citenamefont
  {{Kennedy}}, \citenamefont {{Meiron}}, \citenamefont {{Shukirgaliyev}},
  \citenamefont {{Panamarev}}, \citenamefont {{Berczik}}, \citenamefont
  {{Just}},\ and\ \citenamefont {{Spurzem}}}]{Kennedyetal2016}%
  \BibitemOpen
  \bibfield  {author} {\bibinfo {author} {\bibfnamefont {G.~F.}\ \bibnamefont
  {{Kennedy}}}, \bibinfo {author} {\bibfnamefont {Y.}~\bibnamefont {{Meiron}}},
  \bibinfo {author} {\bibfnamefont {B.}~\bibnamefont {{Shukirgaliyev}}},
  \bibinfo {author} {\bibfnamefont {T.}~\bibnamefont {{Panamarev}}}, \bibinfo
  {author} {\bibfnamefont {P.}~\bibnamefont {{Berczik}}}, \bibinfo {author}
  {\bibfnamefont {A.}~\bibnamefont {{Just}}}, \ and\ \bibinfo {author}
  {\bibfnamefont {R.}~\bibnamefont {{Spurzem}}},\ }\href
  {\doibase10.1093/mnras/stw908} {\bibfield  {journal} {\bibinfo  {journal}
  {MNRAS}\ }\textbf {\bibinfo {volume} {460}},\ \bibinfo {pages} {240--255}
  (\bibinfo {year} {2016})},\ \Eprint {http://arxiv.org/abs/1604.05309}
  {arXiv:1604.05309}\BibitemShut {NoStop}%
\bibitem [{\citenamefont {Panamarev}\ \emph
  {et~al.}(2018{\natexlab{b}})\citenamefont {Panamarev}, \citenamefont
  {Shukirgaliyev}, \citenamefont {Meiron}, \citenamefont {Berczik},
  \citenamefont {Just}, \citenamefont {Spurzem}, \citenamefont {Omarov},\ and\
  \citenamefont {Vilkoviskij}}]{Panamarev:2018who}%
  \BibitemOpen
  \bibfield  {author} {\bibinfo {author} {\bibfnamefont {Taras}\ \bibnamefont
  {Panamarev}}, \bibinfo {author} {\bibfnamefont {Bekdaulet}\ \bibnamefont
  {Shukirgaliyev}}, \bibinfo {author} {\bibfnamefont {Yohai}\ \bibnamefont
  {Meiron}}, \bibinfo {author} {\bibfnamefont {Peter}\ \bibnamefont {Berczik}},
  \bibinfo {author} {\bibfnamefont {Andreas}\ \bibnamefont {Just}}, \bibinfo
  {author} {\bibfnamefont {Rainer}\ \bibnamefont {Spurzem}}, \bibinfo {author}
  {\bibfnamefont {Chingis}\ \bibnamefont {Omarov}}, \ and\ \bibinfo {author}
  {\bibfnamefont {Emmanuil}\ \bibnamefont {Vilkoviskij}},\ }\href
  {\doibase10.1093/mnras/sty459} {\bibfield  {journal} {\bibinfo  {journal}
  {Mon. Not. Roy. Astron. Soc.}\ }\textbf {\bibinfo {volume} {476}},\ \bibinfo
  {pages} {4224--4233} (\bibinfo {year} {2018}{\natexlab{b}})},\ \Eprint
  {http://arxiv.org/abs/1802.03027} {arXiv:1802.03027
  [astro-ph.GA]}\BibitemShut {NoStop}%
\bibitem [{\citenamefont {{Kocsis}}\ and\ \citenamefont
  {{Tremaine}}(2011)}]{Kocsis_Tremaine2011}%
  \BibitemOpen
  \bibfield  {author} {\bibinfo {author} {\bibfnamefont {B.}~\bibnamefont
  {{Kocsis}}}\ and\ \bibinfo {author} {\bibfnamefont {S.}~\bibnamefont
  {{Tremaine}}},\ }\href {\doibase10.1111/j.1365-2966.2010.17897.x} {\bibfield
  {journal} {\bibinfo  {journal} {MNRAS}\ }\textbf {\bibinfo {volume} {412}},\
  \bibinfo {pages} {187--207} (\bibinfo {year} {2011})},\ \Eprint
  {http://arxiv.org/abs/1006.0001} {arXiv:1006.0001 [astro-ph.GA]}\BibitemShut
  {NoStop}%
\bibitem [{\citenamefont {{Perets}}\ \emph {et~al.}(2018)\citenamefont
  {{Perets}}, \citenamefont {{Mastrobuono-Battisti}}, \citenamefont
  {{Meiron}},\ and\ \citenamefont {{Gualandris}}}]{Peretsetal2018}%
  \BibitemOpen
  \bibfield  {author} {\bibinfo {author} {\bibfnamefont {H.~B.}\ \bibnamefont
  {{Perets}}}, \bibinfo {author} {\bibfnamefont {A.}~\bibnamefont
  {{Mastrobuono-Battisti}}}, \bibinfo {author} {\bibfnamefont {Y.}~\bibnamefont
  {{Meiron}}}, \ and\ \bibinfo {author} {\bibfnamefont {A.}~\bibnamefont
  {{Gualandris}}},\ }\href@noop {} {\  (\bibinfo {year} {2018})},\ \Eprint
  {http://arxiv.org/abs/1802.00012} {arXiv:1802.00012}\BibitemShut {NoStop}%
\bibitem [{\citenamefont {{Morscher}}\ \emph {et~al.}(2015)\citenamefont
  {{Morscher}}, \citenamefont {{Pattabiraman}}, \citenamefont {{Rodriguez}},
  \citenamefont {{Rasio}},\ and\ \citenamefont {{Umbreit}}}]{Morscheretal2015}%
  \BibitemOpen
  \bibfield  {author} {\bibinfo {author} {\bibfnamefont {M.}~\bibnamefont
  {{Morscher}}}, \bibinfo {author} {\bibfnamefont {B.}~\bibnamefont
  {{Pattabiraman}}}, \bibinfo {author} {\bibfnamefont {C.}~\bibnamefont
  {{Rodriguez}}}, \bibinfo {author} {\bibfnamefont {F.~A.}\ \bibnamefont
  {{Rasio}}}, \ and\ \bibinfo {author} {\bibfnamefont {S.}~\bibnamefont
  {{Umbreit}}},\ }\href {\doibase10.1088/0004-637X/800/1/9} {\bibfield
  {journal} {\bibinfo  {journal} {Astrophysical Journal}\ }\textbf {\bibinfo
  {volume} {800}},\ \bibinfo {eid} {9} (\bibinfo {year} {2015})},\ \Eprint
  {http://arxiv.org/abs/1409.0866} {arXiv:1409.0866}\BibitemShut {NoStop}%
\bibitem [{\citenamefont {{Arca Sedda}}\ \emph {et~al.}(2018)\citenamefont
  {{Arca Sedda}}, \citenamefont {{Askar}},\ and\ \citenamefont
  {{Giersz}}}]{Arcasedda2018}%
  \BibitemOpen
  \bibfield  {author} {\bibinfo {author} {\bibfnamefont {M.}~\bibnamefont
  {{Arca Sedda}}}, \bibinfo {author} {\bibfnamefont {A.}~\bibnamefont
  {{Askar}}}, \ and\ \bibinfo {author} {\bibfnamefont {M.}~\bibnamefont
  {{Giersz}}},\ }\href@noop {} {\  (\bibinfo {year} {2018})},\ \Eprint
  {http://arxiv.org/abs/1801.00795} {arXiv:1801.00795}\BibitemShut {NoStop}%
\bibitem [{\citenamefont {{Askar}}\ \emph {et~al.}(2018)\citenamefont
  {{Askar}}, \citenamefont {{Sedda}},\ and\ \citenamefont
  {{Giersz}}}]{Askar2018}%
  \BibitemOpen
  \bibfield  {author} {\bibinfo {author} {\bibfnamefont {A.}~\bibnamefont
  {{Askar}}}, \bibinfo {author} {\bibfnamefont {M.~A.}\ \bibnamefont
  {{Sedda}}}, \ and\ \bibinfo {author} {\bibfnamefont {M.}~\bibnamefont
  {{Giersz}}},\ }\href {\doibase10.1093/mnras/sty1186} {\bibfield  {journal}
  {\bibinfo  {journal} {Monthly Notices of The Royal Astronomical Society}\ }
  (\bibinfo {year} {2018}),\ 10.1093/mnras/sty1186},\ \Eprint
  {http://arxiv.org/abs/1802.05284} {arXiv:1802.05284}\BibitemShut {NoStop}%
\bibitem [{\citenamefont {{Chatterjee}}\ \emph {et~al.}(2017)\citenamefont
  {{Chatterjee}}, \citenamefont {{Rodriguez}}, \citenamefont {{Kalogera}},\
  and\ \citenamefont {{Rasio}}}]{Chatterjeeetal2017}%
  \BibitemOpen
  \bibfield  {author} {\bibinfo {author} {\bibfnamefont {S.}~\bibnamefont
  {{Chatterjee}}}, \bibinfo {author} {\bibfnamefont {C.~L.}\ \bibnamefont
  {{Rodriguez}}}, \bibinfo {author} {\bibfnamefont {V.}~\bibnamefont
  {{Kalogera}}}, \ and\ \bibinfo {author} {\bibfnamefont {F.~A.}\ \bibnamefont
  {{Rasio}}},\ }\href {\doibase10.3847/2041-8213/aa5caa} {\bibfield  {journal}
  {\bibinfo  {journal} {Astrophysical Journal}\ }\textbf {\bibinfo {volume}
  {836}},\ \bibinfo {eid} {L26} (\bibinfo {year} {2017})},\ \Eprint
  {http://arxiv.org/abs/1609.06689} {arXiv:1609.06689}\BibitemShut {NoStop}%
\bibitem [{\citenamefont {{Giersz}}\ \emph {et~al.}(2015)\citenamefont
  {{Giersz}}, \citenamefont {{Leigh}}, \citenamefont {{Hypki}}, \citenamefont
  {{L{\"u}tzgendorf}},\ and\ \citenamefont {{Askar}}}]{Gierszetal2015}%
  \BibitemOpen
  \bibfield  {author} {\bibinfo {author} {\bibfnamefont {M.}~\bibnamefont
  {{Giersz}}}, \bibinfo {author} {\bibfnamefont {N.}~\bibnamefont {{Leigh}}},
  \bibinfo {author} {\bibfnamefont {A.}~\bibnamefont {{Hypki}}}, \bibinfo
  {author} {\bibfnamefont {N.}~\bibnamefont {{L{\"u}tzgendorf}}}, \ and\
  \bibinfo {author} {\bibfnamefont {A.}~\bibnamefont {{Askar}}},\ }\href
  {\doibase10.1093/mnras/stv2162} {\bibfield  {journal} {\bibinfo  {journal}
  {MNRAS}\ }\textbf {\bibinfo {volume} {454}},\ \bibinfo {pages} {3150--3165}
  (\bibinfo {year} {2015})},\ \Eprint {http://arxiv.org/abs/1506.05234}
  {arXiv:1506.05234}\BibitemShut {NoStop}%
\bibitem [{\citenamefont {Rodriguez}\ \emph {et~al.}(2018)\citenamefont
  {Rodriguez}, \citenamefont {Amaro-Seoane}, \citenamefont {Chatterjee},\ and\
  \citenamefont {Rasio}}]{Rodriguez:2017pec}%
  \BibitemOpen
  \bibfield  {author} {\bibinfo {author} {\bibfnamefont {Carl~L.}\ \bibnamefont
  {Rodriguez}}, \bibinfo {author} {\bibfnamefont {Pau}\ \bibnamefont
  {Amaro-Seoane}}, \bibinfo {author} {\bibfnamefont {Sourav}\ \bibnamefont
  {Chatterjee}}, \ and\ \bibinfo {author} {\bibfnamefont {Frederic~A.}\
  \bibnamefont {Rasio}},\ }\href {\doibase10.1103/PhysRevLett.120.151101}
  {\bibfield  {journal} {\bibinfo  {journal} {Phys. Rev. Lett.}\ }\textbf
  {\bibinfo {volume} {120}},\ \bibinfo {pages} {151101} (\bibinfo {year}
  {2018})},\ \Eprint {http://arxiv.org/abs/1712.04937} {arXiv:1712.04937
  [astro-ph.HE]}\BibitemShut {NoStop}%
\bibitem [{\citenamefont {{Samsing}}\ \emph {et~al.}(2018)\citenamefont
  {{Samsing}}, \citenamefont {{Askar}},\ and\ \citenamefont
  {{Giersz}}}]{Samsingetal2017}%
  \BibitemOpen
  \bibfield  {author} {\bibinfo {author} {\bibfnamefont {J.}~\bibnamefont
  {{Samsing}}}, \bibinfo {author} {\bibfnamefont {A.}~\bibnamefont {{Askar}}},
  \ and\ \bibinfo {author} {\bibfnamefont {M.}~\bibnamefont {{Giersz}}},\
  }\href {\doibase10.3847/1538-4357/aaab52} {\bibfield  {journal} {\bibinfo
  {journal} {The Astrophysical Journal}\ }\textbf {\bibinfo {volume} {855}},\
  \bibinfo {eid} {124} (\bibinfo {year} {2018})},\ \Eprint
  {http://arxiv.org/abs/1712.06186} {arXiv:1712.06186
  [astro-ph.HE]}\BibitemShut {NoStop}%
\bibitem [{\citenamefont {Samsing}\ and\ \citenamefont
  {D'Orazio}(2018)}]{Samsing:2018isx}%
  \BibitemOpen
  \bibfield  {author} {\bibinfo {author} {\bibfnamefont {Johan}\ \bibnamefont
  {Samsing}}\ and\ \bibinfo {author} {\bibfnamefont {Daniel~J.}\ \bibnamefont
  {D'Orazio}},\ }\href@noop {} {\  (\bibinfo {year} {2018})},\ \Eprint
  {http://arxiv.org/abs/1804.06519} {arXiv:1804.06519
  [astro-ph.HE]}\BibitemShut {NoStop}%
\bibitem [{\citenamefont {Samsing}\ \emph
  {et~al.}(2018{\natexlab{a}})\citenamefont {Samsing}, \citenamefont {Leigh},\
  and\ \citenamefont {Trani}}]{Samsing:2018uoo}%
  \BibitemOpen
  \bibfield  {author} {\bibinfo {author} {\bibfnamefont {Johan}\ \bibnamefont
  {Samsing}}, \bibinfo {author} {\bibfnamefont {Nathan W.~C.}\ \bibnamefont
  {Leigh}}, \ and\ \bibinfo {author} {\bibfnamefont {Alessandro~A.}\
  \bibnamefont {Trani}},\ }\href@noop {} {\  (\bibinfo {year}
  {2018}{\natexlab{a}})},\ \Eprint {http://arxiv.org/abs/1803.08215}
  {arXiv:1803.08215 [astro-ph.HE]}\BibitemShut {NoStop}%
\bibitem [{\citenamefont {{Rauch}}\ and\ \citenamefont
  {{Tremaine}}(1996)}]{Rauch_Tremaine1996}%
  \BibitemOpen
  \bibfield  {author} {\bibinfo {author} {\bibfnamefont {K.~P.}\ \bibnamefont
  {{Rauch}}}\ and\ \bibinfo {author} {\bibfnamefont {S.}~\bibnamefont
  {{Tremaine}}},\ }\href {\doibase10.1016/S1384-1076(96)00012-7} {\bibfield
  {journal} {\bibinfo  {journal} {\na}\ }\textbf {\bibinfo {volume} {1}},\
  \bibinfo {pages} {149--170} (\bibinfo {year} {1996})},\ \Eprint
  {http://arxiv.org/abs/astro-ph/9603018} {astro-ph/9603018}\BibitemShut
  {NoStop}%
\bibitem [{\citenamefont {{Kocsis}}\ and\ \citenamefont
  {{Tremaine}}(2015)}]{Kocsis_Tremaine2015}%
  \BibitemOpen
  \bibfield  {author} {\bibinfo {author} {\bibfnamefont {B.}~\bibnamefont
  {{Kocsis}}}\ and\ \bibinfo {author} {\bibfnamefont {S.}~\bibnamefont
  {{Tremaine}}},\ }\href {\doibase10.1093/mnras/stv057} {\bibfield  {journal}
  {\bibinfo  {journal} {MNRAS}\ }\textbf {\bibinfo {volume} {448}},\ \bibinfo
  {pages} {3265--3296} (\bibinfo {year} {2015})},\ \Eprint
  {http://arxiv.org/abs/1406.1178} {arXiv:1406.1178}\BibitemShut {NoStop}%
\bibitem [{\citenamefont {{Touma}}\ and\ \citenamefont
  {{Tremaine}}(2014)}]{Touma_Tremaine2014}%
  \BibitemOpen
  \bibfield  {author} {\bibinfo {author} {\bibfnamefont {J.}~\bibnamefont
  {{Touma}}}\ and\ \bibinfo {author} {\bibfnamefont {S.}~\bibnamefont
  {{Tremaine}}},\ }\href {\doibase10.1088/1751-8113/47/29/292001} {\bibfield
  {journal} {\bibinfo  {journal} {J. Phys. A: Math. Theor.}\ }\textbf {\bibinfo
  {volume} {47}},\ \bibinfo {pages} {292001} (\bibinfo {year} {2014})},\
  \Eprint {http://arxiv.org/abs/1401.5534} {arXiv:1401.5534}\BibitemShut
  {NoStop}%
\bibitem [{\citenamefont {{Roupas}}\ \emph {et~al.}(2017)\citenamefont
  {{Roupas}}, \citenamefont {{Kocsis}},\ and\ \citenamefont
  {{Tremaine}}}]{Roupasetal2017}%
  \BibitemOpen
  \bibfield  {author} {\bibinfo {author} {\bibfnamefont {Z.}~\bibnamefont
  {{Roupas}}}, \bibinfo {author} {\bibfnamefont {B.}~\bibnamefont {{Kocsis}}},
  \ and\ \bibinfo {author} {\bibfnamefont {S.}~\bibnamefont {{Tremaine}}},\
  }\href {\doibase10.3847/1538-4357/aa7141} {\bibfield  {journal} {\bibinfo
  {journal} {Astrophysical Journal}\ }\textbf {\bibinfo {volume} {842}},\
  \bibinfo {eid} {90} (\bibinfo {year} {2017})},\ \Eprint
  {http://arxiv.org/abs/1701.03271} {arXiv:1701.03271}\BibitemShut {NoStop}%
\bibitem [{\citenamefont {Takács}\ and\ \citenamefont
  {Kocsis}(2018)}]{Takacs:2017wnn}%
  \BibitemOpen
  \bibfield  {author} {\bibinfo {author} {\bibfnamefont {Ádám}\ \bibnamefont
  {Takács}}\ and\ \bibinfo {author} {\bibfnamefont {Bence}\ \bibnamefont
  {Kocsis}},\ }\href {\doibase10.3847/1538-4357/aab268} {\bibfield  {journal}
  {\bibinfo  {journal} {Astrophys. J.}\ }\textbf {\bibinfo {volume} {856}},\
  \bibinfo {pages} {113} (\bibinfo {year} {2018})},\ \Eprint
  {http://arxiv.org/abs/1712.04449} {arXiv:1712.04449
  [astro-ph.GA]}\BibitemShut {NoStop}%
\bibitem [{\citenamefont {{Fouvry}}\ \emph {et~al.}(2018)\citenamefont
  {{Fouvry}}, \citenamefont {{Pichon}},\ and\ \citenamefont
  {{Chavanis}}}]{Fouvry2018}%
  \BibitemOpen
  \bibfield  {author} {\bibinfo {author} {\bibfnamefont {J.-B.}\ \bibnamefont
  {{Fouvry}}}, \bibinfo {author} {\bibfnamefont {C.}~\bibnamefont {{Pichon}}},
  \ and\ \bibinfo {author} {\bibfnamefont {P.-H.}\ \bibnamefont {{Chavanis}}},\
  }\href {\doibase10.1051/0004-6361/201731088} {\bibfield  {journal} {\bibinfo
  {journal} {\aap}\ }\textbf {\bibinfo {volume} {609}},\ \bibinfo {eid} {A38}
  (\bibinfo {year} {2018})},\ \Eprint {http://arxiv.org/abs/1705.01131}
  {arXiv:1705.01131}\BibitemShut {NoStop}%
\bibitem [{\citenamefont {{Portegies Zwart}}\ and\ \citenamefont
  {{McMillan}}(2002)}]{PortegiesZwart_McMillan2002}%
  \BibitemOpen
  \bibfield  {author} {\bibinfo {author} {\bibfnamefont {S.~F.}\ \bibnamefont
  {{Portegies Zwart}}}\ and\ \bibinfo {author} {\bibfnamefont {S.~L.~W.}\
  \bibnamefont {{McMillan}}},\ }\href {\doibase10.1086/341798} {\bibfield
  {journal} {\bibinfo  {journal} {Astrophysical Journal}\ }\textbf {\bibinfo
  {volume} {576}},\ \bibinfo {pages} {899--907} (\bibinfo {year} {2002})},\
  \Eprint {http://arxiv.org/abs/astro-ph/0201055}
  {astro-ph/0201055}\BibitemShut {NoStop}%
\bibitem [{\citenamefont {Fragione}\ \emph {et~al.}(2018)\citenamefont
  {Fragione}, \citenamefont {Ginsburg},\ and\ \citenamefont
  {Kocsis}}]{Fragione:2017blf}%
  \BibitemOpen
  \bibfield  {author} {\bibinfo {author} {\bibfnamefont {Giacomo}\ \bibnamefont
  {Fragione}}, \bibinfo {author} {\bibfnamefont {Idan}\ \bibnamefont
  {Ginsburg}}, \ and\ \bibinfo {author} {\bibfnamefont {Bence}\ \bibnamefont
  {Kocsis}},\ }\href {\doibase10.3847/1538-4357/aab368} {\bibfield  {journal}
  {\bibinfo  {journal} {Astrophys. J.}\ }\textbf {\bibinfo {volume} {856}},\
  \bibinfo {pages} {92} (\bibinfo {year} {2018})},\ \Eprint
  {http://arxiv.org/abs/1711.00483} {arXiv:1711.00483
  [astro-ph.GA]}\BibitemShut {NoStop}%
\bibitem [{\citenamefont {{O'Leary}}\ \emph {et~al.}(2016)\citenamefont
  {{O'Leary}}, \citenamefont {{Meiron}},\ and\ \citenamefont
  {{Kocsis}}}]{OLearyetal2016}%
  \BibitemOpen
  \bibfield  {author} {\bibinfo {author} {\bibfnamefont {R.~M.}\ \bibnamefont
  {{O'Leary}}}, \bibinfo {author} {\bibfnamefont {Y.}~\bibnamefont {{Meiron}}},
  \ and\ \bibinfo {author} {\bibfnamefont {B.}~\bibnamefont {{Kocsis}}},\
  }\href {\doibase10.3847/2041-8205/824/1/L12} {\bibfield  {journal} {\bibinfo
  {journal} {Astrophysical Journal}\ }\textbf {\bibinfo {volume} {824}},\
  \bibinfo {eid} {L12} (\bibinfo {year} {2016})},\ \Eprint
  {http://arxiv.org/abs/1602.02809} {arXiv:1602.02809
  [astro-ph.HE]}\BibitemShut {NoStop}%
\bibitem [{\citenamefont {{Zevin}}\ \emph {et~al.}(2017)\citenamefont
  {{Zevin}}, \citenamefont {{Pankow}}, \citenamefont {{Rodriguez}},
  \citenamefont {{Sampson}}, \citenamefont {{Chase}}, \citenamefont
  {{Kalogera}},\ and\ \citenamefont {{Rasio}}}]{Zevinetal2017}%
  \BibitemOpen
  \bibfield  {author} {\bibinfo {author} {\bibfnamefont {M.}~\bibnamefont
  {{Zevin}}}, \bibinfo {author} {\bibfnamefont {C.}~\bibnamefont {{Pankow}}},
  \bibinfo {author} {\bibfnamefont {C.~L.}\ \bibnamefont {{Rodriguez}}},
  \bibinfo {author} {\bibfnamefont {L.}~\bibnamefont {{Sampson}}}, \bibinfo
  {author} {\bibfnamefont {E.}~\bibnamefont {{Chase}}}, \bibinfo {author}
  {\bibfnamefont {V.}~\bibnamefont {{Kalogera}}}, \ and\ \bibinfo {author}
  {\bibfnamefont {F.~A.}\ \bibnamefont {{Rasio}}},\ }\href
  {\doibase10.3847/1538-4357/aa8408} {\bibfield  {journal} {\bibinfo  {journal}
  {Astrophysical Journal}\ }\textbf {\bibinfo {volume} {846}},\ \bibinfo {eid}
  {82} (\bibinfo {year} {2017})},\ \Eprint {http://arxiv.org/abs/1704.07379}
  {arXiv:1704.07379 [astro-ph.HE]}\BibitemShut {NoStop}%
\bibitem [{\citenamefont {{Kocsis}}\ \emph {et~al.}(2018)\citenamefont
  {{Kocsis}}, \citenamefont {{Suyama}}, \citenamefont {{Tanaka}},\ and\
  \citenamefont {{Yokoyama}}}]{Kocsisetal2018}%
  \BibitemOpen
  \bibfield  {author} {\bibinfo {author} {\bibfnamefont {B.}~\bibnamefont
  {{Kocsis}}}, \bibinfo {author} {\bibfnamefont {T.}~\bibnamefont {{Suyama}}},
  \bibinfo {author} {\bibfnamefont {T.}~\bibnamefont {{Tanaka}}}, \ and\
  \bibinfo {author} {\bibfnamefont {S.}~\bibnamefont {{Yokoyama}}},\ }\href
  {\doibase10.3847/1538-4357/aaa7f4} {\bibfield  {journal} {\bibinfo  {journal}
  {Astrophysical Journal}\ }\textbf {\bibinfo {volume} {854}},\ \bibinfo {eid}
  {41} (\bibinfo {year} {2018})},\ \Eprint {http://arxiv.org/abs/1709.09007}
  {arXiv:1709.09007}\BibitemShut {NoStop}%
\bibitem [{\citenamefont {Chen}\ and\ \citenamefont
  {Huang}(2018)}]{Chen:2018czv}%
  \BibitemOpen
  \bibfield  {author} {\bibinfo {author} {\bibfnamefont {Zu-Cheng}\
  \bibnamefont {Chen}}\ and\ \bibinfo {author} {\bibfnamefont {Qing-Guo}\
  \bibnamefont {Huang}},\ }\href@noop {} {\  (\bibinfo {year} {2018})},\
  \Eprint {http://arxiv.org/abs/1801.10327} {arXiv:1801.10327
  [astro-ph.CO]}\BibitemShut {NoStop}%
\bibitem [{\citenamefont {Amaro-Seoane}\ and\ \citenamefont
  {Chen}(2016)}]{Amaro-Seoane:2015umi}%
  \BibitemOpen
  \bibfield  {author} {\bibinfo {author} {\bibfnamefont {Pau}\ \bibnamefont
  {Amaro-Seoane}}\ and\ \bibinfo {author} {\bibfnamefont {Xian}\ \bibnamefont
  {Chen}},\ }\href {\doibase10.1093/mnras/stw503} {\bibfield  {journal}
  {\bibinfo  {journal} {Mon. Not. Roy. Astron. Soc.}\ }\textbf {\bibinfo
  {volume} {458}},\ \bibinfo {pages} {3075--3082} (\bibinfo {year} {2016})},\
  \Eprint {http://arxiv.org/abs/1512.04897} {arXiv:1512.04897
  [astro-ph.CO]}\BibitemShut {NoStop}%
\bibitem [{\citenamefont {{Fishbach}}\ \emph {et~al.}(2017)\citenamefont
  {{Fishbach}}, \citenamefont {{Holz}},\ and\ \citenamefont
  {{Farr}}}]{Fishbachetal2017}%
  \BibitemOpen
  \bibfield  {author} {\bibinfo {author} {\bibfnamefont {M.}~\bibnamefont
  {{Fishbach}}}, \bibinfo {author} {\bibfnamefont {D.~E.}\ \bibnamefont
  {{Holz}}}, \ and\ \bibinfo {author} {\bibfnamefont {B.}~\bibnamefont
  {{Farr}}},\ }\href {\doibase10.3847/2041-8213/aa7045} {\bibfield  {journal}
  {\bibinfo  {journal} {Astrophysical Journal}\ }\textbf {\bibinfo {volume}
  {840}},\ \bibinfo {eid} {L24} (\bibinfo {year} {2017})},\ \Eprint
  {http://arxiv.org/abs/1703.06869} {arXiv:1703.06869
  [astro-ph.HE]}\BibitemShut {NoStop}%
\bibitem [{\citenamefont {Morawski}\ \emph {et~al.}(2018)\citenamefont
  {Morawski}, \citenamefont {Giersz}, \citenamefont {Askar},\ and\
  \citenamefont {Belczynski}}]{Morawski:2018kfs}%
  \BibitemOpen
  \bibfield  {author} {\bibinfo {author} {\bibfnamefont {J.}~\bibnamefont
  {Morawski}}, \bibinfo {author} {\bibfnamefont {M.}~\bibnamefont {Giersz}},
  \bibinfo {author} {\bibfnamefont {A.}~\bibnamefont {Askar}}, \ and\ \bibinfo
  {author} {\bibfnamefont {K.}~\bibnamefont {Belczynski}},\ }\href@noop {} {\
  (\bibinfo {year} {2018})},\ \Eprint {http://arxiv.org/abs/1802.01192}
  {arXiv:1802.01192 [astro-ph.GA]}\BibitemShut {NoStop}%
\bibitem [{\citenamefont {Brenneman}(2013)}]{Brenneman:2013oba}%
  \BibitemOpen
  \bibfield  {author} {\bibinfo {author} {\bibfnamefont {Laura}\ \bibnamefont
  {Brenneman}},\ }\href {\doibase10.1007/978-1-4614-7771-6} {\  (\bibinfo
  {year} {2013}),\ 10.1007/978-1-4614-7771-6},\ \Eprint
  {http://arxiv.org/abs/1309.6334} {arXiv:1309.6334 [astro-ph.HE]}\BibitemShut
  {NoStop}%
\bibitem [{\citenamefont {{McClintock}}\ \emph {et~al.}(2014)\citenamefont
  {{McClintock}}, \citenamefont {{Narayan}},\ and\ \citenamefont
  {{Steiner}}}]{McClintocketal2014}%
  \BibitemOpen
  \bibfield  {author} {\bibinfo {author} {\bibfnamefont {J.~E.}\ \bibnamefont
  {{McClintock}}}, \bibinfo {author} {\bibfnamefont {R.}~\bibnamefont
  {{Narayan}}}, \ and\ \bibinfo {author} {\bibfnamefont {J.~F.}\ \bibnamefont
  {{Steiner}}},\ }\href {\doibase10.1007/s11214-013-0003-9} {\bibfield
  {journal} {\bibinfo  {journal} {\ssr}\ }\textbf {\bibinfo {volume} {183}},\
  \bibinfo {pages} {295--322} (\bibinfo {year} {2014})},\ \Eprint
  {http://arxiv.org/abs/1303.1583} {arXiv:1303.1583 [astro-ph.HE]}\BibitemShut
  {NoStop}%
\bibitem [{\citenamefont {{Farr}}\ \emph {et~al.}(2017)\citenamefont {{Farr}},
  \citenamefont {{Stevenson}}, \citenamefont {{Miller}}, \citenamefont
  {{Mandel}}, \citenamefont {{Farr}},\ and\ \citenamefont
  {{Vecchio}}}]{Farretal2017}%
  \BibitemOpen
  \bibfield  {author} {\bibinfo {author} {\bibfnamefont {W.~M.}\ \bibnamefont
  {{Farr}}}, \bibinfo {author} {\bibfnamefont {S.}~\bibnamefont {{Stevenson}}},
  \bibinfo {author} {\bibfnamefont {M.~C.}\ \bibnamefont {{Miller}}}, \bibinfo
  {author} {\bibfnamefont {I.}~\bibnamefont {{Mandel}}}, \bibinfo {author}
  {\bibfnamefont {B.}~\bibnamefont {{Farr}}}, \ and\ \bibinfo {author}
  {\bibfnamefont {A.}~\bibnamefont {{Vecchio}}},\ }\href
  {\doibase10.1038/nature23453} {\bibfield  {journal} {\bibinfo  {journal}
  {Nature}\ }\textbf {\bibinfo {volume} {548}},\ \bibinfo {pages} {426--429}
  (\bibinfo {year} {2017})},\ \Eprint {http://arxiv.org/abs/1706.01385}
  {arXiv:1706.01385 [astro-ph.HE]}\BibitemShut {NoStop}%
\bibitem [{\citenamefont {{Farr}}\ \emph {et~al.}(2018)\citenamefont {{Farr}},
  \citenamefont {{Holz}},\ and\ \citenamefont {{Farr}}}]{Farretal2018}%
  \BibitemOpen
  \bibfield  {author} {\bibinfo {author} {\bibfnamefont {B.}~\bibnamefont
  {{Farr}}}, \bibinfo {author} {\bibfnamefont {D.~E.}\ \bibnamefont {{Holz}}},
  \ and\ \bibinfo {author} {\bibfnamefont {W.~M.}\ \bibnamefont {{Farr}}},\
  }\href {\doibase10.3847/2041-8213/aaaa64} {\bibfield  {journal} {\bibinfo
  {journal} {Astrophysical Journal}\ }\textbf {\bibinfo {volume} {854}},\
  \bibinfo {eid} {L9} (\bibinfo {year} {2018})},\ \Eprint
  {http://arxiv.org/abs/1709.07896} {arXiv:1709.07896
  [astro-ph.HE]}\BibitemShut {NoStop}%
\bibitem [{\citenamefont {{O'Leary}}\ \emph {et~al.}(2006)\citenamefont
  {{O'Leary}}, \citenamefont {{Rasio}}, \citenamefont {{Fregeau}},
  \citenamefont {{Ivanova}},\ and\ \citenamefont
  {{O'Shaughnessy}}}]{OLearyetal2006}%
  \BibitemOpen
  \bibfield  {author} {\bibinfo {author} {\bibfnamefont {R.~M.}\ \bibnamefont
  {{O'Leary}}}, \bibinfo {author} {\bibfnamefont {F.~A.}\ \bibnamefont
  {{Rasio}}}, \bibinfo {author} {\bibfnamefont {J.~M.}\ \bibnamefont
  {{Fregeau}}}, \bibinfo {author} {\bibfnamefont {N.}~\bibnamefont
  {{Ivanova}}}, \ and\ \bibinfo {author} {\bibfnamefont {R.}~\bibnamefont
  {{O'Shaughnessy}}},\ }\href {\doibase10.1086/498446} {\bibfield  {journal}
  {\bibinfo  {journal} {Astrophysical Journal}\ }\textbf {\bibinfo {volume}
  {637}},\ \bibinfo {pages} {937--951} (\bibinfo {year} {2006})},\ \Eprint
  {http://arxiv.org/abs/astro-ph/0508224} {astro-ph/0508224}\BibitemShut
  {NoStop}%
\bibitem [{\citenamefont {{Breivik}}\ \emph {et~al.}(2016)\citenamefont
  {{Breivik}}, \citenamefont {{Rodriguez}}, \citenamefont {{Larson}},
  \citenamefont {{Kalogera}},\ and\ \citenamefont {{Rasio}}}]{Breiviketal2016}%
  \BibitemOpen
  \bibfield  {author} {\bibinfo {author} {\bibfnamefont {K.}~\bibnamefont
  {{Breivik}}}, \bibinfo {author} {\bibfnamefont {C.~L.}\ \bibnamefont
  {{Rodriguez}}}, \bibinfo {author} {\bibfnamefont {S.~L.}\ \bibnamefont
  {{Larson}}}, \bibinfo {author} {\bibfnamefont {V.}~\bibnamefont
  {{Kalogera}}}, \ and\ \bibinfo {author} {\bibfnamefont {F.~A.}\ \bibnamefont
  {{Rasio}}},\ }\href {\doibase10.3847/2041-8205/830/1/L18} {\bibfield
  {journal} {\bibinfo  {journal} {Astrophysical Journal}\ }\textbf {\bibinfo
  {volume} {830}},\ \bibinfo {eid} {L18} (\bibinfo {year} {2016})},\ \Eprint
  {http://arxiv.org/abs/1606.09558} {arXiv:1606.09558}\BibitemShut {NoStop}%
\bibitem [{\citenamefont {D'Orazio}\ and\ \citenamefont
  {Samsing}(2018)}]{DOrazio:2018jnv}%
  \BibitemOpen
  \bibfield  {author} {\bibinfo {author} {\bibfnamefont {Daniel~J.}\
  \bibnamefont {D'Orazio}}\ and\ \bibinfo {author} {\bibfnamefont {Johan}\
  \bibnamefont {Samsing}},\ }\href@noop {} {\  (\bibinfo {year} {2018})},\
  \Eprint {http://arxiv.org/abs/1805.06194} {arXiv:1805.06194
  [astro-ph.HE]}\BibitemShut {NoStop}%
\bibitem [{\citenamefont {Samsing}\ \emph
  {et~al.}(2018{\natexlab{b}})\citenamefont {Samsing}, \citenamefont
  {D'Orazio}, \citenamefont {Askar},\ and\ \citenamefont
  {Giersz}}]{Samsing:2018ykz}%
  \BibitemOpen
  \bibfield  {author} {\bibinfo {author} {\bibfnamefont {Johan}\ \bibnamefont
  {Samsing}}, \bibinfo {author} {\bibfnamefont {Daniel~J.}\ \bibnamefont
  {D'Orazio}}, \bibinfo {author} {\bibfnamefont {Abbas}\ \bibnamefont {Askar}},
  \ and\ \bibinfo {author} {\bibfnamefont {Mirek}\ \bibnamefont {Giersz}},\
  }\href@noop {} {\  (\bibinfo {year} {2018}{\natexlab{b}})},\ \Eprint
  {http://arxiv.org/abs/1802.08654} {arXiv:1802.08654
  [astro-ph.HE]}\BibitemShut {NoStop}%
\bibitem [{\citenamefont {Chen}\ and\ \citenamefont
  {Amaro-Seoane}(2017)}]{Chen:2017gfm}%
  \BibitemOpen
  \bibfield  {author} {\bibinfo {author} {\bibfnamefont {Xian}\ \bibnamefont
  {Chen}}\ and\ \bibinfo {author} {\bibfnamefont {Pau}\ \bibnamefont
  {Amaro-Seoane}},\ }\href {\doibase10.3847/2041-8213/aa74ce} {\bibfield
  {journal} {\bibinfo  {journal} {Astrophys. J.}\ }\textbf {\bibinfo {volume}
  {842}},\ \bibinfo {pages} {L2} (\bibinfo {year} {2017})},\ \Eprint
  {http://arxiv.org/abs/1702.08479} {arXiv:1702.08479
  [astro-ph.HE]}\BibitemShut {NoStop}%
\bibitem [{\citenamefont {Gondán}\ \emph
  {et~al.}(2018{\natexlab{a}})\citenamefont {Gondán}, \citenamefont {Kocsis},
  \citenamefont {Raffai},\ and\ \citenamefont {Frei}}]{Gondan:2017wzd}%
  \BibitemOpen
  \bibfield  {author} {\bibinfo {author} {\bibfnamefont {László}\
  \bibnamefont {Gondán}}, \bibinfo {author} {\bibfnamefont {Bence}\
  \bibnamefont {Kocsis}}, \bibinfo {author} {\bibfnamefont {Péter}\
  \bibnamefont {Raffai}}, \ and\ \bibinfo {author} {\bibfnamefont {Zsolt}\
  \bibnamefont {Frei}},\ }\href {\doibase10.3847/1538-4357/aabfee} {\bibfield
  {journal} {\bibinfo  {journal} {Astrophys. J.}\ }\textbf {\bibinfo {volume}
  {860}},\ \bibinfo {pages} {5} (\bibinfo {year} {2018}{\natexlab{a}})},\
  \Eprint {http://arxiv.org/abs/1711.09989} {arXiv:1711.09989
  [astro-ph.HE]}\BibitemShut {NoStop}%
\bibitem [{\citenamefont {Gondán}\ \emph
  {et~al.}(2018{\natexlab{b}})\citenamefont {Gondán}, \citenamefont {Kocsis},
  \citenamefont {Raffai},\ and\ \citenamefont {Frei}}]{Gondan:2017hbp}%
  \BibitemOpen
  \bibfield  {author} {\bibinfo {author} {\bibfnamefont {László}\
  \bibnamefont {Gondán}}, \bibinfo {author} {\bibfnamefont {Bence}\
  \bibnamefont {Kocsis}}, \bibinfo {author} {\bibfnamefont {Péter}\
  \bibnamefont {Raffai}}, \ and\ \bibinfo {author} {\bibfnamefont {Zsolt}\
  \bibnamefont {Frei}},\ }\href {\doibase10.3847/1538-4357/aaad0e} {\bibfield
  {journal} {\bibinfo  {journal} {Astrophys. J.}\ }\textbf {\bibinfo {volume}
  {855}},\ \bibinfo {pages} {34} (\bibinfo {year} {2018}{\natexlab{b}})},\
  \Eprint {http://arxiv.org/abs/1705.10781} {arXiv:1705.10781
  [astro-ph.HE]}\BibitemShut {NoStop}%
\bibitem [{\citenamefont {Kremer}\ \emph {et~al.}(2018)\citenamefont {Kremer},
  \citenamefont {Chatterjee}, \citenamefont {Breivik}, \citenamefont
  {Rodriguez}, \citenamefont {Larson},\ and\ \citenamefont
  {Rasio}}]{Kremer:2018tzm}%
  \BibitemOpen
  \bibfield  {author} {\bibinfo {author} {\bibfnamefont {Kyle}\ \bibnamefont
  {Kremer}}, \bibinfo {author} {\bibfnamefont {Sourav}\ \bibnamefont
  {Chatterjee}}, \bibinfo {author} {\bibfnamefont {Katelyn}\ \bibnamefont
  {Breivik}}, \bibinfo {author} {\bibfnamefont {Carl~L.}\ \bibnamefont
  {Rodriguez}}, \bibinfo {author} {\bibfnamefont {Shane~L.}\ \bibnamefont
  {Larson}}, \ and\ \bibinfo {author} {\bibfnamefont {Frederic~A.}\
  \bibnamefont {Rasio}},\ }\href {\doibase10.1103/PhysRevLett.120.191103}
  {\bibfield  {journal} {\bibinfo  {journal} {Phys. Rev. Lett.}\ }\textbf
  {\bibinfo {volume} {120}},\ \bibinfo {pages} {191103} (\bibinfo {year}
  {2018})},\ \Eprint {http://arxiv.org/abs/1802.05661} {arXiv:1802.05661
  [astro-ph.HE]}\BibitemShut {NoStop}%
\bibitem [{\citenamefont {{Yunes}}\ \emph {et~al.}(2011)\citenamefont
  {{Yunes}}, \citenamefont {{Miller}},\ and\ \citenamefont
  {{Thornburg}}}]{Yunesetal2011}%
  \BibitemOpen
  \bibfield  {author} {\bibinfo {author} {\bibfnamefont {N.}~\bibnamefont
  {{Yunes}}}, \bibinfo {author} {\bibfnamefont {M.~C.}\ \bibnamefont
  {{Miller}}}, \ and\ \bibinfo {author} {\bibfnamefont {J.}~\bibnamefont
  {{Thornburg}}},\ }\href {\doibase10.1103/PhysRevD.83.044030} {\bibfield
  {journal} {\bibinfo  {journal} {Physical Review D}\ }\textbf {\bibinfo
  {volume} {83}},\ \bibinfo {eid} {044030} (\bibinfo {year} {2011})},\ \Eprint
  {http://arxiv.org/abs/1010.1721} {arXiv:1010.1721 [astro-ph.GA]}\BibitemShut
  {NoStop}%
\bibitem [{\citenamefont {{Meiron}}\ \emph {et~al.}(2017)\citenamefont
  {{Meiron}}, \citenamefont {{Kocsis}},\ and\ \citenamefont
  {{Loeb}}}]{Meironetal2017}%
  \BibitemOpen
  \bibfield  {author} {\bibinfo {author} {\bibfnamefont {Y.}~\bibnamefont
  {{Meiron}}}, \bibinfo {author} {\bibfnamefont {B.}~\bibnamefont {{Kocsis}}},
  \ and\ \bibinfo {author} {\bibfnamefont {A.}~\bibnamefont {{Loeb}}},\ }\href
  {\doibase10.3847/1538-4357/834/2/200} {\bibfield  {journal} {\bibinfo
  {journal} {Astrophysical Journal}\ }\textbf {\bibinfo {volume} {834}},\
  \bibinfo {eid} {200} (\bibinfo {year} {2017})},\ \Eprint
  {http://arxiv.org/abs/1604.02148} {arXiv:1604.02148
  [astro-ph.HE]}\BibitemShut {NoStop}%
\bibitem [{\citenamefont {Inayoshi}\ \emph {et~al.}(2017)\citenamefont
  {Inayoshi}, \citenamefont {Tamanini}, \citenamefont {Caprini},\ and\
  \citenamefont {Haiman}}]{Inayoshi:2017hgw}%
  \BibitemOpen
  \bibfield  {author} {\bibinfo {author} {\bibfnamefont {Kohei}\ \bibnamefont
  {Inayoshi}}, \bibinfo {author} {\bibfnamefont {Nicola}\ \bibnamefont
  {Tamanini}}, \bibinfo {author} {\bibfnamefont {Chiara}\ \bibnamefont
  {Caprini}}, \ and\ \bibinfo {author} {\bibfnamefont {Zoltan}\ \bibnamefont
  {Haiman}},\ }\href {\doibase10.1103/PhysRevD.96.063014} {\bibfield  {journal}
  {\bibinfo  {journal} {Phys. Rev.}\ }\textbf {\bibinfo {volume} {D96}},\
  \bibinfo {pages} {063014} (\bibinfo {year} {2017})},\ \Eprint
  {http://arxiv.org/abs/1702.06529} {arXiv:1702.06529
  [astro-ph.HE]}\BibitemShut {NoStop}%
\bibitem [{\citenamefont {{Kocsis}}(2013)}]{Kocsis2013}%
  \BibitemOpen
  \bibfield  {author} {\bibinfo {author} {\bibfnamefont {B.}~\bibnamefont
  {{Kocsis}}},\ }\href {\doibase10.1088/0004-637X/763/2/122} {\bibfield
  {journal} {\bibinfo  {journal} {Astrophysical Journal}\ }\textbf {\bibinfo
  {volume} {763}},\ \bibinfo {eid} {122} (\bibinfo {year} {2013})},\ \Eprint
  {http://arxiv.org/abs/1211.6427} {arXiv:1211.6427 [astro-ph.HE]}\BibitemShut
  {NoStop}%
\bibitem [{\citenamefont {Samsing}\ and\ \citenamefont
  {Ilan}(2017)}]{Samsing:2017xod}%
  \BibitemOpen
  \bibfield  {author} {\bibinfo {author} {\bibfnamefont {Johan}\ \bibnamefont
  {Samsing}}\ and\ \bibinfo {author} {\bibfnamefont {Teva}\ \bibnamefont
  {Ilan}},\ }\href@noop {} {\  (\bibinfo {year} {2017})},\ \Eprint
  {http://arxiv.org/abs/1709.01660} {arXiv:1709.01660
  [astro-ph.HE]}\BibitemShut {NoStop}%
\bibitem [{\citenamefont {Bertone}\ and\ \citenamefont
  {Hooper}(2016)}]{Bertone:2016nfn}%
  \BibitemOpen
  \bibfield  {author} {\bibinfo {author} {\bibfnamefont {Gianfranco}\
  \bibnamefont {Bertone}}\ and\ \bibinfo {author} {\bibfnamefont {Dan}\
  \bibnamefont {Hooper}},\ }\href@noop {} {\bibfield  {journal} {\bibinfo
  {journal} {Submitted to: Rev. Mod. Phys.}\ } (\bibinfo {year} {2016})},\
  \Eprint {http://arxiv.org/abs/1605.04909} {arXiv:1605.04909
  [astro-ph.CO]}\BibitemShut {NoStop}%
\bibitem [{\citenamefont {Silk}\ \emph {et~al.}(2010)\citenamefont {Silk} \emph
  {et~al.}}]{Bertone:2010zza}%
  \BibitemOpen
  \bibfield  {author} {\bibinfo {author} {\bibfnamefont {J.}~\bibnamefont
  {Silk}} \emph {et~al.},\ }\href {\doibase10.1017/CBO9780511770739} {\emph
  {\bibinfo {title} {{Particle Dark Matter: Observations, Models and
  Searches}}}},\ edited by\ \bibinfo {editor} {\bibfnamefont {Gianfranco}\
  \bibnamefont {Bertone}}\ (\bibinfo  {publisher} {Cambridge Univ. Press},\
  \bibinfo {address} {Cambridge},\ \bibinfo {year} {2010})\BibitemShut
  {NoStop}%
\bibitem [{\citenamefont {Bergström}(2000)}]{Bergstrom:2000pn}%
  \BibitemOpen
  \bibfield  {author} {\bibinfo {author} {\bibfnamefont {Lars}\ \bibnamefont
  {Bergström}},\ }\href {\doibase10.1088/0034-4885/63/5/2r3} {\bibfield
  {journal} {\bibinfo  {journal} {Rept. Prog. Phys.}\ }\textbf {\bibinfo
  {volume} {63}},\ \bibinfo {pages} {793} (\bibinfo {year} {2000})},\ \Eprint
  {http://arxiv.org/abs/hep-ph/0002126} {arXiv:hep-ph/0002126
  [hep-ph]}\BibitemShut {NoStop}%
\bibitem [{\citenamefont {Bertone}\ \emph
  {et~al.}(2005{\natexlab{a}})\citenamefont {Bertone}, \citenamefont {Hooper},\
  and\ \citenamefont {Silk}}]{Bertone:2004pz}%
  \BibitemOpen
  \bibfield  {author} {\bibinfo {author} {\bibfnamefont {Gianfranco}\
  \bibnamefont {Bertone}}, \bibinfo {author} {\bibfnamefont {Dan}\ \bibnamefont
  {Hooper}}, \ and\ \bibinfo {author} {\bibfnamefont {Joseph}\ \bibnamefont
  {Silk}},\ }\href {\doibase10.1016/j.physrep.2004.08.031} {\bibfield
  {journal} {\bibinfo  {journal} {Phys. Rept.}\ }\textbf {\bibinfo {volume}
  {405}},\ \bibinfo {pages} {279--390} (\bibinfo {year}
  {2005}{\natexlab{a}})},\ \Eprint {http://arxiv.org/abs/hep-ph/0404175}
  {arXiv:hep-ph/0404175 [hep-ph]}\BibitemShut {NoStop}%
\bibitem [{\citenamefont {Feng}(2010)}]{Feng:2010gw}%
  \BibitemOpen
  \bibfield  {author} {\bibinfo {author} {\bibfnamefont {Jonathan~L.}\
  \bibnamefont {Feng}},\ }\href {\doibase10.1146/annurev-astro-082708-101659}
  {\bibfield  {journal} {\bibinfo  {journal} {Ann. Rev. Astron. Astrophys.}\
  }\textbf {\bibinfo {volume} {48}},\ \bibinfo {pages} {495--545} (\bibinfo
  {year} {2010})},\ \Eprint {http://arxiv.org/abs/1003.0904} {arXiv:1003.0904
  [astro-ph.CO]}\BibitemShut {NoStop}%
\bibitem [{\citenamefont {Carr}\ \emph {et~al.}(2016)\citenamefont {Carr},
  \citenamefont {Kuhnel},\ and\ \citenamefont {Sandstad}}]{Carr:2016drx}%
  \BibitemOpen
  \bibfield  {author} {\bibinfo {author} {\bibfnamefont {Bernard}\ \bibnamefont
  {Carr}}, \bibinfo {author} {\bibfnamefont {Florian}\ \bibnamefont {Kuhnel}},
  \ and\ \bibinfo {author} {\bibfnamefont {Marit}\ \bibnamefont {Sandstad}},\
  }\href {\doibase10.1103/PhysRevD.94.083504} {\bibfield  {journal} {\bibinfo
  {journal} {Phys. Rev.}\ }\textbf {\bibinfo {volume} {D94}},\ \bibinfo {pages}
  {083504} (\bibinfo {year} {2016})},\ \Eprint
  {http://arxiv.org/abs/1607.06077} {arXiv:1607.06077
  [astro-ph.CO]}\BibitemShut {NoStop}%
\bibitem [{\citenamefont {Clesse}\ and\ \citenamefont
  {Garc\'ia-Bellido}(2017{\natexlab{b}})}]{Clesse:2016vqa}%
  \BibitemOpen
  \bibfield  {author} {\bibinfo {author} {\bibfnamefont {Sebastien}\
  \bibnamefont {Clesse}}\ and\ \bibinfo {author} {\bibfnamefont {Juan}\
  \bibnamefont {Garc\'ia-Bellido}},\ }\href
  {\doibase10.1016/j.dark.2016.10.002} {\bibfield  {journal} {\bibinfo
  {journal} {Phys. Dark Univ.}\ }\textbf {\bibinfo {volume} {15}},\ \bibinfo
  {pages} {142--147} (\bibinfo {year} {2017}{\natexlab{b}})},\ \Eprint
  {http://arxiv.org/abs/1603.05234} {arXiv:1603.05234
  [astro-ph.CO]}\BibitemShut {NoStop}%
\bibitem [{\citenamefont {García-Bellido}\ and\ \citenamefont
  {Clesse}(2018)}]{Garcia-Bellido:2017xvr}%
  \BibitemOpen
  \bibfield  {author} {\bibinfo {author} {\bibfnamefont {Juan}\ \bibnamefont
  {García-Bellido}}\ and\ \bibinfo {author} {\bibfnamefont {Sebastien}\
  \bibnamefont {Clesse}},\ }\href {\doibase10.1016/j.dark.2018.01.001}
  {\bibfield  {journal} {\bibinfo  {journal} {Phys. Dark Univ.}\ }\textbf
  {\bibinfo {volume} {19}},\ \bibinfo {pages} {144--148} (\bibinfo {year}
  {2018})},\ \Eprint {http://arxiv.org/abs/1710.04694} {arXiv:1710.04694
  [astro-ph.CO]}\BibitemShut {NoStop}%
\bibitem [{\citenamefont {García-Bellido}(2017)}]{Garcia-Bellido:2017fdg}%
  \BibitemOpen
  \bibfield  {author} {\bibinfo {author} {\bibfnamefont {Juan}\ \bibnamefont
  {García-Bellido}},\ }\bibfield  {booktitle} {\emph {\bibinfo {booktitle}
  {{Proceedings, 11th International LISA Symposium: Zurich, Switzerland,
  September 5-9, 2016}}},\ }\href {\doibase10.1088/1742-6596/840/1/012032}
  {\bibfield  {journal} {\bibinfo  {journal} {J. Phys. Conf. Ser.}\ }\textbf
  {\bibinfo {volume} {840}},\ \bibinfo {pages} {012032} (\bibinfo {year}
  {2017})},\ \Eprint {http://arxiv.org/abs/1702.08275} {arXiv:1702.08275
  [astro-ph.CO]}\BibitemShut {NoStop}%
\bibitem [{\citenamefont {Green}(2015)}]{Green:2014faa}%
  \BibitemOpen
  \bibfield  {author} {\bibinfo {author} {\bibfnamefont {Anne~M.}\ \bibnamefont
  {Green}},\ }\href {\doibase10.1007/978-3-319-10852-0_5} {\bibfield  {journal}
  {\bibinfo  {journal} {Fundam. Theor. Phys.}\ }\textbf {\bibinfo {volume}
  {178}},\ \bibinfo {pages} {129--149} (\bibinfo {year} {2015})},\ \Eprint
  {http://arxiv.org/abs/1403.1198} {arXiv:1403.1198 [gr-qc]}\BibitemShut
  {NoStop}%
\bibitem [{\citenamefont {Garcia-Bellido}\ \emph {et~al.}(1996)\citenamefont
  {Garcia-Bellido}, \citenamefont {Linde},\ and\ \citenamefont
  {Wands}}]{GarciaBellido:1996qt}%
  \BibitemOpen
  \bibfield  {author} {\bibinfo {author} {\bibfnamefont {Juan}\ \bibnamefont
  {Garcia-Bellido}}, \bibinfo {author} {\bibfnamefont {Andrei~D.}\ \bibnamefont
  {Linde}}, \ and\ \bibinfo {author} {\bibfnamefont {David}\ \bibnamefont
  {Wands}},\ }\href {\doibase10.1103/PhysRevD.54.6040} {\bibfield  {journal}
  {\bibinfo  {journal} {Phys. Rev.}\ }\textbf {\bibinfo {volume} {D54}},\
  \bibinfo {pages} {6040--6058} (\bibinfo {year} {1996})},\ \Eprint
  {http://arxiv.org/abs/astro-ph/9605094} {arXiv:astro-ph/9605094
  [astro-ph]}\BibitemShut {NoStop}%
\bibitem [{\citenamefont {Clesse}\ and\ \citenamefont
  {Garc\'ia-Bellido}(2015)}]{Clesse:2015wea}%
  \BibitemOpen
  \bibfield  {author} {\bibinfo {author} {\bibfnamefont {S.}~\bibnamefont
  {Clesse}}\ and\ \bibinfo {author} {\bibfnamefont {J.}~\bibnamefont
  {Garc\'ia-Bellido}},\ }\href {\doibase10.1103/PhysRevD.92.023524} {\bibfield
  {journal} {\bibinfo  {journal} {Phys. Rev.}\ }\textbf {\bibinfo {volume}
  {D92}},\ \bibinfo {pages} {023524} (\bibinfo {year} {2015})},\ \Eprint
  {http://arxiv.org/abs/1501.07565} {arXiv:1501.07565
  [astro-ph.CO]}\BibitemShut {NoStop}%
\bibitem [{\citenamefont {Nakamura}\ \emph {et~al.}(1997)\citenamefont
  {Nakamura}, \citenamefont {Sasaki}, \citenamefont {Tanaka},\ and\
  \citenamefont {Thorne}}]{Nakamura:1997sm}%
  \BibitemOpen
  \bibfield  {author} {\bibinfo {author} {\bibfnamefont {Takashi}\ \bibnamefont
  {Nakamura}}, \bibinfo {author} {\bibfnamefont {Misao}\ \bibnamefont
  {Sasaki}}, \bibinfo {author} {\bibfnamefont {Takahiro}\ \bibnamefont
  {Tanaka}}, \ and\ \bibinfo {author} {\bibfnamefont {Kip~S.}\ \bibnamefont
  {Thorne}},\ }\href {\doibase10.1086/310886} {\bibfield  {journal} {\bibinfo
  {journal} {Astrophys. J.}\ }\textbf {\bibinfo {volume} {487}},\ \bibinfo
  {pages} {L139--L142} (\bibinfo {year} {1997})},\ \Eprint
  {http://arxiv.org/abs/astro-ph/9708060} {arXiv:astro-ph/9708060
  [astro-ph]}\BibitemShut {NoStop}%
\bibitem [{\citenamefont {Ioka}\ \emph {et~al.}(1998)\citenamefont {Ioka},
  \citenamefont {Chiba}, \citenamefont {Tanaka},\ and\ \citenamefont
  {Nakamura}}]{Ioka:1998nz}%
  \BibitemOpen
  \bibfield  {author} {\bibinfo {author} {\bibfnamefont {Kunihito}\
  \bibnamefont {Ioka}}, \bibinfo {author} {\bibfnamefont {Takeshi}\
  \bibnamefont {Chiba}}, \bibinfo {author} {\bibfnamefont {Takahiro}\
  \bibnamefont {Tanaka}}, \ and\ \bibinfo {author} {\bibfnamefont {Takashi}\
  \bibnamefont {Nakamura}},\ }\href {\doibase10.1103/PhysRevD.58.063003}
  {\bibfield  {journal} {\bibinfo  {journal} {Phys. Rev.}\ }\textbf {\bibinfo
  {volume} {D58}},\ \bibinfo {pages} {063003} (\bibinfo {year} {1998})},\
  \Eprint {http://arxiv.org/abs/astro-ph/9807018} {arXiv:astro-ph/9807018
  [astro-ph]}\BibitemShut {NoStop}%
\bibitem [{\citenamefont {Kavanagh}\ \emph {et~al.}(2018)\citenamefont
  {Kavanagh}, \citenamefont {Gaggero},\ and\ \citenamefont
  {Bertone}}]{Kavanagh:2018ggo}%
  \BibitemOpen
  \bibfield  {author} {\bibinfo {author} {\bibfnamefont {Bradley~J.}\
  \bibnamefont {Kavanagh}}, \bibinfo {author} {\bibfnamefont {Daniele}\
  \bibnamefont {Gaggero}}, \ and\ \bibinfo {author} {\bibfnamefont
  {Gianfranco}\ \bibnamefont {Bertone}},\ }\href@noop {} {\  (\bibinfo {year}
  {2018})},\ \Eprint {http://arxiv.org/abs/1805.09034} {arXiv:1805.09034
  [astro-ph.CO]}\BibitemShut {NoStop}%
\bibitem [{\citenamefont {Ali-Haïmoud}(2018)}]{Ali-Haimoud:2018dau}%
  \BibitemOpen
  \bibfield  {author} {\bibinfo {author} {\bibfnamefont {Yacine}\ \bibnamefont
  {Ali-Haïmoud}},\ }\href@noop {} {\  (\bibinfo {year} {2018})},\ \Eprint
  {http://arxiv.org/abs/1805.05912} {arXiv:1805.05912
  [astro-ph.CO]}\BibitemShut {NoStop}%
\bibitem [{\citenamefont {Chiba}\ and\ \citenamefont
  {Yokoyama}(2017)}]{Chiba:2017rvs}%
  \BibitemOpen
  \bibfield  {author} {\bibinfo {author} {\bibfnamefont {Takeshi}\ \bibnamefont
  {Chiba}}\ and\ \bibinfo {author} {\bibfnamefont {Shuichiro}\ \bibnamefont
  {Yokoyama}},\ }\href {\doibase10.1093/ptep/ptx087} {\bibfield  {journal}
  {\bibinfo  {journal} {PTEP}\ }\textbf {\bibinfo {volume} {2017}},\ \bibinfo
  {pages} {083E01} (\bibinfo {year} {2017})},\ \Eprint
  {http://arxiv.org/abs/1704.06573} {arXiv:1704.06573 [gr-qc]}\BibitemShut
  {NoStop}%
\bibitem [{\citenamefont {Ezquiaga}\ \emph {et~al.}(2018)\citenamefont
  {Ezquiaga}, \citenamefont {Garcia-Bellido},\ and\ \citenamefont
  {Ruiz~Morales}}]{Ezquiaga:2017fvi}%
  \BibitemOpen
  \bibfield  {author} {\bibinfo {author} {\bibfnamefont {Jose~Maria}\
  \bibnamefont {Ezquiaga}}, \bibinfo {author} {\bibfnamefont {Juan}\
  \bibnamefont {Garcia-Bellido}}, \ and\ \bibinfo {author} {\bibfnamefont
  {Ester}\ \bibnamefont {Ruiz~Morales}},\ }\href
  {\doibase10.1016/j.physletb.2017.11.039} {\bibfield  {journal} {\bibinfo
  {journal} {Phys. Lett.}\ }\textbf {\bibinfo {volume} {B776}},\ \bibinfo
  {pages} {345--349} (\bibinfo {year} {2018})},\ \Eprint
  {http://arxiv.org/abs/1705.04861} {arXiv:1705.04861
  [astro-ph.CO]}\BibitemShut {NoStop}%
\bibitem [{\citenamefont {Garcia-Bellido}\ and\ \citenamefont
  {Ruiz~Morales}(2017)}]{Garcia-Bellido:2017mdw}%
  \BibitemOpen
  \bibfield  {author} {\bibinfo {author} {\bibfnamefont {Juan}\ \bibnamefont
  {Garcia-Bellido}}\ and\ \bibinfo {author} {\bibfnamefont {Ester}\
  \bibnamefont {Ruiz~Morales}},\ }\href {\doibase10.1016/j.dark.2017.09.007}
  {\bibfield  {journal} {\bibinfo  {journal} {Phys. Dark Univ.}\ }\textbf
  {\bibinfo {volume} {18}},\ \bibinfo {pages} {47--54} (\bibinfo {year}
  {2017})},\ \Eprint {http://arxiv.org/abs/1702.03901} {arXiv:1702.03901
  [astro-ph.CO]}\BibitemShut {NoStop}%
\bibitem [{\citenamefont {Byrnes}\ \emph {et~al.}(2018)\citenamefont {Byrnes},
  \citenamefont {Hindmarsh}, \citenamefont {Young},\ and\ \citenamefont
  {Hawkins}}]{Byrnes:2018clq}%
  \BibitemOpen
  \bibfield  {author} {\bibinfo {author} {\bibfnamefont {Christian~T.}\
  \bibnamefont {Byrnes}}, \bibinfo {author} {\bibfnamefont {Mark}\ \bibnamefont
  {Hindmarsh}}, \bibinfo {author} {\bibfnamefont {Sam}\ \bibnamefont {Young}},
  \ and\ \bibinfo {author} {\bibfnamefont {Michael R.~S.}\ \bibnamefont
  {Hawkins}},\ }\href@noop {} {\  (\bibinfo {year} {2018})},\ \Eprint
  {http://arxiv.org/abs/1801.06138} {arXiv:1801.06138
  [astro-ph.CO]}\BibitemShut {NoStop}%
\bibitem [{\citenamefont {Nakama}\ \emph {et~al.}(2017)\citenamefont {Nakama},
  \citenamefont {Silk},\ and\ \citenamefont {Kamionkowski}}]{Nakama:2016gzw}%
  \BibitemOpen
  \bibfield  {author} {\bibinfo {author} {\bibfnamefont {Tomohiro}\
  \bibnamefont {Nakama}}, \bibinfo {author} {\bibfnamefont {Joseph}\
  \bibnamefont {Silk}}, \ and\ \bibinfo {author} {\bibfnamefont {Marc}\
  \bibnamefont {Kamionkowski}},\ }\href {\doibase10.1103/PhysRevD.95.043511}
  {\bibfield  {journal} {\bibinfo  {journal} {Phys. Rev.}\ }\textbf {\bibinfo
  {volume} {D95}},\ \bibinfo {pages} {043511} (\bibinfo {year} {2017})},\
  \Eprint {http://arxiv.org/abs/1612.06264} {arXiv:1612.06264
  [astro-ph.CO]}\BibitemShut {NoStop}%
\bibitem [{\citenamefont {Garcia-Bellido}\ \emph {et~al.}(2017)\citenamefont
  {Garcia-Bellido}, \citenamefont {Peloso},\ and\ \citenamefont
  {Unal}}]{Garcia-Bellido:2017aan}%
  \BibitemOpen
  \bibfield  {author} {\bibinfo {author} {\bibfnamefont {Juan}\ \bibnamefont
  {Garcia-Bellido}}, \bibinfo {author} {\bibfnamefont {Marco}\ \bibnamefont
  {Peloso}}, \ and\ \bibinfo {author} {\bibfnamefont {Caner}\ \bibnamefont
  {Unal}},\ }\href {\doibase10.1088/1475-7516/2017/09/013} {\bibfield
  {journal} {\bibinfo  {journal} {JCAP}\ }\textbf {\bibinfo {volume} {1709}},\
  \bibinfo {pages} {013} (\bibinfo {year} {2017})},\ \Eprint
  {http://arxiv.org/abs/1707.02441} {arXiv:1707.02441
  [astro-ph.CO]}\BibitemShut {NoStop}%
\bibitem [{\citenamefont {Allsman}\ \emph {et~al.}(2001)\citenamefont {Allsman}
  \emph {et~al.}}]{Allsman:2000kg}%
  \BibitemOpen
  \bibfield  {author} {\bibinfo {author} {\bibfnamefont {R.~A.}\ \bibnamefont
  {Allsman}} \emph {et~al.} (\bibinfo {collaboration} {Macho}),\ }\href
  {\doibase10.1086/319636} {\bibfield  {journal} {\bibinfo  {journal}
  {Astrophys. J.}\ }\textbf {\bibinfo {volume} {550}},\ \bibinfo {pages} {L169}
  (\bibinfo {year} {2001})},\ \Eprint {http://arxiv.org/abs/astro-ph/0011506}
  {arXiv:astro-ph/0011506 [astro-ph]}\BibitemShut {NoStop}%
\bibitem [{\citenamefont {Tisserand}\ \emph {et~al.}(2007)\citenamefont
  {Tisserand} \emph {et~al.}}]{Tisserand:2006zx}%
  \BibitemOpen
  \bibfield  {author} {\bibinfo {author} {\bibfnamefont {P.}~\bibnamefont
  {Tisserand}} \emph {et~al.} (\bibinfo {collaboration} {EROS-2}),\ }\href
  {\doibase10.1051/0004-6361:20066017} {\bibfield  {journal} {\bibinfo
  {journal} {Astron. Astrophys.}\ }\textbf {\bibinfo {volume} {469}},\ \bibinfo
  {pages} {387--404} (\bibinfo {year} {2007})},\ \Eprint
  {http://arxiv.org/abs/astro-ph/0607207} {arXiv:astro-ph/0607207
  [astro-ph]}\BibitemShut {NoStop}%
\bibitem [{\citenamefont {Hawkins}(2015)}]{Hawkins:2015uja}%
  \BibitemOpen
  \bibfield  {author} {\bibinfo {author} {\bibfnamefont {M.~R.~S.}\
  \bibnamefont {Hawkins}},\ }\href {\doibase10.1051/0004-6361/201425400}
  {\bibfield  {journal} {\bibinfo  {journal} {Astron. Astrophys.}\ }\textbf
  {\bibinfo {volume} {575}},\ \bibinfo {pages} {A107} (\bibinfo {year}
  {2015})},\ \Eprint {http://arxiv.org/abs/1503.01935} {arXiv:1503.01935
  [astro-ph.GA]}\BibitemShut {NoStop}%
\bibitem [{\citenamefont {Calchi~Novati}\ \emph {et~al.}(2005)\citenamefont
  {Calchi~Novati} \emph {et~al.}}]{CalchiNovati:2005cd}%
  \BibitemOpen
  \bibfield  {author} {\bibinfo {author} {\bibfnamefont {Sebastiano}\
  \bibnamefont {Calchi~Novati}} \emph {et~al.} (\bibinfo {collaboration}
  {POINT-AGAPE}),\ }\href {\doibase10.1051/0004-6361:20053135} {\bibfield
  {journal} {\bibinfo  {journal} {Astron. Astrophys.}\ }\textbf {\bibinfo
  {volume} {443}},\ \bibinfo {pages} {911} (\bibinfo {year} {2005})},\ \Eprint
  {http://arxiv.org/abs/astro-ph/0504188} {arXiv:astro-ph/0504188
  [astro-ph]}\BibitemShut {NoStop}%
\bibitem [{\citenamefont {Lee}\ \emph {et~al.}(2015)\citenamefont {Lee},
  \citenamefont {Riffeser}, \citenamefont {Seitz}, \citenamefont {Bender},\
  and\ \citenamefont {Koppenhoefer}}]{Lee:2015boa}%
  \BibitemOpen
  \bibfield  {author} {\bibinfo {author} {\bibfnamefont {C.~H.}\ \bibnamefont
  {Lee}}, \bibinfo {author} {\bibfnamefont {A.}~\bibnamefont {Riffeser}},
  \bibinfo {author} {\bibfnamefont {S.}~\bibnamefont {Seitz}}, \bibinfo
  {author} {\bibfnamefont {R.}~\bibnamefont {Bender}}, \ and\ \bibinfo {author}
  {\bibfnamefont {J.}~\bibnamefont {Koppenhoefer}},\ }\href
  {\doibase10.1088/0004-637X/806/2/161} {\bibfield  {journal} {\bibinfo
  {journal} {Astrophys. J.}\ }\textbf {\bibinfo {volume} {806}},\ \bibinfo
  {pages} {161} (\bibinfo {year} {2015})},\ \Eprint
  {http://arxiv.org/abs/1504.07246} {arXiv:1504.07246
  [astro-ph.GA]}\BibitemShut {NoStop}%
\bibitem [{\citenamefont {Zumalacarregui}\ and\ \citenamefont
  {Seljak}(2017)}]{Zumalacarregui:2017qqd}%
  \BibitemOpen
  \bibfield  {author} {\bibinfo {author} {\bibfnamefont {Miguel}\ \bibnamefont
  {Zumalacarregui}}\ and\ \bibinfo {author} {\bibfnamefont {Uros}\ \bibnamefont
  {Seljak}},\ }\href@noop {} {\  (\bibinfo {year} {2017})},\ \Eprint
  {http://arxiv.org/abs/1712.02240} {arXiv:1712.02240
  [astro-ph.CO]}\BibitemShut {NoStop}%
\bibitem [{\citenamefont {Carr}(1981)}]{Carr1981}%
  \BibitemOpen
  \bibfield  {author} {\bibinfo {author} {\bibfnamefont {B.~J.}\ \bibnamefont
  {Carr}},\ }\href {\doibase10.1093/mnras/194.3.639} {\bibfield  {journal}
  {\bibinfo  {journal} {Monthly Notices of the Royal Astronomical Society}\
  }\textbf {\bibinfo {volume} {194}},\ \bibinfo {pages} {639--668} (\bibinfo
  {year} {1981})}\BibitemShut {NoStop}%
\bibitem [{\citenamefont {Ricotti}\ \emph {et~al.}(2008)\citenamefont
  {Ricotti}, \citenamefont {Ostriker},\ and\ \citenamefont
  {Mack}}]{Ricotti:2007au}%
  \BibitemOpen
  \bibfield  {author} {\bibinfo {author} {\bibfnamefont {Massimo}\ \bibnamefont
  {Ricotti}}, \bibinfo {author} {\bibfnamefont {Jeremiah~P.}\ \bibnamefont
  {Ostriker}}, \ and\ \bibinfo {author} {\bibfnamefont {Katherine~J.}\
  \bibnamefont {Mack}},\ }\href {\doibase10.1086/587831} {\bibfield  {journal}
  {\bibinfo  {journal} {Astrophys. J.}\ }\textbf {\bibinfo {volume} {680}},\
  \bibinfo {pages} {829} (\bibinfo {year} {2008})},\ \Eprint
  {http://arxiv.org/abs/0709.0524} {arXiv:0709.0524 [astro-ph]}\BibitemShut
  {NoStop}%
\bibitem [{\citenamefont {Chen}\ \emph {et~al.}(2016)\citenamefont {Chen},
  \citenamefont {Huang},\ and\ \citenamefont {Wang}}]{Chen:2016pud}%
  \BibitemOpen
  \bibfield  {author} {\bibinfo {author} {\bibfnamefont {Lu}~\bibnamefont
  {Chen}}, \bibinfo {author} {\bibfnamefont {Qing-Guo}\ \bibnamefont {Huang}},
  \ and\ \bibinfo {author} {\bibfnamefont {Ke}~\bibnamefont {Wang}},\ }\href
  {\doibase10.1088/1475-7516/2016/12/044} {\bibfield  {journal} {\bibinfo
  {journal} {JCAP}\ }\textbf {\bibinfo {volume} {1612}},\ \bibinfo {pages}
  {044} (\bibinfo {year} {2016})},\ \Eprint {http://arxiv.org/abs/1608.02174}
  {arXiv:1608.02174 [astro-ph.CO]}\BibitemShut {NoStop}%
\bibitem [{\citenamefont {Aloni}\ \emph {et~al.}(2017)\citenamefont {Aloni},
  \citenamefont {Blum},\ and\ \citenamefont {Flauger}}]{Blum:2016cjs}%
  \BibitemOpen
  \bibfield  {author} {\bibinfo {author} {\bibfnamefont {Daniel}\ \bibnamefont
  {Aloni}}, \bibinfo {author} {\bibfnamefont {Kfir}\ \bibnamefont {Blum}}, \
  and\ \bibinfo {author} {\bibfnamefont {Raphael}\ \bibnamefont {Flauger}},\
  }\href {\doibase10.1088/1475-7516/2017/05/017} {\bibfield  {journal}
  {\bibinfo  {journal} {JCAP}\ }\textbf {\bibinfo {volume} {1705}},\ \bibinfo
  {pages} {017} (\bibinfo {year} {2017})},\ \Eprint
  {http://arxiv.org/abs/1612.06811} {arXiv:1612.06811
  [astro-ph.CO]}\BibitemShut {NoStop}%
\bibitem [{\citenamefont {Ali-Haïmoud}\ and\ \citenamefont
  {Kamionkowski}(2017)}]{Ali-Haimoud:2016mbv}%
  \BibitemOpen
  \bibfield  {author} {\bibinfo {author} {\bibfnamefont {Yacine}\ \bibnamefont
  {Ali-Haïmoud}}\ and\ \bibinfo {author} {\bibfnamefont {Marc}\ \bibnamefont
  {Kamionkowski}},\ }\href {\doibase10.1103/PhysRevD.95.043534} {\bibfield
  {journal} {\bibinfo  {journal} {Phys. Rev.}\ }\textbf {\bibinfo {volume}
  {D95}},\ \bibinfo {pages} {043534} (\bibinfo {year} {2017})},\ \Eprint
  {http://arxiv.org/abs/1612.05644} {arXiv:1612.05644
  [astro-ph.CO]}\BibitemShut {NoStop}%
\bibitem [{\citenamefont {Clesse}\ and\ \citenamefont
  {García-Bellido}(2017)}]{Clesse:2016ajp}%
  \BibitemOpen
  \bibfield  {author} {\bibinfo {author} {\bibfnamefont {Sebastien}\
  \bibnamefont {Clesse}}\ and\ \bibinfo {author} {\bibfnamefont {Juan}\
  \bibnamefont {García-Bellido}},\ }\href {\doibase10.1016/j.dark.2017.10.001}
  {\bibfield  {journal} {\bibinfo  {journal} {Phys. Dark Univ.}\ }\textbf
  {\bibinfo {volume} {18}},\ \bibinfo {pages} {105--114} (\bibinfo {year}
  {2017})},\ \Eprint {http://arxiv.org/abs/1610.08479} {arXiv:1610.08479
  [astro-ph.CO]}\BibitemShut {NoStop}%
\bibitem [{\citenamefont {Poulin}\ \emph {et~al.}(2017)\citenamefont {Poulin},
  \citenamefont {Serpico}, \citenamefont {Calore}, \citenamefont {Clesse},\
  and\ \citenamefont {Kohri}}]{Poulin:2017bwe}%
  \BibitemOpen
  \bibfield  {author} {\bibinfo {author} {\bibfnamefont {Vivian}\ \bibnamefont
  {Poulin}}, \bibinfo {author} {\bibfnamefont {Pasquale~D.}\ \bibnamefont
  {Serpico}}, \bibinfo {author} {\bibfnamefont {Francesca}\ \bibnamefont
  {Calore}}, \bibinfo {author} {\bibfnamefont {Sebastien}\ \bibnamefont
  {Clesse}}, \ and\ \bibinfo {author} {\bibfnamefont {Kazunori}\ \bibnamefont
  {Kohri}},\ }\href {\doibase10.1103/PhysRevD.96.083524} {\bibfield  {journal}
  {\bibinfo  {journal} {Phys. Rev.}\ }\textbf {\bibinfo {volume} {D96}},\
  \bibinfo {pages} {083524} (\bibinfo {year} {2017})},\ \Eprint
  {http://arxiv.org/abs/1707.04206} {arXiv:1707.04206
  [astro-ph.CO]}\BibitemShut {NoStop}%
\bibitem [{\citenamefont {Monroy-Rodríguez}\ and\ \citenamefont
  {Allen}(2014)}]{Monroy-Rodriguez:2014ula}%
  \BibitemOpen
  \bibfield  {author} {\bibinfo {author} {\bibfnamefont {Miguel~A.}\
  \bibnamefont {Monroy-Rodríguez}}\ and\ \bibinfo {author} {\bibfnamefont
  {Christine}\ \bibnamefont {Allen}},\ }\href
  {\doibase10.1088/0004-637X/790/2/159} {\bibfield  {journal} {\bibinfo
  {journal} {Astrophys. J.}\ }\textbf {\bibinfo {volume} {790}},\ \bibinfo
  {pages} {159} (\bibinfo {year} {2014})},\ \Eprint
  {http://arxiv.org/abs/1406.5169} {arXiv:1406.5169 [astro-ph.GA]}\BibitemShut
  {NoStop}%
\bibitem [{\citenamefont {Koushiappas}\ and\ \citenamefont
  {Loeb}(2017{\natexlab{a}})}]{Koushiappas:2017chw}%
  \BibitemOpen
  \bibfield  {author} {\bibinfo {author} {\bibfnamefont {Savvas~M.}\
  \bibnamefont {Koushiappas}}\ and\ \bibinfo {author} {\bibfnamefont {Abraham}\
  \bibnamefont {Loeb}},\ }\href {\doibase10.1103/PhysRevLett.119.041102}
  {\bibfield  {journal} {\bibinfo  {journal} {Phys. Rev. Lett.}\ }\textbf
  {\bibinfo {volume} {119}},\ \bibinfo {pages} {041102} (\bibinfo {year}
  {2017}{\natexlab{a}})},\ \Eprint {http://arxiv.org/abs/1704.01668}
  {arXiv:1704.01668 [astro-ph.GA]}\BibitemShut {NoStop}%
\bibitem [{\citenamefont {Li}\ \emph {et~al.}(2017{\natexlab{a}})\citenamefont
  {Li} \emph {et~al.}}]{Li:2016utv}%
  \BibitemOpen
  \bibfield  {author} {\bibinfo {author} {\bibfnamefont {T.~S.}\ \bibnamefont
  {Li}} \emph {et~al.} (\bibinfo {collaboration} {DES}),\ }\href
  {\doibase10.3847/1538-4357/aa6113} {\bibfield  {journal} {\bibinfo  {journal}
  {Astrophys. J.}\ }\textbf {\bibinfo {volume} {838}},\ \bibinfo {pages} {8}
  (\bibinfo {year} {2017}{\natexlab{a}})},\ \Eprint
  {http://arxiv.org/abs/1611.05052} {arXiv:1611.05052
  [astro-ph.GA]}\BibitemShut {NoStop}%
\bibitem [{\citenamefont {Gaggero}\ \emph {et~al.}(2017)\citenamefont
  {Gaggero}, \citenamefont {Bertone}, \citenamefont {Calore}, \citenamefont
  {Connors}, \citenamefont {Lovell}, \citenamefont {Markoff},\ and\
  \citenamefont {Storm}}]{Gaggero:2016dpq}%
  \BibitemOpen
  \bibfield  {author} {\bibinfo {author} {\bibfnamefont {Daniele}\ \bibnamefont
  {Gaggero}}, \bibinfo {author} {\bibfnamefont {Gianfranco}\ \bibnamefont
  {Bertone}}, \bibinfo {author} {\bibfnamefont {Francesca}\ \bibnamefont
  {Calore}}, \bibinfo {author} {\bibfnamefont {Riley M.~T.}\ \bibnamefont
  {Connors}}, \bibinfo {author} {\bibfnamefont {Mark}\ \bibnamefont {Lovell}},
  \bibinfo {author} {\bibfnamefont {Sera}\ \bibnamefont {Markoff}}, \ and\
  \bibinfo {author} {\bibfnamefont {Emma}\ \bibnamefont {Storm}},\ }\href
  {\doibase10.1103/PhysRevLett.118.241101} {\bibfield  {journal} {\bibinfo
  {journal} {Phys. Rev. Lett.}\ }\textbf {\bibinfo {volume} {118}},\ \bibinfo
  {pages} {241101} (\bibinfo {year} {2017})},\ \Eprint
  {http://arxiv.org/abs/1612.00457} {arXiv:1612.00457
  [astro-ph.HE]}\BibitemShut {NoStop}%
\bibitem [{\citenamefont {Green}(2016)}]{Green:2016xgy}%
  \BibitemOpen
  \bibfield  {author} {\bibinfo {author} {\bibfnamefont {Anne~M.}\ \bibnamefont
  {Green}},\ }\href {\doibase10.1103/PhysRevD.94.063530} {\bibfield  {journal}
  {\bibinfo  {journal} {Phys. Rev.}\ }\textbf {\bibinfo {volume} {D94}},\
  \bibinfo {pages} {063530} (\bibinfo {year} {2016})},\ \Eprint
  {http://arxiv.org/abs/1609.01143} {arXiv:1609.01143
  [astro-ph.CO]}\BibitemShut {NoStop}%
\bibitem [{\citenamefont {Bellomo}\ \emph {et~al.}(2018)\citenamefont
  {Bellomo}, \citenamefont {Bernal}, \citenamefont {Raccanelli},\ and\
  \citenamefont {Verde}}]{Bellomo:2017zsr}%
  \BibitemOpen
  \bibfield  {author} {\bibinfo {author} {\bibfnamefont {Nicola}\ \bibnamefont
  {Bellomo}}, \bibinfo {author} {\bibfnamefont {José~Luis}\ \bibnamefont
  {Bernal}}, \bibinfo {author} {\bibfnamefont {Alvise}\ \bibnamefont
  {Raccanelli}}, \ and\ \bibinfo {author} {\bibfnamefont {Licia}\ \bibnamefont
  {Verde}},\ }\href {\doibase10.1088/1475-7516/2018/01/004} {\bibfield
  {journal} {\bibinfo  {journal} {JCAP}\ }\textbf {\bibinfo {volume} {1801}},\
  \bibinfo {pages} {004} (\bibinfo {year} {2018})},\ \Eprint
  {http://arxiv.org/abs/1709.07467} {arXiv:1709.07467
  [astro-ph.CO]}\BibitemShut {NoStop}%
\bibitem [{\citenamefont {Kühnel}\ and\ \citenamefont
  {Freese}(2017)}]{Kuhnel:2017pwq}%
  \BibitemOpen
  \bibfield  {author} {\bibinfo {author} {\bibfnamefont {Florian}\ \bibnamefont
  {Kühnel}}\ and\ \bibinfo {author} {\bibfnamefont {Katherine}\ \bibnamefont
  {Freese}},\ }\href {\doibase10.1103/PhysRevD.95.083508} {\bibfield  {journal}
  {\bibinfo  {journal} {Phys. Rev.}\ }\textbf {\bibinfo {volume} {D95}},\
  \bibinfo {pages} {083508} (\bibinfo {year} {2017})},\ \Eprint
  {http://arxiv.org/abs/1701.07223} {arXiv:1701.07223
  [astro-ph.CO]}\BibitemShut {NoStop}%
\bibitem [{\citenamefont {Carr}\ \emph {et~al.}(2017)\citenamefont {Carr},
  \citenamefont {Raidal}, \citenamefont {Tenkanen}, \citenamefont {Vaskonen},\
  and\ \citenamefont {Veermäe}}]{Carr:2017jsz}%
  \BibitemOpen
  \bibfield  {author} {\bibinfo {author} {\bibfnamefont {Bernard}\ \bibnamefont
  {Carr}}, \bibinfo {author} {\bibfnamefont {Martti}\ \bibnamefont {Raidal}},
  \bibinfo {author} {\bibfnamefont {Tommi}\ \bibnamefont {Tenkanen}}, \bibinfo
  {author} {\bibfnamefont {Ville}\ \bibnamefont {Vaskonen}}, \ and\ \bibinfo
  {author} {\bibfnamefont {Hardi}\ \bibnamefont {Veermäe}},\ }\href
  {\doibase10.1103/PhysRevD.96.023514} {\bibfield  {journal} {\bibinfo
  {journal} {Phys. Rev.}\ }\textbf {\bibinfo {volume} {D96}},\ \bibinfo {pages}
  {023514} (\bibinfo {year} {2017})},\ \Eprint
  {http://arxiv.org/abs/1705.05567} {arXiv:1705.05567
  [astro-ph.CO]}\BibitemShut {NoStop}%
\bibitem [{\citenamefont {Lehmann}\ \emph {et~al.}(2018)\citenamefont
  {Lehmann}, \citenamefont {Profumo},\ and\ \citenamefont
  {Yant}}]{Lehmann:2018ejc}%
  \BibitemOpen
  \bibfield  {author} {\bibinfo {author} {\bibfnamefont {Benjamin~V.}\
  \bibnamefont {Lehmann}}, \bibinfo {author} {\bibfnamefont {Stefano}\
  \bibnamefont {Profumo}}, \ and\ \bibinfo {author} {\bibfnamefont {Jackson}\
  \bibnamefont {Yant}},\ }\href {\doibase10.1088/1475-7516/2018/04/007}
  {\bibfield  {journal} {\bibinfo  {journal} {JCAP}\ }\textbf {\bibinfo
  {volume} {1804}},\ \bibinfo {pages} {007} (\bibinfo {year} {2018})},\ \Eprint
  {http://arxiv.org/abs/1801.00808} {arXiv:1801.00808
  [astro-ph.CO]}\BibitemShut {NoStop}%
\bibitem [{\citenamefont {Koushiappas}\ and\ \citenamefont
  {Loeb}(2017{\natexlab{b}})}]{Koushiappas:2017kqm}%
  \BibitemOpen
  \bibfield  {author} {\bibinfo {author} {\bibfnamefont {Savvas~M.}\
  \bibnamefont {Koushiappas}}\ and\ \bibinfo {author} {\bibfnamefont {Abraham}\
  \bibnamefont {Loeb}},\ }\href {\doibase10.1103/PhysRevLett.119.221104}
  {\bibfield  {journal} {\bibinfo  {journal} {Phys. Rev. Lett.}\ }\textbf
  {\bibinfo {volume} {119}},\ \bibinfo {pages} {221104} (\bibinfo {year}
  {2017}{\natexlab{b}})},\ \Eprint {http://arxiv.org/abs/1708.07380}
  {arXiv:1708.07380 [astro-ph.CO]}\BibitemShut {NoStop}%
\bibitem [{\citenamefont {{Governato}}\ \emph {et~al.}(1994)\citenamefont
  {{Governato}}, \citenamefont {{Colpi}},\ and\ \citenamefont
  {{Maraschi}}}]{1994MNRAS.271..317G}%
  \BibitemOpen
  \bibfield  {author} {\bibinfo {author} {\bibfnamefont {F.}~\bibnamefont
  {{Governato}}}, \bibinfo {author} {\bibfnamefont {M.}~\bibnamefont
  {{Colpi}}}, \ and\ \bibinfo {author} {\bibfnamefont {L.}~\bibnamefont
  {{Maraschi}}},\ }\href {\doibase10.1093/mnras/271.2.317} {\bibfield
  {journal} {\bibinfo  {journal} {MNRAS}\ }\textbf {\bibinfo {volume} {271}}
  (\bibinfo {year} {1994}),\ 10.1093/mnras/271.2.317},\ \Eprint
  {http://arxiv.org/abs/astro-ph/9407018} {astro-ph/9407018}\BibitemShut
  {NoStop}%
\bibitem [{\citenamefont {Khan}\ \emph
  {et~al.}(2016{\natexlab{a}})\citenamefont {Khan}, \citenamefont {Fiacconi},
  \citenamefont {Mayer}, \citenamefont {Berczik},\ and\ \citenamefont
  {Just}}]{Khan:2016vln}%
  \BibitemOpen
  \bibfield  {author} {\bibinfo {author} {\bibfnamefont {Fazeel~M.}\
  \bibnamefont {Khan}}, \bibinfo {author} {\bibfnamefont {Davide}\ \bibnamefont
  {Fiacconi}}, \bibinfo {author} {\bibfnamefont {Lucio}\ \bibnamefont {Mayer}},
  \bibinfo {author} {\bibfnamefont {Peter}\ \bibnamefont {Berczik}}, \ and\
  \bibinfo {author} {\bibfnamefont {Andreas}\ \bibnamefont {Just}},\ }\href
  {\doibase10.3847/0004-637X/828/2/73} {\bibfield  {journal} {\bibinfo
  {journal} {Astrophys. J.}\ }\textbf {\bibinfo {volume} {828}},\ \bibinfo
  {pages} {73} (\bibinfo {year} {2016}{\natexlab{a}})},\ \Eprint
  {http://arxiv.org/abs/1604.00015} {arXiv:1604.00015
  [astro-ph.GA]}\BibitemShut {NoStop}%
\bibitem [{\citenamefont {{Capelo}}\ \emph {et~al.}(2015)\citenamefont
  {{Capelo}}, \citenamefont {{Volonteri}}, \citenamefont {{Dotti}},
  \citenamefont {{Bellovary}}, \citenamefont {{Mayer}},\ and\ \citenamefont
  {{Governato}}}]{2015MNRAS.447.2123C}%
  \BibitemOpen
  \bibfield  {author} {\bibinfo {author} {\bibfnamefont {P.~R.}\ \bibnamefont
  {{Capelo}}}, \bibinfo {author} {\bibfnamefont {M.}~\bibnamefont
  {{Volonteri}}}, \bibinfo {author} {\bibfnamefont {M.}~\bibnamefont
  {{Dotti}}}, \bibinfo {author} {\bibfnamefont {J.~M.}\ \bibnamefont
  {{Bellovary}}}, \bibinfo {author} {\bibfnamefont {L.}~\bibnamefont
  {{Mayer}}}, \ and\ \bibinfo {author} {\bibfnamefont {F.}~\bibnamefont
  {{Governato}}},\ }\href {\doibase10.1093/mnras/stu2500} {\bibfield  {journal}
  {\bibinfo  {journal} {MNRAS}\ }\textbf {\bibinfo {volume} {447}},\ \bibinfo
  {pages} {2123--2143} (\bibinfo {year} {2015})},\ \Eprint
  {http://arxiv.org/abs/1409.0004} {arXiv:1409.0004}\BibitemShut {NoStop}%
\bibitem [{\citenamefont {{Fiacconi}}\ \emph {et~al.}(2013)\citenamefont
  {{Fiacconi}}, \citenamefont {{Mayer}}, \citenamefont {{Ro{\v s}kar}},\ and\
  \citenamefont {{Colpi}}}]{Fiacconi13}%
  \BibitemOpen
  \bibfield  {author} {\bibinfo {author} {\bibfnamefont {D.}~\bibnamefont
  {{Fiacconi}}}, \bibinfo {author} {\bibfnamefont {L.}~\bibnamefont {{Mayer}}},
  \bibinfo {author} {\bibfnamefont {R.}~\bibnamefont {{Ro{\v s}kar}}}, \ and\
  \bibinfo {author} {\bibfnamefont {M.}~\bibnamefont {{Colpi}}},\ }\href
  {\doibase10.1088/2041-8205/777/1/L14} {\bibfield  {journal} {\bibinfo
  {journal} {Astrophysical Journal}\ }\textbf {\bibinfo {volume} {777}},\
  \bibinfo {eid} {L14} (\bibinfo {year} {2013})},\ \Eprint
  {http://arxiv.org/abs/1307.0822} {arXiv:1307.0822}\BibitemShut {NoStop}%
\bibitem [{\citenamefont {{Lupi}}\ \emph
  {et~al.}(2015{\natexlab{a}})\citenamefont {{Lupi}}, \citenamefont
  {{Haardt}},\ and\ \citenamefont {{Dotti}}}]{2015MNRAS.446.1765L}%
  \BibitemOpen
  \bibfield  {author} {\bibinfo {author} {\bibfnamefont {A.}~\bibnamefont
  {{Lupi}}}, \bibinfo {author} {\bibfnamefont {F.}~\bibnamefont {{Haardt}}}, \
  and\ \bibinfo {author} {\bibfnamefont {M.}~\bibnamefont {{Dotti}}},\ }\href
  {\doibase10.1093/mnras/stu2223} {\bibfield  {journal} {\bibinfo  {journal}
  {MNRAS}\ }\textbf {\bibinfo {volume} {446}},\ \bibinfo {pages} {1765--1774}
  (\bibinfo {year} {2015}{\natexlab{a}})},\ \Eprint
  {http://arxiv.org/abs/1410.0959} {arXiv:1410.0959}\BibitemShut {NoStop}%
\bibitem [{\citenamefont {{Tamburello}}\ \emph {et~al.}(2017)\citenamefont
  {{Tamburello}}, \citenamefont {{Capelo}}, \citenamefont {{Mayer}},
  \citenamefont {{Bellovary}},\ and\ \citenamefont
  {{Wadsley}}}]{2017MNRAS.464.2952T}%
  \BibitemOpen
  \bibfield  {author} {\bibinfo {author} {\bibfnamefont {V.}~\bibnamefont
  {{Tamburello}}}, \bibinfo {author} {\bibfnamefont {P.~R.}\ \bibnamefont
  {{Capelo}}}, \bibinfo {author} {\bibfnamefont {L.}~\bibnamefont {{Mayer}}},
  \bibinfo {author} {\bibfnamefont {J.~M.}\ \bibnamefont {{Bellovary}}}, \ and\
  \bibinfo {author} {\bibfnamefont {J.~W.}\ \bibnamefont {{Wadsley}}},\ }\href
  {\doibase10.1093/mnras/stw2561} {\bibfield  {journal} {\bibinfo  {journal}
  {MNRAS}\ }\textbf {\bibinfo {volume} {464}},\ \bibinfo {pages} {2952--2962}
  (\bibinfo {year} {2017})},\ \Eprint {http://arxiv.org/abs/1603.00021}
  {arXiv:1603.00021}\BibitemShut {NoStop}%
\bibitem [{\citenamefont {{Lupi}}\ \emph
  {et~al.}(2015{\natexlab{b}})\citenamefont {{Lupi}}, \citenamefont {{Haardt}},
  \citenamefont {{Dotti}},\ and\ \citenamefont
  {{Colpi}}}]{2015MNRAS.453.3437L}%
  \BibitemOpen
  \bibfield  {author} {\bibinfo {author} {\bibfnamefont {A.}~\bibnamefont
  {{Lupi}}}, \bibinfo {author} {\bibfnamefont {F.}~\bibnamefont {{Haardt}}},
  \bibinfo {author} {\bibfnamefont {M.}~\bibnamefont {{Dotti}}}, \ and\
  \bibinfo {author} {\bibfnamefont {M.}~\bibnamefont {{Colpi}}},\ }\href
  {\doibase10.1093/mnras/stv1920} {\bibfield  {journal} {\bibinfo  {journal}
  {MNRAS}\ }\textbf {\bibinfo {volume} {453}},\ \bibinfo {pages} {3437--3446}
  (\bibinfo {year} {2015}{\natexlab{b}})},\ \Eprint
  {http://arxiv.org/abs/1509.02920} {arXiv:1509.02920}\BibitemShut {NoStop}%
\bibitem [{\citenamefont {{Vasiliev}}\ \emph {et~al.}(2015)\citenamefont
  {{Vasiliev}}, \citenamefont {{Antonini}},\ and\ \citenamefont
  {{Merritt}}}]{2015ApJ...810...49V}%
  \BibitemOpen
  \bibfield  {author} {\bibinfo {author} {\bibfnamefont {E.}~\bibnamefont
  {{Vasiliev}}}, \bibinfo {author} {\bibfnamefont {F.}~\bibnamefont
  {{Antonini}}}, \ and\ \bibinfo {author} {\bibfnamefont {D.}~\bibnamefont
  {{Merritt}}},\ }\href {\doibase10.1088/0004-637X/810/1/49} {\bibfield
  {journal} {\bibinfo  {journal} {Astrophysical Journal}\ }\textbf {\bibinfo
  {volume} {810}},\ \bibinfo {eid} {49} (\bibinfo {year} {2015})},\ \Eprint
  {http://arxiv.org/abs/1505.05480} {arXiv:1505.05480}\BibitemShut {NoStop}%
\bibitem [{\citenamefont {{Roedig}}\ \emph {et~al.}(2012)\citenamefont
  {{Roedig}}, \citenamefont {{Sesana}}, \citenamefont {{Dotti}}, \citenamefont
  {{Cuadra}}, \citenamefont {{Amaro-Seoane}},\ and\ \citenamefont
  {{Haardt}}}]{2012A&A...545A.127R}%
  \BibitemOpen
  \bibfield  {author} {\bibinfo {author} {\bibfnamefont {C.}~\bibnamefont
  {{Roedig}}}, \bibinfo {author} {\bibfnamefont {A.}~\bibnamefont {{Sesana}}},
  \bibinfo {author} {\bibfnamefont {M.}~\bibnamefont {{Dotti}}}, \bibinfo
  {author} {\bibfnamefont {J.}~\bibnamefont {{Cuadra}}}, \bibinfo {author}
  {\bibfnamefont {P.}~\bibnamefont {{Amaro-Seoane}}}, \ and\ \bibinfo {author}
  {\bibfnamefont {F.}~\bibnamefont {{Haardt}}},\ }\href
  {\doibase10.1051/0004-6361/201219986} {\bibfield  {journal} {\bibinfo
  {journal} {\aap}\ }\textbf {\bibinfo {volume} {545}},\ \bibinfo {eid} {A127}
  (\bibinfo {year} {2012})},\ \Eprint {http://arxiv.org/abs/1202.6063}
  {arXiv:1202.6063 [astro-ph.CO]}\BibitemShut {NoStop}%
\bibitem [{\citenamefont {{Callegari}}\ \emph {et~al.}(2009)\citenamefont
  {{Callegari}}, \citenamefont {{Mayer}}, \citenamefont {{Kazantzidis}},
  \citenamefont {{Colpi}}, \citenamefont {{Governato}}, \citenamefont
  {{Quinn}},\ and\ \citenamefont {{Wadsley}}}]{Callegari09}%
  \BibitemOpen
  \bibfield  {author} {\bibinfo {author} {\bibfnamefont {S.}~\bibnamefont
  {{Callegari}}}, \bibinfo {author} {\bibfnamefont {L.}~\bibnamefont
  {{Mayer}}}, \bibinfo {author} {\bibfnamefont {S.}~\bibnamefont
  {{Kazantzidis}}}, \bibinfo {author} {\bibfnamefont {M.}~\bibnamefont
  {{Colpi}}}, \bibinfo {author} {\bibfnamefont {F.}~\bibnamefont
  {{Governato}}}, \bibinfo {author} {\bibfnamefont {T.}~\bibnamefont
  {{Quinn}}}, \ and\ \bibinfo {author} {\bibfnamefont {J.}~\bibnamefont
  {{Wadsley}}},\ }\href {\doibase10.1088/0004-637X/696/1/L89} {\bibfield
  {journal} {\bibinfo  {journal} {Astrophysical Journal}\ }\textbf {\bibinfo
  {volume} {696}},\ \bibinfo {pages} {L89--L92} (\bibinfo {year} {2009})},\
  \Eprint {http://arxiv.org/abs/0811.0615} {arXiv:0811.0615}\BibitemShut
  {NoStop}%
\bibitem [{\citenamefont {{Tremmel}}\ \emph {et~al.}(2018)\citenamefont
  {{Tremmel}}, \citenamefont {{Governato}}, \citenamefont {{Volonteri}},
  \citenamefont {{Quinn}},\ and\ \citenamefont
  {{Pontzen}}}]{2018MNRAS.tmp..160T}%
  \BibitemOpen
  \bibfield  {author} {\bibinfo {author} {\bibfnamefont {M.}~\bibnamefont
  {{Tremmel}}}, \bibinfo {author} {\bibfnamefont {F.}~\bibnamefont
  {{Governato}}}, \bibinfo {author} {\bibfnamefont {M.}~\bibnamefont
  {{Volonteri}}}, \bibinfo {author} {\bibfnamefont {T.~R.}\ \bibnamefont
  {{Quinn}}}, \ and\ \bibinfo {author} {\bibfnamefont {A.}~\bibnamefont
  {{Pontzen}}},\ }\href {\doibase10.1093/mnras/sty139} {\bibfield  {journal}
  {\bibinfo  {journal} {MNRAS}\ } (\bibinfo {year} {2018}),\
  10.1093/mnras/sty139},\ \Eprint {http://arxiv.org/abs/1708.07126}
  {arXiv:1708.07126}\BibitemShut {NoStop}%
\bibitem [{\citenamefont {{Volonteri}}\ and\ \citenamefont
  {{Perna}}(2005)}]{2005MNRAS.358..913V}%
  \BibitemOpen
  \bibfield  {author} {\bibinfo {author} {\bibfnamefont {M.}~\bibnamefont
  {{Volonteri}}}\ and\ \bibinfo {author} {\bibfnamefont {R.}~\bibnamefont
  {{Perna}}},\ }\href {\doibase10.1111/j.1365-2966.2005.08832.x} {\bibfield
  {journal} {\bibinfo  {journal} {MNRAS}\ }\textbf {\bibinfo {volume} {358}},\
  \bibinfo {pages} {913--922} (\bibinfo {year} {2005})},\ \Eprint
  {http://arxiv.org/abs/astro-ph/0501345} {astro-ph/0501345}\BibitemShut
  {NoStop}%
\bibitem [{\citenamefont {{Bellovary}}\ \emph {et~al.}(2010)\citenamefont
  {{Bellovary}}, \citenamefont {{Governato}}, \citenamefont {{Quinn}},
  \citenamefont {{Wadsley}}, \citenamefont {{Shen}},\ and\ \citenamefont
  {{Volonteri}}}]{2010ApJ...721L.148B}%
  \BibitemOpen
  \bibfield  {author} {\bibinfo {author} {\bibfnamefont {J.~M.}\ \bibnamefont
  {{Bellovary}}}, \bibinfo {author} {\bibfnamefont {F.}~\bibnamefont
  {{Governato}}}, \bibinfo {author} {\bibfnamefont {T.~R.}\ \bibnamefont
  {{Quinn}}}, \bibinfo {author} {\bibfnamefont {J.}~\bibnamefont {{Wadsley}}},
  \bibinfo {author} {\bibfnamefont {S.}~\bibnamefont {{Shen}}}, \ and\ \bibinfo
  {author} {\bibfnamefont {M.}~\bibnamefont {{Volonteri}}},\ }\href
  {\doibase10.1088/2041-8205/721/2/L148} {\bibfield  {journal} {\bibinfo
  {journal} {Astrophysical Journal}\ }\textbf {\bibinfo {volume} {721}},\
  \bibinfo {pages} {L148--L152} (\bibinfo {year} {2010})},\ \Eprint
  {http://arxiv.org/abs/1008.5147} {arXiv:1008.5147 [astro-ph.CO]}\BibitemShut
  {NoStop}%
\bibitem [{\citenamefont {{Volonteri}}\ \emph {et~al.}(2016)\citenamefont
  {{Volonteri}}, \citenamefont {{Dubois}}, \citenamefont {{Pichon}},\ and\
  \citenamefont {{Devriendt}}}]{2016MNRAS.460.2979V}%
  \BibitemOpen
  \bibfield  {author} {\bibinfo {author} {\bibfnamefont {M.}~\bibnamefont
  {{Volonteri}}}, \bibinfo {author} {\bibfnamefont {Y.}~\bibnamefont
  {{Dubois}}}, \bibinfo {author} {\bibfnamefont {C.}~\bibnamefont {{Pichon}}},
  \ and\ \bibinfo {author} {\bibfnamefont {J.}~\bibnamefont {{Devriendt}}},\
  }\href {\doibase10.1093/mnras/stw1123} {\bibfield  {journal} {\bibinfo
  {journal} {MNRAS}\ }\textbf {\bibinfo {volume} {460}},\ \bibinfo {pages}
  {2979--2996} (\bibinfo {year} {2016})},\ \Eprint
  {http://arxiv.org/abs/1602.01941} {arXiv:1602.01941}\BibitemShut {NoStop}%
\bibitem [{\citenamefont {{Yu}}(2002)}]{Yu2002}%
  \BibitemOpen
  \bibfield  {author} {\bibinfo {author} {\bibfnamefont {Q.}~\bibnamefont
  {{Yu}}},\ }\href {\doibase10.1046/j.1365-8711.2002.05242.x} {\bibfield
  {journal} {\bibinfo  {journal} {MNRAS}\ }\textbf {\bibinfo {volume} {331}},\
  \bibinfo {pages} {935--958} (\bibinfo {year} {2002})},\ \Eprint
  {http://arxiv.org/abs/astro-ph/0109530} {astro-ph/0109530}\BibitemShut
  {NoStop}%
\bibitem [{\citenamefont {{Mayer}}\ \emph {et~al.}(2007)\citenamefont
  {{Mayer}}, \citenamefont {{Kazantzidis}}, \citenamefont {{Madau}},
  \citenamefont {{Colpi}}, \citenamefont {{Quinn}},\ and\ \citenamefont
  {{Wadsley}}}]{2007Sci...316.1874M}%
  \BibitemOpen
  \bibfield  {author} {\bibinfo {author} {\bibfnamefont {L.}~\bibnamefont
  {{Mayer}}}, \bibinfo {author} {\bibfnamefont {S.}~\bibnamefont
  {{Kazantzidis}}}, \bibinfo {author} {\bibfnamefont {P.}~\bibnamefont
  {{Madau}}}, \bibinfo {author} {\bibfnamefont {M.}~\bibnamefont {{Colpi}}},
  \bibinfo {author} {\bibfnamefont {T.}~\bibnamefont {{Quinn}}}, \ and\
  \bibinfo {author} {\bibfnamefont {J.}~\bibnamefont {{Wadsley}}},\ }\href
  {\doibase10.1126/science.1141858} {\bibfield  {journal} {\bibinfo  {journal}
  {Science}\ }\textbf {\bibinfo {volume} {316}},\ \bibinfo {pages} {1874}
  (\bibinfo {year} {2007})},\ \Eprint {http://arxiv.org/abs/0706.1562}
  {arXiv:0706.1562}\BibitemShut {NoStop}%
\bibitem [{\citenamefont {{Ro{\v s}kar}}\ \emph {et~al.}(2015)\citenamefont
  {{Ro{\v s}kar}}, \citenamefont {{Fiacconi}}, \citenamefont {{Mayer}},
  \citenamefont {{Kazantzidis}}, \citenamefont {{Quinn}},\ and\ \citenamefont
  {{Wadsley}}}]{2015MNRAS.449..494R}%
  \BibitemOpen
  \bibfield  {author} {\bibinfo {author} {\bibfnamefont {R.}~\bibnamefont
  {{Ro{\v s}kar}}}, \bibinfo {author} {\bibfnamefont {D.}~\bibnamefont
  {{Fiacconi}}}, \bibinfo {author} {\bibfnamefont {L.}~\bibnamefont {{Mayer}}},
  \bibinfo {author} {\bibfnamefont {S.}~\bibnamefont {{Kazantzidis}}}, \bibinfo
  {author} {\bibfnamefont {T.~R.}\ \bibnamefont {{Quinn}}}, \ and\ \bibinfo
  {author} {\bibfnamefont {J.}~\bibnamefont {{Wadsley}}},\ }\href
  {\doibase10.1093/mnras/stv312} {\bibfield  {journal} {\bibinfo  {journal}
  {MNRAS}\ }\textbf {\bibinfo {volume} {449}},\ \bibinfo {pages} {494--505}
  (\bibinfo {year} {2015})},\ \Eprint {http://arxiv.org/abs/1406.4505}
  {arXiv:1406.4505}\BibitemShut {NoStop}%
\bibitem [{\citenamefont {{Pfister}}\ \emph {et~al.}(2017)\citenamefont
  {{Pfister}}, \citenamefont {{Lupi}}, \citenamefont {{Capelo}}, \citenamefont
  {{Volonteri}}, \citenamefont {{Bellovary}},\ and\ \citenamefont
  {{Dotti}}}]{2017MNRAS.471.3646P}%
  \BibitemOpen
  \bibfield  {author} {\bibinfo {author} {\bibfnamefont {H.}~\bibnamefont
  {{Pfister}}}, \bibinfo {author} {\bibfnamefont {A.}~\bibnamefont {{Lupi}}},
  \bibinfo {author} {\bibfnamefont {P.~R.}\ \bibnamefont {{Capelo}}}, \bibinfo
  {author} {\bibfnamefont {M.}~\bibnamefont {{Volonteri}}}, \bibinfo {author}
  {\bibfnamefont {J.~M.}\ \bibnamefont {{Bellovary}}}, \ and\ \bibinfo {author}
  {\bibfnamefont {M.}~\bibnamefont {{Dotti}}},\ }\href
  {\doibase10.1093/mnras/stx1853} {\bibfield  {journal} {\bibinfo  {journal}
  {MNRAS}\ }\textbf {\bibinfo {volume} {471}},\ \bibinfo {pages} {3646--3656}
  (\bibinfo {year} {2017})},\ \Eprint {http://arxiv.org/abs/1706.04010}
  {arXiv:1706.04010}\BibitemShut {NoStop}%
\bibitem [{\citenamefont {Volonteri}\ \emph {et~al.}(2003)\citenamefont
  {Volonteri}, \citenamefont {Haardt},\ and\ \citenamefont
  {Madau}}]{Volonteri:2002vz}%
  \BibitemOpen
  \bibfield  {author} {\bibinfo {author} {\bibfnamefont {Marta}\ \bibnamefont
  {Volonteri}}, \bibinfo {author} {\bibfnamefont {Francesco}\ \bibnamefont
  {Haardt}}, \ and\ \bibinfo {author} {\bibfnamefont {Piero}\ \bibnamefont
  {Madau}},\ }\href {\doibase10.1086/344675} {\bibfield  {journal} {\bibinfo
  {journal} {Astrophys. J.}\ }\textbf {\bibinfo {volume} {582}},\ \bibinfo
  {pages} {559--573} (\bibinfo {year} {2003})},\ \Eprint
  {http://arxiv.org/abs/astro-ph/0207276} {arXiv:astro-ph/0207276
  [astro-ph]}\BibitemShut {NoStop}%
\bibitem [{\citenamefont {{Dosopoulou}}\ and\ \citenamefont
  {{Antonini}}(2017)}]{2017ApJ...840...31D}%
  \BibitemOpen
  \bibfield  {author} {\bibinfo {author} {\bibfnamefont {F.}~\bibnamefont
  {{Dosopoulou}}}\ and\ \bibinfo {author} {\bibfnamefont {F.}~\bibnamefont
  {{Antonini}}},\ }\href {\doibase10.3847/1538-4357/aa6b58} {\bibfield
  {journal} {\bibinfo  {journal} {Astrophysical Journal}\ }\textbf {\bibinfo
  {volume} {840}},\ \bibinfo {eid} {31} (\bibinfo {year} {2017})},\ \Eprint
  {http://arxiv.org/abs/1611.06573} {arXiv:1611.06573}\BibitemShut {NoStop}%
\bibitem [{\citenamefont {Begelman}\ \emph {et~al.}(1980)\citenamefont
  {Begelman}, \citenamefont {Blandford},\ and\ \citenamefont
  {Rees}}]{Begelman:1980vb}%
  \BibitemOpen
  \bibfield  {author} {\bibinfo {author} {\bibfnamefont {M.~C.}\ \bibnamefont
  {Begelman}}, \bibinfo {author} {\bibfnamefont {R.~D.}\ \bibnamefont
  {Blandford}}, \ and\ \bibinfo {author} {\bibfnamefont {M.~J.}\ \bibnamefont
  {Rees}},\ }\href {\doibase10.1038/287307a0} {\bibfield  {journal} {\bibinfo
  {journal} {Nature}\ }\textbf {\bibinfo {volume} {287}},\ \bibinfo {pages}
  {307--309} (\bibinfo {year} {1980})}\BibitemShut {NoStop}%
\bibitem [{\citenamefont {{Goldreich}}\ and\ \citenamefont
  {{Tremaine}}(1980)}]{1980ApJ...241..425G}%
  \BibitemOpen
  \bibfield  {author} {\bibinfo {author} {\bibfnamefont {P.}~\bibnamefont
  {{Goldreich}}}\ and\ \bibinfo {author} {\bibfnamefont {S.}~\bibnamefont
  {{Tremaine}}},\ }\href {\doibase10.1086/158356} {\bibfield  {journal}
  {\bibinfo  {journal} {Astrophysical Journal}\ }\textbf {\bibinfo {volume}
  {241}},\ \bibinfo {pages} {425--441} (\bibinfo {year} {1980})}\BibitemShut
  {NoStop}%
\bibitem [{\citenamefont {{Berczik}}\ \emph {et~al.}(2006)\citenamefont
  {{Berczik}}, \citenamefont {{Merritt}}, \citenamefont {{Spurzem}},\ and\
  \citenamefont {{Bischof}}}]{2006ApJ...642L..21B}%
  \BibitemOpen
  \bibfield  {author} {\bibinfo {author} {\bibfnamefont {P.}~\bibnamefont
  {{Berczik}}}, \bibinfo {author} {\bibfnamefont {D.}~\bibnamefont
  {{Merritt}}}, \bibinfo {author} {\bibfnamefont {R.}~\bibnamefont
  {{Spurzem}}}, \ and\ \bibinfo {author} {\bibfnamefont {H.-P.}\ \bibnamefont
  {{Bischof}}},\ }\href {\doibase10.1086/504426} {\bibfield  {journal}
  {\bibinfo  {journal} {Astrophysical Journal}\ }\textbf {\bibinfo {volume}
  {642}},\ \bibinfo {pages} {L21--L24} (\bibinfo {year} {2006})},\ \Eprint
  {http://arxiv.org/abs/astro-ph/0601698} {astro-ph/0601698}\BibitemShut
  {NoStop}%
\bibitem [{\citenamefont {{Sesana}}\ and\ \citenamefont
  {{Khan}}(2015)}]{2015MNRAS.454L..66S}%
  \BibitemOpen
  \bibfield  {author} {\bibinfo {author} {\bibfnamefont {A.}~\bibnamefont
  {{Sesana}}}\ and\ \bibinfo {author} {\bibfnamefont {F.~M.}\ \bibnamefont
  {{Khan}}},\ }\href {\doibase10.1093/mnrasl/slv131} {\bibfield  {journal}
  {\bibinfo  {journal} {MNRAS}\ }\textbf {\bibinfo {volume} {454}},\ \bibinfo
  {pages} {L66--L70} (\bibinfo {year} {2015})},\ \Eprint
  {http://arxiv.org/abs/1505.02062} {arXiv:1505.02062}\BibitemShut {NoStop}%
\bibitem [{\citenamefont {{Shakura}}\ and\ \citenamefont
  {{Sunyaev}}(1973)}]{1973A&A....24..337S}%
  \BibitemOpen
  \bibfield  {author} {\bibinfo {author} {\bibfnamefont {N.~I.}\ \bibnamefont
  {{Shakura}}}\ and\ \bibinfo {author} {\bibfnamefont {R.~A.}\ \bibnamefont
  {{Sunyaev}}},\ }\href@noop {} {\bibfield  {journal} {\bibinfo  {journal}
  {\aap}\ }\textbf {\bibinfo {volume} {24}},\ \bibinfo {pages} {337--355}
  (\bibinfo {year} {1973})}\BibitemShut {NoStop}%
\bibitem [{\citenamefont {{Armitage}}\ and\ \citenamefont
  {{Natarajan}}(2005)}]{ArmitageNarajan2005}%
  \BibitemOpen
  \bibfield  {author} {\bibinfo {author} {\bibfnamefont {P.~J.}\ \bibnamefont
  {{Armitage}}}\ and\ \bibinfo {author} {\bibfnamefont {P.}~\bibnamefont
  {{Natarajan}}},\ }\href {\doibase10.1086/497108} {\bibfield  {journal}
  {\bibinfo  {journal} {{ApJ}}\ }\textbf {\bibinfo {volume} {634}},\ \bibinfo
  {pages} {921--927} (\bibinfo {year} {2005})}\BibitemShut {NoStop}%
\bibitem [{\citenamefont {{MacFadyen}}\ and\ \citenamefont
  {{Milosavljevi{\'c}}}(2008)}]{2008ApJ...672...83M}%
  \BibitemOpen
  \bibfield  {author} {\bibinfo {author} {\bibfnamefont {A.~I.}\ \bibnamefont
  {{MacFadyen}}}\ and\ \bibinfo {author} {\bibfnamefont {M.}~\bibnamefont
  {{Milosavljevi{\'c}}}},\ }\href {\doibase10.1086/523869} {\bibfield
  {journal} {\bibinfo  {journal} {Astrophysical Journal}\ }\textbf {\bibinfo
  {volume} {672}},\ \bibinfo {pages} {83--93} (\bibinfo {year} {2008})},\
  \Eprint {http://arxiv.org/abs/astro-ph/0607467}
  {astro-ph/0607467}\BibitemShut {NoStop}%
\bibitem [{\citenamefont {{Haiman}}\ \emph {et~al.}(2009)\citenamefont
  {{Haiman}}, \citenamefont {{Kocsis}},\ and\ \citenamefont
  {{Menou}}}]{Haiman09}%
  \BibitemOpen
  \bibfield  {author} {\bibinfo {author} {\bibfnamefont {Z.}~\bibnamefont
  {{Haiman}}}, \bibinfo {author} {\bibfnamefont {B.}~\bibnamefont {{Kocsis}}},
  \ and\ \bibinfo {author} {\bibfnamefont {K.}~\bibnamefont {{Menou}}},\ }\href
  {\doibase10.1088/0004-637X/700/2/1952} {\bibfield  {journal} {\bibinfo
  {journal} {Astrophysical Journal}\ }\textbf {\bibinfo {volume} {700}},\
  \bibinfo {pages} {1952--1969} (\bibinfo {year} {2009})},\ \Eprint
  {http://arxiv.org/abs/0904.1383} {arXiv:0904.1383 [astro-ph.CO]}\BibitemShut
  {NoStop}%
\bibitem [{\citenamefont {{Roedig}}\ \emph {et~al.}(2011)\citenamefont
  {{Roedig}}, \citenamefont {{Dotti}}, \citenamefont {{Sesana}}, \citenamefont
  {{Cuadra}},\ and\ \citenamefont {{Colpi}}}]{rds+11}%
  \BibitemOpen
  \bibfield  {author} {\bibinfo {author} {\bibfnamefont {C.}~\bibnamefont
  {{Roedig}}}, \bibinfo {author} {\bibfnamefont {M.}~\bibnamefont {{Dotti}}},
  \bibinfo {author} {\bibfnamefont {A.}~\bibnamefont {{Sesana}}}, \bibinfo
  {author} {\bibfnamefont {J.}~\bibnamefont {{Cuadra}}}, \ and\ \bibinfo
  {author} {\bibfnamefont {M.}~\bibnamefont {{Colpi}}},\ }\href
  {\doibase10.1111/j.1365-2966.2011.18927.x} {\bibfield  {journal} {\bibinfo
  {journal} {Monthly Notices of the Royal Astronomical Society}\ }\textbf
  {\bibinfo {volume} {415}},\ \bibinfo {pages} {3033--3041} (\bibinfo {year}
  {2011})},\ \Eprint {http://arxiv.org/abs/1104.3868} {arXiv:1104.3868
  [astro-ph.CO]}\BibitemShut {NoStop}%
\bibitem [{\citenamefont {{Shi}}\ \emph {et~al.}(2012)\citenamefont {{Shi}},
  \citenamefont {{Krolik}}, \citenamefont {{Lubow}},\ and\ \citenamefont
  {{Hawley}}}]{2012ApJ...749..118S}%
  \BibitemOpen
  \bibfield  {author} {\bibinfo {author} {\bibfnamefont {J.-M.}\ \bibnamefont
  {{Shi}}}, \bibinfo {author} {\bibfnamefont {J.~H.}\ \bibnamefont {{Krolik}}},
  \bibinfo {author} {\bibfnamefont {S.~H.}\ \bibnamefont {{Lubow}}}, \ and\
  \bibinfo {author} {\bibfnamefont {J.~F.}\ \bibnamefont {{Hawley}}},\ }\href
  {\doibase10.1088/0004-637X/749/2/118} {\bibfield  {journal} {\bibinfo
  {journal} {Astrophysical Journal}\ }\textbf {\bibinfo {volume} {749}},\
  \bibinfo {eid} {118} (\bibinfo {year} {2012})},\ \Eprint
  {http://arxiv.org/abs/1110.4866} {arXiv:1110.4866 [astro-ph.HE]}\BibitemShut
  {NoStop}%
\bibitem [{\citenamefont {{D'Orazio}}\ \emph {et~al.}(2013)\citenamefont
  {{D'Orazio}}, \citenamefont {{Haiman}},\ and\ \citenamefont
  {{MacFadyen}}}]{2013MNRAS.436.2997D}%
  \BibitemOpen
  \bibfield  {author} {\bibinfo {author} {\bibfnamefont {D.~J.}\ \bibnamefont
  {{D'Orazio}}}, \bibinfo {author} {\bibfnamefont {Z.}~\bibnamefont
  {{Haiman}}}, \ and\ \bibinfo {author} {\bibfnamefont {A.}~\bibnamefont
  {{MacFadyen}}},\ }\href {\doibase10.1093/mnras/stt1787} {\bibfield  {journal}
  {\bibinfo  {journal} {MNRAS}\ }\textbf {\bibinfo {volume} {436}},\ \bibinfo
  {pages} {2997--3020} (\bibinfo {year} {2013})},\ \Eprint
  {http://arxiv.org/abs/1210.0536} {arXiv:1210.0536 [astro-ph.GA]}\BibitemShut
  {NoStop}%
\bibitem [{\citenamefont {{Shi}}\ and\ \citenamefont
  {{Krolik}}(2015)}]{2015ApJ...807..131S}%
  \BibitemOpen
  \bibfield  {author} {\bibinfo {author} {\bibfnamefont {J.-M.}\ \bibnamefont
  {{Shi}}}\ and\ \bibinfo {author} {\bibfnamefont {J.~H.}\ \bibnamefont
  {{Krolik}}},\ }\href {\doibase10.1088/0004-637X/807/2/131} {\bibfield
  {journal} {\bibinfo  {journal} {Astrophys. J.}\ }\textbf {\bibinfo {volume}
  {807}},\ \bibinfo {eid} {131} (\bibinfo {year} {2015})},\ \Eprint
  {http://arxiv.org/abs/1503.05561} {arXiv:1503.05561
  [astro-ph.HE]}\BibitemShut {NoStop}%
\bibitem [{\citenamefont {{Boylan-Kolchin}}\ \emph {et~al.}(2008)\citenamefont
  {{Boylan-Kolchin}}, \citenamefont {{Ma}},\ and\ \citenamefont
  {{Quataert}}}]{Boylan2008}%
  \BibitemOpen
  \bibfield  {author} {\bibinfo {author} {\bibfnamefont {M.}~\bibnamefont
  {{Boylan-Kolchin}}}, \bibinfo {author} {\bibfnamefont {C.-P.}\ \bibnamefont
  {{Ma}}}, \ and\ \bibinfo {author} {\bibfnamefont {E.}~\bibnamefont
  {{Quataert}}},\ }\href {\doibase10.1111/j.1365-2966.2007.12530.x} {\bibfield
  {journal} {\bibinfo  {journal} {MNRAS}\ }\textbf {\bibinfo {volume} {383}},\
  \bibinfo {pages} {93--101} (\bibinfo {year} {2008})},\ \Eprint
  {http://arxiv.org/abs/0707.2960} {arXiv:0707.2960}\BibitemShut {NoStop}%
\bibitem [{\citenamefont {{McWilliams}}\ \emph {et~al.}(2014)\citenamefont
  {{McWilliams}}, \citenamefont {{Ostriker}},\ and\ \citenamefont
  {{Pretorius}}}]{2014ApJ...789..156M}%
  \BibitemOpen
  \bibfield  {author} {\bibinfo {author} {\bibfnamefont {S.~T.}\ \bibnamefont
  {{McWilliams}}}, \bibinfo {author} {\bibfnamefont {J.~P.}\ \bibnamefont
  {{Ostriker}}}, \ and\ \bibinfo {author} {\bibfnamefont {F.}~\bibnamefont
  {{Pretorius}}},\ }\href {\doibase10.1088/0004-637X/789/2/156} {\bibfield
  {journal} {\bibinfo  {journal} {Astrophysical Journal}\ }\textbf {\bibinfo
  {volume} {789}},\ \bibinfo {eid} {156} (\bibinfo {year} {2014})}\BibitemShut
  {NoStop}%
\bibitem [{\citenamefont {{Sazhin}}(1978)}]{saz78}%
  \BibitemOpen
  \bibfield  {author} {\bibinfo {author} {\bibfnamefont {M.~V.}\ \bibnamefont
  {{Sazhin}}},\ }\href@noop {} {\bibfield  {journal} {\bibinfo  {journal}
  {Soviet Journal of Astronomy}\ }\textbf {\bibinfo {volume} {22}},\ \bibinfo
  {pages} {36--38} (\bibinfo {year} {1978})}\BibitemShut {NoStop}%
\bibitem [{\citenamefont {{Detweiler}}(1979)}]{det79}%
  \BibitemOpen
  \bibfield  {author} {\bibinfo {author} {\bibfnamefont {S.}~\bibnamefont
  {{Detweiler}}},\ }\href {\doibase10.1086/157593} {\bibfield  {journal}
  {\bibinfo  {journal} {Astrophys. J.}\ }\textbf {\bibinfo {volume} {234}},\
  \bibinfo {pages} {1100--1104} (\bibinfo {year} {1979})}\BibitemShut {NoStop}%
\bibitem [{\citenamefont {{Backer}}\ \emph {et~al.}(1982)\citenamefont
  {{Backer}}, \citenamefont {{Kulkarni}}, \citenamefont {{Heiles}},
  \citenamefont {{Davis}},\ and\ \citenamefont {{Goss}}}]{Backer:1982}%
  \BibitemOpen
  \bibfield  {author} {\bibinfo {author} {\bibfnamefont {D.~C.}\ \bibnamefont
  {{Backer}}}, \bibinfo {author} {\bibfnamefont {S.~R.}\ \bibnamefont
  {{Kulkarni}}}, \bibinfo {author} {\bibfnamefont {C.}~\bibnamefont
  {{Heiles}}}, \bibinfo {author} {\bibfnamefont {M.~M.}\ \bibnamefont
  {{Davis}}}, \ and\ \bibinfo {author} {\bibfnamefont {W.~M.}\ \bibnamefont
  {{Goss}}},\ }\href {\doibase10.1038/300615a0} {\bibfield  {journal} {\bibinfo
   {journal} {Nature}\ }\textbf {\bibinfo {volume} {300}},\ \bibinfo {pages}
  {615--618} (\bibinfo {year} {1982})}\BibitemShut {NoStop}%
\bibitem [{\citenamefont {{Hellings}}\ and\ \citenamefont
  {{Downs}}(1983)}]{hd83}%
  \BibitemOpen
  \bibfield  {author} {\bibinfo {author} {\bibfnamefont {R.~W.}\ \bibnamefont
  {{Hellings}}}\ and\ \bibinfo {author} {\bibfnamefont {G.~S.}\ \bibnamefont
  {{Downs}}},\ }\href {\doibase10.1086/183954} {\bibfield  {journal} {\bibinfo
  {journal} {The Astrophysical Journal}\ }\textbf {\bibinfo {volume} {265}},\
  \bibinfo {pages} {L39--L42} (\bibinfo {year} {1983})}\BibitemShut {NoStop}%
\bibitem [{\citenamefont {Romani}(1989)}]{r89}%
  \BibitemOpen
  \bibfield  {author} {\bibinfo {author} {\bibfnamefont {R.W.}\ \bibnamefont
  {Romani}},\ }in\ \href@noop {} {\emph {\bibinfo {booktitle} {Timing Neutron
  Stars}}},\ \bibinfo {series} {NATO ASI Series, Series C}, Vol.\ \bibinfo
  {volume} {262},\ \bibinfo {editor} {edited by\ \bibinfo {editor}
  {\bibfnamefont {H.}~\bibnamefont {{\"{O}}gelman}}\ and\ \bibinfo {editor}
  {\bibfnamefont {E.P.J.}\ \bibnamefont {van~den Heuvel}}}\ (\bibinfo
  {publisher} {Kluwer},\ \bibinfo {address} {Dordrecht, Netherlands; Boston,
  U.S.A.},\ \bibinfo {year} {1989})\ pp.\ \bibinfo {pages}
  {113--117}\BibitemShut {NoStop}%
\bibitem [{\citenamefont {Burke-Spolaor}(2015)}]{Burke-Spolaor:2015xpf}%
  \BibitemOpen
  \bibfield  {author} {\bibinfo {author} {\bibfnamefont {Sarah}\ \bibnamefont
  {Burke-Spolaor}},\ }\href@noop {} {\  (\bibinfo {year} {2015})},\ \Eprint
  {http://arxiv.org/abs/1511.07869} {arXiv:1511.07869
  [astro-ph.IM]}\BibitemShut {NoStop}%
\bibitem [{\citenamefont {{Lommen}}(2015)}]{l15}%
  \BibitemOpen
  \bibfield  {author} {\bibinfo {author} {\bibfnamefont {A.~N.}\ \bibnamefont
  {{Lommen}}},\ }\href {\doibase10.1088/0034-4885/78/12/124901} {\bibfield
  {journal} {\bibinfo  {journal} {Reports on Progress in Physics}\ }\textbf
  {\bibinfo {volume} {78}},\ \bibinfo {eid} {124901} (\bibinfo {year}
  {2015})}\BibitemShut {NoStop}%
\bibitem [{\citenamefont {{Lommen}}(2017)}]{l17}%
  \BibitemOpen
  \bibfield  {author} {\bibinfo {author} {\bibfnamefont {A.~N.}\ \bibnamefont
  {{Lommen}}},\ }\href {\doibase10.1038/s41550-017-0324-9} {\bibfield
  {journal} {\bibinfo  {journal} {Nature Astronomy}\ }\textbf {\bibinfo
  {volume} {1}},\ \bibinfo {pages} {809--811} (\bibinfo {year}
  {2017})}\BibitemShut {NoStop}%
\bibitem [{\citenamefont {{Foster}}\ and\ \citenamefont
  {{Backer}}(1990)}]{fb90}%
  \BibitemOpen
  \bibfield  {author} {\bibinfo {author} {\bibfnamefont {R.~S.}\ \bibnamefont
  {{Foster}}}\ and\ \bibinfo {author} {\bibfnamefont {D.~C.}\ \bibnamefont
  {{Backer}}},\ }\href {\doibase10.1086/169195} {\bibfield  {journal} {\bibinfo
   {journal} {Astrophys. J.}\ }\textbf {\bibinfo {volume} {361}},\ \bibinfo
  {pages} {300--308} (\bibinfo {year} {1990})}\BibitemShut {NoStop}%
\bibitem [{\citenamefont {{Rajagopal}}\ and\ \citenamefont
  {{Romani}}(1995)}]{rm95}%
  \BibitemOpen
  \bibfield  {author} {\bibinfo {author} {\bibfnamefont {M.}~\bibnamefont
  {{Rajagopal}}}\ and\ \bibinfo {author} {\bibfnamefont {R.~W.}\ \bibnamefont
  {{Romani}}},\ }\href {\doibase10.1086/175813} {\bibfield  {journal} {\bibinfo
   {journal} {Astrophys. J.}\ }\textbf {\bibinfo {volume} {446}},\ \bibinfo
  {pages} {543} (\bibinfo {year} {1995})},\ \Eprint
  {http://arxiv.org/abs/astro-ph/9412038} {astro-ph/9412038}\BibitemShut
  {NoStop}%
\bibitem [{\citenamefont {{Jaffe}}\ and\ \citenamefont
  {{Backer}}(2003)}]{jb03}%
  \BibitemOpen
  \bibfield  {author} {\bibinfo {author} {\bibfnamefont {A.~H.}\ \bibnamefont
  {{Jaffe}}}\ and\ \bibinfo {author} {\bibfnamefont {D.~C.}\ \bibnamefont
  {{Backer}}},\ }\href {\doibase10.1086/345443} {\bibfield  {journal} {\bibinfo
   {journal} {Astrophys. J.}\ }\textbf {\bibinfo {volume} {583}},\ \bibinfo
  {pages} {616--631} (\bibinfo {year} {2003})},\ \Eprint
  {http://arxiv.org/abs/astro-ph/0210148} {astro-ph/0210148}\BibitemShut
  {NoStop}%
\bibitem [{\citenamefont {{Wyithe}}\ and\ \citenamefont {{Loeb}}(2003)}]{wl03}%
  \BibitemOpen
  \bibfield  {author} {\bibinfo {author} {\bibfnamefont {J.~S.~B.}\
  \bibnamefont {{Wyithe}}}\ and\ \bibinfo {author} {\bibfnamefont
  {A.}~\bibnamefont {{Loeb}}},\ }\href {\doibase10.1086/375187} {\bibfield
  {journal} {\bibinfo  {journal} {Astrophys. J.}\ }\textbf {\bibinfo {volume}
  {590}},\ \bibinfo {pages} {691--706} (\bibinfo {year} {2003})},\ \Eprint
  {http://arxiv.org/abs/astro-ph/0211556} {astro-ph/0211556}\BibitemShut
  {NoStop}%
\bibitem [{\citenamefont {{Sesana}}\ \emph {et~al.}(2004)\citenamefont
  {{Sesana}}, \citenamefont {{Haardt}}, \citenamefont {{Madau}},\ and\
  \citenamefont {{Volonteri}}}]{shm+04}%
  \BibitemOpen
  \bibfield  {author} {\bibinfo {author} {\bibfnamefont {A.}~\bibnamefont
  {{Sesana}}}, \bibinfo {author} {\bibfnamefont {F.}~\bibnamefont {{Haardt}}},
  \bibinfo {author} {\bibfnamefont {P.}~\bibnamefont {{Madau}}}, \ and\
  \bibinfo {author} {\bibfnamefont {M.}~\bibnamefont {{Volonteri}}},\ }\href
  {\doibase10.1086/422185} {\bibfield  {journal} {\bibinfo  {journal}
  {Astrophys. J.}\ }\textbf {\bibinfo {volume} {611}},\ \bibinfo {pages}
  {623--632} (\bibinfo {year} {2004})},\ \Eprint
  {http://arxiv.org/abs/astro-ph/0401543} {astro-ph/0401543}\BibitemShut
  {NoStop}%
\bibitem [{\citenamefont {{Sesana}}\ \emph {et~al.}(2008)\citenamefont
  {{Sesana}}, \citenamefont {{Vecchio}},\ and\ \citenamefont
  {{Colacino}}}]{svc08}%
  \BibitemOpen
  \bibfield  {author} {\bibinfo {author} {\bibfnamefont {A.}~\bibnamefont
  {{Sesana}}}, \bibinfo {author} {\bibfnamefont {A.}~\bibnamefont {{Vecchio}}},
  \ and\ \bibinfo {author} {\bibfnamefont {C.~N.}\ \bibnamefont {{Colacino}}},\
  }\href {\doibase10.1111/j.1365-2966.2008.13682.x} {\bibfield  {journal}
  {\bibinfo  {journal} {Mon. Not. Roy. Astron. Soc.}\ }\textbf {\bibinfo
  {volume} {390}},\ \bibinfo {pages} {192--209} (\bibinfo {year} {2008})},\
  \Eprint {http://arxiv.org/abs/0804.4476} {arXiv:0804.4476}\BibitemShut
  {NoStop}%
\bibitem [{\citenamefont {{Sesana}}\ \emph {et~al.}(2009)\citenamefont
  {{Sesana}}, \citenamefont {{Vecchio}},\ and\ \citenamefont
  {{Volonteri}}}]{svv09}%
  \BibitemOpen
  \bibfield  {author} {\bibinfo {author} {\bibfnamefont {A.}~\bibnamefont
  {{Sesana}}}, \bibinfo {author} {\bibfnamefont {A.}~\bibnamefont {{Vecchio}}},
  \ and\ \bibinfo {author} {\bibfnamefont {M.}~\bibnamefont {{Volonteri}}},\
  }\href {\doibase10.1111/j.1365-2966.2009.14499.x} {\bibfield  {journal}
  {\bibinfo  {journal} {Mon. Not. Roy. Astron. Soc.}\ }\textbf {\bibinfo
  {volume} {394}},\ \bibinfo {pages} {2255--2265} (\bibinfo {year} {2009})},\
  \Eprint {http://arxiv.org/abs/0809.3412} {arXiv:0809.3412}\BibitemShut
  {NoStop}%
\bibitem [{\citenamefont {{Mingarelli}}\ \emph {et~al.}(2017)\citenamefont
  {{Mingarelli}}, \citenamefont {{Lazio}}, \citenamefont {{Sesana}},
  \citenamefont {{Greene}}, \citenamefont {{Ellis}}, \citenamefont {{Ma}},
  \citenamefont {{Croft}}, \citenamefont {{Burke-Spolaor}},\ and\ \citenamefont
  {{Taylor}}}]{mls+2017}%
  \BibitemOpen
  \bibfield  {author} {\bibinfo {author} {\bibfnamefont {C.~M.~F.}\
  \bibnamefont {{Mingarelli}}}, \bibinfo {author} {\bibfnamefont {T.~J.~W.}\
  \bibnamefont {{Lazio}}}, \bibinfo {author} {\bibfnamefont {A.}~\bibnamefont
  {{Sesana}}}, \bibinfo {author} {\bibfnamefont {J.~E.}\ \bibnamefont
  {{Greene}}}, \bibinfo {author} {\bibfnamefont {J.~A.}\ \bibnamefont
  {{Ellis}}}, \bibinfo {author} {\bibfnamefont {C.-P.}\ \bibnamefont {{Ma}}},
  \bibinfo {author} {\bibfnamefont {S.}~\bibnamefont {{Croft}}}, \bibinfo
  {author} {\bibfnamefont {S.}~\bibnamefont {{Burke-Spolaor}}}, \ and\ \bibinfo
  {author} {\bibfnamefont {S.~R.}\ \bibnamefont {{Taylor}}},\ }\href
  {\doibase10.1038/s41550-017-0299-6} {\bibfield  {journal} {\bibinfo
  {journal} {Nature Astronomy}\ }\textbf {\bibinfo {volume} {1}},\ \bibinfo
  {pages} {886--892} (\bibinfo {year} {2017})},\ \Eprint
  {http://arxiv.org/abs/1708.03491} {arXiv:1708.03491}\BibitemShut {NoStop}%
\bibitem [{\citenamefont {{van Haasteren}}\ and\ \citenamefont
  {{Levin}}(2010)}]{vHL10}%
  \BibitemOpen
  \bibfield  {author} {\bibinfo {author} {\bibfnamefont {R.}~\bibnamefont {{van
  Haasteren}}}\ and\ \bibinfo {author} {\bibfnamefont {Y.}~\bibnamefont
  {{Levin}}},\ }\href {\doibase10.1111/j.1365-2966.2009.15885.x} {\bibfield
  {journal} {\bibinfo  {journal} {Monthly Notices of the Royal Astronomical
  Society}\ }\textbf {\bibinfo {volume} {401}},\ \bibinfo {pages} {2372--2378}
  (\bibinfo {year} {2010})},\ \Eprint {http://arxiv.org/abs/0909.0954}
  {arXiv:0909.0954 [astro-ph.IM]}\BibitemShut {NoStop}%
\bibitem [{\citenamefont {{Madison}}\ \emph {et~al.}(2017)\citenamefont
  {{Madison}}, \citenamefont {{Chernoff}},\ and\ \citenamefont
  {{Cordes}}}]{mcc17}%
  \BibitemOpen
  \bibfield  {author} {\bibinfo {author} {\bibfnamefont {D.~R.}\ \bibnamefont
  {{Madison}}}, \bibinfo {author} {\bibfnamefont {D.~F.}\ \bibnamefont
  {{Chernoff}}}, \ and\ \bibinfo {author} {\bibfnamefont {J.~M.}\ \bibnamefont
  {{Cordes}}},\ }\href {\doibase10.1103/PhysRevD.96.123016} {\bibfield
  {journal} {\bibinfo  {journal} {Phys. Rev. D}\ }\textbf {\bibinfo {volume}
  {96}},\ \bibinfo {eid} {123016} (\bibinfo {year} {2017})},\ \Eprint
  {http://arxiv.org/abs/1710.04974} {arXiv:1710.04974}\BibitemShut {NoStop}%
\bibitem [{\citenamefont {Siemens}\ \emph {et~al.}(2006)\citenamefont
  {Siemens}, \citenamefont {Creighton}, \citenamefont {Maor}, \citenamefont
  {Ray~Majumder}, \citenamefont {Cannon},\ and\ \citenamefont
  {Read}}]{Siemens:2006vk}%
  \BibitemOpen
  \bibfield  {author} {\bibinfo {author} {\bibfnamefont {Xavier}\ \bibnamefont
  {Siemens}}, \bibinfo {author} {\bibfnamefont {Jolien}\ \bibnamefont
  {Creighton}}, \bibinfo {author} {\bibfnamefont {Irit}\ \bibnamefont {Maor}},
  \bibinfo {author} {\bibfnamefont {Saikat}\ \bibnamefont {Ray~Majumder}},
  \bibinfo {author} {\bibfnamefont {Kipp}\ \bibnamefont {Cannon}}, \ and\
  \bibinfo {author} {\bibfnamefont {Jocelyn}\ \bibnamefont {Read}},\ }\href
  {\doibase10.1103/PhysRevD.73.105001} {\bibfield  {journal} {\bibinfo
  {journal} {Phys. Rev.}\ }\textbf {\bibinfo {volume} {D73}},\ \bibinfo {pages}
  {105001} (\bibinfo {year} {2006})},\ \Eprint
  {http://arxiv.org/abs/gr-qc/0603115} {arXiv:gr-qc/0603115
  [gr-qc]}\BibitemShut {NoStop}%
\bibitem [{\citenamefont {{Leblond}}\ \emph {et~al.}(2009)\citenamefont
  {{Leblond}}, \citenamefont {{Shlaer}},\ and\ \citenamefont
  {{Siemens}}}]{lss09}%
  \BibitemOpen
  \bibfield  {author} {\bibinfo {author} {\bibfnamefont {L.}~\bibnamefont
  {{Leblond}}}, \bibinfo {author} {\bibfnamefont {B.}~\bibnamefont {{Shlaer}}},
  \ and\ \bibinfo {author} {\bibfnamefont {X.}~\bibnamefont {{Siemens}}},\
  }\href {\doibase10.1103/PhysRevD.79.123519} {\bibfield  {journal} {\bibinfo
  {journal} {Phys. Rev. D}\ }\textbf {\bibinfo {volume} {79}},\ \bibinfo
  {pages} {123519--+} (\bibinfo {year} {2009})},\ \Eprint
  {http://arxiv.org/abs/0903.4686} {arXiv:0903.4686 [astro-ph.CO]}\BibitemShut
  {NoStop}%
\bibitem [{\citenamefont {{Sanidas}}\ \emph {et~al.}(2013)\citenamefont
  {{Sanidas}}, \citenamefont {{Battye}},\ and\ \citenamefont
  {{Stappers}}}]{sbs13}%
  \BibitemOpen
  \bibfield  {author} {\bibinfo {author} {\bibfnamefont {S.~A.}\ \bibnamefont
  {{Sanidas}}}, \bibinfo {author} {\bibfnamefont {R.~A.}\ \bibnamefont
  {{Battye}}}, \ and\ \bibinfo {author} {\bibfnamefont {B.~W.}\ \bibnamefont
  {{Stappers}}},\ }\href {\doibase10.1088/0004-637X/764/1/108} {\bibfield
  {journal} {\bibinfo  {journal} {Astrophys. J.}\ }\textbf {\bibinfo {volume}
  {764}},\ \bibinfo {eid} {108} (\bibinfo {year} {2013})},\ \Eprint
  {http://arxiv.org/abs/1211.5042} {arXiv:1211.5042}\BibitemShut {NoStop}%
\bibitem [{\citenamefont {Lasky}\ \emph {et~al.}(2016)\citenamefont {Lasky}
  \emph {et~al.}}]{Lasky:2015lej}%
  \BibitemOpen
  \bibfield  {author} {\bibinfo {author} {\bibfnamefont {Paul~D.}\ \bibnamefont
  {Lasky}} \emph {et~al.},\ }\href {\doibase10.1103/PhysRevX.6.011035}
  {\bibfield  {journal} {\bibinfo  {journal} {Phys. Rev.}\ }\textbf {\bibinfo
  {volume} {X6}},\ \bibinfo {pages} {011035} (\bibinfo {year} {2016})},\
  \Eprint {http://arxiv.org/abs/1511.05994} {arXiv:1511.05994
  [astro-ph.CO]}\BibitemShut {NoStop}%
\bibitem [{\citenamefont {{Lee}}\ \emph {et~al.}(2008)\citenamefont {{Lee}},
  \citenamefont {{Jenet}},\ and\ \citenamefont {{Price}}}]{ljp08}%
  \BibitemOpen
  \bibfield  {author} {\bibinfo {author} {\bibfnamefont {K.~J.}\ \bibnamefont
  {{Lee}}}, \bibinfo {author} {\bibfnamefont {F.~A.}\ \bibnamefont {{Jenet}}},
  \ and\ \bibinfo {author} {\bibfnamefont {R.~H.}\ \bibnamefont {{Price}}},\
  }\href {\doibase10.1086/591080} {\bibfield  {journal} {\bibinfo  {journal}
  {Astrophys. J.}\ }\textbf {\bibinfo {volume} {685}},\ \bibinfo {pages}
  {1304--1319} (\bibinfo {year} {2008})}\BibitemShut {NoStop}%
\bibitem [{\citenamefont {{Chamberlin}}\ and\ \citenamefont
  {{Siemens}}(2012)}]{ss12}%
  \BibitemOpen
  \bibfield  {author} {\bibinfo {author} {\bibfnamefont {S.~J.}\ \bibnamefont
  {{Chamberlin}}}\ and\ \bibinfo {author} {\bibfnamefont {X.}~\bibnamefont
  {{Siemens}}},\ }\href {\doibase10.1103/PhysRevD.85.082001} {\bibfield
  {journal} {\bibinfo  {journal} {Physical Review D}\ }\textbf {\bibinfo
  {volume} {85}},\ \bibinfo {eid} {082001} (\bibinfo {year} {2012})},\ \Eprint
  {http://arxiv.org/abs/1111.5661} {arXiv:1111.5661 [astro-ph.HE]}\BibitemShut
  {NoStop}%
\bibitem [{\citenamefont {{Mingarelli}}\ \emph {et~al.}(2013)\citenamefont
  {{Mingarelli}}, \citenamefont {{Sidery}}, \citenamefont {{Mandel}},\ and\
  \citenamefont {{Vecchio}}}]{msmv13}%
  \BibitemOpen
  \bibfield  {author} {\bibinfo {author} {\bibfnamefont {C.~M.~F.}\
  \bibnamefont {{Mingarelli}}}, \bibinfo {author} {\bibfnamefont
  {T.}~\bibnamefont {{Sidery}}}, \bibinfo {author} {\bibfnamefont
  {I.}~\bibnamefont {{Mandel}}}, \ and\ \bibinfo {author} {\bibfnamefont
  {A.}~\bibnamefont {{Vecchio}}},\ }\href {\doibase10.1103/PhysRevD.88.062005}
  {\bibfield  {journal} {\bibinfo  {journal} {Phys. Rev. D}\ }\textbf {\bibinfo
  {volume} {88}},\ \bibinfo {eid} {062005} (\bibinfo {year} {2013})},\ \Eprint
  {http://arxiv.org/abs/1306.5394} {arXiv:1306.5394 [astro-ph.HE]}\BibitemShut
  {NoStop}%
\bibitem [{\citenamefont {{Taylor}}\ and\ \citenamefont {{Gair}}(2013)}]{tg13}%
  \BibitemOpen
  \bibfield  {author} {\bibinfo {author} {\bibfnamefont {S.~R.}\ \bibnamefont
  {{Taylor}}}\ and\ \bibinfo {author} {\bibfnamefont {J.~R.}\ \bibnamefont
  {{Gair}}},\ }\href {\doibase10.1103/PhysRevD.88.084001} {\bibfield  {journal}
  {\bibinfo  {journal} {Phys. Rev. D}\ }\textbf {\bibinfo {volume} {88}},\
  \bibinfo {eid} {084001} (\bibinfo {year} {2013})},\ \Eprint
  {http://arxiv.org/abs/1306.5395} {arXiv:1306.5395 [gr-qc]}\BibitemShut
  {NoStop}%
\bibitem [{\citenamefont {{Mingarelli}}\ and\ \citenamefont
  {{Sidery}}(2014)}]{ms14}%
  \BibitemOpen
  \bibfield  {author} {\bibinfo {author} {\bibfnamefont {C.~M.~F.}\
  \bibnamefont {{Mingarelli}}}\ and\ \bibinfo {author} {\bibfnamefont
  {T.}~\bibnamefont {{Sidery}}},\ }\href {\doibase10.1103/PhysRevD.90.062011}
  {\bibfield  {journal} {\bibinfo  {journal} {Phys. Rev. D}\ }\textbf {\bibinfo
  {volume} {90}},\ \bibinfo {eid} {062011} (\bibinfo {year} {2014})},\ \Eprint
  {http://arxiv.org/abs/1408.6840} {arXiv:1408.6840 [astro-ph.HE]}\BibitemShut
  {NoStop}%
\bibitem [{\citenamefont {Cornish}\ and\ \citenamefont {van
  Haasteren}(2014)}]{Cornish:2014rva}%
  \BibitemOpen
  \bibfield  {author} {\bibinfo {author} {\bibfnamefont {Neil~J.}\ \bibnamefont
  {Cornish}}\ and\ \bibinfo {author} {\bibfnamefont {Rutger}\ \bibnamefont {van
  Haasteren}},\ }\href@noop {} {\  (\bibinfo {year} {2014})},\ \Eprint
  {http://arxiv.org/abs/1406.4511} {arXiv:1406.4511 [gr-qc]}\BibitemShut
  {NoStop}%
\bibitem [{\citenamefont {{Gair}}\ \emph {et~al.}(2014)\citenamefont {{Gair}},
  \citenamefont {{Romano}}, \citenamefont {{Taylor}},\ and\ \citenamefont
  {{Mingarelli}}}]{grt+14}%
  \BibitemOpen
  \bibfield  {author} {\bibinfo {author} {\bibfnamefont {J.}~\bibnamefont
  {{Gair}}}, \bibinfo {author} {\bibfnamefont {J.~D.}\ \bibnamefont
  {{Romano}}}, \bibinfo {author} {\bibfnamefont {S.}~\bibnamefont {{Taylor}}},
  \ and\ \bibinfo {author} {\bibfnamefont {C.~M.~F.}\ \bibnamefont
  {{Mingarelli}}},\ }\href {\doibase10.1103/PhysRevD.90.082001} {\bibfield
  {journal} {\bibinfo  {journal} {Phys. Rev. D}\ }\textbf {\bibinfo {volume}
  {90}},\ \bibinfo {eid} {082001} (\bibinfo {year} {2014})},\ \Eprint
  {http://arxiv.org/abs/1406.4664} {arXiv:1406.4664 [gr-qc]}\BibitemShut
  {NoStop}%
\bibitem [{\citenamefont {{Taylor}}\ \emph {et~al.}(2015)\citenamefont
  {{Taylor}}, \citenamefont {{Mingarelli}}, \citenamefont {{Gair}},
  \citenamefont {{Sesana}}, \citenamefont {{Theureau}}, \citenamefont
  {{Babak}}, \citenamefont {{Bassa}}, \citenamefont {{Brem}}, \citenamefont
  {{Burgay}}, \citenamefont {{Caballero}}, \citenamefont {{Champion}},
  \citenamefont {{Cognard}}, \citenamefont {{Desvignes}}, \citenamefont
  {{Guillemot}}, \citenamefont {{Hessels}}, \citenamefont {{Janssen}},
  \citenamefont {{Karuppusamy}}, \citenamefont {{Kramer}}, \citenamefont
  {{Lassus}}, \citenamefont {{Lazarus}}, \citenamefont {{Lentati}},
  \citenamefont {{Liu}}, \citenamefont {{Os{\l}owski}}, \citenamefont
  {{Perrodin}}, \citenamefont {{Petiteau}}, \citenamefont {{Possenti}},
  \citenamefont {{Purver}}, \citenamefont {{Rosado}}, \citenamefont
  {{Sanidas}}, \citenamefont {{Smits}}, \citenamefont {{Stappers}},
  \citenamefont {{Tiburzi}}, \citenamefont {{van Haasteren}}, \citenamefont
  {{Vecchio}}, \citenamefont {{Verbiest}},\ and\ \citenamefont {{EPTA
  Collaboration}}}]{TaylorEtAl:2015}%
  \BibitemOpen
  \bibfield  {author} {\bibinfo {author} {\bibfnamefont {S.~R.}\ \bibnamefont
  {{Taylor}}}, \bibinfo {author} {\bibfnamefont {C.~M.~F.}\ \bibnamefont
  {{Mingarelli}}}, \bibinfo {author} {\bibfnamefont {J.~R.}\ \bibnamefont
  {{Gair}}}, \bibinfo {author} {\bibfnamefont {A.}~\bibnamefont {{Sesana}}},
  \bibinfo {author} {\bibfnamefont {G.}~\bibnamefont {{Theureau}}}, \bibinfo
  {author} {\bibfnamefont {S.}~\bibnamefont {{Babak}}}, \bibinfo {author}
  {\bibfnamefont {C.~G.}\ \bibnamefont {{Bassa}}}, \bibinfo {author}
  {\bibfnamefont {P.}~\bibnamefont {{Brem}}}, \bibinfo {author} {\bibfnamefont
  {M.}~\bibnamefont {{Burgay}}}, \bibinfo {author} {\bibfnamefont {R.~N.}\
  \bibnamefont {{Caballero}}}, \bibinfo {author} {\bibfnamefont {D.~J.}\
  \bibnamefont {{Champion}}}, \bibinfo {author} {\bibfnamefont
  {I.}~\bibnamefont {{Cognard}}}, \bibinfo {author} {\bibfnamefont
  {G.}~\bibnamefont {{Desvignes}}}, \bibinfo {author} {\bibfnamefont
  {L.}~\bibnamefont {{Guillemot}}}, \bibinfo {author} {\bibfnamefont
  {J.~W.~T.}\ \bibnamefont {{Hessels}}}, \bibinfo {author} {\bibfnamefont
  {G.~H.}\ \bibnamefont {{Janssen}}}, \bibinfo {author} {\bibfnamefont
  {R.}~\bibnamefont {{Karuppusamy}}}, \bibinfo {author} {\bibfnamefont
  {M.}~\bibnamefont {{Kramer}}}, \bibinfo {author} {\bibfnamefont
  {A.}~\bibnamefont {{Lassus}}}, \bibinfo {author} {\bibfnamefont
  {P.}~\bibnamefont {{Lazarus}}}, \bibinfo {author} {\bibfnamefont
  {L.}~\bibnamefont {{Lentati}}}, \bibinfo {author} {\bibfnamefont
  {K.}~\bibnamefont {{Liu}}}, \bibinfo {author} {\bibfnamefont
  {S.}~\bibnamefont {{Os{\l}owski}}}, \bibinfo {author} {\bibfnamefont
  {D.}~\bibnamefont {{Perrodin}}}, \bibinfo {author} {\bibfnamefont
  {A.}~\bibnamefont {{Petiteau}}}, \bibinfo {author} {\bibfnamefont
  {A.}~\bibnamefont {{Possenti}}}, \bibinfo {author} {\bibfnamefont {M.~B.}\
  \bibnamefont {{Purver}}}, \bibinfo {author} {\bibfnamefont {P.~A.}\
  \bibnamefont {{Rosado}}}, \bibinfo {author} {\bibfnamefont {S.~A.}\
  \bibnamefont {{Sanidas}}}, \bibinfo {author} {\bibfnamefont {R.}~\bibnamefont
  {{Smits}}}, \bibinfo {author} {\bibfnamefont {B.}~\bibnamefont {{Stappers}}},
  \bibinfo {author} {\bibfnamefont {C.}~\bibnamefont {{Tiburzi}}}, \bibinfo
  {author} {\bibfnamefont {R.}~\bibnamefont {{van Haasteren}}}, \bibinfo
  {author} {\bibfnamefont {A.}~\bibnamefont {{Vecchio}}}, \bibinfo {author}
  {\bibfnamefont {J.~P.~W.}\ \bibnamefont {{Verbiest}}}, \ and\ \bibinfo
  {author} {\bibnamefont {{EPTA Collaboration}}},\ }\href
  {\doibase10.1103/PhysRevLett.115.041101} {\bibfield  {journal} {\bibinfo
  {journal} {Physical Review Letters}\ }\textbf {\bibinfo {volume} {115}},\
  \bibinfo {eid} {041101} (\bibinfo {year} {2015})},\ \Eprint
  {http://arxiv.org/abs/1506.08817} {arXiv:1506.08817
  [astro-ph.HE]}\BibitemShut {NoStop}%
\bibitem [{\citenamefont {Arzoumanian}\ \emph
  {et~al.}(2018{\natexlab{a}})\citenamefont {Arzoumanian} \emph
  {et~al.}}]{Arzoumanian:2017puf}%
  \BibitemOpen
  \bibfield  {author} {\bibinfo {author} {\bibfnamefont {Zaven}\ \bibnamefont
  {Arzoumanian}} \emph {et~al.} (\bibinfo {collaboration} {NANOGrav}),\ }\href
  {\doibase10.3847/1538-4365/aab5b0} {\bibfield  {journal} {\bibinfo  {journal}
  {Astrophys. J. Suppl.}\ }\textbf {\bibinfo {volume} {235}},\ \bibinfo {pages}
  {37} (\bibinfo {year} {2018}{\natexlab{a}})},\ \Eprint
  {http://arxiv.org/abs/1801.01837} {arXiv:1801.01837
  [astro-ph.HE]}\BibitemShut {NoStop}%
\bibitem [{\citenamefont {{Desvignes}}\ \emph {et~al.}(2016)\citenamefont
  {{Desvignes}}, \citenamefont {{Caballero}}, \citenamefont {{Lentati}},
  \citenamefont {{Verbiest}}, \citenamefont {{Champion}}, \citenamefont
  {{Stappers}}, \citenamefont {{Janssen}}, \citenamefont {{Lazarus}},
  \citenamefont {{Os{\l}owski}}, \citenamefont {{Babak}}, \citenamefont
  {{Bassa}}, \citenamefont {{Brem}}, \citenamefont {{Burgay}}, \citenamefont
  {{Cognard}}, \citenamefont {{Gair}}, \citenamefont {{Graikou}}, \citenamefont
  {{Guillemot}}, \citenamefont {{Hessels}}, \citenamefont {{Jessner}},
  \citenamefont {{Jordan}}, \citenamefont {{Karuppusamy}}, \citenamefont
  {{Kramer}}, \citenamefont {{Lassus}}, \citenamefont {{Lazaridis}},
  \citenamefont {{Lee}}, \citenamefont {{Liu}}, \citenamefont {{Lyne}},
  \citenamefont {{McKee}}, \citenamefont {{Mingarelli}}, \citenamefont
  {{Perrodin}}, \citenamefont {{Petiteau}}, \citenamefont {{Possenti}},
  \citenamefont {{Purver}}, \citenamefont {{Rosado}}, \citenamefont
  {{Sanidas}}, \citenamefont {{Sesana}}, \citenamefont {{Shaifullah}},
  \citenamefont {{Smits}}, \citenamefont {{Taylor}}, \citenamefont
  {{Theureau}}, \citenamefont {{Tiburzi}}, \citenamefont {{van Haasteren}},\
  and\ \citenamefont {{Vecchio}}}]{desvignes+:2016}%
  \BibitemOpen
  \bibfield  {author} {\bibinfo {author} {\bibfnamefont {G.}~\bibnamefont
  {{Desvignes}}}, \bibinfo {author} {\bibfnamefont {R.~N.}\ \bibnamefont
  {{Caballero}}}, \bibinfo {author} {\bibfnamefont {L.}~\bibnamefont
  {{Lentati}}}, \bibinfo {author} {\bibfnamefont {J.~P.~W.}\ \bibnamefont
  {{Verbiest}}}, \bibinfo {author} {\bibfnamefont {D.~J.}\ \bibnamefont
  {{Champion}}}, \bibinfo {author} {\bibfnamefont {B.~W.}\ \bibnamefont
  {{Stappers}}}, \bibinfo {author} {\bibfnamefont {G.~H.}\ \bibnamefont
  {{Janssen}}}, \bibinfo {author} {\bibfnamefont {P.}~\bibnamefont
  {{Lazarus}}}, \bibinfo {author} {\bibfnamefont {S.}~\bibnamefont
  {{Os{\l}owski}}}, \bibinfo {author} {\bibfnamefont {S.}~\bibnamefont
  {{Babak}}}, \bibinfo {author} {\bibfnamefont {C.~G.}\ \bibnamefont
  {{Bassa}}}, \bibinfo {author} {\bibfnamefont {P.}~\bibnamefont {{Brem}}},
  \bibinfo {author} {\bibfnamefont {M.}~\bibnamefont {{Burgay}}}, \bibinfo
  {author} {\bibfnamefont {I.}~\bibnamefont {{Cognard}}}, \bibinfo {author}
  {\bibfnamefont {J.~R.}\ \bibnamefont {{Gair}}}, \bibinfo {author}
  {\bibfnamefont {E.}~\bibnamefont {{Graikou}}}, \bibinfo {author}
  {\bibfnamefont {L.}~\bibnamefont {{Guillemot}}}, \bibinfo {author}
  {\bibfnamefont {J.~W.~T.}\ \bibnamefont {{Hessels}}}, \bibinfo {author}
  {\bibfnamefont {A.}~\bibnamefont {{Jessner}}}, \bibinfo {author}
  {\bibfnamefont {C.}~\bibnamefont {{Jordan}}}, \bibinfo {author}
  {\bibfnamefont {R.}~\bibnamefont {{Karuppusamy}}}, \bibinfo {author}
  {\bibfnamefont {M.}~\bibnamefont {{Kramer}}}, \bibinfo {author}
  {\bibfnamefont {A.}~\bibnamefont {{Lassus}}}, \bibinfo {author}
  {\bibfnamefont {K.}~\bibnamefont {{Lazaridis}}}, \bibinfo {author}
  {\bibfnamefont {K.~J.}\ \bibnamefont {{Lee}}}, \bibinfo {author}
  {\bibfnamefont {K.}~\bibnamefont {{Liu}}}, \bibinfo {author} {\bibfnamefont
  {A.~G.}\ \bibnamefont {{Lyne}}}, \bibinfo {author} {\bibfnamefont
  {J.}~\bibnamefont {{McKee}}}, \bibinfo {author} {\bibfnamefont {C.~M.~F.}\
  \bibnamefont {{Mingarelli}}}, \bibinfo {author} {\bibfnamefont
  {D.}~\bibnamefont {{Perrodin}}}, \bibinfo {author} {\bibfnamefont
  {A.}~\bibnamefont {{Petiteau}}}, \bibinfo {author} {\bibfnamefont
  {A.}~\bibnamefont {{Possenti}}}, \bibinfo {author} {\bibfnamefont {M.~B.}\
  \bibnamefont {{Purver}}}, \bibinfo {author} {\bibfnamefont {P.~A.}\
  \bibnamefont {{Rosado}}}, \bibinfo {author} {\bibfnamefont {S.}~\bibnamefont
  {{Sanidas}}}, \bibinfo {author} {\bibfnamefont {A.}~\bibnamefont {{Sesana}}},
  \bibinfo {author} {\bibfnamefont {G.}~\bibnamefont {{Shaifullah}}}, \bibinfo
  {author} {\bibfnamefont {R.}~\bibnamefont {{Smits}}}, \bibinfo {author}
  {\bibfnamefont {S.~R.}\ \bibnamefont {{Taylor}}}, \bibinfo {author}
  {\bibfnamefont {G.}~\bibnamefont {{Theureau}}}, \bibinfo {author}
  {\bibfnamefont {C.}~\bibnamefont {{Tiburzi}}}, \bibinfo {author}
  {\bibfnamefont {R.}~\bibnamefont {{van Haasteren}}}, \ and\ \bibinfo {author}
  {\bibfnamefont {A.}~\bibnamefont {{Vecchio}}},\ }\href
  {\doibase10.1093/mnras/stw483} {\bibfield  {journal} {\bibinfo  {journal}
  {Monthly Notices of the Royal Astronomical Society}\ }\textbf {\bibinfo
  {volume} {458}},\ \bibinfo {pages} {3341--3380} (\bibinfo {year} {2016})},\
  \Eprint {http://arxiv.org/abs/1602.08511} {arXiv:1602.08511
  [astro-ph.HE]}\BibitemShut {NoStop}%
\bibitem [{\citenamefont {{Shannon}}\ \emph {et~al.}(2015)\citenamefont
  {{Shannon}}, \citenamefont {{Ravi}}, \citenamefont {{Lentati}}, \citenamefont
  {{Lasky}}, \citenamefont {{Hobbs}}, \citenamefont {{Kerr}}, \citenamefont
  {{Manchester}}, \citenamefont {{Coles}}, \citenamefont {{Levin}},
  \citenamefont {{Bailes}}, \citenamefont {{Bhat}}, \citenamefont
  {{Burke-Spolaor}}, \citenamefont {{Dai}}, \citenamefont {{Keith}},
  \citenamefont {{Os{\l}owski}}, \citenamefont {{Reardon}}, \citenamefont {{van
  Straten}}, \citenamefont {{Toomey}}, \citenamefont {{Wang}}, \citenamefont
  {{Wen}}, \citenamefont {{Wyithe}},\ and\ \citenamefont
  {{Zhu}}}]{ShannonEtAl:2015}%
  \BibitemOpen
  \bibfield  {author} {\bibinfo {author} {\bibfnamefont {R.~M.}\ \bibnamefont
  {{Shannon}}}, \bibinfo {author} {\bibfnamefont {V.}~\bibnamefont {{Ravi}}},
  \bibinfo {author} {\bibfnamefont {L.~T.}\ \bibnamefont {{Lentati}}}, \bibinfo
  {author} {\bibfnamefont {P.~D.}\ \bibnamefont {{Lasky}}}, \bibinfo {author}
  {\bibfnamefont {G.}~\bibnamefont {{Hobbs}}}, \bibinfo {author} {\bibfnamefont
  {M.}~\bibnamefont {{Kerr}}}, \bibinfo {author} {\bibfnamefont {R.~N.}\
  \bibnamefont {{Manchester}}}, \bibinfo {author} {\bibfnamefont {W.~A.}\
  \bibnamefont {{Coles}}}, \bibinfo {author} {\bibfnamefont {Y.}~\bibnamefont
  {{Levin}}}, \bibinfo {author} {\bibfnamefont {M.}~\bibnamefont {{Bailes}}},
  \bibinfo {author} {\bibfnamefont {N.~D.~R.}\ \bibnamefont {{Bhat}}}, \bibinfo
  {author} {\bibfnamefont {S.}~\bibnamefont {{Burke-Spolaor}}}, \bibinfo
  {author} {\bibfnamefont {S.}~\bibnamefont {{Dai}}}, \bibinfo {author}
  {\bibfnamefont {M.~J.}\ \bibnamefont {{Keith}}}, \bibinfo {author}
  {\bibfnamefont {S.}~\bibnamefont {{Os{\l}owski}}}, \bibinfo {author}
  {\bibfnamefont {D.~J.}\ \bibnamefont {{Reardon}}}, \bibinfo {author}
  {\bibfnamefont {W.}~\bibnamefont {{van Straten}}}, \bibinfo {author}
  {\bibfnamefont {L.}~\bibnamefont {{Toomey}}}, \bibinfo {author}
  {\bibfnamefont {J.-B.}\ \bibnamefont {{Wang}}}, \bibinfo {author}
  {\bibfnamefont {L.}~\bibnamefont {{Wen}}}, \bibinfo {author} {\bibfnamefont
  {J.~S.~B.}\ \bibnamefont {{Wyithe}}}, \ and\ \bibinfo {author} {\bibfnamefont
  {X.-J.}\ \bibnamefont {{Zhu}}},\ }\href {\doibase10.1126/science.aab1910}
  {\bibfield  {journal} {\bibinfo  {journal} {Science}\ }\textbf {\bibinfo
  {volume} {349}},\ \bibinfo {pages} {1522--1525} (\bibinfo {year} {2015})},\
  \Eprint {http://arxiv.org/abs/1509.07320} {arXiv:1509.07320}\BibitemShut
  {NoStop}%
\bibitem [{\citenamefont {{Verbiest}}\ \emph {et~al.}(2016)\citenamefont
  {{Verbiest}}, \citenamefont {{Lentati}}, \citenamefont {{Hobbs}},
  \citenamefont {{van Haasteren}}, \citenamefont {{Demorest}}, \citenamefont
  {{Janssen}}, \citenamefont {{Wang}}, \citenamefont {{Desvignes}},
  \citenamefont {{Caballero}}, \citenamefont {{Keith}}, \citenamefont
  {{Champion}}, \citenamefont {{Arzoumanian}}, \citenamefont {{Babak}},
  \citenamefont {{Bassa}}, \citenamefont {{Bhat}}, \citenamefont {{Brazier}},
  \citenamefont {{Brem}}, \citenamefont {{Burgay}}, \citenamefont
  {{Burke-Spolaor}}, \citenamefont {{Chamberlin}}, \citenamefont
  {{Chatterjee}}, \citenamefont {{Christy}}, \citenamefont {{Cognard}},
  \citenamefont {{Cordes}}, \citenamefont {{Dai}}, \citenamefont {{Dolch}},
  \citenamefont {{Ellis}}, \citenamefont {{Ferdman}}, \citenamefont
  {{Fonseca}}, \citenamefont {{Gair}}, \citenamefont {{Garver-Daniels}},
  \citenamefont {{Gentile}}, \citenamefont {{Gonzalez}}, \citenamefont
  {{Graikou}}, \citenamefont {{Guillemot}}, \citenamefont {{Hessels}},
  \citenamefont {{Jones}}, \citenamefont {{Karuppusamy}}, \citenamefont
  {{Kerr}}, \citenamefont {{Kramer}}, \citenamefont {{Lam}}, \citenamefont
  {{Lasky}}, \citenamefont {{Lassus}}, \citenamefont {{Lazarus}}, \citenamefont
  {{Lazio}}, \citenamefont {{Lee}}, \citenamefont {{Levin}}, \citenamefont
  {{Liu}}, \citenamefont {{Lynch}}, \citenamefont {{Lyne}}, \citenamefont
  {{Mckee}}, \citenamefont {{McLaughlin}}, \citenamefont {{McWilliams}},
  \citenamefont {{Madison}}, \citenamefont {{Manchester}}, \citenamefont
  {{Mingarelli}}, \citenamefont {{Nice}}, \citenamefont {{Os{\l}owski}},
  \citenamefont {{Palliyaguru}}, \citenamefont {{Pennucci}}, \citenamefont
  {{Perera}}, \citenamefont {{Perrodin}}, \citenamefont {{Possenti}},
  \citenamefont {{Petiteau}}, \citenamefont {{Ransom}}, \citenamefont
  {{Reardon}}, \citenamefont {{Rosado}}, \citenamefont {{Sanidas}},
  \citenamefont {{Sesana}}, \citenamefont {{Shaifullah}}, \citenamefont
  {{Shannon}}, \citenamefont {{Siemens}}, \citenamefont {{Simon}},
  \citenamefont {{Smits}}, \citenamefont {{Spiewak}}, \citenamefont {{Stairs}},
  \citenamefont {{Stappers}}, \citenamefont {{Stinebring}}, \citenamefont
  {{Stovall}}, \citenamefont {{Swiggum}}, \citenamefont {{Taylor}},
  \citenamefont {{Theureau}}, \citenamefont {{Tiburzi}}, \citenamefont
  {{Toomey}}, \citenamefont {{Vallisneri}}, \citenamefont {{van Straten}},
  \citenamefont {{Vecchio}}, \citenamefont {{Wang}}, \citenamefont {{Wen}},
  \citenamefont {{You}}, \citenamefont {{Zhu}},\ and\ \citenamefont
  {{Zhu}}}]{VerbiestEtAl:2016}%
  \BibitemOpen
  \bibfield  {author} {\bibinfo {author} {\bibfnamefont {J.~P.~W.}\
  \bibnamefont {{Verbiest}}}, \bibinfo {author} {\bibfnamefont
  {L.}~\bibnamefont {{Lentati}}}, \bibinfo {author} {\bibfnamefont
  {G.}~\bibnamefont {{Hobbs}}}, \bibinfo {author} {\bibfnamefont
  {R.}~\bibnamefont {{van Haasteren}}}, \bibinfo {author} {\bibfnamefont
  {P.~B.}\ \bibnamefont {{Demorest}}}, \bibinfo {author} {\bibfnamefont
  {G.~H.}\ \bibnamefont {{Janssen}}}, \bibinfo {author} {\bibfnamefont {J.-B.}\
  \bibnamefont {{Wang}}}, \bibinfo {author} {\bibfnamefont {G.}~\bibnamefont
  {{Desvignes}}}, \bibinfo {author} {\bibfnamefont {R.~N.}\ \bibnamefont
  {{Caballero}}}, \bibinfo {author} {\bibfnamefont {M.~J.}\ \bibnamefont
  {{Keith}}}, \bibinfo {author} {\bibfnamefont {D.~J.}\ \bibnamefont
  {{Champion}}}, \bibinfo {author} {\bibfnamefont {Z.}~\bibnamefont
  {{Arzoumanian}}}, \bibinfo {author} {\bibfnamefont {S.}~\bibnamefont
  {{Babak}}}, \bibinfo {author} {\bibfnamefont {C.~G.}\ \bibnamefont
  {{Bassa}}}, \bibinfo {author} {\bibfnamefont {N.~D.~R.}\ \bibnamefont
  {{Bhat}}}, \bibinfo {author} {\bibfnamefont {A.}~\bibnamefont {{Brazier}}},
  \bibinfo {author} {\bibfnamefont {P.}~\bibnamefont {{Brem}}}, \bibinfo
  {author} {\bibfnamefont {M.}~\bibnamefont {{Burgay}}}, \bibinfo {author}
  {\bibfnamefont {S.}~\bibnamefont {{Burke-Spolaor}}}, \bibinfo {author}
  {\bibfnamefont {S.~J.}\ \bibnamefont {{Chamberlin}}}, \bibinfo {author}
  {\bibfnamefont {S.}~\bibnamefont {{Chatterjee}}}, \bibinfo {author}
  {\bibfnamefont {B.}~\bibnamefont {{Christy}}}, \bibinfo {author}
  {\bibfnamefont {I.}~\bibnamefont {{Cognard}}}, \bibinfo {author}
  {\bibfnamefont {J.~M.}\ \bibnamefont {{Cordes}}}, \bibinfo {author}
  {\bibfnamefont {S.}~\bibnamefont {{Dai}}}, \bibinfo {author} {\bibfnamefont
  {T.}~\bibnamefont {{Dolch}}}, \bibinfo {author} {\bibfnamefont {J.~A.}\
  \bibnamefont {{Ellis}}}, \bibinfo {author} {\bibfnamefont {R.~D.}\
  \bibnamefont {{Ferdman}}}, \bibinfo {author} {\bibfnamefont {E.}~\bibnamefont
  {{Fonseca}}}, \bibinfo {author} {\bibfnamefont {J.~R.}\ \bibnamefont
  {{Gair}}}, \bibinfo {author} {\bibfnamefont {N.~E.}\ \bibnamefont
  {{Garver-Daniels}}}, \bibinfo {author} {\bibfnamefont {P.}~\bibnamefont
  {{Gentile}}}, \bibinfo {author} {\bibfnamefont {M.~E.}\ \bibnamefont
  {{Gonzalez}}}, \bibinfo {author} {\bibfnamefont {E.}~\bibnamefont
  {{Graikou}}}, \bibinfo {author} {\bibfnamefont {L.}~\bibnamefont
  {{Guillemot}}}, \bibinfo {author} {\bibfnamefont {J.~W.~T.}\ \bibnamefont
  {{Hessels}}}, \bibinfo {author} {\bibfnamefont {G.}~\bibnamefont {{Jones}}},
  \bibinfo {author} {\bibfnamefont {R.}~\bibnamefont {{Karuppusamy}}}, \bibinfo
  {author} {\bibfnamefont {M.}~\bibnamefont {{Kerr}}}, \bibinfo {author}
  {\bibfnamefont {M.}~\bibnamefont {{Kramer}}}, \bibinfo {author}
  {\bibfnamefont {M.~T.}\ \bibnamefont {{Lam}}}, \bibinfo {author}
  {\bibfnamefont {P.~D.}\ \bibnamefont {{Lasky}}}, \bibinfo {author}
  {\bibfnamefont {A.}~\bibnamefont {{Lassus}}}, \bibinfo {author}
  {\bibfnamefont {P.}~\bibnamefont {{Lazarus}}}, \bibinfo {author}
  {\bibfnamefont {T.~J.~W.}\ \bibnamefont {{Lazio}}}, \bibinfo {author}
  {\bibfnamefont {K.~J.}\ \bibnamefont {{Lee}}}, \bibinfo {author}
  {\bibfnamefont {L.}~\bibnamefont {{Levin}}}, \bibinfo {author} {\bibfnamefont
  {K.}~\bibnamefont {{Liu}}}, \bibinfo {author} {\bibfnamefont {R.~S.}\
  \bibnamefont {{Lynch}}}, \bibinfo {author} {\bibfnamefont {A.~G.}\
  \bibnamefont {{Lyne}}}, \bibinfo {author} {\bibfnamefont {J.}~\bibnamefont
  {{Mckee}}}, \bibinfo {author} {\bibfnamefont {M.~A.}\ \bibnamefont
  {{McLaughlin}}}, \bibinfo {author} {\bibfnamefont {S.~T.}\ \bibnamefont
  {{McWilliams}}}, \bibinfo {author} {\bibfnamefont {D.~R.}\ \bibnamefont
  {{Madison}}}, \bibinfo {author} {\bibfnamefont {R.~N.}\ \bibnamefont
  {{Manchester}}}, \bibinfo {author} {\bibfnamefont {C.~M.~F.}\ \bibnamefont
  {{Mingarelli}}}, \bibinfo {author} {\bibfnamefont {D.~J.}\ \bibnamefont
  {{Nice}}}, \bibinfo {author} {\bibfnamefont {S.}~\bibnamefont
  {{Os{\l}owski}}}, \bibinfo {author} {\bibfnamefont {N.~T.}\ \bibnamefont
  {{Palliyaguru}}}, \bibinfo {author} {\bibfnamefont {T.~T.}\ \bibnamefont
  {{Pennucci}}}, \bibinfo {author} {\bibfnamefont {B.~B.~P.}\ \bibnamefont
  {{Perera}}}, \bibinfo {author} {\bibfnamefont {D.}~\bibnamefont
  {{Perrodin}}}, \bibinfo {author} {\bibfnamefont {A.}~\bibnamefont
  {{Possenti}}}, \bibinfo {author} {\bibfnamefont {A.}~\bibnamefont
  {{Petiteau}}}, \bibinfo {author} {\bibfnamefont {S.~M.}\ \bibnamefont
  {{Ransom}}}, \bibinfo {author} {\bibfnamefont {D.}~\bibnamefont {{Reardon}}},
  \bibinfo {author} {\bibfnamefont {P.~A.}\ \bibnamefont {{Rosado}}}, \bibinfo
  {author} {\bibfnamefont {S.~A.}\ \bibnamefont {{Sanidas}}}, \bibinfo {author}
  {\bibfnamefont {A.}~\bibnamefont {{Sesana}}}, \bibinfo {author}
  {\bibfnamefont {G.}~\bibnamefont {{Shaifullah}}}, \bibinfo {author}
  {\bibfnamefont {R.~M.}\ \bibnamefont {{Shannon}}}, \bibinfo {author}
  {\bibfnamefont {X.}~\bibnamefont {{Siemens}}}, \bibinfo {author}
  {\bibfnamefont {J.}~\bibnamefont {{Simon}}}, \bibinfo {author} {\bibfnamefont
  {R.}~\bibnamefont {{Smits}}}, \bibinfo {author} {\bibfnamefont
  {R.}~\bibnamefont {{Spiewak}}}, \bibinfo {author} {\bibfnamefont {I.~H.}\
  \bibnamefont {{Stairs}}}, \bibinfo {author} {\bibfnamefont {B.~W.}\
  \bibnamefont {{Stappers}}}, \bibinfo {author} {\bibfnamefont {D.~R.}\
  \bibnamefont {{Stinebring}}}, \bibinfo {author} {\bibfnamefont
  {K.}~\bibnamefont {{Stovall}}}, \bibinfo {author} {\bibfnamefont {J.~K.}\
  \bibnamefont {{Swiggum}}}, \bibinfo {author} {\bibfnamefont {S.~R.}\
  \bibnamefont {{Taylor}}}, \bibinfo {author} {\bibfnamefont {G.}~\bibnamefont
  {{Theureau}}}, \bibinfo {author} {\bibfnamefont {C.}~\bibnamefont
  {{Tiburzi}}}, \bibinfo {author} {\bibfnamefont {L.}~\bibnamefont {{Toomey}}},
  \bibinfo {author} {\bibfnamefont {M.}~\bibnamefont {{Vallisneri}}}, \bibinfo
  {author} {\bibfnamefont {W.}~\bibnamefont {{van Straten}}}, \bibinfo {author}
  {\bibfnamefont {A.}~\bibnamefont {{Vecchio}}}, \bibinfo {author}
  {\bibfnamefont {Y.}~\bibnamefont {{Wang}}}, \bibinfo {author} {\bibfnamefont
  {L.}~\bibnamefont {{Wen}}}, \bibinfo {author} {\bibfnamefont {X.~P.}\
  \bibnamefont {{You}}}, \bibinfo {author} {\bibfnamefont {W.~W.}\ \bibnamefont
  {{Zhu}}}, \ and\ \bibinfo {author} {\bibfnamefont {X.-J.}\ \bibnamefont
  {{Zhu}}},\ }\href {\doibase10.1093/mnras/stw347} {\bibfield  {journal}
  {\bibinfo  {journal} {Monthly Notices of the Royal Astronomical Society}\
  }\textbf {\bibinfo {volume} {458}},\ \bibinfo {pages} {1267--1288} (\bibinfo
  {year} {2016})},\ \Eprint {http://arxiv.org/abs/1602.03640} {arXiv:1602.03640
  [astro-ph.IM]}\BibitemShut {NoStop}%
\bibitem [{\citenamefont {Siemens}\ \emph {et~al.}(2013)\citenamefont
  {Siemens}, \citenamefont {Ellis}, \citenamefont {Jenet},\ and\ \citenamefont
  {Romano}}]{Siemens:2013zla}%
  \BibitemOpen
  \bibfield  {author} {\bibinfo {author} {\bibfnamefont {Xavier}\ \bibnamefont
  {Siemens}}, \bibinfo {author} {\bibfnamefont {Justin}\ \bibnamefont {Ellis}},
  \bibinfo {author} {\bibfnamefont {Fredrick}\ \bibnamefont {Jenet}}, \ and\
  \bibinfo {author} {\bibfnamefont {Joseph~D.}\ \bibnamefont {Romano}},\ }\href
  {\doibase10.1088/0264-9381/30/22/224015} {\bibfield  {journal} {\bibinfo
  {journal} {Class. Quant. Grav.}\ }\textbf {\bibinfo {volume} {30}},\ \bibinfo
  {pages} {224015} (\bibinfo {year} {2013})},\ \Eprint
  {http://arxiv.org/abs/1305.3196} {arXiv:1305.3196 [astro-ph.IM]}\BibitemShut
  {NoStop}%
\bibitem [{\citenamefont {{Rosado}}\ \emph {et~al.}(2015)\citenamefont
  {{Rosado}}, \citenamefont {{Sesana}},\ and\ \citenamefont {{Gair}}}]{rsg15}%
  \BibitemOpen
  \bibfield  {author} {\bibinfo {author} {\bibfnamefont {P.~A.}\ \bibnamefont
  {{Rosado}}}, \bibinfo {author} {\bibfnamefont {A.}~\bibnamefont {{Sesana}}},
  \ and\ \bibinfo {author} {\bibfnamefont {J.}~\bibnamefont {{Gair}}},\ }\href
  {\doibase10.1093/mnras/stv1098} {\bibfield  {journal} {\bibinfo  {journal}
  {Mon. Not. Roy. Astron. Soc.}\ }\textbf {\bibinfo {volume} {451}},\ \bibinfo
  {pages} {2417--2433} (\bibinfo {year} {2015})},\ \Eprint
  {http://arxiv.org/abs/1503.04803} {arXiv:1503.04803
  [astro-ph.HE]}\BibitemShut {NoStop}%
\bibitem [{\citenamefont {{Taylor}}\ \emph {et~al.}(2016)\citenamefont
  {{Taylor}}, \citenamefont {{Vallisneri}}, \citenamefont {{Ellis}},
  \citenamefont {{Mingarelli}}, \citenamefont {{Lazio}},\ and\ \citenamefont
  {{van Haasteren}}}]{tve+16}%
  \BibitemOpen
  \bibfield  {author} {\bibinfo {author} {\bibfnamefont {S.~R.}\ \bibnamefont
  {{Taylor}}}, \bibinfo {author} {\bibfnamefont {M.}~\bibnamefont
  {{Vallisneri}}}, \bibinfo {author} {\bibfnamefont {J.~A.}\ \bibnamefont
  {{Ellis}}}, \bibinfo {author} {\bibfnamefont {C.~M.~F.}\ \bibnamefont
  {{Mingarelli}}}, \bibinfo {author} {\bibfnamefont {T.~J.~W.}\ \bibnamefont
  {{Lazio}}}, \ and\ \bibinfo {author} {\bibfnamefont {R.}~\bibnamefont {{van
  Haasteren}}},\ }\href {\doibase10.3847/2041-8205/819/1/L6} {\bibfield
  {journal} {\bibinfo  {journal} {The Astrophysical Journal Letters}\ }\textbf
  {\bibinfo {volume} {819}},\ \bibinfo {eid} {L6} (\bibinfo {year} {2016})},\
  \Eprint {http://arxiv.org/abs/1511.05564} {arXiv:1511.05564
  [astro-ph.IM]}\BibitemShut {NoStop}%
\bibitem [{\citenamefont {Phinney}(2001)}]{Phinney:2001di}%
  \BibitemOpen
  \bibfield  {author} {\bibinfo {author} {\bibfnamefont {E.~S.}\ \bibnamefont
  {Phinney}},\ }\href@noop {} {\  (\bibinfo {year} {2001})},\ \Eprint
  {http://arxiv.org/abs/astro-ph/0108028} {arXiv:astro-ph/0108028
  [astro-ph]}\BibitemShut {NoStop}%
\bibitem [{\citenamefont {{Kocsis}}\ and\ \citenamefont
  {{Sesana}}(2011)}]{ks11}%
  \BibitemOpen
  \bibfield  {author} {\bibinfo {author} {\bibfnamefont {B.}~\bibnamefont
  {{Kocsis}}}\ and\ \bibinfo {author} {\bibfnamefont {A.}~\bibnamefont
  {{Sesana}}},\ }\href {\doibase10.1111/j.1365-2966.2010.17782.x} {\bibfield
  {journal} {\bibinfo  {journal} {Monthly Notices of the Royal Astronomical
  Society}\ }\textbf {\bibinfo {volume} {411}},\ \bibinfo {pages} {1467--1479}
  (\bibinfo {year} {2011})},\ \Eprint {http://arxiv.org/abs/1002.0584}
  {arXiv:1002.0584 [astro-ph.CO]}\BibitemShut {NoStop}%
\bibitem [{\citenamefont {{Ravi}}\ \emph {et~al.}(2014)\citenamefont {{Ravi}},
  \citenamefont {{Wyithe}}, \citenamefont {{Shannon}}, \citenamefont
  {{Hobbs}},\ and\ \citenamefont {{Manchester}}}]{rws+14}%
  \BibitemOpen
  \bibfield  {author} {\bibinfo {author} {\bibfnamefont {V.}~\bibnamefont
  {{Ravi}}}, \bibinfo {author} {\bibfnamefont {J.~S.~B.}\ \bibnamefont
  {{Wyithe}}}, \bibinfo {author} {\bibfnamefont {R.~M.}\ \bibnamefont
  {{Shannon}}}, \bibinfo {author} {\bibfnamefont {G.}~\bibnamefont {{Hobbs}}},
  \ and\ \bibinfo {author} {\bibfnamefont {R.~N.}\ \bibnamefont
  {{Manchester}}},\ }\href {\doibase10.1093/mnras/stu779} {\bibfield  {journal}
  {\bibinfo  {journal} {Mon. Not. Roy. Astron. Soc.}\ }\textbf {\bibinfo
  {volume} {442}},\ \bibinfo {pages} {56--68} (\bibinfo {year} {2014})},\
  \Eprint {http://arxiv.org/abs/1404.5183} {arXiv:1404.5183}\BibitemShut
  {NoStop}%
\bibitem [{\citenamefont {{Sampson}}\ \emph {et~al.}(2015)\citenamefont
  {{Sampson}}, \citenamefont {{Cornish}},\ and\ \citenamefont
  {{McWilliams}}}]{scm15}%
  \BibitemOpen
  \bibfield  {author} {\bibinfo {author} {\bibfnamefont {L.}~\bibnamefont
  {{Sampson}}}, \bibinfo {author} {\bibfnamefont {N.~J.}\ \bibnamefont
  {{Cornish}}}, \ and\ \bibinfo {author} {\bibfnamefont {S.~T.}\ \bibnamefont
  {{McWilliams}}},\ }\href {\doibase10.1103/PhysRevD.91.084055} {\bibfield
  {journal} {\bibinfo  {journal} {Phys. Rev. D}\ }\textbf {\bibinfo {volume}
  {91}},\ \bibinfo {eid} {084055} (\bibinfo {year} {2015})},\ \Eprint
  {http://arxiv.org/abs/1503.02662} {arXiv:1503.02662 [gr-qc]}\BibitemShut
  {NoStop}%
\bibitem [{\citenamefont {{Arzoumanian}}\ \emph {et~al.}(2016)\citenamefont
  {{Arzoumanian}}, \citenamefont {{Brazier}}, \citenamefont {{Burke-Spolaor}},
  \citenamefont {{Chamberlin}}, \citenamefont {{Chatterjee}}, \citenamefont
  {{Christy}}, \citenamefont {{Cordes}}, \citenamefont {{Cornish}},
  \citenamefont {{Crowter}}, \citenamefont {{Demorest}}, \citenamefont
  {{Deng}}, \citenamefont {{Dolch}}, \citenamefont {{Ellis}}, \citenamefont
  {{Ferdman}}, \citenamefont {{Fonseca}}, \citenamefont {{Garver-Daniels}},
  \citenamefont {{Gonzalez}}, \citenamefont {{Jenet}}, \citenamefont {{Jones}},
  \citenamefont {{Jones}}, \citenamefont {{Kaspi}}, \citenamefont {{Koop}},
  \citenamefont {{Lam}}, \citenamefont {{Lazio}}, \citenamefont {{Levin}},
  \citenamefont {{Lommen}}, \citenamefont {{Lorimer}}, \citenamefont {{Luo}},
  \citenamefont {{Lynch}}, \citenamefont {{Madison}}, \citenamefont
  {{McLaughlin}}, \citenamefont {{McWilliams}}, \citenamefont {{Mingarelli}},
  \citenamefont {{Nice}}, \citenamefont {{Palliyaguru}}, \citenamefont
  {{Pennucci}}, \citenamefont {{Ransom}}, \citenamefont {{Sampson}},
  \citenamefont {{Sanidas}}, \citenamefont {{Sesana}}, \citenamefont
  {{Siemens}}, \citenamefont {{Simon}}, \citenamefont {{Stairs}}, \citenamefont
  {{Stinebring}}, \citenamefont {{Stovall}}, \citenamefont {{Swiggum}},
  \citenamefont {{Taylor}}, \citenamefont {{Vallisneri}}, \citenamefont {{van
  Haasteren}}, \citenamefont {{Wang}}, \citenamefont {{Zhu}},\ and\
  \citenamefont {{NANOGrav Collaboration}}}]{ArzoumanianEtAl:2016}%
  \BibitemOpen
  \bibfield  {author} {\bibinfo {author} {\bibfnamefont {Z.}~\bibnamefont
  {{Arzoumanian}}}, \bibinfo {author} {\bibfnamefont {A.}~\bibnamefont
  {{Brazier}}}, \bibinfo {author} {\bibfnamefont {S.}~\bibnamefont
  {{Burke-Spolaor}}}, \bibinfo {author} {\bibfnamefont {S.~J.}\ \bibnamefont
  {{Chamberlin}}}, \bibinfo {author} {\bibfnamefont {S.}~\bibnamefont
  {{Chatterjee}}}, \bibinfo {author} {\bibfnamefont {B.}~\bibnamefont
  {{Christy}}}, \bibinfo {author} {\bibfnamefont {J.~M.}\ \bibnamefont
  {{Cordes}}}, \bibinfo {author} {\bibfnamefont {N.~J.}\ \bibnamefont
  {{Cornish}}}, \bibinfo {author} {\bibfnamefont {K.}~\bibnamefont
  {{Crowter}}}, \bibinfo {author} {\bibfnamefont {P.~B.}\ \bibnamefont
  {{Demorest}}}, \bibinfo {author} {\bibfnamefont {X.}~\bibnamefont {{Deng}}},
  \bibinfo {author} {\bibfnamefont {T.}~\bibnamefont {{Dolch}}}, \bibinfo
  {author} {\bibfnamefont {J.~A.}\ \bibnamefont {{Ellis}}}, \bibinfo {author}
  {\bibfnamefont {R.~D.}\ \bibnamefont {{Ferdman}}}, \bibinfo {author}
  {\bibfnamefont {E.}~\bibnamefont {{Fonseca}}}, \bibinfo {author}
  {\bibfnamefont {N.}~\bibnamefont {{Garver-Daniels}}}, \bibinfo {author}
  {\bibfnamefont {M.~E.}\ \bibnamefont {{Gonzalez}}}, \bibinfo {author}
  {\bibfnamefont {F.}~\bibnamefont {{Jenet}}}, \bibinfo {author} {\bibfnamefont
  {G.}~\bibnamefont {{Jones}}}, \bibinfo {author} {\bibfnamefont {M.~L.}\
  \bibnamefont {{Jones}}}, \bibinfo {author} {\bibfnamefont {V.~M.}\
  \bibnamefont {{Kaspi}}}, \bibinfo {author} {\bibfnamefont {M.}~\bibnamefont
  {{Koop}}}, \bibinfo {author} {\bibfnamefont {M.~T.}\ \bibnamefont {{Lam}}},
  \bibinfo {author} {\bibfnamefont {T.~J.~W.}\ \bibnamefont {{Lazio}}},
  \bibinfo {author} {\bibfnamefont {L.}~\bibnamefont {{Levin}}}, \bibinfo
  {author} {\bibfnamefont {A.~N.}\ \bibnamefont {{Lommen}}}, \bibinfo {author}
  {\bibfnamefont {D.~R.}\ \bibnamefont {{Lorimer}}}, \bibinfo {author}
  {\bibfnamefont {J.}~\bibnamefont {{Luo}}}, \bibinfo {author} {\bibfnamefont
  {R.~S.}\ \bibnamefont {{Lynch}}}, \bibinfo {author} {\bibfnamefont {D.~R.}\
  \bibnamefont {{Madison}}}, \bibinfo {author} {\bibfnamefont {M.~A.}\
  \bibnamefont {{McLaughlin}}}, \bibinfo {author} {\bibfnamefont {S.~T.}\
  \bibnamefont {{McWilliams}}}, \bibinfo {author} {\bibfnamefont {C.~M.~F.}\
  \bibnamefont {{Mingarelli}}}, \bibinfo {author} {\bibfnamefont {D.~J.}\
  \bibnamefont {{Nice}}}, \bibinfo {author} {\bibfnamefont {N.}~\bibnamefont
  {{Palliyaguru}}}, \bibinfo {author} {\bibfnamefont {T.~T.}\ \bibnamefont
  {{Pennucci}}}, \bibinfo {author} {\bibfnamefont {S.~M.}\ \bibnamefont
  {{Ransom}}}, \bibinfo {author} {\bibfnamefont {L.}~\bibnamefont {{Sampson}}},
  \bibinfo {author} {\bibfnamefont {S.~A.}\ \bibnamefont {{Sanidas}}}, \bibinfo
  {author} {\bibfnamefont {A.}~\bibnamefont {{Sesana}}}, \bibinfo {author}
  {\bibfnamefont {X.}~\bibnamefont {{Siemens}}}, \bibinfo {author}
  {\bibfnamefont {J.}~\bibnamefont {{Simon}}}, \bibinfo {author} {\bibfnamefont
  {I.~H.}\ \bibnamefont {{Stairs}}}, \bibinfo {author} {\bibfnamefont {D.~R.}\
  \bibnamefont {{Stinebring}}}, \bibinfo {author} {\bibfnamefont
  {K.}~\bibnamefont {{Stovall}}}, \bibinfo {author} {\bibfnamefont
  {J.}~\bibnamefont {{Swiggum}}}, \bibinfo {author} {\bibfnamefont {S.~R.}\
  \bibnamefont {{Taylor}}}, \bibinfo {author} {\bibfnamefont {M.}~\bibnamefont
  {{Vallisneri}}}, \bibinfo {author} {\bibfnamefont {R.}~\bibnamefont {{van
  Haasteren}}}, \bibinfo {author} {\bibfnamefont {Y.}~\bibnamefont {{Wang}}},
  \bibinfo {author} {\bibfnamefont {W.~W.}\ \bibnamefont {{Zhu}}}, \ and\
  \bibinfo {author} {\bibnamefont {{NANOGrav Collaboration}}},\ }\href
  {\doibase10.3847/0004-637X/821/1/13} {\bibfield  {journal} {\bibinfo
  {journal} {Astrophysical Journal}\ }\textbf {\bibinfo {volume} {821}},\
  \bibinfo {eid} {13} (\bibinfo {year} {2016})},\ \Eprint
  {http://arxiv.org/abs/1508.03024} {arXiv:1508.03024}\BibitemShut {NoStop}%
\bibitem [{\citenamefont {{Sesana}}(2013)}]{Sesana13}%
  \BibitemOpen
  \bibfield  {author} {\bibinfo {author} {\bibfnamefont {A.}~\bibnamefont
  {{Sesana}}},\ }\href {\doibase10.1088/0264-9381/30/22/224014} {\bibfield
  {journal} {\bibinfo  {journal} {Classical and Quantum Gravity}\ }\textbf
  {\bibinfo {volume} {30}},\ \bibinfo {eid} {224014} (\bibinfo {year}
  {2013})},\ \Eprint {http://arxiv.org/abs/1307.2600}
  {arXiv:1307.2600}\BibitemShut {NoStop}%
\bibitem [{\citenamefont {{Taylor}}\ \emph {et~al.}(2017)\citenamefont
  {{Taylor}}, \citenamefont {{Simon}},\ and\ \citenamefont
  {{Sampson}}}]{2017PhRvL.118r1102T}%
  \BibitemOpen
  \bibfield  {author} {\bibinfo {author} {\bibfnamefont {S.~R.}\ \bibnamefont
  {{Taylor}}}, \bibinfo {author} {\bibfnamefont {J.}~\bibnamefont {{Simon}}}, \
  and\ \bibinfo {author} {\bibfnamefont {L.}~\bibnamefont {{Sampson}}},\ }\href
  {\doibase10.1103/PhysRevLett.118.181102} {\bibfield  {journal} {\bibinfo
  {journal} {Physical Review Letters}\ }\textbf {\bibinfo {volume} {118}},\
  \bibinfo {eid} {181102} (\bibinfo {year} {2017})},\ \Eprint
  {http://arxiv.org/abs/1612.02817} {arXiv:1612.02817}\BibitemShut {NoStop}%
\bibitem [{\citenamefont {{Chen}}\ \emph {et~al.}(2017)\citenamefont {{Chen}},
  \citenamefont {{Middleton}}, \citenamefont {{Sesana}}, \citenamefont {{Del
  Pozzo}},\ and\ \citenamefont {{Vecchio}}}]{2017MNRAS.468..404C}%
  \BibitemOpen
  \bibfield  {author} {\bibinfo {author} {\bibfnamefont {S.}~\bibnamefont
  {{Chen}}}, \bibinfo {author} {\bibfnamefont {H.}~\bibnamefont {{Middleton}}},
  \bibinfo {author} {\bibfnamefont {A.}~\bibnamefont {{Sesana}}}, \bibinfo
  {author} {\bibfnamefont {W.}~\bibnamefont {{Del Pozzo}}}, \ and\ \bibinfo
  {author} {\bibfnamefont {A.}~\bibnamefont {{Vecchio}}},\ }\href
  {\doibase10.1093/mnras/stx475} {\bibfield  {journal} {\bibinfo  {journal}
  {Monthly Notices of The Royal Astronomical Society}\ }\textbf {\bibinfo
  {volume} {468}},\ \bibinfo {pages} {404--417} (\bibinfo {year} {2017})},\
  \Eprint {http://arxiv.org/abs/1612.02826} {arXiv:1612.02826
  [astro-ph.HE]}\BibitemShut {NoStop}%
\bibitem [{\citenamefont {{Ryu}}\ \emph {et~al.}(2018)\citenamefont {{Ryu}},
  \citenamefont {{Perna}}, \citenamefont {{Haiman}}, \citenamefont
  {{Ostriker}},\ and\ \citenamefont {{Stone}}}]{rph+18}%
  \BibitemOpen
  \bibfield  {author} {\bibinfo {author} {\bibfnamefont {T.}~\bibnamefont
  {{Ryu}}}, \bibinfo {author} {\bibfnamefont {R.}~\bibnamefont {{Perna}}},
  \bibinfo {author} {\bibfnamefont {Z.}~\bibnamefont {{Haiman}}}, \bibinfo
  {author} {\bibfnamefont {J.~P.}\ \bibnamefont {{Ostriker}}}, \ and\ \bibinfo
  {author} {\bibfnamefont {N.~C.}\ \bibnamefont {{Stone}}},\ }\href
  {\doibase10.1093/mnras/stx2524} {\bibfield  {journal} {\bibinfo  {journal}
  {\mnras}\ }\textbf {\bibinfo {volume} {473}},\ \bibinfo {pages} {3410--3433}
  (\bibinfo {year} {2018})},\ \Eprint {http://arxiv.org/abs/1709.06501}
  {arXiv:1709.06501}\BibitemShut {NoStop}%
\bibitem [{\citenamefont {{Kelley}}\ \emph {et~al.}(2017)\citenamefont
  {{Kelley}}, \citenamefont {{Blecha}}, \citenamefont {{Hernquist}},
  \citenamefont {{Sesana}},\ and\ \citenamefont {{Taylor}}}]{lbh+17}%
  \BibitemOpen
  \bibfield  {author} {\bibinfo {author} {\bibfnamefont {L.~Z.}\ \bibnamefont
  {{Kelley}}}, \bibinfo {author} {\bibfnamefont {L.}~\bibnamefont {{Blecha}}},
  \bibinfo {author} {\bibfnamefont {L.}~\bibnamefont {{Hernquist}}}, \bibinfo
  {author} {\bibfnamefont {A.}~\bibnamefont {{Sesana}}}, \ and\ \bibinfo
  {author} {\bibfnamefont {S.~R.}\ \bibnamefont {{Taylor}}},\ }\href
  {\doibase10.1093/mnras/stx1638} {\bibfield  {journal} {\bibinfo  {journal}
  {\mnras}\ }\textbf {\bibinfo {volume} {471}},\ \bibinfo {pages} {4508--4526}
  (\bibinfo {year} {2017})},\ \Eprint {http://arxiv.org/abs/1702.02180}
  {arXiv:1702.02180 [astro-ph.HE]}\BibitemShut {NoStop}%
\bibitem [{\citenamefont {{Roebber}}\ \emph {et~al.}(2016)\citenamefont
  {{Roebber}}, \citenamefont {{Holder}}, \citenamefont {{Holz}},\ and\
  \citenamefont {{Warren}}}]{rhh+16}%
  \BibitemOpen
  \bibfield  {author} {\bibinfo {author} {\bibfnamefont {E.}~\bibnamefont
  {{Roebber}}}, \bibinfo {author} {\bibfnamefont {G.}~\bibnamefont {{Holder}}},
  \bibinfo {author} {\bibfnamefont {D.~E.}\ \bibnamefont {{Holz}}}, \ and\
  \bibinfo {author} {\bibfnamefont {M.}~\bibnamefont {{Warren}}},\ }\href
  {\doibase10.3847/0004-637X/819/2/163} {\bibfield  {journal} {\bibinfo
  {journal} {\apj}\ }\textbf {\bibinfo {volume} {819}},\ \bibinfo {eid} {163}
  (\bibinfo {year} {2016})},\ \Eprint {http://arxiv.org/abs/1508.07336}
  {arXiv:1508.07336}\BibitemShut {NoStop}%
\bibitem [{\citenamefont {{Inayoshi}}\ \emph {et~al.}(2018)\citenamefont
  {{Inayoshi}}, \citenamefont {{Ichikawa}},\ and\ \citenamefont
  {{Haiman}}}]{iih18}%
  \BibitemOpen
  \bibfield  {author} {\bibinfo {author} {\bibfnamefont {K.}~\bibnamefont
  {{Inayoshi}}}, \bibinfo {author} {\bibfnamefont {K.}~\bibnamefont
  {{Ichikawa}}}, \ and\ \bibinfo {author} {\bibfnamefont {Z.}~\bibnamefont
  {{Haiman}}},\ }\href@noop {} {\bibfield  {journal} {\bibinfo  {journal}
  {ArXiv e-prints}\ } (\bibinfo {year} {2018})},\ \Eprint
  {http://arxiv.org/abs/1805.05334} {arXiv:1805.05334}\BibitemShut {NoStop}%
\bibitem [{\citenamefont {Middleton}\ \emph {et~al.}(2016)\citenamefont
  {Middleton}, \citenamefont {Del~Pozzo}, \citenamefont {Farr}, \citenamefont
  {Sesana},\ and\ \citenamefont {Vecchio}}]{Middleton:2015oda}%
  \BibitemOpen
  \bibfield  {author} {\bibinfo {author} {\bibfnamefont {Hannah}\ \bibnamefont
  {Middleton}}, \bibinfo {author} {\bibfnamefont {Walter}\ \bibnamefont
  {Del~Pozzo}}, \bibinfo {author} {\bibfnamefont {Will~M.}\ \bibnamefont
  {Farr}}, \bibinfo {author} {\bibfnamefont {Alberto}\ \bibnamefont {Sesana}},
  \ and\ \bibinfo {author} {\bibfnamefont {Alberto}\ \bibnamefont {Vecchio}},\
  }\href {\doibase10.1093/mnrasl/slv150} {\bibfield  {journal} {\bibinfo
  {journal} {Mon. Not. Roy. Astron. Soc.}\ }\textbf {\bibinfo {volume} {455}},\
  \bibinfo {pages} {L72--L76} (\bibinfo {year} {2016})},\ \Eprint
  {http://arxiv.org/abs/1507.00992} {arXiv:1507.00992
  [astro-ph.CO]}\BibitemShut {NoStop}%
\bibitem [{\citenamefont {{Kormendy}}\ and\ \citenamefont
  {{Ho}}(2013{\natexlab{b}})}]{kh13}%
  \BibitemOpen
  \bibfield  {author} {\bibinfo {author} {\bibfnamefont {J.}~\bibnamefont
  {{Kormendy}}}\ and\ \bibinfo {author} {\bibfnamefont {L.~C.}\ \bibnamefont
  {{Ho}}},\ }\href {\doibase10.1146/annurev-astro-082708-101811} {\bibfield
  {journal} {\bibinfo  {journal} {Annu. Rev. Astron. Astrophys}\ }\textbf
  {\bibinfo {volume} {51}},\ \bibinfo {pages} {511--653} (\bibinfo {year}
  {2013}{\natexlab{b}})},\ \Eprint {http://arxiv.org/abs/1304.7762}
  {arXiv:1304.7762 [astro-ph.CO]}\BibitemShut {NoStop}%
\bibitem [{\citenamefont {{Middleton}}\ \emph {et~al.}(2018)\citenamefont
  {{Middleton}}, \citenamefont {{Chen}}, \citenamefont {{Del Pozzo}},
  \citenamefont {{Sesana}},\ and\ \citenamefont
  {{Vecchio}}}]{2018NatCo...9..573M}%
  \BibitemOpen
  \bibfield  {author} {\bibinfo {author} {\bibfnamefont {H.}~\bibnamefont
  {{Middleton}}}, \bibinfo {author} {\bibfnamefont {S.}~\bibnamefont {{Chen}}},
  \bibinfo {author} {\bibfnamefont {W.}~\bibnamefont {{Del Pozzo}}}, \bibinfo
  {author} {\bibfnamefont {A.}~\bibnamefont {{Sesana}}}, \ and\ \bibinfo
  {author} {\bibfnamefont {A.}~\bibnamefont {{Vecchio}}},\ }\href
  {\doibase10.1038/s41467-018-02916-7} {\bibfield  {journal} {\bibinfo
  {journal} {Nature Communications}\ }\textbf {\bibinfo {volume} {9}},\
  \bibinfo {eid} {573} (\bibinfo {year} {2018})},\ \Eprint
  {http://arxiv.org/abs/1707.00623} {arXiv:1707.00623}\BibitemShut {NoStop}%
\bibitem [{\citenamefont {Lentati}\ \emph {et~al.}(2015)\citenamefont {Lentati}
  \emph {et~al.}}]{Lentati:2015qwp}%
  \BibitemOpen
  \bibfield  {author} {\bibinfo {author} {\bibfnamefont {L.}~\bibnamefont
  {Lentati}} \emph {et~al.},\ }\href {\doibase10.1093/mnras/stv1538} {\bibfield
   {journal} {\bibinfo  {journal} {Mon. Not. Roy. Astron. Soc.}\ }\textbf
  {\bibinfo {volume} {453}},\ \bibinfo {pages} {2576--2598} (\bibinfo {year}
  {2015})},\ \Eprint {http://arxiv.org/abs/1504.03692} {arXiv:1504.03692
  [astro-ph.CO]}\BibitemShut {NoStop}%
\bibitem [{\citenamefont {Arzoumanian}\ \emph
  {et~al.}(2018{\natexlab{b}})\citenamefont {Arzoumanian} \emph
  {et~al.}}]{Arzoumanian:2018saf}%
  \BibitemOpen
  \bibfield  {author} {\bibinfo {author} {\bibfnamefont {Z.}~\bibnamefont
  {Arzoumanian}} \emph {et~al.} (\bibinfo {collaboration} {NANOGRAV}),\ }\href
  {\doibase10.3847/1538-4357/aabd3b} {\bibfield  {journal} {\bibinfo  {journal}
  {Astrophys. J.}\ }\textbf {\bibinfo {volume} {859}},\ \bibinfo {pages} {47}
  (\bibinfo {year} {2018}{\natexlab{b}})},\ \Eprint
  {http://arxiv.org/abs/1801.02617} {arXiv:1801.02617
  [astro-ph.HE]}\BibitemShut {NoStop}%
\bibitem [{\citenamefont {Janssen}\ \emph {et~al.}(2015)\citenamefont {Janssen}
  \emph {et~al.}}]{Janssen:2014dka}%
  \BibitemOpen
  \bibfield  {author} {\bibinfo {author} {\bibfnamefont {Gemma}\ \bibnamefont
  {Janssen}} \emph {et~al.},\ }\bibfield  {booktitle} {\emph {\bibinfo
  {booktitle} {{Proceedings, Advancing Astrophysics with the Square Kilometre
  Array (AASKA14): Giardini Naxos, Italy, June 9-13, 2014}}},\ }\href
  {\doibase10.22323/1.215.0037} {\bibfield  {journal} {\bibinfo  {journal}
  {PoS}\ }\textbf {\bibinfo {volume} {AASKA14}},\ \bibinfo {pages} {037}
  (\bibinfo {year} {2015})},\ \Eprint {http://arxiv.org/abs/1501.00127}
  {arXiv:1501.00127 [astro-ph.IM]}\BibitemShut {NoStop}%
\bibitem [{\citenamefont {{Tiburzi}}\ \emph {et~al.}(2016)\citenamefont
  {{Tiburzi}}, \citenamefont {{Hobbs}}, \citenamefont {{Kerr}}, \citenamefont
  {{Coles}}, \citenamefont {{Dai}}, \citenamefont {{Manchester}}, \citenamefont
  {{Possenti}}, \citenamefont {{Shannon}},\ and\ \citenamefont
  {{You}}}]{thk+16}%
  \BibitemOpen
  \bibfield  {author} {\bibinfo {author} {\bibfnamefont {C.}~\bibnamefont
  {{Tiburzi}}}, \bibinfo {author} {\bibfnamefont {G.}~\bibnamefont {{Hobbs}}},
  \bibinfo {author} {\bibfnamefont {M.}~\bibnamefont {{Kerr}}}, \bibinfo
  {author} {\bibfnamefont {W.~A.}\ \bibnamefont {{Coles}}}, \bibinfo {author}
  {\bibfnamefont {S.}~\bibnamefont {{Dai}}}, \bibinfo {author} {\bibfnamefont
  {R.~N.}\ \bibnamefont {{Manchester}}}, \bibinfo {author} {\bibfnamefont
  {A.}~\bibnamefont {{Possenti}}}, \bibinfo {author} {\bibfnamefont {R.~M.}\
  \bibnamefont {{Shannon}}}, \ and\ \bibinfo {author} {\bibfnamefont {X.~P.}\
  \bibnamefont {{You}}},\ }\href {\doibase10.1093/mnras/stv2143} {\bibfield
  {journal} {\bibinfo  {journal} {Monthly Notices of The Royal Astronomical
  Society}\ }\textbf {\bibinfo {volume} {455}},\ \bibinfo {pages} {4339--4350}
  (\bibinfo {year} {2016})},\ \Eprint {http://arxiv.org/abs/1510.02363}
  {arXiv:1510.02363 [astro-ph.IM]}\BibitemShut {NoStop}%
\bibitem [{\citenamefont {{Mingarelli}}\ and\ \citenamefont
  {{Mingarelli}}(2018)}]{mm18}%
  \BibitemOpen
  \bibfield  {author} {\bibinfo {author} {\bibfnamefont {C.~M.~F.}\
  \bibnamefont {{Mingarelli}}}\ and\ \bibinfo {author} {\bibfnamefont {A.~B.}\
  \bibnamefont {{Mingarelli}}},\ }\href@noop {} {\bibfield  {journal} {\bibinfo
   {journal} {ArXiv e-prints}\ } (\bibinfo {year} {2018})},\ \Eprint
  {http://arxiv.org/abs/1806.06979} {arXiv:1806.06979
  [astro-ph.IM]}\BibitemShut {NoStop}%
\bibitem [{\citenamefont {Moore}\ \emph {et~al.}(2015)\citenamefont {Moore},
  \citenamefont {Cole},\ and\ \citenamefont {Berry}}]{Moore:2014lga}%
  \BibitemOpen
  \bibfield  {author} {\bibinfo {author} {\bibfnamefont {C.~J.}\ \bibnamefont
  {Moore}}, \bibinfo {author} {\bibfnamefont {R.~H.}\ \bibnamefont {Cole}}, \
  and\ \bibinfo {author} {\bibfnamefont {C.~P.~L.}\ \bibnamefont {Berry}},\
  }\href {\doibase10.1088/0264-9381/32/1/015014} {\bibfield  {journal}
  {\bibinfo  {journal} {Class. Quant. Grav.}\ }\textbf {\bibinfo {volume}
  {32}},\ \bibinfo {pages} {015014} (\bibinfo {year} {2015})},\ \Eprint
  {http://arxiv.org/abs/1408.0740} {arXiv:1408.0740 [gr-qc]}\BibitemShut
  {NoStop}%
\bibitem [{\citenamefont {{Lee}}\ \emph {et~al.}(2011)\citenamefont {{Lee}},
  \citenamefont {{Wex}}, \citenamefont {{Kramer}}, \citenamefont {{Stappers}},
  \citenamefont {{Bassa}}, \citenamefont {{Janssen}}, \citenamefont
  {{Karuppusamy}},\ and\ \citenamefont {{Smits}}}]{lwk+11}%
  \BibitemOpen
  \bibfield  {author} {\bibinfo {author} {\bibfnamefont {K.~J.}\ \bibnamefont
  {{Lee}}}, \bibinfo {author} {\bibfnamefont {N.}~\bibnamefont {{Wex}}},
  \bibinfo {author} {\bibfnamefont {M.}~\bibnamefont {{Kramer}}}, \bibinfo
  {author} {\bibfnamefont {B.~W.}\ \bibnamefont {{Stappers}}}, \bibinfo
  {author} {\bibfnamefont {C.~G.}\ \bibnamefont {{Bassa}}}, \bibinfo {author}
  {\bibfnamefont {G.~H.}\ \bibnamefont {{Janssen}}}, \bibinfo {author}
  {\bibfnamefont {R.}~\bibnamefont {{Karuppusamy}}}, \ and\ \bibinfo {author}
  {\bibfnamefont {R.}~\bibnamefont {{Smits}}},\ }\href
  {\doibase10.1111/j.1365-2966.2011.18622.x} {\bibfield  {journal} {\bibinfo
  {journal} {Mon. Not. Roy. Astron. Soc.}\ }\textbf {\bibinfo {volume} {414}},\
  \bibinfo {pages} {3251--3264} (\bibinfo {year} {2011})},\ \Eprint
  {http://arxiv.org/abs/1103.0115} {arXiv:1103.0115 [astro-ph.HE]}\BibitemShut
  {NoStop}%
\bibitem [{\citenamefont {{Babak}}\ and\ \citenamefont
  {{Sesana}}(2012)}]{2012PhRvD..85d4034B}%
  \BibitemOpen
  \bibfield  {author} {\bibinfo {author} {\bibfnamefont {S.}~\bibnamefont
  {{Babak}}}\ and\ \bibinfo {author} {\bibfnamefont {A.}~\bibnamefont
  {{Sesana}}},\ }\href {\doibase10.1103/PhysRevD.85.044034} {\bibfield
  {journal} {\bibinfo  {journal} {Physical Review D}\ }\textbf {\bibinfo
  {volume} {85}},\ \bibinfo {eid} {044034} (\bibinfo {year} {2012})},\ \Eprint
  {http://arxiv.org/abs/1112.1075} {arXiv:1112.1075}\BibitemShut {NoStop}%
\bibitem [{\citenamefont {{Arzoumanian}}\ \emph {et~al.}(2014)\citenamefont
  {{Arzoumanian}}, \citenamefont {{Brazier}}, \citenamefont {{Burke-Spolaor}},
  \citenamefont {{Chamberlin}}, \citenamefont {{Chatterjee}}, \citenamefont
  {{Cordes}}, \citenamefont {{Demorest}}, \citenamefont {{Deng}}, \citenamefont
  {{Dolch}}, \citenamefont {{Ellis}}, \citenamefont {{Ferdman}}, \citenamefont
  {{Garver-Daniels}}, \citenamefont {{Jenet}}, \citenamefont {{Jones}},
  \citenamefont {{Kaspi}}, \citenamefont {{Koop}}, \citenamefont {{Lam}},
  \citenamefont {{Lazio}}, \citenamefont {{Lommen}}, \citenamefont {{Lorimer}},
  \citenamefont {{Luo}}, \citenamefont {{Lynch}}, \citenamefont {{Madison}},
  \citenamefont {{McLaughlin}}, \citenamefont {{McWilliams}}, \citenamefont
  {{Nice}}, \citenamefont {{Palliyaguru}}, \citenamefont {{Pennucci}},
  \citenamefont {{Ransom}}, \citenamefont {{Sesana}}, \citenamefont
  {{Siemens}}, \citenamefont {{Stairs}}, \citenamefont {{Stinebring}},
  \citenamefont {{Stovall}}, \citenamefont {{Swiggum}}, \citenamefont
  {{Vallisneri}}, \citenamefont {{van Haasteren}}, \citenamefont {{Wang}},
  \citenamefont {{Zhu}},\ and\ \citenamefont {{NANOGrav
  Collaboration}}}]{abb+14}%
  \BibitemOpen
  \bibfield  {author} {\bibinfo {author} {\bibfnamefont {Z.}~\bibnamefont
  {{Arzoumanian}}}, \bibinfo {author} {\bibfnamefont {A.}~\bibnamefont
  {{Brazier}}}, \bibinfo {author} {\bibfnamefont {S.}~\bibnamefont
  {{Burke-Spolaor}}}, \bibinfo {author} {\bibfnamefont {S.~J.}\ \bibnamefont
  {{Chamberlin}}}, \bibinfo {author} {\bibfnamefont {S.}~\bibnamefont
  {{Chatterjee}}}, \bibinfo {author} {\bibfnamefont {J.~M.}\ \bibnamefont
  {{Cordes}}}, \bibinfo {author} {\bibfnamefont {P.~B.}\ \bibnamefont
  {{Demorest}}}, \bibinfo {author} {\bibfnamefont {X.}~\bibnamefont {{Deng}}},
  \bibinfo {author} {\bibfnamefont {T.}~\bibnamefont {{Dolch}}}, \bibinfo
  {author} {\bibfnamefont {J.~A.}\ \bibnamefont {{Ellis}}}, \bibinfo {author}
  {\bibfnamefont {R.~D.}\ \bibnamefont {{Ferdman}}}, \bibinfo {author}
  {\bibfnamefont {N.}~\bibnamefont {{Garver-Daniels}}}, \bibinfo {author}
  {\bibfnamefont {F.}~\bibnamefont {{Jenet}}}, \bibinfo {author} {\bibfnamefont
  {G.}~\bibnamefont {{Jones}}}, \bibinfo {author} {\bibfnamefont {V.~M.}\
  \bibnamefont {{Kaspi}}}, \bibinfo {author} {\bibfnamefont {M.}~\bibnamefont
  {{Koop}}}, \bibinfo {author} {\bibfnamefont {M.~T.}\ \bibnamefont {{Lam}}},
  \bibinfo {author} {\bibfnamefont {T.~J.~W.}\ \bibnamefont {{Lazio}}},
  \bibinfo {author} {\bibfnamefont {A.~N.}\ \bibnamefont {{Lommen}}}, \bibinfo
  {author} {\bibfnamefont {D.~R.}\ \bibnamefont {{Lorimer}}}, \bibinfo {author}
  {\bibfnamefont {J.}~\bibnamefont {{Luo}}}, \bibinfo {author} {\bibfnamefont
  {R.~S.}\ \bibnamefont {{Lynch}}}, \bibinfo {author} {\bibfnamefont {D.~R.}\
  \bibnamefont {{Madison}}}, \bibinfo {author} {\bibfnamefont {M.~A.}\
  \bibnamefont {{McLaughlin}}}, \bibinfo {author} {\bibfnamefont {S.~T.}\
  \bibnamefont {{McWilliams}}}, \bibinfo {author} {\bibfnamefont {D.~J.}\
  \bibnamefont {{Nice}}}, \bibinfo {author} {\bibfnamefont {N.}~\bibnamefont
  {{Palliyaguru}}}, \bibinfo {author} {\bibfnamefont {T.~T.}\ \bibnamefont
  {{Pennucci}}}, \bibinfo {author} {\bibfnamefont {S.~M.}\ \bibnamefont
  {{Ransom}}}, \bibinfo {author} {\bibfnamefont {A.}~\bibnamefont {{Sesana}}},
  \bibinfo {author} {\bibfnamefont {X.}~\bibnamefont {{Siemens}}}, \bibinfo
  {author} {\bibfnamefont {I.~H.}\ \bibnamefont {{Stairs}}}, \bibinfo {author}
  {\bibfnamefont {D.~R.}\ \bibnamefont {{Stinebring}}}, \bibinfo {author}
  {\bibfnamefont {K.}~\bibnamefont {{Stovall}}}, \bibinfo {author}
  {\bibfnamefont {J.}~\bibnamefont {{Swiggum}}}, \bibinfo {author}
  {\bibfnamefont {M.}~\bibnamefont {{Vallisneri}}}, \bibinfo {author}
  {\bibfnamefont {R.}~\bibnamefont {{van Haasteren}}}, \bibinfo {author}
  {\bibfnamefont {Y.}~\bibnamefont {{Wang}}}, \bibinfo {author} {\bibfnamefont
  {W.~W.}\ \bibnamefont {{Zhu}}}, \ and\ \bibinfo {author} {\bibnamefont
  {{NANOGrav Collaboration}}},\ }\href {\doibase10.1088/0004-637X/794/2/141}
  {\bibfield  {journal} {\bibinfo  {journal} {Astrophysical Journal}\ }\textbf
  {\bibinfo {volume} {794}},\ \bibinfo {eid} {141} (\bibinfo {year} {2014})},\
  \Eprint {http://arxiv.org/abs/1404.1267} {arXiv:1404.1267}\BibitemShut
  {NoStop}%
\bibitem [{\citenamefont {Babak}\ \emph {et~al.}(2016)\citenamefont {Babak}
  \emph {et~al.}}]{Babak:2015lua}%
  \BibitemOpen
  \bibfield  {author} {\bibinfo {author} {\bibfnamefont {Stanislav}\
  \bibnamefont {Babak}} \emph {et~al.},\ }\href {\doibase10.1093/mnras/stv2092}
  {\bibfield  {journal} {\bibinfo  {journal} {Mon. Not. Roy. Astron. Soc.}\
  }\textbf {\bibinfo {volume} {455}},\ \bibinfo {pages} {1665--1679} (\bibinfo
  {year} {2016})},\ \Eprint {http://arxiv.org/abs/1509.02165} {arXiv:1509.02165
  [astro-ph.CO]}\BibitemShut {NoStop}%
\bibitem [{\citenamefont {{Sesana}}\ and\ \citenamefont
  {{Vecchio}}(2010)}]{sv10}%
  \BibitemOpen
  \bibfield  {author} {\bibinfo {author} {\bibfnamefont {A.}~\bibnamefont
  {{Sesana}}}\ and\ \bibinfo {author} {\bibfnamefont {A.}~\bibnamefont
  {{Vecchio}}},\ }\href {\doibase10.1103/PhysRevD.81.104008} {\bibfield
  {journal} {\bibinfo  {journal} {Phys. Rev. D}\ }\textbf {\bibinfo {volume}
  {81}},\ \bibinfo {pages} {104008--+} (\bibinfo {year} {2010})},\ \Eprint
  {http://arxiv.org/abs/1003.0677} {arXiv:1003.0677 [astro-ph.CO]}\BibitemShut
  {NoStop}%
\bibitem [{\citenamefont {{Simon}}\ \emph {et~al.}(2014)\citenamefont
  {{Simon}}, \citenamefont {{Polin}}, \citenamefont {{Lommen}}, \citenamefont
  {{Stappers}}, \citenamefont {{Finn}}, \citenamefont {{Jenet}},\ and\
  \citenamefont {{Christy}}}]{spl+14}%
  \BibitemOpen
  \bibfield  {author} {\bibinfo {author} {\bibfnamefont {J.}~\bibnamefont
  {{Simon}}}, \bibinfo {author} {\bibfnamefont {A.}~\bibnamefont {{Polin}}},
  \bibinfo {author} {\bibfnamefont {A.}~\bibnamefont {{Lommen}}}, \bibinfo
  {author} {\bibfnamefont {B.}~\bibnamefont {{Stappers}}}, \bibinfo {author}
  {\bibfnamefont {L.~S.}\ \bibnamefont {{Finn}}}, \bibinfo {author}
  {\bibfnamefont {F.~A.}\ \bibnamefont {{Jenet}}}, \ and\ \bibinfo {author}
  {\bibfnamefont {B.}~\bibnamefont {{Christy}}},\ }\href
  {\doibase10.1088/0004-637X/784/1/60} {\bibfield  {journal} {\bibinfo
  {journal} {\apj}\ }\textbf {\bibinfo {volume} {784}},\ \bibinfo {eid} {60}
  (\bibinfo {year} {2014})},\ \Eprint {http://arxiv.org/abs/1402.1140}
  {arXiv:1402.1140 [astro-ph.CO]}\BibitemShut {NoStop}%
\bibitem [{\citenamefont {{Mingarelli}}\ \emph {et~al.}(2012)\citenamefont
  {{Mingarelli}}, \citenamefont {{Grover}}, \citenamefont {{Sidery}},
  \citenamefont {{Smith}},\ and\ \citenamefont {{Vecchio}}}]{Mingarelli2012}%
  \BibitemOpen
  \bibfield  {author} {\bibinfo {author} {\bibfnamefont {C.~M.~F.}\
  \bibnamefont {{Mingarelli}}}, \bibinfo {author} {\bibfnamefont
  {K.}~\bibnamefont {{Grover}}}, \bibinfo {author} {\bibfnamefont
  {T.}~\bibnamefont {{Sidery}}}, \bibinfo {author} {\bibfnamefont {R.~J.~E.}\
  \bibnamefont {{Smith}}}, \ and\ \bibinfo {author} {\bibfnamefont
  {A.}~\bibnamefont {{Vecchio}}},\ }\href
  {\doibase10.1103/PhysRevLett.109.081104} {\bibfield  {journal} {\bibinfo
  {journal} {Physical Review Letters}\ }\textbf {\bibinfo {volume} {109}},\
  \bibinfo {eid} {081104} (\bibinfo {year} {2012})},\ \Eprint
  {http://arxiv.org/abs/1207.5645} {arXiv:1207.5645 [astro-ph.HE]}\BibitemShut
  {NoStop}%
\bibitem [{\citenamefont {{Sesana}}\ \emph {et~al.}(2012)\citenamefont
  {{Sesana}}, \citenamefont {{Roedig}}, \citenamefont {{Reynolds}},\ and\
  \citenamefont {{Dotti}}}]{srr+12}%
  \BibitemOpen
  \bibfield  {author} {\bibinfo {author} {\bibfnamefont {A.}~\bibnamefont
  {{Sesana}}}, \bibinfo {author} {\bibfnamefont {C.}~\bibnamefont {{Roedig}}},
  \bibinfo {author} {\bibfnamefont {M.~T.}\ \bibnamefont {{Reynolds}}}, \ and\
  \bibinfo {author} {\bibfnamefont {M.}~\bibnamefont {{Dotti}}},\ }\href
  {\doibase10.1111/j.1365-2966.2011.20097.x} {\bibfield  {journal} {\bibinfo
  {journal} {Mon. Not. Roy. Astron. Soc.}\ }\textbf {\bibinfo {volume} {420}},\
  \bibinfo {pages} {860--877} (\bibinfo {year} {2012})},\ \Eprint
  {http://arxiv.org/abs/1107.2927} {arXiv:1107.2927 [astro-ph.CO]}\BibitemShut
  {NoStop}%
\bibitem [{\citenamefont {{Tanaka}}\ \emph {et~al.}(2012)\citenamefont
  {{Tanaka}}, \citenamefont {{Menou}},\ and\ \citenamefont {{Haiman}}}]{tmh12}%
  \BibitemOpen
  \bibfield  {author} {\bibinfo {author} {\bibfnamefont {T.}~\bibnamefont
  {{Tanaka}}}, \bibinfo {author} {\bibfnamefont {K.}~\bibnamefont {{Menou}}}, \
  and\ \bibinfo {author} {\bibfnamefont {Z.}~\bibnamefont {{Haiman}}},\ }\href
  {\doibase10.1111/j.1365-2966.2011.20083.x} {\bibfield  {journal} {\bibinfo
  {journal} {Mon. Not. Roy. Astron. Soc.}\ }\textbf {\bibinfo {volume} {420}},\
  \bibinfo {pages} {705--719} (\bibinfo {year} {2012})},\ \Eprint
  {http://arxiv.org/abs/1107.2937} {arXiv:1107.2937 [astro-ph.CO]}\BibitemShut
  {NoStop}%
\bibitem [{\citenamefont {Burke-Spolaor}(2013)}]{Burke-Spolaor:2013aba}%
  \BibitemOpen
  \bibfield  {author} {\bibinfo {author} {\bibfnamefont {Sarah}\ \bibnamefont
  {Burke-Spolaor}},\ }\href {\doibase10.1088/0264-9381/30/22/224013} {\bibfield
   {journal} {\bibinfo  {journal} {Class. Quant. Grav.}\ }\textbf {\bibinfo
  {volume} {30}},\ \bibinfo {pages} {224013} (\bibinfo {year} {2013})},\
  \Eprint {http://arxiv.org/abs/1308.4408} {arXiv:1308.4408
  [astro-ph.CO]}\BibitemShut {NoStop}%
\bibitem [{\citenamefont {{Moustakas}}(2017)}]{m17}%
  \BibitemOpen
  \bibfield  {author} {\bibinfo {author} {\bibfnamefont {L.~A.}\ \bibnamefont
  {{Moustakas}}},\ }\href {\doibase10.1038/s41550-017-0334-7} {\bibfield
  {journal} {\bibinfo  {journal} {Nature Astronomy}\ }\textbf {\bibinfo
  {volume} {1}},\ \bibinfo {pages} {825--826} (\bibinfo {year}
  {2017})}\BibitemShut {NoStop}%
\bibitem [{\citenamefont {Rosswog}(2015)}]{Rosswog:2015nja}%
  \BibitemOpen
  \bibfield  {author} {\bibinfo {author} {\bibfnamefont {Stephan}\ \bibnamefont
  {Rosswog}},\ }\href {\doibase10.1142/S0218271815300128} {\bibfield  {journal}
  {\bibinfo  {journal} {Int. J. Mod. Phys.}\ }\textbf {\bibinfo {volume}
  {D24}},\ \bibinfo {pages} {1530012} (\bibinfo {year} {2015})},\ \Eprint
  {http://arxiv.org/abs/1501.02081} {arXiv:1501.02081
  [astro-ph.HE]}\BibitemShut {NoStop}%
\bibitem [{\citenamefont {Shibata}\ and\ \citenamefont
  {Taniguchi}(2011)}]{Shibata:2011jka}%
  \BibitemOpen
  \bibfield  {author} {\bibinfo {author} {\bibfnamefont {Masaru}\ \bibnamefont
  {Shibata}}\ and\ \bibinfo {author} {\bibfnamefont {Keisuke}\ \bibnamefont
  {Taniguchi}},\ }\href {\doibase10.12942/lrr-2011-6} {\bibfield  {journal}
  {\bibinfo  {journal} {Living Rev. Rel.}\ }\textbf {\bibinfo {volume} {14}},\
  \bibinfo {pages} {6} (\bibinfo {year} {2011})}\BibitemShut {NoStop}%
\bibitem [{\citenamefont {Baiotti}\ and\ \citenamefont
  {Rezzolla}(2017)}]{Baiotti:2016qnr}%
  \BibitemOpen
  \bibfield  {author} {\bibinfo {author} {\bibfnamefont {Luca}\ \bibnamefont
  {Baiotti}}\ and\ \bibinfo {author} {\bibfnamefont {Luciano}\ \bibnamefont
  {Rezzolla}},\ }\href {\doibase10.1088/1361-6633/aa67bb} {\bibfield  {journal}
  {\bibinfo  {journal} {Rept. Prog. Phys.}\ }\textbf {\bibinfo {volume} {80}},\
  \bibinfo {pages} {096901} (\bibinfo {year} {2017})},\ \Eprint
  {http://arxiv.org/abs/1607.03540} {arXiv:1607.03540 [gr-qc]}\BibitemShut
  {NoStop}%
\bibitem [{\citenamefont {Paschalidis}(2017)}]{Paschalidis:2016agf}%
  \BibitemOpen
  \bibfield  {author} {\bibinfo {author} {\bibfnamefont {Vasileios}\
  \bibnamefont {Paschalidis}},\ }\href {\doibase10.1088/1361-6382/aa61ce}
  {\bibfield  {journal} {\bibinfo  {journal} {Class. Quant. Grav.}\ }\textbf
  {\bibinfo {volume} {34}},\ \bibinfo {pages} {084002} (\bibinfo {year}
  {2017})},\ \Eprint {http://arxiv.org/abs/1611.01519} {arXiv:1611.01519
  [astro-ph.HE]}\BibitemShut {NoStop}%
\bibitem [{\citenamefont {{Eichler}}\ \emph
  {et~al.}(1989{\natexlab{a}})\citenamefont {{Eichler}}, \citenamefont
  {{Livio}}, \citenamefont {{Piran}},\ and\ \citenamefont
  {{Schramm}}}]{1989Natur.340..126E}%
  \BibitemOpen
  \bibfield  {author} {\bibinfo {author} {\bibfnamefont {D.}~\bibnamefont
  {{Eichler}}}, \bibinfo {author} {\bibfnamefont {M.}~\bibnamefont {{Livio}}},
  \bibinfo {author} {\bibfnamefont {T.}~\bibnamefont {{Piran}}}, \ and\
  \bibinfo {author} {\bibfnamefont {D.~N.}\ \bibnamefont {{Schramm}}},\ }\href
  {\doibase10.1038/340126a0} {\bibfield  {journal} {\bibinfo  {journal}
  {Nature}\ }\textbf {\bibinfo {volume} {340}},\ \bibinfo {pages} {126--128}
  (\bibinfo {year} {1989}{\natexlab{a}})}\BibitemShut {NoStop}%
\bibitem [{\citenamefont {{Paczynski}}(1986)}]{1986ApJ...308L..43P}%
  \BibitemOpen
  \bibfield  {author} {\bibinfo {author} {\bibfnamefont {B.}~\bibnamefont
  {{Paczynski}}},\ }\href {\doibase10.1086/184740} {\bibfield  {journal}
  {\bibinfo  {journal} {Astrop. J. Lett.}\ }\textbf {\bibinfo {volume} {308}},\
  \bibinfo {pages} {L43--L46} (\bibinfo {year} {1986})}\BibitemShut {NoStop}%
\bibitem [{\citenamefont {Berger}(2014)}]{Berger:2013jza}%
  \BibitemOpen
  \bibfield  {author} {\bibinfo {author} {\bibfnamefont {Edo}\ \bibnamefont
  {Berger}},\ }\href {\doibase10.1146/annurev-astro-081913-035926} {\bibfield
  {journal} {\bibinfo  {journal} {Ann. Rev. Astron. Astrophys.}\ }\textbf
  {\bibinfo {volume} {52}},\ \bibinfo {pages} {43--105} (\bibinfo {year}
  {2014})},\ \Eprint {http://arxiv.org/abs/1311.2603} {arXiv:1311.2603
  [astro-ph.HE]}\BibitemShut {NoStop}%
\bibitem [{\citenamefont {{Lattimer}}\ and\ \citenamefont
  {{Schramm}}(1974)}]{1974ApJ...192L.145L}%
  \BibitemOpen
  \bibfield  {author} {\bibinfo {author} {\bibfnamefont {J.~M.}\ \bibnamefont
  {{Lattimer}}}\ and\ \bibinfo {author} {\bibfnamefont {D.~N.}\ \bibnamefont
  {{Schramm}}},\ }\href {\doibase10.1086/181612} {\bibfield  {journal}
  {\bibinfo  {journal} {Astrop. J. Lett.}\ }\textbf {\bibinfo {volume} {192}},\
  \bibinfo {pages} {L145--L147} (\bibinfo {year} {1974})}\BibitemShut {NoStop}%
\bibitem [{\citenamefont {{Meyer}}\ \emph {et~al.}(1989)\citenamefont
  {{Meyer}}, \citenamefont {{Welty}},\ and\ \citenamefont
  {{York}}}]{1989ApJ...343L..37M}%
  \BibitemOpen
  \bibfield  {author} {\bibinfo {author} {\bibfnamefont {D.~M.}\ \bibnamefont
  {{Meyer}}}, \bibinfo {author} {\bibfnamefont {D.~E.}\ \bibnamefont
  {{Welty}}}, \ and\ \bibinfo {author} {\bibfnamefont {D.~G.}\ \bibnamefont
  {{York}}},\ }\href {\doibase10.1086/185505} {\bibfield  {journal} {\bibinfo
  {journal} {"Astrop. J. Lett."}\ }\textbf {\bibinfo {volume} {343}},\ \bibinfo
  {pages} {L37--L40} (\bibinfo {year} {1989})}\BibitemShut {NoStop}%
\bibitem [{\citenamefont {{Freiburghaus}}\ \emph {et~al.}(1999)\citenamefont
  {{Freiburghaus}}, \citenamefont {{Rosswog}},\ and\ \citenamefont
  {{Thielemann}}}]{1999ApJ...525L.121F}%
  \BibitemOpen
  \bibfield  {author} {\bibinfo {author} {\bibfnamefont {C.}~\bibnamefont
  {{Freiburghaus}}}, \bibinfo {author} {\bibfnamefont {S.}~\bibnamefont
  {{Rosswog}}}, \ and\ \bibinfo {author} {\bibfnamefont {F.-K.}\ \bibnamefont
  {{Thielemann}}},\ }\href {\doibase10.1086/312343} {\bibfield  {journal}
  {\bibinfo  {journal} {The Astrophysical Journal}\ }\textbf {\bibinfo {volume}
  {525}},\ \bibinfo {pages} {L121--L124} (\bibinfo {year} {1999})}\BibitemShut
  {NoStop}%
\bibitem [{\citenamefont {{Paczynski}}(1998)}]{1998ApJ...494L..45P}%
  \BibitemOpen
  \bibfield  {author} {\bibinfo {author} {\bibfnamefont {B.}~\bibnamefont
  {{Paczynski}}},\ }\href {\doibase10.1086/311148} {\bibfield  {journal}
  {\bibinfo  {journal} {Astrop. J. Lett.}\ }\textbf {\bibinfo {volume} {494}},\
  \bibinfo {pages} {L45} (\bibinfo {year} {1998})},\ \Eprint
  {http://arxiv.org/abs/astro-ph/9710086} {astro-ph/9710086}\BibitemShut
  {NoStop}%
\bibitem [{\citenamefont {Kulkarni}(2005)}]{Kulkarni:2005jw}%
  \BibitemOpen
  \bibfield  {author} {\bibinfo {author} {\bibfnamefont {S.R.}\ \bibnamefont
  {Kulkarni}},\ }\href@noop {} {\  (\bibinfo {year} {2005})},\ \Eprint
  {http://arxiv.org/abs/astro-ph/0510256} {arXiv:astro-ph/0510256
  [astro-ph]}\BibitemShut {NoStop}%
\bibitem [{\citenamefont {Pian}\ \emph {et~al.}(2017)\citenamefont {Pian} \emph
  {et~al.}}]{Pian:2017gtc}%
  \BibitemOpen
  \bibfield  {author} {\bibinfo {author} {\bibfnamefont {E.}~\bibnamefont
  {Pian}} \emph {et~al.},\ }\href {\doibase10.1038/nature24298} {\bibfield
  {journal} {\bibinfo  {journal} {Nature}\ }\textbf {\bibinfo {volume} {551}},\
  \bibinfo {pages} {67--70} (\bibinfo {year} {2017})},\ \Eprint
  {http://arxiv.org/abs/1710.05858} {arXiv:1710.05858
  [astro-ph.HE]}\BibitemShut {NoStop}%
\bibitem [{\citenamefont {Tanvir}\ \emph {et~al.}(2017)\citenamefont {Tanvir}
  \emph {et~al.}}]{Tanvir:2017pws}%
  \BibitemOpen
  \bibfield  {author} {\bibinfo {author} {\bibfnamefont {N.~R.}\ \bibnamefont
  {Tanvir}} \emph {et~al.},\ }\href {\doibase10.3847/2041-8213/aa90b6}
  {\bibfield  {journal} {\bibinfo  {journal} {Astrophys. J.}\ }\textbf
  {\bibinfo {volume} {848}},\ \bibinfo {pages} {L27} (\bibinfo {year}
  {2017})},\ \Eprint {http://arxiv.org/abs/1710.05455} {arXiv:1710.05455
  [astro-ph.HE]}\BibitemShut {NoStop}%
\bibitem [{\citenamefont {Bernuzzi}\ \emph {et~al.}(2015)\citenamefont
  {Bernuzzi}, \citenamefont {Nagar}, \citenamefont {Dietrich},\ and\
  \citenamefont {Damour}}]{Bernuzzi:2014owa}%
  \BibitemOpen
  \bibfield  {author} {\bibinfo {author} {\bibfnamefont {Sebastiano}\
  \bibnamefont {Bernuzzi}}, \bibinfo {author} {\bibfnamefont {Alessandro}\
  \bibnamefont {Nagar}}, \bibinfo {author} {\bibfnamefont {Tim}\ \bibnamefont
  {Dietrich}}, \ and\ \bibinfo {author} {\bibfnamefont {Thibault}\ \bibnamefont
  {Damour}},\ }\href {\doibase10.1103/PhysRevLett.114.161103} {\bibfield
  {journal} {\bibinfo  {journal} {Phys. Rev. Lett.}\ }\textbf {\bibinfo
  {volume} {114}},\ \bibinfo {pages} {161103} (\bibinfo {year} {2015})},\
  \Eprint {http://arxiv.org/abs/1412.4553} {arXiv:1412.4553
  [gr-qc]}\BibitemShut {NoStop}%
\bibitem [{\citenamefont {{Kiuchi}}\ \emph {et~al.}(2017)\citenamefont
  {{Kiuchi}}, \citenamefont {{Kawaguchi}}, \citenamefont {{Kyutoku}},
  \citenamefont {{Sekiguchi}}, \citenamefont {{Shibata}},\ and\ \citenamefont
  {{Taniguchi}}}]{2017PhRvD..96h4060K}%
  \BibitemOpen
  \bibfield  {author} {\bibinfo {author} {\bibfnamefont {K.}~\bibnamefont
  {{Kiuchi}}}, \bibinfo {author} {\bibfnamefont {K.}~\bibnamefont
  {{Kawaguchi}}}, \bibinfo {author} {\bibfnamefont {K.}~\bibnamefont
  {{Kyutoku}}}, \bibinfo {author} {\bibfnamefont {Y.}~\bibnamefont
  {{Sekiguchi}}}, \bibinfo {author} {\bibfnamefont {M.}~\bibnamefont
  {{Shibata}}}, \ and\ \bibinfo {author} {\bibfnamefont {K.}~\bibnamefont
  {{Taniguchi}}},\ }\href {\doibase10.1103/PhysRevD.96.084060} {\bibfield
  {journal} {\bibinfo  {journal} {Physical Review D}\ }\textbf {\bibinfo
  {volume} {96}},\ \bibinfo {eid} {084060} (\bibinfo {year} {2017})},\ \Eprint
  {http://arxiv.org/abs/1708.08926} {arXiv:1708.08926
  [astro-ph.HE]}\BibitemShut {NoStop}%
\bibitem [{\citenamefont {Radice}\ \emph {et~al.}(2014)\citenamefont {Radice},
  \citenamefont {Rezzolla},\ and\ \citenamefont {Galeazzi}}]{Radice:2013hxh}%
  \BibitemOpen
  \bibfield  {author} {\bibinfo {author} {\bibfnamefont {David}\ \bibnamefont
  {Radice}}, \bibinfo {author} {\bibfnamefont {Luciano}\ \bibnamefont
  {Rezzolla}}, \ and\ \bibinfo {author} {\bibfnamefont {Filippo}\ \bibnamefont
  {Galeazzi}},\ }\href {\doibase10.1093/mnrasl/slt137} {\bibfield  {journal}
  {\bibinfo  {journal} {Mon. Not. Roy. Astron. Soc.}\ }\textbf {\bibinfo
  {volume} {437}},\ \bibinfo {pages} {L46--L50} (\bibinfo {year} {2014})},\
  \Eprint {http://arxiv.org/abs/1306.6052} {arXiv:1306.6052
  [gr-qc]}\BibitemShut {NoStop}%
\bibitem [{\citenamefont {Tichy}(2012)}]{Tichy:2012rp}%
  \BibitemOpen
  \bibfield  {author} {\bibinfo {author} {\bibfnamefont {Wolfgang}\
  \bibnamefont {Tichy}},\ }\href {\doibase10.1103/PhysRevD.86.064024}
  {\bibfield  {journal} {\bibinfo  {journal} {Phys. Rev.}\ }\textbf {\bibinfo
  {volume} {D86}},\ \bibinfo {pages} {064024} (\bibinfo {year} {2012})},\
  \Eprint {http://arxiv.org/abs/1209.5336} {arXiv:1209.5336
  [gr-qc]}\BibitemShut {NoStop}%
\bibitem [{\citenamefont {Bernuzzi}\ \emph {et~al.}(2014)\citenamefont
  {Bernuzzi}, \citenamefont {Dietrich}, \citenamefont {Tichy},\ and\
  \citenamefont {Brügmann}}]{Bernuzzi:2013rza}%
  \BibitemOpen
  \bibfield  {author} {\bibinfo {author} {\bibfnamefont {Sebastiano}\
  \bibnamefont {Bernuzzi}}, \bibinfo {author} {\bibfnamefont {Tim}\
  \bibnamefont {Dietrich}}, \bibinfo {author} {\bibfnamefont {Wolfgang}\
  \bibnamefont {Tichy}}, \ and\ \bibinfo {author} {\bibfnamefont {Bernd}\
  \bibnamefont {Brügmann}},\ }\href {\doibase10.1103/PhysRevD.89.104021}
  {\bibfield  {journal} {\bibinfo  {journal} {Phys. Rev.}\ }\textbf {\bibinfo
  {volume} {D89}},\ \bibinfo {pages} {104021} (\bibinfo {year} {2014})},\
  \Eprint {http://arxiv.org/abs/1311.4443} {arXiv:1311.4443
  [gr-qc]}\BibitemShut {NoStop}%
\bibitem [{\citenamefont {{Dietrich}}\ \emph {et~al.}(2015)\citenamefont
  {{Dietrich}}, \citenamefont {{Moldenhauer}}, \citenamefont
  {{Johnson-McDaniel}}, \citenamefont {{Bernuzzi}}, \citenamefont {{Markakis}},
  \citenamefont {{Br{\"u}gmann}},\ and\ \citenamefont
  {{Tichy}}}]{2015PhRvD..92l4007D}%
  \BibitemOpen
  \bibfield  {author} {\bibinfo {author} {\bibfnamefont {T.}~\bibnamefont
  {{Dietrich}}}, \bibinfo {author} {\bibfnamefont {N.}~\bibnamefont
  {{Moldenhauer}}}, \bibinfo {author} {\bibfnamefont {N.~K.}\ \bibnamefont
  {{Johnson-McDaniel}}}, \bibinfo {author} {\bibfnamefont {S.}~\bibnamefont
  {{Bernuzzi}}}, \bibinfo {author} {\bibfnamefont {C.~M.}\ \bibnamefont
  {{Markakis}}}, \bibinfo {author} {\bibfnamefont {B.}~\bibnamefont
  {{Br{\"u}gmann}}}, \ and\ \bibinfo {author} {\bibfnamefont {W.}~\bibnamefont
  {{Tichy}}},\ }\href {\doibase10.1103/PhysRevD.92.124007} {\bibfield
  {journal} {\bibinfo  {journal} {Physical Review D}\ }\textbf {\bibinfo
  {volume} {92}},\ \bibinfo {eid} {124007} (\bibinfo {year} {2015})},\ \Eprint
  {http://arxiv.org/abs/1507.07100} {arXiv:1507.07100 [gr-qc]}\BibitemShut
  {NoStop}%
\bibitem [{\citenamefont {{Dietrich}}\ \emph {et~al.}(2017)\citenamefont
  {{Dietrich}}, \citenamefont {{Bernuzzi}}, \citenamefont {{Ujevic}},\ and\
  \citenamefont {{Tichy}}}]{2017PhRvD..95d4045D}%
  \BibitemOpen
  \bibfield  {author} {\bibinfo {author} {\bibfnamefont {T.}~\bibnamefont
  {{Dietrich}}}, \bibinfo {author} {\bibfnamefont {S.}~\bibnamefont
  {{Bernuzzi}}}, \bibinfo {author} {\bibfnamefont {M.}~\bibnamefont
  {{Ujevic}}}, \ and\ \bibinfo {author} {\bibfnamefont {W.}~\bibnamefont
  {{Tichy}}},\ }\href {\doibase10.1103/PhysRevD.95.044045} {\bibfield
  {journal} {\bibinfo  {journal} {Physical Review D}\ }\textbf {\bibinfo
  {volume} {95}},\ \bibinfo {eid} {044045} (\bibinfo {year} {2017})},\ \Eprint
  {http://arxiv.org/abs/1611.07367} {arXiv:1611.07367 [gr-qc]}\BibitemShut
  {NoStop}%
\bibitem [{\citenamefont {{Hotokezaka}}\ \emph {et~al.}(2016)\citenamefont
  {{Hotokezaka}}, \citenamefont {{Kyutoku}}, \citenamefont {{Sekiguchi}},\ and\
  \citenamefont {{Shibata}}}]{2016PhRvD..93f4082H}%
  \BibitemOpen
  \bibfield  {author} {\bibinfo {author} {\bibfnamefont {K.}~\bibnamefont
  {{Hotokezaka}}}, \bibinfo {author} {\bibfnamefont {K.}~\bibnamefont
  {{Kyutoku}}}, \bibinfo {author} {\bibfnamefont {Y.-i.}\ \bibnamefont
  {{Sekiguchi}}}, \ and\ \bibinfo {author} {\bibfnamefont {M.}~\bibnamefont
  {{Shibata}}},\ }\href {\doibase10.1103/PhysRevD.93.064082} {\bibfield
  {journal} {\bibinfo  {journal} {Physical Review D}\ }\textbf {\bibinfo
  {volume} {93}},\ \bibinfo {eid} {064082} (\bibinfo {year} {2016})},\ \Eprint
  {http://arxiv.org/abs/1603.01286} {arXiv:1603.01286 [gr-qc]}\BibitemShut
  {NoStop}%
\bibitem [{\citenamefont {Hinderer}\ \emph {et~al.}(2016)\citenamefont
  {Hinderer} \emph {et~al.}}]{Hinderer:2016eia}%
  \BibitemOpen
  \bibfield  {author} {\bibinfo {author} {\bibfnamefont {Tanja}\ \bibnamefont
  {Hinderer}} \emph {et~al.},\ }\href {\doibase10.1103/PhysRevLett.116.181101}
  {\bibfield  {journal} {\bibinfo  {journal} {Phys. Rev. Lett.}\ }\textbf
  {\bibinfo {volume} {116}},\ \bibinfo {pages} {181101} (\bibinfo {year}
  {2016})},\ \Eprint {http://arxiv.org/abs/1602.00599} {arXiv:1602.00599
  [gr-qc]}\BibitemShut {NoStop}%
\bibitem [{\citenamefont {{Dietrich}}\ and\ \citenamefont
  {{Ujevic}}(2017)}]{2017CQGra..34j5014D}%
  \BibitemOpen
  \bibfield  {author} {\bibinfo {author} {\bibfnamefont {T.}~\bibnamefont
  {{Dietrich}}}\ and\ \bibinfo {author} {\bibfnamefont {M.}~\bibnamefont
  {{Ujevic}}},\ }\href {\doibase10.1088/1361-6382/aa6bb0} {\bibfield  {journal}
  {\bibinfo  {journal} {Classical and Quantum Gravity}\ }\textbf {\bibinfo
  {volume} {34}},\ \bibinfo {eid} {105014} (\bibinfo {year} {2017})},\ \Eprint
  {http://arxiv.org/abs/1612.03665} {arXiv:1612.03665 [gr-qc]}\BibitemShut
  {NoStop}%
\bibitem [{\citenamefont {{Hotokezaka}}\ \emph {et~al.}(2013)\citenamefont
  {{Hotokezaka}}, \citenamefont {{Kiuchi}}, \citenamefont {{Kyutoku}},
  \citenamefont {{Muranushi}}, \citenamefont {{Sekiguchi}}, \citenamefont
  {{Shibata}},\ and\ \citenamefont {{Taniguchi}}}]{2013PhRvD..88d4026H}%
  \BibitemOpen
  \bibfield  {author} {\bibinfo {author} {\bibfnamefont {K.}~\bibnamefont
  {{Hotokezaka}}}, \bibinfo {author} {\bibfnamefont {K.}~\bibnamefont
  {{Kiuchi}}}, \bibinfo {author} {\bibfnamefont {K.}~\bibnamefont {{Kyutoku}}},
  \bibinfo {author} {\bibfnamefont {T.}~\bibnamefont {{Muranushi}}}, \bibinfo
  {author} {\bibfnamefont {Y.-i.}\ \bibnamefont {{Sekiguchi}}}, \bibinfo
  {author} {\bibfnamefont {M.}~\bibnamefont {{Shibata}}}, \ and\ \bibinfo
  {author} {\bibfnamefont {K.}~\bibnamefont {{Taniguchi}}},\ }\href
  {\doibase10.1103/PhysRevD.88.044026} {\bibfield  {journal} {\bibinfo
  {journal} {Physical Review D}\ }\textbf {\bibinfo {volume} {88}},\ \bibinfo
  {eid} {044026} (\bibinfo {year} {2013})},\ \Eprint
  {http://arxiv.org/abs/1307.5888} {arXiv:1307.5888 [astro-ph.HE]}\BibitemShut
  {NoStop}%
\bibitem [{\citenamefont {Bauswein}\ \emph {et~al.}(2013)\citenamefont
  {Bauswein}, \citenamefont {Baumgarte},\ and\ \citenamefont
  {Janka}}]{Bauswein:2013jpa}%
  \BibitemOpen
  \bibfield  {author} {\bibinfo {author} {\bibfnamefont {A.}~\bibnamefont
  {Bauswein}}, \bibinfo {author} {\bibfnamefont {T.~W.}\ \bibnamefont
  {Baumgarte}}, \ and\ \bibinfo {author} {\bibfnamefont {H.~T.}\ \bibnamefont
  {Janka}},\ }\href {\doibase10.1103/PhysRevLett.111.131101} {\bibfield
  {journal} {\bibinfo  {journal} {Phys. Rev. Lett.}\ }\textbf {\bibinfo
  {volume} {111}},\ \bibinfo {pages} {131101} (\bibinfo {year} {2013})},\
  \Eprint {http://arxiv.org/abs/1307.5191} {arXiv:1307.5191
  [astro-ph.SR]}\BibitemShut {NoStop}%
\bibitem [{\citenamefont {Zappa}\ \emph {et~al.}(2018)\citenamefont {Zappa},
  \citenamefont {Bernuzzi}, \citenamefont {Radice}, \citenamefont {Perego},\
  and\ \citenamefont {Dietrich}}]{Zappa:2017xba}%
  \BibitemOpen
  \bibfield  {author} {\bibinfo {author} {\bibfnamefont {Francesco}\
  \bibnamefont {Zappa}}, \bibinfo {author} {\bibfnamefont {Sebastiano}\
  \bibnamefont {Bernuzzi}}, \bibinfo {author} {\bibfnamefont {David}\
  \bibnamefont {Radice}}, \bibinfo {author} {\bibfnamefont {Albino}\
  \bibnamefont {Perego}}, \ and\ \bibinfo {author} {\bibfnamefont {Tim}\
  \bibnamefont {Dietrich}},\ }\href {\doibase10.1103/PhysRevLett.120.111101}
  {\bibfield  {journal} {\bibinfo  {journal} {Phys. Rev. Lett.}\ }\textbf
  {\bibinfo {volume} {120}},\ \bibinfo {pages} {111101} (\bibinfo {year}
  {2018})},\ \Eprint {http://arxiv.org/abs/1712.04267} {arXiv:1712.04267
  [gr-qc]}\BibitemShut {NoStop}%
\bibitem [{\citenamefont {Bauswein}\ \emph {et~al.}(2016)\citenamefont
  {Bauswein}, \citenamefont {Stergioulas},\ and\ \citenamefont
  {Janka}}]{Bauswein:2015vxa}%
  \BibitemOpen
  \bibfield  {author} {\bibinfo {author} {\bibfnamefont {Andreas}\ \bibnamefont
  {Bauswein}}, \bibinfo {author} {\bibfnamefont {Nikolaos}\ \bibnamefont
  {Stergioulas}}, \ and\ \bibinfo {author} {\bibfnamefont {Hans-Thomas}\
  \bibnamefont {Janka}},\ }\href {\doibase10.1140/epja/i2016-16056-7}
  {\bibfield  {journal} {\bibinfo  {journal} {Eur. Phys. J.}\ }\textbf
  {\bibinfo {volume} {A52}},\ \bibinfo {pages} {56} (\bibinfo {year} {2016})},\
  \Eprint {http://arxiv.org/abs/1508.05493} {arXiv:1508.05493
  [astro-ph.HE]}\BibitemShut {NoStop}%
\bibitem [{\citenamefont {Rezzolla}\ and\ \citenamefont
  {Takami}(2016)}]{Rezzolla:2016nxn}%
  \BibitemOpen
  \bibfield  {author} {\bibinfo {author} {\bibfnamefont {Luciano}\ \bibnamefont
  {Rezzolla}}\ and\ \bibinfo {author} {\bibfnamefont {Kentaro}\ \bibnamefont
  {Takami}},\ }\href {\doibase10.1103/PhysRevD.93.124051} {\bibfield  {journal}
  {\bibinfo  {journal} {Phys. Rev.}\ }\textbf {\bibinfo {volume} {D93}},\
  \bibinfo {pages} {124051} (\bibinfo {year} {2016})},\ \Eprint
  {http://arxiv.org/abs/1604.00246} {arXiv:1604.00246 [gr-qc]}\BibitemShut
  {NoStop}%
\bibitem [{\citenamefont {{Ciolfi}}\ \emph {et~al.}(2017)\citenamefont
  {{Ciolfi}}, \citenamefont {{Kastaun}}, \citenamefont {{Giacomazzo}},
  \citenamefont {{Endrizzi}}, \citenamefont {{Siegel}},\ and\ \citenamefont
  {{Perna}}}]{2017PhRvD..95f3016C}%
  \BibitemOpen
  \bibfield  {author} {\bibinfo {author} {\bibfnamefont {R.}~\bibnamefont
  {{Ciolfi}}}, \bibinfo {author} {\bibfnamefont {W.}~\bibnamefont {{Kastaun}}},
  \bibinfo {author} {\bibfnamefont {B.}~\bibnamefont {{Giacomazzo}}}, \bibinfo
  {author} {\bibfnamefont {A.}~\bibnamefont {{Endrizzi}}}, \bibinfo {author}
  {\bibfnamefont {D.~M.}\ \bibnamefont {{Siegel}}}, \ and\ \bibinfo {author}
  {\bibfnamefont {R.}~\bibnamefont {{Perna}}},\ }\href
  {\doibase10.1103/PhysRevD.95.063016} {\bibfield  {journal} {\bibinfo
  {journal} {Physical Review D}\ }\textbf {\bibinfo {volume} {95}},\ \bibinfo
  {eid} {063016} (\bibinfo {year} {2017})},\ \Eprint
  {http://arxiv.org/abs/1701.08738} {arXiv:1701.08738
  [astro-ph.HE]}\BibitemShut {NoStop}%
\bibitem [{\citenamefont {Paschalidis}\ \emph {et~al.}(2015)\citenamefont
  {Paschalidis}, \citenamefont {East}, \citenamefont {Pretorius},\ and\
  \citenamefont {Shapiro}}]{Paschalidis:2015mla}%
  \BibitemOpen
  \bibfield  {author} {\bibinfo {author} {\bibfnamefont {Vasileios}\
  \bibnamefont {Paschalidis}}, \bibinfo {author} {\bibfnamefont {William~E.}\
  \bibnamefont {East}}, \bibinfo {author} {\bibfnamefont {Frans}\ \bibnamefont
  {Pretorius}}, \ and\ \bibinfo {author} {\bibfnamefont {Stuart~L.}\
  \bibnamefont {Shapiro}},\ }\href {\doibase10.1103/PhysRevD.92.121502}
  {\bibfield  {journal} {\bibinfo  {journal} {Phys. Rev.}\ }\textbf {\bibinfo
  {volume} {D92}},\ \bibinfo {pages} {121502} (\bibinfo {year} {2015})},\
  \Eprint {http://arxiv.org/abs/1510.03432} {arXiv:1510.03432
  [astro-ph.HE]}\BibitemShut {NoStop}%
\bibitem [{\citenamefont {Radice}\ \emph
  {et~al.}(2016{\natexlab{a}})\citenamefont {Radice}, \citenamefont
  {Bernuzzi},\ and\ \citenamefont {Ott}}]{Radice:2016gym}%
  \BibitemOpen
  \bibfield  {author} {\bibinfo {author} {\bibfnamefont {David}\ \bibnamefont
  {Radice}}, \bibinfo {author} {\bibfnamefont {Sebastiano}\ \bibnamefont
  {Bernuzzi}}, \ and\ \bibinfo {author} {\bibfnamefont {Christian~D.}\
  \bibnamefont {Ott}},\ }\href {\doibase10.1103/PhysRevD.94.064011} {\bibfield
  {journal} {\bibinfo  {journal} {Phys. Rev.}\ }\textbf {\bibinfo {volume}
  {D94}},\ \bibinfo {pages} {064011} (\bibinfo {year} {2016}{\natexlab{a}})},\
  \Eprint {http://arxiv.org/abs/1603.05726} {arXiv:1603.05726
  [gr-qc]}\BibitemShut {NoStop}%
\bibitem [{\citenamefont {Lehner}\ \emph {et~al.}(2016)\citenamefont {Lehner},
  \citenamefont {Liebling}, \citenamefont {Palenzuela},\ and\ \citenamefont
  {Motl}}]{Lehner:2016wjg}%
  \BibitemOpen
  \bibfield  {author} {\bibinfo {author} {\bibfnamefont {Luis}\ \bibnamefont
  {Lehner}}, \bibinfo {author} {\bibfnamefont {Steven~L.}\ \bibnamefont
  {Liebling}}, \bibinfo {author} {\bibfnamefont {Carlos}\ \bibnamefont
  {Palenzuela}}, \ and\ \bibinfo {author} {\bibfnamefont {Patrick~M.}\
  \bibnamefont {Motl}},\ }\href {\doibase10.1103/PhysRevD.94.043003} {\bibfield
   {journal} {\bibinfo  {journal} {Phys. Rev.}\ }\textbf {\bibinfo {volume}
  {D94}},\ \bibinfo {pages} {043003} (\bibinfo {year} {2016})},\ \Eprint
  {http://arxiv.org/abs/1605.02369} {arXiv:1605.02369 [gr-qc]}\BibitemShut
  {NoStop}%
\bibitem [{\citenamefont {De~Pietri}\ \emph {et~al.}(2018)\citenamefont
  {De~Pietri}, \citenamefont {Feo}, \citenamefont {Font}, \citenamefont
  {Löffler}, \citenamefont {Maione}, \citenamefont {Pasquali},\ and\
  \citenamefont {Stergioulas}}]{DePietri:2018tpx}%
  \BibitemOpen
  \bibfield  {author} {\bibinfo {author} {\bibfnamefont {Roberto}\ \bibnamefont
  {De~Pietri}}, \bibinfo {author} {\bibfnamefont {Alessandra}\ \bibnamefont
  {Feo}}, \bibinfo {author} {\bibfnamefont {José~A.}\ \bibnamefont {Font}},
  \bibinfo {author} {\bibfnamefont {Frank}\ \bibnamefont {Löffler}}, \bibinfo
  {author} {\bibfnamefont {Francesco}\ \bibnamefont {Maione}}, \bibinfo
  {author} {\bibfnamefont {Michele}\ \bibnamefont {Pasquali}}, \ and\ \bibinfo
  {author} {\bibfnamefont {Nikolaos}\ \bibnamefont {Stergioulas}},\ }\href
  {\doibase10.1103/PhysRevLett.120.221101} {\bibfield  {journal} {\bibinfo
  {journal} {Phys. Rev. Lett.}\ }\textbf {\bibinfo {volume} {120}},\ \bibinfo
  {pages} {221101} (\bibinfo {year} {2018})},\ \Eprint
  {http://arxiv.org/abs/1802.03288} {arXiv:1802.03288 [gr-qc]}\BibitemShut
  {NoStop}%
\bibitem [{\citenamefont {Abbott}\ \emph
  {et~al.}(2017{\natexlab{k}})\citenamefont {Abbott} \emph
  {et~al.}}]{Abbott:2017dke}%
  \BibitemOpen
  \bibfield  {author} {\bibinfo {author} {\bibfnamefont {B.~P.}\ \bibnamefont
  {Abbott}} \emph {et~al.} (\bibinfo {collaboration} {Virgo, LIGO
  Scientific}),\ }\href {\doibase10.3847/2041-8213/aa9a35} {\bibfield
  {journal} {\bibinfo  {journal} {Astrophys. J.}\ }\textbf {\bibinfo {volume}
  {851}},\ \bibinfo {pages} {L16} (\bibinfo {year} {2017}{\natexlab{k}})},\
  \Eprint {http://arxiv.org/abs/1710.09320} {arXiv:1710.09320
  [astro-ph.HE]}\BibitemShut {NoStop}%
\bibitem [{\citenamefont {{Korobkin}}\ \emph {et~al.}(2012)\citenamefont
  {{Korobkin}}, \citenamefont {{Rosswog}}, \citenamefont {{Arcones}},\ and\
  \citenamefont {{Winteler}}}]{2012MNRAS.426.1940K}%
  \BibitemOpen
  \bibfield  {author} {\bibinfo {author} {\bibfnamefont {O.}~\bibnamefont
  {{Korobkin}}}, \bibinfo {author} {\bibfnamefont {S.}~\bibnamefont
  {{Rosswog}}}, \bibinfo {author} {\bibfnamefont {A.}~\bibnamefont
  {{Arcones}}}, \ and\ \bibinfo {author} {\bibfnamefont {C.}~\bibnamefont
  {{Winteler}}},\ }\href {\doibase10.1111/j.1365-2966.2012.21859.x} {\bibfield
  {journal} {\bibinfo  {journal} {MNRAS}\ }\textbf {\bibinfo {volume} {426}},\
  \bibinfo {pages} {1940--1949} (\bibinfo {year} {2012})},\ \Eprint
  {http://arxiv.org/abs/1206.2379} {arXiv:1206.2379 [astro-ph.SR]}\BibitemShut
  {NoStop}%
\bibitem [{\citenamefont {Sekiguchi}\ \emph {et~al.}(2015)\citenamefont
  {Sekiguchi}, \citenamefont {Kiuchi}, \citenamefont {Kyutoku},\ and\
  \citenamefont {Shibata}}]{Sekiguchi:2015dma}%
  \BibitemOpen
  \bibfield  {author} {\bibinfo {author} {\bibfnamefont {Yuichiro}\
  \bibnamefont {Sekiguchi}}, \bibinfo {author} {\bibfnamefont {Kenta}\
  \bibnamefont {Kiuchi}}, \bibinfo {author} {\bibfnamefont {Koutarou}\
  \bibnamefont {Kyutoku}}, \ and\ \bibinfo {author} {\bibfnamefont {Masaru}\
  \bibnamefont {Shibata}},\ }\href {\doibase10.1103/PhysRevD.91.064059}
  {\bibfield  {journal} {\bibinfo  {journal} {Phys. Rev.}\ }\textbf {\bibinfo
  {volume} {D91}},\ \bibinfo {pages} {064059} (\bibinfo {year} {2015})},\
  \Eprint {http://arxiv.org/abs/1502.06660} {arXiv:1502.06660
  [astro-ph.HE]}\BibitemShut {NoStop}%
\bibitem [{\citenamefont {Foucart}\ \emph {et~al.}(2016)\citenamefont
  {Foucart}, \citenamefont {Haas}, \citenamefont {Duez}, \citenamefont
  {O'Connor}, \citenamefont {Ott}, \citenamefont {Roberts}, \citenamefont
  {Kidder}, \citenamefont {Lippuner}, \citenamefont {Pfeiffer},\ and\
  \citenamefont {Scheel}}]{Foucart:2015gaa}%
  \BibitemOpen
  \bibfield  {author} {\bibinfo {author} {\bibfnamefont {Francois}\
  \bibnamefont {Foucart}}, \bibinfo {author} {\bibfnamefont {Roland}\
  \bibnamefont {Haas}}, \bibinfo {author} {\bibfnamefont {Matthew~D.}\
  \bibnamefont {Duez}}, \bibinfo {author} {\bibfnamefont {Evan}\ \bibnamefont
  {O'Connor}}, \bibinfo {author} {\bibfnamefont {Christian~D.}\ \bibnamefont
  {Ott}}, \bibinfo {author} {\bibfnamefont {Luke}\ \bibnamefont {Roberts}},
  \bibinfo {author} {\bibfnamefont {Lawrence~E.}\ \bibnamefont {Kidder}},
  \bibinfo {author} {\bibfnamefont {Jonas}\ \bibnamefont {Lippuner}}, \bibinfo
  {author} {\bibfnamefont {Harald~P.}\ \bibnamefont {Pfeiffer}}, \ and\
  \bibinfo {author} {\bibfnamefont {Mark~A.}\ \bibnamefont {Scheel}},\ }\href
  {\doibase10.1103/PhysRevD.93.044019} {\bibfield  {journal} {\bibinfo
  {journal} {Phys. Rev.}\ }\textbf {\bibinfo {volume} {D93}},\ \bibinfo {pages}
  {044019} (\bibinfo {year} {2016})},\ \Eprint
  {http://arxiv.org/abs/1510.06398} {arXiv:1510.06398
  [astro-ph.HE]}\BibitemShut {NoStop}%
\bibitem [{\citenamefont {Radice}\ \emph
  {et~al.}(2016{\natexlab{b}})\citenamefont {Radice}, \citenamefont {Galeazzi},
  \citenamefont {Lippuner}, \citenamefont {Roberts}, \citenamefont {Ott},\ and\
  \citenamefont {Rezzolla}}]{Radice:2016dwd}%
  \BibitemOpen
  \bibfield  {author} {\bibinfo {author} {\bibfnamefont {David}\ \bibnamefont
  {Radice}}, \bibinfo {author} {\bibfnamefont {Filippo}\ \bibnamefont
  {Galeazzi}}, \bibinfo {author} {\bibfnamefont {Jonas}\ \bibnamefont
  {Lippuner}}, \bibinfo {author} {\bibfnamefont {Luke~F.}\ \bibnamefont
  {Roberts}}, \bibinfo {author} {\bibfnamefont {Christian~D.}\ \bibnamefont
  {Ott}}, \ and\ \bibinfo {author} {\bibfnamefont {Luciano}\ \bibnamefont
  {Rezzolla}},\ }\href {\doibase10.1093/mnras/stw1227} {\bibfield  {journal}
  {\bibinfo  {journal} {Mon. Not. Roy. Astron. Soc.}\ }\textbf {\bibinfo
  {volume} {460}},\ \bibinfo {pages} {3255--3271} (\bibinfo {year}
  {2016}{\natexlab{b}})},\ \Eprint {http://arxiv.org/abs/1601.02426}
  {arXiv:1601.02426 [astro-ph.HE]}\BibitemShut {NoStop}%
\bibitem [{\citenamefont {{Bovard}}\ \emph {et~al.}(2017)\citenamefont
  {{Bovard}}, \citenamefont {{Martin}}, \citenamefont {{Guercilena}},
  \citenamefont {{Arcones}}, \citenamefont {{Rezzolla}},\ and\ \citenamefont
  {{Korobkin}}}]{2017PhRvD..96l4005B}%
  \BibitemOpen
  \bibfield  {author} {\bibinfo {author} {\bibfnamefont {L.}~\bibnamefont
  {{Bovard}}}, \bibinfo {author} {\bibfnamefont {D.}~\bibnamefont {{Martin}}},
  \bibinfo {author} {\bibfnamefont {F.}~\bibnamefont {{Guercilena}}}, \bibinfo
  {author} {\bibfnamefont {A.}~\bibnamefont {{Arcones}}}, \bibinfo {author}
  {\bibfnamefont {L.}~\bibnamefont {{Rezzolla}}}, \ and\ \bibinfo {author}
  {\bibfnamefont {O.}~\bibnamefont {{Korobkin}}},\ }\href
  {\doibase10.1103/PhysRevD.96.124005} {\bibfield  {journal} {\bibinfo
  {journal} {Physical Review D}\ }\textbf {\bibinfo {volume} {96}},\ \bibinfo
  {eid} {124005} (\bibinfo {year} {2017})},\ \Eprint
  {http://arxiv.org/abs/1709.09630} {arXiv:1709.09630 [gr-qc]}\BibitemShut
  {NoStop}%
\bibitem [{\citenamefont {Wanajo}\ \emph {et~al.}(2014)\citenamefont {Wanajo},
  \citenamefont {Sekiguchi}, \citenamefont {Nishimura}, \citenamefont {Kiuchi},
  \citenamefont {Kyutoku},\ and\ \citenamefont {Shibata}}]{Wanajo:2014wha}%
  \BibitemOpen
  \bibfield  {author} {\bibinfo {author} {\bibfnamefont {Shinya}\ \bibnamefont
  {Wanajo}}, \bibinfo {author} {\bibfnamefont {Yuichiro}\ \bibnamefont
  {Sekiguchi}}, \bibinfo {author} {\bibfnamefont {Nobuya}\ \bibnamefont
  {Nishimura}}, \bibinfo {author} {\bibfnamefont {Kenta}\ \bibnamefont
  {Kiuchi}}, \bibinfo {author} {\bibfnamefont {Koutarou}\ \bibnamefont
  {Kyutoku}}, \ and\ \bibinfo {author} {\bibfnamefont {Masaru}\ \bibnamefont
  {Shibata}},\ }\href {\doibase10.1088/2041-8205/789/2/L39} {\bibfield
  {journal} {\bibinfo  {journal} {Astrophys. J.}\ }\textbf {\bibinfo {volume}
  {789}},\ \bibinfo {pages} {L39} (\bibinfo {year} {2014})},\ \Eprint
  {http://arxiv.org/abs/1402.7317} {arXiv:1402.7317 [astro-ph.SR]}\BibitemShut
  {NoStop}%
\bibitem [{\citenamefont {{Goriely}}\ \emph {et~al.}(2015)\citenamefont
  {{Goriely}}, \citenamefont {{Bauswein}}, \citenamefont {{Just}},
  \citenamefont {{Pllumbi}},\ and\ \citenamefont {{Janka}}}]{Goriely:2015}%
  \BibitemOpen
  \bibfield  {author} {\bibinfo {author} {\bibfnamefont {S.}~\bibnamefont
  {{Goriely}}}, \bibinfo {author} {\bibfnamefont {A.}~\bibnamefont
  {{Bauswein}}}, \bibinfo {author} {\bibfnamefont {O.}~\bibnamefont {{Just}}},
  \bibinfo {author} {\bibfnamefont {E.}~\bibnamefont {{Pllumbi}}}, \ and\
  \bibinfo {author} {\bibfnamefont {H.-T.}\ \bibnamefont {{Janka}}},\ }\href
  {\doibase10.1093/mnras/stv1526} {\bibfield  {journal} {\bibinfo  {journal}
  {MNRAS}\ }\textbf {\bibinfo {volume} {452}},\ \bibinfo {pages} {3894--3904}
  (\bibinfo {year} {2015})},\ \Eprint {http://arxiv.org/abs/1504.04377}
  {arXiv:1504.04377 [astro-ph.SR]}\BibitemShut {NoStop}%
\bibitem [{\citenamefont {{Martin}}\ \emph {et~al.}(2018)\citenamefont
  {{Martin}}, \citenamefont {{Perego}}, \citenamefont {{Kastaun}},\ and\
  \citenamefont {{Arcones}}}]{2018CQGra..35c4001M}%
  \BibitemOpen
  \bibfield  {author} {\bibinfo {author} {\bibfnamefont {D.}~\bibnamefont
  {{Martin}}}, \bibinfo {author} {\bibfnamefont {A.}~\bibnamefont {{Perego}}},
  \bibinfo {author} {\bibfnamefont {W.}~\bibnamefont {{Kastaun}}}, \ and\
  \bibinfo {author} {\bibfnamefont {A.}~\bibnamefont {{Arcones}}},\ }\href
  {\doibase10.1088/1361-6382/aa9f5a} {\bibfield  {journal} {\bibinfo  {journal}
  {Classical and Quantum Gravity}\ }\textbf {\bibinfo {volume} {35}},\ \bibinfo
  {eid} {034001} (\bibinfo {year} {2018})},\ \Eprint
  {http://arxiv.org/abs/1710.04900} {arXiv:1710.04900
  [astro-ph.HE]}\BibitemShut {NoStop}%
\bibitem [{\citenamefont {Perego}\ \emph {et~al.}(2014)\citenamefont {Perego},
  \citenamefont {Rosswog}, \citenamefont {Cabezón}, \citenamefont {Korobkin},
  \citenamefont {Käppeli}, \citenamefont {Arcones},\ and\ \citenamefont
  {Liebendörfer}}]{Perego:2014fma}%
  \BibitemOpen
  \bibfield  {author} {\bibinfo {author} {\bibfnamefont {A.}~\bibnamefont
  {Perego}}, \bibinfo {author} {\bibfnamefont {Stephan}\ \bibnamefont
  {Rosswog}}, \bibinfo {author} {\bibfnamefont {Ruben~M.}\ \bibnamefont
  {Cabezón}}, \bibinfo {author} {\bibfnamefont {Oleg}\ \bibnamefont
  {Korobkin}}, \bibinfo {author} {\bibfnamefont {Roger}\ \bibnamefont
  {Käppeli}}, \bibinfo {author} {\bibfnamefont {Almudena}\ \bibnamefont
  {Arcones}}, \ and\ \bibinfo {author} {\bibfnamefont {Matthias}\ \bibnamefont
  {Liebendörfer}},\ }\href {\doibase10.1093/mnras/stu1352} {\bibfield
  {journal} {\bibinfo  {journal} {Mon. Not. Roy. Astron. Soc.}\ }\textbf
  {\bibinfo {volume} {443}},\ \bibinfo {pages} {3134--3156} (\bibinfo {year}
  {2014})},\ \Eprint {http://arxiv.org/abs/1405.6730} {arXiv:1405.6730
  [astro-ph.HE]}\BibitemShut {NoStop}%
\bibitem [{\citenamefont {{Fujibayashi}}\ \emph {et~al.}(2017)\citenamefont
  {{Fujibayashi}}, \citenamefont {{Sekiguchi}}, \citenamefont {{Kiuchi}},\ and\
  \citenamefont {{Shibata}}}]{2017ApJ...846..114F}%
  \BibitemOpen
  \bibfield  {author} {\bibinfo {author} {\bibfnamefont {S.}~\bibnamefont
  {{Fujibayashi}}}, \bibinfo {author} {\bibfnamefont {Y.}~\bibnamefont
  {{Sekiguchi}}}, \bibinfo {author} {\bibfnamefont {K.}~\bibnamefont
  {{Kiuchi}}}, \ and\ \bibinfo {author} {\bibfnamefont {M.}~\bibnamefont
  {{Shibata}}},\ }\href {\doibase10.3847/1538-4357/aa8039} {\bibfield
  {journal} {\bibinfo  {journal} {The Astrophysical Journal}\ }\textbf
  {\bibinfo {volume} {846}},\ \bibinfo {eid} {114} (\bibinfo {year}
  {2017})}\BibitemShut {NoStop}%
\bibitem [{\citenamefont {{Fern{\'a}ndez}}\ and\ \citenamefont
  {{Metzger}}(2013)}]{2013MNRAS.435..502F}%
  \BibitemOpen
  \bibfield  {author} {\bibinfo {author} {\bibfnamefont {R.}~\bibnamefont
  {{Fern{\'a}ndez}}}\ and\ \bibinfo {author} {\bibfnamefont {B.~D.}\
  \bibnamefont {{Metzger}}},\ }\href {\doibase10.1093/mnras/stt1312} {\bibfield
   {journal} {\bibinfo  {journal} {MNRAS}\ }\textbf {\bibinfo {volume} {435}},\
  \bibinfo {pages} {502--517} (\bibinfo {year} {2013})},\ \Eprint
  {http://arxiv.org/abs/1304.6720} {arXiv:1304.6720 [astro-ph.HE]}\BibitemShut
  {NoStop}%
\bibitem [{\citenamefont {Just}\ \emph {et~al.}(2015)\citenamefont {Just},
  \citenamefont {Bauswein}, \citenamefont {Ardevol~Pulpillo}, \citenamefont
  {Goriely},\ and\ \citenamefont {Janka}}]{Just:2015ypa}%
  \BibitemOpen
  \bibfield  {author} {\bibinfo {author} {\bibfnamefont {Oliver}\ \bibnamefont
  {Just}}, \bibinfo {author} {\bibfnamefont {Andreas}\ \bibnamefont
  {Bauswein}}, \bibinfo {author} {\bibfnamefont {Ricard}\ \bibnamefont
  {Ardevol~Pulpillo}}, \bibinfo {author} {\bibfnamefont {Stephane}\
  \bibnamefont {Goriely}}, \ and\ \bibinfo {author} {\bibfnamefont {H.~Thomas}\
  \bibnamefont {Janka}},\ }\bibfield  {booktitle} {\emph {\bibinfo {booktitle}
  {{Proceedings, 13th International Symposium on Nuclei in the Cosmos (NIC
  XIII): Debrecen, Hungary, July 7-11, 2014}}},\ }\href
  {\doibase10.22323/1.204.0103} {\bibfield  {journal} {\bibinfo  {journal}
  {PoS}\ }\textbf {\bibinfo {volume} {NICXIII}},\ \bibinfo {pages} {103}
  (\bibinfo {year} {2015})},\ \Eprint {http://arxiv.org/abs/1504.05448}
  {arXiv:1504.05448 [astro-ph.SR]}\BibitemShut {NoStop}%
\bibitem [{\citenamefont {Siegel}\ and\ \citenamefont
  {Metzger}(2018)}]{Siegel:2017jug}%
  \BibitemOpen
  \bibfield  {author} {\bibinfo {author} {\bibfnamefont {Daniel~M.}\
  \bibnamefont {Siegel}}\ and\ \bibinfo {author} {\bibfnamefont {Brian~D.}\
  \bibnamefont {Metzger}},\ }\href {\doibase10.3847/1538-4357/aabaec}
  {\bibfield  {journal} {\bibinfo  {journal} {Astrophys. J.}\ }\textbf
  {\bibinfo {volume} {858}},\ \bibinfo {pages} {52} (\bibinfo {year} {2018})},\
  \Eprint {http://arxiv.org/abs/1711.00868} {arXiv:1711.00868
  [astro-ph.HE]}\BibitemShut {NoStop}%
\bibitem [{\citenamefont {Fujibayashi}\ \emph {et~al.}(2018)\citenamefont
  {Fujibayashi}, \citenamefont {Kiuchi}, \citenamefont {Nishimura},
  \citenamefont {Sekiguchi},\ and\ \citenamefont
  {Shibata}}]{Fujibayashi:2017puw}%
  \BibitemOpen
  \bibfield  {author} {\bibinfo {author} {\bibfnamefont {Sho}\ \bibnamefont
  {Fujibayashi}}, \bibinfo {author} {\bibfnamefont {Kenta}\ \bibnamefont
  {Kiuchi}}, \bibinfo {author} {\bibfnamefont {Nobuya}\ \bibnamefont
  {Nishimura}}, \bibinfo {author} {\bibfnamefont {Yuichiro}\ \bibnamefont
  {Sekiguchi}}, \ and\ \bibinfo {author} {\bibfnamefont {Masaru}\ \bibnamefont
  {Shibata}},\ }\href {\doibase10.3847/1538-4357/aabafd} {\bibfield  {journal}
  {\bibinfo  {journal} {Astrophys. J.}\ }\textbf {\bibinfo {volume} {860}},\
  \bibinfo {pages} {64} (\bibinfo {year} {2018})},\ \Eprint
  {http://arxiv.org/abs/1711.02093} {arXiv:1711.02093
  [astro-ph.HE]}\BibitemShut {NoStop}%
\bibitem [{\citenamefont {Siegel}\ \emph {et~al.}(2014)\citenamefont {Siegel},
  \citenamefont {Ciolfi},\ and\ \citenamefont {Rezzolla}}]{Siegel:2014ita}%
  \BibitemOpen
  \bibfield  {author} {\bibinfo {author} {\bibfnamefont {Daniel~M.}\
  \bibnamefont {Siegel}}, \bibinfo {author} {\bibfnamefont {Riccardo}\
  \bibnamefont {Ciolfi}}, \ and\ \bibinfo {author} {\bibfnamefont {Luciano}\
  \bibnamefont {Rezzolla}},\ }\href {\doibase10.1088/2041-8205/785/1/L6}
  {\bibfield  {journal} {\bibinfo  {journal} {Astrophys. J.}\ }\textbf
  {\bibinfo {volume} {785}},\ \bibinfo {pages} {L6} (\bibinfo {year} {2014})},\
  \Eprint {http://arxiv.org/abs/1401.4544} {arXiv:1401.4544
  [astro-ph.HE]}\BibitemShut {NoStop}%
\bibitem [{\citenamefont {{Lippuner}}\ \emph {et~al.}(2017)\citenamefont
  {{Lippuner}}, \citenamefont {{Fern{\'a}ndez}}, \citenamefont {{Roberts}},
  \citenamefont {{Foucart}}, \citenamefont {{Kasen}}, \citenamefont
  {{Metzger}},\ and\ \citenamefont {{Ott}}}]{2017MNRAS.472..904L}%
  \BibitemOpen
  \bibfield  {author} {\bibinfo {author} {\bibfnamefont {J.}~\bibnamefont
  {{Lippuner}}}, \bibinfo {author} {\bibfnamefont {R.}~\bibnamefont
  {{Fern{\'a}ndez}}}, \bibinfo {author} {\bibfnamefont {L.~F.}\ \bibnamefont
  {{Roberts}}}, \bibinfo {author} {\bibfnamefont {F.}~\bibnamefont
  {{Foucart}}}, \bibinfo {author} {\bibfnamefont {D.}~\bibnamefont {{Kasen}}},
  \bibinfo {author} {\bibfnamefont {B.~D.}\ \bibnamefont {{Metzger}}}, \ and\
  \bibinfo {author} {\bibfnamefont {C.~D.}\ \bibnamefont {{Ott}}},\ }\href
  {\doibase10.1093/mnras/stx1987} {\bibfield  {journal} {\bibinfo  {journal}
  {MNRAS}\ }\textbf {\bibinfo {volume} {472}},\ \bibinfo {pages} {904--918}
  (\bibinfo {year} {2017})},\ \Eprint {http://arxiv.org/abs/1703.06216}
  {arXiv:1703.06216 [astro-ph.HE]}\BibitemShut {NoStop}%
\bibitem [{\citenamefont {{Wu}}\ \emph {et~al.}(2017)\citenamefont {{Wu}},
  \citenamefont {{Tamborra}}, \citenamefont {{Just}},\ and\ \citenamefont
  {{Janka}}}]{2017PhRvD..96l3015W}%
  \BibitemOpen
  \bibfield  {author} {\bibinfo {author} {\bibfnamefont {M.-R.}\ \bibnamefont
  {{Wu}}}, \bibinfo {author} {\bibfnamefont {I.}~\bibnamefont {{Tamborra}}},
  \bibinfo {author} {\bibfnamefont {O.}~\bibnamefont {{Just}}}, \ and\ \bibinfo
  {author} {\bibfnamefont {H.-T.}\ \bibnamefont {{Janka}}},\ }\href
  {\doibase10.1103/PhysRevD.96.123015} {\bibfield  {journal} {\bibinfo
  {journal} {Physical Review D}\ }\textbf {\bibinfo {volume} {96}},\ \bibinfo
  {eid} {123015} (\bibinfo {year} {2017})},\ \Eprint
  {http://arxiv.org/abs/1711.00477} {arXiv:1711.00477
  [astro-ph.HE]}\BibitemShut {NoStop}%
\bibitem [{\citenamefont {{Fern{\'a}ndez}}\ and\ \citenamefont
  {{Metzger}}(2016)}]{2016ARNPS..66...23F}%
  \BibitemOpen
  \bibfield  {author} {\bibinfo {author} {\bibfnamefont {R.}~\bibnamefont
  {{Fern{\'a}ndez}}}\ and\ \bibinfo {author} {\bibfnamefont {B.~D.}\
  \bibnamefont {{Metzger}}},\ }\href
  {\doibase10.1146/annurev-nucl-102115-044819} {\bibfield  {journal} {\bibinfo
  {journal} {Annual Review of Nuclear and Particle Science}\ }\textbf {\bibinfo
  {volume} {66}},\ \bibinfo {pages} {23--45} (\bibinfo {year} {2016})},\
  \Eprint {http://arxiv.org/abs/1512.05435} {arXiv:1512.05435
  [astro-ph.HE]}\BibitemShut {NoStop}%
\bibitem [{\citenamefont {Metzger}(2017)}]{Metzger:2016pju}%
  \BibitemOpen
  \bibfield  {author} {\bibinfo {author} {\bibfnamefont {Brian~D.}\
  \bibnamefont {Metzger}},\ }\href {\doibase10.1007/s41114-017-0006-z}
  {\bibfield  {journal} {\bibinfo  {journal} {Living Rev. Rel.}\ }\textbf
  {\bibinfo {volume} {20}},\ \bibinfo {pages} {3} (\bibinfo {year} {2017})},\
  \Eprint {http://arxiv.org/abs/1610.09381} {arXiv:1610.09381
  [astro-ph.HE]}\BibitemShut {NoStop}%
\bibitem [{\citenamefont {Barnes}\ and\ \citenamefont
  {Kasen}(2013)}]{Barnes:2013wka}%
  \BibitemOpen
  \bibfield  {author} {\bibinfo {author} {\bibfnamefont {Jennifer}\
  \bibnamefont {Barnes}}\ and\ \bibinfo {author} {\bibfnamefont {Daniel}\
  \bibnamefont {Kasen}},\ }\href {\doibase10.1088/0004-637X/775/1/18}
  {\bibfield  {journal} {\bibinfo  {journal} {Astrophys. J.}\ }\textbf
  {\bibinfo {volume} {775}},\ \bibinfo {pages} {18} (\bibinfo {year} {2013})},\
  \Eprint {http://arxiv.org/abs/1303.5787} {arXiv:1303.5787
  [astro-ph.HE]}\BibitemShut {NoStop}%
\bibitem [{\citenamefont {Tanaka}\ and\ \citenamefont
  {Hotokezaka}(2013)}]{Tanaka:2013ana}%
  \BibitemOpen
  \bibfield  {author} {\bibinfo {author} {\bibfnamefont {Masaomi}\ \bibnamefont
  {Tanaka}}\ and\ \bibinfo {author} {\bibfnamefont {Kenta}\ \bibnamefont
  {Hotokezaka}},\ }\href {\doibase10.1088/0004-637X/775/2/113} {\bibfield
  {journal} {\bibinfo  {journal} {Astrophys. J.}\ }\textbf {\bibinfo {volume}
  {775}},\ \bibinfo {pages} {113} (\bibinfo {year} {2013})},\ \Eprint
  {http://arxiv.org/abs/1306.3742} {arXiv:1306.3742 [astro-ph.HE]}\BibitemShut
  {NoStop}%
\bibitem [{\citenamefont {{Kasen}}\ \emph {et~al.}(2015)\citenamefont
  {{Kasen}}, \citenamefont {{Fern{\'a}ndez}},\ and\ \citenamefont
  {{Metzger}}}]{2015MNRAS.450.1777K}%
  \BibitemOpen
  \bibfield  {author} {\bibinfo {author} {\bibfnamefont {D.}~\bibnamefont
  {{Kasen}}}, \bibinfo {author} {\bibfnamefont {R.}~\bibnamefont
  {{Fern{\'a}ndez}}}, \ and\ \bibinfo {author} {\bibfnamefont {B.~D.}\
  \bibnamefont {{Metzger}}},\ }\href {\doibase10.1093/mnras/stv721} {\bibfield
  {journal} {\bibinfo  {journal} {MNRAS}\ }\textbf {\bibinfo {volume} {450}},\
  \bibinfo {pages} {1777--1786} (\bibinfo {year} {2015})},\ \Eprint
  {http://arxiv.org/abs/1411.3726} {arXiv:1411.3726 [astro-ph.HE]}\BibitemShut
  {NoStop}%
\bibitem [{\citenamefont {Wollaeger}\ \emph {et~al.}(2017)\citenamefont
  {Wollaeger}, \citenamefont {Korobkin}, \citenamefont {Fontes}, \citenamefont
  {Rosswog}, \citenamefont {Even}, \citenamefont {Fryer}, \citenamefont
  {Sollerman}, \citenamefont {Hungerford}, \citenamefont {van Rossum},\ and\
  \citenamefont {Wollaber}}]{Wollaeger:2017ahm}%
  \BibitemOpen
  \bibfield  {author} {\bibinfo {author} {\bibfnamefont {Ryan~T.}\ \bibnamefont
  {Wollaeger}}, \bibinfo {author} {\bibfnamefont {Oleg}\ \bibnamefont
  {Korobkin}}, \bibinfo {author} {\bibfnamefont {Christopher~J.}\ \bibnamefont
  {Fontes}}, \bibinfo {author} {\bibfnamefont {Stephan~K.}\ \bibnamefont
  {Rosswog}}, \bibinfo {author} {\bibfnamefont {Wesley~P.}\ \bibnamefont
  {Even}}, \bibinfo {author} {\bibfnamefont {Christopher~L.}\ \bibnamefont
  {Fryer}}, \bibinfo {author} {\bibfnamefont {Jesper}\ \bibnamefont
  {Sollerman}}, \bibinfo {author} {\bibfnamefont {Aimee~L.}\ \bibnamefont
  {Hungerford}}, \bibinfo {author} {\bibfnamefont {Daniel~R.}\ \bibnamefont
  {van Rossum}}, \ and\ \bibinfo {author} {\bibfnamefont {Allan~B.}\
  \bibnamefont {Wollaber}},\ }\href@noop {} {\  (\bibinfo {year} {2017})},\
  \Eprint {http://arxiv.org/abs/1705.07084} {arXiv:1705.07084
  [astro-ph.HE]}\BibitemShut {NoStop}%
\bibitem [{\citenamefont {{Grossman}}\ \emph {et~al.}(2014)\citenamefont
  {{Grossman}}, \citenamefont {{Korobkin}}, \citenamefont {{Rosswog}},\ and\
  \citenamefont {{Piran}}}]{2014MNRAS.439..757G}%
  \BibitemOpen
  \bibfield  {author} {\bibinfo {author} {\bibfnamefont {D.}~\bibnamefont
  {{Grossman}}}, \bibinfo {author} {\bibfnamefont {O.}~\bibnamefont
  {{Korobkin}}}, \bibinfo {author} {\bibfnamefont {S.}~\bibnamefont
  {{Rosswog}}}, \ and\ \bibinfo {author} {\bibfnamefont {T.}~\bibnamefont
  {{Piran}}},\ }\href {\doibase10.1093/mnras/stt2503} {\bibfield  {journal}
  {\bibinfo  {journal} {MNRAS}\ }\textbf {\bibinfo {volume} {439}},\ \bibinfo
  {pages} {757--770} (\bibinfo {year} {2014})},\ \Eprint
  {http://arxiv.org/abs/1307.2943} {arXiv:1307.2943 [astro-ph.HE]}\BibitemShut
  {NoStop}%
\bibitem [{\citenamefont {Rosswog}\ \emph {et~al.}(2017)\citenamefont
  {Rosswog}, \citenamefont {Feindt}, \citenamefont {Korobkin}, \citenamefont
  {Wu}, \citenamefont {Sollerman}, \citenamefont {Goobar},\ and\ \citenamefont
  {Martinez-Pinedo}}]{Rosswog:2016dhy}%
  \BibitemOpen
  \bibfield  {author} {\bibinfo {author} {\bibfnamefont {S.}~\bibnamefont
  {Rosswog}}, \bibinfo {author} {\bibfnamefont {U.}~\bibnamefont {Feindt}},
  \bibinfo {author} {\bibfnamefont {O.}~\bibnamefont {Korobkin}}, \bibinfo
  {author} {\bibfnamefont {M.~R.}\ \bibnamefont {Wu}}, \bibinfo {author}
  {\bibfnamefont {J.}~\bibnamefont {Sollerman}}, \bibinfo {author}
  {\bibfnamefont {A.}~\bibnamefont {Goobar}}, \ and\ \bibinfo {author}
  {\bibfnamefont {G.}~\bibnamefont {Martinez-Pinedo}},\ }\href
  {\doibase10.1088/1361-6382/aa68a9} {\bibfield  {journal} {\bibinfo  {journal}
  {Class. Quant. Grav.}\ }\textbf {\bibinfo {volume} {34}},\ \bibinfo {pages}
  {104001} (\bibinfo {year} {2017})},\ \Eprint
  {http://arxiv.org/abs/1611.09822} {arXiv:1611.09822
  [astro-ph.HE]}\BibitemShut {NoStop}%
\bibitem [{\citenamefont {{Metzger}}\ and\ \citenamefont
  {{Fern{\'a}ndez}}(2014)}]{2014MNRAS.441.3444M}%
  \BibitemOpen
  \bibfield  {author} {\bibinfo {author} {\bibfnamefont {B.~D.}\ \bibnamefont
  {{Metzger}}}\ and\ \bibinfo {author} {\bibfnamefont {R.}~\bibnamefont
  {{Fern{\'a}ndez}}},\ }\href {\doibase10.1093/mnras/stu802} {\bibfield
  {journal} {\bibinfo  {journal} {MNRAS}\ }\textbf {\bibinfo {volume} {441}},\
  \bibinfo {pages} {3444--3453} (\bibinfo {year} {2014})},\ \Eprint
  {http://arxiv.org/abs/1402.4803} {arXiv:1402.4803 [astro-ph.HE]}\BibitemShut
  {NoStop}%
\bibitem [{\citenamefont {Martin}\ \emph {et~al.}(2015)\citenamefont {Martin},
  \citenamefont {Perego}, \citenamefont {Arcones}, \citenamefont {Thielemann},
  \citenamefont {Korobkin},\ and\ \citenamefont {Rosswog}}]{Martin:2015hxa}%
  \BibitemOpen
  \bibfield  {author} {\bibinfo {author} {\bibfnamefont {Dirk}\ \bibnamefont
  {Martin}}, \bibinfo {author} {\bibfnamefont {Albino}\ \bibnamefont {Perego}},
  \bibinfo {author} {\bibfnamefont {Almudena}\ \bibnamefont {Arcones}},
  \bibinfo {author} {\bibfnamefont {Friedrich-Karl}\ \bibnamefont
  {Thielemann}}, \bibinfo {author} {\bibfnamefont {Oleg}\ \bibnamefont
  {Korobkin}}, \ and\ \bibinfo {author} {\bibfnamefont {Stephan}\ \bibnamefont
  {Rosswog}},\ }\href {\doibase10.1088/0004-637X/813/1/2} {\bibfield  {journal}
  {\bibinfo  {journal} {Astrophys. J.}\ }\textbf {\bibinfo {volume} {813}},\
  \bibinfo {pages} {2} (\bibinfo {year} {2015})},\ \Eprint
  {http://arxiv.org/abs/1506.05048} {arXiv:1506.05048
  [astro-ph.SR]}\BibitemShut {NoStop}%
\bibitem [{\citenamefont {Villar}\ \emph {et~al.}(2017)\citenamefont {Villar}
  \emph {et~al.}}]{Villar:2017wcc}%
  \BibitemOpen
  \bibfield  {author} {\bibinfo {author} {\bibfnamefont {V.~Ashley}\
  \bibnamefont {Villar}} \emph {et~al.},\ }\href
  {\doibase10.3847/2041-8213/aa9c84} {\bibfield  {journal} {\bibinfo  {journal}
  {Astrophys. J.}\ }\textbf {\bibinfo {volume} {851}},\ \bibinfo {pages} {L21}
  (\bibinfo {year} {2017})},\ \Eprint {http://arxiv.org/abs/1710.11576}
  {arXiv:1710.11576 [astro-ph.HE]}\BibitemShut {NoStop}%
\bibitem [{\citenamefont {Tanaka}\ \emph {et~al.}(2017)\citenamefont {Tanaka}
  \emph {et~al.}}]{Tanaka:2017qxj}%
  \BibitemOpen
  \bibfield  {author} {\bibinfo {author} {\bibfnamefont {Masaomi}\ \bibnamefont
  {Tanaka}} \emph {et~al.},\ }\href {\doibase10.1093/pasj/psx121} {\bibfield
  {journal} {\bibinfo  {journal} {Publ. Astron. Soc. Jap.}\ } (\bibinfo {year}
  {2017}),\ 10.1093/pasj/psx121},\ \Eprint {http://arxiv.org/abs/1710.05850}
  {arXiv:1710.05850 [astro-ph.HE]}\BibitemShut {NoStop}%
\bibitem [{\citenamefont {Perego}\ \emph
  {et~al.}(2017{\natexlab{a}})\citenamefont {Perego}, \citenamefont {Radice},\
  and\ \citenamefont {Bernuzzi}}]{Perego:2017wtu}%
  \BibitemOpen
  \bibfield  {author} {\bibinfo {author} {\bibfnamefont {Albino}\ \bibnamefont
  {Perego}}, \bibinfo {author} {\bibfnamefont {David}\ \bibnamefont {Radice}},
  \ and\ \bibinfo {author} {\bibfnamefont {Sebastiano}\ \bibnamefont
  {Bernuzzi}},\ }\href {\doibase10.3847/2041-8213/aa9ab9} {\bibfield  {journal}
  {\bibinfo  {journal} {Astrophys. J.}\ }\textbf {\bibinfo {volume} {850}},\
  \bibinfo {pages} {L37} (\bibinfo {year} {2017}{\natexlab{a}})},\ \Eprint
  {http://arxiv.org/abs/1711.03982} {arXiv:1711.03982
  [astro-ph.HE]}\BibitemShut {NoStop}%
\bibitem [{\citenamefont {{Murguia-Berthier}}\ \emph
  {et~al.}(2017)\citenamefont {{Murguia-Berthier}}, \citenamefont
  {{Ramirez-Ruiz}}, \citenamefont {{Kilpatrick}}, \citenamefont {{Foley}},
  \citenamefont {{Kasen}}, \citenamefont {{Lee}}, \citenamefont {{Piro}},
  \citenamefont {{Coulter}}, \citenamefont {{Drout}}, \citenamefont {{Madore}},
  \citenamefont {{Shappee}}, \citenamefont {{Pan}}, \citenamefont
  {{Prochaska}}, \citenamefont {{Rest}}, \citenamefont {{Rojas-Bravo}},
  \citenamefont {{Siebert}},\ and\ \citenamefont
  {{Simon}}}]{2017ApJ...848L..34M}%
  \BibitemOpen
  \bibfield  {author} {\bibinfo {author} {\bibfnamefont {A.}~\bibnamefont
  {{Murguia-Berthier}}}, \bibinfo {author} {\bibfnamefont {E.}~\bibnamefont
  {{Ramirez-Ruiz}}}, \bibinfo {author} {\bibfnamefont {C.~D.}\ \bibnamefont
  {{Kilpatrick}}}, \bibinfo {author} {\bibfnamefont {R.~J.}\ \bibnamefont
  {{Foley}}}, \bibinfo {author} {\bibfnamefont {D.}~\bibnamefont {{Kasen}}},
  \bibinfo {author} {\bibfnamefont {W.~H.}\ \bibnamefont {{Lee}}}, \bibinfo
  {author} {\bibfnamefont {A.~L.}\ \bibnamefont {{Piro}}}, \bibinfo {author}
  {\bibfnamefont {D.~A.}\ \bibnamefont {{Coulter}}}, \bibinfo {author}
  {\bibfnamefont {M.~R.}\ \bibnamefont {{Drout}}}, \bibinfo {author}
  {\bibfnamefont {B.~F.}\ \bibnamefont {{Madore}}}, \bibinfo {author}
  {\bibfnamefont {B.~J.}\ \bibnamefont {{Shappee}}}, \bibinfo {author}
  {\bibfnamefont {Y.-C.}\ \bibnamefont {{Pan}}}, \bibinfo {author}
  {\bibfnamefont {J.~X.}\ \bibnamefont {{Prochaska}}}, \bibinfo {author}
  {\bibfnamefont {A.}~\bibnamefont {{Rest}}}, \bibinfo {author} {\bibfnamefont
  {C.}~\bibnamefont {{Rojas-Bravo}}}, \bibinfo {author} {\bibfnamefont {M.~R.}\
  \bibnamefont {{Siebert}}}, \ and\ \bibinfo {author} {\bibfnamefont {J.~D.}\
  \bibnamefont {{Simon}}},\ }\href {\doibase10.3847/2041-8213/aa91b3}
  {\bibfield  {journal} {\bibinfo  {journal} {The Astrophysical Journal}\
  }\textbf {\bibinfo {volume} {848}},\ \bibinfo {eid} {L34} (\bibinfo {year}
  {2017})},\ \Eprint {http://arxiv.org/abs/1710.05453} {arXiv:1710.05453
  [astro-ph.HE]}\BibitemShut {NoStop}%
\bibitem [{\citenamefont {{Bauswein}}\ \emph {et~al.}(2017)\citenamefont
  {{Bauswein}}, \citenamefont {{Just}}, \citenamefont {{Janka}},\ and\
  \citenamefont {{Stergioulas}}}]{2017ApJ...850L..34B}%
  \BibitemOpen
  \bibfield  {author} {\bibinfo {author} {\bibfnamefont {A.}~\bibnamefont
  {{Bauswein}}}, \bibinfo {author} {\bibfnamefont {O.}~\bibnamefont {{Just}}},
  \bibinfo {author} {\bibfnamefont {H.-T.}\ \bibnamefont {{Janka}}}, \ and\
  \bibinfo {author} {\bibfnamefont {N.}~\bibnamefont {{Stergioulas}}},\ }\href
  {\doibase10.3847/2041-8213/aa9994} {\bibfield  {journal} {\bibinfo  {journal}
  {The Astrophysical Journal}\ }\textbf {\bibinfo {volume} {850}},\ \bibinfo
  {eid} {L34} (\bibinfo {year} {2017})},\ \Eprint
  {http://arxiv.org/abs/1710.06843} {arXiv:1710.06843
  [astro-ph.HE]}\BibitemShut {NoStop}%
\bibitem [{\citenamefont {Radice}\ \emph {et~al.}(2018)\citenamefont {Radice},
  \citenamefont {Perego}, \citenamefont {Zappa},\ and\ \citenamefont
  {Bernuzzi}}]{Radice:2017lry}%
  \BibitemOpen
  \bibfield  {author} {\bibinfo {author} {\bibfnamefont {David}\ \bibnamefont
  {Radice}}, \bibinfo {author} {\bibfnamefont {Albino}\ \bibnamefont {Perego}},
  \bibinfo {author} {\bibfnamefont {Francesco}\ \bibnamefont {Zappa}}, \ and\
  \bibinfo {author} {\bibfnamefont {Sebastiano}\ \bibnamefont {Bernuzzi}},\
  }\href {\doibase10.3847/2041-8213/aaa402} {\bibfield  {journal} {\bibinfo
  {journal} {Astrophys. J.}\ }\textbf {\bibinfo {volume} {852}},\ \bibinfo
  {pages} {L29} (\bibinfo {year} {2018})},\ \Eprint
  {http://arxiv.org/abs/1711.03647} {arXiv:1711.03647
  [astro-ph.HE]}\BibitemShut {NoStop}%
\bibitem [{\citenamefont {{Margalit}}\ and\ \citenamefont
  {{Metzger}}(2017)}]{2017ApJ...850L..19M}%
  \BibitemOpen
  \bibfield  {author} {\bibinfo {author} {\bibfnamefont {B.}~\bibnamefont
  {{Margalit}}}\ and\ \bibinfo {author} {\bibfnamefont {B.~D.}\ \bibnamefont
  {{Metzger}}},\ }\href {\doibase10.3847/2041-8213/aa991c} {\bibfield
  {journal} {\bibinfo  {journal} {The Astrophysical Journal}\ }\textbf
  {\bibinfo {volume} {850}},\ \bibinfo {eid} {L19} (\bibinfo {year} {2017})},\
  \Eprint {http://arxiv.org/abs/1710.05938} {arXiv:1710.05938
  [astro-ph.HE]}\BibitemShut {NoStop}%
\bibitem [{\citenamefont {{Kyutoku}}\ \emph {et~al.}(2013)\citenamefont
  {{Kyutoku}}, \citenamefont {{Ioka}},\ and\ \citenamefont
  {{Shibata}}}]{2013PhRvD..88d1503K}%
  \BibitemOpen
  \bibfield  {author} {\bibinfo {author} {\bibfnamefont {K.}~\bibnamefont
  {{Kyutoku}}}, \bibinfo {author} {\bibfnamefont {K.}~\bibnamefont {{Ioka}}}, \
  and\ \bibinfo {author} {\bibfnamefont {M.}~\bibnamefont {{Shibata}}},\ }\href
  {\doibase10.1103/PhysRevD.88.041503} {\bibfield  {journal} {\bibinfo
  {journal} {Physical Review D}\ }\textbf {\bibinfo {volume} {88}},\ \bibinfo
  {eid} {041503} (\bibinfo {year} {2013})},\ \Eprint
  {http://arxiv.org/abs/1305.6309} {arXiv:1305.6309 [astro-ph.HE]}\BibitemShut
  {NoStop}%
\bibitem [{\citenamefont {{Deaton}}\ \emph {et~al.}(2013)\citenamefont
  {{Deaton}}, \citenamefont {{Duez}}, \citenamefont {{Foucart}}, \citenamefont
  {{O'Connor}}, \citenamefont {{Ott}}, \citenamefont {{Kidder}}, \citenamefont
  {{Muhlberger}}, \citenamefont {{Scheel}},\ and\ \citenamefont
  {{Szilagyi}}}]{2013ApJ...776...47D}%
  \BibitemOpen
  \bibfield  {author} {\bibinfo {author} {\bibfnamefont {M.~B.}\ \bibnamefont
  {{Deaton}}}, \bibinfo {author} {\bibfnamefont {M.~D.}\ \bibnamefont
  {{Duez}}}, \bibinfo {author} {\bibfnamefont {F.}~\bibnamefont {{Foucart}}},
  \bibinfo {author} {\bibfnamefont {E.}~\bibnamefont {{O'Connor}}}, \bibinfo
  {author} {\bibfnamefont {C.~D.}\ \bibnamefont {{Ott}}}, \bibinfo {author}
  {\bibfnamefont {L.~E.}\ \bibnamefont {{Kidder}}}, \bibinfo {author}
  {\bibfnamefont {C.~D.}\ \bibnamefont {{Muhlberger}}}, \bibinfo {author}
  {\bibfnamefont {M.~A.}\ \bibnamefont {{Scheel}}}, \ and\ \bibinfo {author}
  {\bibfnamefont {B.}~\bibnamefont {{Szilagyi}}},\ }\href
  {\doibase10.1088/0004-637X/776/1/47} {\bibfield  {journal} {\bibinfo
  {journal} {The Astrophysical Journal}\ }\textbf {\bibinfo {volume} {776}},\
  \bibinfo {eid} {47} (\bibinfo {year} {2013})},\ \Eprint
  {http://arxiv.org/abs/1304.3384} {arXiv:1304.3384 [astro-ph.HE]}\BibitemShut
  {NoStop}%
\bibitem [{\citenamefont {{Foucart}}\ \emph {et~al.}(2014)\citenamefont
  {{Foucart}}, \citenamefont {{Deaton}}, \citenamefont {{Duez}}, \citenamefont
  {{O'Connor}}, \citenamefont {{Ott}}, \citenamefont {{Haas}}, \citenamefont
  {{Kidder}}, \citenamefont {{Pfeiffer}}, \citenamefont {{Scheel}},\ and\
  \citenamefont {{Szilagyi}}}]{Foucart:2014}%
  \BibitemOpen
  \bibfield  {author} {\bibinfo {author} {\bibfnamefont {F.}~\bibnamefont
  {{Foucart}}}, \bibinfo {author} {\bibfnamefont {M.~B.}\ \bibnamefont
  {{Deaton}}}, \bibinfo {author} {\bibfnamefont {M.~D.}\ \bibnamefont
  {{Duez}}}, \bibinfo {author} {\bibfnamefont {E.}~\bibnamefont {{O'Connor}}},
  \bibinfo {author} {\bibfnamefont {C.~D.}\ \bibnamefont {{Ott}}}, \bibinfo
  {author} {\bibfnamefont {R.}~\bibnamefont {{Haas}}}, \bibinfo {author}
  {\bibfnamefont {L.~E.}\ \bibnamefont {{Kidder}}}, \bibinfo {author}
  {\bibfnamefont {H.~P.}\ \bibnamefont {{Pfeiffer}}}, \bibinfo {author}
  {\bibfnamefont {M.~A.}\ \bibnamefont {{Scheel}}}, \ and\ \bibinfo {author}
  {\bibfnamefont {B.}~\bibnamefont {{Szilagyi}}},\ }\href
  {\doibase10.1103/PhysRevD.90.024026} {\bibfield  {journal} {\bibinfo
  {journal} {Physical Review D}\ }\textbf {\bibinfo {volume} {90}},\ \bibinfo
  {eid} {024026} (\bibinfo {year} {2014})},\ \Eprint
  {http://arxiv.org/abs/1405.1121} {arXiv:1405.1121 [astro-ph.HE]}\BibitemShut
  {NoStop}%
\bibitem [{\citenamefont {Roberts}\ \emph {et~al.}(2017)\citenamefont
  {Roberts}, \citenamefont {Lippuner}, \citenamefont {Duez}, \citenamefont
  {Faber}, \citenamefont {Foucart}, \citenamefont {Lombardi}, \citenamefont
  {Ning}, \citenamefont {Ott},\ and\ \citenamefont {Ponce}}]{Roberts:2016igt}%
  \BibitemOpen
  \bibfield  {author} {\bibinfo {author} {\bibfnamefont {Luke~F.}\ \bibnamefont
  {Roberts}}, \bibinfo {author} {\bibfnamefont {Jonas}\ \bibnamefont
  {Lippuner}}, \bibinfo {author} {\bibfnamefont {Matthew~D.}\ \bibnamefont
  {Duez}}, \bibinfo {author} {\bibfnamefont {Joshua~A.}\ \bibnamefont {Faber}},
  \bibinfo {author} {\bibfnamefont {Francois}\ \bibnamefont {Foucart}},
  \bibinfo {author} {\bibfnamefont {James~C.}\ \bibnamefont {Lombardi}},
  \bibinfo {author} {\bibfnamefont {Sandra}\ \bibnamefont {Ning}}, \bibinfo
  {author} {\bibfnamefont {Christian~D.}\ \bibnamefont {Ott}}, \ and\ \bibinfo
  {author} {\bibfnamefont {Marcelo}\ \bibnamefont {Ponce}},\ }\href
  {\doibase10.1093/mnras/stw2622} {\bibfield  {journal} {\bibinfo  {journal}
  {Mon. Not. Roy. Astron. Soc.}\ }\textbf {\bibinfo {volume} {464}},\ \bibinfo
  {pages} {3907--3919} (\bibinfo {year} {2017})},\ \Eprint
  {http://arxiv.org/abs/1601.07942} {arXiv:1601.07942
  [astro-ph.HE]}\BibitemShut {NoStop}%
\bibitem [{\citenamefont {{Kiuchi}}\ \emph {et~al.}(2015)\citenamefont
  {{Kiuchi}}, \citenamefont {{Sekiguchi}}, \citenamefont {{Kyutoku}},
  \citenamefont {{Shibata}}, \citenamefont {{Taniguchi}},\ and\ \citenamefont
  {{Wada}}}]{2015PhRvD..92f4034K}%
  \BibitemOpen
  \bibfield  {author} {\bibinfo {author} {\bibfnamefont {K.}~\bibnamefont
  {{Kiuchi}}}, \bibinfo {author} {\bibfnamefont {Y.}~\bibnamefont
  {{Sekiguchi}}}, \bibinfo {author} {\bibfnamefont {K.}~\bibnamefont
  {{Kyutoku}}}, \bibinfo {author} {\bibfnamefont {M.}~\bibnamefont
  {{Shibata}}}, \bibinfo {author} {\bibfnamefont {K.}~\bibnamefont
  {{Taniguchi}}}, \ and\ \bibinfo {author} {\bibfnamefont {T.}~\bibnamefont
  {{Wada}}},\ }\href {\doibase10.1103/PhysRevD.92.064034} {\bibfield  {journal}
  {\bibinfo  {journal} {Physical Review D}\ }\textbf {\bibinfo {volume} {92}},\
  \bibinfo {eid} {064034} (\bibinfo {year} {2015})},\ \Eprint
  {http://arxiv.org/abs/1506.06811} {arXiv:1506.06811
  [astro-ph.HE]}\BibitemShut {NoStop}%
\bibitem [{\citenamefont {{Fern{\'a}ndez}}\ \emph {et~al.}(2017)\citenamefont
  {{Fern{\'a}ndez}}, \citenamefont {{Foucart}}, \citenamefont {{Kasen}},
  \citenamefont {{Lippuner}}, \citenamefont {{Desai}},\ and\ \citenamefont
  {{Roberts}}}]{2017CQGra..34o4001F}%
  \BibitemOpen
  \bibfield  {author} {\bibinfo {author} {\bibfnamefont {R.}~\bibnamefont
  {{Fern{\'a}ndez}}}, \bibinfo {author} {\bibfnamefont {F.}~\bibnamefont
  {{Foucart}}}, \bibinfo {author} {\bibfnamefont {D.}~\bibnamefont {{Kasen}}},
  \bibinfo {author} {\bibfnamefont {J.}~\bibnamefont {{Lippuner}}}, \bibinfo
  {author} {\bibfnamefont {D.}~\bibnamefont {{Desai}}}, \ and\ \bibinfo
  {author} {\bibfnamefont {L.~F.}\ \bibnamefont {{Roberts}}},\ }\href
  {\doibase10.1088/1361-6382/aa7a77} {\bibfield  {journal} {\bibinfo  {journal}
  {Classical and Quantum Gravity}\ }\textbf {\bibinfo {volume} {34}},\ \bibinfo
  {eid} {154001} (\bibinfo {year} {2017})},\ \Eprint
  {http://arxiv.org/abs/1612.04829} {arXiv:1612.04829
  [astro-ph.HE]}\BibitemShut {NoStop}%
\bibitem [{\citenamefont {{Perego}}\ \emph {et~al.}(2016)\citenamefont
  {{Perego}}, \citenamefont {{Cabez{\'o}n}},\ and\ \citenamefont
  {{K{\"a}ppeli}}}]{2016ApJS..223...22P}%
  \BibitemOpen
  \bibfield  {author} {\bibinfo {author} {\bibfnamefont {A.}~\bibnamefont
  {{Perego}}}, \bibinfo {author} {\bibfnamefont {R.~M.}\ \bibnamefont
  {{Cabez{\'o}n}}}, \ and\ \bibinfo {author} {\bibfnamefont {R.}~\bibnamefont
  {{K{\"a}ppeli}}},\ }\href {\doibase10.3847/0067-0049/223/2/22} {\bibfield
  {journal} {\bibinfo  {journal} {The Astrophysical Journal}\ }\textbf
  {\bibinfo {volume} {223}},\ \bibinfo {eid} {22} (\bibinfo {year} {2016})},\
  \Eprint {http://arxiv.org/abs/1511.08519} {arXiv:1511.08519
  [astro-ph.IM]}\BibitemShut {NoStop}%
\bibitem [{\citenamefont {{Foucart}}\ \emph {et~al.}(2016)\citenamefont
  {{Foucart}}, \citenamefont {{O'Connor}}, \citenamefont {{Roberts}},
  \citenamefont {{Kidder}}, \citenamefont {{Pfeiffer}},\ and\ \citenamefont
  {{Scheel}}}]{2016PhRvD..94l3016F}%
  \BibitemOpen
  \bibfield  {author} {\bibinfo {author} {\bibfnamefont {F.}~\bibnamefont
  {{Foucart}}}, \bibinfo {author} {\bibfnamefont {E.}~\bibnamefont
  {{O'Connor}}}, \bibinfo {author} {\bibfnamefont {L.}~\bibnamefont
  {{Roberts}}}, \bibinfo {author} {\bibfnamefont {L.~E.}\ \bibnamefont
  {{Kidder}}}, \bibinfo {author} {\bibfnamefont {H.~P.}\ \bibnamefont
  {{Pfeiffer}}}, \ and\ \bibinfo {author} {\bibfnamefont {M.~A.}\ \bibnamefont
  {{Scheel}}},\ }\href {\doibase10.1103/PhysRevD.94.123016} {\bibfield
  {journal} {\bibinfo  {journal} {Physical Review D}\ }\textbf {\bibinfo
  {volume} {94}},\ \bibinfo {eid} {123016} (\bibinfo {year} {2016})},\ \Eprint
  {http://arxiv.org/abs/1607.07450} {arXiv:1607.07450
  [astro-ph.HE]}\BibitemShut {NoStop}%
\bibitem [{\citenamefont {{Foucart}}(2018)}]{Foucart:2018}%
  \BibitemOpen
  \bibfield  {author} {\bibinfo {author} {\bibfnamefont {F.}~\bibnamefont
  {{Foucart}}},\ }\href {\doibase10.1093/mnras/sty108} {\bibfield  {journal}
  {\bibinfo  {journal} {MNRAS}\ }\textbf {\bibinfo {volume} {475}},\ \bibinfo
  {pages} {4186--4207} (\bibinfo {year} {2018})},\ \Eprint
  {http://arxiv.org/abs/1708.08452} {arXiv:1708.08452
  [astro-ph.HE]}\BibitemShut {NoStop}%
\bibitem [{\citenamefont {{Malkus}}\ \emph {et~al.}(2016)\citenamefont
  {{Malkus}}, \citenamefont {{McLaughlin}},\ and\ \citenamefont
  {{Surman}}}]{2016PhRvD..93d5021M}%
  \BibitemOpen
  \bibfield  {author} {\bibinfo {author} {\bibfnamefont {A.}~\bibnamefont
  {{Malkus}}}, \bibinfo {author} {\bibfnamefont {G.~C.}\ \bibnamefont
  {{McLaughlin}}}, \ and\ \bibinfo {author} {\bibfnamefont {R.}~\bibnamefont
  {{Surman}}},\ }\href {\doibase10.1103/PhysRevD.93.045021} {\bibfield
  {journal} {\bibinfo  {journal} {Physical Review D}\ }\textbf {\bibinfo
  {volume} {93}},\ \bibinfo {eid} {045021} (\bibinfo {year} {2016})},\ \Eprint
  {http://arxiv.org/abs/1507.00946} {arXiv:1507.00946 [hep-ph]}\BibitemShut
  {NoStop}%
\bibitem [{\citenamefont {{Tian}}\ \emph {et~al.}(2017)\citenamefont {{Tian}},
  \citenamefont {{Patwardhan}},\ and\ \citenamefont
  {{Fuller}}}]{2017PhRvD..96d3001T}%
  \BibitemOpen
  \bibfield  {author} {\bibinfo {author} {\bibfnamefont {J.~Y.}\ \bibnamefont
  {{Tian}}}, \bibinfo {author} {\bibfnamefont {A.~V.}\ \bibnamefont
  {{Patwardhan}}}, \ and\ \bibinfo {author} {\bibfnamefont {G.~M.}\
  \bibnamefont {{Fuller}}},\ }\href {\doibase10.1103/PhysRevD.96.043001}
  {\bibfield  {journal} {\bibinfo  {journal} {Physical Review D}\ }\textbf
  {\bibinfo {volume} {96}},\ \bibinfo {eid} {043001} (\bibinfo {year}
  {2017})},\ \Eprint {http://arxiv.org/abs/1703.03039} {arXiv:1703.03039
  [astro-ph.HE]}\BibitemShut {NoStop}%
\bibitem [{\citenamefont {{Zhu}}\ \emph {et~al.}(2016)\citenamefont {{Zhu}},
  \citenamefont {{Perego}},\ and\ \citenamefont
  {{McLaughlin}}}]{2016PhRvD..94j5006Z}%
  \BibitemOpen
  \bibfield  {author} {\bibinfo {author} {\bibfnamefont {Y.~L.}\ \bibnamefont
  {{Zhu}}}, \bibinfo {author} {\bibfnamefont {A.}~\bibnamefont {{Perego}}}, \
  and\ \bibinfo {author} {\bibfnamefont {G.~C.}\ \bibnamefont {{McLaughlin}}},\
  }\href {\doibase10.1103/PhysRevD.94.105006} {\bibfield  {journal} {\bibinfo
  {journal} {Physical Review D}\ }\textbf {\bibinfo {volume} {94}},\ \bibinfo
  {eid} {105006} (\bibinfo {year} {2016})},\ \Eprint
  {http://arxiv.org/abs/1607.04671} {arXiv:1607.04671 [hep-ph]}\BibitemShut
  {NoStop}%
\bibitem [{\citenamefont {{Frensel}}\ \emph {et~al.}(2017)\citenamefont
  {{Frensel}}, \citenamefont {{Wu}}, \citenamefont {{Volpe}},\ and\
  \citenamefont {{Perego}}}]{2017PhRvD..95b3011F}%
  \BibitemOpen
  \bibfield  {author} {\bibinfo {author} {\bibfnamefont {M.}~\bibnamefont
  {{Frensel}}}, \bibinfo {author} {\bibfnamefont {M.-R.}\ \bibnamefont {{Wu}}},
  \bibinfo {author} {\bibfnamefont {C.}~\bibnamefont {{Volpe}}}, \ and\
  \bibinfo {author} {\bibfnamefont {A.}~\bibnamefont {{Perego}}},\ }\href
  {\doibase10.1103/PhysRevD.95.023011} {\bibfield  {journal} {\bibinfo
  {journal} {Physical Review D}\ }\textbf {\bibinfo {volume} {95}},\ \bibinfo
  {eid} {023011} (\bibinfo {year} {2017})},\ \Eprint
  {http://arxiv.org/abs/1607.05938} {arXiv:1607.05938
  [astro-ph.HE]}\BibitemShut {NoStop}%
\bibitem [{\citenamefont {{Chandrasekhar}}(1960)}]{1960PNAS...46..253C}%
  \BibitemOpen
  \bibfield  {author} {\bibinfo {author} {\bibfnamefont {S.}~\bibnamefont
  {{Chandrasekhar}}},\ }\href {\doibase10.1073/pnas.46.2.253} {\bibfield
  {journal} {\bibinfo  {journal} {Proceedings of the National Academy of
  Science}\ }\textbf {\bibinfo {volume} {46}},\ \bibinfo {pages} {253--257}
  (\bibinfo {year} {1960})}\BibitemShut {NoStop}%
\bibitem [{\citenamefont {Kiuchi}\ \emph {et~al.}(2018)\citenamefont {Kiuchi},
  \citenamefont {Kyutoku}, \citenamefont {Sekiguchi},\ and\ \citenamefont
  {Shibata}}]{Kiuchi:2017zzg}%
  \BibitemOpen
  \bibfield  {author} {\bibinfo {author} {\bibfnamefont {Kenta}\ \bibnamefont
  {Kiuchi}}, \bibinfo {author} {\bibfnamefont {Koutarou}\ \bibnamefont
  {Kyutoku}}, \bibinfo {author} {\bibfnamefont {Yuichiro}\ \bibnamefont
  {Sekiguchi}}, \ and\ \bibinfo {author} {\bibfnamefont {Masaru}\ \bibnamefont
  {Shibata}},\ }\href {\doibase10.1103/PhysRevD.97.124039} {\bibfield
  {journal} {\bibinfo  {journal} {Phys. Rev. D}\ }\textbf {\bibinfo {volume}
  {97}},\ \bibinfo {pages} {124039} (\bibinfo {year} {2018})},\ \Eprint
  {http://arxiv.org/abs/1710.01311} {arXiv:1710.01311
  [astro-ph.HE]}\BibitemShut {NoStop}%
\bibitem [{\citenamefont {{Palenzuela}}\ \emph {et~al.}(2015)\citenamefont
  {{Palenzuela}}, \citenamefont {{Liebling}}, \citenamefont {{Neilsen}},
  \citenamefont {{Lehner}}, \citenamefont {{Caballero}}, \citenamefont
  {{O'Connor}},\ and\ \citenamefont {{Anderson}}}]{2015PhRvD..92d4045P}%
  \BibitemOpen
  \bibfield  {author} {\bibinfo {author} {\bibfnamefont {C.}~\bibnamefont
  {{Palenzuela}}}, \bibinfo {author} {\bibfnamefont {S.~L.}\ \bibnamefont
  {{Liebling}}}, \bibinfo {author} {\bibfnamefont {D.}~\bibnamefont
  {{Neilsen}}}, \bibinfo {author} {\bibfnamefont {L.}~\bibnamefont {{Lehner}}},
  \bibinfo {author} {\bibfnamefont {O.~L.}\ \bibnamefont {{Caballero}}},
  \bibinfo {author} {\bibfnamefont {E.}~\bibnamefont {{O'Connor}}}, \ and\
  \bibinfo {author} {\bibfnamefont {M.}~\bibnamefont {{Anderson}}},\ }\href
  {\doibase10.1103/PhysRevD.92.044045} {\bibfield  {journal} {\bibinfo
  {journal} {"Phys. Rev. D"}\ }\textbf {\bibinfo {volume} {92}},\ \bibinfo
  {eid} {044045} (\bibinfo {year} {2015})},\ \Eprint
  {http://arxiv.org/abs/1505.01607} {arXiv:1505.01607 [gr-qc]}\BibitemShut
  {NoStop}%
\bibitem [{\citenamefont {{Endrizzi}}\ \emph {et~al.}(2016)\citenamefont
  {{Endrizzi}}, \citenamefont {{Ciolfi}}, \citenamefont {{Giacomazzo}},
  \citenamefont {{Kastaun}},\ and\ \citenamefont
  {{Kawamura}}}]{2016CQGra..33p4001E}%
  \BibitemOpen
  \bibfield  {author} {\bibinfo {author} {\bibfnamefont {A.}~\bibnamefont
  {{Endrizzi}}}, \bibinfo {author} {\bibfnamefont {R.}~\bibnamefont
  {{Ciolfi}}}, \bibinfo {author} {\bibfnamefont {B.}~\bibnamefont
  {{Giacomazzo}}}, \bibinfo {author} {\bibfnamefont {W.}~\bibnamefont
  {{Kastaun}}}, \ and\ \bibinfo {author} {\bibfnamefont {T.}~\bibnamefont
  {{Kawamura}}},\ }\href {\doibase10.1088/0264-9381/33/16/164001} {\bibfield
  {journal} {\bibinfo  {journal} {Classical and Quantum Gravity}\ }\textbf
  {\bibinfo {volume} {33}},\ \bibinfo {eid} {164001} (\bibinfo {year}
  {2016})},\ \Eprint {http://arxiv.org/abs/1604.03445} {arXiv:1604.03445
  [astro-ph.HE]}\BibitemShut {NoStop}%
\bibitem [{\citenamefont {Radice}(2017)}]{Radice:2017zta}%
  \BibitemOpen
  \bibfield  {author} {\bibinfo {author} {\bibfnamefont {David}\ \bibnamefont
  {Radice}},\ }\href {\doibase10.3847/2041-8213/aa6483} {\bibfield  {journal}
  {\bibinfo  {journal} {Astrophys. J.}\ }\textbf {\bibinfo {volume} {838}},\
  \bibinfo {pages} {L2} (\bibinfo {year} {2017})},\ \Eprint
  {http://arxiv.org/abs/1703.02046} {arXiv:1703.02046
  [astro-ph.HE]}\BibitemShut {NoStop}%
\bibitem [{\citenamefont {Shibata}\ \emph {et~al.}(2017)\citenamefont
  {Shibata}, \citenamefont {Kiuchi},\ and\ \citenamefont
  {Sekiguchi}}]{Shibata:2017jyf}%
  \BibitemOpen
  \bibfield  {author} {\bibinfo {author} {\bibfnamefont {Masaru}\ \bibnamefont
  {Shibata}}, \bibinfo {author} {\bibfnamefont {Kenta}\ \bibnamefont {Kiuchi}},
  \ and\ \bibinfo {author} {\bibfnamefont {Yu-ichiro}\ \bibnamefont
  {Sekiguchi}},\ }\href {\doibase10.1103/PhysRevD.95.083005} {\bibfield
  {journal} {\bibinfo  {journal} {Phys. Rev.}\ }\textbf {\bibinfo {volume}
  {D95}},\ \bibinfo {pages} {083005} (\bibinfo {year} {2017})},\ \Eprint
  {http://arxiv.org/abs/1703.10303} {arXiv:1703.10303
  [astro-ph.HE]}\BibitemShut {NoStop}%
\bibitem [{\citenamefont {Alford}\ \emph {et~al.}(2018)\citenamefont {Alford},
  \citenamefont {Bovard}, \citenamefont {Hanauske}, \citenamefont {Rezzolla},\
  and\ \citenamefont {Schwenzer}}]{Alford:2017rxf}%
  \BibitemOpen
  \bibfield  {author} {\bibinfo {author} {\bibfnamefont {Mark~G.}\ \bibnamefont
  {Alford}}, \bibinfo {author} {\bibfnamefont {Luke}\ \bibnamefont {Bovard}},
  \bibinfo {author} {\bibfnamefont {Matthias}\ \bibnamefont {Hanauske}},
  \bibinfo {author} {\bibfnamefont {Luciano}\ \bibnamefont {Rezzolla}}, \ and\
  \bibinfo {author} {\bibfnamefont {Kai}\ \bibnamefont {Schwenzer}},\ }\href
  {\doibase10.1103/PhysRevLett.120.041101} {\bibfield  {journal} {\bibinfo
  {journal} {Phys. Rev. Lett.}\ }\textbf {\bibinfo {volume} {120}},\ \bibinfo
  {pages} {041101} (\bibinfo {year} {2018})},\ \Eprint
  {http://arxiv.org/abs/1707.09475} {arXiv:1707.09475 [gr-qc]}\BibitemShut
  {NoStop}%
\bibitem [{\citenamefont {Perego}\ \emph
  {et~al.}(2017{\natexlab{b}})\citenamefont {Perego}, \citenamefont {Yasin},\
  and\ \citenamefont {Arcones}}]{Perego:2017fho}%
  \BibitemOpen
  \bibfield  {author} {\bibinfo {author} {\bibfnamefont {Albino}\ \bibnamefont
  {Perego}}, \bibinfo {author} {\bibfnamefont {Hannah}\ \bibnamefont {Yasin}},
  \ and\ \bibinfo {author} {\bibfnamefont {Almudena}\ \bibnamefont {Arcones}},\
  }\href {\doibase10.1088/1361-6471/aa7bdc} {\bibfield  {journal} {\bibinfo
  {journal} {J. Phys.}\ }\textbf {\bibinfo {volume} {G44}},\ \bibinfo {pages}
  {084007} (\bibinfo {year} {2017}{\natexlab{b}})},\ \Eprint
  {http://arxiv.org/abs/1701.02017} {arXiv:1701.02017
  [astro-ph.HE]}\BibitemShut {NoStop}%
\bibitem [{\citenamefont {Kilpatrick}\ \emph {et~al.}(2017)\citenamefont
  {Kilpatrick} \emph {et~al.}}]{Kilpatrick:2017mhz}%
  \BibitemOpen
  \bibfield  {author} {\bibinfo {author} {\bibfnamefont {Charles~D.}\
  \bibnamefont {Kilpatrick}} \emph {et~al.},\ }\href
  {\doibase10.1126/science.aaq0073} {\bibfield  {journal} {\bibinfo  {journal}
  {Science}\ }\textbf {\bibinfo {volume} {358}},\ \bibinfo {pages} {1583--1587}
  (\bibinfo {year} {2017})},\ \Eprint {http://arxiv.org/abs/1710.05434}
  {arXiv:1710.05434 [astro-ph.HE]}\BibitemShut {NoStop}%
\bibitem [{\citenamefont {Drout}\ \emph {et~al.}(2017)\citenamefont {Drout}
  \emph {et~al.}}]{Drout:2017ijr}%
  \BibitemOpen
  \bibfield  {author} {\bibinfo {author} {\bibfnamefont {M.~R.}\ \bibnamefont
  {Drout}} \emph {et~al.},\ }\href {\doibase10.1126/science.aaq0049} {\bibfield
   {journal} {\bibinfo  {journal} {Science}\ }\textbf {\bibinfo {volume}
  {358}},\ \bibinfo {pages} {1570--1574} (\bibinfo {year} {2017})},\ \Eprint
  {http://arxiv.org/abs/1710.05443} {arXiv:1710.05443
  [astro-ph.HE]}\BibitemShut {NoStop}%
\bibitem [{\citenamefont {Smartt}\ \emph {et~al.}(2017)\citenamefont {Smartt}
  \emph {et~al.}}]{Smartt:2017fuw}%
  \BibitemOpen
  \bibfield  {author} {\bibinfo {author} {\bibfnamefont {S.~J.}\ \bibnamefont
  {Smartt}} \emph {et~al.},\ }\href {\doibase10.1038/nature24303} {\bibfield
  {journal} {\bibinfo  {journal} {Nature}\ }\textbf {\bibinfo {volume} {551}},\
  \bibinfo {pages} {75--79} (\bibinfo {year} {2017})},\ \Eprint
  {http://arxiv.org/abs/1710.05841} {arXiv:1710.05841
  [astro-ph.HE]}\BibitemShut {NoStop}%
\bibitem [{\citenamefont {Chornock}\ \emph {et~al.}(2017)\citenamefont
  {Chornock} \emph {et~al.}}]{Chornock:2017sdf}%
  \BibitemOpen
  \bibfield  {author} {\bibinfo {author} {\bibfnamefont {R.}~\bibnamefont
  {Chornock}} \emph {et~al.},\ }\href {\doibase10.3847/2041-8213/aa905c}
  {\bibfield  {journal} {\bibinfo  {journal} {Astrophys. J.}\ }\textbf
  {\bibinfo {volume} {848}},\ \bibinfo {pages} {L19} (\bibinfo {year}
  {2017})},\ \Eprint {http://arxiv.org/abs/1710.05454} {arXiv:1710.05454
  [astro-ph.HE]}\BibitemShut {NoStop}%
\bibitem [{\citenamefont {Arcavi}\ \emph {et~al.}(2017)\citenamefont {Arcavi}
  \emph {et~al.}}]{Arcavi:2017xiz}%
  \BibitemOpen
  \bibfield  {author} {\bibinfo {author} {\bibfnamefont {Iair}\ \bibnamefont
  {Arcavi}} \emph {et~al.},\ }\href {\doibase10.1038/nature24291} {\bibfield
  {journal} {\bibinfo  {journal} {Nature}\ }\textbf {\bibinfo {volume} {551}},\
  \bibinfo {pages} {64} (\bibinfo {year} {2017})},\ \Eprint
  {http://arxiv.org/abs/1710.05843} {arXiv:1710.05843
  [astro-ph.HE]}\BibitemShut {NoStop}%
\bibitem [{\citenamefont {Shappee}\ \emph {et~al.}(2017)\citenamefont {Shappee}
  \emph {et~al.}}]{Shappee:2017zly}%
  \BibitemOpen
  \bibfield  {author} {\bibinfo {author} {\bibfnamefont {B.~J.}\ \bibnamefont
  {Shappee}} \emph {et~al.},\ }\href {\doibase10.1126/science.aaq0186}
  {\bibfield  {journal} {\bibinfo  {journal} {Science}\ }\textbf {\bibinfo
  {volume} {358}},\ \bibinfo {pages} {1574} (\bibinfo {year} {2017})},\ \Eprint
  {http://arxiv.org/abs/1710.05432} {arXiv:1710.05432
  [astro-ph.HE]}\BibitemShut {NoStop}%
\bibitem [{\citenamefont {Kasliwal}\ \emph {et~al.}(2017)\citenamefont
  {Kasliwal} \emph {et~al.}}]{Kasliwal:2017ngb}%
  \BibitemOpen
  \bibfield  {author} {\bibinfo {author} {\bibfnamefont {M.~M.}\ \bibnamefont
  {Kasliwal}} \emph {et~al.},\ }\href {\doibase10.1126/science.aap9455}
  {\bibfield  {journal} {\bibinfo  {journal} {Science}\ }\textbf {\bibinfo
  {volume} {358}},\ \bibinfo {pages} {1559} (\bibinfo {year} {2017})},\ \Eprint
  {http://arxiv.org/abs/1710.05436} {arXiv:1710.05436
  [astro-ph.HE]}\BibitemShut {NoStop}%
\bibitem [{\citenamefont {Andreoni}\ \emph {et~al.}(2017)\citenamefont
  {Andreoni} \emph {et~al.}}]{Andreoni:2017ppd}%
  \BibitemOpen
  \bibfield  {author} {\bibinfo {author} {\bibfnamefont {I.}~\bibnamefont
  {Andreoni}} \emph {et~al.},\ }\href {\doibase10.1017/pasa.2017.65} {\bibfield
   {journal} {\bibinfo  {journal} {Publ. Astron. Soc. Austral.}\ }\textbf
  {\bibinfo {volume} {34}},\ \bibinfo {pages} {e069} (\bibinfo {year}
  {2017})},\ \Eprint {http://arxiv.org/abs/1710.05846} {arXiv:1710.05846
  [astro-ph.HE]}\BibitemShut {NoStop}%
\bibitem [{\citenamefont {McCully}\ \emph {et~al.}(2017)\citenamefont {McCully}
  \emph {et~al.}}]{McCully:2017lgx}%
  \BibitemOpen
  \bibfield  {author} {\bibinfo {author} {\bibfnamefont {Curtis}\ \bibnamefont
  {McCully}} \emph {et~al.},\ }\href {\doibase10.3847/2041-8213/aa9111}
  {\bibfield  {journal} {\bibinfo  {journal} {Astrophys. J.}\ }\textbf
  {\bibinfo {volume} {848}},\ \bibinfo {pages} {L32} (\bibinfo {year}
  {2017})},\ \Eprint {http://arxiv.org/abs/1710.05853} {arXiv:1710.05853
  [astro-ph.HE]}\BibitemShut {NoStop}%
\bibitem [{\citenamefont {Cowperthwaite}\ \emph {et~al.}(2017)\citenamefont
  {Cowperthwaite} \emph {et~al.}}]{Cowperthwaite:2017dyu}%
  \BibitemOpen
  \bibfield  {author} {\bibinfo {author} {\bibfnamefont {P.~S.}\ \bibnamefont
  {Cowperthwaite}} \emph {et~al.},\ }\href {\doibase10.3847/2041-8213/aa8fc7}
  {\bibfield  {journal} {\bibinfo  {journal} {Astrophys. J.}\ }\textbf
  {\bibinfo {volume} {848}},\ \bibinfo {pages} {L17} (\bibinfo {year}
  {2017})},\ \Eprint {http://arxiv.org/abs/1710.05840} {arXiv:1710.05840
  [astro-ph.HE]}\BibitemShut {NoStop}%
\bibitem [{\citenamefont {Troja}\ \emph {et~al.}(2017)\citenamefont {Troja}
  \emph {et~al.}}]{Troja:2017nqp}%
  \BibitemOpen
  \bibfield  {author} {\bibinfo {author} {\bibfnamefont {E.}~\bibnamefont
  {Troja}} \emph {et~al.},\ }\href {\doibase10.1038/nature24290} {\bibfield
  {journal} {\bibinfo  {journal} {Nature}\ }\textbf {\bibinfo {volume} {551}},\
  \bibinfo {pages} {71--74} (\bibinfo {year} {2017})},\ \bibinfo {note}
  {[Nature551,71(2017)]},\ \Eprint {http://arxiv.org/abs/1710.05433}
  {arXiv:1710.05433 [astro-ph.HE]}\BibitemShut {NoStop}%
\bibitem [{\citenamefont {Haggard}\ \emph {et~al.}(2017)\citenamefont
  {Haggard}, \citenamefont {Nynka}, \citenamefont {Ruan}, \citenamefont
  {Kalogera}, \citenamefont {Bradley~Cenko}, \citenamefont {Evans},\ and\
  \citenamefont {Kennea}}]{Haggard:2017qne}%
  \BibitemOpen
  \bibfield  {author} {\bibinfo {author} {\bibfnamefont {Daryl}\ \bibnamefont
  {Haggard}}, \bibinfo {author} {\bibfnamefont {Melania}\ \bibnamefont
  {Nynka}}, \bibinfo {author} {\bibfnamefont {John~J.}\ \bibnamefont {Ruan}},
  \bibinfo {author} {\bibfnamefont {Vicky}\ \bibnamefont {Kalogera}}, \bibinfo
  {author} {\bibfnamefont {S.}~\bibnamefont {Bradley~Cenko}}, \bibinfo {author}
  {\bibfnamefont {Phil}\ \bibnamefont {Evans}}, \ and\ \bibinfo {author}
  {\bibfnamefont {Jamie~A.}\ \bibnamefont {Kennea}},\ }\href
  {\doibase10.3847/2041-8213/aa8ede} {\bibfield  {journal} {\bibinfo  {journal}
  {Astrophys. J.}\ }\textbf {\bibinfo {volume} {848}},\ \bibinfo {pages} {L25}
  (\bibinfo {year} {2017})},\ \Eprint {http://arxiv.org/abs/1710.05852}
  {arXiv:1710.05852 [astro-ph.HE]}\BibitemShut {NoStop}%
\bibitem [{\citenamefont {Margutti}\ \emph {et~al.}(2017)\citenamefont
  {Margutti} \emph {et~al.}}]{Margutti:2017cjl}%
  \BibitemOpen
  \bibfield  {author} {\bibinfo {author} {\bibfnamefont {Raffaella}\
  \bibnamefont {Margutti}} \emph {et~al.},\ }\href
  {\doibase10.3847/2041-8213/aa9057} {\bibfield  {journal} {\bibinfo  {journal}
  {Astrophys. J.}\ }\textbf {\bibinfo {volume} {848}},\ \bibinfo {pages} {L20}
  (\bibinfo {year} {2017})},\ \Eprint {http://arxiv.org/abs/1710.05431}
  {arXiv:1710.05431 [astro-ph.HE]}\BibitemShut {NoStop}%
\bibitem [{\citenamefont {Hallinan}\ \emph {et~al.}(2017)\citenamefont
  {Hallinan} \emph {et~al.}}]{Hallinan:2017woc}%
  \BibitemOpen
  \bibfield  {author} {\bibinfo {author} {\bibfnamefont {G.}~\bibnamefont
  {Hallinan}} \emph {et~al.},\ }\href {\doibase10.1126/science.aap9855}
  {\bibfield  {journal} {\bibinfo  {journal} {Science}\ }\textbf {\bibinfo
  {volume} {358}},\ \bibinfo {pages} {1579} (\bibinfo {year} {2017})},\ \Eprint
  {http://arxiv.org/abs/1710.05435} {arXiv:1710.05435
  [astro-ph.HE]}\BibitemShut {NoStop}%
\bibitem [{\citenamefont {Mooley}\ \emph {et~al.}(2018)\citenamefont {Mooley}
  \emph {et~al.}}]{Mooley:2017enz}%
  \BibitemOpen
  \bibfield  {author} {\bibinfo {author} {\bibfnamefont {K.~P.}\ \bibnamefont
  {Mooley}} \emph {et~al.},\ }\href {\doibase10.1038/nature25452} {\bibfield
  {journal} {\bibinfo  {journal} {Nature}\ }\textbf {\bibinfo {volume} {554}},\
  \bibinfo {pages} {207} (\bibinfo {year} {2018})},\ \Eprint
  {http://arxiv.org/abs/1711.11573} {arXiv:1711.11573
  [astro-ph.HE]}\BibitemShut {NoStop}%
\bibitem [{\citenamefont {Narayan}\ \emph {et~al.}(1992)\citenamefont
  {Narayan}, \citenamefont {Paczynski},\ and\ \citenamefont
  {Piran}}]{Narayan:1992iy}%
  \BibitemOpen
  \bibfield  {author} {\bibinfo {author} {\bibfnamefont {Ramesh}\ \bibnamefont
  {Narayan}}, \bibinfo {author} {\bibfnamefont {Bohdan}\ \bibnamefont
  {Paczynski}}, \ and\ \bibinfo {author} {\bibfnamefont {Tsvi}\ \bibnamefont
  {Piran}},\ }\href {\doibase10.1086/186493} {\bibfield  {journal} {\bibinfo
  {journal} {Astrophys. J.}\ }\textbf {\bibinfo {volume} {395}},\ \bibinfo
  {pages} {L83--L86} (\bibinfo {year} {1992})},\ \Eprint
  {http://arxiv.org/abs/astro-ph/9204001} {arXiv:astro-ph/9204001
  [astro-ph]}\BibitemShut {NoStop}%
\bibitem [{\citenamefont {Goldstein}\ \emph {et~al.}(2017)\citenamefont
  {Goldstein} \emph {et~al.}}]{Goldstein:2017mmi}%
  \BibitemOpen
  \bibfield  {author} {\bibinfo {author} {\bibfnamefont {A.}~\bibnamefont
  {Goldstein}} \emph {et~al.},\ }\href {\doibase10.3847/2041-8213/aa8f41}
  {\bibfield  {journal} {\bibinfo  {journal} {Astrophys. J.}\ }\textbf
  {\bibinfo {volume} {848}},\ \bibinfo {pages} {L14} (\bibinfo {year}
  {2017})},\ \Eprint {http://arxiv.org/abs/1710.05446} {arXiv:1710.05446
  [astro-ph.HE]}\BibitemShut {NoStop}%
\bibitem [{\citenamefont {Savchenko}\ \emph {et~al.}(2017)\citenamefont
  {Savchenko} \emph {et~al.}}]{Savchenko:2017ffs}%
  \BibitemOpen
  \bibfield  {author} {\bibinfo {author} {\bibfnamefont {V.}~\bibnamefont
  {Savchenko}} \emph {et~al.},\ }\href {\doibase10.3847/2041-8213/aa8f94}
  {\bibfield  {journal} {\bibinfo  {journal} {Astrophys. J.}\ }\textbf
  {\bibinfo {volume} {848}},\ \bibinfo {pages} {L15} (\bibinfo {year}
  {2017})},\ \Eprint {http://arxiv.org/abs/1710.05449} {arXiv:1710.05449
  [astro-ph.HE]}\BibitemShut {NoStop}%
\bibitem [{\citenamefont {Margutti}\ \emph {et~al.}(2018)\citenamefont
  {Margutti} \emph {et~al.}}]{Margutti:2018xqd}%
  \BibitemOpen
  \bibfield  {author} {\bibinfo {author} {\bibfnamefont {R.}~\bibnamefont
  {Margutti}} \emph {et~al.},\ }\href {\doibase10.3847/2041-8213/aab2ad}
  {\bibfield  {journal} {\bibinfo  {journal} {Astrophys. J.}\ }\textbf
  {\bibinfo {volume} {856}},\ \bibinfo {pages} {L18} (\bibinfo {year}
  {2018})},\ \Eprint {http://arxiv.org/abs/1801.03531} {arXiv:1801.03531
  [astro-ph.HE]}\BibitemShut {NoStop}%
\bibitem [{\citenamefont {{Eichler}}\ \emph
  {et~al.}(1989{\natexlab{b}})\citenamefont {{Eichler}}, \citenamefont
  {{Livio}}, \citenamefont {{Piran}},\ and\ \citenamefont
  {{Schramm}}}]{Eichler:1989}%
  \BibitemOpen
  \bibfield  {author} {\bibinfo {author} {\bibfnamefont {D.}~\bibnamefont
  {{Eichler}}}, \bibinfo {author} {\bibfnamefont {M.}~\bibnamefont {{Livio}}},
  \bibinfo {author} {\bibfnamefont {T.}~\bibnamefont {{Piran}}}, \ and\
  \bibinfo {author} {\bibfnamefont {D.~N.}\ \bibnamefont {{Schramm}}},\ }\href
  {\doibase10.1038/340126a0} {\bibfield  {journal} {\bibinfo  {journal}
  {Nature}\ }\textbf {\bibinfo {volume} {340}},\ \bibinfo {pages} {126--128}
  (\bibinfo {year} {1989}{\natexlab{b}})}\BibitemShut {NoStop}%
\bibitem [{\citenamefont {Rosswog}\ \emph {et~al.}(1999)\citenamefont
  {Rosswog}, \citenamefont {Liebendoerfer}, \citenamefont {Thielemann},
  \citenamefont {Davies}, \citenamefont {Benz},\ and\ \citenamefont
  {Piran}}]{Rosswog:1998hy}%
  \BibitemOpen
  \bibfield  {author} {\bibinfo {author} {\bibfnamefont {S.}~\bibnamefont
  {Rosswog}}, \bibinfo {author} {\bibfnamefont {M.}~\bibnamefont
  {Liebendoerfer}}, \bibinfo {author} {\bibfnamefont {F.~K.}\ \bibnamefont
  {Thielemann}}, \bibinfo {author} {\bibfnamefont {M.~B.}\ \bibnamefont
  {Davies}}, \bibinfo {author} {\bibfnamefont {W.}~\bibnamefont {Benz}}, \ and\
  \bibinfo {author} {\bibfnamefont {T.}~\bibnamefont {Piran}},\ }\href@noop {}
  {\bibfield  {journal} {\bibinfo  {journal} {Astron. Astrophys.}\ }\textbf
  {\bibinfo {volume} {341}},\ \bibinfo {pages} {499--526} (\bibinfo {year}
  {1999})},\ \Eprint {http://arxiv.org/abs/astro-ph/9811367}
  {arXiv:astro-ph/9811367 [astro-ph]}\BibitemShut {NoStop}%
\bibitem [{\citenamefont {Li}\ and\ \citenamefont
  {Paczynski}(1998)}]{Li:1998bw}%
  \BibitemOpen
  \bibfield  {author} {\bibinfo {author} {\bibfnamefont {Li-Xin}\ \bibnamefont
  {Li}}\ and\ \bibinfo {author} {\bibfnamefont {Bohdan}\ \bibnamefont
  {Paczynski}},\ }\href {\doibase10.1086/311680} {\bibfield  {journal}
  {\bibinfo  {journal} {Astrophys. J.}\ }\textbf {\bibinfo {volume} {507}},\
  \bibinfo {pages} {L59} (\bibinfo {year} {1998})},\ \Eprint
  {http://arxiv.org/abs/astro-ph/9807272} {arXiv:astro-ph/9807272
  [astro-ph]}\BibitemShut {NoStop}%
\bibitem [{\citenamefont {Kasen}\ \emph {et~al.}(2017)\citenamefont {Kasen},
  \citenamefont {Metzger}, \citenamefont {Barnes}, \citenamefont {Quataert},\
  and\ \citenamefont {Ramirez-Ruiz}}]{Kasen:2017sxr}%
  \BibitemOpen
  \bibfield  {author} {\bibinfo {author} {\bibfnamefont {Daniel}\ \bibnamefont
  {Kasen}}, \bibinfo {author} {\bibfnamefont {Brian}\ \bibnamefont {Metzger}},
  \bibinfo {author} {\bibfnamefont {Jennifer}\ \bibnamefont {Barnes}}, \bibinfo
  {author} {\bibfnamefont {Eliot}\ \bibnamefont {Quataert}}, \ and\ \bibinfo
  {author} {\bibfnamefont {Enrico}\ \bibnamefont {Ramirez-Ruiz}},\ }\href
  {\doibase10.1038/nature24453} {\bibfield  {journal} {\bibinfo  {journal}
  {Nature}\ } (\bibinfo {year} {2017}),\ 10.1038/nature24453},\ \bibinfo {note}
  {[Nature551,80(2017)]},\ \Eprint {http://arxiv.org/abs/1710.05463}
  {arXiv:1710.05463 [astro-ph.HE]}\BibitemShut {NoStop}%
\bibitem [{\citenamefont {Nicholl}\ \emph {et~al.}(2017)\citenamefont {Nicholl}
  \emph {et~al.}}]{Nicholl:2017ahq}%
  \BibitemOpen
  \bibfield  {author} {\bibinfo {author} {\bibfnamefont {M.}~\bibnamefont
  {Nicholl}} \emph {et~al.},\ }\href {\doibase10.3847/2041-8213/aa9029}
  {\bibfield  {journal} {\bibinfo  {journal} {Astrophys. J.}\ }\textbf
  {\bibinfo {volume} {848}},\ \bibinfo {pages} {L18} (\bibinfo {year}
  {2017})},\ \Eprint {http://arxiv.org/abs/1710.05456} {arXiv:1710.05456
  [astro-ph.HE]}\BibitemShut {NoStop}%
\bibitem [{\citenamefont {Waxman}\ \emph {et~al.}(2017)\citenamefont {Waxman},
  \citenamefont {Ofek}, \citenamefont {Kushnir},\ and\ \citenamefont
  {Gal-Yam}}]{Waxman:2017sqv}%
  \BibitemOpen
  \bibfield  {author} {\bibinfo {author} {\bibfnamefont {Eli}\ \bibnamefont
  {Waxman}}, \bibinfo {author} {\bibfnamefont {Eran~O.}\ \bibnamefont {Ofek}},
  \bibinfo {author} {\bibfnamefont {Doron}\ \bibnamefont {Kushnir}}, \ and\
  \bibinfo {author} {\bibfnamefont {Avishay}\ \bibnamefont {Gal-Yam}},\
  }\href@noop {} {\  (\bibinfo {year} {2017})},\ \Eprint
  {http://arxiv.org/abs/1711.09638} {arXiv:1711.09638
  [astro-ph.HE]}\BibitemShut {NoStop}%
\bibitem [{\citenamefont {Nakar}\ and\ \citenamefont
  {Piran}(2011)}]{Nakar:2011cw}%
  \BibitemOpen
  \bibfield  {author} {\bibinfo {author} {\bibfnamefont {Ehud}\ \bibnamefont
  {Nakar}}\ and\ \bibinfo {author} {\bibfnamefont {Tsvi}\ \bibnamefont
  {Piran}},\ }\href {\doibase10.1038/nature10365} {\bibfield  {journal}
  {\bibinfo  {journal} {Nature}\ }\textbf {\bibinfo {volume} {478}},\ \bibinfo
  {pages} {82--84} (\bibinfo {year} {2011})},\ \Eprint
  {http://arxiv.org/abs/1102.1020} {arXiv:1102.1020 [astro-ph.HE]}\BibitemShut
  {NoStop}%
\bibitem [{\citenamefont {Fong}\ \emph {et~al.}(2015)\citenamefont {Fong},
  \citenamefont {Berger}, \citenamefont {Margutti},\ and\ \citenamefont
  {Zauderer}}]{Fong:2015oha}%
  \BibitemOpen
  \bibfield  {author} {\bibinfo {author} {\bibfnamefont {Wen-fai}\ \bibnamefont
  {Fong}}, \bibinfo {author} {\bibfnamefont {Edo}\ \bibnamefont {Berger}},
  \bibinfo {author} {\bibfnamefont {Raffaella}\ \bibnamefont {Margutti}}, \
  and\ \bibinfo {author} {\bibfnamefont {B.~Ashley}\ \bibnamefont {Zauderer}},\
  }\href {\doibase10.1088/0004-637X/815/2/102} {\bibfield  {journal} {\bibinfo
  {journal} {Astrophys. J.}\ }\textbf {\bibinfo {volume} {815}},\ \bibinfo
  {pages} {102} (\bibinfo {year} {2015})},\ \Eprint
  {http://arxiv.org/abs/1509.02922} {arXiv:1509.02922
  [astro-ph.HE]}\BibitemShut {NoStop}%
\bibitem [{\citenamefont {Horesh}\ \emph {et~al.}(2016)\citenamefont {Horesh},
  \citenamefont {Hotokezaka}, \citenamefont {Piran}, \citenamefont {Nakar},\
  and\ \citenamefont {Hancock}}]{Horesh:2016dah}%
  \BibitemOpen
  \bibfield  {author} {\bibinfo {author} {\bibfnamefont {Assaf}\ \bibnamefont
  {Horesh}}, \bibinfo {author} {\bibfnamefont {Kenta}\ \bibnamefont
  {Hotokezaka}}, \bibinfo {author} {\bibfnamefont {Tsvi}\ \bibnamefont
  {Piran}}, \bibinfo {author} {\bibfnamefont {Ehud}\ \bibnamefont {Nakar}}, \
  and\ \bibinfo {author} {\bibfnamefont {Paul}\ \bibnamefont {Hancock}},\
  }\href {\doibase10.3847/2041-8205/819/2/L22} {\bibfield  {journal} {\bibinfo
  {journal} {Astrophys. J.}\ }\textbf {\bibinfo {volume} {819}},\ \bibinfo
  {pages} {L22} (\bibinfo {year} {2016})},\ \Eprint
  {http://arxiv.org/abs/1601.01692} {arXiv:1601.01692
  [astro-ph.HE]}\BibitemShut {NoStop}%
\bibitem [{\citenamefont {Fong}\ \emph {et~al.}(2016)\citenamefont {Fong},
  \citenamefont {Metzger}, \citenamefont {Berger},\ and\ \citenamefont
  {Ozel}}]{Fong:2016orv}%
  \BibitemOpen
  \bibfield  {author} {\bibinfo {author} {\bibfnamefont {Wen-fai}\ \bibnamefont
  {Fong}}, \bibinfo {author} {\bibfnamefont {Brian~D.}\ \bibnamefont
  {Metzger}}, \bibinfo {author} {\bibfnamefont {Edo}\ \bibnamefont {Berger}}, \
  and\ \bibinfo {author} {\bibfnamefont {Feryal}\ \bibnamefont {Ozel}},\ }\href
  {\doibase10.3847/0004-637X/831/2/141} {\bibfield  {journal} {\bibinfo
  {journal} {Astrophys. J.}\ }\textbf {\bibinfo {volume} {831}},\ \bibinfo
  {pages} {141} (\bibinfo {year} {2016})},\ \Eprint
  {http://arxiv.org/abs/1607.00416} {arXiv:1607.00416
  [astro-ph.HE]}\BibitemShut {NoStop}%
\bibitem [{\citenamefont {Alexander}\ \emph {et~al.}(2017)\citenamefont
  {Alexander} \emph {et~al.}}]{Alexander:2017aly}%
  \BibitemOpen
  \bibfield  {author} {\bibinfo {author} {\bibfnamefont {K.~D.}\ \bibnamefont
  {Alexander}} \emph {et~al.},\ }\href {\doibase10.3847/2041-8213/aa905d}
  {\bibfield  {journal} {\bibinfo  {journal} {Astrophys. J.}\ }\textbf
  {\bibinfo {volume} {848}},\ \bibinfo {pages} {L21} (\bibinfo {year}
  {2017})},\ \Eprint {http://arxiv.org/abs/1710.05457} {arXiv:1710.05457
  [astro-ph.HE]}\BibitemShut {NoStop}%
\bibitem [{\citenamefont {Kim}\ \emph {et~al.}(2017)\citenamefont {Kim} \emph
  {et~al.}}]{Kim:2017skw}%
  \BibitemOpen
  \bibfield  {author} {\bibinfo {author} {\bibfnamefont {S.}~\bibnamefont
  {Kim}} \emph {et~al.} (\bibinfo {collaboration} {ALMA}),\ }\href
  {\doibase10.3847/2041-8213/aa970b} {\bibfield  {journal} {\bibinfo  {journal}
  {Astrophys. J.}\ }\textbf {\bibinfo {volume} {850}},\ \bibinfo {pages} {L21}
  (\bibinfo {year} {2017})},\ \Eprint {http://arxiv.org/abs/1710.05847}
  {arXiv:1710.05847 [astro-ph.HE]}\BibitemShut {NoStop}%
\bibitem [{\citenamefont {Lamb}\ and\ \citenamefont
  {Kobayashi}(2017)}]{Lamb:2017ych}%
  \BibitemOpen
  \bibfield  {author} {\bibinfo {author} {\bibfnamefont {Gavin~P.}\
  \bibnamefont {Lamb}}\ and\ \bibinfo {author} {\bibfnamefont {Shiho}\
  \bibnamefont {Kobayashi}},\ }\href {\doibase10.1093/mnras/stx2345} {\bibfield
   {journal} {\bibinfo  {journal} {Mon. Not. Roy. Astron. Soc.}\ }\textbf
  {\bibinfo {volume} {472}},\ \bibinfo {pages} {4953--4964} (\bibinfo {year}
  {2017})},\ \Eprint {http://arxiv.org/abs/1706.03000} {arXiv:1706.03000
  [astro-ph.HE]}\BibitemShut {NoStop}%
\bibitem [{\citenamefont {Gottlieb}\ \emph {et~al.}(2017)\citenamefont
  {Gottlieb}, \citenamefont {Nakar}, \citenamefont {Piran},\ and\ \citenamefont
  {Hotokezaka}}]{Gottlieb:2017pju}%
  \BibitemOpen
  \bibfield  {author} {\bibinfo {author} {\bibfnamefont {Ore}\ \bibnamefont
  {Gottlieb}}, \bibinfo {author} {\bibfnamefont {Ehud}\ \bibnamefont {Nakar}},
  \bibinfo {author} {\bibfnamefont {Tsvi}\ \bibnamefont {Piran}}, \ and\
  \bibinfo {author} {\bibfnamefont {Kenta}\ \bibnamefont {Hotokezaka}},\
  }\href@noop {} {\  (\bibinfo {year} {2017})},\ \Eprint
  {http://arxiv.org/abs/1710.05896} {arXiv:1710.05896
  [astro-ph.HE]}\BibitemShut {NoStop}%
\bibitem [{\citenamefont {{Grimm}}\ \emph {et~al.}(2002)\citenamefont
  {{Grimm}}, \citenamefont {{Gilfanov}},\ and\ \citenamefont
  {{Sunyaev}}}]{Grimm2002}%
  \BibitemOpen
  \bibfield  {author} {\bibinfo {author} {\bibfnamefont {H.-J.}\ \bibnamefont
  {{Grimm}}}, \bibinfo {author} {\bibfnamefont {M.}~\bibnamefont {{Gilfanov}}},
  \ and\ \bibinfo {author} {\bibfnamefont {R.}~\bibnamefont {{Sunyaev}}},\
  }\href {\doibase10.1051/0004-6361:20020826} {\bibfield  {journal} {\bibinfo
  {journal} {\aap}\ }\textbf {\bibinfo {volume} {391}},\ \bibinfo {pages}
  {923--944} (\bibinfo {year} {2002})},\ \Eprint
  {http://arxiv.org/abs/astro-ph/0109239} {astro-ph/0109239}\BibitemShut
  {NoStop}%
\bibitem [{\citenamefont {{Liu}}\ \emph {et~al.}(2006)\citenamefont {{Liu}},
  \citenamefont {{van Paradijs}},\ and\ \citenamefont {{van den
  Heuvel}}}]{Liu2006}%
  \BibitemOpen
  \bibfield  {author} {\bibinfo {author} {\bibfnamefont {Q.~Z.}\ \bibnamefont
  {{Liu}}}, \bibinfo {author} {\bibfnamefont {J.}~\bibnamefont {{van
  Paradijs}}}, \ and\ \bibinfo {author} {\bibfnamefont {E.~P.~J.}\ \bibnamefont
  {{van den Heuvel}}},\ }\href {\doibase10.1051/0004-6361:20064987} {\bibfield
  {journal} {\bibinfo  {journal} {\aap}\ }\textbf {\bibinfo {volume} {455}},\
  \bibinfo {pages} {1165--1168} (\bibinfo {year} {2006})},\ \Eprint
  {http://arxiv.org/abs/0707.0549} {arXiv:0707.0549}\BibitemShut {NoStop}%
\bibitem [{\citenamefont {{Liu}}\ \emph {et~al.}(2007)\citenamefont {{Liu}},
  \citenamefont {{van Paradijs}},\ and\ \citenamefont {{van den
  Heuvel}}}]{Liu2007}%
  \BibitemOpen
  \bibfield  {author} {\bibinfo {author} {\bibfnamefont {Q.~Z.}\ \bibnamefont
  {{Liu}}}, \bibinfo {author} {\bibfnamefont {J.}~\bibnamefont {{van
  Paradijs}}}, \ and\ \bibinfo {author} {\bibfnamefont {E.~P.~J.}\ \bibnamefont
  {{van den Heuvel}}},\ }\href {\doibase10.1051/0004-6361:20077303} {\bibfield
  {journal} {\bibinfo  {journal} {\aap}\ }\textbf {\bibinfo {volume} {469}},\
  \bibinfo {pages} {807--810} (\bibinfo {year} {2007})},\ \Eprint
  {http://arxiv.org/abs/0707.0544} {arXiv:0707.0544}\BibitemShut {NoStop}%
\bibitem [{\citenamefont {{McClintock}}\ and\ \citenamefont
  {{Remillard}}(2006)}]{BHrev2006}%
  \BibitemOpen
  \bibfield  {author} {\bibinfo {author} {\bibfnamefont {J.~E.}\ \bibnamefont
  {{McClintock}}}\ and\ \bibinfo {author} {\bibfnamefont {R.~A.}\ \bibnamefont
  {{Remillard}}},\ }\enquote {\bibinfo {title} {{Black hole binaries}},}\ in\
  \href {\doibase10.1017/CBO9780511536281} {\emph {\bibinfo {booktitle}
  {Compact stellar X-ray sources}}},\ \bibinfo {editor} {edited by\ \bibinfo
  {editor} {\bibfnamefont {W.~H.~G.}\ \bibnamefont {{Lewin}}}\ and\ \bibinfo
  {editor} {\bibfnamefont {M.}~\bibnamefont {{van der Klis}}}}\ (\bibinfo
  {publisher} {Cambridge University Press},\ \bibinfo {year} {2006})\
  Chap.~\bibinfo {chapter} {4}, pp.\ \bibinfo {pages} {157--213}\BibitemShut
  {NoStop}%
\bibitem [{\citenamefont {{Tetarenko}}\ \emph {et~al.}(2016)\citenamefont
  {{Tetarenko}}, \citenamefont {{Sivakoff}}, \citenamefont {{Heinke}},\ and\
  \citenamefont {{Gladstone}}}]{BHcat2016}%
  \BibitemOpen
  \bibfield  {author} {\bibinfo {author} {\bibfnamefont {B.~E.}\ \bibnamefont
  {{Tetarenko}}}, \bibinfo {author} {\bibfnamefont {G.~R.}\ \bibnamefont
  {{Sivakoff}}}, \bibinfo {author} {\bibfnamefont {C.~O.}\ \bibnamefont
  {{Heinke}}}, \ and\ \bibinfo {author} {\bibfnamefont {J.~C.}\ \bibnamefont
  {{Gladstone}}},\ }\href {\doibase10.3847/0067-0049/222/2/15} {\bibfield
  {journal} {\bibinfo  {journal} {Astrophysical Journal}\ }\textbf {\bibinfo
  {volume} {222}},\ \bibinfo {eid} {15} (\bibinfo {year} {2016})},\ \Eprint
  {http://arxiv.org/abs/1512.00778} {arXiv:1512.00778
  [astro-ph.HE]}\BibitemShut {NoStop}%
\bibitem [{TeV(accessed June 2018)}]{TeVCat}%
  \BibitemOpen
  \href@noop {} {\enquote {\bibinfo {title} {{TeVCat}},}\ }\bibinfo
  {howpublished} {\url{http://tevcat2.uchicago.edu/}} (\bibinfo {year}
  {accessed June 2018})\BibitemShut {NoStop}%
\bibitem [{\citenamefont {{Dubus}}(2013)}]{dubus_review13}%
  \BibitemOpen
  \bibfield  {author} {\bibinfo {author} {\bibfnamefont {G.}~\bibnamefont
  {{Dubus}}},\ }\href {\doibase10.1007/s00159-013-0064-5} {\bibfield  {journal}
  {\bibinfo  {journal} {\aapr}\ }\textbf {\bibinfo {volume} {21}},\ \bibinfo
  {pages} {64} (\bibinfo {year} {2013})},\ \Eprint
  {http://arxiv.org/abs/1307.7083} {arXiv:1307.7083 [astro-ph.HE]}\BibitemShut
  {NoStop}%
\bibitem [{\citenamefont {{Mirzoyan}}\ and\ \citenamefont
  {{Mukherjee}}(2017)}]{PSRJ2032_ATel_TeV}%
  \BibitemOpen
  \bibfield  {author} {\bibinfo {author} {\bibfnamefont {R.}~\bibnamefont
  {{Mirzoyan}}}\ and\ \bibinfo {author} {\bibfnamefont {R.}~\bibnamefont
  {{Mukherjee}}},\ }\href@noop {} {\bibfield  {journal} {\bibinfo  {journal}
  {The Astronomer's Telegram}\ }\textbf {\bibinfo {volume} {10971}} (\bibinfo
  {year} {2017})}\BibitemShut {NoStop}%
\bibitem [{\citenamefont {{Eger}}\ \emph {et~al.}(2016)\citenamefont {{Eger}},
  \citenamefont {{Laffon}}, \citenamefont {{Bordas}}, \citenamefont {{de
  O{\~n}a Whilhelmi}}, \citenamefont {{Hinton}},\ and\ \citenamefont
  {{P{\"u}hlhofer}}}]{2016MNRAS.457.1753E}%
  \BibitemOpen
  \bibfield  {author} {\bibinfo {author} {\bibfnamefont {P.}~\bibnamefont
  {{Eger}}}, \bibinfo {author} {\bibfnamefont {H.}~\bibnamefont {{Laffon}}},
  \bibinfo {author} {\bibfnamefont {P.}~\bibnamefont {{Bordas}}}, \bibinfo
  {author} {\bibfnamefont {E.}~\bibnamefont {{de O{\~n}a Whilhelmi}}}, \bibinfo
  {author} {\bibfnamefont {J.}~\bibnamefont {{Hinton}}}, \ and\ \bibinfo
  {author} {\bibfnamefont {G.}~\bibnamefont {{P{\"u}hlhofer}}},\ }\href
  {\doibase10.1093/mnras/stw125} {\bibfield  {journal} {\bibinfo  {journal}
  {MNRAS}\ }\textbf {\bibinfo {volume} {457}},\ \bibinfo {pages} {1753--1758}
  (\bibinfo {year} {2016})},\ \Eprint {http://arxiv.org/abs/1601.03208}
  {arXiv:1601.03208 [astro-ph.HE]}\BibitemShut {NoStop}%
\bibitem [{\citenamefont {{Abdo}}\ \emph {et~al.}(2011)\citenamefont {{Abdo}},
  \citenamefont {{Ackermann}}, \citenamefont {{Ajello}}, \citenamefont
  {{Allafort}}, \citenamefont {{Ballet}}, \citenamefont {{Barbiellini}},
  \citenamefont {{Bastieri}}, \citenamefont {{Bechtol}}, \citenamefont
  {{Bellazzini}}, \citenamefont {{Berenji}}, \citenamefont {{Blandford}},
  \citenamefont {{Bonamente}}, \citenamefont {{Borgland}}, \citenamefont
  {{Bregeon}}, \citenamefont {{Brigida}}, \citenamefont {{Bruel}},
  \citenamefont {{Buehler}}, \citenamefont {{Buson}}, \citenamefont
  {{Caliandro}}, \citenamefont {{Cameron}}, \citenamefont {{Camilo}},
  \citenamefont {{Caraveo}}, \citenamefont {{Cecchi}}, \citenamefont
  {{Charles}}, \citenamefont {{Chaty}}, \citenamefont {{Chekhtman}},
  \citenamefont {{Chernyakova}}, \citenamefont {{Cheung}}, \citenamefont
  {{Chiang}}, \citenamefont {{Ciprini}}, \citenamefont {{Claus}}, \citenamefont
  {{Cohen-Tanugi}}, \citenamefont {{Cominsky}}, \citenamefont {{Corbel}},
  \citenamefont {{Cutini}}, \citenamefont {{D'Ammando}}, \citenamefont {{de
  Angelis}}, \citenamefont {{den Hartog}}, \citenamefont {{de Palma}},
  \citenamefont {{Dermer}}, \citenamefont {{Digel}}, \citenamefont {{Silva}},
  \citenamefont {{Dormody}}, \citenamefont {{Drell}}, \citenamefont
  {{Drlica-Wagner}}, \citenamefont {{Dubois}}, \citenamefont {{Dubus}},
  \citenamefont {{Dumora}}, \citenamefont {{Enoto}}, \citenamefont
  {{Espinoza}}, \citenamefont {{Favuzzi}}, \citenamefont {{Fegan}},
  \citenamefont {{Ferrara}}, \citenamefont {{Focke}}, \citenamefont {{Fortin}},
  \citenamefont {{Fukazawa}}, \citenamefont {{Funk}}, \citenamefont {{Fusco}},
  \citenamefont {{Gargano}}, \citenamefont {{Gasparrini}}, \citenamefont
  {{Gehrels}}, \citenamefont {{Germani}}, \citenamefont {{Giglietto}},
  \citenamefont {{Giommi}}, \citenamefont {{Giordano}}, \citenamefont
  {{Giroletti}}, \citenamefont {{Glanzman}}, \citenamefont {{Godfrey}},
  \citenamefont {{Grenier}}, \citenamefont {{Grondin}}, \citenamefont
  {{Grove}}, \citenamefont {{Grundstrom}}, \citenamefont {{Guiriec}},
  \citenamefont {{Gwon}}, \citenamefont {{Hadasch}}, \citenamefont {{Harding}},
  \citenamefont {{Hayashida}}, \citenamefont {{Hays}}, \citenamefont
  {{J{\'o}hannesson}}, \citenamefont {{Johnson}}, \citenamefont {{Johnson}},
  \citenamefont {{Johnston}}, \citenamefont {{Kamae}}, \citenamefont
  {{Katagiri}}, \citenamefont {{Kataoka}}, \citenamefont {{Keith}},
  \citenamefont {{Kerr}}, \citenamefont {{Kn{\"o}dlseder}}, \citenamefont
  {{Kramer}}, \citenamefont {{Kuss}}, \citenamefont {{Lande}}, \citenamefont
  {{Lee}}, \citenamefont {{Lemoine-Goumard}}, \citenamefont {{Longo}},
  \citenamefont {{Loparco}}, \citenamefont {{Lovellette}}, \citenamefont
  {{Lubrano}}, \citenamefont {{Manchester}}, \citenamefont {{Marelli}},
  \citenamefont {{Mazziotta}}, \citenamefont {{Michelson}}, \citenamefont
  {{Mitthumsiri}}, \citenamefont {{Mizuno}}, \citenamefont {{Moiseev}},
  \citenamefont {{Monte}}, \citenamefont {{Monzani}}, \citenamefont
  {{Morselli}}, \citenamefont {{Moskalenko}}, \citenamefont {{Murgia}},
  \citenamefont {{Nakamori}}, \citenamefont {{Naumann-Godo}}, \citenamefont
  {{Neronov}}, \citenamefont {{Nolan}}, \citenamefont {{Norris}}, \citenamefont
  {{Noutsos}}, \citenamefont {{Nuss}}, \citenamefont {{Ohsugi}}, \citenamefont
  {{Okumura}}, \citenamefont {{Omodei}}, \citenamefont {{Orlando}},
  \citenamefont {{Paneque}}, \citenamefont {{Parent}}, \citenamefont
  {{Pesce-Rollins}}, \citenamefont {{Pierbattista}}, \citenamefont {{Piron}},
  \citenamefont {{Porter}}, \citenamefont {{Possenti}}, \citenamefont
  {{Rain{\`o}}}, \citenamefont {{Rando}}, \citenamefont {{Ray}}, \citenamefont
  {{Razzano}}, \citenamefont {{Razzaque}}, \citenamefont {{Reimer}},
  \citenamefont {{Reimer}}, \citenamefont {{Reposeur}}, \citenamefont {{Ritz}},
  \citenamefont {{Sadrozinski}}, \citenamefont {{Scargle}}, \citenamefont
  {{Sgr{\`o}}}, \citenamefont {{Shannon}}, \citenamefont {{Siskind}},
  \citenamefont {{Smith}}, \citenamefont {{Spandre}}, \citenamefont
  {{Spinelli}}, \citenamefont {{Strickman}}, \citenamefont {{Suson}},
  \citenamefont {{Takahashi}}, \citenamefont {{Tanaka}}, \citenamefont
  {{Thayer}}, \citenamefont {{Thayer}}, \citenamefont {{Thompson}},
  \citenamefont {{Thorsett}}, \citenamefont {{Tibaldo}}, \citenamefont
  {{Tibolla}}, \citenamefont {{Torres}}, \citenamefont {{Tosti}}, \citenamefont
  {{Troja}}, \citenamefont {{Uchiyama}}, \citenamefont {{Usher}}, \citenamefont
  {{Vandenbroucke}}, \citenamefont {{Vasileiou}}, \citenamefont {{Vianello}},
  \citenamefont {{Vitale}}, \citenamefont {{Waite}}, \citenamefont {{Wang}},
  \citenamefont {{Winer}}, \citenamefont {{Wolff}}, \citenamefont {{Wood}},
  \citenamefont {{Wood}}, \citenamefont {{Yang}}, \citenamefont {{Ziegler}},\
  and\ \citenamefont {{Zimmer}}}]{Abdo2011}%
  \BibitemOpen
  \bibfield  {author} {\bibinfo {author} {\bibfnamefont {A.~A.}\ \bibnamefont
  {{Abdo}}}, \bibinfo {author} {\bibfnamefont {M.}~\bibnamefont {{Ackermann}}},
  \bibinfo {author} {\bibfnamefont {M.}~\bibnamefont {{Ajello}}}, \bibinfo
  {author} {\bibfnamefont {A.}~\bibnamefont {{Allafort}}}, \bibinfo {author}
  {\bibfnamefont {J.}~\bibnamefont {{Ballet}}}, \bibinfo {author}
  {\bibfnamefont {G.}~\bibnamefont {{Barbiellini}}}, \bibinfo {author}
  {\bibfnamefont {D.}~\bibnamefont {{Bastieri}}}, \bibinfo {author}
  {\bibfnamefont {K.}~\bibnamefont {{Bechtol}}}, \bibinfo {author}
  {\bibfnamefont {R.}~\bibnamefont {{Bellazzini}}}, \bibinfo {author}
  {\bibfnamefont {B.}~\bibnamefont {{Berenji}}}, \bibinfo {author}
  {\bibfnamefont {R.~D.}\ \bibnamefont {{Blandford}}}, \bibinfo {author}
  {\bibfnamefont {E.}~\bibnamefont {{Bonamente}}}, \bibinfo {author}
  {\bibfnamefont {A.~W.}\ \bibnamefont {{Borgland}}}, \bibinfo {author}
  {\bibfnamefont {J.}~\bibnamefont {{Bregeon}}}, \bibinfo {author}
  {\bibfnamefont {M.}~\bibnamefont {{Brigida}}}, \bibinfo {author}
  {\bibfnamefont {P.}~\bibnamefont {{Bruel}}}, \bibinfo {author} {\bibfnamefont
  {R.}~\bibnamefont {{Buehler}}}, \bibinfo {author} {\bibfnamefont
  {S.}~\bibnamefont {{Buson}}}, \bibinfo {author} {\bibfnamefont {G.~A.}\
  \bibnamefont {{Caliandro}}}, \bibinfo {author} {\bibfnamefont {R.~A.}\
  \bibnamefont {{Cameron}}}, \bibinfo {author} {\bibfnamefont {F.}~\bibnamefont
  {{Camilo}}}, \bibinfo {author} {\bibfnamefont {P.~A.}\ \bibnamefont
  {{Caraveo}}}, \bibinfo {author} {\bibfnamefont {C.}~\bibnamefont {{Cecchi}}},
  \bibinfo {author} {\bibfnamefont {E.}~\bibnamefont {{Charles}}}, \bibinfo
  {author} {\bibfnamefont {S.}~\bibnamefont {{Chaty}}}, \bibinfo {author}
  {\bibfnamefont {A.}~\bibnamefont {{Chekhtman}}}, \bibinfo {author}
  {\bibfnamefont {M.}~\bibnamefont {{Chernyakova}}}, \bibinfo {author}
  {\bibfnamefont {C.~C.}\ \bibnamefont {{Cheung}}}, \bibinfo {author}
  {\bibfnamefont {J.}~\bibnamefont {{Chiang}}}, \bibinfo {author}
  {\bibfnamefont {S.}~\bibnamefont {{Ciprini}}}, \bibinfo {author}
  {\bibfnamefont {R.}~\bibnamefont {{Claus}}}, \bibinfo {author} {\bibfnamefont
  {J.}~\bibnamefont {{Cohen-Tanugi}}}, \bibinfo {author} {\bibfnamefont
  {L.~R.}\ \bibnamefont {{Cominsky}}}, \bibinfo {author} {\bibfnamefont
  {S.}~\bibnamefont {{Corbel}}}, \bibinfo {author} {\bibfnamefont
  {S.}~\bibnamefont {{Cutini}}}, \bibinfo {author} {\bibfnamefont
  {F.}~\bibnamefont {{D'Ammando}}}, \bibinfo {author} {\bibfnamefont
  {A.}~\bibnamefont {{de Angelis}}}, \bibinfo {author} {\bibfnamefont {P.~R.}\
  \bibnamefont {{den Hartog}}}, \bibinfo {author} {\bibfnamefont
  {F.}~\bibnamefont {{de Palma}}}, \bibinfo {author} {\bibfnamefont {C.~D.}\
  \bibnamefont {{Dermer}}}, \bibinfo {author} {\bibfnamefont {S.~W.}\
  \bibnamefont {{Digel}}}, \bibinfo {author} {\bibfnamefont {E.~d.~C.~e.}\
  \bibnamefont {{Silva}}}, \bibinfo {author} {\bibfnamefont {M.}~\bibnamefont
  {{Dormody}}}, \bibinfo {author} {\bibfnamefont {P.~S.}\ \bibnamefont
  {{Drell}}}, \bibinfo {author} {\bibfnamefont {A.}~\bibnamefont
  {{Drlica-Wagner}}}, \bibinfo {author} {\bibfnamefont {R.}~\bibnamefont
  {{Dubois}}}, \bibinfo {author} {\bibfnamefont {G.}~\bibnamefont {{Dubus}}},
  \bibinfo {author} {\bibfnamefont {D.}~\bibnamefont {{Dumora}}}, \bibinfo
  {author} {\bibfnamefont {T.}~\bibnamefont {{Enoto}}}, \bibinfo {author}
  {\bibfnamefont {C.~M.}\ \bibnamefont {{Espinoza}}}, \bibinfo {author}
  {\bibfnamefont {C.}~\bibnamefont {{Favuzzi}}}, \bibinfo {author}
  {\bibfnamefont {S.~J.}\ \bibnamefont {{Fegan}}}, \bibinfo {author}
  {\bibfnamefont {E.~C.}\ \bibnamefont {{Ferrara}}}, \bibinfo {author}
  {\bibfnamefont {W.~B.}\ \bibnamefont {{Focke}}}, \bibinfo {author}
  {\bibfnamefont {P.}~\bibnamefont {{Fortin}}}, \bibinfo {author}
  {\bibfnamefont {Y.}~\bibnamefont {{Fukazawa}}}, \bibinfo {author}
  {\bibfnamefont {S.}~\bibnamefont {{Funk}}}, \bibinfo {author} {\bibfnamefont
  {P.}~\bibnamefont {{Fusco}}}, \bibinfo {author} {\bibfnamefont
  {F.}~\bibnamefont {{Gargano}}}, \bibinfo {author} {\bibfnamefont
  {D.}~\bibnamefont {{Gasparrini}}}, \bibinfo {author} {\bibfnamefont
  {N.}~\bibnamefont {{Gehrels}}}, \bibinfo {author} {\bibfnamefont
  {S.}~\bibnamefont {{Germani}}}, \bibinfo {author} {\bibfnamefont
  {N.}~\bibnamefont {{Giglietto}}}, \bibinfo {author} {\bibfnamefont
  {P.}~\bibnamefont {{Giommi}}}, \bibinfo {author} {\bibfnamefont
  {F.}~\bibnamefont {{Giordano}}}, \bibinfo {author} {\bibfnamefont
  {M.}~\bibnamefont {{Giroletti}}}, \bibinfo {author} {\bibfnamefont
  {T.}~\bibnamefont {{Glanzman}}}, \bibinfo {author} {\bibfnamefont
  {G.}~\bibnamefont {{Godfrey}}}, \bibinfo {author} {\bibfnamefont {I.~A.}\
  \bibnamefont {{Grenier}}}, \bibinfo {author} {\bibfnamefont {M.-H.}\
  \bibnamefont {{Grondin}}}, \bibinfo {author} {\bibfnamefont {J.~E.}\
  \bibnamefont {{Grove}}}, \bibinfo {author} {\bibfnamefont {E.}~\bibnamefont
  {{Grundstrom}}}, \bibinfo {author} {\bibfnamefont {S.}~\bibnamefont
  {{Guiriec}}}, \bibinfo {author} {\bibfnamefont {C.}~\bibnamefont {{Gwon}}},
  \bibinfo {author} {\bibfnamefont {D.}~\bibnamefont {{Hadasch}}}, \bibinfo
  {author} {\bibfnamefont {A.~K.}\ \bibnamefont {{Harding}}}, \bibinfo {author}
  {\bibfnamefont {M.}~\bibnamefont {{Hayashida}}}, \bibinfo {author}
  {\bibfnamefont {E.}~\bibnamefont {{Hays}}}, \bibinfo {author} {\bibfnamefont
  {G.}~\bibnamefont {{J{\'o}hannesson}}}, \bibinfo {author} {\bibfnamefont
  {A.~S.}\ \bibnamefont {{Johnson}}}, \bibinfo {author} {\bibfnamefont {T.~J.}\
  \bibnamefont {{Johnson}}}, \bibinfo {author} {\bibfnamefont {S.}~\bibnamefont
  {{Johnston}}}, \bibinfo {author} {\bibfnamefont {T.}~\bibnamefont {{Kamae}}},
  \bibinfo {author} {\bibfnamefont {H.}~\bibnamefont {{Katagiri}}}, \bibinfo
  {author} {\bibfnamefont {J.}~\bibnamefont {{Kataoka}}}, \bibinfo {author}
  {\bibfnamefont {M.}~\bibnamefont {{Keith}}}, \bibinfo {author} {\bibfnamefont
  {M.}~\bibnamefont {{Kerr}}}, \bibinfo {author} {\bibfnamefont
  {J.}~\bibnamefont {{Kn{\"o}dlseder}}}, \bibinfo {author} {\bibfnamefont
  {M.}~\bibnamefont {{Kramer}}}, \bibinfo {author} {\bibfnamefont
  {M.}~\bibnamefont {{Kuss}}}, \bibinfo {author} {\bibfnamefont
  {J.}~\bibnamefont {{Lande}}}, \bibinfo {author} {\bibfnamefont {S.-H.}\
  \bibnamefont {{Lee}}}, \bibinfo {author} {\bibfnamefont {M.}~\bibnamefont
  {{Lemoine-Goumard}}}, \bibinfo {author} {\bibfnamefont {F.}~\bibnamefont
  {{Longo}}}, \bibinfo {author} {\bibfnamefont {F.}~\bibnamefont {{Loparco}}},
  \bibinfo {author} {\bibfnamefont {M.~N.}\ \bibnamefont {{Lovellette}}},
  \bibinfo {author} {\bibfnamefont {P.}~\bibnamefont {{Lubrano}}}, \bibinfo
  {author} {\bibfnamefont {R.~N.}\ \bibnamefont {{Manchester}}}, \bibinfo
  {author} {\bibfnamefont {M.}~\bibnamefont {{Marelli}}}, \bibinfo {author}
  {\bibfnamefont {M.~N.}\ \bibnamefont {{Mazziotta}}}, \bibinfo {author}
  {\bibfnamefont {P.~F.}\ \bibnamefont {{Michelson}}}, \bibinfo {author}
  {\bibfnamefont {W.}~\bibnamefont {{Mitthumsiri}}}, \bibinfo {author}
  {\bibfnamefont {T.}~\bibnamefont {{Mizuno}}}, \bibinfo {author}
  {\bibfnamefont {A.~A.}\ \bibnamefont {{Moiseev}}}, \bibinfo {author}
  {\bibfnamefont {C.}~\bibnamefont {{Monte}}}, \bibinfo {author} {\bibfnamefont
  {M.~E.}\ \bibnamefont {{Monzani}}}, \bibinfo {author} {\bibfnamefont
  {A.}~\bibnamefont {{Morselli}}}, \bibinfo {author} {\bibfnamefont {I.~V.}\
  \bibnamefont {{Moskalenko}}}, \bibinfo {author} {\bibfnamefont
  {S.}~\bibnamefont {{Murgia}}}, \bibinfo {author} {\bibfnamefont
  {T.}~\bibnamefont {{Nakamori}}}, \bibinfo {author} {\bibfnamefont
  {M.}~\bibnamefont {{Naumann-Godo}}}, \bibinfo {author} {\bibfnamefont
  {A.}~\bibnamefont {{Neronov}}}, \bibinfo {author} {\bibfnamefont {P.~L.}\
  \bibnamefont {{Nolan}}}, \bibinfo {author} {\bibfnamefont {J.~P.}\
  \bibnamefont {{Norris}}}, \bibinfo {author} {\bibfnamefont {A.}~\bibnamefont
  {{Noutsos}}}, \bibinfo {author} {\bibfnamefont {E.}~\bibnamefont {{Nuss}}},
  \bibinfo {author} {\bibfnamefont {T.}~\bibnamefont {{Ohsugi}}}, \bibinfo
  {author} {\bibfnamefont {A.}~\bibnamefont {{Okumura}}}, \bibinfo {author}
  {\bibfnamefont {N.}~\bibnamefont {{Omodei}}}, \bibinfo {author}
  {\bibfnamefont {E.}~\bibnamefont {{Orlando}}}, \bibinfo {author}
  {\bibfnamefont {D.}~\bibnamefont {{Paneque}}}, \bibinfo {author}
  {\bibfnamefont {D.}~\bibnamefont {{Parent}}}, \bibinfo {author}
  {\bibfnamefont {M.}~\bibnamefont {{Pesce-Rollins}}}, \bibinfo {author}
  {\bibfnamefont {M.}~\bibnamefont {{Pierbattista}}}, \bibinfo {author}
  {\bibfnamefont {F.}~\bibnamefont {{Piron}}}, \bibinfo {author} {\bibfnamefont
  {T.~A.}\ \bibnamefont {{Porter}}}, \bibinfo {author} {\bibfnamefont
  {A.}~\bibnamefont {{Possenti}}}, \bibinfo {author} {\bibfnamefont
  {S.}~\bibnamefont {{Rain{\`o}}}}, \bibinfo {author} {\bibfnamefont
  {R.}~\bibnamefont {{Rando}}}, \bibinfo {author} {\bibfnamefont {P.~S.}\
  \bibnamefont {{Ray}}}, \bibinfo {author} {\bibfnamefont {M.}~\bibnamefont
  {{Razzano}}}, \bibinfo {author} {\bibfnamefont {S.}~\bibnamefont
  {{Razzaque}}}, \bibinfo {author} {\bibfnamefont {A.}~\bibnamefont
  {{Reimer}}}, \bibinfo {author} {\bibfnamefont {O.}~\bibnamefont {{Reimer}}},
  \bibinfo {author} {\bibfnamefont {T.}~\bibnamefont {{Reposeur}}}, \bibinfo
  {author} {\bibfnamefont {S.}~\bibnamefont {{Ritz}}}, \bibinfo {author}
  {\bibfnamefont {H.~F.-W.}\ \bibnamefont {{Sadrozinski}}}, \bibinfo {author}
  {\bibfnamefont {J.~D.}\ \bibnamefont {{Scargle}}}, \bibinfo {author}
  {\bibfnamefont {C.}~\bibnamefont {{Sgr{\`o}}}}, \bibinfo {author}
  {\bibfnamefont {R.}~\bibnamefont {{Shannon}}}, \bibinfo {author}
  {\bibfnamefont {E.~J.}\ \bibnamefont {{Siskind}}}, \bibinfo {author}
  {\bibfnamefont {P.~D.}\ \bibnamefont {{Smith}}}, \bibinfo {author}
  {\bibfnamefont {G.}~\bibnamefont {{Spandre}}}, \bibinfo {author}
  {\bibfnamefont {P.}~\bibnamefont {{Spinelli}}}, \bibinfo {author}
  {\bibfnamefont {M.~S.}\ \bibnamefont {{Strickman}}}, \bibinfo {author}
  {\bibfnamefont {D.~J.}\ \bibnamefont {{Suson}}}, \bibinfo {author}
  {\bibfnamefont {H.}~\bibnamefont {{Takahashi}}}, \bibinfo {author}
  {\bibfnamefont {T.}~\bibnamefont {{Tanaka}}}, \bibinfo {author}
  {\bibfnamefont {J.~G.}\ \bibnamefont {{Thayer}}}, \bibinfo {author}
  {\bibfnamefont {J.~B.}\ \bibnamefont {{Thayer}}}, \bibinfo {author}
  {\bibfnamefont {D.~J.}\ \bibnamefont {{Thompson}}}, \bibinfo {author}
  {\bibfnamefont {S.~E.}\ \bibnamefont {{Thorsett}}}, \bibinfo {author}
  {\bibfnamefont {L.}~\bibnamefont {{Tibaldo}}}, \bibinfo {author}
  {\bibfnamefont {O.}~\bibnamefont {{Tibolla}}}, \bibinfo {author}
  {\bibfnamefont {D.~F.}\ \bibnamefont {{Torres}}}, \bibinfo {author}
  {\bibfnamefont {G.}~\bibnamefont {{Tosti}}}, \bibinfo {author} {\bibfnamefont
  {E.}~\bibnamefont {{Troja}}}, \bibinfo {author} {\bibfnamefont
  {Y.}~\bibnamefont {{Uchiyama}}}, \bibinfo {author} {\bibfnamefont {T.~L.}\
  \bibnamefont {{Usher}}}, \bibinfo {author} {\bibfnamefont {J.}~\bibnamefont
  {{Vandenbroucke}}}, \bibinfo {author} {\bibfnamefont {V.}~\bibnamefont
  {{Vasileiou}}}, \bibinfo {author} {\bibfnamefont {G.}~\bibnamefont
  {{Vianello}}}, \bibinfo {author} {\bibfnamefont {V.}~\bibnamefont
  {{Vitale}}}, \bibinfo {author} {\bibfnamefont {A.~P.}\ \bibnamefont
  {{Waite}}}, \bibinfo {author} {\bibfnamefont {P.}~\bibnamefont {{Wang}}},
  \bibinfo {author} {\bibfnamefont {B.~L.}\ \bibnamefont {{Winer}}}, \bibinfo
  {author} {\bibfnamefont {M.~T.}\ \bibnamefont {{Wolff}}}, \bibinfo {author}
  {\bibfnamefont {D.~L.}\ \bibnamefont {{Wood}}}, \bibinfo {author}
  {\bibfnamefont {K.~S.}\ \bibnamefont {{Wood}}}, \bibinfo {author}
  {\bibfnamefont {Z.}~\bibnamefont {{Yang}}}, \bibinfo {author} {\bibfnamefont
  {M.}~\bibnamefont {{Ziegler}}}, \ and\ \bibinfo {author} {\bibfnamefont
  {S.}~\bibnamefont {{Zimmer}}},\ }\href {\doibase10.1088/2041-8205/736/1/L11}
  {\bibfield  {journal} {\bibinfo  {journal} {Astrophysical Journal}\ }\textbf
  {\bibinfo {volume} {736}},\ \bibinfo {eid} {L11} (\bibinfo {year} {2011})},\
  \Eprint {http://arxiv.org/abs/1103.4108} {arXiv:1103.4108
  [astro-ph.HE]}\BibitemShut {NoStop}%
\bibitem [{\citenamefont {{Chernyakova}}\ \emph {et~al.}(2015)\citenamefont
  {{Chernyakova}}, \citenamefont {{Neronov}}, \citenamefont {{van Soelen}},
  \citenamefont {{Callanan}}, \citenamefont {{O'Shaughnessy}}, \citenamefont
  {{Babyk}}, \citenamefont {{Tsygankov}}, \citenamefont {{Vovk}}, \citenamefont
  {{Krivonos}}, \citenamefont {{Tomsick}}, \citenamefont {{Malyshev}},
  \citenamefont {{Li}}, \citenamefont {{Wood}}, \citenamefont {{Torres}},
  \citenamefont {{Zhang}}, \citenamefont {{Kretschmar}}, \citenamefont
  {{McSwain}}, \citenamefont {{Buckley}},\ and\ \citenamefont
  {{Koen}}}]{Chernyakova2015}%
  \BibitemOpen
  \bibfield  {author} {\bibinfo {author} {\bibfnamefont {M.}~\bibnamefont
  {{Chernyakova}}}, \bibinfo {author} {\bibfnamefont {A.}~\bibnamefont
  {{Neronov}}}, \bibinfo {author} {\bibfnamefont {B.}~\bibnamefont {{van
  Soelen}}}, \bibinfo {author} {\bibfnamefont {P.}~\bibnamefont {{Callanan}}},
  \bibinfo {author} {\bibfnamefont {L.}~\bibnamefont {{O'Shaughnessy}}},
  \bibinfo {author} {\bibfnamefont {I.}~\bibnamefont {{Babyk}}}, \bibinfo
  {author} {\bibfnamefont {S.}~\bibnamefont {{Tsygankov}}}, \bibinfo {author}
  {\bibfnamefont {I.}~\bibnamefont {{Vovk}}}, \bibinfo {author} {\bibfnamefont
  {R.}~\bibnamefont {{Krivonos}}}, \bibinfo {author} {\bibfnamefont {J.~A.}\
  \bibnamefont {{Tomsick}}}, \bibinfo {author} {\bibfnamefont {D.}~\bibnamefont
  {{Malyshev}}}, \bibinfo {author} {\bibfnamefont {J.}~\bibnamefont {{Li}}},
  \bibinfo {author} {\bibfnamefont {K.}~\bibnamefont {{Wood}}}, \bibinfo
  {author} {\bibfnamefont {D.}~\bibnamefont {{Torres}}}, \bibinfo {author}
  {\bibfnamefont {S.}~\bibnamefont {{Zhang}}}, \bibinfo {author} {\bibfnamefont
  {P.}~\bibnamefont {{Kretschmar}}}, \bibinfo {author} {\bibfnamefont {M.~V.}\
  \bibnamefont {{McSwain}}}, \bibinfo {author} {\bibfnamefont {D.~A.~H.}\
  \bibnamefont {{Buckley}}}, \ and\ \bibinfo {author} {\bibfnamefont
  {C.}~\bibnamefont {{Koen}}},\ }\href {\doibase10.1093/mnras/stv1988}
  {\bibfield  {journal} {\bibinfo  {journal} {MNRAS}\ }\textbf {\bibinfo
  {volume} {454}},\ \bibinfo {pages} {1358--1370} (\bibinfo {year} {2015})},\
  \Eprint {http://arxiv.org/abs/1508.01339} {arXiv:1508.01339
  [astro-ph.HE]}\BibitemShut {NoStop}%
\bibitem [{\citenamefont {{Zdziarski}}\ \emph {et~al.}(2010)\citenamefont
  {{Zdziarski}}, \citenamefont {{Neronov}},\ and\ \citenamefont
  {{Chernyakova}}}]{zdz10}%
  \BibitemOpen
  \bibfield  {author} {\bibinfo {author} {\bibfnamefont {A.~A.}\ \bibnamefont
  {{Zdziarski}}}, \bibinfo {author} {\bibfnamefont {A.}~\bibnamefont
  {{Neronov}}}, \ and\ \bibinfo {author} {\bibfnamefont {M.}~\bibnamefont
  {{Chernyakova}}},\ }\href {\doibase10.1111/j.1365-2966.2010.16263.x}
  {\bibfield  {journal} {\bibinfo  {journal} {MNRAS}\ }\textbf {\bibinfo
  {volume} {403}},\ \bibinfo {pages} {1873--1886} (\bibinfo {year} {2010})},\
  \Eprint {http://arxiv.org/abs/0802.1174} {arXiv:0802.1174}\BibitemShut
  {NoStop}%
\bibitem [{\citenamefont {{Massi}}\ \emph {et~al.}(2017)\citenamefont
  {{Massi}}, \citenamefont {{Migliari}},\ and\ \citenamefont
  {{Chernyakova}}}]{Massi2017}%
  \BibitemOpen
  \bibfield  {author} {\bibinfo {author} {\bibfnamefont {M.}~\bibnamefont
  {{Massi}}}, \bibinfo {author} {\bibfnamefont {S.}~\bibnamefont {{Migliari}}},
  \ and\ \bibinfo {author} {\bibfnamefont {M.}~\bibnamefont {{Chernyakova}}},\
  }\href {\doibase10.1093/mnras/stx778} {\bibfield  {journal} {\bibinfo
  {journal} {MNRAS}\ }\textbf {\bibinfo {volume} {468}},\ \bibinfo {pages}
  {3689--3693} (\bibinfo {year} {2017})},\ \Eprint
  {http://arxiv.org/abs/1704.01335} {arXiv:1704.01335
  [astro-ph.HE]}\BibitemShut {NoStop}%
\bibitem [{\citenamefont {{Yang}}\ \emph {et~al.}(2015)\citenamefont {{Yang}},
  \citenamefont {{Xie}}, \citenamefont {{Yuan}}, \citenamefont {{Zdziarski}},
  \citenamefont {{Gierlinski}}, \citenamefont {{Ho}},\ and\ \citenamefont
  {{Yu}}}]{2015PKAS...30..565Y}%
  \BibitemOpen
  \bibfield  {author} {\bibinfo {author} {\bibfnamefont {Q.-X.}\ \bibnamefont
  {{Yang}}}, \bibinfo {author} {\bibfnamefont {F.-G.}\ \bibnamefont {{Xie}}},
  \bibinfo {author} {\bibfnamefont {F.}~\bibnamefont {{Yuan}}}, \bibinfo
  {author} {\bibfnamefont {A.~A.}\ \bibnamefont {{Zdziarski}}}, \bibinfo
  {author} {\bibfnamefont {M.}~\bibnamefont {{Gierlinski}}}, \bibinfo {author}
  {\bibfnamefont {L.~C.}\ \bibnamefont {{Ho}}}, \ and\ \bibinfo {author}
  {\bibfnamefont {Z.}~\bibnamefont {{Yu}}},\ }\href
  {\doibase10.5303/PKAS.2015.30.2.565} {\bibfield  {journal} {\bibinfo
  {journal} {Publication of Korean Astronomical Society}\ }\textbf {\bibinfo
  {volume} {30}},\ \bibinfo {pages} {565--568} (\bibinfo {year}
  {2015})}\BibitemShut {NoStop}%
\bibitem [{\citenamefont {Bosch-Ramon}\ and\ \citenamefont
  {Paredes}(2005)}]{BoschRamon:2004se}%
  \BibitemOpen
  \bibfield  {author} {\bibinfo {author} {\bibfnamefont {Valenti}\ \bibnamefont
  {Bosch-Ramon}}\ and\ \bibinfo {author} {\bibfnamefont {J.~M.}\ \bibnamefont
  {Paredes}},\ }\bibfield  {booktitle} {\emph {\bibinfo {booktitle} {{5th
  Microquasar Workshop: Microquasars and Related Astrophysics Beijing, China,
  June 7-13, 2004}}},\ }\href {\doibase10.1088/1009-9271/5/S1/133} {\bibfield
  {journal} {\bibinfo  {journal} {Chin. J. Astron. Astrophys.}\ }\textbf
  {\bibinfo {volume} {5}},\ \bibinfo {pages} {133--138} (\bibinfo {year}
  {2005})},\ \Eprint {http://arxiv.org/abs/astro-ph/0411509}
  {arXiv:astro-ph/0411509 [astro-ph]}\BibitemShut {NoStop}%
\bibitem [{\citenamefont {{Massi}}\ and\ \citenamefont
  {{Torricelli-Ciamponi}}(2016)}]{Massi2016}%
  \BibitemOpen
  \bibfield  {author} {\bibinfo {author} {\bibfnamefont {M.}~\bibnamefont
  {{Massi}}}\ and\ \bibinfo {author} {\bibfnamefont {G.}~\bibnamefont
  {{Torricelli-Ciamponi}}},\ }\href {\doibase10.1051/0004-6361/201526938}
  {\bibfield  {journal} {\bibinfo  {journal} {\aap}\ }\textbf {\bibinfo
  {volume} {585}},\ \bibinfo {eid} {A123} (\bibinfo {year} {2016})},\ \Eprint
  {http://arxiv.org/abs/1511.05621} {arXiv:1511.05621
  [astro-ph.HE]}\BibitemShut {NoStop}%
\bibitem [{\citenamefont {{Paredes-Fortuny}}\ \emph {et~al.}(2015)\citenamefont
  {{Paredes-Fortuny}}, \citenamefont {{Rib{\'o}}}, \citenamefont
  {{Bosch-Ramon}}, \citenamefont {{Casares}}, \citenamefont {{Fors}},\ and\
  \citenamefont {{N{\'u}{\~n}ez}}}]{fortuny15}%
  \BibitemOpen
  \bibfield  {author} {\bibinfo {author} {\bibfnamefont {X.}~\bibnamefont
  {{Paredes-Fortuny}}}, \bibinfo {author} {\bibfnamefont {M.}~\bibnamefont
  {{Rib{\'o}}}}, \bibinfo {author} {\bibfnamefont {V.}~\bibnamefont
  {{Bosch-Ramon}}}, \bibinfo {author} {\bibfnamefont {J.}~\bibnamefont
  {{Casares}}}, \bibinfo {author} {\bibfnamefont {O.}~\bibnamefont {{Fors}}}, \
  and\ \bibinfo {author} {\bibfnamefont {J.}~\bibnamefont {{N{\'u}{\~n}ez}}},\
  }\href {\doibase10.1051/0004-6361/201425361} {\bibfield  {journal} {\bibinfo
  {journal} {\aap}\ }\textbf {\bibinfo {volume} {575}},\ \bibinfo {eid} {L6}
  (\bibinfo {year} {2015})},\ \Eprint {http://arxiv.org/abs/1501.02208}
  {arXiv:1501.02208 [astro-ph.HE]}\BibitemShut {NoStop}%
\bibitem [{\citenamefont {{Zanin}}\ \emph {et~al.}(2016)\citenamefont
  {{Zanin}}, \citenamefont {{Fern{\'a}ndez-Barral}}, \citenamefont {{de O{\~n}a
  Wilhelmi}}, \citenamefont {{Aharonian}}, \citenamefont {{Blanch}},
  \citenamefont {{Bosch-Ramon}},\ and\ \citenamefont {{Galindo}}}]{Zanin2016}%
  \BibitemOpen
  \bibfield  {author} {\bibinfo {author} {\bibfnamefont {R.}~\bibnamefont
  {{Zanin}}}, \bibinfo {author} {\bibfnamefont {A.}~\bibnamefont
  {{Fern{\'a}ndez-Barral}}}, \bibinfo {author} {\bibfnamefont {E.}~\bibnamefont
  {{de O{\~n}a Wilhelmi}}}, \bibinfo {author} {\bibfnamefont {F.}~\bibnamefont
  {{Aharonian}}}, \bibinfo {author} {\bibfnamefont {O.}~\bibnamefont
  {{Blanch}}}, \bibinfo {author} {\bibfnamefont {V.}~\bibnamefont
  {{Bosch-Ramon}}}, \ and\ \bibinfo {author} {\bibfnamefont {D.}~\bibnamefont
  {{Galindo}}},\ }\href {\doibase10.1051/0004-6361/201628917} {\bibfield
  {journal} {\bibinfo  {journal} {\aap}\ }\textbf {\bibinfo {volume} {596}},\
  \bibinfo {eid} {A55} (\bibinfo {year} {2016})},\ \Eprint
  {http://arxiv.org/abs/1605.05914} {arXiv:1605.05914
  [astro-ph.HE]}\BibitemShut {NoStop}%
\bibitem [{\citenamefont {{Zdziarski}}\ \emph {et~al.}(2017)\citenamefont
  {{Zdziarski}}, \citenamefont {{Malyshev}}, \citenamefont {{Chernyakova}},\
  and\ \citenamefont {{Pooley}}}]{Zdz_CygX1_2017}%
  \BibitemOpen
  \bibfield  {author} {\bibinfo {author} {\bibfnamefont {A.~A.}\ \bibnamefont
  {{Zdziarski}}}, \bibinfo {author} {\bibfnamefont {D.}~\bibnamefont
  {{Malyshev}}}, \bibinfo {author} {\bibfnamefont {M.}~\bibnamefont
  {{Chernyakova}}}, \ and\ \bibinfo {author} {\bibfnamefont {G.~G.}\
  \bibnamefont {{Pooley}}},\ }\href {\doibase10.1093/mnras/stx1846} {\bibfield
  {journal} {\bibinfo  {journal} {MNRAS}\ }\textbf {\bibinfo {volume} {471}},\
  \bibinfo {pages} {3657--3667} (\bibinfo {year} {2017})},\ \Eprint
  {http://arxiv.org/abs/1607.05059} {arXiv:1607.05059
  [astro-ph.HE]}\BibitemShut {NoStop}%
\bibitem [{\citenamefont {{Cheung}}\ and\ \citenamefont
  {{Loh}}(2016)}]{2016ATel.9502....1C}%
  \BibitemOpen
  \bibfield  {author} {\bibinfo {author} {\bibfnamefont {C.~C.}\ \bibnamefont
  {{Cheung}}}\ and\ \bibinfo {author} {\bibfnamefont {A.}~\bibnamefont
  {{Loh}}},\ }\href@noop {} {\bibfield  {journal} {\bibinfo  {journal} {The
  Astronomer's Telegram}\ }\textbf {\bibinfo {volume} {9502}} (\bibinfo {year}
  {2016})}\BibitemShut {NoStop}%
\bibitem [{\citenamefont {{CTA Consortium}}(2018)}]{CTA}%
  \BibitemOpen
  \bibfield  {author} {\bibinfo {author} {\bibnamefont {{CTA Consortium}}},\
  }\href {\doibase10.1142/10986} {\emph {\bibinfo {title} {{Science with the
  Cherenkov Telescope Array}}}}\ (\bibinfo  {publisher} {World Scientific},\
  \bibinfo {year} {2018})\ \Eprint {http://arxiv.org/abs/1709.07997}
  {arXiv:1709.07997}\BibitemShut {NoStop}%
\bibitem [{\citenamefont {{Artymowicz}}\ and\ \citenamefont
  {{Lubow}}(1994)}]{1994ApJ...421..651A}%
  \BibitemOpen
  \bibfield  {author} {\bibinfo {author} {\bibfnamefont {P.}~\bibnamefont
  {{Artymowicz}}}\ and\ \bibinfo {author} {\bibfnamefont {S.~H.}\ \bibnamefont
  {{Lubow}}},\ }\href {\doibase10.1086/173679} {\bibfield  {journal} {\bibinfo
  {journal} {Astrophys. J.}\ }\textbf {\bibinfo {volume} {421}},\ \bibinfo
  {pages} {651--667} (\bibinfo {year} {1994})}\BibitemShut {NoStop}%
\bibitem [{\citenamefont {{Farris}}\ \emph {et~al.}(2015)\citenamefont
  {{Farris}}, \citenamefont {{Duffell}}, \citenamefont {{MacFadyen}},\ and\
  \citenamefont {{Haiman}}}]{Farris:2014iga}%
  \BibitemOpen
  \bibfield  {author} {\bibinfo {author} {\bibfnamefont {B.~D.}\ \bibnamefont
  {{Farris}}}, \bibinfo {author} {\bibfnamefont {P.}~\bibnamefont {{Duffell}}},
  \bibinfo {author} {\bibfnamefont {A.~I.}\ \bibnamefont {{MacFadyen}}}, \ and\
  \bibinfo {author} {\bibfnamefont {Z.}~\bibnamefont {{Haiman}}},\ }\href
  {\doibase10.1093/mnrasl/slu160} {\bibfield  {journal} {\bibinfo  {journal}
  {MNRAS}\ }\textbf {\bibinfo {volume} {446}},\ \bibinfo {pages} {L36--L40}
  (\bibinfo {year} {2015})},\ \Eprint {http://arxiv.org/abs/1406.0007}
  {arXiv:1406.0007 [astro-ph.HE]}\BibitemShut {NoStop}%
\bibitem [{\citenamefont {{Roedig}}\ \emph {et~al.}(2014)\citenamefont
  {{Roedig}}, \citenamefont {{Krolik}},\ and\ \citenamefont
  {{Miller}}}]{2014ApJ...785..115R}%
  \BibitemOpen
  \bibfield  {author} {\bibinfo {author} {\bibfnamefont {C.}~\bibnamefont
  {{Roedig}}}, \bibinfo {author} {\bibfnamefont {J.~H.}\ \bibnamefont
  {{Krolik}}}, \ and\ \bibinfo {author} {\bibfnamefont {M.~C.}\ \bibnamefont
  {{Miller}}},\ }\href {\doibase10.1088/0004-637X/785/2/115} {\bibfield
  {journal} {\bibinfo  {journal} {Astrophys. J.}\ }\textbf {\bibinfo {volume}
  {785}},\ \bibinfo {eid} {115} (\bibinfo {year} {2014})},\ \Eprint
  {http://arxiv.org/abs/1402.7098} {arXiv:1402.7098 [astro-ph.HE]}\BibitemShut
  {NoStop}%
\bibitem [{\citenamefont {{Ryan}}\ and\ \citenamefont
  {{MacFadyen}}(2017)}]{Ryan:2016vcm}%
  \BibitemOpen
  \bibfield  {author} {\bibinfo {author} {\bibfnamefont {G.}~\bibnamefont
  {{Ryan}}}\ and\ \bibinfo {author} {\bibfnamefont {A.}~\bibnamefont
  {{MacFadyen}}},\ }\href {\doibase10.3847/1538-4357/835/2/199} {\bibfield
  {journal} {\bibinfo  {journal} {The Astrophysical Journal}\ }\textbf
  {\bibinfo {volume} {835}},\ \bibinfo {eid} {199} (\bibinfo {year} {2017})},\
  \Eprint {http://arxiv.org/abs/1611.00341} {arXiv:1611.00341
  [astro-ph.HE]}\BibitemShut {NoStop}%
\bibitem [{\citenamefont {{Komossa}}(2003)}]{Komossa:2003wz}%
  \BibitemOpen
  \bibfield  {author} {\bibinfo {author} {\bibfnamefont {S.}~\bibnamefont
  {{Komossa}}},\ }in\ \href {\doibase10.1063/1.1629428} {\emph {\bibinfo
  {booktitle} {The Astrophysics of Gravitational Wave Sources}}},\ \bibinfo
  {series} {American Institute of Physics Conference Series}, Vol.\ \bibinfo
  {volume} {686},\ \bibinfo {editor} {edited by\ \bibinfo {editor}
  {\bibfnamefont {J.~M.}\ \bibnamefont {{Centrella}}}}\ (\bibinfo {year}
  {2003})\ pp.\ \bibinfo {pages} {161--174},\ \Eprint
  {http://arxiv.org/abs/astro-ph/0306439} {astro-ph/0306439}\BibitemShut
  {NoStop}%
\bibitem [{\citenamefont {{Bogdanovi{\'c}}}\ \emph {et~al.}(2008)\citenamefont
  {{Bogdanovi{\'c}}}, \citenamefont {{Smith}}, \citenamefont {{Sigurdsson}},\
  and\ \citenamefont {{Eracleous}}}]{Bogdanovic:2007bu}%
  \BibitemOpen
  \bibfield  {author} {\bibinfo {author} {\bibfnamefont {T.}~\bibnamefont
  {{Bogdanovi{\'c}}}}, \bibinfo {author} {\bibfnamefont {B.~D.}\ \bibnamefont
  {{Smith}}}, \bibinfo {author} {\bibfnamefont {S.}~\bibnamefont
  {{Sigurdsson}}}, \ and\ \bibinfo {author} {\bibfnamefont {M.}~\bibnamefont
  {{Eracleous}}},\ }\href {\doibase10.1086/521828} {\bibfield  {journal}
  {\bibinfo  {journal} {The Astrophysical Journal}\ }\textbf {\bibinfo {volume}
  {174}},\ \bibinfo {eid} {455-480} (\bibinfo {year} {2008})},\ \Eprint
  {http://arxiv.org/abs/0708.0414} {arXiv:0708.0414}\BibitemShut {NoStop}%
\bibitem [{\citenamefont {{Bogdanovi{\'c}}}(2015)}]{Bogdanovic:2014cua}%
  \BibitemOpen
  \bibfield  {author} {\bibinfo {author} {\bibfnamefont {T.}~\bibnamefont
  {{Bogdanovi{\'c}}}},\ }in\ \href {\doibase10.1007/978-3-319-10488-1_9} {\emph
  {\bibinfo {booktitle} {Gravitational Wave Astrophysics}}},\ \bibinfo {series}
  {Astrophysics and Space Science Proceedings}, Vol.~\bibinfo {volume} {40},\
  \bibinfo {editor} {edited by\ \bibinfo {editor} {\bibfnamefont {C.~F.}\
  \bibnamefont {{Sopuerta}}}}\ (\bibinfo {year} {2015})\ p.\ \bibinfo {pages}
  {103},\ \Eprint {http://arxiv.org/abs/1406.5193} {arXiv:1406.5193
  [astro-ph.HE]}\BibitemShut {NoStop}%
\bibitem [{\citenamefont {{Graham}}\ \emph
  {et~al.}(2015{\natexlab{a}})\citenamefont {{Graham}}, \citenamefont
  {{Djorgovski}}, \citenamefont {{Stern}}, \citenamefont {{Glikman}},
  \citenamefont {{Drake}}, \citenamefont {{Mahabal}}, \citenamefont
  {{Donalek}}, \citenamefont {{Larson}},\ and\ \citenamefont
  {{Christensen}}}]{Graham:2015gma}%
  \BibitemOpen
  \bibfield  {author} {\bibinfo {author} {\bibfnamefont {M.~J.}\ \bibnamefont
  {{Graham}}}, \bibinfo {author} {\bibfnamefont {S.~G.}\ \bibnamefont
  {{Djorgovski}}}, \bibinfo {author} {\bibfnamefont {D.}~\bibnamefont
  {{Stern}}}, \bibinfo {author} {\bibfnamefont {E.}~\bibnamefont {{Glikman}}},
  \bibinfo {author} {\bibfnamefont {A.~J.}\ \bibnamefont {{Drake}}}, \bibinfo
  {author} {\bibfnamefont {A.~A.}\ \bibnamefont {{Mahabal}}}, \bibinfo {author}
  {\bibfnamefont {C.}~\bibnamefont {{Donalek}}}, \bibinfo {author}
  {\bibfnamefont {S.}~\bibnamefont {{Larson}}}, \ and\ \bibinfo {author}
  {\bibfnamefont {E.}~\bibnamefont {{Christensen}}},\ }\href
  {\doibase10.1038/nature14143} {\bibfield  {journal} {\bibinfo  {journal}
  {Nature}\ }\textbf {\bibinfo {volume} {518}},\ \bibinfo {pages} {74--76}
  (\bibinfo {year} {2015}{\natexlab{a}})},\ \Eprint
  {http://arxiv.org/abs/1501.01375} {arXiv:1501.01375}\BibitemShut {NoStop}%
\bibitem [{\citenamefont {{Nguyen}}\ and\ \citenamefont
  {{Bogdanovi{\'c}}}(2016)}]{Nguyen:2016qnk}%
  \BibitemOpen
  \bibfield  {author} {\bibinfo {author} {\bibfnamefont {K.}~\bibnamefont
  {{Nguyen}}}\ and\ \bibinfo {author} {\bibfnamefont {T.}~\bibnamefont
  {{Bogdanovi{\'c}}}},\ }\href {\doibase10.3847/0004-637X/828/2/68} {\bibfield
  {journal} {\bibinfo  {journal} {The Astrophysical Journal}\ }\textbf
  {\bibinfo {volume} {828}},\ \bibinfo {eid} {68} (\bibinfo {year} {2016})},\
  \Eprint {http://arxiv.org/abs/1605.09389} {arXiv:1605.09389
  [astro-ph.HE]}\BibitemShut {NoStop}%
\bibitem [{\citenamefont {{Komossa}}(2015)}]{2015JHEAp...7..148K}%
  \BibitemOpen
  \bibfield  {author} {\bibinfo {author} {\bibfnamefont {S.}~\bibnamefont
  {{Komossa}}},\ }\href {\doibase10.1016/j.jheap.2015.04.006} {\bibfield
  {journal} {\bibinfo  {journal} {Journal of High Energy Astrophysics}\
  }\textbf {\bibinfo {volume} {7}},\ \bibinfo {pages} {148--157} (\bibinfo
  {year} {2015})},\ \Eprint {http://arxiv.org/abs/1505.01093} {arXiv:1505.01093
  [astro-ph.HE]}\BibitemShut {NoStop}%
\bibitem [{\citenamefont {Guo}\ \emph {et~al.}(2014)\citenamefont {Guo},
  \citenamefont {Hu}, \citenamefont {Tao}, \citenamefont {Yin}, \citenamefont
  {Chen},\ and\ \citenamefont {Pan}}]{Guo:2014ora}%
  \BibitemOpen
  \bibfield  {author} {\bibinfo {author} {\bibfnamefont {Di-Fu}\ \bibnamefont
  {Guo}}, \bibinfo {author} {\bibfnamefont {Shao-Ming}\ \bibnamefont {Hu}},
  \bibinfo {author} {\bibfnamefont {Jun}\ \bibnamefont {Tao}}, \bibinfo
  {author} {\bibfnamefont {Hong-Xing}\ \bibnamefont {Yin}}, \bibinfo {author}
  {\bibfnamefont {Xu}~\bibnamefont {Chen}}, \ and\ \bibinfo {author}
  {\bibfnamefont {Hong-Jian}\ \bibnamefont {Pan}},\ }\href
  {\doibase10.1088/1674-4527/14/8/003} {\bibfield  {journal} {\bibinfo
  {journal} {Res. Astron. Astrophys.}\ }\textbf {\bibinfo {volume} {14}},\
  \bibinfo {pages} {923--932} (\bibinfo {year} {2014})},\ \Eprint
  {http://arxiv.org/abs/1405.4636} {arXiv:1405.4636 [astro-ph.HE]}\BibitemShut
  {NoStop}%
\bibitem [{\citenamefont {{Oknyanskij}}\ \emph {et~al.}(2016)\citenamefont
  {{Oknyanskij}}, \citenamefont {{Metlova}}, \citenamefont {{Huseynov}},
  \citenamefont {{Guo}},\ and\ \citenamefont {{Lyuty}}}]{Oknyanskij2016}%
  \BibitemOpen
  \bibfield  {author} {\bibinfo {author} {\bibfnamefont {V.~L.}\ \bibnamefont
  {{Oknyanskij}}}, \bibinfo {author} {\bibfnamefont {N.~V.}\ \bibnamefont
  {{Metlova}}}, \bibinfo {author} {\bibfnamefont {N.~A.}\ \bibnamefont
  {{Huseynov}}}, \bibinfo {author} {\bibfnamefont {D.-F.}\ \bibnamefont
  {{Guo}}}, \ and\ \bibinfo {author} {\bibfnamefont {V.~M.}\ \bibnamefont
  {{Lyuty}}},\ }\href {\doibase10.18524/1810-4215.2016.29.85058} {\bibfield
  {journal} {\bibinfo  {journal} {Odessa Astronomical Publications}\ }\textbf
  {\bibinfo {volume} {29}},\ \bibinfo {pages} {95} (\bibinfo {year}
  {2016})}\BibitemShut {NoStop}%
\bibitem [{\citenamefont {{Netzer}}(2013)}]{Hagai2013}%
  \BibitemOpen
  \bibfield  {author} {\bibinfo {author} {\bibfnamefont {H.}~\bibnamefont
  {{Netzer}}},\ }\href@noop {} {\emph {\bibinfo {title} {{The Physics and
  Evolution of Active Galactic Nuclei}}}}\ (\bibinfo  {publisher} {Cambridge
  University Press},\ \bibinfo {year} {2013})\BibitemShut {NoStop}%
\bibitem [{\citenamefont {Bon}\ \emph {et~al.}(2018)\citenamefont {Bon},
  \citenamefont {Jovanovic}, \citenamefont {Marziani}, \citenamefont {Bon},\
  and\ \citenamefont {Otasevic}}]{Bon:2018ibp}%
  \BibitemOpen
  \bibfield  {author} {\bibinfo {author} {\bibfnamefont {E.}~\bibnamefont
  {Bon}}, \bibinfo {author} {\bibfnamefont {P.}~\bibnamefont {Jovanovic}},
  \bibinfo {author} {\bibfnamefont {P.}~\bibnamefont {Marziani}}, \bibinfo
  {author} {\bibfnamefont {N.}~\bibnamefont {Bon}}, \ and\ \bibinfo {author}
  {\bibfnamefont {A.}~\bibnamefont {Otasevic}},\ }\href
  {\doibase10.3389/fspas.2018.00019} {\  (\bibinfo {year} {2018}),\
  10.3389/fspas.2018.00019},\ \Eprint {http://arxiv.org/abs/1805.07007}
  {arXiv:1805.07007 [astro-ph.GA]}\BibitemShut {NoStop}%
\bibitem [{\citenamefont {{Popovi{\'c}}}(2012)}]{Popovic:2011uy}%
  \BibitemOpen
  \bibfield  {author} {\bibinfo {author} {\bibfnamefont {L.~{\v C}.}\
  \bibnamefont {{Popovi{\'c}}}},\ }\href {\doibase10.1016/j.newar.2011.11.001}
  {\bibfield  {journal} {\bibinfo  {journal} {\nar}\ }\textbf {\bibinfo
  {volume} {56}},\ \bibinfo {pages} {74--91} (\bibinfo {year} {2012})},\
  \Eprint {http://arxiv.org/abs/1109.0710} {arXiv:1109.0710}\BibitemShut
  {NoStop}%
\bibitem [{\citenamefont {{Bon}}\ \emph {et~al.}(2012)\citenamefont {{Bon}},
  \citenamefont {{Jovanovi{\'c}}}, \citenamefont {{Marziani}}, \citenamefont
  {{Shapovalova}}, \citenamefont {{Bon}}, \citenamefont {{Borka
  Jovanovi{\'c}}}, \citenamefont {{Borka}}, \citenamefont {{Sulentic}},\ and\
  \citenamefont {{Popovi{\'c}}}}]{Bon:2012}%
  \BibitemOpen
  \bibfield  {author} {\bibinfo {author} {\bibfnamefont {E.}~\bibnamefont
  {{Bon}}}, \bibinfo {author} {\bibfnamefont {P.}~\bibnamefont
  {{Jovanovi{\'c}}}}, \bibinfo {author} {\bibfnamefont {P.}~\bibnamefont
  {{Marziani}}}, \bibinfo {author} {\bibfnamefont {A.~I.}\ \bibnamefont
  {{Shapovalova}}}, \bibinfo {author} {\bibfnamefont {N.}~\bibnamefont
  {{Bon}}}, \bibinfo {author} {\bibfnamefont {V.}~\bibnamefont {{Borka
  Jovanovi{\'c}}}}, \bibinfo {author} {\bibfnamefont {D.}~\bibnamefont
  {{Borka}}}, \bibinfo {author} {\bibfnamefont {J.}~\bibnamefont {{Sulentic}}},
  \ and\ \bibinfo {author} {\bibfnamefont {L.~{\v C}.}\ \bibnamefont
  {{Popovi{\'c}}}},\ }\href {\doibase10.1088/0004-637X/759/2/118} {\bibfield
  {journal} {\bibinfo  {journal} {The Astrophysical Journal}\ }\textbf
  {\bibinfo {volume} {759}},\ \bibinfo {eid} {118} (\bibinfo {year} {2012})},\
  \Eprint {http://arxiv.org/abs/1209.4524} {arXiv:1209.4524
  [astro-ph.HE]}\BibitemShut {NoStop}%
\bibitem [{\citenamefont {{Bon}}\ \emph {et~al.}(2016)\citenamefont {{Bon}},
  \citenamefont {{Zucker}}, \citenamefont {{Netzer}}, \citenamefont
  {{Marziani}}, \citenamefont {{Bon}}, \citenamefont {{Jovanovi{\'c}}},
  \citenamefont {{Shapovalova}}, \citenamefont {{Komossa}}, \citenamefont
  {{Gaskell}}, \citenamefont {{Popovi{\'c}}}, \citenamefont {{Britzen}},
  \citenamefont {{Chavushyan}}, \citenamefont {{Burenkov}}, \citenamefont
  {{Sergeev}}, \citenamefont {{La Mura}}, \citenamefont {{Vald{\'e}s}},\ and\
  \citenamefont {{Stalevski}}}]{Bon:2016-NGC5548}%
  \BibitemOpen
  \bibfield  {author} {\bibinfo {author} {\bibfnamefont {E.}~\bibnamefont
  {{Bon}}}, \bibinfo {author} {\bibfnamefont {S.}~\bibnamefont {{Zucker}}},
  \bibinfo {author} {\bibfnamefont {H.}~\bibnamefont {{Netzer}}}, \bibinfo
  {author} {\bibfnamefont {P.}~\bibnamefont {{Marziani}}}, \bibinfo {author}
  {\bibfnamefont {N.}~\bibnamefont {{Bon}}}, \bibinfo {author} {\bibfnamefont
  {P.}~\bibnamefont {{Jovanovi{\'c}}}}, \bibinfo {author} {\bibfnamefont
  {A.~I.}\ \bibnamefont {{Shapovalova}}}, \bibinfo {author} {\bibfnamefont
  {S.}~\bibnamefont {{Komossa}}}, \bibinfo {author} {\bibfnamefont {C.~M.}\
  \bibnamefont {{Gaskell}}}, \bibinfo {author} {\bibfnamefont {L.~{\v C}.}\
  \bibnamefont {{Popovi{\'c}}}}, \bibinfo {author} {\bibfnamefont
  {S.}~\bibnamefont {{Britzen}}}, \bibinfo {author} {\bibfnamefont {V.~H.}\
  \bibnamefont {{Chavushyan}}}, \bibinfo {author} {\bibfnamefont {A.~N.}\
  \bibnamefont {{Burenkov}}}, \bibinfo {author} {\bibfnamefont
  {S.}~\bibnamefont {{Sergeev}}}, \bibinfo {author} {\bibfnamefont
  {G.}~\bibnamefont {{La Mura}}}, \bibinfo {author} {\bibfnamefont {J.~R.}\
  \bibnamefont {{Vald{\'e}s}}}, \ and\ \bibinfo {author} {\bibfnamefont
  {M.}~\bibnamefont {{Stalevski}}},\ }\href
  {\doibase10.3847/0067-0049/225/2/29} {\bibfield  {journal} {\bibinfo
  {journal} {The Astrophysical Journal}\ }\textbf {\bibinfo {volume} {225}},\
  \bibinfo {eid} {29} (\bibinfo {year} {2016})},\ \Eprint
  {http://arxiv.org/abs/1606.04606} {arXiv:1606.04606
  [astro-ph.HE]}\BibitemShut {NoStop}%
\bibitem [{\citenamefont {{Li}}\ \emph {et~al.}(2016)\citenamefont {{Li}},
  \citenamefont {{Wang}}, \citenamefont {{Ho}}, \citenamefont {{Lu}},
  \citenamefont {{Qiu}}, \citenamefont {{Du}}, \citenamefont {{Hu}},
  \citenamefont {{Huang}}, \citenamefont {{Zhang}}, \citenamefont {{Wang}},\
  and\ \citenamefont {{Bai}}}]{Li:2016hcm}%
  \BibitemOpen
  \bibfield  {author} {\bibinfo {author} {\bibfnamefont {Y.-R.}\ \bibnamefont
  {{Li}}}, \bibinfo {author} {\bibfnamefont {J.-M.}\ \bibnamefont {{Wang}}},
  \bibinfo {author} {\bibfnamefont {L.~C.}\ \bibnamefont {{Ho}}}, \bibinfo
  {author} {\bibfnamefont {K.-X.}\ \bibnamefont {{Lu}}}, \bibinfo {author}
  {\bibfnamefont {J.}~\bibnamefont {{Qiu}}}, \bibinfo {author} {\bibfnamefont
  {P.}~\bibnamefont {{Du}}}, \bibinfo {author} {\bibfnamefont {C.}~\bibnamefont
  {{Hu}}}, \bibinfo {author} {\bibfnamefont {Y.-K.}\ \bibnamefont {{Huang}}},
  \bibinfo {author} {\bibfnamefont {Z.-X.}\ \bibnamefont {{Zhang}}}, \bibinfo
  {author} {\bibfnamefont {K.}~\bibnamefont {{Wang}}}, \ and\ \bibinfo {author}
  {\bibfnamefont {J.-M.}\ \bibnamefont {{Bai}}},\ }\href
  {\doibase10.3847/0004-637X/822/1/4} {\bibfield  {journal} {\bibinfo
  {journal} {The Astrophysical Journal}\ }\textbf {\bibinfo {volume} {822}},\
  \bibinfo {eid} {4} (\bibinfo {year} {2016})},\ \Eprint
  {http://arxiv.org/abs/1602.05005} {arXiv:1602.05005}\BibitemShut {NoStop}%
\bibitem [{\citenamefont {{Simm}}\ \emph {et~al.}(2016)\citenamefont {{Simm}},
  \citenamefont {{Salvato}}, \citenamefont {{Saglia}}, \citenamefont {{Ponti}},
  \citenamefont {{Lanzuisi}}, \citenamefont {{Trakhtenbrot}}, \citenamefont
  {{Nandra}},\ and\ \citenamefont {{Bender}}}]{Simm:2015ova}%
  \BibitemOpen
  \bibfield  {author} {\bibinfo {author} {\bibfnamefont {T.}~\bibnamefont
  {{Simm}}}, \bibinfo {author} {\bibfnamefont {M.}~\bibnamefont {{Salvato}}},
  \bibinfo {author} {\bibfnamefont {R.}~\bibnamefont {{Saglia}}}, \bibinfo
  {author} {\bibfnamefont {G.}~\bibnamefont {{Ponti}}}, \bibinfo {author}
  {\bibfnamefont {G.}~\bibnamefont {{Lanzuisi}}}, \bibinfo {author}
  {\bibfnamefont {B.}~\bibnamefont {{Trakhtenbrot}}}, \bibinfo {author}
  {\bibfnamefont {K.}~\bibnamefont {{Nandra}}}, \ and\ \bibinfo {author}
  {\bibfnamefont {R.}~\bibnamefont {{Bender}}},\ }\href
  {\doibase10.1051/0004-6361/201527353} {\bibfield  {journal} {\bibinfo
  {journal} {\aap}\ }\textbf {\bibinfo {volume} {585}},\ \bibinfo {eid} {A129}
  (\bibinfo {year} {2016})},\ \Eprint {http://arxiv.org/abs/1510.06737}
  {arXiv:1510.06737}\BibitemShut {NoStop}%
\bibitem [{\citenamefont {{Vaughan}}\ \emph {et~al.}(2016)\citenamefont
  {{Vaughan}}, \citenamefont {{Uttley}}, \citenamefont {{Markowitz}},
  \citenamefont {{Huppenkothen}}, \citenamefont {{Middleton}}, \citenamefont
  {{Alston}}, \citenamefont {{Scargle}},\ and\ \citenamefont
  {{Farr}}}]{Vaughan:2016pyf}%
  \BibitemOpen
  \bibfield  {author} {\bibinfo {author} {\bibfnamefont {S.}~\bibnamefont
  {{Vaughan}}}, \bibinfo {author} {\bibfnamefont {P.}~\bibnamefont {{Uttley}}},
  \bibinfo {author} {\bibfnamefont {A.~G.}\ \bibnamefont {{Markowitz}}},
  \bibinfo {author} {\bibfnamefont {D.}~\bibnamefont {{Huppenkothen}}},
  \bibinfo {author} {\bibfnamefont {M.~J.}\ \bibnamefont {{Middleton}}},
  \bibinfo {author} {\bibfnamefont {W.~N.}\ \bibnamefont {{Alston}}}, \bibinfo
  {author} {\bibfnamefont {J.~D.}\ \bibnamefont {{Scargle}}}, \ and\ \bibinfo
  {author} {\bibfnamefont {W.~M.}\ \bibnamefont {{Farr}}},\ }\href
  {\doibase10.1093/mnras/stw1412} {\bibfield  {journal} {\bibinfo  {journal}
  {MNRAS}\ }\textbf {\bibinfo {volume} {461}},\ \bibinfo {pages} {3145--3152}
  (\bibinfo {year} {2016})},\ \Eprint {http://arxiv.org/abs/1606.02620}
  {arXiv:1606.02620 [astro-ph.IM]}\BibitemShut {NoStop}%
\bibitem [{\citenamefont {{Scargle}}(1982)}]{Scargle:1982bw}%
  \BibitemOpen
  \bibfield  {author} {\bibinfo {author} {\bibfnamefont {J.~D.}\ \bibnamefont
  {{Scargle}}},\ }\href {\doibase10.1086/160554} {\bibfield  {journal}
  {\bibinfo  {journal} {The Astrophysical Journal}\ }\textbf {\bibinfo {volume}
  {263}},\ \bibinfo {pages} {835--853} (\bibinfo {year} {1982})}\BibitemShut
  {NoStop}%
\bibitem [{\citenamefont {{Bon}}\ \emph {et~al.}(2017)\citenamefont {{Bon}},
  \citenamefont {{Marziani}},\ and\ \citenamefont {{Bon}}}]{Bon:2017tgh}%
  \BibitemOpen
  \bibfield  {author} {\bibinfo {author} {\bibfnamefont {E.}~\bibnamefont
  {{Bon}}}, \bibinfo {author} {\bibfnamefont {P.}~\bibnamefont {{Marziani}}}, \
  and\ \bibinfo {author} {\bibfnamefont {N.}~\bibnamefont {{Bon}}},\ }in\ \href
  {\doibase10.1017/S1743921317002277} {\emph {\bibinfo {booktitle} {New
  Frontiers in Black Hole Astrophysics}}},\ \bibinfo {series} {IAU Symposium},
  Vol.\ \bibinfo {volume} {324},\ \bibinfo {editor} {edited by\ \bibinfo
  {editor} {\bibfnamefont {A.}~\bibnamefont {{Gomboc}}}}\ (\bibinfo {year}
  {2017})\ pp.\ \bibinfo {pages} {176--179},\ \Eprint
  {http://arxiv.org/abs/1702.07210} {arXiv:1702.07210}\BibitemShut {NoStop}%
\bibitem [{\citenamefont {{Sillanpaa}}\ \emph {et~al.}(1988)\citenamefont
  {{Sillanpaa}}, \citenamefont {{Haarala}}, \citenamefont {{Valtonen}},
  \citenamefont {{Sundelius}},\ and\ \citenamefont
  {{Byrd}}}]{Sillanpaa:1988zz}%
  \BibitemOpen
  \bibfield  {author} {\bibinfo {author} {\bibfnamefont {A.}~\bibnamefont
  {{Sillanpaa}}}, \bibinfo {author} {\bibfnamefont {S.}~\bibnamefont
  {{Haarala}}}, \bibinfo {author} {\bibfnamefont {M.~J.}\ \bibnamefont
  {{Valtonen}}}, \bibinfo {author} {\bibfnamefont {B.}~\bibnamefont
  {{Sundelius}}}, \ and\ \bibinfo {author} {\bibfnamefont {G.~G.}\ \bibnamefont
  {{Byrd}}},\ }\href {\doibase10.1086/166033} {\bibfield  {journal} {\bibinfo
  {journal} {The Astrophysical Journal}\ }\textbf {\bibinfo {volume} {325}},\
  \bibinfo {pages} {628--634} (\bibinfo {year} {1988})}\BibitemShut {NoStop}%
\bibitem [{\citenamefont {{Graham}}\ \emph
  {et~al.}(2015{\natexlab{b}})\citenamefont {{Graham}}, \citenamefont
  {{Djorgovski}}, \citenamefont {{Stern}}, \citenamefont {{Drake}},
  \citenamefont {{Mahabal}}, \citenamefont {{Donalek}}, \citenamefont
  {{Glikman}}, \citenamefont {{Larson}},\ and\ \citenamefont
  {{Christensen}}}]{Graham:2015tba}%
  \BibitemOpen
  \bibfield  {author} {\bibinfo {author} {\bibfnamefont {M.~J.}\ \bibnamefont
  {{Graham}}}, \bibinfo {author} {\bibfnamefont {S.~G.}\ \bibnamefont
  {{Djorgovski}}}, \bibinfo {author} {\bibfnamefont {D.}~\bibnamefont
  {{Stern}}}, \bibinfo {author} {\bibfnamefont {A.~J.}\ \bibnamefont
  {{Drake}}}, \bibinfo {author} {\bibfnamefont {A.~A.}\ \bibnamefont
  {{Mahabal}}}, \bibinfo {author} {\bibfnamefont {C.}~\bibnamefont
  {{Donalek}}}, \bibinfo {author} {\bibfnamefont {E.}~\bibnamefont
  {{Glikman}}}, \bibinfo {author} {\bibfnamefont {S.}~\bibnamefont {{Larson}}},
  \ and\ \bibinfo {author} {\bibfnamefont {E.}~\bibnamefont {{Christensen}}},\
  }\href {\doibase10.1093/mnras/stv1726} {\bibfield  {journal} {\bibinfo
  {journal} {MNRAS}\ }\textbf {\bibinfo {volume} {453}},\ \bibinfo {pages}
  {1562--1576} (\bibinfo {year} {2015}{\natexlab{b}})},\ \Eprint
  {http://arxiv.org/abs/1507.07603} {arXiv:1507.07603}\BibitemShut {NoStop}%
\bibitem [{\citenamefont {{Charisi}}\ \emph {et~al.}(2016)\citenamefont
  {{Charisi}}, \citenamefont {{Bartos}}, \citenamefont {{Haiman}},
  \citenamefont {{Price-Whelan}}, \citenamefont {{Graham}}, \citenamefont
  {{Bellm}}, \citenamefont {{Laher}},\ and\ \citenamefont
  {{M{\'a}rka}}}]{Charisi:2016fqw}%
  \BibitemOpen
  \bibfield  {author} {\bibinfo {author} {\bibfnamefont {M.}~\bibnamefont
  {{Charisi}}}, \bibinfo {author} {\bibfnamefont {I.}~\bibnamefont {{Bartos}}},
  \bibinfo {author} {\bibfnamefont {Z.}~\bibnamefont {{Haiman}}}, \bibinfo
  {author} {\bibfnamefont {A.~M.}\ \bibnamefont {{Price-Whelan}}}, \bibinfo
  {author} {\bibfnamefont {M.~J.}\ \bibnamefont {{Graham}}}, \bibinfo {author}
  {\bibfnamefont {E.~C.}\ \bibnamefont {{Bellm}}}, \bibinfo {author}
  {\bibfnamefont {R.~R.}\ \bibnamefont {{Laher}}}, \ and\ \bibinfo {author}
  {\bibfnamefont {S.}~\bibnamefont {{M{\'a}rka}}},\ }\href
  {\doibase10.1093/mnras/stw1838} {\bibfield  {journal} {\bibinfo  {journal}
  {MNRAS}\ }\textbf {\bibinfo {volume} {463}},\ \bibinfo {pages} {2145--2171}
  (\bibinfo {year} {2016})},\ \Eprint {http://arxiv.org/abs/1604.01020}
  {arXiv:1604.01020}\BibitemShut {NoStop}%
\bibitem [{\citenamefont {{Liu}}\ \emph {et~al.}(2016)\citenamefont {{Liu}},
  \citenamefont {{Gezari}}, \citenamefont {{Burgett}}, \citenamefont
  {{Chambers}}, \citenamefont {{Draper}}, \citenamefont {{Hodapp}},
  \citenamefont {{Huber}}, \citenamefont {{Kudritzki}}, \citenamefont
  {{Magnier}}, \citenamefont {{Metcalfe}}, \citenamefont {{Tonry}},
  \citenamefont {{Wainscoat}},\ and\ \citenamefont {{Waters}}}]{Liu:2016msr}%
  \BibitemOpen
  \bibfield  {author} {\bibinfo {author} {\bibfnamefont {T.}~\bibnamefont
  {{Liu}}}, \bibinfo {author} {\bibfnamefont {S.}~\bibnamefont {{Gezari}}},
  \bibinfo {author} {\bibfnamefont {W.}~\bibnamefont {{Burgett}}}, \bibinfo
  {author} {\bibfnamefont {K.}~\bibnamefont {{Chambers}}}, \bibinfo {author}
  {\bibfnamefont {P.}~\bibnamefont {{Draper}}}, \bibinfo {author}
  {\bibfnamefont {K.}~\bibnamefont {{Hodapp}}}, \bibinfo {author}
  {\bibfnamefont {M.}~\bibnamefont {{Huber}}}, \bibinfo {author} {\bibfnamefont
  {R.-P.}\ \bibnamefont {{Kudritzki}}}, \bibinfo {author} {\bibfnamefont
  {E.}~\bibnamefont {{Magnier}}}, \bibinfo {author} {\bibfnamefont
  {N.}~\bibnamefont {{Metcalfe}}}, \bibinfo {author} {\bibfnamefont
  {J.}~\bibnamefont {{Tonry}}}, \bibinfo {author} {\bibfnamefont
  {R.}~\bibnamefont {{Wainscoat}}}, \ and\ \bibinfo {author} {\bibfnamefont
  {C.}~\bibnamefont {{Waters}}},\ }\href {\doibase10.3847/0004-637X/833/1/6}
  {\bibfield  {journal} {\bibinfo  {journal} {The Astrophysical Journal}\
  }\textbf {\bibinfo {volume} {833}},\ \bibinfo {eid} {6} (\bibinfo {year}
  {2016})},\ \Eprint {http://arxiv.org/abs/1609.09503} {arXiv:1609.09503
  [astro-ph.HE]}\BibitemShut {NoStop}%
\bibitem [{\citenamefont {{Bhatta}}\ \emph {et~al.}(2016)\citenamefont
  {{Bhatta}}, \citenamefont {{Zola}}, \citenamefont {{Stawarz}}, \citenamefont
  {{Ostrowski}}, \citenamefont {{Winiarski}}, \citenamefont {{Og{\l}oza}},
  \citenamefont {{Dr{\'o}{\.z}d{\.z}}}, \citenamefont {{Siwak}}, \citenamefont
  {{Liakos}}, \citenamefont {{Kozie{\l}-Wierzbowska}}, \citenamefont
  {{Gazeas}}, \citenamefont {{Debski}}, \citenamefont {{Kundera}},
  \citenamefont {{Stachowski}},\ and\ \citenamefont
  {{Paliya}}}]{Bhatta:2016dsn}%
  \BibitemOpen
  \bibfield  {author} {\bibinfo {author} {\bibfnamefont {G.}~\bibnamefont
  {{Bhatta}}}, \bibinfo {author} {\bibfnamefont {S.}~\bibnamefont {{Zola}}},
  \bibinfo {author} {\bibfnamefont {{\L}.}~\bibnamefont {{Stawarz}}}, \bibinfo
  {author} {\bibfnamefont {M.}~\bibnamefont {{Ostrowski}}}, \bibinfo {author}
  {\bibfnamefont {M.}~\bibnamefont {{Winiarski}}}, \bibinfo {author}
  {\bibfnamefont {W.}~\bibnamefont {{Og{\l}oza}}}, \bibinfo {author}
  {\bibfnamefont {M.}~\bibnamefont {{Dr{\'o}{\.z}d{\.z}}}}, \bibinfo {author}
  {\bibfnamefont {M.}~\bibnamefont {{Siwak}}}, \bibinfo {author} {\bibfnamefont
  {A.}~\bibnamefont {{Liakos}}}, \bibinfo {author} {\bibfnamefont
  {D.}~\bibnamefont {{Kozie{\l}-Wierzbowska}}}, \bibinfo {author}
  {\bibfnamefont {K.}~\bibnamefont {{Gazeas}}}, \bibinfo {author}
  {\bibfnamefont {B.}~\bibnamefont {{Debski}}}, \bibinfo {author}
  {\bibfnamefont {T.}~\bibnamefont {{Kundera}}}, \bibinfo {author}
  {\bibfnamefont {G.}~\bibnamefont {{Stachowski}}}, \ and\ \bibinfo {author}
  {\bibfnamefont {V.~S.}\ \bibnamefont {{Paliya}}},\ }\href
  {\doibase10.3847/0004-637X/832/1/47} {\bibfield  {journal} {\bibinfo
  {journal} {The Astrophysical Journal}\ }\textbf {\bibinfo {volume} {832}},\
  \bibinfo {eid} {47} (\bibinfo {year} {2016})},\ \Eprint
  {http://arxiv.org/abs/1609.02388} {arXiv:1609.02388
  [astro-ph.HE]}\BibitemShut {NoStop}%
\bibitem [{\citenamefont {{Kova{\v c}evi{\'c}}}\ \emph
  {et~al.}(2017)\citenamefont {{Kova{\v c}evi{\'c}}}, \citenamefont
  {{Popovi{\'c}}}, \citenamefont {{Shapovalova}},\ and\ \citenamefont
  {{Ili{\'c}}}}]{Kovacevic:2017sjl}%
  \BibitemOpen
  \bibfield  {author} {\bibinfo {author} {\bibfnamefont {A.}~\bibnamefont
  {{Kova{\v c}evi{\'c}}}}, \bibinfo {author} {\bibfnamefont {L.~{\v C}.}\
  \bibnamefont {{Popovi{\'c}}}}, \bibinfo {author} {\bibfnamefont {A.~I.}\
  \bibnamefont {{Shapovalova}}}, \ and\ \bibinfo {author} {\bibfnamefont
  {D.}~\bibnamefont {{Ili{\'c}}}},\ }\href {\doibase10.1007/s10509-017-3009-z}
  {\bibfield  {journal} {\bibinfo  {journal} {\apss}\ }\textbf {\bibinfo
  {volume} {362}},\ \bibinfo {eid} {31} (\bibinfo {year} {2017})},\ \Eprint
  {http://arxiv.org/abs/1701.01566} {arXiv:1701.01566}\BibitemShut {NoStop}%
\bibitem [{\citenamefont {Ackermann}\ \emph {et~al.}(2015)\citenamefont
  {Ackermann} \emph {et~al.}}]{Ackermann:2015wda}%
  \BibitemOpen
  \bibfield  {author} {\bibinfo {author} {\bibfnamefont {M.}~\bibnamefont
  {Ackermann}} \emph {et~al.} (\bibinfo {collaboration} {Fermi-LAT}),\ }\href
  {\doibase10.1088/2041-8205/813/2/L41} {\bibfield  {journal} {\bibinfo
  {journal} {Astrophys. J.}\ }\textbf {\bibinfo {volume} {813}},\ \bibinfo
  {pages} {L41} (\bibinfo {year} {2015})},\ \Eprint
  {http://arxiv.org/abs/1509.02063} {arXiv:1509.02063
  [astro-ph.HE]}\BibitemShut {NoStop}%
\bibitem [{\citenamefont {Li}\ \emph {et~al.}(2017{\natexlab{b}})\citenamefont
  {Li} \emph {et~al.}}]{Li:2017eqf}%
  \BibitemOpen
  \bibfield  {author} {\bibinfo {author} {\bibfnamefont {Yan-Rong}\
  \bibnamefont {Li}} \emph {et~al.},\ }\href@noop {} {\  (\bibinfo {year}
  {2017}{\natexlab{b}})},\ \Eprint {http://arxiv.org/abs/1705.07781}
  {arXiv:1705.07781 [astro-ph.HE]}\BibitemShut {NoStop}%
\bibitem [{\citenamefont {{Cuadra}}\ \emph {et~al.}(2009)\citenamefont
  {{Cuadra}}, \citenamefont {{Armitage}}, \citenamefont {{Alexander}},\ and\
  \citenamefont {{Begelman}}}]{Cuadra:2008xn}%
  \BibitemOpen
  \bibfield  {author} {\bibinfo {author} {\bibfnamefont {J.}~\bibnamefont
  {{Cuadra}}}, \bibinfo {author} {\bibfnamefont {P.~J.}\ \bibnamefont
  {{Armitage}}}, \bibinfo {author} {\bibfnamefont {R.~D.}\ \bibnamefont
  {{Alexander}}}, \ and\ \bibinfo {author} {\bibfnamefont {M.~C.}\ \bibnamefont
  {{Begelman}}},\ }\href {\doibase10.1111/j.1365-2966.2008.14147.x} {\bibfield
  {journal} {\bibinfo  {journal} {MNRAS}\ }\textbf {\bibinfo {volume} {393}},\
  \bibinfo {pages} {1423--1432} (\bibinfo {year} {2009})},\ \Eprint
  {http://arxiv.org/abs/0809.0311} {arXiv:0809.0311}\BibitemShut {NoStop}%
\bibitem [{\citenamefont {{Hayasaki}}\ \emph {et~al.}(2008)\citenamefont
  {{Hayasaki}}, \citenamefont {{Mineshige}},\ and\ \citenamefont
  {{Ho}}}]{Kimitake:2007fs}%
  \BibitemOpen
  \bibfield  {author} {\bibinfo {author} {\bibfnamefont {K.}~\bibnamefont
  {{Hayasaki}}}, \bibinfo {author} {\bibfnamefont {S.}~\bibnamefont
  {{Mineshige}}}, \ and\ \bibinfo {author} {\bibfnamefont {L.~C.}\ \bibnamefont
  {{Ho}}},\ }\href {\doibase10.1086/588837} {\bibfield  {journal} {\bibinfo
  {journal} {The Astrophysical Journal}\ }\textbf {\bibinfo {volume} {682}},\
  \bibinfo {eid} {1134-1140} (\bibinfo {year} {2008})},\ \Eprint
  {http://arxiv.org/abs/0708.2555} {arXiv:0708.2555}\BibitemShut {NoStop}%
\bibitem [{\citenamefont {{Tang}}\ \emph {et~al.}(2017)\citenamefont {{Tang}},
  \citenamefont {{MacFadyen}},\ and\ \citenamefont {{Haiman}}}]{Tang:2017eiz}%
  \BibitemOpen
  \bibfield  {author} {\bibinfo {author} {\bibfnamefont {Y.}~\bibnamefont
  {{Tang}}}, \bibinfo {author} {\bibfnamefont {A.}~\bibnamefont {{MacFadyen}}},
  \ and\ \bibinfo {author} {\bibfnamefont {Z.}~\bibnamefont {{Haiman}}},\
  }\href {\doibase10.1093/mnras/stx1130} {\bibfield  {journal} {\bibinfo
  {journal} {MNRAS}\ }\textbf {\bibinfo {volume} {469}},\ \bibinfo {pages}
  {4258--4267} (\bibinfo {year} {2017})},\ \Eprint
  {http://arxiv.org/abs/1703.03913} {arXiv:1703.03913
  [astro-ph.HE]}\BibitemShut {NoStop}%
\bibitem [{\citenamefont {{Miranda}}\ \emph {et~al.}(2017)\citenamefont
  {{Miranda}}, \citenamefont {{Mu{\~n}oz}},\ and\ \citenamefont
  {{Lai}}}]{2017MNRAS.466.1170M}%
  \BibitemOpen
  \bibfield  {author} {\bibinfo {author} {\bibfnamefont {R.}~\bibnamefont
  {{Miranda}}}, \bibinfo {author} {\bibfnamefont {D.~J.}\ \bibnamefont
  {{Mu{\~n}oz}}}, \ and\ \bibinfo {author} {\bibfnamefont {D.}~\bibnamefont
  {{Lai}}},\ }\href {\doibase10.1093/mnras/stw3189} {\bibfield  {journal}
  {\bibinfo  {journal} {Mon. Not. Roy. Astron. Soc.}\ }\textbf {\bibinfo
  {volume} {466}},\ \bibinfo {pages} {1170--1191} (\bibinfo {year} {2017})},\
  \Eprint {http://arxiv.org/abs/1610.07263} {arXiv:1610.07263
  [astro-ph.SR]}\BibitemShut {NoStop}%
\bibitem [{\citenamefont {{Hayasaki}}\ \emph {et~al.}(2013)\citenamefont
  {{Hayasaki}}, \citenamefont {{Saito}},\ and\ \citenamefont
  {{Mineshige}}}]{2013PASJ...65...86H}%
  \BibitemOpen
  \bibfield  {author} {\bibinfo {author} {\bibfnamefont {K.}~\bibnamefont
  {{Hayasaki}}}, \bibinfo {author} {\bibfnamefont {H.}~\bibnamefont {{Saito}}},
  \ and\ \bibinfo {author} {\bibfnamefont {S.}~\bibnamefont {{Mineshige}}},\
  }\href {\doibase10.1093/pasj/65.4.86} {\bibfield  {journal} {\bibinfo
  {journal} {\pasj}\ }\textbf {\bibinfo {volume} {65}},\ \bibinfo {eid} {86}
  (\bibinfo {year} {2013})},\ \Eprint {http://arxiv.org/abs/1211.5137}
  {arXiv:1211.5137}\BibitemShut {NoStop}%
\bibitem [{\citenamefont {{Sobacchi}}\ \emph {et~al.}(2017)\citenamefont
  {{Sobacchi}}, \citenamefont {{Sormani}},\ and\ \citenamefont
  {{Stamerra}}}]{Sobacchi:2016yez}%
  \BibitemOpen
  \bibfield  {author} {\bibinfo {author} {\bibfnamefont {E.}~\bibnamefont
  {{Sobacchi}}}, \bibinfo {author} {\bibfnamefont {M.~C.}\ \bibnamefont
  {{Sormani}}}, \ and\ \bibinfo {author} {\bibfnamefont {A.}~\bibnamefont
  {{Stamerra}}},\ }\href {\doibase10.1093/mnras/stw2684} {\bibfield  {journal}
  {\bibinfo  {journal} {MNRAS}\ }\textbf {\bibinfo {volume} {465}},\ \bibinfo
  {pages} {161--172} (\bibinfo {year} {2017})},\ \Eprint
  {http://arxiv.org/abs/1610.04709} {arXiv:1610.04709
  [astro-ph.HE]}\BibitemShut {NoStop}%
\bibitem [{\citenamefont {{Rieger}}(2004)}]{Rieger:2004ay}%
  \BibitemOpen
  \bibfield  {author} {\bibinfo {author} {\bibfnamefont {F.~M.}\ \bibnamefont
  {{Rieger}}},\ }\href {\doibase10.1086/426018} {\bibfield  {journal} {\bibinfo
   {journal} {The Astrophysical Journal}\ }\textbf {\bibinfo {volume} {615}},\
  \bibinfo {pages} {L5--L8} (\bibinfo {year} {2004})},\ \Eprint
  {http://arxiv.org/abs/astro-ph/0410188} {astro-ph/0410188}\BibitemShut
  {NoStop}%
\bibitem [{\citenamefont {{Prokhorov}}\ and\ \citenamefont
  {{Moraghan}}(2017)}]{Prokhorov:2017amk}%
  \BibitemOpen
  \bibfield  {author} {\bibinfo {author} {\bibfnamefont {D.~A.}\ \bibnamefont
  {{Prokhorov}}}\ and\ \bibinfo {author} {\bibfnamefont {A.}~\bibnamefont
  {{Moraghan}}},\ }\href {\doibase10.1093/mnras/stx1742} {\bibfield  {journal}
  {\bibinfo  {journal} {MNRAS}\ }\textbf {\bibinfo {volume} {471}},\ \bibinfo
  {pages} {3036--3042} (\bibinfo {year} {2017})},\ \Eprint
  {http://arxiv.org/abs/1707.05829} {arXiv:1707.05829
  [astro-ph.HE]}\BibitemShut {NoStop}%
\bibitem [{\citenamefont {{Sandrinelli}}\ \emph {et~al.}(2016)\citenamefont
  {{Sandrinelli}}, \citenamefont {{Covino}},\ and\ \citenamefont
  {{Treves}}}]{Sandrinelli:2015ijk}%
  \BibitemOpen
  \bibfield  {author} {\bibinfo {author} {\bibfnamefont {A.}~\bibnamefont
  {{Sandrinelli}}}, \bibinfo {author} {\bibfnamefont {S.}~\bibnamefont
  {{Covino}}}, \ and\ \bibinfo {author} {\bibfnamefont {A.}~\bibnamefont
  {{Treves}}},\ }\href {\doibase10.3847/0004-637X/820/1/20} {\bibfield
  {journal} {\bibinfo  {journal} {The Astrophysical Journal}\ }\textbf
  {\bibinfo {volume} {820}},\ \bibinfo {eid} {20} (\bibinfo {year} {2016})},\
  \Eprint {http://arxiv.org/abs/1512.08801} {arXiv:1512.08801}\BibitemShut
  {NoStop}%
\bibitem [{\citenamefont {{Kun}}\ \emph {et~al.}(2014)\citenamefont {{Kun}},
  \citenamefont {{Gab{\'a}nyi}}, \citenamefont {{Karouzos}}, \citenamefont
  {{Britzen}},\ and\ \citenamefont {{Gergely}}}]{Kun:2014tva}%
  \BibitemOpen
  \bibfield  {author} {\bibinfo {author} {\bibfnamefont {E.}~\bibnamefont
  {{Kun}}}, \bibinfo {author} {\bibfnamefont {K.~{\'E}.}\ \bibnamefont
  {{Gab{\'a}nyi}}}, \bibinfo {author} {\bibfnamefont {M.}~\bibnamefont
  {{Karouzos}}}, \bibinfo {author} {\bibfnamefont {S.}~\bibnamefont
  {{Britzen}}}, \ and\ \bibinfo {author} {\bibfnamefont {L.~{\'A}.}\
  \bibnamefont {{Gergely}}},\ }\href {\doibase10.1093/mnras/stu1813} {\bibfield
   {journal} {\bibinfo  {journal} {MNRAS}\ }\textbf {\bibinfo {volume} {445}},\
  \bibinfo {pages} {1370--1382} (\bibinfo {year} {2014})},\ \Eprint
  {http://arxiv.org/abs/1402.2644} {arXiv:1402.2644 [astro-ph.HE]}\BibitemShut
  {NoStop}%
\bibitem [{\citenamefont {{Sesana}}\ \emph {et~al.}(2018)\citenamefont
  {{Sesana}}, \citenamefont {{Haiman}}, \citenamefont {{Kocsis}},\ and\
  \citenamefont {{Kelley}}}]{Sesana:2017lnk}%
  \BibitemOpen
  \bibfield  {author} {\bibinfo {author} {\bibfnamefont {A.}~\bibnamefont
  {{Sesana}}}, \bibinfo {author} {\bibfnamefont {Z.}~\bibnamefont {{Haiman}}},
  \bibinfo {author} {\bibfnamefont {B.}~\bibnamefont {{Kocsis}}}, \ and\
  \bibinfo {author} {\bibfnamefont {L.~Z.}\ \bibnamefont {{Kelley}}},\ }\href
  {\doibase10.3847/1538-4357/aaad0f} {\bibfield  {journal} {\bibinfo  {journal}
  {The Astrophysical Journal}\ }\textbf {\bibinfo {volume} {856}},\ \bibinfo
  {eid} {42} (\bibinfo {year} {2018})},\ \Eprint
  {http://arxiv.org/abs/1703.10611} {arXiv:1703.10611
  [astro-ph.HE]}\BibitemShut {NoStop}%
\bibitem [{\citenamefont {Tamanini}\ \emph {et~al.}(2016)\citenamefont
  {Tamanini}, \citenamefont {Caprini}, \citenamefont {Barausse}, \citenamefont
  {Sesana}, \citenamefont {Klein},\ and\ \citenamefont
  {Petiteau}}]{Tamanini:2016zlh}%
  \BibitemOpen
  \bibfield  {author} {\bibinfo {author} {\bibfnamefont {Nicola}\ \bibnamefont
  {Tamanini}}, \bibinfo {author} {\bibfnamefont {Chiara}\ \bibnamefont
  {Caprini}}, \bibinfo {author} {\bibfnamefont {Enrico}\ \bibnamefont
  {Barausse}}, \bibinfo {author} {\bibfnamefont {Alberto}\ \bibnamefont
  {Sesana}}, \bibinfo {author} {\bibfnamefont {Antoine}\ \bibnamefont {Klein}},
  \ and\ \bibinfo {author} {\bibfnamefont {Antoine}\ \bibnamefont {Petiteau}},\
  }\href {\doibase10.1088/1475-7516/2016/04/002} {\bibfield  {journal}
  {\bibinfo  {journal} {JCAP}\ }\textbf {\bibinfo {volume} {1604}},\ \bibinfo
  {pages} {002} (\bibinfo {year} {2016})},\ \Eprint
  {http://arxiv.org/abs/1601.07112} {arXiv:1601.07112
  [astro-ph.CO]}\BibitemShut {NoStop}%
\bibitem [{\citenamefont {{Tang}}\ \emph {et~al.}(2018)\citenamefont {{Tang}},
  \citenamefont {{Haiman}},\ and\ \citenamefont
  {{MacFadyen}}}]{2018MNRAS.476.2249T}%
  \BibitemOpen
  \bibfield  {author} {\bibinfo {author} {\bibfnamefont {Y.}~\bibnamefont
  {{Tang}}}, \bibinfo {author} {\bibfnamefont {Z.}~\bibnamefont {{Haiman}}}, \
  and\ \bibinfo {author} {\bibfnamefont {A.}~\bibnamefont {{MacFadyen}}},\
  }\href {\doibase10.1093/mnras/sty423} {\bibfield  {journal} {\bibinfo
  {journal} {\mnras}\ }\textbf {\bibinfo {volume} {476}},\ \bibinfo {pages}
  {2249--2257} (\bibinfo {year} {2018})},\ \Eprint
  {http://arxiv.org/abs/1801.02266} {arXiv:1801.02266
  [astro-ph.HE]}\BibitemShut {NoStop}%
\bibitem [{\citenamefont {{Bowen}}\ \emph {et~al.}(2018)\citenamefont
  {{Bowen}}, \citenamefont {{Mewes}}, \citenamefont {{Campanelli}},
  \citenamefont {{Noble}}, \citenamefont {{Krolik}},\ and\ \citenamefont
  {{Zilh{\~a}o}}}]{2018ApJ...853L..17B}%
  \BibitemOpen
  \bibfield  {author} {\bibinfo {author} {\bibfnamefont {D.~B.}\ \bibnamefont
  {{Bowen}}}, \bibinfo {author} {\bibfnamefont {V.}~\bibnamefont {{Mewes}}},
  \bibinfo {author} {\bibfnamefont {M.}~\bibnamefont {{Campanelli}}}, \bibinfo
  {author} {\bibfnamefont {S.~C.}\ \bibnamefont {{Noble}}}, \bibinfo {author}
  {\bibfnamefont {J.~H.}\ \bibnamefont {{Krolik}}}, \ and\ \bibinfo {author}
  {\bibfnamefont {M.}~\bibnamefont {{Zilh{\~a}o}}},\ }\href
  {\doibase10.3847/2041-8213/aaa756} {\bibfield  {journal} {\bibinfo  {journal}
  {Astrophys. J. Letters}\ }\textbf {\bibinfo {volume} {853}},\ \bibinfo {eid}
  {L17} (\bibinfo {year} {2018})},\ \Eprint {http://arxiv.org/abs/1712.05451}
  {arXiv:1712.05451 [astro-ph.HE]}\BibitemShut {NoStop}%
\bibitem [{\citenamefont {{Armitage}}\ and\ \citenamefont
  {{Natarajan}}(2002)}]{armitage}%
  \BibitemOpen
  \bibfield  {author} {\bibinfo {author} {\bibfnamefont {P.~J.}\ \bibnamefont
  {{Armitage}}}\ and\ \bibinfo {author} {\bibfnamefont {P.}~\bibnamefont
  {{Natarajan}}},\ }\href {\doibase10.1086/339770} {\bibfield  {journal}
  {\bibinfo  {journal} {Astrophys. J. Letters}\ }\textbf {\bibinfo {volume}
  {567}},\ \bibinfo {pages} {L9--L12} (\bibinfo {year} {2002})},\ \Eprint
  {http://arxiv.org/abs/astro-ph/0201318} {astro-ph/0201318}\BibitemShut
  {NoStop}%
\bibitem [{\citenamefont {{Cerioli}}\ \emph {et~al.}(2016)\citenamefont
  {{Cerioli}}, \citenamefont {{Lodato}},\ and\ \citenamefont
  {{Price}}}]{2016MNRAS.457..939C}%
  \BibitemOpen
  \bibfield  {author} {\bibinfo {author} {\bibfnamefont {A.}~\bibnamefont
  {{Cerioli}}}, \bibinfo {author} {\bibfnamefont {G.}~\bibnamefont {{Lodato}}},
  \ and\ \bibinfo {author} {\bibfnamefont {D.~J.}\ \bibnamefont {{Price}}},\
  }\href {\doibase10.1093/mnras/stw034} {\bibfield  {journal} {\bibinfo
  {journal} {\mnras}\ }\textbf {\bibinfo {volume} {457}},\ \bibinfo {pages}
  {939--948} (\bibinfo {year} {2016})},\ \Eprint
  {http://arxiv.org/abs/1601.03776} {arXiv:1601.03776
  [astro-ph.HE]}\BibitemShut {NoStop}%
\bibitem [{\citenamefont {{Fontecilla}}\ \emph {et~al.}(2017)\citenamefont
  {{Fontecilla}}, \citenamefont {{Chen}},\ and\ \citenamefont
  {{Cuadra}}}]{2017MNRAS.468L..50F}%
  \BibitemOpen
  \bibfield  {author} {\bibinfo {author} {\bibfnamefont {C.}~\bibnamefont
  {{Fontecilla}}}, \bibinfo {author} {\bibfnamefont {X.}~\bibnamefont
  {{Chen}}}, \ and\ \bibinfo {author} {\bibfnamefont {J.}~\bibnamefont
  {{Cuadra}}},\ }\href {\doibase10.1093/mnrasl/slw258} {\bibfield  {journal}
  {\bibinfo  {journal} {\mnras}\ }\textbf {\bibinfo {volume} {468}},\ \bibinfo
  {pages} {L50--L54} (\bibinfo {year} {2017})},\ \Eprint
  {http://arxiv.org/abs/1610.09382} {arXiv:1610.09382
  [astro-ph.HE]}\BibitemShut {NoStop}%
\bibitem [{\citenamefont {Palenzuela}\ \emph {et~al.}(2010)\citenamefont
  {Palenzuela}, \citenamefont {Lehner},\ and\ \citenamefont
  {Liebling}}]{Palenzuela:2010nf}%
  \BibitemOpen
  \bibfield  {author} {\bibinfo {author} {\bibfnamefont {Carlos}\ \bibnamefont
  {Palenzuela}}, \bibinfo {author} {\bibfnamefont {Luis}\ \bibnamefont
  {Lehner}}, \ and\ \bibinfo {author} {\bibfnamefont {Steven~L.}\ \bibnamefont
  {Liebling}},\ }\href {\doibase10.1126/science.1191766} {\bibfield  {journal}
  {\bibinfo  {journal} {Science}\ }\textbf {\bibinfo {volume} {329}},\ \bibinfo
  {pages} {927} (\bibinfo {year} {2010})},\ \Eprint
  {http://arxiv.org/abs/1005.1067} {arXiv:1005.1067 [astro-ph.HE]}\BibitemShut
  {NoStop}%
\bibitem [{\citenamefont {{Moesta}}\ \emph {et~al.}(2012)\citenamefont
  {{Moesta}}, \citenamefont {{Alic}}, \citenamefont {{Rezzolla}}, \citenamefont
  {{Zanotti}},\ and\ \citenamefont {{Palenzuela}}}]{2012ApJ...749L..32M}%
  \BibitemOpen
  \bibfield  {author} {\bibinfo {author} {\bibfnamefont {P.}~\bibnamefont
  {{Moesta}}}, \bibinfo {author} {\bibfnamefont {D.}~\bibnamefont {{Alic}}},
  \bibinfo {author} {\bibfnamefont {L.}~\bibnamefont {{Rezzolla}}}, \bibinfo
  {author} {\bibfnamefont {O.}~\bibnamefont {{Zanotti}}}, \ and\ \bibinfo
  {author} {\bibfnamefont {C.}~\bibnamefont {{Palenzuela}}},\ }\href
  {\doibase10.1088/2041-8205/749/2/L32} {\bibfield  {journal} {\bibinfo
  {journal} {"Astrop. J. Lett."}\ }\textbf {\bibinfo {volume} {749}},\ \bibinfo
  {eid} {L32} (\bibinfo {year} {2012})},\ \Eprint
  {http://arxiv.org/abs/1109.1177} {arXiv:1109.1177 [gr-qc]}\BibitemShut
  {NoStop}%
\bibitem [{\citenamefont {{Gold}}\ \emph {et~al.}(2014)\citenamefont {{Gold}},
  \citenamefont {{Paschalidis}}, \citenamefont {{Ruiz}}, \citenamefont
  {{Shapiro}}, \citenamefont {{Etienne}},\ and\ \citenamefont
  {{Pfeiffer}}}]{2014PhRvD..90j4030G}%
  \BibitemOpen
  \bibfield  {author} {\bibinfo {author} {\bibfnamefont {R.}~\bibnamefont
  {{Gold}}}, \bibinfo {author} {\bibfnamefont {V.}~\bibnamefont
  {{Paschalidis}}}, \bibinfo {author} {\bibfnamefont {M.}~\bibnamefont
  {{Ruiz}}}, \bibinfo {author} {\bibfnamefont {S.~L.}\ \bibnamefont
  {{Shapiro}}}, \bibinfo {author} {\bibfnamefont {Z.~B.}\ \bibnamefont
  {{Etienne}}}, \ and\ \bibinfo {author} {\bibfnamefont {H.~P.}\ \bibnamefont
  {{Pfeiffer}}},\ }\href {\doibase10.1103/PhysRevD.90.104030} {\bibfield
  {journal} {\bibinfo  {journal} {Phys. Rev. D}\ }\textbf {\bibinfo {volume}
  {90}},\ \bibinfo {eid} {104030} (\bibinfo {year} {2014})},\ \Eprint
  {http://arxiv.org/abs/1410.1543} {arXiv:1410.1543}\BibitemShut {NoStop}%
\bibitem [{\citenamefont {{Fitchett}}(1983)}]{1983MNRAS.203.1049F}%
  \BibitemOpen
  \bibfield  {author} {\bibinfo {author} {\bibfnamefont {M.~J.}\ \bibnamefont
  {{Fitchett}}},\ }\href {\doibase10.1093/mnras/203.4.1049} {\bibfield
  {journal} {\bibinfo  {journal} {\mnras}\ }\textbf {\bibinfo {volume} {203}},\
  \bibinfo {pages} {1049--1062} (\bibinfo {year} {1983})}\BibitemShut {NoStop}%
\bibitem [{\citenamefont {Campanelli}\ \emph
  {et~al.}(2007{\natexlab{a}})\citenamefont {Campanelli}, \citenamefont
  {Lousto}, \citenamefont {Zlochower},\ and\ \citenamefont
  {Merritt}}]{Campanelli:2007ew}%
  \BibitemOpen
  \bibfield  {author} {\bibinfo {author} {\bibfnamefont {Manuela}\ \bibnamefont
  {Campanelli}}, \bibinfo {author} {\bibfnamefont {Carlos~O.}\ \bibnamefont
  {Lousto}}, \bibinfo {author} {\bibfnamefont {Yosef}\ \bibnamefont
  {Zlochower}}, \ and\ \bibinfo {author} {\bibfnamefont {David}\ \bibnamefont
  {Merritt}},\ }\href {\doibase10.1086/516712} {\bibfield  {journal} {\bibinfo
  {journal} {Astrophys. J.}\ }\textbf {\bibinfo {volume} {659}},\ \bibinfo
  {pages} {L5--L8} (\bibinfo {year} {2007}{\natexlab{a}})},\ \Eprint
  {http://arxiv.org/abs/gr-qc/0701164} {arXiv:gr-qc/0701164
  [gr-qc]}\BibitemShut {NoStop}%
\bibitem [{\citenamefont {{Lippai}}\ \emph {et~al.}(2008)\citenamefont
  {{Lippai}}, \citenamefont {{Frei}},\ and\ \citenamefont
  {{Haiman}}}]{2008ApJ...676L...5L}%
  \BibitemOpen
  \bibfield  {author} {\bibinfo {author} {\bibfnamefont {Z.}~\bibnamefont
  {{Lippai}}}, \bibinfo {author} {\bibfnamefont {Z.}~\bibnamefont {{Frei}}}, \
  and\ \bibinfo {author} {\bibfnamefont {Z.}~\bibnamefont {{Haiman}}},\ }\href
  {\doibase10.1086/587034} {\bibfield  {journal} {\bibinfo  {journal}
  {Astrophys. J. Letters}\ }\textbf {\bibinfo {volume} {676}},\ \bibinfo
  {pages} {L5} (\bibinfo {year} {2008})},\ \Eprint
  {http://arxiv.org/abs/0801.0739} {arXiv:0801.0739}\BibitemShut {NoStop}%
\bibitem [{\citenamefont {{Schnittman}}\ and\ \citenamefont
  {{Krolik}}(2008)}]{2008ApJ...684..835S}%
  \BibitemOpen
  \bibfield  {author} {\bibinfo {author} {\bibfnamefont {J.~D.}\ \bibnamefont
  {{Schnittman}}}\ and\ \bibinfo {author} {\bibfnamefont {J.~H.}\ \bibnamefont
  {{Krolik}}},\ }\href {\doibase10.1086/590363} {\bibfield  {journal} {\bibinfo
   {journal} {Astrophys. J.}\ }\textbf {\bibinfo {volume} {684}},\ \bibinfo
  {pages} {835--844} (\bibinfo {year} {2008})},\ \Eprint
  {http://arxiv.org/abs/0802.3556} {arXiv:0802.3556}\BibitemShut {NoStop}%
\bibitem [{\citenamefont {{Shields}}\ and\ \citenamefont
  {{Bonning}}(2008)}]{2008ApJ...682..758S}%
  \BibitemOpen
  \bibfield  {author} {\bibinfo {author} {\bibfnamefont {G.~A.}\ \bibnamefont
  {{Shields}}}\ and\ \bibinfo {author} {\bibfnamefont {E.~W.}\ \bibnamefont
  {{Bonning}}},\ }\href {\doibase10.1086/589427} {\bibfield  {journal}
  {\bibinfo  {journal} {Astrophys. J.}\ }\textbf {\bibinfo {volume} {682}},\
  \bibinfo {pages} {758--766} (\bibinfo {year} {2008})},\ \Eprint
  {http://arxiv.org/abs/0802.3873} {arXiv:0802.3873}\BibitemShut {NoStop}%
\bibitem [{\citenamefont {{Megevand}}\ \emph {et~al.}(2009)\citenamefont
  {{Megevand}}, \citenamefont {{Anderson}}, \citenamefont {{Frank}},
  \citenamefont {{Hirschmann}}, \citenamefont {{Lehner}}, \citenamefont
  {{Liebling}}, \citenamefont {{Motl}},\ and\ \citenamefont
  {{Neilsen}}}]{2009PhRvD..80b4012M}%
  \BibitemOpen
  \bibfield  {author} {\bibinfo {author} {\bibfnamefont {M.}~\bibnamefont
  {{Megevand}}}, \bibinfo {author} {\bibfnamefont {M.}~\bibnamefont
  {{Anderson}}}, \bibinfo {author} {\bibfnamefont {J.}~\bibnamefont {{Frank}}},
  \bibinfo {author} {\bibfnamefont {E.~W.}\ \bibnamefont {{Hirschmann}}},
  \bibinfo {author} {\bibfnamefont {L.}~\bibnamefont {{Lehner}}}, \bibinfo
  {author} {\bibfnamefont {S.~L.}\ \bibnamefont {{Liebling}}}, \bibinfo
  {author} {\bibfnamefont {P.~M.}\ \bibnamefont {{Motl}}}, \ and\ \bibinfo
  {author} {\bibfnamefont {D.}~\bibnamefont {{Neilsen}}},\ }\href
  {\doibase10.1103/PhysRevD.80.024012} {\bibfield  {journal} {\bibinfo
  {journal} {Phys. Rev. D}\ }\textbf {\bibinfo {volume} {80}},\ \bibinfo {eid}
  {024012} (\bibinfo {year} {2009})},\ \Eprint {http://arxiv.org/abs/0905.3390}
  {arXiv:0905.3390 [astro-ph.HE]}\BibitemShut {NoStop}%
\bibitem [{\citenamefont {{Rossi}}\ \emph {et~al.}(2010)\citenamefont
  {{Rossi}}, \citenamefont {{Lodato}}, \citenamefont {{Armitage}},
  \citenamefont {{Pringle}},\ and\ \citenamefont {{King}}}]{rossi}%
  \BibitemOpen
  \bibfield  {author} {\bibinfo {author} {\bibfnamefont {E.~M.}\ \bibnamefont
  {{Rossi}}}, \bibinfo {author} {\bibfnamefont {G.}~\bibnamefont {{Lodato}}},
  \bibinfo {author} {\bibfnamefont {P.~J.}\ \bibnamefont {{Armitage}}},
  \bibinfo {author} {\bibfnamefont {J.~E.}\ \bibnamefont {{Pringle}}}, \ and\
  \bibinfo {author} {\bibfnamefont {A.~R.}\ \bibnamefont {{King}}},\ }\href
  {\doibase10.1111/j.1365-2966.2009.15802.x} {\bibfield  {journal} {\bibinfo
  {journal} {\mnras}\ }\textbf {\bibinfo {volume} {401}},\ \bibinfo {pages}
  {2021--2035} (\bibinfo {year} {2010})},\ \Eprint
  {http://arxiv.org/abs/0910.0002} {arXiv:0910.0002 [astro-ph.HE]}\BibitemShut
  {NoStop}%
\bibitem [{\citenamefont {{Perna}}\ \emph {et~al.}(2016)\citenamefont
  {{Perna}}, \citenamefont {{Lazzati}},\ and\ \citenamefont
  {{Giacomazzo}}}]{perna}%
  \BibitemOpen
  \bibfield  {author} {\bibinfo {author} {\bibfnamefont {R.}~\bibnamefont
  {{Perna}}}, \bibinfo {author} {\bibfnamefont {D.}~\bibnamefont {{Lazzati}}},
  \ and\ \bibinfo {author} {\bibfnamefont {B.}~\bibnamefont {{Giacomazzo}}},\
  }\href {\doibase10.3847/2041-8205/821/1/L18} {\bibfield  {journal} {\bibinfo
  {journal} {Astrophys. J. Letters}\ }\textbf {\bibinfo {volume} {821}},\
  \bibinfo {eid} {L18} (\bibinfo {year} {2016})},\ \Eprint
  {http://arxiv.org/abs/1602.05140} {arXiv:1602.05140
  [astro-ph.HE]}\BibitemShut {NoStop}%
\bibitem [{\citenamefont {{Schnittman}}\ \emph {et~al.}(2016)\citenamefont
  {{Schnittman}}, \citenamefont {{Krolik}},\ and\ \citenamefont
  {{Noble}}}]{2016ApJ...819...48S}%
  \BibitemOpen
  \bibfield  {author} {\bibinfo {author} {\bibfnamefont {J.~D.}\ \bibnamefont
  {{Schnittman}}}, \bibinfo {author} {\bibfnamefont {J.~H.}\ \bibnamefont
  {{Krolik}}}, \ and\ \bibinfo {author} {\bibfnamefont {S.~C.}\ \bibnamefont
  {{Noble}}},\ }\href {\doibase10.3847/0004-637X/819/1/48} {\bibfield
  {journal} {\bibinfo  {journal} {Astrophys. J.}\ }\textbf {\bibinfo {volume}
  {819}},\ \bibinfo {eid} {48} (\bibinfo {year} {2016})},\ \Eprint
  {http://arxiv.org/abs/1512.00729} {arXiv:1512.00729
  [astro-ph.HE]}\BibitemShut {NoStop}%
\bibitem [{\citenamefont {Maggiore}(2007)}]{Maggiore:1900zz}%
  \BibitemOpen
  \bibfield  {author} {\bibinfo {author} {\bibfnamefont {Michele}\ \bibnamefont
  {Maggiore}},\ }\href {http://www.oup.com/uk/catalogue/?ci=9780198570745}
  {\emph {\bibinfo {title} {{Gravitational Waves. Vol. 1: Theory and
  Experiments}}}},\ Oxford Master Series in Physics\ (\bibinfo  {publisher}
  {Oxford University Press},\ \bibinfo {year} {2007})\BibitemShut {NoStop}%
\bibitem [{\citenamefont {Schutz}(1986)}]{Schutz:1986gp}%
  \BibitemOpen
  \bibfield  {author} {\bibinfo {author} {\bibfnamefont {Bernard~F.}\
  \bibnamefont {Schutz}},\ }\href {\doibase10.1038/323310a0} {\bibfield
  {journal} {\bibinfo  {journal} {Nature}\ }\textbf {\bibinfo {volume} {323}},\
  \bibinfo {pages} {310--311} (\bibinfo {year} {1986})}\BibitemShut {NoStop}%
\bibitem [{\citenamefont {Holz}\ and\ \citenamefont
  {Hughes}(2005)}]{Holz:2005df}%
  \BibitemOpen
  \bibfield  {author} {\bibinfo {author} {\bibfnamefont {Daniel~E.}\
  \bibnamefont {Holz}}\ and\ \bibinfo {author} {\bibfnamefont {Scott~A.}\
  \bibnamefont {Hughes}},\ }\href {\doibase10.1086/431341} {\bibfield
  {journal} {\bibinfo  {journal} {Astrophys. J.}\ }\textbf {\bibinfo {volume}
  {629}},\ \bibinfo {pages} {15--22} (\bibinfo {year} {2005})},\ \Eprint
  {http://arxiv.org/abs/astro-ph/0504616} {arXiv:astro-ph/0504616
  [astro-ph]}\BibitemShut {NoStop}%
\bibitem [{\citenamefont {MacLeod}\ and\ \citenamefont
  {Hogan}(2008)}]{MacLeod:2007jd}%
  \BibitemOpen
  \bibfield  {author} {\bibinfo {author} {\bibfnamefont {Chelsea~L.}\
  \bibnamefont {MacLeod}}\ and\ \bibinfo {author} {\bibfnamefont {Craig~J.}\
  \bibnamefont {Hogan}},\ }\href {\doibase10.1103/PhysRevD.77.043512}
  {\bibfield  {journal} {\bibinfo  {journal} {Phys. Rev.}\ }\textbf {\bibinfo
  {volume} {D77}},\ \bibinfo {pages} {043512} (\bibinfo {year} {2008})},\
  \Eprint {http://arxiv.org/abs/0712.0618} {arXiv:0712.0618
  [astro-ph]}\BibitemShut {NoStop}%
\bibitem [{\citenamefont {Jonsson}\ \emph {et~al.}(2007)\citenamefont
  {Jonsson}, \citenamefont {Goobar},\ and\ \citenamefont
  {Mortsell}}]{Jonsson:2006vc}%
  \BibitemOpen
  \bibfield  {author} {\bibinfo {author} {\bibfnamefont {Jakob}\ \bibnamefont
  {Jonsson}}, \bibinfo {author} {\bibfnamefont {Ariel}\ \bibnamefont {Goobar}},
  \ and\ \bibinfo {author} {\bibfnamefont {Edvard}\ \bibnamefont {Mortsell}},\
  }\href {\doibase10.1086/510832} {\bibfield  {journal} {\bibinfo  {journal}
  {Astrophys. J.}\ }\textbf {\bibinfo {volume} {658}},\ \bibinfo {pages}
  {52--59} (\bibinfo {year} {2007})},\ \Eprint
  {http://arxiv.org/abs/astro-ph/0611334} {arXiv:astro-ph/0611334
  [astro-ph]}\BibitemShut {NoStop}%
\bibitem [{\citenamefont {Shapiro}\ \emph {et~al.}(2010)\citenamefont
  {Shapiro}, \citenamefont {Bacon}, \citenamefont {Hendry},\ and\ \citenamefont
  {Hoyle}}]{Shapiro:2009sr}%
  \BibitemOpen
  \bibfield  {author} {\bibinfo {author} {\bibfnamefont {Charles}\ \bibnamefont
  {Shapiro}}, \bibinfo {author} {\bibfnamefont {David}\ \bibnamefont {Bacon}},
  \bibinfo {author} {\bibfnamefont {Martin}\ \bibnamefont {Hendry}}, \ and\
  \bibinfo {author} {\bibfnamefont {Ben}\ \bibnamefont {Hoyle}},\ }\href
  {\doibase10.1111/j.1365-2966.2010.16317.x} {\bibfield  {journal} {\bibinfo
  {journal} {Mon. Not. Roy. Astron. Soc.}\ }\textbf {\bibinfo {volume} {404}},\
  \bibinfo {pages} {858--866} (\bibinfo {year} {2010})},\ \Eprint
  {http://arxiv.org/abs/0907.3635} {arXiv:0907.3635 [astro-ph.CO]}\BibitemShut
  {NoStop}%
\bibitem [{\citenamefont {Hirata}\ \emph {et~al.}(2010)\citenamefont {Hirata},
  \citenamefont {Holz},\ and\ \citenamefont {Cutler}}]{Hirata:2010ba}%
  \BibitemOpen
  \bibfield  {author} {\bibinfo {author} {\bibfnamefont {Christopher~M.}\
  \bibnamefont {Hirata}}, \bibinfo {author} {\bibfnamefont {Daniel~E.}\
  \bibnamefont {Holz}}, \ and\ \bibinfo {author} {\bibfnamefont {Curt}\
  \bibnamefont {Cutler}},\ }\href {\doibase10.1103/PhysRevD.81.124046}
  {\bibfield  {journal} {\bibinfo  {journal} {Phys. Rev.}\ }\textbf {\bibinfo
  {volume} {D81}},\ \bibinfo {pages} {124046} (\bibinfo {year} {2010})},\
  \Eprint {http://arxiv.org/abs/1004.3988} {arXiv:1004.3988
  [astro-ph.CO]}\BibitemShut {NoStop}%
\bibitem [{\citenamefont {Hilbert}\ \emph {et~al.}(2011)\citenamefont
  {Hilbert}, \citenamefont {Gair},\ and\ \citenamefont
  {King}}]{Hilbert:2010am}%
  \BibitemOpen
  \bibfield  {author} {\bibinfo {author} {\bibfnamefont {Stefan}\ \bibnamefont
  {Hilbert}}, \bibinfo {author} {\bibfnamefont {Jonathan~R.}\ \bibnamefont
  {Gair}}, \ and\ \bibinfo {author} {\bibfnamefont {Lindsay~J.}\ \bibnamefont
  {King}},\ }\href {\doibase10.1111/j.1365-2966.2010.17963.x} {\bibfield
  {journal} {\bibinfo  {journal} {Mon. Not. Roy. Astron. Soc.}\ }\textbf
  {\bibinfo {volume} {412}},\ \bibinfo {pages} {1023--1037} (\bibinfo {year}
  {2011})},\ \Eprint {http://arxiv.org/abs/1007.2468} {arXiv:1007.2468
  [astro-ph.CO]}\BibitemShut {NoStop}%
\bibitem [{\citenamefont {Ade}\ \emph {et~al.}(2016)\citenamefont {Ade} \emph
  {et~al.}}]{Ade:2015xua}%
  \BibitemOpen
  \bibfield  {author} {\bibinfo {author} {\bibfnamefont {P.~A.~R.}\
  \bibnamefont {Ade}} \emph {et~al.} (\bibinfo {collaboration} {Planck}),\
  }\href {\doibase10.1051/0004-6361/201525830} {\bibfield  {journal} {\bibinfo
  {journal} {Astron. Astrophys.}\ }\textbf {\bibinfo {volume} {594}},\ \bibinfo
  {pages} {A13} (\bibinfo {year} {2016})},\ \Eprint
  {http://arxiv.org/abs/1502.01589} {arXiv:1502.01589
  [astro-ph.CO]}\BibitemShut {NoStop}%
\bibitem [{\citenamefont {Riess}\ \emph {et~al.}(2016)\citenamefont {Riess}
  \emph {et~al.}}]{Riess:2016jrr}%
  \BibitemOpen
  \bibfield  {author} {\bibinfo {author} {\bibfnamefont {Adam~G.}\ \bibnamefont
  {Riess}} \emph {et~al.},\ }\href {\doibase10.3847/0004-637X/826/1/56}
  {\bibfield  {journal} {\bibinfo  {journal} {Astrophys. J.}\ }\textbf
  {\bibinfo {volume} {826}},\ \bibinfo {pages} {56} (\bibinfo {year} {2016})},\
  \Eprint {http://arxiv.org/abs/1604.01424} {arXiv:1604.01424
  [astro-ph.CO]}\BibitemShut {NoStop}%
\bibitem [{\citenamefont {Dalal}\ \emph {et~al.}(2006)\citenamefont {Dalal},
  \citenamefont {Holz}, \citenamefont {Hughes},\ and\ \citenamefont
  {Jain}}]{Dalal:2006qt}%
  \BibitemOpen
  \bibfield  {author} {\bibinfo {author} {\bibfnamefont {Neal}\ \bibnamefont
  {Dalal}}, \bibinfo {author} {\bibfnamefont {Daniel~E.}\ \bibnamefont {Holz}},
  \bibinfo {author} {\bibfnamefont {Scott~A.}\ \bibnamefont {Hughes}}, \ and\
  \bibinfo {author} {\bibfnamefont {Bhuvnesh}\ \bibnamefont {Jain}},\ }\href
  {\doibase10.1103/PhysRevD.74.063006} {\bibfield  {journal} {\bibinfo
  {journal} {Phys. Rev.}\ }\textbf {\bibinfo {volume} {D74}},\ \bibinfo {pages}
  {063006} (\bibinfo {year} {2006})},\ \Eprint
  {http://arxiv.org/abs/astro-ph/0601275} {arXiv:astro-ph/0601275
  [astro-ph]}\BibitemShut {NoStop}%
\bibitem [{\citenamefont {Nissanke}\ \emph {et~al.}(2010)\citenamefont
  {Nissanke}, \citenamefont {Holz}, \citenamefont {Hughes}, \citenamefont
  {Dalal},\ and\ \citenamefont {Sievers}}]{Nissanke:2009kt}%
  \BibitemOpen
  \bibfield  {author} {\bibinfo {author} {\bibfnamefont {Samaya}\ \bibnamefont
  {Nissanke}}, \bibinfo {author} {\bibfnamefont {Daniel~E.}\ \bibnamefont
  {Holz}}, \bibinfo {author} {\bibfnamefont {Scott~A.}\ \bibnamefont {Hughes}},
  \bibinfo {author} {\bibfnamefont {Neal}\ \bibnamefont {Dalal}}, \ and\
  \bibinfo {author} {\bibfnamefont {Jonathan~L.}\ \bibnamefont {Sievers}},\
  }\href {\doibase10.1088/0004-637X/725/1/496} {\bibfield  {journal} {\bibinfo
  {journal} {Astrophys. J.}\ }\textbf {\bibinfo {volume} {725}},\ \bibinfo
  {pages} {496--514} (\bibinfo {year} {2010})},\ \Eprint
  {http://arxiv.org/abs/0904.1017} {arXiv:0904.1017 [astro-ph.CO]}\BibitemShut
  {NoStop}%
\bibitem [{\citenamefont {Del~Pozzo}(2012)}]{DelPozzo:2011yh}%
  \BibitemOpen
  \bibfield  {author} {\bibinfo {author} {\bibfnamefont {Walter}\ \bibnamefont
  {Del~Pozzo}},\ }\href {\doibase10.1103/PhysRevD.86.043011} {\bibfield
  {journal} {\bibinfo  {journal} {Phys. Rev.}\ }\textbf {\bibinfo {volume}
  {D86}},\ \bibinfo {pages} {043011} (\bibinfo {year} {2012})},\ \Eprint
  {http://arxiv.org/abs/1108.1317} {arXiv:1108.1317 [astro-ph.CO]}\BibitemShut
  {NoStop}%
\bibitem [{\citenamefont {Nissanke}\ \emph {et~al.}(2013)\citenamefont
  {Nissanke}, \citenamefont {Holz}, \citenamefont {Dalal}, \citenamefont
  {Hughes}, \citenamefont {Sievers},\ and\ \citenamefont
  {Hirata}}]{Nissanke:2013fka}%
  \BibitemOpen
  \bibfield  {author} {\bibinfo {author} {\bibfnamefont {Samaya}\ \bibnamefont
  {Nissanke}}, \bibinfo {author} {\bibfnamefont {Daniel~E.}\ \bibnamefont
  {Holz}}, \bibinfo {author} {\bibfnamefont {Neal}\ \bibnamefont {Dalal}},
  \bibinfo {author} {\bibfnamefont {Scott~A.}\ \bibnamefont {Hughes}}, \bibinfo
  {author} {\bibfnamefont {Jonathan~L.}\ \bibnamefont {Sievers}}, \ and\
  \bibinfo {author} {\bibfnamefont {Christopher~M.}\ \bibnamefont {Hirata}},\
  }\href@noop {} {\  (\bibinfo {year} {2013})},\ \Eprint
  {http://arxiv.org/abs/1307.2638} {arXiv:1307.2638 [astro-ph.CO]}\BibitemShut
  {NoStop}%
\bibitem [{\citenamefont {Menou}\ \emph {et~al.}(2008)\citenamefont {Menou},
  \citenamefont {Haiman},\ and\ \citenamefont {Kocsis}}]{Menou:2008dv}%
  \BibitemOpen
  \bibfield  {author} {\bibinfo {author} {\bibfnamefont {Kristen}\ \bibnamefont
  {Menou}}, \bibinfo {author} {\bibfnamefont {Zoltan}\ \bibnamefont {Haiman}},
  \ and\ \bibinfo {author} {\bibfnamefont {Bence}\ \bibnamefont {Kocsis}},\
  }\href {\doibase10.1016/j.newar.2008.03.020} {\bibfield  {journal} {\bibinfo
  {journal} {New Astron. Rev.}\ }\textbf {\bibinfo {volume} {51}},\ \bibinfo
  {pages} {884--890} (\bibinfo {year} {2008})},\ \Eprint
  {http://arxiv.org/abs/0803.3627} {arXiv:0803.3627 [astro-ph]}\BibitemShut
  {NoStop}%
\bibitem [{\citenamefont {Stavridis}\ \emph {et~al.}(2009)\citenamefont
  {Stavridis}, \citenamefont {Arun},\ and\ \citenamefont
  {Will}}]{Stavridis:2009ys}%
  \BibitemOpen
  \bibfield  {author} {\bibinfo {author} {\bibfnamefont {Adamantios}\
  \bibnamefont {Stavridis}}, \bibinfo {author} {\bibfnamefont {K.~G.}\
  \bibnamefont {Arun}}, \ and\ \bibinfo {author} {\bibfnamefont {Clifford~M.}\
  \bibnamefont {Will}},\ }\href {\doibase10.1103/PhysRevD.80.067501} {\bibfield
   {journal} {\bibinfo  {journal} {Phys. Rev.}\ }\textbf {\bibinfo {volume}
  {D80}},\ \bibinfo {pages} {067501} (\bibinfo {year} {2009})},\ \Eprint
  {http://arxiv.org/abs/0907.4686} {arXiv:0907.4686 [gr-qc]}\BibitemShut
  {NoStop}%
\bibitem [{\citenamefont {Van Den~Broeck}\ \emph {et~al.}(2010)\citenamefont
  {Van Den~Broeck}, \citenamefont {Trias}, \citenamefont {Sathyaprakash},\ and\
  \citenamefont {Sintes}}]{VanDenBroeck:2010fp}%
  \BibitemOpen
  \bibfield  {author} {\bibinfo {author} {\bibfnamefont {Chris}\ \bibnamefont
  {Van Den~Broeck}}, \bibinfo {author} {\bibfnamefont {M.}~\bibnamefont
  {Trias}}, \bibinfo {author} {\bibfnamefont {B.~S.}\ \bibnamefont
  {Sathyaprakash}}, \ and\ \bibinfo {author} {\bibfnamefont {A.~M.}\
  \bibnamefont {Sintes}},\ }\href {\doibase10.1103/PhysRevD.81.124031}
  {\bibfield  {journal} {\bibinfo  {journal} {Phys. Rev.}\ }\textbf {\bibinfo
  {volume} {D81}},\ \bibinfo {pages} {124031} (\bibinfo {year} {2010})},\
  \Eprint {http://arxiv.org/abs/1001.3099} {arXiv:1001.3099
  [gr-qc]}\BibitemShut {NoStop}%
\bibitem [{\citenamefont {Shang}\ and\ \citenamefont
  {Haiman}(2011)}]{Shang:2010ta}%
  \BibitemOpen
  \bibfield  {author} {\bibinfo {author} {\bibfnamefont {Cien}\ \bibnamefont
  {Shang}}\ and\ \bibinfo {author} {\bibfnamefont {Zoltan}\ \bibnamefont
  {Haiman}},\ }\href {\doibase10.1111/j.1365-2966.2010.17607.x} {\bibfield
  {journal} {\bibinfo  {journal} {Mon. Not. Roy. Astron. Soc.}\ }\textbf
  {\bibinfo {volume} {411}},\ \bibinfo {pages} {9} (\bibinfo {year} {2011})},\
  \Eprint {http://arxiv.org/abs/1004.3562} {arXiv:1004.3562
  [astro-ph.CO]}\BibitemShut {NoStop}%
\bibitem [{\citenamefont {Sereno}\ \emph {et~al.}(2011)\citenamefont {Sereno},
  \citenamefont {Jetzer}, \citenamefont {Sesana},\ and\ \citenamefont
  {Volonteri}}]{Sereno:2011ty}%
  \BibitemOpen
  \bibfield  {author} {\bibinfo {author} {\bibfnamefont {M.}~\bibnamefont
  {Sereno}}, \bibinfo {author} {\bibfnamefont {Ph.}\ \bibnamefont {Jetzer}},
  \bibinfo {author} {\bibfnamefont {A.}~\bibnamefont {Sesana}}, \ and\ \bibinfo
  {author} {\bibfnamefont {M.}~\bibnamefont {Volonteri}},\ }\href
  {\doibase10.1111/j.1365-2966.2011.18895.x} {\bibfield  {journal} {\bibinfo
  {journal} {Mon. Not. Roy. Astron. Soc.}\ }\textbf {\bibinfo {volume} {415}},\
  \bibinfo {pages} {2773} (\bibinfo {year} {2011})},\ \Eprint
  {http://arxiv.org/abs/1104.1977} {arXiv:1104.1977 [astro-ph.CO]}\BibitemShut
  {NoStop}%
\bibitem [{\citenamefont {Petiteau}\ \emph {et~al.}(2011)\citenamefont
  {Petiteau}, \citenamefont {Babak},\ and\ \citenamefont
  {Sesana}}]{Petiteau:2011we}%
  \BibitemOpen
  \bibfield  {author} {\bibinfo {author} {\bibfnamefont {Antoine}\ \bibnamefont
  {Petiteau}}, \bibinfo {author} {\bibfnamefont {Stanislav}\ \bibnamefont
  {Babak}}, \ and\ \bibinfo {author} {\bibfnamefont {Alberto}\ \bibnamefont
  {Sesana}},\ }\href {\doibase10.1088/0004-637X/732/2/82} {\bibfield  {journal}
  {\bibinfo  {journal} {Astrophys. J.}\ }\textbf {\bibinfo {volume} {732}},\
  \bibinfo {pages} {82} (\bibinfo {year} {2011})},\ \Eprint
  {http://arxiv.org/abs/1102.0769} {arXiv:1102.0769 [astro-ph.CO]}\BibitemShut
  {NoStop}%
\bibitem [{\citenamefont {Chen}\ \emph {et~al.}(2017)\citenamefont {Chen},
  \citenamefont {Fishbach},\ and\ \citenamefont {Holz}}]{Chen:2017rfc}%
  \BibitemOpen
  \bibfield  {author} {\bibinfo {author} {\bibfnamefont {Hsin-Yu}\ \bibnamefont
  {Chen}}, \bibinfo {author} {\bibfnamefont {Maya}\ \bibnamefont {Fishbach}}, \
  and\ \bibinfo {author} {\bibfnamefont {Daniel~E.}\ \bibnamefont {Holz}},\
  }\href@noop {} {\  (\bibinfo {year} {2017})},\ \Eprint
  {http://arxiv.org/abs/1712.06531} {arXiv:1712.06531
  [astro-ph.CO]}\BibitemShut {NoStop}%
\bibitem [{\citenamefont {Seto}\ and\ \citenamefont
  {Kyutoku}(2018)}]{Seto:2017swx}%
  \BibitemOpen
  \bibfield  {author} {\bibinfo {author} {\bibfnamefont {Naoki}\ \bibnamefont
  {Seto}}\ and\ \bibinfo {author} {\bibfnamefont {Koutarou}\ \bibnamefont
  {Kyutoku}},\ }\href {\doibase10.1093/mnras/sty090} {\bibfield  {journal}
  {\bibinfo  {journal} {Mon. Not. Roy. Astron. Soc.}\ }\textbf {\bibinfo
  {volume} {475}},\ \bibinfo {pages} {4133--4139} (\bibinfo {year} {2018})},\
  \Eprint {http://arxiv.org/abs/1710.06424} {arXiv:1710.06424
  [astro-ph.CO]}\BibitemShut {NoStop}%
\bibitem [{\citenamefont {Nishizawa}(2017)}]{Nishizawa:2016ood}%
  \BibitemOpen
  \bibfield  {author} {\bibinfo {author} {\bibfnamefont {Atsushi}\ \bibnamefont
  {Nishizawa}},\ }\href {\doibase10.1103/PhysRevD.96.101303} {\bibfield
  {journal} {\bibinfo  {journal} {Phys. Rev.}\ }\textbf {\bibinfo {volume}
  {D96}},\ \bibinfo {pages} {101303} (\bibinfo {year} {2017})},\ \Eprint
  {http://arxiv.org/abs/1612.06060} {arXiv:1612.06060
  [astro-ph.CO]}\BibitemShut {NoStop}%
\bibitem [{\citenamefont {Sathyaprakash}\ \emph {et~al.}(2010)\citenamefont
  {Sathyaprakash}, \citenamefont {Schutz},\ and\ \citenamefont {Van
  Den~Broeck}}]{Sathyaprakash:2009xt}%
  \BibitemOpen
  \bibfield  {author} {\bibinfo {author} {\bibfnamefont {B.~S.}\ \bibnamefont
  {Sathyaprakash}}, \bibinfo {author} {\bibfnamefont {B.~F.}\ \bibnamefont
  {Schutz}}, \ and\ \bibinfo {author} {\bibfnamefont {C.}~\bibnamefont {Van
  Den~Broeck}},\ }\href {\doibase10.1088/0264-9381/27/21/215006} {\bibfield
  {journal} {\bibinfo  {journal} {Class. Quant. Grav.}\ }\textbf {\bibinfo
  {volume} {27}},\ \bibinfo {pages} {215006} (\bibinfo {year} {2010})},\
  \Eprint {http://arxiv.org/abs/0906.4151} {arXiv:0906.4151
  [astro-ph.CO]}\BibitemShut {NoStop}%
\bibitem [{\citenamefont {Taylor}\ and\ \citenamefont
  {Gair}(2012)}]{Taylor:2012db}%
  \BibitemOpen
  \bibfield  {author} {\bibinfo {author} {\bibfnamefont {Stephen~R.}\
  \bibnamefont {Taylor}}\ and\ \bibinfo {author} {\bibfnamefont {Jonathan~R.}\
  \bibnamefont {Gair}},\ }\href {\doibase10.1103/PhysRevD.86.023502} {\bibfield
   {journal} {\bibinfo  {journal} {Phys. Rev.}\ }\textbf {\bibinfo {volume}
  {D86}},\ \bibinfo {pages} {023502} (\bibinfo {year} {2012})},\ \Eprint
  {http://arxiv.org/abs/1204.6739} {arXiv:1204.6739 [astro-ph.CO]}\BibitemShut
  {NoStop}%
\bibitem [{\citenamefont {Cai}\ and\ \citenamefont {Yang}(2017)}]{Cai:2016sby}%
  \BibitemOpen
  \bibfield  {author} {\bibinfo {author} {\bibfnamefont {Rong-Gen}\
  \bibnamefont {Cai}}\ and\ \bibinfo {author} {\bibfnamefont {Tao}\
  \bibnamefont {Yang}},\ }\href {\doibase10.1103/PhysRevD.95.044024} {\bibfield
   {journal} {\bibinfo  {journal} {Phys. Rev.}\ }\textbf {\bibinfo {volume}
  {D95}},\ \bibinfo {pages} {044024} (\bibinfo {year} {2017})},\ \Eprint
  {http://arxiv.org/abs/1608.08008} {arXiv:1608.08008
  [astro-ph.CO]}\BibitemShut {NoStop}%
\bibitem [{\citenamefont {Messenger}\ and\ \citenamefont
  {Read}(2012)}]{Messenger:2011gi}%
  \BibitemOpen
  \bibfield  {author} {\bibinfo {author} {\bibfnamefont {C.}~\bibnamefont
  {Messenger}}\ and\ \bibinfo {author} {\bibfnamefont {J.}~\bibnamefont
  {Read}},\ }\href {\doibase10.1103/PhysRevLett.108.091101} {\bibfield
  {journal} {\bibinfo  {journal} {Phys. Rev. Lett.}\ }\textbf {\bibinfo
  {volume} {108}},\ \bibinfo {pages} {091101} (\bibinfo {year} {2012})},\
  \Eprint {http://arxiv.org/abs/1107.5725} {arXiv:1107.5725
  [gr-qc]}\BibitemShut {NoStop}%
\bibitem [{\citenamefont {Taylor}\ \emph {et~al.}(2012)\citenamefont {Taylor},
  \citenamefont {Gair},\ and\ \citenamefont {Mandel}}]{Taylor:2011fs}%
  \BibitemOpen
  \bibfield  {author} {\bibinfo {author} {\bibfnamefont {Stephen~R.}\
  \bibnamefont {Taylor}}, \bibinfo {author} {\bibfnamefont {Jonathan~R.}\
  \bibnamefont {Gair}}, \ and\ \bibinfo {author} {\bibfnamefont {Ilya}\
  \bibnamefont {Mandel}},\ }\href {\doibase10.1103/PhysRevD.85.023535}
  {\bibfield  {journal} {\bibinfo  {journal} {Phys. Rev.}\ }\textbf {\bibinfo
  {volume} {D85}},\ \bibinfo {pages} {023535} (\bibinfo {year} {2012})},\
  \Eprint {http://arxiv.org/abs/1108.5161} {arXiv:1108.5161
  [gr-qc]}\BibitemShut {NoStop}%
\bibitem [{\citenamefont {Del~Pozzo}\ \emph {et~al.}(2017)\citenamefont
  {Del~Pozzo}, \citenamefont {Li},\ and\ \citenamefont
  {Messenger}}]{DelPozzo:2015bna}%
  \BibitemOpen
  \bibfield  {author} {\bibinfo {author} {\bibfnamefont {Walter}\ \bibnamefont
  {Del~Pozzo}}, \bibinfo {author} {\bibfnamefont {Tjonnie G.~F.}\ \bibnamefont
  {Li}}, \ and\ \bibinfo {author} {\bibfnamefont {Chris}\ \bibnamefont
  {Messenger}},\ }\href {\doibase10.1103/PhysRevD.95.043502} {\bibfield
  {journal} {\bibinfo  {journal} {Phys. Rev.}\ }\textbf {\bibinfo {volume}
  {D95}},\ \bibinfo {pages} {043502} (\bibinfo {year} {2017})},\ \Eprint
  {http://arxiv.org/abs/1506.06590} {arXiv:1506.06590 [gr-qc]}\BibitemShut
  {NoStop}%
\bibitem [{\citenamefont {Kyutoku}\ and\ \citenamefont
  {Seto}(2017)}]{Kyutoku:2016zxn}%
  \BibitemOpen
  \bibfield  {author} {\bibinfo {author} {\bibfnamefont {Koutarou}\
  \bibnamefont {Kyutoku}}\ and\ \bibinfo {author} {\bibfnamefont {Naoki}\
  \bibnamefont {Seto}},\ }\href {\doibase10.1103/PhysRevD.95.083525} {\bibfield
   {journal} {\bibinfo  {journal} {Phys. Rev.}\ }\textbf {\bibinfo {volume}
  {D95}},\ \bibinfo {pages} {083525} (\bibinfo {year} {2017})},\ \Eprint
  {http://arxiv.org/abs/1609.07142} {arXiv:1609.07142
  [astro-ph.CO]}\BibitemShut {NoStop}%
\bibitem [{\citenamefont {Del~Pozzo}\ \emph {et~al.}(2018)\citenamefont
  {Del~Pozzo}, \citenamefont {Sesana},\ and\ \citenamefont
  {Klein}}]{DelPozzo:2017kme}%
  \BibitemOpen
  \bibfield  {author} {\bibinfo {author} {\bibfnamefont {Walter}\ \bibnamefont
  {Del~Pozzo}}, \bibinfo {author} {\bibfnamefont {Alberto}\ \bibnamefont
  {Sesana}}, \ and\ \bibinfo {author} {\bibfnamefont {Antoine}\ \bibnamefont
  {Klein}},\ }\href {\doibase10.1093/mnras/sty057} {\bibfield  {journal}
  {\bibinfo  {journal} {Mon. Not. Roy. Astron. Soc.}\ }\textbf {\bibinfo
  {volume} {475}},\ \bibinfo {pages} {3485--3492} (\bibinfo {year} {2018})},\
  \Eprint {http://arxiv.org/abs/1703.01300} {arXiv:1703.01300
  [astro-ph.CO]}\BibitemShut {NoStop}%
\bibitem [{\citenamefont {Caprini}\ and\ \citenamefont
  {Tamanini}(2016)}]{Caprini:2016qxs}%
  \BibitemOpen
  \bibfield  {author} {\bibinfo {author} {\bibfnamefont {Chiara}\ \bibnamefont
  {Caprini}}\ and\ \bibinfo {author} {\bibfnamefont {Nicola}\ \bibnamefont
  {Tamanini}},\ }\href {\doibase10.1088/1475-7516/2016/10/006} {\bibfield
  {journal} {\bibinfo  {journal} {JCAP}\ }\textbf {\bibinfo {volume} {1610}},\
  \bibinfo {pages} {006} (\bibinfo {year} {2016})},\ \Eprint
  {http://arxiv.org/abs/1607.08755} {arXiv:1607.08755
  [astro-ph.CO]}\BibitemShut {NoStop}%
\bibitem [{\citenamefont {Cai}\ \emph {et~al.}(2017)\citenamefont {Cai},
  \citenamefont {Tamanini},\ and\ \citenamefont {Yang}}]{Cai:2017yww}%
  \BibitemOpen
  \bibfield  {author} {\bibinfo {author} {\bibfnamefont {Rong-Gen}\
  \bibnamefont {Cai}}, \bibinfo {author} {\bibfnamefont {Nicola}\ \bibnamefont
  {Tamanini}}, \ and\ \bibinfo {author} {\bibfnamefont {Tao}\ \bibnamefont
  {Yang}},\ }\href {\doibase10.1088/1475-7516/2017/05/031} {\bibfield
  {journal} {\bibinfo  {journal} {JCAP}\ }\textbf {\bibinfo {volume} {1705}},\
  \bibinfo {pages} {031} (\bibinfo {year} {2017})},\ \Eprint
  {http://arxiv.org/abs/1703.07323} {arXiv:1703.07323
  [astro-ph.CO]}\BibitemShut {NoStop}%
\bibitem [{\citenamefont {Cutler}\ and\ \citenamefont
  {Holz}(2009)}]{Cutler:2009qv}%
  \BibitemOpen
  \bibfield  {author} {\bibinfo {author} {\bibfnamefont {Curt}\ \bibnamefont
  {Cutler}}\ and\ \bibinfo {author} {\bibfnamefont {Daniel~E.}\ \bibnamefont
  {Holz}},\ }\href {\doibase10.1103/PhysRevD.80.104009} {\bibfield  {journal}
  {\bibinfo  {journal} {Phys. Rev.}\ }\textbf {\bibinfo {volume} {D80}},\
  \bibinfo {pages} {104009} (\bibinfo {year} {2009})},\ \Eprint
  {http://arxiv.org/abs/0906.3752} {arXiv:0906.3752 [astro-ph.CO]}\BibitemShut
  {NoStop}%
\bibitem [{\citenamefont {Nishizawa}\ \emph {et~al.}(2011)\citenamefont
  {Nishizawa}, \citenamefont {Taruya},\ and\ \citenamefont
  {Saito}}]{Nishizawa:2010xx}%
  \BibitemOpen
  \bibfield  {author} {\bibinfo {author} {\bibfnamefont {Atsushi}\ \bibnamefont
  {Nishizawa}}, \bibinfo {author} {\bibfnamefont {Atsushi}\ \bibnamefont
  {Taruya}}, \ and\ \bibinfo {author} {\bibfnamefont {Shun}\ \bibnamefont
  {Saito}},\ }\href {\doibase10.1103/PhysRevD.83.084045} {\bibfield  {journal}
  {\bibinfo  {journal} {Phys. Rev.}\ }\textbf {\bibinfo {volume} {D83}},\
  \bibinfo {pages} {084045} (\bibinfo {year} {2011})},\ \Eprint
  {http://arxiv.org/abs/1011.5000} {arXiv:1011.5000 [astro-ph.CO]}\BibitemShut
  {NoStop}%
\bibitem [{\citenamefont {Kawamura}\ \emph {et~al.}(2011)\citenamefont
  {Kawamura} \emph {et~al.}}]{Kawamura:2011zz}%
  \BibitemOpen
  \bibfield  {author} {\bibinfo {author} {\bibfnamefont {Seiji}\ \bibnamefont
  {Kawamura}} \emph {et~al.},\ }\bibfield  {booktitle} {\emph {\bibinfo
  {booktitle} {{Laser interferometer space antenna. Proceedings, 8th
  International LISA Symposium, Stanford, USA, June 28-July 2, 2010}}},\ }\href
  {\doibase10.1088/0264-9381/28/9/094011} {\bibfield  {journal} {\bibinfo
  {journal} {Class. Quant. Grav.}\ }\textbf {\bibinfo {volume} {28}},\ \bibinfo
  {pages} {094011} (\bibinfo {year} {2011})}\BibitemShut {NoStop}%
\bibitem [{\citenamefont {Arabsalmani}\ \emph {et~al.}(2013)\citenamefont
  {Arabsalmani}, \citenamefont {Sahni},\ and\ \citenamefont
  {Saini}}]{Arabsalmani:2013bj}%
  \BibitemOpen
  \bibfield  {author} {\bibinfo {author} {\bibfnamefont {Maryam}\ \bibnamefont
  {Arabsalmani}}, \bibinfo {author} {\bibfnamefont {Varun}\ \bibnamefont
  {Sahni}}, \ and\ \bibinfo {author} {\bibfnamefont {Tarun~Deep}\ \bibnamefont
  {Saini}},\ }\href {\doibase10.1103/PhysRevD.87.083001} {\bibfield  {journal}
  {\bibinfo  {journal} {Phys. Rev.}\ }\textbf {\bibinfo {volume} {D87}},\
  \bibinfo {pages} {083001} (\bibinfo {year} {2013})},\ \Eprint
  {http://arxiv.org/abs/1301.5779} {arXiv:1301.5779 [astro-ph.CO]}\BibitemShut
  {NoStop}%
\bibitem [{\citenamefont {Seto}\ \emph {et~al.}(2001)\citenamefont {Seto},
  \citenamefont {Kawamura},\ and\ \citenamefont {Nakamura}}]{Seto:2001qf}%
  \BibitemOpen
  \bibfield  {author} {\bibinfo {author} {\bibfnamefont {Naoki}\ \bibnamefont
  {Seto}}, \bibinfo {author} {\bibfnamefont {Seiji}\ \bibnamefont {Kawamura}},
  \ and\ \bibinfo {author} {\bibfnamefont {Takashi}\ \bibnamefont {Nakamura}},\
  }\href {\doibase10.1103/PhysRevLett.87.221103} {\bibfield  {journal}
  {\bibinfo  {journal} {Phys. Rev. Lett.}\ }\textbf {\bibinfo {volume} {87}},\
  \bibinfo {pages} {221103} (\bibinfo {year} {2001})},\ \Eprint
  {http://arxiv.org/abs/astro-ph/0108011} {arXiv:astro-ph/0108011
  [astro-ph]}\BibitemShut {NoStop}%
\bibitem [{\citenamefont {Nishizawa}\ \emph {et~al.}(2012)\citenamefont
  {Nishizawa}, \citenamefont {Yagi}, \citenamefont {Taruya},\ and\
  \citenamefont {Tanaka}}]{Nishizawa:2011eq}%
  \BibitemOpen
  \bibfield  {author} {\bibinfo {author} {\bibfnamefont {Atsushi}\ \bibnamefont
  {Nishizawa}}, \bibinfo {author} {\bibfnamefont {Kent}\ \bibnamefont {Yagi}},
  \bibinfo {author} {\bibfnamefont {Atsushi}\ \bibnamefont {Taruya}}, \ and\
  \bibinfo {author} {\bibfnamefont {Takahiro}\ \bibnamefont {Tanaka}},\ }\href
  {\doibase10.1103/PhysRevD.85.044047} {\bibfield  {journal} {\bibinfo
  {journal} {Phys. Rev.}\ }\textbf {\bibinfo {volume} {D85}},\ \bibinfo {pages}
  {044047} (\bibinfo {year} {2012})},\ \Eprint {http://arxiv.org/abs/1110.2865}
  {arXiv:1110.2865 [astro-ph.CO]}\BibitemShut {NoStop}%
\bibitem [{\citenamefont {Bonvin}\ \emph {et~al.}(2017)\citenamefont {Bonvin},
  \citenamefont {Caprini}, \citenamefont {Sturani},\ and\ \citenamefont
  {Tamanini}}]{Bonvin:2016qxr}%
  \BibitemOpen
  \bibfield  {author} {\bibinfo {author} {\bibfnamefont {Camille}\ \bibnamefont
  {Bonvin}}, \bibinfo {author} {\bibfnamefont {Chiara}\ \bibnamefont
  {Caprini}}, \bibinfo {author} {\bibfnamefont {Riccardo}\ \bibnamefont
  {Sturani}}, \ and\ \bibinfo {author} {\bibfnamefont {Nicola}\ \bibnamefont
  {Tamanini}},\ }\href {\doibase10.1103/PhysRevD.95.044029} {\bibfield
  {journal} {\bibinfo  {journal} {Phys. Rev.}\ }\textbf {\bibinfo {volume}
  {D95}},\ \bibinfo {pages} {044029} (\bibinfo {year} {2017})},\ \Eprint
  {http://arxiv.org/abs/1609.08093} {arXiv:1609.08093
  [astro-ph.CO]}\BibitemShut {NoStop}%
\bibitem [{\citenamefont {Vitale}\ and\ \citenamefont
  {Chen}(2018)}]{Vitale:2018wlg}%
  \BibitemOpen
  \bibfield  {author} {\bibinfo {author} {\bibfnamefont {Salvatore}\
  \bibnamefont {Vitale}}\ and\ \bibinfo {author} {\bibfnamefont {Hsin-Yu}\
  \bibnamefont {Chen}},\ }\href@noop {} {\  (\bibinfo {year} {2018})},\ \Eprint
  {http://arxiv.org/abs/1804.07337} {arXiv:1804.07337
  [astro-ph.CO]}\BibitemShut {NoStop}%
\bibitem [{\citenamefont {Babak}\ \emph
  {et~al.}(2017{\natexlab{a}})\citenamefont {Babak}, \citenamefont {Gair},
  \citenamefont {Sesana}, \citenamefont {Barausse}, \citenamefont {Sopuerta},
  \citenamefont {Berry}, \citenamefont {Berti}, \citenamefont {Amaro-Seoane},
  \citenamefont {Petiteau},\ and\ \citenamefont {Klein}}]{Babak:2017tow}%
  \BibitemOpen
  \bibfield  {author} {\bibinfo {author} {\bibfnamefont {Stanislav}\
  \bibnamefont {Babak}}, \bibinfo {author} {\bibfnamefont {Jonathan}\
  \bibnamefont {Gair}}, \bibinfo {author} {\bibfnamefont {Alberto}\
  \bibnamefont {Sesana}}, \bibinfo {author} {\bibfnamefont {Enrico}\
  \bibnamefont {Barausse}}, \bibinfo {author} {\bibfnamefont {Carlos~F.}\
  \bibnamefont {Sopuerta}}, \bibinfo {author} {\bibfnamefont {Christopher
  P.~L.}\ \bibnamefont {Berry}}, \bibinfo {author} {\bibfnamefont {Emanuele}\
  \bibnamefont {Berti}}, \bibinfo {author} {\bibfnamefont {Pau}\ \bibnamefont
  {Amaro-Seoane}}, \bibinfo {author} {\bibfnamefont {Antoine}\ \bibnamefont
  {Petiteau}}, \ and\ \bibinfo {author} {\bibfnamefont {Antoine}\ \bibnamefont
  {Klein}},\ }\href {\doibase10.1103/PhysRevD.95.103012} {\bibfield  {journal}
  {\bibinfo  {journal} {Phys. Rev.}\ }\textbf {\bibinfo {volume} {D95}},\
  \bibinfo {pages} {103012} (\bibinfo {year} {2017}{\natexlab{a}})},\ \Eprint
  {http://arxiv.org/abs/1703.09722} {arXiv:1703.09722 [gr-qc]}\BibitemShut
  {NoStop}%
\bibitem [{\citenamefont {Tamanini}(2017)}]{Tamanini:2016uin}%
  \BibitemOpen
  \bibfield  {author} {\bibinfo {author} {\bibfnamefont {Nicola}\ \bibnamefont
  {Tamanini}},\ }\bibfield  {booktitle} {\emph {\bibinfo {booktitle}
  {{Proceedings, 11th International LISA Symposium: Zurich, Switzerland,
  September 5-9, 2016}}},\ }\href {\doibase10.1088/1742-6596/840/1/012029}
  {\bibfield  {journal} {\bibinfo  {journal} {J. Phys. Conf. Ser.}\ }\textbf
  {\bibinfo {volume} {840}},\ \bibinfo {pages} {012029} (\bibinfo {year}
  {2017})},\ \Eprint {http://arxiv.org/abs/1612.02634} {arXiv:1612.02634
  [astro-ph.CO]}\BibitemShut {NoStop}%
\bibitem [{\citenamefont {Creminelli}\ and\ \citenamefont
  {Vernizzi}(2017)}]{Creminelli:2017sry}%
  \BibitemOpen
  \bibfield  {author} {\bibinfo {author} {\bibfnamefont {Paolo}\ \bibnamefont
  {Creminelli}}\ and\ \bibinfo {author} {\bibfnamefont {Filippo}\ \bibnamefont
  {Vernizzi}},\ }\href {\doibase10.1103/PhysRevLett.119.251302} {\bibfield
  {journal} {\bibinfo  {journal} {Phys. Rev. Lett.}\ }\textbf {\bibinfo
  {volume} {119}},\ \bibinfo {pages} {251302} (\bibinfo {year} {2017})},\
  \Eprint {http://arxiv.org/abs/1710.05877} {arXiv:1710.05877
  [astro-ph.CO]}\BibitemShut {NoStop}%
\bibitem [{\citenamefont {Ezquiaga}\ and\ \citenamefont
  {Zumalacárregui}(2017)}]{Ezquiaga:2017ekz}%
  \BibitemOpen
  \bibfield  {author} {\bibinfo {author} {\bibfnamefont {Jose~María}\
  \bibnamefont {Ezquiaga}}\ and\ \bibinfo {author} {\bibfnamefont {Miguel}\
  \bibnamefont {Zumalacárregui}},\ }\href
  {\doibase10.1103/PhysRevLett.119.251304} {\bibfield  {journal} {\bibinfo
  {journal} {Phys. Rev. Lett.}\ }\textbf {\bibinfo {volume} {119}},\ \bibinfo
  {pages} {251304} (\bibinfo {year} {2017})},\ \Eprint
  {http://arxiv.org/abs/1710.05901} {arXiv:1710.05901
  [astro-ph.CO]}\BibitemShut {NoStop}%
\bibitem [{\citenamefont {Baker}\ \emph {et~al.}(2017)\citenamefont {Baker},
  \citenamefont {Bellini}, \citenamefont {Ferreira}, \citenamefont {Lagos},
  \citenamefont {Noller},\ and\ \citenamefont {Sawicki}}]{Baker:2017hug}%
  \BibitemOpen
  \bibfield  {author} {\bibinfo {author} {\bibfnamefont {T.}~\bibnamefont
  {Baker}}, \bibinfo {author} {\bibfnamefont {E.}~\bibnamefont {Bellini}},
  \bibinfo {author} {\bibfnamefont {P.~G.}\ \bibnamefont {Ferreira}}, \bibinfo
  {author} {\bibfnamefont {M.}~\bibnamefont {Lagos}}, \bibinfo {author}
  {\bibfnamefont {J.}~\bibnamefont {Noller}}, \ and\ \bibinfo {author}
  {\bibfnamefont {I.}~\bibnamefont {Sawicki}},\ }\href
  {\doibase10.1103/PhysRevLett.119.251301} {\bibfield  {journal} {\bibinfo
  {journal} {Phys. Rev. Lett.}\ }\textbf {\bibinfo {volume} {119}},\ \bibinfo
  {pages} {251301} (\bibinfo {year} {2017})},\ \Eprint
  {http://arxiv.org/abs/1710.06394} {arXiv:1710.06394
  [astro-ph.CO]}\BibitemShut {NoStop}%
\bibitem [{\citenamefont {Sakstein}\ and\ \citenamefont
  {Jain}(2017)}]{Sakstein:2017xjx}%
  \BibitemOpen
  \bibfield  {author} {\bibinfo {author} {\bibfnamefont {Jeremy}\ \bibnamefont
  {Sakstein}}\ and\ \bibinfo {author} {\bibfnamefont {Bhuvnesh}\ \bibnamefont
  {Jain}},\ }\href {\doibase10.1103/PhysRevLett.119.251303} {\bibfield
  {journal} {\bibinfo  {journal} {Phys. Rev. Lett.}\ }\textbf {\bibinfo
  {volume} {119}},\ \bibinfo {pages} {251303} (\bibinfo {year} {2017})},\
  \Eprint {http://arxiv.org/abs/1710.05893} {arXiv:1710.05893
  [astro-ph.CO]}\BibitemShut {NoStop}%
\bibitem [{\citenamefont {Belgacem}\ \emph {et~al.}(2018)\citenamefont
  {Belgacem}, \citenamefont {Dirian}, \citenamefont {Foffa},\ and\
  \citenamefont {Maggiore}}]{Belgacem:2017ihm}%
  \BibitemOpen
  \bibfield  {author} {\bibinfo {author} {\bibfnamefont {Enis}\ \bibnamefont
  {Belgacem}}, \bibinfo {author} {\bibfnamefont {Yves}\ \bibnamefont {Dirian}},
  \bibinfo {author} {\bibfnamefont {Stefano}\ \bibnamefont {Foffa}}, \ and\
  \bibinfo {author} {\bibfnamefont {Michele}\ \bibnamefont {Maggiore}},\ }\href
  {\doibase10.1103/PhysRevD.97.104066} {\bibfield  {journal} {\bibinfo
  {journal} {Phys. Rev.}\ }\textbf {\bibinfo {volume} {D97}},\ \bibinfo {pages}
  {104066} (\bibinfo {year} {2018})},\ \Eprint
  {http://arxiv.org/abs/1712.08108} {arXiv:1712.08108
  [astro-ph.CO]}\BibitemShut {NoStop}%
\bibitem [{\citenamefont {Amendola}\ \emph {et~al.}(2017)\citenamefont
  {Amendola}, \citenamefont {Sawicki}, \citenamefont {Kunz},\ and\
  \citenamefont {Saltas}}]{Amendola:2017ovw}%
  \BibitemOpen
  \bibfield  {author} {\bibinfo {author} {\bibfnamefont {Luca}\ \bibnamefont
  {Amendola}}, \bibinfo {author} {\bibfnamefont {Ignacy}\ \bibnamefont
  {Sawicki}}, \bibinfo {author} {\bibfnamefont {Martin}\ \bibnamefont {Kunz}},
  \ and\ \bibinfo {author} {\bibfnamefont {Ippocratis~D.}\ \bibnamefont
  {Saltas}},\ }\href@noop {} {\  (\bibinfo {year} {2017})},\ \Eprint
  {http://arxiv.org/abs/1712.08623} {arXiv:1712.08623
  [astro-ph.CO]}\BibitemShut {NoStop}%
\bibitem [{\citenamefont {Linder}(2018)}]{Linder:2018jil}%
  \BibitemOpen
  \bibfield  {author} {\bibinfo {author} {\bibfnamefont {Eric~V.}\ \bibnamefont
  {Linder}},\ }\href {\doibase10.1088/1475-7516/2018/03/005} {\bibfield
  {journal} {\bibinfo  {journal} {JCAP}\ }\textbf {\bibinfo {volume} {1803}},\
  \bibinfo {pages} {005} (\bibinfo {year} {2018})},\ \Eprint
  {http://arxiv.org/abs/1801.01503} {arXiv:1801.01503
  [astro-ph.CO]}\BibitemShut {NoStop}%
\bibitem [{\citenamefont {Pardo}\ \emph {et~al.}(2018)\citenamefont {Pardo},
  \citenamefont {Fishbach}, \citenamefont {Holz},\ and\ \citenamefont
  {Spergel}}]{Pardo:2018ipy}%
  \BibitemOpen
  \bibfield  {author} {\bibinfo {author} {\bibfnamefont {Kris}\ \bibnamefont
  {Pardo}}, \bibinfo {author} {\bibfnamefont {Maya}\ \bibnamefont {Fishbach}},
  \bibinfo {author} {\bibfnamefont {Daniel~E.}\ \bibnamefont {Holz}}, \ and\
  \bibinfo {author} {\bibfnamefont {David~N.}\ \bibnamefont {Spergel}},\
  }\href@noop {} {\  (\bibinfo {year} {2018})},\ \Eprint
  {http://arxiv.org/abs/1801.08160} {arXiv:1801.08160 [gr-qc]}\BibitemShut
  {NoStop}%
\bibitem [{\citenamefont {Yagi}\ \emph
  {et~al.}(2012{\natexlab{a}})\citenamefont {Yagi}, \citenamefont {Nishizawa},\
  and\ \citenamefont {Yoo}}]{Yagi:2011bt}%
  \BibitemOpen
  \bibfield  {author} {\bibinfo {author} {\bibfnamefont {Kent}\ \bibnamefont
  {Yagi}}, \bibinfo {author} {\bibfnamefont {Atsushi}\ \bibnamefont
  {Nishizawa}}, \ and\ \bibinfo {author} {\bibfnamefont {Chul-Moon}\
  \bibnamefont {Yoo}},\ }\href {\doibase10.1088/1475-7516/2012/04/031}
  {\bibfield  {journal} {\bibinfo  {journal} {JCAP}\ }\textbf {\bibinfo
  {volume} {1204}},\ \bibinfo {pages} {031} (\bibinfo {year}
  {2012}{\natexlab{a}})},\ \Eprint {http://arxiv.org/abs/1112.6040}
  {arXiv:1112.6040 [astro-ph.CO]}\BibitemShut {NoStop}%
\bibitem [{\citenamefont {Namikawa}\ \emph {et~al.}(2016)\citenamefont
  {Namikawa}, \citenamefont {Nishizawa},\ and\ \citenamefont
  {Taruya}}]{Namikawa:2015prh}%
  \BibitemOpen
  \bibfield  {author} {\bibinfo {author} {\bibfnamefont {Toshiya}\ \bibnamefont
  {Namikawa}}, \bibinfo {author} {\bibfnamefont {Atsushi}\ \bibnamefont
  {Nishizawa}}, \ and\ \bibinfo {author} {\bibfnamefont {Atsushi}\ \bibnamefont
  {Taruya}},\ }\href {\doibase10.1103/PhysRevLett.116.121302} {\bibfield
  {journal} {\bibinfo  {journal} {Phys. Rev. Lett.}\ }\textbf {\bibinfo
  {volume} {116}},\ \bibinfo {pages} {121302} (\bibinfo {year} {2016})},\
  \Eprint {http://arxiv.org/abs/1511.04638} {arXiv:1511.04638
  [astro-ph.CO]}\BibitemShut {NoStop}%
\bibitem [{\citenamefont {Cai}\ \emph {et~al.}(2018)\citenamefont {Cai},
  \citenamefont {Liu}, \citenamefont {Liu}, \citenamefont {Wang},\ and\
  \citenamefont {Yang}}]{Cai:2017aea}%
  \BibitemOpen
  \bibfield  {author} {\bibinfo {author} {\bibfnamefont {Rong-Gen}\
  \bibnamefont {Cai}}, \bibinfo {author} {\bibfnamefont {Tong-Bo}\ \bibnamefont
  {Liu}}, \bibinfo {author} {\bibfnamefont {Xue-Wen}\ \bibnamefont {Liu}},
  \bibinfo {author} {\bibfnamefont {Shao-Jiang}\ \bibnamefont {Wang}}, \ and\
  \bibinfo {author} {\bibfnamefont {Tao}\ \bibnamefont {Yang}},\ }\href
  {\doibase10.1103/PhysRevD.97.103005} {\bibfield  {journal} {\bibinfo
  {journal} {Phys. Rev.}\ }\textbf {\bibinfo {volume} {D97}},\ \bibinfo {pages}
  {103005} (\bibinfo {year} {2018})},\ \Eprint
  {http://arxiv.org/abs/1712.00952} {arXiv:1712.00952
  [astro-ph.CO]}\BibitemShut {NoStop}%
\bibitem [{\citenamefont {Camera}\ and\ \citenamefont
  {Nishizawa}(2013)}]{Camera:2013xfa}%
  \BibitemOpen
  \bibfield  {author} {\bibinfo {author} {\bibfnamefont {Stefano}\ \bibnamefont
  {Camera}}\ and\ \bibinfo {author} {\bibfnamefont {Atsushi}\ \bibnamefont
  {Nishizawa}},\ }\href {\doibase10.1103/PhysRevLett.110.151103} {\bibfield
  {journal} {\bibinfo  {journal} {Phys. Rev. Lett.}\ }\textbf {\bibinfo
  {volume} {110}},\ \bibinfo {pages} {151103} (\bibinfo {year} {2013})},\
  \Eprint {http://arxiv.org/abs/1303.5446} {arXiv:1303.5446
  [astro-ph.CO]}\BibitemShut {NoStop}%
\bibitem [{\citenamefont {Oguri}(2016)}]{Oguri:2016dgk}%
  \BibitemOpen
  \bibfield  {author} {\bibinfo {author} {\bibfnamefont {Masamune}\
  \bibnamefont {Oguri}},\ }\href {\doibase10.1103/PhysRevD.93.083511}
  {\bibfield  {journal} {\bibinfo  {journal} {Phys. Rev.}\ }\textbf {\bibinfo
  {volume} {D93}},\ \bibinfo {pages} {083511} (\bibinfo {year} {2016})},\
  \Eprint {http://arxiv.org/abs/1603.02356} {arXiv:1603.02356
  [astro-ph.CO]}\BibitemShut {NoStop}%
\bibitem [{\citenamefont {Raccanelli}(2017)}]{Raccanelli:2016fmc}%
  \BibitemOpen
  \bibfield  {author} {\bibinfo {author} {\bibfnamefont {Alvise}\ \bibnamefont
  {Raccanelli}},\ }\href {\doibase10.1093/mnras/stx835} {\bibfield  {journal}
  {\bibinfo  {journal} {Mon. Not. Roy. Astron. Soc.}\ }\textbf {\bibinfo
  {volume} {469}},\ \bibinfo {pages} {656--670} (\bibinfo {year} {2017})},\
  \Eprint {http://arxiv.org/abs/1609.09377} {arXiv:1609.09377
  [astro-ph.CO]}\BibitemShut {NoStop}%
\bibitem [{\citenamefont {Romano}\ and\ \citenamefont
  {Cornish}(2017)}]{Romano:2016dpx}%
  \BibitemOpen
  \bibfield  {author} {\bibinfo {author} {\bibfnamefont {Joseph~D.}\
  \bibnamefont {Romano}}\ and\ \bibinfo {author} {\bibfnamefont {Neil~J.}\
  \bibnamefont {Cornish}},\ }\href {\doibase10.1007/s41114-017-0004-1}
  {\bibfield  {journal} {\bibinfo  {journal} {Living Rev. Rel.}\ }\textbf
  {\bibinfo {volume} {20}},\ \bibinfo {pages} {2} (\bibinfo {year} {2017})},\
  \Eprint {http://arxiv.org/abs/1608.06889} {arXiv:1608.06889
  [gr-qc]}\BibitemShut {NoStop}%
\bibitem [{\citenamefont {Caprini}\ and\ \citenamefont
  {Figueroa}(2018)}]{Caprini:2018mtu}%
  \BibitemOpen
  \bibfield  {author} {\bibinfo {author} {\bibfnamefont {Chiara}\ \bibnamefont
  {Caprini}}\ and\ \bibinfo {author} {\bibfnamefont {Daniel~G.}\ \bibnamefont
  {Figueroa}},\ }\href@noop {} {\  (\bibinfo {year} {2018})},\ \Eprint
  {http://arxiv.org/abs/1801.04268} {arXiv:1801.04268
  [astro-ph.CO]}\BibitemShut {NoStop}%
\bibitem [{\citenamefont {Abbott}\ \emph
  {et~al.}(2017{\natexlab{l}})\citenamefont {Abbott} \emph
  {et~al.}}]{TheLIGOScientific:2016dpb}%
  \BibitemOpen
  \bibfield  {author} {\bibinfo {author} {\bibfnamefont {Benjamin~P.}\
  \bibnamefont {Abbott}} \emph {et~al.} (\bibinfo {collaboration} {Virgo, LIGO
  Scientific}),\ }\href {\doibase10.1103/PhysRevLett.118.121101} {\bibfield
  {journal} {\bibinfo  {journal} {Phys. Rev. Lett.}\ }\textbf {\bibinfo
  {volume} {118}},\ \bibinfo {pages} {121101} (\bibinfo {year}
  {2017}{\natexlab{l}})},\ \bibinfo {note} {[Erratum: Phys. Rev.
  Lett.119,no.2,029901(2017)]},\ \Eprint {http://arxiv.org/abs/1612.02029}
  {arXiv:1612.02029 [gr-qc]}\BibitemShut {NoStop}%
\bibitem [{\citenamefont {Abbott}\ \emph
  {et~al.}(2018{\natexlab{b}})\citenamefont {Abbott} \emph
  {et~al.}}]{Abbott:2017mem}%
  \BibitemOpen
  \bibfield  {author} {\bibinfo {author} {\bibfnamefont {B.~P.}\ \bibnamefont
  {Abbott}} \emph {et~al.} (\bibinfo {collaboration} {Virgo, LIGO
  Scientific}),\ }\href {\doibase10.1103/PhysRevD.97.102002} {\bibfield
  {journal} {\bibinfo  {journal} {Phys. Rev.}\ }\textbf {\bibinfo {volume}
  {D97}},\ \bibinfo {pages} {102002} (\bibinfo {year} {2018}{\natexlab{b}})},\
  \Eprint {http://arxiv.org/abs/1712.01168} {arXiv:1712.01168
  [gr-qc]}\BibitemShut {NoStop}%
\bibitem [{\citenamefont {Abbott}\ \emph
  {et~al.}(2018{\natexlab{c}})\citenamefont {Abbott} \emph
  {et~al.}}]{Abbott:2017xzg}%
  \BibitemOpen
  \bibfield  {author} {\bibinfo {author} {\bibfnamefont {Benjamin~P.}\
  \bibnamefont {Abbott}} \emph {et~al.} (\bibinfo {collaboration} {Virgo, LIGO
  Scientific}),\ }\href {\doibase10.1103/PhysRevLett.120.091101} {\bibfield
  {journal} {\bibinfo  {journal} {Phys. Rev. Lett.}\ }\textbf {\bibinfo
  {volume} {120}},\ \bibinfo {pages} {091101} (\bibinfo {year}
  {2018}{\natexlab{c}})},\ \Eprint {http://arxiv.org/abs/1710.05837}
  {arXiv:1710.05837 [gr-qc]}\BibitemShut {NoStop}%
\bibitem [{\citenamefont {Caprini}\ \emph {et~al.}()\citenamefont {Caprini}
  \emph {et~al.}}]{CosWGpreparation}%
  \BibitemOpen
  \bibfield  {author} {\bibinfo {author} {\bibfnamefont {Chiara}\ \bibnamefont
  {Caprini}} \emph {et~al.} (\bibinfo {collaboration} {LISA Cosmology Working
  Group}),\ }\href@noop {} {\ }\bibinfo {note} {(in preparation)}\BibitemShut
  {NoStop}%
\bibitem [{\citenamefont {Adams}\ and\ \citenamefont
  {Cornish}(2014)}]{Adams:2013qma}%
  \BibitemOpen
  \bibfield  {author} {\bibinfo {author} {\bibfnamefont {Matthew~R.}\
  \bibnamefont {Adams}}\ and\ \bibinfo {author} {\bibfnamefont {Neil~J.}\
  \bibnamefont {Cornish}},\ }\href {\doibase10.1103/PhysRevD.89.022001}
  {\bibfield  {journal} {\bibinfo  {journal} {Phys. Rev.}\ }\textbf {\bibinfo
  {volume} {D89}},\ \bibinfo {pages} {022001} (\bibinfo {year} {2014})},\
  \Eprint {http://arxiv.org/abs/1307.4116} {arXiv:1307.4116
  [gr-qc]}\BibitemShut {NoStop}%
\bibitem [{\citenamefont {Regimbau}\ \emph {et~al.}(2017)\citenamefont
  {Regimbau}, \citenamefont {Evans}, \citenamefont {Christensen}, \citenamefont
  {Katsavounidis}, \citenamefont {Sathyaprakash},\ and\ \citenamefont
  {Vitale}}]{Regimbau:2016ike}%
  \BibitemOpen
  \bibfield  {author} {\bibinfo {author} {\bibfnamefont {T.}~\bibnamefont
  {Regimbau}}, \bibinfo {author} {\bibfnamefont {M.}~\bibnamefont {Evans}},
  \bibinfo {author} {\bibfnamefont {N.}~\bibnamefont {Christensen}}, \bibinfo
  {author} {\bibfnamefont {E.}~\bibnamefont {Katsavounidis}}, \bibinfo {author}
  {\bibfnamefont {B.}~\bibnamefont {Sathyaprakash}}, \ and\ \bibinfo {author}
  {\bibfnamefont {S.}~\bibnamefont {Vitale}},\ }\href
  {\doibase10.1103/PhysRevLett.118.151105} {\bibfield  {journal} {\bibinfo
  {journal} {Phys. Rev. Lett.}\ }\textbf {\bibinfo {volume} {118}},\ \bibinfo
  {pages} {151105} (\bibinfo {year} {2017})},\ \Eprint
  {http://arxiv.org/abs/1611.08943} {arXiv:1611.08943
  [astro-ph.CO]}\BibitemShut {NoStop}%
\bibitem [{\citenamefont {Sesana}(2016)}]{Sesana:2016ljz}%
  \BibitemOpen
  \bibfield  {author} {\bibinfo {author} {\bibfnamefont {Alberto}\ \bibnamefont
  {Sesana}},\ }\href {\doibase10.1103/PhysRevLett.116.231102} {\bibfield
  {journal} {\bibinfo  {journal} {Phys. Rev. Lett.}\ }\textbf {\bibinfo
  {volume} {116}},\ \bibinfo {pages} {231102} (\bibinfo {year} {2016})},\
  \Eprint {http://arxiv.org/abs/1602.06951} {arXiv:1602.06951
  [gr-qc]}\BibitemShut {NoStop}%
\bibitem [{\citenamefont {Magorrian}\ \emph {et~al.}(1998)\citenamefont
  {Magorrian}, \citenamefont {Tremaine}, \citenamefont {Richstone},
  \citenamefont {Bender}, \citenamefont {Bower} \emph
  {et~al.}}]{Magorrian:1997hw}%
  \BibitemOpen
  \bibfield  {author} {\bibinfo {author} {\bibfnamefont {John}\ \bibnamefont
  {Magorrian}}, \bibinfo {author} {\bibfnamefont {Scott}\ \bibnamefont
  {Tremaine}}, \bibinfo {author} {\bibfnamefont {Douglas}\ \bibnamefont
  {Richstone}}, \bibinfo {author} {\bibfnamefont {Ralf}\ \bibnamefont
  {Bender}}, \bibinfo {author} {\bibfnamefont {Gary}\ \bibnamefont {Bower}},
  \emph {et~al.},\ }\href {\doibase10.1086/300353} {\bibfield  {journal}
  {\bibinfo  {journal} {Astron. J.}\ }\textbf {\bibinfo {volume} {115}},\
  \bibinfo {pages} {2285} (\bibinfo {year} {1998})},\ \Eprint
  {http://arxiv.org/abs/astro-ph/9708072} {arXiv:astro-ph/9708072
  [astro-ph]}\BibitemShut {NoStop}%
\bibitem [{\citenamefont {Gair}\ \emph {et~al.}(2004)\citenamefont {Gair},
  \citenamefont {Barack}, \citenamefont {Creighton}, \citenamefont {Cutler},
  \citenamefont {Larson} \emph {et~al.}}]{Gair:2004iv}%
  \BibitemOpen
  \bibfield  {author} {\bibinfo {author} {\bibfnamefont {Jonathan~R.}\
  \bibnamefont {Gair}}, \bibinfo {author} {\bibfnamefont {Leor}\ \bibnamefont
  {Barack}}, \bibinfo {author} {\bibfnamefont {Teviet}\ \bibnamefont
  {Creighton}}, \bibinfo {author} {\bibfnamefont {Curt}\ \bibnamefont
  {Cutler}}, \bibinfo {author} {\bibfnamefont {Shane~L.}\ \bibnamefont
  {Larson}},  \emph {et~al.},\ }\href {\doibase10.1088/0264-9381/21/20/003}
  {\bibfield  {journal} {\bibinfo  {journal} {Class. Quant. Grav.}\ }\textbf
  {\bibinfo {volume} {21}},\ \bibinfo {pages} {S1595--S1606} (\bibinfo {year}
  {2004})},\ \Eprint {http://arxiv.org/abs/gr-qc/0405137} {arXiv:gr-qc/0405137
  [gr-qc]}\BibitemShut {NoStop}%
\bibitem [{\citenamefont {Amaro-Seoane}\ \emph {et~al.}(2007)\citenamefont
  {Amaro-Seoane}, \citenamefont {Gair}, \citenamefont {Freitag}, \citenamefont
  {Coleman~Miller}, \citenamefont {Mandel}, \citenamefont {Cutler},\ and\
  \citenamefont {Babak}}]{AmaroSeoane:2007aw}%
  \BibitemOpen
  \bibfield  {author} {\bibinfo {author} {\bibfnamefont {Pau}\ \bibnamefont
  {Amaro-Seoane}}, \bibinfo {author} {\bibfnamefont {Jonathan~R.}\ \bibnamefont
  {Gair}}, \bibinfo {author} {\bibfnamefont {Marc}\ \bibnamefont {Freitag}},
  \bibinfo {author} {\bibfnamefont {M.}~\bibnamefont {Coleman~Miller}},
  \bibinfo {author} {\bibfnamefont {Ilya}\ \bibnamefont {Mandel}}, \bibinfo
  {author} {\bibfnamefont {Curt~J.}\ \bibnamefont {Cutler}}, \ and\ \bibinfo
  {author} {\bibfnamefont {Stanislav}\ \bibnamefont {Babak}},\ }\href
  {\doibase10.1088/0264-9381/24/17/R01} {\bibfield  {journal} {\bibinfo
  {journal} {Class. Quant. Grav.}\ }\textbf {\bibinfo {volume} {24}},\ \bibinfo
  {pages} {R113--R169} (\bibinfo {year} {2007})},\ \Eprint
  {http://arxiv.org/abs/astro-ph/0703495} {arXiv:astro-ph/0703495
  [ASTRO-PH]}\BibitemShut {NoStop}%
\bibitem [{\citenamefont {Gair}(2009)}]{Gair:2008bx}%
  \BibitemOpen
  \bibfield  {author} {\bibinfo {author} {\bibfnamefont {Jonathan~R.}\
  \bibnamefont {Gair}},\ }\href {\doibase10.1088/0264-9381/26/9/094034}
  {\bibfield  {journal} {\bibinfo  {journal} {Class. Quant. Grav.}\ }\textbf
  {\bibinfo {volume} {26}},\ \bibinfo {pages} {094034} (\bibinfo {year}
  {2009})},\ \Eprint {http://arxiv.org/abs/0811.0188} {arXiv:0811.0188
  [gr-qc]}\BibitemShut {NoStop}%
\bibitem [{\citenamefont {Amaro-Seoane}\ \emph {et~al.}(2013)\citenamefont
  {Amaro-Seoane} \emph {et~al.}}]{AmaroSeoane:2012km}%
  \BibitemOpen
  \bibfield  {author} {\bibinfo {author} {\bibfnamefont {Pau}\ \bibnamefont
  {Amaro-Seoane}} \emph {et~al.},\ }\href@noop {} {\bibfield  {journal}
  {\bibinfo  {journal} {GW Notes}\ }\textbf {\bibinfo {volume} {6}},\ \bibinfo
  {pages} {4--110} (\bibinfo {year} {2013})},\ \Eprint
  {http://arxiv.org/abs/1201.3621} {arXiv:1201.3621 [astro-ph.CO]}\BibitemShut
  {NoStop}%
\bibitem [{\citenamefont {Hinderer}\ and\ \citenamefont
  {Flanagan}(2008)}]{Hinderer:2008dm}%
  \BibitemOpen
  \bibfield  {author} {\bibinfo {author} {\bibfnamefont {Tanja}\ \bibnamefont
  {Hinderer}}\ and\ \bibinfo {author} {\bibfnamefont {Eanna~E.}\ \bibnamefont
  {Flanagan}},\ }\href {\doibase10.1103/PhysRevD.78.064028} {\bibfield
  {journal} {\bibinfo  {journal} {Phys. Rev.}\ }\textbf {\bibinfo {volume}
  {D78}},\ \bibinfo {pages} {064028} (\bibinfo {year} {2008})},\ \Eprint
  {http://arxiv.org/abs/0805.3337} {arXiv:0805.3337 [gr-qc]}\BibitemShut
  {NoStop}%
\bibitem [{\citenamefont {Isoyama}\ \emph {et~al.}(2013)\citenamefont
  {Isoyama}, \citenamefont {Fujita}, \citenamefont {Sago}, \citenamefont
  {Tagoshi},\ and\ \citenamefont {Tanaka}}]{Isoyama:2012bx}%
  \BibitemOpen
  \bibfield  {author} {\bibinfo {author} {\bibfnamefont {Soichiro}\
  \bibnamefont {Isoyama}}, \bibinfo {author} {\bibfnamefont {Ryuichi}\
  \bibnamefont {Fujita}}, \bibinfo {author} {\bibfnamefont {Norichika}\
  \bibnamefont {Sago}}, \bibinfo {author} {\bibfnamefont {Hideyuki}\
  \bibnamefont {Tagoshi}}, \ and\ \bibinfo {author} {\bibfnamefont {Takahiro}\
  \bibnamefont {Tanaka}},\ }\href {\doibase10.1103/PhysRevD.87.024010}
  {\bibfield  {journal} {\bibinfo  {journal} {Phys. Rev.}\ }\textbf {\bibinfo
  {volume} {D87}},\ \bibinfo {pages} {024010} (\bibinfo {year} {2013})},\
  \Eprint {http://arxiv.org/abs/1210.2569} {arXiv:1210.2569
  [gr-qc]}\BibitemShut {NoStop}%
\bibitem [{\citenamefont {Burko}\ and\ \citenamefont
  {Khanna}(2013)}]{Burko:2013cca}%
  \BibitemOpen
  \bibfield  {author} {\bibinfo {author} {\bibfnamefont {Lior~M.}\ \bibnamefont
  {Burko}}\ and\ \bibinfo {author} {\bibfnamefont {Gaurav}\ \bibnamefont
  {Khanna}},\ }\href {\doibase10.1103/PhysRevD.88.024002} {\bibfield  {journal}
  {\bibinfo  {journal} {Phys. Rev.}\ }\textbf {\bibinfo {volume} {D88}},\
  \bibinfo {pages} {024002} (\bibinfo {year} {2013})},\ \Eprint
  {http://arxiv.org/abs/1304.5296} {arXiv:1304.5296 [gr-qc]}\BibitemShut
  {NoStop}%
\bibitem [{LISA Mission web site()}]{LISA}%
  \BibitemOpen
  LISA Mission web site,\ \href@noop {} {}\bibinfo {howpublished}
  {\url{https://www.lisamission.org}} (\bibinfo {year} {accessed June
  2018})\BibitemShut {NoStop}%
\bibitem [{\citenamefont {Barack}\ and\ \citenamefont
  {Cutler}(2004)}]{Barack:2003fp}%
  \BibitemOpen
  \bibfield  {author} {\bibinfo {author} {\bibfnamefont {Leor}\ \bibnamefont
  {Barack}}\ and\ \bibinfo {author} {\bibfnamefont {Curt}\ \bibnamefont
  {Cutler}},\ }\href {\doibase10.1103/PhysRevD.69.082005} {\bibfield  {journal}
  {\bibinfo  {journal} {Phys. Rev.}\ }\textbf {\bibinfo {volume} {D69}},\
  \bibinfo {pages} {082005} (\bibinfo {year} {2004})},\ \Eprint
  {http://arxiv.org/abs/gr-qc/0310125} {arXiv:gr-qc/0310125
  [gr-qc]}\BibitemShut {NoStop}%
\bibitem [{\citenamefont {Babak}\ \emph {et~al.}(2011)\citenamefont {Babak},
  \citenamefont {Gair}, \citenamefont {Petiteau},\ and\ \citenamefont
  {Sesana}}]{Babak:2010ej}%
  \BibitemOpen
  \bibfield  {author} {\bibinfo {author} {\bibfnamefont {Stanislav}\
  \bibnamefont {Babak}}, \bibinfo {author} {\bibfnamefont {Jonathan~R.}\
  \bibnamefont {Gair}}, \bibinfo {author} {\bibfnamefont {Antoine}\
  \bibnamefont {Petiteau}}, \ and\ \bibinfo {author} {\bibfnamefont {Alberto}\
  \bibnamefont {Sesana}},\ }\href {\doibase10.1088/0264-9381/28/11/114001}
  {\bibfield  {journal} {\bibinfo  {journal} {Class. Quant. Grav.}\ }\textbf
  {\bibinfo {volume} {28}},\ \bibinfo {pages} {114001} (\bibinfo {year}
  {2011})},\ \Eprint {http://arxiv.org/abs/1011.2062} {arXiv:1011.2062
  [gr-qc]}\BibitemShut {NoStop}%
\bibitem [{\citenamefont {Gair}\ \emph
  {et~al.}(2011{\natexlab{a}})\citenamefont {Gair}, \citenamefont {Sesana},
  \citenamefont {Berti},\ and\ \citenamefont {Volonteri}}]{Gair:2010bx}%
  \BibitemOpen
  \bibfield  {author} {\bibinfo {author} {\bibfnamefont {Jonathan~R.}\
  \bibnamefont {Gair}}, \bibinfo {author} {\bibfnamefont {Alberto}\
  \bibnamefont {Sesana}}, \bibinfo {author} {\bibfnamefont {Emanuele}\
  \bibnamefont {Berti}}, \ and\ \bibinfo {author} {\bibfnamefont {Marta}\
  \bibnamefont {Volonteri}},\ }\href {\doibase10.1088/0264-9381/28/9/094018}
  {\bibfield  {journal} {\bibinfo  {journal} {Class. Quant. Grav.}\ }\textbf
  {\bibinfo {volume} {28}},\ \bibinfo {pages} {094018} (\bibinfo {year}
  {2011}{\natexlab{a}})},\ \Eprint {http://arxiv.org/abs/1009.6172}
  {arXiv:1009.6172 [gr-qc]}\BibitemShut {NoStop}%
\bibitem [{\citenamefont {Abbott}\ \emph
  {et~al.}(2017{\natexlab{m}})\citenamefont {Abbott} \emph
  {et~al.}}]{Abbott:2017iws}%
  \BibitemOpen
  \bibfield  {author} {\bibinfo {author} {\bibfnamefont {Benjamin~P.}\
  \bibnamefont {Abbott}} \emph {et~al.} (\bibinfo {collaboration} {Virgo, LIGO
  Scientific}),\ }\href {\doibase10.1103/PhysRevD.96.022001} {\bibfield
  {journal} {\bibinfo  {journal} {Phys. Rev.}\ }\textbf {\bibinfo {volume}
  {D96}},\ \bibinfo {pages} {022001} (\bibinfo {year} {2017}{\natexlab{m}})},\
  \Eprint {http://arxiv.org/abs/1704.04628} {arXiv:1704.04628
  [gr-qc]}\BibitemShut {NoStop}%
\bibitem [{\citenamefont {Haster}\ \emph {et~al.}(2016)\citenamefont {Haster},
  \citenamefont {Wang}, \citenamefont {Berry}, \citenamefont {Stevenson},
  \citenamefont {Veitch},\ and\ \citenamefont {Mandel}}]{Haster:2015cnn}%
  \BibitemOpen
  \bibfield  {author} {\bibinfo {author} {\bibfnamefont {Carl-Johan}\
  \bibnamefont {Haster}}, \bibinfo {author} {\bibfnamefont {Zhilu}\
  \bibnamefont {Wang}}, \bibinfo {author} {\bibfnamefont {Christopher P.~L.}\
  \bibnamefont {Berry}}, \bibinfo {author} {\bibfnamefont {Simon}\ \bibnamefont
  {Stevenson}}, \bibinfo {author} {\bibfnamefont {John}\ \bibnamefont
  {Veitch}}, \ and\ \bibinfo {author} {\bibfnamefont {Ilya}\ \bibnamefont
  {Mandel}},\ }\href {\doibase10.1093/mnras/stw233} {\bibfield  {journal}
  {\bibinfo  {journal} {Mon. Not. Roy. Astron. Soc.}\ }\textbf {\bibinfo
  {volume} {457}},\ \bibinfo {pages} {4499--4506} (\bibinfo {year} {2016})},\
  \Eprint {http://arxiv.org/abs/1511.01431} {arXiv:1511.01431
  [astro-ph.HE]}\BibitemShut {NoStop}%
\bibitem [{\citenamefont {Glampedakis}\ and\ \citenamefont
  {Kennefick}(2002)}]{Glampedakis:2002ya}%
  \BibitemOpen
  \bibfield  {author} {\bibinfo {author} {\bibfnamefont {Kostas}\ \bibnamefont
  {Glampedakis}}\ and\ \bibinfo {author} {\bibfnamefont {Daniel}\ \bibnamefont
  {Kennefick}},\ }\href {\doibase10.1103/PhysRevD.66.044002} {\bibfield
  {journal} {\bibinfo  {journal} {Phys. Rev.}\ }\textbf {\bibinfo {volume}
  {D66}},\ \bibinfo {pages} {044002} (\bibinfo {year} {2002})},\ \Eprint
  {http://arxiv.org/abs/gr-qc/0203086} {arXiv:gr-qc/0203086
  [gr-qc]}\BibitemShut {NoStop}%
\bibitem [{\citenamefont {Mino}(2003)}]{Mino:2003yg}%
  \BibitemOpen
  \bibfield  {author} {\bibinfo {author} {\bibfnamefont {Yasushi}\ \bibnamefont
  {Mino}},\ }\href {\doibase10.1103/PhysRevD.67.084027} {\bibfield  {journal}
  {\bibinfo  {journal} {Phys. Rev.}\ }\textbf {\bibinfo {volume} {D67}},\
  \bibinfo {pages} {084027} (\bibinfo {year} {2003})},\ \Eprint
  {http://arxiv.org/abs/gr-qc/0302075} {arXiv:gr-qc/0302075
  [gr-qc]}\BibitemShut {NoStop}%
\bibitem [{\citenamefont {Hughes}\ \emph {et~al.}(2005)\citenamefont {Hughes},
  \citenamefont {Drasco}, \citenamefont {Flanagan},\ and\ \citenamefont
  {Franklin}}]{Hughes:2005qb}%
  \BibitemOpen
  \bibfield  {author} {\bibinfo {author} {\bibfnamefont {Scott~A.}\
  \bibnamefont {Hughes}}, \bibinfo {author} {\bibfnamefont {Steve}\
  \bibnamefont {Drasco}}, \bibinfo {author} {\bibfnamefont {Eanna~E.}\
  \bibnamefont {Flanagan}}, \ and\ \bibinfo {author} {\bibfnamefont {Joel}\
  \bibnamefont {Franklin}},\ }\href {\doibase10.1103/PhysRevLett.94.221101}
  {\bibfield  {journal} {\bibinfo  {journal} {Phys. Rev. Lett.}\ }\textbf
  {\bibinfo {volume} {94}},\ \bibinfo {pages} {221101} (\bibinfo {year}
  {2005})},\ \Eprint {http://arxiv.org/abs/gr-qc/0504015} {arXiv:gr-qc/0504015
  [gr-qc]}\BibitemShut {NoStop}%
\bibitem [{\citenamefont {Sago}\ \emph {et~al.}(2006)\citenamefont {Sago},
  \citenamefont {Tanaka}, \citenamefont {Hikida}, \citenamefont {Ganz},\ and\
  \citenamefont {Nakano}}]{Sago:2005fn}%
  \BibitemOpen
  \bibfield  {author} {\bibinfo {author} {\bibfnamefont {Norichika}\
  \bibnamefont {Sago}}, \bibinfo {author} {\bibfnamefont {Takahiro}\
  \bibnamefont {Tanaka}}, \bibinfo {author} {\bibfnamefont {Wataru}\
  \bibnamefont {Hikida}}, \bibinfo {author} {\bibfnamefont {Katsuhiko}\
  \bibnamefont {Ganz}}, \ and\ \bibinfo {author} {\bibfnamefont {Hiroyuki}\
  \bibnamefont {Nakano}},\ }\href {\doibase10.1143/PTP.115.873} {\bibfield
  {journal} {\bibinfo  {journal} {Prog. Theor. Phys.}\ }\textbf {\bibinfo
  {volume} {115}},\ \bibinfo {pages} {873--907} (\bibinfo {year} {2006})},\
  \Eprint {http://arxiv.org/abs/gr-qc/0511151} {arXiv:gr-qc/0511151
  [gr-qc]}\BibitemShut {NoStop}%
\bibitem [{\citenamefont {Drasco}\ \emph {et~al.}(2005)\citenamefont {Drasco},
  \citenamefont {Flanagan},\ and\ \citenamefont {Hughes}}]{Drasco:2005is}%
  \BibitemOpen
  \bibfield  {author} {\bibinfo {author} {\bibfnamefont {Steve}\ \bibnamefont
  {Drasco}}, \bibinfo {author} {\bibfnamefont {Eanna~E.}\ \bibnamefont
  {Flanagan}}, \ and\ \bibinfo {author} {\bibfnamefont {Scott~A.}\ \bibnamefont
  {Hughes}},\ }\href {\doibase10.1088/0264-9381/22/15/011} {\bibfield
  {journal} {\bibinfo  {journal} {Class. Quant. Grav.}\ }\textbf {\bibinfo
  {volume} {22}},\ \bibinfo {pages} {S801--846} (\bibinfo {year} {2005})},\
  \Eprint {http://arxiv.org/abs/gr-qc/0505075} {arXiv:gr-qc/0505075
  [gr-qc]}\BibitemShut {NoStop}%
\bibitem [{\citenamefont {Drasco}\ and\ \citenamefont
  {Hughes}(2006)}]{Drasco:2005kz}%
  \BibitemOpen
  \bibfield  {author} {\bibinfo {author} {\bibfnamefont {Steve}\ \bibnamefont
  {Drasco}}\ and\ \bibinfo {author} {\bibfnamefont {Scott~A.}\ \bibnamefont
  {Hughes}},\ }\href {\doibase10.1103/PhysRevD.73.024027} {\bibfield  {journal}
  {\bibinfo  {journal} {Phys. Rev.}\ }\textbf {\bibinfo {volume} {D73}},\
  \bibinfo {pages} {024027} (\bibinfo {year} {2006})},\ \bibinfo {note}
  {[Erratum: Phys. Rev.D90,no.10,109905(2014)]},\ \Eprint
  {http://arxiv.org/abs/gr-qc/0509101} {arXiv:gr-qc/0509101
  [gr-qc]}\BibitemShut {NoStop}%
\bibitem [{\citenamefont {Sundararajan}\ \emph {et~al.}(2007)\citenamefont
  {Sundararajan}, \citenamefont {Khanna},\ and\ \citenamefont
  {Hughes}}]{Sundararajan:2007jg}%
  \BibitemOpen
  \bibfield  {author} {\bibinfo {author} {\bibfnamefont {Pranesh~A.}\
  \bibnamefont {Sundararajan}}, \bibinfo {author} {\bibfnamefont {Gaurav}\
  \bibnamefont {Khanna}}, \ and\ \bibinfo {author} {\bibfnamefont {Scott~A.}\
  \bibnamefont {Hughes}},\ }\href {\doibase10.1103/PhysRevD.76.104005}
  {\bibfield  {journal} {\bibinfo  {journal} {Phys. Rev.}\ }\textbf {\bibinfo
  {volume} {D76}},\ \bibinfo {pages} {104005} (\bibinfo {year} {2007})},\
  \Eprint {http://arxiv.org/abs/gr-qc/0703028} {arXiv:gr-qc/0703028
  [gr-qc]}\BibitemShut {NoStop}%
\bibitem [{\citenamefont {Fujita}\ \emph {et~al.}(2009)\citenamefont {Fujita},
  \citenamefont {Hikida},\ and\ \citenamefont {Tagoshi}}]{Fujita:2009us}%
  \BibitemOpen
  \bibfield  {author} {\bibinfo {author} {\bibfnamefont {Ryuichi}\ \bibnamefont
  {Fujita}}, \bibinfo {author} {\bibfnamefont {Wataru}\ \bibnamefont {Hikida}},
  \ and\ \bibinfo {author} {\bibfnamefont {Hideyuki}\ \bibnamefont {Tagoshi}},\
  }\href {\doibase10.1143/PTP.121.843} {\bibfield  {journal} {\bibinfo
  {journal} {Prog. Theor. Phys.}\ }\textbf {\bibinfo {volume} {121}},\ \bibinfo
  {pages} {843--874} (\bibinfo {year} {2009})},\ \Eprint
  {http://arxiv.org/abs/0904.3810} {arXiv:0904.3810 [gr-qc]}\BibitemShut
  {NoStop}%
\bibitem [{\citenamefont {Poisson}\ \emph {et~al.}(2011)\citenamefont
  {Poisson}, \citenamefont {Pound},\ and\ \citenamefont
  {Vega}}]{Poisson:2011nh}%
  \BibitemOpen
  \bibfield  {author} {\bibinfo {author} {\bibfnamefont {Eric}\ \bibnamefont
  {Poisson}}, \bibinfo {author} {\bibfnamefont {Adam}\ \bibnamefont {Pound}}, \
  and\ \bibinfo {author} {\bibfnamefont {Ian}\ \bibnamefont {Vega}},\ }\href
  {\doibase10.12942/lrr-2011-7} {\bibfield  {journal} {\bibinfo  {journal}
  {Living Rev. Rel.}\ }\textbf {\bibinfo {volume} {14}},\ \bibinfo {pages} {7}
  (\bibinfo {year} {2011})},\ \Eprint {http://arxiv.org/abs/1102.0529}
  {arXiv:1102.0529 [gr-qc]}\BibitemShut {NoStop}%
\bibitem [{\citenamefont {Barack}\ and\ \citenamefont
  {Pound}(2018)}]{Barack:2018yvs}%
  \BibitemOpen
  \bibfield  {author} {\bibinfo {author} {\bibfnamefont {Leor}\ \bibnamefont
  {Barack}}\ and\ \bibinfo {author} {\bibfnamefont {Adam}\ \bibnamefont
  {Pound}},\ }\href@noop {} {\  (\bibinfo {year} {2018})},\ \Eprint
  {http://arxiv.org/abs/1805.10385} {arXiv:1805.10385 [gr-qc]}\BibitemShut
  {NoStop}%
\bibitem [{\citenamefont {DeWitt}\ and\ \citenamefont
  {Brehme}(1960)}]{DeWitt:1960fc}%
  \BibitemOpen
  \bibfield  {author} {\bibinfo {author} {\bibfnamefont {Bryce~S.}\
  \bibnamefont {DeWitt}}\ and\ \bibinfo {author} {\bibfnamefont {Robert~W.}\
  \bibnamefont {Brehme}},\ }\href {\doibase10.1016/0003-4916(60)90030-0}
  {\bibfield  {journal} {\bibinfo  {journal} {Annals Phys.}\ }\textbf {\bibinfo
  {volume} {9}},\ \bibinfo {pages} {220--259} (\bibinfo {year}
  {1960})}\BibitemShut {NoStop}%
\bibitem [{\citenamefont {Hobbs}(1968)}]{Hobbs:1968a}%
  \BibitemOpen
  \bibfield  {author} {\bibinfo {author} {\bibfnamefont {J.~M.}\ \bibnamefont
  {Hobbs}},\ }\href {\doibase10.1016/0003-4916(68)90231-5} {\bibfield
  {journal} {\bibinfo  {journal} {Ann. Phys.}\ }\textbf {\bibinfo {volume}
  {47}},\ \bibinfo {pages} {141} (\bibinfo {year} {1968})}\BibitemShut
  {NoStop}%
\bibitem [{\citenamefont {Mino}\ \emph {et~al.}(1997)\citenamefont {Mino},
  \citenamefont {Sasaki},\ and\ \citenamefont {Tanaka}}]{Mino:1996nk}%
  \BibitemOpen
  \bibfield  {author} {\bibinfo {author} {\bibfnamefont {Yasushi}\ \bibnamefont
  {Mino}}, \bibinfo {author} {\bibfnamefont {Misao}\ \bibnamefont {Sasaki}}, \
  and\ \bibinfo {author} {\bibfnamefont {Takahiro}\ \bibnamefont {Tanaka}},\
  }\href {\doibase10.1103/PhysRevD.55.3457} {\bibfield  {journal} {\bibinfo
  {journal} {Phys. Rev.}\ }\textbf {\bibinfo {volume} {D55}},\ \bibinfo {pages}
  {3457--3476} (\bibinfo {year} {1997})},\ \Eprint
  {http://arxiv.org/abs/gr-qc/9606018} {arXiv:gr-qc/9606018
  [gr-qc]}\BibitemShut {NoStop}%
\bibitem [{\citenamefont {Quinn}\ and\ \citenamefont
  {Wald}(1997)}]{Quinn:1996am}%
  \BibitemOpen
  \bibfield  {author} {\bibinfo {author} {\bibfnamefont {Theodore~C.}\
  \bibnamefont {Quinn}}\ and\ \bibinfo {author} {\bibfnamefont {Robert~M.}\
  \bibnamefont {Wald}},\ }\href {\doibase10.1103/PhysRevD.56.3381} {\bibfield
  {journal} {\bibinfo  {journal} {Phys. Rev.}\ }\textbf {\bibinfo {volume}
  {D56}},\ \bibinfo {pages} {3381--3394} (\bibinfo {year} {1997})},\ \Eprint
  {http://arxiv.org/abs/gr-qc/9610053} {arXiv:gr-qc/9610053
  [gr-qc]}\BibitemShut {NoStop}%
\bibitem [{\citenamefont {Detweiler}\ and\ \citenamefont
  {Whiting}(2003)}]{Detweiler:2002mi}%
  \BibitemOpen
  \bibfield  {author} {\bibinfo {author} {\bibfnamefont {Steven~L.}\
  \bibnamefont {Detweiler}}\ and\ \bibinfo {author} {\bibfnamefont
  {Bernard~F.}\ \bibnamefont {Whiting}},\ }\href
  {\doibase10.1103/PhysRevD.67.024025} {\bibfield  {journal} {\bibinfo
  {journal} {Phys. Rev.}\ }\textbf {\bibinfo {volume} {D67}},\ \bibinfo {pages}
  {024025} (\bibinfo {year} {2003})},\ \Eprint
  {http://arxiv.org/abs/gr-qc/0202086} {arXiv:gr-qc/0202086
  [gr-qc]}\BibitemShut {NoStop}%
\bibitem [{\citenamefont {Gralla}\ and\ \citenamefont
  {Wald}(2008)}]{Gralla:2008fg}%
  \BibitemOpen
  \bibfield  {author} {\bibinfo {author} {\bibfnamefont {Samuel~E.}\
  \bibnamefont {Gralla}}\ and\ \bibinfo {author} {\bibfnamefont {Robert~M.}\
  \bibnamefont {Wald}},\ }\href {\doibase10.1088/0264-9381/25/20/205009,
  10.1088/0264-9381/28/15/159501} {\bibfield  {journal} {\bibinfo  {journal}
  {Class. Quant. Grav.}\ }\textbf {\bibinfo {volume} {25}},\ \bibinfo {pages}
  {205009} (\bibinfo {year} {2008})},\ \Eprint {http://arxiv.org/abs/0806.3293}
  {arXiv:0806.3293 [gr-qc]}\BibitemShut {NoStop}%
\bibitem [{\citenamefont {Pound}(2010{\natexlab{a}})}]{Pound:2009sm}%
  \BibitemOpen
  \bibfield  {author} {\bibinfo {author} {\bibfnamefont {Adam}\ \bibnamefont
  {Pound}},\ }\href {\doibase10.1103/PhysRevD.81.024023} {\bibfield  {journal}
  {\bibinfo  {journal} {Phys. Rev.}\ }\textbf {\bibinfo {volume} {D81}},\
  \bibinfo {pages} {024023} (\bibinfo {year} {2010}{\natexlab{a}})},\ \Eprint
  {http://arxiv.org/abs/0907.5197} {arXiv:0907.5197 [gr-qc]}\BibitemShut
  {NoStop}%
\bibitem [{\citenamefont {Pound}(2010{\natexlab{b}})}]{Pound:2010pj}%
  \BibitemOpen
  \bibfield  {author} {\bibinfo {author} {\bibfnamefont {Adam}\ \bibnamefont
  {Pound}},\ }\href {\doibase10.1103/PhysRevD.81.124009} {\bibfield  {journal}
  {\bibinfo  {journal} {Phys. Rev.}\ }\textbf {\bibinfo {volume} {D81}},\
  \bibinfo {pages} {124009} (\bibinfo {year} {2010}{\natexlab{b}})},\ \Eprint
  {http://arxiv.org/abs/1003.3954} {arXiv:1003.3954 [gr-qc]}\BibitemShut
  {NoStop}%
\bibitem [{\citenamefont {Pound}(2012{\natexlab{a}})}]{Pound:2012dk}%
  \BibitemOpen
  \bibfield  {author} {\bibinfo {author} {\bibfnamefont {Adam}\ \bibnamefont
  {Pound}},\ }\href {\doibase10.1103/PhysRevD.86.084019} {\bibfield  {journal}
  {\bibinfo  {journal} {Phys. Rev.}\ }\textbf {\bibinfo {volume} {D86}},\
  \bibinfo {pages} {084019} (\bibinfo {year} {2012}{\natexlab{a}})},\ \Eprint
  {http://arxiv.org/abs/1206.6538} {arXiv:1206.6538 [gr-qc]}\BibitemShut
  {NoStop}%
\bibitem [{\citenamefont {Rosenthal}(2005)}]{Rosenthal:2005it}%
  \BibitemOpen
  \bibfield  {author} {\bibinfo {author} {\bibfnamefont {Eran}\ \bibnamefont
  {Rosenthal}},\ }\href {\doibase10.1103/PhysRevD.72.121503} {\bibfield
  {journal} {\bibinfo  {journal} {Phys. Rev.}\ }\textbf {\bibinfo {volume}
  {D72}},\ \bibinfo {pages} {121503} (\bibinfo {year} {2005})},\ \Eprint
  {http://arxiv.org/abs/gr-qc/0508050} {arXiv:gr-qc/0508050
  [gr-qc]}\BibitemShut {NoStop}%
\bibitem [{\citenamefont {Rosenthal}(2006)}]{Rosenthal:2006iy}%
  \BibitemOpen
  \bibfield  {author} {\bibinfo {author} {\bibfnamefont {Eran}\ \bibnamefont
  {Rosenthal}},\ }\href {\doibase10.1103/PhysRevD.74.084018} {\bibfield
  {journal} {\bibinfo  {journal} {Phys. Rev.}\ }\textbf {\bibinfo {volume}
  {D74}},\ \bibinfo {pages} {084018} (\bibinfo {year} {2006})},\ \Eprint
  {http://arxiv.org/abs/gr-qc/0609069} {arXiv:gr-qc/0609069
  [gr-qc]}\BibitemShut {NoStop}%
\bibitem [{\citenamefont {Detweiler}(2012)}]{Detweiler:2011tt}%
  \BibitemOpen
  \bibfield  {author} {\bibinfo {author} {\bibfnamefont {Steven}\ \bibnamefont
  {Detweiler}},\ }\href {\doibase10.1103/PhysRevD.85.044048} {\bibfield
  {journal} {\bibinfo  {journal} {Phys. Rev.}\ }\textbf {\bibinfo {volume}
  {D85}},\ \bibinfo {pages} {044048} (\bibinfo {year} {2012})},\ \Eprint
  {http://arxiv.org/abs/1107.2098} {arXiv:1107.2098 [gr-qc]}\BibitemShut
  {NoStop}%
\bibitem [{\citenamefont {Gralla}(2012)}]{Gralla:2012db}%
  \BibitemOpen
  \bibfield  {author} {\bibinfo {author} {\bibfnamefont {Samuel~E.}\
  \bibnamefont {Gralla}},\ }\href {\doibase10.1103/PhysRevD.85.124011}
  {\bibfield  {journal} {\bibinfo  {journal} {Phys. Rev.}\ }\textbf {\bibinfo
  {volume} {D85}},\ \bibinfo {pages} {124011} (\bibinfo {year} {2012})},\
  \Eprint {http://arxiv.org/abs/1203.3189} {arXiv:1203.3189
  [gr-qc]}\BibitemShut {NoStop}%
\bibitem [{\citenamefont {Pound}(2012{\natexlab{b}})}]{Pound:2012nt}%
  \BibitemOpen
  \bibfield  {author} {\bibinfo {author} {\bibfnamefont {Adam}\ \bibnamefont
  {Pound}},\ }\href {\doibase10.1103/PhysRevLett.109.051101} {\bibfield
  {journal} {\bibinfo  {journal} {Phys. Rev. Lett.}\ }\textbf {\bibinfo
  {volume} {109}},\ \bibinfo {pages} {051101} (\bibinfo {year}
  {2012}{\natexlab{b}})},\ \Eprint {http://arxiv.org/abs/1201.5089}
  {arXiv:1201.5089 [gr-qc]}\BibitemShut {NoStop}%
\bibitem [{\citenamefont {Pound}\ and\ \citenamefont
  {Miller}(2014)}]{Pound:2014xva}%
  \BibitemOpen
  \bibfield  {author} {\bibinfo {author} {\bibfnamefont {Adam}\ \bibnamefont
  {Pound}}\ and\ \bibinfo {author} {\bibfnamefont {Jeremy}\ \bibnamefont
  {Miller}},\ }\href {\doibase10.1103/PhysRevD.89.104020} {\bibfield  {journal}
  {\bibinfo  {journal} {Phys. Rev.}\ }\textbf {\bibinfo {volume} {D89}},\
  \bibinfo {pages} {104020} (\bibinfo {year} {2014})},\ \Eprint
  {http://arxiv.org/abs/1403.1843} {arXiv:1403.1843 [gr-qc]}\BibitemShut
  {NoStop}%
\bibitem [{\citenamefont {Harte}(2008)}]{Harte:2008xq}%
  \BibitemOpen
  \bibfield  {author} {\bibinfo {author} {\bibfnamefont {Abraham~I.}\
  \bibnamefont {Harte}},\ }\href {\doibase10.1088/0264-9381/25/23/235020}
  {\bibfield  {journal} {\bibinfo  {journal} {Class. Quant. Grav.}\ }\textbf
  {\bibinfo {volume} {25}},\ \bibinfo {pages} {235020} (\bibinfo {year}
  {2008})},\ \Eprint {http://arxiv.org/abs/0807.1150} {arXiv:0807.1150
  [gr-qc]}\BibitemShut {NoStop}%
\bibitem [{\citenamefont {Harte}(2010)}]{Harte:2009yr}%
  \BibitemOpen
  \bibfield  {author} {\bibinfo {author} {\bibfnamefont {Abraham~I.}\
  \bibnamefont {Harte}},\ }\href {\doibase10.1088/0264-9381/27/13/135002}
  {\bibfield  {journal} {\bibinfo  {journal} {Class. Quant. Grav.}\ }\textbf
  {\bibinfo {volume} {27}},\ \bibinfo {pages} {135002} (\bibinfo {year}
  {2010})},\ \Eprint {http://arxiv.org/abs/0910.4614} {arXiv:0910.4614
  [gr-qc]}\BibitemShut {NoStop}%
\bibitem [{\citenamefont {Harte}(2012)}]{Harte:2011ku}%
  \BibitemOpen
  \bibfield  {author} {\bibinfo {author} {\bibfnamefont {Abraham~I.}\
  \bibnamefont {Harte}},\ }\href {\doibase10.1088/0264-9381/29/5/055012}
  {\bibfield  {journal} {\bibinfo  {journal} {Class. Quant. Grav.}\ }\textbf
  {\bibinfo {volume} {29}},\ \bibinfo {pages} {055012} (\bibinfo {year}
  {2012})},\ \Eprint {http://arxiv.org/abs/1103.0543} {arXiv:1103.0543
  [gr-qc]}\BibitemShut {NoStop}%
\bibitem [{\citenamefont {{Gal'tsov}}(1982)}]{1982JPhA...15.3737G}%
  \BibitemOpen
  \bibfield  {author} {\bibinfo {author} {\bibfnamefont {D.~V.}\ \bibnamefont
  {{Gal'tsov}}},\ }\href {\doibase10.1088/0305-4470/15/12/025} {\bibfield
  {journal} {\bibinfo  {journal} {Journal of Physics A Mathematical General}\
  }\textbf {\bibinfo {volume} {15}},\ \bibinfo {pages} {3737--3749} (\bibinfo
  {year} {1982})}\BibitemShut {NoStop}%
\bibitem [{\citenamefont {Poisson}\ and\ \citenamefont
  {Wiseman}(1998)}]{Poisson:Wiseman:1998}%
  \BibitemOpen
  \bibfield  {author} {\bibinfo {author} {\bibfnamefont {E.}~\bibnamefont
  {Poisson}}\ and\ \bibinfo {author} {\bibfnamefont {A.~G.}\ \bibnamefont
  {Wiseman}},\ }\href@noop {} {\enquote {\bibinfo {title} {{Suggestion at the
  1st Capra Ranch meeting on Radiation Reaction}},}\ } (\bibinfo {year}
  {1998})\BibitemShut {NoStop}%
\bibitem [{\citenamefont {Barack}\ and\ \citenamefont
  {Ori}(2000)}]{Barack:1999wf}%
  \BibitemOpen
  \bibfield  {author} {\bibinfo {author} {\bibfnamefont {Leor}\ \bibnamefont
  {Barack}}\ and\ \bibinfo {author} {\bibfnamefont {Amos}\ \bibnamefont
  {Ori}},\ }\href {\doibase10.1103/PhysRevD.61.061502} {\bibfield  {journal}
  {\bibinfo  {journal} {Phys. Rev.}\ }\textbf {\bibinfo {volume} {D61}},\
  \bibinfo {pages} {061502} (\bibinfo {year} {2000})},\ \Eprint
  {http://arxiv.org/abs/gr-qc/9912010} {arXiv:gr-qc/9912010
  [gr-qc]}\BibitemShut {NoStop}%
\bibitem [{\citenamefont {Barack}\ \emph {et~al.}(2002)\citenamefont {Barack},
  \citenamefont {Mino}, \citenamefont {Nakano}, \citenamefont {Ori},\ and\
  \citenamefont {Sasaki}}]{Barack:2001gx}%
  \BibitemOpen
  \bibfield  {author} {\bibinfo {author} {\bibfnamefont {Leor}\ \bibnamefont
  {Barack}}, \bibinfo {author} {\bibfnamefont {Yasushi}\ \bibnamefont {Mino}},
  \bibinfo {author} {\bibfnamefont {Hiroyuki}\ \bibnamefont {Nakano}}, \bibinfo
  {author} {\bibfnamefont {Amos}\ \bibnamefont {Ori}}, \ and\ \bibinfo {author}
  {\bibfnamefont {Misao}\ \bibnamefont {Sasaki}},\ }\href
  {\doibase10.1103/PhysRevLett.88.091101} {\bibfield  {journal} {\bibinfo
  {journal} {Phys. Rev. Lett.}\ }\textbf {\bibinfo {volume} {88}},\ \bibinfo
  {pages} {091101} (\bibinfo {year} {2002})},\ \Eprint
  {http://arxiv.org/abs/gr-qc/0111001} {arXiv:gr-qc/0111001
  [gr-qc]}\BibitemShut {NoStop}%
\bibitem [{\citenamefont {Anderson}\ and\ \citenamefont
  {Wiseman}(2005)}]{Anderson:2005gb}%
  \BibitemOpen
  \bibfield  {author} {\bibinfo {author} {\bibfnamefont {Warren~G.}\
  \bibnamefont {Anderson}}\ and\ \bibinfo {author} {\bibfnamefont {Alan~G.}\
  \bibnamefont {Wiseman}},\ }\href {\doibase10.1088/0264-9381/22/15/010}
  {\bibfield  {journal} {\bibinfo  {journal} {Class. Quant. Grav.}\ }\textbf
  {\bibinfo {volume} {22}},\ \bibinfo {pages} {S783--S800} (\bibinfo {year}
  {2005})},\ \Eprint {http://arxiv.org/abs/gr-qc/0506136} {arXiv:gr-qc/0506136
  [gr-qc]}\BibitemShut {NoStop}%
\bibitem [{\citenamefont {Barack}\ and\ \citenamefont
  {Golbourn}(2007)}]{Barack:2007jh}%
  \BibitemOpen
  \bibfield  {author} {\bibinfo {author} {\bibfnamefont {Leor}\ \bibnamefont
  {Barack}}\ and\ \bibinfo {author} {\bibfnamefont {Darren~A.}\ \bibnamefont
  {Golbourn}},\ }\href {\doibase10.1103/PhysRevD.76.044020} {\bibfield
  {journal} {\bibinfo  {journal} {Phys. Rev.}\ }\textbf {\bibinfo {volume}
  {D76}},\ \bibinfo {pages} {044020} (\bibinfo {year} {2007})},\ \Eprint
  {http://arxiv.org/abs/0705.3620} {arXiv:0705.3620 [gr-qc]}\BibitemShut
  {NoStop}%
\bibitem [{\citenamefont {Vega}\ and\ \citenamefont
  {Detweiler}(2008)}]{Vega:2007mc}%
  \BibitemOpen
  \bibfield  {author} {\bibinfo {author} {\bibfnamefont {Ian}\ \bibnamefont
  {Vega}}\ and\ \bibinfo {author} {\bibfnamefont {Steven~L.}\ \bibnamefont
  {Detweiler}},\ }\href {\doibase10.1103/PhysRevD.77.084008} {\bibfield
  {journal} {\bibinfo  {journal} {Phys. Rev.}\ }\textbf {\bibinfo {volume}
  {D77}},\ \bibinfo {pages} {084008} (\bibinfo {year} {2008})},\ \Eprint
  {http://arxiv.org/abs/0712.4405} {arXiv:0712.4405 [gr-qc]}\BibitemShut
  {NoStop}%
\bibitem [{\citenamefont {van~de Meent}(2018)}]{vandeMeent:2017bcc}%
  \BibitemOpen
  \bibfield  {author} {\bibinfo {author} {\bibfnamefont {Maarten}\ \bibnamefont
  {van~de Meent}},\ }\href {\doibase10.1103/PhysRevD.97.104033} {\bibfield
  {journal} {\bibinfo  {journal} {Phys. Rev.}\ }\textbf {\bibinfo {volume}
  {D97}},\ \bibinfo {pages} {104033} (\bibinfo {year} {2018})},\ \Eprint
  {http://arxiv.org/abs/1711.09607} {arXiv:1711.09607 [gr-qc]}\BibitemShut
  {NoStop}%
\bibitem [{\citenamefont {Wardell}\ and\ \citenamefont
  {Gopakumar}(2015)}]{Wardell:2015kea}%
  \BibitemOpen
  \bibfield  {author} {\bibinfo {author} {\bibfnamefont {Barry}\ \bibnamefont
  {Wardell}}\ and\ \bibinfo {author} {\bibfnamefont {Achamveedu}\ \bibnamefont
  {Gopakumar}},\ }\bibfield  {booktitle} {\emph {\bibinfo {booktitle}
  {{Proceedings, 524th WE-Heraeus-Seminar: Equations of Motion in Relativistic
  Gravity (EOM 2013): Bad Honnef, Germany, February 17-23, 2013}}},\ }\href
  {\doibase10.1007/978-3-319-18335-0_14} {\bibfield  {journal} {\bibinfo
  {journal} {Fund. Theor. Phys.}\ }\textbf {\bibinfo {volume} {179}},\ \bibinfo
  {pages} {487--522} (\bibinfo {year} {2015})},\ \Eprint
  {http://arxiv.org/abs/1501.07322} {arXiv:1501.07322 [gr-qc]}\BibitemShut
  {NoStop}%
\bibitem [{\citenamefont {Barack}\ and\ \citenamefont
  {Burko}(2000)}]{Barack:2000zq}%
  \BibitemOpen
  \bibfield  {author} {\bibinfo {author} {\bibfnamefont {Leor}\ \bibnamefont
  {Barack}}\ and\ \bibinfo {author} {\bibfnamefont {Lior~M.}\ \bibnamefont
  {Burko}},\ }\href {\doibase10.1103/PhysRevD.62.084040} {\bibfield  {journal}
  {\bibinfo  {journal} {Phys. Rev.}\ }\textbf {\bibinfo {volume} {D62}},\
  \bibinfo {pages} {084040} (\bibinfo {year} {2000})},\ \Eprint
  {http://arxiv.org/abs/gr-qc/0007033} {arXiv:gr-qc/0007033
  [gr-qc]}\BibitemShut {NoStop}%
\bibitem [{\citenamefont {Burko}(2000)}]{Burko:2000xx}%
  \BibitemOpen
  \bibfield  {author} {\bibinfo {author} {\bibfnamefont {Lior~M.}\ \bibnamefont
  {Burko}},\ }\href {\doibase10.1103/PhysRevLett.84.4529} {\bibfield  {journal}
  {\bibinfo  {journal} {Phys. Rev. Lett.}\ }\textbf {\bibinfo {volume} {84}},\
  \bibinfo {pages} {4529} (\bibinfo {year} {2000})},\ \Eprint
  {http://arxiv.org/abs/gr-qc/0003074} {arXiv:gr-qc/0003074
  [gr-qc]}\BibitemShut {NoStop}%
\bibitem [{\citenamefont {Burko}\ and\ \citenamefont
  {Liu}(2001)}]{Burko:2001kr}%
  \BibitemOpen
  \bibfield  {author} {\bibinfo {author} {\bibfnamefont {Lior~M.}\ \bibnamefont
  {Burko}}\ and\ \bibinfo {author} {\bibfnamefont {Yuk~Tung}\ \bibnamefont
  {Liu}},\ }\href {\doibase10.1103/PhysRevD.64.024006} {\bibfield  {journal}
  {\bibinfo  {journal} {Phys. Rev.}\ }\textbf {\bibinfo {volume} {D64}},\
  \bibinfo {pages} {024006} (\bibinfo {year} {2001})},\ \Eprint
  {http://arxiv.org/abs/gr-qc/0103008} {arXiv:gr-qc/0103008
  [gr-qc]}\BibitemShut {NoStop}%
\bibitem [{\citenamefont {Detweiler}\ \emph {et~al.}(2003)\citenamefont
  {Detweiler}, \citenamefont {Messaritaki},\ and\ \citenamefont
  {Whiting}}]{Detweiler:2002gi}%
  \BibitemOpen
  \bibfield  {author} {\bibinfo {author} {\bibfnamefont {Steven~L.}\
  \bibnamefont {Detweiler}}, \bibinfo {author} {\bibfnamefont {Eirini}\
  \bibnamefont {Messaritaki}}, \ and\ \bibinfo {author} {\bibfnamefont
  {Bernard~F.}\ \bibnamefont {Whiting}},\ }\href
  {\doibase10.1103/PhysRevD.67.104016} {\bibfield  {journal} {\bibinfo
  {journal} {Phys. Rev.}\ }\textbf {\bibinfo {volume} {D67}},\ \bibinfo {pages}
  {104016} (\bibinfo {year} {2003})},\ \Eprint
  {http://arxiv.org/abs/gr-qc/0205079} {arXiv:gr-qc/0205079
  [gr-qc]}\BibitemShut {NoStop}%
\bibitem [{\citenamefont {Diaz-Rivera}\ \emph {et~al.}(2004)\citenamefont
  {Diaz-Rivera}, \citenamefont {Messaritaki}, \citenamefont {Whiting},\ and\
  \citenamefont {Detweiler}}]{DiazRivera:2004ik}%
  \BibitemOpen
  \bibfield  {author} {\bibinfo {author} {\bibfnamefont {Luz~Maria}\
  \bibnamefont {Diaz-Rivera}}, \bibinfo {author} {\bibfnamefont {Eirini}\
  \bibnamefont {Messaritaki}}, \bibinfo {author} {\bibfnamefont {Bernard~F.}\
  \bibnamefont {Whiting}}, \ and\ \bibinfo {author} {\bibfnamefont {Steven~L.}\
  \bibnamefont {Detweiler}},\ }\href {\doibase10.1103/PhysRevD.70.124018}
  {\bibfield  {journal} {\bibinfo  {journal} {Phys. Rev.}\ }\textbf {\bibinfo
  {volume} {D70}},\ \bibinfo {pages} {124018} (\bibinfo {year} {2004})},\
  \Eprint {http://arxiv.org/abs/gr-qc/0410011} {arXiv:gr-qc/0410011
  [gr-qc]}\BibitemShut {NoStop}%
\bibitem [{\citenamefont {Haas}(2007)}]{Haas:2007kz}%
  \BibitemOpen
  \bibfield  {author} {\bibinfo {author} {\bibfnamefont {Roland}\ \bibnamefont
  {Haas}},\ }\href {\doibase10.1103/PhysRevD.75.124011} {\bibfield  {journal}
  {\bibinfo  {journal} {Phys. Rev.}\ }\textbf {\bibinfo {volume} {D75}},\
  \bibinfo {pages} {124011} (\bibinfo {year} {2007})},\ \Eprint
  {http://arxiv.org/abs/0704.0797} {arXiv:0704.0797 [gr-qc]}\BibitemShut
  {NoStop}%
\bibitem [{\citenamefont {Ottewill}\ and\ \citenamefont
  {Wardell}(2008)}]{Ottewill:2007mz}%
  \BibitemOpen
  \bibfield  {author} {\bibinfo {author} {\bibfnamefont {Adrian~C.}\
  \bibnamefont {Ottewill}}\ and\ \bibinfo {author} {\bibfnamefont {Barry}\
  \bibnamefont {Wardell}},\ }\href {\doibase10.1103/PhysRevD.77.104002}
  {\bibfield  {journal} {\bibinfo  {journal} {Phys. Rev.}\ }\textbf {\bibinfo
  {volume} {D77}},\ \bibinfo {pages} {104002} (\bibinfo {year} {2008})},\
  \Eprint {http://arxiv.org/abs/0711.2469} {arXiv:0711.2469
  [gr-qc]}\BibitemShut {NoStop}%
\bibitem [{\citenamefont {Ottewill}\ and\ \citenamefont
  {Wardell}(2009)}]{Ottewill:2008uu}%
  \BibitemOpen
  \bibfield  {author} {\bibinfo {author} {\bibfnamefont {Adrian~C.}\
  \bibnamefont {Ottewill}}\ and\ \bibinfo {author} {\bibfnamefont {Barry}\
  \bibnamefont {Wardell}},\ }\href {\doibase10.1103/PhysRevD.79.024031}
  {\bibfield  {journal} {\bibinfo  {journal} {Phys. Rev.}\ }\textbf {\bibinfo
  {volume} {D79}},\ \bibinfo {pages} {024031} (\bibinfo {year} {2009})},\
  \Eprint {http://arxiv.org/abs/0810.1961} {arXiv:0810.1961
  [gr-qc]}\BibitemShut {NoStop}%
\bibitem [{\citenamefont {Lousto}\ and\ \citenamefont
  {Nakano}(2008)}]{Lousto:2008mb}%
  \BibitemOpen
  \bibfield  {author} {\bibinfo {author} {\bibfnamefont {Carlos~O.}\
  \bibnamefont {Lousto}}\ and\ \bibinfo {author} {\bibfnamefont {Hiroyuki}\
  \bibnamefont {Nakano}},\ }\href {\doibase10.1088/0264-9381/25/14/145018}
  {\bibfield  {journal} {\bibinfo  {journal} {Class. Quant. Grav.}\ }\textbf
  {\bibinfo {volume} {25}},\ \bibinfo {pages} {145018} (\bibinfo {year}
  {2008})},\ \Eprint {http://arxiv.org/abs/0802.4277} {arXiv:0802.4277
  [gr-qc]}\BibitemShut {NoStop}%
\bibitem [{\citenamefont {Vega}\ \emph {et~al.}(2009)\citenamefont {Vega},
  \citenamefont {Diener}, \citenamefont {Tichy},\ and\ \citenamefont
  {Detweiler}}]{Vega:2009qb}%
  \BibitemOpen
  \bibfield  {author} {\bibinfo {author} {\bibfnamefont {Ian}\ \bibnamefont
  {Vega}}, \bibinfo {author} {\bibfnamefont {Peter}\ \bibnamefont {Diener}},
  \bibinfo {author} {\bibfnamefont {Wolfgang}\ \bibnamefont {Tichy}}, \ and\
  \bibinfo {author} {\bibfnamefont {Steven~L.}\ \bibnamefont {Detweiler}},\
  }\href {\doibase10.1103/PhysRevD.80.084021} {\bibfield  {journal} {\bibinfo
  {journal} {Phys. Rev.}\ }\textbf {\bibinfo {volume} {D80}},\ \bibinfo {pages}
  {084021} (\bibinfo {year} {2009})},\ \Eprint {http://arxiv.org/abs/0908.2138}
  {arXiv:0908.2138 [gr-qc]}\BibitemShut {NoStop}%
\bibitem [{\citenamefont {Canizares}\ and\ \citenamefont
  {Sopuerta}(2009)}]{Canizares:2009ay}%
  \BibitemOpen
  \bibfield  {author} {\bibinfo {author} {\bibfnamefont {Priscilla}\
  \bibnamefont {Canizares}}\ and\ \bibinfo {author} {\bibfnamefont {Carlos~F.}\
  \bibnamefont {Sopuerta}},\ }\href {\doibase10.1103/PhysRevD.79.084020}
  {\bibfield  {journal} {\bibinfo  {journal} {Phys. Rev.}\ }\textbf {\bibinfo
  {volume} {D79}},\ \bibinfo {pages} {084020} (\bibinfo {year} {2009})},\
  \Eprint {http://arxiv.org/abs/0903.0505} {arXiv:0903.0505
  [gr-qc]}\BibitemShut {NoStop}%
\bibitem [{\citenamefont {Dolan}\ and\ \citenamefont
  {Barack}(2011)}]{Dolan:2010mt}%
  \BibitemOpen
  \bibfield  {author} {\bibinfo {author} {\bibfnamefont {Sam~R.}\ \bibnamefont
  {Dolan}}\ and\ \bibinfo {author} {\bibfnamefont {Leor}\ \bibnamefont
  {Barack}},\ }\href {\doibase10.1103/PhysRevD.83.024019} {\bibfield  {journal}
  {\bibinfo  {journal} {Phys. Rev.}\ }\textbf {\bibinfo {volume} {D83}},\
  \bibinfo {pages} {024019} (\bibinfo {year} {2011})},\ \Eprint
  {http://arxiv.org/abs/1010.5255} {arXiv:1010.5255 [gr-qc]}\BibitemShut
  {NoStop}%
\bibitem [{\citenamefont {Thornburg}(2010)}]{Thornburg:2010tq}%
  \BibitemOpen
  \bibfield  {author} {\bibinfo {author} {\bibfnamefont {Jonathan}\
  \bibnamefont {Thornburg}},\ }\href@noop {} {\  (\bibinfo {year} {2010})},\
  \Eprint {http://arxiv.org/abs/1006.3788} {arXiv:1006.3788
  [gr-qc]}\BibitemShut {NoStop}%
\bibitem [{\citenamefont {Canizares}\ \emph {et~al.}(2010)\citenamefont
  {Canizares}, \citenamefont {Sopuerta},\ and\ \citenamefont
  {Jaramillo}}]{Canizares:2010yx}%
  \BibitemOpen
  \bibfield  {author} {\bibinfo {author} {\bibfnamefont {Priscilla}\
  \bibnamefont {Canizares}}, \bibinfo {author} {\bibfnamefont {Carlos~F.}\
  \bibnamefont {Sopuerta}}, \ and\ \bibinfo {author} {\bibfnamefont
  {Jose~Luis}\ \bibnamefont {Jaramillo}},\ }\href
  {\doibase10.1103/PhysRevD.82.044023} {\bibfield  {journal} {\bibinfo
  {journal} {Phys. Rev.}\ }\textbf {\bibinfo {volume} {D82}},\ \bibinfo {pages}
  {044023} (\bibinfo {year} {2010})},\ \Eprint {http://arxiv.org/abs/1006.3201}
  {arXiv:1006.3201 [gr-qc]}\BibitemShut {NoStop}%
\bibitem [{\citenamefont {Warburton}\ and\ \citenamefont
  {Barack}(2010)}]{Warburton:2010eq}%
  \BibitemOpen
  \bibfield  {author} {\bibinfo {author} {\bibfnamefont {Niels}\ \bibnamefont
  {Warburton}}\ and\ \bibinfo {author} {\bibfnamefont {Leor}\ \bibnamefont
  {Barack}},\ }\href {\doibase10.1103/PhysRevD.81.084039} {\bibfield  {journal}
  {\bibinfo  {journal} {Phys. Rev.}\ }\textbf {\bibinfo {volume} {D81}},\
  \bibinfo {pages} {084039} (\bibinfo {year} {2010})},\ \Eprint
  {http://arxiv.org/abs/1003.1860} {arXiv:1003.1860 [gr-qc]}\BibitemShut
  {NoStop}%
\bibitem [{\citenamefont {Warburton}\ and\ \citenamefont
  {Barack}(2011)}]{Warburton:2011hp}%
  \BibitemOpen
  \bibfield  {author} {\bibinfo {author} {\bibfnamefont {Niels}\ \bibnamefont
  {Warburton}}\ and\ \bibinfo {author} {\bibfnamefont {Leor}\ \bibnamefont
  {Barack}},\ }\href {\doibase10.1103/PhysRevD.83.124038} {\bibfield  {journal}
  {\bibinfo  {journal} {Phys. Rev.}\ }\textbf {\bibinfo {volume} {D83}},\
  \bibinfo {pages} {124038} (\bibinfo {year} {2011})},\ \Eprint
  {http://arxiv.org/abs/1103.0287} {arXiv:1103.0287 [gr-qc]}\BibitemShut
  {NoStop}%
\bibitem [{\citenamefont {Diener}\ \emph {et~al.}(2012)\citenamefont {Diener},
  \citenamefont {Vega}, \citenamefont {Wardell},\ and\ \citenamefont
  {Detweiler}}]{Diener:2011cc}%
  \BibitemOpen
  \bibfield  {author} {\bibinfo {author} {\bibfnamefont {Peter}\ \bibnamefont
  {Diener}}, \bibinfo {author} {\bibfnamefont {Ian}\ \bibnamefont {Vega}},
  \bibinfo {author} {\bibfnamefont {Barry}\ \bibnamefont {Wardell}}, \ and\
  \bibinfo {author} {\bibfnamefont {Steven}\ \bibnamefont {Detweiler}},\ }\href
  {\doibase10.1103/PhysRevLett.108.191102} {\bibfield  {journal} {\bibinfo
  {journal} {Phys. Rev. Lett.}\ }\textbf {\bibinfo {volume} {108}},\ \bibinfo
  {pages} {191102} (\bibinfo {year} {2012})},\ \Eprint
  {http://arxiv.org/abs/1112.4821} {arXiv:1112.4821 [gr-qc]}\BibitemShut
  {NoStop}%
\bibitem [{\citenamefont {Dolan}\ \emph {et~al.}(2011)\citenamefont {Dolan},
  \citenamefont {Barack},\ and\ \citenamefont {Wardell}}]{Dolan:2011dx}%
  \BibitemOpen
  \bibfield  {author} {\bibinfo {author} {\bibfnamefont {Sam~R.}\ \bibnamefont
  {Dolan}}, \bibinfo {author} {\bibfnamefont {Leor}\ \bibnamefont {Barack}}, \
  and\ \bibinfo {author} {\bibfnamefont {Barry}\ \bibnamefont {Wardell}},\
  }\href {\doibase10.1103/PhysRevD.84.084001} {\bibfield  {journal} {\bibinfo
  {journal} {Phys. Rev.}\ }\textbf {\bibinfo {volume} {D84}},\ \bibinfo {pages}
  {084001} (\bibinfo {year} {2011})},\ \Eprint {http://arxiv.org/abs/1107.0012}
  {arXiv:1107.0012 [gr-qc]}\BibitemShut {NoStop}%
\bibitem [{\citenamefont {Casals}\ \emph {et~al.}(2012)\citenamefont {Casals},
  \citenamefont {Poisson},\ and\ \citenamefont {Vega}}]{Casals:2012qq}%
  \BibitemOpen
  \bibfield  {author} {\bibinfo {author} {\bibfnamefont {Marc}\ \bibnamefont
  {Casals}}, \bibinfo {author} {\bibfnamefont {Eric}\ \bibnamefont {Poisson}},
  \ and\ \bibinfo {author} {\bibfnamefont {Ian}\ \bibnamefont {Vega}},\ }\href
  {\doibase10.1103/PhysRevD.86.064033} {\bibfield  {journal} {\bibinfo
  {journal} {Phys. Rev.}\ }\textbf {\bibinfo {volume} {D86}},\ \bibinfo {pages}
  {064033} (\bibinfo {year} {2012})},\ \Eprint {http://arxiv.org/abs/1206.3772}
  {arXiv:1206.3772 [gr-qc]}\BibitemShut {NoStop}%
\bibitem [{\citenamefont {Ottewill}\ and\ \citenamefont
  {Taylor}(2012)}]{Ottewill:2012aj}%
  \BibitemOpen
  \bibfield  {author} {\bibinfo {author} {\bibfnamefont {Adrian~C.}\
  \bibnamefont {Ottewill}}\ and\ \bibinfo {author} {\bibfnamefont {Peter}\
  \bibnamefont {Taylor}},\ }\href {\doibase10.1103/PhysRevD.86.024036}
  {\bibfield  {journal} {\bibinfo  {journal} {Phys. Rev.}\ }\textbf {\bibinfo
  {volume} {D86}},\ \bibinfo {pages} {024036} (\bibinfo {year} {2012})},\
  \Eprint {http://arxiv.org/abs/1205.5587} {arXiv:1205.5587
  [gr-qc]}\BibitemShut {NoStop}%
\bibitem [{\citenamefont {Casals}\ \emph {et~al.}(2013)\citenamefont {Casals},
  \citenamefont {Dolan}, \citenamefont {Ottewill},\ and\ \citenamefont
  {Wardell}}]{Casals:2013mpa}%
  \BibitemOpen
  \bibfield  {author} {\bibinfo {author} {\bibfnamefont {Marc}\ \bibnamefont
  {Casals}}, \bibinfo {author} {\bibfnamefont {Sam}\ \bibnamefont {Dolan}},
  \bibinfo {author} {\bibfnamefont {Adrian~C.}\ \bibnamefont {Ottewill}}, \
  and\ \bibinfo {author} {\bibfnamefont {Barry}\ \bibnamefont {Wardell}},\
  }\href {\doibase10.1103/PhysRevD.88.044022} {\bibfield  {journal} {\bibinfo
  {journal} {Phys. Rev.}\ }\textbf {\bibinfo {volume} {D88}},\ \bibinfo {pages}
  {044022} (\bibinfo {year} {2013})},\ \Eprint {http://arxiv.org/abs/1306.0884}
  {arXiv:1306.0884 [gr-qc]}\BibitemShut {NoStop}%
\bibitem [{\citenamefont {Warburton}\ and\ \citenamefont
  {Wardell}(2014)}]{Warburton:2013lea}%
  \BibitemOpen
  \bibfield  {author} {\bibinfo {author} {\bibfnamefont {Niels}\ \bibnamefont
  {Warburton}}\ and\ \bibinfo {author} {\bibfnamefont {Barry}\ \bibnamefont
  {Wardell}},\ }\href {\doibase10.1103/PhysRevD.89.044046} {\bibfield
  {journal} {\bibinfo  {journal} {Phys. Rev.}\ }\textbf {\bibinfo {volume}
  {D89}},\ \bibinfo {pages} {044046} (\bibinfo {year} {2014})},\ \Eprint
  {http://arxiv.org/abs/1311.3104} {arXiv:1311.3104 [gr-qc]}\BibitemShut
  {NoStop}%
\bibitem [{\citenamefont {Vega}\ \emph {et~al.}(2013)\citenamefont {Vega},
  \citenamefont {Wardell}, \citenamefont {Diener}, \citenamefont {Cupp},\ and\
  \citenamefont {Haas}}]{Vega:2013wxa}%
  \BibitemOpen
  \bibfield  {author} {\bibinfo {author} {\bibfnamefont {Ian}\ \bibnamefont
  {Vega}}, \bibinfo {author} {\bibfnamefont {Barry}\ \bibnamefont {Wardell}},
  \bibinfo {author} {\bibfnamefont {Peter}\ \bibnamefont {Diener}}, \bibinfo
  {author} {\bibfnamefont {Samuel}\ \bibnamefont {Cupp}}, \ and\ \bibinfo
  {author} {\bibfnamefont {Roland}\ \bibnamefont {Haas}},\ }\href
  {\doibase10.1103/PhysRevD.88.084021} {\bibfield  {journal} {\bibinfo
  {journal} {Phys. Rev.}\ }\textbf {\bibinfo {volume} {D88}},\ \bibinfo {pages}
  {084021} (\bibinfo {year} {2013})},\ \Eprint {http://arxiv.org/abs/1307.3476}
  {arXiv:1307.3476 [gr-qc]}\BibitemShut {NoStop}%
\bibitem [{\citenamefont {Wardell}\ \emph {et~al.}(2014)\citenamefont
  {Wardell}, \citenamefont {Galley}, \citenamefont {Zenginoglu}, \citenamefont
  {Casals}, \citenamefont {Dolan} \emph {et~al.}}]{Wardell:2014kea}%
  \BibitemOpen
  \bibfield  {author} {\bibinfo {author} {\bibfnamefont {Barry}\ \bibnamefont
  {Wardell}}, \bibinfo {author} {\bibfnamefont {Chad~R.}\ \bibnamefont
  {Galley}}, \bibinfo {author} {\bibfnamefont {Anil}\ \bibnamefont
  {Zenginoglu}}, \bibinfo {author} {\bibfnamefont {Marc}\ \bibnamefont
  {Casals}}, \bibinfo {author} {\bibfnamefont {Sam~R.}\ \bibnamefont {Dolan}},
  \emph {et~al.},\ }\href {\doibase10.1103/PhysRevD.89.084021} {\bibfield
  {journal} {\bibinfo  {journal} {Phys. Rev.}\ }\textbf {\bibinfo {volume}
  {D89}},\ \bibinfo {pages} {084021} (\bibinfo {year} {2014})},\ \Eprint
  {http://arxiv.org/abs/1401.1506} {arXiv:1401.1506 [gr-qc]}\BibitemShut
  {NoStop}%
\bibitem [{\citenamefont {Warburton}(2015)}]{Warburton:2014bya}%
  \BibitemOpen
  \bibfield  {author} {\bibinfo {author} {\bibfnamefont {Niels}\ \bibnamefont
  {Warburton}},\ }\href {\doibase10.1103/PhysRevD.91.024045} {\bibfield
  {journal} {\bibinfo  {journal} {Phys. Rev.}\ }\textbf {\bibinfo {volume}
  {D91}},\ \bibinfo {pages} {024045} (\bibinfo {year} {2015})},\ \Eprint
  {http://arxiv.org/abs/1408.2885} {arXiv:1408.2885 [gr-qc]}\BibitemShut
  {NoStop}%
\bibitem [{\citenamefont {Gralla}\ \emph {et~al.}(2015)\citenamefont {Gralla},
  \citenamefont {Porfyriadis},\ and\ \citenamefont
  {Warburton}}]{Gralla:2015rpa}%
  \BibitemOpen
  \bibfield  {author} {\bibinfo {author} {\bibfnamefont {Samuel~E.}\
  \bibnamefont {Gralla}}, \bibinfo {author} {\bibfnamefont {Achilleas~P.}\
  \bibnamefont {Porfyriadis}}, \ and\ \bibinfo {author} {\bibfnamefont {Niels}\
  \bibnamefont {Warburton}},\ }\href {\doibase10.1103/PhysRevD.92.064029}
  {\bibfield  {journal} {\bibinfo  {journal} {Phys. Rev.}\ }\textbf {\bibinfo
  {volume} {D92}},\ \bibinfo {pages} {064029} (\bibinfo {year} {2015})},\
  \Eprint {http://arxiv.org/abs/1506.08496} {arXiv:1506.08496
  [gr-qc]}\BibitemShut {NoStop}%
\bibitem [{\citenamefont {Thornburg}\ and\ \citenamefont
  {Wardell}(2017)}]{Thornburg:2016msc}%
  \BibitemOpen
  \bibfield  {author} {\bibinfo {author} {\bibfnamefont {Jonathan}\
  \bibnamefont {Thornburg}}\ and\ \bibinfo {author} {\bibfnamefont {Barry}\
  \bibnamefont {Wardell}},\ }\href {\doibase10.1103/PhysRevD.95.084043}
  {\bibfield  {journal} {\bibinfo  {journal} {Phys. Rev.}\ }\textbf {\bibinfo
  {volume} {D95}},\ \bibinfo {pages} {084043} (\bibinfo {year} {2017})},\
  \Eprint {http://arxiv.org/abs/1610.09319} {arXiv:1610.09319
  [gr-qc]}\BibitemShut {NoStop}%
\bibitem [{\citenamefont {Heffernan}\ \emph {et~al.}(2017)\citenamefont
  {Heffernan}, \citenamefont {Ottewill}, \citenamefont {Warburton},
  \citenamefont {Wardell},\ and\ \citenamefont {Diener}}]{Heffernan:2017cad}%
  \BibitemOpen
  \bibfield  {author} {\bibinfo {author} {\bibfnamefont {Anna}\ \bibnamefont
  {Heffernan}}, \bibinfo {author} {\bibfnamefont {Adrian~C.}\ \bibnamefont
  {Ottewill}}, \bibinfo {author} {\bibfnamefont {Niels}\ \bibnamefont
  {Warburton}}, \bibinfo {author} {\bibfnamefont {Barry}\ \bibnamefont
  {Wardell}}, \ and\ \bibinfo {author} {\bibfnamefont {Peter}\ \bibnamefont
  {Diener}},\ }\href@noop {} {\  (\bibinfo {year} {2017})},\ \Eprint
  {http://arxiv.org/abs/1712.01098} {arXiv:1712.01098 [gr-qc]}\BibitemShut
  {NoStop}%
\bibitem [{\citenamefont {Barack}\ and\ \citenamefont
  {Lousto}(2002)}]{Barack:2002ku}%
  \BibitemOpen
  \bibfield  {author} {\bibinfo {author} {\bibfnamefont {Leor}\ \bibnamefont
  {Barack}}\ and\ \bibinfo {author} {\bibfnamefont {Carlos~O.}\ \bibnamefont
  {Lousto}},\ }\href {\doibase10.1103/PhysRevD.66.061502} {\bibfield  {journal}
  {\bibinfo  {journal} {Phys. Rev.}\ }\textbf {\bibinfo {volume} {D66}},\
  \bibinfo {pages} {061502} (\bibinfo {year} {2002})},\ \Eprint
  {http://arxiv.org/abs/gr-qc/0205043} {arXiv:gr-qc/0205043
  [gr-qc]}\BibitemShut {NoStop}%
\bibitem [{\citenamefont {Barack}\ and\ \citenamefont
  {Lousto}(2005)}]{Barack:2005nr}%
  \BibitemOpen
  \bibfield  {author} {\bibinfo {author} {\bibfnamefont {Leor}\ \bibnamefont
  {Barack}}\ and\ \bibinfo {author} {\bibfnamefont {Carlos~O.}\ \bibnamefont
  {Lousto}},\ }\href {\doibase10.1103/PhysRevD.72.104026} {\bibfield  {journal}
  {\bibinfo  {journal} {Phys. Rev.}\ }\textbf {\bibinfo {volume} {D72}},\
  \bibinfo {pages} {104026} (\bibinfo {year} {2005})},\ \Eprint
  {http://arxiv.org/abs/gr-qc/0510019} {arXiv:gr-qc/0510019
  [gr-qc]}\BibitemShut {NoStop}%
\bibitem [{\citenamefont {Barack}\ and\ \citenamefont
  {Sago}(2007)}]{Barack:2007tm}%
  \BibitemOpen
  \bibfield  {author} {\bibinfo {author} {\bibfnamefont {Leor}\ \bibnamefont
  {Barack}}\ and\ \bibinfo {author} {\bibfnamefont {Norichika}\ \bibnamefont
  {Sago}},\ }\href {\doibase10.1103/PhysRevD.75.064021} {\bibfield  {journal}
  {\bibinfo  {journal} {Phys. Rev.}\ }\textbf {\bibinfo {volume} {D75}},\
  \bibinfo {pages} {064021} (\bibinfo {year} {2007})},\ \Eprint
  {http://arxiv.org/abs/gr-qc/0701069} {arXiv:gr-qc/0701069
  [gr-qc]}\BibitemShut {NoStop}%
\bibitem [{\citenamefont {Sago}(2009)}]{Sago:2009zz}%
  \BibitemOpen
  \bibfield  {author} {\bibinfo {author} {\bibfnamefont {Norichika}\
  \bibnamefont {Sago}},\ }\href {\doibase10.1088/0264-9381/26/9/094025}
  {\bibfield  {journal} {\bibinfo  {journal} {Class. Quant. Grav.}\ }\textbf
  {\bibinfo {volume} {26}},\ \bibinfo {pages} {094025} (\bibinfo {year}
  {2009})}\BibitemShut {NoStop}%
\bibitem [{\citenamefont {Shah}\ \emph {et~al.}(2011)\citenamefont {Shah},
  \citenamefont {Keidl}, \citenamefont {Friedman}, \citenamefont {Kim},\ and\
  \citenamefont {Price}}]{Shah:2010bi}%
  \BibitemOpen
  \bibfield  {author} {\bibinfo {author} {\bibfnamefont {Abhay~G.}\
  \bibnamefont {Shah}}, \bibinfo {author} {\bibfnamefont {Tobias~S.}\
  \bibnamefont {Keidl}}, \bibinfo {author} {\bibfnamefont {John~L.}\
  \bibnamefont {Friedman}}, \bibinfo {author} {\bibfnamefont {Dong-Hoon}\
  \bibnamefont {Kim}}, \ and\ \bibinfo {author} {\bibfnamefont {Larry~R.}\
  \bibnamefont {Price}},\ }\href {\doibase10.1103/PhysRevD.83.064018}
  {\bibfield  {journal} {\bibinfo  {journal} {Phys. Rev.}\ }\textbf {\bibinfo
  {volume} {D83}},\ \bibinfo {pages} {064018} (\bibinfo {year} {2011})},\
  \Eprint {http://arxiv.org/abs/1009.4876} {arXiv:1009.4876
  [gr-qc]}\BibitemShut {NoStop}%
\bibitem [{\citenamefont {Akcay}(2011)}]{Akcay:2010dx}%
  \BibitemOpen
  \bibfield  {author} {\bibinfo {author} {\bibfnamefont {Sarp}\ \bibnamefont
  {Akcay}},\ }\href {\doibase10.1103/PhysRevD.83.124026} {\bibfield  {journal}
  {\bibinfo  {journal} {Phys. Rev.}\ }\textbf {\bibinfo {volume} {D83}},\
  \bibinfo {pages} {124026} (\bibinfo {year} {2011})},\ \Eprint
  {http://arxiv.org/abs/1012.5860} {arXiv:1012.5860 [gr-qc]}\BibitemShut
  {NoStop}%
\bibitem [{\citenamefont {Barack}\ and\ \citenamefont
  {Sago}(2010)}]{Barack:2010tm}%
  \BibitemOpen
  \bibfield  {author} {\bibinfo {author} {\bibfnamefont {Leor}\ \bibnamefont
  {Barack}}\ and\ \bibinfo {author} {\bibfnamefont {Norichika}\ \bibnamefont
  {Sago}},\ }\href {\doibase10.1103/PhysRevD.81.084021} {\bibfield  {journal}
  {\bibinfo  {journal} {Phys. Rev.}\ }\textbf {\bibinfo {volume} {D81}},\
  \bibinfo {pages} {084021} (\bibinfo {year} {2010})},\ \Eprint
  {http://arxiv.org/abs/1002.2386} {arXiv:1002.2386 [gr-qc]}\BibitemShut
  {NoStop}%
\bibitem [{\citenamefont {Keidl}\ \emph {et~al.}(2010)\citenamefont {Keidl},
  \citenamefont {Shah}, \citenamefont {Friedman}, \citenamefont {Kim},\ and\
  \citenamefont {Price}}]{Keidl:2010pm}%
  \BibitemOpen
  \bibfield  {author} {\bibinfo {author} {\bibfnamefont {Tobias~S.}\
  \bibnamefont {Keidl}}, \bibinfo {author} {\bibfnamefont {Abhay~G.}\
  \bibnamefont {Shah}}, \bibinfo {author} {\bibfnamefont {John~L.}\
  \bibnamefont {Friedman}}, \bibinfo {author} {\bibfnamefont {Dong-Hoon}\
  \bibnamefont {Kim}}, \ and\ \bibinfo {author} {\bibfnamefont {Larry~R.}\
  \bibnamefont {Price}},\ }\href {\doibase10.1103/PhysRevD.82.124012}
  {\bibfield  {journal} {\bibinfo  {journal} {Phys. Rev.}\ }\textbf {\bibinfo
  {volume} {D82}},\ \bibinfo {pages} {124012} (\bibinfo {year} {2010})},\
  \Eprint {http://arxiv.org/abs/1004.2276} {arXiv:1004.2276
  [gr-qc]}\BibitemShut {NoStop}%
\bibitem [{\citenamefont {Warburton}\ \emph {et~al.}(2012)\citenamefont
  {Warburton}, \citenamefont {Akcay}, \citenamefont {Barack}, \citenamefont
  {Gair},\ and\ \citenamefont {Sago}}]{Warburton:2011fk}%
  \BibitemOpen
  \bibfield  {author} {\bibinfo {author} {\bibfnamefont {Niels}\ \bibnamefont
  {Warburton}}, \bibinfo {author} {\bibfnamefont {Sarp}\ \bibnamefont {Akcay}},
  \bibinfo {author} {\bibfnamefont {Leor}\ \bibnamefont {Barack}}, \bibinfo
  {author} {\bibfnamefont {Jonathan~R.}\ \bibnamefont {Gair}}, \ and\ \bibinfo
  {author} {\bibfnamefont {Norichika}\ \bibnamefont {Sago}},\ }\href
  {\doibase10.1103/PhysRevD.85.061501} {\bibfield  {journal} {\bibinfo
  {journal} {Phys. Rev.}\ }\textbf {\bibinfo {volume} {D85}},\ \bibinfo {pages}
  {061501} (\bibinfo {year} {2012})},\ \Eprint {http://arxiv.org/abs/1111.6908}
  {arXiv:1111.6908 [gr-qc]}\BibitemShut {NoStop}%
\bibitem [{\citenamefont {Dolan}\ and\ \citenamefont
  {Barack}(2013)}]{Dolan:2012jg}%
  \BibitemOpen
  \bibfield  {author} {\bibinfo {author} {\bibfnamefont {Sam~R.}\ \bibnamefont
  {Dolan}}\ and\ \bibinfo {author} {\bibfnamefont {Leor}\ \bibnamefont
  {Barack}},\ }\href {\doibase10.1103/PhysRevD.87.084066} {\bibfield  {journal}
  {\bibinfo  {journal} {Phys. Rev.}\ }\textbf {\bibinfo {volume} {D87}},\
  \bibinfo {pages} {084066} (\bibinfo {year} {2013})},\ \Eprint
  {http://arxiv.org/abs/1211.4586} {arXiv:1211.4586 [gr-qc]}\BibitemShut
  {NoStop}%
\bibitem [{\citenamefont {Shah}\ \emph {et~al.}(2012)\citenamefont {Shah},
  \citenamefont {Friedman},\ and\ \citenamefont {Keidl}}]{Shah:2012gu}%
  \BibitemOpen
  \bibfield  {author} {\bibinfo {author} {\bibfnamefont {Abhay~G.}\
  \bibnamefont {Shah}}, \bibinfo {author} {\bibfnamefont {John~L.}\
  \bibnamefont {Friedman}}, \ and\ \bibinfo {author} {\bibfnamefont
  {Tobias~S.}\ \bibnamefont {Keidl}},\ }\href
  {\doibase10.1103/PhysRevD.86.084059} {\bibfield  {journal} {\bibinfo
  {journal} {Phys. Rev.}\ }\textbf {\bibinfo {volume} {D86}},\ \bibinfo {pages}
  {084059} (\bibinfo {year} {2012})},\ \Eprint {http://arxiv.org/abs/1207.5595}
  {arXiv:1207.5595 [gr-qc]}\BibitemShut {NoStop}%
\bibitem [{\citenamefont {Dolan}(2013)}]{Dolan:Capra16}%
  \BibitemOpen
  \bibfield  {author} {\bibinfo {author} {\bibfnamefont {Sam}\ \bibnamefont
  {Dolan}},\ }\href@noop {} {\enquote {\bibinfo {title} {{Approaches to
  Self-Force Calculations on Kerr Spacetime}},}\ } (\bibinfo {year} {2013}),\
  \bibinfo {note} {{16th Capra Meeting on Radiation Reaction in General
  Relativity}, \url{http://maths.ucd.ie/capra16/talks/Dolan.pdf}}\BibitemShut
  {NoStop}%
\bibitem [{\citenamefont {Akcay}\ \emph {et~al.}(2013)\citenamefont {Akcay},
  \citenamefont {Warburton},\ and\ \citenamefont {Barack}}]{Akcay:2013wfa}%
  \BibitemOpen
  \bibfield  {author} {\bibinfo {author} {\bibfnamefont {Sarp}\ \bibnamefont
  {Akcay}}, \bibinfo {author} {\bibfnamefont {Niels}\ \bibnamefont
  {Warburton}}, \ and\ \bibinfo {author} {\bibfnamefont {Leor}\ \bibnamefont
  {Barack}},\ }\href {\doibase10.1103/PhysRevD.88.104009} {\bibfield  {journal}
  {\bibinfo  {journal} {Phys. Rev.}\ }\textbf {\bibinfo {volume} {D88}},\
  \bibinfo {pages} {104009} (\bibinfo {year} {2013})},\ \Eprint
  {http://arxiv.org/abs/1308.5223} {arXiv:1308.5223 [gr-qc]}\BibitemShut
  {NoStop}%
\bibitem [{\citenamefont {Pound}\ \emph {et~al.}(2014)\citenamefont {Pound},
  \citenamefont {Merlin},\ and\ \citenamefont {Barack}}]{Pound:2013faa}%
  \BibitemOpen
  \bibfield  {author} {\bibinfo {author} {\bibfnamefont {Adam}\ \bibnamefont
  {Pound}}, \bibinfo {author} {\bibfnamefont {Cesar}\ \bibnamefont {Merlin}}, \
  and\ \bibinfo {author} {\bibfnamefont {Leor}\ \bibnamefont {Barack}},\ }\href
  {\doibase10.1103/PhysRevD.89.024009} {\bibfield  {journal} {\bibinfo
  {journal} {Phys. Rev.}\ }\textbf {\bibinfo {volume} {D89}},\ \bibinfo {pages}
  {024009} (\bibinfo {year} {2014})},\ \Eprint {http://arxiv.org/abs/1310.1513}
  {arXiv:1310.1513 [gr-qc]}\BibitemShut {NoStop}%
\bibitem [{\citenamefont {Isoyama}\ \emph {et~al.}(2014)\citenamefont
  {Isoyama}, \citenamefont {Barack}, \citenamefont {Dolan}, \citenamefont
  {Le~Tiec}, \citenamefont {Nakano} \emph {et~al.}}]{Isoyama:2014mja}%
  \BibitemOpen
  \bibfield  {author} {\bibinfo {author} {\bibfnamefont {Soichiro}\
  \bibnamefont {Isoyama}}, \bibinfo {author} {\bibfnamefont {Leor}\
  \bibnamefont {Barack}}, \bibinfo {author} {\bibfnamefont {Sam~R.}\
  \bibnamefont {Dolan}}, \bibinfo {author} {\bibfnamefont {Alexandre}\
  \bibnamefont {Le~Tiec}}, \bibinfo {author} {\bibfnamefont {Hiroyuki}\
  \bibnamefont {Nakano}},  \emph {et~al.},\ }\href
  {\doibase10.1103/PhysRevLett.113.161101} {\bibfield  {journal} {\bibinfo
  {journal} {Phys. Rev. Lett.}\ }\textbf {\bibinfo {volume} {113}},\ \bibinfo
  {pages} {161101} (\bibinfo {year} {2014})},\ \Eprint
  {http://arxiv.org/abs/1404.6133} {arXiv:1404.6133 [gr-qc]}\BibitemShut
  {NoStop}%
\bibitem [{\citenamefont {Merlin}\ and\ \citenamefont
  {Shah}(2015)}]{Merlin:2014qda}%
  \BibitemOpen
  \bibfield  {author} {\bibinfo {author} {\bibfnamefont {Cesar}\ \bibnamefont
  {Merlin}}\ and\ \bibinfo {author} {\bibfnamefont {Abhay~G.}\ \bibnamefont
  {Shah}},\ }\href {\doibase10.1103/PhysRevD.91.024005} {\bibfield  {journal}
  {\bibinfo  {journal} {Phys. Rev.}\ }\textbf {\bibinfo {volume} {D91}},\
  \bibinfo {pages} {024005} (\bibinfo {year} {2015})},\ \Eprint
  {http://arxiv.org/abs/1410.2998} {arXiv:1410.2998 [gr-qc]}\BibitemShut
  {NoStop}%
\bibitem [{\citenamefont {Osburn}\ \emph {et~al.}(2016)\citenamefont {Osburn},
  \citenamefont {Warburton},\ and\ \citenamefont {Evans}}]{Osburn:2015duj}%
  \BibitemOpen
  \bibfield  {author} {\bibinfo {author} {\bibfnamefont {Thomas}\ \bibnamefont
  {Osburn}}, \bibinfo {author} {\bibfnamefont {Niels}\ \bibnamefont
  {Warburton}}, \ and\ \bibinfo {author} {\bibfnamefont {Charles~R.}\
  \bibnamefont {Evans}},\ }\href {\doibase10.1103/PhysRevD.93.064024}
  {\bibfield  {journal} {\bibinfo  {journal} {Phys. Rev.}\ }\textbf {\bibinfo
  {volume} {D93}},\ \bibinfo {pages} {064024} (\bibinfo {year} {2016})},\
  \Eprint {http://arxiv.org/abs/1511.01498} {arXiv:1511.01498
  [gr-qc]}\BibitemShut {NoStop}%
\bibitem [{\citenamefont {Gralla}\ \emph {et~al.}(2016)\citenamefont {Gralla},
  \citenamefont {Hughes},\ and\ \citenamefont {Warburton}}]{Gralla:2016qfw}%
  \BibitemOpen
  \bibfield  {author} {\bibinfo {author} {\bibfnamefont {Samuel~E.}\
  \bibnamefont {Gralla}}, \bibinfo {author} {\bibfnamefont {Scott~A.}\
  \bibnamefont {Hughes}}, \ and\ \bibinfo {author} {\bibfnamefont {Niels}\
  \bibnamefont {Warburton}},\ }\href {\doibase10.1088/0264-9381/33/15/155002}
  {\bibfield  {journal} {\bibinfo  {journal} {Class. Quant. Grav.}\ }\textbf
  {\bibinfo {volume} {33}},\ \bibinfo {pages} {155002} (\bibinfo {year}
  {2016})},\ \Eprint {http://arxiv.org/abs/1603.01221} {arXiv:1603.01221
  [gr-qc]}\BibitemShut {NoStop}%
\bibitem [{\citenamefont {Hopper}\ and\ \citenamefont
  {Evans}(2013)}]{Hopper:2012ty}%
  \BibitemOpen
  \bibfield  {author} {\bibinfo {author} {\bibfnamefont {Seth}\ \bibnamefont
  {Hopper}}\ and\ \bibinfo {author} {\bibfnamefont {Charles~R.}\ \bibnamefont
  {Evans}},\ }\href {\doibase10.1103/PhysRevD.87.064008} {\bibfield  {journal}
  {\bibinfo  {journal} {Phys. Rev.}\ }\textbf {\bibinfo {volume} {D87}},\
  \bibinfo {pages} {064008} (\bibinfo {year} {2013})},\ \Eprint
  {http://arxiv.org/abs/1210.7969} {arXiv:1210.7969 [gr-qc]}\BibitemShut
  {NoStop}%
\bibitem [{\citenamefont {Osburn}\ \emph {et~al.}(2014)\citenamefont {Osburn},
  \citenamefont {Forseth}, \citenamefont {Evans},\ and\ \citenamefont
  {Hopper}}]{Osburn:2014hoa}%
  \BibitemOpen
  \bibfield  {author} {\bibinfo {author} {\bibfnamefont {Thomas}\ \bibnamefont
  {Osburn}}, \bibinfo {author} {\bibfnamefont {Erik}\ \bibnamefont {Forseth}},
  \bibinfo {author} {\bibfnamefont {Charles~R.}\ \bibnamefont {Evans}}, \ and\
  \bibinfo {author} {\bibfnamefont {Seth}\ \bibnamefont {Hopper}},\ }\href
  {\doibase10.1103/PhysRevD.90.104031} {\bibfield  {journal} {\bibinfo
  {journal} {Phys. Rev.}\ }\textbf {\bibinfo {volume} {D90}},\ \bibinfo {pages}
  {104031} (\bibinfo {year} {2014})},\ \Eprint {http://arxiv.org/abs/1409.4419}
  {arXiv:1409.4419 [gr-qc]}\BibitemShut {NoStop}%
\bibitem [{\citenamefont {Pound}(2015{\natexlab{a}})}]{Pound:2015fma}%
  \BibitemOpen
  \bibfield  {author} {\bibinfo {author} {\bibfnamefont {Adam}\ \bibnamefont
  {Pound}},\ }\href {\doibase10.1103/PhysRevD.92.044021} {\bibfield  {journal}
  {\bibinfo  {journal} {Phys. Rev.}\ }\textbf {\bibinfo {volume} {D92}},\
  \bibinfo {pages} {044021} (\bibinfo {year} {2015}{\natexlab{a}})},\ \Eprint
  {http://arxiv.org/abs/1506.02894} {arXiv:1506.02894 [gr-qc]}\BibitemShut
  {NoStop}%
\bibitem [{\citenamefont {Shah}\ \emph {et~al.}(2016)\citenamefont {Shah},
  \citenamefont {Whiting}, \citenamefont {Aksteiner}, \citenamefont
  {Andersson},\ and\ \citenamefont {Backdähl}}]{Shah:2016juc}%
  \BibitemOpen
  \bibfield  {author} {\bibinfo {author} {\bibfnamefont {Abhay~G.}\
  \bibnamefont {Shah}}, \bibinfo {author} {\bibfnamefont {Bernard~F.}\
  \bibnamefont {Whiting}}, \bibinfo {author} {\bibfnamefont {Steffen}\
  \bibnamefont {Aksteiner}}, \bibinfo {author} {\bibfnamefont {Lars}\
  \bibnamefont {Andersson}}, \ and\ \bibinfo {author} {\bibfnamefont {Thomas}\
  \bibnamefont {Backdähl}},\ }\href@noop {} {\  (\bibinfo {year} {2016})},\
  \Eprint {http://arxiv.org/abs/1611.08291} {arXiv:1611.08291
  [gr-qc]}\BibitemShut {NoStop}%
\bibitem [{\citenamefont {Thompson}\ \emph {et~al.}(2017)\citenamefont
  {Thompson}, \citenamefont {Thompson}, \citenamefont {Whiting},\ and\
  \citenamefont {Chen}}]{Chen:2016plo}%
  \BibitemOpen
  \bibfield  {author} {\bibinfo {author} {\bibfnamefont {Jonathan~E.}\
  \bibnamefont {Thompson}}, \bibinfo {author} {\bibfnamefont {Jonathan}\
  \bibnamefont {Thompson}}, \bibinfo {author} {\bibfnamefont {Bernard~F.}\
  \bibnamefont {Whiting}}, \ and\ \bibinfo {author} {\bibfnamefont {Hector}\
  \bibnamefont {Chen}},\ }\href {\doibase10.1088/1361-6382/aa7f5b} {\bibfield
  {journal} {\bibinfo  {journal} {Class. Quant. Grav.}\ }\textbf {\bibinfo
  {volume} {34}},\ \bibinfo {pages} {174001} (\bibinfo {year} {2017})},\
  \Eprint {http://arxiv.org/abs/1611.06214} {arXiv:1611.06214
  [gr-qc]}\BibitemShut {NoStop}%
\bibitem [{\citenamefont {Haas}\ and\ \citenamefont
  {Poisson}(2006)}]{Haas:2006ne}%
  \BibitemOpen
  \bibfield  {author} {\bibinfo {author} {\bibfnamefont {Roland}\ \bibnamefont
  {Haas}}\ and\ \bibinfo {author} {\bibfnamefont {Eric}\ \bibnamefont
  {Poisson}},\ }\href {\doibase10.1103/PhysRevD.74.044009} {\bibfield
  {journal} {\bibinfo  {journal} {Phys. Rev.}\ }\textbf {\bibinfo {volume}
  {D74}},\ \bibinfo {pages} {044009} (\bibinfo {year} {2006})},\ \Eprint
  {http://arxiv.org/abs/gr-qc/0605077} {arXiv:gr-qc/0605077
  [gr-qc]}\BibitemShut {NoStop}%
\bibitem [{\citenamefont {Barack}\ \emph {et~al.}(2007)\citenamefont {Barack},
  \citenamefont {Golbourn},\ and\ \citenamefont {Sago}}]{Barack:2007we}%
  \BibitemOpen
  \bibfield  {author} {\bibinfo {author} {\bibfnamefont {Leor}\ \bibnamefont
  {Barack}}, \bibinfo {author} {\bibfnamefont {Darren~A.}\ \bibnamefont
  {Golbourn}}, \ and\ \bibinfo {author} {\bibfnamefont {Norichika}\
  \bibnamefont {Sago}},\ }\href {\doibase10.1103/PhysRevD.76.124036} {\bibfield
   {journal} {\bibinfo  {journal} {Phys. Rev.}\ }\textbf {\bibinfo {volume}
  {D76}},\ \bibinfo {pages} {124036} (\bibinfo {year} {2007})},\ \Eprint
  {http://arxiv.org/abs/0709.4588} {arXiv:0709.4588 [gr-qc]}\BibitemShut
  {NoStop}%
\bibitem [{\citenamefont {Wardell}\ \emph {et~al.}(2012)\citenamefont
  {Wardell}, \citenamefont {Vega}, \citenamefont {Thornburg},\ and\
  \citenamefont {Diener}}]{Wardell:2011gb}%
  \BibitemOpen
  \bibfield  {author} {\bibinfo {author} {\bibfnamefont {Barry}\ \bibnamefont
  {Wardell}}, \bibinfo {author} {\bibfnamefont {Ian}\ \bibnamefont {Vega}},
  \bibinfo {author} {\bibfnamefont {Jonathan}\ \bibnamefont {Thornburg}}, \
  and\ \bibinfo {author} {\bibfnamefont {Peter}\ \bibnamefont {Diener}},\
  }\href {\doibase10.1103/PhysRevD.85.104044} {\bibfield  {journal} {\bibinfo
  {journal} {Phys. Rev.}\ }\textbf {\bibinfo {volume} {D85}},\ \bibinfo {pages}
  {104044} (\bibinfo {year} {2012})},\ \Eprint {http://arxiv.org/abs/1112.6355}
  {arXiv:1112.6355 [gr-qc]}\BibitemShut {NoStop}%
\bibitem [{\citenamefont {Vega}\ \emph {et~al.}(2011)\citenamefont {Vega},
  \citenamefont {Wardell},\ and\ \citenamefont {Diener}}]{Vega:2011wf}%
  \BibitemOpen
  \bibfield  {author} {\bibinfo {author} {\bibfnamefont {Ian}\ \bibnamefont
  {Vega}}, \bibinfo {author} {\bibfnamefont {Barry}\ \bibnamefont {Wardell}}, \
  and\ \bibinfo {author} {\bibfnamefont {Peter}\ \bibnamefont {Diener}},\
  }\href {\doibase10.1088/0264-9381/28/13/134010} {\bibfield  {journal}
  {\bibinfo  {journal} {Class. Quant. Grav.}\ }\textbf {\bibinfo {volume}
  {28}},\ \bibinfo {pages} {134010} (\bibinfo {year} {2011})},\ \Eprint
  {http://arxiv.org/abs/1101.2925} {arXiv:1101.2925 [gr-qc]}\BibitemShut
  {NoStop}%
\bibitem [{\citenamefont {Heffernan}\ \emph {et~al.}(2012)\citenamefont
  {Heffernan}, \citenamefont {Ottewill},\ and\ \citenamefont
  {Wardell}}]{Heffernan:2012su}%
  \BibitemOpen
  \bibfield  {author} {\bibinfo {author} {\bibfnamefont {Anna}\ \bibnamefont
  {Heffernan}}, \bibinfo {author} {\bibfnamefont {Adrian}\ \bibnamefont
  {Ottewill}}, \ and\ \bibinfo {author} {\bibfnamefont {Barry}\ \bibnamefont
  {Wardell}},\ }\href {\doibase10.1103/PhysRevD.86.104023} {\bibfield
  {journal} {\bibinfo  {journal} {Phys. Rev.}\ }\textbf {\bibinfo {volume}
  {D86}},\ \bibinfo {pages} {104023} (\bibinfo {year} {2012})},\ \Eprint
  {http://arxiv.org/abs/1204.0794} {arXiv:1204.0794 [gr-qc]}\BibitemShut
  {NoStop}%
\bibitem [{\citenamefont {Heffernan}\ \emph {et~al.}(2014)\citenamefont
  {Heffernan}, \citenamefont {Ottewill},\ and\ \citenamefont
  {Wardell}}]{Heffernan:2012vj}%
  \BibitemOpen
  \bibfield  {author} {\bibinfo {author} {\bibfnamefont {Anna}\ \bibnamefont
  {Heffernan}}, \bibinfo {author} {\bibfnamefont {Adrian}\ \bibnamefont
  {Ottewill}}, \ and\ \bibinfo {author} {\bibfnamefont {Barry}\ \bibnamefont
  {Wardell}},\ }\href {\doibase10.1103/PhysRevD.89.024030} {\bibfield
  {journal} {\bibinfo  {journal} {Phys. Rev.}\ }\textbf {\bibinfo {volume}
  {D89}},\ \bibinfo {pages} {024030} (\bibinfo {year} {2014})},\ \Eprint
  {http://arxiv.org/abs/1211.6446} {arXiv:1211.6446 [gr-qc]}\BibitemShut
  {NoStop}%
\bibitem [{\citenamefont {Linz}\ \emph {et~al.}(2014)\citenamefont {Linz},
  \citenamefont {Friedman},\ and\ \citenamefont {Wiseman}}]{Linz:2014pka}%
  \BibitemOpen
  \bibfield  {author} {\bibinfo {author} {\bibfnamefont {Thomas~M.}\
  \bibnamefont {Linz}}, \bibinfo {author} {\bibfnamefont {John~L.}\
  \bibnamefont {Friedman}}, \ and\ \bibinfo {author} {\bibfnamefont {Alan~G.}\
  \bibnamefont {Wiseman}},\ }\href {\doibase10.1103/PhysRevD.90.024064}
  {\bibfield  {journal} {\bibinfo  {journal} {Phys. Rev.}\ }\textbf {\bibinfo
  {volume} {D90}},\ \bibinfo {pages} {024064} (\bibinfo {year} {2014})},\
  \Eprint {http://arxiv.org/abs/1404.7039} {arXiv:1404.7039
  [gr-qc]}\BibitemShut {NoStop}%
\bibitem [{\citenamefont {Wardell}\ and\ \citenamefont
  {Warburton}(2015)}]{Wardell:2015ada}%
  \BibitemOpen
  \bibfield  {author} {\bibinfo {author} {\bibfnamefont {Barry}\ \bibnamefont
  {Wardell}}\ and\ \bibinfo {author} {\bibfnamefont {Niels}\ \bibnamefont
  {Warburton}},\ }\href {\doibase10.1103/PhysRevD.92.084019} {\bibfield
  {journal} {\bibinfo  {journal} {Phys. Rev.}\ }\textbf {\bibinfo {volume}
  {D92}},\ \bibinfo {pages} {084019} (\bibinfo {year} {2015})},\ \Eprint
  {http://arxiv.org/abs/1505.07841} {arXiv:1505.07841 [gr-qc]}\BibitemShut
  {NoStop}%
\bibitem [{\citenamefont {Barack}\ \emph {et~al.}(2008)\citenamefont {Barack},
  \citenamefont {Ori},\ and\ \citenamefont {Sago}}]{Barack:2008ms}%
  \BibitemOpen
  \bibfield  {author} {\bibinfo {author} {\bibfnamefont {Leor}\ \bibnamefont
  {Barack}}, \bibinfo {author} {\bibfnamefont {Amos}\ \bibnamefont {Ori}}, \
  and\ \bibinfo {author} {\bibfnamefont {Norichika}\ \bibnamefont {Sago}},\
  }\href {\doibase10.1103/PhysRevD.78.084021} {\bibfield  {journal} {\bibinfo
  {journal} {Phys. Rev.}\ }\textbf {\bibinfo {volume} {D78}},\ \bibinfo {pages}
  {084021} (\bibinfo {year} {2008})},\ \Eprint {http://arxiv.org/abs/0808.2315}
  {arXiv:0808.2315 [gr-qc]}\BibitemShut {NoStop}%
\bibitem [{\citenamefont {Field}\ \emph {et~al.}(2009)\citenamefont {Field},
  \citenamefont {Hesthaven},\ and\ \citenamefont {Lau}}]{Field:2009kk}%
  \BibitemOpen
  \bibfield  {author} {\bibinfo {author} {\bibfnamefont {Scott~E.}\
  \bibnamefont {Field}}, \bibinfo {author} {\bibfnamefont {Jan~S.}\
  \bibnamefont {Hesthaven}}, \ and\ \bibinfo {author} {\bibfnamefont
  {Stephen~R.}\ \bibnamefont {Lau}},\ }\href
  {\doibase10.1088/0264-9381/26/16/165010} {\bibfield  {journal} {\bibinfo
  {journal} {Class. Quant. Grav.}\ }\textbf {\bibinfo {volume} {26}},\ \bibinfo
  {pages} {165010} (\bibinfo {year} {2009})},\ \Eprint
  {http://arxiv.org/abs/0902.1287} {arXiv:0902.1287 [gr-qc]}\BibitemShut
  {NoStop}%
\bibitem [{\citenamefont {Field}\ \emph {et~al.}(2010)\citenamefont {Field},
  \citenamefont {Hesthaven},\ and\ \citenamefont {Lau}}]{Field:2010xn}%
  \BibitemOpen
  \bibfield  {author} {\bibinfo {author} {\bibfnamefont {Scott~E.}\
  \bibnamefont {Field}}, \bibinfo {author} {\bibfnamefont {Jan~S.}\
  \bibnamefont {Hesthaven}}, \ and\ \bibinfo {author} {\bibfnamefont
  {Stephen~R.}\ \bibnamefont {Lau}},\ }\href
  {\doibase10.1103/PhysRevD.81.124030} {\bibfield  {journal} {\bibinfo
  {journal} {Phys. Rev.}\ }\textbf {\bibinfo {volume} {D81}},\ \bibinfo {pages}
  {124030} (\bibinfo {year} {2010})},\ \Eprint {http://arxiv.org/abs/1001.2578}
  {arXiv:1001.2578 [gr-qc]}\BibitemShut {NoStop}%
\bibitem [{\citenamefont {Hopper}\ and\ \citenamefont
  {Evans}(2010)}]{Hopper:2010uv}%
  \BibitemOpen
  \bibfield  {author} {\bibinfo {author} {\bibfnamefont {Seth}\ \bibnamefont
  {Hopper}}\ and\ \bibinfo {author} {\bibfnamefont {Charles~R.}\ \bibnamefont
  {Evans}},\ }\href {\doibase10.1103/PhysRevD.82.084010} {\bibfield  {journal}
  {\bibinfo  {journal} {Phys. Rev.}\ }\textbf {\bibinfo {volume} {D82}},\
  \bibinfo {pages} {084010} (\bibinfo {year} {2010})},\ \Eprint
  {http://arxiv.org/abs/1006.4907} {arXiv:1006.4907 [gr-qc]}\BibitemShut
  {NoStop}%
\bibitem [{\citenamefont {Hopper}\ \emph {et~al.}(2015)\citenamefont {Hopper},
  \citenamefont {Forseth}, \citenamefont {Osburn},\ and\ \citenamefont
  {Evans}}]{Hopper:2015jxa}%
  \BibitemOpen
  \bibfield  {author} {\bibinfo {author} {\bibfnamefont {Seth}\ \bibnamefont
  {Hopper}}, \bibinfo {author} {\bibfnamefont {Erik}\ \bibnamefont {Forseth}},
  \bibinfo {author} {\bibfnamefont {Thomas}\ \bibnamefont {Osburn}}, \ and\
  \bibinfo {author} {\bibfnamefont {Charles~R.}\ \bibnamefont {Evans}},\ }\href
  {\doibase10.1103/PhysRevD.92.044048} {\bibfield  {journal} {\bibinfo
  {journal} {Phys. Rev.}\ }\textbf {\bibinfo {volume} {D92}},\ \bibinfo {pages}
  {044048} (\bibinfo {year} {2015})},\ \Eprint
  {http://arxiv.org/abs/1506.04742} {arXiv:1506.04742 [gr-qc]}\BibitemShut
  {NoStop}%
\bibitem [{\citenamefont {Hopper}(2018)}]{Hopper:2017iyq}%
  \BibitemOpen
  \bibfield  {author} {\bibinfo {author} {\bibfnamefont {Seth}\ \bibnamefont
  {Hopper}},\ }\href {\doibase10.1103/PhysRevD.97.064007} {\bibfield  {journal}
  {\bibinfo  {journal} {Phys. Rev.}\ }\textbf {\bibinfo {volume} {D97}},\
  \bibinfo {pages} {064007} (\bibinfo {year} {2018})},\ \Eprint
  {http://arxiv.org/abs/1706.05455} {arXiv:1706.05455 [gr-qc]}\BibitemShut
  {NoStop}%
\bibitem [{\citenamefont {Barack}\ and\ \citenamefont
  {Giudice}(2017)}]{Barack:2017oir}%
  \BibitemOpen
  \bibfield  {author} {\bibinfo {author} {\bibfnamefont {Leor}\ \bibnamefont
  {Barack}}\ and\ \bibinfo {author} {\bibfnamefont {Paco}\ \bibnamefont
  {Giudice}},\ }\href {\doibase10.1103/PhysRevD.95.104033} {\bibfield
  {journal} {\bibinfo  {journal} {Phys. Rev.}\ }\textbf {\bibinfo {volume}
  {D95}},\ \bibinfo {pages} {104033} (\bibinfo {year} {2017})},\ \Eprint
  {http://arxiv.org/abs/1702.04204} {arXiv:1702.04204 [gr-qc]}\BibitemShut
  {NoStop}%
\bibitem [{\citenamefont {Barack}\ and\ \citenamefont
  {Ori}(2001)}]{Barack:2001ph}%
  \BibitemOpen
  \bibfield  {author} {\bibinfo {author} {\bibfnamefont {Leor}\ \bibnamefont
  {Barack}}\ and\ \bibinfo {author} {\bibfnamefont {Amos}\ \bibnamefont
  {Ori}},\ }\href {\doibase10.1103/PhysRevD.64.124003} {\bibfield  {journal}
  {\bibinfo  {journal} {Phys. Rev.}\ }\textbf {\bibinfo {volume} {D64}},\
  \bibinfo {pages} {124003} (\bibinfo {year} {2001})},\ \Eprint
  {http://arxiv.org/abs/gr-qc/0107056} {arXiv:gr-qc/0107056
  [gr-qc]}\BibitemShut {NoStop}%
\bibitem [{\citenamefont {Gralla}(2011)}]{Gralla:2011zr}%
  \BibitemOpen
  \bibfield  {author} {\bibinfo {author} {\bibfnamefont {Samuel~E.}\
  \bibnamefont {Gralla}},\ }\href {\doibase10.1103/PhysRevD.84.084050}
  {\bibfield  {journal} {\bibinfo  {journal} {Phys. Rev.}\ }\textbf {\bibinfo
  {volume} {D84}},\ \bibinfo {pages} {084050} (\bibinfo {year} {2011})},\
  \Eprint {http://arxiv.org/abs/1104.5635} {arXiv:1104.5635
  [gr-qc]}\BibitemShut {NoStop}%
\bibitem [{\citenamefont {Detweiler}(2008)}]{Detweiler:2008ft}%
  \BibitemOpen
  \bibfield  {author} {\bibinfo {author} {\bibfnamefont {Steven~L.}\
  \bibnamefont {Detweiler}},\ }\href {\doibase10.1103/PhysRevD.77.124026}
  {\bibfield  {journal} {\bibinfo  {journal} {Phys. Rev.}\ }\textbf {\bibinfo
  {volume} {D77}},\ \bibinfo {pages} {124026} (\bibinfo {year} {2008})},\
  \Eprint {http://arxiv.org/abs/0804.3529} {arXiv:0804.3529
  [gr-qc]}\BibitemShut {NoStop}%
\bibitem [{\citenamefont {Barack}\ and\ \citenamefont
  {Sago}(2009)}]{Barack:2009ey}%
  \BibitemOpen
  \bibfield  {author} {\bibinfo {author} {\bibfnamefont {Leor}\ \bibnamefont
  {Barack}}\ and\ \bibinfo {author} {\bibfnamefont {Norichika}\ \bibnamefont
  {Sago}},\ }\href {\doibase10.1103/PhysRevLett.102.191101} {\bibfield
  {journal} {\bibinfo  {journal} {Phys. Rev. Lett.}\ }\textbf {\bibinfo
  {volume} {102}},\ \bibinfo {pages} {191101} (\bibinfo {year} {2009})},\
  \Eprint {http://arxiv.org/abs/0902.0573} {arXiv:0902.0573
  [gr-qc]}\BibitemShut {NoStop}%
\bibitem [{\citenamefont {Barack}\ \emph {et~al.}(2010)\citenamefont {Barack},
  \citenamefont {Damour},\ and\ \citenamefont {Sago}}]{Barack:2010ny}%
  \BibitemOpen
  \bibfield  {author} {\bibinfo {author} {\bibfnamefont {Leor}\ \bibnamefont
  {Barack}}, \bibinfo {author} {\bibfnamefont {Thibault}\ \bibnamefont
  {Damour}}, \ and\ \bibinfo {author} {\bibfnamefont {Norichika}\ \bibnamefont
  {Sago}},\ }\href {\doibase10.1103/PhysRevD.82.084036} {\bibfield  {journal}
  {\bibinfo  {journal} {Phys. Rev.}\ }\textbf {\bibinfo {volume} {D82}},\
  \bibinfo {pages} {084036} (\bibinfo {year} {2010})},\ \Eprint
  {http://arxiv.org/abs/1008.0935} {arXiv:1008.0935 [gr-qc]}\BibitemShut
  {NoStop}%
\bibitem [{\citenamefont {van~de Meent}(2017)}]{vandeMeent:2016hel}%
  \BibitemOpen
  \bibfield  {author} {\bibinfo {author} {\bibfnamefont {Maarten}\ \bibnamefont
  {van~de Meent}},\ }\href {\doibase10.1103/PhysRevLett.118.011101} {\bibfield
  {journal} {\bibinfo  {journal} {Phys. Rev. Lett.}\ }\textbf {\bibinfo
  {volume} {118}},\ \bibinfo {pages} {011101} (\bibinfo {year} {2017})},\
  \Eprint {http://arxiv.org/abs/1610.03497} {arXiv:1610.03497
  [gr-qc]}\BibitemShut {NoStop}%
\bibitem [{\citenamefont {Dolan}\ \emph {et~al.}(2014)\citenamefont {Dolan},
  \citenamefont {Warburton}, \citenamefont {Harte}, \citenamefont {Le~Tiec},
  \citenamefont {Wardell} \emph {et~al.}}]{Dolan:2013roa}%
  \BibitemOpen
  \bibfield  {author} {\bibinfo {author} {\bibfnamefont {Sam~R.}\ \bibnamefont
  {Dolan}}, \bibinfo {author} {\bibfnamefont {Niels}\ \bibnamefont
  {Warburton}}, \bibinfo {author} {\bibfnamefont {Abraham~I.}\ \bibnamefont
  {Harte}}, \bibinfo {author} {\bibfnamefont {Alexandre}\ \bibnamefont
  {Le~Tiec}}, \bibinfo {author} {\bibfnamefont {Barry}\ \bibnamefont
  {Wardell}},  \emph {et~al.},\ }\href {\doibase10.1103/PhysRevD.89.064011}
  {\bibfield  {journal} {\bibinfo  {journal} {Phys. Rev.}\ }\textbf {\bibinfo
  {volume} {D89}},\ \bibinfo {pages} {064011} (\bibinfo {year} {2014})},\
  \Eprint {http://arxiv.org/abs/1312.0775} {arXiv:1312.0775
  [gr-qc]}\BibitemShut {NoStop}%
\bibitem [{\citenamefont {Shah}\ and\ \citenamefont
  {Pound}(2015)}]{Shah:2015nva}%
  \BibitemOpen
  \bibfield  {author} {\bibinfo {author} {\bibfnamefont {Abhay~G.}\
  \bibnamefont {Shah}}\ and\ \bibinfo {author} {\bibfnamefont {Adam}\
  \bibnamefont {Pound}},\ }\href {\doibase10.1103/PhysRevD.91.124022}
  {\bibfield  {journal} {\bibinfo  {journal} {Phys. Rev.}\ }\textbf {\bibinfo
  {volume} {D91}},\ \bibinfo {pages} {124022} (\bibinfo {year} {2015})},\
  \Eprint {http://arxiv.org/abs/1503.02414} {arXiv:1503.02414
  [gr-qc]}\BibitemShut {NoStop}%
\bibitem [{\citenamefont {Akcay}\ \emph {et~al.}(2017)\citenamefont {Akcay},
  \citenamefont {Dempsey},\ and\ \citenamefont {Dolan}}]{Akcay:2016dku}%
  \BibitemOpen
  \bibfield  {author} {\bibinfo {author} {\bibfnamefont {Sarp}\ \bibnamefont
  {Akcay}}, \bibinfo {author} {\bibfnamefont {David}\ \bibnamefont {Dempsey}},
  \ and\ \bibinfo {author} {\bibfnamefont {Sam~R.}\ \bibnamefont {Dolan}},\
  }\href {\doibase10.1088/1361-6382/aa61d6} {\bibfield  {journal} {\bibinfo
  {journal} {Class. Quant. Grav.}\ }\textbf {\bibinfo {volume} {34}},\ \bibinfo
  {pages} {084001} (\bibinfo {year} {2017})},\ \Eprint
  {http://arxiv.org/abs/1608.04811} {arXiv:1608.04811 [gr-qc]}\BibitemShut
  {NoStop}%
\bibitem [{\citenamefont {Akcay}(2017)}]{Akcay:2017azq}%
  \BibitemOpen
  \bibfield  {author} {\bibinfo {author} {\bibfnamefont {Sarp}\ \bibnamefont
  {Akcay}},\ }\href {\doibase10.1103/PhysRevD.96.044024} {\bibfield  {journal}
  {\bibinfo  {journal} {Phys. Rev.}\ }\textbf {\bibinfo {volume} {D96}},\
  \bibinfo {pages} {044024} (\bibinfo {year} {2017})},\ \Eprint
  {http://arxiv.org/abs/1705.03282} {arXiv:1705.03282 [gr-qc]}\BibitemShut
  {NoStop}%
\bibitem [{\citenamefont {Dolan}\ \emph {et~al.}(2015)\citenamefont {Dolan},
  \citenamefont {Nolan}, \citenamefont {Ottewill}, \citenamefont {Warburton},\
  and\ \citenamefont {Wardell}}]{Dolan:2014pja}%
  \BibitemOpen
  \bibfield  {author} {\bibinfo {author} {\bibfnamefont {Sam~R}\ \bibnamefont
  {Dolan}}, \bibinfo {author} {\bibfnamefont {Patrick}\ \bibnamefont {Nolan}},
  \bibinfo {author} {\bibfnamefont {Adrian~C}\ \bibnamefont {Ottewill}},
  \bibinfo {author} {\bibfnamefont {Niels}\ \bibnamefont {Warburton}}, \ and\
  \bibinfo {author} {\bibfnamefont {Barry}\ \bibnamefont {Wardell}},\ }\href
  {\doibase10.1103/PhysRevD.91.023009} {\bibfield  {journal} {\bibinfo
  {journal} {Phys. Rev.}\ }\textbf {\bibinfo {volume} {D91}},\ \bibinfo {pages}
  {023009} (\bibinfo {year} {2015})},\ \Eprint {http://arxiv.org/abs/1406.4890}
  {arXiv:1406.4890 [gr-qc]}\BibitemShut {NoStop}%
\bibitem [{\citenamefont {Nolan}\ \emph {et~al.}(2015)\citenamefont {Nolan},
  \citenamefont {Kavanagh}, \citenamefont {Dolan}, \citenamefont {Ottewill},
  \citenamefont {Warburton},\ and\ \citenamefont {Wardell}}]{Nolan:2015vpa}%
  \BibitemOpen
  \bibfield  {author} {\bibinfo {author} {\bibfnamefont {Patrick}\ \bibnamefont
  {Nolan}}, \bibinfo {author} {\bibfnamefont {Chris}\ \bibnamefont {Kavanagh}},
  \bibinfo {author} {\bibfnamefont {Sam~R.}\ \bibnamefont {Dolan}}, \bibinfo
  {author} {\bibfnamefont {Adrian~C.}\ \bibnamefont {Ottewill}}, \bibinfo
  {author} {\bibfnamefont {Niels}\ \bibnamefont {Warburton}}, \ and\ \bibinfo
  {author} {\bibfnamefont {Barry}\ \bibnamefont {Wardell}},\ }\href
  {\doibase10.1103/PhysRevD.92.123008} {\bibfield  {journal} {\bibinfo
  {journal} {Phys. Rev.}\ }\textbf {\bibinfo {volume} {D92}},\ \bibinfo {pages}
  {123008} (\bibinfo {year} {2015})},\ \Eprint
  {http://arxiv.org/abs/1505.04447} {arXiv:1505.04447 [gr-qc]}\BibitemShut
  {NoStop}%
\bibitem [{\citenamefont {Sago}\ \emph {et~al.}(2008)\citenamefont {Sago},
  \citenamefont {Barack},\ and\ \citenamefont {Detweiler}}]{Sago:2008id}%
  \BibitemOpen
  \bibfield  {author} {\bibinfo {author} {\bibfnamefont {Norichika}\
  \bibnamefont {Sago}}, \bibinfo {author} {\bibfnamefont {Leor}\ \bibnamefont
  {Barack}}, \ and\ \bibinfo {author} {\bibfnamefont {Steven~L.}\ \bibnamefont
  {Detweiler}},\ }\href {\doibase10.1103/PhysRevD.78.124024} {\bibfield
  {journal} {\bibinfo  {journal} {Phys. Rev.}\ }\textbf {\bibinfo {volume}
  {D78}},\ \bibinfo {pages} {124024} (\bibinfo {year} {2008})},\ \Eprint
  {http://arxiv.org/abs/0810.2530} {arXiv:0810.2530 [gr-qc]}\BibitemShut
  {NoStop}%
\bibitem [{\citenamefont {Blanchet}\ \emph
  {et~al.}(2010{\natexlab{a}})\citenamefont {Blanchet}, \citenamefont
  {Detweiler}, \citenamefont {Le~Tiec},\ and\ \citenamefont
  {Whiting}}]{Blanchet:2010zd}%
  \BibitemOpen
  \bibfield  {author} {\bibinfo {author} {\bibfnamefont {Luc}\ \bibnamefont
  {Blanchet}}, \bibinfo {author} {\bibfnamefont {Steven~L.}\ \bibnamefont
  {Detweiler}}, \bibinfo {author} {\bibfnamefont {Alexandre}\ \bibnamefont
  {Le~Tiec}}, \ and\ \bibinfo {author} {\bibfnamefont {Bernard~F.}\
  \bibnamefont {Whiting}},\ }\href {\doibase10.1103/PhysRevD.81.084033}
  {\bibfield  {journal} {\bibinfo  {journal} {Phys. Rev.}\ }\textbf {\bibinfo
  {volume} {D81}},\ \bibinfo {pages} {084033} (\bibinfo {year}
  {2010}{\natexlab{a}})},\ \Eprint {http://arxiv.org/abs/1002.0726}
  {arXiv:1002.0726 [gr-qc]}\BibitemShut {NoStop}%
\bibitem [{\citenamefont {Blanchet}\ \emph
  {et~al.}(2010{\natexlab{b}})\citenamefont {Blanchet}, \citenamefont
  {Detweiler}, \citenamefont {Le~Tiec},\ and\ \citenamefont
  {Whiting}}]{Blanchet:2009sd}%
  \BibitemOpen
  \bibfield  {author} {\bibinfo {author} {\bibfnamefont {Luc}\ \bibnamefont
  {Blanchet}}, \bibinfo {author} {\bibfnamefont {Steven~L.}\ \bibnamefont
  {Detweiler}}, \bibinfo {author} {\bibfnamefont {Alexandre}\ \bibnamefont
  {Le~Tiec}}, \ and\ \bibinfo {author} {\bibfnamefont {Bernard~F.}\
  \bibnamefont {Whiting}},\ }\href {\doibase10.1103/PhysRevD.81.064004}
  {\bibfield  {journal} {\bibinfo  {journal} {Phys. Rev.}\ }\textbf {\bibinfo
  {volume} {D81}},\ \bibinfo {pages} {064004} (\bibinfo {year}
  {2010}{\natexlab{b}})},\ \Eprint {http://arxiv.org/abs/0910.0207}
  {arXiv:0910.0207 [gr-qc]}\BibitemShut {NoStop}%
\bibitem [{\citenamefont {Akcay}\ \emph {et~al.}(2012)\citenamefont {Akcay},
  \citenamefont {Barack}, \citenamefont {Damour},\ and\ \citenamefont
  {Sago}}]{Akcay:2012ea}%
  \BibitemOpen
  \bibfield  {author} {\bibinfo {author} {\bibfnamefont {Sarp}\ \bibnamefont
  {Akcay}}, \bibinfo {author} {\bibfnamefont {Leor}\ \bibnamefont {Barack}},
  \bibinfo {author} {\bibfnamefont {Thibault}\ \bibnamefont {Damour}}, \ and\
  \bibinfo {author} {\bibfnamefont {Norichika}\ \bibnamefont {Sago}},\ }\href
  {\doibase10.1103/PhysRevD.86.104041} {\bibfield  {journal} {\bibinfo
  {journal} {Phys. Rev.}\ }\textbf {\bibinfo {volume} {D86}},\ \bibinfo {pages}
  {104041} (\bibinfo {year} {2012})},\ \Eprint {http://arxiv.org/abs/1209.0964}
  {arXiv:1209.0964 [gr-qc]}\BibitemShut {NoStop}%
\bibitem [{\citenamefont {Bini}\ and\ \citenamefont
  {Damour}(2014{\natexlab{a}})}]{Bini:2014zxa}%
  \BibitemOpen
  \bibfield  {author} {\bibinfo {author} {\bibfnamefont {Donato}\ \bibnamefont
  {Bini}}\ and\ \bibinfo {author} {\bibfnamefont {Thibault}\ \bibnamefont
  {Damour}},\ }\href {\doibase10.1103/PhysRevD.90.124037} {\bibfield  {journal}
  {\bibinfo  {journal} {Phys. Rev.}\ }\textbf {\bibinfo {volume} {D90}},\
  \bibinfo {pages} {124037} (\bibinfo {year} {2014}{\natexlab{a}})},\ \Eprint
  {http://arxiv.org/abs/1409.6933} {arXiv:1409.6933 [gr-qc]}\BibitemShut
  {NoStop}%
\bibitem [{\citenamefont {Blanchet}\ \emph
  {et~al.}(2014{\natexlab{a}})\citenamefont {Blanchet}, \citenamefont {Faye},\
  and\ \citenamefont {Whiting}}]{Blanchet:2014bza}%
  \BibitemOpen
  \bibfield  {author} {\bibinfo {author} {\bibfnamefont {Luc}\ \bibnamefont
  {Blanchet}}, \bibinfo {author} {\bibfnamefont {Guillaume}\ \bibnamefont
  {Faye}}, \ and\ \bibinfo {author} {\bibfnamefont {Bernard~F.}\ \bibnamefont
  {Whiting}},\ }\href {\doibase10.1103/PhysRevD.90.044017} {\bibfield
  {journal} {\bibinfo  {journal} {Phys. Rev.}\ }\textbf {\bibinfo {volume}
  {D90}},\ \bibinfo {pages} {044017} (\bibinfo {year} {2014}{\natexlab{a}})},\
  \Eprint {http://arxiv.org/abs/1405.5151} {arXiv:1405.5151
  [gr-qc]}\BibitemShut {NoStop}%
\bibitem [{\citenamefont {Bini}\ \emph {et~al.}(2015)\citenamefont {Bini},
  \citenamefont {Damour},\ and\ \citenamefont {Geralico}}]{Bini:2015xua}%
  \BibitemOpen
  \bibfield  {author} {\bibinfo {author} {\bibfnamefont {Donato}\ \bibnamefont
  {Bini}}, \bibinfo {author} {\bibfnamefont {Thibault}\ \bibnamefont {Damour}},
  \ and\ \bibinfo {author} {\bibfnamefont {Andrea}\ \bibnamefont {Geralico}},\
  }\href {\doibase10.1103/PhysRevD.92.124058} {\bibfield  {journal} {\bibinfo
  {journal} {Phys. Rev.}\ }\textbf {\bibinfo {volume} {D92}},\ \bibinfo {pages}
  {124058} (\bibinfo {year} {2015})},\ \bibinfo {note} {[Erratum: Phys.
  Rev.D93,no.10,109902(2016)]},\ \Eprint {http://arxiv.org/abs/1510.06230}
  {arXiv:1510.06230 [gr-qc]}\BibitemShut {NoStop}%
\bibitem [{\citenamefont {Bini}\ \emph
  {et~al.}(2016{\natexlab{a}})\citenamefont {Bini}, \citenamefont {Damour},\
  and\ \citenamefont {Geralico}}]{Bini:2015bfb}%
  \BibitemOpen
  \bibfield  {author} {\bibinfo {author} {\bibfnamefont {Donato}\ \bibnamefont
  {Bini}}, \bibinfo {author} {\bibfnamefont {Thibault}\ \bibnamefont {Damour}},
  \ and\ \bibinfo {author} {\bibfnamefont {Andrea}\ \bibnamefont {Geralico}},\
  }\href {\doibase10.1103/PhysRevD.93.064023} {\bibfield  {journal} {\bibinfo
  {journal} {Phys. Rev.}\ }\textbf {\bibinfo {volume} {D93}},\ \bibinfo {pages}
  {064023} (\bibinfo {year} {2016}{\natexlab{a}})},\ \Eprint
  {http://arxiv.org/abs/1511.04533} {arXiv:1511.04533 [gr-qc]}\BibitemShut
  {NoStop}%
\bibitem [{\citenamefont {Kavanagh}\ \emph {et~al.}(2015)\citenamefont
  {Kavanagh}, \citenamefont {Ottewill},\ and\ \citenamefont
  {Wardell}}]{Kavanagh:2015lva}%
  \BibitemOpen
  \bibfield  {author} {\bibinfo {author} {\bibfnamefont {Chris}\ \bibnamefont
  {Kavanagh}}, \bibinfo {author} {\bibfnamefont {Adrian~C.}\ \bibnamefont
  {Ottewill}}, \ and\ \bibinfo {author} {\bibfnamefont {Barry}\ \bibnamefont
  {Wardell}},\ }\href {\doibase10.1103/PhysRevD.92.084025} {\bibfield
  {journal} {\bibinfo  {journal} {Phys. Rev.}\ }\textbf {\bibinfo {volume}
  {D92}},\ \bibinfo {pages} {084025} (\bibinfo {year} {2015})},\ \Eprint
  {http://arxiv.org/abs/1503.02334} {arXiv:1503.02334 [gr-qc]}\BibitemShut
  {NoStop}%
\bibitem [{\citenamefont {Akcay}\ and\ \citenamefont {van~de
  Meent}(2016)}]{Akcay:2015pjz}%
  \BibitemOpen
  \bibfield  {author} {\bibinfo {author} {\bibfnamefont {Sarp}\ \bibnamefont
  {Akcay}}\ and\ \bibinfo {author} {\bibfnamefont {Maarten}\ \bibnamefont
  {van~de Meent}},\ }\href {\doibase10.1103/PhysRevD.93.064063} {\bibfield
  {journal} {\bibinfo  {journal} {Phys. Rev.}\ }\textbf {\bibinfo {volume}
  {D93}},\ \bibinfo {pages} {064063} (\bibinfo {year} {2016})},\ \Eprint
  {http://arxiv.org/abs/1512.03392} {arXiv:1512.03392 [gr-qc]}\BibitemShut
  {NoStop}%
\bibitem [{\citenamefont {Akcay}\ \emph {et~al.}(2015)\citenamefont {Akcay},
  \citenamefont {Le~Tiec}, \citenamefont {Barack}, \citenamefont {Sago},\ and\
  \citenamefont {Warburton}}]{Akcay:2015pza}%
  \BibitemOpen
  \bibfield  {author} {\bibinfo {author} {\bibfnamefont {Sarp}\ \bibnamefont
  {Akcay}}, \bibinfo {author} {\bibfnamefont {Alexandre}\ \bibnamefont
  {Le~Tiec}}, \bibinfo {author} {\bibfnamefont {Leor}\ \bibnamefont {Barack}},
  \bibinfo {author} {\bibfnamefont {Norichika}\ \bibnamefont {Sago}}, \ and\
  \bibinfo {author} {\bibfnamefont {Niels}\ \bibnamefont {Warburton}},\ }\href
  {\doibase10.1103/PhysRevD.91.124014} {\bibfield  {journal} {\bibinfo
  {journal} {Phys. Rev.}\ }\textbf {\bibinfo {volume} {D91}},\ \bibinfo {pages}
  {124014} (\bibinfo {year} {2015})},\ \Eprint
  {http://arxiv.org/abs/1503.01374} {arXiv:1503.01374 [gr-qc]}\BibitemShut
  {NoStop}%
\bibitem [{\citenamefont {Hopper}\ \emph {et~al.}(2016)\citenamefont {Hopper},
  \citenamefont {Kavanagh},\ and\ \citenamefont {Ottewill}}]{Hopper:2015icj}%
  \BibitemOpen
  \bibfield  {author} {\bibinfo {author} {\bibfnamefont {Seth}\ \bibnamefont
  {Hopper}}, \bibinfo {author} {\bibfnamefont {Chris}\ \bibnamefont
  {Kavanagh}}, \ and\ \bibinfo {author} {\bibfnamefont {Adrian~C.}\
  \bibnamefont {Ottewill}},\ }\href {\doibase10.1103/PhysRevD.93.044010}
  {\bibfield  {journal} {\bibinfo  {journal} {Phys. Rev.}\ }\textbf {\bibinfo
  {volume} {D93}},\ \bibinfo {pages} {044010} (\bibinfo {year} {2016})},\
  \Eprint {http://arxiv.org/abs/1512.01556} {arXiv:1512.01556
  [gr-qc]}\BibitemShut {NoStop}%
\bibitem [{\citenamefont {Kavanagh}\ \emph {et~al.}(2016)\citenamefont
  {Kavanagh}, \citenamefont {Ottewill},\ and\ \citenamefont
  {Wardell}}]{Kavanagh:2016idg}%
  \BibitemOpen
  \bibfield  {author} {\bibinfo {author} {\bibfnamefont {Chris}\ \bibnamefont
  {Kavanagh}}, \bibinfo {author} {\bibfnamefont {Adrian~C.}\ \bibnamefont
  {Ottewill}}, \ and\ \bibinfo {author} {\bibfnamefont {Barry}\ \bibnamefont
  {Wardell}},\ }\href {\doibase10.1103/PhysRevD.93.124038} {\bibfield
  {journal} {\bibinfo  {journal} {Phys. Rev.}\ }\textbf {\bibinfo {volume}
  {D93}},\ \bibinfo {pages} {124038} (\bibinfo {year} {2016})},\ \Eprint
  {http://arxiv.org/abs/1601.03394} {arXiv:1601.03394 [gr-qc]}\BibitemShut
  {NoStop}%
\bibitem [{\citenamefont {Bini}\ \emph
  {et~al.}(2016{\natexlab{b}})\citenamefont {Bini}, \citenamefont {Damour},\
  and\ \citenamefont {Geralico}}]{Bini:2016qtx}%
  \BibitemOpen
  \bibfield  {author} {\bibinfo {author} {\bibfnamefont {Donato}\ \bibnamefont
  {Bini}}, \bibinfo {author} {\bibfnamefont {Thibault}\ \bibnamefont {Damour}},
  \ and\ \bibinfo {author} {\bibfnamefont {andrea}\ \bibnamefont {Geralico}},\
  }\href {\doibase10.1103/PhysRevD.93.104017} {\bibfield  {journal} {\bibinfo
  {journal} {Phys. Rev.}\ }\textbf {\bibinfo {volume} {D93}},\ \bibinfo {pages}
  {104017} (\bibinfo {year} {2016}{\natexlab{b}})},\ \Eprint
  {http://arxiv.org/abs/1601.02988} {arXiv:1601.02988 [gr-qc]}\BibitemShut
  {NoStop}%
\bibitem [{\citenamefont {Bini}\ \emph
  {et~al.}(2016{\natexlab{c}})\citenamefont {Bini}, \citenamefont {Damour},\
  and\ \citenamefont {Geralico}}]{Bini:2016dvs}%
  \BibitemOpen
  \bibfield  {author} {\bibinfo {author} {\bibfnamefont {Donato}\ \bibnamefont
  {Bini}}, \bibinfo {author} {\bibfnamefont {Thibault}\ \bibnamefont {Damour}},
  \ and\ \bibinfo {author} {\bibfnamefont {Andrea}\ \bibnamefont {Geralico}},\
  }\href {\doibase10.1103/PhysRevD.93.124058} {\bibfield  {journal} {\bibinfo
  {journal} {Phys. Rev.}\ }\textbf {\bibinfo {volume} {D93}},\ \bibinfo {pages}
  {124058} (\bibinfo {year} {2016}{\natexlab{c}})},\ \Eprint
  {http://arxiv.org/abs/1602.08282} {arXiv:1602.08282 [gr-qc]}\BibitemShut
  {NoStop}%
\bibitem [{\citenamefont {Kavanagh}\ \emph {et~al.}(2017)\citenamefont
  {Kavanagh}, \citenamefont {Bini}, \citenamefont {Damour}, \citenamefont
  {Hopper}, \citenamefont {Ottewill},\ and\ \citenamefont
  {Wardell}}]{Kavanagh:2017wot}%
  \BibitemOpen
  \bibfield  {author} {\bibinfo {author} {\bibfnamefont {Chris}\ \bibnamefont
  {Kavanagh}}, \bibinfo {author} {\bibfnamefont {Donato}\ \bibnamefont {Bini}},
  \bibinfo {author} {\bibfnamefont {Thibault}\ \bibnamefont {Damour}}, \bibinfo
  {author} {\bibfnamefont {Seth}\ \bibnamefont {Hopper}}, \bibinfo {author}
  {\bibfnamefont {Adrian~C.}\ \bibnamefont {Ottewill}}, \ and\ \bibinfo
  {author} {\bibfnamefont {Barry}\ \bibnamefont {Wardell}},\ }\href
  {\doibase10.1103/PhysRevD.96.064012} {\bibfield  {journal} {\bibinfo
  {journal} {Phys. Rev.}\ }\textbf {\bibinfo {volume} {D96}},\ \bibinfo {pages}
  {064012} (\bibinfo {year} {2017})},\ \Eprint
  {http://arxiv.org/abs/1706.00459} {arXiv:1706.00459 [gr-qc]}\BibitemShut
  {NoStop}%
\bibitem [{\citenamefont {Le~Tiec}\ \emph {et~al.}(2011)\citenamefont
  {Le~Tiec}, \citenamefont {Mroue}, \citenamefont {Barack}, \citenamefont
  {Buonanno}, \citenamefont {Pfeiffer} \emph {et~al.}}]{LeTiec:2011bk}%
  \BibitemOpen
  \bibfield  {author} {\bibinfo {author} {\bibfnamefont {Alexandre}\
  \bibnamefont {Le~Tiec}}, \bibinfo {author} {\bibfnamefont {Abdul~H.}\
  \bibnamefont {Mroue}}, \bibinfo {author} {\bibfnamefont {Leor}\ \bibnamefont
  {Barack}}, \bibinfo {author} {\bibfnamefont {Alessandra}\ \bibnamefont
  {Buonanno}}, \bibinfo {author} {\bibfnamefont {Harald~P.}\ \bibnamefont
  {Pfeiffer}},  \emph {et~al.},\ }\href
  {\doibase10.1103/PhysRevLett.107.141101} {\bibfield  {journal} {\bibinfo
  {journal} {Phys. Rev. Lett.}\ }\textbf {\bibinfo {volume} {107}},\ \bibinfo
  {pages} {141101} (\bibinfo {year} {2011})},\ \Eprint
  {http://arxiv.org/abs/1106.3278} {arXiv:1106.3278 [gr-qc]}\BibitemShut
  {NoStop}%
\bibitem [{\citenamefont {Le~Tiec}\ \emph
  {et~al.}(2012{\natexlab{a}})\citenamefont {Le~Tiec}, \citenamefont
  {Barausse},\ and\ \citenamefont {Buonanno}}]{LeTiec:2011dp}%
  \BibitemOpen
  \bibfield  {author} {\bibinfo {author} {\bibfnamefont {Alexandre}\
  \bibnamefont {Le~Tiec}}, \bibinfo {author} {\bibfnamefont {Enrico}\
  \bibnamefont {Barausse}}, \ and\ \bibinfo {author} {\bibfnamefont
  {Alessandra}\ \bibnamefont {Buonanno}},\ }\href
  {\doibase10.1103/PhysRevLett.108.131103} {\bibfield  {journal} {\bibinfo
  {journal} {Phys. Rev. Lett.}\ }\textbf {\bibinfo {volume} {108}},\ \bibinfo
  {pages} {131103} (\bibinfo {year} {2012}{\natexlab{a}})},\ \Eprint
  {http://arxiv.org/abs/1111.5609} {arXiv:1111.5609 [gr-qc]}\BibitemShut
  {NoStop}%
\bibitem [{\citenamefont {Le~Tiec}\ \emph {et~al.}(2013)\citenamefont
  {Le~Tiec}, \citenamefont {Buonanno}, \citenamefont {Mroué}, \citenamefont
  {Pfeiffer}, \citenamefont {Hemberger} \emph {et~al.}}]{Tiec:2013twa}%
  \BibitemOpen
  \bibfield  {author} {\bibinfo {author} {\bibfnamefont {Alexandre}\
  \bibnamefont {Le~Tiec}}, \bibinfo {author} {\bibfnamefont {Alessandra}\
  \bibnamefont {Buonanno}}, \bibinfo {author} {\bibfnamefont {Abdul~H.}\
  \bibnamefont {Mroué}}, \bibinfo {author} {\bibfnamefont {Harald~P.}\
  \bibnamefont {Pfeiffer}}, \bibinfo {author} {\bibfnamefont {Daniel~A.}\
  \bibnamefont {Hemberger}},  \emph {et~al.},\ }\href
  {\doibase10.1103/PhysRevD.88.124027} {\bibfield  {journal} {\bibinfo
  {journal} {Phys. Rev.}\ }\textbf {\bibinfo {volume} {D88}},\ \bibinfo {pages}
  {124027} (\bibinfo {year} {2013})},\ \Eprint {http://arxiv.org/abs/1309.0541}
  {arXiv:1309.0541 [gr-qc]}\BibitemShut {NoStop}%
\bibitem [{\citenamefont {Zimmerman}\ \emph {et~al.}(2016)\citenamefont
  {Zimmerman}, \citenamefont {Lewis},\ and\ \citenamefont
  {Pfeiffer}}]{Zimmerman:2016ajr}%
  \BibitemOpen
  \bibfield  {author} {\bibinfo {author} {\bibfnamefont {Aaron}\ \bibnamefont
  {Zimmerman}}, \bibinfo {author} {\bibfnamefont {Adam G.~M.}\ \bibnamefont
  {Lewis}}, \ and\ \bibinfo {author} {\bibfnamefont {Harald~P.}\ \bibnamefont
  {Pfeiffer}},\ }\href {\doibase10.1103/PhysRevLett.117.191101} {\bibfield
  {journal} {\bibinfo  {journal} {Phys. Rev. Lett.}\ }\textbf {\bibinfo
  {volume} {117}},\ \bibinfo {pages} {191101} (\bibinfo {year} {2016})},\
  \Eprint {http://arxiv.org/abs/1606.08056} {arXiv:1606.08056
  [gr-qc]}\BibitemShut {NoStop}%
\bibitem [{\citenamefont {Le~Tiec}\ and\ \citenamefont
  {Grandclément}(2017)}]{LeTiec:2017ebm}%
  \BibitemOpen
  \bibfield  {author} {\bibinfo {author} {\bibfnamefont {Alexandre}\
  \bibnamefont {Le~Tiec}}\ and\ \bibinfo {author} {\bibfnamefont {Philippe}\
  \bibnamefont {Grandclément}},\ }\href@noop {} {\  (\bibinfo {year}
  {2017})},\ \Eprint {http://arxiv.org/abs/1710.03673} {arXiv:1710.03673
  [gr-qc]}\BibitemShut {NoStop}%
\bibitem [{\citenamefont {Barack}\ and\ \citenamefont
  {Sago}(2011)}]{Barack:2011ed}%
  \BibitemOpen
  \bibfield  {author} {\bibinfo {author} {\bibfnamefont {Leor}\ \bibnamefont
  {Barack}}\ and\ \bibinfo {author} {\bibfnamefont {Norichika}\ \bibnamefont
  {Sago}},\ }\href {\doibase10.1103/PhysRevD.83.084023} {\bibfield  {journal}
  {\bibinfo  {journal} {Phys. Rev.}\ }\textbf {\bibinfo {volume} {D83}},\
  \bibinfo {pages} {084023} (\bibinfo {year} {2011})},\ \Eprint
  {http://arxiv.org/abs/1101.3331} {arXiv:1101.3331 [gr-qc]}\BibitemShut
  {NoStop}%
\bibitem [{\citenamefont {Detweiler}(2005)}]{Detweiler:2005kq}%
  \BibitemOpen
  \bibfield  {author} {\bibinfo {author} {\bibfnamefont {Steven~L.}\
  \bibnamefont {Detweiler}},\ }\href {\doibase10.1088/0264-9381/22/15/006}
  {\bibfield  {journal} {\bibinfo  {journal} {Class. Quant. Grav.}\ }\textbf
  {\bibinfo {volume} {22}},\ \bibinfo {pages} {S681--S716} (\bibinfo {year}
  {2005})},\ \Eprint {http://arxiv.org/abs/gr-qc/0501004} {arXiv:gr-qc/0501004
  [gr-qc]}\BibitemShut {NoStop}%
\bibitem [{\citenamefont {Hopper}\ and\ \citenamefont
  {Cardoso}(2018)}]{Hopper:2017qus}%
  \BibitemOpen
  \bibfield  {author} {\bibinfo {author} {\bibfnamefont {Seth}\ \bibnamefont
  {Hopper}}\ and\ \bibinfo {author} {\bibfnamefont {Vitor}\ \bibnamefont
  {Cardoso}},\ }\href {\doibase10.1103/PhysRevD.97.044031} {\bibfield
  {journal} {\bibinfo  {journal} {Phys. Rev.}\ }\textbf {\bibinfo {volume}
  {D97}},\ \bibinfo {pages} {044031} (\bibinfo {year} {2018})},\ \Eprint
  {http://arxiv.org/abs/1706.02791} {arXiv:1706.02791 [gr-qc]}\BibitemShut
  {NoStop}%
\bibitem [{\citenamefont {Teukolsky}(1972)}]{Teukolsky:1972my}%
  \BibitemOpen
  \bibfield  {author} {\bibinfo {author} {\bibfnamefont {S.A.}\ \bibnamefont
  {Teukolsky}},\ }\href {\doibase10.1103/PhysRevLett.29.1114} {\bibfield
  {journal} {\bibinfo  {journal} {Phys. Rev. Lett.}\ }\textbf {\bibinfo
  {volume} {29}},\ \bibinfo {pages} {1114--1118} (\bibinfo {year}
  {1972})}\BibitemShut {NoStop}%
\bibitem [{\citenamefont {Teukolsky}(1973)}]{Teukolsky:1973ha}%
  \BibitemOpen
  \bibfield  {author} {\bibinfo {author} {\bibfnamefont {Saul~A.}\ \bibnamefont
  {Teukolsky}},\ }\href {\doibase10.1086/152444} {\bibfield  {journal}
  {\bibinfo  {journal} {Astrophys. J.}\ }\textbf {\bibinfo {volume} {185}},\
  \bibinfo {pages} {635--647} (\bibinfo {year} {1973})}\BibitemShut {NoStop}%
\bibitem [{\citenamefont {Merlin}\ \emph {et~al.}(2016)\citenamefont {Merlin},
  \citenamefont {Ori}, \citenamefont {Barack}, \citenamefont {Pound},\ and\
  \citenamefont {van~de Meent}}]{Merlin:2016boc}%
  \BibitemOpen
  \bibfield  {author} {\bibinfo {author} {\bibfnamefont {Cesar}\ \bibnamefont
  {Merlin}}, \bibinfo {author} {\bibfnamefont {Amos}\ \bibnamefont {Ori}},
  \bibinfo {author} {\bibfnamefont {Leor}\ \bibnamefont {Barack}}, \bibinfo
  {author} {\bibfnamefont {Adam}\ \bibnamefont {Pound}}, \ and\ \bibinfo
  {author} {\bibfnamefont {Maarten}\ \bibnamefont {van~de Meent}},\ }\href
  {\doibase10.1103/PhysRevD.94.104066} {\bibfield  {journal} {\bibinfo
  {journal} {Phys. Rev.}\ }\textbf {\bibinfo {volume} {D94}},\ \bibinfo {pages}
  {104066} (\bibinfo {year} {2016})},\ \Eprint
  {http://arxiv.org/abs/1609.01227} {arXiv:1609.01227 [gr-qc]}\BibitemShut
  {NoStop}%
\bibitem [{\citenamefont {van De~Meent}(2017)}]{vandeMeent:2017fqk}%
  \BibitemOpen
  \bibfield  {author} {\bibinfo {author} {\bibfnamefont {Maarten}\ \bibnamefont
  {van De~Meent}},\ }\href {\doibase10.1088/1361-6382/aa71c3} {\bibfield
  {journal} {\bibinfo  {journal} {Class. Quant. Grav.}\ }\textbf {\bibinfo
  {volume} {34}},\ \bibinfo {pages} {124003} (\bibinfo {year} {2017})},\
  \Eprint {http://arxiv.org/abs/1702.00969} {arXiv:1702.00969
  [gr-qc]}\BibitemShut {NoStop}%
\bibitem [{\citenamefont {van~de Meent}\ and\ \citenamefont
  {Shah}(2015)}]{vandeMeent:2015lxa}%
  \BibitemOpen
  \bibfield  {author} {\bibinfo {author} {\bibfnamefont {Maarten}\ \bibnamefont
  {van~de Meent}}\ and\ \bibinfo {author} {\bibfnamefont {Abhay~G.}\
  \bibnamefont {Shah}},\ }\href {\doibase10.1103/PhysRevD.92.064025} {\bibfield
   {journal} {\bibinfo  {journal} {Phys. Rev.}\ }\textbf {\bibinfo {volume}
  {D92}},\ \bibinfo {pages} {064025} (\bibinfo {year} {2015})},\ \Eprint
  {http://arxiv.org/abs/1506.04755} {arXiv:1506.04755 [gr-qc]}\BibitemShut
  {NoStop}%
\bibitem [{\citenamefont {van~de Meent}(2016)}]{vandeMeent:2016pee}%
  \BibitemOpen
  \bibfield  {author} {\bibinfo {author} {\bibfnamefont {Maarten}\ \bibnamefont
  {van~de Meent}},\ }\href {\doibase10.1103/PhysRevD.94.044034} {\bibfield
  {journal} {\bibinfo  {journal} {Phys. Rev.}\ }\textbf {\bibinfo {volume}
  {D94}},\ \bibinfo {pages} {044034} (\bibinfo {year} {2016})},\ \Eprint
  {http://arxiv.org/abs/1606.06297} {arXiv:1606.06297 [gr-qc]}\BibitemShut
  {NoStop}%
\bibitem [{\citenamefont {{Leaver}}(1986)}]{Leaver:1986JMP}%
  \BibitemOpen
  \bibfield  {author} {\bibinfo {author} {\bibfnamefont {E.~W.}\ \bibnamefont
  {{Leaver}}},\ }\href {\doibase10.1063/1.527130} {\bibfield  {journal}
  {\bibinfo  {journal} {Journal of Mathematical Physics}\ }\textbf {\bibinfo
  {volume} {27}},\ \bibinfo {pages} {1238--1265} (\bibinfo {year}
  {1986})}\BibitemShut {NoStop}%
\bibitem [{\citenamefont {Mano}\ \emph {et~al.}(1996)\citenamefont {Mano},
  \citenamefont {Suzuki},\ and\ \citenamefont {Takasugi}}]{Mano:1996vt}%
  \BibitemOpen
  \bibfield  {author} {\bibinfo {author} {\bibfnamefont {Shuhei}\ \bibnamefont
  {Mano}}, \bibinfo {author} {\bibfnamefont {Hisao}\ \bibnamefont {Suzuki}}, \
  and\ \bibinfo {author} {\bibfnamefont {Eiichi}\ \bibnamefont {Takasugi}},\
  }\href {\doibase10.1143/PTP.95.1079} {\bibfield  {journal} {\bibinfo
  {journal} {Prog. Theor. Phys.}\ }\textbf {\bibinfo {volume} {95}},\ \bibinfo
  {pages} {1079--1096} (\bibinfo {year} {1996})},\ \Eprint
  {http://arxiv.org/abs/gr-qc/9603020} {arXiv:gr-qc/9603020
  [gr-qc]}\BibitemShut {NoStop}%
\bibitem [{\citenamefont {Sasaki}\ and\ \citenamefont
  {Tagoshi}(2003)}]{Sasaki:2003xr}%
  \BibitemOpen
  \bibfield  {author} {\bibinfo {author} {\bibfnamefont {Misao}\ \bibnamefont
  {Sasaki}}\ and\ \bibinfo {author} {\bibfnamefont {Hideyuki}\ \bibnamefont
  {Tagoshi}},\ }\href {\doibase10.12942/lrr-2003-6} {\bibfield  {journal}
  {\bibinfo  {journal} {Living Rev. Rel.}\ }\textbf {\bibinfo {volume} {6}},\
  \bibinfo {pages} {6} (\bibinfo {year} {2003})},\ \Eprint
  {http://arxiv.org/abs/gr-qc/0306120} {arXiv:gr-qc/0306120
  [gr-qc]}\BibitemShut {NoStop}%
\bibitem [{\citenamefont {Shah}\ \emph {et~al.}(2014)\citenamefont {Shah},
  \citenamefont {Friedman},\ and\ \citenamefont {Whiting}}]{Shah:2013uya}%
  \BibitemOpen
  \bibfield  {author} {\bibinfo {author} {\bibfnamefont {Abhay~G}\ \bibnamefont
  {Shah}}, \bibinfo {author} {\bibfnamefont {John~L}\ \bibnamefont {Friedman}},
  \ and\ \bibinfo {author} {\bibfnamefont {Bernard~F}\ \bibnamefont
  {Whiting}},\ }\href {\doibase10.1103/PhysRevD.89.064042} {\bibfield
  {journal} {\bibinfo  {journal} {Phys. Rev.}\ }\textbf {\bibinfo {volume}
  {D89}},\ \bibinfo {pages} {064042} (\bibinfo {year} {2014})},\ \Eprint
  {http://arxiv.org/abs/1312.1952} {arXiv:1312.1952 [gr-qc]}\BibitemShut
  {NoStop}%
\bibitem [{\citenamefont {Shah}(2014)}]{Shah:2014tka}%
  \BibitemOpen
  \bibfield  {author} {\bibinfo {author} {\bibfnamefont {Abhay~G.}\
  \bibnamefont {Shah}},\ }\href {\doibase10.1103/PhysRevD.90.044025} {\bibfield
   {journal} {\bibinfo  {journal} {Phys. Rev.}\ }\textbf {\bibinfo {volume}
  {D90}},\ \bibinfo {pages} {044025} (\bibinfo {year} {2014})},\ \Eprint
  {http://arxiv.org/abs/1403.2697} {arXiv:1403.2697 [gr-qc]}\BibitemShut
  {NoStop}%
\bibitem [{\citenamefont {Johnson-McDaniel}\ \emph {et~al.}(2015)\citenamefont
  {Johnson-McDaniel}, \citenamefont {Shah},\ and\ \citenamefont
  {Whiting}}]{Johnson-McDaniel:2015vva}%
  \BibitemOpen
  \bibfield  {author} {\bibinfo {author} {\bibfnamefont {Nathan~K.}\
  \bibnamefont {Johnson-McDaniel}}, \bibinfo {author} {\bibfnamefont
  {Abhay~G.}\ \bibnamefont {Shah}}, \ and\ \bibinfo {author} {\bibfnamefont
  {Bernard~F.}\ \bibnamefont {Whiting}},\ }\href
  {\doibase10.1103/PhysRevD.92.044007} {\bibfield  {journal} {\bibinfo
  {journal} {Phys. Rev.}\ }\textbf {\bibinfo {volume} {D92}},\ \bibinfo {pages}
  {044007} (\bibinfo {year} {2015})},\ \Eprint
  {http://arxiv.org/abs/1503.02638} {arXiv:1503.02638 [gr-qc]}\BibitemShut
  {NoStop}%
\bibitem [{\citenamefont {Bini}\ and\ \citenamefont
  {Damour}(2015{\natexlab{a}})}]{Bini:2015bla}%
  \BibitemOpen
  \bibfield  {author} {\bibinfo {author} {\bibfnamefont {Donato}\ \bibnamefont
  {Bini}}\ and\ \bibinfo {author} {\bibfnamefont {Thibault}\ \bibnamefont
  {Damour}},\ }\href {\doibase10.1103/PhysRevD.91.064050} {\bibfield  {journal}
  {\bibinfo  {journal} {Phys. Rev.}\ }\textbf {\bibinfo {volume} {D91}},\
  \bibinfo {pages} {064050} (\bibinfo {year} {2015}{\natexlab{a}})},\ \Eprint
  {http://arxiv.org/abs/1502.02450} {arXiv:1502.02450 [gr-qc]}\BibitemShut
  {NoStop}%
\bibitem [{\citenamefont {Bini}\ and\ \citenamefont
  {Damour}(2015{\natexlab{b}})}]{Bini:2015mza}%
  \BibitemOpen
  \bibfield  {author} {\bibinfo {author} {\bibfnamefont {Donato}\ \bibnamefont
  {Bini}}\ and\ \bibinfo {author} {\bibfnamefont {Thibault}\ \bibnamefont
  {Damour}},\ }\href {\doibase10.1103/PhysRevD.91.064064} {\bibfield  {journal}
  {\bibinfo  {journal} {Phys. Rev.}\ }\textbf {\bibinfo {volume} {D91}},\
  \bibinfo {pages} {064064} (\bibinfo {year} {2015}{\natexlab{b}})},\ \Eprint
  {http://arxiv.org/abs/1503.01272} {arXiv:1503.01272 [gr-qc]}\BibitemShut
  {NoStop}%
\bibitem [{\citenamefont {Porfyriadis}\ and\ \citenamefont
  {Strominger}(2014)}]{Porfyriadis:2014fja}%
  \BibitemOpen
  \bibfield  {author} {\bibinfo {author} {\bibfnamefont {Achilleas~P.}\
  \bibnamefont {Porfyriadis}}\ and\ \bibinfo {author} {\bibfnamefont {Andrew}\
  \bibnamefont {Strominger}},\ }\href {\doibase10.1103/PhysRevD.90.044038}
  {\bibfield  {journal} {\bibinfo  {journal} {Phys. Rev.}\ }\textbf {\bibinfo
  {volume} {D90}},\ \bibinfo {pages} {044038} (\bibinfo {year} {2014})},\
  \Eprint {http://arxiv.org/abs/1401.3746} {arXiv:1401.3746
  [hep-th]}\BibitemShut {NoStop}%
\bibitem [{\citenamefont {Hadar}\ \emph {et~al.}(2014)\citenamefont {Hadar},
  \citenamefont {Porfyriadis},\ and\ \citenamefont
  {Strominger}}]{Hadar:2014dpa}%
  \BibitemOpen
  \bibfield  {author} {\bibinfo {author} {\bibfnamefont {Shahar}\ \bibnamefont
  {Hadar}}, \bibinfo {author} {\bibfnamefont {Achilleas~P.}\ \bibnamefont
  {Porfyriadis}}, \ and\ \bibinfo {author} {\bibfnamefont {Andrew}\
  \bibnamefont {Strominger}},\ }\href {\doibase10.1103/PhysRevD.90.064045}
  {\bibfield  {journal} {\bibinfo  {journal} {Phys. Rev.}\ }\textbf {\bibinfo
  {volume} {D90}},\ \bibinfo {pages} {064045} (\bibinfo {year} {2014})},\
  \Eprint {http://arxiv.org/abs/1403.2797} {arXiv:1403.2797
  [hep-th]}\BibitemShut {NoStop}%
\bibitem [{\citenamefont {Hadar}\ \emph {et~al.}(2015)\citenamefont {Hadar},
  \citenamefont {Porfyriadis},\ and\ \citenamefont
  {Strominger}}]{Hadar:2015xpa}%
  \BibitemOpen
  \bibfield  {author} {\bibinfo {author} {\bibfnamefont {Shahar}\ \bibnamefont
  {Hadar}}, \bibinfo {author} {\bibfnamefont {Achilleas~P.}\ \bibnamefont
  {Porfyriadis}}, \ and\ \bibinfo {author} {\bibfnamefont {Andrew}\
  \bibnamefont {Strominger}},\ }\href {\doibase10.1007/JHEP07(2015)078}
  {\bibfield  {journal} {\bibinfo  {journal} {JHEP}\ }\textbf {\bibinfo
  {volume} {07}},\ \bibinfo {pages} {078} (\bibinfo {year} {2015})},\ \Eprint
  {http://arxiv.org/abs/1504.07650} {arXiv:1504.07650 [hep-th]}\BibitemShut
  {NoStop}%
\bibitem [{\citenamefont {Hadar}\ and\ \citenamefont
  {Porfyriadis}(2017)}]{Hadar:2016vmk}%
  \BibitemOpen
  \bibfield  {author} {\bibinfo {author} {\bibfnamefont {Shahar}\ \bibnamefont
  {Hadar}}\ and\ \bibinfo {author} {\bibfnamefont {Achilleas~P.}\ \bibnamefont
  {Porfyriadis}},\ }\href {\doibase10.1007/JHEP03(2017)014} {\bibfield
  {journal} {\bibinfo  {journal} {JHEP}\ }\textbf {\bibinfo {volume} {03}},\
  \bibinfo {pages} {014} (\bibinfo {year} {2017})},\ \Eprint
  {http://arxiv.org/abs/1611.09834} {arXiv:1611.09834 [hep-th]}\BibitemShut
  {NoStop}%
\bibitem [{\citenamefont {Comp\`ere}\ \emph {et~al.}(2018)\citenamefont
  {Comp\`ere}, \citenamefont {Fransen}, \citenamefont {Hertog},\ and\
  \citenamefont {Long}}]{Compere:2017hsi}%
  \BibitemOpen
  \bibfield  {author} {\bibinfo {author} {\bibfnamefont {Geoffrey}\
  \bibnamefont {Comp\`ere}}, \bibinfo {author} {\bibfnamefont {Kwinten}\
  \bibnamefont {Fransen}}, \bibinfo {author} {\bibfnamefont {Thomas}\
  \bibnamefont {Hertog}}, \ and\ \bibinfo {author} {\bibfnamefont {Jiang}\
  \bibnamefont {Long}},\ }\href {\doibase10.1088/1361-6382/aab99e} {\bibfield
  {journal} {\bibinfo  {journal} {Class. Quant. Grav.}\ }\textbf {\bibinfo
  {volume} {35}},\ \bibinfo {pages} {104002} (\bibinfo {year} {2018})},\
  \Eprint {http://arxiv.org/abs/1712.07130} {arXiv:1712.07130
  [gr-qc]}\BibitemShut {NoStop}%
\bibitem [{\citenamefont {Jacobson}\ and\ \citenamefont
  {Sotiriou}(2009)}]{Jacobson:2009kt}%
  \BibitemOpen
  \bibfield  {author} {\bibinfo {author} {\bibfnamefont {Ted}\ \bibnamefont
  {Jacobson}}\ and\ \bibinfo {author} {\bibfnamefont {Thomas~P.}\ \bibnamefont
  {Sotiriou}},\ }\href {\doibase10.1103/PhysRevLett.103.141101} {\bibfield
  {journal} {\bibinfo  {journal} {Phys. Rev. Lett.}\ }\textbf {\bibinfo
  {volume} {103}},\ \bibinfo {pages} {141101} (\bibinfo {year} {2009})},\
  \bibinfo {note} {[Erratum: Phys. Rev. Lett.103,209903(2009)]},\ \Eprint
  {http://arxiv.org/abs/0907.4146} {arXiv:0907.4146 [gr-qc]}\BibitemShut
  {NoStop}%
\bibitem [{\citenamefont {Hubeny}(1999)}]{Hubeny:1998ga}%
  \BibitemOpen
  \bibfield  {author} {\bibinfo {author} {\bibfnamefont {Veronika~E.}\
  \bibnamefont {Hubeny}},\ }\href {\doibase10.1103/PhysRevD.59.064013}
  {\bibfield  {journal} {\bibinfo  {journal} {Phys. Rev.}\ }\textbf {\bibinfo
  {volume} {D59}},\ \bibinfo {pages} {064013} (\bibinfo {year} {1999})},\
  \Eprint {http://arxiv.org/abs/gr-qc/9808043} {arXiv:gr-qc/9808043
  [gr-qc]}\BibitemShut {NoStop}%
\bibitem [{\citenamefont {Saa}\ and\ \citenamefont
  {Santarelli}(2011)}]{Saa:2011wq}%
  \BibitemOpen
  \bibfield  {author} {\bibinfo {author} {\bibfnamefont {Alberto}\ \bibnamefont
  {Saa}}\ and\ \bibinfo {author} {\bibfnamefont {Raphael}\ \bibnamefont
  {Santarelli}},\ }\href {\doibase10.1103/PhysRevD.84.027501} {\bibfield
  {journal} {\bibinfo  {journal} {Phys. Rev.}\ }\textbf {\bibinfo {volume}
  {D84}},\ \bibinfo {pages} {027501} (\bibinfo {year} {2011})},\ \Eprint
  {http://arxiv.org/abs/1105.3950} {arXiv:1105.3950 [gr-qc]}\BibitemShut
  {NoStop}%
\bibitem [{\citenamefont {Gao}\ and\ \citenamefont {Zhang}(2013)}]{Gao:2012ca}%
  \BibitemOpen
  \bibfield  {author} {\bibinfo {author} {\bibfnamefont {Sijie}\ \bibnamefont
  {Gao}}\ and\ \bibinfo {author} {\bibfnamefont {Yuan}\ \bibnamefont {Zhang}},\
  }\href {\doibase10.1103/PhysRevD.87.044028} {\bibfield  {journal} {\bibinfo
  {journal} {Phys. Rev.}\ }\textbf {\bibinfo {volume} {D87}},\ \bibinfo {pages}
  {044028} (\bibinfo {year} {2013})},\ \Eprint {http://arxiv.org/abs/1211.2631}
  {arXiv:1211.2631 [gr-qc]}\BibitemShut {NoStop}%
\bibitem [{\citenamefont {Barausse}\ \emph {et~al.}(2010)\citenamefont
  {Barausse}, \citenamefont {Cardoso},\ and\ \citenamefont
  {Khanna}}]{Barausse:2010ka}%
  \BibitemOpen
  \bibfield  {author} {\bibinfo {author} {\bibfnamefont {Enrico}\ \bibnamefont
  {Barausse}}, \bibinfo {author} {\bibfnamefont {Vitor}\ \bibnamefont
  {Cardoso}}, \ and\ \bibinfo {author} {\bibfnamefont {Gaurav}\ \bibnamefont
  {Khanna}},\ }\href {\doibase10.1103/PhysRevLett.105.261102} {\bibfield
  {journal} {\bibinfo  {journal} {Phys. Rev. Lett.}\ }\textbf {\bibinfo
  {volume} {105}},\ \bibinfo {pages} {261102} (\bibinfo {year} {2010})},\
  \Eprint {http://arxiv.org/abs/1008.5159} {arXiv:1008.5159
  [gr-qc]}\BibitemShut {NoStop}%
\bibitem [{\citenamefont {Zimmerman}\ \emph {et~al.}(2013)\citenamefont
  {Zimmerman}, \citenamefont {Vega}, \citenamefont {Poisson},\ and\
  \citenamefont {Haas}}]{Zimmerman:2012zu}%
  \BibitemOpen
  \bibfield  {author} {\bibinfo {author} {\bibfnamefont {Peter}\ \bibnamefont
  {Zimmerman}}, \bibinfo {author} {\bibfnamefont {Ian}\ \bibnamefont {Vega}},
  \bibinfo {author} {\bibfnamefont {Eric}\ \bibnamefont {Poisson}}, \ and\
  \bibinfo {author} {\bibfnamefont {Roland}\ \bibnamefont {Haas}},\ }\href
  {\doibase10.1103/PhysRevD.87.041501} {\bibfield  {journal} {\bibinfo
  {journal} {Phys. Rev.}\ }\textbf {\bibinfo {volume} {D87}},\ \bibinfo {pages}
  {041501} (\bibinfo {year} {2013})},\ \Eprint {http://arxiv.org/abs/1211.3889}
  {arXiv:1211.3889 [gr-qc]}\BibitemShut {NoStop}%
\bibitem [{\citenamefont {Revelar}\ and\ \citenamefont
  {Vega}(2017)}]{Revelar:2017sem}%
  \BibitemOpen
  \bibfield  {author} {\bibinfo {author} {\bibfnamefont {Karl~Simon}\
  \bibnamefont {Revelar}}\ and\ \bibinfo {author} {\bibfnamefont {Ian}\
  \bibnamefont {Vega}},\ }\href {\doibase10.1103/PhysRevD.96.064010} {\bibfield
   {journal} {\bibinfo  {journal} {Phys. Rev.}\ }\textbf {\bibinfo {volume}
  {D96}},\ \bibinfo {pages} {064010} (\bibinfo {year} {2017})},\ \Eprint
  {http://arxiv.org/abs/1706.07190} {arXiv:1706.07190 [gr-qc]}\BibitemShut
  {NoStop}%
\bibitem [{\citenamefont {Colleoni}\ and\ \citenamefont
  {Barack}(2015)}]{Colleoni:2015afa}%
  \BibitemOpen
  \bibfield  {author} {\bibinfo {author} {\bibfnamefont {Marta}\ \bibnamefont
  {Colleoni}}\ and\ \bibinfo {author} {\bibfnamefont {Leor}\ \bibnamefont
  {Barack}},\ }\href {\doibase10.1103/PhysRevD.91.104024} {\bibfield  {journal}
  {\bibinfo  {journal} {Phys. Rev.}\ }\textbf {\bibinfo {volume} {D91}},\
  \bibinfo {pages} {104024} (\bibinfo {year} {2015})},\ \Eprint
  {http://arxiv.org/abs/1501.07330} {arXiv:1501.07330 [gr-qc]}\BibitemShut
  {NoStop}%
\bibitem [{\citenamefont {Colleoni}\ \emph {et~al.}(2015)\citenamefont
  {Colleoni}, \citenamefont {Barack}, \citenamefont {Shah},\ and\ \citenamefont
  {van~de Meent}}]{Colleoni:2015ena}%
  \BibitemOpen
  \bibfield  {author} {\bibinfo {author} {\bibfnamefont {Marta}\ \bibnamefont
  {Colleoni}}, \bibinfo {author} {\bibfnamefont {Leor}\ \bibnamefont {Barack}},
  \bibinfo {author} {\bibfnamefont {Abhay~G.}\ \bibnamefont {Shah}}, \ and\
  \bibinfo {author} {\bibfnamefont {Maarten}\ \bibnamefont {van~de Meent}},\
  }\href {\doibase10.1103/PhysRevD.92.084044} {\bibfield  {journal} {\bibinfo
  {journal} {Phys. Rev.}\ }\textbf {\bibinfo {volume} {D92}},\ \bibinfo {pages}
  {084044} (\bibinfo {year} {2015})},\ \Eprint
  {http://arxiv.org/abs/1508.04031} {arXiv:1508.04031 [gr-qc]}\BibitemShut
  {NoStop}%
\bibitem [{\citenamefont {Sorce}\ and\ \citenamefont
  {Wald}(2017)}]{Sorce:2017dst}%
  \BibitemOpen
  \bibfield  {author} {\bibinfo {author} {\bibfnamefont {Jonathan}\
  \bibnamefont {Sorce}}\ and\ \bibinfo {author} {\bibfnamefont {Robert~M.}\
  \bibnamefont {Wald}},\ }\href {\doibase10.1103/PhysRevD.96.104014} {\bibfield
   {journal} {\bibinfo  {journal} {Phys. Rev.}\ }\textbf {\bibinfo {volume}
  {D96}},\ \bibinfo {pages} {104014} (\bibinfo {year} {2017})},\ \Eprint
  {http://arxiv.org/abs/1707.05862} {arXiv:1707.05862 [gr-qc]}\BibitemShut
  {NoStop}%
\bibitem [{\citenamefont {Warburton}\ \emph {et~al.}(2017)\citenamefont
  {Warburton}, \citenamefont {Osburn},\ and\ \citenamefont
  {Evans}}]{Warburton:2017sxk}%
  \BibitemOpen
  \bibfield  {author} {\bibinfo {author} {\bibfnamefont {Niels}\ \bibnamefont
  {Warburton}}, \bibinfo {author} {\bibfnamefont {Thomas}\ \bibnamefont
  {Osburn}}, \ and\ \bibinfo {author} {\bibfnamefont {Charles~R.}\ \bibnamefont
  {Evans}},\ }\href {\doibase10.1103/PhysRevD.96.084057} {\bibfield  {journal}
  {\bibinfo  {journal} {Phys. Rev.}\ }\textbf {\bibinfo {volume} {D96}},\
  \bibinfo {pages} {084057} (\bibinfo {year} {2017})},\ \Eprint
  {http://arxiv.org/abs/1708.03720} {arXiv:1708.03720 [gr-qc]}\BibitemShut
  {NoStop}%
\bibitem [{\citenamefont {Sundararajan}\ \emph {et~al.}(2008)\citenamefont
  {Sundararajan}, \citenamefont {Khanna}, \citenamefont {Hughes},\ and\
  \citenamefont {Drasco}}]{Sundararajan:2008zm}%
  \BibitemOpen
  \bibfield  {author} {\bibinfo {author} {\bibfnamefont {Pranesh~A.}\
  \bibnamefont {Sundararajan}}, \bibinfo {author} {\bibfnamefont {Gaurav}\
  \bibnamefont {Khanna}}, \bibinfo {author} {\bibfnamefont {Scott~A.}\
  \bibnamefont {Hughes}}, \ and\ \bibinfo {author} {\bibfnamefont {Steve}\
  \bibnamefont {Drasco}},\ }\href {\doibase10.1103/PhysRevD.78.024022}
  {\bibfield  {journal} {\bibinfo  {journal} {Phys. Rev.}\ }\textbf {\bibinfo
  {volume} {D78}},\ \bibinfo {pages} {024022} (\bibinfo {year} {2008})},\
  \Eprint {http://arxiv.org/abs/0803.0317} {arXiv:0803.0317
  [gr-qc]}\BibitemShut {NoStop}%
\bibitem [{\citenamefont {Chua}\ \emph {et~al.}(2017)\citenamefont {Chua},
  \citenamefont {Moore},\ and\ \citenamefont {Gair}}]{Chua:2017ujo}%
  \BibitemOpen
  \bibfield  {author} {\bibinfo {author} {\bibfnamefont {Alvin J.~K.}\
  \bibnamefont {Chua}}, \bibinfo {author} {\bibfnamefont {Christopher~J.}\
  \bibnamefont {Moore}}, \ and\ \bibinfo {author} {\bibfnamefont {Jonathan~R.}\
  \bibnamefont {Gair}},\ }\href {\doibase10.1103/PhysRevD.96.044005} {\bibfield
   {journal} {\bibinfo  {journal} {Phys. Rev.}\ }\textbf {\bibinfo {volume}
  {D96}},\ \bibinfo {pages} {044005} (\bibinfo {year} {2017})},\ \Eprint
  {http://arxiv.org/abs/1705.04259} {arXiv:1705.04259 [gr-qc]}\BibitemShut
  {NoStop}%
\bibitem [{\citenamefont {Berry}\ and\ \citenamefont
  {Gair}(2013)}]{Berry:2012im}%
  \BibitemOpen
  \bibfield  {author} {\bibinfo {author} {\bibfnamefont {C.~P.~L.}\
  \bibnamefont {Berry}}\ and\ \bibinfo {author} {\bibfnamefont {J.~R.}\
  \bibnamefont {Gair}},\ }\href {\doibase10.1093/mnras/sts360} {\bibfield
  {journal} {\bibinfo  {journal} {Mon. Not. Roy. Astron. Soc.}\ }\textbf
  {\bibinfo {volume} {429}},\ \bibinfo {pages} {589--612} (\bibinfo {year}
  {2013})},\ \Eprint {http://arxiv.org/abs/1210.2778} {arXiv:1210.2778
  [astro-ph.HE]}\BibitemShut {NoStop}%
\bibitem [{\citenamefont {Van De~Meent}\ and\ \citenamefont
  {Warburton}(2018)}]{vandeMeent:2018rms}%
  \BibitemOpen
  \bibfield  {author} {\bibinfo {author} {\bibfnamefont {Maarten}\ \bibnamefont
  {Van De~Meent}}\ and\ \bibinfo {author} {\bibfnamefont {Niels}\ \bibnamefont
  {Warburton}},\ }\href@noop {} {\  (\bibinfo {year} {2018})},\ \Eprint
  {http://arxiv.org/abs/1802.05281} {arXiv:1802.05281 [gr-qc]}\BibitemShut
  {NoStop}%
\bibitem [{\citenamefont {Galley}\ and\ \citenamefont
  {Rothstein}(2017)}]{Galley:2016zee}%
  \BibitemOpen
  \bibfield  {author} {\bibinfo {author} {\bibfnamefont {Chad~R.}\ \bibnamefont
  {Galley}}\ and\ \bibinfo {author} {\bibfnamefont {Ira~Z.}\ \bibnamefont
  {Rothstein}},\ }\href {\doibase10.1103/PhysRevD.95.104054} {\bibfield
  {journal} {\bibinfo  {journal} {Phys. Rev.}\ }\textbf {\bibinfo {volume}
  {D95}},\ \bibinfo {pages} {104054} (\bibinfo {year} {2017})},\ \Eprint
  {http://arxiv.org/abs/1609.08268} {arXiv:1609.08268 [gr-qc]}\BibitemShut
  {NoStop}%
\bibitem [{\citenamefont {Moxon}\ and\ \citenamefont
  {Flanagan}(2018)}]{Moxon:2017ozd}%
  \BibitemOpen
  \bibfield  {author} {\bibinfo {author} {\bibfnamefont {Jordan}\ \bibnamefont
  {Moxon}}\ and\ \bibinfo {author} {\bibfnamefont {Éanna}\ \bibnamefont
  {Flanagan}},\ }\href {\doibase10.1103/PhysRevD.97.105001} {\bibfield
  {journal} {\bibinfo  {journal} {Phys. Rev.}\ }\textbf {\bibinfo {volume}
  {D97}},\ \bibinfo {pages} {105001} (\bibinfo {year} {2018})},\ \Eprint
  {http://arxiv.org/abs/1711.05212} {arXiv:1711.05212 [gr-qc]}\BibitemShut
  {NoStop}%
\bibitem [{\citenamefont {Flanagan}\ and\ \citenamefont
  {Hinderer}(2012)}]{Flanagan:2010cd}%
  \BibitemOpen
  \bibfield  {author} {\bibinfo {author} {\bibfnamefont {Eanna~E.}\
  \bibnamefont {Flanagan}}\ and\ \bibinfo {author} {\bibfnamefont {Tanja}\
  \bibnamefont {Hinderer}},\ }\href {\doibase10.1103/PhysRevLett.109.071102}
  {\bibfield  {journal} {\bibinfo  {journal} {Phys. Rev. Lett.}\ }\textbf
  {\bibinfo {volume} {109}},\ \bibinfo {pages} {071102} (\bibinfo {year}
  {2012})},\ \Eprint {http://arxiv.org/abs/1009.4923} {arXiv:1009.4923
  [gr-qc]}\BibitemShut {NoStop}%
\bibitem [{\citenamefont {Flanagan}\ \emph {et~al.}(2014)\citenamefont
  {Flanagan}, \citenamefont {Hughes},\ and\ \citenamefont
  {Ruangsri}}]{Flanagan:2012kg}%
  \BibitemOpen
  \bibfield  {author} {\bibinfo {author} {\bibfnamefont {Éanna~E.}\
  \bibnamefont {Flanagan}}, \bibinfo {author} {\bibfnamefont {Scott~A.}\
  \bibnamefont {Hughes}}, \ and\ \bibinfo {author} {\bibfnamefont {Uchupol}\
  \bibnamefont {Ruangsri}},\ }\href {\doibase10.1103/PhysRevD.89.084028}
  {\bibfield  {journal} {\bibinfo  {journal} {Phys. Rev.}\ }\textbf {\bibinfo
  {volume} {D89}},\ \bibinfo {pages} {084028} (\bibinfo {year} {2014})},\
  \Eprint {http://arxiv.org/abs/1208.3906} {arXiv:1208.3906
  [gr-qc]}\BibitemShut {NoStop}%
\bibitem [{\citenamefont {van~de
  Meent}(2014{\natexlab{a}})}]{vandeMeent:2013sza}%
  \BibitemOpen
  \bibfield  {author} {\bibinfo {author} {\bibfnamefont {Maarten}\ \bibnamefont
  {van~de Meent}},\ }\href {\doibase10.1103/PhysRevD.89.084033} {\bibfield
  {journal} {\bibinfo  {journal} {Phys. Rev.}\ }\textbf {\bibinfo {volume}
  {D89}},\ \bibinfo {pages} {084033} (\bibinfo {year} {2014}{\natexlab{a}})},\
  \Eprint {http://arxiv.org/abs/1311.4457} {arXiv:1311.4457
  [gr-qc]}\BibitemShut {NoStop}%
\bibitem [{\citenamefont {van~de
  Meent}(2014{\natexlab{b}})}]{vandeMeent:2014raa}%
  \BibitemOpen
  \bibfield  {author} {\bibinfo {author} {\bibfnamefont {Maarten}\ \bibnamefont
  {van~de Meent}},\ }\href {\doibase10.1103/PhysRevD.90.044027} {\bibfield
  {journal} {\bibinfo  {journal} {Phys. Rev.}\ }\textbf {\bibinfo {volume}
  {D90}},\ \bibinfo {pages} {044027} (\bibinfo {year} {2014}{\natexlab{b}})},\
  \Eprint {http://arxiv.org/abs/1406.2594} {arXiv:1406.2594
  [gr-qc]}\BibitemShut {NoStop}%
\bibitem [{\citenamefont {Brink}\ \emph {et~al.}(2015)\citenamefont {Brink},
  \citenamefont {Geyer},\ and\ \citenamefont {Hinderer}}]{Brink:2013nna}%
  \BibitemOpen
  \bibfield  {author} {\bibinfo {author} {\bibfnamefont {Jeandrew}\
  \bibnamefont {Brink}}, \bibinfo {author} {\bibfnamefont {Marisa}\
  \bibnamefont {Geyer}}, \ and\ \bibinfo {author} {\bibfnamefont {Tanja}\
  \bibnamefont {Hinderer}},\ }\href {\doibase10.1103/PhysRevLett.114.081102}
  {\bibfield  {journal} {\bibinfo  {journal} {Phys. Rev. Lett.}\ }\textbf
  {\bibinfo {volume} {114}},\ \bibinfo {pages} {081102} (\bibinfo {year}
  {2015})},\ \Eprint {http://arxiv.org/abs/1304.0330} {arXiv:1304.0330
  [gr-qc]}\BibitemShut {NoStop}%
\bibitem [{\citenamefont {Berry}\ \emph {et~al.}(2016)\citenamefont {Berry},
  \citenamefont {Cole}, \citenamefont {Cañizares},\ and\ \citenamefont
  {Gair}}]{Berry:2016bit}%
  \BibitemOpen
  \bibfield  {author} {\bibinfo {author} {\bibfnamefont {Christopher P.~L.}\
  \bibnamefont {Berry}}, \bibinfo {author} {\bibfnamefont {Robert~H.}\
  \bibnamefont {Cole}}, \bibinfo {author} {\bibfnamefont {Priscilla}\
  \bibnamefont {Cañizares}}, \ and\ \bibinfo {author} {\bibfnamefont
  {Jonathan~R.}\ \bibnamefont {Gair}},\ }\href
  {\doibase10.1103/PhysRevD.94.124042} {\bibfield  {journal} {\bibinfo
  {journal} {Phys. Rev.}\ }\textbf {\bibinfo {volume} {D94}},\ \bibinfo {pages}
  {124042} (\bibinfo {year} {2016})},\ \Eprint
  {http://arxiv.org/abs/1608.08951} {arXiv:1608.08951 [gr-qc]}\BibitemShut
  {NoStop}%
\bibitem [{\citenamefont {Berry}\ \emph {et~al.}(2017)\citenamefont {Berry},
  \citenamefont {Cole}, \citenamefont {Cañizares},\ and\ \citenamefont
  {Gair}}]{Berry:2017cty}%
  \BibitemOpen
  \bibfield  {author} {\bibinfo {author} {\bibfnamefont {C.~P.~L.}\
  \bibnamefont {Berry}}, \bibinfo {author} {\bibfnamefont {R.~H.}\ \bibnamefont
  {Cole}}, \bibinfo {author} {\bibfnamefont {P.}~\bibnamefont {Cañizares}}, \
  and\ \bibinfo {author} {\bibfnamefont {J.~R.}\ \bibnamefont {Gair}},\
  }\bibfield  {booktitle} {\emph {\bibinfo {booktitle} {{Proceedings, 11th
  International LISA Symposium: Zurich, Switzerland, September 5-9, 2016}}},\
  }\href {\doibase10.1088/1742-6596/840/1/012052} {\bibfield  {journal}
  {\bibinfo  {journal} {J. Phys. Conf. Ser.}\ }\textbf {\bibinfo {volume}
  {840}},\ \bibinfo {pages} {012052} (\bibinfo {year} {2017})},\ \Eprint
  {http://arxiv.org/abs/1702.05481} {arXiv:1702.05481 [gr-qc]}\BibitemShut
  {NoStop}%
\bibitem [{\citenamefont {Pound}(2014)}]{Pound:2014koa}%
  \BibitemOpen
  \bibfield  {author} {\bibinfo {author} {\bibfnamefont {Adam}\ \bibnamefont
  {Pound}},\ }\href {\doibase10.1103/PhysRevD.90.084039} {\bibfield  {journal}
  {\bibinfo  {journal} {Phys. Rev.}\ }\textbf {\bibinfo {volume} {D90}},\
  \bibinfo {pages} {084039} (\bibinfo {year} {2014})},\ \Eprint
  {http://arxiv.org/abs/1404.1543} {arXiv:1404.1543 [gr-qc]}\BibitemShut
  {NoStop}%
\bibitem [{\citenamefont {Pound}(2015{\natexlab{b}})}]{Pound:2015wva}%
  \BibitemOpen
  \bibfield  {author} {\bibinfo {author} {\bibfnamefont {Adam}\ \bibnamefont
  {Pound}},\ }\href {\doibase10.1103/PhysRevD.92.104047} {\bibfield  {journal}
  {\bibinfo  {journal} {Phys. Rev.}\ }\textbf {\bibinfo {volume} {D92}},\
  \bibinfo {pages} {104047} (\bibinfo {year} {2015}{\natexlab{b}})},\ \Eprint
  {http://arxiv.org/abs/1510.05172} {arXiv:1510.05172 [gr-qc]}\BibitemShut
  {NoStop}%
\bibitem [{\citenamefont {Miller}\ \emph {et~al.}(2016)\citenamefont {Miller},
  \citenamefont {Wardell},\ and\ \citenamefont {Pound}}]{Miller:2016hjv}%
  \BibitemOpen
  \bibfield  {author} {\bibinfo {author} {\bibfnamefont {Jeremy}\ \bibnamefont
  {Miller}}, \bibinfo {author} {\bibfnamefont {Barry}\ \bibnamefont {Wardell}},
  \ and\ \bibinfo {author} {\bibfnamefont {Adam}\ \bibnamefont {Pound}},\
  }\href {\doibase10.1103/PhysRevD.94.104018} {\bibfield  {journal} {\bibinfo
  {journal} {Phys. Rev.}\ }\textbf {\bibinfo {volume} {D94}},\ \bibinfo {pages}
  {104018} (\bibinfo {year} {2016})},\ \Eprint
  {http://arxiv.org/abs/1608.06783} {arXiv:1608.06783 [gr-qc]}\BibitemShut
  {NoStop}%
\bibitem [{\citenamefont {Pound}(2017)}]{Pound:2017psq}%
  \BibitemOpen
  \bibfield  {author} {\bibinfo {author} {\bibfnamefont {Adam}\ \bibnamefont
  {Pound}},\ }\href {\doibase10.1103/PhysRevD.95.104056} {\bibfield  {journal}
  {\bibinfo  {journal} {Phys. Rev.}\ }\textbf {\bibinfo {volume} {D95}},\
  \bibinfo {pages} {104056} (\bibinfo {year} {2017})},\ \Eprint
  {http://arxiv.org/abs/1703.02836} {arXiv:1703.02836 [gr-qc]}\BibitemShut
  {NoStop}%
\bibitem [{\citenamefont {Ruangsri}\ \emph {et~al.}(2016)\citenamefont
  {Ruangsri}, \citenamefont {Vigeland},\ and\ \citenamefont
  {Hughes}}]{Ruangsri:2015cvg}%
  \BibitemOpen
  \bibfield  {author} {\bibinfo {author} {\bibfnamefont {Uchupol}\ \bibnamefont
  {Ruangsri}}, \bibinfo {author} {\bibfnamefont {Sarah~J.}\ \bibnamefont
  {Vigeland}}, \ and\ \bibinfo {author} {\bibfnamefont {Scott~A.}\ \bibnamefont
  {Hughes}},\ }\href {\doibase10.1103/PhysRevD.94.044008} {\bibfield  {journal}
  {\bibinfo  {journal} {Phys. Rev.}\ }\textbf {\bibinfo {volume} {D94}},\
  \bibinfo {pages} {044008} (\bibinfo {year} {2016})},\ \Eprint
  {http://arxiv.org/abs/1512.00376} {arXiv:1512.00376 [gr-qc]}\BibitemShut
  {NoStop}%
\bibitem [{\citenamefont {Faye}\ \emph {et~al.}(2006)\citenamefont {Faye},
  \citenamefont {Blanchet},\ and\ \citenamefont {Buonanno}}]{Faye:2006gx}%
  \BibitemOpen
  \bibfield  {author} {\bibinfo {author} {\bibfnamefont {Guillaume}\
  \bibnamefont {Faye}}, \bibinfo {author} {\bibfnamefont {Luc}\ \bibnamefont
  {Blanchet}}, \ and\ \bibinfo {author} {\bibfnamefont {Alessandra}\
  \bibnamefont {Buonanno}},\ }\href {\doibase10.1103/PhysRevD.74.104033}
  {\bibfield  {journal} {\bibinfo  {journal} {Phys. Rev.}\ }\textbf {\bibinfo
  {volume} {D74}},\ \bibinfo {pages} {104033} (\bibinfo {year} {2006})},\
  \Eprint {http://arxiv.org/abs/gr-qc/0605139} {arXiv:gr-qc/0605139
  [gr-qc]}\BibitemShut {NoStop}%
\bibitem [{\citenamefont {Han}(2010)}]{Han:2010tp}%
  \BibitemOpen
  \bibfield  {author} {\bibinfo {author} {\bibfnamefont {Wen-Biao}\
  \bibnamefont {Han}},\ }\href {\doibase10.1103/PhysRevD.82.084013} {\bibfield
  {journal} {\bibinfo  {journal} {Phys. Rev.}\ }\textbf {\bibinfo {volume}
  {D82}},\ \bibinfo {pages} {084013} (\bibinfo {year} {2010})},\ \Eprint
  {http://arxiv.org/abs/1008.3324} {arXiv:1008.3324 [gr-qc]}\BibitemShut
  {NoStop}%
\bibitem [{\citenamefont {Harms}\ \emph {et~al.}(2016)\citenamefont {Harms},
  \citenamefont {Lukes-Gerakopoulos}, \citenamefont {Bernuzzi},\ and\
  \citenamefont {Nagar}}]{Harms:2016ctx}%
  \BibitemOpen
  \bibfield  {author} {\bibinfo {author} {\bibfnamefont {Enno}\ \bibnamefont
  {Harms}}, \bibinfo {author} {\bibfnamefont {Georgios}\ \bibnamefont
  {Lukes-Gerakopoulos}}, \bibinfo {author} {\bibfnamefont {Sebastiano}\
  \bibnamefont {Bernuzzi}}, \ and\ \bibinfo {author} {\bibfnamefont
  {Alessandro}\ \bibnamefont {Nagar}},\ }\href
  {\doibase10.1103/PhysRevD.94.104010} {\bibfield  {journal} {\bibinfo
  {journal} {Phys. Rev.}\ }\textbf {\bibinfo {volume} {D94}},\ \bibinfo {pages}
  {104010} (\bibinfo {year} {2016})},\ \Eprint
  {http://arxiv.org/abs/1609.00356} {arXiv:1609.00356 [gr-qc]}\BibitemShut
  {NoStop}%
\bibitem [{\citenamefont {Maia}\ \emph
  {et~al.}(2017{\natexlab{a}})\citenamefont {Maia}, \citenamefont {Galley},
  \citenamefont {Leibovich},\ and\ \citenamefont {Porto}}]{Maia:2017gxn}%
  \BibitemOpen
  \bibfield  {author} {\bibinfo {author} {\bibfnamefont {Natalia~T.}\
  \bibnamefont {Maia}}, \bibinfo {author} {\bibfnamefont {Chad~R.}\
  \bibnamefont {Galley}}, \bibinfo {author} {\bibfnamefont {Adam~K.}\
  \bibnamefont {Leibovich}}, \ and\ \bibinfo {author} {\bibfnamefont
  {Rafael~A.}\ \bibnamefont {Porto}},\ }\href
  {\doibase10.1103/PhysRevD.96.084064} {\bibfield  {journal} {\bibinfo
  {journal} {Phys. Rev.}\ }\textbf {\bibinfo {volume} {D96}},\ \bibinfo {pages}
  {084064} (\bibinfo {year} {2017}{\natexlab{a}})},\ \Eprint
  {http://arxiv.org/abs/1705.07934} {arXiv:1705.07934 [gr-qc]}\BibitemShut
  {NoStop}%
\bibitem [{\citenamefont {Maia}\ \emph
  {et~al.}(2017{\natexlab{b}})\citenamefont {Maia}, \citenamefont {Galley},
  \citenamefont {Leibovich},\ and\ \citenamefont {Porto}}]{Maia:2017yok}%
  \BibitemOpen
  \bibfield  {author} {\bibinfo {author} {\bibfnamefont {Natalia~T.}\
  \bibnamefont {Maia}}, \bibinfo {author} {\bibfnamefont {Chad~R.}\
  \bibnamefont {Galley}}, \bibinfo {author} {\bibfnamefont {Adam~K.}\
  \bibnamefont {Leibovich}}, \ and\ \bibinfo {author} {\bibfnamefont
  {Rafael~A.}\ \bibnamefont {Porto}},\ }\href
  {\doibase10.1103/PhysRevD.96.084065} {\bibfield  {journal} {\bibinfo
  {journal} {Phys. Rev.}\ }\textbf {\bibinfo {volume} {D96}},\ \bibinfo {pages}
  {084065} (\bibinfo {year} {2017}{\natexlab{b}})},\ \Eprint
  {http://arxiv.org/abs/1705.07938} {arXiv:1705.07938 [gr-qc]}\BibitemShut
  {NoStop}%
\bibitem [{\citenamefont {Lukes-Gerakopoulos}\ \emph
  {et~al.}(2017)\citenamefont {Lukes-Gerakopoulos}, \citenamefont {Harms},
  \citenamefont {Bernuzzi},\ and\ \citenamefont
  {Nagar}}]{Lukes-Gerakopoulos:2017vkj}%
  \BibitemOpen
  \bibfield  {author} {\bibinfo {author} {\bibfnamefont {Georgios}\
  \bibnamefont {Lukes-Gerakopoulos}}, \bibinfo {author} {\bibfnamefont {Enno}\
  \bibnamefont {Harms}}, \bibinfo {author} {\bibfnamefont {Sebastiano}\
  \bibnamefont {Bernuzzi}}, \ and\ \bibinfo {author} {\bibfnamefont
  {Alessandro}\ \bibnamefont {Nagar}},\ }\href
  {\doibase10.1103/PhysRevD.96.064051} {\bibfield  {journal} {\bibinfo
  {journal} {Phys. Rev.}\ }\textbf {\bibinfo {volume} {D96}},\ \bibinfo {pages}
  {064051} (\bibinfo {year} {2017})},\ \Eprint
  {http://arxiv.org/abs/1707.07537} {arXiv:1707.07537 [gr-qc]}\BibitemShut
  {NoStop}%
\bibitem [{\citenamefont {Zimmerman}(2015)}]{Zimmerman:2015hua}%
  \BibitemOpen
  \bibfield  {author} {\bibinfo {author} {\bibfnamefont {Peter}\ \bibnamefont
  {Zimmerman}},\ }\href {\doibase10.1103/PhysRevD.92.064051} {\bibfield
  {journal} {\bibinfo  {journal} {Phys. Rev.}\ }\textbf {\bibinfo {volume}
  {D92}},\ \bibinfo {pages} {064051} (\bibinfo {year} {2015})},\ \Eprint
  {http://arxiv.org/abs/1507.04076} {arXiv:1507.04076 [gr-qc]}\BibitemShut
  {NoStop}%
\bibitem [{\citenamefont {Cardoso}\ and\ \citenamefont {other}(accessed June
  2018)}]{Cardosoweb}%
  \BibitemOpen
  \bibfield  {author} {\bibinfo {author} {\bibfnamefont {Vitor}\ \bibnamefont
  {Cardoso}}\ and\ \bibinfo {author} {\bibnamefont {other}},\ }\href@noop {}
  {\enquote {\bibinfo {title} {Gravitation in técnico: files},}\ }\bibinfo
  {howpublished} {\url{https://centra.tecnico.ulisboa.pt/network/grit/files/}}
  (\bibinfo {year} {accessed June 2018})\BibitemShut {NoStop}%
\bibitem [{\citenamefont {Berti}(2017 (accessed June 2018))}]{Bertiweb}%
  \BibitemOpen
  \bibfield  {author} {\bibinfo {author} {\bibfnamefont {Emanuele}\
  \bibnamefont {Berti}},\ }\href@noop {} {\enquote {\bibinfo {title} {Ringdown
  resources},}\ }\bibinfo {howpublished}
  {\url{http://www.phy.olemiss.edu/~berti/ringdown/}} (\bibinfo {year} {2017
  (accessed June 2018)})\BibitemShut {NoStop}%
\bibitem [{\citenamefont {Chua}\ and\ \citenamefont {Gair}(2017)}]{EKS}%
  \BibitemOpen
  \bibfield  {author} {\bibinfo {author} {\bibfnamefont {Alvin}\ \bibnamefont
  {Chua}}\ and\ \bibinfo {author} {\bibfnamefont {Jonathan}\ \bibnamefont
  {Gair}},\ }\href@noop {} {\enquote {\bibinfo {title} {{EMRI Kludge Suite}},}\
  }\bibinfo {howpublished}
  {\url{https://github.com/alvincjk/EMRI\_Kludge\_Suite}} (\bibinfo {year}
  {2017})\BibitemShut {NoStop}%
\bibitem [{\citenamefont {Fujita}\ \emph
  {et~al.}(2017{\natexlab{a}})\citenamefont {Fujita}, \citenamefont {Nakano},
  \citenamefont {Sago}, \citenamefont {Sasaki},\ and\ \citenamefont
  {Tanaka}}]{BHPC}%
  \BibitemOpen
  \bibfield  {author} {\bibinfo {author} {\bibfnamefont {Ryuichi}\ \bibnamefont
  {Fujita}}, \bibinfo {author} {\bibfnamefont {Hiroyuki}\ \bibnamefont
  {Nakano}}, \bibinfo {author} {\bibfnamefont {Norichika}\ \bibnamefont
  {Sago}}, \bibinfo {author} {\bibfnamefont {Misao}\ \bibnamefont {Sasaki}}, \
  and\ \bibinfo {author} {\bibfnamefont {Takahiro}\ \bibnamefont {Tanaka}},\
  }\href@noop {} {\enquote {\bibinfo {title} {Black hole perturbation club},}\
  }\bibinfo {howpublished}
  {\url{http://www2.yukawa.kyoto-u.ac.jp/~misao.sasaki/BHPC/}} (\bibinfo {year}
  {2017}{\natexlab{a}})\BibitemShut {NoStop}%
\bibitem [{BHP(2017)}]{BHPT}%
  \BibitemOpen
  \href@noop {} {\enquote {\bibinfo {title} {Black hole perturbation
  toolkit},}\ }\bibinfo {howpublished} {\url{http://bhptoolkit.org/}} (\bibinfo
  {year} {2017})\BibitemShut {NoStop}%
\bibitem [{\citenamefont {Futamase}\ and\ \citenamefont
  {Itoh}(2007)}]{FuIt.07}%
  \BibitemOpen
  \bibfield  {author} {\bibinfo {author} {\bibfnamefont {T.}~\bibnamefont
  {Futamase}}\ and\ \bibinfo {author} {\bibfnamefont {Y.}~\bibnamefont
  {Itoh}},\ }\href {\doibase10.12942/lrr-2007-2} {\bibfield  {journal}
  {\bibinfo  {journal} {Living Rev. Relativity}\ }\textbf {\bibinfo {volume}
  {10}},\ \bibinfo {pages} {2} (\bibinfo {year} {2007})}\BibitemShut {NoStop}%
\bibitem [{\citenamefont {Blanchet}(2011)}]{Blanchet:2011wga}%
  \BibitemOpen
  \bibfield  {author} {\bibinfo {author} {\bibfnamefont {Luc}\ \bibnamefont
  {Blanchet}},\ }\bibfield  {booktitle} {\emph {\bibinfo {booktitle} {{Mass and
  motion in general relativity. Proceedings, School on Mass, Orleans, France,
  June 23-25, 2008}}},\ }\href {\doibase10.1007/978-90-481-3015-3_5} {\bibfield
   {journal} {\bibinfo  {journal} {Fundam. Theor. Phys.}\ }\textbf {\bibinfo
  {volume} {162}},\ \bibinfo {pages} {125--166} (\bibinfo {year} {2011})},\
  \bibinfo {note} {[,125(2009)]},\ \Eprint {http://arxiv.org/abs/0907.3596}
  {arXiv:0907.3596 [gr-qc]}\BibitemShut {NoStop}%
\bibitem [{\citenamefont {Sch{\"a}fer}(2010)}]{Schafer:2009dq}%
  \BibitemOpen
  \bibfield  {author} {\bibinfo {author} {\bibfnamefont {Gerhard}\ \bibnamefont
  {Sch{\"a}fer}},\ }\enquote {\bibinfo {title} {{Post-Newtonian methods:
  Analytic results on the binary problem}},}\ in\ \href@noop {} {\emph
  {\bibinfo {booktitle} {Mass and Motion in General Relativity (Fundamental
  Theories of Physics, Vol. 162)}}}\ (\bibinfo  {publisher} {Springer},\
  \bibinfo {address} {New York},\ \bibinfo {year} {2010})\ Chap.~\bibinfo
  {chapter} {6},\ \Eprint {http://arxiv.org/abs/0910.2857} {arXiv:0910.2857
  [gr-qc]}\BibitemShut {NoStop}%
\bibitem [{\citenamefont {Foffa}\ and\ \citenamefont
  {Sturani}(2014)}]{Foffa:2013qca}%
  \BibitemOpen
  \bibfield  {author} {\bibinfo {author} {\bibfnamefont {Stefano}\ \bibnamefont
  {Foffa}}\ and\ \bibinfo {author} {\bibfnamefont {Riccardo}\ \bibnamefont
  {Sturani}},\ }\href {\doibase10.1088/0264-9381/31/4/043001} {\bibfield
  {journal} {\bibinfo  {journal} {Class. Quant. Grav.}\ }\textbf {\bibinfo
  {volume} {31}},\ \bibinfo {pages} {043001} (\bibinfo {year} {2014})},\
  \Eprint {http://arxiv.org/abs/1309.3474} {arXiv:1309.3474
  [gr-qc]}\BibitemShut {NoStop}%
\bibitem [{\citenamefont {Rothstein}(2014)}]{Rothstein:2014sra}%
  \BibitemOpen
  \bibfield  {author} {\bibinfo {author} {\bibfnamefont {Ira~Z.}\ \bibnamefont
  {Rothstein}},\ }\href {\doibase10.1007/s10714-014-1726-y} {\bibfield
  {journal} {\bibinfo  {journal} {Gen. Rel. Grav.}\ }\textbf {\bibinfo {volume}
  {46}},\ \bibinfo {pages} {1726} (\bibinfo {year} {2014})}\BibitemShut
  {NoStop}%
\bibitem [{\citenamefont {Blanchet}(2014)}]{Blanchet:2013haa}%
  \BibitemOpen
  \bibfield  {author} {\bibinfo {author} {\bibfnamefont {Luc}\ \bibnamefont
  {Blanchet}},\ }\href {\doibase10.12942/lrr-2014-2} {\bibfield  {journal}
  {\bibinfo  {journal} {Living Rev. Rel.}\ }\textbf {\bibinfo {volume} {17}},\
  \bibinfo {pages} {2} (\bibinfo {year} {2014})},\ \Eprint
  {http://arxiv.org/abs/1310.1528} {arXiv:1310.1528 [gr-qc]}\BibitemShut
  {NoStop}%
\bibitem [{\citenamefont {Porto}(2016{\natexlab{a}})}]{Porto:2016pyg}%
  \BibitemOpen
  \bibfield  {author} {\bibinfo {author} {\bibfnamefont {Rafael~A.}\
  \bibnamefont {Porto}},\ }\href {\doibase10.1016/j.physrep.2016.04.003}
  {\bibfield  {journal} {\bibinfo  {journal} {Phys. Rept.}\ }\textbf {\bibinfo
  {volume} {633}},\ \bibinfo {pages} {1--104} (\bibinfo {year}
  {2016}{\natexlab{a}})},\ \Eprint {http://arxiv.org/abs/1601.04914}
  {arXiv:1601.04914 [hep-th]}\BibitemShut {NoStop}%
\bibitem [{\citenamefont {Schäfer}\ and\ \citenamefont
  {Jaranowski}(2018)}]{Schafer:2018kuf}%
  \BibitemOpen
  \bibfield  {author} {\bibinfo {author} {\bibfnamefont {Gerhard}\ \bibnamefont
  {Schäfer}}\ and\ \bibinfo {author} {\bibfnamefont {Piotr}\ \bibnamefont
  {Jaranowski}},\ }\href@noop {} {\  (\bibinfo {year} {2018})},\ \Eprint
  {http://arxiv.org/abs/1805.07240} {arXiv:1805.07240 [gr-qc]}\BibitemShut
  {NoStop}%
\bibitem [{\citenamefont {Poisson}\ and\ \citenamefont
  {Will}(2014)}]{PoissonWill}%
  \BibitemOpen
  \bibfield  {author} {\bibinfo {author} {\bibfnamefont {E.}~\bibnamefont
  {Poisson}}\ and\ \bibinfo {author} {\bibfnamefont {C.M.}\ \bibnamefont
  {Will}},\ }\href@noop {} {\emph {\bibinfo {title} {{Gravity: Newtonian,
  Post-Newtonian, Relativistic}}}}\ (\bibinfo  {publisher} {Cambridge
  University Press},\ \bibinfo {address} {Cambridge, UK},\ \bibinfo {year}
  {2014})\BibitemShut {NoStop}%
\bibitem [{\citenamefont {Blanchet}\ \emph {et~al.}(1998)\citenamefont
  {Blanchet}, \citenamefont {Faye},\ and\ \citenamefont
  {Ponsot}}]{Blanchet:1998vx}%
  \BibitemOpen
  \bibfield  {author} {\bibinfo {author} {\bibfnamefont {Luc}\ \bibnamefont
  {Blanchet}}, \bibinfo {author} {\bibfnamefont {Guillaume}\ \bibnamefont
  {Faye}}, \ and\ \bibinfo {author} {\bibfnamefont {Benedicte}\ \bibnamefont
  {Ponsot}},\ }\href {\doibase10.1103/PhysRevD.58.124002} {\bibfield  {journal}
  {\bibinfo  {journal} {Phys. Rev.}\ }\textbf {\bibinfo {volume} {D58}},\
  \bibinfo {pages} {124002} (\bibinfo {year} {1998})},\ \Eprint
  {http://arxiv.org/abs/gr-qc/9804079} {arXiv:gr-qc/9804079
  [gr-qc]}\BibitemShut {NoStop}%
\bibitem [{\citenamefont {Blanchet}\ and\ \citenamefont
  {Faye}(2001)}]{Blanchet:2000ub}%
  \BibitemOpen
  \bibfield  {author} {\bibinfo {author} {\bibfnamefont {Luc}\ \bibnamefont
  {Blanchet}}\ and\ \bibinfo {author} {\bibfnamefont {Guillaume}\ \bibnamefont
  {Faye}},\ }\href {\doibase10.1103/PhysRevD.63.062005} {\bibfield  {journal}
  {\bibinfo  {journal} {Phys. Rev.}\ }\textbf {\bibinfo {volume} {D63}},\
  \bibinfo {pages} {062005} (\bibinfo {year} {2001})},\ \Eprint
  {http://arxiv.org/abs/gr-qc/0007051} {arXiv:gr-qc/0007051
  [gr-qc]}\BibitemShut {NoStop}%
\bibitem [{\citenamefont {de~Andrade}\ \emph {et~al.}(2001)\citenamefont
  {de~Andrade}, \citenamefont {Blanchet},\ and\ \citenamefont
  {Faye}}]{deAndrade:2000gf}%
  \BibitemOpen
  \bibfield  {author} {\bibinfo {author} {\bibfnamefont {Vanessa~C.}\
  \bibnamefont {de~Andrade}}, \bibinfo {author} {\bibfnamefont {Luc}\
  \bibnamefont {Blanchet}}, \ and\ \bibinfo {author} {\bibfnamefont
  {Guillaume}\ \bibnamefont {Faye}},\ }\href
  {\doibase10.1088/0264-9381/18/5/301} {\bibfield  {journal} {\bibinfo
  {journal} {Class. Quant. Grav.}\ }\textbf {\bibinfo {volume} {18}},\ \bibinfo
  {pages} {753--778} (\bibinfo {year} {2001})},\ \Eprint
  {http://arxiv.org/abs/gr-qc/0011063} {arXiv:gr-qc/0011063
  [gr-qc]}\BibitemShut {NoStop}%
\bibitem [{\citenamefont {Blanchet}\ \emph
  {et~al.}(2004{\natexlab{a}})\citenamefont {Blanchet}, \citenamefont
  {Damour},\ and\ \citenamefont {Esposito-Farese}}]{Blanchet:2003gy}%
  \BibitemOpen
  \bibfield  {author} {\bibinfo {author} {\bibfnamefont {Luc}\ \bibnamefont
  {Blanchet}}, \bibinfo {author} {\bibfnamefont {Thibault}\ \bibnamefont
  {Damour}}, \ and\ \bibinfo {author} {\bibfnamefont {Gilles}\ \bibnamefont
  {Esposito-Farese}},\ }\href {\doibase10.1103/PhysRevD.69.124007} {\bibfield
  {journal} {\bibinfo  {journal} {Phys. Rev.}\ }\textbf {\bibinfo {volume}
  {D69}},\ \bibinfo {pages} {124007} (\bibinfo {year} {2004}{\natexlab{a}})},\
  \Eprint {http://arxiv.org/abs/gr-qc/0311052} {arXiv:gr-qc/0311052
  [gr-qc]}\BibitemShut {NoStop}%
\bibitem [{\citenamefont {Mitchell}\ and\ \citenamefont
  {Will}(2007)}]{Mitchell:2007ea}%
  \BibitemOpen
  \bibfield  {author} {\bibinfo {author} {\bibfnamefont {Thomas}\ \bibnamefont
  {Mitchell}}\ and\ \bibinfo {author} {\bibfnamefont {Clifford~M.}\
  \bibnamefont {Will}},\ }\href {\doibase10.1103/PhysRevD.75.124025} {\bibfield
   {journal} {\bibinfo  {journal} {Phys. Rev.}\ }\textbf {\bibinfo {volume}
  {D75}},\ \bibinfo {pages} {124025} (\bibinfo {year} {2007})},\ \Eprint
  {http://arxiv.org/abs/0704.2243} {arXiv:0704.2243 [gr-qc]}\BibitemShut
  {NoStop}%
\bibitem [{\citenamefont {Jaranowski}\ and\ \citenamefont
  {Schaefer}(1998)}]{Jaranowski:1997ky}%
  \BibitemOpen
  \bibfield  {author} {\bibinfo {author} {\bibfnamefont {Piotr}\ \bibnamefont
  {Jaranowski}}\ and\ \bibinfo {author} {\bibfnamefont {Gerhard}\ \bibnamefont
  {Schaefer}},\ }\href {\doibase10.1103/PhysRevD.57.7274} {\bibfield  {journal}
  {\bibinfo  {journal} {Phys. Rev.}\ }\textbf {\bibinfo {volume} {D57}},\
  \bibinfo {pages} {7274--7291} (\bibinfo {year} {1998})},\ \bibinfo {note}
  {[Erratum: Phys. Rev.D63,029902(2001)]},\ \Eprint
  {http://arxiv.org/abs/gr-qc/9712075} {arXiv:gr-qc/9712075
  [gr-qc]}\BibitemShut {NoStop}%
\bibitem [{\citenamefont {Damour}\ \emph {et~al.}(2000)\citenamefont {Damour},
  \citenamefont {Jaranowski},\ and\ \citenamefont {Schaefer}}]{Damour:2000kk}%
  \BibitemOpen
  \bibfield  {author} {\bibinfo {author} {\bibfnamefont {Thibault}\
  \bibnamefont {Damour}}, \bibinfo {author} {\bibfnamefont {Piotr}\
  \bibnamefont {Jaranowski}}, \ and\ \bibinfo {author} {\bibfnamefont
  {Gerhard}\ \bibnamefont {Schaefer}},\ }\href
  {\doibase10.1103/PhysRevD.62.021501} {\bibfield  {journal} {\bibinfo
  {journal} {Phys. Rev.}\ }\textbf {\bibinfo {volume} {D62}},\ \bibinfo {pages}
  {021501} (\bibinfo {year} {2000})},\ \bibinfo {note} {[Erratum: Phys.
  Rev.D63,029903(2001)]},\ \Eprint {http://arxiv.org/abs/gr-qc/0003051}
  {arXiv:gr-qc/0003051 [gr-qc]}\BibitemShut {NoStop}%
\bibitem [{\citenamefont {Damour}\ \emph {et~al.}(2001)\citenamefont {Damour},
  \citenamefont {Jaranowski},\ and\ \citenamefont {Schaefer}}]{Damour:2001bu}%
  \BibitemOpen
  \bibfield  {author} {\bibinfo {author} {\bibfnamefont {Thibault}\
  \bibnamefont {Damour}}, \bibinfo {author} {\bibfnamefont {Piotr}\
  \bibnamefont {Jaranowski}}, \ and\ \bibinfo {author} {\bibfnamefont
  {Gerhard}\ \bibnamefont {Schaefer}},\ }\href
  {\doibase10.1016/S0370-2693(01)00642-6} {\bibfield  {journal} {\bibinfo
  {journal} {Phys. Lett.}\ }\textbf {\bibinfo {volume} {B513}},\ \bibinfo
  {pages} {147--155} (\bibinfo {year} {2001})},\ \Eprint
  {http://arxiv.org/abs/gr-qc/0105038} {arXiv:gr-qc/0105038
  [gr-qc]}\BibitemShut {NoStop}%
\bibitem [{\citenamefont {Itoh}\ \emph {et~al.}(2001)\citenamefont {Itoh},
  \citenamefont {Futamase},\ and\ \citenamefont {Asada}}]{Itoh:2001np}%
  \BibitemOpen
  \bibfield  {author} {\bibinfo {author} {\bibfnamefont {Yousuke}\ \bibnamefont
  {Itoh}}, \bibinfo {author} {\bibfnamefont {Toshifumi}\ \bibnamefont
  {Futamase}}, \ and\ \bibinfo {author} {\bibfnamefont {Hideki}\ \bibnamefont
  {Asada}},\ }\href {\doibase10.1103/PhysRevD.63.064038} {\bibfield  {journal}
  {\bibinfo  {journal} {Phys. Rev.}\ }\textbf {\bibinfo {volume} {D63}},\
  \bibinfo {pages} {064038} (\bibinfo {year} {2001})},\ \Eprint
  {http://arxiv.org/abs/gr-qc/0101114} {arXiv:gr-qc/0101114
  [gr-qc]}\BibitemShut {NoStop}%
\bibitem [{\citenamefont {Itoh}\ and\ \citenamefont
  {Futamase}(2003)}]{Itoh:2003fy}%
  \BibitemOpen
  \bibfield  {author} {\bibinfo {author} {\bibfnamefont {Yousuke}\ \bibnamefont
  {Itoh}}\ and\ \bibinfo {author} {\bibfnamefont {Toshifumi}\ \bibnamefont
  {Futamase}},\ }\href {\doibase10.1103/PhysRevD.68.121501} {\bibfield
  {journal} {\bibinfo  {journal} {Phys. Rev.}\ }\textbf {\bibinfo {volume}
  {D68}},\ \bibinfo {pages} {121501} (\bibinfo {year} {2003})},\ \Eprint
  {http://arxiv.org/abs/gr-qc/0310028} {arXiv:gr-qc/0310028
  [gr-qc]}\BibitemShut {NoStop}%
\bibitem [{\citenamefont {Itoh}(2004)}]{Itoh:2003fz}%
  \BibitemOpen
  \bibfield  {author} {\bibinfo {author} {\bibfnamefont {Yousuke}\ \bibnamefont
  {Itoh}},\ }\href {\doibase10.1103/PhysRevD.69.064018} {\bibfield  {journal}
  {\bibinfo  {journal} {Phys. Rev.}\ }\textbf {\bibinfo {volume} {D69}},\
  \bibinfo {pages} {064018} (\bibinfo {year} {2004})},\ \Eprint
  {http://arxiv.org/abs/gr-qc/0310029} {arXiv:gr-qc/0310029
  [gr-qc]}\BibitemShut {NoStop}%
\bibitem [{\citenamefont {Goldberger}\ and\ \citenamefont
  {Rothstein}(2006)}]{Goldberger:2004jt}%
  \BibitemOpen
  \bibfield  {author} {\bibinfo {author} {\bibfnamefont {Walter~D.}\
  \bibnamefont {Goldberger}}\ and\ \bibinfo {author} {\bibfnamefont {Ira~Z.}\
  \bibnamefont {Rothstein}},\ }\href {\doibase10.1103/PhysRevD.73.104029}
  {\bibfield  {journal} {\bibinfo  {journal} {Phys. Rev.}\ }\textbf {\bibinfo
  {volume} {D73}},\ \bibinfo {pages} {104029} (\bibinfo {year} {2006})},\
  \Eprint {http://arxiv.org/abs/hep-th/0409156} {arXiv:hep-th/0409156
  [hep-th]}\BibitemShut {NoStop}%
\bibitem [{\citenamefont {Foffa}\ and\ \citenamefont
  {Sturani}(2011)}]{Foffa:2011ub}%
  \BibitemOpen
  \bibfield  {author} {\bibinfo {author} {\bibfnamefont {Stefano}\ \bibnamefont
  {Foffa}}\ and\ \bibinfo {author} {\bibfnamefont {Riccardo}\ \bibnamefont
  {Sturani}},\ }\href {\doibase10.1103/PhysRevD.84.044031} {\bibfield
  {journal} {\bibinfo  {journal} {Phys. Rev.}\ }\textbf {\bibinfo {volume}
  {D84}},\ \bibinfo {pages} {044031} (\bibinfo {year} {2011})},\ \Eprint
  {http://arxiv.org/abs/1104.1122} {arXiv:1104.1122 [gr-qc]}\BibitemShut
  {NoStop}%
\bibitem [{\citenamefont {Iyer}\ and\ \citenamefont
  {Will}(1993)}]{Iyer:1993xi}%
  \BibitemOpen
  \bibfield  {author} {\bibinfo {author} {\bibfnamefont {Bala~R.}\ \bibnamefont
  {Iyer}}\ and\ \bibinfo {author} {\bibfnamefont {C.~M.}\ \bibnamefont
  {Will}},\ }\href {\doibase10.1103/PhysRevLett.70.113} {\bibfield  {journal}
  {\bibinfo  {journal} {Phys. Rev. Lett.}\ }\textbf {\bibinfo {volume} {70}},\
  \bibinfo {pages} {113--116} (\bibinfo {year} {1993})}\BibitemShut {NoStop}%
\bibitem [{\citenamefont {Iyer}\ and\ \citenamefont
  {Will}(1995)}]{Iyer:1995rn}%
  \BibitemOpen
  \bibfield  {author} {\bibinfo {author} {\bibfnamefont {Bala~R.}\ \bibnamefont
  {Iyer}}\ and\ \bibinfo {author} {\bibfnamefont {C.~M.}\ \bibnamefont
  {Will}},\ }\href {\doibase10.1103/PhysRevD.52.6882} {\bibfield  {journal}
  {\bibinfo  {journal} {Phys. Rev.}\ }\textbf {\bibinfo {volume} {D52}},\
  \bibinfo {pages} {6882--6893} (\bibinfo {year} {1995})}\BibitemShut {NoStop}%
\bibitem [{\citenamefont {Jaranowski}\ and\ \citenamefont
  {Schaefer}(1997)}]{Jaranowski:1996nv}%
  \BibitemOpen
  \bibfield  {author} {\bibinfo {author} {\bibfnamefont {Piotr}\ \bibnamefont
  {Jaranowski}}\ and\ \bibinfo {author} {\bibfnamefont {Gerhard}\ \bibnamefont
  {Schaefer}},\ }\href {\doibase10.1103/PhysRevD.55.4712} {\bibfield  {journal}
  {\bibinfo  {journal} {Phys. Rev.}\ }\textbf {\bibinfo {volume} {D55}},\
  \bibinfo {pages} {4712--4722} (\bibinfo {year} {1997})}\BibitemShut {NoStop}%
\bibitem [{\citenamefont {Pati}\ and\ \citenamefont
  {Will}(2002)}]{Pati:2002ux}%
  \BibitemOpen
  \bibfield  {author} {\bibinfo {author} {\bibfnamefont {Michael~E.}\
  \bibnamefont {Pati}}\ and\ \bibinfo {author} {\bibfnamefont {Clifford~M.}\
  \bibnamefont {Will}},\ }\href {\doibase10.1103/PhysRevD.65.104008} {\bibfield
   {journal} {\bibinfo  {journal} {Phys. Rev.}\ }\textbf {\bibinfo {volume}
  {D65}},\ \bibinfo {pages} {104008} (\bibinfo {year} {2002})},\ \Eprint
  {http://arxiv.org/abs/gr-qc/0201001} {arXiv:gr-qc/0201001
  [gr-qc]}\BibitemShut {NoStop}%
\bibitem [{\citenamefont {Konigsdorffer}\ \emph {et~al.}(2003)\citenamefont
  {Konigsdorffer}, \citenamefont {Faye},\ and\ \citenamefont
  {Schaefer}}]{Konigsdorffer:2003ue}%
  \BibitemOpen
  \bibfield  {author} {\bibinfo {author} {\bibfnamefont {Christian}\
  \bibnamefont {Konigsdorffer}}, \bibinfo {author} {\bibfnamefont {Guillaume}\
  \bibnamefont {Faye}}, \ and\ \bibinfo {author} {\bibfnamefont {Gerhard}\
  \bibnamefont {Schaefer}},\ }\href {\doibase10.1103/PhysRevD.68.044004}
  {\bibfield  {journal} {\bibinfo  {journal} {Phys. Rev.}\ }\textbf {\bibinfo
  {volume} {D68}},\ \bibinfo {pages} {044004} (\bibinfo {year} {2003})},\
  \Eprint {http://arxiv.org/abs/gr-qc/0305048} {arXiv:gr-qc/0305048
  [gr-qc]}\BibitemShut {NoStop}%
\bibitem [{\citenamefont {Nissanke}\ and\ \citenamefont
  {Blanchet}(2005)}]{Nissanke:2004er}%
  \BibitemOpen
  \bibfield  {author} {\bibinfo {author} {\bibfnamefont {Samaya}\ \bibnamefont
  {Nissanke}}\ and\ \bibinfo {author} {\bibfnamefont {Luc}\ \bibnamefont
  {Blanchet}},\ }\href {\doibase10.1088/0264-9381/22/6/008} {\bibfield
  {journal} {\bibinfo  {journal} {Class. Quant. Grav.}\ }\textbf {\bibinfo
  {volume} {22}},\ \bibinfo {pages} {1007--1032} (\bibinfo {year} {2005})},\
  \Eprint {http://arxiv.org/abs/gr-qc/0412018} {arXiv:gr-qc/0412018
  [gr-qc]}\BibitemShut {NoStop}%
\bibitem [{\citenamefont {Blanchet}\ and\ \citenamefont
  {Damour}(1986)}]{Blanchet:1985sp}%
  \BibitemOpen
  \bibfield  {author} {\bibinfo {author} {\bibfnamefont {Luc}\ \bibnamefont
  {Blanchet}}\ and\ \bibinfo {author} {\bibfnamefont {Thibault}\ \bibnamefont
  {Damour}},\ }\href {\doibase10.1098/rsta.1986.0125} {\bibfield  {journal}
  {\bibinfo  {journal} {Phil. Trans. Roy. Soc. Lond.}\ }\textbf {\bibinfo
  {volume} {A320}},\ \bibinfo {pages} {379--430} (\bibinfo {year}
  {1986})}\BibitemShut {NoStop}%
\bibitem [{\citenamefont {Blanchet}(1987)}]{Blanchet:1986dk}%
  \BibitemOpen
  \bibfield  {author} {\bibinfo {author} {\bibfnamefont {Luc}\ \bibnamefont
  {Blanchet}},\ }\href {\doibase10.1098/rspa.1987.0022} {\bibfield  {journal}
  {\bibinfo  {journal} {Proc. Roy. Soc. Lond.}\ }\textbf {\bibinfo {volume}
  {A409}},\ \bibinfo {pages} {383--399} (\bibinfo {year} {1987})}\BibitemShut
  {NoStop}%
\bibitem [{\citenamefont {Blanchet}(1998)}]{Blanchet:1998in}%
  \BibitemOpen
  \bibfield  {author} {\bibinfo {author} {\bibfnamefont {Luc}\ \bibnamefont
  {Blanchet}},\ }\href {\doibase10.1088/0264-9381/15/7/013} {\bibfield
  {journal} {\bibinfo  {journal} {Class. Quant. Grav.}\ }\textbf {\bibinfo
  {volume} {15}},\ \bibinfo {pages} {1971--1999} (\bibinfo {year} {1998})},\
  \Eprint {http://arxiv.org/abs/gr-qc/9801101} {arXiv:gr-qc/9801101
  [gr-qc]}\BibitemShut {NoStop}%
\bibitem [{\citenamefont {Will}\ and\ \citenamefont
  {Wiseman}(1996)}]{Will:1996zj}%
  \BibitemOpen
  \bibfield  {author} {\bibinfo {author} {\bibfnamefont {Clifford~M.}\
  \bibnamefont {Will}}\ and\ \bibinfo {author} {\bibfnamefont {Alan~G.}\
  \bibnamefont {Wiseman}},\ }\href {\doibase10.1103/PhysRevD.54.4813}
  {\bibfield  {journal} {\bibinfo  {journal} {Phys. Rev.}\ }\textbf {\bibinfo
  {volume} {D54}},\ \bibinfo {pages} {4813--4848} (\bibinfo {year} {1996})},\
  \Eprint {http://arxiv.org/abs/gr-qc/9608012} {arXiv:gr-qc/9608012
  [gr-qc]}\BibitemShut {NoStop}%
\bibitem [{\citenamefont {Pati}\ and\ \citenamefont
  {Will}(2000)}]{Pati:2000vt}%
  \BibitemOpen
  \bibfield  {author} {\bibinfo {author} {\bibfnamefont {Michael~E.}\
  \bibnamefont {Pati}}\ and\ \bibinfo {author} {\bibfnamefont {Clifford~M.}\
  \bibnamefont {Will}},\ }\href {\doibase10.1103/PhysRevD.62.124015} {\bibfield
   {journal} {\bibinfo  {journal} {Phys. Rev.}\ }\textbf {\bibinfo {volume}
  {D62}},\ \bibinfo {pages} {124015} (\bibinfo {year} {2000})},\ \Eprint
  {http://arxiv.org/abs/gr-qc/0007087} {arXiv:gr-qc/0007087
  [gr-qc]}\BibitemShut {NoStop}%
\bibitem [{\citenamefont {Goldberger}\ and\ \citenamefont
  {Ross}(2010)}]{Goldberger:2009qd}%
  \BibitemOpen
  \bibfield  {author} {\bibinfo {author} {\bibfnamefont {Walter~D.}\
  \bibnamefont {Goldberger}}\ and\ \bibinfo {author} {\bibfnamefont {Andreas}\
  \bibnamefont {Ross}},\ }\href {\doibase10.1103/PhysRevD.81.124015} {\bibfield
   {journal} {\bibinfo  {journal} {Phys. Rev.}\ }\textbf {\bibinfo {volume}
  {D81}},\ \bibinfo {pages} {124015} (\bibinfo {year} {2010})},\ \Eprint
  {http://arxiv.org/abs/0912.4254} {arXiv:0912.4254 [gr-qc]}\BibitemShut
  {NoStop}%
\bibitem [{\citenamefont {Ross}(2012)}]{Ross:2012fc}%
  \BibitemOpen
  \bibfield  {author} {\bibinfo {author} {\bibfnamefont {Andreas}\ \bibnamefont
  {Ross}},\ }\href {\doibase10.1103/PhysRevD.85.125033} {\bibfield  {journal}
  {\bibinfo  {journal} {Phys. Rev.}\ }\textbf {\bibinfo {volume} {D85}},\
  \bibinfo {pages} {125033} (\bibinfo {year} {2012})},\ \Eprint
  {http://arxiv.org/abs/1202.4750} {arXiv:1202.4750 [gr-qc]}\BibitemShut
  {NoStop}%
\bibitem [{\citenamefont {Blanchet}\ \emph
  {et~al.}(2002{\natexlab{a}})\citenamefont {Blanchet}, \citenamefont {Iyer},\
  and\ \citenamefont {Joguet}}]{Blanchet:2001aw}%
  \BibitemOpen
  \bibfield  {author} {\bibinfo {author} {\bibfnamefont {Luc}\ \bibnamefont
  {Blanchet}}, \bibinfo {author} {\bibfnamefont {Bala~R.}\ \bibnamefont
  {Iyer}}, \ and\ \bibinfo {author} {\bibfnamefont {Benoit}\ \bibnamefont
  {Joguet}},\ }\href {\doibase10.1103/PhysRevD.65.064005} {\bibfield  {journal}
  {\bibinfo  {journal} {Phys. Rev.}\ }\textbf {\bibinfo {volume} {D65}},\
  \bibinfo {pages} {064005} (\bibinfo {year} {2002}{\natexlab{a}})},\ \bibinfo
  {note} {[Erratum: Phys. Rev.D71,129903(2005)]},\ \Eprint
  {http://arxiv.org/abs/gr-qc/0105098} {arXiv:gr-qc/0105098
  [gr-qc]}\BibitemShut {NoStop}%
\bibitem [{\citenamefont {Blanchet}\ \emph
  {et~al.}(2002{\natexlab{b}})\citenamefont {Blanchet}, \citenamefont {Faye},
  \citenamefont {Iyer},\ and\ \citenamefont {Joguet}}]{Blanchet:2001ax}%
  \BibitemOpen
  \bibfield  {author} {\bibinfo {author} {\bibfnamefont {Luc}\ \bibnamefont
  {Blanchet}}, \bibinfo {author} {\bibfnamefont {Guillaume}\ \bibnamefont
  {Faye}}, \bibinfo {author} {\bibfnamefont {Bala~R.}\ \bibnamefont {Iyer}}, \
  and\ \bibinfo {author} {\bibfnamefont {Benoit}\ \bibnamefont {Joguet}},\
  }\href {\doibase10.1103/PhysRevD.65.061501} {\bibfield  {journal} {\bibinfo
  {journal} {Phys. Rev.}\ }\textbf {\bibinfo {volume} {D65}},\ \bibinfo {pages}
  {061501} (\bibinfo {year} {2002}{\natexlab{b}})},\ \bibinfo {note} {[Erratum:
  Phys. Rev.D71,129902(2005)]},\ \Eprint {http://arxiv.org/abs/gr-qc/0105099}
  {arXiv:gr-qc/0105099 [gr-qc]}\BibitemShut {NoStop}%
\bibitem [{\citenamefont {Blanchet}\ \emph
  {et~al.}(2004{\natexlab{b}})\citenamefont {Blanchet}, \citenamefont {Damour},
  \citenamefont {Esposito-Farese},\ and\ \citenamefont
  {Iyer}}]{Blanchet:2004ek}%
  \BibitemOpen
  \bibfield  {author} {\bibinfo {author} {\bibfnamefont {Luc}\ \bibnamefont
  {Blanchet}}, \bibinfo {author} {\bibfnamefont {Thibault}\ \bibnamefont
  {Damour}}, \bibinfo {author} {\bibfnamefont {Gilles}\ \bibnamefont
  {Esposito-Farese}}, \ and\ \bibinfo {author} {\bibfnamefont {Bala~R.}\
  \bibnamefont {Iyer}},\ }\href {\doibase10.1103/PhysRevLett.93.091101}
  {\bibfield  {journal} {\bibinfo  {journal} {Phys. Rev. Lett.}\ }\textbf
  {\bibinfo {volume} {93}},\ \bibinfo {pages} {091101} (\bibinfo {year}
  {2004}{\natexlab{b}})},\ \Eprint {http://arxiv.org/abs/gr-qc/0406012}
  {arXiv:gr-qc/0406012 [gr-qc]}\BibitemShut {NoStop}%
\bibitem [{\citenamefont {Blanchet}\ \emph {et~al.}(2005)\citenamefont
  {Blanchet}, \citenamefont {Damour}, \citenamefont {Esposito-Farese},\ and\
  \citenamefont {Iyer}}]{Blanchet:2005tk}%
  \BibitemOpen
  \bibfield  {author} {\bibinfo {author} {\bibfnamefont {Luc}\ \bibnamefont
  {Blanchet}}, \bibinfo {author} {\bibfnamefont {Thibault}\ \bibnamefont
  {Damour}}, \bibinfo {author} {\bibfnamefont {Gilles}\ \bibnamefont
  {Esposito-Farese}}, \ and\ \bibinfo {author} {\bibfnamefont {Bala~R.}\
  \bibnamefont {Iyer}},\ }\href {\doibase10.1103/PhysRevD.71.124004} {\bibfield
   {journal} {\bibinfo  {journal} {Phys. Rev.}\ }\textbf {\bibinfo {volume}
  {D71}},\ \bibinfo {pages} {124004} (\bibinfo {year} {2005})},\ \Eprint
  {http://arxiv.org/abs/gr-qc/0503044} {arXiv:gr-qc/0503044
  [gr-qc]}\BibitemShut {NoStop}%
\bibitem [{\citenamefont {Arun}\ \emph
  {et~al.}(2008{\natexlab{a}})\citenamefont {Arun}, \citenamefont {Blanchet},
  \citenamefont {Iyer},\ and\ \citenamefont {Qusailah}}]{Arun:2007rg}%
  \BibitemOpen
  \bibfield  {author} {\bibinfo {author} {\bibfnamefont {K.~G.}\ \bibnamefont
  {Arun}}, \bibinfo {author} {\bibfnamefont {Luc}\ \bibnamefont {Blanchet}},
  \bibinfo {author} {\bibfnamefont {Bala~R.}\ \bibnamefont {Iyer}}, \ and\
  \bibinfo {author} {\bibfnamefont {Moh'd S.~S.}\ \bibnamefont {Qusailah}},\
  }\href {\doibase10.1103/PhysRevD.77.064034} {\bibfield  {journal} {\bibinfo
  {journal} {Phys. Rev.}\ }\textbf {\bibinfo {volume} {D77}},\ \bibinfo {pages}
  {064034} (\bibinfo {year} {2008}{\natexlab{a}})},\ \Eprint
  {http://arxiv.org/abs/0711.0250} {arXiv:0711.0250 [gr-qc]}\BibitemShut
  {NoStop}%
\bibitem [{\citenamefont {Arun}\ \emph
  {et~al.}(2008{\natexlab{b}})\citenamefont {Arun}, \citenamefont {Blanchet},
  \citenamefont {Iyer},\ and\ \citenamefont {Qusailah}}]{Arun:2007sg}%
  \BibitemOpen
  \bibfield  {author} {\bibinfo {author} {\bibfnamefont {K.~G.}\ \bibnamefont
  {Arun}}, \bibinfo {author} {\bibfnamefont {Luc}\ \bibnamefont {Blanchet}},
  \bibinfo {author} {\bibfnamefont {Bala~R.}\ \bibnamefont {Iyer}}, \ and\
  \bibinfo {author} {\bibfnamefont {Moh'd S.~S.}\ \bibnamefont {Qusailah}},\
  }\href {\doibase10.1103/PhysRevD.77.064035} {\bibfield  {journal} {\bibinfo
  {journal} {Phys. Rev.}\ }\textbf {\bibinfo {volume} {D77}},\ \bibinfo {pages}
  {064035} (\bibinfo {year} {2008}{\natexlab{b}})},\ \Eprint
  {http://arxiv.org/abs/0711.0302} {arXiv:0711.0302 [gr-qc]}\BibitemShut
  {NoStop}%
\bibitem [{\citenamefont {Arun}\ \emph {et~al.}(2009)\citenamefont {Arun},
  \citenamefont {Blanchet}, \citenamefont {Iyer},\ and\ \citenamefont
  {Sinha}}]{Arun:2009mc}%
  \BibitemOpen
  \bibfield  {author} {\bibinfo {author} {\bibfnamefont {K.~G.}\ \bibnamefont
  {Arun}}, \bibinfo {author} {\bibfnamefont {Luc}\ \bibnamefont {Blanchet}},
  \bibinfo {author} {\bibfnamefont {Bala~R.}\ \bibnamefont {Iyer}}, \ and\
  \bibinfo {author} {\bibfnamefont {Siddhartha}\ \bibnamefont {Sinha}},\ }\href
  {\doibase10.1103/PhysRevD.80.124018} {\bibfield  {journal} {\bibinfo
  {journal} {Phys. Rev.}\ }\textbf {\bibinfo {volume} {D80}},\ \bibinfo {pages}
  {124018} (\bibinfo {year} {2009})},\ \Eprint {http://arxiv.org/abs/0908.3854}
  {arXiv:0908.3854 [gr-qc]}\BibitemShut {NoStop}%
\bibitem [{\citenamefont {Moore}\ \emph {et~al.}(2016)\citenamefont {Moore},
  \citenamefont {Favata}, \citenamefont {Arun},\ and\ \citenamefont
  {Mishra}}]{Moore:2016qxz}%
  \BibitemOpen
  \bibfield  {author} {\bibinfo {author} {\bibfnamefont {Blake}\ \bibnamefont
  {Moore}}, \bibinfo {author} {\bibfnamefont {Marc}\ \bibnamefont {Favata}},
  \bibinfo {author} {\bibfnamefont {K.~G.}\ \bibnamefont {Arun}}, \ and\
  \bibinfo {author} {\bibfnamefont {Chandra~Kant}\ \bibnamefont {Mishra}},\
  }\href {\doibase10.1103/PhysRevD.93.124061} {\bibfield  {journal} {\bibinfo
  {journal} {Phys. Rev.}\ }\textbf {\bibinfo {volume} {D93}},\ \bibinfo {pages}
  {124061} (\bibinfo {year} {2016})},\ \Eprint
  {http://arxiv.org/abs/1605.00304} {arXiv:1605.00304 [gr-qc]}\BibitemShut
  {NoStop}%
\bibitem [{\citenamefont {Arun}\ \emph {et~al.}(2004)\citenamefont {Arun},
  \citenamefont {Blanchet}, \citenamefont {Iyer},\ and\ \citenamefont
  {Qusailah}}]{Arun:2004ff}%
  \BibitemOpen
  \bibfield  {author} {\bibinfo {author} {\bibfnamefont {K.~G.}\ \bibnamefont
  {Arun}}, \bibinfo {author} {\bibfnamefont {Luc}\ \bibnamefont {Blanchet}},
  \bibinfo {author} {\bibfnamefont {Bala~R.}\ \bibnamefont {Iyer}}, \ and\
  \bibinfo {author} {\bibfnamefont {Moh'd S.~S.}\ \bibnamefont {Qusailah}},\
  }\href {\doibase10.1088/0264-9381/21/15/010} {\bibfield  {journal} {\bibinfo
  {journal} {Class. Quant. Grav.}\ }\textbf {\bibinfo {volume} {21}},\ \bibinfo
  {pages} {3771--3802} (\bibinfo {year} {2004})},\ \bibinfo {note} {[Erratum:
  Class. Quant. Grav.22,3115(2005)]},\ \Eprint
  {http://arxiv.org/abs/gr-qc/0404085} {arXiv:gr-qc/0404085
  [gr-qc]}\BibitemShut {NoStop}%
\bibitem [{\citenamefont {Kidder}\ \emph {et~al.}(2007)\citenamefont {Kidder},
  \citenamefont {Blanchet},\ and\ \citenamefont {Iyer}}]{Kidder:2007gz}%
  \BibitemOpen
  \bibfield  {author} {\bibinfo {author} {\bibfnamefont {Lawrence~E.}\
  \bibnamefont {Kidder}}, \bibinfo {author} {\bibfnamefont {Luc}\ \bibnamefont
  {Blanchet}}, \ and\ \bibinfo {author} {\bibfnamefont {Bala~R.}\ \bibnamefont
  {Iyer}},\ }\href {\doibase10.1088/0264-9381/24/20/N01} {\bibfield  {journal}
  {\bibinfo  {journal} {Class. Quant. Grav.}\ }\textbf {\bibinfo {volume}
  {24}},\ \bibinfo {pages} {5307--5312} (\bibinfo {year} {2007})},\ \Eprint
  {http://arxiv.org/abs/0706.0726} {arXiv:0706.0726 [gr-qc]}\BibitemShut
  {NoStop}%
\bibitem [{\citenamefont {Kidder}(2008)}]{Kidder:2007rt}%
  \BibitemOpen
  \bibfield  {author} {\bibinfo {author} {\bibfnamefont {Lawrence~E.}\
  \bibnamefont {Kidder}},\ }\href {\doibase10.1103/PhysRevD.77.044016}
  {\bibfield  {journal} {\bibinfo  {journal} {Phys. Rev.}\ }\textbf {\bibinfo
  {volume} {D77}},\ \bibinfo {pages} {044016} (\bibinfo {year} {2008})},\
  \Eprint {http://arxiv.org/abs/0710.0614} {arXiv:0710.0614
  [gr-qc]}\BibitemShut {NoStop}%
\bibitem [{\citenamefont {Blanchet}\ \emph {et~al.}(2008)\citenamefont
  {Blanchet}, \citenamefont {Faye}, \citenamefont {Iyer},\ and\ \citenamefont
  {Sinha}}]{Blanchet:2008je}%
  \BibitemOpen
  \bibfield  {author} {\bibinfo {author} {\bibfnamefont {Luc}\ \bibnamefont
  {Blanchet}}, \bibinfo {author} {\bibfnamefont {Guillaume}\ \bibnamefont
  {Faye}}, \bibinfo {author} {\bibfnamefont {Bala~R.}\ \bibnamefont {Iyer}}, \
  and\ \bibinfo {author} {\bibfnamefont {Siddhartha}\ \bibnamefont {Sinha}},\
  }\href {\doibase10.1088/0264-9381/25/16/165003} {\bibfield  {journal}
  {\bibinfo  {journal} {Class. Quant. Grav.}\ }\textbf {\bibinfo {volume}
  {25}},\ \bibinfo {pages} {165003} (\bibinfo {year} {2008})},\ \bibinfo {note}
  {[Erratum: Class. Quant. Grav.29,239501(2012)]},\ \Eprint
  {http://arxiv.org/abs/0802.1249} {arXiv:0802.1249 [gr-qc]}\BibitemShut
  {NoStop}%
\bibitem [{\citenamefont {Faye}\ \emph {et~al.}(2012)\citenamefont {Faye},
  \citenamefont {Marsat}, \citenamefont {Blanchet},\ and\ \citenamefont
  {Iyer}}]{Faye:2012we}%
  \BibitemOpen
  \bibfield  {author} {\bibinfo {author} {\bibfnamefont {Guillaume}\
  \bibnamefont {Faye}}, \bibinfo {author} {\bibfnamefont {Sylvain}\
  \bibnamefont {Marsat}}, \bibinfo {author} {\bibfnamefont {Luc}\ \bibnamefont
  {Blanchet}}, \ and\ \bibinfo {author} {\bibfnamefont {Bala~R.}\ \bibnamefont
  {Iyer}},\ }\href {\doibase10.1088/0264-9381/29/17/175004} {\bibfield
  {journal} {\bibinfo  {journal} {Class. Quant. Grav.}\ }\textbf {\bibinfo
  {volume} {29}},\ \bibinfo {pages} {175004} (\bibinfo {year} {2012})},\
  \Eprint {http://arxiv.org/abs/1204.1043} {arXiv:1204.1043
  [gr-qc]}\BibitemShut {NoStop}%
\bibitem [{\citenamefont {Mishra}\ \emph {et~al.}(2015)\citenamefont {Mishra},
  \citenamefont {Arun},\ and\ \citenamefont {Iyer}}]{Mishra:2015bqa}%
  \BibitemOpen
  \bibfield  {author} {\bibinfo {author} {\bibfnamefont {Chandra~Kant}\
  \bibnamefont {Mishra}}, \bibinfo {author} {\bibfnamefont {K.~G.}\
  \bibnamefont {Arun}}, \ and\ \bibinfo {author} {\bibfnamefont {Bala~R.}\
  \bibnamefont {Iyer}},\ }\href {\doibase10.1103/PhysRevD.91.084040} {\bibfield
   {journal} {\bibinfo  {journal} {Phys. Rev.}\ }\textbf {\bibinfo {volume}
  {D91}},\ \bibinfo {pages} {084040} (\bibinfo {year} {2015})},\ \Eprint
  {http://arxiv.org/abs/1501.07096} {arXiv:1501.07096 [gr-qc]}\BibitemShut
  {NoStop}%
\bibitem [{\citenamefont {Foffa}(2014)}]{Foffa:2013gja}%
  \BibitemOpen
  \bibfield  {author} {\bibinfo {author} {\bibfnamefont {Stefano}\ \bibnamefont
  {Foffa}},\ }\href {\doibase10.1103/PhysRevD.89.024019} {\bibfield  {journal}
  {\bibinfo  {journal} {Phys. Rev.}\ }\textbf {\bibinfo {volume} {D89}},\
  \bibinfo {pages} {024019} (\bibinfo {year} {2014})},\ \Eprint
  {http://arxiv.org/abs/1309.3956} {arXiv:1309.3956 [gr-qc]}\BibitemShut
  {NoStop}%
\bibitem [{\citenamefont {Jaranowski}\ and\ \citenamefont
  {Schafer}(2012)}]{Jaranowski:2012eb}%
  \BibitemOpen
  \bibfield  {author} {\bibinfo {author} {\bibfnamefont {Piotr}\ \bibnamefont
  {Jaranowski}}\ and\ \bibinfo {author} {\bibfnamefont {Gerhard}\ \bibnamefont
  {Schafer}},\ }\href {\doibase10.1103/PhysRevD.86.061503} {\bibfield
  {journal} {\bibinfo  {journal} {Phys. Rev.}\ }\textbf {\bibinfo {volume}
  {D86}},\ \bibinfo {pages} {061503} (\bibinfo {year} {2012})},\ \Eprint
  {http://arxiv.org/abs/1207.5448} {arXiv:1207.5448 [gr-qc]}\BibitemShut
  {NoStop}%
\bibitem [{\citenamefont {Jaranowski}\ and\ \citenamefont
  {Schäfer}(2013)}]{Jaranowski:2013lca}%
  \BibitemOpen
  \bibfield  {author} {\bibinfo {author} {\bibfnamefont {Piotr}\ \bibnamefont
  {Jaranowski}}\ and\ \bibinfo {author} {\bibfnamefont {Gerhard}\ \bibnamefont
  {Schäfer}},\ }\href {\doibase10.1103/PhysRevD.87.081503} {\bibfield
  {journal} {\bibinfo  {journal} {Phys. Rev.}\ }\textbf {\bibinfo {volume}
  {D87}},\ \bibinfo {pages} {081503} (\bibinfo {year} {2013})},\ \Eprint
  {http://arxiv.org/abs/1303.3225} {arXiv:1303.3225 [gr-qc]}\BibitemShut
  {NoStop}%
\bibitem [{\citenamefont {Damour}\ \emph
  {et~al.}(2014{\natexlab{a}})\citenamefont {Damour}, \citenamefont
  {Jaranowski},\ and\ \citenamefont {Schäfer}}]{Damour:2014jta}%
  \BibitemOpen
  \bibfield  {author} {\bibinfo {author} {\bibfnamefont {Thibault}\
  \bibnamefont {Damour}}, \bibinfo {author} {\bibfnamefont {Piotr}\
  \bibnamefont {Jaranowski}}, \ and\ \bibinfo {author} {\bibfnamefont
  {Gerhard}\ \bibnamefont {Schäfer}},\ }\href
  {\doibase10.1103/PhysRevD.89.064058} {\bibfield  {journal} {\bibinfo
  {journal} {Phys. Rev.}\ }\textbf {\bibinfo {volume} {D89}},\ \bibinfo {pages}
  {064058} (\bibinfo {year} {2014}{\natexlab{a}})},\ \Eprint
  {http://arxiv.org/abs/1401.4548} {arXiv:1401.4548 [gr-qc]}\BibitemShut
  {NoStop}%
\bibitem [{\citenamefont {Jaranowski}\ and\ \citenamefont
  {Schäfer}(2015)}]{Jaranowski:2015lha}%
  \BibitemOpen
  \bibfield  {author} {\bibinfo {author} {\bibfnamefont {Piotr}\ \bibnamefont
  {Jaranowski}}\ and\ \bibinfo {author} {\bibfnamefont {Gerhard}\ \bibnamefont
  {Schäfer}},\ }\href {\doibase10.1103/PhysRevD.92.124043} {\bibfield
  {journal} {\bibinfo  {journal} {Phys. Rev.}\ }\textbf {\bibinfo {volume}
  {D92}},\ \bibinfo {pages} {124043} (\bibinfo {year} {2015})},\ \Eprint
  {http://arxiv.org/abs/1508.01016} {arXiv:1508.01016 [gr-qc]}\BibitemShut
  {NoStop}%
\bibitem [{\citenamefont {Damour}\ \emph {et~al.}(2016)\citenamefont {Damour},
  \citenamefont {Jaranowski},\ and\ \citenamefont {Schäfer}}]{Damour:2016abl}%
  \BibitemOpen
  \bibfield  {author} {\bibinfo {author} {\bibfnamefont {Thibault}\
  \bibnamefont {Damour}}, \bibinfo {author} {\bibfnamefont {Piotr}\
  \bibnamefont {Jaranowski}}, \ and\ \bibinfo {author} {\bibfnamefont
  {Gerhard}\ \bibnamefont {Schäfer}},\ }\href
  {\doibase10.1103/PhysRevD.93.084014} {\bibfield  {journal} {\bibinfo
  {journal} {Phys. Rev.}\ }\textbf {\bibinfo {volume} {D93}},\ \bibinfo {pages}
  {084014} (\bibinfo {year} {2016})},\ \Eprint
  {http://arxiv.org/abs/1601.01283} {arXiv:1601.01283 [gr-qc]}\BibitemShut
  {NoStop}%
\bibitem [{\citenamefont {Damour}\ and\ \citenamefont
  {Jaranowski}(2017)}]{Damour:2017ced}%
  \BibitemOpen
  \bibfield  {author} {\bibinfo {author} {\bibfnamefont {Thibault}\
  \bibnamefont {Damour}}\ and\ \bibinfo {author} {\bibfnamefont {Piotr}\
  \bibnamefont {Jaranowski}},\ }\href {\doibase10.1103/PhysRevD.95.084005}
  {\bibfield  {journal} {\bibinfo  {journal} {Phys. Rev.}\ }\textbf {\bibinfo
  {volume} {D95}},\ \bibinfo {pages} {084005} (\bibinfo {year} {2017})},\
  \Eprint {http://arxiv.org/abs/1701.02645} {arXiv:1701.02645
  [gr-qc]}\BibitemShut {NoStop}%
\bibitem [{\citenamefont {Bernard}\ \emph {et~al.}(2016)\citenamefont
  {Bernard}, \citenamefont {Blanchet}, \citenamefont {Bohé}, \citenamefont
  {Faye},\ and\ \citenamefont {Marsat}}]{Bernard:2015njp}%
  \BibitemOpen
  \bibfield  {author} {\bibinfo {author} {\bibfnamefont {Laura}\ \bibnamefont
  {Bernard}}, \bibinfo {author} {\bibfnamefont {Luc}\ \bibnamefont {Blanchet}},
  \bibinfo {author} {\bibfnamefont {Alejandro}\ \bibnamefont {Bohé}}, \bibinfo
  {author} {\bibfnamefont {Guillaume}\ \bibnamefont {Faye}}, \ and\ \bibinfo
  {author} {\bibfnamefont {Sylvain}\ \bibnamefont {Marsat}},\ }\href
  {\doibase10.1103/PhysRevD.93.084037} {\bibfield  {journal} {\bibinfo
  {journal} {Phys. Rev.}\ }\textbf {\bibinfo {volume} {D93}},\ \bibinfo {pages}
  {084037} (\bibinfo {year} {2016})},\ \Eprint
  {http://arxiv.org/abs/1512.02876} {arXiv:1512.02876 [gr-qc]}\BibitemShut
  {NoStop}%
\bibitem [{\citenamefont {Bernard}\ \emph
  {et~al.}(2017{\natexlab{a}})\citenamefont {Bernard}, \citenamefont
  {Blanchet}, \citenamefont {Bohé}, \citenamefont {Faye},\ and\ \citenamefont
  {Marsat}}]{Bernard:2016wrg}%
  \BibitemOpen
  \bibfield  {author} {\bibinfo {author} {\bibfnamefont {Laura}\ \bibnamefont
  {Bernard}}, \bibinfo {author} {\bibfnamefont {Luc}\ \bibnamefont {Blanchet}},
  \bibinfo {author} {\bibfnamefont {Alejandro}\ \bibnamefont {Bohé}}, \bibinfo
  {author} {\bibfnamefont {Guillaume}\ \bibnamefont {Faye}}, \ and\ \bibinfo
  {author} {\bibfnamefont {Sylvain}\ \bibnamefont {Marsat}},\ }\href
  {\doibase10.1103/PhysRevD.95.044026} {\bibfield  {journal} {\bibinfo
  {journal} {Phys. Rev.}\ }\textbf {\bibinfo {volume} {D95}},\ \bibinfo {pages}
  {044026} (\bibinfo {year} {2017}{\natexlab{a}})},\ \Eprint
  {http://arxiv.org/abs/1610.07934} {arXiv:1610.07934 [gr-qc]}\BibitemShut
  {NoStop}%
\bibitem [{\citenamefont {Bernard}\ \emph
  {et~al.}(2017{\natexlab{b}})\citenamefont {Bernard}, \citenamefont
  {Blanchet}, \citenamefont {Bohé}, \citenamefont {Faye},\ and\ \citenamefont
  {Marsat}}]{Bernard:2017bvn}%
  \BibitemOpen
  \bibfield  {author} {\bibinfo {author} {\bibfnamefont {Laura}\ \bibnamefont
  {Bernard}}, \bibinfo {author} {\bibfnamefont {Luc}\ \bibnamefont {Blanchet}},
  \bibinfo {author} {\bibfnamefont {Alejandro}\ \bibnamefont {Bohé}}, \bibinfo
  {author} {\bibfnamefont {Guillaume}\ \bibnamefont {Faye}}, \ and\ \bibinfo
  {author} {\bibfnamefont {Sylvain}\ \bibnamefont {Marsat}},\ }\href
  {\doibase10.1103/PhysRevD.96.104043} {\bibfield  {journal} {\bibinfo
  {journal} {Phys. Rev.}\ }\textbf {\bibinfo {volume} {D96}},\ \bibinfo {pages}
  {104043} (\bibinfo {year} {2017}{\natexlab{b}})},\ \Eprint
  {http://arxiv.org/abs/1706.08480} {arXiv:1706.08480 [gr-qc]}\BibitemShut
  {NoStop}%
\bibitem [{\citenamefont {Marchand}\ \emph {et~al.}(2018)\citenamefont
  {Marchand}, \citenamefont {Bernard}, \citenamefont {Blanchet},\ and\
  \citenamefont {Faye}}]{Marchand:2017pir}%
  \BibitemOpen
  \bibfield  {author} {\bibinfo {author} {\bibfnamefont {Tanguy}\ \bibnamefont
  {Marchand}}, \bibinfo {author} {\bibfnamefont {Laura}\ \bibnamefont
  {Bernard}}, \bibinfo {author} {\bibfnamefont {Luc}\ \bibnamefont {Blanchet}},
  \ and\ \bibinfo {author} {\bibfnamefont {Guillaume}\ \bibnamefont {Faye}},\
  }\href {\doibase10.1103/PhysRevD.97.044023} {\bibfield  {journal} {\bibinfo
  {journal} {Phys. Rev.}\ }\textbf {\bibinfo {volume} {D97}},\ \bibinfo {pages}
  {044023} (\bibinfo {year} {2018})},\ \Eprint
  {http://arxiv.org/abs/1707.09289} {arXiv:1707.09289 [gr-qc]}\BibitemShut
  {NoStop}%
\bibitem [{\citenamefont {Bernard}\ \emph {et~al.}(2018)\citenamefont
  {Bernard}, \citenamefont {Blanchet}, \citenamefont {Faye},\ and\
  \citenamefont {Marchand}}]{Bernard:2017ktp}%
  \BibitemOpen
  \bibfield  {author} {\bibinfo {author} {\bibfnamefont {Laura}\ \bibnamefont
  {Bernard}}, \bibinfo {author} {\bibfnamefont {Luc}\ \bibnamefont {Blanchet}},
  \bibinfo {author} {\bibfnamefont {Guillaume}\ \bibnamefont {Faye}}, \ and\
  \bibinfo {author} {\bibfnamefont {Tanguy}\ \bibnamefont {Marchand}},\ }\href
  {\doibase10.1103/PhysRevD.97.044037} {\bibfield  {journal} {\bibinfo
  {journal} {Phys. Rev.}\ }\textbf {\bibinfo {volume} {D97}},\ \bibinfo {pages}
  {044037} (\bibinfo {year} {2018})},\ \Eprint
  {http://arxiv.org/abs/1711.00283} {arXiv:1711.00283 [gr-qc]}\BibitemShut
  {NoStop}%
\bibitem [{\citenamefont {Foffa}\ and\ \citenamefont
  {Sturani}(2013{\natexlab{a}})}]{Foffa:2012rn}%
  \BibitemOpen
  \bibfield  {author} {\bibinfo {author} {\bibfnamefont {Stefano}\ \bibnamefont
  {Foffa}}\ and\ \bibinfo {author} {\bibfnamefont {Riccardo}\ \bibnamefont
  {Sturani}},\ }\href {\doibase10.1103/PhysRevD.87.064011} {\bibfield
  {journal} {\bibinfo  {journal} {Phys. Rev.}\ }\textbf {\bibinfo {volume}
  {D87}},\ \bibinfo {pages} {064011} (\bibinfo {year} {2013}{\natexlab{a}})},\
  \Eprint {http://arxiv.org/abs/1206.7087} {arXiv:1206.7087
  [gr-qc]}\BibitemShut {NoStop}%
\bibitem [{\citenamefont {Foffa}\ \emph {et~al.}(2017)\citenamefont {Foffa},
  \citenamefont {Mastrolia}, \citenamefont {Sturani},\ and\ \citenamefont
  {Sturm}}]{Foffa:2016rgu}%
  \BibitemOpen
  \bibfield  {author} {\bibinfo {author} {\bibfnamefont {Stefano}\ \bibnamefont
  {Foffa}}, \bibinfo {author} {\bibfnamefont {Pierpaolo}\ \bibnamefont
  {Mastrolia}}, \bibinfo {author} {\bibfnamefont {Riccardo}\ \bibnamefont
  {Sturani}}, \ and\ \bibinfo {author} {\bibfnamefont {Christian}\ \bibnamefont
  {Sturm}},\ }\href {\doibase10.1103/PhysRevD.95.104009} {\bibfield  {journal}
  {\bibinfo  {journal} {Phys. Rev.}\ }\textbf {\bibinfo {volume} {D95}},\
  \bibinfo {pages} {104009} (\bibinfo {year} {2017})},\ \Eprint
  {http://arxiv.org/abs/1612.00482} {arXiv:1612.00482 [gr-qc]}\BibitemShut
  {NoStop}%
\bibitem [{\citenamefont {Porto}\ and\ \citenamefont
  {Rothstein}(2017)}]{Porto:2017dgs}%
  \BibitemOpen
  \bibfield  {author} {\bibinfo {author} {\bibfnamefont {Rafael~A.}\
  \bibnamefont {Porto}}\ and\ \bibinfo {author} {\bibfnamefont {Ira~Z.}\
  \bibnamefont {Rothstein}},\ }\href {\doibase10.1103/PhysRevD.96.024062}
  {\bibfield  {journal} {\bibinfo  {journal} {Phys. Rev.}\ }\textbf {\bibinfo
  {volume} {D96}},\ \bibinfo {pages} {024062} (\bibinfo {year} {2017})},\
  \Eprint {http://arxiv.org/abs/1703.06433} {arXiv:1703.06433
  [gr-qc]}\BibitemShut {NoStop}%
\bibitem [{\citenamefont {Damour}\ \emph {et~al.}(2015)\citenamefont {Damour},
  \citenamefont {Jaranowski},\ and\ \citenamefont {Schäfer}}]{Damour:2015isa}%
  \BibitemOpen
  \bibfield  {author} {\bibinfo {author} {\bibfnamefont {Thibault}\
  \bibnamefont {Damour}}, \bibinfo {author} {\bibfnamefont {Piotr}\
  \bibnamefont {Jaranowski}}, \ and\ \bibinfo {author} {\bibfnamefont
  {Gerhard}\ \bibnamefont {Schäfer}},\ }\href
  {\doibase10.1103/PhysRevD.91.084024} {\bibfield  {journal} {\bibinfo
  {journal} {Phys. Rev.}\ }\textbf {\bibinfo {volume} {D91}},\ \bibinfo {pages}
  {084024} (\bibinfo {year} {2015})},\ \Eprint
  {http://arxiv.org/abs/1502.07245} {arXiv:1502.07245 [gr-qc]}\BibitemShut
  {NoStop}%
\bibitem [{\citenamefont {Bini}\ and\ \citenamefont
  {Damour}(2013)}]{Bini:2013zaa}%
  \BibitemOpen
  \bibfield  {author} {\bibinfo {author} {\bibfnamefont {Donato}\ \bibnamefont
  {Bini}}\ and\ \bibinfo {author} {\bibfnamefont {Thibault}\ \bibnamefont
  {Damour}},\ }\href {\doibase10.1103/PhysRevD.87.121501} {\bibfield  {journal}
  {\bibinfo  {journal} {Phys. Rev.}\ }\textbf {\bibinfo {volume} {D87}},\
  \bibinfo {pages} {121501} (\bibinfo {year} {2013})},\ \Eprint
  {http://arxiv.org/abs/1305.4884} {arXiv:1305.4884 [gr-qc]}\BibitemShut
  {NoStop}%
\bibitem [{\citenamefont {Blanchet}\ and\ \citenamefont
  {Damour}(1988)}]{Blanchet:1987wq}%
  \BibitemOpen
  \bibfield  {author} {\bibinfo {author} {\bibfnamefont {Luc}\ \bibnamefont
  {Blanchet}}\ and\ \bibinfo {author} {\bibfnamefont {Thibault}\ \bibnamefont
  {Damour}},\ }\href {\doibase10.1103/PhysRevD.37.1410} {\bibfield  {journal}
  {\bibinfo  {journal} {Phys. Rev.}\ }\textbf {\bibinfo {volume} {D37}},\
  \bibinfo {pages} {1410} (\bibinfo {year} {1988})}\BibitemShut {NoStop}%
\bibitem [{\citenamefont {Foffa}\ and\ \citenamefont
  {Sturani}(2013{\natexlab{b}})}]{Foffa:2011np}%
  \BibitemOpen
  \bibfield  {author} {\bibinfo {author} {\bibfnamefont {S.}~\bibnamefont
  {Foffa}}\ and\ \bibinfo {author} {\bibfnamefont {Riccardo}\ \bibnamefont
  {Sturani}},\ }\href {\doibase10.1103/PhysRevD.87.044056} {\bibfield
  {journal} {\bibinfo  {journal} {Phys. Rev.}\ }\textbf {\bibinfo {volume}
  {D87}},\ \bibinfo {pages} {044056} (\bibinfo {year} {2013}{\natexlab{b}})},\
  \Eprint {http://arxiv.org/abs/1111.5488} {arXiv:1111.5488
  [gr-qc]}\BibitemShut {NoStop}%
\bibitem [{\citenamefont {Galley}\ \emph {et~al.}(2016)\citenamefont {Galley},
  \citenamefont {Leibovich}, \citenamefont {Porto},\ and\ \citenamefont
  {Ross}}]{Galley:2015kus}%
  \BibitemOpen
  \bibfield  {author} {\bibinfo {author} {\bibfnamefont {Chad~R.}\ \bibnamefont
  {Galley}}, \bibinfo {author} {\bibfnamefont {Adam~K.}\ \bibnamefont
  {Leibovich}}, \bibinfo {author} {\bibfnamefont {Rafael~A.}\ \bibnamefont
  {Porto}}, \ and\ \bibinfo {author} {\bibfnamefont {Andreas}\ \bibnamefont
  {Ross}},\ }\href {\doibase10.1103/PhysRevD.93.124010} {\bibfield  {journal}
  {\bibinfo  {journal} {Phys. Rev.}\ }\textbf {\bibinfo {volume} {D93}},\
  \bibinfo {pages} {124010} (\bibinfo {year} {2016})},\ \Eprint
  {http://arxiv.org/abs/1511.07379} {arXiv:1511.07379 [gr-qc]}\BibitemShut
  {NoStop}%
\bibitem [{\citenamefont {Reynolds}(2013)}]{Reynolds:2013rva}%
  \BibitemOpen
  \bibfield  {author} {\bibinfo {author} {\bibfnamefont {Christopher~S.}\
  \bibnamefont {Reynolds}},\ }\href {\doibase10.1088/0264-9381/30/24/244004}
  {\bibfield  {journal} {\bibinfo  {journal} {Class. Quant. Grav.}\ }\textbf
  {\bibinfo {volume} {30}},\ \bibinfo {pages} {244004} (\bibinfo {year}
  {2013})},\ \Eprint {http://arxiv.org/abs/1307.3246} {arXiv:1307.3246
  [astro-ph.HE]}\BibitemShut {NoStop}%
\bibitem [{\citenamefont {Miller}\ and\ \citenamefont
  {Miller}(2014)}]{Miller:2014aaa}%
  \BibitemOpen
  \bibfield  {author} {\bibinfo {author} {\bibfnamefont {M.~Coleman}\
  \bibnamefont {Miller}}\ and\ \bibinfo {author} {\bibfnamefont {Jon~M.}\
  \bibnamefont {Miller}},\ }\href {\doibase10.1016/j.physrep.2014.09.003}
  {\bibfield  {journal} {\bibinfo  {journal} {Phys. Rept.}\ }\textbf {\bibinfo
  {volume} {548}},\ \bibinfo {pages} {1--34} (\bibinfo {year} {2014})},\
  \Eprint {http://arxiv.org/abs/1408.4145} {arXiv:1408.4145
  [astro-ph.HE]}\BibitemShut {NoStop}%
\bibitem [{\citenamefont {Hartung}\ and\ \citenamefont
  {Steinhoff}(2011{\natexlab{a}})}]{Hartung:2011te}%
  \BibitemOpen
  \bibfield  {author} {\bibinfo {author} {\bibfnamefont {Johannes}\
  \bibnamefont {Hartung}}\ and\ \bibinfo {author} {\bibfnamefont {Jan}\
  \bibnamefont {Steinhoff}},\ }\href {\doibase10.1002/andp.201100094}
  {\bibfield  {journal} {\bibinfo  {journal} {Annalen Phys.}\ }\textbf
  {\bibinfo {volume} {523}},\ \bibinfo {pages} {783--790} (\bibinfo {year}
  {2011}{\natexlab{a}})},\ \Eprint {http://arxiv.org/abs/1104.3079}
  {arXiv:1104.3079 [gr-qc]}\BibitemShut {NoStop}%
\bibitem [{\citenamefont {Hartung}\ \emph {et~al.}(2013)\citenamefont
  {Hartung}, \citenamefont {Steinhoff},\ and\ \citenamefont
  {Schafer}}]{Hartung:2013dza}%
  \BibitemOpen
  \bibfield  {author} {\bibinfo {author} {\bibfnamefont {Johannes}\
  \bibnamefont {Hartung}}, \bibinfo {author} {\bibfnamefont {Jan}\ \bibnamefont
  {Steinhoff}}, \ and\ \bibinfo {author} {\bibfnamefont {Gerhard}\ \bibnamefont
  {Schafer}},\ }\href {\doibase10.1002/andp.201200271} {\bibfield  {journal}
  {\bibinfo  {journal} {Annalen Phys.}\ }\textbf {\bibinfo {volume} {525}},\
  \bibinfo {pages} {359--394} (\bibinfo {year} {2013})},\ \Eprint
  {http://arxiv.org/abs/1302.6723} {arXiv:1302.6723 [gr-qc]}\BibitemShut
  {NoStop}%
\bibitem [{\citenamefont {Marsat}\ \emph {et~al.}(2013)\citenamefont {Marsat},
  \citenamefont {Bohe}, \citenamefont {Faye},\ and\ \citenamefont
  {Blanchet}}]{Marsat:2012fn}%
  \BibitemOpen
  \bibfield  {author} {\bibinfo {author} {\bibfnamefont {Sylvain}\ \bibnamefont
  {Marsat}}, \bibinfo {author} {\bibfnamefont {Alejandro}\ \bibnamefont
  {Bohe}}, \bibinfo {author} {\bibfnamefont {Guillaume}\ \bibnamefont {Faye}},
  \ and\ \bibinfo {author} {\bibfnamefont {Luc}\ \bibnamefont {Blanchet}},\
  }\href {\doibase10.1088/0264-9381/30/5/055007} {\bibfield  {journal}
  {\bibinfo  {journal} {Class. Quant. Grav.}\ }\textbf {\bibinfo {volume}
  {30}},\ \bibinfo {pages} {055007} (\bibinfo {year} {2013})},\ \Eprint
  {http://arxiv.org/abs/1210.4143} {arXiv:1210.4143 [gr-qc]}\BibitemShut
  {NoStop}%
\bibitem [{\citenamefont {Bohe}\ \emph {et~al.}(2013)\citenamefont {Bohe},
  \citenamefont {Marsat}, \citenamefont {Faye},\ and\ \citenamefont
  {Blanchet}}]{Bohe:2012mr}%
  \BibitemOpen
  \bibfield  {author} {\bibinfo {author} {\bibfnamefont {Alejandro}\
  \bibnamefont {Bohe}}, \bibinfo {author} {\bibfnamefont {Sylvain}\
  \bibnamefont {Marsat}}, \bibinfo {author} {\bibfnamefont {Guillaume}\
  \bibnamefont {Faye}}, \ and\ \bibinfo {author} {\bibfnamefont {Luc}\
  \bibnamefont {Blanchet}},\ }\href {\doibase10.1088/0264-9381/30/7/075017}
  {\bibfield  {journal} {\bibinfo  {journal} {Class. Quant. Grav.}\ }\textbf
  {\bibinfo {volume} {30}},\ \bibinfo {pages} {075017} (\bibinfo {year}
  {2013})},\ \Eprint {http://arxiv.org/abs/1212.5520} {arXiv:1212.5520
  [gr-qc]}\BibitemShut {NoStop}%
\bibitem [{\citenamefont {Levi}\ and\ \citenamefont
  {Steinhoff}(2016{\natexlab{a}})}]{Levi:2015uxa}%
  \BibitemOpen
  \bibfield  {author} {\bibinfo {author} {\bibfnamefont {Michele}\ \bibnamefont
  {Levi}}\ and\ \bibinfo {author} {\bibfnamefont {Jan}\ \bibnamefont
  {Steinhoff}},\ }\href {\doibase10.1088/1475-7516/2016/01/011} {\bibfield
  {journal} {\bibinfo  {journal} {JCAP}\ }\textbf {\bibinfo {volume} {1601}},\
  \bibinfo {pages} {011} (\bibinfo {year} {2016}{\natexlab{a}})},\ \Eprint
  {http://arxiv.org/abs/1506.05056} {arXiv:1506.05056 [gr-qc]}\BibitemShut
  {NoStop}%
\bibitem [{\citenamefont {Hartung}\ and\ \citenamefont
  {Steinhoff}(2011{\natexlab{b}})}]{Hartung:2011ea}%
  \BibitemOpen
  \bibfield  {author} {\bibinfo {author} {\bibfnamefont {Johannes}\
  \bibnamefont {Hartung}}\ and\ \bibinfo {author} {\bibfnamefont {Jan}\
  \bibnamefont {Steinhoff}},\ }\href {\doibase10.1002/andp.201100163}
  {\bibfield  {journal} {\bibinfo  {journal} {Annalen Phys.}\ }\textbf
  {\bibinfo {volume} {523}},\ \bibinfo {pages} {919--924} (\bibinfo {year}
  {2011}{\natexlab{b}})},\ \Eprint {http://arxiv.org/abs/1107.4294}
  {arXiv:1107.4294 [gr-qc]}\BibitemShut {NoStop}%
\bibitem [{\citenamefont {Levi}(2012)}]{Levi:2011eq}%
  \BibitemOpen
  \bibfield  {author} {\bibinfo {author} {\bibfnamefont {Michele}\ \bibnamefont
  {Levi}},\ }\href {\doibase10.1103/PhysRevD.85.064043} {\bibfield  {journal}
  {\bibinfo  {journal} {Phys. Rev.}\ }\textbf {\bibinfo {volume} {D85}},\
  \bibinfo {pages} {064043} (\bibinfo {year} {2012})},\ \Eprint
  {http://arxiv.org/abs/1107.4322} {arXiv:1107.4322 [gr-qc]}\BibitemShut
  {NoStop}%
\bibitem [{\citenamefont {Levi}\ and\ \citenamefont
  {Steinhoff}(2014)}]{Levi:2014sba}%
  \BibitemOpen
  \bibfield  {author} {\bibinfo {author} {\bibfnamefont {Michele}\ \bibnamefont
  {Levi}}\ and\ \bibinfo {author} {\bibfnamefont {Jan}\ \bibnamefont
  {Steinhoff}},\ }\href {\doibase10.1088/1475-7516/2014/12/003} {\bibfield
  {journal} {\bibinfo  {journal} {JCAP}\ }\textbf {\bibinfo {volume} {1412}},\
  \bibinfo {pages} {003} (\bibinfo {year} {2014})},\ \Eprint
  {http://arxiv.org/abs/1408.5762} {arXiv:1408.5762 [gr-qc]}\BibitemShut
  {NoStop}%
\bibitem [{\citenamefont {Levi}\ and\ \citenamefont
  {Steinhoff}(2016{\natexlab{b}})}]{Levi:2015ixa}%
  \BibitemOpen
  \bibfield  {author} {\bibinfo {author} {\bibfnamefont {Michele}\ \bibnamefont
  {Levi}}\ and\ \bibinfo {author} {\bibfnamefont {Jan}\ \bibnamefont
  {Steinhoff}},\ }\href {\doibase10.1088/1475-7516/2016/01/008} {\bibfield
  {journal} {\bibinfo  {journal} {JCAP}\ }\textbf {\bibinfo {volume} {1601}},\
  \bibinfo {pages} {008} (\bibinfo {year} {2016}{\natexlab{b}})},\ \Eprint
  {http://arxiv.org/abs/1506.05794} {arXiv:1506.05794 [gr-qc]}\BibitemShut
  {NoStop}%
\bibitem [{\citenamefont {Levi}\ and\ \citenamefont
  {Steinhoff}(2016{\natexlab{c}})}]{Levi:2016ofk}%
  \BibitemOpen
  \bibfield  {author} {\bibinfo {author} {\bibfnamefont {Michele}\ \bibnamefont
  {Levi}}\ and\ \bibinfo {author} {\bibfnamefont {Jan}\ \bibnamefont
  {Steinhoff}},\ }\href@noop {} {\  (\bibinfo {year} {2016}{\natexlab{c}})},\
  \Eprint {http://arxiv.org/abs/1607.04252} {arXiv:1607.04252
  [gr-qc]}\BibitemShut {NoStop}%
\bibitem [{\citenamefont {Hergt}\ and\ \citenamefont
  {Schaefer}(2008{\natexlab{a}})}]{Hergt:2007ha}%
  \BibitemOpen
  \bibfield  {author} {\bibinfo {author} {\bibfnamefont {Steven}\ \bibnamefont
  {Hergt}}\ and\ \bibinfo {author} {\bibfnamefont {Gerhard}\ \bibnamefont
  {Schaefer}},\ }\href {\doibase10.1103/PhysRevD.77.104001} {\bibfield
  {journal} {\bibinfo  {journal} {Phys. Rev.}\ }\textbf {\bibinfo {volume}
  {D77}},\ \bibinfo {pages} {104001} (\bibinfo {year} {2008}{\natexlab{a}})},\
  \Eprint {http://arxiv.org/abs/0712.1515} {arXiv:0712.1515
  [gr-qc]}\BibitemShut {NoStop}%
\bibitem [{\citenamefont {Hergt}\ and\ \citenamefont
  {Schaefer}(2008{\natexlab{b}})}]{Hergt:2008jn}%
  \BibitemOpen
  \bibfield  {author} {\bibinfo {author} {\bibfnamefont {Steven}\ \bibnamefont
  {Hergt}}\ and\ \bibinfo {author} {\bibfnamefont {Gerhard}\ \bibnamefont
  {Schaefer}},\ }\href {\doibase10.1103/PhysRevD.78.124004} {\bibfield
  {journal} {\bibinfo  {journal} {Phys. Rev.}\ }\textbf {\bibinfo {volume}
  {D78}},\ \bibinfo {pages} {124004} (\bibinfo {year} {2008}{\natexlab{b}})},\
  \Eprint {http://arxiv.org/abs/0809.2208} {arXiv:0809.2208
  [gr-qc]}\BibitemShut {NoStop}%
\bibitem [{\citenamefont {Marsat}(2015)}]{Marsat:2014xea}%
  \BibitemOpen
  \bibfield  {author} {\bibinfo {author} {\bibfnamefont {Sylvain}\ \bibnamefont
  {Marsat}},\ }\href {\doibase10.1088/0264-9381/32/8/085008} {\bibfield
  {journal} {\bibinfo  {journal} {Class. Quant. Grav.}\ }\textbf {\bibinfo
  {volume} {32}},\ \bibinfo {pages} {085008} (\bibinfo {year} {2015})},\
  \Eprint {http://arxiv.org/abs/1411.4118} {arXiv:1411.4118
  [gr-qc]}\BibitemShut {NoStop}%
\bibitem [{\citenamefont {Levi}\ and\ \citenamefont
  {Steinhoff}(2015)}]{Levi:2014gsa}%
  \BibitemOpen
  \bibfield  {author} {\bibinfo {author} {\bibfnamefont {Michele}\ \bibnamefont
  {Levi}}\ and\ \bibinfo {author} {\bibfnamefont {Jan}\ \bibnamefont
  {Steinhoff}},\ }\href {\doibase10.1007/JHEP06(2015)059} {\bibfield  {journal}
  {\bibinfo  {journal} {JHEP}\ }\textbf {\bibinfo {volume} {06}},\ \bibinfo
  {pages} {059} (\bibinfo {year} {2015})},\ \Eprint
  {http://arxiv.org/abs/1410.2601} {arXiv:1410.2601 [gr-qc]}\BibitemShut
  {NoStop}%
\bibitem [{\citenamefont {Vaidya}(2015)}]{Vaidya:2014kza}%
  \BibitemOpen
  \bibfield  {author} {\bibinfo {author} {\bibfnamefont {Varun}\ \bibnamefont
  {Vaidya}},\ }\href {\doibase10.1103/PhysRevD.91.024017} {\bibfield  {journal}
  {\bibinfo  {journal} {Phys. Rev.}\ }\textbf {\bibinfo {volume} {D91}},\
  \bibinfo {pages} {024017} (\bibinfo {year} {2015})},\ \Eprint
  {http://arxiv.org/abs/1410.5348} {arXiv:1410.5348 [hep-th]}\BibitemShut
  {NoStop}%
\bibitem [{\citenamefont {Vines}\ and\ \citenamefont
  {Steinhoff}(2018)}]{Vines:2016qwa}%
  \BibitemOpen
  \bibfield  {author} {\bibinfo {author} {\bibfnamefont {Justin}\ \bibnamefont
  {Vines}}\ and\ \bibinfo {author} {\bibfnamefont {Jan}\ \bibnamefont
  {Steinhoff}},\ }\href {\doibase10.1103/PhysRevD.97.064010} {\bibfield
  {journal} {\bibinfo  {journal} {Phys. Rev.}\ }\textbf {\bibinfo {volume}
  {D97}},\ \bibinfo {pages} {064010} (\bibinfo {year} {2018})},\ \Eprint
  {http://arxiv.org/abs/1606.08832} {arXiv:1606.08832 [gr-qc]}\BibitemShut
  {NoStop}%
\bibitem [{\citenamefont {Gerosa}\ \emph
  {et~al.}(2015{\natexlab{a}})\citenamefont {Gerosa}, \citenamefont {Kesden},
  \citenamefont {Sperhake}, \citenamefont {Berti},\ and\ \citenamefont
  {O'Shaughnessy}}]{Gerosa:2015tea}%
  \BibitemOpen
  \bibfield  {author} {\bibinfo {author} {\bibfnamefont {Davide}\ \bibnamefont
  {Gerosa}}, \bibinfo {author} {\bibfnamefont {Michael}\ \bibnamefont
  {Kesden}}, \bibinfo {author} {\bibfnamefont {Ulrich}\ \bibnamefont
  {Sperhake}}, \bibinfo {author} {\bibfnamefont {Emanuele}\ \bibnamefont
  {Berti}}, \ and\ \bibinfo {author} {\bibfnamefont {Richard}\ \bibnamefont
  {O'Shaughnessy}},\ }\href {\doibase10.1103/PhysRevD.92.064016} {\bibfield
  {journal} {\bibinfo  {journal} {Phys. Rev.}\ }\textbf {\bibinfo {volume}
  {D92}},\ \bibinfo {pages} {064016} (\bibinfo {year} {2015}{\natexlab{a}})},\
  \Eprint {http://arxiv.org/abs/1506.03492} {arXiv:1506.03492
  [gr-qc]}\BibitemShut {NoStop}%
\bibitem [{\citenamefont {Kesden}\ \emph {et~al.}(2015)\citenamefont {Kesden},
  \citenamefont {Gerosa}, \citenamefont {O'Shaughnessy}, \citenamefont
  {Berti},\ and\ \citenamefont {Sperhake}}]{Kesden:2014sla}%
  \BibitemOpen
  \bibfield  {author} {\bibinfo {author} {\bibfnamefont {Michael}\ \bibnamefont
  {Kesden}}, \bibinfo {author} {\bibfnamefont {Davide}\ \bibnamefont {Gerosa}},
  \bibinfo {author} {\bibfnamefont {Richard}\ \bibnamefont {O'Shaughnessy}},
  \bibinfo {author} {\bibfnamefont {Emanuele}\ \bibnamefont {Berti}}, \ and\
  \bibinfo {author} {\bibfnamefont {Ulrich}\ \bibnamefont {Sperhake}},\ }\href
  {\doibase10.1103/PhysRevLett.114.081103} {\bibfield  {journal} {\bibinfo
  {journal} {Phys. Rev. Lett.}\ }\textbf {\bibinfo {volume} {114}},\ \bibinfo
  {pages} {081103} (\bibinfo {year} {2015})},\ \Eprint
  {http://arxiv.org/abs/1411.0674} {arXiv:1411.0674 [gr-qc]}\BibitemShut
  {NoStop}%
\bibitem [{\citenamefont {Gerosa}\ \emph
  {et~al.}(2015{\natexlab{b}})\citenamefont {Gerosa}, \citenamefont {Kesden},
  \citenamefont {O'Shaughnessy}, \citenamefont {Klein}, \citenamefont {Berti},
  \citenamefont {Sperhake},\ and\ \citenamefont {Trifirò}}]{Gerosa:2015hba}%
  \BibitemOpen
  \bibfield  {author} {\bibinfo {author} {\bibfnamefont {Davide}\ \bibnamefont
  {Gerosa}}, \bibinfo {author} {\bibfnamefont {Michael}\ \bibnamefont
  {Kesden}}, \bibinfo {author} {\bibfnamefont {Richard}\ \bibnamefont
  {O'Shaughnessy}}, \bibinfo {author} {\bibfnamefont {Antoine}\ \bibnamefont
  {Klein}}, \bibinfo {author} {\bibfnamefont {Emanuele}\ \bibnamefont {Berti}},
  \bibinfo {author} {\bibfnamefont {Ulrich}\ \bibnamefont {Sperhake}}, \ and\
  \bibinfo {author} {\bibfnamefont {Daniele}\ \bibnamefont {Trifirò}},\ }\href
  {\doibase10.1103/PhysRevLett.115.141102} {\bibfield  {journal} {\bibinfo
  {journal} {Phys. Rev. Lett.}\ }\textbf {\bibinfo {volume} {115}},\ \bibinfo
  {pages} {141102} (\bibinfo {year} {2015}{\natexlab{b}})},\ \Eprint
  {http://arxiv.org/abs/1506.09116} {arXiv:1506.09116 [gr-qc]}\BibitemShut
  {NoStop}%
\bibitem [{\citenamefont {Zhao}\ \emph {et~al.}(2017)\citenamefont {Zhao},
  \citenamefont {Kesden},\ and\ \citenamefont {Gerosa}}]{Zhao:2017tro}%
  \BibitemOpen
  \bibfield  {author} {\bibinfo {author} {\bibfnamefont {Xinyu}\ \bibnamefont
  {Zhao}}, \bibinfo {author} {\bibfnamefont {Michael}\ \bibnamefont {Kesden}},
  \ and\ \bibinfo {author} {\bibfnamefont {Davide}\ \bibnamefont {Gerosa}},\
  }\href {\doibase10.1103/PhysRevD.96.024007} {\bibfield  {journal} {\bibinfo
  {journal} {Phys. Rev.}\ }\textbf {\bibinfo {volume} {D96}},\ \bibinfo {pages}
  {024007} (\bibinfo {year} {2017})},\ \Eprint
  {http://arxiv.org/abs/1705.02369} {arXiv:1705.02369 [gr-qc]}\BibitemShut
  {NoStop}%
\bibitem [{\citenamefont {Bohé}\ \emph {et~al.}(2013)\citenamefont {Bohé},
  \citenamefont {Marsat},\ and\ \citenamefont {Blanchet}}]{Bohe:2013cla}%
  \BibitemOpen
  \bibfield  {author} {\bibinfo {author} {\bibfnamefont {Alejandro}\
  \bibnamefont {Bohé}}, \bibinfo {author} {\bibfnamefont {Sylvain}\
  \bibnamefont {Marsat}}, \ and\ \bibinfo {author} {\bibfnamefont {Luc}\
  \bibnamefont {Blanchet}},\ }\href {\doibase10.1088/0264-9381/30/13/135009}
  {\bibfield  {journal} {\bibinfo  {journal} {Class. Quant. Grav.}\ }\textbf
  {\bibinfo {volume} {30}},\ \bibinfo {pages} {135009} (\bibinfo {year}
  {2013})},\ \Eprint {http://arxiv.org/abs/1303.7412} {arXiv:1303.7412
  [gr-qc]}\BibitemShut {NoStop}%
\bibitem [{\citenamefont {Marsat}\ \emph {et~al.}(2014)\citenamefont {Marsat},
  \citenamefont {Bohé}, \citenamefont {Blanchet},\ and\ \citenamefont
  {Buonanno}}]{Marsat:2013caa}%
  \BibitemOpen
  \bibfield  {author} {\bibinfo {author} {\bibfnamefont {Sylvain}\ \bibnamefont
  {Marsat}}, \bibinfo {author} {\bibfnamefont {Alejandro}\ \bibnamefont
  {Bohé}}, \bibinfo {author} {\bibfnamefont {Luc}\ \bibnamefont {Blanchet}}, \
  and\ \bibinfo {author} {\bibfnamefont {Alessandra}\ \bibnamefont
  {Buonanno}},\ }\href {\doibase10.1088/0264-9381/31/2/025023} {\bibfield
  {journal} {\bibinfo  {journal} {Class. Quant. Grav.}\ }\textbf {\bibinfo
  {volume} {31}},\ \bibinfo {pages} {025023} (\bibinfo {year} {2014})},\
  \Eprint {http://arxiv.org/abs/1307.6793} {arXiv:1307.6793
  [gr-qc]}\BibitemShut {NoStop}%
\bibitem [{\citenamefont {Buonanno}\ \emph {et~al.}(2013)\citenamefont
  {Buonanno}, \citenamefont {Faye},\ and\ \citenamefont
  {Hinderer}}]{Buonanno:2012rv}%
  \BibitemOpen
  \bibfield  {author} {\bibinfo {author} {\bibfnamefont {Alessandra}\
  \bibnamefont {Buonanno}}, \bibinfo {author} {\bibfnamefont {Guillaume}\
  \bibnamefont {Faye}}, \ and\ \bibinfo {author} {\bibfnamefont {Tanja}\
  \bibnamefont {Hinderer}},\ }\href {\doibase10.1103/PhysRevD.87.044009}
  {\bibfield  {journal} {\bibinfo  {journal} {Phys. Rev.}\ }\textbf {\bibinfo
  {volume} {D87}},\ \bibinfo {pages} {044009} (\bibinfo {year} {2013})},\
  \Eprint {http://arxiv.org/abs/1209.6349} {arXiv:1209.6349
  [gr-qc]}\BibitemShut {NoStop}%
\bibitem [{\citenamefont {Mishra}\ \emph {et~al.}(2016)\citenamefont {Mishra},
  \citenamefont {Kela}, \citenamefont {Arun},\ and\ \citenamefont
  {Faye}}]{Mishra:2016whh}%
  \BibitemOpen
  \bibfield  {author} {\bibinfo {author} {\bibfnamefont {Chandra~Kant}\
  \bibnamefont {Mishra}}, \bibinfo {author} {\bibfnamefont {Aditya}\
  \bibnamefont {Kela}}, \bibinfo {author} {\bibfnamefont {K.~G.}\ \bibnamefont
  {Arun}}, \ and\ \bibinfo {author} {\bibfnamefont {Guillaume}\ \bibnamefont
  {Faye}},\ }\href {\doibase10.1103/PhysRevD.93.084054} {\bibfield  {journal}
  {\bibinfo  {journal} {Phys. Rev.}\ }\textbf {\bibinfo {volume} {D93}},\
  \bibinfo {pages} {084054} (\bibinfo {year} {2016})},\ \Eprint
  {http://arxiv.org/abs/1601.05588} {arXiv:1601.05588 [gr-qc]}\BibitemShut
  {NoStop}%
\bibitem [{\citenamefont {Porto}\ \emph {et~al.}(2011)\citenamefont {Porto},
  \citenamefont {Ross},\ and\ \citenamefont {Rothstein}}]{Porto:2010zg}%
  \BibitemOpen
  \bibfield  {author} {\bibinfo {author} {\bibfnamefont {Rafael~A.}\
  \bibnamefont {Porto}}, \bibinfo {author} {\bibfnamefont {Andreas}\
  \bibnamefont {Ross}}, \ and\ \bibinfo {author} {\bibfnamefont {Ira~Z.}\
  \bibnamefont {Rothstein}},\ }\href {\doibase10.1088/1475-7516/2011/03/009}
  {\bibfield  {journal} {\bibinfo  {journal} {JCAP}\ }\textbf {\bibinfo
  {volume} {1103}},\ \bibinfo {pages} {009} (\bibinfo {year} {2011})},\ \Eprint
  {http://arxiv.org/abs/1007.1312} {arXiv:1007.1312 [gr-qc]}\BibitemShut
  {NoStop}%
\bibitem [{\citenamefont {Bohé}\ \emph {et~al.}(2015)\citenamefont {Bohé},
  \citenamefont {Faye}, \citenamefont {Marsat},\ and\ \citenamefont
  {Porter}}]{Bohe:2015ana}%
  \BibitemOpen
  \bibfield  {author} {\bibinfo {author} {\bibfnamefont {Alejandro}\
  \bibnamefont {Bohé}}, \bibinfo {author} {\bibfnamefont {Guillaume}\
  \bibnamefont {Faye}}, \bibinfo {author} {\bibfnamefont {Sylvain}\
  \bibnamefont {Marsat}}, \ and\ \bibinfo {author} {\bibfnamefont {Edward~K.}\
  \bibnamefont {Porter}},\ }\href {\doibase10.1088/0264-9381/32/19/195010}
  {\bibfield  {journal} {\bibinfo  {journal} {Class. Quant. Grav.}\ }\textbf
  {\bibinfo {volume} {32}},\ \bibinfo {pages} {195010} (\bibinfo {year}
  {2015})},\ \Eprint {http://arxiv.org/abs/1501.01529} {arXiv:1501.01529
  [gr-qc]}\BibitemShut {NoStop}%
\bibitem [{\citenamefont {Porto}\ \emph {et~al.}(2012)\citenamefont {Porto},
  \citenamefont {Ross},\ and\ \citenamefont {Rothstein}}]{Porto:2012as}%
  \BibitemOpen
  \bibfield  {author} {\bibinfo {author} {\bibfnamefont {Rafael~A.}\
  \bibnamefont {Porto}}, \bibinfo {author} {\bibfnamefont {Andreas}\
  \bibnamefont {Ross}}, \ and\ \bibinfo {author} {\bibfnamefont {Ira~Z.}\
  \bibnamefont {Rothstein}},\ }\href {\doibase10.1088/1475-7516/2012/09/028}
  {\bibfield  {journal} {\bibinfo  {journal} {JCAP}\ }\textbf {\bibinfo
  {volume} {1209}},\ \bibinfo {pages} {028} (\bibinfo {year} {2012})},\ \Eprint
  {http://arxiv.org/abs/1203.2962} {arXiv:1203.2962 [gr-qc]}\BibitemShut
  {NoStop}%
\bibitem [{\citenamefont {Vines}(2018)}]{Vines:2017hyw}%
  \BibitemOpen
  \bibfield  {author} {\bibinfo {author} {\bibfnamefont {Justin}\ \bibnamefont
  {Vines}},\ }\href {\doibase10.1088/1361-6382/aaa3a8} {\bibfield  {journal}
  {\bibinfo  {journal} {Class. Quant. Grav.}\ }\textbf {\bibinfo {volume}
  {35}},\ \bibinfo {pages} {084002} (\bibinfo {year} {2018})},\ \Eprint
  {http://arxiv.org/abs/1709.06016} {arXiv:1709.06016 [gr-qc]}\BibitemShut
  {NoStop}%
\bibitem [{\citenamefont {Damour}(2010)}]{Damour:2009sm}%
  \BibitemOpen
  \bibfield  {author} {\bibinfo {author} {\bibfnamefont {Thibault}\
  \bibnamefont {Damour}},\ }\href {\doibase10.1103/PhysRevD.81.024017}
  {\bibfield  {journal} {\bibinfo  {journal} {Phys. Rev.}\ }\textbf {\bibinfo
  {volume} {D81}},\ \bibinfo {pages} {024017} (\bibinfo {year} {2010})},\
  \Eprint {http://arxiv.org/abs/0910.5533} {arXiv:0910.5533
  [gr-qc]}\BibitemShut {NoStop}%
\bibitem [{\citenamefont {Blanchet}\ \emph
  {et~al.}(2014{\natexlab{b}})\citenamefont {Blanchet}, \citenamefont {Faye},\
  and\ \citenamefont {Whiting}}]{Blanchet:2013txa}%
  \BibitemOpen
  \bibfield  {author} {\bibinfo {author} {\bibfnamefont {Luc}\ \bibnamefont
  {Blanchet}}, \bibinfo {author} {\bibfnamefont {Guillaume}\ \bibnamefont
  {Faye}}, \ and\ \bibinfo {author} {\bibfnamefont {Bernard~F.}\ \bibnamefont
  {Whiting}},\ }\href {\doibase10.1103/PhysRevD.89.064026} {\bibfield
  {journal} {\bibinfo  {journal} {Phys. Rev.}\ }\textbf {\bibinfo {volume}
  {D89}},\ \bibinfo {pages} {064026} (\bibinfo {year} {2014}{\natexlab{b}})},\
  \Eprint {http://arxiv.org/abs/1312.2975} {arXiv:1312.2975
  [gr-qc]}\BibitemShut {NoStop}%
\bibitem [{\citenamefont {Mano}\ and\ \citenamefont
  {Takasugi}(1997)}]{Mano:1996gn}%
  \BibitemOpen
  \bibfield  {author} {\bibinfo {author} {\bibfnamefont {Shuhei}\ \bibnamefont
  {Mano}}\ and\ \bibinfo {author} {\bibfnamefont {Eiichi}\ \bibnamefont
  {Takasugi}},\ }\href {\doibase10.1143/PTP.97.213} {\bibfield  {journal}
  {\bibinfo  {journal} {Prog. Theor. Phys.}\ }\textbf {\bibinfo {volume}
  {97}},\ \bibinfo {pages} {213--232} (\bibinfo {year} {1997})},\ \Eprint
  {http://arxiv.org/abs/gr-qc/9611014} {arXiv:gr-qc/9611014
  [gr-qc]}\BibitemShut {NoStop}%
\bibitem [{\citenamefont {Bini}\ and\ \citenamefont
  {Damour}(2014{\natexlab{b}})}]{Bini:2013rfa}%
  \BibitemOpen
  \bibfield  {author} {\bibinfo {author} {\bibfnamefont {Donato}\ \bibnamefont
  {Bini}}\ and\ \bibinfo {author} {\bibfnamefont {Thibault}\ \bibnamefont
  {Damour}},\ }\href {\doibase10.1103/PhysRevD.89.064063} {\bibfield  {journal}
  {\bibinfo  {journal} {Phys. Rev.}\ }\textbf {\bibinfo {volume} {D89}},\
  \bibinfo {pages} {064063} (\bibinfo {year} {2014}{\natexlab{b}})},\ \Eprint
  {http://arxiv.org/abs/1312.2503} {arXiv:1312.2503 [gr-qc]}\BibitemShut
  {NoStop}%
\bibitem [{\citenamefont {Bini}\ and\ \citenamefont
  {Damour}(2014{\natexlab{c}})}]{Bini:2014nfa}%
  \BibitemOpen
  \bibfield  {author} {\bibinfo {author} {\bibfnamefont {Donato}\ \bibnamefont
  {Bini}}\ and\ \bibinfo {author} {\bibfnamefont {Thibault}\ \bibnamefont
  {Damour}},\ }\href {\doibase10.1103/PhysRevD.89.104047} {\bibfield  {journal}
  {\bibinfo  {journal} {Phys. Rev.}\ }\textbf {\bibinfo {volume} {D89}},\
  \bibinfo {pages} {104047} (\bibinfo {year} {2014}{\natexlab{c}})},\ \Eprint
  {http://arxiv.org/abs/1403.2366} {arXiv:1403.2366 [gr-qc]}\BibitemShut
  {NoStop}%
\bibitem [{\citenamefont {Bini}\ and\ \citenamefont
  {Damour}(2014{\natexlab{d}})}]{Bini:2014ica}%
  \BibitemOpen
  \bibfield  {author} {\bibinfo {author} {\bibfnamefont {Donato}\ \bibnamefont
  {Bini}}\ and\ \bibinfo {author} {\bibfnamefont {Thibault}\ \bibnamefont
  {Damour}},\ }\href {\doibase10.1103/PhysRevD.90.024039} {\bibfield  {journal}
  {\bibinfo  {journal} {Phys. Rev.}\ }\textbf {\bibinfo {volume} {D90}},\
  \bibinfo {pages} {024039} (\bibinfo {year} {2014}{\natexlab{d}})},\ \Eprint
  {http://arxiv.org/abs/1404.2747} {arXiv:1404.2747 [gr-qc]}\BibitemShut
  {NoStop}%
\bibitem [{\citenamefont {Le~Tiec}\ \emph
  {et~al.}(2012{\natexlab{b}})\citenamefont {Le~Tiec}, \citenamefont
  {Blanchet},\ and\ \citenamefont {Whiting}}]{LeTiec:2011ab}%
  \BibitemOpen
  \bibfield  {author} {\bibinfo {author} {\bibfnamefont {Alexandre}\
  \bibnamefont {Le~Tiec}}, \bibinfo {author} {\bibfnamefont {Luc}\ \bibnamefont
  {Blanchet}}, \ and\ \bibinfo {author} {\bibfnamefont {Bernard~F.}\
  \bibnamefont {Whiting}},\ }\href {\doibase10.1103/PhysRevD.85.064039}
  {\bibfield  {journal} {\bibinfo  {journal} {Phys. Rev.}\ }\textbf {\bibinfo
  {volume} {D85}},\ \bibinfo {pages} {064039} (\bibinfo {year}
  {2012}{\natexlab{b}})},\ \Eprint {http://arxiv.org/abs/1111.5378}
  {arXiv:1111.5378 [gr-qc]}\BibitemShut {NoStop}%
\bibitem [{\citenamefont {Friedman}\ \emph {et~al.}(2002)\citenamefont
  {Friedman}, \citenamefont {Uryu},\ and\ \citenamefont
  {Shibata}}]{Friedman:2001pf}%
  \BibitemOpen
  \bibfield  {author} {\bibinfo {author} {\bibfnamefont {John~L.}\ \bibnamefont
  {Friedman}}, \bibinfo {author} {\bibfnamefont {Koji}\ \bibnamefont {Uryu}}, \
  and\ \bibinfo {author} {\bibfnamefont {Masaru}\ \bibnamefont {Shibata}},\
  }\href {\doibase10.1103/PhysRevD.65.064035} {\bibfield  {journal} {\bibinfo
  {journal} {Phys. Rev.}\ }\textbf {\bibinfo {volume} {D65}},\ \bibinfo {pages}
  {064035} (\bibinfo {year} {2002})},\ \bibinfo {note} {[Erratum: Phys.
  Rev.D70,129904(2004)]},\ \Eprint {http://arxiv.org/abs/gr-qc/0108070}
  {arXiv:gr-qc/0108070 [gr-qc]}\BibitemShut {NoStop}%
\bibitem [{\citenamefont {Blanchet}\ \emph {et~al.}(2013)\citenamefont
  {Blanchet}, \citenamefont {Buonanno},\ and\ \citenamefont
  {Le~Tiec}}]{Blanchet:2012at}%
  \BibitemOpen
  \bibfield  {author} {\bibinfo {author} {\bibfnamefont {Luc}\ \bibnamefont
  {Blanchet}}, \bibinfo {author} {\bibfnamefont {Alessandra}\ \bibnamefont
  {Buonanno}}, \ and\ \bibinfo {author} {\bibfnamefont {Alexandre}\
  \bibnamefont {Le~Tiec}},\ }\href {\doibase10.1103/PhysRevD.87.024030}
  {\bibfield  {journal} {\bibinfo  {journal} {Phys. Rev.}\ }\textbf {\bibinfo
  {volume} {D87}},\ \bibinfo {pages} {024030} (\bibinfo {year} {2013})},\
  \Eprint {http://arxiv.org/abs/1211.1060} {arXiv:1211.1060
  [gr-qc]}\BibitemShut {NoStop}%
\bibitem [{\citenamefont {Le~Tiec}(2015)}]{Tiec:2015cxa}%
  \BibitemOpen
  \bibfield  {author} {\bibinfo {author} {\bibfnamefont {Alexandre}\
  \bibnamefont {Le~Tiec}},\ }\href {\doibase10.1103/PhysRevD.92.084021}
  {\bibfield  {journal} {\bibinfo  {journal} {Phys. Rev.}\ }\textbf {\bibinfo
  {volume} {D92}},\ \bibinfo {pages} {084021} (\bibinfo {year} {2015})},\
  \Eprint {http://arxiv.org/abs/1506.05648} {arXiv:1506.05648
  [gr-qc]}\BibitemShut {NoStop}%
\bibitem [{\citenamefont {Blanchet}\ and\ \citenamefont
  {Le~Tiec}(2017)}]{Blanchet:2017rcn}%
  \BibitemOpen
  \bibfield  {author} {\bibinfo {author} {\bibfnamefont {Luc}\ \bibnamefont
  {Blanchet}}\ and\ \bibinfo {author} {\bibfnamefont {Alexandre}\ \bibnamefont
  {Le~Tiec}},\ }\href {\doibase10.1088/1361-6382/aa79d7} {\bibfield  {journal}
  {\bibinfo  {journal} {Class. Quant. Grav.}\ }\textbf {\bibinfo {volume}
  {34}},\ \bibinfo {pages} {164001} (\bibinfo {year} {2017})},\ \Eprint
  {http://arxiv.org/abs/1702.06839} {arXiv:1702.06839 [gr-qc]}\BibitemShut
  {NoStop}%
\bibitem [{\citenamefont {Gralla}\ and\ \citenamefont
  {Le~Tiec}(2013)}]{Gralla:2012dm}%
  \BibitemOpen
  \bibfield  {author} {\bibinfo {author} {\bibfnamefont {Samuel~E.}\
  \bibnamefont {Gralla}}\ and\ \bibinfo {author} {\bibfnamefont {Alexandre}\
  \bibnamefont {Le~Tiec}},\ }\href {\doibase10.1103/PhysRevD.88.044021}
  {\bibfield  {journal} {\bibinfo  {journal} {Phys. Rev.}\ }\textbf {\bibinfo
  {volume} {D88}},\ \bibinfo {pages} {044021} (\bibinfo {year} {2013})},\
  \Eprint {http://arxiv.org/abs/1210.8444} {arXiv:1210.8444
  [gr-qc]}\BibitemShut {NoStop}%
\bibitem [{\citenamefont {Le~Tiec}(2014)}]{Tiec:2013kua}%
  \BibitemOpen
  \bibfield  {author} {\bibinfo {author} {\bibfnamefont {Alexandre}\
  \bibnamefont {Le~Tiec}},\ }\href {\doibase10.1088/0264-9381/31/9/097001}
  {\bibfield  {journal} {\bibinfo  {journal} {Class. Quant. Grav.}\ }\textbf
  {\bibinfo {volume} {31}},\ \bibinfo {pages} {097001} (\bibinfo {year}
  {2014})},\ \Eprint {http://arxiv.org/abs/1311.3836} {arXiv:1311.3836
  [gr-qc]}\BibitemShut {NoStop}%
\bibitem [{\citenamefont {Fujita}\ \emph
  {et~al.}(2017{\natexlab{b}})\citenamefont {Fujita}, \citenamefont {Isoyama},
  \citenamefont {Le~Tiec}, \citenamefont {Nakano}, \citenamefont {Sago},\ and\
  \citenamefont {Tanaka}}]{Fujita:2016igj}%
  \BibitemOpen
  \bibfield  {author} {\bibinfo {author} {\bibfnamefont {Ryuichi}\ \bibnamefont
  {Fujita}}, \bibinfo {author} {\bibfnamefont {Soichiro}\ \bibnamefont
  {Isoyama}}, \bibinfo {author} {\bibfnamefont {Alexandre}\ \bibnamefont
  {Le~Tiec}}, \bibinfo {author} {\bibfnamefont {Hiroyuki}\ \bibnamefont
  {Nakano}}, \bibinfo {author} {\bibfnamefont {Norichika}\ \bibnamefont
  {Sago}}, \ and\ \bibinfo {author} {\bibfnamefont {Takahiro}\ \bibnamefont
  {Tanaka}},\ }\href {\doibase10.1088/1361-6382/aa7342} {\bibfield  {journal}
  {\bibinfo  {journal} {Class. Quant. Grav.}\ }\textbf {\bibinfo {volume}
  {34}},\ \bibinfo {pages} {134001} (\bibinfo {year} {2017}{\natexlab{b}})},\
  \Eprint {http://arxiv.org/abs/1612.02504} {arXiv:1612.02504
  [gr-qc]}\BibitemShut {NoStop}%
\bibitem [{\citenamefont {Barausse}\ \emph
  {et~al.}(2012{\natexlab{b}})\citenamefont {Barausse}, \citenamefont
  {Buonanno},\ and\ \citenamefont {Le~Tiec}}]{Barausse:2011dq}%
  \BibitemOpen
  \bibfield  {author} {\bibinfo {author} {\bibfnamefont {Enrico}\ \bibnamefont
  {Barausse}}, \bibinfo {author} {\bibfnamefont {Alessandra}\ \bibnamefont
  {Buonanno}}, \ and\ \bibinfo {author} {\bibfnamefont {Alexandre}\
  \bibnamefont {Le~Tiec}},\ }\href {\doibase10.1103/PhysRevD.85.064010}
  {\bibfield  {journal} {\bibinfo  {journal} {Phys. Rev.}\ }\textbf {\bibinfo
  {volume} {D85}},\ \bibinfo {pages} {064010} (\bibinfo {year}
  {2012}{\natexlab{b}})},\ \Eprint {http://arxiv.org/abs/1111.5610}
  {arXiv:1111.5610 [gr-qc]}\BibitemShut {NoStop}%
\bibitem [{\citenamefont {Marchand}\ \emph {et~al.}(2016)\citenamefont
  {Marchand}, \citenamefont {Blanchet},\ and\ \citenamefont
  {Faye}}]{Marchand:2016vox}%
  \BibitemOpen
  \bibfield  {author} {\bibinfo {author} {\bibfnamefont {Tanguy}\ \bibnamefont
  {Marchand}}, \bibinfo {author} {\bibfnamefont {Luc}\ \bibnamefont
  {Blanchet}}, \ and\ \bibinfo {author} {\bibfnamefont {Guillaume}\
  \bibnamefont {Faye}},\ }\href {\doibase10.1088/0264-9381/33/24/244003}
  {\bibfield  {journal} {\bibinfo  {journal} {Class. Quant. Grav.}\ }\textbf
  {\bibinfo {volume} {33}},\ \bibinfo {pages} {244003} (\bibinfo {year}
  {2016})},\ \Eprint {http://arxiv.org/abs/1607.07601} {arXiv:1607.07601
  [gr-qc]}\BibitemShut {NoStop}%
\bibitem [{\citenamefont {Abbott}\ \emph
  {et~al.}(2017{\natexlab{n}})\citenamefont {Abbott} \emph
  {et~al.}}]{Abbott:2016wiq}%
  \BibitemOpen
  \bibfield  {author} {\bibinfo {author} {\bibfnamefont {Benjamin~P.}\
  \bibnamefont {Abbott}} \emph {et~al.} (\bibinfo {collaboration} {Virgo, LIGO
  Scientific}),\ }\href {\doibase10.1088/1361-6382/aa6854} {\bibfield
  {journal} {\bibinfo  {journal} {Class. Quant. Grav.}\ }\textbf {\bibinfo
  {volume} {34}},\ \bibinfo {pages} {104002} (\bibinfo {year}
  {2017}{\natexlab{n}})},\ \Eprint {http://arxiv.org/abs/1611.07531}
  {arXiv:1611.07531 [gr-qc]}\BibitemShut {NoStop}%
\bibitem [{\citenamefont {Damour}\ \emph
  {et~al.}(2014{\natexlab{b}})\citenamefont {Damour}, \citenamefont
  {Guercilena}, \citenamefont {Hinder}, \citenamefont {Hopper}, \citenamefont
  {Nagar},\ and\ \citenamefont {Rezzolla}}]{Damour:2014afa}%
  \BibitemOpen
  \bibfield  {author} {\bibinfo {author} {\bibfnamefont {T.}~\bibnamefont
  {Damour}}, \bibinfo {author} {\bibfnamefont {F.}~\bibnamefont {Guercilena}},
  \bibinfo {author} {\bibfnamefont {I.}~\bibnamefont {Hinder}}, \bibinfo
  {author} {\bibfnamefont {S.}~\bibnamefont {Hopper}}, \bibinfo {author}
  {\bibfnamefont {A.}~\bibnamefont {Nagar}}, \ and\ \bibinfo {author}
  {\bibfnamefont {L.}~\bibnamefont {Rezzolla}},\ }\href
  {\doibase10.1103/PhysRevD.89.081503} {\bibfield  {journal} {\bibinfo
  {journal} {Phys. Rev. D}\ }\textbf {\bibinfo {volume} {89}},\ \bibinfo
  {pages} {081503} (\bibinfo {year} {2014}{\natexlab{b}})},\ \Eprint
  {http://arxiv.org/abs/1402.7307} {1402.7307}\BibitemShut {NoStop}%
\bibitem [{\citenamefont {Bini}\ and\ \citenamefont
  {Damour}(2018)}]{Bini:2018ywr}%
  \BibitemOpen
  \bibfield  {author} {\bibinfo {author} {\bibfnamefont {Donato}\ \bibnamefont
  {Bini}}\ and\ \bibinfo {author} {\bibfnamefont {Thibault}\ \bibnamefont
  {Damour}},\ }\href@noop {} {\  (\bibinfo {year} {2018})},\ \Eprint
  {http://arxiv.org/abs/1805.10809} {arXiv:1805.10809 [gr-qc]}\BibitemShut
  {NoStop}%
\bibitem [{\citenamefont {Bini}\ and\ \citenamefont
  {Damour}(2017{\natexlab{a}})}]{Bini:2017wfr}%
  \BibitemOpen
  \bibfield  {author} {\bibinfo {author} {\bibfnamefont {Donato}\ \bibnamefont
  {Bini}}\ and\ \bibinfo {author} {\bibfnamefont {Thibault}\ \bibnamefont
  {Damour}},\ }\href {\doibase10.1103/PhysRevD.96.064021} {\bibfield  {journal}
  {\bibinfo  {journal} {Phys. Rev.}\ }\textbf {\bibinfo {volume} {D96}},\
  \bibinfo {pages} {064021} (\bibinfo {year} {2017}{\natexlab{a}})},\ \Eprint
  {http://arxiv.org/abs/1706.06877} {arXiv:1706.06877 [gr-qc]}\BibitemShut
  {NoStop}%
\bibitem [{\citenamefont {Mirshekari}\ and\ \citenamefont
  {Will}(2013)}]{Mirshekari:2013vb}%
  \BibitemOpen
  \bibfield  {author} {\bibinfo {author} {\bibfnamefont {Saeed}\ \bibnamefont
  {Mirshekari}}\ and\ \bibinfo {author} {\bibfnamefont {Clifford~M.}\
  \bibnamefont {Will}},\ }\href {\doibase10.1103/PhysRevD.87.084070} {\bibfield
   {journal} {\bibinfo  {journal} {Phys. Rev.}\ }\textbf {\bibinfo {volume}
  {D87}},\ \bibinfo {pages} {084070} (\bibinfo {year} {2013})},\ \Eprint
  {http://arxiv.org/abs/1301.4680} {arXiv:1301.4680 [gr-qc]}\BibitemShut
  {NoStop}%
\bibitem [{\citenamefont {Lang}(2014)}]{Lang:2013fna}%
  \BibitemOpen
  \bibfield  {author} {\bibinfo {author} {\bibfnamefont {Ryan~N.}\ \bibnamefont
  {Lang}},\ }\href {\doibase10.1103/PhysRevD.89.084014} {\bibfield  {journal}
  {\bibinfo  {journal} {Phys. Rev.}\ }\textbf {\bibinfo {volume} {D89}},\
  \bibinfo {pages} {084014} (\bibinfo {year} {2014})},\ \Eprint
  {http://arxiv.org/abs/1310.3320} {arXiv:1310.3320 [gr-qc]}\BibitemShut
  {NoStop}%
\bibitem [{\citenamefont {Lang}(2015)}]{Lang:2014osa}%
  \BibitemOpen
  \bibfield  {author} {\bibinfo {author} {\bibfnamefont {Ryan~N.}\ \bibnamefont
  {Lang}},\ }\href {\doibase10.1103/PhysRevD.91.084027} {\bibfield  {journal}
  {\bibinfo  {journal} {Phys. Rev.}\ }\textbf {\bibinfo {volume} {D91}},\
  \bibinfo {pages} {084027} (\bibinfo {year} {2015})},\ \Eprint
  {http://arxiv.org/abs/1411.3073} {arXiv:1411.3073 [gr-qc]}\BibitemShut
  {NoStop}%
\bibitem [{\citenamefont {Sennett}\ \emph {et~al.}(2016)\citenamefont
  {Sennett}, \citenamefont {Marsat},\ and\ \citenamefont
  {Buonanno}}]{Sennett:2016klh}%
  \BibitemOpen
  \bibfield  {author} {\bibinfo {author} {\bibfnamefont {Noah}\ \bibnamefont
  {Sennett}}, \bibinfo {author} {\bibfnamefont {Sylvain}\ \bibnamefont
  {Marsat}}, \ and\ \bibinfo {author} {\bibfnamefont {Alessandra}\ \bibnamefont
  {Buonanno}},\ }\href {\doibase10.1103/PhysRevD.94.084003} {\bibfield
  {journal} {\bibinfo  {journal} {Phys. Rev.}\ }\textbf {\bibinfo {volume}
  {D94}},\ \bibinfo {pages} {084003} (\bibinfo {year} {2016})},\ \Eprint
  {http://arxiv.org/abs/1607.01420} {arXiv:1607.01420 [gr-qc]}\BibitemShut
  {NoStop}%
\bibitem [{\citenamefont {Bernard}(2018)}]{Bernard:2018hta}%
  \BibitemOpen
  \bibfield  {author} {\bibinfo {author} {\bibfnamefont {L.}~\bibnamefont
  {Bernard}},\ }\href@noop {} {\  (\bibinfo {year} {2018})},\ \Eprint
  {http://arxiv.org/abs/1802.10201} {arXiv:1802.10201 [gr-qc]}\BibitemShut
  {NoStop}%
\bibitem [{\citenamefont {Yagi}\ \emph
  {et~al.}(2012{\natexlab{b}})\citenamefont {Yagi}, \citenamefont {Stein},
  \citenamefont {Yunes},\ and\ \citenamefont {Tanaka}}]{Yagi:2011xp}%
  \BibitemOpen
  \bibfield  {author} {\bibinfo {author} {\bibfnamefont {Kent}\ \bibnamefont
  {Yagi}}, \bibinfo {author} {\bibfnamefont {Leo~C.}\ \bibnamefont {Stein}},
  \bibinfo {author} {\bibfnamefont {Nicolás}\ \bibnamefont {Yunes}}, \ and\
  \bibinfo {author} {\bibfnamefont {Takahiro}\ \bibnamefont {Tanaka}},\ }\href
  {\doibase10.1103/PhysRevD.85.064022} {\bibfield  {journal} {\bibinfo
  {journal} {Phys. Rev.}\ }\textbf {\bibinfo {volume} {D85}},\ \bibinfo {pages}
  {064022} (\bibinfo {year} {2012}{\natexlab{b}})},\ \bibinfo {note} {[Erratum:
  Phys. Rev.D93,no.2,029902(2016)]},\ \Eprint {http://arxiv.org/abs/1110.5950}
  {arXiv:1110.5950 [gr-qc]}\BibitemShut {NoStop}%
\bibitem [{\citenamefont {Yagi}\ \emph {et~al.}(2013)\citenamefont {Yagi},
  \citenamefont {Stein}, \citenamefont {Yunes},\ and\ \citenamefont
  {Tanaka}}]{Yagi:2013mbt}%
  \BibitemOpen
  \bibfield  {author} {\bibinfo {author} {\bibfnamefont {Kent}\ \bibnamefont
  {Yagi}}, \bibinfo {author} {\bibfnamefont {Leo~C.}\ \bibnamefont {Stein}},
  \bibinfo {author} {\bibfnamefont {Nicolas}\ \bibnamefont {Yunes}}, \ and\
  \bibinfo {author} {\bibfnamefont {Takahiro}\ \bibnamefont {Tanaka}},\ }\href
  {\doibase10.1103/PhysRevD.87.084058} {\bibfield  {journal} {\bibinfo
  {journal} {Phys. Rev.}\ }\textbf {\bibinfo {volume} {D87}},\ \bibinfo {pages}
  {084058} (\bibinfo {year} {2013})},\ \bibinfo {note} {[Erratum: Phys.
  Rev.D93,no.8,089909(2016)]},\ \Eprint {http://arxiv.org/abs/1302.1918}
  {arXiv:1302.1918 [gr-qc]}\BibitemShut {NoStop}%
\bibitem [{\citenamefont {Damour}(2016)}]{Damour:2016gwp}%
  \BibitemOpen
  \bibfield  {author} {\bibinfo {author} {\bibfnamefont {Thibault}\
  \bibnamefont {Damour}},\ }\href {\doibase10.1103/PhysRevD.94.104015}
  {\bibfield  {journal} {\bibinfo  {journal} {Phys. Rev.}\ }\textbf {\bibinfo
  {volume} {D94}},\ \bibinfo {pages} {104015} (\bibinfo {year} {2016})},\
  \Eprint {http://arxiv.org/abs/1609.00354} {arXiv:1609.00354
  [gr-qc]}\BibitemShut {NoStop}%
\bibitem [{\citenamefont {Bini}\ and\ \citenamefont
  {Damour}(2017{\natexlab{b}})}]{Bini:2017xzy}%
  \BibitemOpen
  \bibfield  {author} {\bibinfo {author} {\bibfnamefont {Donato}\ \bibnamefont
  {Bini}}\ and\ \bibinfo {author} {\bibfnamefont {Thibault}\ \bibnamefont
  {Damour}},\ }\href {\doibase10.1103/PhysRevD.96.104038} {\bibfield  {journal}
  {\bibinfo  {journal} {Phys. Rev.}\ }\textbf {\bibinfo {volume} {D96}},\
  \bibinfo {pages} {104038} (\bibinfo {year} {2017}{\natexlab{b}})},\ \Eprint
  {http://arxiv.org/abs/1709.00590} {arXiv:1709.00590 [gr-qc]}\BibitemShut
  {NoStop}%
\bibitem [{\citenamefont {Damour}(2018)}]{Damour:2017zjx}%
  \BibitemOpen
  \bibfield  {author} {\bibinfo {author} {\bibfnamefont {Thibault}\
  \bibnamefont {Damour}},\ }\href {\doibase10.1103/PhysRevD.97.044038}
  {\bibfield  {journal} {\bibinfo  {journal} {Phys. Rev.}\ }\textbf {\bibinfo
  {volume} {D97}},\ \bibinfo {pages} {044038} (\bibinfo {year} {2018})},\
  \Eprint {http://arxiv.org/abs/1710.10599} {arXiv:1710.10599
  [gr-qc]}\BibitemShut {NoStop}%
\bibitem [{\citenamefont {Pretorius}(2005{\natexlab{a}})}]{Pretorius:2005gq}%
  \BibitemOpen
  \bibfield  {author} {\bibinfo {author} {\bibfnamefont {Frans}\ \bibnamefont
  {Pretorius}},\ }\href {\doibase10.1103/PhysRevLett.95.121101} {\bibfield
  {journal} {\bibinfo  {journal} {Phys. Rev. Lett.}\ }\textbf {\bibinfo
  {volume} {95}},\ \bibinfo {pages} {121101} (\bibinfo {year}
  {2005}{\natexlab{a}})},\ \Eprint {http://arxiv.org/abs/gr-qc/0507014}
  {arXiv:gr-qc/0507014 [gr-qc]}\BibitemShut {NoStop}%
\bibitem [{\citenamefont {Campanelli}\ \emph
  {et~al.}(2006{\natexlab{a}})\citenamefont {Campanelli}, \citenamefont
  {Lousto}, \citenamefont {Marronetti},\ and\ \citenamefont
  {Zlochower}}]{Campanelli:2005dd}%
  \BibitemOpen
  \bibfield  {author} {\bibinfo {author} {\bibfnamefont {Manuela}\ \bibnamefont
  {Campanelli}}, \bibinfo {author} {\bibfnamefont {C.~O.}\ \bibnamefont
  {Lousto}}, \bibinfo {author} {\bibfnamefont {P.}~\bibnamefont {Marronetti}},
  \ and\ \bibinfo {author} {\bibfnamefont {Y.}~\bibnamefont {Zlochower}},\
  }\href {\doibase10.1103/PhysRevLett.96.111101} {\bibfield  {journal}
  {\bibinfo  {journal} {Phys. Rev. Lett.}\ }\textbf {\bibinfo {volume} {96}},\
  \bibinfo {pages} {111101} (\bibinfo {year} {2006}{\natexlab{a}})},\ \Eprint
  {http://arxiv.org/abs/gr-qc/0511048} {arXiv:gr-qc/0511048
  [gr-qc]}\BibitemShut {NoStop}%
\bibitem [{\citenamefont {Baker}\ \emph
  {et~al.}(2006{\natexlab{a}})\citenamefont {Baker}, \citenamefont {Centrella},
  \citenamefont {Choi}, \citenamefont {Koppitz},\ and\ \citenamefont {van
  Meter}}]{Baker:2005vv}%
  \BibitemOpen
  \bibfield  {author} {\bibinfo {author} {\bibfnamefont {John~G.}\ \bibnamefont
  {Baker}}, \bibinfo {author} {\bibfnamefont {Joan}\ \bibnamefont {Centrella}},
  \bibinfo {author} {\bibfnamefont {Dae-Il}\ \bibnamefont {Choi}}, \bibinfo
  {author} {\bibfnamefont {Michael}\ \bibnamefont {Koppitz}}, \ and\ \bibinfo
  {author} {\bibfnamefont {James}\ \bibnamefont {van Meter}},\ }\href
  {\doibase10.1103/PhysRevLett.96.111102} {\bibfield  {journal} {\bibinfo
  {journal} {Phys. Rev. Lett.}\ }\textbf {\bibinfo {volume} {96}},\ \bibinfo
  {pages} {111102} (\bibinfo {year} {2006}{\natexlab{a}})},\ \Eprint
  {http://arxiv.org/abs/gr-qc/0511103} {arXiv:gr-qc/0511103
  [gr-qc]}\BibitemShut {NoStop}%
\bibitem [{\citenamefont {Sperhake}(2015)}]{Sperhake:2014wpa}%
  \BibitemOpen
  \bibfield  {author} {\bibinfo {author} {\bibfnamefont {Ulrich}\ \bibnamefont
  {Sperhake}},\ }\href {\doibase10.1088/0264-9381/32/12/124011} {\bibfield
  {journal} {\bibinfo  {journal} {Class. Quant. Grav.}\ }\textbf {\bibinfo
  {volume} {32}},\ \bibinfo {pages} {124011} (\bibinfo {year} {2015})},\
  \Eprint {http://arxiv.org/abs/1411.3997} {arXiv:1411.3997
  [gr-qc]}\BibitemShut {NoStop}%
\bibitem [{\citenamefont {Campanelli}\ \emph
  {et~al.}(2006{\natexlab{b}})\citenamefont {Campanelli}, \citenamefont
  {Lousto},\ and\ \citenamefont {Zlochower}}]{Campanelli:2006uy}%
  \BibitemOpen
  \bibfield  {author} {\bibinfo {author} {\bibfnamefont {Manuela}\ \bibnamefont
  {Campanelli}}, \bibinfo {author} {\bibfnamefont {C.~O.}\ \bibnamefont
  {Lousto}}, \ and\ \bibinfo {author} {\bibfnamefont {Y.}~\bibnamefont
  {Zlochower}},\ }\href {\doibase10.1103/PhysRevD.74.041501} {\bibfield
  {journal} {\bibinfo  {journal} {Phys. Rev. D}\ }\textbf {\bibinfo {volume}
  {74}},\ \bibinfo {pages} {041501} (\bibinfo {year} {2006}{\natexlab{b}})},\
  \Eprint {http://arxiv.org/abs/gr-qc/0604012} {arXiv:gr-qc/0604012
  [gr-qc]}\BibitemShut {NoStop}%
\bibitem [{\citenamefont {Hannam}\ \emph
  {et~al.}(2008{\natexlab{a}})\citenamefont {Hannam}, \citenamefont {Husa},
  \citenamefont {Bruegmann},\ and\ \citenamefont {Gopakumar}}]{Hannam:2007wf}%
  \BibitemOpen
  \bibfield  {author} {\bibinfo {author} {\bibfnamefont {Mark}\ \bibnamefont
  {Hannam}}, \bibinfo {author} {\bibfnamefont {Sascha}\ \bibnamefont {Husa}},
  \bibinfo {author} {\bibfnamefont {Bernd}\ \bibnamefont {Bruegmann}}, \ and\
  \bibinfo {author} {\bibfnamefont {Achamveedu}\ \bibnamefont {Gopakumar}},\
  }\href {\doibase10.1103/PhysRevD.78.104007} {\bibfield  {journal} {\bibinfo
  {journal} {Phys. Rev.}\ }\textbf {\bibinfo {volume} {D78}},\ \bibinfo {pages}
  {104007} (\bibinfo {year} {2008}{\natexlab{a}})},\ \Eprint
  {http://arxiv.org/abs/0712.3787} {arXiv:0712.3787 [gr-qc]}\BibitemShut
  {NoStop}%
\bibitem [{\citenamefont {Dain}\ \emph {et~al.}(2008)\citenamefont {Dain},
  \citenamefont {Lousto},\ and\ \citenamefont {Zlochower}}]{Dain:2008ck}%
  \BibitemOpen
  \bibfield  {author} {\bibinfo {author} {\bibfnamefont {Sergio}\ \bibnamefont
  {Dain}}, \bibinfo {author} {\bibfnamefont {Carlos~O.}\ \bibnamefont
  {Lousto}}, \ and\ \bibinfo {author} {\bibfnamefont {Yosef}\ \bibnamefont
  {Zlochower}},\ }\href {\doibase10.1103/PhysRevD.78.024039} {\bibfield
  {journal} {\bibinfo  {journal} {Phys. Rev.}\ }\textbf {\bibinfo {volume}
  {D78}},\ \bibinfo {pages} {024039} (\bibinfo {year} {2008})},\ \Eprint
  {http://arxiv.org/abs/0803.0351} {arXiv:0803.0351 [gr-qc]}\BibitemShut
  {NoStop}%
\bibitem [{\citenamefont {Campanelli}\ \emph
  {et~al.}(2007{\natexlab{b}})\citenamefont {Campanelli}, \citenamefont
  {Lousto}, \citenamefont {Zlochower}, \citenamefont {Krishnan},\ and\
  \citenamefont {Merritt}}]{Campanelli:2006fy}%
  \BibitemOpen
  \bibfield  {author} {\bibinfo {author} {\bibfnamefont {Manuela}\ \bibnamefont
  {Campanelli}}, \bibinfo {author} {\bibfnamefont {Carlos~O.}\ \bibnamefont
  {Lousto}}, \bibinfo {author} {\bibfnamefont {Yosef}\ \bibnamefont
  {Zlochower}}, \bibinfo {author} {\bibfnamefont {Badri}\ \bibnamefont
  {Krishnan}}, \ and\ \bibinfo {author} {\bibfnamefont {David}\ \bibnamefont
  {Merritt}},\ }\href {\doibase10.1103/PhysRevD.75.064030} {\bibfield
  {journal} {\bibinfo  {journal} {Phys. Rev.}\ }\textbf {\bibinfo {volume}
  {D75}},\ \bibinfo {pages} {064030} (\bibinfo {year} {2007}{\natexlab{b}})},\
  \Eprint {http://arxiv.org/abs/gr-qc/0612076} {arXiv:gr-qc/0612076
  [gr-qc]}\BibitemShut {NoStop}%
\bibitem [{\citenamefont {Campanelli}\ \emph {et~al.}(2009)\citenamefont
  {Campanelli}, \citenamefont {Lousto}, \citenamefont {Nakano},\ and\
  \citenamefont {Zlochower}}]{Campanelli:2008nk}%
  \BibitemOpen
  \bibfield  {author} {\bibinfo {author} {\bibfnamefont {Manuela}\ \bibnamefont
  {Campanelli}}, \bibinfo {author} {\bibfnamefont {Carlos~O.}\ \bibnamefont
  {Lousto}}, \bibinfo {author} {\bibfnamefont {Hiroyuki}\ \bibnamefont
  {Nakano}}, \ and\ \bibinfo {author} {\bibfnamefont {Yosef}\ \bibnamefont
  {Zlochower}},\ }\href {\doibase10.1103/PhysRevD.79.084010} {\bibfield
  {journal} {\bibinfo  {journal} {Phys. Rev.}\ }\textbf {\bibinfo {volume}
  {D79}},\ \bibinfo {pages} {084010} (\bibinfo {year} {2009})},\ \Eprint
  {http://arxiv.org/abs/0808.0713} {arXiv:0808.0713 [gr-qc]}\BibitemShut
  {NoStop}%
\bibitem [{\citenamefont {Schmidt}\ \emph {et~al.}(2011)\citenamefont
  {Schmidt}, \citenamefont {Hannam}, \citenamefont {Husa},\ and\ \citenamefont
  {Ajith}}]{Schmidt:2010it}%
  \BibitemOpen
  \bibfield  {author} {\bibinfo {author} {\bibfnamefont {Patricia}\
  \bibnamefont {Schmidt}}, \bibinfo {author} {\bibfnamefont {Mark}\
  \bibnamefont {Hannam}}, \bibinfo {author} {\bibfnamefont {Sascha}\
  \bibnamefont {Husa}}, \ and\ \bibinfo {author} {\bibfnamefont
  {P.}~\bibnamefont {Ajith}},\ }\href {\doibase10.1103/PhysRevD.84.024046}
  {\bibfield  {journal} {\bibinfo  {journal} {Phys. Rev.}\ }\textbf {\bibinfo
  {volume} {D84}},\ \bibinfo {pages} {024046} (\bibinfo {year} {2011})},\
  \Eprint {http://arxiv.org/abs/1012.2879} {arXiv:1012.2879
  [gr-qc]}\BibitemShut {NoStop}%
\bibitem [{\citenamefont {Sperhake}\ \emph
  {et~al.}(2008{\natexlab{a}})\citenamefont {Sperhake}, \citenamefont {Berti},
  \citenamefont {Cardoso}, \citenamefont {Gonzalez}, \citenamefont
  {Bruegmann},\ and\ \citenamefont {Ansorg}}]{Sperhake:2007gu}%
  \BibitemOpen
  \bibfield  {author} {\bibinfo {author} {\bibfnamefont {Ulrich}\ \bibnamefont
  {Sperhake}}, \bibinfo {author} {\bibfnamefont {Emanuele}\ \bibnamefont
  {Berti}}, \bibinfo {author} {\bibfnamefont {Vitor}\ \bibnamefont {Cardoso}},
  \bibinfo {author} {\bibfnamefont {Jose~A.}\ \bibnamefont {Gonzalez}},
  \bibinfo {author} {\bibfnamefont {Bernd}\ \bibnamefont {Bruegmann}}, \ and\
  \bibinfo {author} {\bibfnamefont {Marcus}\ \bibnamefont {Ansorg}},\ }\href
  {\doibase10.1103/PhysRevD.78.064069} {\bibfield  {journal} {\bibinfo
  {journal} {Phys. Rev.}\ }\textbf {\bibinfo {volume} {D78}},\ \bibinfo {pages}
  {064069} (\bibinfo {year} {2008}{\natexlab{a}})},\ \Eprint
  {http://arxiv.org/abs/0710.3823} {arXiv:0710.3823 [gr-qc]}\BibitemShut
  {NoStop}%
\bibitem [{\citenamefont {Hinder}\ \emph {et~al.}(2008)\citenamefont {Hinder},
  \citenamefont {Vaishnav}, \citenamefont {Herrmann}, \citenamefont
  {Shoemaker},\ and\ \citenamefont {Laguna}}]{Hinder:2007qu}%
  \BibitemOpen
  \bibfield  {author} {\bibinfo {author} {\bibfnamefont {Ian}\ \bibnamefont
  {Hinder}}, \bibinfo {author} {\bibfnamefont {Birjoo}\ \bibnamefont
  {Vaishnav}}, \bibinfo {author} {\bibfnamefont {Frank}\ \bibnamefont
  {Herrmann}}, \bibinfo {author} {\bibfnamefont {Deirdre}\ \bibnamefont
  {Shoemaker}}, \ and\ \bibinfo {author} {\bibfnamefont {Pablo}\ \bibnamefont
  {Laguna}},\ }\href {\doibase10.1103/PhysRevD.77.081502} {\bibfield  {journal}
  {\bibinfo  {journal} {Phys. Rev.}\ }\textbf {\bibinfo {volume} {D77}},\
  \bibinfo {pages} {081502} (\bibinfo {year} {2008})},\ \Eprint
  {http://arxiv.org/abs/0710.5167} {arXiv:0710.5167 [gr-qc]}\BibitemShut
  {NoStop}%
\bibitem [{\citenamefont {Szilágyi}\ \emph {et~al.}(2015)\citenamefont
  {Szilágyi}, \citenamefont {Blackman}, \citenamefont {Buonanno},
  \citenamefont {Taracchini}, \citenamefont {Pfeiffer}, \citenamefont {Scheel},
  \citenamefont {Chu}, \citenamefont {Kidder},\ and\ \citenamefont
  {Pan}}]{Szilagyi:2015rwa}%
  \BibitemOpen
  \bibfield  {author} {\bibinfo {author} {\bibfnamefont {Béla}\ \bibnamefont
  {Szilágyi}}, \bibinfo {author} {\bibfnamefont {Jonathan}\ \bibnamefont
  {Blackman}}, \bibinfo {author} {\bibfnamefont {Alessandra}\ \bibnamefont
  {Buonanno}}, \bibinfo {author} {\bibfnamefont {Andrea}\ \bibnamefont
  {Taracchini}}, \bibinfo {author} {\bibfnamefont {Harald~P.}\ \bibnamefont
  {Pfeiffer}}, \bibinfo {author} {\bibfnamefont {Mark~A.}\ \bibnamefont
  {Scheel}}, \bibinfo {author} {\bibfnamefont {Tony}\ \bibnamefont {Chu}},
  \bibinfo {author} {\bibfnamefont {Lawrence~E.}\ \bibnamefont {Kidder}}, \
  and\ \bibinfo {author} {\bibfnamefont {Yi}~\bibnamefont {Pan}},\ }\href
  {\doibase10.1103/PhysRevLett.115.031102} {\bibfield  {journal} {\bibinfo
  {journal} {Phys. Rev. Lett.}\ }\textbf {\bibinfo {volume} {115}},\ \bibinfo
  {pages} {031102} (\bibinfo {year} {2015})},\ \Eprint
  {http://arxiv.org/abs/1502.04953} {arXiv:1502.04953 [gr-qc]}\BibitemShut
  {NoStop}%
\bibitem [{\citenamefont {Bruegmann}\ \emph
  {et~al.}(2008{\natexlab{a}})\citenamefont {Bruegmann}, \citenamefont
  {Gonzalez}, \citenamefont {Hannam}, \citenamefont {Husa}, \citenamefont
  {Sperhake},\ and\ \citenamefont {Tichy}}]{Bruegmann:2006at}%
  \BibitemOpen
  \bibfield  {author} {\bibinfo {author} {\bibfnamefont {Bernd}\ \bibnamefont
  {Bruegmann}}, \bibinfo {author} {\bibfnamefont {Jose~A.}\ \bibnamefont
  {Gonzalez}}, \bibinfo {author} {\bibfnamefont {Mark}\ \bibnamefont {Hannam}},
  \bibinfo {author} {\bibfnamefont {Sascha}\ \bibnamefont {Husa}}, \bibinfo
  {author} {\bibfnamefont {Ulrich}\ \bibnamefont {Sperhake}}, \ and\ \bibinfo
  {author} {\bibfnamefont {Wolfgang}\ \bibnamefont {Tichy}},\ }\href
  {\doibase10.1103/PhysRevD.77.024027} {\bibfield  {journal} {\bibinfo
  {journal} {Phys. Rev.}\ }\textbf {\bibinfo {volume} {D77}},\ \bibinfo {pages}
  {024027} (\bibinfo {year} {2008}{\natexlab{a}})},\ \Eprint
  {http://arxiv.org/abs/gr-qc/0610128} {arXiv:gr-qc/0610128
  [gr-qc]}\BibitemShut {NoStop}%
\bibitem [{\citenamefont {Loffler}\ \emph {et~al.}(2012)\citenamefont {Loffler}
  \emph {et~al.}}]{Loffler:2011ay}%
  \BibitemOpen
  \bibfield  {author} {\bibinfo {author} {\bibfnamefont {Frank}\ \bibnamefont
  {Loffler}} \emph {et~al.},\ }\href {\doibase10.1088/0264-9381/29/11/115001}
  {\bibfield  {journal} {\bibinfo  {journal} {Class. Quant. Grav.}\ }\textbf
  {\bibinfo {volume} {29}},\ \bibinfo {pages} {115001} (\bibinfo {year}
  {2012})},\ \Eprint {http://arxiv.org/abs/1111.3344} {arXiv:1111.3344
  [gr-qc]}\BibitemShut {NoStop}%
\bibitem [{\citenamefont {Campanelli}\ \emph
  {et~al.}(2006{\natexlab{c}})\citenamefont {Campanelli}, \citenamefont
  {Lousto},\ and\ \citenamefont {Zlochower}}]{Campanelli:2006gf}%
  \BibitemOpen
  \bibfield  {author} {\bibinfo {author} {\bibfnamefont {Manuela}\ \bibnamefont
  {Campanelli}}, \bibinfo {author} {\bibfnamefont {C.~O.}\ \bibnamefont
  {Lousto}}, \ and\ \bibinfo {author} {\bibfnamefont {Y.}~\bibnamefont
  {Zlochower}},\ }\href {\doibase10.1103/PhysRevD.73.061501} {\bibfield
  {journal} {\bibinfo  {journal} {Phys. Rev.}\ }\textbf {\bibinfo {volume}
  {D73}},\ \bibinfo {pages} {061501} (\bibinfo {year} {2006}{\natexlab{c}})},\
  \Eprint {http://arxiv.org/abs/gr-qc/0601091} {arXiv:gr-qc/0601091
  [gr-qc]}\BibitemShut {NoStop}%
\bibitem [{\citenamefont {Vaishnav}\ \emph {et~al.}(2007)\citenamefont
  {Vaishnav}, \citenamefont {Hinder}, \citenamefont {Herrmann},\ and\
  \citenamefont {Shoemaker}}]{Vaishnav:2007nm}%
  \BibitemOpen
  \bibfield  {author} {\bibinfo {author} {\bibfnamefont {Birjoo}\ \bibnamefont
  {Vaishnav}}, \bibinfo {author} {\bibfnamefont {Ian}\ \bibnamefont {Hinder}},
  \bibinfo {author} {\bibfnamefont {Frank}\ \bibnamefont {Herrmann}}, \ and\
  \bibinfo {author} {\bibfnamefont {Deirdre}\ \bibnamefont {Shoemaker}},\
  }\href {\doibase10.1103/PhysRevD.76.084020} {\bibfield  {journal} {\bibinfo
  {journal} {Phys. Rev.}\ }\textbf {\bibinfo {volume} {D76}},\ \bibinfo {pages}
  {084020} (\bibinfo {year} {2007})},\ \Eprint {http://arxiv.org/abs/0705.3829}
  {arXiv:0705.3829 [gr-qc]}\BibitemShut {NoStop}%
\bibitem [{\citenamefont {Sperhake}(2007)}]{Sperhake:2006cy}%
  \BibitemOpen
  \bibfield  {author} {\bibinfo {author} {\bibfnamefont {Ulrich}\ \bibnamefont
  {Sperhake}},\ }\href {\doibase10.1103/PhysRevD.76.104015} {\bibfield
  {journal} {\bibinfo  {journal} {Phys. Rev.}\ }\textbf {\bibinfo {volume}
  {D76}},\ \bibinfo {pages} {104015} (\bibinfo {year} {2007})},\ \Eprint
  {http://arxiv.org/abs/gr-qc/0606079} {arXiv:gr-qc/0606079
  [gr-qc]}\BibitemShut {NoStop}%
\bibitem [{\citenamefont {Baker}\ \emph
  {et~al.}(2006{\natexlab{b}})\citenamefont {Baker}, \citenamefont {Centrella},
  \citenamefont {Choi}, \citenamefont {Koppitz},\ and\ \citenamefont {van
  Meter}}]{Baker:2006yw}%
  \BibitemOpen
  \bibfield  {author} {\bibinfo {author} {\bibfnamefont {John~G.}\ \bibnamefont
  {Baker}}, \bibinfo {author} {\bibfnamefont {Joan}\ \bibnamefont {Centrella}},
  \bibinfo {author} {\bibfnamefont {Dae-Il}\ \bibnamefont {Choi}}, \bibinfo
  {author} {\bibfnamefont {Michael}\ \bibnamefont {Koppitz}}, \ and\ \bibinfo
  {author} {\bibfnamefont {James}\ \bibnamefont {van Meter}},\ }\href
  {\doibase10.1103/PhysRevD.73.104002} {\bibfield  {journal} {\bibinfo
  {journal} {Phys. Rev.}\ }\textbf {\bibinfo {volume} {D73}},\ \bibinfo {pages}
  {104002} (\bibinfo {year} {2006}{\natexlab{b}})},\ \Eprint
  {http://arxiv.org/abs/gr-qc/0602026} {arXiv:gr-qc/0602026
  [gr-qc]}\BibitemShut {NoStop}%
\bibitem [{\citenamefont {Pretorius}(2005{\natexlab{b}})}]{Pretorius:2004jg}%
  \BibitemOpen
  \bibfield  {author} {\bibinfo {author} {\bibfnamefont {Frans}\ \bibnamefont
  {Pretorius}},\ }\href {\doibase10.1088/0264-9381/22/2/014} {\bibfield
  {journal} {\bibinfo  {journal} {Class. Quant. Grav.}\ }\textbf {\bibinfo
  {volume} {22}},\ \bibinfo {pages} {425--452} (\bibinfo {year}
  {2005}{\natexlab{b}})},\ \Eprint {http://arxiv.org/abs/gr-qc/0407110}
  {arXiv:gr-qc/0407110 [gr-qc]}\BibitemShut {NoStop}%
\bibitem [{\citenamefont {Scheel}\ \emph {et~al.}(2006)\citenamefont {Scheel},
  \citenamefont {Pfeiffer}, \citenamefont {Lindblom}, \citenamefont {Kidder},
  \citenamefont {Rinne},\ and\ \citenamefont {Teukolsky}}]{Scheel:2006gg}%
  \BibitemOpen
  \bibfield  {author} {\bibinfo {author} {\bibfnamefont {Mark~A.}\ \bibnamefont
  {Scheel}}, \bibinfo {author} {\bibfnamefont {Harald~P.}\ \bibnamefont
  {Pfeiffer}}, \bibinfo {author} {\bibfnamefont {Lee}\ \bibnamefont
  {Lindblom}}, \bibinfo {author} {\bibfnamefont {Lawrence~E.}\ \bibnamefont
  {Kidder}}, \bibinfo {author} {\bibfnamefont {Oliver}\ \bibnamefont {Rinne}},
  \ and\ \bibinfo {author} {\bibfnamefont {Saul~A.}\ \bibnamefont
  {Teukolsky}},\ }\href {\doibase10.1103/PhysRevD.74.104006} {\bibfield
  {journal} {\bibinfo  {journal} {Phys. Rev.}\ }\textbf {\bibinfo {volume}
  {D74}},\ \bibinfo {pages} {104006} (\bibinfo {year} {2006})},\ \Eprint
  {http://arxiv.org/abs/gr-qc/0607056} {arXiv:gr-qc/0607056
  [gr-qc]}\BibitemShut {NoStop}%
\bibitem [{\citenamefont {Hilditch}\ \emph {et~al.}(2016)\citenamefont
  {Hilditch}, \citenamefont {Weyhausen},\ and\ \citenamefont
  {Brügmann}}]{Hilditch:2015aba}%
  \BibitemOpen
  \bibfield  {author} {\bibinfo {author} {\bibfnamefont {David}\ \bibnamefont
  {Hilditch}}, \bibinfo {author} {\bibfnamefont {Andreas}\ \bibnamefont
  {Weyhausen}}, \ and\ \bibinfo {author} {\bibfnamefont {Bernd}\ \bibnamefont
  {Brügmann}},\ }\href {\doibase10.1103/PhysRevD.93.063006} {\bibfield
  {journal} {\bibinfo  {journal} {Phys. Rev.}\ }\textbf {\bibinfo {volume}
  {D93}},\ \bibinfo {pages} {063006} (\bibinfo {year} {2016})},\ \Eprint
  {http://arxiv.org/abs/1504.04732} {arXiv:1504.04732 [gr-qc]}\BibitemShut
  {NoStop}%
\bibitem [{\citenamefont {Clough}\ \emph {et~al.}(2015)\citenamefont {Clough},
  \citenamefont {Figueras}, \citenamefont {Finkel}, \citenamefont {Kunesch},
  \citenamefont {Lim},\ and\ \citenamefont {Tunyasuvunakool}}]{Clough:2015sqa}%
  \BibitemOpen
  \bibfield  {author} {\bibinfo {author} {\bibfnamefont {Katy}\ \bibnamefont
  {Clough}}, \bibinfo {author} {\bibfnamefont {Pau}\ \bibnamefont {Figueras}},
  \bibinfo {author} {\bibfnamefont {Hal}\ \bibnamefont {Finkel}}, \bibinfo
  {author} {\bibfnamefont {Markus}\ \bibnamefont {Kunesch}}, \bibinfo {author}
  {\bibfnamefont {Eugene~A.}\ \bibnamefont {Lim}}, \ and\ \bibinfo {author}
  {\bibfnamefont {Saran}\ \bibnamefont {Tunyasuvunakool}},\ }\href
  {\doibase10.1088/0264-9381/32/24/245011} {\bibfield  {journal} {\bibinfo
  {journal} {Class. Quant. Grav.}\ }\textbf {\bibinfo {volume} {32}},\ \bibinfo
  {pages} {245011} (\bibinfo {year} {2015})},\ \bibinfo {note} {[Class. Quant.
  Grav.32,24(2015)]},\ \Eprint {http://arxiv.org/abs/1503.03436}
  {arXiv:1503.03436 [gr-qc]}\BibitemShut {NoStop}%
\bibitem [{\citenamefont {Mroue}\ \emph {et~al.}(2013)\citenamefont {Mroue}
  \emph {et~al.}}]{Mroue:2013xna}%
  \BibitemOpen
  \bibfield  {author} {\bibinfo {author} {\bibfnamefont {Abdul~H.}\
  \bibnamefont {Mroue}} \emph {et~al.},\ }\href
  {\doibase10.1103/PhysRevLett.111.241104} {\bibfield  {journal} {\bibinfo
  {journal} {Phys. Rev. Lett.}\ }\textbf {\bibinfo {volume} {111}},\ \bibinfo
  {pages} {241104} (\bibinfo {year} {2013})},\ \Eprint
  {http://arxiv.org/abs/1304.6077} {arXiv:1304.6077 [gr-qc]}\BibitemShut
  {NoStop}%
\bibitem [{\citenamefont {Chu}\ \emph {et~al.}(2016)\citenamefont {Chu},
  \citenamefont {Fong}, \citenamefont {Kumar}, \citenamefont {Pfeiffer},
  \citenamefont {Boyle}, \citenamefont {Hemberger}, \citenamefont {Kidder},
  \citenamefont {Scheel},\ and\ \citenamefont {Szilagyi}}]{Chu:2015kft}%
  \BibitemOpen
  \bibfield  {author} {\bibinfo {author} {\bibfnamefont {Tony}\ \bibnamefont
  {Chu}}, \bibinfo {author} {\bibfnamefont {Heather}\ \bibnamefont {Fong}},
  \bibinfo {author} {\bibfnamefont {Prayush}\ \bibnamefont {Kumar}}, \bibinfo
  {author} {\bibfnamefont {Harald~P.}\ \bibnamefont {Pfeiffer}}, \bibinfo
  {author} {\bibfnamefont {Michael}\ \bibnamefont {Boyle}}, \bibinfo {author}
  {\bibfnamefont {Daniel~A.}\ \bibnamefont {Hemberger}}, \bibinfo {author}
  {\bibfnamefont {Lawrence~E.}\ \bibnamefont {Kidder}}, \bibinfo {author}
  {\bibfnamefont {Mark~A.}\ \bibnamefont {Scheel}}, \ and\ \bibinfo {author}
  {\bibfnamefont {Bela}\ \bibnamefont {Szilagyi}},\ }\href
  {\doibase10.1088/0264-9381/33/16/165001} {\bibfield  {journal} {\bibinfo
  {journal} {Class. Quant. Grav.}\ }\textbf {\bibinfo {volume} {33}},\ \bibinfo
  {pages} {165001} (\bibinfo {year} {2016})},\ \Eprint
  {http://arxiv.org/abs/1512.06800} {arXiv:1512.06800 [gr-qc]}\BibitemShut
  {NoStop}%
\bibitem [{\citenamefont {Jani}\ \emph {et~al.}(2016)\citenamefont {Jani},
  \citenamefont {Healy}, \citenamefont {Clark}, \citenamefont {London},
  \citenamefont {Laguna},\ and\ \citenamefont {Shoemaker}}]{Jani:2016wkt}%
  \BibitemOpen
  \bibfield  {author} {\bibinfo {author} {\bibfnamefont {Karan}\ \bibnamefont
  {Jani}}, \bibinfo {author} {\bibfnamefont {James}\ \bibnamefont {Healy}},
  \bibinfo {author} {\bibfnamefont {James~A.}\ \bibnamefont {Clark}}, \bibinfo
  {author} {\bibfnamefont {Lionel}\ \bibnamefont {London}}, \bibinfo {author}
  {\bibfnamefont {Pablo}\ \bibnamefont {Laguna}}, \ and\ \bibinfo {author}
  {\bibfnamefont {Deirdre}\ \bibnamefont {Shoemaker}},\ }\href
  {\doibase10.1088/0264-9381/33/20/204001} {\bibfield  {journal} {\bibinfo
  {journal} {Class. Quant. Grav.}\ }\textbf {\bibinfo {volume} {33}},\ \bibinfo
  {pages} {204001} (\bibinfo {year} {2016})},\ \Eprint
  {http://arxiv.org/abs/1605.03204} {arXiv:1605.03204 [gr-qc]}\BibitemShut
  {NoStop}%
\bibitem [{\citenamefont {Healy}\ \emph {et~al.}(2017)\citenamefont {Healy},
  \citenamefont {Lousto}, \citenamefont {Zlochower},\ and\ \citenamefont
  {Campanelli}}]{Healy:2017psd}%
  \BibitemOpen
  \bibfield  {author} {\bibinfo {author} {\bibfnamefont {James}\ \bibnamefont
  {Healy}}, \bibinfo {author} {\bibfnamefont {Carlos~O.}\ \bibnamefont
  {Lousto}}, \bibinfo {author} {\bibfnamefont {Yosef}\ \bibnamefont
  {Zlochower}}, \ and\ \bibinfo {author} {\bibfnamefont {Manuela}\ \bibnamefont
  {Campanelli}},\ }\href {\doibase10.1088/1361-6382/aa91b1} {\bibfield
  {journal} {\bibinfo  {journal} {Class. Quant. Grav.}\ }\textbf {\bibinfo
  {volume} {34}},\ \bibinfo {pages} {224001} (\bibinfo {year} {2017})},\
  \Eprint {http://arxiv.org/abs/1703.03423} {arXiv:1703.03423
  [gr-qc]}\BibitemShut {NoStop}%
\bibitem [{\citenamefont {Thornburg}(1987)}]{0264-9381-4-5-013}%
  \BibitemOpen
  \bibfield  {author} {\bibinfo {author} {\bibfnamefont {J}~\bibnamefont
  {Thornburg}},\ }\href {\doibase10.1088/0264-9381/4/5/013} {\bibfield
  {journal} {\bibinfo  {journal} {Classical and Quantum Gravity}\ }\textbf
  {\bibinfo {volume} {4}},\ \bibinfo {pages} {1119} (\bibinfo {year}
  {1987})}\BibitemShut {NoStop}%
\bibitem [{\citenamefont {Cook}\ and\ \citenamefont
  {Pfeiffer}(2004)}]{Cook:2004kt}%
  \BibitemOpen
  \bibfield  {author} {\bibinfo {author} {\bibfnamefont {Gregory~B.}\
  \bibnamefont {Cook}}\ and\ \bibinfo {author} {\bibfnamefont {Harald~P.}\
  \bibnamefont {Pfeiffer}},\ }\href {\doibase10.1103/PhysRevD.70.104016}
  {\bibfield  {journal} {\bibinfo  {journal} {Phys. Rev.}\ }\textbf {\bibinfo
  {volume} {D70}},\ \bibinfo {pages} {104016} (\bibinfo {year} {2004})},\
  \Eprint {http://arxiv.org/abs/gr-qc/0407078} {arXiv:gr-qc/0407078
  [gr-qc]}\BibitemShut {NoStop}%
\bibitem [{\citenamefont {Szilágyi}(2014)}]{Szilagyi:2014fna}%
  \BibitemOpen
  \bibfield  {author} {\bibinfo {author} {\bibfnamefont {Béla}\ \bibnamefont
  {Szilágyi}},\ }\href {\doibase10.1142/S0218271814300146} {\bibfield
  {journal} {\bibinfo  {journal} {Int. J. Mod. Phys.}\ }\textbf {\bibinfo
  {volume} {D23}},\ \bibinfo {pages} {1430014} (\bibinfo {year} {2014})},\
  \Eprint {http://arxiv.org/abs/1405.3693} {arXiv:1405.3693
  [gr-qc]}\BibitemShut {NoStop}%
\bibitem [{\citenamefont {Kidder}\ \emph {et~al.}(2000)\citenamefont {Kidder},
  \citenamefont {Scheel}, \citenamefont {Teukolsky}, \citenamefont {Carlson},\
  and\ \citenamefont {Cook}}]{Kidder:2000yq}%
  \BibitemOpen
  \bibfield  {author} {\bibinfo {author} {\bibfnamefont {Lawrence~E.}\
  \bibnamefont {Kidder}}, \bibinfo {author} {\bibfnamefont {Mark~A.}\
  \bibnamefont {Scheel}}, \bibinfo {author} {\bibfnamefont {Saul~A.}\
  \bibnamefont {Teukolsky}}, \bibinfo {author} {\bibfnamefont {Eric~D.}\
  \bibnamefont {Carlson}}, \ and\ \bibinfo {author} {\bibfnamefont
  {Gregory~B.}\ \bibnamefont {Cook}},\ }\href
  {\doibase10.1103/PhysRevD.62.084032} {\bibfield  {journal} {\bibinfo
  {journal} {Phys. Rev.}\ }\textbf {\bibinfo {volume} {D62}},\ \bibinfo {pages}
  {084032} (\bibinfo {year} {2000})},\ \Eprint
  {http://arxiv.org/abs/gr-qc/0005056} {arXiv:gr-qc/0005056
  [gr-qc]}\BibitemShut {NoStop}%
\bibitem [{\citenamefont {Brandt}\ and\ \citenamefont
  {Bruegmann}(1997)}]{Brandt:1997tf}%
  \BibitemOpen
  \bibfield  {author} {\bibinfo {author} {\bibfnamefont {Steven}\ \bibnamefont
  {Brandt}}\ and\ \bibinfo {author} {\bibfnamefont {Bernd}\ \bibnamefont
  {Bruegmann}},\ }\href {\doibase10.1103/PhysRevLett.78.3606} {\bibfield
  {journal} {\bibinfo  {journal} {Phys. Rev. Lett.}\ }\textbf {\bibinfo
  {volume} {78}},\ \bibinfo {pages} {3606--3609} (\bibinfo {year} {1997})},\
  \Eprint {http://arxiv.org/abs/gr-qc/9703066} {arXiv:gr-qc/9703066
  [gr-qc]}\BibitemShut {NoStop}%
\bibitem [{\citenamefont {Hannam}\ \emph
  {et~al.}(2008{\natexlab{b}})\citenamefont {Hannam}, \citenamefont {Husa},
  \citenamefont {Ohme}, \citenamefont {Bruegmann},\ and\ \citenamefont
  {O'Murchadha}}]{Hannam:2008sg}%
  \BibitemOpen
  \bibfield  {author} {\bibinfo {author} {\bibfnamefont {Mark}\ \bibnamefont
  {Hannam}}, \bibinfo {author} {\bibfnamefont {Sascha}\ \bibnamefont {Husa}},
  \bibinfo {author} {\bibfnamefont {Frank}\ \bibnamefont {Ohme}}, \bibinfo
  {author} {\bibfnamefont {Bernd}\ \bibnamefont {Bruegmann}}, \ and\ \bibinfo
  {author} {\bibfnamefont {Niall}\ \bibnamefont {O'Murchadha}},\ }\href
  {\doibase10.1103/PhysRevD.78.064020} {\bibfield  {journal} {\bibinfo
  {journal} {Phys. Rev.}\ }\textbf {\bibinfo {volume} {D78}},\ \bibinfo {pages}
  {064020} (\bibinfo {year} {2008}{\natexlab{b}})},\ \Eprint
  {http://arxiv.org/abs/0804.0628} {arXiv:0804.0628 [gr-qc]}\BibitemShut
  {NoStop}%
\bibitem [{\citenamefont {van Meter}\ \emph {et~al.}(2006)\citenamefont {van
  Meter}, \citenamefont {Baker}, \citenamefont {Koppitz},\ and\ \citenamefont
  {Choi}}]{vanMeter:2006vi}%
  \BibitemOpen
  \bibfield  {author} {\bibinfo {author} {\bibfnamefont {James~R.}\
  \bibnamefont {van Meter}}, \bibinfo {author} {\bibfnamefont {John~G.}\
  \bibnamefont {Baker}}, \bibinfo {author} {\bibfnamefont {Michael}\
  \bibnamefont {Koppitz}}, \ and\ \bibinfo {author} {\bibfnamefont {Dae-Il}\
  \bibnamefont {Choi}},\ }\href {\doibase10.1103/PhysRevD.73.124011} {\bibfield
   {journal} {\bibinfo  {journal} {Phys. Rev.}\ }\textbf {\bibinfo {volume}
  {D73}},\ \bibinfo {pages} {124011} (\bibinfo {year} {2006})},\ \Eprint
  {http://arxiv.org/abs/gr-qc/0605030} {arXiv:gr-qc/0605030
  [gr-qc]}\BibitemShut {NoStop}%
\bibitem [{\citenamefont {Gundlach}\ and\ \citenamefont
  {Martin-Garcia}(2006)}]{Gundlach:2006tw}%
  \BibitemOpen
  \bibfield  {author} {\bibinfo {author} {\bibfnamefont {Carsten}\ \bibnamefont
  {Gundlach}}\ and\ \bibinfo {author} {\bibfnamefont {Jose~M.}\ \bibnamefont
  {Martin-Garcia}},\ }\href {\doibase10.1103/PhysRevD.74.024016} {\bibfield
  {journal} {\bibinfo  {journal} {Phys. Rev.}\ }\textbf {\bibinfo {volume}
  {D74}},\ \bibinfo {pages} {024016} (\bibinfo {year} {2006})},\ \Eprint
  {http://arxiv.org/abs/gr-qc/0604035} {arXiv:gr-qc/0604035
  [gr-qc]}\BibitemShut {NoStop}%
\bibitem [{\citenamefont {Bowen}\ and\ \citenamefont
  {York}(1980)}]{Bowen:1980yu}%
  \BibitemOpen
  \bibfield  {author} {\bibinfo {author} {\bibfnamefont {Jeffrey~M.}\
  \bibnamefont {Bowen}}\ and\ \bibinfo {author} {\bibfnamefont {James~W.}\
  \bibnamefont {York}, \bibfnamefont {Jr.}},\ }\href
  {\doibase10.1103/PhysRevD.21.2047} {\bibfield  {journal} {\bibinfo  {journal}
  {Phys. Rev.}\ }\textbf {\bibinfo {volume} {D21}},\ \bibinfo {pages}
  {2047--2056} (\bibinfo {year} {1980})}\BibitemShut {NoStop}%
\bibitem [{\citenamefont {Dain}\ \emph {et~al.}(2002)\citenamefont {Dain},
  \citenamefont {Lousto},\ and\ \citenamefont {Takahashi}}]{Dain:2002ee}%
  \BibitemOpen
  \bibfield  {author} {\bibinfo {author} {\bibfnamefont {Sergio}\ \bibnamefont
  {Dain}}, \bibinfo {author} {\bibfnamefont {Carlos~O.}\ \bibnamefont
  {Lousto}}, \ and\ \bibinfo {author} {\bibfnamefont {Ryoji}\ \bibnamefont
  {Takahashi}},\ }\href {\doibase10.1103/PhysRevD.65.104038} {\bibfield
  {journal} {\bibinfo  {journal} {Phys. Rev.}\ }\textbf {\bibinfo {volume}
  {D65}},\ \bibinfo {pages} {104038} (\bibinfo {year} {2002})},\ \Eprint
  {http://arxiv.org/abs/gr-qc/0201062} {arXiv:gr-qc/0201062
  [gr-qc]}\BibitemShut {NoStop}%
\bibitem [{\citenamefont {York}(1999)}]{York:1998hy}%
  \BibitemOpen
  \bibfield  {author} {\bibinfo {author} {\bibfnamefont {James~W.}\
  \bibnamefont {York}, \bibfnamefont {Jr.}},\ }\href
  {\doibase10.1103/PhysRevLett.82.1350} {\bibfield  {journal} {\bibinfo
  {journal} {Phys. Rev. Lett.}\ }\textbf {\bibinfo {volume} {82}},\ \bibinfo
  {pages} {1350--1353} (\bibinfo {year} {1999})},\ \Eprint
  {http://arxiv.org/abs/gr-qc/9810051} {arXiv:gr-qc/9810051
  [gr-qc]}\BibitemShut {NoStop}%
\bibitem [{\citenamefont {Gourgoulhon}\ \emph {et~al.}(2002)\citenamefont
  {Gourgoulhon}, \citenamefont {Grandclement},\ and\ \citenamefont
  {Bonazzola}}]{Gourgoulhon:2001ec}%
  \BibitemOpen
  \bibfield  {author} {\bibinfo {author} {\bibfnamefont {Eric}\ \bibnamefont
  {Gourgoulhon}}, \bibinfo {author} {\bibfnamefont {Philippe}\ \bibnamefont
  {Grandclement}}, \ and\ \bibinfo {author} {\bibfnamefont {Silvano}\
  \bibnamefont {Bonazzola}},\ }\href {\doibase10.1103/PhysRevD.65.044020}
  {\bibfield  {journal} {\bibinfo  {journal} {Phys. Rev.}\ }\textbf {\bibinfo
  {volume} {D65}},\ \bibinfo {pages} {044020} (\bibinfo {year} {2002})},\
  \Eprint {http://arxiv.org/abs/gr-qc/0106015} {arXiv:gr-qc/0106015
  [gr-qc]}\BibitemShut {NoStop}%
\bibitem [{\citenamefont {Grandclement}\ \emph {et~al.}(2002)\citenamefont
  {Grandclement}, \citenamefont {Gourgoulhon},\ and\ \citenamefont
  {Bonazzola}}]{Grandclement:2001ed}%
  \BibitemOpen
  \bibfield  {author} {\bibinfo {author} {\bibfnamefont {Philippe}\
  \bibnamefont {Grandclement}}, \bibinfo {author} {\bibfnamefont {Eric}\
  \bibnamefont {Gourgoulhon}}, \ and\ \bibinfo {author} {\bibfnamefont
  {Silvano}\ \bibnamefont {Bonazzola}},\ }\href
  {\doibase10.1103/PhysRevD.65.044021} {\bibfield  {journal} {\bibinfo
  {journal} {Phys. Rev.}\ }\textbf {\bibinfo {volume} {D65}},\ \bibinfo {pages}
  {044021} (\bibinfo {year} {2002})},\ \Eprint
  {http://arxiv.org/abs/gr-qc/0106016} {arXiv:gr-qc/0106016
  [gr-qc]}\BibitemShut {NoStop}%
\bibitem [{\citenamefont {Cook}(2002)}]{Cook:2001wi}%
  \BibitemOpen
  \bibfield  {author} {\bibinfo {author} {\bibfnamefont {Gregory~B.}\
  \bibnamefont {Cook}},\ }\href {\doibase10.1103/PhysRevD.65.084003} {\bibfield
   {journal} {\bibinfo  {journal} {Phys. Rev.}\ }\textbf {\bibinfo {volume}
  {D65}},\ \bibinfo {pages} {084003} (\bibinfo {year} {2002})},\ \Eprint
  {http://arxiv.org/abs/gr-qc/0108076} {arXiv:gr-qc/0108076
  [gr-qc]}\BibitemShut {NoStop}%
\bibitem [{\citenamefont {Pfeiffer}\ and\ \citenamefont
  {York}(2003)}]{Pfeiffer:2002iy}%
  \BibitemOpen
  \bibfield  {author} {\bibinfo {author} {\bibfnamefont {Harald~P.}\
  \bibnamefont {Pfeiffer}}\ and\ \bibinfo {author} {\bibfnamefont {James~W.}\
  \bibnamefont {York}, \bibfnamefont {Jr.}},\ }\href
  {\doibase10.1103/PhysRevD.67.044022} {\bibfield  {journal} {\bibinfo
  {journal} {Phys. Rev.}\ }\textbf {\bibinfo {volume} {D67}},\ \bibinfo {pages}
  {044022} (\bibinfo {year} {2003})},\ \Eprint
  {http://arxiv.org/abs/gr-qc/0207095} {arXiv:gr-qc/0207095
  [gr-qc]}\BibitemShut {NoStop}%
\bibitem [{\citenamefont {Lovelace}\ \emph {et~al.}(2008)\citenamefont
  {Lovelace}, \citenamefont {Owen}, \citenamefont {Pfeiffer},\ and\
  \citenamefont {Chu}}]{Lovelace:2008tw}%
  \BibitemOpen
  \bibfield  {author} {\bibinfo {author} {\bibfnamefont {Geoffrey}\
  \bibnamefont {Lovelace}}, \bibinfo {author} {\bibfnamefont {Robert}\
  \bibnamefont {Owen}}, \bibinfo {author} {\bibfnamefont {Harald~P.}\
  \bibnamefont {Pfeiffer}}, \ and\ \bibinfo {author} {\bibfnamefont {Tony}\
  \bibnamefont {Chu}},\ }\href {\doibase10.1103/PhysRevD.78.084017} {\bibfield
  {journal} {\bibinfo  {journal} {Phys. Rev.}\ }\textbf {\bibinfo {volume}
  {D78}},\ \bibinfo {pages} {084017} (\bibinfo {year} {2008})},\ \Eprint
  {http://arxiv.org/abs/0805.4192} {arXiv:0805.4192 [gr-qc]}\BibitemShut
  {NoStop}%
\bibitem [{\citenamefont {Ruchlin}\ \emph {et~al.}(2017)\citenamefont
  {Ruchlin}, \citenamefont {Healy}, \citenamefont {Lousto},\ and\ \citenamefont
  {Zlochower}}]{Ruchlin:2014zva}%
  \BibitemOpen
  \bibfield  {author} {\bibinfo {author} {\bibfnamefont {Ian}\ \bibnamefont
  {Ruchlin}}, \bibinfo {author} {\bibfnamefont {James}\ \bibnamefont {Healy}},
  \bibinfo {author} {\bibfnamefont {Carlos~O.}\ \bibnamefont {Lousto}}, \ and\
  \bibinfo {author} {\bibfnamefont {Yosef}\ \bibnamefont {Zlochower}},\ }\href
  {\doibase10.1103/PhysRevD.95.024033} {\bibfield  {journal} {\bibinfo
  {journal} {Phys. Rev.}\ }\textbf {\bibinfo {volume} {D95}},\ \bibinfo {pages}
  {024033} (\bibinfo {year} {2017})},\ \Eprint {http://arxiv.org/abs/1410.8607}
  {arXiv:1410.8607 [gr-qc]}\BibitemShut {NoStop}%
\bibitem [{\citenamefont {Zlochower}\ \emph {et~al.}(2017)\citenamefont
  {Zlochower}, \citenamefont {Healy}, \citenamefont {Lousto},\ and\
  \citenamefont {Ruchlin}}]{Zlochower:2017bbg}%
  \BibitemOpen
  \bibfield  {author} {\bibinfo {author} {\bibfnamefont {Yosef}\ \bibnamefont
  {Zlochower}}, \bibinfo {author} {\bibfnamefont {James}\ \bibnamefont
  {Healy}}, \bibinfo {author} {\bibfnamefont {Carlos~O.}\ \bibnamefont
  {Lousto}}, \ and\ \bibinfo {author} {\bibfnamefont {Ian}\ \bibnamefont
  {Ruchlin}},\ }\href {\doibase10.1103/PhysRevD.96.044002} {\bibfield
  {journal} {\bibinfo  {journal} {Phys. Rev.}\ }\textbf {\bibinfo {volume}
  {D96}},\ \bibinfo {pages} {044002} (\bibinfo {year} {2017})},\ \Eprint
  {http://arxiv.org/abs/1706.01980} {arXiv:1706.01980 [gr-qc]}\BibitemShut
  {NoStop}%
\bibitem [{\citenamefont {Friedrich}\ and\ \citenamefont
  {Rendall}(2000)}]{Friedrich:2000qv}%
  \BibitemOpen
  \bibfield  {author} {\bibinfo {author} {\bibfnamefont {Helmut}\ \bibnamefont
  {Friedrich}}\ and\ \bibinfo {author} {\bibfnamefont {Alan~D.}\ \bibnamefont
  {Rendall}},\ }\bibfield  {booktitle} {\emph {\bibinfo {booktitle} {{In
  *Schmidt, B.G. (ed.): Einstein's field equations and their physical
  implications* 127-224.}}},\ }\href {\doibase10.1007/3-540-46580-4_2}
  {\bibfield  {journal} {\bibinfo  {journal} {Lect. Notes Phys.}\ }\textbf
  {\bibinfo {volume} {540}},\ \bibinfo {pages} {127--224} (\bibinfo {year}
  {2000})},\ \Eprint {http://arxiv.org/abs/gr-qc/0002074} {arXiv:gr-qc/0002074
  [gr-qc]}\BibitemShut {NoStop}%
\bibitem [{\citenamefont {Lindblom}\ \emph {et~al.}(2006)\citenamefont
  {Lindblom}, \citenamefont {Scheel}, \citenamefont {Kidder}, \citenamefont
  {Owen},\ and\ \citenamefont {Rinne}}]{Lindblom:2005qh}%
  \BibitemOpen
  \bibfield  {author} {\bibinfo {author} {\bibfnamefont {Lee}\ \bibnamefont
  {Lindblom}}, \bibinfo {author} {\bibfnamefont {Mark~A.}\ \bibnamefont
  {Scheel}}, \bibinfo {author} {\bibfnamefont {Lawrence~E.}\ \bibnamefont
  {Kidder}}, \bibinfo {author} {\bibfnamefont {Robert}\ \bibnamefont {Owen}}, \
  and\ \bibinfo {author} {\bibfnamefont {Oliver}\ \bibnamefont {Rinne}},\
  }\href {\doibase10.1088/0264-9381/23/16/S09} {\bibfield  {journal} {\bibinfo
  {journal} {Class. Quant. Grav.}\ }\textbf {\bibinfo {volume} {23}},\ \bibinfo
  {pages} {S447--S462} (\bibinfo {year} {2006})},\ \Eprint
  {http://arxiv.org/abs/gr-qc/0512093} {arXiv:gr-qc/0512093
  [gr-qc]}\BibitemShut {NoStop}%
\bibitem [{\citenamefont {Shibata}\ and\ \citenamefont
  {Nakamura}(1995)}]{PhysRevD.52.5428}%
  \BibitemOpen
  \bibfield  {author} {\bibinfo {author} {\bibfnamefont {Masaru}\ \bibnamefont
  {Shibata}}\ and\ \bibinfo {author} {\bibfnamefont {Takashi}\ \bibnamefont
  {Nakamura}},\ }\href {\doibase10.1103/PhysRevD.52.5428} {\bibfield  {journal}
  {\bibinfo  {journal} {Phys. Rev. D}\ }\textbf {\bibinfo {volume} {52}},\
  \bibinfo {pages} {5428--5444} (\bibinfo {year} {1995})}\BibitemShut {NoStop}%
\bibitem [{\citenamefont {Baumgarte}\ and\ \citenamefont
  {Shapiro}(1999)}]{Baumgarte:1998te}%
  \BibitemOpen
  \bibfield  {author} {\bibinfo {author} {\bibfnamefont {Thomas~W.}\
  \bibnamefont {Baumgarte}}\ and\ \bibinfo {author} {\bibfnamefont {Stuart~L.}\
  \bibnamefont {Shapiro}},\ }\href {\doibase10.1103/PhysRevD.59.024007}
  {\bibfield  {journal} {\bibinfo  {journal} {Phys. Rev.}\ }\textbf {\bibinfo
  {volume} {D59}},\ \bibinfo {pages} {024007} (\bibinfo {year} {1999})},\
  \Eprint {http://arxiv.org/abs/gr-qc/9810065} {arXiv:gr-qc/9810065
  [gr-qc]}\BibitemShut {NoStop}%
\bibitem [{\citenamefont {Bona}\ \emph {et~al.}(2003)\citenamefont {Bona},
  \citenamefont {Ledvinka}, \citenamefont {Palenzuela},\ and\ \citenamefont
  {Zacek}}]{Bona:2003fj}%
  \BibitemOpen
  \bibfield  {author} {\bibinfo {author} {\bibfnamefont {C.}~\bibnamefont
  {Bona}}, \bibinfo {author} {\bibfnamefont {T.}~\bibnamefont {Ledvinka}},
  \bibinfo {author} {\bibfnamefont {C.}~\bibnamefont {Palenzuela}}, \ and\
  \bibinfo {author} {\bibfnamefont {M.}~\bibnamefont {Zacek}},\ }\href
  {\doibase10.1103/PhysRevD.67.104005} {\bibfield  {journal} {\bibinfo
  {journal} {Phys. Rev.}\ }\textbf {\bibinfo {volume} {D67}},\ \bibinfo {pages}
  {104005} (\bibinfo {year} {2003})},\ \Eprint
  {http://arxiv.org/abs/gr-qc/0302083} {arXiv:gr-qc/0302083
  [gr-qc]}\BibitemShut {NoStop}%
\bibitem [{\citenamefont {Bona}\ and\ \citenamefont
  {Palenzuela}(2004)}]{Bona:2004yp}%
  \BibitemOpen
  \bibfield  {author} {\bibinfo {author} {\bibfnamefont {C.}~\bibnamefont
  {Bona}}\ and\ \bibinfo {author} {\bibfnamefont {C.}~\bibnamefont
  {Palenzuela}},\ }\href {\doibase10.1103/PhysRevD.69.104003} {\bibfield
  {journal} {\bibinfo  {journal} {Phys. Rev.}\ }\textbf {\bibinfo {volume}
  {D69}},\ \bibinfo {pages} {104003} (\bibinfo {year} {2004})},\ \Eprint
  {http://arxiv.org/abs/gr-qc/0401019} {arXiv:gr-qc/0401019
  [gr-qc]}\BibitemShut {NoStop}%
\bibitem [{\citenamefont {Gundlach}\ \emph {et~al.}(2005)\citenamefont
  {Gundlach}, \citenamefont {Martin-Garcia}, \citenamefont {Calabrese},\ and\
  \citenamefont {Hinder}}]{Gundlach:2005eh}%
  \BibitemOpen
  \bibfield  {author} {\bibinfo {author} {\bibfnamefont {Carsten}\ \bibnamefont
  {Gundlach}}, \bibinfo {author} {\bibfnamefont {Jose~M.}\ \bibnamefont
  {Martin-Garcia}}, \bibinfo {author} {\bibfnamefont {Gioel}\ \bibnamefont
  {Calabrese}}, \ and\ \bibinfo {author} {\bibfnamefont {Ian}\ \bibnamefont
  {Hinder}},\ }\href {\doibase10.1088/0264-9381/22/17/025} {\bibfield
  {journal} {\bibinfo  {journal} {Class. Quant. Grav.}\ }\textbf {\bibinfo
  {volume} {22}},\ \bibinfo {pages} {3767--3774} (\bibinfo {year} {2005})},\
  \Eprint {http://arxiv.org/abs/gr-qc/0504114} {arXiv:gr-qc/0504114
  [gr-qc]}\BibitemShut {NoStop}%
\bibitem [{\citenamefont {Hilditch}\ \emph {et~al.}(2013)\citenamefont
  {Hilditch}, \citenamefont {Bernuzzi}, \citenamefont {Thierfelder},
  \citenamefont {Cao}, \citenamefont {Tichy},\ and\ \citenamefont
  {Bruegmann}}]{Hilditch:2012fp}%
  \BibitemOpen
  \bibfield  {author} {\bibinfo {author} {\bibfnamefont {David}\ \bibnamefont
  {Hilditch}}, \bibinfo {author} {\bibfnamefont {Sebastiano}\ \bibnamefont
  {Bernuzzi}}, \bibinfo {author} {\bibfnamefont {Marcus}\ \bibnamefont
  {Thierfelder}}, \bibinfo {author} {\bibfnamefont {Zhoujian}\ \bibnamefont
  {Cao}}, \bibinfo {author} {\bibfnamefont {Wolfgang}\ \bibnamefont {Tichy}}, \
  and\ \bibinfo {author} {\bibfnamefont {Bernd}\ \bibnamefont {Bruegmann}},\
  }\href {\doibase10.1103/PhysRevD.88.084057} {\bibfield  {journal} {\bibinfo
  {journal} {Phys. Rev.}\ }\textbf {\bibinfo {volume} {D88}},\ \bibinfo {pages}
  {084057} (\bibinfo {year} {2013})},\ \Eprint {http://arxiv.org/abs/1212.2901}
  {arXiv:1212.2901 [gr-qc]}\BibitemShut {NoStop}%
\bibitem [{\citenamefont {Bernuzzi}\ and\ \citenamefont
  {Hilditch}(2010)}]{Bernuzzi:2009ex}%
  \BibitemOpen
  \bibfield  {author} {\bibinfo {author} {\bibfnamefont {Sebastiano}\
  \bibnamefont {Bernuzzi}}\ and\ \bibinfo {author} {\bibfnamefont {David}\
  \bibnamefont {Hilditch}},\ }\href {\doibase10.1103/PhysRevD.81.084003}
  {\bibfield  {journal} {\bibinfo  {journal} {Phys. Rev.}\ }\textbf {\bibinfo
  {volume} {D81}},\ \bibinfo {pages} {084003} (\bibinfo {year} {2010})},\
  \Eprint {http://arxiv.org/abs/0912.2920} {arXiv:0912.2920
  [gr-qc]}\BibitemShut {NoStop}%
\bibitem [{\citenamefont {Alic}\ \emph {et~al.}(2012)\citenamefont {Alic},
  \citenamefont {Bona-Casas}, \citenamefont {Bona}, \citenamefont {Rezzolla},\
  and\ \citenamefont {Palenzuela}}]{Alic:2011gg}%
  \BibitemOpen
  \bibfield  {author} {\bibinfo {author} {\bibfnamefont {Daniela}\ \bibnamefont
  {Alic}}, \bibinfo {author} {\bibfnamefont {Carles}\ \bibnamefont
  {Bona-Casas}}, \bibinfo {author} {\bibfnamefont {Carles}\ \bibnamefont
  {Bona}}, \bibinfo {author} {\bibfnamefont {Luciano}\ \bibnamefont
  {Rezzolla}}, \ and\ \bibinfo {author} {\bibfnamefont {Carlos}\ \bibnamefont
  {Palenzuela}},\ }\href {\doibase10.1103/PhysRevD.85.064040} {\bibfield
  {journal} {\bibinfo  {journal} {Phys. Rev.}\ }\textbf {\bibinfo {volume}
  {D85}},\ \bibinfo {pages} {064040} (\bibinfo {year} {2012})},\ \Eprint
  {http://arxiv.org/abs/1106.2254} {arXiv:1106.2254 [gr-qc]}\BibitemShut
  {NoStop}%
\bibitem [{\citenamefont {Dumbser}\ \emph {et~al.}(2018)\citenamefont
  {Dumbser}, \citenamefont {Guercilena}, \citenamefont {Köppel}, \citenamefont
  {Rezzolla},\ and\ \citenamefont {Zanotti}}]{Dumbser:2017okk}%
  \BibitemOpen
  \bibfield  {author} {\bibinfo {author} {\bibfnamefont {Michael}\ \bibnamefont
  {Dumbser}}, \bibinfo {author} {\bibfnamefont {Federico}\ \bibnamefont
  {Guercilena}}, \bibinfo {author} {\bibfnamefont {Sven}\ \bibnamefont
  {Köppel}}, \bibinfo {author} {\bibfnamefont {Luciano}\ \bibnamefont
  {Rezzolla}}, \ and\ \bibinfo {author} {\bibfnamefont {Olindo}\ \bibnamefont
  {Zanotti}},\ }\href {\doibase10.1103/PhysRevD.97.084053} {\bibfield
  {journal} {\bibinfo  {journal} {Phys. Rev.}\ }\textbf {\bibinfo {volume}
  {D97}},\ \bibinfo {pages} {084053} (\bibinfo {year} {2018})},\ \Eprint
  {http://arxiv.org/abs/1707.09910} {arXiv:1707.09910 [gr-qc]}\BibitemShut
  {NoStop}%
\bibitem [{\citenamefont {Newman}\ and\ \citenamefont
  {Penrose}(1962)}]{Newman:1961qr}%
  \BibitemOpen
  \bibfield  {author} {\bibinfo {author} {\bibfnamefont {Ezra}\ \bibnamefont
  {Newman}}\ and\ \bibinfo {author} {\bibfnamefont {Roger}\ \bibnamefont
  {Penrose}},\ }\href {\doibase10.1063/1.1724257} {\bibfield  {journal}
  {\bibinfo  {journal} {J. Math. Phys.}\ }\textbf {\bibinfo {volume} {3}},\
  \bibinfo {pages} {566--578} (\bibinfo {year} {1962})}\BibitemShut {NoStop}%
\bibitem [{\citenamefont {Stewart}(1990)}]{Stewart:1990aa}%
  \BibitemOpen
  \bibfield  {author} {\bibinfo {author} {\bibfnamefont {J.~M.}\ \bibnamefont
  {Stewart}},\ }\href@noop {} {\emph {\bibinfo {title} {{Advanced General
  Relativity}}}}\ (\bibinfo  {publisher} {Cambridge University Press},\
  \bibinfo {year} {1990})\BibitemShut {NoStop}%
\bibitem [{\citenamefont {Bishop}\ \emph {et~al.}(1996)\citenamefont {Bishop},
  \citenamefont {Gomez}, \citenamefont {Lehner},\ and\ \citenamefont
  {Winicour}}]{Bishop:1996gt}%
  \BibitemOpen
  \bibfield  {author} {\bibinfo {author} {\bibfnamefont {Nigel~T.}\
  \bibnamefont {Bishop}}, \bibinfo {author} {\bibfnamefont {Roberto}\
  \bibnamefont {Gomez}}, \bibinfo {author} {\bibfnamefont {Luis}\ \bibnamefont
  {Lehner}}, \ and\ \bibinfo {author} {\bibfnamefont {Jeffrey}\ \bibnamefont
  {Winicour}},\ }\href {\doibase10.1103/PhysRevD.54.6153} {\bibfield  {journal}
  {\bibinfo  {journal} {Phys. Rev.}\ }\textbf {\bibinfo {volume} {D54}},\
  \bibinfo {pages} {6153--6165} (\bibinfo {year} {1996})}\BibitemShut {NoStop}%
\bibitem [{\citenamefont {Winicour}(1999)}]{Winicour:1999ba}%
  \BibitemOpen
  \bibfield  {author} {\bibinfo {author} {\bibfnamefont {Jeffrey}\ \bibnamefont
  {Winicour}},\ }\bibfield  {booktitle} {\emph {\bibinfo {booktitle}
  {{Proceedings, 9th Yukawa International Seminar on Black Holes and
  Gravitational Waves: New Eyes in the 21st Century (YKIS 99): Kyoto, Japan,
  June 28-July 2, 1999}}},\ }\href {\doibase10.1143/PTPS.136.57} {\bibfield
  {journal} {\bibinfo  {journal} {Prog. Theor. Phys. Suppl.}\ }\textbf
  {\bibinfo {volume} {136}},\ \bibinfo {pages} {57--71} (\bibinfo {year}
  {1999})},\ \Eprint {http://arxiv.org/abs/gr-qc/9911106} {arXiv:gr-qc/9911106
  [gr-qc]}\BibitemShut {NoStop}%
\bibitem [{\citenamefont {Bondi}\ \emph {et~al.}(1962)\citenamefont {Bondi},
  \citenamefont {van~der Burg},\ and\ \citenamefont {Metzner}}]{Bondi21}%
  \BibitemOpen
  \bibfield  {author} {\bibinfo {author} {\bibfnamefont {H.}~\bibnamefont
  {Bondi}}, \bibinfo {author} {\bibfnamefont {M.~G.~J.}\ \bibnamefont {van~der
  Burg}}, \ and\ \bibinfo {author} {\bibfnamefont {A.~W.~K.}\ \bibnamefont
  {Metzner}},\ }\href {\doibase10.1098/rspa.1962.0161} {\bibfield  {journal}
  {\bibinfo  {journal} {Proc. Roy. Soc. Lond.}\ }\textbf {\bibinfo {volume}
  {269}},\ \bibinfo {pages} {21--52} (\bibinfo {year} {1962})}\BibitemShut
  {NoStop}%
\bibitem [{\citenamefont {Sachs}(1962)}]{Sachs103}%
  \BibitemOpen
  \bibfield  {author} {\bibinfo {author} {\bibfnamefont {Rainer~K.}\
  \bibnamefont {Sachs}},\ }\href {\doibase10.1098/rspa.1962.0206} {\bibfield
  {journal} {\bibinfo  {journal} {Proc. Roy. Soc. Lond.}\ }\textbf {\bibinfo
  {volume} {270}},\ \bibinfo {pages} {103--126} (\bibinfo {year}
  {1962})}\BibitemShut {NoStop}%
\bibitem [{\citenamefont {Husa}\ \emph {et~al.}(2016)\citenamefont {Husa},
  \citenamefont {Khan}, \citenamefont {Hannam}, \citenamefont {Puerrer},
  \citenamefont {Ohme}, \citenamefont {Forteza},\ and\ \citenamefont
  {Bohe}}]{Husa:2015iqa}%
  \BibitemOpen
  \bibfield  {author} {\bibinfo {author} {\bibfnamefont {Sascha}\ \bibnamefont
  {Husa}}, \bibinfo {author} {\bibfnamefont {Sebastian}\ \bibnamefont {Khan}},
  \bibinfo {author} {\bibfnamefont {Mark}\ \bibnamefont {Hannam}}, \bibinfo
  {author} {\bibfnamefont {Michael}\ \bibnamefont {Puerrer}}, \bibinfo {author}
  {\bibfnamefont {Frank}\ \bibnamefont {Ohme}}, \bibinfo {author}
  {\bibfnamefont {Xisco~Jimenez}\ \bibnamefont {Forteza}}, \ and\ \bibinfo
  {author} {\bibfnamefont {Alejandro}\ \bibnamefont {Bohe}},\ }\href
  {\doibase10.1103/PhysRevD.93.044006} {\bibfield  {journal} {\bibinfo
  {journal} {Phys. Rev. D}\ }\textbf {\bibinfo {volume} {93}},\ \bibinfo
  {pages} {044006} (\bibinfo {year} {2016})},\ \Eprint
  {http://arxiv.org/abs/1508.07250} {arXiv:1508.07250 [gr-qc]}\BibitemShut
  {NoStop}%
\bibitem [{\citenamefont {Scheel}\ \emph {et~al.}(2015)\citenamefont {Scheel},
  \citenamefont {Giesler}, \citenamefont {Hemberger}, \citenamefont {Lovelace},
  \citenamefont {Kuper}, \citenamefont {Boyle}, \citenamefont {Szilágyi},\
  and\ \citenamefont {Kidder}}]{Scheel:2014ina}%
  \BibitemOpen
  \bibfield  {author} {\bibinfo {author} {\bibfnamefont {Mark~A.}\ \bibnamefont
  {Scheel}}, \bibinfo {author} {\bibfnamefont {Matthew}\ \bibnamefont
  {Giesler}}, \bibinfo {author} {\bibfnamefont {Daniel~A.}\ \bibnamefont
  {Hemberger}}, \bibinfo {author} {\bibfnamefont {Geoffrey}\ \bibnamefont
  {Lovelace}}, \bibinfo {author} {\bibfnamefont {Kevin}\ \bibnamefont {Kuper}},
  \bibinfo {author} {\bibfnamefont {Michael}\ \bibnamefont {Boyle}}, \bibinfo
  {author} {\bibfnamefont {B.}~\bibnamefont {Szilágyi}}, \ and\ \bibinfo
  {author} {\bibfnamefont {Lawrence~E.}\ \bibnamefont {Kidder}},\ }\href
  {\doibase10.1088/0264-9381/32/10/105009} {\bibfield  {journal} {\bibinfo
  {journal} {Class. Quant. Grav.}\ }\textbf {\bibinfo {volume} {32}},\ \bibinfo
  {pages} {105009} (\bibinfo {year} {2015})},\ \Eprint
  {http://arxiv.org/abs/1412.1803} {arXiv:1412.1803 [gr-qc]}\BibitemShut
  {NoStop}%
\bibitem [{\citenamefont {Thorne}(1974)}]{Thorne:1974ve}%
  \BibitemOpen
  \bibfield  {author} {\bibinfo {author} {\bibfnamefont {Kip~S.}\ \bibnamefont
  {Thorne}},\ }\href {\doibase10.1086/152991} {\bibfield  {journal} {\bibinfo
  {journal} {Astrophys. J.}\ }\textbf {\bibinfo {volume} {191}},\ \bibinfo
  {pages} {507--520} (\bibinfo {year} {1974})}\BibitemShut {NoStop}%
\bibitem [{\citenamefont {Lousto}\ and\ \citenamefont
  {Zlochower}(2011{\natexlab{a}})}]{Lousto:2010ut}%
  \BibitemOpen
  \bibfield  {author} {\bibinfo {author} {\bibfnamefont {Carlos~O.}\
  \bibnamefont {Lousto}}\ and\ \bibinfo {author} {\bibfnamefont {Yosef}\
  \bibnamefont {Zlochower}},\ }\href {\doibase10.1103/PhysRevLett.106.041101}
  {\bibfield  {journal} {\bibinfo  {journal} {Phys. Rev. Lett.}\ }\textbf
  {\bibinfo {volume} {106}},\ \bibinfo {pages} {041101} (\bibinfo {year}
  {2011}{\natexlab{a}})},\ \Eprint {http://arxiv.org/abs/1009.0292}
  {arXiv:1009.0292 [gr-qc]}\BibitemShut {NoStop}%
\bibitem [{\citenamefont {Lau}\ \emph {et~al.}(2011)\citenamefont {Lau},
  \citenamefont {Lovelace},\ and\ \citenamefont {Pfeiffer}}]{Lau:2011we}%
  \BibitemOpen
  \bibfield  {author} {\bibinfo {author} {\bibfnamefont {Stephen~R.}\
  \bibnamefont {Lau}}, \bibinfo {author} {\bibfnamefont {Geoffrey}\
  \bibnamefont {Lovelace}}, \ and\ \bibinfo {author} {\bibfnamefont
  {Harald~P.}\ \bibnamefont {Pfeiffer}},\ }\href
  {\doibase10.1103/PhysRevD.84.084023} {\bibfield  {journal} {\bibinfo
  {journal} {Phys. Rev.}\ }\textbf {\bibinfo {volume} {D84}},\ \bibinfo {pages}
  {084023} (\bibinfo {year} {2011})},\ \Eprint {http://arxiv.org/abs/1105.3922}
  {arXiv:1105.3922 [gr-qc]}\BibitemShut {NoStop}%
\bibitem [{\citenamefont {Hannam}(2009)}]{Hannam:2009rd}%
  \BibitemOpen
  \bibfield  {author} {\bibinfo {author} {\bibfnamefont {Mark}\ \bibnamefont
  {Hannam}},\ }\bibfield  {booktitle} {\emph {\bibinfo {booktitle} {{Numerical
  relativity data analysis. Proceedings, 2nd Meeting, NRDA 2008, Syracuse, USA,
  August 11-14, 2008}}},\ }\href {\doibase10.1088/0264-9381/26/11/114001}
  {\bibfield  {journal} {\bibinfo  {journal} {Class. Quant. Grav.}\ }\textbf
  {\bibinfo {volume} {26}},\ \bibinfo {pages} {114001} (\bibinfo {year}
  {2009})},\ \Eprint {http://arxiv.org/abs/0901.2931} {arXiv:0901.2931
  [gr-qc]}\BibitemShut {NoStop}%
\bibitem [{\citenamefont {Berti}\ \emph {et~al.}(2013)\citenamefont {Berti},
  \citenamefont {Cardoso}, \citenamefont {Gualtieri}, \citenamefont
  {Horbatsch},\ and\ \citenamefont {Sperhake}}]{Berti:2013gfa}%
  \BibitemOpen
  \bibfield  {author} {\bibinfo {author} {\bibfnamefont {Emanuele}\
  \bibnamefont {Berti}}, \bibinfo {author} {\bibfnamefont {Vitor}\ \bibnamefont
  {Cardoso}}, \bibinfo {author} {\bibfnamefont {Leonardo}\ \bibnamefont
  {Gualtieri}}, \bibinfo {author} {\bibfnamefont {Michael}\ \bibnamefont
  {Horbatsch}}, \ and\ \bibinfo {author} {\bibfnamefont {Ulrich}\ \bibnamefont
  {Sperhake}},\ }\href {\doibase10.1103/PhysRevD.87.124020} {\bibfield
  {journal} {\bibinfo  {journal} {Phys. Rev.}\ }\textbf {\bibinfo {volume}
  {D87}},\ \bibinfo {pages} {124020} (\bibinfo {year} {2013})},\ \Eprint
  {http://arxiv.org/abs/1304.2836} {arXiv:1304.2836 [gr-qc]}\BibitemShut
  {NoStop}%
\bibitem [{\citenamefont {Okounkova}\ \emph {et~al.}(2017)\citenamefont
  {Okounkova}, \citenamefont {Stein}, \citenamefont {Scheel},\ and\
  \citenamefont {Hemberger}}]{Okounkova:2017yby}%
  \BibitemOpen
  \bibfield  {author} {\bibinfo {author} {\bibfnamefont {M.}~\bibnamefont
  {Okounkova}}, \bibinfo {author} {\bibfnamefont {L.~C.}\ \bibnamefont
  {Stein}}, \bibinfo {author} {\bibfnamefont {M.~A.}\ \bibnamefont {Scheel}}, \
  and\ \bibinfo {author} {\bibfnamefont {D.~A.}\ \bibnamefont {Hemberger}},\
  }\href {\doibase10.1103/PhysRevD.96.044020} {\bibfield  {journal} {\bibinfo
  {journal} {Phys. Rev. D}\ }\textbf {\bibinfo {volume} {96}},\ \bibinfo
  {pages} {044020} (\bibinfo {year} {2017})},\ \Eprint
  {http://arxiv.org/abs/1705.07924} {1705.07924}\BibitemShut {NoStop}%
\bibitem [{\citenamefont {Papallo}\ and\ \citenamefont
  {Reall}(2017)}]{Papallo:2017qvl}%
  \BibitemOpen
  \bibfield  {author} {\bibinfo {author} {\bibfnamefont {Giuseppe}\
  \bibnamefont {Papallo}}\ and\ \bibinfo {author} {\bibfnamefont {Harvey~S.}\
  \bibnamefont {Reall}},\ }\href {\doibase10.1103/PhysRevD.96.044019}
  {\bibfield  {journal} {\bibinfo  {journal} {Phys. Rev.}\ }\textbf {\bibinfo
  {volume} {D96}},\ \bibinfo {pages} {044019} (\bibinfo {year} {2017})},\
  \Eprint {http://arxiv.org/abs/1705.04370} {arXiv:1705.04370
  [gr-qc]}\BibitemShut {NoStop}%
\bibitem [{\citenamefont {Delsate}\ \emph {et~al.}(2015)\citenamefont
  {Delsate}, \citenamefont {Hilditch},\ and\ \citenamefont
  {Witek}}]{Delsate:2014hba}%
  \BibitemOpen
  \bibfield  {author} {\bibinfo {author} {\bibfnamefont {Térence}\
  \bibnamefont {Delsate}}, \bibinfo {author} {\bibfnamefont {David}\
  \bibnamefont {Hilditch}}, \ and\ \bibinfo {author} {\bibfnamefont {Helvi}\
  \bibnamefont {Witek}},\ }\href {\doibase10.1103/PhysRevD.91.024027}
  {\bibfield  {journal} {\bibinfo  {journal} {Phys. Rev.}\ }\textbf {\bibinfo
  {volume} {D91}},\ \bibinfo {pages} {024027} (\bibinfo {year} {2015})},\
  \Eprint {http://arxiv.org/abs/1407.6727} {arXiv:1407.6727
  [gr-qc]}\BibitemShut {NoStop}%
\bibitem [{\citenamefont {Khan}\ \emph
  {et~al.}(2016{\natexlab{b}})\citenamefont {Khan}, \citenamefont {Husa},
  \citenamefont {Hannam}, \citenamefont {Ohme}, \citenamefont {Puerrer},
  \citenamefont {Jimenez~Forteza},\ and\ \citenamefont {Bohe}}]{Khan:2015jqa}%
  \BibitemOpen
  \bibfield  {author} {\bibinfo {author} {\bibfnamefont {Sebastian}\
  \bibnamefont {Khan}}, \bibinfo {author} {\bibfnamefont {Sascha}\ \bibnamefont
  {Husa}}, \bibinfo {author} {\bibfnamefont {Mark}\ \bibnamefont {Hannam}},
  \bibinfo {author} {\bibfnamefont {Frank}\ \bibnamefont {Ohme}}, \bibinfo
  {author} {\bibfnamefont {Michael}\ \bibnamefont {Puerrer}}, \bibinfo {author}
  {\bibfnamefont {Xisco}\ \bibnamefont {Jimenez~Forteza}}, \ and\ \bibinfo
  {author} {\bibfnamefont {Alejandro}\ \bibnamefont {Bohe}},\ }\href
  {\doibase10.1103/PhysRevD.93.044007} {\bibfield  {journal} {\bibinfo
  {journal} {Phys. Rev.}\ }\textbf {\bibinfo {volume} {D93}},\ \bibinfo {pages}
  {044007} (\bibinfo {year} {2016}{\natexlab{b}})},\ \Eprint
  {http://arxiv.org/abs/1508.07253} {arXiv:1508.07253 [gr-qc]}\BibitemShut
  {NoStop}%
\bibitem [{\citenamefont {Boh\'{e}}\ \emph {et~al.}(2017)\citenamefont
  {Boh\'{e}} \emph {et~al.}}]{Bohe:2016gbl}%
  \BibitemOpen
  \bibfield  {author} {\bibinfo {author} {\bibfnamefont {Alejandro}\
  \bibnamefont {Boh\'{e}}} \emph {et~al.},\ }\href
  {\doibase10.1103/PhysRevD.95.044028} {\bibfield  {journal} {\bibinfo
  {journal} {Phys. Rev. D}\ }\textbf {\bibinfo {volume} {95}},\ \bibinfo
  {pages} {044028} (\bibinfo {year} {2017})},\ \Eprint
  {http://arxiv.org/abs/1611.03703} {arXiv:1611.03703 [gr-qc]}\BibitemShut
  {NoStop}%
\bibitem [{\citenamefont {Apostolatos}\ \emph {et~al.}(1994)\citenamefont
  {Apostolatos}, \citenamefont {Cutler}, \citenamefont {Sussman},\ and\
  \citenamefont {Thorne}}]{Apostolatos:1994mx}%
  \BibitemOpen
  \bibfield  {author} {\bibinfo {author} {\bibfnamefont {Theocharis~A.}\
  \bibnamefont {Apostolatos}}, \bibinfo {author} {\bibfnamefont {Curt}\
  \bibnamefont {Cutler}}, \bibinfo {author} {\bibfnamefont {Gerald~J.}\
  \bibnamefont {Sussman}}, \ and\ \bibinfo {author} {\bibfnamefont {Kip~S.}\
  \bibnamefont {Thorne}},\ }\href {\doibase10.1103/PhysRevD.49.6274} {\bibfield
   {journal} {\bibinfo  {journal} {Phys. Rev.}\ }\textbf {\bibinfo {volume}
  {D49}},\ \bibinfo {pages} {6274--6297} (\bibinfo {year} {1994})}\BibitemShut
  {NoStop}%
\bibitem [{\citenamefont {Kidder}(1995)}]{Kidder:1995zr}%
  \BibitemOpen
  \bibfield  {author} {\bibinfo {author} {\bibfnamefont {Lawrence~E.}\
  \bibnamefont {Kidder}},\ }\href {\doibase10.1103/PhysRevD.52.821} {\bibfield
  {journal} {\bibinfo  {journal} {Phys. Rev.}\ }\textbf {\bibinfo {volume}
  {D52}},\ \bibinfo {pages} {821--847} (\bibinfo {year} {1995})},\ \Eprint
  {http://arxiv.org/abs/gr-qc/9506022} {arXiv:gr-qc/9506022
  [gr-qc]}\BibitemShut {NoStop}%
\bibitem [{\citenamefont {Hannam}\ \emph {et~al.}(2014)\citenamefont {Hannam},
  \citenamefont {Schmidt}, \citenamefont {Bohe}, \citenamefont {Haegel},
  \citenamefont {Husa}, \citenamefont {Ohme}, \citenamefont {Pratten},\ and\
  \citenamefont {Puerrer}}]{Hannam:2013oca}%
  \BibitemOpen
  \bibfield  {author} {\bibinfo {author} {\bibfnamefont {Mark}\ \bibnamefont
  {Hannam}}, \bibinfo {author} {\bibfnamefont {Patricia}\ \bibnamefont
  {Schmidt}}, \bibinfo {author} {\bibfnamefont {Alejandro}\ \bibnamefont
  {Bohe}}, \bibinfo {author} {\bibfnamefont {Leila}\ \bibnamefont {Haegel}},
  \bibinfo {author} {\bibfnamefont {Sascha}\ \bibnamefont {Husa}}, \bibinfo
  {author} {\bibfnamefont {Frank}\ \bibnamefont {Ohme}}, \bibinfo {author}
  {\bibfnamefont {Geraint}\ \bibnamefont {Pratten}}, \ and\ \bibinfo {author}
  {\bibfnamefont {Michael}\ \bibnamefont {Puerrer}},\ }\href
  {\doibase10.1103/PhysRevLett.113.151101} {\bibfield  {journal} {\bibinfo
  {journal} {Phys. Rev. Lett.}\ }\textbf {\bibinfo {volume} {113}},\ \bibinfo
  {pages} {151101} (\bibinfo {year} {2014})},\ \Eprint
  {http://arxiv.org/abs/1308.3271} {arXiv:1308.3271 [gr-qc]}\BibitemShut
  {NoStop}%
\bibitem [{\citenamefont {Pan}\ \emph {et~al.}(2014)\citenamefont {Pan},
  \citenamefont {Buonanno}, \citenamefont {Taracchini}, \citenamefont {Kidder},
  \citenamefont {Mrou{\'e}}, \citenamefont {Pfeiffer}, \citenamefont {Scheel},\
  and\ \citenamefont {Szil{\'a}gyi}}]{Pan:2013rra}%
  \BibitemOpen
  \bibfield  {author} {\bibinfo {author} {\bibfnamefont {Yi}~\bibnamefont
  {Pan}}, \bibinfo {author} {\bibfnamefont {Alessandra}\ \bibnamefont
  {Buonanno}}, \bibinfo {author} {\bibfnamefont {Andrea}\ \bibnamefont
  {Taracchini}}, \bibinfo {author} {\bibfnamefont {Lawrence~E.}\ \bibnamefont
  {Kidder}}, \bibinfo {author} {\bibfnamefont {Abdul~H.}\ \bibnamefont
  {Mrou{\'e}}}, \bibinfo {author} {\bibfnamefont {Harald~P.}\ \bibnamefont
  {Pfeiffer}}, \bibinfo {author} {\bibfnamefont {Mark~A.}\ \bibnamefont
  {Scheel}}, \ and\ \bibinfo {author} {\bibfnamefont {B{\'e}la}\ \bibnamefont
  {Szil{\'a}gyi}},\ }\href {\doibase10.1103/PhysRevD.89.084006} {\bibfield
  {journal} {\bibinfo  {journal} {Phys. Rev. D}\ }\textbf {\bibinfo {volume}
  {89}},\ \bibinfo {pages} {084006} (\bibinfo {year} {2014})},\ \Eprint
  {http://arxiv.org/abs/1307.6232} {arXiv:1307.6232 [gr-qc]}\BibitemShut
  {NoStop}%
\bibitem [{\citenamefont {Abbott}\ \emph
  {et~al.}(2016{\natexlab{n}})\citenamefont {Abbott} \emph
  {et~al.}}]{Abbott:2016apu}%
  \BibitemOpen
  \bibfield  {author} {\bibinfo {author} {\bibfnamefont {B.~P.}\ \bibnamefont
  {Abbott}} \emph {et~al.} (\bibinfo {collaboration} {Virgo, LIGO
  Scientific}),\ }\href {\doibase10.1103/PhysRevD.94.064035} {\bibfield
  {journal} {\bibinfo  {journal} {Phys. Rev.}\ }\textbf {\bibinfo {volume}
  {D94}},\ \bibinfo {pages} {064035} (\bibinfo {year} {2016}{\natexlab{n}})},\
  \Eprint {http://arxiv.org/abs/1606.01262} {arXiv:1606.01262
  [gr-qc]}\BibitemShut {NoStop}%
\bibitem [{\citenamefont {Lange}\ \emph {et~al.}(2017)\citenamefont {Lange}
  \emph {et~al.}}]{Lange:2017wki}%
  \BibitemOpen
  \bibfield  {author} {\bibinfo {author} {\bibfnamefont {J.}~\bibnamefont
  {Lange}} \emph {et~al.},\ }\href {\doibase10.1103/PhysRevD.96.104041}
  {\bibfield  {journal} {\bibinfo  {journal} {Phys. Rev.}\ }\textbf {\bibinfo
  {volume} {D96}},\ \bibinfo {pages} {104041} (\bibinfo {year} {2017})},\
  \Eprint {http://arxiv.org/abs/1705.09833} {arXiv:1705.09833
  [gr-qc]}\BibitemShut {NoStop}%
\bibitem [{\citenamefont {Blackman}\ \emph {et~al.}(2015)\citenamefont
  {Blackman}, \citenamefont {Field}, \citenamefont {Galley}, \citenamefont
  {Szilágyi}, \citenamefont {Scheel}, \citenamefont {Tiglio},\ and\
  \citenamefont {Hemberger}}]{Blackman:2015pia}%
  \BibitemOpen
  \bibfield  {author} {\bibinfo {author} {\bibfnamefont {Jonathan}\
  \bibnamefont {Blackman}}, \bibinfo {author} {\bibfnamefont {Scott~E.}\
  \bibnamefont {Field}}, \bibinfo {author} {\bibfnamefont {Chad~R.}\
  \bibnamefont {Galley}}, \bibinfo {author} {\bibfnamefont {Béla}\
  \bibnamefont {Szilágyi}}, \bibinfo {author} {\bibfnamefont {Mark~A.}\
  \bibnamefont {Scheel}}, \bibinfo {author} {\bibfnamefont {Manuel}\
  \bibnamefont {Tiglio}}, \ and\ \bibinfo {author} {\bibfnamefont {Daniel~A.}\
  \bibnamefont {Hemberger}},\ }\href {\doibase10.1103/PhysRevLett.115.121102}
  {\bibfield  {journal} {\bibinfo  {journal} {Phys. Rev. Lett.}\ }\textbf
  {\bibinfo {volume} {115}},\ \bibinfo {pages} {121102} (\bibinfo {year}
  {2015})},\ \Eprint {http://arxiv.org/abs/1502.07758} {arXiv:1502.07758
  [gr-qc]}\BibitemShut {NoStop}%
\bibitem [{\citenamefont {Blackman}\ \emph
  {et~al.}(2017{\natexlab{a}})\citenamefont {Blackman}, \citenamefont {Field},
  \citenamefont {Scheel}, \citenamefont {Galley}, \citenamefont {Hemberger},
  \citenamefont {Schmidt},\ and\ \citenamefont {Smith}}]{Blackman:2017dfb}%
  \BibitemOpen
  \bibfield  {author} {\bibinfo {author} {\bibfnamefont {Jonathan}\
  \bibnamefont {Blackman}}, \bibinfo {author} {\bibfnamefont {Scott~E.}\
  \bibnamefont {Field}}, \bibinfo {author} {\bibfnamefont {Mark~A.}\
  \bibnamefont {Scheel}}, \bibinfo {author} {\bibfnamefont {Chad~R.}\
  \bibnamefont {Galley}}, \bibinfo {author} {\bibfnamefont {Daniel~A.}\
  \bibnamefont {Hemberger}}, \bibinfo {author} {\bibfnamefont {Patricia}\
  \bibnamefont {Schmidt}}, \ and\ \bibinfo {author} {\bibfnamefont {Rory}\
  \bibnamefont {Smith}},\ }\href {\doibase10.1103/PhysRevD.95.104023}
  {\bibfield  {journal} {\bibinfo  {journal} {Phys. Rev.}\ }\textbf {\bibinfo
  {volume} {D95}},\ \bibinfo {pages} {104023} (\bibinfo {year}
  {2017}{\natexlab{a}})},\ \Eprint {http://arxiv.org/abs/1701.00550}
  {arXiv:1701.00550 [gr-qc]}\BibitemShut {NoStop}%
\bibitem [{\citenamefont {Blackman}\ \emph
  {et~al.}(2017{\natexlab{b}})\citenamefont {Blackman}, \citenamefont {Field},
  \citenamefont {Scheel}, \citenamefont {Galley}, \citenamefont {Ott},
  \citenamefont {Boyle}, \citenamefont {Kidder}, \citenamefont {Pfeiffer},\
  and\ \citenamefont {Szilágyi}}]{Blackman:2017pcm}%
  \BibitemOpen
  \bibfield  {author} {\bibinfo {author} {\bibfnamefont {Jonathan}\
  \bibnamefont {Blackman}}, \bibinfo {author} {\bibfnamefont {Scott~E.}\
  \bibnamefont {Field}}, \bibinfo {author} {\bibfnamefont {Mark~A.}\
  \bibnamefont {Scheel}}, \bibinfo {author} {\bibfnamefont {Chad~R.}\
  \bibnamefont {Galley}}, \bibinfo {author} {\bibfnamefont {Christian~D.}\
  \bibnamefont {Ott}}, \bibinfo {author} {\bibfnamefont {Michael}\ \bibnamefont
  {Boyle}}, \bibinfo {author} {\bibfnamefont {Lawrence~E.}\ \bibnamefont
  {Kidder}}, \bibinfo {author} {\bibfnamefont {Harald~P.}\ \bibnamefont
  {Pfeiffer}}, \ and\ \bibinfo {author} {\bibfnamefont {Béla}\ \bibnamefont
  {Szilágyi}},\ }\href {\doibase10.1103/PhysRevD.96.024058} {\bibfield
  {journal} {\bibinfo  {journal} {Phys. Rev.}\ }\textbf {\bibinfo {volume}
  {D96}},\ \bibinfo {pages} {024058} (\bibinfo {year} {2017}{\natexlab{b}})},\
  \Eprint {http://arxiv.org/abs/1705.07089} {arXiv:1705.07089
  [gr-qc]}\BibitemShut {NoStop}%
\bibitem [{\citenamefont {Kokkotas}\ and\ \citenamefont
  {Schmidt}(1999)}]{Kokkotas:1999bd}%
  \BibitemOpen
  \bibfield  {author} {\bibinfo {author} {\bibfnamefont {Kostas~D.}\
  \bibnamefont {Kokkotas}}\ and\ \bibinfo {author} {\bibfnamefont {Bernd~G.}\
  \bibnamefont {Schmidt}},\ }\href {\doibase10.12942/lrr-1999-2} {\bibfield
  {journal} {\bibinfo  {journal} {Living Rev. Rel.}\ }\textbf {\bibinfo
  {volume} {2}},\ \bibinfo {pages} {2} (\bibinfo {year} {1999})},\ \Eprint
  {http://arxiv.org/abs/gr-qc/9909058} {arXiv:gr-qc/9909058
  [gr-qc]}\BibitemShut {NoStop}%
\bibitem [{\citenamefont {Kamaretsos}\ \emph
  {et~al.}(2012{\natexlab{a}})\citenamefont {Kamaretsos}, \citenamefont
  {Hannam}, \citenamefont {Husa},\ and\ \citenamefont
  {Sathyaprakash}}]{Kamaretsos:2011um}%
  \BibitemOpen
  \bibfield  {author} {\bibinfo {author} {\bibfnamefont {Ioannis}\ \bibnamefont
  {Kamaretsos}}, \bibinfo {author} {\bibfnamefont {Mark}\ \bibnamefont
  {Hannam}}, \bibinfo {author} {\bibfnamefont {Sascha}\ \bibnamefont {Husa}}, \
  and\ \bibinfo {author} {\bibfnamefont {B.~S.}\ \bibnamefont
  {Sathyaprakash}},\ }\href {\doibase10.1103/PhysRevD.85.024018} {\bibfield
  {journal} {\bibinfo  {journal} {Phys. Rev.}\ }\textbf {\bibinfo {volume}
  {D85}},\ \bibinfo {pages} {024018} (\bibinfo {year} {2012}{\natexlab{a}})},\
  \Eprint {http://arxiv.org/abs/1107.0854} {arXiv:1107.0854
  [gr-qc]}\BibitemShut {NoStop}%
\bibitem [{\citenamefont {Kamaretsos}\ \emph
  {et~al.}(2012{\natexlab{b}})\citenamefont {Kamaretsos}, \citenamefont
  {Hannam},\ and\ \citenamefont {Sathyaprakash}}]{Kamaretsos:2012bs}%
  \BibitemOpen
  \bibfield  {author} {\bibinfo {author} {\bibfnamefont {Ioannis}\ \bibnamefont
  {Kamaretsos}}, \bibinfo {author} {\bibfnamefont {Mark}\ \bibnamefont
  {Hannam}}, \ and\ \bibinfo {author} {\bibfnamefont {B.}~\bibnamefont
  {Sathyaprakash}},\ }\href {\doibase10.1103/PhysRevLett.109.141102} {\bibfield
   {journal} {\bibinfo  {journal} {Phys. Rev. Lett.}\ }\textbf {\bibinfo
  {volume} {109}},\ \bibinfo {pages} {141102} (\bibinfo {year}
  {2012}{\natexlab{b}})},\ \Eprint {http://arxiv.org/abs/1207.0399}
  {arXiv:1207.0399 [gr-qc]}\BibitemShut {NoStop}%
\bibitem [{\citenamefont {Berti}\ \emph {et~al.}(2006)\citenamefont {Berti},
  \citenamefont {Cardoso},\ and\ \citenamefont {Will}}]{Berti:2005ys}%
  \BibitemOpen
  \bibfield  {author} {\bibinfo {author} {\bibfnamefont {Emanuele}\
  \bibnamefont {Berti}}, \bibinfo {author} {\bibfnamefont {Vitor}\ \bibnamefont
  {Cardoso}}, \ and\ \bibinfo {author} {\bibfnamefont {Clifford~M.}\
  \bibnamefont {Will}},\ }\href {\doibase10.1103/PhysRevD.73.064030} {\bibfield
   {journal} {\bibinfo  {journal} {Phys. Rev.}\ }\textbf {\bibinfo {volume}
  {D73}},\ \bibinfo {pages} {064030} (\bibinfo {year} {2006})},\ \Eprint
  {http://arxiv.org/abs/gr-qc/0512160} {arXiv:gr-qc/0512160
  [gr-qc]}\BibitemShut {NoStop}%
\bibitem [{\citenamefont {Barausse}\ and\ \citenamefont
  {Rezzolla}(2009)}]{Barausse:2009uz}%
  \BibitemOpen
  \bibfield  {author} {\bibinfo {author} {\bibfnamefont {Enrico}\ \bibnamefont
  {Barausse}}\ and\ \bibinfo {author} {\bibfnamefont {Luciano}\ \bibnamefont
  {Rezzolla}},\ }\href {\doibase10.1088/0004-637X/704/1/L40} {\bibfield
  {journal} {\bibinfo  {journal} {Astrophys. J.}\ }\textbf {\bibinfo {volume}
  {704}},\ \bibinfo {pages} {L40--L44} (\bibinfo {year} {2009})},\ \Eprint
  {http://arxiv.org/abs/0904.2577} {arXiv:0904.2577 [gr-qc]}\BibitemShut
  {NoStop}%
\bibitem [{\citenamefont {Hemberger}\ \emph {et~al.}(2013)\citenamefont
  {Hemberger}, \citenamefont {Lovelace}, \citenamefont {Loredo}, \citenamefont
  {Kidder}, \citenamefont {Scheel}, \citenamefont {Szilágyi}, \citenamefont
  {Taylor},\ and\ \citenamefont {Teukolsky}}]{Hemberger:2013hsa}%
  \BibitemOpen
  \bibfield  {author} {\bibinfo {author} {\bibfnamefont {Daniel~A.}\
  \bibnamefont {Hemberger}}, \bibinfo {author} {\bibfnamefont {Geoffrey}\
  \bibnamefont {Lovelace}}, \bibinfo {author} {\bibfnamefont {Thomas~J.}\
  \bibnamefont {Loredo}}, \bibinfo {author} {\bibfnamefont {Lawrence~E.}\
  \bibnamefont {Kidder}}, \bibinfo {author} {\bibfnamefont {Mark~A.}\
  \bibnamefont {Scheel}}, \bibinfo {author} {\bibfnamefont {Béla}\
  \bibnamefont {Szilágyi}}, \bibinfo {author} {\bibfnamefont {Nicholas~W.}\
  \bibnamefont {Taylor}}, \ and\ \bibinfo {author} {\bibfnamefont {Saul~A.}\
  \bibnamefont {Teukolsky}},\ }\href {\doibase10.1103/PhysRevD.88.064014}
  {\bibfield  {journal} {\bibinfo  {journal} {Phys. Rev.}\ }\textbf {\bibinfo
  {volume} {D88}},\ \bibinfo {pages} {064014} (\bibinfo {year} {2013})},\
  \Eprint {http://arxiv.org/abs/1305.5991} {arXiv:1305.5991
  [gr-qc]}\BibitemShut {NoStop}%
\bibitem [{\citenamefont {Hofmann}\ \emph {et~al.}(2016)\citenamefont
  {Hofmann}, \citenamefont {Barausse},\ and\ \citenamefont
  {Rezzolla}}]{Hofmann:2016yih}%
  \BibitemOpen
  \bibfield  {author} {\bibinfo {author} {\bibfnamefont {Fabian}\ \bibnamefont
  {Hofmann}}, \bibinfo {author} {\bibfnamefont {Enrico}\ \bibnamefont
  {Barausse}}, \ and\ \bibinfo {author} {\bibfnamefont {Luciano}\ \bibnamefont
  {Rezzolla}},\ }\href {\doibase10.3847/2041-8205/825/2/L19} {\bibfield
  {journal} {\bibinfo  {journal} {Astrophys. J.}\ }\textbf {\bibinfo {volume}
  {825}},\ \bibinfo {pages} {L19} (\bibinfo {year} {2016})},\ \Eprint
  {http://arxiv.org/abs/1605.01938} {arXiv:1605.01938 [gr-qc]}\BibitemShut
  {NoStop}%
\bibitem [{\citenamefont {Healy}\ and\ \citenamefont
  {Lousto}(2017)}]{Healy:2016lce}%
  \BibitemOpen
  \bibfield  {author} {\bibinfo {author} {\bibfnamefont {James}\ \bibnamefont
  {Healy}}\ and\ \bibinfo {author} {\bibfnamefont {Carlos~O.}\ \bibnamefont
  {Lousto}},\ }\href {\doibase10.1103/PhysRevD.95.024037} {\bibfield  {journal}
  {\bibinfo  {journal} {Phys. Rev.}\ }\textbf {\bibinfo {volume} {D95}},\
  \bibinfo {pages} {024037} (\bibinfo {year} {2017})},\ \Eprint
  {http://arxiv.org/abs/1610.09713} {arXiv:1610.09713 [gr-qc]}\BibitemShut
  {NoStop}%
\bibitem [{\citenamefont {Jiménez-Forteza}\ \emph {et~al.}(2017)\citenamefont
  {Jiménez-Forteza}, \citenamefont {Keitel}, \citenamefont {Husa},
  \citenamefont {Hannam}, \citenamefont {Khan},\ and\ \citenamefont
  {Pürrer}}]{Jimenez-Forteza:2016oae}%
  \BibitemOpen
  \bibfield  {author} {\bibinfo {author} {\bibfnamefont {Xisco}\ \bibnamefont
  {Jiménez-Forteza}}, \bibinfo {author} {\bibfnamefont {David}\ \bibnamefont
  {Keitel}}, \bibinfo {author} {\bibfnamefont {Sascha}\ \bibnamefont {Husa}},
  \bibinfo {author} {\bibfnamefont {Mark}\ \bibnamefont {Hannam}}, \bibinfo
  {author} {\bibfnamefont {Sebastian}\ \bibnamefont {Khan}}, \ and\ \bibinfo
  {author} {\bibfnamefont {Michael}\ \bibnamefont {Pürrer}},\ }\href
  {\doibase10.1103/PhysRevD.95.064024} {\bibfield  {journal} {\bibinfo
  {journal} {Phys. Rev.}\ }\textbf {\bibinfo {volume} {D95}},\ \bibinfo {pages}
  {064024} (\bibinfo {year} {2017})},\ \Eprint
  {http://arxiv.org/abs/1611.00332} {arXiv:1611.00332 [gr-qc]}\BibitemShut
  {NoStop}%
\bibitem [{\citenamefont {Buonanno}\ \emph {et~al.}(2008)\citenamefont
  {Buonanno}, \citenamefont {Kidder},\ and\ \citenamefont
  {Lehner}}]{Buonanno:2007sv}%
  \BibitemOpen
  \bibfield  {author} {\bibinfo {author} {\bibfnamefont {Alessandra}\
  \bibnamefont {Buonanno}}, \bibinfo {author} {\bibfnamefont {Lawrence~E.}\
  \bibnamefont {Kidder}}, \ and\ \bibinfo {author} {\bibfnamefont {Luis}\
  \bibnamefont {Lehner}},\ }\href {\doibase10.1103/PhysRevD.77.026004}
  {\bibfield  {journal} {\bibinfo  {journal} {Phys. Rev.}\ }\textbf {\bibinfo
  {volume} {D77}},\ \bibinfo {pages} {026004} (\bibinfo {year} {2008})},\
  \Eprint {http://arxiv.org/abs/0709.3839} {arXiv:0709.3839
  [astro-ph]}\BibitemShut {NoStop}%
\bibitem [{\citenamefont {Rezzolla}\ \emph {et~al.}(2008)\citenamefont
  {Rezzolla}, \citenamefont {Barausse}, \citenamefont {Dorband}, \citenamefont
  {Pollney}, \citenamefont {Reisswig}, \citenamefont {Seiler},\ and\
  \citenamefont {Husa}}]{Rezzolla:2007rz}%
  \BibitemOpen
  \bibfield  {author} {\bibinfo {author} {\bibfnamefont {Luciano}\ \bibnamefont
  {Rezzolla}}, \bibinfo {author} {\bibfnamefont {Enrico}\ \bibnamefont
  {Barausse}}, \bibinfo {author} {\bibfnamefont {Ernst~Nils}\ \bibnamefont
  {Dorband}}, \bibinfo {author} {\bibfnamefont {Denis}\ \bibnamefont
  {Pollney}}, \bibinfo {author} {\bibfnamefont {Christian}\ \bibnamefont
  {Reisswig}}, \bibinfo {author} {\bibfnamefont {Jennifer}\ \bibnamefont
  {Seiler}}, \ and\ \bibinfo {author} {\bibfnamefont {Sascha}\ \bibnamefont
  {Husa}},\ }\href {\doibase10.1103/PhysRevD.78.044002} {\bibfield  {journal}
  {\bibinfo  {journal} {Phys. Rev.}\ }\textbf {\bibinfo {volume} {D78}},\
  \bibinfo {pages} {044002} (\bibinfo {year} {2008})},\ \Eprint
  {http://arxiv.org/abs/0712.3541} {arXiv:0712.3541 [gr-qc]}\BibitemShut
  {NoStop}%
\bibitem [{\citenamefont {Lousto}\ \emph {et~al.}(2010)\citenamefont {Lousto},
  \citenamefont {Campanelli}, \citenamefont {Zlochower},\ and\ \citenamefont
  {Nakano}}]{Lousto:2009mf}%
  \BibitemOpen
  \bibfield  {author} {\bibinfo {author} {\bibfnamefont {Carlos~O.}\
  \bibnamefont {Lousto}}, \bibinfo {author} {\bibfnamefont {Manuela}\
  \bibnamefont {Campanelli}}, \bibinfo {author} {\bibfnamefont {Yosef}\
  \bibnamefont {Zlochower}}, \ and\ \bibinfo {author} {\bibfnamefont
  {Hiroyuki}\ \bibnamefont {Nakano}},\ }\bibfield  {booktitle} {\emph {\bibinfo
  {booktitle} {{Numerical relativity and data analysis. Proceedings, 3rd Annual
  Meeting, NRDA 2009, Potsdam, Germany, July 6-9, 2009}}},\ }\href
  {\doibase10.1088/0264-9381/27/11/114006} {\bibfield  {journal} {\bibinfo
  {journal} {Class. Quant. Grav.}\ }\textbf {\bibinfo {volume} {27}},\ \bibinfo
  {pages} {114006} (\bibinfo {year} {2010})},\ \Eprint
  {http://arxiv.org/abs/0904.3541} {arXiv:0904.3541 [gr-qc]}\BibitemShut
  {NoStop}%
\bibitem [{\citenamefont {Ghosh}\ \emph {et~al.}(2016)\citenamefont {Ghosh}
  \emph {et~al.}}]{Ghosh:2016qgn}%
  \BibitemOpen
  \bibfield  {author} {\bibinfo {author} {\bibfnamefont {Abhirup}\ \bibnamefont
  {Ghosh}} \emph {et~al.},\ }\href {\doibase10.1103/PhysRevD.94.021101}
  {\bibfield  {journal} {\bibinfo  {journal} {Phys. Rev.}\ }\textbf {\bibinfo
  {volume} {D94}},\ \bibinfo {pages} {021101} (\bibinfo {year} {2016})},\
  \Eprint {http://arxiv.org/abs/1602.02453} {arXiv:1602.02453
  [gr-qc]}\BibitemShut {NoStop}%
\bibitem [{\citenamefont {Gonzalez}\ \emph
  {et~al.}(2007{\natexlab{a}})\citenamefont {Gonzalez}, \citenamefont
  {Sperhake}, \citenamefont {Bruegmann}, \citenamefont {Hannam},\ and\
  \citenamefont {Husa}}]{Gonzalez:2006md}%
  \BibitemOpen
  \bibfield  {author} {\bibinfo {author} {\bibfnamefont {Jose~A.}\ \bibnamefont
  {Gonzalez}}, \bibinfo {author} {\bibfnamefont {Ulrich}\ \bibnamefont
  {Sperhake}}, \bibinfo {author} {\bibfnamefont {Bernd}\ \bibnamefont
  {Bruegmann}}, \bibinfo {author} {\bibfnamefont {Mark}\ \bibnamefont
  {Hannam}}, \ and\ \bibinfo {author} {\bibfnamefont {Sascha}\ \bibnamefont
  {Husa}},\ }\href {\doibase10.1103/PhysRevLett.98.091101} {\bibfield
  {journal} {\bibinfo  {journal} {Phys. Rev. Lett.}\ }\textbf {\bibinfo
  {volume} {98}},\ \bibinfo {pages} {091101} (\bibinfo {year}
  {2007}{\natexlab{a}})},\ \Eprint {http://arxiv.org/abs/gr-qc/0610154}
  {arXiv:gr-qc/0610154 [gr-qc]}\BibitemShut {NoStop}%
\bibitem [{\citenamefont {Gonzalez}\ \emph
  {et~al.}(2007{\natexlab{b}})\citenamefont {Gonzalez}, \citenamefont {Hannam},
  \citenamefont {Sperhake}, \citenamefont {Bruegmann},\ and\ \citenamefont
  {Husa}}]{Gonzalez:2007hi}%
  \BibitemOpen
  \bibfield  {author} {\bibinfo {author} {\bibfnamefont {J.~A.}\ \bibnamefont
  {Gonzalez}}, \bibinfo {author} {\bibfnamefont {M.~D.}\ \bibnamefont
  {Hannam}}, \bibinfo {author} {\bibfnamefont {U.}~\bibnamefont {Sperhake}},
  \bibinfo {author} {\bibfnamefont {Bernd}\ \bibnamefont {Bruegmann}}, \ and\
  \bibinfo {author} {\bibfnamefont {S.}~\bibnamefont {Husa}},\ }\href
  {\doibase10.1103/PhysRevLett.98.231101} {\bibfield  {journal} {\bibinfo
  {journal} {Phys. Rev. Lett.}\ }\textbf {\bibinfo {volume} {98}},\ \bibinfo
  {pages} {231101} (\bibinfo {year} {2007}{\natexlab{b}})},\ \Eprint
  {http://arxiv.org/abs/gr-qc/0702052} {arXiv:gr-qc/0702052
  [GR-QC]}\BibitemShut {NoStop}%
\bibitem [{\citenamefont {Bruegmann}\ \emph
  {et~al.}(2008{\natexlab{b}})\citenamefont {Bruegmann}, \citenamefont
  {Gonzalez}, \citenamefont {Hannam}, \citenamefont {Husa},\ and\ \citenamefont
  {Sperhake}}]{Brugmann:2007zj}%
  \BibitemOpen
  \bibfield  {author} {\bibinfo {author} {\bibfnamefont {Bernd}\ \bibnamefont
  {Bruegmann}}, \bibinfo {author} {\bibfnamefont {Jose~A.}\ \bibnamefont
  {Gonzalez}}, \bibinfo {author} {\bibfnamefont {Mark}\ \bibnamefont {Hannam}},
  \bibinfo {author} {\bibfnamefont {Sascha}\ \bibnamefont {Husa}}, \ and\
  \bibinfo {author} {\bibfnamefont {Ulrich}\ \bibnamefont {Sperhake}},\ }\href
  {\doibase10.1103/PhysRevD.77.124047} {\bibfield  {journal} {\bibinfo
  {journal} {Phys. Rev.}\ }\textbf {\bibinfo {volume} {D77}},\ \bibinfo {pages}
  {124047} (\bibinfo {year} {2008}{\natexlab{b}})},\ \Eprint
  {http://arxiv.org/abs/0707.0135} {arXiv:0707.0135 [gr-qc]}\BibitemShut
  {NoStop}%
\bibitem [{\citenamefont {Lousto}\ and\ \citenamefont
  {Zlochower}(2011{\natexlab{b}})}]{Lousto:2011kp}%
  \BibitemOpen
  \bibfield  {author} {\bibinfo {author} {\bibfnamefont {Carlos~O.}\
  \bibnamefont {Lousto}}\ and\ \bibinfo {author} {\bibfnamefont {Yosef}\
  \bibnamefont {Zlochower}},\ }\href {\doibase10.1103/PhysRevLett.107.231102}
  {\bibfield  {journal} {\bibinfo  {journal} {Phys. Rev. Lett.}\ }\textbf
  {\bibinfo {volume} {107}},\ \bibinfo {pages} {231102} (\bibinfo {year}
  {2011}{\natexlab{b}})},\ \Eprint {http://arxiv.org/abs/1108.2009}
  {arXiv:1108.2009 [gr-qc]}\BibitemShut {NoStop}%
\bibitem [{\citenamefont {Gerosa}\ \emph {et~al.}(2018)\citenamefont {Gerosa},
  \citenamefont {Hébert},\ and\ \citenamefont {Stein}}]{Gerosa:2018qay}%
  \BibitemOpen
  \bibfield  {author} {\bibinfo {author} {\bibfnamefont {Davide}\ \bibnamefont
  {Gerosa}}, \bibinfo {author} {\bibfnamefont {François}\ \bibnamefont
  {Hébert}}, \ and\ \bibinfo {author} {\bibfnamefont {Leo~C.}\ \bibnamefont
  {Stein}},\ }\href {\doibase10.1103/PhysRevD.97.104049} {\bibfield  {journal}
  {\bibinfo  {journal} {Phys. Rev.}\ }\textbf {\bibinfo {volume} {D97}},\
  \bibinfo {pages} {104049} (\bibinfo {year} {2018})},\ \Eprint
  {http://arxiv.org/abs/1802.04276} {arXiv:1802.04276 [gr-qc]}\BibitemShut
  {NoStop}%
\bibitem [{\citenamefont {Gerosa}\ and\ \citenamefont
  {Moore}(2016)}]{Gerosa:2016vip}%
  \BibitemOpen
  \bibfield  {author} {\bibinfo {author} {\bibfnamefont {Davide}\ \bibnamefont
  {Gerosa}}\ and\ \bibinfo {author} {\bibfnamefont {Christopher~J.}\
  \bibnamefont {Moore}},\ }\href {\doibase10.1103/PhysRevLett.117.011101}
  {\bibfield  {journal} {\bibinfo  {journal} {Phys. Rev. Lett.}\ }\textbf
  {\bibinfo {volume} {117}},\ \bibinfo {pages} {011101} (\bibinfo {year}
  {2016})},\ \Eprint {http://arxiv.org/abs/1606.04226} {arXiv:1606.04226
  [gr-qc]}\BibitemShut {NoStop}%
\bibitem [{\citenamefont {Lousto}\ \emph {et~al.}(2016)\citenamefont {Lousto},
  \citenamefont {Healy},\ and\ \citenamefont {Nakano}}]{Lousto:2015uwa}%
  \BibitemOpen
  \bibfield  {author} {\bibinfo {author} {\bibfnamefont {Carlos~O.}\
  \bibnamefont {Lousto}}, \bibinfo {author} {\bibfnamefont {James}\
  \bibnamefont {Healy}}, \ and\ \bibinfo {author} {\bibfnamefont {Hiroyuki}\
  \bibnamefont {Nakano}},\ }\href {\doibase10.1103/PhysRevD.93.044031}
  {\bibfield  {journal} {\bibinfo  {journal} {Phys. Rev.}\ }\textbf {\bibinfo
  {volume} {D93}},\ \bibinfo {pages} {044031} (\bibinfo {year} {2016})},\
  \Eprint {http://arxiv.org/abs/1506.04768} {arXiv:1506.04768
  [gr-qc]}\BibitemShut {NoStop}%
\bibitem [{\citenamefont {Berti}(2006)}]{Berti:2006ew}%
  \BibitemOpen
  \bibfield  {author} {\bibinfo {author} {\bibfnamefont {Emanuele}\
  \bibnamefont {Berti}},\ }\bibfield  {booktitle} {\emph {\bibinfo {booktitle}
  {{Gravitational wave data analysis. Proceedings, 10th Workshop, GWDAW-10,
  Brownsville, USA, December 14-17, 2005}}},\ }\href
  {\doibase10.1088/0264-9381/23/19/S17} {\bibfield  {journal} {\bibinfo
  {journal} {Class. Quant. Grav.}\ }\textbf {\bibinfo {volume} {23}},\ \bibinfo
  {pages} {S785--S798} (\bibinfo {year} {2006})},\ \Eprint
  {http://arxiv.org/abs/astro-ph/0602470} {arXiv:astro-ph/0602470
  [astro-ph]}\BibitemShut {NoStop}%
\bibitem [{\citenamefont {Cutler}\ and\ \citenamefont
  {Vallisneri}(2007)}]{Cutler:2007mi}%
  \BibitemOpen
  \bibfield  {author} {\bibinfo {author} {\bibfnamefont {Curt}\ \bibnamefont
  {Cutler}}\ and\ \bibinfo {author} {\bibfnamefont {Michele}\ \bibnamefont
  {Vallisneri}},\ }\href {\doibase10.1103/PhysRevD.76.104018} {\bibfield
  {journal} {\bibinfo  {journal} {Phys. Rev.}\ }\textbf {\bibinfo {volume}
  {D76}},\ \bibinfo {pages} {104018} (\bibinfo {year} {2007})},\ \Eprint
  {http://arxiv.org/abs/0707.2982} {arXiv:0707.2982 [gr-qc]}\BibitemShut
  {NoStop}%
\bibitem [{\citenamefont {Kowalski}\ \emph {et~al.}(2008)\citenamefont
  {Kowalski} \emph {et~al.}}]{Kowalski:2008ez}%
  \BibitemOpen
  \bibfield  {author} {\bibinfo {author} {\bibfnamefont {M.}~\bibnamefont
  {Kowalski}} \emph {et~al.} (\bibinfo {collaboration} {Supernova Cosmology
  Project}),\ }\href {\doibase10.1086/589937} {\bibfield  {journal} {\bibinfo
  {journal} {Astrophys. J.}\ }\textbf {\bibinfo {volume} {686}},\ \bibinfo
  {pages} {749--778} (\bibinfo {year} {2008})},\ \Eprint
  {http://arxiv.org/abs/0804.4142} {0804.4142}\BibitemShut {NoStop}%
\bibitem [{\citenamefont {Choptuik}(1993)}]{Choptuik:1992jv}%
  \BibitemOpen
  \bibfield  {author} {\bibinfo {author} {\bibfnamefont {M.~W.}\ \bibnamefont
  {Choptuik}},\ }\href {\doibase10.1103/PhysRevLett.70.9} {\bibfield  {journal}
  {\bibinfo  {journal} {Phys. Rev. Lett.}\ }\textbf {\bibinfo {volume} {70}},\
  \bibinfo {pages} {9--12} (\bibinfo {year} {1993})}\BibitemShut {NoStop}%
\bibitem [{\citenamefont {Khlopov}(2017)}]{Khlopov:2017vcj}%
  \BibitemOpen
  \bibfield  {author} {\bibinfo {author} {\bibfnamefont {M.~{\relax Yu}.}\
  \bibnamefont {Khlopov}},\ }\href {\doibase10.1142/S0217732317020011}
  {\bibfield  {journal} {\bibinfo  {journal} {Mod. Phys. Lett.}\ }\textbf
  {\bibinfo {volume} {A32}},\ \bibinfo {pages} {1702001} (\bibinfo {year}
  {2017})},\ \Eprint {http://arxiv.org/abs/1704.06511} {1704.06511}\BibitemShut
  {NoStop}%
\bibitem [{\citenamefont {Arvanitaki}\ and\ \citenamefont
  {Dubovsky}(2011)}]{Arvanitaki:2010sy}%
  \BibitemOpen
  \bibfield  {author} {\bibinfo {author} {\bibfnamefont {Asimina}\ \bibnamefont
  {Arvanitaki}}\ and\ \bibinfo {author} {\bibfnamefont {Sergei}\ \bibnamefont
  {Dubovsky}},\ }\href {\doibase10.1103/PhysRevD.83.044026} {\bibfield
  {journal} {\bibinfo  {journal} {Phys. Rev.}\ }\textbf {\bibinfo {volume}
  {D83}},\ \bibinfo {pages} {044026} (\bibinfo {year} {2011})},\ \Eprint
  {http://arxiv.org/abs/1004.3558} {arXiv:1004.3558 [hep-th]}\BibitemShut
  {NoStop}%
\bibitem [{\citenamefont {Ackerman}\ \emph {et~al.}(2009)\citenamefont
  {Ackerman}, \citenamefont {Buckley}, \citenamefont {Carroll},\ and\
  \citenamefont {Kamionkowski}}]{Ackerman:mha}%
  \BibitemOpen
  \bibfield  {author} {\bibinfo {author} {\bibfnamefont {Lotty}\ \bibnamefont
  {Ackerman}}, \bibinfo {author} {\bibfnamefont {Matthew~R.}\ \bibnamefont
  {Buckley}}, \bibinfo {author} {\bibfnamefont {Sean~M.}\ \bibnamefont
  {Carroll}}, \ and\ \bibinfo {author} {\bibfnamefont {Marc}\ \bibnamefont
  {Kamionkowski}},\ }\bibfield  {booktitle} {\emph {\bibinfo {booktitle}
  {{Proceedings, 7th International Heidelberg Conference on Dark Matter in
  Astro and Particle Physics (DARK 2009): Christchurch, New Zealand, January
  18-24, 2009}}},\ }\href {\doibase10.1103/PhysRevD.79.023519} {\bibfield
  {journal} {\bibinfo  {journal} {Phys. Rev.}\ }\textbf {\bibinfo {volume}
  {D79}},\ \bibinfo {pages} {023519} (\bibinfo {year} {2009})},\ \Eprint
  {http://arxiv.org/abs/0810.5126} {arXiv:0810.5126 [hep-ph]}\BibitemShut
  {NoStop}%
\bibitem [{\citenamefont {Klasen}\ \emph {et~al.}(2015)\citenamefont {Klasen},
  \citenamefont {Pohl},\ and\ \citenamefont {Sigl}}]{Klasen:2015uma}%
  \BibitemOpen
  \bibfield  {author} {\bibinfo {author} {\bibfnamefont {M.}~\bibnamefont
  {Klasen}}, \bibinfo {author} {\bibfnamefont {M.}~\bibnamefont {Pohl}}, \ and\
  \bibinfo {author} {\bibfnamefont {G.}~\bibnamefont {Sigl}},\ }\href
  {\doibase10.1016/j.ppnp.2015.07.001} {\bibfield  {journal} {\bibinfo
  {journal} {Prog. Part. Nucl. Phys.}\ }\textbf {\bibinfo {volume} {85}},\
  \bibinfo {pages} {1--32} (\bibinfo {year} {2015})},\ \Eprint
  {http://arxiv.org/abs/1507.03800} {1507.03800}\BibitemShut {NoStop}%
\bibitem [{\citenamefont {Witek}\ \emph {et~al.}(2013)\citenamefont {Witek},
  \citenamefont {Cardoso}, \citenamefont {Ishibashi},\ and\ \citenamefont
  {Sperhake}}]{Witek:2012tr}%
  \BibitemOpen
  \bibfield  {author} {\bibinfo {author} {\bibfnamefont {Helvi}\ \bibnamefont
  {Witek}}, \bibinfo {author} {\bibfnamefont {Vitor}\ \bibnamefont {Cardoso}},
  \bibinfo {author} {\bibfnamefont {Akihiro}\ \bibnamefont {Ishibashi}}, \ and\
  \bibinfo {author} {\bibfnamefont {Ulrich}\ \bibnamefont {Sperhake}},\ }\href
  {\doibase10.1103/PhysRevD.87.043513} {\bibfield  {journal} {\bibinfo
  {journal} {Phys. Rev.}\ }\textbf {\bibinfo {volume} {D87}},\ \bibinfo {pages}
  {043513} (\bibinfo {year} {2013})},\ \Eprint {http://arxiv.org/abs/1212.0551}
  {arXiv:1212.0551 [gr-qc]}\BibitemShut {NoStop}%
\bibitem [{\citenamefont {Okawa}\ \emph {et~al.}(2014)\citenamefont {Okawa},
  \citenamefont {Witek},\ and\ \citenamefont {Cardoso}}]{Okawa:2014nda}%
  \BibitemOpen
  \bibfield  {author} {\bibinfo {author} {\bibfnamefont {H.}~\bibnamefont
  {Okawa}}, \bibinfo {author} {\bibfnamefont {H.}~\bibnamefont {Witek}}, \ and\
  \bibinfo {author} {\bibfnamefont {V.}~\bibnamefont {Cardoso}},\ }\href
  {\doibase10.1103/PhysRevD.89.104032} {\bibfield  {journal} {\bibinfo
  {journal} {Phys. Rev. D}\ }\textbf {\bibinfo {volume} {89}},\ \bibinfo
  {pages} {104032} (\bibinfo {year} {2014})},\ \Eprint
  {http://arxiv.org/abs/1401.1548} {1401.1548}\BibitemShut {NoStop}%
\bibitem [{\citenamefont {Zilh{\~a}o}\ \emph {et~al.}(2015)\citenamefont
  {Zilh{\~a}o}, \citenamefont {Witek},\ and\ \citenamefont
  {Cardoso}}]{Zilhao:2015tya}%
  \BibitemOpen
  \bibfield  {author} {\bibinfo {author} {\bibfnamefont {M.}~\bibnamefont
  {Zilh{\~a}o}}, \bibinfo {author} {\bibfnamefont {H.}~\bibnamefont {Witek}}, \
  and\ \bibinfo {author} {\bibfnamefont {V.}~\bibnamefont {Cardoso}},\ }\href
  {\doibase10.1088/0264-9381/32/23/234003} {\bibfield  {journal} {\bibinfo
  {journal} {Class. Quant. Grav.}\ }\textbf {\bibinfo {volume} {32}},\ \bibinfo
  {pages} {234003} (\bibinfo {year} {2015})},\ \Eprint
  {http://arxiv.org/abs/1505.00797} {1505.00797}\BibitemShut {NoStop}%
\bibitem [{\citenamefont {Zel'dovich}(1971)}]{Zeldovich:1971}%
  \BibitemOpen
  \bibfield  {author} {\bibinfo {author} {\bibfnamefont {Ya.~B.}\ \bibnamefont
  {Zel'dovich}},\ }\href
  {http://www.jetpletters.ac.ru/ps/1604/article_24607.shtml} {\bibfield
  {journal} {\bibinfo  {journal} {Pis'ma Zh. Eksp. Teor. Fiz.}\ }\textbf
  {\bibinfo {volume} {14}},\ \bibinfo {pages} {270} (\bibinfo {year}
  {1971})}\BibitemShut {NoStop}%
\bibitem [{\citenamefont {Zel'dovich}(1972)}]{Zeldovich1972}%
  \BibitemOpen
  \bibfield  {author} {\bibinfo {author} {\bibfnamefont {Ya.~B.}\ \bibnamefont
  {Zel'dovich}},\ }\href
  {http://www.jetp.ac.ru/cgi-bin/e/index/e/35/6/p1085?a=list} {\bibfield
  {journal} {\bibinfo  {journal} {Zh. Eksp. Teor. Fiz}\ }\textbf {\bibinfo
  {volume} {62}},\ \bibinfo {pages} {2076} (\bibinfo {year}
  {1972})}\BibitemShut {NoStop}%
\bibitem [{\citenamefont {Misner}(1972)}]{Misner:1972kx}%
  \BibitemOpen
  \bibfield  {author} {\bibinfo {author} {\bibfnamefont {C.~W.}\ \bibnamefont
  {Misner}},\ }\href {\doibase10.1103/PhysRevLett.28.994} {\bibfield  {journal}
  {\bibinfo  {journal} {Phys. Rev. Lett.}\ }\textbf {\bibinfo {volume} {28}},\
  \bibinfo {pages} {994--997} (\bibinfo {year} {1972})}\BibitemShut {NoStop}%
\bibitem [{\citenamefont {Brito}\ \emph
  {et~al.}(2015{\natexlab{a}})\citenamefont {Brito}, \citenamefont {Cardoso},\
  and\ \citenamefont {Pani}}]{Brito:2015oca}%
  \BibitemOpen
  \bibfield  {author} {\bibinfo {author} {\bibfnamefont {Richard}\ \bibnamefont
  {Brito}}, \bibinfo {author} {\bibfnamefont {Vitor}\ \bibnamefont {Cardoso}},
  \ and\ \bibinfo {author} {\bibfnamefont {Paolo}\ \bibnamefont {Pani}},\
  }\href {\doibase10.1007/978-3-319-19000-6} {\bibfield  {journal} {\bibinfo
  {journal} {Lect. Notes Phys.}\ }\textbf {\bibinfo {volume} {906}},\ \bibinfo
  {pages} {pp.1--237} (\bibinfo {year} {2015}{\natexlab{a}})},\ \Eprint
  {http://arxiv.org/abs/1501.06570} {arXiv:1501.06570 [gr-qc]}\BibitemShut
  {NoStop}%
\bibitem [{\citenamefont {Penrose}\ and\ \citenamefont
  {Floyd}(1971)}]{Penrose:1971uk}%
  \BibitemOpen
  \bibfield  {author} {\bibinfo {author} {\bibfnamefont {R.}~\bibnamefont
  {Penrose}}\ and\ \bibinfo {author} {\bibfnamefont {R.~M.}\ \bibnamefont
  {Floyd}},\ }\href {\doibase10.1038/physci229177a0} {\bibfield  {journal}
  {\bibinfo  {journal} {Nature}\ }\textbf {\bibinfo {volume} {229}},\ \bibinfo
  {pages} {177--179} (\bibinfo {year} {1971})}\BibitemShut {NoStop}%
\bibitem [{\citenamefont {Press}\ and\ \citenamefont
  {Teukolsky}(1972)}]{Press:1972zz}%
  \BibitemOpen
  \bibfield  {author} {\bibinfo {author} {\bibfnamefont {W.~H.}\ \bibnamefont
  {Press}}\ and\ \bibinfo {author} {\bibfnamefont {S.~A.}\ \bibnamefont
  {Teukolsky}},\ }\href {\doibase10.1038/238211a0} {\bibfield  {journal}
  {\bibinfo  {journal} {Nature}\ }\textbf {\bibinfo {volume} {238}},\ \bibinfo
  {pages} {211--212} (\bibinfo {year} {1972})}\BibitemShut {NoStop}%
\bibitem [{\citenamefont {Cardoso}\ \emph {et~al.}(2004)\citenamefont
  {Cardoso}, \citenamefont {Dias}, \citenamefont {Lemos},\ and\ \citenamefont
  {Yoshida}}]{Cardoso:2004nk}%
  \BibitemOpen
  \bibfield  {author} {\bibinfo {author} {\bibfnamefont {Vitor}\ \bibnamefont
  {Cardoso}}, \bibinfo {author} {\bibfnamefont {Oscar J.~C.}\ \bibnamefont
  {Dias}}, \bibinfo {author} {\bibfnamefont {Jose P.~S.}\ \bibnamefont
  {Lemos}}, \ and\ \bibinfo {author} {\bibfnamefont {Shijun}\ \bibnamefont
  {Yoshida}},\ }\href {\doibase10.1103/PhysRevD.70.044039} {\bibfield
  {journal} {\bibinfo  {journal} {Phys. Rev.}\ }\textbf {\bibinfo {volume}
  {D70}},\ \bibinfo {pages} {044039} (\bibinfo {year} {2004})},\ \Eprint
  {http://arxiv.org/abs/hep-th/0404096} {arXiv:hep-th/0404096}\BibitemShut
  {NoStop}%
\bibitem [{\citenamefont {Cardoso}\ \emph {et~al.}(2011)\citenamefont
  {Cardoso}, \citenamefont {Chakrabarti}, \citenamefont {Pani}, \citenamefont
  {Berti},\ and\ \citenamefont {Gualtieri}}]{Cardoso:2011xi}%
  \BibitemOpen
  \bibfield  {author} {\bibinfo {author} {\bibfnamefont {Vitor}\ \bibnamefont
  {Cardoso}}, \bibinfo {author} {\bibfnamefont {Sayan}\ \bibnamefont
  {Chakrabarti}}, \bibinfo {author} {\bibfnamefont {Paolo}\ \bibnamefont
  {Pani}}, \bibinfo {author} {\bibfnamefont {Emanuele}\ \bibnamefont {Berti}},
  \ and\ \bibinfo {author} {\bibfnamefont {Leonardo}\ \bibnamefont
  {Gualtieri}},\ }\href {\doibase10.1103/PhysRevLett.107.241101} {\bibfield
  {journal} {\bibinfo  {journal} {Phys. Rev. Lett.}\ }\textbf {\bibinfo
  {volume} {107}},\ \bibinfo {pages} {241101} (\bibinfo {year} {2011})},\
  \Eprint {http://arxiv.org/abs/1109.6021} {arXiv:1109.6021
  [gr-qc]}\BibitemShut {NoStop}%
\bibitem [{\citenamefont {Fujita}\ and\ \citenamefont
  {Cardoso}(2017)}]{Fujita:2016yav}%
  \BibitemOpen
  \bibfield  {author} {\bibinfo {author} {\bibfnamefont {Ryuichi}\ \bibnamefont
  {Fujita}}\ and\ \bibinfo {author} {\bibfnamefont {Vitor}\ \bibnamefont
  {Cardoso}},\ }\href {\doibase10.1103/PhysRevD.95.044016} {\bibfield
  {journal} {\bibinfo  {journal} {Phys. Rev.}\ }\textbf {\bibinfo {volume}
  {D95}},\ \bibinfo {pages} {044016} (\bibinfo {year} {2017})},\ \Eprint
  {http://arxiv.org/abs/1612.00978} {arXiv:1612.00978 [gr-qc]}\BibitemShut
  {NoStop}%
\bibitem [{\citenamefont {East}\ \emph {et~al.}(2014)\citenamefont {East},
  \citenamefont {Ramazanoğlu},\ and\ \citenamefont
  {Pretorius}}]{East:2013mfa}%
  \BibitemOpen
  \bibfield  {author} {\bibinfo {author} {\bibfnamefont {William~E.}\
  \bibnamefont {East}}, \bibinfo {author} {\bibfnamefont {Fethi~M.}\
  \bibnamefont {Ramazanoğlu}}, \ and\ \bibinfo {author} {\bibfnamefont
  {Frans}\ \bibnamefont {Pretorius}},\ }\href
  {\doibase10.1103/PhysRevD.89.061503} {\bibfield  {journal} {\bibinfo
  {journal} {Phys. Rev.}\ }\textbf {\bibinfo {volume} {D89}},\ \bibinfo {pages}
  {061503} (\bibinfo {year} {2014})},\ \Eprint {http://arxiv.org/abs/1312.4529}
  {arXiv:1312.4529 [gr-qc]}\BibitemShut {NoStop}%
\bibitem [{\citenamefont {Bosch}\ \emph {et~al.}(2016)\citenamefont {Bosch},
  \citenamefont {Green},\ and\ \citenamefont {Lehner}}]{Bosch:2016vcp}%
  \BibitemOpen
  \bibfield  {author} {\bibinfo {author} {\bibfnamefont {P.}~\bibnamefont
  {Bosch}}, \bibinfo {author} {\bibfnamefont {S.~R.}\ \bibnamefont {Green}}, \
  and\ \bibinfo {author} {\bibfnamefont {L.}~\bibnamefont {Lehner}},\ }\href
  {\doibase10.1103/PhysRevLett.116.141102} {\bibfield  {journal} {\bibinfo
  {journal} {Phys. Rev. Lett.}\ }\textbf {\bibinfo {volume} {116}},\ \bibinfo
  {pages} {141102} (\bibinfo {year} {2016})},\ \Eprint
  {http://arxiv.org/abs/1601.01384} {1601.01384}\BibitemShut {NoStop}%
\bibitem [{\citenamefont {Sanchis-Gual}\ \emph {et~al.}(2016)\citenamefont
  {Sanchis-Gual}, \citenamefont {Degollado}, \citenamefont {Herdeiro},
  \citenamefont {Font},\ and\ \citenamefont {Montero}}]{Sanchis-Gual:2016tcm}%
  \BibitemOpen
  \bibfield  {author} {\bibinfo {author} {\bibfnamefont {Nicolas}\ \bibnamefont
  {Sanchis-Gual}}, \bibinfo {author} {\bibfnamefont {Juan~Carlos}\ \bibnamefont
  {Degollado}}, \bibinfo {author} {\bibfnamefont {Carlos}\ \bibnamefont
  {Herdeiro}}, \bibinfo {author} {\bibfnamefont {José~A.}\ \bibnamefont
  {Font}}, \ and\ \bibinfo {author} {\bibfnamefont {Pedro~J.}\ \bibnamefont
  {Montero}},\ }\href {\doibase10.1103/PhysRevD.94.044061} {\bibfield
  {journal} {\bibinfo  {journal} {Phys. Rev.}\ }\textbf {\bibinfo {volume}
  {D94}},\ \bibinfo {pages} {044061} (\bibinfo {year} {2016})},\ \Eprint
  {http://arxiv.org/abs/1607.06304} {arXiv:1607.06304 [gr-qc]}\BibitemShut
  {NoStop}%
\bibitem [{\citenamefont {Chesler}\ and\ \citenamefont
  {Lowe}(2018)}]{Chesler:2018txn}%
  \BibitemOpen
  \bibfield  {author} {\bibinfo {author} {\bibfnamefont {Paul~M.}\ \bibnamefont
  {Chesler}}\ and\ \bibinfo {author} {\bibfnamefont {David~A.}\ \bibnamefont
  {Lowe}},\ }\href@noop {} {\  (\bibinfo {year} {2018})},\ \Eprint
  {http://arxiv.org/abs/1801.09711} {arXiv:1801.09711 [gr-qc]}\BibitemShut
  {NoStop}%
\bibitem [{\citenamefont {East}\ and\ \citenamefont
  {Pretorius}(2017)}]{East:2017ovw}%
  \BibitemOpen
  \bibfield  {author} {\bibinfo {author} {\bibfnamefont {William~E.}\
  \bibnamefont {East}}\ and\ \bibinfo {author} {\bibfnamefont {Frans}\
  \bibnamefont {Pretorius}},\ }\href {\doibase10.1103/PhysRevLett.119.041101}
  {\bibfield  {journal} {\bibinfo  {journal} {Phys. Rev. Lett.}\ }\textbf
  {\bibinfo {volume} {119}},\ \bibinfo {pages} {041101} (\bibinfo {year}
  {2017})},\ \Eprint {http://arxiv.org/abs/1704.04791} {arXiv:1704.04791
  [gr-qc]}\BibitemShut {NoStop}%
\bibitem [{\citenamefont {Brito}\ \emph
  {et~al.}(2015{\natexlab{b}})\citenamefont {Brito}, \citenamefont {Cardoso},\
  and\ \citenamefont {Okawa}}]{Brito:2015yga}%
  \BibitemOpen
  \bibfield  {author} {\bibinfo {author} {\bibfnamefont {Richard}\ \bibnamefont
  {Brito}}, \bibinfo {author} {\bibfnamefont {Vitor}\ \bibnamefont {Cardoso}},
  \ and\ \bibinfo {author} {\bibfnamefont {Hirotada}\ \bibnamefont {Okawa}},\
  }\href {\doibase10.1103/PhysRevLett.115.111301} {\bibfield  {journal}
  {\bibinfo  {journal} {Phys. Rev. Lett.}\ }\textbf {\bibinfo {volume} {115}},\
  \bibinfo {pages} {111301} (\bibinfo {year} {2015}{\natexlab{b}})},\ \Eprint
  {http://arxiv.org/abs/1508.04773} {arXiv:1508.04773 [gr-qc]}\BibitemShut
  {NoStop}%
\bibitem [{\citenamefont {Herdeiro}\ and\ \citenamefont
  {Radu}(2014{\natexlab{a}})}]{Herdeiro:2014goa}%
  \BibitemOpen
  \bibfield  {author} {\bibinfo {author} {\bibfnamefont {Carlos A.~R.}\
  \bibnamefont {Herdeiro}}\ and\ \bibinfo {author} {\bibfnamefont {Eugen}\
  \bibnamefont {Radu}},\ }\href {\doibase10.1103/PhysRevLett.112.221101}
  {\bibfield  {journal} {\bibinfo  {journal} {Phys. Rev. Lett.}\ }\textbf
  {\bibinfo {volume} {112}},\ \bibinfo {pages} {221101} (\bibinfo {year}
  {2014}{\natexlab{a}})},\ \Eprint {http://arxiv.org/abs/1403.2757}
  {arXiv:1403.2757 [gr-qc]}\BibitemShut {NoStop}%
\bibitem [{\citenamefont {Brihaye}\ \emph {et~al.}(2014)\citenamefont
  {Brihaye}, \citenamefont {Herdeiro},\ and\ \citenamefont
  {Radu}}]{Brihaye:2014nba}%
  \BibitemOpen
  \bibfield  {author} {\bibinfo {author} {\bibfnamefont {Y.}~\bibnamefont
  {Brihaye}}, \bibinfo {author} {\bibfnamefont {C.}~\bibnamefont {Herdeiro}}, \
  and\ \bibinfo {author} {\bibfnamefont {E.}~\bibnamefont {Radu}},\ }\href
  {\doibase10.1016/j.physletb.2014.10.019} {\bibfield  {journal} {\bibinfo
  {journal} {Phys. Lett. B}\ }\textbf {\bibinfo {volume} {739}},\ \bibinfo
  {pages} {1--7} (\bibinfo {year} {2014})},\ \Eprint
  {http://arxiv.org/abs/1408.5581} {1408.5581}\BibitemShut {NoStop}%
\bibitem [{\citenamefont {Benone}\ \emph {et~al.}(2014)\citenamefont {Benone},
  \citenamefont {Crispino}, \citenamefont {Herdeiro},\ and\ \citenamefont
  {Radu}}]{Benone:2014ssa}%
  \BibitemOpen
  \bibfield  {author} {\bibinfo {author} {\bibfnamefont {C.~L.}\ \bibnamefont
  {Benone}}, \bibinfo {author} {\bibfnamefont {L.~C.~B.}\ \bibnamefont
  {Crispino}}, \bibinfo {author} {\bibfnamefont {C.}~\bibnamefont {Herdeiro}},
  \ and\ \bibinfo {author} {\bibfnamefont {E.}~\bibnamefont {Radu}},\ }\href
  {\doibase10.1103/PhysRevD.90.104024} {\bibfield  {journal} {\bibinfo
  {journal} {Phys. Rev. D}\ }\textbf {\bibinfo {volume} {90}},\ \bibinfo
  {pages} {104024} (\bibinfo {year} {2014})},\ \Eprint
  {http://arxiv.org/abs/1409.1593} {1409.1593}\BibitemShut {NoStop}%
\bibitem [{\citenamefont {Herdeiro}\ \emph {et~al.}(2016)\citenamefont
  {Herdeiro}, \citenamefont {Radu},\ and\ \citenamefont
  {Runarsson}}]{Herdeiro:2016tmi}%
  \BibitemOpen
  \bibfield  {author} {\bibinfo {author} {\bibfnamefont {Carlos}\ \bibnamefont
  {Herdeiro}}, \bibinfo {author} {\bibfnamefont {Eugen}\ \bibnamefont {Radu}},
  \ and\ \bibinfo {author} {\bibfnamefont {Helgi}\ \bibnamefont {Runarsson}},\
  }\href {\doibase10.1088/0264-9381/33/15/154001} {\bibfield  {journal}
  {\bibinfo  {journal} {Class. Quant. Grav.}\ }\textbf {\bibinfo {volume}
  {33}},\ \bibinfo {pages} {154001} (\bibinfo {year} {2016})},\ \Eprint
  {http://arxiv.org/abs/1603.02687} {arXiv:1603.02687 [gr-qc]}\BibitemShut
  {NoStop}%
\bibitem [{\citenamefont {Chodosh}\ and\ \citenamefont
  {Shlapentokh-Rothman}(2017)}]{Chodosh:2015oma}%
  \BibitemOpen
  \bibfield  {author} {\bibinfo {author} {\bibfnamefont {Otis}\ \bibnamefont
  {Chodosh}}\ and\ \bibinfo {author} {\bibfnamefont {Yakov}\ \bibnamefont
  {Shlapentokh-Rothman}},\ }\href {\doibase10.1007/s00220-017-2998-3}
  {\bibfield  {journal} {\bibinfo  {journal} {Commun. Math. Phys.}\ }\textbf
  {\bibinfo {volume} {356}},\ \bibinfo {pages} {1155--1250} (\bibinfo {year}
  {2017})},\ \Eprint {http://arxiv.org/abs/1510.08025} {arXiv:1510.08025
  [gr-qc]}\BibitemShut {NoStop}%
\bibitem [{\citenamefont {Degollado}\ \emph {et~al.}(2018)\citenamefont
  {Degollado}, \citenamefont {Herdeiro},\ and\ \citenamefont
  {Radu}}]{Degollado:2018ypf}%
  \BibitemOpen
  \bibfield  {author} {\bibinfo {author} {\bibfnamefont {Juan~Carlos}\
  \bibnamefont {Degollado}}, \bibinfo {author} {\bibfnamefont {Carlos A.~R.}\
  \bibnamefont {Herdeiro}}, \ and\ \bibinfo {author} {\bibfnamefont {Eugen}\
  \bibnamefont {Radu}},\ }\href {\doibase10.1016/j.physletb.2018.04.052}
  {\bibfield  {journal} {\bibinfo  {journal} {Phys. Lett.}\ }\textbf {\bibinfo
  {volume} {B781}},\ \bibinfo {pages} {651--655} (\bibinfo {year} {2018})},\
  \Eprint {http://arxiv.org/abs/1802.07266} {arXiv:1802.07266
  [gr-qc]}\BibitemShut {NoStop}%
\bibitem [{\citenamefont {Wheeler}(1955)}]{Wheeler:1955zz}%
  \BibitemOpen
  \bibfield  {author} {\bibinfo {author} {\bibfnamefont {J.~A.}\ \bibnamefont
  {Wheeler}},\ }\href {\doibase10.1103/PhysRev.97.511} {\bibfield  {journal}
  {\bibinfo  {journal} {Phys. Rev.}\ }\textbf {\bibinfo {volume} {97}},\
  \bibinfo {pages} {511--536} (\bibinfo {year} {1955})}\BibitemShut {NoStop}%
\bibitem [{\citenamefont {Kaup}(1968)}]{Kaup:1968zz}%
  \BibitemOpen
  \bibfield  {author} {\bibinfo {author} {\bibfnamefont {David~J.}\
  \bibnamefont {Kaup}},\ }\href {\doibase10.1103/PhysRev.172.1331} {\bibfield
  {journal} {\bibinfo  {journal} {Phys. Rev.}\ }\textbf {\bibinfo {volume}
  {172}},\ \bibinfo {pages} {1331--1342} (\bibinfo {year} {1968})}\BibitemShut
  {NoStop}%
\bibitem [{\citenamefont {Jetzer}(1992)}]{Jetzer:1991jr}%
  \BibitemOpen
  \bibfield  {author} {\bibinfo {author} {\bibfnamefont {Philippe}\
  \bibnamefont {Jetzer}},\ }\href {\doibase10.1016/0370-1573(92)90123-H}
  {\bibfield  {journal} {\bibinfo  {journal} {Phys. Rept.}\ }\textbf {\bibinfo
  {volume} {220}},\ \bibinfo {pages} {163} (\bibinfo {year}
  {1992})}\BibitemShut {NoStop}%
\bibitem [{\citenamefont {Mundim}(2010)}]{Mundim:2010hi}%
  \BibitemOpen
  \bibfield  {author} {\bibinfo {author} {\bibfnamefont {B.~C.}\ \bibnamefont
  {Mundim}},\ }\emph {\bibinfo {title} {{A Numerical Study of Boson Star
  Binaries}}},\ \href
  {https://inspirehep.net/record/847127/files/arXiv:1003.0239.pdf} {Ph.D.
  thesis},\ \bibinfo  {school} {British Columbia U.} (\bibinfo {year} {2010}),\
  \Eprint {http://arxiv.org/abs/1003.0239} {1003.0239}\BibitemShut {NoStop}%
\bibitem [{\citenamefont {Liebling}\ and\ \citenamefont
  {Palenzuela}(2012)}]{Liebling:2012fv}%
  \BibitemOpen
  \bibfield  {author} {\bibinfo {author} {\bibfnamefont {Steven~L.}\
  \bibnamefont {Liebling}}\ and\ \bibinfo {author} {\bibfnamefont {Carlos}\
  \bibnamefont {Palenzuela}},\ }\href {\doibase10.12942/lrr-2012-6} {\bibfield
  {journal} {\bibinfo  {journal} {Living Rev. Rel.}\ }\textbf {\bibinfo
  {volume} {15}},\ \bibinfo {pages} {6} (\bibinfo {year} {2012})},\ \Eprint
  {http://arxiv.org/abs/1202.5809} {1202.5809}\BibitemShut {NoStop}%
\bibitem [{\citenamefont {Mielke}\ and\ \citenamefont
  {Schunck}(1997)}]{Mielke:1997re}%
  \BibitemOpen
  \bibfield  {author} {\bibinfo {author} {\bibfnamefont {E.~W.}\ \bibnamefont
  {Mielke}}\ and\ \bibinfo {author} {\bibfnamefont {F.~E.}\ \bibnamefont
  {Schunck}},\ }in\ \href@noop {} {\emph {\bibinfo {booktitle} {{Recent
  developments in theoretical and experimental general relativity, gravitation,
  and relativistic field theories. Proceedings, 8th Marcel Grossmann meeting,
  MG8, Jerusalem, Israel, June 22-27, 1997. Pts. A, B}}}}\ (\bibinfo {year}
  {1997})\ pp.\ \bibinfo {pages} {1607--1626},\ \Eprint
  {http://arxiv.org/abs/gr-qc/9801063} {gr-qc/9801063}\BibitemShut {NoStop}%
\bibitem [{\citenamefont {Colpi}\ \emph {et~al.}(1986)\citenamefont {Colpi},
  \citenamefont {Shapiro},\ and\ \citenamefont {Wasserman}}]{Colpi:1986ye}%
  \BibitemOpen
  \bibfield  {author} {\bibinfo {author} {\bibfnamefont {M.}~\bibnamefont
  {Colpi}}, \bibinfo {author} {\bibfnamefont {S.~L.}\ \bibnamefont {Shapiro}},
  \ and\ \bibinfo {author} {\bibfnamefont {I.}~\bibnamefont {Wasserman}},\
  }\href {\doibase10.1103/PhysRevLett.57.2485} {\bibfield  {journal} {\bibinfo
  {journal} {Phys. Rev. Lett.}\ }\textbf {\bibinfo {volume} {57}},\ \bibinfo
  {pages} {2485--2488} (\bibinfo {year} {1986})}\BibitemShut {NoStop}%
\bibitem [{\citenamefont {Helfer}\ \emph {et~al.}(2017)\citenamefont {Helfer},
  \citenamefont {Marsh}, \citenamefont {Clough}, \citenamefont {Fairbairn},
  \citenamefont {Lim},\ and\ \citenamefont {Becerril}}]{Helfer:2016ljl}%
  \BibitemOpen
  \bibfield  {author} {\bibinfo {author} {\bibfnamefont {T.}~\bibnamefont
  {Helfer}}, \bibinfo {author} {\bibfnamefont {D.~J.~E.}\ \bibnamefont
  {Marsh}}, \bibinfo {author} {\bibfnamefont {K.}~\bibnamefont {Clough}},
  \bibinfo {author} {\bibfnamefont {M.}~\bibnamefont {Fairbairn}}, \bibinfo
  {author} {\bibfnamefont {E.~A.}\ \bibnamefont {Lim}}, \ and\ \bibinfo
  {author} {\bibfnamefont {R.}~\bibnamefont {Becerril}},\ }\href
  {\doibase10.1088/1475-7516/2017/03/055} {\bibfield  {journal} {\bibinfo
  {journal} {JCAP}\ }\textbf {\bibinfo {volume} {1703}},\ \bibinfo {pages}
  {055} (\bibinfo {year} {2017})},\ \Eprint {http://arxiv.org/abs/1609.04724}
  {1609.04724}\BibitemShut {NoStop}%
\bibitem [{\citenamefont {Kobayashi}\ \emph {et~al.}(1994)\citenamefont
  {Kobayashi}, \citenamefont {Kasai},\ and\ \citenamefont
  {Futamase}}]{Kobayashi:1994qi}%
  \BibitemOpen
  \bibfield  {author} {\bibinfo {author} {\bibfnamefont {Y.}~\bibnamefont
  {Kobayashi}}, \bibinfo {author} {\bibfnamefont {M.}~\bibnamefont {Kasai}}, \
  and\ \bibinfo {author} {\bibfnamefont {T.}~\bibnamefont {Futamase}},\ }\href
  {\doibase10.1103/PhysRevD.50.7721} {\bibfield  {journal} {\bibinfo  {journal}
  {Phys. Rev. D}\ }\textbf {\bibinfo {volume} {50}},\ \bibinfo {pages}
  {7721--7724} (\bibinfo {year} {1994})}\BibitemShut {NoStop}%
\bibitem [{\citenamefont {Brito}\ \emph
  {et~al.}(2016{\natexlab{a}})\citenamefont {Brito}, \citenamefont {Cardoso},
  \citenamefont {Herdeiro},\ and\ \citenamefont {Radu}}]{Brito:2015pxa}%
  \BibitemOpen
  \bibfield  {author} {\bibinfo {author} {\bibfnamefont {R.}~\bibnamefont
  {Brito}}, \bibinfo {author} {\bibfnamefont {V.}~\bibnamefont {Cardoso}},
  \bibinfo {author} {\bibfnamefont {C.~A.~R.}\ \bibnamefont {Herdeiro}}, \ and\
  \bibinfo {author} {\bibfnamefont {E.}~\bibnamefont {Radu}},\ }\href
  {\doibase10.1016/j.physletb.2015.11.051} {\bibfield  {journal} {\bibinfo
  {journal} {Phys. Lett.}\ }\textbf {\bibinfo {volume} {B752}},\ \bibinfo
  {pages} {291--295} (\bibinfo {year} {2016}{\natexlab{a}})},\ \Eprint
  {http://arxiv.org/abs/1508.05395} {1508.05395}\BibitemShut {NoStop}%
\bibitem [{\citenamefont {Seidel}\ and\ \citenamefont
  {Suen}(1991)}]{Seidel:1991zh}%
  \BibitemOpen
  \bibfield  {author} {\bibinfo {author} {\bibfnamefont {E.}~\bibnamefont
  {Seidel}}\ and\ \bibinfo {author} {\bibfnamefont {W.~M.}\ \bibnamefont
  {Suen}},\ }\href {\doibase10.1103/PhysRevLett.66.1659} {\bibfield  {journal}
  {\bibinfo  {journal} {Phys. Rev. Lett.}\ }\textbf {\bibinfo {volume} {66}},\
  \bibinfo {pages} {1659--1662} (\bibinfo {year} {1991})}\BibitemShut {NoStop}%
\bibitem [{\citenamefont {Guzman}\ and\ \citenamefont
  {Rueda-Becerril}(2009)}]{Guzman:2009zz}%
  \BibitemOpen
  \bibfield  {author} {\bibinfo {author} {\bibfnamefont {F.~S.}\ \bibnamefont
  {Guzman}}\ and\ \bibinfo {author} {\bibfnamefont {J.~M.}\ \bibnamefont
  {Rueda-Becerril}},\ }\href {\doibase10.1103/PhysRevD.80.084023} {\bibfield
  {journal} {\bibinfo  {journal} {Phys. Rev. D}\ }\textbf {\bibinfo {volume}
  {80}},\ \bibinfo {pages} {084023} (\bibinfo {year} {2009})},\ \Eprint
  {http://arxiv.org/abs/1009.1250} {1009.1250}\BibitemShut {NoStop}%
\bibitem [{\citenamefont {Cardoso}\ \emph
  {et~al.}(2016{\natexlab{a}})\citenamefont {Cardoso}, \citenamefont {Macedo},
  \citenamefont {Pani},\ and\ \citenamefont {Ferrari}}]{Cardoso:2016olt}%
  \BibitemOpen
  \bibfield  {author} {\bibinfo {author} {\bibfnamefont {Vitor}\ \bibnamefont
  {Cardoso}}, \bibinfo {author} {\bibfnamefont {Caio F.~B.}\ \bibnamefont
  {Macedo}}, \bibinfo {author} {\bibfnamefont {Paolo}\ \bibnamefont {Pani}}, \
  and\ \bibinfo {author} {\bibfnamefont {Valeria}\ \bibnamefont {Ferrari}},\
  }\href {\doibase10.1088/1475-7516/2016/05/054} {\bibfield  {journal}
  {\bibinfo  {journal} {JCAP}\ }\textbf {\bibinfo {volume} {1605}},\ \bibinfo
  {pages} {054} (\bibinfo {year} {2016}{\natexlab{a}})},\ \Eprint
  {http://arxiv.org/abs/1604.07845} {arXiv:1604.07845 [hep-ph]}\BibitemShut
  {NoStop}%
\bibitem [{\citenamefont {Sennett}\ \emph {et~al.}(2017)\citenamefont
  {Sennett}, \citenamefont {Hinderer}, \citenamefont {Steinhoff}, \citenamefont
  {Buonanno},\ and\ \citenamefont {Ossokine}}]{Sennett:2017etc}%
  \BibitemOpen
  \bibfield  {author} {\bibinfo {author} {\bibfnamefont {N.}~\bibnamefont
  {Sennett}}, \bibinfo {author} {\bibfnamefont {T.}~\bibnamefont {Hinderer}},
  \bibinfo {author} {\bibfnamefont {J.}~\bibnamefont {Steinhoff}}, \bibinfo
  {author} {\bibfnamefont {A.}~\bibnamefont {Buonanno}}, \ and\ \bibinfo
  {author} {\bibfnamefont {S.}~\bibnamefont {Ossokine}},\ }\href
  {\doibase10.1103/PhysRevD.96.024002} {\bibfield  {journal} {\bibinfo
  {journal} {Phys. Rev. D}\ }\textbf {\bibinfo {volume} {96}},\ \bibinfo
  {pages} {024002} (\bibinfo {year} {2017})},\ \Eprint
  {http://arxiv.org/abs/1704.08651} {1704.08651}\BibitemShut {NoStop}%
\bibitem [{\citenamefont {Balakrishna}\ \emph {et~al.}(1998)\citenamefont
  {Balakrishna}, \citenamefont {Seidel},\ and\ \citenamefont
  {Suen}}]{Balakrishna:1997ej}%
  \BibitemOpen
  \bibfield  {author} {\bibinfo {author} {\bibfnamefont {J.}~\bibnamefont
  {Balakrishna}}, \bibinfo {author} {\bibfnamefont {E.}~\bibnamefont {Seidel}},
  \ and\ \bibinfo {author} {\bibfnamefont {W.-M.}\ \bibnamefont {Suen}},\
  }\href {\doibase10.1103/PhysRevD.58.104004} {\bibfield  {journal} {\bibinfo
  {journal} {Phys. Rev. D}\ }\textbf {\bibinfo {volume} {58}},\ \bibinfo
  {pages} {104004} (\bibinfo {year} {1998})},\ \Eprint
  {http://arxiv.org/abs/gr-qc/9712064} {gr-qc/9712064}\BibitemShut {NoStop}%
\bibitem [{\citenamefont {Valdez-Alvarado}\ \emph {et~al.}(2013)\citenamefont
  {Valdez-Alvarado}, \citenamefont {Palenzuela}, \citenamefont {Alic},\ and\
  \citenamefont {Ure{\~n}a-L{\'o}pez}}]{ValdezAlvarado:2012xc}%
  \BibitemOpen
  \bibfield  {author} {\bibinfo {author} {\bibfnamefont {S.}~\bibnamefont
  {Valdez-Alvarado}}, \bibinfo {author} {\bibfnamefont {C.}~\bibnamefont
  {Palenzuela}}, \bibinfo {author} {\bibfnamefont {D.}~\bibnamefont {Alic}}, \
  and\ \bibinfo {author} {\bibfnamefont {L.~A.}\ \bibnamefont
  {Ure{\~n}a-L{\'o}pez}},\ }\href {\doibase10.1103/PhysRevD.87.084040}
  {\bibfield  {journal} {\bibinfo  {journal} {Phys. Rev. D}\ }\textbf {\bibinfo
  {volume} {87}},\ \bibinfo {pages} {084040} (\bibinfo {year} {2013})},\
  \Eprint {http://arxiv.org/abs/1210.2299} {1210.2299}\BibitemShut {NoStop}%
\bibitem [{\citenamefont {Collodel}\ \emph {et~al.}(2017)\citenamefont
  {Collodel}, \citenamefont {Kleihaus},\ and\ \citenamefont
  {Kunz}}]{Collodel:2017biu}%
  \BibitemOpen
  \bibfield  {author} {\bibinfo {author} {\bibfnamefont {Lucas~G.}\
  \bibnamefont {Collodel}}, \bibinfo {author} {\bibfnamefont {Burkhard}\
  \bibnamefont {Kleihaus}}, \ and\ \bibinfo {author} {\bibfnamefont {Jutta}\
  \bibnamefont {Kunz}},\ }\href {\doibase10.1103/PhysRevD.96.084066} {\bibfield
   {journal} {\bibinfo  {journal} {Phys. Rev.}\ }\textbf {\bibinfo {volume}
  {D96}},\ \bibinfo {pages} {084066} (\bibinfo {year} {2017})},\ \Eprint
  {http://arxiv.org/abs/1708.02057} {arXiv:1708.02057 [gr-qc]}\BibitemShut
  {NoStop}%
\bibitem [{\citenamefont {Sanchis-Gual}\ \emph {et~al.}(2017)\citenamefont
  {Sanchis-Gual}, \citenamefont {Herdeiro}, \citenamefont {Radu}, \citenamefont
  {Degollado},\ and\ \citenamefont {Font}}]{Sanchis-Gual:2017bhw}%
  \BibitemOpen
  \bibfield  {author} {\bibinfo {author} {\bibfnamefont {Nicolas}\ \bibnamefont
  {Sanchis-Gual}}, \bibinfo {author} {\bibfnamefont {Carlos}\ \bibnamefont
  {Herdeiro}}, \bibinfo {author} {\bibfnamefont {Eugen}\ \bibnamefont {Radu}},
  \bibinfo {author} {\bibfnamefont {Juan~Carlos}\ \bibnamefont {Degollado}}, \
  and\ \bibinfo {author} {\bibfnamefont {José~A.}\ \bibnamefont {Font}},\
  }\href {\doibase10.1103/PhysRevD.95.104028} {\bibfield  {journal} {\bibinfo
  {journal} {Phys. Rev.}\ }\textbf {\bibinfo {volume} {D95}},\ \bibinfo {pages}
  {104028} (\bibinfo {year} {2017})},\ \Eprint
  {http://arxiv.org/abs/1702.04532} {arXiv:1702.04532 [gr-qc]}\BibitemShut
  {NoStop}%
\bibitem [{\citenamefont {Palenzuela}\ \emph {et~al.}(2007)\citenamefont
  {Palenzuela}, \citenamefont {Olabarrieta}, \citenamefont {Lehner},\ and\
  \citenamefont {Liebling}}]{Palenzuela:2006wp}%
  \BibitemOpen
  \bibfield  {author} {\bibinfo {author} {\bibfnamefont {C.}~\bibnamefont
  {Palenzuela}}, \bibinfo {author} {\bibfnamefont {I.}~\bibnamefont
  {Olabarrieta}}, \bibinfo {author} {\bibfnamefont {L.}~\bibnamefont {Lehner}},
  \ and\ \bibinfo {author} {\bibfnamefont {S.~L.}\ \bibnamefont {Liebling}},\
  }\href {\doibase10.1103/PhysRevD.75.064005} {\bibfield  {journal} {\bibinfo
  {journal} {Phys. Rev. D}\ }\textbf {\bibinfo {volume} {75}},\ \bibinfo
  {pages} {064005} (\bibinfo {year} {2007})},\ \Eprint
  {http://arxiv.org/abs/gr-qc/0612067} {gr-qc/0612067}\BibitemShut {NoStop}%
\bibitem [{\citenamefont {Bezares}\ \emph {et~al.}(2017)\citenamefont
  {Bezares}, \citenamefont {Palenzuela},\ and\ \citenamefont
  {Bona}}]{Bezares:2017mzk}%
  \BibitemOpen
  \bibfield  {author} {\bibinfo {author} {\bibfnamefont {Miguel}\ \bibnamefont
  {Bezares}}, \bibinfo {author} {\bibfnamefont {Carlos}\ \bibnamefont
  {Palenzuela}}, \ and\ \bibinfo {author} {\bibfnamefont {Carles}\ \bibnamefont
  {Bona}},\ }\href {\doibase10.1103/PhysRevD.95.124005} {\bibfield  {journal}
  {\bibinfo  {journal} {Phys. Rev. D}\ }\textbf {\bibinfo {volume} {95}},\
  \bibinfo {pages} {124005} (\bibinfo {year} {2017})},\ \Eprint
  {http://arxiv.org/abs/1705.01071} {1705.01071}\BibitemShut {NoStop}%
\bibitem [{\citenamefont {Choptuik}\ and\ \citenamefont
  {Pretorius}(2010)}]{Choptuik:2009ww}%
  \BibitemOpen
  \bibfield  {author} {\bibinfo {author} {\bibfnamefont {Matthew~W.}\
  \bibnamefont {Choptuik}}\ and\ \bibinfo {author} {\bibfnamefont {Frans}\
  \bibnamefont {Pretorius}},\ }\href {\doibase10.1103/PhysRevLett.104.111101}
  {\bibfield  {journal} {\bibinfo  {journal} {Phys. Rev. Lett.}\ }\textbf
  {\bibinfo {volume} {104}},\ \bibinfo {pages} {111101} (\bibinfo {year}
  {2010})},\ \Eprint {http://arxiv.org/abs/0908.1780} {arXiv:0908.1780
  [gr-qc]}\BibitemShut {NoStop}%
\bibitem [{\citenamefont {Palenzuela}\ \emph {et~al.}(2008)\citenamefont
  {Palenzuela}, \citenamefont {Lehner},\ and\ \citenamefont
  {Liebling}}]{Palenzuela:2007dm}%
  \BibitemOpen
  \bibfield  {author} {\bibinfo {author} {\bibfnamefont {C.}~\bibnamefont
  {Palenzuela}}, \bibinfo {author} {\bibfnamefont {L.}~\bibnamefont {Lehner}},
  \ and\ \bibinfo {author} {\bibfnamefont {S.~L.}\ \bibnamefont {Liebling}},\
  }\href {\doibase10.1103/PhysRevD.77.044036} {\bibfield  {journal} {\bibinfo
  {journal} {Phys. Rev. D}\ }\textbf {\bibinfo {volume} {77}},\ \bibinfo
  {pages} {044036} (\bibinfo {year} {2008})},\ \Eprint
  {http://arxiv.org/abs/0706.2435} {0706.2435}\BibitemShut {NoStop}%
\bibitem [{\citenamefont {Palenzuela}\ \emph {et~al.}(2017)\citenamefont
  {Palenzuela}, \citenamefont {Pani}, \citenamefont {Bezares}, \citenamefont
  {Cardoso}, \citenamefont {Lehner},\ and\ \citenamefont
  {Liebling}}]{Palenzuela:2017kcg}%
  \BibitemOpen
  \bibfield  {author} {\bibinfo {author} {\bibfnamefont {Carlos}\ \bibnamefont
  {Palenzuela}}, \bibinfo {author} {\bibfnamefont {Paolo}\ \bibnamefont
  {Pani}}, \bibinfo {author} {\bibfnamefont {Miguel}\ \bibnamefont {Bezares}},
  \bibinfo {author} {\bibfnamefont {Vitor}\ \bibnamefont {Cardoso}}, \bibinfo
  {author} {\bibfnamefont {Luis}\ \bibnamefont {Lehner}}, \ and\ \bibinfo
  {author} {\bibfnamefont {Steven}\ \bibnamefont {Liebling}},\ }\href
  {\doibase10.1103/PhysRevD.96.104058} {\bibfield  {journal} {\bibinfo
  {journal} {Phys. Rev.}\ }\textbf {\bibinfo {volume} {D96}},\ \bibinfo {pages}
  {104058} (\bibinfo {year} {2017})},\ \Eprint
  {http://arxiv.org/abs/1710.09432} {arXiv:1710.09432 [gr-qc]}\BibitemShut
  {NoStop}%
\bibitem [{\citenamefont {Berti}\ \emph {et~al.}(2015)\citenamefont {Berti}
  \emph {et~al.}}]{Berti:2015itd}%
  \BibitemOpen
  \bibfield  {author} {\bibinfo {author} {\bibfnamefont {Emanuele}\
  \bibnamefont {Berti}} \emph {et~al.},\ }\href
  {\doibase10.1088/0264-9381/32/24/243001} {\bibfield  {journal} {\bibinfo
  {journal} {Class. Quant. Grav.}\ }\textbf {\bibinfo {volume} {32}},\ \bibinfo
  {pages} {243001} (\bibinfo {year} {2015})},\ \Eprint
  {http://arxiv.org/abs/1501.07274} {arXiv:1501.07274 [gr-qc]}\BibitemShut
  {NoStop}%
\bibitem [{\citenamefont {Fujii}\ and\ \citenamefont
  {Maeda}(2007)}]{Fujii:2003pa}%
  \BibitemOpen
  \bibfield  {author} {\bibinfo {author} {\bibfnamefont {Y.}~\bibnamefont
  {Fujii}}\ and\ \bibinfo {author} {\bibfnamefont {K.}~\bibnamefont {Maeda}},\
  }\href {http://www.cambridge.org/uk/catalogue/catalogue.asp?isbn=0521811597}
  {\emph {\bibinfo {title} {{The scalar-tensor theory of gravitation}}}}\
  (\bibinfo  {publisher} {Cambridge University Press},\ \bibinfo {year}
  {2007})\BibitemShut {NoStop}%
\bibitem [{\citenamefont {Brans}\ and\ \citenamefont
  {Dicke}(1961)}]{Brans:1961sx}%
  \BibitemOpen
  \bibfield  {author} {\bibinfo {author} {\bibfnamefont {C.}~\bibnamefont
  {Brans}}\ and\ \bibinfo {author} {\bibfnamefont {R.H.}\ \bibnamefont
  {Dicke}},\ }\href {\doibase10.1103/PhysRev.124.925} {\bibfield  {journal}
  {\bibinfo  {journal} {Phys. Rev.}\ }\textbf {\bibinfo {volume} {124}},\
  \bibinfo {pages} {925--935} (\bibinfo {year} {1961})}\BibitemShut {NoStop}%
\bibitem [{\citenamefont {De~Felice}\ and\ \citenamefont
  {Tsujikawa}(2010)}]{DeFelice:2010aj}%
  \BibitemOpen
  \bibfield  {author} {\bibinfo {author} {\bibfnamefont {Antonio}\ \bibnamefont
  {De~Felice}}\ and\ \bibinfo {author} {\bibfnamefont {Shinji}\ \bibnamefont
  {Tsujikawa}},\ }\href {\doibase10.12942/lrr-2010-3} {\bibfield  {journal}
  {\bibinfo  {journal} {Living Rev. Rel.}\ }\textbf {\bibinfo {volume} {13}},\
  \bibinfo {pages} {3} (\bibinfo {year} {2010})},\ \Eprint
  {http://arxiv.org/abs/1002.4928} {arXiv:1002.4928 [gr-qc]}\BibitemShut
  {NoStop}%
\bibitem [{\citenamefont {Damour}\ and\ \citenamefont
  {Esposito-Far{\'e}se}(1992)}]{Damour:1992we}%
  \BibitemOpen
  \bibfield  {author} {\bibinfo {author} {\bibfnamefont {T.}~\bibnamefont
  {Damour}}\ and\ \bibinfo {author} {\bibfnamefont {G.}~\bibnamefont
  {Esposito-Far{\'e}se}},\ }\href {\doibase10.1088/0264-9381/9/9/015}
  {\bibfield  {journal} {\bibinfo  {journal} {Class. Quant. Grav.}\ }\textbf
  {\bibinfo {volume} {9}},\ \bibinfo {pages} {2093--2176} (\bibinfo {year}
  {1992})}\BibitemShut {NoStop}%
\bibitem [{\citenamefont {Salgado}(2006)}]{Salgado:2005hx}%
  \BibitemOpen
  \bibfield  {author} {\bibinfo {author} {\bibfnamefont {Marcelo}\ \bibnamefont
  {Salgado}},\ }\href {\doibase10.1088/0264-9381/23/14/010} {\bibfield
  {journal} {\bibinfo  {journal} {Class. Quant. Grav.}\ }\textbf {\bibinfo
  {volume} {23}},\ \bibinfo {pages} {4719--4742} (\bibinfo {year} {2006})},\
  \Eprint {http://arxiv.org/abs/gr-qc/0509001} {arXiv:gr-qc/0509001
  [gr-qc]}\BibitemShut {NoStop}%
\bibitem [{\citenamefont {Salgado}\ \emph {et~al.}(2008)\citenamefont
  {Salgado}, \citenamefont {Martinez-del Rio}, \citenamefont {Alcubierre},\
  and\ \citenamefont {Nunez}}]{Salgado:2008xh}%
  \BibitemOpen
  \bibfield  {author} {\bibinfo {author} {\bibfnamefont {Marcelo}\ \bibnamefont
  {Salgado}}, \bibinfo {author} {\bibfnamefont {David}\ \bibnamefont
  {Martinez-del Rio}}, \bibinfo {author} {\bibfnamefont {Miguel}\ \bibnamefont
  {Alcubierre}}, \ and\ \bibinfo {author} {\bibfnamefont {Dario}\ \bibnamefont
  {Nunez}},\ }\href {\doibase10.1103/PhysRevD.77.104010} {\bibfield  {journal}
  {\bibinfo  {journal} {Phys. Rev.}\ }\textbf {\bibinfo {volume} {D77}},\
  \bibinfo {pages} {104010} (\bibinfo {year} {2008})},\ \Eprint
  {http://arxiv.org/abs/0801.2372} {arXiv:0801.2372 [gr-qc]}\BibitemShut
  {NoStop}%
\bibitem [{\citenamefont {Jakubiec}\ and\ \citenamefont
  {Kijowski}(1989)}]{KijowskiJakubiecUniversality}%
  \BibitemOpen
  \bibfield  {author} {\bibinfo {author} {\bibfnamefont {A.}~\bibnamefont
  {Jakubiec}}\ and\ \bibinfo {author} {\bibfnamefont {J.}~\bibnamefont
  {Kijowski}},\ }\href {\doibase10.1063/1.528377} {\bibfield  {journal}
  {\bibinfo  {journal} {Jour.\ Math.\ Phys.}\ }\textbf {\bibinfo {volume}
  {30}},\ \bibinfo {pages} {1073--1076} (\bibinfo {year} {1989})}\BibitemShut
  {NoStop}%
\bibitem [{\citenamefont {Damour}\ and\ \citenamefont
  {Esposito-Far{\`e}se}(1993)}]{Damour:1993hw}%
  \BibitemOpen
  \bibfield  {author} {\bibinfo {author} {\bibfnamefont {T.}~\bibnamefont
  {Damour}}\ and\ \bibinfo {author} {\bibfnamefont {G.}~\bibnamefont
  {Esposito-Far{\`e}se}},\ }\href {\doibase10.1103/PhysRevLett.70.2220}
  {\bibfield  {journal} {\bibinfo  {journal} {Phys. Rev. Lett.}\ }\textbf
  {\bibinfo {volume} {70}},\ \bibinfo {pages} {2220--2223} (\bibinfo {year}
  {1993})}\BibitemShut {NoStop}%
\bibitem [{\citenamefont {Ramazano\u{g}lu}(2017)}]{Ramazanoglu:2017xbl}%
  \BibitemOpen
  \bibfield  {author} {\bibinfo {author} {\bibfnamefont {F.~M.}\ \bibnamefont
  {Ramazano\u{g}lu}},\ }\href {\doibase10.1103/PhysRevD.96.064009} {\bibfield
  {journal} {\bibinfo  {journal} {Phys. Rev. D}\ }\textbf {\bibinfo {volume}
  {96}},\ \bibinfo {pages} {064009} (\bibinfo {year} {2017})},\ \Eprint
  {http://arxiv.org/abs/1706.01056} {1706.01056}\BibitemShut {NoStop}%
\bibitem [{\citenamefont {Bertotti}\ \emph {et~al.}(2003)\citenamefont
  {Bertotti}, \citenamefont {Iess},\ and\ \citenamefont
  {Tortora}}]{Bertotti:2003rm}%
  \BibitemOpen
  \bibfield  {author} {\bibinfo {author} {\bibfnamefont {B.}~\bibnamefont
  {Bertotti}}, \bibinfo {author} {\bibfnamefont {L.}~\bibnamefont {Iess}}, \
  and\ \bibinfo {author} {\bibfnamefont {P.}~\bibnamefont {Tortora}},\ }\href
  {\doibase10.1038/nature01997} {\bibfield  {journal} {\bibinfo  {journal}
  {Nature}\ }\textbf {\bibinfo {volume} {425}},\ \bibinfo {pages} {374--376}
  (\bibinfo {year} {2003})}\BibitemShut {NoStop}%
\bibitem [{\citenamefont {Doneva}\ \emph {et~al.}(2013)\citenamefont {Doneva},
  \citenamefont {Yazadjiev}, \citenamefont {Stergioulas},\ and\ \citenamefont
  {Kokkotas}}]{Doneva:2013qva}%
  \BibitemOpen
  \bibfield  {author} {\bibinfo {author} {\bibfnamefont {Daniela~D.}\
  \bibnamefont {Doneva}}, \bibinfo {author} {\bibfnamefont {Stoytcho~S.}\
  \bibnamefont {Yazadjiev}}, \bibinfo {author} {\bibfnamefont {Nikolaos}\
  \bibnamefont {Stergioulas}}, \ and\ \bibinfo {author} {\bibfnamefont
  {Kostas~D.}\ \bibnamefont {Kokkotas}},\ }\href
  {\doibase10.1103/PhysRevD.88.084060} {\bibfield  {journal} {\bibinfo
  {journal} {Phys. Rev.}\ }\textbf {\bibinfo {volume} {D88}},\ \bibinfo {pages}
  {084060} (\bibinfo {year} {2013})},\ \Eprint {http://arxiv.org/abs/1309.0605}
  {arXiv:1309.0605 [gr-qc]}\BibitemShut {NoStop}%
\bibitem [{\citenamefont {Mendes}(2015)}]{Mendes:2014ufa}%
  \BibitemOpen
  \bibfield  {author} {\bibinfo {author} {\bibfnamefont {Raissa F.~P.}\
  \bibnamefont {Mendes}},\ }\href {\doibase10.1103/PhysRevD.91.064024}
  {\bibfield  {journal} {\bibinfo  {journal} {Phys. Rev.}\ }\textbf {\bibinfo
  {volume} {D91}},\ \bibinfo {pages} {064024} (\bibinfo {year} {2015})},\
  \Eprint {http://arxiv.org/abs/1412.6789} {arXiv:1412.6789
  [gr-qc]}\BibitemShut {NoStop}%
\bibitem [{\citenamefont {Silva}\ \emph {et~al.}(2015)\citenamefont {Silva},
  \citenamefont {Macedo}, \citenamefont {Berti},\ and\ \citenamefont
  {Crispino}}]{Silva:2014fca}%
  \BibitemOpen
  \bibfield  {author} {\bibinfo {author} {\bibfnamefont {Hector~O.}\
  \bibnamefont {Silva}}, \bibinfo {author} {\bibfnamefont {Caio F.~B.}\
  \bibnamefont {Macedo}}, \bibinfo {author} {\bibfnamefont {Emanuele}\
  \bibnamefont {Berti}}, \ and\ \bibinfo {author} {\bibfnamefont {Luís C.~B.}\
  \bibnamefont {Crispino}},\ }\href {\doibase10.1088/0264-9381/32/14/145008}
  {\bibfield  {journal} {\bibinfo  {journal} {Class. Quant. Grav.}\ }\textbf
  {\bibinfo {volume} {32}},\ \bibinfo {pages} {145008} (\bibinfo {year}
  {2015})},\ \Eprint {http://arxiv.org/abs/1411.6286} {arXiv:1411.6286
  [gr-qc]}\BibitemShut {NoStop}%
\bibitem [{\citenamefont {Horbatsch}\ \emph {et~al.}(2015)\citenamefont
  {Horbatsch}, \citenamefont {Silva}, \citenamefont {Gerosa}, \citenamefont
  {Pani}, \citenamefont {Berti}, \citenamefont {Gualtieri},\ and\ \citenamefont
  {Sperhake}}]{Horbatsch:2015bua}%
  \BibitemOpen
  \bibfield  {author} {\bibinfo {author} {\bibfnamefont {Michael}\ \bibnamefont
  {Horbatsch}}, \bibinfo {author} {\bibfnamefont {Hector~O.}\ \bibnamefont
  {Silva}}, \bibinfo {author} {\bibfnamefont {Davide}\ \bibnamefont {Gerosa}},
  \bibinfo {author} {\bibfnamefont {Paolo}\ \bibnamefont {Pani}}, \bibinfo
  {author} {\bibfnamefont {Emanuele}\ \bibnamefont {Berti}}, \bibinfo {author}
  {\bibfnamefont {Leonardo}\ \bibnamefont {Gualtieri}}, \ and\ \bibinfo
  {author} {\bibfnamefont {Ulrich}\ \bibnamefont {Sperhake}},\ }\href
  {\doibase10.1088/0264-9381/32/20/204001} {\bibfield  {journal} {\bibinfo
  {journal} {Class. Quant. Grav.}\ }\textbf {\bibinfo {volume} {32}},\ \bibinfo
  {pages} {204001} (\bibinfo {year} {2015})},\ \Eprint
  {http://arxiv.org/abs/1505.07462} {arXiv:1505.07462 [gr-qc]}\BibitemShut
  {NoStop}%
\bibitem [{\citenamefont {Palenzuela}\ and\ \citenamefont
  {Liebling}(2016)}]{Palenzuela:2015ima}%
  \BibitemOpen
  \bibfield  {author} {\bibinfo {author} {\bibfnamefont {C.}~\bibnamefont
  {Palenzuela}}\ and\ \bibinfo {author} {\bibfnamefont {S.~L.}\ \bibnamefont
  {Liebling}},\ }\href {\doibase10.1103/PhysRevD.93.044009} {\bibfield
  {journal} {\bibinfo  {journal} {Phys. Rev. D}\ }\textbf {\bibinfo {volume}
  {93}},\ \bibinfo {pages} {044009} (\bibinfo {year} {2016})},\ \Eprint
  {http://arxiv.org/abs/1510.03471} {1510.03471}\BibitemShut {NoStop}%
\bibitem [{\citenamefont {Ramazanoğlu}\ and\ \citenamefont
  {Pretorius}(2016)}]{Ramazanoglu:2016kul}%
  \BibitemOpen
  \bibfield  {author} {\bibinfo {author} {\bibfnamefont {Fethi~M.}\
  \bibnamefont {Ramazanoğlu}}\ and\ \bibinfo {author} {\bibfnamefont {Frans}\
  \bibnamefont {Pretorius}},\ }\href {\doibase10.1103/PhysRevD.93.064005}
  {\bibfield  {journal} {\bibinfo  {journal} {Phys. Rev.}\ }\textbf {\bibinfo
  {volume} {D93}},\ \bibinfo {pages} {064005} (\bibinfo {year} {2016})},\
  \Eprint {http://arxiv.org/abs/1601.07475} {arXiv:1601.07475
  [gr-qc]}\BibitemShut {NoStop}%
\bibitem [{\citenamefont {Morisaki}\ and\ \citenamefont
  {Suyama}(2017)}]{Morisaki:2017nit}%
  \BibitemOpen
  \bibfield  {author} {\bibinfo {author} {\bibfnamefont {Soichiro}\
  \bibnamefont {Morisaki}}\ and\ \bibinfo {author} {\bibfnamefont {Teruaki}\
  \bibnamefont {Suyama}},\ }\href {\doibase10.1103/PhysRevD.96.084026}
  {\bibfield  {journal} {\bibinfo  {journal} {Phys. Rev. D}\ }\textbf {\bibinfo
  {volume} {96}},\ \bibinfo {pages} {084026} (\bibinfo {year} {2017})},\
  \Eprint {http://arxiv.org/abs/1707.02809} {1707.02809}\BibitemShut {NoStop}%
\bibitem [{\citenamefont {Shibata}\ \emph {et~al.}(1994)\citenamefont
  {Shibata}, \citenamefont {Nakao},\ and\ \citenamefont
  {Nakamura}}]{Shibata:1994qd}%
  \BibitemOpen
  \bibfield  {author} {\bibinfo {author} {\bibfnamefont {M.}~\bibnamefont
  {Shibata}}, \bibinfo {author} {\bibfnamefont {K.}~\bibnamefont {Nakao}}, \
  and\ \bibinfo {author} {\bibfnamefont {T.}~\bibnamefont {Nakamura}},\ }\href
  {\doibase10.1103/PhysRevD.50.7304} {\bibfield  {journal} {\bibinfo  {journal}
  {Phys. Rev. D}\ }\textbf {\bibinfo {volume} {50}},\ \bibinfo {pages}
  {7304--7317} (\bibinfo {year} {1994})}\BibitemShut {NoStop}%
\bibitem [{\citenamefont {Scheel}\ \emph
  {et~al.}(1995{\natexlab{a}})\citenamefont {Scheel}, \citenamefont {Shapiro},\
  and\ \citenamefont {Teukolsky}}]{Scheel:1994yr}%
  \BibitemOpen
  \bibfield  {author} {\bibinfo {author} {\bibfnamefont {M.~A.}\ \bibnamefont
  {Scheel}}, \bibinfo {author} {\bibfnamefont {S.~L.}\ \bibnamefont {Shapiro}},
  \ and\ \bibinfo {author} {\bibfnamefont {S.~A.}\ \bibnamefont {Teukolsky}},\
  }\href {\doibase10.1103/PhysRevD.51.4208} {\bibfield  {journal} {\bibinfo
  {journal} {Phys. Rev. D}\ }\textbf {\bibinfo {volume} {51}},\ \bibinfo
  {pages} {4208--4235} (\bibinfo {year} {1995}{\natexlab{a}})}\BibitemShut
  {NoStop}%
\bibitem [{\citenamefont {Scheel}\ \emph
  {et~al.}(1995{\natexlab{b}})\citenamefont {Scheel}, \citenamefont {Shapiro},\
  and\ \citenamefont {Teukolsky}}]{Scheel:1994yn}%
  \BibitemOpen
  \bibfield  {author} {\bibinfo {author} {\bibfnamefont {M.~A.}\ \bibnamefont
  {Scheel}}, \bibinfo {author} {\bibfnamefont {S.~L.}\ \bibnamefont {Shapiro}},
  \ and\ \bibinfo {author} {\bibfnamefont {S.~A.}\ \bibnamefont {Teukolsky}},\
  }\href {\doibase10.1103/PhysRevD.51.4236} {\bibfield  {journal} {\bibinfo
  {journal} {Phys. Rev. D}\ }\textbf {\bibinfo {volume} {51}},\ \bibinfo
  {pages} {4236--4249} (\bibinfo {year} {1995}{\natexlab{b}})}\BibitemShut
  {NoStop}%
\bibitem [{\citenamefont {Harada}\ \emph {et~al.}(1997)\citenamefont {Harada},
  \citenamefont {Chiba}, \citenamefont {Nakao},\ and\ \citenamefont
  {Nakamura}}]{Harada:1996wt}%
  \BibitemOpen
  \bibfield  {author} {\bibinfo {author} {\bibfnamefont {T.}~\bibnamefont
  {Harada}}, \bibinfo {author} {\bibfnamefont {T.}~\bibnamefont {Chiba}},
  \bibinfo {author} {\bibfnamefont {K.-i.}\ \bibnamefont {Nakao}}, \ and\
  \bibinfo {author} {\bibfnamefont {T.}~\bibnamefont {Nakamura}},\ }\href
  {\doibase10.1103/PhysRevD.55.2024} {\bibfield  {journal} {\bibinfo  {journal}
  {Phys. Rev. D}\ }\textbf {\bibinfo {volume} {55}},\ \bibinfo {pages}
  {2024--2037} (\bibinfo {year} {1997})},\ \Eprint
  {http://arxiv.org/abs/gr-qc/9611031} {gr-qc/9611031}\BibitemShut {NoStop}%
\bibitem [{\citenamefont {Novak}(1998{\natexlab{a}})}]{Novak:1997hw}%
  \BibitemOpen
  \bibfield  {author} {\bibinfo {author} {\bibfnamefont {Jerome}\ \bibnamefont
  {Novak}},\ }\href {\doibase10.1103/PhysRevD.57.4789} {\bibfield  {journal}
  {\bibinfo  {journal} {Phys. Rev.}\ }\textbf {\bibinfo {volume} {D57}},\
  \bibinfo {pages} {4789--4801} (\bibinfo {year} {1998}{\natexlab{a}})},\
  \Eprint {http://arxiv.org/abs/gr-qc/9707041} {arXiv:gr-qc/9707041
  [gr-qc]}\BibitemShut {NoStop}%
\bibitem [{\citenamefont {Novak}(1998{\natexlab{b}})}]{Novak:1998rk}%
  \BibitemOpen
  \bibfield  {author} {\bibinfo {author} {\bibfnamefont {Jerome}\ \bibnamefont
  {Novak}},\ }\href {\doibase10.1103/PhysRevD.58.064019} {\bibfield  {journal}
  {\bibinfo  {journal} {Phys. Rev.}\ }\textbf {\bibinfo {volume} {D58}},\
  \bibinfo {pages} {064019} (\bibinfo {year} {1998}{\natexlab{b}})},\ \Eprint
  {http://arxiv.org/abs/gr-qc/9806022} {arXiv:gr-qc/9806022
  [gr-qc]}\BibitemShut {NoStop}%
\bibitem [{\citenamefont {Novak}\ and\ \citenamefont
  {Ibanez}(2000)}]{Novak:1999jg}%
  \BibitemOpen
  \bibfield  {author} {\bibinfo {author} {\bibfnamefont {Jerome}\ \bibnamefont
  {Novak}}\ and\ \bibinfo {author} {\bibfnamefont {Jose~M.}\ \bibnamefont
  {Ibanez}},\ }\href {\doibase10.1086/308627} {\bibfield  {journal} {\bibinfo
  {journal} {Astrophys. J.}\ }\textbf {\bibinfo {volume} {533}},\ \bibinfo
  {pages} {392--405} (\bibinfo {year} {2000})},\ \Eprint
  {http://arxiv.org/abs/astro-ph/9911298} {arXiv:astro-ph/9911298
  [astro-ph]}\BibitemShut {NoStop}%
\bibitem [{\citenamefont {Gerosa}\ \emph {et~al.}(2016)\citenamefont {Gerosa},
  \citenamefont {Sperhake},\ and\ \citenamefont {Ott}}]{Gerosa:2016fri}%
  \BibitemOpen
  \bibfield  {author} {\bibinfo {author} {\bibfnamefont {Davide}\ \bibnamefont
  {Gerosa}}, \bibinfo {author} {\bibfnamefont {Ulrich}\ \bibnamefont
  {Sperhake}}, \ and\ \bibinfo {author} {\bibfnamefont {Christian~D.}\
  \bibnamefont {Ott}},\ }\href {\doibase10.1088/0264-9381/33/13/135002}
  {\bibfield  {journal} {\bibinfo  {journal} {Class. Quant. Grav.}\ }\textbf
  {\bibinfo {volume} {33}},\ \bibinfo {pages} {135002} (\bibinfo {year}
  {2016})},\ \Eprint {http://arxiv.org/abs/1602.06952} {arXiv:1602.06952
  [gr-qc]}\BibitemShut {NoStop}%
\bibitem [{\citenamefont {Alsing}\ \emph {et~al.}(2012)\citenamefont {Alsing},
  \citenamefont {Berti}, \citenamefont {Will},\ and\ \citenamefont
  {Zaglauer}}]{Alsing:2011er}%
  \BibitemOpen
  \bibfield  {author} {\bibinfo {author} {\bibfnamefont {Justin}\ \bibnamefont
  {Alsing}}, \bibinfo {author} {\bibfnamefont {Emanuele}\ \bibnamefont
  {Berti}}, \bibinfo {author} {\bibfnamefont {Clifford~M.}\ \bibnamefont
  {Will}}, \ and\ \bibinfo {author} {\bibfnamefont {Helmut}\ \bibnamefont
  {Zaglauer}},\ }\href {\doibase10.1103/PhysRevD.85.064041} {\bibfield
  {journal} {\bibinfo  {journal} {Phys. Rev.}\ }\textbf {\bibinfo {volume}
  {D85}},\ \bibinfo {pages} {064041} (\bibinfo {year} {2012})},\ \Eprint
  {http://arxiv.org/abs/1112.4903} {arXiv:1112.4903 [gr-qc]}\BibitemShut
  {NoStop}%
\bibitem [{\citenamefont {Sperhake}\ \emph {et~al.}(2017)\citenamefont
  {Sperhake}, \citenamefont {Moore}, \citenamefont {Rosca}, \citenamefont
  {Agathos}, \citenamefont {Gerosa},\ and\ \citenamefont
  {Ott}}]{Sperhake:2017itk}%
  \BibitemOpen
  \bibfield  {author} {\bibinfo {author} {\bibfnamefont {Ulrich}\ \bibnamefont
  {Sperhake}}, \bibinfo {author} {\bibfnamefont {Christopher~J.}\ \bibnamefont
  {Moore}}, \bibinfo {author} {\bibfnamefont {Roxana}\ \bibnamefont {Rosca}},
  \bibinfo {author} {\bibfnamefont {Michalis}\ \bibnamefont {Agathos}},
  \bibinfo {author} {\bibfnamefont {Davide}\ \bibnamefont {Gerosa}}, \ and\
  \bibinfo {author} {\bibfnamefont {Christian~D.}\ \bibnamefont {Ott}},\ }\href
  {\doibase10.1103/PhysRevLett.119.201103} {\bibfield  {journal} {\bibinfo
  {journal} {Phys. Rev. Lett.}\ }\textbf {\bibinfo {volume} {119}},\ \bibinfo
  {pages} {201103} (\bibinfo {year} {2017})},\ \Eprint
  {http://arxiv.org/abs/1708.03651} {arXiv:1708.03651 [gr-qc]}\BibitemShut
  {NoStop}%
\bibitem [{\citenamefont {Healy}\ \emph {et~al.}(2012)\citenamefont {Healy},
  \citenamefont {Bode}, \citenamefont {Haas}, \citenamefont {Pazos},
  \citenamefont {Laguna}, \citenamefont {Shoemaker},\ and\ \citenamefont
  {Yunes}}]{Healy:2011ef}%
  \BibitemOpen
  \bibfield  {author} {\bibinfo {author} {\bibfnamefont {James}\ \bibnamefont
  {Healy}}, \bibinfo {author} {\bibfnamefont {Tanja}\ \bibnamefont {Bode}},
  \bibinfo {author} {\bibfnamefont {Roland}\ \bibnamefont {Haas}}, \bibinfo
  {author} {\bibfnamefont {Enrique}\ \bibnamefont {Pazos}}, \bibinfo {author}
  {\bibfnamefont {Pablo}\ \bibnamefont {Laguna}}, \bibinfo {author}
  {\bibfnamefont {Deirdre~M.}\ \bibnamefont {Shoemaker}}, \ and\ \bibinfo
  {author} {\bibfnamefont {Nicolas}\ \bibnamefont {Yunes}},\ }\href
  {\doibase10.1088/0264-9381/29/23/232002} {\bibfield  {journal} {\bibinfo
  {journal} {Class. Quant. Grav.}\ }\textbf {\bibinfo {volume} {29}},\ \bibinfo
  {pages} {232002} (\bibinfo {year} {2012})},\ \Eprint
  {http://arxiv.org/abs/1112.3928} {arXiv:1112.3928 [gr-qc]}\BibitemShut
  {NoStop}%
\bibitem [{\citenamefont {Barausse}\ \emph {et~al.}(2013)\citenamefont
  {Barausse}, \citenamefont {Palenzuela}, \citenamefont {Ponce},\ and\
  \citenamefont {Lehner}}]{Barausse:2012da}%
  \BibitemOpen
  \bibfield  {author} {\bibinfo {author} {\bibfnamefont {Enrico}\ \bibnamefont
  {Barausse}}, \bibinfo {author} {\bibfnamefont {Carlos}\ \bibnamefont
  {Palenzuela}}, \bibinfo {author} {\bibfnamefont {Marcelo}\ \bibnamefont
  {Ponce}}, \ and\ \bibinfo {author} {\bibfnamefont {Luis}\ \bibnamefont
  {Lehner}},\ }\href {\doibase10.1103/PhysRevD.87.081506} {\bibfield  {journal}
  {\bibinfo  {journal} {Phys. Rev.}\ }\textbf {\bibinfo {volume} {D87}},\
  \bibinfo {pages} {081506} (\bibinfo {year} {2013})},\ \Eprint
  {http://arxiv.org/abs/1212.5053} {arXiv:1212.5053 [gr-qc]}\BibitemShut
  {NoStop}%
\bibitem [{\citenamefont {Palenzuela}\ \emph {et~al.}(2014)\citenamefont
  {Palenzuela}, \citenamefont {Barausse}, \citenamefont {Ponce},\ and\
  \citenamefont {Lehner}}]{Palenzuela:2013hsa}%
  \BibitemOpen
  \bibfield  {author} {\bibinfo {author} {\bibfnamefont {Carlos}\ \bibnamefont
  {Palenzuela}}, \bibinfo {author} {\bibfnamefont {Enrico}\ \bibnamefont
  {Barausse}}, \bibinfo {author} {\bibfnamefont {Marcelo}\ \bibnamefont
  {Ponce}}, \ and\ \bibinfo {author} {\bibfnamefont {Luis}\ \bibnamefont
  {Lehner}},\ }\href {\doibase10.1103/PhysRevD.89.044024} {\bibfield  {journal}
  {\bibinfo  {journal} {Phys. Rev.}\ }\textbf {\bibinfo {volume} {D89}},\
  \bibinfo {pages} {044024} (\bibinfo {year} {2014})},\ \Eprint
  {http://arxiv.org/abs/1310.4481} {arXiv:1310.4481 [gr-qc]}\BibitemShut
  {NoStop}%
\bibitem [{\citenamefont {Antoniadis}\ \emph {et~al.}(1998)\citenamefont
  {Antoniadis}, \citenamefont {Arkani-Hamed}, \citenamefont {Dimopoulos},\ and\
  \citenamefont {Dvali}}]{Antoniadis:1998ig}%
  \BibitemOpen
  \bibfield  {author} {\bibinfo {author} {\bibfnamefont {Ignatios}\
  \bibnamefont {Antoniadis}}, \bibinfo {author} {\bibfnamefont {Nima}\
  \bibnamefont {Arkani-Hamed}}, \bibinfo {author} {\bibfnamefont {Savas}\
  \bibnamefont {Dimopoulos}}, \ and\ \bibinfo {author} {\bibfnamefont {G.~R.}\
  \bibnamefont {Dvali}},\ }\href {\doibase10.1016/S0370-2693(98)00860-0}
  {\bibfield  {journal} {\bibinfo  {journal} {Phys. Lett.}\ }\textbf {\bibinfo
  {volume} {B436}},\ \bibinfo {pages} {257--263} (\bibinfo {year} {1998})},\
  \Eprint {http://arxiv.org/abs/hep-ph/9804398} {arXiv:hep-ph/9804398
  [hep-ph]}\BibitemShut {NoStop}%
\bibitem [{\citenamefont {Arkani-Hamed}\ \emph {et~al.}(1998)\citenamefont
  {Arkani-Hamed}, \citenamefont {Dimopoulos},\ and\ \citenamefont
  {Dvali}}]{ArkaniHamed:1998rs}%
  \BibitemOpen
  \bibfield  {author} {\bibinfo {author} {\bibfnamefont {Nima}\ \bibnamefont
  {Arkani-Hamed}}, \bibinfo {author} {\bibfnamefont {Savas}\ \bibnamefont
  {Dimopoulos}}, \ and\ \bibinfo {author} {\bibfnamefont {G.~R.}\ \bibnamefont
  {Dvali}},\ }\href {\doibase10.1016/S0370-2693(98)00466-3} {\bibfield
  {journal} {\bibinfo  {journal} {Phys. Lett.}\ }\textbf {\bibinfo {volume}
  {B429}},\ \bibinfo {pages} {263--272} (\bibinfo {year} {1998})},\ \Eprint
  {http://arxiv.org/abs/hep-ph/9803315} {arXiv:hep-ph/9803315
  [hep-ph]}\BibitemShut {NoStop}%
\bibitem [{\citenamefont {Randall}\ and\ \citenamefont
  {Sundrum}(1999{\natexlab{a}})}]{Randall:1999ee}%
  \BibitemOpen
  \bibfield  {author} {\bibinfo {author} {\bibfnamefont {Lisa}\ \bibnamefont
  {Randall}}\ and\ \bibinfo {author} {\bibfnamefont {Raman}\ \bibnamefont
  {Sundrum}},\ }\href {\doibase10.1103/PhysRevLett.83.3370} {\bibfield
  {journal} {\bibinfo  {journal} {Phys. Rev. Lett.}\ }\textbf {\bibinfo
  {volume} {83}},\ \bibinfo {pages} {3370--3373} (\bibinfo {year}
  {1999}{\natexlab{a}})},\ \Eprint {http://arxiv.org/abs/hep-ph/9905221}
  {arXiv:hep-ph/9905221 [hep-ph]}\BibitemShut {NoStop}%
\bibitem [{\citenamefont {Randall}\ and\ \citenamefont
  {Sundrum}(1999{\natexlab{b}})}]{Randall:1999vf}%
  \BibitemOpen
  \bibfield  {author} {\bibinfo {author} {\bibfnamefont {Lisa}\ \bibnamefont
  {Randall}}\ and\ \bibinfo {author} {\bibfnamefont {Raman}\ \bibnamefont
  {Sundrum}},\ }\href {\doibase10.1103/PhysRevLett.83.4690} {\bibfield
  {journal} {\bibinfo  {journal} {Phys. Rev. Lett.}\ }\textbf {\bibinfo
  {volume} {83}},\ \bibinfo {pages} {4690--4693} (\bibinfo {year}
  {1999}{\natexlab{b}})},\ \Eprint {http://arxiv.org/abs/hep-th/9906064}
  {arXiv:hep-th/9906064 [hep-th]}\BibitemShut {NoStop}%
\bibitem [{\citenamefont {Banks}\ and\ \citenamefont
  {Fischler}(1999)}]{Banks:1999gd}%
  \BibitemOpen
  \bibfield  {author} {\bibinfo {author} {\bibfnamefont {Tom}\ \bibnamefont
  {Banks}}\ and\ \bibinfo {author} {\bibfnamefont {Willy}\ \bibnamefont
  {Fischler}},\ }\href@noop {} {\  (\bibinfo {year} {1999})},\ \Eprint
  {http://arxiv.org/abs/hep-th/9906038} {arXiv:hep-th/9906038
  [hep-th]}\BibitemShut {NoStop}%
\bibitem [{\citenamefont {Dimopoulos}\ and\ \citenamefont
  {Landsberg}(2001)}]{Dimopoulos:2001hw}%
  \BibitemOpen
  \bibfield  {author} {\bibinfo {author} {\bibfnamefont {Savas}\ \bibnamefont
  {Dimopoulos}}\ and\ \bibinfo {author} {\bibfnamefont {Greg~L.}\ \bibnamefont
  {Landsberg}},\ }\href {\doibase10.1103/PhysRevLett.87.161602} {\bibfield
  {journal} {\bibinfo  {journal} {Phys. Rev. Lett.}\ }\textbf {\bibinfo
  {volume} {87}},\ \bibinfo {pages} {161602} (\bibinfo {year} {2001})},\
  \Eprint {http://arxiv.org/abs/hep-ph/0106295} {arXiv:hep-ph/0106295
  [hep-ph]}\BibitemShut {NoStop}%
\bibitem [{\citenamefont {Giddings}\ and\ \citenamefont
  {Thomas}(2002)}]{Giddings:2001bu}%
  \BibitemOpen
  \bibfield  {author} {\bibinfo {author} {\bibfnamefont {Steven~B.}\
  \bibnamefont {Giddings}}\ and\ \bibinfo {author} {\bibfnamefont {Scott~D.}\
  \bibnamefont {Thomas}},\ }\href {\doibase10.1103/PhysRevD.65.056010}
  {\bibfield  {journal} {\bibinfo  {journal} {Phys. Rev.}\ }\textbf {\bibinfo
  {volume} {D65}},\ \bibinfo {pages} {056010} (\bibinfo {year} {2002})},\
  \Eprint {http://arxiv.org/abs/hep-ph/0106219} {arXiv:hep-ph/0106219
  [hep-ph]}\BibitemShut {NoStop}%
\bibitem [{\citenamefont {Frost}\ \emph {et~al.}(2009)\citenamefont {Frost},
  \citenamefont {Gaunt}, \citenamefont {Sampaio}, \citenamefont {Casals},
  \citenamefont {Dolan}, \citenamefont {Parker},\ and\ \citenamefont
  {Webber}}]{Frost:2009cf}%
  \BibitemOpen
  \bibfield  {author} {\bibinfo {author} {\bibfnamefont {James~A.}\
  \bibnamefont {Frost}}, \bibinfo {author} {\bibfnamefont {Jonathan~R.}\
  \bibnamefont {Gaunt}}, \bibinfo {author} {\bibfnamefont {Marco O.~P.}\
  \bibnamefont {Sampaio}}, \bibinfo {author} {\bibfnamefont {Marc}\
  \bibnamefont {Casals}}, \bibinfo {author} {\bibfnamefont {Sam~R.}\
  \bibnamefont {Dolan}}, \bibinfo {author} {\bibfnamefont {Michael~Andrew}\
  \bibnamefont {Parker}}, \ and\ \bibinfo {author} {\bibfnamefont {Bryan~R.}\
  \bibnamefont {Webber}},\ }\href {\doibase10.1088/1126-6708/2009/10/014}
  {\bibfield  {journal} {\bibinfo  {journal} {JHEP}\ }\textbf {\bibinfo
  {volume} {10}},\ \bibinfo {pages} {014} (\bibinfo {year} {2009})},\ \Eprint
  {http://arxiv.org/abs/0904.0979} {arXiv:0904.0979 [hep-ph]}\BibitemShut
  {NoStop}%
\bibitem [{\citenamefont {Sperhake}\ \emph
  {et~al.}(2008{\natexlab{b}})\citenamefont {Sperhake}, \citenamefont
  {Cardoso}, \citenamefont {Pretorius}, \citenamefont {Berti},\ and\
  \citenamefont {Gonzalez}}]{Sperhake:2008ga}%
  \BibitemOpen
  \bibfield  {author} {\bibinfo {author} {\bibfnamefont {Ulrich}\ \bibnamefont
  {Sperhake}}, \bibinfo {author} {\bibfnamefont {Vitor}\ \bibnamefont
  {Cardoso}}, \bibinfo {author} {\bibfnamefont {Frans}\ \bibnamefont
  {Pretorius}}, \bibinfo {author} {\bibfnamefont {Emanuele}\ \bibnamefont
  {Berti}}, \ and\ \bibinfo {author} {\bibfnamefont {Jose~A.}\ \bibnamefont
  {Gonzalez}},\ }\href {\doibase10.1103/PhysRevLett.101.161101} {\bibfield
  {journal} {\bibinfo  {journal} {Phys. Rev. Lett.}\ }\textbf {\bibinfo
  {volume} {101}},\ \bibinfo {pages} {161101} (\bibinfo {year}
  {2008}{\natexlab{b}})},\ \Eprint {http://arxiv.org/abs/0806.1738}
  {arXiv:0806.1738 [gr-qc]}\BibitemShut {NoStop}%
\bibitem [{\citenamefont {Healy}\ \emph {et~al.}(2016)\citenamefont {Healy},
  \citenamefont {Ruchlin}, \citenamefont {Lousto},\ and\ \citenamefont
  {Zlochower}}]{Healy:2015mla}%
  \BibitemOpen
  \bibfield  {author} {\bibinfo {author} {\bibfnamefont {J.}~\bibnamefont
  {Healy}}, \bibinfo {author} {\bibfnamefont {I.}~\bibnamefont {Ruchlin}},
  \bibinfo {author} {\bibfnamefont {C.~O.}\ \bibnamefont {Lousto}}, \ and\
  \bibinfo {author} {\bibfnamefont {Y.}~\bibnamefont {Zlochower}},\ }\href
  {\doibase10.1103/PhysRevD.94.104020} {\bibfield  {journal} {\bibinfo
  {journal} {Phys. Rev. D}\ }\textbf {\bibinfo {volume} {94}},\ \bibinfo
  {pages} {104020} (\bibinfo {year} {2016})},\ \Eprint
  {http://arxiv.org/abs/1506.06153} {1506.06153}\BibitemShut {NoStop}%
\bibitem [{\citenamefont {Shibata}\ \emph {et~al.}(2008)\citenamefont
  {Shibata}, \citenamefont {Okawa},\ and\ \citenamefont
  {Yamamoto}}]{Shibata:2008rq}%
  \BibitemOpen
  \bibfield  {author} {\bibinfo {author} {\bibfnamefont {Masaru}\ \bibnamefont
  {Shibata}}, \bibinfo {author} {\bibfnamefont {Hirotada}\ \bibnamefont
  {Okawa}}, \ and\ \bibinfo {author} {\bibfnamefont {Tetsuro}\ \bibnamefont
  {Yamamoto}},\ }\href {\doibase10.1103/PhysRevD.78.101501} {\bibfield
  {journal} {\bibinfo  {journal} {Phys. Rev.}\ }\textbf {\bibinfo {volume}
  {D78}},\ \bibinfo {pages} {101501} (\bibinfo {year} {2008})},\ \Eprint
  {http://arxiv.org/abs/0810.4735} {arXiv:0810.4735 [gr-qc]}\BibitemShut
  {NoStop}%
\bibitem [{\citenamefont {Pretorius}\ and\ \citenamefont
  {Khurana}(2007)}]{Pretorius:2007jn}%
  \BibitemOpen
  \bibfield  {author} {\bibinfo {author} {\bibfnamefont {Frans}\ \bibnamefont
  {Pretorius}}\ and\ \bibinfo {author} {\bibfnamefont {Deepak}\ \bibnamefont
  {Khurana}},\ }\bibfield  {booktitle} {\emph {\bibinfo {booktitle} {{New
  frontiers in numerical relativity. Proceedings, International Meeting, NFNR
  2006, Potsdam, Germany, July 17-21, 2006}}},\ }\href
  {\doibase10.1088/0264-9381/24/12/S07} {\bibfield  {journal} {\bibinfo
  {journal} {Class. Quant. Grav.}\ }\textbf {\bibinfo {volume} {24}},\ \bibinfo
  {pages} {S83--S108} (\bibinfo {year} {2007})},\ \Eprint
  {http://arxiv.org/abs/gr-qc/0702084} {arXiv:gr-qc/0702084
  [GR-QC]}\BibitemShut {NoStop}%
\bibitem [{\citenamefont {Sperhake}\ \emph {et~al.}(2009)\citenamefont
  {Sperhake}, \citenamefont {Cardoso}, \citenamefont {Pretorius}, \citenamefont
  {Berti}, \citenamefont {Hinderer},\ and\ \citenamefont
  {Yunes}}]{Sperhake:2009jz}%
  \BibitemOpen
  \bibfield  {author} {\bibinfo {author} {\bibfnamefont {Ulrich}\ \bibnamefont
  {Sperhake}}, \bibinfo {author} {\bibfnamefont {Vitor}\ \bibnamefont
  {Cardoso}}, \bibinfo {author} {\bibfnamefont {Frans}\ \bibnamefont
  {Pretorius}}, \bibinfo {author} {\bibfnamefont {Emanuele}\ \bibnamefont
  {Berti}}, \bibinfo {author} {\bibfnamefont {Tanja}\ \bibnamefont {Hinderer}},
  \ and\ \bibinfo {author} {\bibfnamefont {Nicolas}\ \bibnamefont {Yunes}},\
  }\href {\doibase10.1103/PhysRevLett.103.131102} {\bibfield  {journal}
  {\bibinfo  {journal} {Phys. Rev. Lett.}\ }\textbf {\bibinfo {volume} {103}},\
  \bibinfo {pages} {131102} (\bibinfo {year} {2009})},\ \Eprint
  {http://arxiv.org/abs/0907.1252} {arXiv:0907.1252 [gr-qc]}\BibitemShut
  {NoStop}%
\bibitem [{\citenamefont {Sperhake}\ \emph {et~al.}(2013)\citenamefont
  {Sperhake}, \citenamefont {Berti}, \citenamefont {Cardoso},\ and\
  \citenamefont {Pretorius}}]{Sperhake:2012me}%
  \BibitemOpen
  \bibfield  {author} {\bibinfo {author} {\bibfnamefont {Ulrich}\ \bibnamefont
  {Sperhake}}, \bibinfo {author} {\bibfnamefont {Emanuele}\ \bibnamefont
  {Berti}}, \bibinfo {author} {\bibfnamefont {Vitor}\ \bibnamefont {Cardoso}},
  \ and\ \bibinfo {author} {\bibfnamefont {Frans}\ \bibnamefont {Pretorius}},\
  }\href {\doibase10.1103/PhysRevLett.111.041101} {\bibfield  {journal}
  {\bibinfo  {journal} {Phys. Rev. Lett.}\ }\textbf {\bibinfo {volume} {111}},\
  \bibinfo {pages} {041101} (\bibinfo {year} {2013})},\ \Eprint
  {http://arxiv.org/abs/1211.6114} {arXiv:1211.6114 [gr-qc]}\BibitemShut
  {NoStop}%
\bibitem [{\citenamefont {East}\ and\ \citenamefont
  {Pretorius}(2013)}]{East:2012mb}%
  \BibitemOpen
  \bibfield  {author} {\bibinfo {author} {\bibfnamefont {William~E.}\
  \bibnamefont {East}}\ and\ \bibinfo {author} {\bibfnamefont {Frans}\
  \bibnamefont {Pretorius}},\ }\href {\doibase10.1103/PhysRevLett.110.101101}
  {\bibfield  {journal} {\bibinfo  {journal} {Phys. Rev. Lett.}\ }\textbf
  {\bibinfo {volume} {110}},\ \bibinfo {pages} {101101} (\bibinfo {year}
  {2013})},\ \Eprint {http://arxiv.org/abs/1210.0443} {arXiv:1210.0443
  [gr-qc]}\BibitemShut {NoStop}%
\bibitem [{\citenamefont {Rezzolla}\ and\ \citenamefont
  {Takami}(2013)}]{Rezzolla:2012nr}%
  \BibitemOpen
  \bibfield  {author} {\bibinfo {author} {\bibfnamefont {Luciano}\ \bibnamefont
  {Rezzolla}}\ and\ \bibinfo {author} {\bibfnamefont {Kentaro}\ \bibnamefont
  {Takami}},\ }\href {\doibase10.1088/0264-9381/30/1/012001} {\bibfield
  {journal} {\bibinfo  {journal} {Class. Quant. Grav.}\ }\textbf {\bibinfo
  {volume} {30}},\ \bibinfo {pages} {012001} (\bibinfo {year} {2013})},\
  \Eprint {http://arxiv.org/abs/1209.6138} {arXiv:1209.6138
  [gr-qc]}\BibitemShut {NoStop}%
\bibitem [{\citenamefont {Sperhake}\ \emph {et~al.}(2016)\citenamefont
  {Sperhake}, \citenamefont {Berti}, \citenamefont {Cardoso},\ and\
  \citenamefont {Pretorius}}]{Sperhake:2015siy}%
  \BibitemOpen
  \bibfield  {author} {\bibinfo {author} {\bibfnamefont {Ulrich}\ \bibnamefont
  {Sperhake}}, \bibinfo {author} {\bibfnamefont {Emanuele}\ \bibnamefont
  {Berti}}, \bibinfo {author} {\bibfnamefont {Vitor}\ \bibnamefont {Cardoso}},
  \ and\ \bibinfo {author} {\bibfnamefont {Frans}\ \bibnamefont {Pretorius}},\
  }\href {\doibase10.1103/PhysRevD.93.044012} {\bibfield  {journal} {\bibinfo
  {journal} {Phys. Rev.}\ }\textbf {\bibinfo {volume} {D93}},\ \bibinfo {pages}
  {044012} (\bibinfo {year} {2016})},\ \Eprint
  {http://arxiv.org/abs/1511.08209} {arXiv:1511.08209 [gr-qc]}\BibitemShut
  {NoStop}%
\bibitem [{\citenamefont {Galtsov}\ \emph {et~al.}(2010)\citenamefont
  {Galtsov}, \citenamefont {Kofinas}, \citenamefont {Spirin},\ and\
  \citenamefont {Tomaras}}]{Galtsov:2010vtu}%
  \BibitemOpen
  \bibfield  {author} {\bibinfo {author} {\bibfnamefont {Dmitry~V.}\
  \bibnamefont {Galtsov}}, \bibinfo {author} {\bibfnamefont {Georgios}\
  \bibnamefont {Kofinas}}, \bibinfo {author} {\bibfnamefont {Pavel}\
  \bibnamefont {Spirin}}, \ and\ \bibinfo {author} {\bibfnamefont
  {Theodore~N.}\ \bibnamefont {Tomaras}},\ }\href
  {\doibase10.1007/JHEP05(2010)055} {\bibfield  {journal} {\bibinfo  {journal}
  {JHEP}\ }\textbf {\bibinfo {volume} {05}},\ \bibinfo {pages} {055} (\bibinfo
  {year} {2010})},\ \Eprint {http://arxiv.org/abs/1003.2982} {arXiv:1003.2982
  [hep-th]}\BibitemShut {NoStop}%
\bibitem [{\citenamefont {Berti}\ \emph {et~al.}(2011)\citenamefont {Berti},
  \citenamefont {Cardoso},\ and\ \citenamefont {Kipapa}}]{Berti:2010gx}%
  \BibitemOpen
  \bibfield  {author} {\bibinfo {author} {\bibfnamefont {Emanuele}\
  \bibnamefont {Berti}}, \bibinfo {author} {\bibfnamefont {Vitor}\ \bibnamefont
  {Cardoso}}, \ and\ \bibinfo {author} {\bibfnamefont {Barnabas}\ \bibnamefont
  {Kipapa}},\ }\href {\doibase10.1103/PhysRevD.83.084018} {\bibfield  {journal}
  {\bibinfo  {journal} {Phys. Rev.}\ }\textbf {\bibinfo {volume} {D83}},\
  \bibinfo {pages} {084018} (\bibinfo {year} {2011})},\ \Eprint
  {http://arxiv.org/abs/1010.3874} {arXiv:1010.3874 [gr-qc]}\BibitemShut
  {NoStop}%
\bibitem [{\citenamefont {Gal'tsov}\ \emph {et~al.}(2013)\citenamefont
  {Gal'tsov}, \citenamefont {Spirin},\ and\ \citenamefont
  {Tomaras}}]{Galtsov:2012pcw}%
  \BibitemOpen
  \bibfield  {author} {\bibinfo {author} {\bibfnamefont {Dmitry}\ \bibnamefont
  {Gal'tsov}}, \bibinfo {author} {\bibfnamefont {Pavel}\ \bibnamefont
  {Spirin}}, \ and\ \bibinfo {author} {\bibfnamefont {Theodore~N.}\
  \bibnamefont {Tomaras}},\ }\href {\doibase10.1007/JHEP01(2013)087} {\bibfield
   {journal} {\bibinfo  {journal} {JHEP}\ }\textbf {\bibinfo {volume} {01}},\
  \bibinfo {pages} {087} (\bibinfo {year} {2013})},\ \Eprint
  {http://arxiv.org/abs/1210.6976} {arXiv:1210.6976 [hep-th]}\BibitemShut
  {NoStop}%
\bibitem [{\citenamefont {Zilhao}\ \emph {et~al.}(2010)\citenamefont {Zilhao},
  \citenamefont {Witek}, \citenamefont {Sperhake}, \citenamefont {Cardoso},
  \citenamefont {Gualtieri}, \citenamefont {Herdeiro},\ and\ \citenamefont
  {Nerozzi}}]{Zilhao:2010sr}%
  \BibitemOpen
  \bibfield  {author} {\bibinfo {author} {\bibfnamefont {Miguel}\ \bibnamefont
  {Zilhao}}, \bibinfo {author} {\bibfnamefont {Helvi}\ \bibnamefont {Witek}},
  \bibinfo {author} {\bibfnamefont {Ulrich}\ \bibnamefont {Sperhake}}, \bibinfo
  {author} {\bibfnamefont {Vitor}\ \bibnamefont {Cardoso}}, \bibinfo {author}
  {\bibfnamefont {Leonardo}\ \bibnamefont {Gualtieri}}, \bibinfo {author}
  {\bibfnamefont {Carlos}\ \bibnamefont {Herdeiro}}, \ and\ \bibinfo {author}
  {\bibfnamefont {Andrea}\ \bibnamefont {Nerozzi}},\ }\href
  {\doibase10.1103/PhysRevD.81.084052} {\bibfield  {journal} {\bibinfo
  {journal} {Phys. Rev.}\ }\textbf {\bibinfo {volume} {D81}},\ \bibinfo {pages}
  {084052} (\bibinfo {year} {2010})},\ \Eprint {http://arxiv.org/abs/1001.2302}
  {arXiv:1001.2302 [gr-qc]}\BibitemShut {NoStop}%
\bibitem [{\citenamefont {Yoshino}\ and\ \citenamefont
  {Shibata}(2011{\natexlab{a}})}]{Yoshino:2011zz}%
  \BibitemOpen
  \bibfield  {author} {\bibinfo {author} {\bibfnamefont {H.}~\bibnamefont
  {Yoshino}}\ and\ \bibinfo {author} {\bibfnamefont {M.}~\bibnamefont
  {Shibata}},\ }\href {\doibase10.1143/PTPS.189.269} {\bibfield  {journal}
  {\bibinfo  {journal} {Prog. Theor. Phys. Suppl.}\ }\textbf {\bibinfo {volume}
  {189}},\ \bibinfo {pages} {269--310} (\bibinfo {year}
  {2011}{\natexlab{a}})}\BibitemShut {NoStop}%
\bibitem [{\citenamefont {Yoshino}\ and\ \citenamefont
  {Shibata}(2011{\natexlab{b}})}]{Yoshino:2011zza}%
  \BibitemOpen
  \bibfield  {author} {\bibinfo {author} {\bibfnamefont {H.}~\bibnamefont
  {Yoshino}}\ and\ \bibinfo {author} {\bibfnamefont {M.}~\bibnamefont
  {Shibata}},\ }\href {\doibase10.1143/PTPS.190.282} {\bibfield  {journal}
  {\bibinfo  {journal} {Prog. Theor. Phys. Suppl.}\ }\textbf {\bibinfo {volume}
  {190}},\ \bibinfo {pages} {282--303} (\bibinfo {year}
  {2011}{\natexlab{b}})}\BibitemShut {NoStop}%
\bibitem [{\citenamefont {Cook}\ \emph {et~al.}(2016)\citenamefont {Cook},
  \citenamefont {Figueras}, \citenamefont {Kunesch}, \citenamefont {Sperhake},\
  and\ \citenamefont {Tunyasuvunakool}}]{Cook:2016soy}%
  \BibitemOpen
  \bibfield  {author} {\bibinfo {author} {\bibfnamefont {William~G.}\
  \bibnamefont {Cook}}, \bibinfo {author} {\bibfnamefont {Pau}\ \bibnamefont
  {Figueras}}, \bibinfo {author} {\bibfnamefont {Markus}\ \bibnamefont
  {Kunesch}}, \bibinfo {author} {\bibfnamefont {Ulrich}\ \bibnamefont
  {Sperhake}}, \ and\ \bibinfo {author} {\bibfnamefont {Saran}\ \bibnamefont
  {Tunyasuvunakool}},\ }\bibfield  {booktitle} {\emph {\bibinfo {booktitle}
  {{Proceedings, 3rd Amazonian Symposium on Physics: Belem, Brazil, September
  28-October 2, 2015}}},\ }\href {\doibase10.1142/S0218271816410133} {\bibfield
   {journal} {\bibinfo  {journal} {Int. J. Mod. Phys.}\ }\textbf {\bibinfo
  {volume} {D25}},\ \bibinfo {pages} {1641013} (\bibinfo {year} {2016})},\
  \Eprint {http://arxiv.org/abs/1603.00362} {arXiv:1603.00362
  [gr-qc]}\BibitemShut {NoStop}%
\bibitem [{\citenamefont {Kodama}\ and\ \citenamefont
  {Ishibashi}(2003)}]{Kodama:2003jz}%
  \BibitemOpen
  \bibfield  {author} {\bibinfo {author} {\bibfnamefont {Hideo}\ \bibnamefont
  {Kodama}}\ and\ \bibinfo {author} {\bibfnamefont {Akihiro}\ \bibnamefont
  {Ishibashi}},\ }\href {\doibase10.1143/PTP.110.701} {\bibfield  {journal}
  {\bibinfo  {journal} {Prog. Theor. Phys.}\ }\textbf {\bibinfo {volume}
  {110}},\ \bibinfo {pages} {701--722} (\bibinfo {year} {2003})},\ \Eprint
  {http://arxiv.org/abs/hep-th/0305147} {arXiv:hep-th/0305147
  [hep-th]}\BibitemShut {NoStop}%
\bibitem [{\citenamefont {Kodama}\ and\ \citenamefont
  {Ishibashi}(2004)}]{Kodama:2003kk}%
  \BibitemOpen
  \bibfield  {author} {\bibinfo {author} {\bibfnamefont {Hideo}\ \bibnamefont
  {Kodama}}\ and\ \bibinfo {author} {\bibfnamefont {Akihiro}\ \bibnamefont
  {Ishibashi}},\ }\href {\doibase10.1143/PTP.111.29} {\bibfield  {journal}
  {\bibinfo  {journal} {Prog. Theor. Phys.}\ }\textbf {\bibinfo {volume}
  {111}},\ \bibinfo {pages} {29--73} (\bibinfo {year} {2004})},\ \Eprint
  {http://arxiv.org/abs/hep-th/0308128} {arXiv:hep-th/0308128
  [hep-th]}\BibitemShut {NoStop}%
\bibitem [{\citenamefont {Witek}\ \emph {et~al.}(2010)\citenamefont {Witek},
  \citenamefont {Zilhao}, \citenamefont {Gualtieri}, \citenamefont {Cardoso},
  \citenamefont {Herdeiro}, \citenamefont {Nerozzi},\ and\ \citenamefont
  {Sperhake}}]{Witek:2010xi}%
  \BibitemOpen
  \bibfield  {author} {\bibinfo {author} {\bibfnamefont {Helvi}\ \bibnamefont
  {Witek}}, \bibinfo {author} {\bibfnamefont {Miguel}\ \bibnamefont {Zilhao}},
  \bibinfo {author} {\bibfnamefont {Leonardo}\ \bibnamefont {Gualtieri}},
  \bibinfo {author} {\bibfnamefont {Vitor}\ \bibnamefont {Cardoso}}, \bibinfo
  {author} {\bibfnamefont {Carlos}\ \bibnamefont {Herdeiro}}, \bibinfo {author}
  {\bibfnamefont {Andrea}\ \bibnamefont {Nerozzi}}, \ and\ \bibinfo {author}
  {\bibfnamefont {Ulrich}\ \bibnamefont {Sperhake}},\ }\href
  {\doibase10.1103/PhysRevD.82.104014} {\bibfield  {journal} {\bibinfo
  {journal} {Phys. Rev.}\ }\textbf {\bibinfo {volume} {D82}},\ \bibinfo {pages}
  {104014} (\bibinfo {year} {2010})},\ \Eprint {http://arxiv.org/abs/1006.3081}
  {arXiv:1006.3081 [gr-qc]}\BibitemShut {NoStop}%
\bibitem [{\citenamefont {Yoshino}\ and\ \citenamefont
  {Shibata}(2009)}]{Yoshino:2009xp}%
  \BibitemOpen
  \bibfield  {author} {\bibinfo {author} {\bibfnamefont {Hirotaka}\
  \bibnamefont {Yoshino}}\ and\ \bibinfo {author} {\bibfnamefont {Masaru}\
  \bibnamefont {Shibata}},\ }\href {\doibase10.1103/PhysRevD.80.084025}
  {\bibfield  {journal} {\bibinfo  {journal} {Phys. Rev.}\ }\textbf {\bibinfo
  {volume} {D80}},\ \bibinfo {pages} {084025} (\bibinfo {year} {2009})},\
  \Eprint {http://arxiv.org/abs/0907.2760} {arXiv:0907.2760
  [gr-qc]}\BibitemShut {NoStop}%
\bibitem [{\citenamefont {Godazgar}\ and\ \citenamefont
  {Reall}(2012)}]{Godazgar:2012zq}%
  \BibitemOpen
  \bibfield  {author} {\bibinfo {author} {\bibfnamefont {Mahdi}\ \bibnamefont
  {Godazgar}}\ and\ \bibinfo {author} {\bibfnamefont {Harvey~S.}\ \bibnamefont
  {Reall}},\ }\href {\doibase10.1103/PhysRevD.85.084021} {\bibfield  {journal}
  {\bibinfo  {journal} {Phys. Rev.}\ }\textbf {\bibinfo {volume} {D85}},\
  \bibinfo {pages} {084021} (\bibinfo {year} {2012})},\ \Eprint
  {http://arxiv.org/abs/1201.4373} {arXiv:1201.4373 [gr-qc]}\BibitemShut
  {NoStop}%
\bibitem [{\citenamefont {Cook}\ and\ \citenamefont
  {Sperhake}(2017)}]{Cook:2016qnt}%
  \BibitemOpen
  \bibfield  {author} {\bibinfo {author} {\bibfnamefont {William~G.}\
  \bibnamefont {Cook}}\ and\ \bibinfo {author} {\bibfnamefont {Ulrich}\
  \bibnamefont {Sperhake}},\ }\href {\doibase10.1088/1361-6382/aa5294}
  {\bibfield  {journal} {\bibinfo  {journal} {Class. Quant. Grav.}\ }\textbf
  {\bibinfo {volume} {34}},\ \bibinfo {pages} {035010} (\bibinfo {year}
  {2017})},\ \Eprint {http://arxiv.org/abs/1609.01292} {arXiv:1609.01292
  [gr-qc]}\BibitemShut {NoStop}%
\bibitem [{\citenamefont {Zilhao}\ \emph {et~al.}(2011)\citenamefont {Zilhao},
  \citenamefont {Ansorg}, \citenamefont {Cardoso}, \citenamefont {Gualtieri},
  \citenamefont {Herdeiro}, \citenamefont {Sperhake},\ and\ \citenamefont
  {Witek}}]{Zilhao:2011yc}%
  \BibitemOpen
  \bibfield  {author} {\bibinfo {author} {\bibfnamefont {Miguel}\ \bibnamefont
  {Zilhao}}, \bibinfo {author} {\bibfnamefont {Marcus}\ \bibnamefont {Ansorg}},
  \bibinfo {author} {\bibfnamefont {Vitor}\ \bibnamefont {Cardoso}}, \bibinfo
  {author} {\bibfnamefont {Leonardo}\ \bibnamefont {Gualtieri}}, \bibinfo
  {author} {\bibfnamefont {Carlos}\ \bibnamefont {Herdeiro}}, \bibinfo {author}
  {\bibfnamefont {Ulrich}\ \bibnamefont {Sperhake}}, \ and\ \bibinfo {author}
  {\bibfnamefont {Helvi}\ \bibnamefont {Witek}},\ }\href
  {\doibase10.1103/PhysRevD.84.084039} {\bibfield  {journal} {\bibinfo
  {journal} {Phys. Rev.}\ }\textbf {\bibinfo {volume} {D84}},\ \bibinfo {pages}
  {084039} (\bibinfo {year} {2011})},\ \Eprint {http://arxiv.org/abs/1109.2149}
  {arXiv:1109.2149 [gr-qc]}\BibitemShut {NoStop}%
\bibitem [{\citenamefont {Okawa}\ \emph {et~al.}(2011)\citenamefont {Okawa},
  \citenamefont {Nakao},\ and\ \citenamefont {Shibata}}]{Okawa:2011fv}%
  \BibitemOpen
  \bibfield  {author} {\bibinfo {author} {\bibfnamefont {Hirotada}\
  \bibnamefont {Okawa}}, \bibinfo {author} {\bibfnamefont {Ken-ichi}\
  \bibnamefont {Nakao}}, \ and\ \bibinfo {author} {\bibfnamefont {Masaru}\
  \bibnamefont {Shibata}},\ }\href {\doibase10.1103/PhysRevD.83.121501}
  {\bibfield  {journal} {\bibinfo  {journal} {Phys. Rev.}\ }\textbf {\bibinfo
  {volume} {D83}},\ \bibinfo {pages} {121501} (\bibinfo {year} {2011})},\
  \Eprint {http://arxiv.org/abs/1105.3331} {arXiv:1105.3331
  [gr-qc]}\BibitemShut {NoStop}%
\bibitem [{\citenamefont {Cook}\ \emph {et~al.}(2017)\citenamefont {Cook},
  \citenamefont {Sperhake}, \citenamefont {Berti},\ and\ \citenamefont
  {Cardoso}}]{Cook:2017fec}%
  \BibitemOpen
  \bibfield  {author} {\bibinfo {author} {\bibfnamefont {William~G.}\
  \bibnamefont {Cook}}, \bibinfo {author} {\bibfnamefont {Ulrich}\ \bibnamefont
  {Sperhake}}, \bibinfo {author} {\bibfnamefont {Emanuele}\ \bibnamefont
  {Berti}}, \ and\ \bibinfo {author} {\bibfnamefont {Vitor}\ \bibnamefont
  {Cardoso}},\ }\href {\doibase10.1103/PhysRevD.96.124006} {\bibfield
  {journal} {\bibinfo  {journal} {Phys. Rev.}\ }\textbf {\bibinfo {volume}
  {D96}},\ \bibinfo {pages} {124006} (\bibinfo {year} {2017})},\ \Eprint
  {http://arxiv.org/abs/1709.10514} {arXiv:1709.10514 [gr-qc]}\BibitemShut
  {NoStop}%
\bibitem [{\citenamefont {Witek}\ \emph {et~al.}(2011)\citenamefont {Witek},
  \citenamefont {Cardoso}, \citenamefont {Gualtieri}, \citenamefont {Herdeiro},
  \citenamefont {Sperhake},\ and\ \citenamefont {Zilhao}}]{Witek:2010az}%
  \BibitemOpen
  \bibfield  {author} {\bibinfo {author} {\bibfnamefont {Helvi}\ \bibnamefont
  {Witek}}, \bibinfo {author} {\bibfnamefont {Vitor}\ \bibnamefont {Cardoso}},
  \bibinfo {author} {\bibfnamefont {Leonardo}\ \bibnamefont {Gualtieri}},
  \bibinfo {author} {\bibfnamefont {Carlos}\ \bibnamefont {Herdeiro}}, \bibinfo
  {author} {\bibfnamefont {Ulrich}\ \bibnamefont {Sperhake}}, \ and\ \bibinfo
  {author} {\bibfnamefont {Miguel}\ \bibnamefont {Zilhao}},\ }\href
  {\doibase10.1103/PhysRevD.83.044017} {\bibfield  {journal} {\bibinfo
  {journal} {Phys. Rev.}\ }\textbf {\bibinfo {volume} {D83}},\ \bibinfo {pages}
  {044017} (\bibinfo {year} {2011})},\ \Eprint {http://arxiv.org/abs/1011.0742}
  {arXiv:1011.0742 [gr-qc]}\BibitemShut {NoStop}%
\bibitem [{\citenamefont {Witek}\ \emph {et~al.}(2014)\citenamefont {Witek},
  \citenamefont {Okawa}, \citenamefont {Cardoso}, \citenamefont {Gualtieri},
  \citenamefont {Herdeiro}, \citenamefont {Shibata}, \citenamefont {Sperhake},\
  and\ \citenamefont {Zilhao}}]{Witek:2014mha}%
  \BibitemOpen
  \bibfield  {author} {\bibinfo {author} {\bibfnamefont {Helvi}\ \bibnamefont
  {Witek}}, \bibinfo {author} {\bibfnamefont {Hirotada}\ \bibnamefont {Okawa}},
  \bibinfo {author} {\bibfnamefont {Vitor}\ \bibnamefont {Cardoso}}, \bibinfo
  {author} {\bibfnamefont {Leonardo}\ \bibnamefont {Gualtieri}}, \bibinfo
  {author} {\bibfnamefont {Carlos}\ \bibnamefont {Herdeiro}}, \bibinfo {author}
  {\bibfnamefont {Masaru}\ \bibnamefont {Shibata}}, \bibinfo {author}
  {\bibfnamefont {Ulrich}\ \bibnamefont {Sperhake}}, \ and\ \bibinfo {author}
  {\bibfnamefont {Miguel}\ \bibnamefont {Zilhao}},\ }\href
  {\doibase10.1103/PhysRevD.90.084014} {\bibfield  {journal} {\bibinfo
  {journal} {Phys. Rev.}\ }\textbf {\bibinfo {volume} {D90}},\ \bibinfo {pages}
  {084014} (\bibinfo {year} {2014})},\ \Eprint {http://arxiv.org/abs/1406.2703}
  {arXiv:1406.2703 [gr-qc]}\BibitemShut {NoStop}%
\bibitem [{\citenamefont {Sperhake}(2013)}]{Sperhake:2013qa}%
  \BibitemOpen
  \bibfield  {author} {\bibinfo {author} {\bibfnamefont {Ulrich}\ \bibnamefont
  {Sperhake}},\ }\href {\doibase10.1142/S021827181330005X} {\bibfield
  {journal} {\bibinfo  {journal} {Int. J. Mod. Phys.}\ }\textbf {\bibinfo
  {volume} {D22}},\ \bibinfo {pages} {1330005} (\bibinfo {year} {2013})},\
  \Eprint {http://arxiv.org/abs/1301.3772} {arXiv:1301.3772
  [gr-qc]}\BibitemShut {NoStop}%
\bibitem [{\citenamefont {Cardoso}\ \emph {et~al.}(2015)\citenamefont
  {Cardoso}, \citenamefont {Gualtieri}, \citenamefont {Herdeiro},\ and\
  \citenamefont {Sperhake}}]{Cardoso:2014uka}%
  \BibitemOpen
  \bibfield  {author} {\bibinfo {author} {\bibfnamefont {Vitor}\ \bibnamefont
  {Cardoso}}, \bibinfo {author} {\bibfnamefont {Leonardo}\ \bibnamefont
  {Gualtieri}}, \bibinfo {author} {\bibfnamefont {Carlos}\ \bibnamefont
  {Herdeiro}}, \ and\ \bibinfo {author} {\bibfnamefont {Ulrich}\ \bibnamefont
  {Sperhake}},\ }\href {\doibase10.1007/lrr-2015-1} {\bibfield  {journal}
  {\bibinfo  {journal} {Living Rev. Relativity}\ }\textbf {\bibinfo {volume}
  {18}},\ \bibinfo {pages} {1} (\bibinfo {year} {2015})},\ \Eprint
  {http://arxiv.org/abs/1409.0014} {arXiv:1409.0014 [gr-qc]}\BibitemShut
  {NoStop}%
\bibitem [{\citenamefont {Emparan}\ and\ \citenamefont
  {Reall}(2008)}]{Emparan:2008eg}%
  \BibitemOpen
  \bibfield  {author} {\bibinfo {author} {\bibfnamefont {Roberto}\ \bibnamefont
  {Emparan}}\ and\ \bibinfo {author} {\bibfnamefont {Harvey~S.}\ \bibnamefont
  {Reall}},\ }\href {\doibase10.12942/lrr-2008-6} {\bibfield  {journal}
  {\bibinfo  {journal} {Living Rev. Rel.}\ }\textbf {\bibinfo {volume} {11}},\
  \bibinfo {pages} {6} (\bibinfo {year} {2008})},\ \Eprint
  {http://arxiv.org/abs/0801.3471} {arXiv:0801.3471 [hep-th]}\BibitemShut
  {NoStop}%
\bibitem [{\citenamefont {Gregory}\ and\ \citenamefont
  {Laflamme}(1993)}]{Gregory:1993vy}%
  \BibitemOpen
  \bibfield  {author} {\bibinfo {author} {\bibfnamefont {R.}~\bibnamefont
  {Gregory}}\ and\ \bibinfo {author} {\bibfnamefont {R.}~\bibnamefont
  {Laflamme}},\ }\href {\doibase10.1103/PhysRevLett.70.2837} {\bibfield
  {journal} {\bibinfo  {journal} {Phys. Rev. Lett.}\ }\textbf {\bibinfo
  {volume} {70}},\ \bibinfo {pages} {2837--2840} (\bibinfo {year} {1993})},\
  \Eprint {http://arxiv.org/abs/hep-th/9301052} {arXiv:hep-th/9301052
  [hep-th]}\BibitemShut {NoStop}%
\bibitem [{\citenamefont {Figueras}\ \emph {et~al.}(2016)\citenamefont
  {Figueras}, \citenamefont {Kunesch},\ and\ \citenamefont
  {Tunyasuvunakool}}]{Figueras:2015hkb}%
  \BibitemOpen
  \bibfield  {author} {\bibinfo {author} {\bibfnamefont {Pau}\ \bibnamefont
  {Figueras}}, \bibinfo {author} {\bibfnamefont {Markus}\ \bibnamefont
  {Kunesch}}, \ and\ \bibinfo {author} {\bibfnamefont {Saran}\ \bibnamefont
  {Tunyasuvunakool}},\ }\href {\doibase10.1103/PhysRevLett.116.071102}
  {\bibfield  {journal} {\bibinfo  {journal} {Phys. Rev. Lett.}\ }\textbf
  {\bibinfo {volume} {116}},\ \bibinfo {pages} {071102} (\bibinfo {year}
  {2016})},\ \Eprint {http://arxiv.org/abs/1512.04532} {arXiv:1512.04532
  [hep-th]}\BibitemShut {NoStop}%
\bibitem [{\citenamefont {Lehner}\ and\ \citenamefont
  {Pretorius}(2010)}]{Lehner:2010pn}%
  \BibitemOpen
  \bibfield  {author} {\bibinfo {author} {\bibfnamefont {Luis}\ \bibnamefont
  {Lehner}}\ and\ \bibinfo {author} {\bibfnamefont {Frans}\ \bibnamefont
  {Pretorius}},\ }\href {\doibase10.1103/PhysRevLett.105.101102} {\bibfield
  {journal} {\bibinfo  {journal} {Phys. Rev. Lett.}\ }\textbf {\bibinfo
  {volume} {105}},\ \bibinfo {pages} {101102} (\bibinfo {year} {2010})},\
  \Eprint {http://arxiv.org/abs/1006.5960} {arXiv:1006.5960
  [hep-th]}\BibitemShut {NoStop}%
\bibitem [{\citenamefont {Shibata}\ and\ \citenamefont
  {Yoshino}(2010)}]{Shibata:2010wz}%
  \BibitemOpen
  \bibfield  {author} {\bibinfo {author} {\bibfnamefont {Masaru}\ \bibnamefont
  {Shibata}}\ and\ \bibinfo {author} {\bibfnamefont {Hirotaka}\ \bibnamefont
  {Yoshino}},\ }\href {\doibase10.1103/PhysRevD.81.104035} {\bibfield
  {journal} {\bibinfo  {journal} {Phys. Rev.}\ }\textbf {\bibinfo {volume}
  {D81}},\ \bibinfo {pages} {104035} (\bibinfo {year} {2010})},\ \Eprint
  {http://arxiv.org/abs/1004.4970} {arXiv:1004.4970 [gr-qc]}\BibitemShut
  {NoStop}%
\bibitem [{\citenamefont {Figueras}\ \emph {et~al.}(2017)\citenamefont
  {Figueras}, \citenamefont {Kunesch}, \citenamefont {Lehner},\ and\
  \citenamefont {Tunyasuvunakool}}]{Figueras:2017zwa}%
  \BibitemOpen
  \bibfield  {author} {\bibinfo {author} {\bibfnamefont {P.}~\bibnamefont
  {Figueras}}, \bibinfo {author} {\bibfnamefont {M.}~\bibnamefont {Kunesch}},
  \bibinfo {author} {\bibfnamefont {L.}~\bibnamefont {Lehner}}, \ and\ \bibinfo
  {author} {\bibfnamefont {S.}~\bibnamefont {Tunyasuvunakool}},\ }\href
  {\doibase10.1103/PhysRevLett.118.151103} {\bibfield  {journal} {\bibinfo
  {journal} {Phys. Rev. Lett.}\ }\textbf {\bibinfo {volume} {118}},\ \bibinfo
  {pages} {151103} (\bibinfo {year} {2017})},\ \Eprint
  {http://arxiv.org/abs/1702.01755} {1702.01755}\BibitemShut {NoStop}%
\bibitem [{\citenamefont {Bizon}\ and\ \citenamefont
  {Rostworowski}(2011)}]{Bizon:2011gg}%
  \BibitemOpen
  \bibfield  {author} {\bibinfo {author} {\bibfnamefont {Piotr}\ \bibnamefont
  {Bizon}}\ and\ \bibinfo {author} {\bibfnamefont {Andrzej}\ \bibnamefont
  {Rostworowski}},\ }\href {\doibase10.1103/PhysRevLett.107.031102} {\bibfield
  {journal} {\bibinfo  {journal} {Phys. Rev. Lett.}\ }\textbf {\bibinfo
  {volume} {107}},\ \bibinfo {pages} {031102} (\bibinfo {year} {2011})},\
  \Eprint {http://arxiv.org/abs/1104.3702} {arXiv:1104.3702
  [gr-qc]}\BibitemShut {NoStop}%
\bibitem [{\citenamefont {Jalmuzna}\ \emph {et~al.}(2011)\citenamefont
  {Jalmuzna}, \citenamefont {Rostworowski},\ and\ \citenamefont
  {Bizon}}]{Jalmuzna:2011qw}%
  \BibitemOpen
  \bibfield  {author} {\bibinfo {author} {\bibfnamefont {Joanna}\ \bibnamefont
  {Jalmuzna}}, \bibinfo {author} {\bibfnamefont {Andrzej}\ \bibnamefont
  {Rostworowski}}, \ and\ \bibinfo {author} {\bibfnamefont {Piotr}\
  \bibnamefont {Bizon}},\ }\href {\doibase10.1103/PhysRevD.84.085021}
  {\bibfield  {journal} {\bibinfo  {journal} {Phys. Rev.}\ }\textbf {\bibinfo
  {volume} {D84}},\ \bibinfo {pages} {085021} (\bibinfo {year} {2011})},\
  \Eprint {http://arxiv.org/abs/1108.4539} {arXiv:1108.4539
  [gr-qc]}\BibitemShut {NoStop}%
\bibitem [{\citenamefont {Buchel}\ \emph {et~al.}(2012)\citenamefont {Buchel},
  \citenamefont {Lehner},\ and\ \citenamefont {Liebling}}]{Buchel:2012uh}%
  \BibitemOpen
  \bibfield  {author} {\bibinfo {author} {\bibfnamefont {Alex}\ \bibnamefont
  {Buchel}}, \bibinfo {author} {\bibfnamefont {Luis}\ \bibnamefont {Lehner}}, \
  and\ \bibinfo {author} {\bibfnamefont {Steven~L.}\ \bibnamefont {Liebling}},\
  }\href {\doibase10.1103/PhysRevD.86.123011} {\bibfield  {journal} {\bibinfo
  {journal} {Phys. Rev.}\ }\textbf {\bibinfo {volume} {D86}},\ \bibinfo {pages}
  {123011} (\bibinfo {year} {2012})},\ \Eprint {http://arxiv.org/abs/1210.0890}
  {arXiv:1210.0890 [gr-qc]}\BibitemShut {NoStop}%
\bibitem [{\citenamefont {Bizoń}\ and\ \citenamefont
  {Jałmużna}(2013)}]{Bizon:2013xha}%
  \BibitemOpen
  \bibfield  {author} {\bibinfo {author} {\bibfnamefont {Piotr}\ \bibnamefont
  {Bizoń}}\ and\ \bibinfo {author} {\bibfnamefont {Joanna}\ \bibnamefont
  {Jałmużna}},\ }\href {\doibase10.1103/PhysRevLett.111.041102} {\bibfield
  {journal} {\bibinfo  {journal} {Phys. Rev. Lett.}\ }\textbf {\bibinfo
  {volume} {111}},\ \bibinfo {pages} {041102} (\bibinfo {year} {2013})},\
  \Eprint {http://arxiv.org/abs/1306.0317} {arXiv:1306.0317
  [gr-qc]}\BibitemShut {NoStop}%
\bibitem [{\citenamefont {Santos-Oliv{\'a}n}\ and\ \citenamefont
  {Sopuerta}(2016)}]{Olivan:2015fmy}%
  \BibitemOpen
  \bibfield  {author} {\bibinfo {author} {\bibfnamefont {D.}~\bibnamefont
  {Santos-Oliv{\'a}n}}\ and\ \bibinfo {author} {\bibfnamefont {C.~F.}\
  \bibnamefont {Sopuerta}},\ }\href {\doibase10.1103/PhysRevLett.116.041101}
  {\bibfield  {journal} {\bibinfo  {journal} {Phys. Rev. Lett.}\ }\textbf
  {\bibinfo {volume} {116}},\ \bibinfo {pages} {041101} (\bibinfo {year}
  {2016})},\ \Eprint {http://arxiv.org/abs/1511.04344} {1511.04344}\BibitemShut
  {NoStop}%
\bibitem [{\citenamefont {Bantilan}\ \emph {et~al.}(2017)\citenamefont
  {Bantilan}, \citenamefont {Figueras}, \citenamefont {Kunesch},\ and\
  \citenamefont {Romatschke}}]{Bantilan:2017kok}%
  \BibitemOpen
  \bibfield  {author} {\bibinfo {author} {\bibfnamefont {Hans}\ \bibnamefont
  {Bantilan}}, \bibinfo {author} {\bibfnamefont {Pau}\ \bibnamefont
  {Figueras}}, \bibinfo {author} {\bibfnamefont {Markus}\ \bibnamefont
  {Kunesch}}, \ and\ \bibinfo {author} {\bibfnamefont {Paul}\ \bibnamefont
  {Romatschke}},\ }\href {\doibase10.1103/PhysRevLett.119.191103} {\bibfield
  {journal} {\bibinfo  {journal} {Phys. Rev. Lett.}\ }\textbf {\bibinfo
  {volume} {119}},\ \bibinfo {pages} {191103} (\bibinfo {year} {2017})},\
  \Eprint {http://arxiv.org/abs/1706.04199} {arXiv:1706.04199
  [hep-th]}\BibitemShut {NoStop}%
\bibitem [{\citenamefont {Bantilan}\ \emph {et~al.}(2012)\citenamefont
  {Bantilan}, \citenamefont {Pretorius},\ and\ \citenamefont
  {Gubser}}]{Bantilan:2012vu}%
  \BibitemOpen
  \bibfield  {author} {\bibinfo {author} {\bibfnamefont {Hans}\ \bibnamefont
  {Bantilan}}, \bibinfo {author} {\bibfnamefont {Frans}\ \bibnamefont
  {Pretorius}}, \ and\ \bibinfo {author} {\bibfnamefont {Steven~S.}\
  \bibnamefont {Gubser}},\ }\href {\doibase10.1103/PhysRevD.85.084038}
  {\bibfield  {journal} {\bibinfo  {journal} {Phys. Rev.}\ }\textbf {\bibinfo
  {volume} {D85}},\ \bibinfo {pages} {084038} (\bibinfo {year} {2012})},\
  \Eprint {http://arxiv.org/abs/1201.2132} {arXiv:1201.2132
  [hep-th]}\BibitemShut {NoStop}%
\bibitem [{\citenamefont {Bantilan}\ and\ \citenamefont
  {Romatschke}(2015)}]{Bantilan:2014sra}%
  \BibitemOpen
  \bibfield  {author} {\bibinfo {author} {\bibfnamefont {Hans}\ \bibnamefont
  {Bantilan}}\ and\ \bibinfo {author} {\bibfnamefont {Paul}\ \bibnamefont
  {Romatschke}},\ }\href {\doibase10.1103/PhysRevLett.114.081601} {\bibfield
  {journal} {\bibinfo  {journal} {Phys. Rev. Lett.}\ }\textbf {\bibinfo
  {volume} {114}},\ \bibinfo {pages} {081601} (\bibinfo {year} {2015})},\
  \Eprint {http://arxiv.org/abs/1410.4799} {arXiv:1410.4799
  [hep-th]}\BibitemShut {NoStop}%
\bibitem [{\citenamefont {Bentivegna}\ and\ \citenamefont
  {Korzynski}(2012)}]{Bentivegna:2012ei}%
  \BibitemOpen
  \bibfield  {author} {\bibinfo {author} {\bibfnamefont {Eloisa}\ \bibnamefont
  {Bentivegna}}\ and\ \bibinfo {author} {\bibfnamefont {Mikolaj}\ \bibnamefont
  {Korzynski}},\ }\href {\doibase10.1088/0264-9381/29/16/165007} {\bibfield
  {journal} {\bibinfo  {journal} {Class. Quant. Grav.}\ }\textbf {\bibinfo
  {volume} {29}},\ \bibinfo {pages} {165007} (\bibinfo {year} {2012})},\
  \Eprint {http://arxiv.org/abs/1204.3568} {arXiv:1204.3568
  [gr-qc]}\BibitemShut {NoStop}%
\bibitem [{\citenamefont {Yoo}\ \emph {et~al.}(2013)\citenamefont {Yoo},
  \citenamefont {Okawa},\ and\ \citenamefont {Nakao}}]{Yoo:2013yea}%
  \BibitemOpen
  \bibfield  {author} {\bibinfo {author} {\bibfnamefont {Chul-Moon}\
  \bibnamefont {Yoo}}, \bibinfo {author} {\bibfnamefont {Hirotada}\
  \bibnamefont {Okawa}}, \ and\ \bibinfo {author} {\bibfnamefont {Ken-ichi}\
  \bibnamefont {Nakao}},\ }\href {\doibase10.1103/PhysRevLett.111.161102}
  {\bibfield  {journal} {\bibinfo  {journal} {Phys. Rev. Lett.}\ }\textbf
  {\bibinfo {volume} {111}},\ \bibinfo {pages} {161102} (\bibinfo {year}
  {2013})},\ \Eprint {http://arxiv.org/abs/1306.1389} {arXiv:1306.1389
  [gr-qc]}\BibitemShut {NoStop}%
\bibitem [{\citenamefont {Yoo}\ and\ \citenamefont
  {Okawa}(2014)}]{Yoo:2014boa}%
  \BibitemOpen
  \bibfield  {author} {\bibinfo {author} {\bibfnamefont {C.-M.}\ \bibnamefont
  {Yoo}}\ and\ \bibinfo {author} {\bibfnamefont {H.}~\bibnamefont {Okawa}},\
  }\href {\doibase10.1103/PhysRevD.89.123502} {\bibfield  {journal} {\bibinfo
  {journal} {Phys. Rev. D}\ }\textbf {\bibinfo {volume} {89}},\ \bibinfo
  {pages} {123502} (\bibinfo {year} {2014})},\ \Eprint
  {http://arxiv.org/abs/1404.1435} {1404.1435}\BibitemShut {NoStop}%
\bibitem [{\citenamefont {Bentivegna}\ and\ \citenamefont
  {Bruni}(2016)}]{Bentivegna:2015flc}%
  \BibitemOpen
  \bibfield  {author} {\bibinfo {author} {\bibfnamefont {E.}~\bibnamefont
  {Bentivegna}}\ and\ \bibinfo {author} {\bibfnamefont {M.}~\bibnamefont
  {Bruni}},\ }\href {\doibase10.1103/PhysRevLett.116.251302} {\bibfield
  {journal} {\bibinfo  {journal} {Phys. Rev. Lett.}\ }\textbf {\bibinfo
  {volume} {116}},\ \bibinfo {pages} {251302} (\bibinfo {year} {2016})},\
  \Eprint {http://arxiv.org/abs/1511.05124} {1511.05124}\BibitemShut {NoStop}%
\bibitem [{\citenamefont {Mertens}\ \emph {et~al.}(2016)\citenamefont
  {Mertens}, \citenamefont {Giblin},\ and\ \citenamefont
  {Starkman}}]{Mertens:2015ttp}%
  \BibitemOpen
  \bibfield  {author} {\bibinfo {author} {\bibfnamefont {J.~B.}\ \bibnamefont
  {Mertens}}, \bibinfo {author} {\bibfnamefont {J.~T.}\ \bibnamefont {Giblin}},
  \ and\ \bibinfo {author} {\bibfnamefont {G.~D.}\ \bibnamefont {Starkman}},\
  }\href {\doibase10.1103/PhysRevD.93.124059} {\bibfield  {journal} {\bibinfo
  {journal} {Phys. Rev. D}\ }\textbf {\bibinfo {volume} {93}},\ \bibinfo
  {pages} {124059} (\bibinfo {year} {2016})},\ \Eprint
  {http://arxiv.org/abs/1511.01106} {1511.01106}\BibitemShut {NoStop}%
\bibitem [{\citenamefont {Clough}\ \emph {et~al.}(2017)\citenamefont {Clough},
  \citenamefont {Lim}, \citenamefont {DiNunno}, \citenamefont {Fischler},
  \citenamefont {Flauger},\ and\ \citenamefont {Paban}}]{Clough:2016ymm}%
  \BibitemOpen
  \bibfield  {author} {\bibinfo {author} {\bibfnamefont {K.}~\bibnamefont
  {Clough}}, \bibinfo {author} {\bibfnamefont {E.~A.}\ \bibnamefont {Lim}},
  \bibinfo {author} {\bibfnamefont {B.~S.}\ \bibnamefont {DiNunno}}, \bibinfo
  {author} {\bibfnamefont {W.}~\bibnamefont {Fischler}}, \bibinfo {author}
  {\bibfnamefont {R.}~\bibnamefont {Flauger}}, \ and\ \bibinfo {author}
  {\bibfnamefont {S.}~\bibnamefont {Paban}},\ }\href
  {\doibase10.1088/1475-7516/2017/09/025} {\bibfield  {journal} {\bibinfo
  {journal} {JCAP}\ }\textbf {\bibinfo {volume} {1709}},\ \bibinfo {pages}
  {025} (\bibinfo {year} {2017})},\ \Eprint {http://arxiv.org/abs/1608.04408}
  {1608.04408}\BibitemShut {NoStop}%
\bibitem [{\citenamefont {Bentivegna}\ \emph {et~al.}(2017)\citenamefont
  {Bentivegna}, \citenamefont {Korzy{\'n}ski}, \citenamefont {Hinder},\ and\
  \citenamefont {Gerlicher}}]{Bentivegna:2016fls}%
  \BibitemOpen
  \bibfield  {author} {\bibinfo {author} {\bibfnamefont {E.}~\bibnamefont
  {Bentivegna}}, \bibinfo {author} {\bibfnamefont {M.}~\bibnamefont
  {Korzy{\'n}ski}}, \bibinfo {author} {\bibfnamefont {I.}~\bibnamefont
  {Hinder}}, \ and\ \bibinfo {author} {\bibfnamefont {D.}~\bibnamefont
  {Gerlicher}},\ }\href {\doibase10.1088/1475-7516/2017/03/014} {\bibfield
  {journal} {\bibinfo  {journal} {JCAP}\ }\textbf {\bibinfo {volume} {1703}},\
  \bibinfo {pages} {014} (\bibinfo {year} {2017})},\ \Eprint
  {http://arxiv.org/abs/1611.09275} {1611.09275}\BibitemShut {NoStop}%
\bibitem [{\citenamefont {Zilhao}\ \emph {et~al.}(2012)\citenamefont {Zilhao},
  \citenamefont {Cardoso}, \citenamefont {Gualtieri}, \citenamefont {Herdeiro},
  \citenamefont {Sperhake},\ and\ \citenamefont {Witek}}]{Zilhao:2012bb}%
  \BibitemOpen
  \bibfield  {author} {\bibinfo {author} {\bibfnamefont {Miguel}\ \bibnamefont
  {Zilhao}}, \bibinfo {author} {\bibfnamefont {Vitor}\ \bibnamefont {Cardoso}},
  \bibinfo {author} {\bibfnamefont {Leonardo}\ \bibnamefont {Gualtieri}},
  \bibinfo {author} {\bibfnamefont {Carlos}\ \bibnamefont {Herdeiro}}, \bibinfo
  {author} {\bibfnamefont {Ulrich}\ \bibnamefont {Sperhake}}, \ and\ \bibinfo
  {author} {\bibfnamefont {Helvi}\ \bibnamefont {Witek}},\ }\href
  {\doibase10.1103/PhysRevD.85.104039} {\bibfield  {journal} {\bibinfo
  {journal} {Phys. Rev.}\ }\textbf {\bibinfo {volume} {D85}},\ \bibinfo {pages}
  {104039} (\bibinfo {year} {2012})},\ \Eprint {http://arxiv.org/abs/1204.2019}
  {arXiv:1204.2019 [gr-qc]}\BibitemShut {NoStop}%
\bibitem [{\citenamefont {Buonanno}\ and\ \citenamefont
  {Sathyaprakash}(2015)}]{Buonanno:2014aza}%
  \BibitemOpen
  \bibfield  {author} {\bibinfo {author} {\bibfnamefont {Alessandra}\
  \bibnamefont {Buonanno}}\ and\ \bibinfo {author} {\bibfnamefont {B.S.}\
  \bibnamefont {Sathyaprakash}},\ }in\ \href@noop {} {\emph {\bibinfo
  {booktitle} {General Relativity and Gravitation: A Centennial
  Perspective}}},\ \bibinfo {editor} {edited by\ \bibinfo {editor}
  {\bibfnamefont {A.}~\bibnamefont {Ashtekar}}, \bibinfo {editor}
  {\bibfnamefont {B.}~\bibnamefont {Berger}}, \bibinfo {editor} {\bibfnamefont
  {J.}~\bibnamefont {Isenberg}}, \ and\ \bibinfo {editor} {\bibfnamefont
  {M.~A.~H.}\ \bibnamefont {MacCallum}}}\ (\bibinfo  {publisher} {Cambridge
  University Press},\ \bibinfo {year} {2015})\ \Eprint
  {http://arxiv.org/abs/1410.7832} {arXiv:1410.7832 [gr-qc]}\BibitemShut
  {NoStop}%
\bibitem [{\citenamefont {Hannam}(2014)}]{Hannam:2013pra}%
  \BibitemOpen
  \bibfield  {author} {\bibinfo {author} {\bibfnamefont {Mark}\ \bibnamefont
  {Hannam}},\ }\href {\doibase10.1007/s10714-014-1767-2} {\bibfield  {journal}
  {\bibinfo  {journal} {Gen. Rel. Grav.}\ }\textbf {\bibinfo {volume} {46}},\
  \bibinfo {pages} {1767} (\bibinfo {year} {2014})},\ \Eprint
  {http://arxiv.org/abs/1312.3641} {arXiv:1312.3641 [gr-qc]}\BibitemShut
  {NoStop}%
\bibitem [{\citenamefont {Damour}(2008)}]{Damour:2008yg}%
  \BibitemOpen
  \bibfield  {author} {\bibinfo {author} {\bibfnamefont {Thibault}\
  \bibnamefont {Damour}},\ }\bibfield  {booktitle} {\emph {\bibinfo {booktitle}
  {{Relativistic field theories. Proceedings, 2nd Stueckelberg Workshop,
  Pescara, Italy, September 3-8, 2007}}},\ }\href
  {\doibase10.1142/S0217751X08039992} {\bibfield  {journal} {\bibinfo
  {journal} {Int. J. Mod. Phys.}\ }\textbf {\bibinfo {volume} {A23}},\ \bibinfo
  {pages} {1130--1148} (\bibinfo {year} {2008})},\ \Eprint
  {http://arxiv.org/abs/0802.4047} {arXiv:0802.4047 [gr-qc]}\BibitemShut
  {NoStop}%
\bibitem [{\citenamefont {Buonanno}\ and\ \citenamefont
  {Damour}(1999)}]{Buonanno:1998gg}%
  \BibitemOpen
  \bibfield  {author} {\bibinfo {author} {\bibfnamefont {A.}~\bibnamefont
  {Buonanno}}\ and\ \bibinfo {author} {\bibfnamefont {T.}~\bibnamefont
  {Damour}},\ }\href {\doibase10.1103/PhysRevD.59.084006} {\bibfield  {journal}
  {\bibinfo  {journal} {Phys. Rev.}\ }\textbf {\bibinfo {volume} {D59}},\
  \bibinfo {pages} {084006} (\bibinfo {year} {1999})},\ \Eprint
  {http://arxiv.org/abs/gr-qc/9811091} {arXiv:gr-qc/9811091
  [gr-qc]}\BibitemShut {NoStop}%
\bibitem [{\citenamefont {Buonanno}\ and\ \citenamefont
  {Damour}(2000)}]{Buonanno:2000ef}%
  \BibitemOpen
  \bibfield  {author} {\bibinfo {author} {\bibfnamefont {Alessandra}\
  \bibnamefont {Buonanno}}\ and\ \bibinfo {author} {\bibfnamefont {Thibault}\
  \bibnamefont {Damour}},\ }\href {\doibase10.1103/PhysRevD.62.064015}
  {\bibfield  {journal} {\bibinfo  {journal} {Phys. Rev.}\ }\textbf {\bibinfo
  {volume} {D62}},\ \bibinfo {pages} {064015} (\bibinfo {year} {2000})},\
  \Eprint {http://arxiv.org/abs/gr-qc/0001013} {arXiv:gr-qc/0001013
  [gr-qc]}\BibitemShut {NoStop}%
\bibitem [{\citenamefont {Maheshwari}\ \emph {et~al.}(1981)\citenamefont
  {Maheshwari}, \citenamefont {Nissimov},\ and\ \citenamefont
  {Todorov}}]{Maheshwari:2016edp}%
  \BibitemOpen
  \bibfield  {author} {\bibinfo {author} {\bibfnamefont {Amar}\ \bibnamefont
  {Maheshwari}}, \bibinfo {author} {\bibfnamefont {Emil}\ \bibnamefont
  {Nissimov}}, \ and\ \bibinfo {author} {\bibfnamefont {Ivan}\ \bibnamefont
  {Todorov}},\ }\href {\doibase10.1007/BF02285306} {\bibfield  {journal}
  {\bibinfo  {journal} {Lett. Math. Phys.}\ }\textbf {\bibinfo {volume} {5}},\
  \bibinfo {pages} {359--366} (\bibinfo {year} {1981})},\ \Eprint
  {http://arxiv.org/abs/1611.02943} {arXiv:1611.02943 [gr-qc]}\BibitemShut
  {NoStop}%
\bibitem [{\citenamefont {Brezin}\ \emph {et~al.}(1970)\citenamefont {Brezin},
  \citenamefont {Itzykson},\ and\ \citenamefont {Zinn-Justin}}]{Brezin:1970zr}%
  \BibitemOpen
  \bibfield  {author} {\bibinfo {author} {\bibfnamefont {E.}~\bibnamefont
  {Brezin}}, \bibinfo {author} {\bibfnamefont {C.}~\bibnamefont {Itzykson}}, \
  and\ \bibinfo {author} {\bibfnamefont {Jean}\ \bibnamefont {Zinn-Justin}},\
  }\href {\doibase10.1103/PhysRevD.1.2349} {\bibfield  {journal} {\bibinfo
  {journal} {Phys. Rev.}\ }\textbf {\bibinfo {volume} {D1}},\ \bibinfo {pages}
  {2349--2355} (\bibinfo {year} {1970})}\BibitemShut {NoStop}%
\bibitem [{\citenamefont {Barausse}\ \emph {et~al.}(2009)\citenamefont
  {Barausse}, \citenamefont {Racine},\ and\ \citenamefont
  {Buonanno}}]{Barausse:2009aa}%
  \BibitemOpen
  \bibfield  {author} {\bibinfo {author} {\bibfnamefont {Enrico}\ \bibnamefont
  {Barausse}}, \bibinfo {author} {\bibfnamefont {Etienne}\ \bibnamefont
  {Racine}}, \ and\ \bibinfo {author} {\bibfnamefont {Alessandra}\ \bibnamefont
  {Buonanno}},\ }\href {\doibase10.1103/PhysRevD.80.104025} {\bibfield
  {journal} {\bibinfo  {journal} {Phys. Rev.}\ }\textbf {\bibinfo {volume}
  {D80}},\ \bibinfo {pages} {104025} (\bibinfo {year} {2009})},\ \bibinfo
  {note} {[Erratum: Phys. Rev.D85,069904(2012)]},\ \Eprint
  {http://arxiv.org/abs/0907.4745} {arXiv:0907.4745 [gr-qc]}\BibitemShut
  {NoStop}%
\bibitem [{\citenamefont {Barausse}\ and\ \citenamefont
  {Buonanno}(2010)}]{Barausse:2009xi}%
  \BibitemOpen
  \bibfield  {author} {\bibinfo {author} {\bibfnamefont {Enrico}\ \bibnamefont
  {Barausse}}\ and\ \bibinfo {author} {\bibfnamefont {Alessandra}\ \bibnamefont
  {Buonanno}},\ }\href {\doibase10.1103/PhysRevD.81.084024} {\bibfield
  {journal} {\bibinfo  {journal} {Phys. Rev.}\ }\textbf {\bibinfo {volume}
  {D81}},\ \bibinfo {pages} {084024} (\bibinfo {year} {2010})},\ \Eprint
  {http://arxiv.org/abs/0912.3517} {arXiv:0912.3517 [gr-qc]}\BibitemShut
  {NoStop}%
\bibitem [{\citenamefont {Barausse}\ and\ \citenamefont
  {Buonanno}(2011)}]{Barausse:2011ys}%
  \BibitemOpen
  \bibfield  {author} {\bibinfo {author} {\bibfnamefont {Enrico}\ \bibnamefont
  {Barausse}}\ and\ \bibinfo {author} {\bibfnamefont {Alessandra}\ \bibnamefont
  {Buonanno}},\ }\href {\doibase10.1103/PhysRevD.84.104027} {\bibfield
  {journal} {\bibinfo  {journal} {Phys. Rev.}\ }\textbf {\bibinfo {volume}
  {D84}},\ \bibinfo {pages} {104027} (\bibinfo {year} {2011})},\ \Eprint
  {http://arxiv.org/abs/1107.2904} {arXiv:1107.2904 [gr-qc]}\BibitemShut
  {NoStop}%
\bibitem [{\citenamefont {Taracchini}\ \emph {et~al.}(2014)\citenamefont
  {Taracchini} \emph {et~al.}}]{Taracchini:2013rva}%
  \BibitemOpen
  \bibfield  {author} {\bibinfo {author} {\bibfnamefont {Andrea}\ \bibnamefont
  {Taracchini}} \emph {et~al.},\ }\href {\doibase10.1103/PhysRevD.89.061502}
  {\bibfield  {journal} {\bibinfo  {journal} {Phys. Rev.}\ }\textbf {\bibinfo
  {volume} {D89}},\ \bibinfo {pages} {061502} (\bibinfo {year} {2014})},\
  \Eprint {http://arxiv.org/abs/1311.2544} {arXiv:1311.2544
  [gr-qc]}\BibitemShut {NoStop}%
\bibitem [{\citenamefont {Damour}(2001)}]{Damour:2001tu}%
  \BibitemOpen
  \bibfield  {author} {\bibinfo {author} {\bibfnamefont {Thibault}\
  \bibnamefont {Damour}},\ }\href {\doibase10.1103/PhysRevD.64.124013}
  {\bibfield  {journal} {\bibinfo  {journal} {Phys. Rev.}\ }\textbf {\bibinfo
  {volume} {D64}},\ \bibinfo {pages} {124013} (\bibinfo {year} {2001})},\
  \Eprint {http://arxiv.org/abs/gr-qc/0103018} {arXiv:gr-qc/0103018
  [gr-qc]}\BibitemShut {NoStop}%
\bibitem [{\citenamefont {Damour}\ \emph {et~al.}(2008)\citenamefont {Damour},
  \citenamefont {Jaranowski},\ and\ \citenamefont {Schaefer}}]{Damour:2008qf}%
  \BibitemOpen
  \bibfield  {author} {\bibinfo {author} {\bibfnamefont {Thibault}\
  \bibnamefont {Damour}}, \bibinfo {author} {\bibfnamefont {Piotr}\
  \bibnamefont {Jaranowski}}, \ and\ \bibinfo {author} {\bibfnamefont
  {Gerhard}\ \bibnamefont {Schaefer}},\ }\href
  {\doibase10.1103/PhysRevD.78.024009} {\bibfield  {journal} {\bibinfo
  {journal} {Phys. Rev.}\ }\textbf {\bibinfo {volume} {D78}},\ \bibinfo {pages}
  {024009} (\bibinfo {year} {2008})},\ \Eprint {http://arxiv.org/abs/0803.0915}
  {arXiv:0803.0915 [gr-qc]}\BibitemShut {NoStop}%
\bibitem [{\citenamefont {Nagar}(2011)}]{Nagar:2011fx}%
  \BibitemOpen
  \bibfield  {author} {\bibinfo {author} {\bibfnamefont {Alessandro}\
  \bibnamefont {Nagar}},\ }\href {\doibase10.1103/PhysRevD.84.084028}
  {\bibfield  {journal} {\bibinfo  {journal} {Phys. Rev.}\ }\textbf {\bibinfo
  {volume} {D84}},\ \bibinfo {pages} {084028} (\bibinfo {year} {2011})},\
  \bibinfo {note} {[Erratum: Phys. Rev.D88,no.8,089901(2013)]},\ \Eprint
  {http://arxiv.org/abs/1106.4349} {arXiv:1106.4349 [gr-qc]}\BibitemShut
  {NoStop}%
\bibitem [{\citenamefont {Balmelli}\ and\ \citenamefont
  {Jetzer}(2015)}]{Balmelli:2015lva}%
  \BibitemOpen
  \bibfield  {author} {\bibinfo {author} {\bibfnamefont {Simone}\ \bibnamefont
  {Balmelli}}\ and\ \bibinfo {author} {\bibfnamefont {Philippe}\ \bibnamefont
  {Jetzer}},\ }\href {\doibase10.1103/PhysRevD.91.064011} {\bibfield  {journal}
  {\bibinfo  {journal} {Phys. Rev.}\ }\textbf {\bibinfo {volume} {D91}},\
  \bibinfo {pages} {064011} (\bibinfo {year} {2015})},\ \Eprint
  {http://arxiv.org/abs/1502.01343} {arXiv:1502.01343 [gr-qc]}\BibitemShut
  {NoStop}%
\bibitem [{\citenamefont {Damour}\ and\ \citenamefont
  {Nagar}(2014{\natexlab{a}})}]{Damour:2014sva}%
  \BibitemOpen
  \bibfield  {author} {\bibinfo {author} {\bibfnamefont {Thibault}\
  \bibnamefont {Damour}}\ and\ \bibinfo {author} {\bibfnamefont {Alessandro}\
  \bibnamefont {Nagar}},\ }\href {\doibase10.1103/PhysRevD.90.044018}
  {\bibfield  {journal} {\bibinfo  {journal} {Phys. Rev.}\ }\textbf {\bibinfo
  {volume} {D90}},\ \bibinfo {pages} {044018} (\bibinfo {year}
  {2014}{\natexlab{a}})},\ \Eprint {http://arxiv.org/abs/1406.6913}
  {arXiv:1406.6913 [gr-qc]}\BibitemShut {NoStop}%
\bibitem [{\citenamefont {Damour}\ and\ \citenamefont
  {Nagar}(2007)}]{Damour:2007xr}%
  \BibitemOpen
  \bibfield  {author} {\bibinfo {author} {\bibfnamefont {Thibault}\
  \bibnamefont {Damour}}\ and\ \bibinfo {author} {\bibfnamefont {Alessandro}\
  \bibnamefont {Nagar}},\ }\href {\doibase10.1103/PhysRevD.76.064028}
  {\bibfield  {journal} {\bibinfo  {journal} {Phys. Rev.}\ }\textbf {\bibinfo
  {volume} {D76}},\ \bibinfo {pages} {064028} (\bibinfo {year} {2007})},\
  \Eprint {http://arxiv.org/abs/0705.2519} {arXiv:0705.2519
  [gr-qc]}\BibitemShut {NoStop}%
\bibitem [{\citenamefont {Damour}\ \emph {et~al.}(2009)\citenamefont {Damour},
  \citenamefont {Iyer},\ and\ \citenamefont {Nagar}}]{Damour:2008gu}%
  \BibitemOpen
  \bibfield  {author} {\bibinfo {author} {\bibfnamefont {Thibault}\
  \bibnamefont {Damour}}, \bibinfo {author} {\bibfnamefont {Bala~R.}\
  \bibnamefont {Iyer}}, \ and\ \bibinfo {author} {\bibfnamefont {Alessandro}\
  \bibnamefont {Nagar}},\ }\href {\doibase10.1103/PhysRevD.79.064004}
  {\bibfield  {journal} {\bibinfo  {journal} {Phys. Rev.}\ }\textbf {\bibinfo
  {volume} {D79}},\ \bibinfo {pages} {064004} (\bibinfo {year} {2009})},\
  \Eprint {http://arxiv.org/abs/0811.2069} {arXiv:0811.2069
  [gr-qc]}\BibitemShut {NoStop}%
\bibitem [{\citenamefont {Damour}\ \emph {et~al.}(2003)\citenamefont {Damour},
  \citenamefont {Iyer}, \citenamefont {Jaranowski},\ and\ \citenamefont
  {Sathyaprakash}}]{Damour:2002vi}%
  \BibitemOpen
  \bibfield  {author} {\bibinfo {author} {\bibfnamefont {Thibault}\
  \bibnamefont {Damour}}, \bibinfo {author} {\bibfnamefont {Bala~R.}\
  \bibnamefont {Iyer}}, \bibinfo {author} {\bibfnamefont {Piotr}\ \bibnamefont
  {Jaranowski}}, \ and\ \bibinfo {author} {\bibfnamefont {B.~S.}\ \bibnamefont
  {Sathyaprakash}},\ }\href {\doibase10.1103/PhysRevD.67.064028} {\bibfield
  {journal} {\bibinfo  {journal} {Phys. Rev.}\ }\textbf {\bibinfo {volume}
  {D67}},\ \bibinfo {pages} {064028} (\bibinfo {year} {2003})},\ \Eprint
  {http://arxiv.org/abs/gr-qc/0211041} {arXiv:gr-qc/0211041
  [gr-qc]}\BibitemShut {NoStop}%
\bibitem [{\citenamefont {Taracchini}\ \emph {et~al.}(2012)\citenamefont
  {Taracchini}, \citenamefont {Pan}, \citenamefont {Buonanno}, \citenamefont
  {Barausse}, \citenamefont {Boyle}, \citenamefont {Chu}, \citenamefont
  {Lovelace}, \citenamefont {Pfeiffer},\ and\ \citenamefont
  {Scheel}}]{Taracchini:2012ig}%
  \BibitemOpen
  \bibfield  {author} {\bibinfo {author} {\bibfnamefont {Andrea}\ \bibnamefont
  {Taracchini}}, \bibinfo {author} {\bibfnamefont {Yi}~\bibnamefont {Pan}},
  \bibinfo {author} {\bibfnamefont {Alessandra}\ \bibnamefont {Buonanno}},
  \bibinfo {author} {\bibfnamefont {Enrico}\ \bibnamefont {Barausse}}, \bibinfo
  {author} {\bibfnamefont {Michael}\ \bibnamefont {Boyle}}, \bibinfo {author}
  {\bibfnamefont {Tony}\ \bibnamefont {Chu}}, \bibinfo {author} {\bibfnamefont
  {Geoffrey}\ \bibnamefont {Lovelace}}, \bibinfo {author} {\bibfnamefont
  {Harald~P.}\ \bibnamefont {Pfeiffer}}, \ and\ \bibinfo {author}
  {\bibfnamefont {Mark~A.}\ \bibnamefont {Scheel}},\ }\href
  {\doibase10.1103/PhysRevD.86.024011} {\bibfield  {journal} {\bibinfo
  {journal} {Phys. Rev.}\ }\textbf {\bibinfo {volume} {D86}},\ \bibinfo {pages}
  {024011} (\bibinfo {year} {2012})},\ \Eprint {http://arxiv.org/abs/1202.0790}
  {arXiv:1202.0790 [gr-qc]}\BibitemShut {NoStop}%
\bibitem [{\citenamefont {Damour}\ \emph {et~al.}(2013)\citenamefont {Damour},
  \citenamefont {Nagar},\ and\ \citenamefont {Bernuzzi}}]{Damour:2012ky}%
  \BibitemOpen
  \bibfield  {author} {\bibinfo {author} {\bibfnamefont {Thibault}\
  \bibnamefont {Damour}}, \bibinfo {author} {\bibfnamefont {Alessandro}\
  \bibnamefont {Nagar}}, \ and\ \bibinfo {author} {\bibfnamefont {Sebastiano}\
  \bibnamefont {Bernuzzi}},\ }\href {\doibase10.1103/PhysRevD.87.084035}
  {\bibfield  {journal} {\bibinfo  {journal} {Phys. Rev.}\ }\textbf {\bibinfo
  {volume} {D87}},\ \bibinfo {pages} {084035} (\bibinfo {year} {2013})},\
  \Eprint {http://arxiv.org/abs/1212.4357} {arXiv:1212.4357
  [gr-qc]}\BibitemShut {NoStop}%
\bibitem [{\citenamefont {Price}\ and\ \citenamefont
  {Pullin}(1994)}]{Price:1994pm}%
  \BibitemOpen
  \bibfield  {author} {\bibinfo {author} {\bibfnamefont {Richard~H.}\
  \bibnamefont {Price}}\ and\ \bibinfo {author} {\bibfnamefont {Jorge}\
  \bibnamefont {Pullin}},\ }\href {\doibase10.1103/PhysRevLett.72.3297}
  {\bibfield  {journal} {\bibinfo  {journal} {Phys. Rev. Lett.}\ }\textbf
  {\bibinfo {volume} {72}},\ \bibinfo {pages} {3297--3300} (\bibinfo {year}
  {1994})},\ \Eprint {http://arxiv.org/abs/gr-qc/9402039} {arXiv:gr-qc/9402039
  [gr-qc]}\BibitemShut {NoStop}%
\bibitem [{\citenamefont {Baker}\ \emph {et~al.}(2008)\citenamefont {Baker},
  \citenamefont {Boggs}, \citenamefont {Centrella}, \citenamefont {Kelly},
  \citenamefont {McWilliams},\ and\ \citenamefont {van Meter}}]{Baker:2008mj}%
  \BibitemOpen
  \bibfield  {author} {\bibinfo {author} {\bibfnamefont {John~G.}\ \bibnamefont
  {Baker}}, \bibinfo {author} {\bibfnamefont {William~D.}\ \bibnamefont
  {Boggs}}, \bibinfo {author} {\bibfnamefont {Joan}\ \bibnamefont {Centrella}},
  \bibinfo {author} {\bibfnamefont {Bernard~J.}\ \bibnamefont {Kelly}},
  \bibinfo {author} {\bibfnamefont {Sean~T.}\ \bibnamefont {McWilliams}}, \
  and\ \bibinfo {author} {\bibfnamefont {James~R.}\ \bibnamefont {van Meter}},\
  }\href {\doibase10.1103/PhysRevD.78.044046} {\bibfield  {journal} {\bibinfo
  {journal} {Phys. Rev.}\ }\textbf {\bibinfo {volume} {D78}},\ \bibinfo {pages}
  {044046} (\bibinfo {year} {2008})},\ \Eprint {http://arxiv.org/abs/0805.1428}
  {arXiv:0805.1428 [gr-qc]}\BibitemShut {NoStop}%
\bibitem [{\citenamefont {Damour}\ and\ \citenamefont
  {Nagar}(2014{\natexlab{b}})}]{Damour:2014yha}%
  \BibitemOpen
  \bibfield  {author} {\bibinfo {author} {\bibfnamefont {Thibault}\
  \bibnamefont {Damour}}\ and\ \bibinfo {author} {\bibfnamefont {Alessandro}\
  \bibnamefont {Nagar}},\ }\href {\doibase10.1103/PhysRevD.90.024054}
  {\bibfield  {journal} {\bibinfo  {journal} {Phys. Rev.}\ }\textbf {\bibinfo
  {volume} {D90}},\ \bibinfo {pages} {024054} (\bibinfo {year}
  {2014}{\natexlab{b}})},\ \Eprint {http://arxiv.org/abs/1406.0401}
  {arXiv:1406.0401 [gr-qc]}\BibitemShut {NoStop}%
\bibitem [{\citenamefont {Pan}\ \emph {et~al.}(2011{\natexlab{a}})\citenamefont
  {Pan}, \citenamefont {Buonanno}, \citenamefont {Fujita}, \citenamefont
  {Racine},\ and\ \citenamefont {Tagoshi}}]{Pan:2010hz}%
  \BibitemOpen
  \bibfield  {author} {\bibinfo {author} {\bibfnamefont {Yi}~\bibnamefont
  {Pan}}, \bibinfo {author} {\bibfnamefont {Alessandra}\ \bibnamefont
  {Buonanno}}, \bibinfo {author} {\bibfnamefont {Ryuichi}\ \bibnamefont
  {Fujita}}, \bibinfo {author} {\bibfnamefont {Etienne}\ \bibnamefont
  {Racine}}, \ and\ \bibinfo {author} {\bibfnamefont {Hideyuki}\ \bibnamefont
  {Tagoshi}},\ }\href {\doibase10.1103/PhysRevD.83.064003} {\bibfield
  {journal} {\bibinfo  {journal} {Phys. Rev.}\ }\textbf {\bibinfo {volume}
  {D83}},\ \bibinfo {pages} {064003} (\bibinfo {year} {2011}{\natexlab{a}})},\
  \bibinfo {note} {[Erratum: Phys. Rev.D87,no.10,109901(2013)]},\ \Eprint
  {http://arxiv.org/abs/1006.0431} {arXiv:1006.0431 [gr-qc]}\BibitemShut
  {NoStop}%
\bibitem [{\citenamefont {Babak}\ \emph
  {et~al.}(2017{\natexlab{b}})\citenamefont {Babak}, \citenamefont
  {Taracchini},\ and\ \citenamefont {Buonanno}}]{Babak:2016tgq}%
  \BibitemOpen
  \bibfield  {author} {\bibinfo {author} {\bibfnamefont {Stanislav}\
  \bibnamefont {Babak}}, \bibinfo {author} {\bibfnamefont {Andrea}\
  \bibnamefont {Taracchini}}, \ and\ \bibinfo {author} {\bibfnamefont
  {Alessandra}\ \bibnamefont {Buonanno}},\ }\href
  {\doibase10.1103/PhysRevD.95.024010} {\bibfield  {journal} {\bibinfo
  {journal} {Phys. Rev.}\ }\textbf {\bibinfo {volume} {D95}},\ \bibinfo {pages}
  {024010} (\bibinfo {year} {2017}{\natexlab{b}})},\ \Eprint
  {http://arxiv.org/abs/1607.05661} {arXiv:1607.05661 [gr-qc]}\BibitemShut
  {NoStop}%
\bibitem [{\citenamefont {Nagar}\ \emph {et~al.}(2018)\citenamefont {Nagar}
  \emph {et~al.}}]{Nagar:2018zoe}%
  \BibitemOpen
  \bibfield  {author} {\bibinfo {author} {\bibfnamefont {Alessandro}\
  \bibnamefont {Nagar}} \emph {et~al.},\ }\href@noop {} {\  (\bibinfo {year}
  {2018})},\ \Eprint {http://arxiv.org/abs/1806.01772} {arXiv:1806.01772
  [gr-qc]}\BibitemShut {NoStop}%
\bibitem [{\citenamefont {Damour}\ and\ \citenamefont
  {Nagar}(2010)}]{Damour:2009wj}%
  \BibitemOpen
  \bibfield  {author} {\bibinfo {author} {\bibfnamefont {Thibault}\
  \bibnamefont {Damour}}\ and\ \bibinfo {author} {\bibfnamefont {Alessandro}\
  \bibnamefont {Nagar}},\ }\href {\doibase10.1103/PhysRevD.81.084016}
  {\bibfield  {journal} {\bibinfo  {journal} {Phys. Rev.}\ }\textbf {\bibinfo
  {volume} {D81}},\ \bibinfo {pages} {084016} (\bibinfo {year} {2010})},\
  \Eprint {http://arxiv.org/abs/0911.5041} {arXiv:0911.5041
  [gr-qc]}\BibitemShut {NoStop}%
\bibitem [{\citenamefont {Bini}\ \emph {et~al.}(2012)\citenamefont {Bini},
  \citenamefont {Damour},\ and\ \citenamefont {Faye}}]{Bini:2012gu}%
  \BibitemOpen
  \bibfield  {author} {\bibinfo {author} {\bibfnamefont {Donato}\ \bibnamefont
  {Bini}}, \bibinfo {author} {\bibfnamefont {Thibault}\ \bibnamefont {Damour}},
  \ and\ \bibinfo {author} {\bibfnamefont {Guillaume}\ \bibnamefont {Faye}},\
  }\href {\doibase10.1103/PhysRevD.85.124034} {\bibfield  {journal} {\bibinfo
  {journal} {Phys. Rev.}\ }\textbf {\bibinfo {volume} {D85}},\ \bibinfo {pages}
  {124034} (\bibinfo {year} {2012})},\ \Eprint {http://arxiv.org/abs/1202.3565}
  {arXiv:1202.3565 [gr-qc]}\BibitemShut {NoStop}%
\bibitem [{\citenamefont {Steinhoff}\ \emph {et~al.}(2016)\citenamefont
  {Steinhoff}, \citenamefont {Hinderer}, \citenamefont {Buonanno},\ and\
  \citenamefont {Taracchini}}]{Steinhoff:2016rfi}%
  \BibitemOpen
  \bibfield  {author} {\bibinfo {author} {\bibfnamefont {Jan}\ \bibnamefont
  {Steinhoff}}, \bibinfo {author} {\bibfnamefont {Tanja}\ \bibnamefont
  {Hinderer}}, \bibinfo {author} {\bibfnamefont {Alessandra}\ \bibnamefont
  {Buonanno}}, \ and\ \bibinfo {author} {\bibfnamefont {Andrea}\ \bibnamefont
  {Taracchini}},\ }\href {\doibase10.1103/PhysRevD.94.104028} {\bibfield
  {journal} {\bibinfo  {journal} {Phys. Rev.}\ }\textbf {\bibinfo {volume}
  {D94}},\ \bibinfo {pages} {104028} (\bibinfo {year} {2016})},\ \Eprint
  {http://arxiv.org/abs/1608.01907} {arXiv:1608.01907 [gr-qc]}\BibitemShut
  {NoStop}%
\bibitem [{\citenamefont {Ajith}\ \emph {et~al.}(2007)\citenamefont {Ajith}
  \emph {et~al.}}]{Ajith:2007qp}%
  \BibitemOpen
  \bibfield  {author} {\bibinfo {author} {\bibfnamefont {Parameswaran}\
  \bibnamefont {Ajith}} \emph {et~al.},\ }\bibfield  {booktitle} {\emph
  {\bibinfo {booktitle} {{Gravitational wave data analysis. Proceedings: 11th
  Workshop, GWDAW-11, Potsdam, Germany, Dec 18-21, 2006}}},\ }\href
  {\doibase10.1088/0264-9381/24/19/S31} {\bibfield  {journal} {\bibinfo
  {journal} {Class. Quant. Grav.}\ }\textbf {\bibinfo {volume} {24}},\ \bibinfo
  {pages} {S689--S700} (\bibinfo {year} {2007})},\ \Eprint
  {http://arxiv.org/abs/0704.3764} {arXiv:0704.3764 [gr-qc]}\BibitemShut
  {NoStop}%
\bibitem [{\citenamefont {Ajith}\ \emph {et~al.}(2008)\citenamefont {Ajith}
  \emph {et~al.}}]{Ajith:2007kx}%
  \BibitemOpen
  \bibfield  {author} {\bibinfo {author} {\bibfnamefont {P.}~\bibnamefont
  {Ajith}} \emph {et~al.},\ }\href {\doibase10.1103/PhysRevD.77.104017}
  {\bibfield  {journal} {\bibinfo  {journal} {Phys. Rev.}\ }\textbf {\bibinfo
  {volume} {D77}},\ \bibinfo {pages} {104017} (\bibinfo {year} {2008})},\
  \bibinfo {note} {[Erratum: Phys. Rev.D79,129901(2009)]},\ \Eprint
  {http://arxiv.org/abs/0710.2335} {arXiv:0710.2335 [gr-qc]}\BibitemShut
  {NoStop}%
\bibitem [{\citenamefont {Ajith}\ \emph {et~al.}(2011)\citenamefont {Ajith}
  \emph {et~al.}}]{Ajith:2009bn}%
  \BibitemOpen
  \bibfield  {author} {\bibinfo {author} {\bibfnamefont {P.}~\bibnamefont
  {Ajith}} \emph {et~al.},\ }\href {\doibase10.1103/PhysRevLett.106.241101}
  {\bibfield  {journal} {\bibinfo  {journal} {Phys. Rev. Lett.}\ }\textbf
  {\bibinfo {volume} {106}},\ \bibinfo {pages} {241101} (\bibinfo {year}
  {2011})},\ \Eprint {http://arxiv.org/abs/0909.2867} {arXiv:0909.2867
  [gr-qc]}\BibitemShut {NoStop}%
\bibitem [{\citenamefont {Santamaria}\ \emph {et~al.}(2010)\citenamefont
  {Santamaria} \emph {et~al.}}]{Santamaria:2010yb}%
  \BibitemOpen
  \bibfield  {author} {\bibinfo {author} {\bibfnamefont {L.}~\bibnamefont
  {Santamaria}} \emph {et~al.},\ }\href {\doibase10.1103/PhysRevD.82.064016}
  {\bibfield  {journal} {\bibinfo  {journal} {Phys. Rev.}\ }\textbf {\bibinfo
  {volume} {D82}},\ \bibinfo {pages} {064016} (\bibinfo {year} {2010})},\
  \Eprint {http://arxiv.org/abs/1005.3306} {arXiv:1005.3306
  [gr-qc]}\BibitemShut {NoStop}%
\bibitem [{\citenamefont {Schmidt}\ \emph {et~al.}(2012)\citenamefont
  {Schmidt}, \citenamefont {Hannam},\ and\ \citenamefont
  {Husa}}]{Schmidt:2012rh}%
  \BibitemOpen
  \bibfield  {author} {\bibinfo {author} {\bibfnamefont {Patricia}\
  \bibnamefont {Schmidt}}, \bibinfo {author} {\bibfnamefont {Mark}\
  \bibnamefont {Hannam}}, \ and\ \bibinfo {author} {\bibfnamefont {Sascha}\
  \bibnamefont {Husa}},\ }\href {\doibase10.1103/PhysRevD.86.104063} {\bibfield
   {journal} {\bibinfo  {journal} {Phys. Rev.}\ }\textbf {\bibinfo {volume}
  {D86}},\ \bibinfo {pages} {104063} (\bibinfo {year} {2012})},\ \Eprint
  {http://arxiv.org/abs/1207.3088} {arXiv:1207.3088 [gr-qc]}\BibitemShut
  {NoStop}%
\bibitem [{\citenamefont {Schmidt}\ \emph {et~al.}(2015)\citenamefont
  {Schmidt}, \citenamefont {Ohme},\ and\ \citenamefont
  {Hannam}}]{Schmidt:2014iyl}%
  \BibitemOpen
  \bibfield  {author} {\bibinfo {author} {\bibfnamefont {Patricia}\
  \bibnamefont {Schmidt}}, \bibinfo {author} {\bibfnamefont {Frank}\
  \bibnamefont {Ohme}}, \ and\ \bibinfo {author} {\bibfnamefont {Mark}\
  \bibnamefont {Hannam}},\ }\href {\doibase10.1103/PhysRevD.91.024043}
  {\bibfield  {journal} {\bibinfo  {journal} {Phys. Rev.}\ }\textbf {\bibinfo
  {volume} {D91}},\ \bibinfo {pages} {024043} (\bibinfo {year} {2015})},\
  \Eprint {http://arxiv.org/abs/1408.1810} {arXiv:1408.1810
  [gr-qc]}\BibitemShut {NoStop}%
\bibitem [{\citenamefont {Krishnendu}\ \emph {et~al.}(2017)\citenamefont
  {Krishnendu}, \citenamefont {Arun},\ and\ \citenamefont
  {Mishra}}]{Krishnendu:2017shb}%
  \BibitemOpen
  \bibfield  {author} {\bibinfo {author} {\bibfnamefont {N.~V.}\ \bibnamefont
  {Krishnendu}}, \bibinfo {author} {\bibfnamefont {K.~G.}\ \bibnamefont
  {Arun}}, \ and\ \bibinfo {author} {\bibfnamefont {Chandra~Kant}\ \bibnamefont
  {Mishra}},\ }\href {\doibase10.1103/PhysRevLett.119.091101} {\bibfield
  {journal} {\bibinfo  {journal} {Phys. Rev. Lett.}\ }\textbf {\bibinfo
  {volume} {119}},\ \bibinfo {pages} {091101} (\bibinfo {year} {2017})},\
  \Eprint {http://arxiv.org/abs/1701.06318} {arXiv:1701.06318
  [gr-qc]}\BibitemShut {NoStop}%
\bibitem [{\citenamefont {Vines}\ \emph {et~al.}(2011)\citenamefont {Vines},
  \citenamefont {Flanagan},\ and\ \citenamefont {Hinderer}}]{Vines:2011ud}%
  \BibitemOpen
  \bibfield  {author} {\bibinfo {author} {\bibfnamefont {Justin}\ \bibnamefont
  {Vines}}, \bibinfo {author} {\bibfnamefont {Eanna~E.}\ \bibnamefont
  {Flanagan}}, \ and\ \bibinfo {author} {\bibfnamefont {Tanja}\ \bibnamefont
  {Hinderer}},\ }\href {\doibase10.1103/PhysRevD.83.084051} {\bibfield
  {journal} {\bibinfo  {journal} {Phys. Rev.}\ }\textbf {\bibinfo {volume}
  {D83}},\ \bibinfo {pages} {084051} (\bibinfo {year} {2011})},\ \Eprint
  {http://arxiv.org/abs/1101.1673} {arXiv:1101.1673 [gr-qc]}\BibitemShut
  {NoStop}%
\bibitem [{\citenamefont {Damour}\ \emph {et~al.}(2012)\citenamefont {Damour},
  \citenamefont {Nagar},\ and\ \citenamefont {Villain}}]{Damour:2012yf}%
  \BibitemOpen
  \bibfield  {author} {\bibinfo {author} {\bibfnamefont {Thibault}\
  \bibnamefont {Damour}}, \bibinfo {author} {\bibfnamefont {Alessandro}\
  \bibnamefont {Nagar}}, \ and\ \bibinfo {author} {\bibfnamefont {Loic}\
  \bibnamefont {Villain}},\ }\href {\doibase10.1103/PhysRevD.85.123007}
  {\bibfield  {journal} {\bibinfo  {journal} {Phys. Rev.}\ }\textbf {\bibinfo
  {volume} {D85}},\ \bibinfo {pages} {123007} (\bibinfo {year} {2012})},\
  \Eprint {http://arxiv.org/abs/1203.4352} {arXiv:1203.4352
  [gr-qc]}\BibitemShut {NoStop}%
\bibitem [{\citenamefont {Dietrich}\ \emph {et~al.}(2017)\citenamefont
  {Dietrich}, \citenamefont {Bernuzzi},\ and\ \citenamefont
  {Tichy}}]{Dietrich:2017aum}%
  \BibitemOpen
  \bibfield  {author} {\bibinfo {author} {\bibfnamefont {Tim}\ \bibnamefont
  {Dietrich}}, \bibinfo {author} {\bibfnamefont {Sebastiano}\ \bibnamefont
  {Bernuzzi}}, \ and\ \bibinfo {author} {\bibfnamefont {Wolfgang}\ \bibnamefont
  {Tichy}},\ }\href {\doibase10.1103/PhysRevD.96.121501} {\bibfield  {journal}
  {\bibinfo  {journal} {Phys. Rev.}\ }\textbf {\bibinfo {volume} {D96}},\
  \bibinfo {pages} {121501} (\bibinfo {year} {2017})},\ \Eprint
  {http://arxiv.org/abs/1706.02969} {arXiv:1706.02969 [gr-qc]}\BibitemShut
  {NoStop}%
\bibitem [{\citenamefont {Dietrich}\ \emph {et~al.}(2018)\citenamefont
  {Dietrich} \emph {et~al.}}]{Dietrich:2018uni}%
  \BibitemOpen
  \bibfield  {author} {\bibinfo {author} {\bibfnamefont {Tim}\ \bibnamefont
  {Dietrich}} \emph {et~al.},\ }\href@noop {} {\  (\bibinfo {year} {2018})},\
  \Eprint {http://arxiv.org/abs/1804.02235} {arXiv:1804.02235
  [gr-qc]}\BibitemShut {NoStop}%
\bibitem [{\citenamefont {Pan}\ \emph {et~al.}(2011{\natexlab{b}})\citenamefont
  {Pan}, \citenamefont {Buonanno}, \citenamefont {Boyle}, \citenamefont
  {Buchman}, \citenamefont {Kidder}, \citenamefont {Pfeiffer},\ and\
  \citenamefont {Scheel}}]{Pan:2011gk}%
  \BibitemOpen
  \bibfield  {author} {\bibinfo {author} {\bibfnamefont {Yi}~\bibnamefont
  {Pan}}, \bibinfo {author} {\bibfnamefont {Alessandra}\ \bibnamefont
  {Buonanno}}, \bibinfo {author} {\bibfnamefont {Michael}\ \bibnamefont
  {Boyle}}, \bibinfo {author} {\bibfnamefont {Luisa~T.}\ \bibnamefont
  {Buchman}}, \bibinfo {author} {\bibfnamefont {Lawrence~E.}\ \bibnamefont
  {Kidder}}, \bibinfo {author} {\bibfnamefont {Harald~P.}\ \bibnamefont
  {Pfeiffer}}, \ and\ \bibinfo {author} {\bibfnamefont {Mark~A.}\ \bibnamefont
  {Scheel}},\ }\href {\doibase10.1103/PhysRevD.84.124052} {\bibfield  {journal}
  {\bibinfo  {journal} {Phys. Rev.}\ }\textbf {\bibinfo {volume} {D84}},\
  \bibinfo {pages} {124052} (\bibinfo {year} {2011}{\natexlab{b}})},\ \Eprint
  {http://arxiv.org/abs/1106.1021} {arXiv:1106.1021 [gr-qc]}\BibitemShut
  {NoStop}%
\bibitem [{\citenamefont {Cotesta}\ \emph {et~al.}(2018)\citenamefont
  {Cotesta}, \citenamefont {Buonanno}, \citenamefont {Bohé}, \citenamefont
  {Taracchini}, \citenamefont {Hinder},\ and\ \citenamefont
  {Ossokine}}]{Cotesta:2018fcv}%
  \BibitemOpen
  \bibfield  {author} {\bibinfo {author} {\bibfnamefont {Roberto}\ \bibnamefont
  {Cotesta}}, \bibinfo {author} {\bibfnamefont {Alessandra}\ \bibnamefont
  {Buonanno}}, \bibinfo {author} {\bibfnamefont {Alejandro}\ \bibnamefont
  {Bohé}}, \bibinfo {author} {\bibfnamefont {Andrea}\ \bibnamefont
  {Taracchini}}, \bibinfo {author} {\bibfnamefont {Ian}\ \bibnamefont
  {Hinder}}, \ and\ \bibinfo {author} {\bibfnamefont {Serguei}\ \bibnamefont
  {Ossokine}},\ }\href@noop {} {\  (\bibinfo {year} {2018})},\ \Eprint
  {http://arxiv.org/abs/1803.10701} {arXiv:1803.10701 [gr-qc]}\BibitemShut
  {NoStop}%
\bibitem [{\citenamefont {Calderon~Bustillo}\ \emph {et~al.}(2015)\citenamefont
  {Calderon~Bustillo}, \citenamefont {Bohe}, \citenamefont {Husa},
  \citenamefont {Sintes}, \citenamefont {Hannam},\ and\ \citenamefont
  {Puerrer}}]{Bustillo:2015ova}%
  \BibitemOpen
  \bibfield  {author} {\bibinfo {author} {\bibfnamefont {Juan}\ \bibnamefont
  {Calderon~Bustillo}}, \bibinfo {author} {\bibfnamefont {Alejandro}\
  \bibnamefont {Bohe}}, \bibinfo {author} {\bibfnamefont {Sascha}\ \bibnamefont
  {Husa}}, \bibinfo {author} {\bibfnamefont {Alicia~M.}\ \bibnamefont
  {Sintes}}, \bibinfo {author} {\bibfnamefont {Mark}\ \bibnamefont {Hannam}}, \
  and\ \bibinfo {author} {\bibfnamefont {Michael}\ \bibnamefont {Puerrer}},\
  }\href@noop {} {\  (\bibinfo {year} {2015})},\ \Eprint
  {http://arxiv.org/abs/1501.00918} {arXiv:1501.00918 [gr-qc]}\BibitemShut
  {NoStop}%
\bibitem [{\citenamefont {Huerta}\ \emph {et~al.}(2017)\citenamefont {Huerta}
  \emph {et~al.}}]{Huerta:2016rwp}%
  \BibitemOpen
  \bibfield  {author} {\bibinfo {author} {\bibfnamefont {E.~A.}\ \bibnamefont
  {Huerta}} \emph {et~al.},\ }\href {\doibase10.1103/PhysRevD.95.024038}
  {\bibfield  {journal} {\bibinfo  {journal} {Phys. Rev.}\ }\textbf {\bibinfo
  {volume} {D95}},\ \bibinfo {pages} {024038} (\bibinfo {year} {2017})},\
  \Eprint {http://arxiv.org/abs/1609.05933} {arXiv:1609.05933
  [gr-qc]}\BibitemShut {NoStop}%
\bibitem [{\citenamefont {Hinder}\ \emph {et~al.}(2017)\citenamefont {Hinder},
  \citenamefont {Kidder},\ and\ \citenamefont {Pfeiffer}}]{Hinder:2017sxy}%
  \BibitemOpen
  \bibfield  {author} {\bibinfo {author} {\bibfnamefont {Ian}\ \bibnamefont
  {Hinder}}, \bibinfo {author} {\bibfnamefont {Lawrence~E.}\ \bibnamefont
  {Kidder}}, \ and\ \bibinfo {author} {\bibfnamefont {Harald~P.}\ \bibnamefont
  {Pfeiffer}},\ }\href@noop {} {\  (\bibinfo {year} {2017})},\ \Eprint
  {http://arxiv.org/abs/1709.02007} {arXiv:1709.02007 [gr-qc]}\BibitemShut
  {NoStop}%
\bibitem [{\citenamefont {Hinderer}\ and\ \citenamefont
  {Babak}(2017)}]{Hinderer:2017jcs}%
  \BibitemOpen
  \bibfield  {author} {\bibinfo {author} {\bibfnamefont {Tanja}\ \bibnamefont
  {Hinderer}}\ and\ \bibinfo {author} {\bibfnamefont {Stanislav}\ \bibnamefont
  {Babak}},\ }\href {\doibase10.1103/PhysRevD.96.104048} {\bibfield  {journal}
  {\bibinfo  {journal} {Phys. Rev.}\ }\textbf {\bibinfo {volume} {D96}},\
  \bibinfo {pages} {104048} (\bibinfo {year} {2017})},\ \Eprint
  {http://arxiv.org/abs/1707.08426} {arXiv:1707.08426 [gr-qc]}\BibitemShut
  {NoStop}%
\bibitem [{\citenamefont {Puerrer}(2016)}]{Purrer:2015tud}%
  \BibitemOpen
  \bibfield  {author} {\bibinfo {author} {\bibfnamefont {Michael}\ \bibnamefont
  {Puerrer}},\ }\href {\doibase10.1103/PhysRevD.93.064041} {\bibfield
  {journal} {\bibinfo  {journal} {Phys. Rev.}\ }\textbf {\bibinfo {volume}
  {D93}},\ \bibinfo {pages} {064041} (\bibinfo {year} {2016})},\ \Eprint
  {http://arxiv.org/abs/1512.02248} {arXiv:1512.02248 [gr-qc]}\BibitemShut
  {NoStop}%
\bibitem [{\citenamefont {Agathos}\ \emph {et~al.}(2014)\citenamefont
  {Agathos}, \citenamefont {Del~Pozzo}, \citenamefont {Li}, \citenamefont {Van
  Den~Broeck}, \citenamefont {Veitch},\ and\ \citenamefont
  {Vitale}}]{Agathos:2013upa}%
  \BibitemOpen
  \bibfield  {author} {\bibinfo {author} {\bibfnamefont {Michalis}\
  \bibnamefont {Agathos}}, \bibinfo {author} {\bibfnamefont {Walter}\
  \bibnamefont {Del~Pozzo}}, \bibinfo {author} {\bibfnamefont {Tjonnie G.~F.}\
  \bibnamefont {Li}}, \bibinfo {author} {\bibfnamefont {Chris}\ \bibnamefont
  {Van Den~Broeck}}, \bibinfo {author} {\bibfnamefont {John}\ \bibnamefont
  {Veitch}}, \ and\ \bibinfo {author} {\bibfnamefont {Salvatore}\ \bibnamefont
  {Vitale}},\ }\href {\doibase10.1103/PhysRevD.89.082001} {\bibfield  {journal}
  {\bibinfo  {journal} {Phys. Rev.}\ }\textbf {\bibinfo {volume} {D89}},\
  \bibinfo {pages} {082001} (\bibinfo {year} {2014})},\ \Eprint
  {http://arxiv.org/abs/1311.0420} {arXiv:1311.0420 [gr-qc]}\BibitemShut
  {NoStop}%
\bibitem [{\citenamefont {Brito}\ \emph {et~al.}(2018)\citenamefont {Brito},
  \citenamefont {Buonanno},\ and\ \citenamefont {Raymond}}]{Brito:2018rfr}%
  \BibitemOpen
  \bibfield  {author} {\bibinfo {author} {\bibfnamefont {Richard}\ \bibnamefont
  {Brito}}, \bibinfo {author} {\bibfnamefont {Alessandra}\ \bibnamefont
  {Buonanno}}, \ and\ \bibinfo {author} {\bibfnamefont {Vivien}\ \bibnamefont
  {Raymond}},\ }\href@noop {} {\  (\bibinfo {year} {2018})},\ \Eprint
  {http://arxiv.org/abs/1805.00293} {arXiv:1805.00293 [gr-qc]}\BibitemShut
  {NoStop}%
\bibitem [{\citenamefont {{Klimenko}}\ \emph {et~al.}(2008)\citenamefont
  {{Klimenko}}, \citenamefont {{Yakushin}}, \citenamefont {{Mercer}},\ and\
  \citenamefont {{Mitselmakher}}}]{2008CQGra..25k4029K}%
  \BibitemOpen
  \bibfield  {author} {\bibinfo {author} {\bibfnamefont {S.}~\bibnamefont
  {{Klimenko}}}, \bibinfo {author} {\bibfnamefont {I.}~\bibnamefont
  {{Yakushin}}}, \bibinfo {author} {\bibfnamefont {A.}~\bibnamefont
  {{Mercer}}}, \ and\ \bibinfo {author} {\bibfnamefont {G.}~\bibnamefont
  {{Mitselmakher}}},\ }\href {\doibase10.1088/0264-9381/25/11/114029}
  {\bibfield  {journal} {\bibinfo  {journal} {Classical and Quantum Gravity}\
  }\textbf {\bibinfo {volume} {25}},\ \bibinfo {eid} {114029} (\bibinfo {year}
  {2008})},\ \Eprint {http://arxiv.org/abs/0802.3232} {arXiv:0802.3232
  [gr-qc]}\BibitemShut {NoStop}%
\bibitem [{\citenamefont {{Sutton}}\ \emph {et~al.}(2010)\citenamefont
  {{Sutton}}, \citenamefont {{Jones}}, \citenamefont {{Chatterji}},
  \citenamefont {{Kalmus}}, \citenamefont {{Leonor}}, \citenamefont
  {{Poprocki}}, \citenamefont {{Rollins}}, \citenamefont {{Searle}},
  \citenamefont {{Stein}}, \citenamefont {{Tinto}},\ and\ \citenamefont
  {{Was}}}]{2010NJPh...12e3034S}%
  \BibitemOpen
  \bibfield  {author} {\bibinfo {author} {\bibfnamefont {P.~J.}\ \bibnamefont
  {{Sutton}}}, \bibinfo {author} {\bibfnamefont {G.}~\bibnamefont {{Jones}}},
  \bibinfo {author} {\bibfnamefont {S.}~\bibnamefont {{Chatterji}}}, \bibinfo
  {author} {\bibfnamefont {P.}~\bibnamefont {{Kalmus}}}, \bibinfo {author}
  {\bibfnamefont {I.}~\bibnamefont {{Leonor}}}, \bibinfo {author}
  {\bibfnamefont {S.}~\bibnamefont {{Poprocki}}}, \bibinfo {author}
  {\bibfnamefont {J.}~\bibnamefont {{Rollins}}}, \bibinfo {author}
  {\bibfnamefont {A.}~\bibnamefont {{Searle}}}, \bibinfo {author}
  {\bibfnamefont {L.}~\bibnamefont {{Stein}}}, \bibinfo {author} {\bibfnamefont
  {M.}~\bibnamefont {{Tinto}}}, \ and\ \bibinfo {author} {\bibfnamefont
  {M.}~\bibnamefont {{Was}}},\ }\href {\doibase10.1088/1367-2630/12/5/053034}
  {\bibfield  {journal} {\bibinfo  {journal} {New Journal of Physics}\ }\textbf
  {\bibinfo {volume} {12}},\ \bibinfo {eid} {053034} (\bibinfo {year}
  {2010})},\ \Eprint {http://arxiv.org/abs/0908.3665} {arXiv:0908.3665
  [gr-qc]}\BibitemShut {NoStop}%
\bibitem [{\citenamefont {{W{\c a}s}}\ \emph {et~al.}(2012)\citenamefont {{W{\c
  a}s}}, \citenamefont {{Sutton}}, \citenamefont {{Jones}},\ and\ \citenamefont
  {{Leonor}}}]{2012PhRvD..86b2003W}%
  \BibitemOpen
  \bibfield  {author} {\bibinfo {author} {\bibfnamefont {M.}~\bibnamefont
  {{W{\c a}s}}}, \bibinfo {author} {\bibfnamefont {P.~J.}\ \bibnamefont
  {{Sutton}}}, \bibinfo {author} {\bibfnamefont {G.}~\bibnamefont {{Jones}}}, \
  and\ \bibinfo {author} {\bibfnamefont {I.}~\bibnamefont {{Leonor}}},\ }\href
  {\doibase10.1103/PhysRevD.86.022003} {\bibfield  {journal} {\bibinfo
  {journal} {Physical Review D}\ }\textbf {\bibinfo {volume} {86}},\ \bibinfo
  {eid} {022003} (\bibinfo {year} {2012})},\ \Eprint
  {http://arxiv.org/abs/1201.5599} {arXiv:1201.5599 [gr-qc]}\BibitemShut
  {NoStop}%
\bibitem [{\citenamefont {{Cornish}}\ and\ \citenamefont
  {{Littenberg}}(2015)}]{2015CQGra..32m5012C}%
  \BibitemOpen
  \bibfield  {author} {\bibinfo {author} {\bibfnamefont {N.~J.}\ \bibnamefont
  {{Cornish}}}\ and\ \bibinfo {author} {\bibfnamefont {T.~B.}\ \bibnamefont
  {{Littenberg}}},\ }\href {\doibase10.1088/0264-9381/32/13/135012} {\bibfield
  {journal} {\bibinfo  {journal} {Classical and Quantum Gravity}\ }\textbf
  {\bibinfo {volume} {32}},\ \bibinfo {eid} {135012} (\bibinfo {year}
  {2015})},\ \Eprint {http://arxiv.org/abs/1410.3835} {arXiv:1410.3835
  [gr-qc]}\BibitemShut {NoStop}%
\bibitem [{\citenamefont {Dal~Canton}\ \emph {et~al.}(2014)\citenamefont
  {Dal~Canton} \emph {et~al.}}]{Canton:2014ena}%
  \BibitemOpen
  \bibfield  {author} {\bibinfo {author} {\bibfnamefont {Tito}\ \bibnamefont
  {Dal~Canton}} \emph {et~al.},\ }\href {\doibase10.1103/PhysRevD.90.082004}
  {\bibfield  {journal} {\bibinfo  {journal} {Phys. Rev.}\ }\textbf {\bibinfo
  {volume} {D90}},\ \bibinfo {pages} {082004} (\bibinfo {year} {2014})},\
  \Eprint {http://arxiv.org/abs/1405.6731} {arXiv:1405.6731
  [gr-qc]}\BibitemShut {NoStop}%
\bibitem [{\citenamefont {Usman}\ \emph {et~al.}(2016)\citenamefont {Usman}
  \emph {et~al.}}]{Usman:2015kfa}%
  \BibitemOpen
  \bibfield  {author} {\bibinfo {author} {\bibfnamefont {Samantha~A.}\
  \bibnamefont {Usman}} \emph {et~al.},\ }\href
  {\doibase10.1088/0264-9381/33/21/215004} {\bibfield  {journal} {\bibinfo
  {journal} {Class. Quant. Grav.}\ }\textbf {\bibinfo {volume} {33}},\ \bibinfo
  {pages} {215004} (\bibinfo {year} {2016})},\ \Eprint
  {http://arxiv.org/abs/1508.02357} {arXiv:1508.02357 [gr-qc]}\BibitemShut
  {NoStop}%
\bibitem [{\citenamefont {{Cannon}}\ \emph {et~al.}(2012)\citenamefont
  {{Cannon}}, \citenamefont {{Cariou}}, \citenamefont {{Chapman}},
  \citenamefont {{Crispin-Ortuzar}}, \citenamefont {{Fotopoulos}},
  \citenamefont {{Frei}}, \citenamefont {{Hanna}}, \citenamefont {{Kara}},
  \citenamefont {{Keppel}}, \citenamefont {{Liao}}, \citenamefont
  {{Privitera}}, \citenamefont {{Searle}}, \citenamefont {{Singer}},\ and\
  \citenamefont {{Weinstein}}}]{2012ApJ...748..136C}%
  \BibitemOpen
  \bibfield  {author} {\bibinfo {author} {\bibfnamefont {K.}~\bibnamefont
  {{Cannon}}}, \bibinfo {author} {\bibfnamefont {R.}~\bibnamefont {{Cariou}}},
  \bibinfo {author} {\bibfnamefont {A.}~\bibnamefont {{Chapman}}}, \bibinfo
  {author} {\bibfnamefont {M.}~\bibnamefont {{Crispin-Ortuzar}}}, \bibinfo
  {author} {\bibfnamefont {N.}~\bibnamefont {{Fotopoulos}}}, \bibinfo {author}
  {\bibfnamefont {M.}~\bibnamefont {{Frei}}}, \bibinfo {author} {\bibfnamefont
  {C.}~\bibnamefont {{Hanna}}}, \bibinfo {author} {\bibfnamefont
  {E.}~\bibnamefont {{Kara}}}, \bibinfo {author} {\bibfnamefont
  {D.}~\bibnamefont {{Keppel}}}, \bibinfo {author} {\bibfnamefont
  {L.}~\bibnamefont {{Liao}}}, \bibinfo {author} {\bibfnamefont
  {S.}~\bibnamefont {{Privitera}}}, \bibinfo {author} {\bibfnamefont
  {A.}~\bibnamefont {{Searle}}}, \bibinfo {author} {\bibfnamefont
  {L.}~\bibnamefont {{Singer}}}, \ and\ \bibinfo {author} {\bibfnamefont
  {A.}~\bibnamefont {{Weinstein}}},\ }\href
  {\doibase10.1088/0004-637X/748/2/136} {\bibfield  {journal} {\bibinfo
  {journal} {The Astrophysical Journal}\ }\textbf {\bibinfo {volume} {748}},\
  \bibinfo {eid} {136} (\bibinfo {year} {2012})},\ \Eprint
  {http://arxiv.org/abs/1107.2665} {arXiv:1107.2665 [astro-ph.IM]}\BibitemShut
  {NoStop}%
\bibitem [{\citenamefont {{Privitera}}\ \emph {et~al.}(2014)\citenamefont
  {{Privitera}}, \citenamefont {{Mohapatra}}, \citenamefont {{Ajith}},
  \citenamefont {{Cannon}}, \citenamefont {{Fotopoulos}}, \citenamefont
  {{Frei}}, \citenamefont {{Hanna}}, \citenamefont {{Weinstein}},\ and\
  \citenamefont {{Whelan}}}]{2014PhRvD..89b4003P}%
  \BibitemOpen
  \bibfield  {author} {\bibinfo {author} {\bibfnamefont {S.}~\bibnamefont
  {{Privitera}}}, \bibinfo {author} {\bibfnamefont {S.~R.~P.}\ \bibnamefont
  {{Mohapatra}}}, \bibinfo {author} {\bibfnamefont {P.}~\bibnamefont
  {{Ajith}}}, \bibinfo {author} {\bibfnamefont {K.}~\bibnamefont {{Cannon}}},
  \bibinfo {author} {\bibfnamefont {N.}~\bibnamefont {{Fotopoulos}}}, \bibinfo
  {author} {\bibfnamefont {M.~A.}\ \bibnamefont {{Frei}}}, \bibinfo {author}
  {\bibfnamefont {C.}~\bibnamefont {{Hanna}}}, \bibinfo {author} {\bibfnamefont
  {A.~J.}\ \bibnamefont {{Weinstein}}}, \ and\ \bibinfo {author} {\bibfnamefont
  {J.~T.}\ \bibnamefont {{Whelan}}},\ }\href
  {\doibase10.1103/PhysRevD.89.024003} {\bibfield  {journal} {\bibinfo
  {journal} {Physical Review D}\ }\textbf {\bibinfo {volume} {89}},\ \bibinfo
  {eid} {024003} (\bibinfo {year} {2014})},\ \Eprint
  {http://arxiv.org/abs/1310.5633} {arXiv:1310.5633 [gr-qc]}\BibitemShut
  {NoStop}%
\bibitem [{\citenamefont {{Veitch}}\ \emph {et~al.}(2015)\citenamefont
  {{Veitch}}, \citenamefont {{Raymond}}, \citenamefont {{Farr}}, \citenamefont
  {{Farr}}, \citenamefont {{Graff}}, \citenamefont {{Vitale}}, \citenamefont
  {{Aylott}}, \citenamefont {{Blackburn}}, \citenamefont {{Christensen}},
  \citenamefont {{Coughlin}}, \citenamefont {{Del Pozzo}}, \citenamefont
  {{Feroz}}, \citenamefont {{Gair}}, \citenamefont {{Haster}}, \citenamefont
  {{Kalogera}}, \citenamefont {{Littenberg}}, \citenamefont {{Mandel}},
  \citenamefont {{O'Shaughnessy}}, \citenamefont {{Pitkin}}, \citenamefont
  {{Rodriguez}}, \citenamefont {{R{\"o}ver}}, \citenamefont {{Sidery}},
  \citenamefont {{Smith}}, \citenamefont {{Van Der Sluys}}, \citenamefont
  {{Vecchio}}, \citenamefont {{Vousden}},\ and\ \citenamefont
  {{Wade}}}]{2015PhRvD..91d2003V}%
  \BibitemOpen
  \bibfield  {author} {\bibinfo {author} {\bibfnamefont {J.}~\bibnamefont
  {{Veitch}}}, \bibinfo {author} {\bibfnamefont {V.}~\bibnamefont {{Raymond}}},
  \bibinfo {author} {\bibfnamefont {B.}~\bibnamefont {{Farr}}}, \bibinfo
  {author} {\bibfnamefont {W.}~\bibnamefont {{Farr}}}, \bibinfo {author}
  {\bibfnamefont {P.}~\bibnamefont {{Graff}}}, \bibinfo {author} {\bibfnamefont
  {S.}~\bibnamefont {{Vitale}}}, \bibinfo {author} {\bibfnamefont
  {B.}~\bibnamefont {{Aylott}}}, \bibinfo {author} {\bibfnamefont
  {K.}~\bibnamefont {{Blackburn}}}, \bibinfo {author} {\bibfnamefont
  {N.}~\bibnamefont {{Christensen}}}, \bibinfo {author} {\bibfnamefont
  {M.}~\bibnamefont {{Coughlin}}}, \bibinfo {author} {\bibfnamefont
  {W.}~\bibnamefont {{Del Pozzo}}}, \bibinfo {author} {\bibfnamefont
  {F.}~\bibnamefont {{Feroz}}}, \bibinfo {author} {\bibfnamefont
  {J.}~\bibnamefont {{Gair}}}, \bibinfo {author} {\bibfnamefont {C.-J.}\
  \bibnamefont {{Haster}}}, \bibinfo {author} {\bibfnamefont {V.}~\bibnamefont
  {{Kalogera}}}, \bibinfo {author} {\bibfnamefont {T.}~\bibnamefont
  {{Littenberg}}}, \bibinfo {author} {\bibfnamefont {I.}~\bibnamefont
  {{Mandel}}}, \bibinfo {author} {\bibfnamefont {R.}~\bibnamefont
  {{O'Shaughnessy}}}, \bibinfo {author} {\bibfnamefont {M.}~\bibnamefont
  {{Pitkin}}}, \bibinfo {author} {\bibfnamefont {C.}~\bibnamefont
  {{Rodriguez}}}, \bibinfo {author} {\bibfnamefont {C.}~\bibnamefont
  {{R{\"o}ver}}}, \bibinfo {author} {\bibfnamefont {T.}~\bibnamefont
  {{Sidery}}}, \bibinfo {author} {\bibfnamefont {R.}~\bibnamefont {{Smith}}},
  \bibinfo {author} {\bibfnamefont {M.}~\bibnamefont {{Van Der Sluys}}},
  \bibinfo {author} {\bibfnamefont {A.}~\bibnamefont {{Vecchio}}}, \bibinfo
  {author} {\bibfnamefont {W.}~\bibnamefont {{Vousden}}}, \ and\ \bibinfo
  {author} {\bibfnamefont {L.}~\bibnamefont {{Wade}}},\ }\href
  {\doibase10.1103/PhysRevD.91.042003} {\bibfield  {journal} {\bibinfo
  {journal} {Physical Review D}\ }\textbf {\bibinfo {volume} {91}},\ \bibinfo
  {eid} {042003} (\bibinfo {year} {2015})},\ \Eprint
  {http://arxiv.org/abs/1409.7215} {arXiv:1409.7215 [gr-qc]}\BibitemShut
  {NoStop}%
\bibitem [{\citenamefont {Pankow}\ \emph {et~al.}(2015)\citenamefont {Pankow},
  \citenamefont {Brady}, \citenamefont {Ochsner},\ and\ \citenamefont
  {O'Shaughnessy}}]{Pankow:2015cra}%
  \BibitemOpen
  \bibfield  {author} {\bibinfo {author} {\bibfnamefont {C.}~\bibnamefont
  {Pankow}}, \bibinfo {author} {\bibfnamefont {P.}~\bibnamefont {Brady}},
  \bibinfo {author} {\bibfnamefont {E.}~\bibnamefont {Ochsner}}, \ and\
  \bibinfo {author} {\bibfnamefont {R.}~\bibnamefont {O'Shaughnessy}},\ }\href
  {\doibase10.1103/PhysRevD.92.023002} {\bibfield  {journal} {\bibinfo
  {journal} {Phys. Rev.}\ }\textbf {\bibinfo {volume} {D92}},\ \bibinfo {pages}
  {023002} (\bibinfo {year} {2015})},\ \Eprint
  {http://arxiv.org/abs/1502.04370} {arXiv:1502.04370 [gr-qc]}\BibitemShut
  {NoStop}%
\bibitem [{\citenamefont {Lange}\ \emph {et~al.}(2018)\citenamefont {Lange},
  \citenamefont {O'Shaughnessy},\ and\ \citenamefont {Rizzo}}]{Lange:2018pyp}%
  \BibitemOpen
  \bibfield  {author} {\bibinfo {author} {\bibfnamefont {Jacob}\ \bibnamefont
  {Lange}}, \bibinfo {author} {\bibfnamefont {Richard}\ \bibnamefont
  {O'Shaughnessy}}, \ and\ \bibinfo {author} {\bibfnamefont {Monica}\
  \bibnamefont {Rizzo}},\ }\href@noop {} {\  (\bibinfo {year} {2018})},\
  \Eprint {http://arxiv.org/abs/1805.10457} {arXiv:1805.10457
  [gr-qc]}\BibitemShut {NoStop}%
\bibitem [{\citenamefont {{Veitch}}\ and\ \citenamefont
  {{Vecchio}}(2010)}]{2010PhRvD..81f2003V}%
  \BibitemOpen
  \bibfield  {author} {\bibinfo {author} {\bibfnamefont {J.}~\bibnamefont
  {{Veitch}}}\ and\ \bibinfo {author} {\bibfnamefont {A.}~\bibnamefont
  {{Vecchio}}},\ }\href {\doibase10.1103/PhysRevD.81.062003} {\bibfield
  {journal} {\bibinfo  {journal} {Physical Review D}\ }\textbf {\bibinfo
  {volume} {81}},\ \bibinfo {eid} {062003} (\bibinfo {year} {2010})},\ \Eprint
  {http://arxiv.org/abs/0911.3820} {arXiv:0911.3820}\BibitemShut {NoStop}%
\bibitem [{\citenamefont {{van der Sluys}}\ \emph {et~al.}(2008)\citenamefont
  {{van der Sluys}}, \citenamefont {{R{\"o}ver}}, \citenamefont {{Stroeer}},
  \citenamefont {{Raymond}}, \citenamefont {{Mandel}}, \citenamefont
  {{Christensen}}, \citenamefont {{Kalogera}}, \citenamefont {{Meyer}},\ and\
  \citenamefont {{Vecchio}}}]{2008ApJ...688L..61V}%
  \BibitemOpen
  \bibfield  {author} {\bibinfo {author} {\bibfnamefont {M.~V.}\ \bibnamefont
  {{van der Sluys}}}, \bibinfo {author} {\bibfnamefont {C.}~\bibnamefont
  {{R{\"o}ver}}}, \bibinfo {author} {\bibfnamefont {A.}~\bibnamefont
  {{Stroeer}}}, \bibinfo {author} {\bibfnamefont {V.}~\bibnamefont
  {{Raymond}}}, \bibinfo {author} {\bibfnamefont {I.}~\bibnamefont {{Mandel}}},
  \bibinfo {author} {\bibfnamefont {N.}~\bibnamefont {{Christensen}}}, \bibinfo
  {author} {\bibfnamefont {V.}~\bibnamefont {{Kalogera}}}, \bibinfo {author}
  {\bibfnamefont {R.}~\bibnamefont {{Meyer}}}, \ and\ \bibinfo {author}
  {\bibfnamefont {A.}~\bibnamefont {{Vecchio}}},\ }\href
  {\doibase10.1086/595279} {\bibfield  {journal} {\bibinfo  {journal}
  {Astrophys. J. Lett.}\ }\textbf {\bibinfo {volume} {688}},\ \bibinfo {eid}
  {L61} (\bibinfo {year} {2008})},\ \Eprint {http://arxiv.org/abs/0710.1897}
  {arXiv:0710.1897}\BibitemShut {NoStop}%
\bibitem [{\citenamefont {Abbott}\ \emph
  {et~al.}(2018{\natexlab{d}})\citenamefont {Abbott} \emph
  {et~al.}}]{Abbott:2018wiz}%
  \BibitemOpen
  \bibfield  {author} {\bibinfo {author} {\bibfnamefont {B.~P.}\ \bibnamefont
  {Abbott}} \emph {et~al.} (\bibinfo {collaboration} {Virgo, LIGO
  Scientific}),\ }\href@noop {} {\  (\bibinfo {year} {2018}{\natexlab{d}})},\
  \Eprint {http://arxiv.org/abs/1805.11579} {arXiv:1805.11579
  [gr-qc]}\BibitemShut {NoStop}%
\bibitem [{\citenamefont {{Brown}}\ \emph {et~al.}(2007)\citenamefont
  {{Brown}}, \citenamefont {{Brink}}, \citenamefont {{Fang}}, \citenamefont
  {{Gair}}, \citenamefont {{Li}}, \citenamefont {{Lovelace}}, \citenamefont
  {{Mandel}},\ and\ \citenamefont {{Thorne}}}]{brown:2007}%
  \BibitemOpen
  \bibfield  {author} {\bibinfo {author} {\bibfnamefont {D.~A.}\ \bibnamefont
  {{Brown}}}, \bibinfo {author} {\bibfnamefont {J.}~\bibnamefont {{Brink}}},
  \bibinfo {author} {\bibfnamefont {H.}~\bibnamefont {{Fang}}}, \bibinfo
  {author} {\bibfnamefont {J.~R.}\ \bibnamefont {{Gair}}}, \bibinfo {author}
  {\bibfnamefont {C.}~\bibnamefont {{Li}}}, \bibinfo {author} {\bibfnamefont
  {G.}~\bibnamefont {{Lovelace}}}, \bibinfo {author} {\bibfnamefont
  {I.}~\bibnamefont {{Mandel}}}, \ and\ \bibinfo {author} {\bibfnamefont
  {K.~S.}\ \bibnamefont {{Thorne}}},\ }\href
  {\doibase10.1103/PhysRevLett.99.201102} {\bibfield  {journal} {\bibinfo
  {journal} {Physical Review Letters}\ }\textbf {\bibinfo {volume} {99}},\
  \bibinfo {pages} {201102--+} (\bibinfo {year} {2007})},\ \Eprint
  {http://arxiv.org/abs/gr-qc/0612060} {gr-qc/0612060}\BibitemShut {NoStop}%
\bibitem [{\citenamefont {Gair}\ \emph
  {et~al.}(2011{\natexlab{b}})\citenamefont {Gair}, \citenamefont {Mandel},
  \citenamefont {Miller},\ and\ \citenamefont {Volonteri}}]{Gair:2010dx}%
  \BibitemOpen
  \bibfield  {author} {\bibinfo {author} {\bibfnamefont {Jonathan~R.}\
  \bibnamefont {Gair}}, \bibinfo {author} {\bibfnamefont {Ilya}\ \bibnamefont
  {Mandel}}, \bibinfo {author} {\bibfnamefont {M.~Coleman}\ \bibnamefont
  {Miller}}, \ and\ \bibinfo {author} {\bibfnamefont {Marta}\ \bibnamefont
  {Volonteri}},\ }\href {\doibase10.1007/s10714-010-1104-3} {\bibfield
  {journal} {\bibinfo  {journal} {Gen. Rel. Grav.}\ }\textbf {\bibinfo {volume}
  {43}},\ \bibinfo {pages} {485--518} (\bibinfo {year} {2011}{\natexlab{b}})},\
  \Eprint {http://arxiv.org/abs/0907.5450} {arXiv:0907.5450
  [astro-ph.CO]}\BibitemShut {NoStop}%
\bibitem [{\citenamefont {{Huerta}}\ and\ \citenamefont
  {{Gair}}(2011{\natexlab{a}})}]{Huerta:2011a}%
  \BibitemOpen
  \bibfield  {author} {\bibinfo {author} {\bibfnamefont {E.~A.}\ \bibnamefont
  {{Huerta}}}\ and\ \bibinfo {author} {\bibfnamefont {J.~R.}\ \bibnamefont
  {{Gair}}},\ }\href {\doibase10.1103/PhysRevD.83.044020} {\bibfield  {journal}
  {\bibinfo  {journal} {Phys. Rev. D}\ }\textbf {\bibinfo {volume} {83}},\
  \bibinfo {pages} {044020--+} (\bibinfo {year} {2011}{\natexlab{a}})},\
  \Eprint {http://arxiv.org/abs/1009.1985} {arXiv:1009.1985
  [gr-qc]}\BibitemShut {NoStop}%
\bibitem [{\citenamefont {{Huerta}}\ and\ \citenamefont
  {{Gair}}(2011{\natexlab{b}})}]{Huerta:2011b}%
  \BibitemOpen
  \bibfield  {author} {\bibinfo {author} {\bibfnamefont {E.~A.}\ \bibnamefont
  {{Huerta}}}\ and\ \bibinfo {author} {\bibfnamefont {J.~R.}\ \bibnamefont
  {{Gair}}},\ }\href {\doibase10.1103/PhysRevD.83.044021} {\bibfield  {journal}
  {\bibinfo  {journal} {Phys. Rev. D}\ }\textbf {\bibinfo {volume} {83}},\
  \bibinfo {pages} {044021--+} (\bibinfo {year} {2011}{\natexlab{b}})},\
  \Eprint {http://arxiv.org/abs/1011.0421} {arXiv:1011.0421
  [gr-qc]}\BibitemShut {NoStop}%
\bibitem [{\citenamefont {{Huerta}}\ \emph
  {et~al.}(2012{\natexlab{a}})\citenamefont {{Huerta}}, \citenamefont
  {{Gair}},\ and\ \citenamefont {{Brown}}}]{Huerta:2012a}%
  \BibitemOpen
  \bibfield  {author} {\bibinfo {author} {\bibfnamefont {E.~A.}\ \bibnamefont
  {{Huerta}}}, \bibinfo {author} {\bibfnamefont {J.~R.}\ \bibnamefont
  {{Gair}}}, \ and\ \bibinfo {author} {\bibfnamefont {D.~A.}\ \bibnamefont
  {{Brown}}},\ }\href {\doibase10.1103/PhysRevD.85.064023} {\bibfield
  {journal} {\bibinfo  {journal} {Physical Review D}\ }\textbf {\bibinfo
  {volume} {85}},\ \bibinfo {eid} {064023} (\bibinfo {year}
  {2012}{\natexlab{a}})},\ \Eprint {http://arxiv.org/abs/1111.3243}
  {arXiv:1111.3243 [gr-qc]}\BibitemShut {NoStop}%
\bibitem [{\citenamefont {{Huerta}}\ \emph
  {et~al.}(2012{\natexlab{b}})\citenamefont {{Huerta}}, \citenamefont
  {{Kumar}},\ and\ \citenamefont {{Brown}}}]{Huerta:2012b}%
  \BibitemOpen
  \bibfield  {author} {\bibinfo {author} {\bibfnamefont {E.~A.}\ \bibnamefont
  {{Huerta}}}, \bibinfo {author} {\bibfnamefont {P.}~\bibnamefont {{Kumar}}}, \
  and\ \bibinfo {author} {\bibfnamefont {D.~A.}\ \bibnamefont {{Brown}}},\
  }\href {\doibase10.1103/PhysRevD.86.024024} {\bibfield  {journal} {\bibinfo
  {journal} {Physical Review D}\ }\textbf {\bibinfo {volume} {86}},\ \bibinfo
  {eid} {024024} (\bibinfo {year} {2012}{\natexlab{b}})},\ \Eprint
  {http://arxiv.org/abs/1205.5562} {arXiv:1205.5562 [gr-qc]}\BibitemShut
  {NoStop}%
\bibitem [{\citenamefont {Barausse}(2012)}]{Barausse:2012fy}%
  \BibitemOpen
  \bibfield  {author} {\bibinfo {author} {\bibfnamefont {Enrico}\ \bibnamefont
  {Barausse}},\ }\href {\doibase10.1111/j.1365-2966.2012.21057.x} {\bibfield
  {journal} {\bibinfo  {journal} {Mon. Not. Roy. Astron. Soc.}\ }\textbf
  {\bibinfo {volume} {423}},\ \bibinfo {pages} {2533--2557} (\bibinfo {year}
  {2012})},\ \Eprint {http://arxiv.org/abs/1201.5888} {arXiv:1201.5888
  [astro-ph.CO]}\BibitemShut {NoStop}%
\bibitem [{\citenamefont {Wysocki}\ \emph
  {et~al.}(2018{\natexlab{a}})\citenamefont {Wysocki}, \citenamefont {Gerosa},
  \citenamefont {O'Shaughnessy}, \citenamefont {Belczynski}, \citenamefont
  {Gladysz}, \citenamefont {Berti}, \citenamefont {Kesden},\ and\ \citenamefont
  {Holz}}]{Wysocki:2017isg}%
  \BibitemOpen
  \bibfield  {author} {\bibinfo {author} {\bibfnamefont {Daniel}\ \bibnamefont
  {Wysocki}}, \bibinfo {author} {\bibfnamefont {Davide}\ \bibnamefont
  {Gerosa}}, \bibinfo {author} {\bibfnamefont {Richard}\ \bibnamefont
  {O'Shaughnessy}}, \bibinfo {author} {\bibfnamefont {Krzysztof}\ \bibnamefont
  {Belczynski}}, \bibinfo {author} {\bibfnamefont {Wojciech}\ \bibnamefont
  {Gladysz}}, \bibinfo {author} {\bibfnamefont {Emanuele}\ \bibnamefont
  {Berti}}, \bibinfo {author} {\bibfnamefont {Michael}\ \bibnamefont {Kesden}},
  \ and\ \bibinfo {author} {\bibfnamefont {Daniel~E.}\ \bibnamefont {Holz}},\
  }\href {\doibase10.1103/PhysRevD.97.043014} {\bibfield  {journal} {\bibinfo
  {journal} {Phys. Rev.}\ }\textbf {\bibinfo {volume} {D97}},\ \bibinfo {pages}
  {043014} (\bibinfo {year} {2018}{\natexlab{a}})},\ \Eprint
  {http://arxiv.org/abs/1709.01943} {arXiv:1709.01943
  [astro-ph.HE]}\BibitemShut {NoStop}%
\bibitem [{\citenamefont {Wysocki}\ \emph
  {et~al.}(2018{\natexlab{b}})\citenamefont {Wysocki}, \citenamefont {Lange},\
  and\ \citenamefont {O.~'shaughnessy}}]{Wysocki:2018mpo}%
  \BibitemOpen
  \bibfield  {author} {\bibinfo {author} {\bibfnamefont {Daniel}\ \bibnamefont
  {Wysocki}}, \bibinfo {author} {\bibfnamefont {Jacob}\ \bibnamefont {Lange}},
  \ and\ \bibinfo {author} {\bibfnamefont {Richard}\ \bibnamefont
  {O.~'shaughnessy}},\ }\href@noop {} {\  (\bibinfo {year}
  {2018}{\natexlab{b}})},\ \Eprint {http://arxiv.org/abs/1805.06442}
  {arXiv:1805.06442 [gr-qc]}\BibitemShut {NoStop}%
\bibitem [{\citenamefont {Amaro-Seoane}(2018)}]{AmaroSeoane:2012tx}%
  \BibitemOpen
  \bibfield  {author} {\bibinfo {author} {\bibfnamefont {Pau}\ \bibnamefont
  {Amaro-Seoane}},\ }\href {\doibase10.1007/s41114-018-0013-8} {\bibfield
  {journal} {\bibinfo  {journal} {Living Rev. Rel.}\ }\textbf {\bibinfo
  {volume} {21}},\ \bibinfo {pages} {4} (\bibinfo {year} {2018})},\ \Eprint
  {http://arxiv.org/abs/1205.5240} {arXiv:1205.5240 [astro-ph.CO]}\BibitemShut
  {NoStop}%
\bibitem [{\citenamefont {Gair}\ \emph {et~al.}(2013)\citenamefont {Gair},
  \citenamefont {Vallisneri}, \citenamefont {Larson},\ and\ \citenamefont
  {Baker}}]{Gair:2012nm}%
  \BibitemOpen
  \bibfield  {author} {\bibinfo {author} {\bibfnamefont {Jonathan~R.}\
  \bibnamefont {Gair}}, \bibinfo {author} {\bibfnamefont {Michele}\
  \bibnamefont {Vallisneri}}, \bibinfo {author} {\bibfnamefont {Shane~L.}\
  \bibnamefont {Larson}}, \ and\ \bibinfo {author} {\bibfnamefont {John~G.}\
  \bibnamefont {Baker}},\ }\href {\doibase10.12942/lrr-2013-7} {\bibfield
  {journal} {\bibinfo  {journal} {Living Rev. Rel.}\ }\textbf {\bibinfo
  {volume} {16}},\ \bibinfo {pages} {7} (\bibinfo {year} {2013})},\ \Eprint
  {http://arxiv.org/abs/1212.5575} {arXiv:1212.5575 [gr-qc]}\BibitemShut
  {NoStop}%
\bibitem [{\citenamefont {{P{\"u}rrer}}(2014)}]{2014CQGra..31s5010P}%
  \BibitemOpen
  \bibfield  {author} {\bibinfo {author} {\bibfnamefont {M.}~\bibnamefont
  {{P{\"u}rrer}}},\ }\href {\doibase10.1088/0264-9381/31/19/195010} {\bibfield
  {journal} {\bibinfo  {journal} {Classical and Quantum Gravity}\ }\textbf
  {\bibinfo {volume} {31}},\ \bibinfo {eid} {195010} (\bibinfo {year}
  {2014})},\ \Eprint {http://arxiv.org/abs/1402.4146} {arXiv:1402.4146
  [gr-qc]}\BibitemShut {NoStop}%
\bibitem [{\citenamefont {{Field}}\ \emph {et~al.}(2011)\citenamefont
  {{Field}}, \citenamefont {{Galley}}, \citenamefont {{Herrmann}},
  \citenamefont {{Hesthaven}}, \citenamefont {{Ochsner}},\ and\ \citenamefont
  {{Tiglio}}}]{2011PhRvL.106v1102F}%
  \BibitemOpen
  \bibfield  {author} {\bibinfo {author} {\bibfnamefont {S.~E.}\ \bibnamefont
  {{Field}}}, \bibinfo {author} {\bibfnamefont {C.~R.}\ \bibnamefont
  {{Galley}}}, \bibinfo {author} {\bibfnamefont {F.}~\bibnamefont
  {{Herrmann}}}, \bibinfo {author} {\bibfnamefont {J.~S.}\ \bibnamefont
  {{Hesthaven}}}, \bibinfo {author} {\bibfnamefont {E.}~\bibnamefont
  {{Ochsner}}}, \ and\ \bibinfo {author} {\bibfnamefont {M.}~\bibnamefont
  {{Tiglio}}},\ }\href {\doibase10.1103/PhysRevLett.106.221102} {\bibfield
  {journal} {\bibinfo  {journal} {Physical Review Letters}\ }\textbf {\bibinfo
  {volume} {106}},\ \bibinfo {eid} {221102} (\bibinfo {year} {2011})},\ \Eprint
  {http://arxiv.org/abs/1101.3765} {arXiv:1101.3765 [gr-qc]}\BibitemShut
  {NoStop}%
\bibitem [{\citenamefont {{Field}}\ \emph {et~al.}(2012)\citenamefont
  {{Field}}, \citenamefont {{Galley}},\ and\ \citenamefont
  {{Ochsner}}}]{2012PhRvD..86h4046F}%
  \BibitemOpen
  \bibfield  {author} {\bibinfo {author} {\bibfnamefont {S.~E.}\ \bibnamefont
  {{Field}}}, \bibinfo {author} {\bibfnamefont {C.~R.}\ \bibnamefont
  {{Galley}}}, \ and\ \bibinfo {author} {\bibfnamefont {E.}~\bibnamefont
  {{Ochsner}}},\ }\href {\doibase10.1103/PhysRevD.86.084046} {\bibfield
  {journal} {\bibinfo  {journal} {Physical Review D}\ }\textbf {\bibinfo
  {volume} {86}},\ \bibinfo {eid} {084046} (\bibinfo {year} {2012})},\ \Eprint
  {http://arxiv.org/abs/1205.6009} {arXiv:1205.6009 [gr-qc]}\BibitemShut
  {NoStop}%
\bibitem [{\citenamefont {{Canizares}}\ \emph {et~al.}(2013)\citenamefont
  {{Canizares}}, \citenamefont {{Field}}, \citenamefont {{Gair}},\ and\
  \citenamefont {{Tiglio}}}]{2013PhRvD..87l4005C}%
  \BibitemOpen
  \bibfield  {author} {\bibinfo {author} {\bibfnamefont {P.}~\bibnamefont
  {{Canizares}}}, \bibinfo {author} {\bibfnamefont {S.~E.}\ \bibnamefont
  {{Field}}}, \bibinfo {author} {\bibfnamefont {J.~R.}\ \bibnamefont {{Gair}}},
  \ and\ \bibinfo {author} {\bibfnamefont {M.}~\bibnamefont {{Tiglio}}},\
  }\href {\doibase10.1103/PhysRevD.87.124005} {\bibfield  {journal} {\bibinfo
  {journal} {Physical Review D}\ }\textbf {\bibinfo {volume} {87}},\ \bibinfo
  {eid} {124005} (\bibinfo {year} {2013})},\ \Eprint
  {http://arxiv.org/abs/1304.0462} {arXiv:1304.0462 [gr-qc]}\BibitemShut
  {NoStop}%
\bibitem [{\citenamefont {{Field}}\ \emph {et~al.}(2014)\citenamefont
  {{Field}}, \citenamefont {{Galley}}, \citenamefont {{Hesthaven}},
  \citenamefont {{Kaye}},\ and\ \citenamefont
  {{Tiglio}}}]{2014PhRvX...4c1006F}%
  \BibitemOpen
  \bibfield  {author} {\bibinfo {author} {\bibfnamefont {S.~E.}\ \bibnamefont
  {{Field}}}, \bibinfo {author} {\bibfnamefont {C.~R.}\ \bibnamefont
  {{Galley}}}, \bibinfo {author} {\bibfnamefont {J.~S.}\ \bibnamefont
  {{Hesthaven}}}, \bibinfo {author} {\bibfnamefont {J.}~\bibnamefont {{Kaye}}},
  \ and\ \bibinfo {author} {\bibfnamefont {M.}~\bibnamefont {{Tiglio}}},\
  }\href {\doibase10.1103/PhysRevX.4.031006} {\bibfield  {journal} {\bibinfo
  {journal} {Physical Review X}\ }\textbf {\bibinfo {volume} {4}},\ \bibinfo
  {eid} {031006} (\bibinfo {year} {2014})},\ \Eprint
  {http://arxiv.org/abs/1308.3565} {arXiv:1308.3565 [gr-qc]}\BibitemShut
  {NoStop}%
\bibitem [{\citenamefont {{Canizares}}\ \emph {et~al.}(2015)\citenamefont
  {{Canizares}}, \citenamefont {{Field}}, \citenamefont {{Gair}}, \citenamefont
  {{Raymond}}, \citenamefont {{Smith}},\ and\ \citenamefont
  {{Tiglio}}}]{2015PhRvL.114g1104C}%
  \BibitemOpen
  \bibfield  {author} {\bibinfo {author} {\bibfnamefont {P.}~\bibnamefont
  {{Canizares}}}, \bibinfo {author} {\bibfnamefont {S.~E.}\ \bibnamefont
  {{Field}}}, \bibinfo {author} {\bibfnamefont {J.}~\bibnamefont {{Gair}}},
  \bibinfo {author} {\bibfnamefont {V.}~\bibnamefont {{Raymond}}}, \bibinfo
  {author} {\bibfnamefont {R.}~\bibnamefont {{Smith}}}, \ and\ \bibinfo
  {author} {\bibfnamefont {M.}~\bibnamefont {{Tiglio}}},\ }\href
  {\doibase10.1103/PhysRevLett.114.071104} {\bibfield  {journal} {\bibinfo
  {journal} {Physical Review Letters}\ }\textbf {\bibinfo {volume} {114}},\
  \bibinfo {eid} {071104} (\bibinfo {year} {2015})},\ \Eprint
  {http://arxiv.org/abs/1404.6284} {arXiv:1404.6284 [gr-qc]}\BibitemShut
  {NoStop}%
\bibitem [{\citenamefont {{Smith}}\ \emph {et~al.}(2016)\citenamefont
  {{Smith}}, \citenamefont {{Field}}, \citenamefont {{Blackburn}},
  \citenamefont {{Haster}}, \citenamefont {{P{\"u}rrer}}, \citenamefont
  {{Raymond}},\ and\ \citenamefont {{Schmidt}}}]{2016PhRvD..94d4031S}%
  \BibitemOpen
  \bibfield  {author} {\bibinfo {author} {\bibfnamefont {R.}~\bibnamefont
  {{Smith}}}, \bibinfo {author} {\bibfnamefont {S.~E.}\ \bibnamefont
  {{Field}}}, \bibinfo {author} {\bibfnamefont {K.}~\bibnamefont
  {{Blackburn}}}, \bibinfo {author} {\bibfnamefont {C.-J.}\ \bibnamefont
  {{Haster}}}, \bibinfo {author} {\bibfnamefont {M.}~\bibnamefont
  {{P{\"u}rrer}}}, \bibinfo {author} {\bibfnamefont {V.}~\bibnamefont
  {{Raymond}}}, \ and\ \bibinfo {author} {\bibfnamefont {P.}~\bibnamefont
  {{Schmidt}}},\ }\href {\doibase10.1103/PhysRevD.94.044031} {\bibfield
  {journal} {\bibinfo  {journal} {Physical Review D}\ }\textbf {\bibinfo
  {volume} {94}},\ \bibinfo {eid} {044031} (\bibinfo {year} {2016})},\ \Eprint
  {http://arxiv.org/abs/1604.08253} {arXiv:1604.08253 [gr-qc]}\BibitemShut
  {NoStop}%
\bibitem [{\citenamefont {{Peters}}\ and\ \citenamefont
  {{Mathews}}(1963)}]{1963PhRv..131..435P}%
  \BibitemOpen
  \bibfield  {author} {\bibinfo {author} {\bibfnamefont {P.~C.}\ \bibnamefont
  {{Peters}}}\ and\ \bibinfo {author} {\bibfnamefont {J.}~\bibnamefont
  {{Mathews}}},\ }\href {\doibase10.1103/PhysRev.131.435} {\bibfield  {journal}
  {\bibinfo  {journal} {Physical Review}\ }\textbf {\bibinfo {volume} {131}},\
  \bibinfo {pages} {435--440} (\bibinfo {year} {1963})}\BibitemShut {NoStop}%
\bibitem [{\citenamefont {{Arnaud}}\ \emph {et~al.}(2007)\citenamefont
  {{Arnaud}}, \citenamefont {{Babak}}, \citenamefont {{Baker}}, \citenamefont
  {{Benacquista}}, \citenamefont {{Cornish}}, \citenamefont {{Cutler}},
  \citenamefont {{Finn}}, \citenamefont {{Larson}}, \citenamefont
  {{Littenberg}}, \citenamefont {{Porter}}, \citenamefont {{Vallisneri}},
  \citenamefont {{Vecchio}}, \citenamefont {{Vinet}},\ and\ \citenamefont
  {{Data Challenge Task Force}}}]{2007CQGra..24S.551A}%
  \BibitemOpen
  \bibfield  {author} {\bibinfo {author} {\bibfnamefont {K.~A.}\ \bibnamefont
  {{Arnaud}}}, \bibinfo {author} {\bibfnamefont {S.}~\bibnamefont {{Babak}}},
  \bibinfo {author} {\bibfnamefont {J.~G.}\ \bibnamefont {{Baker}}}, \bibinfo
  {author} {\bibfnamefont {M.~J.}\ \bibnamefont {{Benacquista}}}, \bibinfo
  {author} {\bibfnamefont {N.~J.}\ \bibnamefont {{Cornish}}}, \bibinfo {author}
  {\bibfnamefont {C.}~\bibnamefont {{Cutler}}}, \bibinfo {author}
  {\bibfnamefont {L.~S.}\ \bibnamefont {{Finn}}}, \bibinfo {author}
  {\bibfnamefont {S.~L.}\ \bibnamefont {{Larson}}}, \bibinfo {author}
  {\bibfnamefont {T.}~\bibnamefont {{Littenberg}}}, \bibinfo {author}
  {\bibfnamefont {E.~K.}\ \bibnamefont {{Porter}}}, \bibinfo {author}
  {\bibfnamefont {M.}~\bibnamefont {{Vallisneri}}}, \bibinfo {author}
  {\bibfnamefont {A.}~\bibnamefont {{Vecchio}}}, \bibinfo {author}
  {\bibfnamefont {J.-Y.}\ \bibnamefont {{Vinet}}}, \ and\ \bibinfo {author}
  {\bibfnamefont {T.~M.~L.}\ \bibnamefont {{Data Challenge Task Force}}},\
  }\href {\doibase10.1088/0264-9381/24/19/S18} {\bibfield  {journal} {\bibinfo
  {journal} {Classical and Quantum Gravity}\ }\textbf {\bibinfo {volume}
  {24}},\ \bibinfo {pages} {S551--S564} (\bibinfo {year} {2007})},\ \Eprint
  {http://arxiv.org/abs/gr-qc/0701170} {gr-qc/0701170}\BibitemShut {NoStop}%
\bibitem [{\citenamefont {Babak}\ \emph {et~al.}(2008)\citenamefont {Babak}
  \emph {et~al.}}]{Babak:2007zd}%
  \BibitemOpen
  \bibfield  {author} {\bibinfo {author} {\bibfnamefont {Stanislav}\
  \bibnamefont {Babak}} \emph {et~al.} (\bibinfo {collaboration} {Mock LISA
  Data Challenge Task Force}),\ }\bibfield  {booktitle} {\emph {\bibinfo
  {booktitle} {{Proceedings, 18th International Conference on General
  Relativity and Gravitation (GRG18) and 7th Edoardo Amaldi Conference on
  Gravitational Waves (Amaldi7), Sydney, Australia, July 2007}}},\ }\href
  {\doibase10.1088/0264-9381/25/11/114037} {\bibfield  {journal} {\bibinfo
  {journal} {Class. Quant. Grav.}\ }\textbf {\bibinfo {volume} {25}},\ \bibinfo
  {pages} {114037} (\bibinfo {year} {2008})},\ \Eprint
  {http://arxiv.org/abs/0711.2667} {arXiv:0711.2667 [gr-qc]}\BibitemShut
  {NoStop}%
\bibitem [{\citenamefont {Barack}\ and\ \citenamefont
  {Cutler}(2007)}]{Barack:2006pq}%
  \BibitemOpen
  \bibfield  {author} {\bibinfo {author} {\bibfnamefont {Leor}\ \bibnamefont
  {Barack}}\ and\ \bibinfo {author} {\bibfnamefont {Curt}\ \bibnamefont
  {Cutler}},\ }\href {\doibase10.1103/PhysRevD.75.042003} {\bibfield  {journal}
  {\bibinfo  {journal} {Phys. Rev.}\ }\textbf {\bibinfo {volume} {D75}},\
  \bibinfo {pages} {042003} (\bibinfo {year} {2007})},\ \Eprint
  {http://arxiv.org/abs/gr-qc/0612029} {arXiv:gr-qc/0612029
  [gr-qc]}\BibitemShut {NoStop}%
\bibitem [{\citenamefont {Gair}\ and\ \citenamefont
  {Yunes}(2011)}]{Gair:2011ym}%
  \BibitemOpen
  \bibfield  {author} {\bibinfo {author} {\bibfnamefont {Jonathan}\
  \bibnamefont {Gair}}\ and\ \bibinfo {author} {\bibfnamefont {Nicolas}\
  \bibnamefont {Yunes}},\ }\href {\doibase10.1103/PhysRevD.84.064016}
  {\bibfield  {journal} {\bibinfo  {journal} {Phys. Rev.}\ }\textbf {\bibinfo
  {volume} {D84}},\ \bibinfo {pages} {064016} (\bibinfo {year} {2011})},\
  \Eprint {http://arxiv.org/abs/1106.6313} {arXiv:1106.6313
  [gr-qc]}\BibitemShut {NoStop}%
\bibitem [{\citenamefont {Gair}\ and\ \citenamefont
  {Glampedakis}(2006)}]{Gair:2005ih}%
  \BibitemOpen
  \bibfield  {author} {\bibinfo {author} {\bibfnamefont {Jonathan~R}\
  \bibnamefont {Gair}}\ and\ \bibinfo {author} {\bibfnamefont {Kostas}\
  \bibnamefont {Glampedakis}},\ }\href {\doibase10.1103/PhysRevD.73.064037}
  {\bibfield  {journal} {\bibinfo  {journal} {Phys. Rev.}\ }\textbf {\bibinfo
  {volume} {D73}},\ \bibinfo {pages} {064037} (\bibinfo {year} {2006})},\
  \Eprint {http://arxiv.org/abs/gr-qc/0510129}
  {arXiv:gr-qc/0510129}\BibitemShut {NoStop}%
\bibitem [{\citenamefont {Babak}\ \emph {et~al.}(2007)\citenamefont {Babak},
  \citenamefont {Fang}, \citenamefont {Gair}, \citenamefont {Glampedakis},\
  and\ \citenamefont {Hughes}}]{Babak:2006uv}%
  \BibitemOpen
  \bibfield  {author} {\bibinfo {author} {\bibfnamefont {Stanislav}\
  \bibnamefont {Babak}}, \bibinfo {author} {\bibfnamefont {Hua}\ \bibnamefont
  {Fang}}, \bibinfo {author} {\bibfnamefont {Jonathan~R.}\ \bibnamefont
  {Gair}}, \bibinfo {author} {\bibfnamefont {Kostas}\ \bibnamefont
  {Glampedakis}}, \ and\ \bibinfo {author} {\bibfnamefont {Scott~A.}\
  \bibnamefont {Hughes}},\ }\href {\doibase10.1103/PhysRevD.75.024005}
  {\bibfield  {journal} {\bibinfo  {journal} {Phys. Rev.}\ }\textbf {\bibinfo
  {volume} {D75}},\ \bibinfo {pages} {024005} (\bibinfo {year} {2007})},\
  \bibinfo {note} {[Erratum: Phys. Rev.D77,04990(2008)]},\ \Eprint
  {http://arxiv.org/abs/gr-qc/0607007} {arXiv:gr-qc/0607007
  [gr-qc]}\BibitemShut {NoStop}%
\bibitem [{\citenamefont {{Huerta}}\ and\ \citenamefont
  {{Gair}}(2009)}]{2009PhRvD..79h4021H}%
  \BibitemOpen
  \bibfield  {author} {\bibinfo {author} {\bibfnamefont {E.~A.}\ \bibnamefont
  {{Huerta}}}\ and\ \bibinfo {author} {\bibfnamefont {J.~R.}\ \bibnamefont
  {{Gair}}},\ }\href {\doibase10.1103/PhysRevD.79.084021} {\bibfield  {journal}
  {\bibinfo  {journal} {Physical Review D}\ }\textbf {\bibinfo {volume} {79}},\
  \bibinfo {pages} {084021--+} (\bibinfo {year} {2009})},\ \Eprint
  {http://arxiv.org/abs/0812.4208} {arXiv:0812.4208 [gr-qc]}\BibitemShut
  {NoStop}%
\bibitem [{\citenamefont {Pound}\ and\ \citenamefont
  {Poisson}(2008)}]{Pound:2007th}%
  \BibitemOpen
  \bibfield  {author} {\bibinfo {author} {\bibfnamefont {Adam}\ \bibnamefont
  {Pound}}\ and\ \bibinfo {author} {\bibfnamefont {Eric}\ \bibnamefont
  {Poisson}},\ }\href {\doibase10.1103/PhysRevD.77.044013} {\bibfield
  {journal} {\bibinfo  {journal} {Phys. Rev.}\ }\textbf {\bibinfo {volume}
  {D77}},\ \bibinfo {pages} {044013} (\bibinfo {year} {2008})},\ \Eprint
  {http://arxiv.org/abs/0708.3033} {arXiv:0708.3033 [gr-qc]}\BibitemShut
  {NoStop}%
\bibitem [{\citenamefont {Gair}\ \emph
  {et~al.}(2011{\natexlab{c}})\citenamefont {Gair}, \citenamefont {Flanagan},
  \citenamefont {Drasco}, \citenamefont {Hinderer},\ and\ \citenamefont
  {Babak}}]{Gair:2010iv}%
  \BibitemOpen
  \bibfield  {author} {\bibinfo {author} {\bibfnamefont {Jonathan~R.}\
  \bibnamefont {Gair}}, \bibinfo {author} {\bibfnamefont {Eanna~E.}\
  \bibnamefont {Flanagan}}, \bibinfo {author} {\bibfnamefont {Steve}\
  \bibnamefont {Drasco}}, \bibinfo {author} {\bibfnamefont {Tanja}\
  \bibnamefont {Hinderer}}, \ and\ \bibinfo {author} {\bibfnamefont
  {Stanislav}\ \bibnamefont {Babak}},\ }\href
  {\doibase10.1103/PhysRevD.83.044037} {\bibfield  {journal} {\bibinfo
  {journal} {Phys. Rev.}\ }\textbf {\bibinfo {volume} {D83}},\ \bibinfo {pages}
  {044037} (\bibinfo {year} {2011}{\natexlab{c}})}\BibitemShut {NoStop}%
\bibitem [{\citenamefont {{Chua}}\ and\ \citenamefont
  {{Gair}}(2015)}]{2015CQGra..32w2002C}%
  \BibitemOpen
  \bibfield  {author} {\bibinfo {author} {\bibfnamefont {A.~J.~K.}\
  \bibnamefont {{Chua}}}\ and\ \bibinfo {author} {\bibfnamefont {J.~R.}\
  \bibnamefont {{Gair}}},\ }\href {\doibase10.1088/0264-9381/32/23/232002}
  {\bibfield  {journal} {\bibinfo  {journal} {Classical and Quantum Gravity}\
  }\textbf {\bibinfo {volume} {32}},\ \bibinfo {eid} {232002} (\bibinfo {year}
  {2015})},\ \Eprint {http://arxiv.org/abs/1510.06245} {arXiv:1510.06245
  [gr-qc]}\BibitemShut {NoStop}%
\bibitem [{\citenamefont {Chrusciel}\ \emph {et~al.}(2012)\citenamefont
  {Chrusciel}, \citenamefont {Lopes~Costa},\ and\ \citenamefont
  {Heusler}}]{Chrusciel:2012jk}%
  \BibitemOpen
  \bibfield  {author} {\bibinfo {author} {\bibfnamefont {Piotr~T.}\
  \bibnamefont {Chrusciel}}, \bibinfo {author} {\bibfnamefont {Joao}\
  \bibnamefont {Lopes~Costa}}, \ and\ \bibinfo {author} {\bibfnamefont
  {Markus}\ \bibnamefont {Heusler}},\ }\href {\doibase10.12942/lrr-2012-7}
  {\bibfield  {journal} {\bibinfo  {journal} {Living Rev. Rel.}\ }\textbf
  {\bibinfo {volume} {15}},\ \bibinfo {pages} {7} (\bibinfo {year} {2012})},\
  \Eprint {http://arxiv.org/abs/1205.6112} {arXiv:1205.6112
  [gr-qc]}\BibitemShut {NoStop}%
\bibitem [{\citenamefont {Alexakis}\ \emph {et~al.}(2010)\citenamefont
  {Alexakis}, \citenamefont {Ionescu},\ and\ \citenamefont
  {Klainerman}}]{Alexakis:2009ch}%
  \BibitemOpen
  \bibfield  {author} {\bibinfo {author} {\bibfnamefont {S.}~\bibnamefont
  {Alexakis}}, \bibinfo {author} {\bibfnamefont {A.~D.}\ \bibnamefont
  {Ionescu}}, \ and\ \bibinfo {author} {\bibfnamefont {S.}~\bibnamefont
  {Klainerman}},\ }\href {\doibase10.1007/s00220-010-1072-1} {\bibfield
  {journal} {\bibinfo  {journal} {Commun. Math. Phys.}\ }\textbf {\bibinfo
  {volume} {299}},\ \bibinfo {pages} {89--127} (\bibinfo {year} {2010})},\
  \Eprint {http://arxiv.org/abs/0904.0982} {arXiv:0904.0982
  [gr-qc]}\BibitemShut {NoStop}%
\bibitem [{\citenamefont {Alexakis}\ \emph {et~al.}(2014)\citenamefont
  {Alexakis}, \citenamefont {Ionescu},\ and\ \citenamefont
  {Klainerman}}]{AIK3}%
  \BibitemOpen
  \bibfield  {author} {\bibinfo {author} {\bibfnamefont {S.}~\bibnamefont
  {Alexakis}}, \bibinfo {author} {\bibfnamefont {A.D.}\ \bibnamefont
  {Ionescu}}, \ and\ \bibinfo {author} {\bibfnamefont {S.}~\bibnamefont
  {Klainerman}},\ }\href {\doibase10.1215/00127094-2819517} {\bibfield
  {journal} {\bibinfo  {journal} {Duke Math.\ J.}\ }\textbf {\bibinfo {volume}
  {163}},\ \bibinfo {pages} {2603--2615} (\bibinfo {year} {2014})},\ \Eprint
  {http://arxiv.org/abs/1304.0487} {arXiv:1304.0487 [gr-qc]}\BibitemShut
  {NoStop}%
\bibitem [{\citenamefont {Reynolds}(2014)}]{Reynolds:2013qqa}%
  \BibitemOpen
  \bibfield  {author} {\bibinfo {author} {\bibfnamefont {Christopher~S.}\
  \bibnamefont {Reynolds}},\ }\href {\doibase10.1007/s11214-013-0006-6}
  {\bibfield  {journal} {\bibinfo  {journal} {Space Sci. Rev.}\ }\textbf
  {\bibinfo {volume} {183}},\ \bibinfo {pages} {277--294} (\bibinfo {year}
  {2014})},\ \Eprint {http://arxiv.org/abs/1302.3260} {arXiv:1302.3260
  [astro-ph.HE]}\BibitemShut {NoStop}%
\bibitem [{\citenamefont {Dyatlov}(2011)}]{Dyatlov}%
  \BibitemOpen
  \bibfield  {author} {\bibinfo {author} {\bibfnamefont {Semyon}\ \bibnamefont
  {Dyatlov}},\ }\href {\doibase10.1007/s00220-011-1286-x} {\bibfield  {journal}
  {\bibinfo  {journal} {Commun.\ Math.\ Phys.}\ }\textbf {\bibinfo {volume}
  {306}},\ \bibinfo {pages} {119--163} (\bibinfo {year} {2011})}\BibitemShut
  {NoStop}%
\bibitem [{\citenamefont {Dyatlov}(2015)}]{Dyatlov2}%
  \BibitemOpen
  \bibfield  {author} {\bibinfo {author} {\bibfnamefont {Semyon}\ \bibnamefont
  {Dyatlov}},\ }\href {\doibase10.1007/s00220-014-2255-y} {\bibfield  {journal}
  {\bibinfo  {journal} {Commun.\ Math.\ Phys.}\ }\textbf {\bibinfo {volume}
  {335}},\ \bibinfo {pages} {1445--1485} (\bibinfo {year} {2015})}\BibitemShut
  {NoStop}%
\bibitem [{\citenamefont {Dafermos}\ \emph {et~al.}(2018)\citenamefont
  {Dafermos}, \citenamefont {Holzegel},\ and\ \citenamefont
  {Rodnianski}}]{DHRT}%
  \BibitemOpen
  \bibfield  {author} {\bibinfo {author} {\bibfnamefont {Mihalis}\ \bibnamefont
  {Dafermos}}, \bibinfo {author} {\bibfnamefont {Gustav}\ \bibnamefont
  {Holzegel}}, \ and\ \bibinfo {author} {\bibfnamefont {Igor}\ \bibnamefont
  {Rodnianski}},\ }\href@noop {} {\  (\bibinfo {year} {2018})},\ \Eprint
  {http://arxiv.org/abs/1601.06467} {arXiv:1601.06467 [gr-qc]}\BibitemShut
  {NoStop}%
\bibitem [{\citenamefont {Klainerman}\ and\ \citenamefont
  {Szeftel}(2017)}]{KlainermanSzeftel}%
  \BibitemOpen
  \bibfield  {author} {\bibinfo {author} {\bibfnamefont {Sergiu}\ \bibnamefont
  {Klainerman}}\ and\ \bibinfo {author} {\bibfnamefont {Jeremie}\ \bibnamefont
  {Szeftel}},\ }\href@noop {} {\  (\bibinfo {year} {2017})},\ \Eprint
  {http://arxiv.org/abs/1711.07597} {arXiv:1711.07597 [gr-qc]}\BibitemShut
  {NoStop}%
\bibitem [{\citenamefont {Szabados}(2009)}]{Szabados:2009eka}%
  \BibitemOpen
  \bibfield  {author} {\bibinfo {author} {\bibfnamefont {László~B.}\
  \bibnamefont {Szabados}},\ }\href {\doibase10.12942/lrr-2009-4} {\bibfield
  {journal} {\bibinfo  {journal} {Living Rev. Rel.}\ }\textbf {\bibinfo
  {volume} {12}},\ \bibinfo {pages} {4} (\bibinfo {year} {2009})}\BibitemShut
  {NoStop}%
\bibitem [{\citenamefont {Chen}\ \emph {et~al.}(2018)\citenamefont {Chen},
  \citenamefont {Wang},\ and\ \citenamefont {Yau}}]{ChenWangYau1804}%
  \BibitemOpen
  \bibfield  {author} {\bibinfo {author} {\bibfnamefont {Po-Ning}\ \bibnamefont
  {Chen}}, \bibinfo {author} {\bibfnamefont {Mu-Tao}\ \bibnamefont {Wang}}, \
  and\ \bibinfo {author} {\bibfnamefont {Shing-Tung}\ \bibnamefont {Yau}},\
  }\href@noop {} {\  (\bibinfo {year} {2018})},\ \Eprint
  {http://arxiv.org/abs/1804.08216} {arXiv:1804.08216 [math.DG]}\BibitemShut
  {NoStop}%
\bibitem [{\citenamefont {Chen}\ \emph {et~al.}(2019)\citenamefont {Chen},
  \citenamefont {Wang}, \citenamefont {Wang},\ and\ \citenamefont
  {Yau}}]{Chen:2019obg}%
  \BibitemOpen
  \bibfield  {author} {\bibinfo {author} {\bibfnamefont {Po-Ning}\ \bibnamefont
  {Chen}}, \bibinfo {author} {\bibfnamefont {Mu-Tao}\ \bibnamefont {Wang}},
  \bibinfo {author} {\bibfnamefont {Ye-Kai}\ \bibnamefont {Wang}}, \ and\
  \bibinfo {author} {\bibfnamefont {Shing-Tung}\ \bibnamefont {Yau}},\
  }\href@noop {} {\  (\bibinfo {year} {2019})},\ \Eprint
  {http://arxiv.org/abs/1901.06952} {arXiv:1901.06952 [gr-qc]}\BibitemShut
  {NoStop}%
\bibitem [{\citenamefont {Gustafsson}\ \emph {et~al.}(2013)\citenamefont
  {Gustafsson}, \citenamefont {Kreiss},\ and\ \citenamefont {Oliger}}]{KGO}%
  \BibitemOpen
  \bibfield  {author} {\bibinfo {author} {\bibfnamefont {B.}~\bibnamefont
  {Gustafsson}}, \bibinfo {author} {\bibfnamefont {H.-O.}\ \bibnamefont
  {Kreiss}}, \ and\ \bibinfo {author} {\bibfnamefont {J.}~\bibnamefont
  {Oliger}},\ }\href {\doibase10.1002/9781118548448} {\emph {\bibinfo {title}
  {Time-dependent problems and difference methods}}},\ \bibinfo {edition}
  {2nd}\ ed.,\ Pure and Applied Mathematics (Hoboken)\ (\bibinfo  {publisher}
  {John Wiley \& Sons, Inc., Hoboken, NJ},\ \bibinfo {year} {2013})\ pp.\
  \bibinfo {pages} {xiv+509}\BibitemShut {NoStop}%
\bibitem [{\citenamefont {Sarbach}\ and\ \citenamefont
  {Tiglio}(2012)}]{Sarbach:2012pr}%
  \BibitemOpen
  \bibfield  {author} {\bibinfo {author} {\bibfnamefont {Olivier}\ \bibnamefont
  {Sarbach}}\ and\ \bibinfo {author} {\bibfnamefont {Manuel}\ \bibnamefont
  {Tiglio}},\ }\href {\doibase10.12942/lrr-2012-9} {\bibfield  {journal}
  {\bibinfo  {journal} {Living Rev. Rel.}\ }\textbf {\bibinfo {volume} {15}},\
  \bibinfo {pages} {9} (\bibinfo {year} {2012})},\ \Eprint
  {http://arxiv.org/abs/1203.6443} {arXiv:1203.6443 [gr-qc]}\BibitemShut
  {NoStop}%
\bibitem [{\citenamefont {Sotiriou}\ \emph {et~al.}(2008)\citenamefont
  {Sotiriou}, \citenamefont {Faraoni},\ and\ \citenamefont
  {Liberati}}]{Sotiriou:2007zu}%
  \BibitemOpen
  \bibfield  {author} {\bibinfo {author} {\bibfnamefont {Thomas~P}\
  \bibnamefont {Sotiriou}}, \bibinfo {author} {\bibfnamefont {Valerio}\
  \bibnamefont {Faraoni}}, \ and\ \bibinfo {author} {\bibfnamefont {Stefano}\
  \bibnamefont {Liberati}},\ }\href {\doibase10.1142/S0218271808012097}
  {\bibfield  {journal} {\bibinfo  {journal} {Int. J. Mod. Phys.}\ }\textbf
  {\bibinfo {volume} {D17}},\ \bibinfo {pages} {399--423} (\bibinfo {year}
  {2008})},\ \Eprint {http://arxiv.org/abs/0707.2748} {arXiv:0707.2748
  [gr-qc]}\BibitemShut {NoStop}%
\bibitem [{\citenamefont {Perivolaropoulos}(2010)}]{Perivolaropoulos:2009ak}%
  \BibitemOpen
  \bibfield  {author} {\bibinfo {author} {\bibfnamefont {L.}~\bibnamefont
  {Perivolaropoulos}},\ }\href {\doibase10.1103/PhysRevD.81.047501} {\bibfield
  {journal} {\bibinfo  {journal} {Phys. Rev.}\ }\textbf {\bibinfo {volume}
  {D81}},\ \bibinfo {pages} {047501} (\bibinfo {year} {2010})},\ \Eprint
  {http://arxiv.org/abs/0911.3401} {arXiv:0911.3401 [gr-qc]}\BibitemShut
  {NoStop}%
\bibitem [{\citenamefont {Damour}\ and\ \citenamefont
  {Esposito-Farese}(1996)}]{Damour:1996ke}%
  \BibitemOpen
  \bibfield  {author} {\bibinfo {author} {\bibfnamefont {Thibault}\
  \bibnamefont {Damour}}\ and\ \bibinfo {author} {\bibfnamefont {Gilles}\
  \bibnamefont {Esposito-Farese}},\ }\href {\doibase10.1103/PhysRevD.54.1474}
  {\bibfield  {journal} {\bibinfo  {journal} {Phys. Rev.}\ }\textbf {\bibinfo
  {volume} {D54}},\ \bibinfo {pages} {1474--1491} (\bibinfo {year} {1996})},\
  \Eprint {http://arxiv.org/abs/gr-qc/9602056} {arXiv:gr-qc/9602056
  [gr-qc]}\BibitemShut {NoStop}%
\bibitem [{\citenamefont {Will}(2014)}]{Will:2014kxa}%
  \BibitemOpen
  \bibfield  {author} {\bibinfo {author} {\bibfnamefont {Clifford~M.}\
  \bibnamefont {Will}},\ }\href {\doibase10.12942/lrr-2014-4} {\bibfield
  {journal} {\bibinfo  {journal} {Living Rev. Rel.}\ }\textbf {\bibinfo
  {volume} {17}},\ \bibinfo {pages} {4} (\bibinfo {year} {2014})},\ \Eprint
  {http://arxiv.org/abs/1403.7377} {arXiv:1403.7377 [gr-qc]}\BibitemShut
  {NoStop}%
\bibitem [{\citenamefont {Freire}\ \emph {et~al.}(2012)\citenamefont {Freire},
  \citenamefont {Wex}, \citenamefont {Esposito-Farese}, \citenamefont
  {Verbiest}, \citenamefont {Bailes}, \citenamefont {Jacoby}, \citenamefont
  {Kramer}, \citenamefont {Stairs}, \citenamefont {Antoniadis},\ and\
  \citenamefont {Janssen}}]{Freire:2012mg}%
  \BibitemOpen
  \bibfield  {author} {\bibinfo {author} {\bibfnamefont {Paulo C.~C.}\
  \bibnamefont {Freire}}, \bibinfo {author} {\bibfnamefont {Norbert}\
  \bibnamefont {Wex}}, \bibinfo {author} {\bibfnamefont {Gilles}\ \bibnamefont
  {Esposito-Farese}}, \bibinfo {author} {\bibfnamefont {Joris P.~W.}\
  \bibnamefont {Verbiest}}, \bibinfo {author} {\bibfnamefont {Matthew}\
  \bibnamefont {Bailes}}, \bibinfo {author} {\bibfnamefont {Bryan~A.}\
  \bibnamefont {Jacoby}}, \bibinfo {author} {\bibfnamefont {Michael}\
  \bibnamefont {Kramer}}, \bibinfo {author} {\bibfnamefont {Ingrid~H.}\
  \bibnamefont {Stairs}}, \bibinfo {author} {\bibfnamefont {John}\ \bibnamefont
  {Antoniadis}}, \ and\ \bibinfo {author} {\bibfnamefont {Gemma~H.}\
  \bibnamefont {Janssen}},\ }\href {\doibase10.1111/j.1365-2966.2012.21253.x}
  {\bibfield  {journal} {\bibinfo  {journal} {Mon. Not. Roy. Astron. Soc.}\
  }\textbf {\bibinfo {volume} {423}},\ \bibinfo {pages} {3328} (\bibinfo {year}
  {2012})},\ \Eprint {http://arxiv.org/abs/1205.1450} {arXiv:1205.1450
  [astro-ph.GA]}\BibitemShut {NoStop}%
\bibitem [{\citenamefont {Shao}\ \emph {et~al.}(2017)\citenamefont {Shao},
  \citenamefont {Sennett}, \citenamefont {Buonanno}, \citenamefont {Kramer},\
  and\ \citenamefont {Wex}}]{Shao:2017gwu}%
  \BibitemOpen
  \bibfield  {author} {\bibinfo {author} {\bibfnamefont {Lijing}\ \bibnamefont
  {Shao}}, \bibinfo {author} {\bibfnamefont {Noah}\ \bibnamefont {Sennett}},
  \bibinfo {author} {\bibfnamefont {Alessandra}\ \bibnamefont {Buonanno}},
  \bibinfo {author} {\bibfnamefont {Michael}\ \bibnamefont {Kramer}}, \ and\
  \bibinfo {author} {\bibfnamefont {Norbert}\ \bibnamefont {Wex}},\ }\href
  {\doibase10.1103/PhysRevX.7.041025} {\bibfield  {journal} {\bibinfo
  {journal} {Phys. Rev.}\ }\textbf {\bibinfo {volume} {X7}},\ \bibinfo {pages}
  {041025} (\bibinfo {year} {2017})},\ \Eprint
  {http://arxiv.org/abs/1704.07561} {arXiv:1704.07561 [gr-qc]}\BibitemShut
  {NoStop}%
\bibitem [{\citenamefont {Shibata}\ \emph {et~al.}(2014)\citenamefont
  {Shibata}, \citenamefont {Taniguchi}, \citenamefont {Okawa},\ and\
  \citenamefont {Buonanno}}]{Shibata:2013pra}%
  \BibitemOpen
  \bibfield  {author} {\bibinfo {author} {\bibfnamefont {Masaru}\ \bibnamefont
  {Shibata}}, \bibinfo {author} {\bibfnamefont {Keisuke}\ \bibnamefont
  {Taniguchi}}, \bibinfo {author} {\bibfnamefont {Hirotada}\ \bibnamefont
  {Okawa}}, \ and\ \bibinfo {author} {\bibfnamefont {Alessandra}\ \bibnamefont
  {Buonanno}},\ }\href {\doibase10.1103/PhysRevD.89.084005} {\bibfield
  {journal} {\bibinfo  {journal} {Phys. Rev.}\ }\textbf {\bibinfo {volume}
  {D89}},\ \bibinfo {pages} {084005} (\bibinfo {year} {2014})},\ \Eprint
  {http://arxiv.org/abs/1310.0627} {arXiv:1310.0627 [gr-qc]}\BibitemShut
  {NoStop}%
\bibitem [{\citenamefont {Taniguchi}\ \emph {et~al.}(2015)\citenamefont
  {Taniguchi}, \citenamefont {Shibata},\ and\ \citenamefont
  {Buonanno}}]{Taniguchi:2014fqa}%
  \BibitemOpen
  \bibfield  {author} {\bibinfo {author} {\bibfnamefont {Keisuke}\ \bibnamefont
  {Taniguchi}}, \bibinfo {author} {\bibfnamefont {Masaru}\ \bibnamefont
  {Shibata}}, \ and\ \bibinfo {author} {\bibfnamefont {Alessandra}\
  \bibnamefont {Buonanno}},\ }\href {\doibase10.1103/PhysRevD.91.024033}
  {\bibfield  {journal} {\bibinfo  {journal} {Phys. Rev.}\ }\textbf {\bibinfo
  {volume} {D91}},\ \bibinfo {pages} {024033} (\bibinfo {year} {2015})},\
  \Eprint {http://arxiv.org/abs/1410.0738} {arXiv:1410.0738
  [gr-qc]}\BibitemShut {NoStop}%
\bibitem [{\citenamefont {Sennett}\ and\ \citenamefont
  {Buonanno}(2016)}]{Sennett:2016rwa}%
  \BibitemOpen
  \bibfield  {author} {\bibinfo {author} {\bibfnamefont {Noah}\ \bibnamefont
  {Sennett}}\ and\ \bibinfo {author} {\bibfnamefont {Alessandra}\ \bibnamefont
  {Buonanno}},\ }\href {\doibase10.1103/PhysRevD.93.124004} {\bibfield
  {journal} {\bibinfo  {journal} {Phys. Rev.}\ }\textbf {\bibinfo {volume}
  {D93}},\ \bibinfo {pages} {124004} (\bibinfo {year} {2016})},\ \Eprint
  {http://arxiv.org/abs/1603.03300} {arXiv:1603.03300 [gr-qc]}\BibitemShut
  {NoStop}%
\bibitem [{\citenamefont {Sampson}\ \emph {et~al.}(2014)\citenamefont
  {Sampson}, \citenamefont {Yunes}, \citenamefont {Cornish}, \citenamefont
  {Ponce}, \citenamefont {Barausse}, \citenamefont {Klein}, \citenamefont
  {Palenzuela},\ and\ \citenamefont {Lehner}}]{Sampson:2014qqa}%
  \BibitemOpen
  \bibfield  {author} {\bibinfo {author} {\bibfnamefont {Laura}\ \bibnamefont
  {Sampson}}, \bibinfo {author} {\bibfnamefont {Nicolás}\ \bibnamefont
  {Yunes}}, \bibinfo {author} {\bibfnamefont {Neil}\ \bibnamefont {Cornish}},
  \bibinfo {author} {\bibfnamefont {Marcelo}\ \bibnamefont {Ponce}}, \bibinfo
  {author} {\bibfnamefont {Enrico}\ \bibnamefont {Barausse}}, \bibinfo {author}
  {\bibfnamefont {Antoine}\ \bibnamefont {Klein}}, \bibinfo {author}
  {\bibfnamefont {Carlos}\ \bibnamefont {Palenzuela}}, \ and\ \bibinfo {author}
  {\bibfnamefont {Luis}\ \bibnamefont {Lehner}},\ }\href
  {\doibase10.1103/PhysRevD.90.124091} {\bibfield  {journal} {\bibinfo
  {journal} {Phys. Rev.}\ }\textbf {\bibinfo {volume} {D90}},\ \bibinfo {pages}
  {124091} (\bibinfo {year} {2014})},\ \Eprint {http://arxiv.org/abs/1407.7038}
  {arXiv:1407.7038 [gr-qc]}\BibitemShut {NoStop}%
\bibitem [{\citenamefont {Anderson}\ and\ \citenamefont
  {Yunes}(2017)}]{Anderson:2017phb}%
  \BibitemOpen
  \bibfield  {author} {\bibinfo {author} {\bibfnamefont {David}\ \bibnamefont
  {Anderson}}\ and\ \bibinfo {author} {\bibfnamefont {Nicolás}\ \bibnamefont
  {Yunes}},\ }\href {\doibase10.1103/PhysRevD.96.064037} {\bibfield  {journal}
  {\bibinfo  {journal} {Phys. Rev.}\ }\textbf {\bibinfo {volume} {D96}},\
  \bibinfo {pages} {064037} (\bibinfo {year} {2017})},\ \Eprint
  {http://arxiv.org/abs/1705.06351} {arXiv:1705.06351 [gr-qc]}\BibitemShut
  {NoStop}%
\bibitem [{\citenamefont {Hawking}(1972{\natexlab{a}})}]{Hawking:1972qk}%
  \BibitemOpen
  \bibfield  {author} {\bibinfo {author} {\bibfnamefont {S.~W.}\ \bibnamefont
  {Hawking}},\ }\href {\doibase10.1007/BF01877518} {\bibfield  {journal}
  {\bibinfo  {journal} {Commun. Math. Phys.}\ }\textbf {\bibinfo {volume}
  {25}},\ \bibinfo {pages} {167--171} (\bibinfo {year}
  {1972}{\natexlab{a}})}\BibitemShut {NoStop}%
\bibitem [{\citenamefont {Bekenstein}(1995)}]{Bekenstein:1995un}%
  \BibitemOpen
  \bibfield  {author} {\bibinfo {author} {\bibfnamefont {J.~D.}\ \bibnamefont
  {Bekenstein}},\ }\href {\doibase10.1103/PhysRevD.51.R6608} {\bibfield
  {journal} {\bibinfo  {journal} {Phys. Rev.}\ }\textbf {\bibinfo {volume}
  {D51}},\ \bibinfo {pages} {R6608} (\bibinfo {year} {1995})}\BibitemShut
  {NoStop}%
\bibitem [{\citenamefont {Sotiriou}\ and\ \citenamefont
  {Faraoni}(2012)}]{Sotiriou:2011dz}%
  \BibitemOpen
  \bibfield  {author} {\bibinfo {author} {\bibfnamefont {Thomas~P.}\
  \bibnamefont {Sotiriou}}\ and\ \bibinfo {author} {\bibfnamefont {Valerio}\
  \bibnamefont {Faraoni}},\ }\href {\doibase10.1103/PhysRevLett.108.081103}
  {\bibfield  {journal} {\bibinfo  {journal} {Phys. Rev. Lett.}\ }\textbf
  {\bibinfo {volume} {108}},\ \bibinfo {pages} {081103} (\bibinfo {year}
  {2012})},\ \Eprint {http://arxiv.org/abs/1109.6324} {arXiv:1109.6324
  [gr-qc]}\BibitemShut {NoStop}%
\bibitem [{\citenamefont {Sotiriou}(2015{\natexlab{a}})}]{Sotiriou:2015pka}%
  \BibitemOpen
  \bibfield  {author} {\bibinfo {author} {\bibfnamefont {Thomas~P.}\
  \bibnamefont {Sotiriou}},\ }\href {\doibase10.1088/0264-9381/32/21/214002}
  {\bibfield  {journal} {\bibinfo  {journal} {Class. Quant. Grav.}\ }\textbf
  {\bibinfo {volume} {32}},\ \bibinfo {pages} {214002} (\bibinfo {year}
  {2015}{\natexlab{a}})},\ \Eprint {http://arxiv.org/abs/1505.00248}
  {arXiv:1505.00248 [gr-qc]}\BibitemShut {NoStop}%
\bibitem [{\citenamefont {Horndeski}(1974)}]{Horndeski:1974wa}%
  \BibitemOpen
  \bibfield  {author} {\bibinfo {author} {\bibfnamefont {Gregory~Walter}\
  \bibnamefont {Horndeski}},\ }\href {\doibase10.1007/BF01807638} {\bibfield
  {journal} {\bibinfo  {journal} {Int. J. Theor. Phys.}\ }\textbf {\bibinfo
  {volume} {10}},\ \bibinfo {pages} {363--384} (\bibinfo {year}
  {1974})}\BibitemShut {NoStop}%
\bibitem [{\citenamefont {Deffayet}\ \emph {et~al.}(2009)\citenamefont
  {Deffayet}, \citenamefont {Deser},\ and\ \citenamefont
  {Esposito-Farese}}]{Deffayet:2009mn}%
  \BibitemOpen
  \bibfield  {author} {\bibinfo {author} {\bibfnamefont {C.}~\bibnamefont
  {Deffayet}}, \bibinfo {author} {\bibfnamefont {S.}~\bibnamefont {Deser}}, \
  and\ \bibinfo {author} {\bibfnamefont {G.}~\bibnamefont {Esposito-Farese}},\
  }\href {\doibase10.1103/PhysRevD.80.064015} {\bibfield  {journal} {\bibinfo
  {journal} {Phys. Rev.}\ }\textbf {\bibinfo {volume} {D80}},\ \bibinfo {pages}
  {064015} (\bibinfo {year} {2009})},\ \Eprint {http://arxiv.org/abs/0906.1967}
  {arXiv:0906.1967 [gr-qc]}\BibitemShut {NoStop}%
\bibitem [{\citenamefont {Nicolis}\ \emph {et~al.}(2009)\citenamefont
  {Nicolis}, \citenamefont {Rattazzi},\ and\ \citenamefont
  {Trincherini}}]{Nicolis:2008in}%
  \BibitemOpen
  \bibfield  {author} {\bibinfo {author} {\bibfnamefont {Alberto}\ \bibnamefont
  {Nicolis}}, \bibinfo {author} {\bibfnamefont {Riccardo}\ \bibnamefont
  {Rattazzi}}, \ and\ \bibinfo {author} {\bibfnamefont {Enrico}\ \bibnamefont
  {Trincherini}},\ }\href {\doibase10.1103/PhysRevD.79.064036} {\bibfield
  {journal} {\bibinfo  {journal} {Phys. Rev.}\ }\textbf {\bibinfo {volume}
  {D79}},\ \bibinfo {pages} {064036} (\bibinfo {year} {2009})},\ \Eprint
  {http://arxiv.org/abs/0811.2197} {arXiv:0811.2197 [hep-th]}\BibitemShut
  {NoStop}%
\bibitem [{\citenamefont {Hui}\ and\ \citenamefont
  {Nicolis}(2013)}]{Hui:2012qt}%
  \BibitemOpen
  \bibfield  {author} {\bibinfo {author} {\bibfnamefont {Lam}\ \bibnamefont
  {Hui}}\ and\ \bibinfo {author} {\bibfnamefont {Alberto}\ \bibnamefont
  {Nicolis}},\ }\href {\doibase10.1103/PhysRevLett.110.241104} {\bibfield
  {journal} {\bibinfo  {journal} {Phys. Rev. Lett.}\ }\textbf {\bibinfo
  {volume} {110}},\ \bibinfo {pages} {241104} (\bibinfo {year} {2013})},\
  \Eprint {http://arxiv.org/abs/1202.1296} {arXiv:1202.1296
  [hep-th]}\BibitemShut {NoStop}%
\bibitem [{\citenamefont {Sotiriou}\ and\ \citenamefont
  {Zhou}(2014{\natexlab{a}})}]{Sotiriou:2013qea}%
  \BibitemOpen
  \bibfield  {author} {\bibinfo {author} {\bibfnamefont {Thomas~P.}\
  \bibnamefont {Sotiriou}}\ and\ \bibinfo {author} {\bibfnamefont
  {Shuang-Yong}\ \bibnamefont {Zhou}},\ }\href
  {\doibase10.1103/PhysRevLett.112.251102} {\bibfield  {journal} {\bibinfo
  {journal} {Phys. Rev. Lett.}\ }\textbf {\bibinfo {volume} {112}},\ \bibinfo
  {pages} {251102} (\bibinfo {year} {2014}{\natexlab{a}})},\ \Eprint
  {http://arxiv.org/abs/1312.3622} {arXiv:1312.3622 [gr-qc]}\BibitemShut
  {NoStop}%
\bibitem [{\citenamefont {Kobayashi}\ \emph {et~al.}(2011)\citenamefont
  {Kobayashi}, \citenamefont {Yamaguchi},\ and\ \citenamefont
  {Yokoyama}}]{Kobayashi:2011nu}%
  \BibitemOpen
  \bibfield  {author} {\bibinfo {author} {\bibfnamefont {Tsutomu}\ \bibnamefont
  {Kobayashi}}, \bibinfo {author} {\bibfnamefont {Masahide}\ \bibnamefont
  {Yamaguchi}}, \ and\ \bibinfo {author} {\bibfnamefont {Jun'ichi}\
  \bibnamefont {Yokoyama}},\ }\href {\doibase10.1143/PTP.126.511} {\bibfield
  {journal} {\bibinfo  {journal} {Prog. Theor. Phys.}\ }\textbf {\bibinfo
  {volume} {126}},\ \bibinfo {pages} {511--529} (\bibinfo {year} {2011})},\
  \Eprint {http://arxiv.org/abs/1105.5723} {arXiv:1105.5723
  [hep-th]}\BibitemShut {NoStop}%
\bibitem [{\citenamefont {Yunes}\ and\ \citenamefont
  {Stein}(2011)}]{Yunes:2011we}%
  \BibitemOpen
  \bibfield  {author} {\bibinfo {author} {\bibfnamefont {Nicolas}\ \bibnamefont
  {Yunes}}\ and\ \bibinfo {author} {\bibfnamefont {Leo~C.}\ \bibnamefont
  {Stein}},\ }\href {\doibase10.1103/PhysRevD.83.104002} {\bibfield  {journal}
  {\bibinfo  {journal} {Phys. Rev.}\ }\textbf {\bibinfo {volume} {D83}},\
  \bibinfo {pages} {104002} (\bibinfo {year} {2011})},\ \Eprint
  {http://arxiv.org/abs/1101.2921} {arXiv:1101.2921 [gr-qc]}\BibitemShut
  {NoStop}%
\bibitem [{\citenamefont {Sotiriou}\ and\ \citenamefont
  {Zhou}(2014{\natexlab{b}})}]{Sotiriou:2014pfa}%
  \BibitemOpen
  \bibfield  {author} {\bibinfo {author} {\bibfnamefont {Thomas~P.}\
  \bibnamefont {Sotiriou}}\ and\ \bibinfo {author} {\bibfnamefont
  {Shuang-Yong}\ \bibnamefont {Zhou}},\ }\href
  {\doibase10.1103/PhysRevD.90.124063} {\bibfield  {journal} {\bibinfo
  {journal} {Phys. Rev.}\ }\textbf {\bibinfo {volume} {D90}},\ \bibinfo {pages}
  {124063} (\bibinfo {year} {2014}{\natexlab{b}})},\ \Eprint
  {http://arxiv.org/abs/1408.1698} {arXiv:1408.1698 [gr-qc]}\BibitemShut
  {NoStop}%
\bibitem [{\citenamefont {Campbell}\ \emph {et~al.}(1992)\citenamefont
  {Campbell}, \citenamefont {Kaloper},\ and\ \citenamefont
  {Olive}}]{Campbell:1991kz}%
  \BibitemOpen
  \bibfield  {author} {\bibinfo {author} {\bibfnamefont {Bruce~A.}\
  \bibnamefont {Campbell}}, \bibinfo {author} {\bibfnamefont {Nemanja}\
  \bibnamefont {Kaloper}}, \ and\ \bibinfo {author} {\bibfnamefont {Keith~A.}\
  \bibnamefont {Olive}},\ }\href {\doibase10.1016/0370-2693(92)91452-F}
  {\bibfield  {journal} {\bibinfo  {journal} {Phys. Lett.}\ }\textbf {\bibinfo
  {volume} {B285}},\ \bibinfo {pages} {199--205} (\bibinfo {year}
  {1992})}\BibitemShut {NoStop}%
\bibitem [{\citenamefont {Kanti}\ \emph {et~al.}(1996)\citenamefont {Kanti},
  \citenamefont {Mavromatos}, \citenamefont {Rizos}, \citenamefont {Tamvakis},\
  and\ \citenamefont {Winstanley}}]{Kanti:1995vq}%
  \BibitemOpen
  \bibfield  {author} {\bibinfo {author} {\bibfnamefont {P.}~\bibnamefont
  {Kanti}}, \bibinfo {author} {\bibfnamefont {N.~E.}\ \bibnamefont
  {Mavromatos}}, \bibinfo {author} {\bibfnamefont {J.}~\bibnamefont {Rizos}},
  \bibinfo {author} {\bibfnamefont {K.}~\bibnamefont {Tamvakis}}, \ and\
  \bibinfo {author} {\bibfnamefont {E.}~\bibnamefont {Winstanley}},\ }\href
  {\doibase10.1103/PhysRevD.54.5049} {\bibfield  {journal} {\bibinfo  {journal}
  {Phys. Rev.}\ }\textbf {\bibinfo {volume} {D54}},\ \bibinfo {pages}
  {5049--5058} (\bibinfo {year} {1996})},\ \Eprint
  {http://arxiv.org/abs/hep-th/9511071} {arXiv:hep-th/9511071
  [hep-th]}\BibitemShut {NoStop}%
\bibitem [{\citenamefont {Metsaev}\ and\ \citenamefont
  {Tseytlin}(1987)}]{Metsaev:1987zx}%
  \BibitemOpen
  \bibfield  {author} {\bibinfo {author} {\bibfnamefont {R.~R.}\ \bibnamefont
  {Metsaev}}\ and\ \bibinfo {author} {\bibfnamefont {Arkady~A.}\ \bibnamefont
  {Tseytlin}},\ }\href {\doibase10.1016/0550-3213(87)90077-0} {\bibfield
  {journal} {\bibinfo  {journal} {Nucl. Phys.}\ }\textbf {\bibinfo {volume}
  {B293}},\ \bibinfo {pages} {385--419} (\bibinfo {year} {1987})}\BibitemShut
  {NoStop}%
\bibitem [{\citenamefont {Maeda}\ \emph {et~al.}(2009)\citenamefont {Maeda},
  \citenamefont {Ohta},\ and\ \citenamefont {Sasagawa}}]{Maeda:2009uy}%
  \BibitemOpen
  \bibfield  {author} {\bibinfo {author} {\bibfnamefont {Kei-ichi}\
  \bibnamefont {Maeda}}, \bibinfo {author} {\bibfnamefont {Nobuyoshi}\
  \bibnamefont {Ohta}}, \ and\ \bibinfo {author} {\bibfnamefont {Yukinori}\
  \bibnamefont {Sasagawa}},\ }\href {\doibase10.1103/PhysRevD.80.104032}
  {\bibfield  {journal} {\bibinfo  {journal} {Phys. Rev.}\ }\textbf {\bibinfo
  {volume} {D80}},\ \bibinfo {pages} {104032} (\bibinfo {year} {2009})},\
  \Eprint {http://arxiv.org/abs/0908.4151} {arXiv:0908.4151
  [hep-th]}\BibitemShut {NoStop}%
\bibitem [{\citenamefont {Silva}\ \emph {et~al.}(2018)\citenamefont {Silva},
  \citenamefont {Sakstein}, \citenamefont {Gualtieri}, \citenamefont
  {Sotiriou},\ and\ \citenamefont {Berti}}]{Silva:2017uqg}%
  \BibitemOpen
  \bibfield  {author} {\bibinfo {author} {\bibfnamefont {Hector~O.}\
  \bibnamefont {Silva}}, \bibinfo {author} {\bibfnamefont {Jeremy}\
  \bibnamefont {Sakstein}}, \bibinfo {author} {\bibfnamefont {Leonardo}\
  \bibnamefont {Gualtieri}}, \bibinfo {author} {\bibfnamefont {Thomas~P.}\
  \bibnamefont {Sotiriou}}, \ and\ \bibinfo {author} {\bibfnamefont {Emanuele}\
  \bibnamefont {Berti}},\ }\href {\doibase10.1103/PhysRevLett.120.131104}
  {\bibfield  {journal} {\bibinfo  {journal} {Phys. Rev. Lett.}\ }\textbf
  {\bibinfo {volume} {120}},\ \bibinfo {pages} {131104} (\bibinfo {year}
  {2018})},\ \Eprint {http://arxiv.org/abs/1711.02080} {arXiv:1711.02080
  [gr-qc]}\BibitemShut {NoStop}%
\bibitem [{\citenamefont {Antoniou}\ \emph
  {et~al.}(2018{\natexlab{a}})\citenamefont {Antoniou}, \citenamefont
  {Bakopoulos},\ and\ \citenamefont {Kanti}}]{Antoniou:2017acq}%
  \BibitemOpen
  \bibfield  {author} {\bibinfo {author} {\bibfnamefont {G.}~\bibnamefont
  {Antoniou}}, \bibinfo {author} {\bibfnamefont {A.}~\bibnamefont
  {Bakopoulos}}, \ and\ \bibinfo {author} {\bibfnamefont {P.}~\bibnamefont
  {Kanti}},\ }\href {\doibase10.1103/PhysRevLett.120.131102} {\bibfield
  {journal} {\bibinfo  {journal} {Phys. Rev. Lett.}\ }\textbf {\bibinfo
  {volume} {120}},\ \bibinfo {pages} {131102} (\bibinfo {year}
  {2018}{\natexlab{a}})},\ \Eprint {http://arxiv.org/abs/1711.03390}
  {arXiv:1711.03390 [hep-th]}\BibitemShut {NoStop}%
\bibitem [{\citenamefont {Doneva}\ and\ \citenamefont
  {Yazadjiev}(2018)}]{Doneva:2017bvd}%
  \BibitemOpen
  \bibfield  {author} {\bibinfo {author} {\bibfnamefont {Daniela~D.}\
  \bibnamefont {Doneva}}\ and\ \bibinfo {author} {\bibfnamefont {Stoytcho~S.}\
  \bibnamefont {Yazadjiev}},\ }\href {\doibase10.1103/PhysRevLett.120.131103}
  {\bibfield  {journal} {\bibinfo  {journal} {Phys. Rev. Lett.}\ }\textbf
  {\bibinfo {volume} {120}},\ \bibinfo {pages} {131103} (\bibinfo {year}
  {2018})},\ \Eprint {http://arxiv.org/abs/1711.01187} {arXiv:1711.01187
  [gr-qc]}\BibitemShut {NoStop}%
\bibitem [{\citenamefont {Antoniou}\ \emph
  {et~al.}(2018{\natexlab{b}})\citenamefont {Antoniou}, \citenamefont
  {Bakopoulos},\ and\ \citenamefont {Kanti}}]{Antoniou:2017hxj}%
  \BibitemOpen
  \bibfield  {author} {\bibinfo {author} {\bibfnamefont {G.}~\bibnamefont
  {Antoniou}}, \bibinfo {author} {\bibfnamefont {A.}~\bibnamefont
  {Bakopoulos}}, \ and\ \bibinfo {author} {\bibfnamefont {P.}~\bibnamefont
  {Kanti}},\ }\href {\doibase10.1103/PhysRevD.97.084037} {\bibfield  {journal}
  {\bibinfo  {journal} {Phys. Rev.}\ }\textbf {\bibinfo {volume} {D97}},\
  \bibinfo {pages} {084037} (\bibinfo {year} {2018}{\natexlab{b}})},\ \Eprint
  {http://arxiv.org/abs/1711.07431} {arXiv:1711.07431 [hep-th]}\BibitemShut
  {NoStop}%
\bibitem [{\citenamefont {Blázquez-Salcedo}\ \emph {et~al.}(2018)\citenamefont
  {Blázquez-Salcedo}, \citenamefont {Doneva}, \citenamefont {Kunz},\ and\
  \citenamefont {Yazadjiev}}]{Blazquez-Salcedo:2018jnn}%
  \BibitemOpen
  \bibfield  {author} {\bibinfo {author} {\bibfnamefont {Jose~Luis}\
  \bibnamefont {Blázquez-Salcedo}}, \bibinfo {author} {\bibfnamefont
  {Daniela~D.}\ \bibnamefont {Doneva}}, \bibinfo {author} {\bibfnamefont
  {Jutta}\ \bibnamefont {Kunz}}, \ and\ \bibinfo {author} {\bibfnamefont
  {Stoytcho~S.}\ \bibnamefont {Yazadjiev}},\ }\href@noop {} {\  (\bibinfo
  {year} {2018})},\ \Eprint {http://arxiv.org/abs/1805.05755} {arXiv:1805.05755
  [gr-qc]}\BibitemShut {NoStop}%
\bibitem [{\citenamefont {Barausse}\ and\ \citenamefont
  {Yagi}(2015)}]{Barausse:2015wia}%
  \BibitemOpen
  \bibfield  {author} {\bibinfo {author} {\bibfnamefont {Enrico}\ \bibnamefont
  {Barausse}}\ and\ \bibinfo {author} {\bibfnamefont {Kent}\ \bibnamefont
  {Yagi}},\ }\href {\doibase10.1103/PhysRevLett.115.211105} {\bibfield
  {journal} {\bibinfo  {journal} {Phys. Rev. Lett.}\ }\textbf {\bibinfo
  {volume} {115}},\ \bibinfo {pages} {211105} (\bibinfo {year} {2015})},\
  \Eprint {http://arxiv.org/abs/1509.04539} {arXiv:1509.04539
  [gr-qc]}\BibitemShut {NoStop}%
\bibitem [{\citenamefont {Barausse}(2017)}]{Barausse:2017gip}%
  \BibitemOpen
  \bibfield  {author} {\bibinfo {author} {\bibfnamefont {Enrico}\ \bibnamefont
  {Barausse}},\ }\bibfield  {booktitle} {\emph {\bibinfo {booktitle}
  {{Proceedings, 3rd International Symposium on Quest for the Origin of
  Particles and the Universe (KMI2017): Nagoya, Japan, January 5-7, 2017}}},\
  }\href {\doibase10.22323/1.294.0029} {\bibfield  {journal} {\bibinfo
  {journal} {PoS}\ }\textbf {\bibinfo {volume} {KMI2017}},\ \bibinfo {pages}
  {029} (\bibinfo {year} {2017})},\ \Eprint {http://arxiv.org/abs/1703.05699}
  {arXiv:1703.05699 [gr-qc]}\BibitemShut {NoStop}%
\bibitem [{\citenamefont {Lehébel}\ \emph {et~al.}(2017)\citenamefont
  {Lehébel}, \citenamefont {Babichev},\ and\ \citenamefont
  {Charmousis}}]{Lehebel:2017fag}%
  \BibitemOpen
  \bibfield  {author} {\bibinfo {author} {\bibfnamefont {Antoine}\ \bibnamefont
  {Lehébel}}, \bibinfo {author} {\bibfnamefont {Eugeny}\ \bibnamefont
  {Babichev}}, \ and\ \bibinfo {author} {\bibfnamefont {Christos}\ \bibnamefont
  {Charmousis}},\ }\href {\doibase10.1088/1475-7516/2017/07/037} {\bibfield
  {journal} {\bibinfo  {journal} {JCAP}\ }\textbf {\bibinfo {volume} {1707}},\
  \bibinfo {pages} {037} (\bibinfo {year} {2017})},\ \Eprint
  {http://arxiv.org/abs/1706.04989} {arXiv:1706.04989 [gr-qc]}\BibitemShut
  {NoStop}%
\bibitem [{\citenamefont {Yagi}\ \emph {et~al.}(2016)\citenamefont {Yagi},
  \citenamefont {Stein},\ and\ \citenamefont {Yunes}}]{Yagi:2015oca}%
  \BibitemOpen
  \bibfield  {author} {\bibinfo {author} {\bibfnamefont {Kent}\ \bibnamefont
  {Yagi}}, \bibinfo {author} {\bibfnamefont {Leo~C.}\ \bibnamefont {Stein}}, \
  and\ \bibinfo {author} {\bibfnamefont {Nicolas}\ \bibnamefont {Yunes}},\
  }\href {\doibase10.1103/PhysRevD.93.024010} {\bibfield  {journal} {\bibinfo
  {journal} {Phys. Rev.}\ }\textbf {\bibinfo {volume} {D93}},\ \bibinfo {pages}
  {024010} (\bibinfo {year} {2016})},\ \Eprint
  {http://arxiv.org/abs/1510.02152} {arXiv:1510.02152 [gr-qc]}\BibitemShut
  {NoStop}%
\bibitem [{\citenamefont {Lombriser}\ and\ \citenamefont
  {Lima}(2017)}]{Lombriser:2016yzn}%
  \BibitemOpen
  \bibfield  {author} {\bibinfo {author} {\bibfnamefont {Lucas}\ \bibnamefont
  {Lombriser}}\ and\ \bibinfo {author} {\bibfnamefont {Nelson~A.}\ \bibnamefont
  {Lima}},\ }\href {\doibase10.1016/j.physletb.2016.12.048} {\bibfield
  {journal} {\bibinfo  {journal} {Phys. Lett.}\ }\textbf {\bibinfo {volume}
  {B765}},\ \bibinfo {pages} {382--385} (\bibinfo {year} {2017})},\ \Eprint
  {http://arxiv.org/abs/1602.07670} {arXiv:1602.07670
  [astro-ph.CO]}\BibitemShut {NoStop}%
\bibitem [{\citenamefont {Lombriser}\ and\ \citenamefont
  {Taylor}(2016)}]{Lombriser:2015sxa}%
  \BibitemOpen
  \bibfield  {author} {\bibinfo {author} {\bibfnamefont {Lucas}\ \bibnamefont
  {Lombriser}}\ and\ \bibinfo {author} {\bibfnamefont {Andy}\ \bibnamefont
  {Taylor}},\ }\href {\doibase10.1088/1475-7516/2016/03/031} {\bibfield
  {journal} {\bibinfo  {journal} {JCAP}\ }\textbf {\bibinfo {volume} {1603}},\
  \bibinfo {pages} {031} (\bibinfo {year} {2016})},\ \Eprint
  {http://arxiv.org/abs/1509.08458} {arXiv:1509.08458
  [astro-ph.CO]}\BibitemShut {NoStop}%
\bibitem [{\citenamefont {Zumalacárregui}\ and\ \citenamefont
  {García-Bellido}(2014)}]{Zumalacarregui:2013pma}%
  \BibitemOpen
  \bibfield  {author} {\bibinfo {author} {\bibfnamefont {Miguel}\ \bibnamefont
  {Zumalacárregui}}\ and\ \bibinfo {author} {\bibfnamefont {Juan}\
  \bibnamefont {García-Bellido}},\ }\href {\doibase10.1103/PhysRevD.89.064046}
  {\bibfield  {journal} {\bibinfo  {journal} {Phys. Rev.}\ }\textbf {\bibinfo
  {volume} {D89}},\ \bibinfo {pages} {064046} (\bibinfo {year} {2014})},\
  \Eprint {http://arxiv.org/abs/1308.4685} {arXiv:1308.4685
  [gr-qc]}\BibitemShut {NoStop}%
\bibitem [{\citenamefont {Gleyzes}\ \emph
  {et~al.}(2015{\natexlab{a}})\citenamefont {Gleyzes}, \citenamefont
  {Langlois}, \citenamefont {Piazza},\ and\ \citenamefont
  {Vernizzi}}]{Gleyzes:2014dya}%
  \BibitemOpen
  \bibfield  {author} {\bibinfo {author} {\bibfnamefont {Jérôme}\
  \bibnamefont {Gleyzes}}, \bibinfo {author} {\bibfnamefont {David}\
  \bibnamefont {Langlois}}, \bibinfo {author} {\bibfnamefont {Federico}\
  \bibnamefont {Piazza}}, \ and\ \bibinfo {author} {\bibfnamefont {Filippo}\
  \bibnamefont {Vernizzi}},\ }\href {\doibase10.1103/PhysRevLett.114.211101}
  {\bibfield  {journal} {\bibinfo  {journal} {Phys. Rev. Lett.}\ }\textbf
  {\bibinfo {volume} {114}},\ \bibinfo {pages} {211101} (\bibinfo {year}
  {2015}{\natexlab{a}})},\ \Eprint {http://arxiv.org/abs/1404.6495}
  {arXiv:1404.6495 [hep-th]}\BibitemShut {NoStop}%
\bibitem [{\citenamefont {Gleyzes}\ \emph
  {et~al.}(2015{\natexlab{b}})\citenamefont {Gleyzes}, \citenamefont
  {Langlois}, \citenamefont {Piazza},\ and\ \citenamefont
  {Vernizzi}}]{Gleyzes:2014qga}%
  \BibitemOpen
  \bibfield  {author} {\bibinfo {author} {\bibfnamefont {Jérôme}\
  \bibnamefont {Gleyzes}}, \bibinfo {author} {\bibfnamefont {David}\
  \bibnamefont {Langlois}}, \bibinfo {author} {\bibfnamefont {Federico}\
  \bibnamefont {Piazza}}, \ and\ \bibinfo {author} {\bibfnamefont {Filippo}\
  \bibnamefont {Vernizzi}},\ }\href {\doibase10.1088/1475-7516/2015/02/018}
  {\bibfield  {journal} {\bibinfo  {journal} {JCAP}\ }\textbf {\bibinfo
  {volume} {1502}},\ \bibinfo {pages} {018} (\bibinfo {year}
  {2015}{\natexlab{b}})},\ \Eprint {http://arxiv.org/abs/1408.1952}
  {arXiv:1408.1952 [astro-ph.CO]}\BibitemShut {NoStop}%
\bibitem [{\citenamefont {Langlois}\ and\ \citenamefont
  {Noui}(2016)}]{Langlois:2015cwa}%
  \BibitemOpen
  \bibfield  {author} {\bibinfo {author} {\bibfnamefont {David}\ \bibnamefont
  {Langlois}}\ and\ \bibinfo {author} {\bibfnamefont {Karim}\ \bibnamefont
  {Noui}},\ }\href {\doibase10.1088/1475-7516/2016/02/034} {\bibfield
  {journal} {\bibinfo  {journal} {JCAP}\ }\textbf {\bibinfo {volume} {1602}},\
  \bibinfo {pages} {034} (\bibinfo {year} {2016})},\ \Eprint
  {http://arxiv.org/abs/1510.06930} {arXiv:1510.06930 [gr-qc]}\BibitemShut
  {NoStop}%
\bibitem [{\citenamefont {Crisostomi}\ \emph {et~al.}(2016)\citenamefont
  {Crisostomi}, \citenamefont {Koyama},\ and\ \citenamefont
  {Tasinato}}]{Crisostomi:2016czh}%
  \BibitemOpen
  \bibfield  {author} {\bibinfo {author} {\bibfnamefont {Marco}\ \bibnamefont
  {Crisostomi}}, \bibinfo {author} {\bibfnamefont {Kazuya}\ \bibnamefont
  {Koyama}}, \ and\ \bibinfo {author} {\bibfnamefont {Gianmassimo}\
  \bibnamefont {Tasinato}},\ }\href {\doibase10.1088/1475-7516/2016/04/044}
  {\bibfield  {journal} {\bibinfo  {journal} {JCAP}\ }\textbf {\bibinfo
  {volume} {1604}},\ \bibinfo {pages} {044} (\bibinfo {year} {2016})},\ \Eprint
  {http://arxiv.org/abs/1602.03119} {arXiv:1602.03119 [hep-th]}\BibitemShut
  {NoStop}%
\bibitem [{\citenamefont {Ben~Achour}\ \emph {et~al.}(2016)\citenamefont
  {Ben~Achour}, \citenamefont {Langlois},\ and\ \citenamefont
  {Noui}}]{Achour:2016rkg}%
  \BibitemOpen
  \bibfield  {author} {\bibinfo {author} {\bibfnamefont {Jibril}\ \bibnamefont
  {Ben~Achour}}, \bibinfo {author} {\bibfnamefont {David}\ \bibnamefont
  {Langlois}}, \ and\ \bibinfo {author} {\bibfnamefont {Karim}\ \bibnamefont
  {Noui}},\ }\href {\doibase10.1103/PhysRevD.93.124005} {\bibfield  {journal}
  {\bibinfo  {journal} {Phys. Rev.}\ }\textbf {\bibinfo {volume} {D93}},\
  \bibinfo {pages} {124005} (\bibinfo {year} {2016})},\ \Eprint
  {http://arxiv.org/abs/1602.08398} {arXiv:1602.08398 [gr-qc]}\BibitemShut
  {NoStop}%
\bibitem [{\citenamefont {Jackiw}\ and\ \citenamefont
  {Pi}(2003)}]{Jackiw:2003pm}%
  \BibitemOpen
  \bibfield  {author} {\bibinfo {author} {\bibfnamefont {R.}~\bibnamefont
  {Jackiw}}\ and\ \bibinfo {author} {\bibfnamefont {S.~Y.}\ \bibnamefont
  {Pi}},\ }\href {\doibase10.1103/PhysRevD.68.104012} {\bibfield  {journal}
  {\bibinfo  {journal} {Phys. Rev.}\ }\textbf {\bibinfo {volume} {D68}},\
  \bibinfo {pages} {104012} (\bibinfo {year} {2003})},\ \Eprint
  {http://arxiv.org/abs/gr-qc/0308071} {arXiv:gr-qc/0308071
  [gr-qc]}\BibitemShut {NoStop}%
\bibitem [{\citenamefont {Alexander}\ and\ \citenamefont
  {Yunes}(2009)}]{Alexander:2009tp}%
  \BibitemOpen
  \bibfield  {author} {\bibinfo {author} {\bibfnamefont {Stephon}\ \bibnamefont
  {Alexander}}\ and\ \bibinfo {author} {\bibfnamefont {Nicolas}\ \bibnamefont
  {Yunes}},\ }\href {\doibase10.1016/j.physrep.2009.07.002} {\bibfield
  {journal} {\bibinfo  {journal} {Phys. Rept.}\ }\textbf {\bibinfo {volume}
  {480}},\ \bibinfo {pages} {1--55} (\bibinfo {year} {2009})},\ \Eprint
  {http://arxiv.org/abs/0907.2562} {arXiv:0907.2562 [hep-th]}\BibitemShut
  {NoStop}%
\bibitem [{\citenamefont {Weinberg}(2013)}]{Weinberg:1996kr}%
  \BibitemOpen
  \bibfield  {author} {\bibinfo {author} {\bibfnamefont {Steven}\ \bibnamefont
  {Weinberg}},\ }\href@noop {} {\emph {\bibinfo {title} {{The quantum theory of
  fields. Vol. 2: Modern applications}}}}\ (\bibinfo  {publisher} {Cambridge
  University Press},\ \bibinfo {year} {2013})\BibitemShut {NoStop}%
\bibitem [{\citenamefont {Green}\ \emph {et~al.}(1988)\citenamefont {Green},
  \citenamefont {Schwarz},\ and\ \citenamefont {Witten}}]{Green:1987mn}%
  \BibitemOpen
  \bibfield  {author} {\bibinfo {author} {\bibfnamefont {Michael~B.}\
  \bibnamefont {Green}}, \bibinfo {author} {\bibfnamefont {J.~H.}\ \bibnamefont
  {Schwarz}}, \ and\ \bibinfo {author} {\bibfnamefont {Edward}\ \bibnamefont
  {Witten}},\ }\href
  {http://www.cambridge.org/us/academic/subjects/physics/theoretical-physics-and-mathematical-physics/superstring-theory-volume-2}
  {\emph {\bibinfo {title} {{Superstring Theory. Vol. 2: Loop Amplitudes,
  Anomalies And Phenomenology}}}}\ (\bibinfo  {publisher} {Cambridge University
  Press},\ \bibinfo {year} {1988})\BibitemShut {NoStop}%
\bibitem [{\citenamefont {Ashtekar}\ \emph {et~al.}(1989)\citenamefont
  {Ashtekar}, \citenamefont {Balachandran},\ and\ \citenamefont
  {Jo}}]{Ashtekar:1988sw}%
  \BibitemOpen
  \bibfield  {author} {\bibinfo {author} {\bibfnamefont {Abhay}\ \bibnamefont
  {Ashtekar}}, \bibinfo {author} {\bibfnamefont {A.~P.}\ \bibnamefont
  {Balachandran}}, \ and\ \bibinfo {author} {\bibfnamefont {Sang}\ \bibnamefont
  {Jo}},\ }\href {\doibase10.1142/S0217751X89000649} {\bibfield  {journal}
  {\bibinfo  {journal} {Int. J. Mod. Phys.}\ }\textbf {\bibinfo {volume}
  {A4}},\ \bibinfo {pages} {1493} (\bibinfo {year} {1989})}\BibitemShut
  {NoStop}%
\bibitem [{\citenamefont {Alexander}\ and\ \citenamefont
  {Gates}(2006)}]{Alexander:2004xd}%
  \BibitemOpen
  \bibfield  {author} {\bibinfo {author} {\bibfnamefont {Stephon H.~S.}\
  \bibnamefont {Alexander}}\ and\ \bibinfo {author} {\bibfnamefont {S.~James}\
  \bibnamefont {Gates}, \bibfnamefont {Jr.}},\ }\href
  {\doibase10.1088/1475-7516/2006/06/018} {\bibfield  {journal} {\bibinfo
  {journal} {JCAP}\ }\textbf {\bibinfo {volume} {0606}},\ \bibinfo {pages}
  {018} (\bibinfo {year} {2006})},\ \Eprint
  {http://arxiv.org/abs/hep-th/0409014} {arXiv:hep-th/0409014
  [hep-th]}\BibitemShut {NoStop}%
\bibitem [{\citenamefont {Taveras}\ and\ \citenamefont
  {Yunes}(2008)}]{Taveras:2008yf}%
  \BibitemOpen
  \bibfield  {author} {\bibinfo {author} {\bibfnamefont {Victor}\ \bibnamefont
  {Taveras}}\ and\ \bibinfo {author} {\bibfnamefont {Nicolas}\ \bibnamefont
  {Yunes}},\ }\href {\doibase10.1103/PhysRevD.78.064070} {\bibfield  {journal}
  {\bibinfo  {journal} {Phys. Rev.}\ }\textbf {\bibinfo {volume} {D78}},\
  \bibinfo {pages} {064070} (\bibinfo {year} {2008})},\ \Eprint
  {http://arxiv.org/abs/0807.2652} {arXiv:0807.2652 [gr-qc]}\BibitemShut
  {NoStop}%
\bibitem [{\citenamefont {Calcagni}\ and\ \citenamefont
  {Mercuri}(2009)}]{Calcagni:2009xz}%
  \BibitemOpen
  \bibfield  {author} {\bibinfo {author} {\bibfnamefont {Gianluca}\
  \bibnamefont {Calcagni}}\ and\ \bibinfo {author} {\bibfnamefont {Simone}\
  \bibnamefont {Mercuri}},\ }\href {\doibase10.1103/PhysRevD.79.084004}
  {\bibfield  {journal} {\bibinfo  {journal} {Phys. Rev.}\ }\textbf {\bibinfo
  {volume} {D79}},\ \bibinfo {pages} {084004} (\bibinfo {year} {2009})},\
  \Eprint {http://arxiv.org/abs/0902.0957} {arXiv:0902.0957
  [gr-qc]}\BibitemShut {NoStop}%
\bibitem [{\citenamefont {Gates}\ \emph {et~al.}(2009)\citenamefont {Gates},
  \citenamefont {Ketov},\ and\ \citenamefont {Yunes}}]{Gates:2009pt}%
  \BibitemOpen
  \bibfield  {author} {\bibinfo {author} {\bibfnamefont {Jr.}\ \bibnamefont
  {Gates}, \bibfnamefont {S.James}}, \bibinfo {author} {\bibfnamefont
  {Sergei~V.}\ \bibnamefont {Ketov}}, \ and\ \bibinfo {author} {\bibfnamefont
  {Nicolas}\ \bibnamefont {Yunes}},\ }\href
  {\doibase10.1103/PhysRevD.80.065003} {\bibfield  {journal} {\bibinfo
  {journal} {Phys. Rev.}\ }\textbf {\bibinfo {volume} {D80}},\ \bibinfo {pages}
  {065003} (\bibinfo {year} {2009})},\ \Eprint {http://arxiv.org/abs/0906.4978}
  {arXiv:0906.4978 [hep-th]}\BibitemShut {NoStop}%
\bibitem [{\citenamefont {Weinberg}(2008)}]{Weinberg:2008hq}%
  \BibitemOpen
  \bibfield  {author} {\bibinfo {author} {\bibfnamefont {Steven}\ \bibnamefont
  {Weinberg}},\ }\href {\doibase10.1103/PhysRevD.77.123541} {\bibfield
  {journal} {\bibinfo  {journal} {Phys. Rev.}\ }\textbf {\bibinfo {volume}
  {D77}},\ \bibinfo {pages} {123541} (\bibinfo {year} {2008})},\ \Eprint
  {http://arxiv.org/abs/0804.4291} {arXiv:0804.4291 [hep-th]}\BibitemShut
  {NoStop}%
\bibitem [{\citenamefont {Crisostomi}\ \emph {et~al.}(2018)\citenamefont
  {Crisostomi}, \citenamefont {Noui}, \citenamefont {Charmousis},\ and\
  \citenamefont {Langlois}}]{Crisostomi:2017ugk}%
  \BibitemOpen
  \bibfield  {author} {\bibinfo {author} {\bibfnamefont {Marco}\ \bibnamefont
  {Crisostomi}}, \bibinfo {author} {\bibfnamefont {Karim}\ \bibnamefont
  {Noui}}, \bibinfo {author} {\bibfnamefont {Christos}\ \bibnamefont
  {Charmousis}}, \ and\ \bibinfo {author} {\bibfnamefont {David}\ \bibnamefont
  {Langlois}},\ }\href {\doibase10.1103/PhysRevD.97.044034} {\bibfield
  {journal} {\bibinfo  {journal} {Phys. Rev.}\ }\textbf {\bibinfo {volume}
  {D97}},\ \bibinfo {pages} {044034} (\bibinfo {year} {2018})},\ \Eprint
  {http://arxiv.org/abs/1710.04531} {arXiv:1710.04531 [hep-th]}\BibitemShut
  {NoStop}%
\bibitem [{\citenamefont {Yunes}\ and\ \citenamefont
  {Pretorius}(2009{\natexlab{a}})}]{Yunes:2009hc}%
  \BibitemOpen
  \bibfield  {author} {\bibinfo {author} {\bibfnamefont {Nicolas}\ \bibnamefont
  {Yunes}}\ and\ \bibinfo {author} {\bibfnamefont {Frans}\ \bibnamefont
  {Pretorius}},\ }\href {\doibase10.1103/PhysRevD.79.084043} {\bibfield
  {journal} {\bibinfo  {journal} {Phys. Rev.}\ }\textbf {\bibinfo {volume}
  {D79}},\ \bibinfo {pages} {084043} (\bibinfo {year} {2009}{\natexlab{a}})},\
  \Eprint {http://arxiv.org/abs/0902.4669} {arXiv:0902.4669
  [gr-qc]}\BibitemShut {NoStop}%
\bibitem [{\citenamefont {Konno}\ \emph {et~al.}(2009)\citenamefont {Konno},
  \citenamefont {Matsuyama},\ and\ \citenamefont {Tanda}}]{Konno:2009kg}%
  \BibitemOpen
  \bibfield  {author} {\bibinfo {author} {\bibfnamefont {Kohkichi}\
  \bibnamefont {Konno}}, \bibinfo {author} {\bibfnamefont {Toyoki}\
  \bibnamefont {Matsuyama}}, \ and\ \bibinfo {author} {\bibfnamefont {Satoshi}\
  \bibnamefont {Tanda}},\ }\href {\doibase10.1143/PTP.122.561} {\bibfield
  {journal} {\bibinfo  {journal} {Prog. Theor. Phys.}\ }\textbf {\bibinfo
  {volume} {122}},\ \bibinfo {pages} {561--568} (\bibinfo {year} {2009})},\
  \Eprint {http://arxiv.org/abs/0902.4767} {arXiv:0902.4767
  [gr-qc]}\BibitemShut {NoStop}%
\bibitem [{\citenamefont {Yagi}\ \emph
  {et~al.}(2012{\natexlab{c}})\citenamefont {Yagi}, \citenamefont {Yunes},\
  and\ \citenamefont {Tanaka}}]{Yagi:2012ya}%
  \BibitemOpen
  \bibfield  {author} {\bibinfo {author} {\bibfnamefont {Kent}\ \bibnamefont
  {Yagi}}, \bibinfo {author} {\bibfnamefont {Nicolas}\ \bibnamefont {Yunes}}, \
  and\ \bibinfo {author} {\bibfnamefont {Takahiro}\ \bibnamefont {Tanaka}},\
  }\href {\doibase10.1103/PhysRevD.86.044037} {\bibfield  {journal} {\bibinfo
  {journal} {Phys. Rev.}\ }\textbf {\bibinfo {volume} {D86}},\ \bibinfo {pages}
  {044037} (\bibinfo {year} {2012}{\natexlab{c}})},\ \bibinfo {note} {[Erratum:
  Phys. Rev.D89,049902(2014)]},\ \Eprint {http://arxiv.org/abs/1206.6130}
  {arXiv:1206.6130 [gr-qc]}\BibitemShut {NoStop}%
\bibitem [{\citenamefont {Stein}(2014)}]{Stein:2014xba}%
  \BibitemOpen
  \bibfield  {author} {\bibinfo {author} {\bibfnamefont {Leo~C.}\ \bibnamefont
  {Stein}},\ }\href {\doibase10.1103/PhysRevD.90.044061} {\bibfield  {journal}
  {\bibinfo  {journal} {Phys. Rev.}\ }\textbf {\bibinfo {volume} {D90}},\
  \bibinfo {pages} {044061} (\bibinfo {year} {2014})},\ \Eprint
  {http://arxiv.org/abs/1407.2350} {arXiv:1407.2350 [gr-qc]}\BibitemShut
  {NoStop}%
\bibitem [{\citenamefont {McNees}\ \emph {et~al.}(2016)\citenamefont {McNees},
  \citenamefont {Stein},\ and\ \citenamefont {Yunes}}]{McNees:2015srl}%
  \BibitemOpen
  \bibfield  {author} {\bibinfo {author} {\bibfnamefont {Robert}\ \bibnamefont
  {McNees}}, \bibinfo {author} {\bibfnamefont {Leo~C.}\ \bibnamefont {Stein}},
  \ and\ \bibinfo {author} {\bibfnamefont {Nicolás}\ \bibnamefont {Yunes}},\
  }\href {\doibase10.1088/0264-9381/33/23/235013} {\bibfield  {journal}
  {\bibinfo  {journal} {Class. Quant. Grav.}\ }\textbf {\bibinfo {volume}
  {33}},\ \bibinfo {pages} {235013} (\bibinfo {year} {2016})},\ \Eprint
  {http://arxiv.org/abs/1512.05453} {arXiv:1512.05453 [gr-qc]}\BibitemShut
  {NoStop}%
\bibitem [{\citenamefont {Chen}\ and\ \citenamefont
  {Stein}(2018)}]{Chen:2018jed}%
  \BibitemOpen
  \bibfield  {author} {\bibinfo {author} {\bibfnamefont {Baoyi}\ \bibnamefont
  {Chen}}\ and\ \bibinfo {author} {\bibfnamefont {Leo~C.}\ \bibnamefont
  {Stein}},\ }\href {\doibase10.1103/PhysRevD.97.084012} {\bibfield  {journal}
  {\bibinfo  {journal} {Phys. Rev.}\ }\textbf {\bibinfo {volume} {D97}},\
  \bibinfo {pages} {084012} (\bibinfo {year} {2018})},\ \Eprint
  {http://arxiv.org/abs/1802.02159} {arXiv:1802.02159 [gr-qc]}\BibitemShut
  {NoStop}%
\bibitem [{\citenamefont {Sopuerta}\ and\ \citenamefont
  {Yunes}(2009)}]{Sopuerta:2009iy}%
  \BibitemOpen
  \bibfield  {author} {\bibinfo {author} {\bibfnamefont {Carlos~F.}\
  \bibnamefont {Sopuerta}}\ and\ \bibinfo {author} {\bibfnamefont {Nicolas}\
  \bibnamefont {Yunes}},\ }\href {\doibase10.1103/PhysRevD.80.064006}
  {\bibfield  {journal} {\bibinfo  {journal} {Phys. Rev.}\ }\textbf {\bibinfo
  {volume} {D80}},\ \bibinfo {pages} {064006} (\bibinfo {year} {2009})},\
  \Eprint {http://arxiv.org/abs/0904.4501} {arXiv:0904.4501
  [gr-qc]}\BibitemShut {NoStop}%
\bibitem [{\citenamefont {Pani}\ \emph {et~al.}(2011)\citenamefont {Pani},
  \citenamefont {Cardoso},\ and\ \citenamefont {Gualtieri}}]{Pani:2011xj}%
  \BibitemOpen
  \bibfield  {author} {\bibinfo {author} {\bibfnamefont {Paolo}\ \bibnamefont
  {Pani}}, \bibinfo {author} {\bibfnamefont {Vitor}\ \bibnamefont {Cardoso}}, \
  and\ \bibinfo {author} {\bibfnamefont {Leonardo}\ \bibnamefont {Gualtieri}},\
  }\href {\doibase10.1103/PhysRevD.83.104048} {\bibfield  {journal} {\bibinfo
  {journal} {Phys. Rev.}\ }\textbf {\bibinfo {volume} {D83}},\ \bibinfo {pages}
  {104048} (\bibinfo {year} {2011})},\ \Eprint {http://arxiv.org/abs/1104.1183}
  {arXiv:1104.1183 [gr-qc]}\BibitemShut {NoStop}%
\bibitem [{\citenamefont {Yagi}\ \emph
  {et~al.}(2012{\natexlab{d}})\citenamefont {Yagi}, \citenamefont {Yunes},\
  and\ \citenamefont {Tanaka}}]{Yagi:2012vf}%
  \BibitemOpen
  \bibfield  {author} {\bibinfo {author} {\bibfnamefont {Kent}\ \bibnamefont
  {Yagi}}, \bibinfo {author} {\bibfnamefont {Nicolas}\ \bibnamefont {Yunes}}, \
  and\ \bibinfo {author} {\bibfnamefont {Takahiro}\ \bibnamefont {Tanaka}},\
  }\href {\doibase10.1103/PhysRevLett.109.251105} {\bibfield  {journal}
  {\bibinfo  {journal} {Phys. Rev. Lett.}\ }\textbf {\bibinfo {volume} {109}},\
  \bibinfo {pages} {251105} (\bibinfo {year} {2012}{\natexlab{d}})},\ \bibinfo
  {note} {[Erratum: Phys. Rev. Lett.116,no.16,169902(2016)]},\ \Eprint
  {http://arxiv.org/abs/1208.5102} {arXiv:1208.5102 [gr-qc]}\BibitemShut
  {NoStop}%
\bibitem [{\citenamefont {Canizares}\ \emph {et~al.}(2012)\citenamefont
  {Canizares}, \citenamefont {Gair},\ and\ \citenamefont
  {Sopuerta}}]{Canizares:2012is}%
  \BibitemOpen
  \bibfield  {author} {\bibinfo {author} {\bibfnamefont {Priscilla}\
  \bibnamefont {Canizares}}, \bibinfo {author} {\bibfnamefont {Jonathan~R.}\
  \bibnamefont {Gair}}, \ and\ \bibinfo {author} {\bibfnamefont {Carlos~F.}\
  \bibnamefont {Sopuerta}},\ }\href {\doibase10.1103/PhysRevD.86.044010}
  {\bibfield  {journal} {\bibinfo  {journal} {Phys. Rev.}\ }\textbf {\bibinfo
  {volume} {D86}},\ \bibinfo {pages} {044010} (\bibinfo {year} {2012})},\
  \Eprint {http://arxiv.org/abs/1205.1253} {arXiv:1205.1253
  [gr-qc]}\BibitemShut {NoStop}%
\bibitem [{\citenamefont {Ali-Haimoud}\ and\ \citenamefont
  {Chen}(2011)}]{AliHaimoud:2011fw}%
  \BibitemOpen
  \bibfield  {author} {\bibinfo {author} {\bibfnamefont {Yacine}\ \bibnamefont
  {Ali-Haimoud}}\ and\ \bibinfo {author} {\bibfnamefont {Yanbei}\ \bibnamefont
  {Chen}},\ }\href {\doibase10.1103/PhysRevD.84.124033} {\bibfield  {journal}
  {\bibinfo  {journal} {Phys. Rev.}\ }\textbf {\bibinfo {volume} {D84}},\
  \bibinfo {pages} {124033} (\bibinfo {year} {2011})},\ \Eprint
  {http://arxiv.org/abs/1110.5329} {arXiv:1110.5329 [astro-ph.HE]}\BibitemShut
  {NoStop}%
\bibitem [{\citenamefont {Alexander}\ \emph {et~al.}(2008)\citenamefont
  {Alexander}, \citenamefont {Finn},\ and\ \citenamefont
  {Yunes}}]{Alexander:2007kv}%
  \BibitemOpen
  \bibfield  {author} {\bibinfo {author} {\bibfnamefont {Stephon}\ \bibnamefont
  {Alexander}}, \bibinfo {author} {\bibfnamefont {Lee~Samuel}\ \bibnamefont
  {Finn}}, \ and\ \bibinfo {author} {\bibfnamefont {Nicolas}\ \bibnamefont
  {Yunes}},\ }\href {\doibase10.1103/PhysRevD.78.066005} {\bibfield  {journal}
  {\bibinfo  {journal} {Phys. Rev.}\ }\textbf {\bibinfo {volume} {D78}},\
  \bibinfo {pages} {066005} (\bibinfo {year} {2008})},\ \Eprint
  {http://arxiv.org/abs/0712.2542} {arXiv:0712.2542 [gr-qc]}\BibitemShut
  {NoStop}%
\bibitem [{\citenamefont {Yunes}\ and\ \citenamefont
  {Finn}(2009)}]{Yunes:2008bu}%
  \BibitemOpen
  \bibfield  {author} {\bibinfo {author} {\bibfnamefont {Nicolas}\ \bibnamefont
  {Yunes}}\ and\ \bibinfo {author} {\bibfnamefont {Lee~Samuel}\ \bibnamefont
  {Finn}},\ }\bibfield  {booktitle} {\emph {\bibinfo {booktitle} {{Laser
  Interferometer Space Antenna. Proceedings, 7th international LISA Symposium,
  Barcelona, Spain, June 16-20, 2008}}},\ }\href
  {\doibase10.1088/1742-6596/154/1/012041} {\bibfield  {journal} {\bibinfo
  {journal} {J. Phys. Conf. Ser.}\ }\textbf {\bibinfo {volume} {154}},\
  \bibinfo {pages} {012041} (\bibinfo {year} {2009})},\ \Eprint
  {http://arxiv.org/abs/0811.0181} {arXiv:0811.0181 [gr-qc]}\BibitemShut
  {NoStop}%
\bibitem [{\citenamefont {Yunes}\ \emph {et~al.}(2010)\citenamefont {Yunes},
  \citenamefont {O'Shaughnessy}, \citenamefont {Owen},\ and\ \citenamefont
  {Alexander}}]{Yunes:2010yf}%
  \BibitemOpen
  \bibfield  {author} {\bibinfo {author} {\bibfnamefont {Nicolas}\ \bibnamefont
  {Yunes}}, \bibinfo {author} {\bibfnamefont {Richard}\ \bibnamefont
  {O'Shaughnessy}}, \bibinfo {author} {\bibfnamefont {Benjamin~J.}\
  \bibnamefont {Owen}}, \ and\ \bibinfo {author} {\bibfnamefont {Stephon}\
  \bibnamefont {Alexander}},\ }\href {\doibase10.1103/PhysRevD.82.064017}
  {\bibfield  {journal} {\bibinfo  {journal} {Phys. Rev.}\ }\textbf {\bibinfo
  {volume} {D82}},\ \bibinfo {pages} {064017} (\bibinfo {year} {2010})},\
  \Eprint {http://arxiv.org/abs/1005.3310} {arXiv:1005.3310
  [gr-qc]}\BibitemShut {NoStop}%
\bibitem [{\citenamefont {Yagi}\ and\ \citenamefont
  {Yang}(2018)}]{Yagi:2017zhb}%
  \BibitemOpen
  \bibfield  {author} {\bibinfo {author} {\bibfnamefont {Kent}\ \bibnamefont
  {Yagi}}\ and\ \bibinfo {author} {\bibfnamefont {Huan}\ \bibnamefont {Yang}},\
  }\href {\doibase10.1103/PhysRevD.97.104018} {\bibfield  {journal} {\bibinfo
  {journal} {Phys. Rev.}\ }\textbf {\bibinfo {volume} {D97}},\ \bibinfo {pages}
  {104018} (\bibinfo {year} {2018})},\ \Eprint
  {http://arxiv.org/abs/1712.00682} {arXiv:1712.00682 [gr-qc]}\BibitemShut
  {NoStop}%
\bibitem [{\citenamefont {Smith}\ \emph {et~al.}(2008)\citenamefont {Smith},
  \citenamefont {Erickcek}, \citenamefont {Caldwell},\ and\ \citenamefont
  {Kamionkowski}}]{Smith:2007jm}%
  \BibitemOpen
  \bibfield  {author} {\bibinfo {author} {\bibfnamefont {Tristan~L.}\
  \bibnamefont {Smith}}, \bibinfo {author} {\bibfnamefont {Adrienne~L.}\
  \bibnamefont {Erickcek}}, \bibinfo {author} {\bibfnamefont {Robert~R.}\
  \bibnamefont {Caldwell}}, \ and\ \bibinfo {author} {\bibfnamefont {Marc}\
  \bibnamefont {Kamionkowski}},\ }\href {\doibase10.1103/PhysRevD.77.024015}
  {\bibfield  {journal} {\bibinfo  {journal} {Phys. Rev.}\ }\textbf {\bibinfo
  {volume} {D77}},\ \bibinfo {pages} {024015} (\bibinfo {year} {2008})},\
  \Eprint {http://arxiv.org/abs/0708.0001} {arXiv:0708.0001
  [astro-ph]}\BibitemShut {NoStop}%
\bibitem [{\citenamefont {Yunes}\ and\ \citenamefont
  {Spergel}(2009)}]{Yunes:2008ua}%
  \BibitemOpen
  \bibfield  {author} {\bibinfo {author} {\bibfnamefont {Nicolas}\ \bibnamefont
  {Yunes}}\ and\ \bibinfo {author} {\bibfnamefont {David~N.}\ \bibnamefont
  {Spergel}},\ }\href {\doibase10.1103/PhysRevD.80.042004} {\bibfield
  {journal} {\bibinfo  {journal} {Phys. Rev.}\ }\textbf {\bibinfo {volume}
  {D80}},\ \bibinfo {pages} {042004} (\bibinfo {year} {2009})},\ \Eprint
  {http://arxiv.org/abs/0810.5541} {arXiv:0810.5541 [gr-qc]}\BibitemShut
  {NoStop}%
\bibitem [{\citenamefont {Ali-Haimoud}(2011)}]{AliHaimoud:2011bk}%
  \BibitemOpen
  \bibfield  {author} {\bibinfo {author} {\bibfnamefont {Yacine}\ \bibnamefont
  {Ali-Haimoud}},\ }\href {\doibase10.1103/PhysRevD.83.124050} {\bibfield
  {journal} {\bibinfo  {journal} {Phys. Rev.}\ }\textbf {\bibinfo {volume}
  {D83}},\ \bibinfo {pages} {124050} (\bibinfo {year} {2011})},\ \Eprint
  {http://arxiv.org/abs/1105.0009} {arXiv:1105.0009 [astro-ph.HE]}\BibitemShut
  {NoStop}%
\bibitem [{\citenamefont {{Faraoni}}(2004)}]{2004cstg.book.....F}%
  \BibitemOpen
  \bibfield  {author} {\bibinfo {author} {\bibfnamefont {V.}~\bibnamefont
  {{Faraoni}}},\ }\href@noop {} {\emph {\bibinfo {title} {{Cosmology in
  Scalar-Tensor Gravity}}}}\ (\bibinfo  {publisher} {Kluwer Academic
  Publishers},\ \bibinfo {year} {2004})\BibitemShut {NoStop}%
\bibitem [{\citenamefont {Deffayet}\ and\ \citenamefont
  {Steer}(2013)}]{Deffayet:2013lga}%
  \BibitemOpen
  \bibfield  {author} {\bibinfo {author} {\bibfnamefont {Cédric}\ \bibnamefont
  {Deffayet}}\ and\ \bibinfo {author} {\bibfnamefont {Danièle~A.}\
  \bibnamefont {Steer}},\ }\href {\doibase10.1088/0264-9381/30/21/214006}
  {\bibfield  {journal} {\bibinfo  {journal} {Class. Quant. Grav.}\ }\textbf
  {\bibinfo {volume} {30}},\ \bibinfo {pages} {214006} (\bibinfo {year}
  {2013})},\ \Eprint {http://arxiv.org/abs/1307.2450} {arXiv:1307.2450
  [hep-th]}\BibitemShut {NoStop}%
\bibitem [{\citenamefont {Sotiriou}(2015{\natexlab{b}})}]{Sotiriou:2015lxa}%
  \BibitemOpen
  \bibfield  {author} {\bibinfo {author} {\bibfnamefont {Thomas~P.}\
  \bibnamefont {Sotiriou}},\ }\bibfield  {booktitle} {\emph {\bibinfo
  {booktitle} {{Proceedings of the 7th Aegean Summer School : Beyond Einstein's
  theory of gravity. Modifications of Einstein's Theory of Gravity at Large
  Distances.: Paros, Greece, September 23-28, 2013}}},\ }\href
  {\doibase10.1007/978-3-319-10070-8_1} {\bibfield  {journal} {\bibinfo
  {journal} {Lect. Notes Phys.}\ }\textbf {\bibinfo {volume} {892}},\ \bibinfo
  {pages} {3--24} (\bibinfo {year} {2015}{\natexlab{b}})},\ \Eprint
  {http://arxiv.org/abs/1404.2955} {arXiv:1404.2955 [gr-qc]}\BibitemShut
  {NoStop}%
\bibitem [{\citenamefont {Jacobson}\ and\ \citenamefont
  {Mattingly}(2001)}]{Jacobson:2000xp}%
  \BibitemOpen
  \bibfield  {author} {\bibinfo {author} {\bibfnamefont {Ted}\ \bibnamefont
  {Jacobson}}\ and\ \bibinfo {author} {\bibfnamefont {David}\ \bibnamefont
  {Mattingly}},\ }\href {\doibase10.1103/PhysRevD.64.024028} {\bibfield
  {journal} {\bibinfo  {journal} {Phys. Rev.}\ }\textbf {\bibinfo {volume}
  {D64}},\ \bibinfo {pages} {024028} (\bibinfo {year} {2001})},\ \Eprint
  {http://arxiv.org/abs/gr-qc/0007031} {arXiv:gr-qc/0007031
  [gr-qc]}\BibitemShut {NoStop}%
\bibitem [{\citenamefont {Withers}(2009)}]{Withers:2009qg}%
  \BibitemOpen
  \bibfield  {author} {\bibinfo {author} {\bibfnamefont {Benjamin}\
  \bibnamefont {Withers}},\ }\bibfield  {booktitle} {\emph {\bibinfo
  {booktitle} {{Strings, Supergravity and Gauge Theories. Proceedings, CERN
  Winter School, CERN, Geneva, Switzerland, February 9-13 2009}}},\ }\href
  {\doibase10.1088/0264-9381/26/22/225009} {\bibfield  {journal} {\bibinfo
  {journal} {Class. Quant. Grav.}\ }\textbf {\bibinfo {volume} {26}},\ \bibinfo
  {pages} {225009} (\bibinfo {year} {2009})},\ \Eprint
  {http://arxiv.org/abs/0905.2446} {arXiv:0905.2446 [gr-qc]}\BibitemShut
  {NoStop}%
\bibitem [{\citenamefont {Jacobson}(2010)}]{Jacobson:2010mx}%
  \BibitemOpen
  \bibfield  {author} {\bibinfo {author} {\bibfnamefont {Ted}\ \bibnamefont
  {Jacobson}},\ }\href {\doibase10.1103/PhysRevD.81.101502} {\bibfield
  {journal} {\bibinfo  {journal} {Phys. Rev.}\ }\textbf {\bibinfo {volume}
  {D81}},\ \bibinfo {pages} {101502} (\bibinfo {year} {2010})},\ \bibinfo
  {note} {[Erratum: Phys. Rev.D82,129901(2010)]},\ \Eprint
  {http://arxiv.org/abs/1001.4823} {arXiv:1001.4823 [hep-th]}\BibitemShut
  {NoStop}%
\bibitem [{\citenamefont {Horava}(2009)}]{Horava:2009uw}%
  \BibitemOpen
  \bibfield  {author} {\bibinfo {author} {\bibfnamefont {Petr}\ \bibnamefont
  {Horava}},\ }\href {\doibase10.1103/PhysRevD.79.084008} {\bibfield  {journal}
  {\bibinfo  {journal} {Phys. Rev.}\ }\textbf {\bibinfo {volume} {D79}},\
  \bibinfo {pages} {084008} (\bibinfo {year} {2009})},\ \Eprint
  {http://arxiv.org/abs/0901.3775} {arXiv:0901.3775 [hep-th]}\BibitemShut
  {NoStop}%
\bibitem [{\citenamefont {Blas}\ \emph {et~al.}(2010)\citenamefont {Blas},
  \citenamefont {Pujolas},\ and\ \citenamefont {Sibiryakov}}]{Blas:2009qj}%
  \BibitemOpen
  \bibfield  {author} {\bibinfo {author} {\bibfnamefont {D.}~\bibnamefont
  {Blas}}, \bibinfo {author} {\bibfnamefont {O.}~\bibnamefont {Pujolas}}, \
  and\ \bibinfo {author} {\bibfnamefont {S.}~\bibnamefont {Sibiryakov}},\
  }\href {\doibase10.1103/PhysRevLett.104.181302} {\bibfield  {journal}
  {\bibinfo  {journal} {Phys. Rev. Lett.}\ }\textbf {\bibinfo {volume} {104}},\
  \bibinfo {pages} {181302} (\bibinfo {year} {2010})},\ \Eprint
  {http://arxiv.org/abs/0909.3525} {arXiv:0909.3525 [hep-th]}\BibitemShut
  {NoStop}%
\bibitem [{\citenamefont {Visser}(2009{\natexlab{a}})}]{Visser:2009fg}%
  \BibitemOpen
  \bibfield  {author} {\bibinfo {author} {\bibfnamefont {Matt}\ \bibnamefont
  {Visser}},\ }\href {\doibase10.1103/PhysRevD.80.025011} {\bibfield  {journal}
  {\bibinfo  {journal} {Phys. Rev.}\ }\textbf {\bibinfo {volume} {D80}},\
  \bibinfo {pages} {025011} (\bibinfo {year} {2009}{\natexlab{a}})},\ \Eprint
  {http://arxiv.org/abs/0902.0590} {arXiv:0902.0590 [hep-th]}\BibitemShut
  {NoStop}%
\bibitem [{\citenamefont {Visser}(2009{\natexlab{b}})}]{Visser:2009ys}%
  \BibitemOpen
  \bibfield  {author} {\bibinfo {author} {\bibfnamefont {Matt}\ \bibnamefont
  {Visser}},\ }\href@noop {} {\  (\bibinfo {year} {2009}{\natexlab{b}})},\
  \Eprint {http://arxiv.org/abs/0912.4757} {arXiv:0912.4757
  [hep-th]}\BibitemShut {NoStop}%
\bibitem [{\citenamefont {Sotiriou}\ \emph
  {et~al.}(2009{\natexlab{a}})\citenamefont {Sotiriou}, \citenamefont
  {Visser},\ and\ \citenamefont {Weinfurtner}}]{Sotiriou:2009gy}%
  \BibitemOpen
  \bibfield  {author} {\bibinfo {author} {\bibfnamefont {Thomas~P.}\
  \bibnamefont {Sotiriou}}, \bibinfo {author} {\bibfnamefont {Matt}\
  \bibnamefont {Visser}}, \ and\ \bibinfo {author} {\bibfnamefont {Silke}\
  \bibnamefont {Weinfurtner}},\ }\href {\doibase10.1103/PhysRevLett.102.251601}
  {\bibfield  {journal} {\bibinfo  {journal} {Phys. Rev. Lett.}\ }\textbf
  {\bibinfo {volume} {102}},\ \bibinfo {pages} {251601} (\bibinfo {year}
  {2009}{\natexlab{a}})},\ \Eprint {http://arxiv.org/abs/0904.4464}
  {arXiv:0904.4464 [hep-th]}\BibitemShut {NoStop}%
\bibitem [{\citenamefont {Barvinsky}\ \emph {et~al.}(2016)\citenamefont
  {Barvinsky}, \citenamefont {Blas}, \citenamefont {Herrero-Valea},
  \citenamefont {Sibiryakov},\ and\ \citenamefont
  {Steinwachs}}]{Barvinsky:2015kil}%
  \BibitemOpen
  \bibfield  {author} {\bibinfo {author} {\bibfnamefont {Andrei~O.}\
  \bibnamefont {Barvinsky}}, \bibinfo {author} {\bibfnamefont {Diego}\
  \bibnamefont {Blas}}, \bibinfo {author} {\bibfnamefont {Mario}\ \bibnamefont
  {Herrero-Valea}}, \bibinfo {author} {\bibfnamefont {Sergey~M.}\ \bibnamefont
  {Sibiryakov}}, \ and\ \bibinfo {author} {\bibfnamefont {Christian~F.}\
  \bibnamefont {Steinwachs}},\ }\href {\doibase10.1103/PhysRevD.93.064022}
  {\bibfield  {journal} {\bibinfo  {journal} {Phys. Rev.}\ }\textbf {\bibinfo
  {volume} {D93}},\ \bibinfo {pages} {064022} (\bibinfo {year} {2016})},\
  \Eprint {http://arxiv.org/abs/1512.02250} {arXiv:1512.02250
  [hep-th]}\BibitemShut {NoStop}%
\bibitem [{\citenamefont {Sotiriou}\ \emph {et~al.}(2011)\citenamefont
  {Sotiriou}, \citenamefont {Visser},\ and\ \citenamefont
  {Weinfurtner}}]{Sotiriou:2011dr}%
  \BibitemOpen
  \bibfield  {author} {\bibinfo {author} {\bibfnamefont {Thomas~P.}\
  \bibnamefont {Sotiriou}}, \bibinfo {author} {\bibfnamefont {Matt}\
  \bibnamefont {Visser}}, \ and\ \bibinfo {author} {\bibfnamefont {Silke}\
  \bibnamefont {Weinfurtner}},\ }\href {\doibase10.1103/PhysRevD.83.124021}
  {\bibfield  {journal} {\bibinfo  {journal} {Phys. Rev.}\ }\textbf {\bibinfo
  {volume} {D83}},\ \bibinfo {pages} {124021} (\bibinfo {year} {2011})},\
  \Eprint {http://arxiv.org/abs/1103.3013} {arXiv:1103.3013
  [hep-th]}\BibitemShut {NoStop}%
\bibitem [{\citenamefont {Barvinsky}\ \emph {et~al.}(2017)\citenamefont
  {Barvinsky}, \citenamefont {Blas}, \citenamefont {Herrero-Valea},
  \citenamefont {Sibiryakov},\ and\ \citenamefont
  {Steinwachs}}]{Barvinsky:2017kob}%
  \BibitemOpen
  \bibfield  {author} {\bibinfo {author} {\bibfnamefont {Andrei~O.}\
  \bibnamefont {Barvinsky}}, \bibinfo {author} {\bibfnamefont {Diego}\
  \bibnamefont {Blas}}, \bibinfo {author} {\bibfnamefont {Mario}\ \bibnamefont
  {Herrero-Valea}}, \bibinfo {author} {\bibfnamefont {Sergey~M.}\ \bibnamefont
  {Sibiryakov}}, \ and\ \bibinfo {author} {\bibfnamefont {Christian~F.}\
  \bibnamefont {Steinwachs}},\ }\href {\doibase10.1103/PhysRevLett.119.211301}
  {\bibfield  {journal} {\bibinfo  {journal} {Phys. Rev. Lett.}\ }\textbf
  {\bibinfo {volume} {119}},\ \bibinfo {pages} {211301} (\bibinfo {year}
  {2017})},\ \Eprint {http://arxiv.org/abs/1706.06809} {arXiv:1706.06809
  [hep-th]}\BibitemShut {NoStop}%
\bibitem [{\citenamefont {Sotiriou}\ \emph
  {et~al.}(2009{\natexlab{b}})\citenamefont {Sotiriou}, \citenamefont
  {Visser},\ and\ \citenamefont {Weinfurtner}}]{Sotiriou:2009bx}%
  \BibitemOpen
  \bibfield  {author} {\bibinfo {author} {\bibfnamefont {Thomas~P.}\
  \bibnamefont {Sotiriou}}, \bibinfo {author} {\bibfnamefont {Matt}\
  \bibnamefont {Visser}}, \ and\ \bibinfo {author} {\bibfnamefont {Silke}\
  \bibnamefont {Weinfurtner}},\ }\href {\doibase10.1088/1126-6708/2009/10/033}
  {\bibfield  {journal} {\bibinfo  {journal} {JHEP}\ }\textbf {\bibinfo
  {volume} {10}},\ \bibinfo {pages} {033} (\bibinfo {year}
  {2009}{\natexlab{b}})},\ \Eprint {http://arxiv.org/abs/0905.2798}
  {arXiv:0905.2798 [hep-th]}\BibitemShut {NoStop}%
\bibitem [{\citenamefont {Charmousis}\ \emph {et~al.}(2009)\citenamefont
  {Charmousis}, \citenamefont {Niz}, \citenamefont {Padilla},\ and\
  \citenamefont {Saffin}}]{Charmousis:2009tc}%
  \BibitemOpen
  \bibfield  {author} {\bibinfo {author} {\bibfnamefont {Christos}\
  \bibnamefont {Charmousis}}, \bibinfo {author} {\bibfnamefont {Gustavo}\
  \bibnamefont {Niz}}, \bibinfo {author} {\bibfnamefont {Antonio}\ \bibnamefont
  {Padilla}}, \ and\ \bibinfo {author} {\bibfnamefont {Paul~M.}\ \bibnamefont
  {Saffin}},\ }\href {\doibase10.1088/1126-6708/2009/08/070} {\bibfield
  {journal} {\bibinfo  {journal} {JHEP}\ }\textbf {\bibinfo {volume} {08}},\
  \bibinfo {pages} {070} (\bibinfo {year} {2009})},\ \Eprint
  {http://arxiv.org/abs/0905.2579} {arXiv:0905.2579 [hep-th]}\BibitemShut
  {NoStop}%
\bibitem [{\citenamefont {Blas}\ \emph {et~al.}(2009)\citenamefont {Blas},
  \citenamefont {Pujolas},\ and\ \citenamefont {Sibiryakov}}]{Blas:2009yd}%
  \BibitemOpen
  \bibfield  {author} {\bibinfo {author} {\bibfnamefont {D.}~\bibnamefont
  {Blas}}, \bibinfo {author} {\bibfnamefont {O.}~\bibnamefont {Pujolas}}, \
  and\ \bibinfo {author} {\bibfnamefont {S.}~\bibnamefont {Sibiryakov}},\
  }\href {\doibase10.1088/1126-6708/2009/10/029} {\bibfield  {journal}
  {\bibinfo  {journal} {JHEP}\ }\textbf {\bibinfo {volume} {10}},\ \bibinfo
  {pages} {029} (\bibinfo {year} {2009})},\ \Eprint
  {http://arxiv.org/abs/0906.3046} {arXiv:0906.3046 [hep-th]}\BibitemShut
  {NoStop}%
\bibitem [{\citenamefont {Koyama}\ and\ \citenamefont
  {Arroja}(2010)}]{Koyama:2009hc}%
  \BibitemOpen
  \bibfield  {author} {\bibinfo {author} {\bibfnamefont {Kazuya}\ \bibnamefont
  {Koyama}}\ and\ \bibinfo {author} {\bibfnamefont {Frederico}\ \bibnamefont
  {Arroja}},\ }\href {\doibase10.1007/JHEP03(2010)061} {\bibfield  {journal}
  {\bibinfo  {journal} {JHEP}\ }\textbf {\bibinfo {volume} {03}},\ \bibinfo
  {pages} {061} (\bibinfo {year} {2010})},\ \Eprint
  {http://arxiv.org/abs/0910.1998} {arXiv:0910.1998 [hep-th]}\BibitemShut
  {NoStop}%
\bibitem [{\citenamefont {Sotiriou}(2018)}]{Sotiriou:2017obf}%
  \BibitemOpen
  \bibfield  {author} {\bibinfo {author} {\bibfnamefont {Thomas~P.}\
  \bibnamefont {Sotiriou}},\ }\href {\doibase10.1103/PhysRevLett.120.041104}
  {\bibfield  {journal} {\bibinfo  {journal} {Phys. Rev. Lett.}\ }\textbf
  {\bibinfo {volume} {120}},\ \bibinfo {pages} {041104} (\bibinfo {year}
  {2018})},\ \Eprint {http://arxiv.org/abs/1709.00940} {arXiv:1709.00940
  [gr-qc]}\BibitemShut {NoStop}%
\bibitem [{\citenamefont {Blas}\ and\ \citenamefont
  {Sibiryakov}(2011)}]{Blas:2011ni}%
  \BibitemOpen
  \bibfield  {author} {\bibinfo {author} {\bibfnamefont {D.}~\bibnamefont
  {Blas}}\ and\ \bibinfo {author} {\bibfnamefont {S.}~\bibnamefont
  {Sibiryakov}},\ }\href {\doibase10.1103/PhysRevD.84.124043} {\bibfield
  {journal} {\bibinfo  {journal} {Phys. Rev.}\ }\textbf {\bibinfo {volume}
  {D84}},\ \bibinfo {pages} {124043} (\bibinfo {year} {2011})},\ \Eprint
  {http://arxiv.org/abs/1110.2195} {arXiv:1110.2195 [hep-th]}\BibitemShut
  {NoStop}%
\bibitem [{\citenamefont {Bhattacharyya}\ \emph
  {et~al.}(2016{\natexlab{a}})\citenamefont {Bhattacharyya}, \citenamefont
  {Coates}, \citenamefont {Colombo},\ and\ \citenamefont
  {Sotiriou}}]{Bhattacharyya:2015uxt}%
  \BibitemOpen
  \bibfield  {author} {\bibinfo {author} {\bibfnamefont {Jishnu}\ \bibnamefont
  {Bhattacharyya}}, \bibinfo {author} {\bibfnamefont {Andrew}\ \bibnamefont
  {Coates}}, \bibinfo {author} {\bibfnamefont {Mattia}\ \bibnamefont
  {Colombo}}, \ and\ \bibinfo {author} {\bibfnamefont {Thomas~P.}\ \bibnamefont
  {Sotiriou}},\ }\href {\doibase10.1103/PhysRevD.93.064056} {\bibfield
  {journal} {\bibinfo  {journal} {Phys. Rev.}\ }\textbf {\bibinfo {volume}
  {D93}},\ \bibinfo {pages} {064056} (\bibinfo {year} {2016}{\natexlab{a}})},\
  \Eprint {http://arxiv.org/abs/1512.04899} {arXiv:1512.04899
  [gr-qc]}\BibitemShut {NoStop}%
\bibitem [{\citenamefont {Bhattacharyya}\ \emph
  {et~al.}(2016{\natexlab{b}})\citenamefont {Bhattacharyya}, \citenamefont
  {Colombo},\ and\ \citenamefont {Sotiriou}}]{Bhattacharyya:2015gwa}%
  \BibitemOpen
  \bibfield  {author} {\bibinfo {author} {\bibfnamefont {Jishnu}\ \bibnamefont
  {Bhattacharyya}}, \bibinfo {author} {\bibfnamefont {Mattia}\ \bibnamefont
  {Colombo}}, \ and\ \bibinfo {author} {\bibfnamefont {Thomas~P.}\ \bibnamefont
  {Sotiriou}},\ }\href {\doibase10.1088/0264-9381/33/23/235003} {\bibfield
  {journal} {\bibinfo  {journal} {Class. Quant. Grav.}\ }\textbf {\bibinfo
  {volume} {33}},\ \bibinfo {pages} {235003} (\bibinfo {year}
  {2016}{\natexlab{b}})},\ \Eprint {http://arxiv.org/abs/1509.01558}
  {arXiv:1509.01558 [gr-qc]}\BibitemShut {NoStop}%
\bibitem [{\citenamefont {Barausse}\ and\ \citenamefont
  {Sotiriou}(2013{\natexlab{a}})}]{Barausse:2013nwa}%
  \BibitemOpen
  \bibfield  {author} {\bibinfo {author} {\bibfnamefont {Enrico}\ \bibnamefont
  {Barausse}}\ and\ \bibinfo {author} {\bibfnamefont {Thomas~P.}\ \bibnamefont
  {Sotiriou}},\ }\href {\doibase10.1088/0264-9381/30/24/244010} {\bibfield
  {journal} {\bibinfo  {journal} {Class. Quant. Grav.}\ }\textbf {\bibinfo
  {volume} {30}},\ \bibinfo {pages} {244010} (\bibinfo {year}
  {2013}{\natexlab{a}})},\ \Eprint {http://arxiv.org/abs/1307.3359}
  {arXiv:1307.3359 [gr-qc]}\BibitemShut {NoStop}%
\bibitem [{\citenamefont {Jacobson}(2007)}]{Jacobson:2008aj}%
  \BibitemOpen
  \bibfield  {author} {\bibinfo {author} {\bibfnamefont {Ted}\ \bibnamefont
  {Jacobson}},\ }\bibfield  {booktitle} {\emph {\bibinfo {booktitle}
  {{Proceedings, Workshop on From quantum to emergent gravity: Theory and
  phenomenology (QG-Ph): Trieste, Italy, June 11-15, 2007}}},\ }\href
  {\doibase10.22323/1.043.0020} {\bibfield  {journal} {\bibinfo  {journal}
  {PoS}\ }\textbf {\bibinfo {volume} {QG-PH}},\ \bibinfo {pages} {020}
  (\bibinfo {year} {2007})},\ \Eprint {http://arxiv.org/abs/0801.1547}
  {arXiv:0801.1547 [gr-qc]}\BibitemShut {NoStop}%
\bibitem [{\citenamefont {Sotiriou}(2011)}]{Sotiriou:2010wn}%
  \BibitemOpen
  \bibfield  {author} {\bibinfo {author} {\bibfnamefont {Thomas~P.}\
  \bibnamefont {Sotiriou}},\ }\bibfield  {booktitle} {\emph {\bibinfo
  {booktitle} {{Proceedings, 14th Conference on Recent developments in gravity
  (NEB 14): Ioannina, Greece, June 8-11, 2010}}},\ }\href
  {\doibase10.1088/1742-6596/283/1/012034} {\bibfield  {journal} {\bibinfo
  {journal} {J. Phys. Conf. Ser.}\ }\textbf {\bibinfo {volume} {283}},\
  \bibinfo {pages} {012034} (\bibinfo {year} {2011})},\ \Eprint
  {http://arxiv.org/abs/1010.3218} {arXiv:1010.3218 [hep-th]}\BibitemShut
  {NoStop}%
\bibitem [{\citenamefont {Emir~Gumrukcuoglu}\ \emph {et~al.}(2018)\citenamefont
  {Emir~Gumrukcuoglu}, \citenamefont {Saravani},\ and\ \citenamefont
  {Sotiriou}}]{Gumrukcuoglu:2017ijh}%
  \BibitemOpen
  \bibfield  {author} {\bibinfo {author} {\bibfnamefont {A.}~\bibnamefont
  {Emir~Gumrukcuoglu}}, \bibinfo {author} {\bibfnamefont {Mehdi}\ \bibnamefont
  {Saravani}}, \ and\ \bibinfo {author} {\bibfnamefont {Thomas~P.}\
  \bibnamefont {Sotiriou}},\ }\href {\doibase10.1103/PhysRevD.97.024032}
  {\bibfield  {journal} {\bibinfo  {journal} {Phys. Rev.}\ }\textbf {\bibinfo
  {volume} {D97}},\ \bibinfo {pages} {024032} (\bibinfo {year} {2018})},\
  \Eprint {http://arxiv.org/abs/1711.08845} {arXiv:1711.08845
  [gr-qc]}\BibitemShut {NoStop}%
\bibitem [{\citenamefont {Oost}\ \emph {et~al.}(2018)\citenamefont {Oost},
  \citenamefont {Mukohyama},\ and\ \citenamefont {Wang}}]{Oost:2018tcv}%
  \BibitemOpen
  \bibfield  {author} {\bibinfo {author} {\bibfnamefont {Jacob}\ \bibnamefont
  {Oost}}, \bibinfo {author} {\bibfnamefont {Shinji}\ \bibnamefont
  {Mukohyama}}, \ and\ \bibinfo {author} {\bibfnamefont {Anzhong}\ \bibnamefont
  {Wang}},\ }\href@noop {} {\bibfield  {journal} {\bibinfo  {journal} {Phys.
  Rev.}\ }\textbf {\bibinfo {volume} {D97}},\ \bibinfo {pages} {124023}
  (\bibinfo {year} {2018})},\ \Eprint {http://arxiv.org/abs/1802.04303}
  {arXiv:1802.04303 [gr-qc]}\BibitemShut {NoStop}%
\bibitem [{\citenamefont {de~Rham}\ \emph {et~al.}(2011)\citenamefont
  {de~Rham}, \citenamefont {Gabadadze},\ and\ \citenamefont
  {Tolley}}]{deRham:2010kj}%
  \BibitemOpen
  \bibfield  {author} {\bibinfo {author} {\bibfnamefont {Claudia}\ \bibnamefont
  {de~Rham}}, \bibinfo {author} {\bibfnamefont {Gregory}\ \bibnamefont
  {Gabadadze}}, \ and\ \bibinfo {author} {\bibfnamefont {Andrew~J.}\
  \bibnamefont {Tolley}},\ }\href {\doibase10.1103/PhysRevLett.106.231101}
  {\bibfield  {journal} {\bibinfo  {journal} {Phys. Rev. Lett.}\ }\textbf
  {\bibinfo {volume} {106}},\ \bibinfo {pages} {231101} (\bibinfo {year}
  {2011})},\ \Eprint {http://arxiv.org/abs/1011.1232} {arXiv:1011.1232
  [hep-th]}\BibitemShut {NoStop}%
\bibitem [{\citenamefont {Hassan}\ and\ \citenamefont
  {Rosen}(2012{\natexlab{a}})}]{Hassan:2011hr}%
  \BibitemOpen
  \bibfield  {author} {\bibinfo {author} {\bibfnamefont {S.~F.}\ \bibnamefont
  {Hassan}}\ and\ \bibinfo {author} {\bibfnamefont {Rachel~A.}\ \bibnamefont
  {Rosen}},\ }\href {\doibase10.1103/PhysRevLett.108.041101} {\bibfield
  {journal} {\bibinfo  {journal} {Phys. Rev. Lett.}\ }\textbf {\bibinfo
  {volume} {108}},\ \bibinfo {pages} {041101} (\bibinfo {year}
  {2012}{\natexlab{a}})},\ \Eprint {http://arxiv.org/abs/1106.3344}
  {arXiv:1106.3344 [hep-th]}\BibitemShut {NoStop}%
\bibitem [{\citenamefont {Hassan}\ and\ \citenamefont
  {Rosen}(2012{\natexlab{b}})}]{Hassan:2011zd}%
  \BibitemOpen
  \bibfield  {author} {\bibinfo {author} {\bibfnamefont {S.~F.}\ \bibnamefont
  {Hassan}}\ and\ \bibinfo {author} {\bibfnamefont {Rachel~A.}\ \bibnamefont
  {Rosen}},\ }\href {\doibase10.1007/JHEP02(2012)126} {\bibfield  {journal}
  {\bibinfo  {journal} {JHEP}\ }\textbf {\bibinfo {volume} {02}},\ \bibinfo
  {pages} {126} (\bibinfo {year} {2012}{\natexlab{b}})},\ \Eprint
  {http://arxiv.org/abs/1109.3515} {arXiv:1109.3515 [hep-th]}\BibitemShut
  {NoStop}%
\bibitem [{\citenamefont {Schmidt-May}\ and\ \citenamefont {von
  Strauss}(2016)}]{Schmidt-May:2015vnx}%
  \BibitemOpen
  \bibfield  {author} {\bibinfo {author} {\bibfnamefont {Angnis}\ \bibnamefont
  {Schmidt-May}}\ and\ \bibinfo {author} {\bibfnamefont {Mikael}\ \bibnamefont
  {von Strauss}},\ }\href {\doibase10.1088/1751-8113/49/18/183001} {\bibfield
  {journal} {\bibinfo  {journal} {J. Phys.}\ }\textbf {\bibinfo {volume}
  {A49}},\ \bibinfo {pages} {183001} (\bibinfo {year} {2016})},\ \Eprint
  {http://arxiv.org/abs/1512.00021} {arXiv:1512.00021 [hep-th]}\BibitemShut
  {NoStop}%
\bibitem [{\citenamefont {de~Rham}\ \emph {et~al.}(2017)\citenamefont
  {de~Rham}, \citenamefont {Deskins}, \citenamefont {Tolley},\ and\
  \citenamefont {Zhou}}]{deRham:2016nuf}%
  \BibitemOpen
  \bibfield  {author} {\bibinfo {author} {\bibfnamefont {Claudia}\ \bibnamefont
  {de~Rham}}, \bibinfo {author} {\bibfnamefont {J.~Tate}\ \bibnamefont
  {Deskins}}, \bibinfo {author} {\bibfnamefont {Andrew~J.}\ \bibnamefont
  {Tolley}}, \ and\ \bibinfo {author} {\bibfnamefont {Shuang-Yong}\
  \bibnamefont {Zhou}},\ }\href {\doibase10.1103/RevModPhys.89.025004}
  {\bibfield  {journal} {\bibinfo  {journal} {Rev. Mod. Phys.}\ }\textbf
  {\bibinfo {volume} {89}},\ \bibinfo {pages} {025004} (\bibinfo {year}
  {2017})},\ \Eprint {http://arxiv.org/abs/1606.08462} {arXiv:1606.08462
  [astro-ph.CO]}\BibitemShut {NoStop}%
\bibitem [{\citenamefont {Akrami}\ \emph {et~al.}(2015)\citenamefont {Akrami},
  \citenamefont {Hassan}, \citenamefont {Könnig}, \citenamefont
  {Schmidt-May},\ and\ \citenamefont {Solomon}}]{Akrami:2015qga}%
  \BibitemOpen
  \bibfield  {author} {\bibinfo {author} {\bibfnamefont {Yashar}\ \bibnamefont
  {Akrami}}, \bibinfo {author} {\bibfnamefont {S.~F.}\ \bibnamefont {Hassan}},
  \bibinfo {author} {\bibfnamefont {Frank}\ \bibnamefont {Könnig}}, \bibinfo
  {author} {\bibfnamefont {Angnis}\ \bibnamefont {Schmidt-May}}, \ and\
  \bibinfo {author} {\bibfnamefont {Adam~R.}\ \bibnamefont {Solomon}},\ }\href
  {\doibase10.1016/j.physletb.2015.06.062} {\bibfield  {journal} {\bibinfo
  {journal} {Phys. Lett.}\ }\textbf {\bibinfo {volume} {B748}},\ \bibinfo
  {pages} {37--44} (\bibinfo {year} {2015})},\ \Eprint
  {http://arxiv.org/abs/1503.07521} {arXiv:1503.07521 [gr-qc]}\BibitemShut
  {NoStop}%
\bibitem [{\citenamefont {Aoki}\ and\ \citenamefont
  {Maeda}(2018)}]{Aoki:2017cnz}%
  \BibitemOpen
  \bibfield  {author} {\bibinfo {author} {\bibfnamefont {Katsuki}\ \bibnamefont
  {Aoki}}\ and\ \bibinfo {author} {\bibfnamefont {Kei-ichi}\ \bibnamefont
  {Maeda}},\ }\href {\doibase10.1103/PhysRevD.97.044002} {\bibfield  {journal}
  {\bibinfo  {journal} {Phys. Rev.}\ }\textbf {\bibinfo {volume} {D97}},\
  \bibinfo {pages} {044002} (\bibinfo {year} {2018})},\ \Eprint
  {http://arxiv.org/abs/1707.05003} {arXiv:1707.05003 [hep-th]}\BibitemShut
  {NoStop}%
\bibitem [{\citenamefont {Babichev}\ \emph {et~al.}(2016)\citenamefont
  {Babichev}, \citenamefont {Marzola}, \citenamefont {Raidal}, \citenamefont
  {Schmidt-May}, \citenamefont {Urban}, \citenamefont {Veermäe},\ and\
  \citenamefont {von Strauss}}]{Babichev:2016bxi}%
  \BibitemOpen
  \bibfield  {author} {\bibinfo {author} {\bibfnamefont {Eugeny}\ \bibnamefont
  {Babichev}}, \bibinfo {author} {\bibfnamefont {Luca}\ \bibnamefont
  {Marzola}}, \bibinfo {author} {\bibfnamefont {Martti}\ \bibnamefont
  {Raidal}}, \bibinfo {author} {\bibfnamefont {Angnis}\ \bibnamefont
  {Schmidt-May}}, \bibinfo {author} {\bibfnamefont {Federico}\ \bibnamefont
  {Urban}}, \bibinfo {author} {\bibfnamefont {Hardi}\ \bibnamefont {Veermäe}},
  \ and\ \bibinfo {author} {\bibfnamefont {Mikael}\ \bibnamefont {von
  Strauss}},\ }\href {\doibase10.1088/1475-7516/2016/09/016} {\bibfield
  {journal} {\bibinfo  {journal} {JCAP}\ }\textbf {\bibinfo {volume} {1609}},\
  \bibinfo {pages} {016} (\bibinfo {year} {2016})},\ \Eprint
  {http://arxiv.org/abs/1607.03497} {arXiv:1607.03497 [hep-th]}\BibitemShut
  {NoStop}%
\bibitem [{\citenamefont {Hassan}\ and\ \citenamefont
  {Kocic}(2018)}]{Hassan:2017ugh}%
  \BibitemOpen
  \bibfield  {author} {\bibinfo {author} {\bibfnamefont {S.~F.}\ \bibnamefont
  {Hassan}}\ and\ \bibinfo {author} {\bibfnamefont {Mikica}\ \bibnamefont
  {Kocic}},\ }\href {\doibase10.1007/JHEP05(2018)099} {\bibfield  {journal}
  {\bibinfo  {journal} {JHEP}\ }\textbf {\bibinfo {volume} {05}},\ \bibinfo
  {pages} {099} (\bibinfo {year} {2018})},\ \Eprint
  {http://arxiv.org/abs/1706.07806} {arXiv:1706.07806 [hep-th]}\BibitemShut
  {NoStop}%
\bibitem [{\citenamefont {Kocic}(2018)}]{Kocic:2018yvr}%
  \BibitemOpen
  \bibfield  {author} {\bibinfo {author} {\bibfnamefont {Mikica}\ \bibnamefont
  {Kocic}},\ }\href@noop {} {\  (\bibinfo {year} {2018})},\ \Eprint
  {http://arxiv.org/abs/1804.03659} {arXiv:1804.03659 [hep-th]}\BibitemShut
  {NoStop}%
\bibitem [{\citenamefont {Yagi}\ \emph {et~al.}(2014)\citenamefont {Yagi},
  \citenamefont {Blas}, \citenamefont {Yunes},\ and\ \citenamefont
  {Barausse}}]{Yagi:2013qpa}%
  \BibitemOpen
  \bibfield  {author} {\bibinfo {author} {\bibfnamefont {Kent}\ \bibnamefont
  {Yagi}}, \bibinfo {author} {\bibfnamefont {Diego}\ \bibnamefont {Blas}},
  \bibinfo {author} {\bibfnamefont {Nicolás}\ \bibnamefont {Yunes}}, \ and\
  \bibinfo {author} {\bibfnamefont {Enrico}\ \bibnamefont {Barausse}},\ }\href
  {\doibase10.1103/PhysRevLett.112.161101} {\bibfield  {journal} {\bibinfo
  {journal} {Phys. Rev. Lett.}\ }\textbf {\bibinfo {volume} {112}},\ \bibinfo
  {pages} {161101} (\bibinfo {year} {2014})},\ \Eprint
  {http://arxiv.org/abs/1307.6219} {arXiv:1307.6219 [gr-qc]}\BibitemShut
  {NoStop}%
\bibitem [{\citenamefont {Mirshekari}\ \emph {et~al.}(2012)\citenamefont
  {Mirshekari}, \citenamefont {Yunes},\ and\ \citenamefont
  {Will}}]{Mirshekari:2011yq}%
  \BibitemOpen
  \bibfield  {author} {\bibinfo {author} {\bibfnamefont {Saeed}\ \bibnamefont
  {Mirshekari}}, \bibinfo {author} {\bibfnamefont {Nicolas}\ \bibnamefont
  {Yunes}}, \ and\ \bibinfo {author} {\bibfnamefont {Clifford~M.}\ \bibnamefont
  {Will}},\ }\href {\doibase10.1103/PhysRevD.85.024041} {\bibfield  {journal}
  {\bibinfo  {journal} {Phys. Rev.}\ }\textbf {\bibinfo {volume} {D85}},\
  \bibinfo {pages} {024041} (\bibinfo {year} {2012})},\ \Eprint
  {http://arxiv.org/abs/1110.2720} {arXiv:1110.2720 [gr-qc]}\BibitemShut
  {NoStop}%
\bibitem [{\citenamefont {Kostelecký}\ and\ \citenamefont
  {Mewes}(2016)}]{Kostelecky:2016kfm}%
  \BibitemOpen
  \bibfield  {author} {\bibinfo {author} {\bibfnamefont {V.~Alan}\ \bibnamefont
  {Kostelecký}}\ and\ \bibinfo {author} {\bibfnamefont {Matthew}\ \bibnamefont
  {Mewes}},\ }\href {\doibase10.1016/j.physletb.2016.04.040} {\bibfield
  {journal} {\bibinfo  {journal} {Phys. Lett.}\ }\textbf {\bibinfo {volume}
  {B757}},\ \bibinfo {pages} {510--514} (\bibinfo {year} {2016})},\ \Eprint
  {http://arxiv.org/abs/1602.04782} {arXiv:1602.04782 [gr-qc]}\BibitemShut
  {NoStop}%
\bibitem [{\citenamefont {Yunes}\ \emph {et~al.}(2016)\citenamefont {Yunes},
  \citenamefont {Yagi},\ and\ \citenamefont {Pretorius}}]{Yunes:2016jcc}%
  \BibitemOpen
  \bibfield  {author} {\bibinfo {author} {\bibfnamefont {Nicolas}\ \bibnamefont
  {Yunes}}, \bibinfo {author} {\bibfnamefont {Kent}\ \bibnamefont {Yagi}}, \
  and\ \bibinfo {author} {\bibfnamefont {Frans}\ \bibnamefont {Pretorius}},\
  }\href {\doibase10.1103/PhysRevD.94.084002} {\bibfield  {journal} {\bibinfo
  {journal} {Phys. Rev.}\ }\textbf {\bibinfo {volume} {D94}},\ \bibinfo {pages}
  {084002} (\bibinfo {year} {2016})},\ \Eprint
  {http://arxiv.org/abs/1603.08955} {arXiv:1603.08955 [gr-qc]}\BibitemShut
  {NoStop}%
\bibitem [{\citenamefont {Will}(2018)}]{Will:2018gku}%
  \BibitemOpen
  \bibfield  {author} {\bibinfo {author} {\bibfnamefont {Clifford~M.}\
  \bibnamefont {Will}},\ }\href@noop {} {\  (\bibinfo {year} {2018})},\ \Eprint
  {http://arxiv.org/abs/1805.10523} {arXiv:1805.10523 [gr-qc]}\BibitemShut
  {NoStop}%
\bibitem [{\citenamefont {Arun}\ \emph
  {et~al.}(2006{\natexlab{a}})\citenamefont {Arun}, \citenamefont {Iyer},
  \citenamefont {Qusailah},\ and\ \citenamefont {Sathyaprakash}}]{Arun:2006yw}%
  \BibitemOpen
  \bibfield  {author} {\bibinfo {author} {\bibfnamefont {K.~G.}\ \bibnamefont
  {Arun}}, \bibinfo {author} {\bibfnamefont {Bala~R.}\ \bibnamefont {Iyer}},
  \bibinfo {author} {\bibfnamefont {M.~S.~S.}\ \bibnamefont {Qusailah}}, \ and\
  \bibinfo {author} {\bibfnamefont {B.~S.}\ \bibnamefont {Sathyaprakash}},\
  }\href {\doibase10.1088/0264-9381/23/9/L01} {\bibfield  {journal} {\bibinfo
  {journal} {Class. Quant. Grav.}\ }\textbf {\bibinfo {volume} {23}},\ \bibinfo
  {pages} {L37--L43} (\bibinfo {year} {2006}{\natexlab{a}})},\ \Eprint
  {http://arxiv.org/abs/gr-qc/0604018} {arXiv:gr-qc/0604018
  [gr-qc]}\BibitemShut {NoStop}%
\bibitem [{\citenamefont {Arun}\ \emph
  {et~al.}(2006{\natexlab{b}})\citenamefont {Arun}, \citenamefont {Iyer},
  \citenamefont {Qusailah},\ and\ \citenamefont {Sathyaprakash}}]{Arun:2006hn}%
  \BibitemOpen
  \bibfield  {author} {\bibinfo {author} {\bibfnamefont {K.~G.}\ \bibnamefont
  {Arun}}, \bibinfo {author} {\bibfnamefont {Bala~R.}\ \bibnamefont {Iyer}},
  \bibinfo {author} {\bibfnamefont {M.~S.~S.}\ \bibnamefont {Qusailah}}, \ and\
  \bibinfo {author} {\bibfnamefont {B.~S.}\ \bibnamefont {Sathyaprakash}},\
  }\href {\doibase10.1103/PhysRevD.74.024006} {\bibfield  {journal} {\bibinfo
  {journal} {Phys. Rev.}\ }\textbf {\bibinfo {volume} {D74}},\ \bibinfo {pages}
  {024006} (\bibinfo {year} {2006}{\natexlab{b}})},\ \Eprint
  {http://arxiv.org/abs/gr-qc/0604067} {arXiv:gr-qc/0604067
  [gr-qc]}\BibitemShut {NoStop}%
\bibitem [{\citenamefont {Mishra}\ \emph {et~al.}(2010)\citenamefont {Mishra},
  \citenamefont {Arun}, \citenamefont {Iyer},\ and\ \citenamefont
  {Sathyaprakash}}]{Mishra:2010tp}%
  \BibitemOpen
  \bibfield  {author} {\bibinfo {author} {\bibfnamefont {Chandra~Kant}\
  \bibnamefont {Mishra}}, \bibinfo {author} {\bibfnamefont {K.~G.}\
  \bibnamefont {Arun}}, \bibinfo {author} {\bibfnamefont {Bala~R.}\
  \bibnamefont {Iyer}}, \ and\ \bibinfo {author} {\bibfnamefont {B.~S.}\
  \bibnamefont {Sathyaprakash}},\ }\href {\doibase10.1103/PhysRevD.82.064010}
  {\bibfield  {journal} {\bibinfo  {journal} {Phys. Rev.}\ }\textbf {\bibinfo
  {volume} {D82}},\ \bibinfo {pages} {064010} (\bibinfo {year} {2010})},\
  \Eprint {http://arxiv.org/abs/1005.0304} {arXiv:1005.0304
  [gr-qc]}\BibitemShut {NoStop}%
\bibitem [{\citenamefont {Stairs}(2003)}]{Stairs:2003eg}%
  \BibitemOpen
  \bibfield  {author} {\bibinfo {author} {\bibfnamefont {Ingrid~H.}\
  \bibnamefont {Stairs}},\ }\href {\doibase10.12942/lrr-2003-5} {\bibfield
  {journal} {\bibinfo  {journal} {Living Rev. Rel.}\ }\textbf {\bibinfo
  {volume} {6}},\ \bibinfo {pages} {5} (\bibinfo {year} {2003})},\ \Eprint
  {http://arxiv.org/abs/astro-ph/0307536} {arXiv:astro-ph/0307536
  [astro-ph]}\BibitemShut {NoStop}%
\bibitem [{\citenamefont {Perrodin}\ and\ \citenamefont
  {Sesana}(2017)}]{Perrodin:2017bxr}%
  \BibitemOpen
  \bibfield  {author} {\bibinfo {author} {\bibfnamefont {Delphine}\
  \bibnamefont {Perrodin}}\ and\ \bibinfo {author} {\bibfnamefont {Alberto}\
  \bibnamefont {Sesana}},\ }\href@noop {} {\  (\bibinfo {year} {2017})},\
  \Eprint {http://arxiv.org/abs/1709.02816} {arXiv:1709.02816
  [astro-ph.HE]}\BibitemShut {NoStop}%
\bibitem [{\citenamefont {Yunes}\ and\ \citenamefont
  {Pretorius}(2009{\natexlab{b}})}]{Yunes:2009ke}%
  \BibitemOpen
  \bibfield  {author} {\bibinfo {author} {\bibfnamefont {Nicolas}\ \bibnamefont
  {Yunes}}\ and\ \bibinfo {author} {\bibfnamefont {Frans}\ \bibnamefont
  {Pretorius}},\ }\href {\doibase10.1103/PhysRevD.80.122003} {\bibfield
  {journal} {\bibinfo  {journal} {Phys. Rev.}\ }\textbf {\bibinfo {volume}
  {D80}},\ \bibinfo {pages} {122003} (\bibinfo {year} {2009}{\natexlab{b}})},\
  \Eprint {http://arxiv.org/abs/0909.3328} {arXiv:0909.3328
  [gr-qc]}\BibitemShut {NoStop}%
\bibitem [{\citenamefont {Jai-akson}\ \emph {et~al.}(2017)\citenamefont
  {Jai-akson}, \citenamefont {Chatrabhuti}, \citenamefont {Evnin},\ and\
  \citenamefont {Lehner}}]{Jai-akson:2017ldo}%
  \BibitemOpen
  \bibfield  {author} {\bibinfo {author} {\bibfnamefont {Puttarak}\
  \bibnamefont {Jai-akson}}, \bibinfo {author} {\bibfnamefont {Auttakit}\
  \bibnamefont {Chatrabhuti}}, \bibinfo {author} {\bibfnamefont {Oleg}\
  \bibnamefont {Evnin}}, \ and\ \bibinfo {author} {\bibfnamefont {Luis}\
  \bibnamefont {Lehner}},\ }\href {\doibase10.1103/PhysRevD.96.044031}
  {\bibfield  {journal} {\bibinfo  {journal} {Phys. Rev.}\ }\textbf {\bibinfo
  {volume} {D96}},\ \bibinfo {pages} {044031} (\bibinfo {year} {2017})},\
  \Eprint {http://arxiv.org/abs/1706.06519} {arXiv:1706.06519
  [gr-qc]}\BibitemShut {NoStop}%
\bibitem [{\citenamefont {Hirschmann}\ \emph {et~al.}(2018)\citenamefont
  {Hirschmann}, \citenamefont {Lehner}, \citenamefont {Liebling},\ and\
  \citenamefont {Palenzuela}}]{Hirschmann:2017psw}%
  \BibitemOpen
  \bibfield  {author} {\bibinfo {author} {\bibfnamefont {Eric~W.}\ \bibnamefont
  {Hirschmann}}, \bibinfo {author} {\bibfnamefont {Luis}\ \bibnamefont
  {Lehner}}, \bibinfo {author} {\bibfnamefont {Steven~L.}\ \bibnamefont
  {Liebling}}, \ and\ \bibinfo {author} {\bibfnamefont {Carlos}\ \bibnamefont
  {Palenzuela}},\ }\href {\doibase10.1103/PhysRevD.97.064032} {\bibfield
  {journal} {\bibinfo  {journal} {Phys. Rev.}\ }\textbf {\bibinfo {volume}
  {D97}},\ \bibinfo {pages} {064032} (\bibinfo {year} {2018})},\ \Eprint
  {http://arxiv.org/abs/1706.09875} {arXiv:1706.09875 [gr-qc]}\BibitemShut
  {NoStop}%
\bibitem [{\citenamefont {Ferrari}\ and\ \citenamefont
  {Gualtieri}(2008)}]{Ferrari:2007dd}%
  \BibitemOpen
  \bibfield  {author} {\bibinfo {author} {\bibfnamefont {Valeria}\ \bibnamefont
  {Ferrari}}\ and\ \bibinfo {author} {\bibfnamefont {Leonardo}\ \bibnamefont
  {Gualtieri}},\ }\href {\doibase10.1007/s10714-007-0585-1} {\bibfield
  {journal} {\bibinfo  {journal} {Gen. Rel. Grav.}\ }\textbf {\bibinfo {volume}
  {40}},\ \bibinfo {pages} {945--970} (\bibinfo {year} {2008})},\ \Eprint
  {http://arxiv.org/abs/0709.0657} {arXiv:0709.0657 [gr-qc]}\BibitemShut
  {NoStop}%
\bibitem [{\citenamefont {Berti}\ \emph {et~al.}(2009)\citenamefont {Berti},
  \citenamefont {Cardoso},\ and\ \citenamefont {Starinets}}]{Berti:2009kk}%
  \BibitemOpen
  \bibfield  {author} {\bibinfo {author} {\bibfnamefont {Emanuele}\
  \bibnamefont {Berti}}, \bibinfo {author} {\bibfnamefont {Vitor}\ \bibnamefont
  {Cardoso}}, \ and\ \bibinfo {author} {\bibfnamefont {Andrei~O.}\ \bibnamefont
  {Starinets}},\ }\href {\doibase10.1088/0264-9381/26/16/163001} {\bibfield
  {journal} {\bibinfo  {journal} {Class. Quantum Grav.}\ }\textbf {\bibinfo
  {volume} {26}},\ \bibinfo {pages} {163001} (\bibinfo {year} {2009})},\
  \Eprint {http://arxiv.org/abs/0905.2975} {arXiv:0905.2975
  [gr-qc]}\BibitemShut {NoStop}%
\bibitem [{\citenamefont {Konoplya}(2003)}]{Konoplya:2003ii}%
  \BibitemOpen
  \bibfield  {author} {\bibinfo {author} {\bibfnamefont {R.~A.}\ \bibnamefont
  {Konoplya}},\ }\href {\doibase10.1103/PhysRevD.68.024018} {\bibfield
  {journal} {\bibinfo  {journal} {Phys. Rev.}\ }\textbf {\bibinfo {volume}
  {D68}},\ \bibinfo {pages} {024018} (\bibinfo {year} {2003})},\ \Eprint
  {http://arxiv.org/abs/gr-qc/0303052} {arXiv:gr-qc/0303052
  [gr-qc]}\BibitemShut {NoStop}%
\bibitem [{\citenamefont {Cardoso}\ and\ \citenamefont
  {Gualtieri}(2016)}]{Cardoso:2016ryw}%
  \BibitemOpen
  \bibfield  {author} {\bibinfo {author} {\bibfnamefont {Vitor}\ \bibnamefont
  {Cardoso}}\ and\ \bibinfo {author} {\bibfnamefont {Leonardo}\ \bibnamefont
  {Gualtieri}},\ }\href {\doibase10.1088/0264-9381/33/17/174001} {\bibfield
  {journal} {\bibinfo  {journal} {Class. Quant. Grav.}\ }\textbf {\bibinfo
  {volume} {33}},\ \bibinfo {pages} {174001} (\bibinfo {year} {2016})},\
  \Eprint {http://arxiv.org/abs/1607.03133} {arXiv:1607.03133
  [gr-qc]}\BibitemShut {NoStop}%
\bibitem [{\citenamefont {Carter}(1971)}]{Carter:1971zc}%
  \BibitemOpen
  \bibfield  {author} {\bibinfo {author} {\bibfnamefont {B.}~\bibnamefont
  {Carter}},\ }\href {\doibase10.1103/PhysRevLett.26.331} {\bibfield  {journal}
  {\bibinfo  {journal} {Phys. Rev. Lett.}\ }\textbf {\bibinfo {volume} {26}},\
  \bibinfo {pages} {331--333} (\bibinfo {year} {1971})}\BibitemShut {NoStop}%
\bibitem [{\citenamefont {Robinson}(2009)}]{Robinson}%
  \BibitemOpen
  \bibfield  {author} {\bibinfo {author} {\bibfnamefont {D.}~\bibnamefont
  {Robinson}},\ }\enquote {\bibinfo {title} {{Four decades of black holes
  uniqueness theorems}},}\ in\ \href@noop {} {\emph {\bibinfo {booktitle} {The
  Kerr spacetime: Rotating black holes in general relativity}}},\ \bibinfo
  {editor} {edited by\ \bibinfo {editor} {\bibfnamefont {David}\ \bibnamefont
  {Wiltshire}}, \bibinfo {editor} {\bibfnamefont {Matt}\ \bibnamefont
  {Visser}}, \ and\ \bibinfo {editor} {\bibfnamefont {Susan}\ \bibnamefont
  {Scott}}}\ (\bibinfo  {publisher} {Cambridge University Press},\ \bibinfo
  {year} {2009})\BibitemShut {NoStop}%
\bibitem [{\citenamefont {Gossan}\ \emph {et~al.}(2012)\citenamefont {Gossan},
  \citenamefont {Veitch},\ and\ \citenamefont {Sathyaprakash}}]{Gossan:2011ha}%
  \BibitemOpen
  \bibfield  {author} {\bibinfo {author} {\bibfnamefont {S.}~\bibnamefont
  {Gossan}}, \bibinfo {author} {\bibfnamefont {J.}~\bibnamefont {Veitch}}, \
  and\ \bibinfo {author} {\bibfnamefont {B.~S.}\ \bibnamefont
  {Sathyaprakash}},\ }\href {\doibase10.1103/PhysRevD.85.124056} {\bibfield
  {journal} {\bibinfo  {journal} {Phys. Rev.}\ }\textbf {\bibinfo {volume}
  {D85}},\ \bibinfo {pages} {124056} (\bibinfo {year} {2012})},\ \Eprint
  {http://arxiv.org/abs/1111.5819} {arXiv:1111.5819 [gr-qc]}\BibitemShut
  {NoStop}%
\bibitem [{\citenamefont {Berti}\ \emph {et~al.}(2016)\citenamefont {Berti},
  \citenamefont {Sesana}, \citenamefont {Barausse}, \citenamefont {Cardoso},\
  and\ \citenamefont {Belczynski}}]{Berti:2016lat}%
  \BibitemOpen
  \bibfield  {author} {\bibinfo {author} {\bibfnamefont {Emanuele}\
  \bibnamefont {Berti}}, \bibinfo {author} {\bibfnamefont {Alberto}\
  \bibnamefont {Sesana}}, \bibinfo {author} {\bibfnamefont {Enrico}\
  \bibnamefont {Barausse}}, \bibinfo {author} {\bibfnamefont {Vitor}\
  \bibnamefont {Cardoso}}, \ and\ \bibinfo {author} {\bibfnamefont {Krzysztof}\
  \bibnamefont {Belczynski}},\ }\href {\doibase10.1103/PhysRevLett.117.101102}
  {\bibfield  {journal} {\bibinfo  {journal} {Phys. Rev. Lett.}\ }\textbf
  {\bibinfo {volume} {117}},\ \bibinfo {pages} {101102} (\bibinfo {year}
  {2016})},\ \Eprint {http://arxiv.org/abs/1605.09286} {arXiv:1605.09286
  [gr-qc]}\BibitemShut {NoStop}%
\bibitem [{\citenamefont {Meidam}\ \emph {et~al.}(2014)\citenamefont {Meidam},
  \citenamefont {Agathos}, \citenamefont {Van Den~Broeck}, \citenamefont
  {Veitch},\ and\ \citenamefont {Sathyaprakash}}]{Meidam:2014jpa}%
  \BibitemOpen
  \bibfield  {author} {\bibinfo {author} {\bibfnamefont {J.}~\bibnamefont
  {Meidam}}, \bibinfo {author} {\bibfnamefont {M.}~\bibnamefont {Agathos}},
  \bibinfo {author} {\bibfnamefont {C.}~\bibnamefont {Van Den~Broeck}},
  \bibinfo {author} {\bibfnamefont {J.}~\bibnamefont {Veitch}}, \ and\ \bibinfo
  {author} {\bibfnamefont {B.~S.}\ \bibnamefont {Sathyaprakash}},\ }\href
  {\doibase10.1103/PhysRevD.90.064009} {\bibfield  {journal} {\bibinfo
  {journal} {Phys. Rev.}\ }\textbf {\bibinfo {volume} {D90}},\ \bibinfo {pages}
  {064009} (\bibinfo {year} {2014})},\ \Eprint {http://arxiv.org/abs/1406.3201}
  {arXiv:1406.3201 [gr-qc]}\BibitemShut {NoStop}%
\bibitem [{\citenamefont {Vishveshwara}(1970)}]{Vishveshwara}%
  \BibitemOpen
  \bibfield  {author} {\bibinfo {author} {\bibfnamefont {C.~V.}\ \bibnamefont
  {Vishveshwara}},\ }\href {\doibase10.1103/PhysRevD.1.2870} {\bibfield
  {journal} {\bibinfo  {journal} {Phys. Rev. D}\ }\textbf {\bibinfo {volume}
  {1}},\ \bibinfo {pages} {2870--2879} (\bibinfo {year} {1970})}\BibitemShut
  {NoStop}%
\bibitem [{\citenamefont {Regge}\ and\ \citenamefont
  {Wheeler}(1957)}]{Regge:1957td}%
  \BibitemOpen
  \bibfield  {author} {\bibinfo {author} {\bibfnamefont {Tullio}\ \bibnamefont
  {Regge}}\ and\ \bibinfo {author} {\bibfnamefont {John~A.}\ \bibnamefont
  {Wheeler}},\ }\href {\doibase10.1103/PhysRev.108.1063} {\bibfield  {journal}
  {\bibinfo  {journal} {Phys. Rev.}\ }\textbf {\bibinfo {volume} {108}},\
  \bibinfo {pages} {1063--1069} (\bibinfo {year} {1957})}\BibitemShut {NoStop}%
\bibitem [{\citenamefont {Zerilli}(1970)}]{Zerilli:1970se}%
  \BibitemOpen
  \bibfield  {author} {\bibinfo {author} {\bibfnamefont {Frank~J.}\
  \bibnamefont {Zerilli}},\ }\href {\doibase10.1103/PhysRevLett.24.737}
  {\bibfield  {journal} {\bibinfo  {journal} {Phys. Rev. Lett.}\ }\textbf
  {\bibinfo {volume} {24}},\ \bibinfo {pages} {737--738} (\bibinfo {year}
  {1970})}\BibitemShut {NoStop}%
\bibitem [{\citenamefont {Chandrasekhar}(1998)}]{Chandra}%
  \BibitemOpen
  \bibfield  {author} {\bibinfo {author} {\bibfnamefont {Subrahmanyan}\
  \bibnamefont {Chandrasekhar}},\ }\href@noop {} {\emph {\bibinfo {title} {{The
  mathematical theory of black holes}}}}\ (\bibinfo  {publisher} {Clarendon
  Press},\ \bibinfo {year} {1998})\BibitemShut {NoStop}%
\bibitem [{\citenamefont {Pani}\ \emph
  {et~al.}(2012{\natexlab{a}})\citenamefont {Pani}, \citenamefont {Cardoso},
  \citenamefont {Gualtieri}, \citenamefont {Berti},\ and\ \citenamefont
  {Ishibashi}}]{Pani:2012bp}%
  \BibitemOpen
  \bibfield  {author} {\bibinfo {author} {\bibfnamefont {Paolo}\ \bibnamefont
  {Pani}}, \bibinfo {author} {\bibfnamefont {Vitor}\ \bibnamefont {Cardoso}},
  \bibinfo {author} {\bibfnamefont {Leonardo}\ \bibnamefont {Gualtieri}},
  \bibinfo {author} {\bibfnamefont {Emanuele}\ \bibnamefont {Berti}}, \ and\
  \bibinfo {author} {\bibfnamefont {Akihiro}\ \bibnamefont {Ishibashi}},\
  }\href {\doibase10.1103/PhysRevD.86.104017} {\bibfield  {journal} {\bibinfo
  {journal} {Phys. Rev.}\ }\textbf {\bibinfo {volume} {D86}},\ \bibinfo {pages}
  {104017} (\bibinfo {year} {2012}{\natexlab{a}})},\ \Eprint
  {http://arxiv.org/abs/1209.0773} {arXiv:1209.0773 [gr-qc]}\BibitemShut
  {NoStop}%
\bibitem [{\citenamefont {Pani}(2013)}]{Pani:2013pma}%
  \BibitemOpen
  \bibfield  {author} {\bibinfo {author} {\bibfnamefont {Paolo}\ \bibnamefont
  {Pani}},\ }\bibfield  {booktitle} {\emph {\bibinfo {booktitle} {{Proceedings,
  Spring School on Numerical Relativity and High Energy Physics (NR/HEP2)}}},\
  }\href {\doibase10.1142/S0217751X13400186} {\bibfield  {journal} {\bibinfo
  {journal} {Int. J. Mod. Phys.}\ }\textbf {\bibinfo {volume} {A28}},\ \bibinfo
  {pages} {1340018} (\bibinfo {year} {2013})},\ \Eprint
  {http://arxiv.org/abs/1305.6759} {arXiv:1305.6759 [gr-qc]}\BibitemShut
  {NoStop}%
\bibitem [{\citenamefont {Dias}\ \emph {et~al.}(2016)\citenamefont {Dias},
  \citenamefont {Santos},\ and\ \citenamefont {Way}}]{Dias:2015nua}%
  \BibitemOpen
  \bibfield  {author} {\bibinfo {author} {\bibfnamefont {Óscar J.~C.}\
  \bibnamefont {Dias}}, \bibinfo {author} {\bibfnamefont {Jorge~E.}\
  \bibnamefont {Santos}}, \ and\ \bibinfo {author} {\bibfnamefont {Benson}\
  \bibnamefont {Way}},\ }\href {\doibase10.1088/0264-9381/33/13/133001}
  {\bibfield  {journal} {\bibinfo  {journal} {Class. Quant. Grav.}\ }\textbf
  {\bibinfo {volume} {33}},\ \bibinfo {pages} {133001} (\bibinfo {year}
  {2016})},\ \Eprint {http://arxiv.org/abs/1510.02804} {arXiv:1510.02804
  [hep-th]}\BibitemShut {NoStop}%
\bibitem [{\citenamefont {Barausse}\ \emph {et~al.}(2014)\citenamefont
  {Barausse}, \citenamefont {Cardoso},\ and\ \citenamefont
  {Pani}}]{Barausse:2014tra}%
  \BibitemOpen
  \bibfield  {author} {\bibinfo {author} {\bibfnamefont {Enrico}\ \bibnamefont
  {Barausse}}, \bibinfo {author} {\bibfnamefont {Vitor}\ \bibnamefont
  {Cardoso}}, \ and\ \bibinfo {author} {\bibfnamefont {Paolo}\ \bibnamefont
  {Pani}},\ }\href {\doibase10.1103/PhysRevD.89.104059} {\bibfield  {journal}
  {\bibinfo  {journal} {Phys. Rev.}\ }\textbf {\bibinfo {volume} {D89}},\
  \bibinfo {pages} {104059} (\bibinfo {year} {2014})},\ \Eprint
  {http://arxiv.org/abs/1404.7149} {arXiv:1404.7149 [gr-qc]}\BibitemShut
  {NoStop}%
\bibitem [{\citenamefont {Ferrari}\ and\ \citenamefont
  {Mashhoon}(1984)}]{Ferrari:1984zz}%
  \BibitemOpen
  \bibfield  {author} {\bibinfo {author} {\bibfnamefont {Valeria}\ \bibnamefont
  {Ferrari}}\ and\ \bibinfo {author} {\bibfnamefont {Bahram}\ \bibnamefont
  {Mashhoon}},\ }\href {\doibase10.1103/PhysRevD.30.295} {\bibfield  {journal}
  {\bibinfo  {journal} {Phys. Rev.}\ }\textbf {\bibinfo {volume} {D30}},\
  \bibinfo {pages} {295--304} (\bibinfo {year} {1984})}\BibitemShut {NoStop}%
\bibitem [{\citenamefont {Cardoso}\ \emph {et~al.}(2009)\citenamefont
  {Cardoso}, \citenamefont {Miranda}, \citenamefont {Berti}, \citenamefont
  {Witek},\ and\ \citenamefont {Zanchin}}]{Cardoso:2008bp}%
  \BibitemOpen
  \bibfield  {author} {\bibinfo {author} {\bibfnamefont {Vitor}\ \bibnamefont
  {Cardoso}}, \bibinfo {author} {\bibfnamefont {Alex~S.}\ \bibnamefont
  {Miranda}}, \bibinfo {author} {\bibfnamefont {Emanuele}\ \bibnamefont
  {Berti}}, \bibinfo {author} {\bibfnamefont {Helvi}\ \bibnamefont {Witek}}, \
  and\ \bibinfo {author} {\bibfnamefont {Vilson~T.}\ \bibnamefont {Zanchin}},\
  }\href {\doibase10.1103/PhysRevD.79.064016} {\bibfield  {journal} {\bibinfo
  {journal} {Phys. Rev.}\ }\textbf {\bibinfo {volume} {D79}},\ \bibinfo {pages}
  {064016} (\bibinfo {year} {2009})},\ \Eprint {http://arxiv.org/abs/0812.1806}
  {arXiv:0812.1806 [hep-th]}\BibitemShut {NoStop}%
\bibitem [{\citenamefont {Yang}\ \emph {et~al.}(2012)\citenamefont {Yang},
  \citenamefont {Nichols}, \citenamefont {Zhang}, \citenamefont {Zimmerman},
  \citenamefont {Zhang},\ and\ \citenamefont {Chen}}]{Yang:2012he}%
  \BibitemOpen
  \bibfield  {author} {\bibinfo {author} {\bibfnamefont {Huan}\ \bibnamefont
  {Yang}}, \bibinfo {author} {\bibfnamefont {David~A.}\ \bibnamefont
  {Nichols}}, \bibinfo {author} {\bibfnamefont {Fan}\ \bibnamefont {Zhang}},
  \bibinfo {author} {\bibfnamefont {Aaron}\ \bibnamefont {Zimmerman}}, \bibinfo
  {author} {\bibfnamefont {Zhongyang}\ \bibnamefont {Zhang}}, \ and\ \bibinfo
  {author} {\bibfnamefont {Yanbei}\ \bibnamefont {Chen}},\ }\href
  {\doibase10.1103/PhysRevD.86.104006} {\bibfield  {journal} {\bibinfo
  {journal} {Phys. Rev.}\ }\textbf {\bibinfo {volume} {D86}},\ \bibinfo {pages}
  {104006} (\bibinfo {year} {2012})},\ \Eprint {http://arxiv.org/abs/1207.4253}
  {arXiv:1207.4253 [gr-qc]}\BibitemShut {NoStop}%
\bibitem [{\citenamefont {Blazquez-Salcedo}\ \emph {et~al.}(2016)\citenamefont
  {Blazquez-Salcedo}, \citenamefont {Macedo}, \citenamefont {Cardoso},
  \citenamefont {Ferrari}, \citenamefont {Gualtieri}, \citenamefont {Khoo},
  \citenamefont {Kunz},\ and\ \citenamefont {Pani}}]{Blazquez-Salcedo:2016enn}%
  \BibitemOpen
  \bibfield  {author} {\bibinfo {author} {\bibfnamefont {Jose~Luis}\
  \bibnamefont {Blazquez-Salcedo}}, \bibinfo {author} {\bibfnamefont {Caio
  F.~B.}\ \bibnamefont {Macedo}}, \bibinfo {author} {\bibfnamefont {Vitor}\
  \bibnamefont {Cardoso}}, \bibinfo {author} {\bibfnamefont {Valeria}\
  \bibnamefont {Ferrari}}, \bibinfo {author} {\bibfnamefont {Leonardo}\
  \bibnamefont {Gualtieri}}, \bibinfo {author} {\bibfnamefont {Fech~Scen}\
  \bibnamefont {Khoo}}, \bibinfo {author} {\bibfnamefont {Jutta}\ \bibnamefont
  {Kunz}}, \ and\ \bibinfo {author} {\bibfnamefont {Paolo}\ \bibnamefont
  {Pani}},\ }\href {\doibase10.1103/PhysRevD.94.104024} {\bibfield  {journal}
  {\bibinfo  {journal} {Phys. Rev.}\ }\textbf {\bibinfo {volume} {D94}},\
  \bibinfo {pages} {104024} (\bibinfo {year} {2016})},\ \Eprint
  {http://arxiv.org/abs/1609.01286} {arXiv:1609.01286 [gr-qc]}\BibitemShut
  {NoStop}%
\bibitem [{\citenamefont {Glampedakis}\ \emph {et~al.}(2017)\citenamefont
  {Glampedakis}, \citenamefont {Pappas}, \citenamefont {Silva},\ and\
  \citenamefont {Berti}}]{Glampedakis:2017dvb}%
  \BibitemOpen
  \bibfield  {author} {\bibinfo {author} {\bibfnamefont {Kostas}\ \bibnamefont
  {Glampedakis}}, \bibinfo {author} {\bibfnamefont {George}\ \bibnamefont
  {Pappas}}, \bibinfo {author} {\bibfnamefont {Hector~O.}\ \bibnamefont
  {Silva}}, \ and\ \bibinfo {author} {\bibfnamefont {Emanuele}\ \bibnamefont
  {Berti}},\ }\href {\doibase10.1103/PhysRevD.96.064054} {\bibfield  {journal}
  {\bibinfo  {journal} {Phys. Rev.}\ }\textbf {\bibinfo {volume} {D96}},\
  \bibinfo {pages} {064054} (\bibinfo {year} {2017})},\ \Eprint
  {http://arxiv.org/abs/1706.07658} {arXiv:1706.07658 [gr-qc]}\BibitemShut
  {NoStop}%
\bibitem [{\citenamefont {Konoplya}\ and\ \citenamefont
  {Stuchlík}(2017)}]{Konoplya:2017wot}%
  \BibitemOpen
  \bibfield  {author} {\bibinfo {author} {\bibfnamefont {R.~A.}\ \bibnamefont
  {Konoplya}}\ and\ \bibinfo {author} {\bibfnamefont {Z.}~\bibnamefont
  {Stuchlík}},\ }\href {\doibase10.1016/j.physletb.2017.06.015} {\bibfield
  {journal} {\bibinfo  {journal} {Phys. Lett.}\ }\textbf {\bibinfo {volume}
  {B771}},\ \bibinfo {pages} {597--602} (\bibinfo {year} {2017})},\ \Eprint
  {http://arxiv.org/abs/1705.05928} {arXiv:1705.05928 [gr-qc]}\BibitemShut
  {NoStop}%
\bibitem [{\citenamefont {Barausse}\ and\ \citenamefont
  {Sotiriou}(2008)}]{Barausse:2008xv}%
  \BibitemOpen
  \bibfield  {author} {\bibinfo {author} {\bibfnamefont {Enrico}\ \bibnamefont
  {Barausse}}\ and\ \bibinfo {author} {\bibfnamefont {Thomas~P.}\ \bibnamefont
  {Sotiriou}},\ }\href {\doibase10.1103/PhysRevLett.101.099001} {\bibfield
  {journal} {\bibinfo  {journal} {Phys. Rev. Lett.}\ }\textbf {\bibinfo
  {volume} {101}},\ \bibinfo {pages} {099001} (\bibinfo {year} {2008})},\
  \Eprint {http://arxiv.org/abs/0803.3433} {arXiv:0803.3433
  [gr-qc]}\BibitemShut {NoStop}%
\bibitem [{\citenamefont {Buonanno}\ \emph {et~al.}(2007)\citenamefont
  {Buonanno}, \citenamefont {Cook},\ and\ \citenamefont
  {Pretorius}}]{Buonanno:2006ui}%
  \BibitemOpen
  \bibfield  {author} {\bibinfo {author} {\bibfnamefont {Alessandra}\
  \bibnamefont {Buonanno}}, \bibinfo {author} {\bibfnamefont {Gregory~B.}\
  \bibnamefont {Cook}}, \ and\ \bibinfo {author} {\bibfnamefont {Frans}\
  \bibnamefont {Pretorius}},\ }\href {\doibase10.1103/PhysRevD.75.124018}
  {\bibfield  {journal} {\bibinfo  {journal} {Phys. Rev.}\ }\textbf {\bibinfo
  {volume} {D75}},\ \bibinfo {pages} {124018} (\bibinfo {year} {2007})},\
  \Eprint {http://arxiv.org/abs/gr-qc/0610122} {arXiv:gr-qc/0610122
  [gr-qc]}\BibitemShut {NoStop}%
\bibitem [{\citenamefont {Berti}\ \emph
  {et~al.}(2007{\natexlab{a}})\citenamefont {Berti}, \citenamefont {Cardoso},
  \citenamefont {Gonzalez}, \citenamefont {Sperhake}, \citenamefont {Hannam},
  \citenamefont {Husa},\ and\ \citenamefont {Bruegmann}}]{Berti:2007fi}%
  \BibitemOpen
  \bibfield  {author} {\bibinfo {author} {\bibfnamefont {Emanuele}\
  \bibnamefont {Berti}}, \bibinfo {author} {\bibfnamefont {Vitor}\ \bibnamefont
  {Cardoso}}, \bibinfo {author} {\bibfnamefont {Jose~A.}\ \bibnamefont
  {Gonzalez}}, \bibinfo {author} {\bibfnamefont {Ulrich}\ \bibnamefont
  {Sperhake}}, \bibinfo {author} {\bibfnamefont {Mark}\ \bibnamefont {Hannam}},
  \bibinfo {author} {\bibfnamefont {Sascha}\ \bibnamefont {Husa}}, \ and\
  \bibinfo {author} {\bibfnamefont {Bernd}\ \bibnamefont {Bruegmann}},\ }\href
  {\doibase10.1103/PhysRevD.76.064034} {\bibfield  {journal} {\bibinfo
  {journal} {Phys. Rev.}\ }\textbf {\bibinfo {volume} {D76}},\ \bibinfo {pages}
  {064034} (\bibinfo {year} {2007}{\natexlab{a}})},\ \Eprint
  {http://arxiv.org/abs/gr-qc/0703053} {arXiv:gr-qc/0703053
  [GR-QC]}\BibitemShut {NoStop}%
\bibitem [{\citenamefont {Barausse}\ \emph
  {et~al.}(2012{\natexlab{c}})\citenamefont {Barausse}, \citenamefont
  {Buonanno}, \citenamefont {Hughes}, \citenamefont {Khanna}, \citenamefont
  {O'Sullivan},\ and\ \citenamefont {Pan}}]{Barausse:2011kb}%
  \BibitemOpen
  \bibfield  {author} {\bibinfo {author} {\bibfnamefont {Enrico}\ \bibnamefont
  {Barausse}}, \bibinfo {author} {\bibfnamefont {Alessandra}\ \bibnamefont
  {Buonanno}}, \bibinfo {author} {\bibfnamefont {Scott~A.}\ \bibnamefont
  {Hughes}}, \bibinfo {author} {\bibfnamefont {Gaurav}\ \bibnamefont {Khanna}},
  \bibinfo {author} {\bibfnamefont {Stephen}\ \bibnamefont {O'Sullivan}}, \
  and\ \bibinfo {author} {\bibfnamefont {Yi}~\bibnamefont {Pan}},\ }\href
  {\doibase10.1103/PhysRevD.85.024046} {\bibfield  {journal} {\bibinfo
  {journal} {Phys. Rev.}\ }\textbf {\bibinfo {volume} {D85}},\ \bibinfo {pages}
  {024046} (\bibinfo {year} {2012}{\natexlab{c}})},\ \Eprint
  {http://arxiv.org/abs/1110.3081} {arXiv:1110.3081 [gr-qc]}\BibitemShut
  {NoStop}%
\bibitem [{\citenamefont {Baibhav}\ \emph {et~al.}(2018)\citenamefont
  {Baibhav}, \citenamefont {Berti}, \citenamefont {Cardoso},\ and\
  \citenamefont {Khanna}}]{Baibhav:2017jhs}%
  \BibitemOpen
  \bibfield  {author} {\bibinfo {author} {\bibfnamefont {Vishal}\ \bibnamefont
  {Baibhav}}, \bibinfo {author} {\bibfnamefont {Emanuele}\ \bibnamefont
  {Berti}}, \bibinfo {author} {\bibfnamefont {Vitor}\ \bibnamefont {Cardoso}},
  \ and\ \bibinfo {author} {\bibfnamefont {Gaurav}\ \bibnamefont {Khanna}},\
  }\href {\doibase10.1103/PhysRevD.97.044048} {\bibfield  {journal} {\bibinfo
  {journal} {Phys. Rev.}\ }\textbf {\bibinfo {volume} {D97}},\ \bibinfo {pages}
  {044048} (\bibinfo {year} {2018})},\ \Eprint
  {http://arxiv.org/abs/1710.02156} {arXiv:1710.02156 [gr-qc]}\BibitemShut
  {NoStop}%
\bibitem [{\citenamefont {Berti}\ \emph {et~al.}(2005)\citenamefont {Berti},
  \citenamefont {Buonanno},\ and\ \citenamefont {Will}}]{Berti:2004bd}%
  \BibitemOpen
  \bibfield  {author} {\bibinfo {author} {\bibfnamefont {Emanuele}\
  \bibnamefont {Berti}}, \bibinfo {author} {\bibfnamefont {Alessandra}\
  \bibnamefont {Buonanno}}, \ and\ \bibinfo {author} {\bibfnamefont
  {Clifford~M.}\ \bibnamefont {Will}},\ }\href
  {\doibase10.1103/PhysRevD.71.084025} {\bibfield  {journal} {\bibinfo
  {journal} {Phys. Rev.}\ }\textbf {\bibinfo {volume} {D71}},\ \bibinfo {pages}
  {084025} (\bibinfo {year} {2005})},\ \Eprint
  {http://arxiv.org/abs/gr-qc/0411129} {arXiv:gr-qc/0411129
  [gr-qc]}\BibitemShut {NoStop}%
\bibitem [{\citenamefont {Berti}\ \emph
  {et~al.}(2007{\natexlab{b}})\citenamefont {Berti}, \citenamefont {Cardoso},
  \citenamefont {Cardoso},\ and\ \citenamefont {Cavaglia}}]{Berti:2007zu}%
  \BibitemOpen
  \bibfield  {author} {\bibinfo {author} {\bibfnamefont {Emanuele}\
  \bibnamefont {Berti}}, \bibinfo {author} {\bibfnamefont {Jaime}\ \bibnamefont
  {Cardoso}}, \bibinfo {author} {\bibfnamefont {Vitor}\ \bibnamefont
  {Cardoso}}, \ and\ \bibinfo {author} {\bibfnamefont {Marco}\ \bibnamefont
  {Cavaglia}},\ }\href {\doibase10.1103/PhysRevD.76.104044} {\bibfield
  {journal} {\bibinfo  {journal} {Phys. Rev.}\ }\textbf {\bibinfo {volume}
  {D76}},\ \bibinfo {pages} {104044} (\bibinfo {year} {2007}{\natexlab{b}})},\
  \Eprint {http://arxiv.org/abs/0707.1202} {arXiv:0707.1202
  [gr-qc]}\BibitemShut {NoStop}%
\bibitem [{\citenamefont {Yang}\ \emph {et~al.}(2017)\citenamefont {Yang},
  \citenamefont {Yagi}, \citenamefont {Blackman}, \citenamefont {Lehner},
  \citenamefont {Paschalidis}, \citenamefont {Pretorius},\ and\ \citenamefont
  {Yunes}}]{Yang:2017zxs}%
  \BibitemOpen
  \bibfield  {author} {\bibinfo {author} {\bibfnamefont {Huan}\ \bibnamefont
  {Yang}}, \bibinfo {author} {\bibfnamefont {Kent}\ \bibnamefont {Yagi}},
  \bibinfo {author} {\bibfnamefont {Jonathan}\ \bibnamefont {Blackman}},
  \bibinfo {author} {\bibfnamefont {Luis}\ \bibnamefont {Lehner}}, \bibinfo
  {author} {\bibfnamefont {Vasileios}\ \bibnamefont {Paschalidis}}, \bibinfo
  {author} {\bibfnamefont {Frans}\ \bibnamefont {Pretorius}}, \ and\ \bibinfo
  {author} {\bibfnamefont {Nicolás}\ \bibnamefont {Yunes}},\ }\href
  {\doibase10.1103/PhysRevLett.118.161101} {\bibfield  {journal} {\bibinfo
  {journal} {Phys. Rev. Lett.}\ }\textbf {\bibinfo {volume} {118}},\ \bibinfo
  {pages} {161101} (\bibinfo {year} {2017})},\ \Eprint
  {http://arxiv.org/abs/1701.05808} {arXiv:1701.05808 [gr-qc]}\BibitemShut
  {NoStop}%
\bibitem [{\citenamefont {Bhagwat}\ \emph {et~al.}(2018)\citenamefont
  {Bhagwat}, \citenamefont {Okounkova}, \citenamefont {Ballmer}, \citenamefont
  {Brown}, \citenamefont {Giesler}, \citenamefont {Scheel},\ and\ \citenamefont
  {Teukolsky}}]{Bhagwat:2017tkm}%
  \BibitemOpen
  \bibfield  {author} {\bibinfo {author} {\bibfnamefont {Swetha}\ \bibnamefont
  {Bhagwat}}, \bibinfo {author} {\bibfnamefont {Maria}\ \bibnamefont
  {Okounkova}}, \bibinfo {author} {\bibfnamefont {Stefan~W.}\ \bibnamefont
  {Ballmer}}, \bibinfo {author} {\bibfnamefont {Duncan~A.}\ \bibnamefont
  {Brown}}, \bibinfo {author} {\bibfnamefont {Matthew}\ \bibnamefont
  {Giesler}}, \bibinfo {author} {\bibfnamefont {Mark~A.}\ \bibnamefont
  {Scheel}}, \ and\ \bibinfo {author} {\bibfnamefont {Saul~A.}\ \bibnamefont
  {Teukolsky}},\ }\href {\doibase10.1103/PhysRevD.97.104065} {\bibfield
  {journal} {\bibinfo  {journal} {Phys. Rev.}\ }\textbf {\bibinfo {volume}
  {D97}},\ \bibinfo {pages} {104065} (\bibinfo {year} {2018})},\ \Eprint
  {http://arxiv.org/abs/1711.00926} {arXiv:1711.00926 [gr-qc]}\BibitemShut
  {NoStop}%
\bibitem [{\citenamefont {Pani}\ \emph {et~al.}(2009)\citenamefont {Pani},
  \citenamefont {Berti}, \citenamefont {Cardoso}, \citenamefont {Chen},\ and\
  \citenamefont {Norte}}]{Pani:2009ss}%
  \BibitemOpen
  \bibfield  {author} {\bibinfo {author} {\bibfnamefont {Paolo}\ \bibnamefont
  {Pani}}, \bibinfo {author} {\bibfnamefont {Emanuele}\ \bibnamefont {Berti}},
  \bibinfo {author} {\bibfnamefont {Vitor}\ \bibnamefont {Cardoso}}, \bibinfo
  {author} {\bibfnamefont {Yanbei}\ \bibnamefont {Chen}}, \ and\ \bibinfo
  {author} {\bibfnamefont {Richard}\ \bibnamefont {Norte}},\ }\href
  {\doibase10.1103/PhysRevD.80.124047} {\bibfield  {journal} {\bibinfo
  {journal} {Phys. Rev.}\ }\textbf {\bibinfo {volume} {D80}},\ \bibinfo {pages}
  {124047} (\bibinfo {year} {2009})},\ \Eprint {http://arxiv.org/abs/0909.0287}
  {arXiv:0909.0287 [gr-qc]}\BibitemShut {NoStop}%
\bibitem [{\citenamefont {Macedo}\ \emph
  {et~al.}(2013{\natexlab{a}})\citenamefont {Macedo}, \citenamefont {Pani},
  \citenamefont {Cardoso},\ and\ \citenamefont {Crispino}}]{Macedo:2013jja}%
  \BibitemOpen
  \bibfield  {author} {\bibinfo {author} {\bibfnamefont {Caio F.~B.}\
  \bibnamefont {Macedo}}, \bibinfo {author} {\bibfnamefont {Paolo}\
  \bibnamefont {Pani}}, \bibinfo {author} {\bibfnamefont {Vitor}\ \bibnamefont
  {Cardoso}}, \ and\ \bibinfo {author} {\bibfnamefont {Luís C.~B.}\
  \bibnamefont {Crispino}},\ }\href {\doibase10.1103/PhysRevD.88.064046}
  {\bibfield  {journal} {\bibinfo  {journal} {Phys. Rev.}\ }\textbf {\bibinfo
  {volume} {D88}},\ \bibinfo {pages} {064046} (\bibinfo {year}
  {2013}{\natexlab{a}})},\ \Eprint {http://arxiv.org/abs/1307.4812}
  {arXiv:1307.4812 [gr-qc]}\BibitemShut {NoStop}%
\bibitem [{\citenamefont {Chirenti}\ and\ \citenamefont
  {Rezzolla}(2016)}]{Chirenti:2016hzd}%
  \BibitemOpen
  \bibfield  {author} {\bibinfo {author} {\bibfnamefont {Cecilia}\ \bibnamefont
  {Chirenti}}\ and\ \bibinfo {author} {\bibfnamefont {Luciano}\ \bibnamefont
  {Rezzolla}},\ }\href {\doibase10.1103/PhysRevD.94.084016} {\bibfield
  {journal} {\bibinfo  {journal} {Phys. Rev.}\ }\textbf {\bibinfo {volume}
  {D94}},\ \bibinfo {pages} {084016} (\bibinfo {year} {2016})},\ \Eprint
  {http://arxiv.org/abs/1602.08759} {arXiv:1602.08759 [gr-qc]}\BibitemShut
  {NoStop}%
\bibitem [{\citenamefont {Cardoso}\ \emph
  {et~al.}(2016{\natexlab{b}})\citenamefont {Cardoso}, \citenamefont
  {Franzin},\ and\ \citenamefont {Pani}}]{Cardoso:2016rao}%
  \BibitemOpen
  \bibfield  {author} {\bibinfo {author} {\bibfnamefont {Vitor}\ \bibnamefont
  {Cardoso}}, \bibinfo {author} {\bibfnamefont {Edgardo}\ \bibnamefont
  {Franzin}}, \ and\ \bibinfo {author} {\bibfnamefont {Paolo}\ \bibnamefont
  {Pani}},\ }\href {\doibase10.1103/PhysRevLett.116.171101} {\bibfield
  {journal} {\bibinfo  {journal} {Phys. Rev. Lett.}\ }\textbf {\bibinfo
  {volume} {116}},\ \bibinfo {pages} {171101} (\bibinfo {year}
  {2016}{\natexlab{b}})},\ \bibinfo {note} {[Erratum: Phys. Rev.
  Lett.117,no.8,089902(2016)]},\ \Eprint {http://arxiv.org/abs/1602.07309}
  {arXiv:1602.07309 [gr-qc]}\BibitemShut {NoStop}%
\bibitem [{\citenamefont {Cardoso}\ \emph
  {et~al.}(2016{\natexlab{c}})\citenamefont {Cardoso}, \citenamefont {Hopper},
  \citenamefont {Macedo}, \citenamefont {Palenzuela},\ and\ \citenamefont
  {Pani}}]{Cardoso:2016oxy}%
  \BibitemOpen
  \bibfield  {author} {\bibinfo {author} {\bibfnamefont {Vitor}\ \bibnamefont
  {Cardoso}}, \bibinfo {author} {\bibfnamefont {Seth}\ \bibnamefont {Hopper}},
  \bibinfo {author} {\bibfnamefont {Caio F.~B.}\ \bibnamefont {Macedo}},
  \bibinfo {author} {\bibfnamefont {Carlos}\ \bibnamefont {Palenzuela}}, \ and\
  \bibinfo {author} {\bibfnamefont {Paolo}\ \bibnamefont {Pani}},\ }\href
  {\doibase10.1103/PhysRevD.94.084031} {\bibfield  {journal} {\bibinfo
  {journal} {Phys. Rev.}\ }\textbf {\bibinfo {volume} {D94}},\ \bibinfo {pages}
  {084031} (\bibinfo {year} {2016}{\natexlab{c}})},\ \Eprint
  {http://arxiv.org/abs/1608.08637} {arXiv:1608.08637 [gr-qc]}\BibitemShut
  {NoStop}%
\bibitem [{\citenamefont {Cardoso}\ and\ \citenamefont
  {Pani}(2017{\natexlab{a}})}]{Cardoso:2017cqb}%
  \BibitemOpen
  \bibfield  {author} {\bibinfo {author} {\bibfnamefont {Vitor}\ \bibnamefont
  {Cardoso}}\ and\ \bibinfo {author} {\bibfnamefont {Paolo}\ \bibnamefont
  {Pani}},\ }\href {\doibase10.1038/s41550-017-0225-y} {\bibfield  {journal}
  {\bibinfo  {journal} {Nat. Astron.}\ }\textbf {\bibinfo {volume} {1}},\
  \bibinfo {pages} {586--591} (\bibinfo {year} {2017}{\natexlab{a}})},\ \Eprint
  {http://arxiv.org/abs/1709.01525} {arXiv:1709.01525 [gr-qc]}\BibitemShut
  {NoStop}%
\bibitem [{\citenamefont {Tattersall}\ \emph {et~al.}(2018)\citenamefont
  {Tattersall}, \citenamefont {Ferreira},\ and\ \citenamefont
  {Lagos}}]{Tattersall:2017erk}%
  \BibitemOpen
  \bibfield  {author} {\bibinfo {author} {\bibfnamefont {Oliver~J.}\
  \bibnamefont {Tattersall}}, \bibinfo {author} {\bibfnamefont {Pedro~G.}\
  \bibnamefont {Ferreira}}, \ and\ \bibinfo {author} {\bibfnamefont {Macarena}\
  \bibnamefont {Lagos}},\ }\href {\doibase10.1103/PhysRevD.97.044021}
  {\bibfield  {journal} {\bibinfo  {journal} {Phys. Rev.}\ }\textbf {\bibinfo
  {volume} {D97}},\ \bibinfo {pages} {044021} (\bibinfo {year} {2018})},\
  \Eprint {http://arxiv.org/abs/1711.01992} {arXiv:1711.01992
  [gr-qc]}\BibitemShut {NoStop}%
\bibitem [{\citenamefont {Tattersall}\ and\ \citenamefont
  {Ferreira}(2018)}]{Tattersall:2018nve}%
  \BibitemOpen
  \bibfield  {author} {\bibinfo {author} {\bibfnamefont {Oliver~J.}\
  \bibnamefont {Tattersall}}\ and\ \bibinfo {author} {\bibfnamefont {Pedro~G.}\
  \bibnamefont {Ferreira}},\ }\href {\doibase10.1103/PhysRevD.97.104047}
  {\bibfield  {journal} {\bibinfo  {journal} {Phys. Rev.}\ }\textbf {\bibinfo
  {volume} {D97}},\ \bibinfo {pages} {104047} (\bibinfo {year} {2018})},\
  \Eprint {http://arxiv.org/abs/1804.08950} {arXiv:1804.08950
  [gr-qc]}\BibitemShut {NoStop}%
\bibitem [{\citenamefont {Brustein}\ \emph
  {et~al.}(2017{\natexlab{a}})\citenamefont {Brustein}, \citenamefont
  {Medved},\ and\ \citenamefont {Yagi}}]{Brustein:2017koc}%
  \BibitemOpen
  \bibfield  {author} {\bibinfo {author} {\bibfnamefont {Ram}\ \bibnamefont
  {Brustein}}, \bibinfo {author} {\bibfnamefont {A.~J.~M.}\ \bibnamefont
  {Medved}}, \ and\ \bibinfo {author} {\bibfnamefont {K.}~\bibnamefont
  {Yagi}},\ }\href {\doibase10.1103/PhysRevD.96.064033} {\bibfield  {journal}
  {\bibinfo  {journal} {Phys. Rev.}\ }\textbf {\bibinfo {volume} {D96}},\
  \bibinfo {pages} {064033} (\bibinfo {year} {2017}{\natexlab{a}})},\ \Eprint
  {http://arxiv.org/abs/1704.05789} {arXiv:1704.05789 [gr-qc]}\BibitemShut
  {NoStop}%
\bibitem [{\citenamefont {Cardoso}\ \emph
  {et~al.}(2013{\natexlab{a}})\citenamefont {Cardoso}, \citenamefont {Carucci},
  \citenamefont {Pani},\ and\ \citenamefont {Sotiriou}}]{Cardoso:2013fwa}%
  \BibitemOpen
  \bibfield  {author} {\bibinfo {author} {\bibfnamefont {Vitor}\ \bibnamefont
  {Cardoso}}, \bibinfo {author} {\bibfnamefont {Isabella~P.}\ \bibnamefont
  {Carucci}}, \bibinfo {author} {\bibfnamefont {Paolo}\ \bibnamefont {Pani}}, \
  and\ \bibinfo {author} {\bibfnamefont {Thomas~P.}\ \bibnamefont {Sotiriou}},\
  }\href {\doibase10.1103/PhysRevLett.111.111101} {\bibfield  {journal}
  {\bibinfo  {journal} {Phys. Rev. Lett.}\ }\textbf {\bibinfo {volume} {111}},\
  \bibinfo {pages} {111101} (\bibinfo {year} {2013}{\natexlab{a}})},\ \Eprint
  {http://arxiv.org/abs/1308.6587} {arXiv:1308.6587 [gr-qc]}\BibitemShut
  {NoStop}%
\bibitem [{\citenamefont {Cardoso}\ \emph
  {et~al.}(2013{\natexlab{b}})\citenamefont {Cardoso}, \citenamefont {Carucci},
  \citenamefont {Pani},\ and\ \citenamefont {Sotiriou}}]{Cardoso:2013opa}%
  \BibitemOpen
  \bibfield  {author} {\bibinfo {author} {\bibfnamefont {Vitor}\ \bibnamefont
  {Cardoso}}, \bibinfo {author} {\bibfnamefont {Isabella~P.}\ \bibnamefont
  {Carucci}}, \bibinfo {author} {\bibfnamefont {Paolo}\ \bibnamefont {Pani}}, \
  and\ \bibinfo {author} {\bibfnamefont {Thomas~P.}\ \bibnamefont {Sotiriou}},\
  }\href {\doibase10.1103/PhysRevD.88.044056} {\bibfield  {journal} {\bibinfo
  {journal} {Phys. Rev.}\ }\textbf {\bibinfo {volume} {D88}},\ \bibinfo {pages}
  {044056} (\bibinfo {year} {2013}{\natexlab{b}})},\ \Eprint
  {http://arxiv.org/abs/1305.6936} {arXiv:1305.6936 [gr-qc]}\BibitemShut
  {NoStop}%
\bibitem [{\citenamefont {Babichev}\ and\ \citenamefont
  {Fabbri}(2013)}]{Babichev:2013una}%
  \BibitemOpen
  \bibfield  {author} {\bibinfo {author} {\bibfnamefont {Eugeny}\ \bibnamefont
  {Babichev}}\ and\ \bibinfo {author} {\bibfnamefont {Alessandro}\ \bibnamefont
  {Fabbri}},\ }\href {\doibase10.1088/0264-9381/30/15/152001} {\bibfield
  {journal} {\bibinfo  {journal} {Class. Quant. Grav.}\ }\textbf {\bibinfo
  {volume} {30}},\ \bibinfo {pages} {152001} (\bibinfo {year} {2013})},\
  \Eprint {http://arxiv.org/abs/1304.5992} {arXiv:1304.5992
  [gr-qc]}\BibitemShut {NoStop}%
\bibitem [{\citenamefont {Brito}\ \emph
  {et~al.}(2013{\natexlab{a}})\citenamefont {Brito}, \citenamefont {Cardoso},\
  and\ \citenamefont {Pani}}]{Brito:2013wya}%
  \BibitemOpen
  \bibfield  {author} {\bibinfo {author} {\bibfnamefont {Richard}\ \bibnamefont
  {Brito}}, \bibinfo {author} {\bibfnamefont {Vitor}\ \bibnamefont {Cardoso}},
  \ and\ \bibinfo {author} {\bibfnamefont {Paolo}\ \bibnamefont {Pani}},\
  }\href {\doibase10.1103/PhysRevD.88.023514} {\bibfield  {journal} {\bibinfo
  {journal} {Phys. Rev.}\ }\textbf {\bibinfo {volume} {D88}},\ \bibinfo {pages}
  {023514} (\bibinfo {year} {2013}{\natexlab{a}})},\ \Eprint
  {http://arxiv.org/abs/1304.6725} {arXiv:1304.6725 [gr-qc]}\BibitemShut
  {NoStop}%
\bibitem [{\citenamefont {Cunha}\ \emph {et~al.}(2017)\citenamefont {Cunha},
  \citenamefont {Berti},\ and\ \citenamefont {Herdeiro}}]{Cunha:2017qtt}%
  \BibitemOpen
  \bibfield  {author} {\bibinfo {author} {\bibfnamefont {Pedro V.~P.}\
  \bibnamefont {Cunha}}, \bibinfo {author} {\bibfnamefont {Emanuele}\
  \bibnamefont {Berti}}, \ and\ \bibinfo {author} {\bibfnamefont {Carlos
  A.~R.}\ \bibnamefont {Herdeiro}},\ }\href
  {\doibase10.1103/PhysRevLett.119.251102} {\bibfield  {journal} {\bibinfo
  {journal} {Phys. Rev. Lett.}\ }\textbf {\bibinfo {volume} {119}},\ \bibinfo
  {pages} {251102} (\bibinfo {year} {2017})},\ \Eprint
  {http://arxiv.org/abs/1708.04211} {arXiv:1708.04211 [gr-qc]}\BibitemShut
  {NoStop}%
\bibitem [{\citenamefont {Collins}\ and\ \citenamefont
  {Hughes}(2004)}]{Collins:2004ex}%
  \BibitemOpen
  \bibfield  {author} {\bibinfo {author} {\bibfnamefont {Nathan~A.}\
  \bibnamefont {Collins}}\ and\ \bibinfo {author} {\bibfnamefont {Scott~A.}\
  \bibnamefont {Hughes}},\ }\href {\doibase10.1103/PhysRevD.69.124022}
  {\bibfield  {journal} {\bibinfo  {journal} {Phys. Rev.}\ }\textbf {\bibinfo
  {volume} {D69}},\ \bibinfo {pages} {124022} (\bibinfo {year} {2004})},\
  \Eprint {http://arxiv.org/abs/gr-qc/0402063} {arXiv:gr-qc/0402063
  [gr-qc]}\BibitemShut {NoStop}%
\bibitem [{\citenamefont {Vigeland}\ and\ \citenamefont
  {Hughes}(2010)}]{Vigeland:2009pr}%
  \BibitemOpen
  \bibfield  {author} {\bibinfo {author} {\bibfnamefont {Sarah~J.}\
  \bibnamefont {Vigeland}}\ and\ \bibinfo {author} {\bibfnamefont {Scott~A.}\
  \bibnamefont {Hughes}},\ }\href {\doibase10.1103/PhysRevD.81.024030}
  {\bibfield  {journal} {\bibinfo  {journal} {Phys. Rev.}\ }\textbf {\bibinfo
  {volume} {D81}},\ \bibinfo {pages} {024030} (\bibinfo {year} {2010})},\
  \Eprint {http://arxiv.org/abs/0911.1756} {arXiv:0911.1756
  [gr-qc]}\BibitemShut {NoStop}%
\bibitem [{\citenamefont {Vigeland}(2010)}]{Vigeland:2010xe}%
  \BibitemOpen
  \bibfield  {author} {\bibinfo {author} {\bibfnamefont {Sarah~J.}\
  \bibnamefont {Vigeland}},\ }\href {\doibase10.1103/PhysRevD.82.104041}
  {\bibfield  {journal} {\bibinfo  {journal} {Phys. Rev.}\ }\textbf {\bibinfo
  {volume} {D82}},\ \bibinfo {pages} {104041} (\bibinfo {year} {2010})},\
  \Eprint {http://arxiv.org/abs/1008.1278} {arXiv:1008.1278
  [gr-qc]}\BibitemShut {NoStop}%
\bibitem [{\citenamefont {Newman}\ and\ \citenamefont
  {Janis}(1965)}]{Newman:1965tw}%
  \BibitemOpen
  \bibfield  {author} {\bibinfo {author} {\bibfnamefont {E.~T.}\ \bibnamefont
  {Newman}}\ and\ \bibinfo {author} {\bibfnamefont {A.~I.}\ \bibnamefont
  {Janis}},\ }\href {\doibase10.1063/1.1704350} {\bibfield  {journal} {\bibinfo
   {journal} {J. Math. Phys.}\ }\textbf {\bibinfo {volume} {6}},\ \bibinfo
  {pages} {915--917} (\bibinfo {year} {1965})}\BibitemShut {NoStop}%
\bibitem [{\citenamefont {Hartle}\ and\ \citenamefont
  {Thorne}(1968)}]{Hartle:1968si}%
  \BibitemOpen
  \bibfield  {author} {\bibinfo {author} {\bibfnamefont {James~B.}\
  \bibnamefont {Hartle}}\ and\ \bibinfo {author} {\bibfnamefont {Kip~S.}\
  \bibnamefont {Thorne}},\ }\href {\doibase10.1086/149707} {\bibfield
  {journal} {\bibinfo  {journal} {Astrophys. J.}\ }\textbf {\bibinfo {volume}
  {153}},\ \bibinfo {pages} {807} (\bibinfo {year} {1968})}\BibitemShut
  {NoStop}%
\bibitem [{\citenamefont {Glampedakis}\ and\ \citenamefont
  {Babak}(2006)}]{Glampedakis:2005cf}%
  \BibitemOpen
  \bibfield  {author} {\bibinfo {author} {\bibfnamefont {Kostas}\ \bibnamefont
  {Glampedakis}}\ and\ \bibinfo {author} {\bibfnamefont {Stanislav}\
  \bibnamefont {Babak}},\ }\href {\doibase10.1088/0264-9381/23/12/013}
  {\bibfield  {journal} {\bibinfo  {journal} {Class. Quant. Grav.}\ }\textbf
  {\bibinfo {volume} {23}},\ \bibinfo {pages} {4167--4188} (\bibinfo {year}
  {2006})},\ \Eprint {http://arxiv.org/abs/gr-qc/0510057} {arXiv:gr-qc/0510057
  [gr-qc]}\BibitemShut {NoStop}%
\bibitem [{\citenamefont {Vigeland}\ \emph {et~al.}(2011)\citenamefont
  {Vigeland}, \citenamefont {Yunes},\ and\ \citenamefont
  {Stein}}]{Vigeland:2011ji}%
  \BibitemOpen
  \bibfield  {author} {\bibinfo {author} {\bibfnamefont {Sarah}\ \bibnamefont
  {Vigeland}}, \bibinfo {author} {\bibfnamefont {Nicolas}\ \bibnamefont
  {Yunes}}, \ and\ \bibinfo {author} {\bibfnamefont {Leo}\ \bibnamefont
  {Stein}},\ }\href {\doibase10.1103/PhysRevD.83.104027} {\bibfield  {journal}
  {\bibinfo  {journal} {Phys. Rev.}\ }\textbf {\bibinfo {volume} {D83}},\
  \bibinfo {pages} {104027} (\bibinfo {year} {2011})},\ \Eprint
  {http://arxiv.org/abs/1102.3706} {arXiv:1102.3706 [gr-qc]}\BibitemShut
  {NoStop}%
\bibitem [{\citenamefont {Johannsen}(2013{\natexlab{a}})}]{Johannsen:2015pca}%
  \BibitemOpen
  \bibfield  {author} {\bibinfo {author} {\bibfnamefont {Tim}\ \bibnamefont
  {Johannsen}},\ }\href {\doibase10.1103/PhysRevD.88.044002} {\bibfield
  {journal} {\bibinfo  {journal} {Phys. Rev.}\ }\textbf {\bibinfo {volume}
  {D88}},\ \bibinfo {pages} {044002} (\bibinfo {year} {2013}{\natexlab{a}})},\
  \Eprint {http://arxiv.org/abs/1501.02809} {arXiv:1501.02809
  [gr-qc]}\BibitemShut {NoStop}%
\bibitem [{\citenamefont {Johannsen}\ and\ \citenamefont
  {Psaltis}(2011)}]{Johannsen:2011dh}%
  \BibitemOpen
  \bibfield  {author} {\bibinfo {author} {\bibfnamefont {Tim}\ \bibnamefont
  {Johannsen}}\ and\ \bibinfo {author} {\bibfnamefont {Dimitrios}\ \bibnamefont
  {Psaltis}},\ }\href {\doibase10.1103/PhysRevD.83.124015} {\bibfield
  {journal} {\bibinfo  {journal} {Phys. Rev.}\ }\textbf {\bibinfo {volume}
  {D83}},\ \bibinfo {pages} {124015} (\bibinfo {year} {2011})},\ \Eprint
  {http://arxiv.org/abs/1105.3191} {arXiv:1105.3191 [gr-qc]}\BibitemShut
  {NoStop}%
\bibitem [{\citenamefont {Cardoso}\ \emph
  {et~al.}(2014{\natexlab{a}})\citenamefont {Cardoso}, \citenamefont {Pani},\
  and\ \citenamefont {Rico}}]{Cardoso:2014rha}%
  \BibitemOpen
  \bibfield  {author} {\bibinfo {author} {\bibfnamefont {Vitor}\ \bibnamefont
  {Cardoso}}, \bibinfo {author} {\bibfnamefont {Paolo}\ \bibnamefont {Pani}}, \
  and\ \bibinfo {author} {\bibfnamefont {Joao}\ \bibnamefont {Rico}},\ }\href
  {\doibase10.1103/PhysRevD.89.064007} {\bibfield  {journal} {\bibinfo
  {journal} {Phys. Rev.}\ }\textbf {\bibinfo {volume} {D89}},\ \bibinfo {pages}
  {064007} (\bibinfo {year} {2014}{\natexlab{a}})},\ \Eprint
  {http://arxiv.org/abs/1401.0528} {arXiv:1401.0528 [gr-qc]}\BibitemShut
  {NoStop}%
\bibitem [{\citenamefont {Rezzolla}\ and\ \citenamefont
  {Zhidenko}(2014)}]{Rezzolla:2014mua}%
  \BibitemOpen
  \bibfield  {author} {\bibinfo {author} {\bibfnamefont {Luciano}\ \bibnamefont
  {Rezzolla}}\ and\ \bibinfo {author} {\bibfnamefont {Alexander}\ \bibnamefont
  {Zhidenko}},\ }\href {\doibase10.1103/PhysRevD.90.084009} {\bibfield
  {journal} {\bibinfo  {journal} {Phys. Rev.}\ }\textbf {\bibinfo {volume}
  {D90}},\ \bibinfo {pages} {084009} (\bibinfo {year} {2014})},\ \Eprint
  {http://arxiv.org/abs/1407.3086} {arXiv:1407.3086 [gr-qc]}\BibitemShut
  {NoStop}%
\bibitem [{\citenamefont {Konoplya}\ \emph {et~al.}(2016)\citenamefont
  {Konoplya}, \citenamefont {Rezzolla},\ and\ \citenamefont
  {Zhidenko}}]{Konoplya:2016jvv}%
  \BibitemOpen
  \bibfield  {author} {\bibinfo {author} {\bibfnamefont {Roman}\ \bibnamefont
  {Konoplya}}, \bibinfo {author} {\bibfnamefont {Luciano}\ \bibnamefont
  {Rezzolla}}, \ and\ \bibinfo {author} {\bibfnamefont {Alexander}\
  \bibnamefont {Zhidenko}},\ }\href {\doibase10.1103/PhysRevD.93.064015}
  {\bibfield  {journal} {\bibinfo  {journal} {Phys. Rev.}\ }\textbf {\bibinfo
  {volume} {D93}},\ \bibinfo {pages} {064015} (\bibinfo {year} {2016})},\
  \Eprint {http://arxiv.org/abs/1602.02378} {arXiv:1602.02378
  [gr-qc]}\BibitemShut {NoStop}%
\bibitem [{\citenamefont {Lin}\ \emph {et~al.}(2015)\citenamefont {Lin},
  \citenamefont {Tsukamoto}, \citenamefont {Ghasemi-Nodehi},\ and\
  \citenamefont {Bambi}}]{Lin:2015oan}%
  \BibitemOpen
  \bibfield  {author} {\bibinfo {author} {\bibfnamefont {Nan}\ \bibnamefont
  {Lin}}, \bibinfo {author} {\bibfnamefont {Naoki}\ \bibnamefont {Tsukamoto}},
  \bibinfo {author} {\bibfnamefont {M.}~\bibnamefont {Ghasemi-Nodehi}}, \ and\
  \bibinfo {author} {\bibfnamefont {Cosimo}\ \bibnamefont {Bambi}},\ }\href
  {\doibase10.1140/epjc/s10052-015-3837-3} {\bibfield  {journal} {\bibinfo
  {journal} {Eur. Phys. J.}\ }\textbf {\bibinfo {volume} {C75}},\ \bibinfo
  {pages} {599} (\bibinfo {year} {2015})},\ \Eprint
  {http://arxiv.org/abs/1512.00724} {arXiv:1512.00724 [gr-qc]}\BibitemShut
  {NoStop}%
\bibitem [{\citenamefont {Ghasemi-Nodehi}\ and\ \citenamefont
  {Bambi}(2016)}]{Ghasemi-Nodehi:2016wao}%
  \BibitemOpen
  \bibfield  {author} {\bibinfo {author} {\bibfnamefont {M.}~\bibnamefont
  {Ghasemi-Nodehi}}\ and\ \bibinfo {author} {\bibfnamefont {Cosimo}\
  \bibnamefont {Bambi}},\ }\href {\doibase10.1140/epjc/s10052-016-4137-2}
  {\bibfield  {journal} {\bibinfo  {journal} {Eur. Phys. J.}\ }\textbf
  {\bibinfo {volume} {C76}},\ \bibinfo {pages} {290} (\bibinfo {year}
  {2016})},\ \Eprint {http://arxiv.org/abs/1604.07032} {arXiv:1604.07032
  [gr-qc]}\BibitemShut {NoStop}%
\bibitem [{\citenamefont {Johannsen}(2013{\natexlab{b}})}]{Johannsen:2013rqa}%
  \BibitemOpen
  \bibfield  {author} {\bibinfo {author} {\bibfnamefont {Tim}\ \bibnamefont
  {Johannsen}},\ }\href {\doibase10.1103/PhysRevD.87.124017} {\bibfield
  {journal} {\bibinfo  {journal} {Phys. Rev.}\ }\textbf {\bibinfo {volume}
  {D87}},\ \bibinfo {pages} {124017} (\bibinfo {year} {2013}{\natexlab{b}})},\
  \Eprint {http://arxiv.org/abs/1304.7786} {arXiv:1304.7786
  [gr-qc]}\BibitemShut {NoStop}%
\bibitem [{\citenamefont {Yagi}\ and\ \citenamefont
  {Stein}(2016)}]{Yagi:2016jml}%
  \BibitemOpen
  \bibfield  {author} {\bibinfo {author} {\bibfnamefont {Kent}\ \bibnamefont
  {Yagi}}\ and\ \bibinfo {author} {\bibfnamefont {Leo~C.}\ \bibnamefont
  {Stein}},\ }\href {\doibase10.1088/0264-9381/33/5/054001} {\bibfield
  {journal} {\bibinfo  {journal} {Class. Quant. Grav.}\ }\textbf {\bibinfo
  {volume} {33}},\ \bibinfo {pages} {054001} (\bibinfo {year} {2016})},\
  \Eprint {http://arxiv.org/abs/1602.02413} {arXiv:1602.02413
  [gr-qc]}\BibitemShut {NoStop}%
\bibitem [{\citenamefont {Moore}\ \emph {et~al.}(2017)\citenamefont {Moore},
  \citenamefont {Chua},\ and\ \citenamefont {Gair}}]{Moore:2017lxy}%
  \BibitemOpen
  \bibfield  {author} {\bibinfo {author} {\bibfnamefont {Christopher~J.}\
  \bibnamefont {Moore}}, \bibinfo {author} {\bibfnamefont {Alvin J.~K.}\
  \bibnamefont {Chua}}, \ and\ \bibinfo {author} {\bibfnamefont {Jonathan~R.}\
  \bibnamefont {Gair}},\ }\href {\doibase10.1088/1361-6382/aa85fa} {\bibfield
  {journal} {\bibinfo  {journal} {Class. Quant. Grav.}\ }\textbf {\bibinfo
  {volume} {34}},\ \bibinfo {pages} {195009} (\bibinfo {year} {2017})},\
  \Eprint {http://arxiv.org/abs/1707.00712} {arXiv:1707.00712
  [gr-qc]}\BibitemShut {NoStop}%
\bibitem [{\citenamefont {{Four{\`e}s-Bruhat}}(1952)}]{1952AcM....88..141F}%
  \BibitemOpen
  \bibfield  {author} {\bibinfo {author} {\bibfnamefont {Y.}~\bibnamefont
  {{Four{\`e}s-Bruhat}}},\ }\href {\doibase10.1007/BF02392131} {\bibfield
  {journal} {\bibinfo  {journal} {Acta Mathematica, Volume 88, Issue 1, pp
  141-225}\ }\textbf {\bibinfo {volume} {88}} (\bibinfo {year} {1952}),\
  10.1007/BF02392131}\BibitemShut {NoStop}%
\bibitem [{\citenamefont {Papallo}(2017)}]{Papallo:2017ddx}%
  \BibitemOpen
  \bibfield  {author} {\bibinfo {author} {\bibfnamefont {Giuseppe}\
  \bibnamefont {Papallo}},\ }\href {\doibase10.1103/PhysRevD.96.124036}
  {\bibfield  {journal} {\bibinfo  {journal} {Phys. Rev.}\ }\textbf {\bibinfo
  {volume} {D96}},\ \bibinfo {pages} {124036} (\bibinfo {year} {2017})},\
  \Eprint {http://arxiv.org/abs/1710.10155} {arXiv:1710.10155
  [gr-qc]}\BibitemShut {NoStop}%
\bibitem [{\citenamefont {Israel}(1976)}]{Israel:1976tn}%
  \BibitemOpen
  \bibfield  {author} {\bibinfo {author} {\bibfnamefont {W.}~\bibnamefont
  {Israel}},\ }\href {\doibase10.1016/0003-4916(76)90064-6} {\bibfield
  {journal} {\bibinfo  {journal} {Annals Phys.}\ }\textbf {\bibinfo {volume}
  {100}},\ \bibinfo {pages} {310--331} (\bibinfo {year} {1976})}\BibitemShut
  {NoStop}%
\bibitem [{\citenamefont {{Israel}}\ and\ \citenamefont
  {{Stewart}}(1976)}]{1976PhLA...58..213I}%
  \BibitemOpen
  \bibfield  {author} {\bibinfo {author} {\bibfnamefont {W.}~\bibnamefont
  {{Israel}}}\ and\ \bibinfo {author} {\bibfnamefont {J.~M.}\ \bibnamefont
  {{Stewart}}},\ }\href {\doibase10.1016/0375-9601(76)90075-X} {\bibfield
  {journal} {\bibinfo  {journal} {Physics Letters A}\ }\textbf {\bibinfo
  {volume} {58}},\ \bibinfo {pages} {213--215} (\bibinfo {year}
  {1976})}\BibitemShut {NoStop}%
\bibitem [{\citenamefont {Israel}\ and\ \citenamefont
  {Stewart}(1979)}]{Israel:1979wp}%
  \BibitemOpen
  \bibfield  {author} {\bibinfo {author} {\bibfnamefont {W.}~\bibnamefont
  {Israel}}\ and\ \bibinfo {author} {\bibfnamefont {J.~M.}\ \bibnamefont
  {Stewart}},\ }\href {\doibase10.1016/0003-4916(79)90130-1} {\bibfield
  {journal} {\bibinfo  {journal} {Annals Phys.}\ }\textbf {\bibinfo {volume}
  {118}},\ \bibinfo {pages} {341--372} (\bibinfo {year} {1979})}\BibitemShut
  {NoStop}%
\bibitem [{\citenamefont {{Cayuso}}\ \emph {et~al.}(2017)\citenamefont
  {{Cayuso}}, \citenamefont {{Ortiz}},\ and\ \citenamefont
  {{Lehner}}}]{2017PhRvD..96h4043C}%
  \BibitemOpen
  \bibfield  {author} {\bibinfo {author} {\bibfnamefont {J.}~\bibnamefont
  {{Cayuso}}}, \bibinfo {author} {\bibfnamefont {N.}~\bibnamefont {{Ortiz}}}, \
  and\ \bibinfo {author} {\bibfnamefont {L.}~\bibnamefont {{Lehner}}},\ }\href
  {\doibase10.1103/PhysRevD.96.084043} {\bibfield  {journal} {\bibinfo
  {journal} {"Phys. Rev. D"}\ }\textbf {\bibinfo {volume} {96}},\ \bibinfo
  {eid} {084043} (\bibinfo {year} {2017})},\ \Eprint
  {http://arxiv.org/abs/1706.07421} {arXiv:1706.07421 [gr-qc]}\BibitemShut
  {NoStop}%
\bibitem [{\citenamefont {Bel}\ and\ \citenamefont {Zia}(1985)}]{Bel:1985zz}%
  \BibitemOpen
  \bibfield  {author} {\bibinfo {author} {\bibfnamefont {L.}~\bibnamefont
  {Bel}}\ and\ \bibinfo {author} {\bibfnamefont {H.~Sirousse}\ \bibnamefont
  {Zia}},\ }\href {\doibase10.1103/PhysRevD.32.3128} {\bibfield  {journal}
  {\bibinfo  {journal} {Phys. Rev.}\ }\textbf {\bibinfo {volume} {D32}},\
  \bibinfo {pages} {3128--3135} (\bibinfo {year} {1985})}\BibitemShut {NoStop}%
\bibitem [{\citenamefont {Simon}(1990)}]{Simon:1990ic}%
  \BibitemOpen
  \bibfield  {author} {\bibinfo {author} {\bibfnamefont {Jonathan~Z.}\
  \bibnamefont {Simon}},\ }\href {\doibase10.1103/PhysRevD.41.3720} {\bibfield
  {journal} {\bibinfo  {journal} {Phys. Rev.}\ }\textbf {\bibinfo {volume}
  {D41}},\ \bibinfo {pages} {3720} (\bibinfo {year} {1990})}\BibitemShut
  {NoStop}%
\bibitem [{\citenamefont {Sotiriou}(2009)}]{Sotiriou:2008ya}%
  \BibitemOpen
  \bibfield  {author} {\bibinfo {author} {\bibfnamefont {Thomas~P.}\
  \bibnamefont {Sotiriou}},\ }\href {\doibase10.1103/PhysRevD.79.044035}
  {\bibfield  {journal} {\bibinfo  {journal} {Phys. Rev.}\ }\textbf {\bibinfo
  {volume} {D79}},\ \bibinfo {pages} {044035} (\bibinfo {year} {2009})},\
  \Eprint {http://arxiv.org/abs/0811.1799} {arXiv:0811.1799
  [gr-qc]}\BibitemShut {NoStop}%
\bibitem [{\citenamefont {Benkel}\ \emph {et~al.}(2016)\citenamefont {Benkel},
  \citenamefont {Sotiriou},\ and\ \citenamefont {Witek}}]{Benkel:2016kcq}%
  \BibitemOpen
  \bibfield  {author} {\bibinfo {author} {\bibfnamefont {Robert}\ \bibnamefont
  {Benkel}}, \bibinfo {author} {\bibfnamefont {Thomas~P.}\ \bibnamefont
  {Sotiriou}}, \ and\ \bibinfo {author} {\bibfnamefont {Helvi}\ \bibnamefont
  {Witek}},\ }\href {\doibase10.1103/PhysRevD.94.121503} {\bibfield  {journal}
  {\bibinfo  {journal} {Phys. Rev.}\ }\textbf {\bibinfo {volume} {D94}},\
  \bibinfo {pages} {121503} (\bibinfo {year} {2016})},\ \Eprint
  {http://arxiv.org/abs/1612.08184} {arXiv:1612.08184 [gr-qc]}\BibitemShut
  {NoStop}%
\bibitem [{\citenamefont {Benkel}\ \emph {et~al.}(2017)\citenamefont {Benkel},
  \citenamefont {Sotiriou},\ and\ \citenamefont {Witek}}]{Benkel:2016rlz}%
  \BibitemOpen
  \bibfield  {author} {\bibinfo {author} {\bibfnamefont {Robert}\ \bibnamefont
  {Benkel}}, \bibinfo {author} {\bibfnamefont {Thomas~P.}\ \bibnamefont
  {Sotiriou}}, \ and\ \bibinfo {author} {\bibfnamefont {Helvi}\ \bibnamefont
  {Witek}},\ }\href {\doibase10.1088/1361-6382/aa5ce7} {\bibfield  {journal}
  {\bibinfo  {journal} {Class. Quant. Grav.}\ }\textbf {\bibinfo {volume}
  {34}},\ \bibinfo {pages} {064001} (\bibinfo {year} {2017})},\ \Eprint
  {http://arxiv.org/abs/1610.09168} {arXiv:1610.09168 [gr-qc]}\BibitemShut
  {NoStop}%
\bibitem [{\citenamefont {Sotiriou}\ and\ \citenamefont
  {Faraoni}(2010)}]{Sotiriou:2008rp}%
  \BibitemOpen
  \bibfield  {author} {\bibinfo {author} {\bibfnamefont {Thomas~P.}\
  \bibnamefont {Sotiriou}}\ and\ \bibinfo {author} {\bibfnamefont {Valerio}\
  \bibnamefont {Faraoni}},\ }\href {\doibase10.1103/RevModPhys.82.451}
  {\bibfield  {journal} {\bibinfo  {journal} {Rev. Mod. Phys.}\ }\textbf
  {\bibinfo {volume} {82}},\ \bibinfo {pages} {451--497} (\bibinfo {year}
  {2010})},\ \Eprint {http://arxiv.org/abs/0805.1726} {arXiv:0805.1726
  [gr-qc]}\BibitemShut {NoStop}%
\bibitem [{\citenamefont {Lanahan-Tremblay}\ and\ \citenamefont
  {Faraoni}(2007)}]{LanahanTremblay:2007sg}%
  \BibitemOpen
  \bibfield  {author} {\bibinfo {author} {\bibfnamefont {Nicolas}\ \bibnamefont
  {Lanahan-Tremblay}}\ and\ \bibinfo {author} {\bibfnamefont {Valerio}\
  \bibnamefont {Faraoni}},\ }\href {\doibase10.1088/0264-9381/24/22/024}
  {\bibfield  {journal} {\bibinfo  {journal} {Class. Quant. Grav.}\ }\textbf
  {\bibinfo {volume} {24}},\ \bibinfo {pages} {5667--5680} (\bibinfo {year}
  {2007})},\ \Eprint {http://arxiv.org/abs/0709.4414} {arXiv:0709.4414
  [gr-qc]}\BibitemShut {NoStop}%
\bibitem [{\citenamefont {Sagunski}\ \emph {et~al.}(2018)\citenamefont
  {Sagunski}, \citenamefont {Zhang}, \citenamefont {Johnson}, \citenamefont
  {Lehner}, \citenamefont {Sakellariadou}, \citenamefont {Liebling},
  \citenamefont {Palenzuela},\ and\ \citenamefont
  {Neilsen}}]{Sagunski:2017nzb}%
  \BibitemOpen
  \bibfield  {author} {\bibinfo {author} {\bibfnamefont {Laura}\ \bibnamefont
  {Sagunski}}, \bibinfo {author} {\bibfnamefont {Jun}\ \bibnamefont {Zhang}},
  \bibinfo {author} {\bibfnamefont {Matthew~C.}\ \bibnamefont {Johnson}},
  \bibinfo {author} {\bibfnamefont {Luis}\ \bibnamefont {Lehner}}, \bibinfo
  {author} {\bibfnamefont {Mairi}\ \bibnamefont {Sakellariadou}}, \bibinfo
  {author} {\bibfnamefont {Steven~L.}\ \bibnamefont {Liebling}}, \bibinfo
  {author} {\bibfnamefont {Carlos}\ \bibnamefont {Palenzuela}}, \ and\ \bibinfo
  {author} {\bibfnamefont {David}\ \bibnamefont {Neilsen}},\ }\href
  {\doibase10.1103/PhysRevD.97.064016} {\bibfield  {journal} {\bibinfo
  {journal} {Phys. Rev.}\ }\textbf {\bibinfo {volume} {D97}},\ \bibinfo {pages}
  {064016} (\bibinfo {year} {2018})},\ \Eprint
  {http://arxiv.org/abs/1709.06634} {arXiv:1709.06634 [gr-qc]}\BibitemShut
  {NoStop}%
\bibitem [{\citenamefont {Schwarzschild}(1916)}]{Schwarzschild:1916uq}%
  \BibitemOpen
  \bibfield  {author} {\bibinfo {author} {\bibfnamefont {Karl}\ \bibnamefont
  {Schwarzschild}},\ }\href@noop {} {\bibfield  {journal} {\bibinfo  {journal}
  {Sitzungsber. Preuss. Akad. Wiss. Berlin (Math. Phys.)}\ }\textbf {\bibinfo
  {volume} {1916}},\ \bibinfo {pages} {189--196} (\bibinfo {year} {1916})},\
  \Eprint {http://arxiv.org/abs/physics/9905030} {arXiv:physics/9905030
  [physics]}\BibitemShut {NoStop}%
\bibitem [{\citenamefont {Einstein}(1915)}]{Einstein:1915ca}%
  \BibitemOpen
  \bibfield  {author} {\bibinfo {author} {\bibfnamefont {Albert}\ \bibnamefont
  {Einstein}},\ }\href@noop {} {\bibfield  {journal} {\bibinfo  {journal}
  {Sitzungsber. Preuss. Akad. Wiss. Berlin (Math. Phys.)}\ }\textbf {\bibinfo
  {volume} {1915}},\ \bibinfo {pages} {844--847} (\bibinfo {year}
  {1915})}\BibitemShut {NoStop}%
\bibitem [{\citenamefont {Kerr}(1963)}]{Kerr:1963ud}%
  \BibitemOpen
  \bibfield  {author} {\bibinfo {author} {\bibfnamefont {Roy~P.}\ \bibnamefont
  {Kerr}},\ }\href {\doibase10.1103/PhysRevLett.11.237} {\bibfield  {journal}
  {\bibinfo  {journal} {Phys. Rev. Lett.}\ }\textbf {\bibinfo {volume} {11}},\
  \bibinfo {pages} {237--238} (\bibinfo {year} {1963})}\BibitemShut {NoStop}%
\bibitem [{\citenamefont {Kerr}(2007)}]{Kerr:2007dk}%
  \BibitemOpen
  \bibfield  {author} {\bibinfo {author} {\bibfnamefont {Roy~P.}\ \bibnamefont
  {Kerr}},\ }in\ \href
  {http://inspirehep.net/record/752664/files/arXiv:0706.1109.pdf} {\emph
  {\bibinfo {booktitle} {{Kerr Fest: Black Holes in Astrophysics, General
  Relativity and Quantum Gravity Christchurch, New Zealand, August 26-28,
  2004}}}}\ (\bibinfo {year} {2007})\ \Eprint {http://arxiv.org/abs/0706.1109}
  {arXiv:0706.1109 [gr-qc]}\BibitemShut {NoStop}%
\bibitem [{\citenamefont {Wheeler}\ and\ \citenamefont
  {Ford}(1998)}]{Wheeler:1998vs}%
  \BibitemOpen
  \bibfield  {author} {\bibinfo {author} {\bibfnamefont {J.~A.}\ \bibnamefont
  {Wheeler}}\ and\ \bibinfo {author} {\bibfnamefont {K.}~\bibnamefont {Ford}},\
  }\href@noop {} {\emph {\bibinfo {title} {{Geons, black holes, and quantum
  foam: A life in physics}}}}\ (\bibinfo  {publisher} {W. W. Norton \&
  Company},\ \bibinfo {year} {1998})\BibitemShut {NoStop}%
\bibitem [{\citenamefont {Israel}(1967)}]{Israel:1967wq}%
  \BibitemOpen
  \bibfield  {author} {\bibinfo {author} {\bibfnamefont {Werner}\ \bibnamefont
  {Israel}},\ }\href {\doibase10.1103/PhysRev.164.1776} {\bibfield  {journal}
  {\bibinfo  {journal} {Phys. Rev.}\ }\textbf {\bibinfo {volume} {164}},\
  \bibinfo {pages} {1776--1779} (\bibinfo {year} {1967})}\BibitemShut {NoStop}%
\bibitem [{\citenamefont {Robinson}(1975)}]{Robinson:1975bv}%
  \BibitemOpen
  \bibfield  {author} {\bibinfo {author} {\bibfnamefont {D.~C.}\ \bibnamefont
  {Robinson}},\ }\href {\doibase10.1103/PhysRevLett.34.905} {\bibfield
  {journal} {\bibinfo  {journal} {Phys. Rev. Lett.}\ }\textbf {\bibinfo
  {volume} {34}},\ \bibinfo {pages} {905--906} (\bibinfo {year}
  {1975})}\BibitemShut {NoStop}%
\bibitem [{\citenamefont {Carter}(1997)}]{Carter:1997im}%
  \BibitemOpen
  \bibfield  {author} {\bibinfo {author} {\bibfnamefont {B.}~\bibnamefont
  {Carter}},\ }in\ \href@noop {} {\emph {\bibinfo {booktitle} {{Recent
  developments in theoretical and experimental general relativity, gravitation,
  and relativistic field theories. Proceedings, 8th Marcel Grossmann meeting,
  MG8, Jerusalem, Israel, June 22-27, 1997. Pts. A, B}}}}\ (\bibinfo {year}
  {1997})\ pp.\ \bibinfo {pages} {136--155},\ \Eprint
  {http://arxiv.org/abs/gr-qc/9712038} {arXiv:gr-qc/9712038
  [gr-qc]}\BibitemShut {NoStop}%
\bibitem [{\citenamefont {Hawking}(1972{\natexlab{b}})}]{Hawking:1971vc}%
  \BibitemOpen
  \bibfield  {author} {\bibinfo {author} {\bibfnamefont {S.~W.}\ \bibnamefont
  {Hawking}},\ }\href {\doibase10.1007/BF01877517} {\bibfield  {journal}
  {\bibinfo  {journal} {Commun. Math. Phys.}\ }\textbf {\bibinfo {volume}
  {25}},\ \bibinfo {pages} {152--166} (\bibinfo {year}
  {1972}{\natexlab{b}})}\BibitemShut {NoStop}%
\bibitem [{\citenamefont {Hansen}(1974)}]{Hansen:1974zz}%
  \BibitemOpen
  \bibfield  {author} {\bibinfo {author} {\bibfnamefont {R.~O.}\ \bibnamefont
  {Hansen}},\ }\href {\doibase10.1063/1.1666501} {\bibfield  {journal}
  {\bibinfo  {journal} {J. Math. Phys.}\ }\textbf {\bibinfo {volume} {15}},\
  \bibinfo {pages} {46--52} (\bibinfo {year} {1974})}\BibitemShut {NoStop}%
\bibitem [{\citenamefont {Ruffini}\ and\ \citenamefont
  {Wheeler}(1971)}]{Ruffini:1971bza}%
  \BibitemOpen
  \bibfield  {author} {\bibinfo {author} {\bibfnamefont {Remo}\ \bibnamefont
  {Ruffini}}\ and\ \bibinfo {author} {\bibfnamefont {John~A.}\ \bibnamefont
  {Wheeler}},\ }\href {\doibase10.1063/1.3022513} {\bibfield  {journal}
  {\bibinfo  {journal} {Phys. Today}\ }\textbf {\bibinfo {volume} {24}},\
  \bibinfo {pages} {30} (\bibinfo {year} {1971})}\BibitemShut {NoStop}%
\bibitem [{\citenamefont {Narayan}\ and\ \citenamefont
  {McClintock}(2013)}]{Narayan:2013gca}%
  \BibitemOpen
  \bibfield  {author} {\bibinfo {author} {\bibfnamefont {Ramesh}\ \bibnamefont
  {Narayan}}\ and\ \bibinfo {author} {\bibfnamefont {Jeffrey~E.}\ \bibnamefont
  {McClintock}},\ }\href@noop {} {\  (\bibinfo {year} {2013})},\ \Eprint
  {http://arxiv.org/abs/1312.6698} {arXiv:1312.6698 [astro-ph.HE]}\BibitemShut
  {NoStop}%
\bibitem [{\citenamefont {Schucking}(1989)}]{Schucking:1989}%
  \BibitemOpen
  \bibfield  {author} {\bibinfo {author} {\bibfnamefont {E.~L.}\ \bibnamefont
  {Schucking}},\ }\href {\doibase10.1063/1.881214} {\bibfield  {journal}
  {\bibinfo  {journal} {Phys. Today}\ }\textbf {\bibinfo {volume} {42}},\
  \bibinfo {pages} {46--52} (\bibinfo {year} {1989})}\BibitemShut {NoStop}%
\bibitem [{\citenamefont {Misner}\ \emph {et~al.}(1973)\citenamefont {Misner},
  \citenamefont {Thorne},\ and\ \citenamefont {Wheeler}}]{Misner:1974qy}%
  \BibitemOpen
  \bibfield  {author} {\bibinfo {author} {\bibfnamefont {Charles~W.}\
  \bibnamefont {Misner}}, \bibinfo {author} {\bibfnamefont {K.S.}\ \bibnamefont
  {Thorne}}, \ and\ \bibinfo {author} {\bibfnamefont {J.A.}\ \bibnamefont
  {Wheeler}},\ }\href@noop {} {\emph {\bibinfo {title} {{Gravitation}}}}\
  (\bibinfo  {publisher} {{Freeman}},\ \bibinfo {address} {San Francisco},\
  \bibinfo {year} {1973})\BibitemShut {NoStop}%
\bibitem [{\citenamefont {Chandrasekhar}(1987)}]{Chandrasekhar:1989}%
  \BibitemOpen
  \bibfield  {author} {\bibinfo {author} {\bibfnamefont {S.}~\bibnamefont
  {Chandrasekhar}},\ }\href@noop {} {\emph {\bibinfo {title} {{Truth and
  Beauty}}}}\ (\bibinfo  {publisher} {University Of Chicago Press},\ \bibinfo
  {address} {Chicago},\ \bibinfo {year} {1987})\BibitemShut {NoStop}%
\bibitem [{\citenamefont {Bekenstein}(1996)}]{Bekenstein:1996pn}%
  \BibitemOpen
  \bibfield  {author} {\bibinfo {author} {\bibfnamefont {Jacob~D.}\
  \bibnamefont {Bekenstein}},\ }in\ \href@noop {} {\emph {\bibinfo {booktitle}
  {{Physics. Proceedings, 2nd International A.D. Sakharov Conference, Moscow,
  Russia, May 20-24, 1996}}}}\ (\bibinfo {year} {1996})\ pp.\ \bibinfo {pages}
  {216--219},\ \Eprint {http://arxiv.org/abs/gr-qc/9605059}
  {arXiv:gr-qc/9605059 [gr-qc]}\BibitemShut {NoStop}%
\bibitem [{\citenamefont {Herdeiro}\ and\ \citenamefont
  {Radu}(2015{\natexlab{a}})}]{Herdeiro:2015waa}%
  \BibitemOpen
  \bibfield  {author} {\bibinfo {author} {\bibfnamefont {Carlos A.~R.}\
  \bibnamefont {Herdeiro}}\ and\ \bibinfo {author} {\bibfnamefont {Eugen}\
  \bibnamefont {Radu}},\ }\bibfield  {booktitle} {\emph {\bibinfo {booktitle}
  {{Proceedings, 7th Black Holes Workshop 2014: Aveiro, Portugal, December
  18-19, 2014}}},\ }\href {\doibase10.1142/S0218271815420146} {\bibfield
  {journal} {\bibinfo  {journal} {Int. J. Mod. Phys.}\ }\textbf {\bibinfo
  {volume} {D24}},\ \bibinfo {pages} {1542014} (\bibinfo {year}
  {2015}{\natexlab{a}})},\ \Eprint {http://arxiv.org/abs/1504.08209}
  {arXiv:1504.08209 [gr-qc]}\BibitemShut {NoStop}%
\bibitem [{\citenamefont {Volkov}(2017)}]{Volkov:2016ehx}%
  \BibitemOpen
  \bibfield  {author} {\bibinfo {author} {\bibfnamefont {Mikhail~S.}\
  \bibnamefont {Volkov}},\ }in\ \href {\doibase10.1142/9789813226609_0184}
  {\emph {\bibinfo {booktitle} {{Proceedings, 14th Marcel Grossmann Meeting on
  Recent Developments in Theoretical and Experimental General Relativity,
  Astrophysics, and Relativistic Field Theories (MG14) (In 4 Volumes): Rome,
  Italy, July 12-18, 2015}}}},\ Vol.~\bibinfo {volume} {2}\ (\bibinfo {year}
  {2017})\ pp.\ \bibinfo {pages} {1779--1798},\ \Eprint
  {http://arxiv.org/abs/1601.08230} {arXiv:1601.08230 [gr-qc]}\BibitemShut
  {NoStop}%
\bibitem [{\citenamefont {Chase}(1970)}]{Chase:1970}%
  \BibitemOpen
  \bibfield  {author} {\bibinfo {author} {\bibfnamefont {J.~E.}\ \bibnamefont
  {Chase}},\ }\href {\doibase10.1007/BF01646635} {\bibfield  {journal}
  {\bibinfo  {journal} {Commun. Math. Phys.}\ }\textbf {\bibinfo {volume}
  {19}},\ \bibinfo {pages} {276--288} (\bibinfo {year} {1970})}\BibitemShut
  {NoStop}%
\bibitem [{\citenamefont {Penney}(1968)}]{Penney:1968zz}%
  \BibitemOpen
  \bibfield  {author} {\bibinfo {author} {\bibfnamefont {R.}~\bibnamefont
  {Penney}},\ }\href {\doibase10.1103/PhysRev.174.1578} {\bibfield  {journal}
  {\bibinfo  {journal} {Phys. Rev.}\ }\textbf {\bibinfo {volume} {174}},\
  \bibinfo {pages} {1578--1579} (\bibinfo {year} {1968})}\BibitemShut {NoStop}%
\bibitem [{\citenamefont {Bekenstein}(1972{\natexlab{a}})}]{Bekenstein:1971hc}%
  \BibitemOpen
  \bibfield  {author} {\bibinfo {author} {\bibfnamefont {Jacob~D.}\
  \bibnamefont {Bekenstein}},\ }\href {\doibase10.1103/PhysRevD.5.1239}
  {\bibfield  {journal} {\bibinfo  {journal} {Phys. Rev.}\ }\textbf {\bibinfo
  {volume} {D5}},\ \bibinfo {pages} {1239--1246} (\bibinfo {year}
  {1972}{\natexlab{a}})}\BibitemShut {NoStop}%
\bibitem [{\citenamefont {Bekenstein}(1972{\natexlab{b}})}]{Bekenstein:1972ky}%
  \BibitemOpen
  \bibfield  {author} {\bibinfo {author} {\bibfnamefont {J.~D.}\ \bibnamefont
  {Bekenstein}},\ }\href {\doibase10.1103/PhysRevD.5.2403} {\bibfield
  {journal} {\bibinfo  {journal} {Phys. Rev.}\ }\textbf {\bibinfo {volume}
  {D5}},\ \bibinfo {pages} {2403--2412} (\bibinfo {year}
  {1972}{\natexlab{b}})}\BibitemShut {NoStop}%
\bibitem [{\citenamefont {Bekenstein}(1972{\natexlab{c}})}]{Bekenstein:1972ny}%
  \BibitemOpen
  \bibfield  {author} {\bibinfo {author} {\bibfnamefont {J.~D.}\ \bibnamefont
  {Bekenstein}},\ }\href {\doibase10.1103/PhysRevLett.28.452} {\bibfield
  {journal} {\bibinfo  {journal} {Phys. Rev. Lett.}\ }\textbf {\bibinfo
  {volume} {28}},\ \bibinfo {pages} {452--455} (\bibinfo {year}
  {1972}{\natexlab{c}})}\BibitemShut {NoStop}%
\bibitem [{\citenamefont {Bizon}(1994)}]{Bizon:1994dh}%
  \BibitemOpen
  \bibfield  {author} {\bibinfo {author} {\bibfnamefont {Piotr}\ \bibnamefont
  {Bizon}},\ }\href@noop {} {\bibfield  {journal} {\bibinfo  {journal} {Acta
  Phys. Polon.}\ }\textbf {\bibinfo {volume} {B25}},\ \bibinfo {pages}
  {877--898} (\bibinfo {year} {1994})},\ \Eprint
  {http://arxiv.org/abs/gr-qc/9402016} {arXiv:gr-qc/9402016
  [gr-qc]}\BibitemShut {NoStop}%
\bibitem [{\citenamefont {Volkov}\ and\ \citenamefont
  {Gal'tsov}(1999)}]{Volkov:1998cc}%
  \BibitemOpen
  \bibfield  {author} {\bibinfo {author} {\bibfnamefont {Mikhail~S.}\
  \bibnamefont {Volkov}}\ and\ \bibinfo {author} {\bibfnamefont {Dmitri~V.}\
  \bibnamefont {Gal'tsov}},\ }\href {\doibase10.1016/S0370-1573(99)00010-1}
  {\bibfield  {journal} {\bibinfo  {journal} {Phys. Rept.}\ }\textbf {\bibinfo
  {volume} {319}},\ \bibinfo {pages} {1--83} (\bibinfo {year} {1999})},\
  \Eprint {http://arxiv.org/abs/hep-th/9810070} {arXiv:hep-th/9810070
  [hep-th]}\BibitemShut {NoStop}%
\bibitem [{\citenamefont {Volkov}\ and\ \citenamefont
  {Galtsov}(1989)}]{Volkov:1989fi}%
  \BibitemOpen
  \bibfield  {author} {\bibinfo {author} {\bibfnamefont {M.~S.}\ \bibnamefont
  {Volkov}}\ and\ \bibinfo {author} {\bibfnamefont {D.~V.}\ \bibnamefont
  {Galtsov}},\ }\href
  {http://www.jetpletters.ac.ru/ps/1130/article_17118.shtml} {\bibfield
  {journal} {\bibinfo  {journal} {JETP Lett.}\ }\textbf {\bibinfo {volume}
  {50}},\ \bibinfo {pages} {346--350} (\bibinfo {year} {1989})},\ \bibinfo
  {note} {[Pisma Zh. Eksp. Teor. Fiz.50,312(1989)]}\BibitemShut {NoStop}%
\bibitem [{\citenamefont {Volkov}\ and\ \citenamefont
  {Galtsov}(1990)}]{Volkov:1990sva}%
  \BibitemOpen
  \bibfield  {author} {\bibinfo {author} {\bibfnamefont {M.~S.}\ \bibnamefont
  {Volkov}}\ and\ \bibinfo {author} {\bibfnamefont {D.~V.}\ \bibnamefont
  {Galtsov}},\ }\href@noop {} {\bibfield  {journal} {\bibinfo  {journal} {Sov.
  J. Nucl. Phys.}\ }\textbf {\bibinfo {volume} {51}},\ \bibinfo {pages}
  {747--753} (\bibinfo {year} {1990})},\ \bibinfo {note} {[Yad.
  Fiz.51,1171(1990)]}\BibitemShut {NoStop}%
\bibitem [{\citenamefont {{K{\"u}nzle}}\ and\ \citenamefont
  {{Masood-ul-Alam}}(1990)}]{1990JMP....31..928K}%
  \BibitemOpen
  \bibfield  {author} {\bibinfo {author} {\bibfnamefont {H.~P.}\ \bibnamefont
  {{K{\"u}nzle}}}\ and\ \bibinfo {author} {\bibfnamefont {A.~K.~M.}\
  \bibnamefont {{Masood-ul-Alam}}},\ }\href {\doibase10.1063/1.528773}
  {\bibfield  {journal} {\bibinfo  {journal} {Journal of Mathematical Physics}\
  }\textbf {\bibinfo {volume} {31}},\ \bibinfo {pages} {928--935} (\bibinfo
  {year} {1990})}\BibitemShut {NoStop}%
\bibitem [{\citenamefont {Bizon}(1990)}]{Bizon:1990sr}%
  \BibitemOpen
  \bibfield  {author} {\bibinfo {author} {\bibfnamefont {P.}~\bibnamefont
  {Bizon}},\ }\href {\doibase10.1103/PhysRevLett.64.2844} {\bibfield  {journal}
  {\bibinfo  {journal} {Phys. Rev. Lett.}\ }\textbf {\bibinfo {volume} {64}},\
  \bibinfo {pages} {2844--2847} (\bibinfo {year} {1990})}\BibitemShut {NoStop}%
\bibitem [{\citenamefont {Torii}\ \emph {et~al.}(2001)\citenamefont {Torii},
  \citenamefont {Maeda},\ and\ \citenamefont {Narita}}]{Torii:2001pg}%
  \BibitemOpen
  \bibfield  {author} {\bibinfo {author} {\bibfnamefont {Takashi}\ \bibnamefont
  {Torii}}, \bibinfo {author} {\bibfnamefont {Kengo}\ \bibnamefont {Maeda}}, \
  and\ \bibinfo {author} {\bibfnamefont {Makoto}\ \bibnamefont {Narita}},\
  }\href {\doibase10.1103/PhysRevD.64.044007} {\bibfield  {journal} {\bibinfo
  {journal} {Phys. Rev.}\ }\textbf {\bibinfo {volume} {D64}},\ \bibinfo {pages}
  {044007} (\bibinfo {year} {2001})}\BibitemShut {NoStop}%
\bibitem [{\citenamefont {Jacobson}(1999)}]{Jacobson:1999vr}%
  \BibitemOpen
  \bibfield  {author} {\bibinfo {author} {\bibfnamefont {Ted}\ \bibnamefont
  {Jacobson}},\ }\href {\doibase10.1103/PhysRevLett.83.2699} {\bibfield
  {journal} {\bibinfo  {journal} {Phys. Rev. Lett.}\ }\textbf {\bibinfo
  {volume} {83}},\ \bibinfo {pages} {2699--2702} (\bibinfo {year} {1999})},\
  \Eprint {http://arxiv.org/abs/astro-ph/9905303} {arXiv:astro-ph/9905303
  [astro-ph]}\BibitemShut {NoStop}%
\bibitem [{\citenamefont {Horbatsch}\ and\ \citenamefont
  {Burgess}(2012)}]{Horbatsch:2011ye}%
  \BibitemOpen
  \bibfield  {author} {\bibinfo {author} {\bibfnamefont {M.~W.}\ \bibnamefont
  {Horbatsch}}\ and\ \bibinfo {author} {\bibfnamefont {C.~P.}\ \bibnamefont
  {Burgess}},\ }\href {\doibase10.1088/1475-7516/2012/05/010} {\bibfield
  {journal} {\bibinfo  {journal} {JCAP}\ }\textbf {\bibinfo {volume} {1205}},\
  \bibinfo {pages} {010} (\bibinfo {year} {2012})},\ \Eprint
  {http://arxiv.org/abs/1111.4009} {arXiv:1111.4009 [gr-qc]}\BibitemShut
  {NoStop}%
\bibitem [{\citenamefont {Smolić}(2017)}]{Smolic:2016dmh}%
  \BibitemOpen
  \bibfield  {author} {\bibinfo {author} {\bibfnamefont {Ivica}\ \bibnamefont
  {Smolić}},\ }\href {\doibase10.1103/PhysRevD.95.024016} {\bibfield
  {journal} {\bibinfo  {journal} {Phys. Rev.}\ }\textbf {\bibinfo {volume}
  {D95}},\ \bibinfo {pages} {024016} (\bibinfo {year} {2017})},\ \Eprint
  {http://arxiv.org/abs/1609.04013} {arXiv:1609.04013 [gr-qc]}\BibitemShut
  {NoStop}%
\bibitem [{\citenamefont {Herdeiro}\ and\ \citenamefont
  {Radu}(2015{\natexlab{b}})}]{Herdeiro:2015gia}%
  \BibitemOpen
  \bibfield  {author} {\bibinfo {author} {\bibfnamefont {Carlos}\ \bibnamefont
  {Herdeiro}}\ and\ \bibinfo {author} {\bibfnamefont {Eugen}\ \bibnamefont
  {Radu}},\ }\href {\doibase10.1088/0264-9381/32/14/144001} {\bibfield
  {journal} {\bibinfo  {journal} {Class. Quant. Grav.}\ }\textbf {\bibinfo
  {volume} {32}},\ \bibinfo {pages} {144001} (\bibinfo {year}
  {2015}{\natexlab{b}})},\ \Eprint {http://arxiv.org/abs/1501.04319}
  {arXiv:1501.04319 [gr-qc]}\BibitemShut {NoStop}%
\bibitem [{\citenamefont {Kleihaus}\ \emph {et~al.}(2015)\citenamefont
  {Kleihaus}, \citenamefont {Kunz},\ and\ \citenamefont
  {Yazadjiev}}]{Kleihaus:2015iea}%
  \BibitemOpen
  \bibfield  {author} {\bibinfo {author} {\bibfnamefont {Burkhard}\
  \bibnamefont {Kleihaus}}, \bibinfo {author} {\bibfnamefont {Jutta}\
  \bibnamefont {Kunz}}, \ and\ \bibinfo {author} {\bibfnamefont {Stoytcho}\
  \bibnamefont {Yazadjiev}},\ }\href {\doibase10.1016/j.physletb.2015.04.014}
  {\bibfield  {journal} {\bibinfo  {journal} {Phys. Lett.}\ }\textbf {\bibinfo
  {volume} {B744}},\ \bibinfo {pages} {406--412} (\bibinfo {year} {2015})},\
  \Eprint {http://arxiv.org/abs/1503.01672} {arXiv:1503.01672
  [gr-qc]}\BibitemShut {NoStop}%
\bibitem [{\citenamefont {Herdeiro}\ \emph {et~al.}(2015)\citenamefont
  {Herdeiro}, \citenamefont {Radu},\ and\ \citenamefont
  {Rúnarsson}}]{Herdeiro:2015tia}%
  \BibitemOpen
  \bibfield  {author} {\bibinfo {author} {\bibfnamefont {Carlos A.~R.}\
  \bibnamefont {Herdeiro}}, \bibinfo {author} {\bibfnamefont {Eugen}\
  \bibnamefont {Radu}}, \ and\ \bibinfo {author} {\bibfnamefont {Helgi}\
  \bibnamefont {Rúnarsson}},\ }\href {\doibase10.1103/PhysRevD.92.084059}
  {\bibfield  {journal} {\bibinfo  {journal} {Phys. Rev.}\ }\textbf {\bibinfo
  {volume} {D92}},\ \bibinfo {pages} {084059} (\bibinfo {year} {2015})},\
  \Eprint {http://arxiv.org/abs/1509.02923} {arXiv:1509.02923
  [gr-qc]}\BibitemShut {NoStop}%
\bibitem [{\citenamefont {Arvanitaki}\ \emph {et~al.}(2010)\citenamefont
  {Arvanitaki}, \citenamefont {Dimopoulos}, \citenamefont {Dubovsky},
  \citenamefont {Kaloper},\ and\ \citenamefont
  {March-Russell}}]{Arvanitaki:2009fg}%
  \BibitemOpen
  \bibfield  {author} {\bibinfo {author} {\bibfnamefont {Asimina}\ \bibnamefont
  {Arvanitaki}}, \bibinfo {author} {\bibfnamefont {Savas}\ \bibnamefont
  {Dimopoulos}}, \bibinfo {author} {\bibfnamefont {Sergei}\ \bibnamefont
  {Dubovsky}}, \bibinfo {author} {\bibfnamefont {Nemanja}\ \bibnamefont
  {Kaloper}}, \ and\ \bibinfo {author} {\bibfnamefont {John}\ \bibnamefont
  {March-Russell}},\ }\href {\doibase10.1103/PhysRevD.81.123530} {\bibfield
  {journal} {\bibinfo  {journal} {Phys. Rev.}\ }\textbf {\bibinfo {volume}
  {D81}},\ \bibinfo {pages} {123530} (\bibinfo {year} {2010})},\ \Eprint
  {http://arxiv.org/abs/0905.4720} {arXiv:0905.4720 [hep-th]}\BibitemShut
  {NoStop}%
\bibitem [{\citenamefont {Herdeiro}\ and\ \citenamefont
  {Radu}(2017)}]{Herdeiro:2017phl}%
  \BibitemOpen
  \bibfield  {author} {\bibinfo {author} {\bibfnamefont {Carlos A.~R.}\
  \bibnamefont {Herdeiro}}\ and\ \bibinfo {author} {\bibfnamefont {Eugen}\
  \bibnamefont {Radu}},\ }\href {\doibase10.1103/PhysRevLett.119.261101}
  {\bibfield  {journal} {\bibinfo  {journal} {Phys. Rev. Lett.}\ }\textbf
  {\bibinfo {volume} {119}},\ \bibinfo {pages} {261101} (\bibinfo {year}
  {2017})},\ \Eprint {http://arxiv.org/abs/1706.06597} {arXiv:1706.06597
  [gr-qc]}\BibitemShut {NoStop}%
\bibitem [{\citenamefont {Herdeiro}\ and\ \citenamefont
  {Radu}(2014{\natexlab{b}})}]{Herdeiro:2014jaa}%
  \BibitemOpen
  \bibfield  {author} {\bibinfo {author} {\bibfnamefont {Carlos}\ \bibnamefont
  {Herdeiro}}\ and\ \bibinfo {author} {\bibfnamefont {Eugen}\ \bibnamefont
  {Radu}},\ }\href {\doibase10.1103/PhysRevD.89.124018} {\bibfield  {journal}
  {\bibinfo  {journal} {Phys. Rev.}\ }\textbf {\bibinfo {volume} {D89}},\
  \bibinfo {pages} {124018} (\bibinfo {year} {2014}{\natexlab{b}})},\ \Eprint
  {http://arxiv.org/abs/1406.1225} {arXiv:1406.1225 [gr-qc]}\BibitemShut
  {NoStop}%
\bibitem [{\citenamefont {Ganchev}\ and\ \citenamefont
  {Santos}(2018)}]{Ganchev:2017uuo}%
  \BibitemOpen
  \bibfield  {author} {\bibinfo {author} {\bibfnamefont {Bogdan}\ \bibnamefont
  {Ganchev}}\ and\ \bibinfo {author} {\bibfnamefont {Jorge~E.}\ \bibnamefont
  {Santos}},\ }\href {\doibase10.1103/PhysRevLett.120.171101} {\bibfield
  {journal} {\bibinfo  {journal} {Phys. Rev. Lett.}\ }\textbf {\bibinfo
  {volume} {120}},\ \bibinfo {pages} {171101} (\bibinfo {year} {2018})},\
  \Eprint {http://arxiv.org/abs/1711.08464} {arXiv:1711.08464
  [gr-qc]}\BibitemShut {NoStop}%
\bibitem [{\citenamefont {Babichev}\ and\ \citenamefont
  {Charmousis}(2014)}]{Babichev:2013cya}%
  \BibitemOpen
  \bibfield  {author} {\bibinfo {author} {\bibfnamefont {Eugeny}\ \bibnamefont
  {Babichev}}\ and\ \bibinfo {author} {\bibfnamefont {Christos}\ \bibnamefont
  {Charmousis}},\ }\href {\doibase10.1007/JHEP08(2014)106} {\bibfield
  {journal} {\bibinfo  {journal} {JHEP}\ }\textbf {\bibinfo {volume} {08}},\
  \bibinfo {pages} {106} (\bibinfo {year} {2014})},\ \Eprint
  {http://arxiv.org/abs/1312.3204} {arXiv:1312.3204 [gr-qc]}\BibitemShut
  {NoStop}%
\bibitem [{\citenamefont {Babichev}\ \emph
  {et~al.}(2017{\natexlab{a}})\citenamefont {Babichev}, \citenamefont
  {Charmousis},\ and\ \citenamefont {Lehébel}}]{Babichev:2017guv}%
  \BibitemOpen
  \bibfield  {author} {\bibinfo {author} {\bibfnamefont {Eugeny}\ \bibnamefont
  {Babichev}}, \bibinfo {author} {\bibfnamefont {Christos}\ \bibnamefont
  {Charmousis}}, \ and\ \bibinfo {author} {\bibfnamefont {Antoine}\
  \bibnamefont {Lehébel}},\ }\href {\doibase10.1088/1475-7516/2017/04/027}
  {\bibfield  {journal} {\bibinfo  {journal} {JCAP}\ }\textbf {\bibinfo
  {volume} {1704}},\ \bibinfo {pages} {027} (\bibinfo {year}
  {2017}{\natexlab{a}})},\ \Eprint {http://arxiv.org/abs/1702.01938}
  {arXiv:1702.01938 [gr-qc]}\BibitemShut {NoStop}%
\bibitem [{\citenamefont {Ogawa}\ \emph {et~al.}(2016)\citenamefont {Ogawa},
  \citenamefont {Kobayashi},\ and\ \citenamefont {Suyama}}]{Ogawa:2015pea}%
  \BibitemOpen
  \bibfield  {author} {\bibinfo {author} {\bibfnamefont {Hiromu}\ \bibnamefont
  {Ogawa}}, \bibinfo {author} {\bibfnamefont {Tsutomu}\ \bibnamefont
  {Kobayashi}}, \ and\ \bibinfo {author} {\bibfnamefont {Teruaki}\ \bibnamefont
  {Suyama}},\ }\href {\doibase10.1103/PhysRevD.93.064078} {\bibfield  {journal}
  {\bibinfo  {journal} {Phys. Rev.}\ }\textbf {\bibinfo {volume} {D93}},\
  \bibinfo {pages} {064078} (\bibinfo {year} {2016})},\ \Eprint
  {http://arxiv.org/abs/1510.07400} {arXiv:1510.07400 [gr-qc]}\BibitemShut
  {NoStop}%
\bibitem [{\citenamefont {Babichev}\ \emph
  {et~al.}(2018{\natexlab{a}})\citenamefont {Babichev}, \citenamefont
  {Charmousis}, \citenamefont {Esposito-Farèse},\ and\ \citenamefont
  {Lehébel}}]{Babichev:2017lmw}%
  \BibitemOpen
  \bibfield  {author} {\bibinfo {author} {\bibfnamefont {Eugeny}\ \bibnamefont
  {Babichev}}, \bibinfo {author} {\bibfnamefont {Christos}\ \bibnamefont
  {Charmousis}}, \bibinfo {author} {\bibfnamefont {Gilles}\ \bibnamefont
  {Esposito-Farèse}}, \ and\ \bibinfo {author} {\bibfnamefont {Antoine}\
  \bibnamefont {Lehébel}},\ }\href {\doibase10.1103/PhysRevLett.120.241101}
  {\bibfield  {journal} {\bibinfo  {journal} {Phys. Rev. Lett.}\ }\textbf
  {\bibinfo {volume} {120}},\ \bibinfo {pages} {241101} (\bibinfo {year}
  {2018}{\natexlab{a}})},\ \Eprint {http://arxiv.org/abs/1712.04398}
  {arXiv:1712.04398 [gr-qc]}\BibitemShut {NoStop}%
\bibitem [{\citenamefont {Babichev}\ \emph
  {et~al.}(2018{\natexlab{b}})\citenamefont {Babichev}, \citenamefont
  {Charmousis}, \citenamefont {Esposito-Farese},\ and\ \citenamefont
  {Lehebel}}]{Babichev:2018uiw}%
  \BibitemOpen
  \bibfield  {author} {\bibinfo {author} {\bibfnamefont {Eugeny}\ \bibnamefont
  {Babichev}}, \bibinfo {author} {\bibfnamefont {Christos}\ \bibnamefont
  {Charmousis}}, \bibinfo {author} {\bibfnamefont {Gilles}\ \bibnamefont
  {Esposito-Farese}}, \ and\ \bibinfo {author} {\bibfnamefont {Antoine}\
  \bibnamefont {Lehebel}},\ }\href@noop {} {\  (\bibinfo {year}
  {2018}{\natexlab{b}})},\ \Eprint {http://arxiv.org/abs/1803.11444}
  {arXiv:1803.11444 [gr-qc]}\BibitemShut {NoStop}%
\bibitem [{\citenamefont {Heisenberg}(2014)}]{Heisenberg:2014rta}%
  \BibitemOpen
  \bibfield  {author} {\bibinfo {author} {\bibfnamefont {Lavinia}\ \bibnamefont
  {Heisenberg}},\ }\href {\doibase10.1088/1475-7516/2014/05/015} {\bibfield
  {journal} {\bibinfo  {journal} {J. Cosm. Astropart.}\ }\textbf {\bibinfo
  {volume} {1405}},\ \bibinfo {pages} {015} (\bibinfo {year} {2014})},\ \Eprint
  {http://arxiv.org/abs/1402.7026} {arXiv:1402.7026 [hep-th]}\BibitemShut
  {NoStop}%
\bibitem [{\citenamefont {Beltran~Jimenez}\ and\ \citenamefont
  {Heisenberg}(2016)}]{Jimenez:2016isa}%
  \BibitemOpen
  \bibfield  {author} {\bibinfo {author} {\bibfnamefont {Jose}\ \bibnamefont
  {Beltran~Jimenez}}\ and\ \bibinfo {author} {\bibfnamefont {Lavinia}\
  \bibnamefont {Heisenberg}},\ }\href {\doibase10.1016/j.physletb.2016.04.017}
  {\bibfield  {journal} {\bibinfo  {journal} {Phys. Lett.}\ }\textbf {\bibinfo
  {volume} {B757}},\ \bibinfo {pages} {405--411} (\bibinfo {year} {2016})},\
  \Eprint {http://arxiv.org/abs/1602.03410} {arXiv:1602.03410
  [hep-th]}\BibitemShut {NoStop}%
\bibitem [{\citenamefont {Chagoya}\ \emph {et~al.}(2016)\citenamefont
  {Chagoya}, \citenamefont {Niz},\ and\ \citenamefont
  {Tasinato}}]{Chagoya:2016aar}%
  \BibitemOpen
  \bibfield  {author} {\bibinfo {author} {\bibfnamefont {Javier}\ \bibnamefont
  {Chagoya}}, \bibinfo {author} {\bibfnamefont {Gustavo}\ \bibnamefont {Niz}},
  \ and\ \bibinfo {author} {\bibfnamefont {Gianmassimo}\ \bibnamefont
  {Tasinato}},\ }\href {\doibase10.1088/0264-9381/33/17/175007} {\bibfield
  {journal} {\bibinfo  {journal} {Class. Quant. Grav.}\ }\textbf {\bibinfo
  {volume} {33}},\ \bibinfo {pages} {175007} (\bibinfo {year} {2016})},\
  \Eprint {http://arxiv.org/abs/1602.08697} {arXiv:1602.08697
  [hep-th]}\BibitemShut {NoStop}%
\bibitem [{\citenamefont {Minamitsuji}(2016)}]{Minamitsuji:2016ydr}%
  \BibitemOpen
  \bibfield  {author} {\bibinfo {author} {\bibfnamefont {Masato}\ \bibnamefont
  {Minamitsuji}},\ }\href {\doibase10.1103/PhysRevD.94.084039} {\bibfield
  {journal} {\bibinfo  {journal} {Phys. Rev.}\ }\textbf {\bibinfo {volume}
  {D94}},\ \bibinfo {pages} {084039} (\bibinfo {year} {2016})},\ \Eprint
  {http://arxiv.org/abs/1607.06278} {arXiv:1607.06278 [gr-qc]}\BibitemShut
  {NoStop}%
\bibitem [{\citenamefont {Heisenberg}\ \emph
  {et~al.}(2017{\natexlab{a}})\citenamefont {Heisenberg}, \citenamefont {Kase},
  \citenamefont {Minamitsuji},\ and\ \citenamefont
  {Tsujikawa}}]{Heisenberg:2017xda}%
  \BibitemOpen
  \bibfield  {author} {\bibinfo {author} {\bibfnamefont {Lavinia}\ \bibnamefont
  {Heisenberg}}, \bibinfo {author} {\bibfnamefont {Ryotaro}\ \bibnamefont
  {Kase}}, \bibinfo {author} {\bibfnamefont {Masato}\ \bibnamefont
  {Minamitsuji}}, \ and\ \bibinfo {author} {\bibfnamefont {Shinji}\
  \bibnamefont {Tsujikawa}},\ }\href {\doibase10.1103/PhysRevD.96.084049}
  {\bibfield  {journal} {\bibinfo  {journal} {Phys. Rev.}\ }\textbf {\bibinfo
  {volume} {D96}},\ \bibinfo {pages} {084049} (\bibinfo {year}
  {2017}{\natexlab{a}})},\ \Eprint {http://arxiv.org/abs/1705.09662}
  {arXiv:1705.09662 [gr-qc]}\BibitemShut {NoStop}%
\bibitem [{\citenamefont {Heisenberg}\ \emph
  {et~al.}(2017{\natexlab{b}})\citenamefont {Heisenberg}, \citenamefont {Kase},
  \citenamefont {Minamitsuji},\ and\ \citenamefont
  {Tsujikawa}}]{Heisenberg:2017hwb}%
  \BibitemOpen
  \bibfield  {author} {\bibinfo {author} {\bibfnamefont {Lavinia}\ \bibnamefont
  {Heisenberg}}, \bibinfo {author} {\bibfnamefont {Ryotaro}\ \bibnamefont
  {Kase}}, \bibinfo {author} {\bibfnamefont {Masato}\ \bibnamefont
  {Minamitsuji}}, \ and\ \bibinfo {author} {\bibfnamefont {Shinji}\
  \bibnamefont {Tsujikawa}},\ }\href {\doibase10.1088/1475-7516/2017/08/024}
  {\bibfield  {journal} {\bibinfo  {journal} {JCAP}\ }\textbf {\bibinfo
  {volume} {1708}},\ \bibinfo {pages} {024} (\bibinfo {year}
  {2017}{\natexlab{b}})},\ \Eprint {http://arxiv.org/abs/1706.05115}
  {arXiv:1706.05115 [gr-qc]}\BibitemShut {NoStop}%
\bibitem [{\citenamefont {Fan}(2016)}]{Fan:2016jnz}%
  \BibitemOpen
  \bibfield  {author} {\bibinfo {author} {\bibfnamefont {Zhong-Ying}\
  \bibnamefont {Fan}},\ }\href {\doibase10.1007/JHEP09(2016)039} {\bibfield
  {journal} {\bibinfo  {journal} {JHEP}\ }\textbf {\bibinfo {volume} {09}},\
  \bibinfo {pages} {039} (\bibinfo {year} {2016})},\ \Eprint
  {http://arxiv.org/abs/1606.00684} {arXiv:1606.00684 [hep-th]}\BibitemShut
  {NoStop}%
\bibitem [{\citenamefont {Cisterna}\ \emph {et~al.}(2016)\citenamefont
  {Cisterna}, \citenamefont {Hassaine}, \citenamefont {Oliva},\ and\
  \citenamefont {Rinaldi}}]{Cisterna:2016nwq}%
  \BibitemOpen
  \bibfield  {author} {\bibinfo {author} {\bibfnamefont {Adolfo}\ \bibnamefont
  {Cisterna}}, \bibinfo {author} {\bibfnamefont {Mokhtar}\ \bibnamefont
  {Hassaine}}, \bibinfo {author} {\bibfnamefont {Julio}\ \bibnamefont {Oliva}},
  \ and\ \bibinfo {author} {\bibfnamefont {Massimiliano}\ \bibnamefont
  {Rinaldi}},\ }\href {\doibase10.1103/PhysRevD.94.104039} {\bibfield
  {journal} {\bibinfo  {journal} {Phys. Rev.}\ }\textbf {\bibinfo {volume}
  {D94}},\ \bibinfo {pages} {104039} (\bibinfo {year} {2016})},\ \Eprint
  {http://arxiv.org/abs/1609.03430} {arXiv:1609.03430 [gr-qc]}\BibitemShut
  {NoStop}%
\bibitem [{\citenamefont {Babichev}\ \emph
  {et~al.}(2017{\natexlab{b}})\citenamefont {Babichev}, \citenamefont
  {Charmousis},\ and\ \citenamefont {Hassaine}}]{Babichev:2017rti}%
  \BibitemOpen
  \bibfield  {author} {\bibinfo {author} {\bibfnamefont {Eugeny}\ \bibnamefont
  {Babichev}}, \bibinfo {author} {\bibfnamefont {Christos}\ \bibnamefont
  {Charmousis}}, \ and\ \bibinfo {author} {\bibfnamefont {Mokhtar}\
  \bibnamefont {Hassaine}},\ }\href {\doibase10.1007/JHEP05(2017)114}
  {\bibfield  {journal} {\bibinfo  {journal} {JHEP}\ }\textbf {\bibinfo
  {volume} {05}},\ \bibinfo {pages} {114} (\bibinfo {year}
  {2017}{\natexlab{b}})},\ \Eprint {http://arxiv.org/abs/1703.07676}
  {arXiv:1703.07676 [gr-qc]}\BibitemShut {NoStop}%
\bibitem [{\citenamefont {Heisenberg}(2018)}]{Heisenberg:2018acv}%
  \BibitemOpen
  \bibfield  {author} {\bibinfo {author} {\bibfnamefont {Lavinia}\ \bibnamefont
  {Heisenberg}},\ }\href@noop {} {\  (\bibinfo {year} {2018})},\ \Eprint
  {http://arxiv.org/abs/1801.01523} {arXiv:1801.01523 [gr-qc]}\BibitemShut
  {NoStop}%
\bibitem [{\citenamefont {Heisenberg}\ and\ \citenamefont
  {Tsujikawa}(2018)}]{Heisenberg:2018vti}%
  \BibitemOpen
  \bibfield  {author} {\bibinfo {author} {\bibfnamefont {Lavinia}\ \bibnamefont
  {Heisenberg}}\ and\ \bibinfo {author} {\bibfnamefont {Shinji}\ \bibnamefont
  {Tsujikawa}},\ }\href {\doibase10.1016/j.physletb.2018.03.059} {\bibfield
  {journal} {\bibinfo  {journal} {Phys. Lett.}\ }\textbf {\bibinfo {volume}
  {B780}},\ \bibinfo {pages} {638--646} (\bibinfo {year} {2018})},\ \Eprint
  {http://arxiv.org/abs/1802.07035} {arXiv:1802.07035 [gr-qc]}\BibitemShut
  {NoStop}%
\bibitem [{\citenamefont {Heisenberg}\ \emph {et~al.}(2018)\citenamefont
  {Heisenberg}, \citenamefont {Kase},\ and\ \citenamefont
  {Tsujikawa}}]{Heisenberg:2018mgr}%
  \BibitemOpen
  \bibfield  {author} {\bibinfo {author} {\bibfnamefont {Lavinia}\ \bibnamefont
  {Heisenberg}}, \bibinfo {author} {\bibfnamefont {Ryotaro}\ \bibnamefont
  {Kase}}, \ and\ \bibinfo {author} {\bibfnamefont {Shinji}\ \bibnamefont
  {Tsujikawa}},\ }\href@noop {} {\  (\bibinfo {year} {2018})},\ \Eprint
  {http://arxiv.org/abs/1804.00535} {arXiv:1804.00535 [gr-qc]}\BibitemShut
  {NoStop}%
\bibitem [{\citenamefont {Volkov}(2012)}]{Volkov:2012wp}%
  \BibitemOpen
  \bibfield  {author} {\bibinfo {author} {\bibfnamefont {Mikhail~S.}\
  \bibnamefont {Volkov}},\ }\href {\doibase10.1103/PhysRevD.85.124043}
  {\bibfield  {journal} {\bibinfo  {journal} {Phys. Rev.}\ }\textbf {\bibinfo
  {volume} {D85}},\ \bibinfo {pages} {124043} (\bibinfo {year} {2012})},\
  \Eprint {http://arxiv.org/abs/1202.6682} {arXiv:1202.6682
  [hep-th]}\BibitemShut {NoStop}%
\bibitem [{\citenamefont {Volkov}(2013)}]{Volkov:2013roa}%
  \BibitemOpen
  \bibfield  {author} {\bibinfo {author} {\bibfnamefont {Mikhail~S.}\
  \bibnamefont {Volkov}},\ }\href {\doibase10.1088/0264-9381/30/18/184009}
  {\bibfield  {journal} {\bibinfo  {journal} {Class. Quant. Grav.}\ }\textbf
  {\bibinfo {volume} {30}},\ \bibinfo {pages} {184009} (\bibinfo {year}
  {2013})},\ \Eprint {http://arxiv.org/abs/1304.0238} {arXiv:1304.0238
  [hep-th]}\BibitemShut {NoStop}%
\bibitem [{\citenamefont {Brito}\ \emph
  {et~al.}(2013{\natexlab{b}})\citenamefont {Brito}, \citenamefont {Cardoso},\
  and\ \citenamefont {Pani}}]{Brito:2013xaa}%
  \BibitemOpen
  \bibfield  {author} {\bibinfo {author} {\bibfnamefont {Richard}\ \bibnamefont
  {Brito}}, \bibinfo {author} {\bibfnamefont {Vitor}\ \bibnamefont {Cardoso}},
  \ and\ \bibinfo {author} {\bibfnamefont {Paolo}\ \bibnamefont {Pani}},\
  }\href {\doibase10.1103/PhysRevD.88.064006} {\bibfield  {journal} {\bibinfo
  {journal} {Phys. Rev.}\ }\textbf {\bibinfo {volume} {D88}},\ \bibinfo {pages}
  {064006} (\bibinfo {year} {2013}{\natexlab{b}})},\ \Eprint
  {http://arxiv.org/abs/1309.0818} {arXiv:1309.0818 [gr-qc]}\BibitemShut
  {NoStop}%
\bibitem [{\citenamefont {Volkov}(2015)}]{Volkov:2014ooa}%
  \BibitemOpen
  \bibfield  {author} {\bibinfo {author} {\bibfnamefont {Mikhail~S.}\
  \bibnamefont {Volkov}},\ }\bibfield  {booktitle} {\emph {\bibinfo {booktitle}
  {{Proceedings of the 7th Aegean Summer School : Beyond Einstein's theory of
  gravity. Modifications of Einstein's Theory of Gravity at Large Distances.:
  Paros, Greece, September 23-28, 2013}}},\ }\href
  {\doibase10.1007/978-3-319-10070-8_6} {\bibfield  {journal} {\bibinfo
  {journal} {Lect. Notes Phys.}\ }\textbf {\bibinfo {volume} {892}},\ \bibinfo
  {pages} {161--180} (\bibinfo {year} {2015})},\ \Eprint
  {http://arxiv.org/abs/1405.1742} {arXiv:1405.1742 [hep-th]}\BibitemShut
  {NoStop}%
\bibitem [{\citenamefont {Babichev}\ and\ \citenamefont
  {Brito}(2015)}]{Babichev:2015xha}%
  \BibitemOpen
  \bibfield  {author} {\bibinfo {author} {\bibfnamefont {Eugeny}\ \bibnamefont
  {Babichev}}\ and\ \bibinfo {author} {\bibfnamefont {Richard}\ \bibnamefont
  {Brito}},\ }\href {\doibase10.1088/0264-9381/32/15/154001} {\bibfield
  {journal} {\bibinfo  {journal} {Class. Quant. Grav.}\ }\textbf {\bibinfo
  {volume} {32}},\ \bibinfo {pages} {154001} (\bibinfo {year} {2015})},\
  \Eprint {http://arxiv.org/abs/1503.07529} {arXiv:1503.07529
  [gr-qc]}\BibitemShut {NoStop}%
\bibitem [{\citenamefont {Torsello}\ \emph {et~al.}(2017)\citenamefont
  {Torsello}, \citenamefont {Kocic},\ and\ \citenamefont
  {Mortsell}}]{Torsello:2017cmz}%
  \BibitemOpen
  \bibfield  {author} {\bibinfo {author} {\bibfnamefont {Francesco}\
  \bibnamefont {Torsello}}, \bibinfo {author} {\bibfnamefont {Mikica}\
  \bibnamefont {Kocic}}, \ and\ \bibinfo {author} {\bibfnamefont {Edvard}\
  \bibnamefont {Mortsell}},\ }\href {\doibase10.1103/PhysRevD.96.064003}
  {\bibfield  {journal} {\bibinfo  {journal} {Phys. Rev.}\ }\textbf {\bibinfo
  {volume} {D96}},\ \bibinfo {pages} {064003} (\bibinfo {year} {2017})},\
  \Eprint {http://arxiv.org/abs/1703.07787} {arXiv:1703.07787
  [gr-qc]}\BibitemShut {NoStop}%
\bibitem [{\citenamefont {Deffayet}\ and\ \citenamefont
  {Jacobson}(2012)}]{Deffayet:2011rh}%
  \BibitemOpen
  \bibfield  {author} {\bibinfo {author} {\bibfnamefont {C.}~\bibnamefont
  {Deffayet}}\ and\ \bibinfo {author} {\bibfnamefont {Ted}\ \bibnamefont
  {Jacobson}},\ }\href {\doibase10.1088/0264-9381/29/6/065009} {\bibfield
  {journal} {\bibinfo  {journal} {Class. Quant. Grav.}\ }\textbf {\bibinfo
  {volume} {29}},\ \bibinfo {pages} {065009} (\bibinfo {year} {2012})},\
  \Eprint {http://arxiv.org/abs/1107.4978} {arXiv:1107.4978
  [gr-qc]}\BibitemShut {NoStop}%
\bibitem [{\citenamefont {Babichev}\ and\ \citenamefont
  {Fabbri}(2014)}]{Babichev:2014oua}%
  \BibitemOpen
  \bibfield  {author} {\bibinfo {author} {\bibfnamefont {Eugeny}\ \bibnamefont
  {Babichev}}\ and\ \bibinfo {author} {\bibfnamefont {Alessandro}\ \bibnamefont
  {Fabbri}},\ }\href {\doibase10.1103/PhysRevD.89.081502} {\bibfield  {journal}
  {\bibinfo  {journal} {Phys. Rev.}\ }\textbf {\bibinfo {volume} {D89}},\
  \bibinfo {pages} {081502} (\bibinfo {year} {2014})},\ \Eprint
  {http://arxiv.org/abs/1401.6871} {arXiv:1401.6871 [gr-qc]}\BibitemShut
  {NoStop}%
\bibitem [{\citenamefont {Rosen}(2017)}]{Rosen:2017dvn}%
  \BibitemOpen
  \bibfield  {author} {\bibinfo {author} {\bibfnamefont {Rachel~A}\
  \bibnamefont {Rosen}},\ }\href {\doibase10.1007/JHEP10(2017)206} {\bibfield
  {journal} {\bibinfo  {journal} {JHEP}\ }\textbf {\bibinfo {volume} {10}},\
  \bibinfo {pages} {206} (\bibinfo {year} {2017})},\ \Eprint
  {http://arxiv.org/abs/1702.06543} {arXiv:1702.06543 [hep-th]}\BibitemShut
  {NoStop}%
\bibitem [{\citenamefont {Eling}\ and\ \citenamefont
  {Jacobson}(2006)}]{Eling:2006ec}%
  \BibitemOpen
  \bibfield  {author} {\bibinfo {author} {\bibfnamefont {Christopher}\
  \bibnamefont {Eling}}\ and\ \bibinfo {author} {\bibfnamefont {Ted}\
  \bibnamefont {Jacobson}},\ }\href {\doibase10.1088/0264-9381/23/18/009}
  {\bibfield  {journal} {\bibinfo  {journal} {Class. Quant. Grav.}\ }\textbf
  {\bibinfo {volume} {23}},\ \bibinfo {pages} {5643--5660} (\bibinfo {year}
  {2006})},\ \bibinfo {note} {[Erratum: Class. Quant. Grav.27,049802(2010)]},\
  \Eprint {http://arxiv.org/abs/gr-qc/0604088} {arXiv:gr-qc/0604088
  [gr-qc]}\BibitemShut {NoStop}%
\bibitem [{\citenamefont {Barausse}\ \emph {et~al.}(2011)\citenamefont
  {Barausse}, \citenamefont {Jacobson},\ and\ \citenamefont
  {Sotiriou}}]{Barausse:2011pu}%
  \BibitemOpen
  \bibfield  {author} {\bibinfo {author} {\bibfnamefont {Enrico}\ \bibnamefont
  {Barausse}}, \bibinfo {author} {\bibfnamefont {Ted}\ \bibnamefont
  {Jacobson}}, \ and\ \bibinfo {author} {\bibfnamefont {Thomas~P.}\
  \bibnamefont {Sotiriou}},\ }\href {\doibase10.1103/PhysRevD.83.124043}
  {\bibfield  {journal} {\bibinfo  {journal} {Phys. Rev.}\ }\textbf {\bibinfo
  {volume} {D83}},\ \bibinfo {pages} {124043} (\bibinfo {year} {2011})},\
  \Eprint {http://arxiv.org/abs/1104.2889} {arXiv:1104.2889
  [gr-qc]}\BibitemShut {NoStop}%
\bibitem [{\citenamefont {Barausse}\ and\ \citenamefont
  {Sotiriou}(2013{\natexlab{b}})}]{Barausse:2012qh}%
  \BibitemOpen
  \bibfield  {author} {\bibinfo {author} {\bibfnamefont {Enrico}\ \bibnamefont
  {Barausse}}\ and\ \bibinfo {author} {\bibfnamefont {Thomas~P.}\ \bibnamefont
  {Sotiriou}},\ }\href {\doibase10.1103/PhysRevD.87.087504} {\bibfield
  {journal} {\bibinfo  {journal} {Phys. Rev.}\ }\textbf {\bibinfo {volume}
  {D87}},\ \bibinfo {pages} {087504} (\bibinfo {year} {2013}{\natexlab{b}})},\
  \Eprint {http://arxiv.org/abs/1212.1334} {arXiv:1212.1334
  [gr-qc]}\BibitemShut {NoStop}%
\bibitem [{\citenamefont {Sotiriou}\ \emph {et~al.}(2014)\citenamefont
  {Sotiriou}, \citenamefont {Vega},\ and\ \citenamefont
  {Vernieri}}]{Sotiriou:2014gna}%
  \BibitemOpen
  \bibfield  {author} {\bibinfo {author} {\bibfnamefont {Thomas~P.}\
  \bibnamefont {Sotiriou}}, \bibinfo {author} {\bibfnamefont {Ian}\
  \bibnamefont {Vega}}, \ and\ \bibinfo {author} {\bibfnamefont {Daniele}\
  \bibnamefont {Vernieri}},\ }\href {\doibase10.1103/PhysRevD.90.044046}
  {\bibfield  {journal} {\bibinfo  {journal} {Phys. Rev.}\ }\textbf {\bibinfo
  {volume} {D90}},\ \bibinfo {pages} {044046} (\bibinfo {year} {2014})},\
  \Eprint {http://arxiv.org/abs/1405.3715} {arXiv:1405.3715
  [gr-qc]}\BibitemShut {NoStop}%
\bibitem [{\citenamefont {Barausse}\ \emph {et~al.}(2016)\citenamefont
  {Barausse}, \citenamefont {Sotiriou},\ and\ \citenamefont
  {Vega}}]{Barausse:2015frm}%
  \BibitemOpen
  \bibfield  {author} {\bibinfo {author} {\bibfnamefont {Enrico}\ \bibnamefont
  {Barausse}}, \bibinfo {author} {\bibfnamefont {Thomas~P.}\ \bibnamefont
  {Sotiriou}}, \ and\ \bibinfo {author} {\bibfnamefont {Ian}\ \bibnamefont
  {Vega}},\ }\href {\doibase10.1103/PhysRevD.93.044044} {\bibfield  {journal}
  {\bibinfo  {journal} {Phys. Rev.}\ }\textbf {\bibinfo {volume} {D93}},\
  \bibinfo {pages} {044044} (\bibinfo {year} {2016})},\ \Eprint
  {http://arxiv.org/abs/1512.05894} {arXiv:1512.05894 [gr-qc]}\BibitemShut
  {NoStop}%
\bibitem [{\citenamefont {Abramowicz}\ \emph {et~al.}(2002)\citenamefont
  {Abramowicz}, \citenamefont {Kluzniak},\ and\ \citenamefont
  {Lasota}}]{Abramowicz:2002vt}%
  \BibitemOpen
  \bibfield  {author} {\bibinfo {author} {\bibfnamefont {Marek~A.}\
  \bibnamefont {Abramowicz}}, \bibinfo {author} {\bibfnamefont {Wlodek}\
  \bibnamefont {Kluzniak}}, \ and\ \bibinfo {author} {\bibfnamefont
  {Jean-Pierre}\ \bibnamefont {Lasota}},\ }\href
  {\doibase10.1051/0004-6361:20021645} {\bibfield  {journal} {\bibinfo
  {journal} {Astron. Astrophys.}\ }\textbf {\bibinfo {volume} {396}},\ \bibinfo
  {pages} {L31--L34} (\bibinfo {year} {2002})},\ \Eprint
  {http://arxiv.org/abs/astro-ph/0207270} {arXiv:astro-ph/0207270
  [astro-ph]}\BibitemShut {NoStop}%
\bibitem [{\citenamefont {Cardoso}\ and\ \citenamefont
  {Pani}(2017{\natexlab{b}})}]{Cardoso:2017njb}%
  \BibitemOpen
  \bibfield  {author} {\bibinfo {author} {\bibfnamefont {Vitor}\ \bibnamefont
  {Cardoso}}\ and\ \bibinfo {author} {\bibfnamefont {Paolo}\ \bibnamefont
  {Pani}},\ }\href {\doibase10.1038/s41550-017-0225-y} {\bibfield  {journal}
  {\bibinfo  {journal} {Nature Astronomy}\ }\textbf {\bibinfo {volume} {1}},\
  \bibinfo {pages} {586} (\bibinfo {year} {2017}{\natexlab{b}})},\ \Eprint
  {http://arxiv.org/abs/1707.03021} {arXiv:1707.03021 [gr-qc]}\BibitemShut
  {NoStop}%
\bibitem [{\citenamefont {Mazur}\ and\ \citenamefont
  {Mottola}(2004)}]{Mazur:2004fk}%
  \BibitemOpen
  \bibfield  {author} {\bibinfo {author} {\bibfnamefont {Pawel~O.}\
  \bibnamefont {Mazur}}\ and\ \bibinfo {author} {\bibfnamefont {Emil}\
  \bibnamefont {Mottola}},\ }\href {\doibase10.1073/pnas.0402717101} {\bibfield
   {journal} {\bibinfo  {journal} {Proc. Nat. Acad. Sci.}\ }\textbf {\bibinfo
  {volume} {101}},\ \bibinfo {pages} {9545--9550} (\bibinfo {year} {2004})},\
  \Eprint {http://arxiv.org/abs/gr-qc/0407075} {arXiv:gr-qc/0407075
  [gr-qc]}\BibitemShut {NoStop}%
\bibitem [{\citenamefont {Mathur}(2005)}]{Mathur:2005zp}%
  \BibitemOpen
  \bibfield  {author} {\bibinfo {author} {\bibfnamefont {Samir~D.}\
  \bibnamefont {Mathur}},\ }\href {\doibase10.1002/prop.200410203} {\bibfield
  {journal} {\bibinfo  {journal} {Fortsch. Phys.}\ }\textbf {\bibinfo {volume}
  {53}},\ \bibinfo {pages} {793--827} (\bibinfo {year} {2005})},\ \Eprint
  {http://arxiv.org/abs/hep-th/0502050} {arXiv:hep-th/0502050
  [hep-th]}\BibitemShut {NoStop}%
\bibitem [{\citenamefont {Mathur}(2009)}]{Mathur:2008nj}%
  \BibitemOpen
  \bibfield  {author} {\bibinfo {author} {\bibfnamefont {Samir~D.}\
  \bibnamefont {Mathur}},\ }\href {\doibase10.1166/asl.2009.1021} {\bibfield
  {journal} {\bibinfo  {journal} {Adv. Sci. Lett.}\ }\textbf {\bibinfo {volume}
  {2}},\ \bibinfo {pages} {133--150} (\bibinfo {year} {2009})},\ \Eprint
  {http://arxiv.org/abs/0810.4525} {arXiv:0810.4525 [hep-th]}\BibitemShut
  {NoStop}%
\bibitem [{\citenamefont {Barcel\'o}\ \emph {et~al.}(2016)\citenamefont
  {Barcel\'o}, \citenamefont {Carballo-Rubio},\ and\ \citenamefont
  {Garay}}]{Barcelo:2015noa}%
  \BibitemOpen
  \bibfield  {author} {\bibinfo {author} {\bibfnamefont {Carlos}\ \bibnamefont
  {Barcel\'o}}, \bibinfo {author} {\bibfnamefont {Ra\'ul}\ \bibnamefont
  {Carballo-Rubio}}, \ and\ \bibinfo {author} {\bibfnamefont {Luis~J.}\
  \bibnamefont {Garay}},\ }\href {\doibase10.3390/universe2020007} {\bibfield
  {journal} {\bibinfo  {journal} {Universe}\ }\textbf {\bibinfo {volume} {2}},\
  \bibinfo {pages} {7} (\bibinfo {year} {2016})},\ \Eprint
  {http://arxiv.org/abs/1510.04957} {arXiv:1510.04957 [gr-qc]}\BibitemShut
  {NoStop}%
\bibitem [{\citenamefont {Danielsson}\ \emph {et~al.}(2017)\citenamefont
  {Danielsson}, \citenamefont {Dibitetto},\ and\ \citenamefont
  {Giri}}]{Danielsson:2017riq}%
  \BibitemOpen
  \bibfield  {author} {\bibinfo {author} {\bibfnamefont {U.~H.}\ \bibnamefont
  {Danielsson}}, \bibinfo {author} {\bibfnamefont {G.}~\bibnamefont
  {Dibitetto}}, \ and\ \bibinfo {author} {\bibfnamefont {S.}~\bibnamefont
  {Giri}},\ }\href {\doibase10.1007/JHEP10(2017)171} {\bibfield  {journal}
  {\bibinfo  {journal} {JHEP}\ }\textbf {\bibinfo {volume} {10}},\ \bibinfo
  {pages} {171} (\bibinfo {year} {2017})},\ \Eprint
  {http://arxiv.org/abs/1705.10172} {arXiv:1705.10172 [hep-th]}\BibitemShut
  {NoStop}%
\bibitem [{\citenamefont {Danielsson}\ and\ \citenamefont
  {Giri}(2018)}]{Danielsson:2017pvl}%
  \BibitemOpen
  \bibfield  {author} {\bibinfo {author} {\bibfnamefont {Ulf}\ \bibnamefont
  {Danielsson}}\ and\ \bibinfo {author} {\bibfnamefont {Suvendu}\ \bibnamefont
  {Giri}},\ }\href {\doibase10.1007/JHEP07(2018)070} {\bibfield  {journal}
  {\bibinfo  {journal} {JHEP}\ }\textbf {\bibinfo {volume} {07}},\ \bibinfo
  {pages} {070} (\bibinfo {year} {2018})},\ \Eprint
  {http://arxiv.org/abs/1712.00511} {arXiv:1712.00511 [hep-th]}\BibitemShut
  {NoStop}%
\bibitem [{\citenamefont {Berthiere}\ \emph {et~al.}(2017)\citenamefont
  {Berthiere}, \citenamefont {Sarkar},\ and\ \citenamefont
  {Solodukhin}}]{Berthiere:2017tms}%
  \BibitemOpen
  \bibfield  {author} {\bibinfo {author} {\bibfnamefont {Clement}\ \bibnamefont
  {Berthiere}}, \bibinfo {author} {\bibfnamefont {Debajyoti}\ \bibnamefont
  {Sarkar}}, \ and\ \bibinfo {author} {\bibfnamefont {Sergey~N.}\ \bibnamefont
  {Solodukhin}},\ }\href@noop {} {\  (\bibinfo {year} {2017})},\ \Eprint
  {http://arxiv.org/abs/1712.09914} {arXiv:1712.09914 [hep-th]}\BibitemShut
  {NoStop}%
\bibitem [{\citenamefont {Giddings}(2013)}]{Giddings:2013kcj}%
  \BibitemOpen
  \bibfield  {author} {\bibinfo {author} {\bibfnamefont {Steven~B.}\
  \bibnamefont {Giddings}},\ }\href {\doibase10.1103/PhysRevD.88.024018}
  {\bibfield  {journal} {\bibinfo  {journal} {Phys. Rev.}\ }\textbf {\bibinfo
  {volume} {D88}},\ \bibinfo {pages} {024018} (\bibinfo {year} {2013})},\
  \Eprint {http://arxiv.org/abs/1302.2613} {arXiv:1302.2613
  [hep-th]}\BibitemShut {NoStop}%
\bibitem [{\citenamefont {Giddings}(2014)}]{Giddings:2014nla}%
  \BibitemOpen
  \bibfield  {author} {\bibinfo {author} {\bibfnamefont {Steven~B.}\
  \bibnamefont {Giddings}},\ }\href {\doibase10.1016/j.physletb.2014.08.070}
  {\bibfield  {journal} {\bibinfo  {journal} {Phys. Lett.}\ }\textbf {\bibinfo
  {volume} {B738}},\ \bibinfo {pages} {92--96} (\bibinfo {year} {2014})},\
  \Eprint {http://arxiv.org/abs/1401.5804} {arXiv:1401.5804
  [hep-th]}\BibitemShut {NoStop}%
\bibitem [{\citenamefont {Giddings}(2017)}]{Giddings:2017mym}%
  \BibitemOpen
  \bibfield  {author} {\bibinfo {author} {\bibfnamefont {Steven~B.}\
  \bibnamefont {Giddings}},\ }\href {\doibase10.1007/JHEP12(2017)047}
  {\bibfield  {journal} {\bibinfo  {journal} {JHEP}\ }\textbf {\bibinfo
  {volume} {12}},\ \bibinfo {pages} {047} (\bibinfo {year} {2017})},\ \Eprint
  {http://arxiv.org/abs/1701.08765} {arXiv:1701.08765 [hep-th]}\BibitemShut
  {NoStop}%
\bibitem [{\citenamefont {Feinblum}\ and\ \citenamefont
  {McKinley}(1968)}]{feinblum68}%
  \BibitemOpen
  \bibfield  {author} {\bibinfo {author} {\bibfnamefont {David~A.}\
  \bibnamefont {Feinblum}}\ and\ \bibinfo {author} {\bibfnamefont {William~A.}\
  \bibnamefont {McKinley}},\ }\href {\doibase10.1103/PhysRev.168.1445}
  {\bibfield  {journal} {\bibinfo  {journal} {Phys. Rev.}\ }\textbf {\bibinfo
  {volume} {168}},\ \bibinfo {pages} {1445--1450} (\bibinfo {year}
  {1968})}\BibitemShut {NoStop}%
\bibitem [{\citenamefont {Ruffini}\ and\ \citenamefont
  {Bonazzola}(1969)}]{ruffini69}%
  \BibitemOpen
  \bibfield  {author} {\bibinfo {author} {\bibfnamefont {R.}~\bibnamefont
  {Ruffini}}\ and\ \bibinfo {author} {\bibfnamefont {S.}~\bibnamefont
  {Bonazzola}},\ }\href {\doibase10.1103/PhysRev.187.1767} {\bibfield
  {journal} {\bibinfo  {journal} {Phys. Rev.}\ }\textbf {\bibinfo {volume}
  {187}},\ \bibinfo {pages} {1767--1783} (\bibinfo {year} {1969})}\BibitemShut
  {NoStop}%
\bibitem [{\citenamefont {{Liebling}}\ and\ \citenamefont
  {{Palenzuela}}(2017)}]{liebling17}%
  \BibitemOpen
  \bibfield  {author} {\bibinfo {author} {\bibfnamefont {S.~L.}\ \bibnamefont
  {{Liebling}}}\ and\ \bibinfo {author} {\bibfnamefont {C.}~\bibnamefont
  {{Palenzuela}}},\ }\href {\doibase10.1007/s41114-017-0007-y} {\bibfield
  {journal} {\bibinfo  {journal} {Living Reviews in Relativity}\ }\textbf
  {\bibinfo {volume} {20}},\ \bibinfo {eid} {5} (\bibinfo {year}
  {2017})}\BibitemShut {NoStop}%
\bibitem [{\citenamefont {Mazur}\ and\ \citenamefont
  {Mottola}(2001)}]{Mazur:2001fv}%
  \BibitemOpen
  \bibfield  {author} {\bibinfo {author} {\bibfnamefont {Pawel~O.}\
  \bibnamefont {Mazur}}\ and\ \bibinfo {author} {\bibfnamefont {Emil}\
  \bibnamefont {Mottola}},\ }\href@noop {} {\  (\bibinfo {year} {2001})},\
  \Eprint {http://arxiv.org/abs/gr-qc/0109035} {arXiv:gr-qc/0109035
  [gr-qc]}\BibitemShut {NoStop}%
\bibitem [{\citenamefont {Cattoen}\ \emph {et~al.}(2005)\citenamefont
  {Cattoen}, \citenamefont {Faber},\ and\ \citenamefont
  {Visser}}]{Cattoen:2005he}%
  \BibitemOpen
  \bibfield  {author} {\bibinfo {author} {\bibfnamefont {Celine}\ \bibnamefont
  {Cattoen}}, \bibinfo {author} {\bibfnamefont {Tristan}\ \bibnamefont
  {Faber}}, \ and\ \bibinfo {author} {\bibfnamefont {Matt}\ \bibnamefont
  {Visser}},\ }\href {\doibase10.1088/0264-9381/22/20/002} {\bibfield
  {journal} {\bibinfo  {journal} {Class. Quant. Grav.}\ }\textbf {\bibinfo
  {volume} {22}},\ \bibinfo {pages} {4189--4202} (\bibinfo {year} {2005})},\
  \Eprint {http://arxiv.org/abs/gr-qc/0505137} {arXiv:gr-qc/0505137
  [gr-qc]}\BibitemShut {NoStop}%
\bibitem [{\citenamefont {Chirenti}\ and\ \citenamefont
  {Rezzolla}(2007)}]{Chirenti:2007mk}%
  \BibitemOpen
  \bibfield  {author} {\bibinfo {author} {\bibfnamefont {Cecilia B. M.~H.}\
  \bibnamefont {Chirenti}}\ and\ \bibinfo {author} {\bibfnamefont {Luciano}\
  \bibnamefont {Rezzolla}},\ }\href {\doibase10.1088/0264-9381/24/16/013}
  {\bibfield  {journal} {\bibinfo  {journal} {Class. Quant. Grav.}\ }\textbf
  {\bibinfo {volume} {24}},\ \bibinfo {pages} {4191--4206} (\bibinfo {year}
  {2007})},\ \Eprint {http://arxiv.org/abs/0706.1513} {arXiv:0706.1513
  [gr-qc]}\BibitemShut {NoStop}%
\bibitem [{\citenamefont {Mottola}\ and\ \citenamefont
  {Vaulin}(2006)}]{Mottola:2006ew}%
  \BibitemOpen
  \bibfield  {author} {\bibinfo {author} {\bibfnamefont {Emil}\ \bibnamefont
  {Mottola}}\ and\ \bibinfo {author} {\bibfnamefont {Ruslan}\ \bibnamefont
  {Vaulin}},\ }\href {\doibase10.1103/PhysRevD.74.064004} {\bibfield  {journal}
  {\bibinfo  {journal} {Phys. Rev.}\ }\textbf {\bibinfo {volume} {D74}},\
  \bibinfo {pages} {064004} (\bibinfo {year} {2006})},\ \Eprint
  {http://arxiv.org/abs/gr-qc/0604051} {arXiv:gr-qc/0604051
  [gr-qc]}\BibitemShut {NoStop}%
\bibitem [{\citenamefont {Barceló}\ \emph {et~al.}(2009)\citenamefont
  {Barceló}, \citenamefont {Liberati}, \citenamefont {Sonego},\ and\
  \citenamefont {Visser}}]{Barcelo:2009tpa}%
  \BibitemOpen
  \bibfield  {author} {\bibinfo {author} {\bibfnamefont {Carlos}\ \bibnamefont
  {Barceló}}, \bibinfo {author} {\bibfnamefont {Stefano}\ \bibnamefont
  {Liberati}}, \bibinfo {author} {\bibfnamefont {Sebastiano}\ \bibnamefont
  {Sonego}}, \ and\ \bibinfo {author} {\bibfnamefont {Matt}\ \bibnamefont
  {Visser}},\ }\href {\doibase10.1038/scientificamerican1009-38} {\bibfield
  {journal} {\bibinfo  {journal} {Sci. Am.}\ }\textbf {\bibinfo {volume}
  {301}},\ \bibinfo {pages} {38--45} (\bibinfo {year} {2009})}\BibitemShut
  {NoStop}%
\bibitem [{\citenamefont {Gimon}\ and\ \citenamefont
  {Horava}(2009)}]{Gimon:2007ur}%
  \BibitemOpen
  \bibfield  {author} {\bibinfo {author} {\bibfnamefont {Eric~G.}\ \bibnamefont
  {Gimon}}\ and\ \bibinfo {author} {\bibfnamefont {Petr}\ \bibnamefont
  {Horava}},\ }\href {\doibase10.1016/j.physletb.2009.01.026} {\bibfield
  {journal} {\bibinfo  {journal} {Phys. Lett.}\ }\textbf {\bibinfo {volume}
  {B672}},\ \bibinfo {pages} {299--302} (\bibinfo {year} {2009})},\ \Eprint
  {http://arxiv.org/abs/0706.2873} {arXiv:0706.2873 [hep-th]}\BibitemShut
  {NoStop}%
\bibitem [{\citenamefont {Brustein}\ and\ \citenamefont
  {Medved}(2017)}]{Brustein:2016msz}%
  \BibitemOpen
  \bibfield  {author} {\bibinfo {author} {\bibfnamefont {Ram}\ \bibnamefont
  {Brustein}}\ and\ \bibinfo {author} {\bibfnamefont {A.~J.~M.}\ \bibnamefont
  {Medved}},\ }\href {\doibase10.1002/prop.201600114} {\bibfield  {journal}
  {\bibinfo  {journal} {Fortsch. Phys.}\ }\textbf {\bibinfo {volume} {65}},\
  \bibinfo {pages} {0114} (\bibinfo {year} {2017})},\ \Eprint
  {http://arxiv.org/abs/1602.07706} {arXiv:1602.07706 [hep-th]}\BibitemShut
  {NoStop}%
\bibitem [{\citenamefont {Brustein}\ \emph
  {et~al.}(2017{\natexlab{b}})\citenamefont {Brustein}, \citenamefont
  {Medved},\ and\ \citenamefont {Yagi}}]{Brustein:2017kcj}%
  \BibitemOpen
  \bibfield  {author} {\bibinfo {author} {\bibfnamefont {Ram}\ \bibnamefont
  {Brustein}}, \bibinfo {author} {\bibfnamefont {A.~J.~M.}\ \bibnamefont
  {Medved}}, \ and\ \bibinfo {author} {\bibfnamefont {K.}~\bibnamefont
  {Yagi}},\ }\href {\doibase10.1103/PhysRevD.96.124021} {\bibfield  {journal}
  {\bibinfo  {journal} {Phys. Rev.}\ }\textbf {\bibinfo {volume} {D96}},\
  \bibinfo {pages} {124021} (\bibinfo {year} {2017}{\natexlab{b}})},\ \Eprint
  {http://arxiv.org/abs/1701.07444} {arXiv:1701.07444 [gr-qc]}\BibitemShut
  {NoStop}%
\bibitem [{\citenamefont {Holdom}\ and\ \citenamefont
  {Ren}(2017)}]{Holdom:2016nek}%
  \BibitemOpen
  \bibfield  {author} {\bibinfo {author} {\bibfnamefont {Bob}\ \bibnamefont
  {Holdom}}\ and\ \bibinfo {author} {\bibfnamefont {Jing}\ \bibnamefont
  {Ren}},\ }\href {\doibase10.1103/PhysRevD.95.084034} {\bibfield  {journal}
  {\bibinfo  {journal} {Phys. Rev.}\ }\textbf {\bibinfo {volume} {D95}},\
  \bibinfo {pages} {084034} (\bibinfo {year} {2017})},\ \Eprint
  {http://arxiv.org/abs/1612.04889} {arXiv:1612.04889 [gr-qc]}\BibitemShut
  {NoStop}%
\bibitem [{\citenamefont {Friedman}(1978)}]{Friedman:1978wla}%
  \BibitemOpen
  \bibfield  {author} {\bibinfo {author} {\bibfnamefont {John~L.}\ \bibnamefont
  {Friedman}},\ }\href {\doibase10.1007/BF01202527} {\bibfield  {journal}
  {\bibinfo  {journal} {Commun. Math. Phys.}\ }\textbf {\bibinfo {volume}
  {62}},\ \bibinfo {pages} {247--278} (\bibinfo {year} {1978})}\BibitemShut
  {NoStop}%
\bibitem [{\citenamefont {{Comins}}\ and\ \citenamefont
  {{Schutz}}(1978)}]{CominsSchutz}%
  \BibitemOpen
  \bibfield  {author} {\bibinfo {author} {\bibfnamefont {N.}~\bibnamefont
  {{Comins}}}\ and\ \bibinfo {author} {\bibfnamefont {B.~F.}\ \bibnamefont
  {{Schutz}}},\ }\href {\doibase10.1098/rspa.1978.0196} {\bibfield  {journal}
  {\bibinfo  {journal} {Proceedings of the Royal Society of London Series A}\
  }\textbf {\bibinfo {volume} {364}},\ \bibinfo {pages} {211--226} (\bibinfo
  {year} {1978})}\BibitemShut {NoStop}%
\bibitem [{\citenamefont {{Yoshida}}\ and\ \citenamefont
  {{Eriguchi}}(1996)}]{YoshidaEriguchi}%
  \BibitemOpen
  \bibfield  {author} {\bibinfo {author} {\bibfnamefont {S.}~\bibnamefont
  {{Yoshida}}}\ and\ \bibinfo {author} {\bibfnamefont {Y.}~\bibnamefont
  {{Eriguchi}}},\ }\href {\doibase10.1093/mnras/282.2.580} {\bibfield
  {journal} {\bibinfo  {journal} {Mon. Not. Roy. Astron. Soc.}\ }\textbf
  {\bibinfo {volume} {282}},\ \bibinfo {pages} {580--586} (\bibinfo {year}
  {1996})}\BibitemShut {NoStop}%
\bibitem [{\citenamefont {{Kokkotas}}\ \emph {et~al.}(2004)\citenamefont
  {{Kokkotas}}, \citenamefont {{Ruoff}},\ and\ \citenamefont
  {{Andersson}}}]{KokkotasRuoffAndersson}%
  \BibitemOpen
  \bibfield  {author} {\bibinfo {author} {\bibfnamefont {K.~D.}\ \bibnamefont
  {{Kokkotas}}}, \bibinfo {author} {\bibfnamefont {J.}~\bibnamefont {{Ruoff}}},
  \ and\ \bibinfo {author} {\bibfnamefont {N.}~\bibnamefont {{Andersson}}},\
  }\href {\doibase10.1103/PhysRevD.70.043003} {\bibfield  {journal} {\bibinfo
  {journal} {Physical Review D}\ }\textbf {\bibinfo {volume} {70}},\ \bibinfo
  {eid} {043003} (\bibinfo {year} {2004})},\ \Eprint
  {http://arxiv.org/abs/astro-ph/0212429} {astro-ph/0212429}\BibitemShut
  {NoStop}%
\bibitem [{\citenamefont {Moschidis}(2018)}]{Moschidis:2016zjy}%
  \BibitemOpen
  \bibfield  {author} {\bibinfo {author} {\bibfnamefont {Georgios}\
  \bibnamefont {Moschidis}},\ }\href {\doibase10.1007/s00220-017-3010-y}
  {\bibfield  {journal} {\bibinfo  {journal} {Commun. Math. Phys.}\ }\textbf
  {\bibinfo {volume} {358}},\ \bibinfo {pages} {437--520} (\bibinfo {year}
  {2018})},\ \Eprint {http://arxiv.org/abs/1608.02035} {arXiv:1608.02035
  [math.AP]}\BibitemShut {NoStop}%
\bibitem [{\citenamefont {Cardoso}\ \emph {et~al.}(2008)\citenamefont
  {Cardoso}, \citenamefont {Pani}, \citenamefont {Cadoni},\ and\ \citenamefont
  {Cavaglia}}]{Cardoso:2007az}%
  \BibitemOpen
  \bibfield  {author} {\bibinfo {author} {\bibfnamefont {Vitor}\ \bibnamefont
  {Cardoso}}, \bibinfo {author} {\bibfnamefont {Paolo}\ \bibnamefont {Pani}},
  \bibinfo {author} {\bibfnamefont {Mariano}\ \bibnamefont {Cadoni}}, \ and\
  \bibinfo {author} {\bibfnamefont {Marco}\ \bibnamefont {Cavaglia}},\ }\href
  {\doibase10.1103/PhysRevD.77.124044} {\bibfield  {journal} {\bibinfo
  {journal} {Phys. Rev.}\ }\textbf {\bibinfo {volume} {D77}},\ \bibinfo {pages}
  {124044} (\bibinfo {year} {2008})},\ \Eprint {http://arxiv.org/abs/0709.0532}
  {arXiv:0709.0532 [gr-qc]}\BibitemShut {NoStop}%
\bibitem [{\citenamefont {Chirenti}\ and\ \citenamefont
  {Rezzolla}(2008)}]{Chirenti:2008pf}%
  \BibitemOpen
  \bibfield  {author} {\bibinfo {author} {\bibfnamefont {Cecilia B. M.~H.}\
  \bibnamefont {Chirenti}}\ and\ \bibinfo {author} {\bibfnamefont {Luciano}\
  \bibnamefont {Rezzolla}},\ }\href {\doibase10.1103/PhysRevD.78.084011}
  {\bibfield  {journal} {\bibinfo  {journal} {Phys. Rev.}\ }\textbf {\bibinfo
  {volume} {D78}},\ \bibinfo {pages} {084011} (\bibinfo {year} {2008})},\
  \Eprint {http://arxiv.org/abs/0808.4080} {arXiv:0808.4080
  [gr-qc]}\BibitemShut {NoStop}%
\bibitem [{\citenamefont {Cardoso}\ \emph
  {et~al.}(2014{\natexlab{b}})\citenamefont {Cardoso}, \citenamefont
  {Crispino}, \citenamefont {Macedo}, \citenamefont {Okawa},\ and\
  \citenamefont {Pani}}]{Cardoso:2014sna}%
  \BibitemOpen
  \bibfield  {author} {\bibinfo {author} {\bibfnamefont {Vitor}\ \bibnamefont
  {Cardoso}}, \bibinfo {author} {\bibfnamefont {Luís C.~B.}\ \bibnamefont
  {Crispino}}, \bibinfo {author} {\bibfnamefont {Caio F.~B.}\ \bibnamefont
  {Macedo}}, \bibinfo {author} {\bibfnamefont {Hirotada}\ \bibnamefont
  {Okawa}}, \ and\ \bibinfo {author} {\bibfnamefont {Paolo}\ \bibnamefont
  {Pani}},\ }\href {\doibase10.1103/PhysRevD.90.044069} {\bibfield  {journal}
  {\bibinfo  {journal} {Phys. Rev.}\ }\textbf {\bibinfo {volume} {D90}},\
  \bibinfo {pages} {044069} (\bibinfo {year} {2014}{\natexlab{b}})},\ \Eprint
  {http://arxiv.org/abs/1406.5510} {arXiv:1406.5510 [gr-qc]}\BibitemShut
  {NoStop}%
\bibitem [{\citenamefont {Maggio}\ \emph {et~al.}(2017)\citenamefont {Maggio},
  \citenamefont {Pani},\ and\ \citenamefont {Ferrari}}]{Maggio:2017ivp}%
  \BibitemOpen
  \bibfield  {author} {\bibinfo {author} {\bibfnamefont {Elisa}\ \bibnamefont
  {Maggio}}, \bibinfo {author} {\bibfnamefont {Paolo}\ \bibnamefont {Pani}}, \
  and\ \bibinfo {author} {\bibfnamefont {Valeria}\ \bibnamefont {Ferrari}},\
  }\href {\doibase10.1103/PhysRevD.96.104047} {\bibfield  {journal} {\bibinfo
  {journal} {Phys. Rev.}\ }\textbf {\bibinfo {volume} {D96}},\ \bibinfo {pages}
  {104047} (\bibinfo {year} {2017})},\ \Eprint
  {http://arxiv.org/abs/1703.03696} {arXiv:1703.03696 [gr-qc]}\BibitemShut
  {NoStop}%
\bibitem [{\citenamefont {{Bardeen}}(1973)}]{Bardeen1973}%
  \BibitemOpen
  \bibfield  {author} {\bibinfo {author} {\bibfnamefont {J.~M.}\ \bibnamefont
  {{Bardeen}}},\ }in\ \href@noop {} {\emph {\bibinfo {booktitle} {Black Holes
  (Les Astres Occlus)}}},\ \bibinfo {editor} {edited by\ \bibinfo {editor}
  {\bibfnamefont {C.}~\bibnamefont {{Dewitt}}}\ and\ \bibinfo {editor}
  {\bibfnamefont {B.~S.}\ \bibnamefont {{Dewitt}}}}\ (\bibinfo {year} {1973})\
  pp.\ \bibinfo {pages} {215--239}\BibitemShut {NoStop}%
\bibitem [{\citenamefont {{Falcke}}\ \emph {et~al.}(2000)\citenamefont
  {{Falcke}}, \citenamefont {{Melia}},\ and\ \citenamefont
  {{Agol}}}]{FalckeMeliaAgol2000}%
  \BibitemOpen
  \bibfield  {author} {\bibinfo {author} {\bibfnamefont {H.}~\bibnamefont
  {{Falcke}}}, \bibinfo {author} {\bibfnamefont {F.}~\bibnamefont {{Melia}}}, \
  and\ \bibinfo {author} {\bibfnamefont {E.}~\bibnamefont {{Agol}}},\ }\href
  {\doibase10.1086/312423} {\bibfield  {journal} {\bibinfo  {journal} {The
  Astrophysical Journal}\ }\textbf {\bibinfo {volume} {528}},\ \bibinfo {pages}
  {L13--L16} (\bibinfo {year} {2000})},\ \Eprint
  {http://arxiv.org/abs/astro-ph/9912263} {astro-ph/9912263}\BibitemShut
  {NoStop}%
\bibitem [{\citenamefont {Doeleman}\ \emph {et~al.}(2009)\citenamefont
  {Doeleman} \emph {et~al.}}]{Doeleman:2009te}%
  \BibitemOpen
  \bibfield  {author} {\bibinfo {author} {\bibfnamefont {Sheperd}\ \bibnamefont
  {Doeleman}} \emph {et~al.},\ }\href@noop {} {\  (\bibinfo {year} {2009})},\
  \Eprint {http://arxiv.org/abs/0906.3899} {arXiv:0906.3899
  [astro-ph.CO]}\BibitemShut {NoStop}%
\bibitem [{\citenamefont {Cunha}\ and\ \citenamefont
  {Herdeiro}(2018)}]{Cunha:2018acu}%
  \BibitemOpen
  \bibfield  {author} {\bibinfo {author} {\bibfnamefont {Pedro V.~P.}\
  \bibnamefont {Cunha}}\ and\ \bibinfo {author} {\bibfnamefont {Carlos A.~R.}\
  \bibnamefont {Herdeiro}},\ }\href {\doibase10.1007/s10714-018-2361-9}
  {\bibfield  {journal} {\bibinfo  {journal} {Gen. Rel. Grav.}\ }\textbf
  {\bibinfo {volume} {50}},\ \bibinfo {pages} {42} (\bibinfo {year} {2018})},\
  \Eprint {http://arxiv.org/abs/1801.00860} {arXiv:1801.00860
  [gr-qc]}\BibitemShut {NoStop}%
\bibitem [{\citenamefont {{Grenzebach}}\ \emph {et~al.}(2014)\citenamefont
  {{Grenzebach}}, \citenamefont {{Perlick}},\ and\ \citenamefont
  {{L{\"a}mmerzahl}}}]{grenzebach14}%
  \BibitemOpen
  \bibfield  {author} {\bibinfo {author} {\bibfnamefont {A.}~\bibnamefont
  {{Grenzebach}}}, \bibinfo {author} {\bibfnamefont {V.}~\bibnamefont
  {{Perlick}}}, \ and\ \bibinfo {author} {\bibfnamefont {C.}~\bibnamefont
  {{L{\"a}mmerzahl}}},\ }\href {\doibase10.1103/PhysRevD.89.124004} {\bibfield
  {journal} {\bibinfo  {journal} {Physical Review D}\ }\textbf {\bibinfo
  {volume} {89}},\ \bibinfo {eid} {124004} (\bibinfo {year} {2014})},\ \Eprint
  {http://arxiv.org/abs/1403.5234} {arXiv:1403.5234 [gr-qc]}\BibitemShut
  {NoStop}%
\bibitem [{\citenamefont {Ghasemi-Nodehi}\ \emph {et~al.}(2015)\citenamefont
  {Ghasemi-Nodehi}, \citenamefont {Li},\ and\ \citenamefont
  {Bambi}}]{Ghasemi-Nodehi:2015raa}%
  \BibitemOpen
  \bibfield  {author} {\bibinfo {author} {\bibfnamefont {Masoumeh}\
  \bibnamefont {Ghasemi-Nodehi}}, \bibinfo {author} {\bibfnamefont {Zilong}\
  \bibnamefont {Li}}, \ and\ \bibinfo {author} {\bibfnamefont {Cosimo}\
  \bibnamefont {Bambi}},\ }\href {\doibase10.1140/epjc/s10052-015-3539-x}
  {\bibfield  {journal} {\bibinfo  {journal} {Eur. Phys. J.}\ }\textbf
  {\bibinfo {volume} {C75}},\ \bibinfo {pages} {315} (\bibinfo {year}
  {2015})},\ \Eprint {http://arxiv.org/abs/1506.02627} {arXiv:1506.02627
  [gr-qc]}\BibitemShut {NoStop}%
\bibitem [{\citenamefont {{Wang}}\ \emph {et~al.}(2017)\citenamefont {{Wang}},
  \citenamefont {{Chen}},\ and\ \citenamefont {{Jing}}}]{wang17}%
  \BibitemOpen
  \bibfield  {author} {\bibinfo {author} {\bibfnamefont {M.}~\bibnamefont
  {{Wang}}}, \bibinfo {author} {\bibfnamefont {S.}~\bibnamefont {{Chen}}}, \
  and\ \bibinfo {author} {\bibfnamefont {J.}~\bibnamefont {{Jing}}},\ }\href
  {\doibase10.1088/1475-7516/2017/10/051} {\bibfield  {journal} {\bibinfo
  {journal} {JCAP}\ }\textbf {\bibinfo {volume} {10}},\ \bibinfo {eid} {051}
  (\bibinfo {year} {2017})},\ \Eprint {http://arxiv.org/abs/1707.09451}
  {arXiv:1707.09451 [gr-qc]}\BibitemShut {NoStop}%
\bibitem [{\citenamefont {{Amarilla}}\ and\ \citenamefont
  {{Eiroa}}(2012)}]{amarilla12}%
  \BibitemOpen
  \bibfield  {author} {\bibinfo {author} {\bibfnamefont {L.}~\bibnamefont
  {{Amarilla}}}\ and\ \bibinfo {author} {\bibfnamefont {E.~F.}\ \bibnamefont
  {{Eiroa}}},\ }\href {\doibase10.1103/PhysRevD.85.064019} {\bibfield
  {journal} {\bibinfo  {journal} {Physical Review D}\ }\textbf {\bibinfo
  {volume} {85}},\ \bibinfo {eid} {064019} (\bibinfo {year} {2012})},\ \Eprint
  {http://arxiv.org/abs/1112.6349} {arXiv:1112.6349 [gr-qc]}\BibitemShut
  {NoStop}%
\bibitem [{\citenamefont {{Amarilla}}\ and\ \citenamefont
  {{Eiroa}}(2013)}]{amarilla13}%
  \BibitemOpen
  \bibfield  {author} {\bibinfo {author} {\bibfnamefont {L.}~\bibnamefont
  {{Amarilla}}}\ and\ \bibinfo {author} {\bibfnamefont {E.~F.}\ \bibnamefont
  {{Eiroa}}},\ }\href {\doibase10.1103/PhysRevD.87.044057} {\bibfield
  {journal} {\bibinfo  {journal} {Physical Review D}\ }\textbf {\bibinfo
  {volume} {87}},\ \bibinfo {eid} {044057} (\bibinfo {year} {2013})},\ \Eprint
  {http://arxiv.org/abs/1301.0532} {arXiv:1301.0532 [gr-qc]}\BibitemShut
  {NoStop}%
\bibitem [{\citenamefont {{Wei}}\ and\ \citenamefont {{Liu}}(2013)}]{wei13}%
  \BibitemOpen
  \bibfield  {author} {\bibinfo {author} {\bibfnamefont {S.-W.}\ \bibnamefont
  {{Wei}}}\ and\ \bibinfo {author} {\bibfnamefont {Y.-X.}\ \bibnamefont
  {{Liu}}},\ }\href {\doibase10.1088/1475-7516/2013/11/063} {\bibfield
  {journal} {\bibinfo  {journal} {JCAP}\ }\textbf {\bibinfo {volume} {11}},\
  \bibinfo {eid} {063} (\bibinfo {year} {2013})},\ \Eprint
  {http://arxiv.org/abs/1311.4251} {arXiv:1311.4251 [gr-qc]}\BibitemShut
  {NoStop}%
\bibitem [{\citenamefont {{Vincent}}(2014)}]{vincent14}%
  \BibitemOpen
  \bibfield  {author} {\bibinfo {author} {\bibfnamefont {F.~H.}\ \bibnamefont
  {{Vincent}}},\ }\href {\doibase10.1088/0264-9381/31/2/025010} {\bibfield
  {journal} {\bibinfo  {journal} {Classical and Quantum Gravity}\ }\textbf
  {\bibinfo {volume} {31}},\ \bibinfo {eid} {025010} (\bibinfo {year}
  {2014})},\ \Eprint {http://arxiv.org/abs/1311.3251} {arXiv:1311.3251
  [astro-ph.HE]}\BibitemShut {NoStop}%
\bibitem [{\citenamefont {{Tinchev}}\ and\ \citenamefont
  {{Yazadjiev}}(2014)}]{tinchev14}%
  \BibitemOpen
  \bibfield  {author} {\bibinfo {author} {\bibfnamefont {V.~K.}\ \bibnamefont
  {{Tinchev}}}\ and\ \bibinfo {author} {\bibfnamefont {S.~S.}\ \bibnamefont
  {{Yazadjiev}}},\ }\href {\doibase10.1142/S0218271814500606} {\bibfield
  {journal} {\bibinfo  {journal} {International Journal of Modern Physics D}\
  }\textbf {\bibinfo {volume} {23}},\ \bibinfo {eid} {1450060} (\bibinfo {year}
  {2014})},\ \Eprint {http://arxiv.org/abs/1311.1353} {arXiv:1311.1353
  [gr-qc]}\BibitemShut {NoStop}%
\bibitem [{\citenamefont {{Moffat}}(2015)}]{moffat15}%
  \BibitemOpen
  \bibfield  {author} {\bibinfo {author} {\bibfnamefont {J.~W.}\ \bibnamefont
  {{Moffat}}},\ }\href {\doibase10.1140/epjc/s10052-015-3352-6} {\bibfield
  {journal} {\bibinfo  {journal} {European Physical Journal C}\ }\textbf
  {\bibinfo {volume} {75}},\ \bibinfo {eid} {130} (\bibinfo {year} {2015})},\
  \Eprint {http://arxiv.org/abs/1502.01677} {arXiv:1502.01677
  [gr-qc]}\BibitemShut {NoStop}%
\bibitem [{\citenamefont {{Schee}}\ and\ \citenamefont
  {{Stuchl{\'{\i}}k}}(2016)}]{schee16}%
  \BibitemOpen
  \bibfield  {author} {\bibinfo {author} {\bibfnamefont {J.}~\bibnamefont
  {{Schee}}}\ and\ \bibinfo {author} {\bibfnamefont {Z.}~\bibnamefont
  {{Stuchl{\'{\i}}k}}},\ }\href {\doibase10.1140/epjc/s10052-016-4511-0}
  {\bibfield  {journal} {\bibinfo  {journal} {European Physical Journal C}\
  }\textbf {\bibinfo {volume} {76}},\ \bibinfo {eid} {643} (\bibinfo {year}
  {2016})},\ \Eprint {http://arxiv.org/abs/1606.09037} {arXiv:1606.09037
  [astro-ph.HE]}\BibitemShut {NoStop}%
\bibitem [{\citenamefont {{Cunha}}\ \emph {et~al.}(2017)\citenamefont
  {{Cunha}}, \citenamefont {{Herdeiro}}, \citenamefont {{Kleihaus}},
  \citenamefont {{Kunz}},\ and\ \citenamefont {{Radu}}}]{cunha17}%
  \BibitemOpen
  \bibfield  {author} {\bibinfo {author} {\bibfnamefont {P.~V.~P.}\
  \bibnamefont {{Cunha}}}, \bibinfo {author} {\bibfnamefont {C.~A.~R.}\
  \bibnamefont {{Herdeiro}}}, \bibinfo {author} {\bibfnamefont
  {B.}~\bibnamefont {{Kleihaus}}}, \bibinfo {author} {\bibfnamefont
  {J.}~\bibnamefont {{Kunz}}}, \ and\ \bibinfo {author} {\bibfnamefont
  {E.}~\bibnamefont {{Radu}}},\ }\href {\doibase10.1016/j.physletb.2017.03.020}
  {\bibfield  {journal} {\bibinfo  {journal} {Physics Letters B}\ }\textbf
  {\bibinfo {volume} {768}},\ \bibinfo {pages} {373--379} (\bibinfo {year}
  {2017})},\ \Eprint {http://arxiv.org/abs/1701.00079} {arXiv:1701.00079
  [gr-qc]}\BibitemShut {NoStop}%
\bibitem [{\citenamefont {Hioki}\ and\ \citenamefont {Maeda}(2009)}]{hioki09}%
  \BibitemOpen
  \bibfield  {author} {\bibinfo {author} {\bibfnamefont {Kenta}\ \bibnamefont
  {Hioki}}\ and\ \bibinfo {author} {\bibfnamefont {Kei-ichi}\ \bibnamefont
  {Maeda}},\ }\href {\doibase10.1103/PhysRevD.80.024042} {\bibfield  {journal}
  {\bibinfo  {journal} {Phys. Rev. D}\ }\textbf {\bibinfo {volume} {80}},\
  \bibinfo {pages} {024042} (\bibinfo {year} {2009})}\BibitemShut {NoStop}%
\bibitem [{\citenamefont {{Nedkova}}\ \emph {et~al.}(2013)\citenamefont
  {{Nedkova}}, \citenamefont {{Tinchev}},\ and\ \citenamefont
  {{Yazadjiev}}}]{nedkova13}%
  \BibitemOpen
  \bibfield  {author} {\bibinfo {author} {\bibfnamefont {P.~G.}\ \bibnamefont
  {{Nedkova}}}, \bibinfo {author} {\bibfnamefont {V.~K.}\ \bibnamefont
  {{Tinchev}}}, \ and\ \bibinfo {author} {\bibfnamefont {S.~S.}\ \bibnamefont
  {{Yazadjiev}}},\ }\href {\doibase10.1103/PhysRevD.88.124019} {\bibfield
  {journal} {\bibinfo  {journal} {Physical Review D}\ }\textbf {\bibinfo
  {volume} {88}},\ \bibinfo {eid} {124019} (\bibinfo {year} {2013})},\ \Eprint
  {http://arxiv.org/abs/1307.7647} {arXiv:1307.7647 [gr-qc]}\BibitemShut
  {NoStop}%
\bibitem [{\citenamefont {{Cunha}}\ \emph {et~al.}(2015)\citenamefont
  {{Cunha}}, \citenamefont {{Herdeiro}}, \citenamefont {{Radu}},\ and\
  \citenamefont {{R{\'u}narsson}}}]{cunha15}%
  \BibitemOpen
  \bibfield  {author} {\bibinfo {author} {\bibfnamefont {P.~V.~P.}\
  \bibnamefont {{Cunha}}}, \bibinfo {author} {\bibfnamefont {C.~A.~R.}\
  \bibnamefont {{Herdeiro}}}, \bibinfo {author} {\bibfnamefont
  {E.}~\bibnamefont {{Radu}}}, \ and\ \bibinfo {author} {\bibfnamefont {H.~F.}\
  \bibnamefont {{R{\'u}narsson}}},\ }\href
  {\doibase10.1103/PhysRevLett.115.211102} {\bibfield  {journal} {\bibinfo
  {journal} {Physical Review Letters}\ }\textbf {\bibinfo {volume} {115}},\
  \bibinfo {eid} {211102} (\bibinfo {year} {2015})},\ \Eprint
  {http://arxiv.org/abs/1509.00021} {arXiv:1509.00021 [gr-qc]}\BibitemShut
  {NoStop}%
\bibitem [{\citenamefont {{Vincent}}\ \emph {et~al.}(2016)\citenamefont
  {{Vincent}}, \citenamefont {{Meliani}}, \citenamefont {{Grandcl{\'e}ment}},
  \citenamefont {{Gourgoulhon}},\ and\ \citenamefont {{Straub}}}]{vincent16}%
  \BibitemOpen
  \bibfield  {author} {\bibinfo {author} {\bibfnamefont {F.~H.}\ \bibnamefont
  {{Vincent}}}, \bibinfo {author} {\bibfnamefont {Z.}~\bibnamefont
  {{Meliani}}}, \bibinfo {author} {\bibfnamefont {P.}~\bibnamefont
  {{Grandcl{\'e}ment}}}, \bibinfo {author} {\bibfnamefont {E.}~\bibnamefont
  {{Gourgoulhon}}}, \ and\ \bibinfo {author} {\bibfnamefont {O.}~\bibnamefont
  {{Straub}}},\ }\href {\doibase10.1088/0264-9381/33/10/105015} {\bibfield
  {journal} {\bibinfo  {journal} {Classical and Quantum Gravity}\ }\textbf
  {\bibinfo {volume} {33}},\ \bibinfo {eid} {105015} (\bibinfo {year}
  {2016})},\ \Eprint {http://arxiv.org/abs/1510.04170} {arXiv:1510.04170
  [gr-qc]}\BibitemShut {NoStop}%
\bibitem [{\citenamefont {{Psaltis}}\ \emph {et~al.}(2015)\citenamefont
  {{Psaltis}}, \citenamefont {{{\"O}zel}}, \citenamefont {{Chan}},\ and\
  \citenamefont {{Marrone}}}]{2015ApJ...814..115P}%
  \BibitemOpen
  \bibfield  {author} {\bibinfo {author} {\bibfnamefont {D.}~\bibnamefont
  {{Psaltis}}}, \bibinfo {author} {\bibfnamefont {F.}~\bibnamefont
  {{{\"O}zel}}}, \bibinfo {author} {\bibfnamefont {C.-K.}\ \bibnamefont
  {{Chan}}}, \ and\ \bibinfo {author} {\bibfnamefont {D.~P.}\ \bibnamefont
  {{Marrone}}},\ }\href {\doibase10.1088/0004-637X/814/2/115} {\bibfield
  {journal} {\bibinfo  {journal} {The Astrophysical Journal}\ }\textbf
  {\bibinfo {volume} {814}},\ \bibinfo {eid} {115} (\bibinfo {year} {2015})},\
  \Eprint {http://arxiv.org/abs/1411.1454} {arXiv:1411.1454
  [astro-ph.HE]}\BibitemShut {NoStop}%
\bibitem [{\citenamefont {Mizuno}\ \emph {et~al.}(2018)\citenamefont {Mizuno},
  \citenamefont {Younsi}, \citenamefont {Fromm}, \citenamefont {Porth},
  \citenamefont {De~Laurentis}, \citenamefont {Olivares}, \citenamefont
  {Falcke}, \citenamefont {Kramer},\ and\ \citenamefont
  {Rezzolla}}]{Mizuno:2018lxz}%
  \BibitemOpen
  \bibfield  {author} {\bibinfo {author} {\bibfnamefont {Yosuke}\ \bibnamefont
  {Mizuno}}, \bibinfo {author} {\bibfnamefont {Ziri}\ \bibnamefont {Younsi}},
  \bibinfo {author} {\bibfnamefont {Christian~M.}\ \bibnamefont {Fromm}},
  \bibinfo {author} {\bibfnamefont {Oliver}\ \bibnamefont {Porth}}, \bibinfo
  {author} {\bibfnamefont {Mariafelicia}\ \bibnamefont {De~Laurentis}},
  \bibinfo {author} {\bibfnamefont {Hector}\ \bibnamefont {Olivares}}, \bibinfo
  {author} {\bibfnamefont {Heino}\ \bibnamefont {Falcke}}, \bibinfo {author}
  {\bibfnamefont {Michael}\ \bibnamefont {Kramer}}, \ and\ \bibinfo {author}
  {\bibfnamefont {Luciano}\ \bibnamefont {Rezzolla}},\ }\href
  {\doibase10.1038/s41550-018-0449-5} {\  (\bibinfo {year} {2018}),\
  10.1038/s41550-018-0449-5},\ \Eprint {http://arxiv.org/abs/1804.05812}
  {arXiv:1804.05812 [astro-ph.GA]}\BibitemShut {NoStop}%
\bibitem [{\citenamefont {{Reynolds}}\ and\ \citenamefont
  {{Nowak}}(2003)}]{reynolds03}%
  \BibitemOpen
  \bibfield  {author} {\bibinfo {author} {\bibfnamefont {C.~S.}\ \bibnamefont
  {{Reynolds}}}\ and\ \bibinfo {author} {\bibfnamefont {M.~A.}\ \bibnamefont
  {{Nowak}}},\ }\href {\doibase10.1016/S0370-1573(02)00584-7} {\bibfield
  {journal} {\bibinfo  {journal} {"Physics Reports"}\ }\textbf {\bibinfo
  {volume} {377}},\ \bibinfo {pages} {389--466} (\bibinfo {year} {2003})},\
  \Eprint {http://arxiv.org/abs/astro-ph/0212065}
  {astro-ph/0212065}\BibitemShut {NoStop}%
\bibitem [{\citenamefont {{McClintock}}\ \emph {et~al.}(2011)\citenamefont
  {{McClintock}}, \citenamefont {{Narayan}}, \citenamefont {{Davis}},
  \citenamefont {{Gou}}, \citenamefont {{Kulkarni}}, \citenamefont {{Orosz}},
  \citenamefont {{Penna}}, \citenamefont {{Remillard}},\ and\ \citenamefont
  {{Steiner}}}]{mcclintock11}%
  \BibitemOpen
  \bibfield  {author} {\bibinfo {author} {\bibfnamefont {J.~E.}\ \bibnamefont
  {{McClintock}}}, \bibinfo {author} {\bibfnamefont {R.}~\bibnamefont
  {{Narayan}}}, \bibinfo {author} {\bibfnamefont {S.~W.}\ \bibnamefont
  {{Davis}}}, \bibinfo {author} {\bibfnamefont {L.}~\bibnamefont {{Gou}}},
  \bibinfo {author} {\bibfnamefont {A.}~\bibnamefont {{Kulkarni}}}, \bibinfo
  {author} {\bibfnamefont {J.~A.}\ \bibnamefont {{Orosz}}}, \bibinfo {author}
  {\bibfnamefont {R.~F.}\ \bibnamefont {{Penna}}}, \bibinfo {author}
  {\bibfnamefont {R.~A.}\ \bibnamefont {{Remillard}}}, \ and\ \bibinfo {author}
  {\bibfnamefont {J.~F.}\ \bibnamefont {{Steiner}}},\ }\href
  {\doibase10.1088/0264-9381/28/11/114009} {\bibfield  {journal} {\bibinfo
  {journal} {Classical and Quantum Gravity}\ }\textbf {\bibinfo {volume}
  {28}},\ \bibinfo {eid} {114009} (\bibinfo {year} {2011})},\ \Eprint
  {http://arxiv.org/abs/1101.0811} {arXiv:1101.0811 [astro-ph.HE]}\BibitemShut
  {NoStop}%
\bibitem [{\citenamefont {Bambi}(2017)}]{Bambi:2015kza}%
  \BibitemOpen
  \bibfield  {author} {\bibinfo {author} {\bibfnamefont {Cosimo}\ \bibnamefont
  {Bambi}},\ }\href {\doibase10.1103/RevModPhys.89.025001} {\bibfield
  {journal} {\bibinfo  {journal} {Rev. Mod. Phys.}\ }\textbf {\bibinfo {volume}
  {89}},\ \bibinfo {pages} {025001} (\bibinfo {year} {2017})},\ \Eprint
  {http://arxiv.org/abs/1509.03884} {arXiv:1509.03884 [gr-qc]}\BibitemShut
  {NoStop}%
\bibitem [{\citenamefont {{Johannsen}}\ and\ \citenamefont
  {{Psaltis}}(2013)}]{johannsen13b}%
  \BibitemOpen
  \bibfield  {author} {\bibinfo {author} {\bibfnamefont {T.}~\bibnamefont
  {{Johannsen}}}\ and\ \bibinfo {author} {\bibfnamefont {D.}~\bibnamefont
  {{Psaltis}}},\ }\href {\doibase10.1088/0004-637X/773/1/57} {\bibfield
  {journal} {\bibinfo  {journal} {The Astrophysical Journal}\ }\textbf
  {\bibinfo {volume} {773}},\ \bibinfo {eid} {57} (\bibinfo {year} {2013})},\
  \Eprint {http://arxiv.org/abs/1202.6069} {arXiv:1202.6069
  [astro-ph.HE]}\BibitemShut {NoStop}%
\bibitem [{\citenamefont {{Moore}}\ and\ \citenamefont
  {{Gair}}(2015)}]{moore15}%
  \BibitemOpen
  \bibfield  {author} {\bibinfo {author} {\bibfnamefont {C.~J.}\ \bibnamefont
  {{Moore}}}\ and\ \bibinfo {author} {\bibfnamefont {J.~R.}\ \bibnamefont
  {{Gair}}},\ }\href {\doibase10.1103/PhysRevD.92.024039} {\bibfield  {journal}
  {\bibinfo  {journal} {Physical Review D}\ }\textbf {\bibinfo {volume} {92}},\
  \bibinfo {eid} {024039} (\bibinfo {year} {2015})},\ \Eprint
  {http://arxiv.org/abs/1507.02998} {arXiv:1507.02998 [gr-qc]}\BibitemShut
  {NoStop}%
\bibitem [{\citenamefont {{Ni}}\ \emph
  {et~al.}(2016{\natexlab{a}})\citenamefont {{Ni}}, \citenamefont {{Jiang}},\
  and\ \citenamefont {{Bambi}}}]{ni16}%
  \BibitemOpen
  \bibfield  {author} {\bibinfo {author} {\bibfnamefont {Y.}~\bibnamefont
  {{Ni}}}, \bibinfo {author} {\bibfnamefont {J.}~\bibnamefont {{Jiang}}}, \
  and\ \bibinfo {author} {\bibfnamefont {C.}~\bibnamefont {{Bambi}}},\ }\href
  {\doibase10.1088/1475-7516/2016/09/014} {\bibfield  {journal} {\bibinfo
  {journal} {JCAP}\ }\textbf {\bibinfo {volume} {9}},\ \bibinfo {eid} {014}
  (\bibinfo {year} {2016}{\natexlab{a}})},\ \Eprint
  {http://arxiv.org/abs/1607.04893} {arXiv:1607.04893 [gr-qc]}\BibitemShut
  {NoStop}%
\bibitem [{\citenamefont {{Harko}}\ \emph
  {et~al.}(2009{\natexlab{a}})\citenamefont {{Harko}}, \citenamefont
  {{Kov{\'a}cs}},\ and\ \citenamefont {{Lobo}}}]{harko09}%
  \BibitemOpen
  \bibfield  {author} {\bibinfo {author} {\bibfnamefont {T.}~\bibnamefont
  {{Harko}}}, \bibinfo {author} {\bibfnamefont {Z.}~\bibnamefont
  {{Kov{\'a}cs}}}, \ and\ \bibinfo {author} {\bibfnamefont {F.~S.~N.}\
  \bibnamefont {{Lobo}}},\ }\href {\doibase10.1088/0264-9381/26/21/215006}
  {\bibfield  {journal} {\bibinfo  {journal} {Classical and Quantum Gravity}\
  }\textbf {\bibinfo {volume} {26}},\ \bibinfo {eid} {215006} (\bibinfo {year}
  {2009}{\natexlab{a}})},\ \Eprint {http://arxiv.org/abs/0905.1355}
  {arXiv:0905.1355 [gr-qc]}\BibitemShut {NoStop}%
\bibitem [{\citenamefont {{Harko}}\ \emph
  {et~al.}(2009{\natexlab{b}})\citenamefont {{Harko}}, \citenamefont
  {{Kov{\'a}cs}},\ and\ \citenamefont {{Lobo}}}]{harko09b}%
  \BibitemOpen
  \bibfield  {author} {\bibinfo {author} {\bibfnamefont {T.}~\bibnamefont
  {{Harko}}}, \bibinfo {author} {\bibfnamefont {Z.}~\bibnamefont
  {{Kov{\'a}cs}}}, \ and\ \bibinfo {author} {\bibfnamefont {F.~S.~N.}\
  \bibnamefont {{Lobo}}},\ }\href {\doibase10.1103/PhysRevD.80.044021}
  {\bibfield  {journal} {\bibinfo  {journal} {Physical Review D}\ }\textbf
  {\bibinfo {volume} {80}},\ \bibinfo {eid} {044021} (\bibinfo {year}
  {2009}{\natexlab{b}})},\ \Eprint {http://arxiv.org/abs/0907.1449}
  {arXiv:0907.1449 [gr-qc]}\BibitemShut {NoStop}%
\bibitem [{\citenamefont {{Schee}}\ and\ \citenamefont
  {{Stuchl{\'{\i}}k}}(2009)}]{schee09}%
  \BibitemOpen
  \bibfield  {author} {\bibinfo {author} {\bibfnamefont {J.}~\bibnamefont
  {{Schee}}}\ and\ \bibinfo {author} {\bibfnamefont {Z.}~\bibnamefont
  {{Stuchl{\'{\i}}k}}},\ }\href {\doibase10.1007/s10714-008-0753-y} {\bibfield
  {journal} {\bibinfo  {journal} {General Relativity and Gravitation}\ }\textbf
  {\bibinfo {volume} {41}},\ \bibinfo {pages} {1795--1818} (\bibinfo {year}
  {2009})},\ \Eprint {http://arxiv.org/abs/0812.3017}
  {arXiv:0812.3017}\BibitemShut {NoStop}%
\bibitem [{\citenamefont {{Cao}}\ \emph {et~al.}(2016)\citenamefont {{Cao}},
  \citenamefont {{C{\'a}rdenas-Avenda{\~n}o}}, \citenamefont {{Zhou}},
  \citenamefont {{Bambi}}, \citenamefont {{Herdeiro}},\ and\ \citenamefont
  {{Radu}}}]{cao16}%
  \BibitemOpen
  \bibfield  {author} {\bibinfo {author} {\bibfnamefont {Z.}~\bibnamefont
  {{Cao}}}, \bibinfo {author} {\bibfnamefont {A.}~\bibnamefont
  {{C{\'a}rdenas-Avenda{\~n}o}}}, \bibinfo {author} {\bibfnamefont
  {M.}~\bibnamefont {{Zhou}}}, \bibinfo {author} {\bibfnamefont
  {C.}~\bibnamefont {{Bambi}}}, \bibinfo {author} {\bibfnamefont {C.~A.~R.}\
  \bibnamefont {{Herdeiro}}}, \ and\ \bibinfo {author} {\bibfnamefont
  {E.}~\bibnamefont {{Radu}}},\ }\href {\doibase10.1088/1475-7516/2016/10/003}
  {\bibfield  {journal} {\bibinfo  {journal} {JCAP}\ }\textbf {\bibinfo
  {volume} {10}},\ \bibinfo {eid} {003} (\bibinfo {year} {2016})},\ \Eprint
  {http://arxiv.org/abs/1609.00901} {arXiv:1609.00901 [gr-qc]}\BibitemShut
  {NoStop}%
\bibitem [{\citenamefont {{Ni}}\ \emph
  {et~al.}(2016{\natexlab{b}})\citenamefont {{Ni}}, \citenamefont {{Zhou}},
  \citenamefont {{C{\'a}rdenas-Avenda{\~n}o}}, \citenamefont {{Bambi}},
  \citenamefont {{Herdeiro}},\ and\ \citenamefont {{Radu}}}]{ni16b}%
  \BibitemOpen
  \bibfield  {author} {\bibinfo {author} {\bibfnamefont {Y.}~\bibnamefont
  {{Ni}}}, \bibinfo {author} {\bibfnamefont {M.}~\bibnamefont {{Zhou}}},
  \bibinfo {author} {\bibfnamefont {A.}~\bibnamefont
  {{C{\'a}rdenas-Avenda{\~n}o}}}, \bibinfo {author} {\bibfnamefont
  {C.}~\bibnamefont {{Bambi}}}, \bibinfo {author} {\bibfnamefont {C.~A.~R.}\
  \bibnamefont {{Herdeiro}}}, \ and\ \bibinfo {author} {\bibfnamefont
  {E.}~\bibnamefont {{Radu}}},\ }\href {\doibase10.1088/1475-7516/2016/07/049}
  {\bibfield  {journal} {\bibinfo  {journal} {JCAP}\ }\textbf {\bibinfo
  {volume} {7}},\ \bibinfo {eid} {049} (\bibinfo {year}
  {2016}{\natexlab{b}})},\ \Eprint {http://arxiv.org/abs/1606.04654}
  {arXiv:1606.04654 [gr-qc]}\BibitemShut {NoStop}%
\bibitem [{\citenamefont {{Remillard}}\ and\ \citenamefont
  {{McClintock}}(2006)}]{remillard06}%
  \BibitemOpen
  \bibfield  {author} {\bibinfo {author} {\bibfnamefont {R.~A.}\ \bibnamefont
  {{Remillard}}}\ and\ \bibinfo {author} {\bibfnamefont {J.~E.}\ \bibnamefont
  {{McClintock}}},\ }\href {\doibase10.1146/annurev.astro.44.051905.092532}
  {\bibfield  {journal} {\bibinfo  {journal} {"Ann. Rev. Astron. Astrophys"}\
  }\textbf {\bibinfo {volume} {44}},\ \bibinfo {pages} {49--92} (\bibinfo
  {year} {2006})},\ \Eprint {http://arxiv.org/abs/astro-ph/0606352}
  {astro-ph/0606352}\BibitemShut {NoStop}%
\bibitem [{\citenamefont {{Johannsen}}\ and\ \citenamefont
  {{Psaltis}}(2011)}]{johannsen11}%
  \BibitemOpen
  \bibfield  {author} {\bibinfo {author} {\bibfnamefont {T.}~\bibnamefont
  {{Johannsen}}}\ and\ \bibinfo {author} {\bibfnamefont {D.}~\bibnamefont
  {{Psaltis}}},\ }\href {\doibase10.1088/0004-637X/726/1/11} {\bibfield
  {journal} {\bibinfo  {journal} {The Astrophysical Journal}\ }\textbf
  {\bibinfo {volume} {726}},\ \bibinfo {eid} {11} (\bibinfo {year} {2011})},\
  \Eprint {http://arxiv.org/abs/1010.1000} {arXiv:1010.1000
  [astro-ph.HE]}\BibitemShut {NoStop}%
\bibitem [{\citenamefont {{Bambi}}\ and\ \citenamefont
  {{Nampalliwar}}(2016)}]{bambi16}%
  \BibitemOpen
  \bibfield  {author} {\bibinfo {author} {\bibfnamefont {C.}~\bibnamefont
  {{Bambi}}}\ and\ \bibinfo {author} {\bibfnamefont {S.}~\bibnamefont
  {{Nampalliwar}}},\ }\href {\doibase10.1209/0295-5075/116/30006} {\bibfield
  {journal} {\bibinfo  {journal} {EPL (Europhysics Letters)}\ }\textbf
  {\bibinfo {volume} {116}},\ \bibinfo {pages} {30006} (\bibinfo {year}
  {2016})},\ \Eprint {http://arxiv.org/abs/1604.02643} {arXiv:1604.02643
  [gr-qc]}\BibitemShut {NoStop}%
\bibitem [{\citenamefont {{Maselli}}\ \emph {et~al.}(2015)\citenamefont
  {{Maselli}}, \citenamefont {{Gualtieri}}, \citenamefont {{Pani}},
  \citenamefont {{Stella}},\ and\ \citenamefont {{Ferrari}}}]{maselli15}%
  \BibitemOpen
  \bibfield  {author} {\bibinfo {author} {\bibfnamefont {A.}~\bibnamefont
  {{Maselli}}}, \bibinfo {author} {\bibfnamefont {L.}~\bibnamefont
  {{Gualtieri}}}, \bibinfo {author} {\bibfnamefont {P.}~\bibnamefont {{Pani}}},
  \bibinfo {author} {\bibfnamefont {L.}~\bibnamefont {{Stella}}}, \ and\
  \bibinfo {author} {\bibfnamefont {V.}~\bibnamefont {{Ferrari}}},\ }\href
  {\doibase10.1088/0004-637X/801/2/115} {\bibfield  {journal} {\bibinfo
  {journal} {The Astrophysical Journal}\ }\textbf {\bibinfo {volume} {801}},\
  \bibinfo {eid} {115} (\bibinfo {year} {2015})},\ \Eprint
  {http://arxiv.org/abs/1412.3473} {arXiv:1412.3473 [astro-ph.HE]}\BibitemShut
  {NoStop}%
\bibitem [{\citenamefont {{Chen}}\ \emph {et~al.}(2016)\citenamefont {{Chen}},
  \citenamefont {{Wang}},\ and\ \citenamefont {{Jing}}}]{chen16}%
  \BibitemOpen
  \bibfield  {author} {\bibinfo {author} {\bibfnamefont {S.}~\bibnamefont
  {{Chen}}}, \bibinfo {author} {\bibfnamefont {M.}~\bibnamefont {{Wang}}}, \
  and\ \bibinfo {author} {\bibfnamefont {J.}~\bibnamefont {{Jing}}},\ }\href
  {\doibase10.1088/0264-9381/33/19/195002} {\bibfield  {journal} {\bibinfo
  {journal} {Classical and Quantum Gravity}\ }\textbf {\bibinfo {volume}
  {33}},\ \bibinfo {eid} {195002} (\bibinfo {year} {2016})},\ \Eprint
  {http://arxiv.org/abs/1604.07106} {arXiv:1604.07106 [gr-qc]}\BibitemShut
  {NoStop}%
\bibitem [{\citenamefont {{Franchini}}\ \emph {et~al.}(2017)\citenamefont
  {{Franchini}}, \citenamefont {{Pani}}, \citenamefont {{Maselli}},
  \citenamefont {{Gualtieri}}, \citenamefont {{Herdeiro}}, \citenamefont
  {{Radu}},\ and\ \citenamefont {{Ferrari}}}]{franchini17}%
  \BibitemOpen
  \bibfield  {author} {\bibinfo {author} {\bibfnamefont {N.}~\bibnamefont
  {{Franchini}}}, \bibinfo {author} {\bibfnamefont {P.}~\bibnamefont {{Pani}}},
  \bibinfo {author} {\bibfnamefont {A.}~\bibnamefont {{Maselli}}}, \bibinfo
  {author} {\bibfnamefont {L.}~\bibnamefont {{Gualtieri}}}, \bibinfo {author}
  {\bibfnamefont {C.~A.~R.}\ \bibnamefont {{Herdeiro}}}, \bibinfo {author}
  {\bibfnamefont {E.}~\bibnamefont {{Radu}}}, \ and\ \bibinfo {author}
  {\bibfnamefont {V.}~\bibnamefont {{Ferrari}}},\ }\href
  {\doibase10.1103/PhysRevD.95.124025} {\bibfield  {journal} {\bibinfo
  {journal} {Physical Review D}\ }\textbf {\bibinfo {volume} {95}},\ \bibinfo
  {eid} {124025} (\bibinfo {year} {2017})},\ \Eprint
  {http://arxiv.org/abs/1612.00038} {arXiv:1612.00038
  [astro-ph.HE]}\BibitemShut {NoStop}%
\bibitem [{\citenamefont {{Taylor}}(1994)}]{tay94}%
  \BibitemOpen
  \bibfield  {author} {\bibinfo {author} {\bibfnamefont {J.~H.}\ \bibnamefont
  {{Taylor}}, \bibfnamefont {Jr.}},\ }\href {\doibase10.1103/RevModPhys.66.711}
  {\bibfield  {journal} {\bibinfo  {journal} {Reviews of Modern Physics}\
  }\textbf {\bibinfo {volume} {66}},\ \bibinfo {pages} {711--719} (\bibinfo
  {year} {1994})}\BibitemShut {NoStop}%
\bibitem [{\citenamefont {Kramer}\ \emph {et~al.}(2006)\citenamefont {Kramer}
  \emph {et~al.}}]{ksm+06}%
  \BibitemOpen
  \bibfield  {author} {\bibinfo {author} {\bibfnamefont {M.}~\bibnamefont
  {Kramer}} \emph {et~al.},\ }\href {\doibase10.1126/science.1132305}
  {\bibfield  {journal} {\bibinfo  {journal} {Science}\ }\textbf {\bibinfo
  {volume} {314}},\ \bibinfo {pages} {97--102} (\bibinfo {year} {2006})},\
  \Eprint {http://arxiv.org/abs/astro-ph/0609417} {arXiv:astro-ph/0609417
  [astro-ph]}\BibitemShut {NoStop}%
\bibitem [{\citenamefont {Kramer}(1998)}]{kra98}%
  \BibitemOpen
  \bibfield  {author} {\bibinfo {author} {\bibfnamefont {Michael}\ \bibnamefont
  {Kramer}},\ }\href {\doibase10.1086/306535} {\bibfield  {journal} {\bibinfo
  {journal} {Astrophys. J.}\ }\textbf {\bibinfo {volume} {509}},\ \bibinfo
  {pages} {856} (\bibinfo {year} {1998})},\ \Eprint
  {http://arxiv.org/abs/astro-ph/9808127} {arXiv:astro-ph/9808127
  [astro-ph]}\BibitemShut {NoStop}%
\bibitem [{\citenamefont {Breton}\ \emph {et~al.}(2008)\citenamefont {Breton},
  \citenamefont {Kaspi}, \citenamefont {Kramer}, \citenamefont {McLaughlin},
  \citenamefont {Lyutikov}, \citenamefont {Ransom}, \citenamefont {Stairs},
  \citenamefont {Ferdman}, \citenamefont {Camilo},\ and\ \citenamefont
  {Possenti}}]{bkk+08}%
  \BibitemOpen
  \bibfield  {author} {\bibinfo {author} {\bibfnamefont {Rene~P.}\ \bibnamefont
  {Breton}}, \bibinfo {author} {\bibfnamefont {Victoria~M.}\ \bibnamefont
  {Kaspi}}, \bibinfo {author} {\bibfnamefont {Michael}\ \bibnamefont {Kramer}},
  \bibinfo {author} {\bibfnamefont {Maura~A.}\ \bibnamefont {McLaughlin}},
  \bibinfo {author} {\bibfnamefont {Maxim}\ \bibnamefont {Lyutikov}}, \bibinfo
  {author} {\bibfnamefont {Scott~M.}\ \bibnamefont {Ransom}}, \bibinfo {author}
  {\bibfnamefont {Ingrid~H.}\ \bibnamefont {Stairs}}, \bibinfo {author}
  {\bibfnamefont {Robert~D.}\ \bibnamefont {Ferdman}}, \bibinfo {author}
  {\bibfnamefont {Fernando}\ \bibnamefont {Camilo}}, \ and\ \bibinfo {author}
  {\bibfnamefont {Andrea}\ \bibnamefont {Possenti}},\ }\href
  {\doibase10.1126/science.1159295} {\bibfield  {journal} {\bibinfo  {journal}
  {Science}\ }\textbf {\bibinfo {volume} {321}},\ \bibinfo {pages} {104--107}
  (\bibinfo {year} {2008})},\ \Eprint {http://arxiv.org/abs/0807.2644}
  {arXiv:0807.2644 [astro-ph]}\BibitemShut {NoStop}%
\bibitem [{\citenamefont {Wex}\ and\ \citenamefont {Kopeikin}(1999)}]{wk99}%
  \BibitemOpen
  \bibfield  {author} {\bibinfo {author} {\bibfnamefont {N.}~\bibnamefont
  {Wex}}\ and\ \bibinfo {author} {\bibfnamefont {S.}~\bibnamefont {Kopeikin}},\
  }\href {\doibase10.1086/306933} {\bibfield  {journal} {\bibinfo  {journal}
  {Astrophys. J.}\ }\textbf {\bibinfo {volume} {514}},\ \bibinfo {pages} {388}
  (\bibinfo {year} {1999})},\ \Eprint {http://arxiv.org/abs/astro-ph/9811052}
  {arXiv:astro-ph/9811052 [astro-ph]}\BibitemShut {NoStop}%
\bibitem [{\citenamefont {Wharton}\ \emph {et~al.}(2012)\citenamefont
  {Wharton}, \citenamefont {Chatterjee}, \citenamefont {Cordes}, \citenamefont
  {Deneva},\ and\ \citenamefont {Lazio}}]{Wharton:2011dv}%
  \BibitemOpen
  \bibfield  {author} {\bibinfo {author} {\bibfnamefont {R.~S.}\ \bibnamefont
  {Wharton}}, \bibinfo {author} {\bibfnamefont {S.}~\bibnamefont {Chatterjee}},
  \bibinfo {author} {\bibfnamefont {J.~M.}\ \bibnamefont {Cordes}}, \bibinfo
  {author} {\bibfnamefont {J.~S.}\ \bibnamefont {Deneva}}, \ and\ \bibinfo
  {author} {\bibfnamefont {T.~J.~W.}\ \bibnamefont {Lazio}},\ }\href
  {\doibase10.1088/0004-637X/753/2/108} {\bibfield  {journal} {\bibinfo
  {journal} {Astrophys. J.}\ }\textbf {\bibinfo {volume} {753}},\ \bibinfo
  {pages} {108} (\bibinfo {year} {2012})},\ \Eprint
  {http://arxiv.org/abs/1111.4216} {arXiv:1111.4216 [astro-ph.HE]}\BibitemShut
  {NoStop}%
\bibitem [{\citenamefont {Liu}\ \emph {et~al.}(2012)\citenamefont {Liu},
  \citenamefont {Wex}, \citenamefont {Kramer}, \citenamefont {Cordes},\ and\
  \citenamefont {Lazio}}]{Liu:2011ae}%
  \BibitemOpen
  \bibfield  {author} {\bibinfo {author} {\bibfnamefont {K.}~\bibnamefont
  {Liu}}, \bibinfo {author} {\bibfnamefont {N.}~\bibnamefont {Wex}}, \bibinfo
  {author} {\bibfnamefont {M.}~\bibnamefont {Kramer}}, \bibinfo {author}
  {\bibfnamefont {J.~M.}\ \bibnamefont {Cordes}}, \ and\ \bibinfo {author}
  {\bibfnamefont {T.~J.~W.}\ \bibnamefont {Lazio}},\ }\href
  {\doibase10.1088/0004-637X/747/1/1} {\bibfield  {journal} {\bibinfo
  {journal} {Astrophys. J.}\ }\textbf {\bibinfo {volume} {747}},\ \bibinfo
  {pages} {1} (\bibinfo {year} {2012})},\ \Eprint
  {http://arxiv.org/abs/1112.2151} {arXiv:1112.2151 [astro-ph.HE]}\BibitemShut
  {NoStop}%
\bibitem [{\citenamefont {Liu}\ \emph {et~al.}(2014)\citenamefont {Liu},
  \citenamefont {Eatough}, \citenamefont {Wex},\ and\ \citenamefont
  {Kramer}}]{Liu:2014uka}%
  \BibitemOpen
  \bibfield  {author} {\bibinfo {author} {\bibfnamefont {K.}~\bibnamefont
  {Liu}}, \bibinfo {author} {\bibfnamefont {R.~P.}\ \bibnamefont {Eatough}},
  \bibinfo {author} {\bibfnamefont {N.}~\bibnamefont {Wex}}, \ and\ \bibinfo
  {author} {\bibfnamefont {M.}~\bibnamefont {Kramer}},\ }\href
  {\doibase10.1093/mnras/stu1913} {\bibfield  {journal} {\bibinfo  {journal}
  {Mon. Not. Roy. Astron. Soc.}\ }\textbf {\bibinfo {volume} {445}},\ \bibinfo
  {pages} {3115--3132} (\bibinfo {year} {2014})},\ \Eprint
  {http://arxiv.org/abs/1409.3882} {arXiv:1409.3882 [astro-ph.GA]}\BibitemShut
  {NoStop}%
\bibitem [{\citenamefont {Zhang}\ and\ \citenamefont
  {Saha}(2017)}]{Zhang:2017qbb}%
  \BibitemOpen
  \bibfield  {author} {\bibinfo {author} {\bibfnamefont {Fupeng}\ \bibnamefont
  {Zhang}}\ and\ \bibinfo {author} {\bibfnamefont {Prasenjit}\ \bibnamefont
  {Saha}},\ }\href {\doibase10.3847/1538-4357/aa8f47} {\bibfield  {journal}
  {\bibinfo  {journal} {Astrophys. J.}\ }\textbf {\bibinfo {volume} {849}},\
  \bibinfo {pages} {33} (\bibinfo {year} {2017})},\ \Eprint
  {http://arxiv.org/abs/1709.08341} {arXiv:1709.08341
  [astro-ph.GA]}\BibitemShut {NoStop}%
\bibitem [{\citenamefont {Psaltis}\ \emph {et~al.}(2016)\citenamefont
  {Psaltis}, \citenamefont {Wex},\ and\ \citenamefont
  {Kramer}}]{Psaltis:2015uza}%
  \BibitemOpen
  \bibfield  {author} {\bibinfo {author} {\bibfnamefont {Dimitrios}\
  \bibnamefont {Psaltis}}, \bibinfo {author} {\bibfnamefont {Norbert}\
  \bibnamefont {Wex}}, \ and\ \bibinfo {author} {\bibfnamefont {Michael}\
  \bibnamefont {Kramer}},\ }\href {\doibase10.3847/0004-637X/818/2/121}
  {\bibfield  {journal} {\bibinfo  {journal} {Astrophys. J.}\ }\textbf
  {\bibinfo {volume} {818}},\ \bibinfo {pages} {121} (\bibinfo {year}
  {2016})},\ \Eprint {http://arxiv.org/abs/1510.00394} {arXiv:1510.00394
  [astro-ph.HE]}\BibitemShut {NoStop}%
\bibitem [{\citenamefont {Simonetti}\ \emph {et~al.}(2011)\citenamefont
  {Simonetti}, \citenamefont {Kavic}, \citenamefont {Minic}, \citenamefont
  {Surani},\ and\ \citenamefont {Vejayan}}]{Simonetti:2010mk}%
  \BibitemOpen
  \bibfield  {author} {\bibinfo {author} {\bibfnamefont {John~H.}\ \bibnamefont
  {Simonetti}}, \bibinfo {author} {\bibfnamefont {Michael}\ \bibnamefont
  {Kavic}}, \bibinfo {author} {\bibfnamefont {Djordje}\ \bibnamefont {Minic}},
  \bibinfo {author} {\bibfnamefont {Umair}\ \bibnamefont {Surani}}, \ and\
  \bibinfo {author} {\bibfnamefont {Vipin}\ \bibnamefont {Vejayan}},\ }\href
  {\doibase10.1088/2041-8205/737/2/L28} {\bibfield  {journal} {\bibinfo
  {journal} {Astrophys. J.}\ }\textbf {\bibinfo {volume} {737}},\ \bibinfo
  {pages} {L28} (\bibinfo {year} {2011})},\ \Eprint
  {http://arxiv.org/abs/1010.5245} {arXiv:1010.5245 [astro-ph.HE]}\BibitemShut
  {NoStop}%
\bibitem [{\citenamefont {Estes}\ \emph {et~al.}(2017)\citenamefont {Estes},
  \citenamefont {Kavic}, \citenamefont {Lippert},\ and\ \citenamefont
  {Simonetti}}]{Estes:2016wgv}%
  \BibitemOpen
  \bibfield  {author} {\bibinfo {author} {\bibfnamefont {John}\ \bibnamefont
  {Estes}}, \bibinfo {author} {\bibfnamefont {Michael}\ \bibnamefont {Kavic}},
  \bibinfo {author} {\bibfnamefont {Matthew}\ \bibnamefont {Lippert}}, \ and\
  \bibinfo {author} {\bibfnamefont {John~H.}\ \bibnamefont {Simonetti}},\
  }\href {\doibase10.3847/1538-4357/aa610e} {\bibfield  {journal} {\bibinfo
  {journal} {Astrophys. J.}\ }\textbf {\bibinfo {volume} {837}},\ \bibinfo
  {pages} {87} (\bibinfo {year} {2017})},\ \Eprint
  {http://arxiv.org/abs/1607.00018} {arXiv:1607.00018 [hep-th]}\BibitemShut
  {NoStop}%
\bibitem [{\citenamefont {Cameron}\ \emph {et~al.}(2018)\citenamefont {Cameron}
  \emph {et~al.}}]{cck+18}%
  \BibitemOpen
  \bibfield  {author} {\bibinfo {author} {\bibfnamefont {A.~D.}\ \bibnamefont
  {Cameron}} \emph {et~al.},\ }\href {\doibase10.1093/mnrasl/sly003} {\bibfield
   {journal} {\bibinfo  {journal} {Mon. Not. Roy. Astron. Soc.}\ }\textbf
  {\bibinfo {volume} {475}},\ \bibinfo {pages} {L57--L61} (\bibinfo {year}
  {2018})},\ \Eprint {http://arxiv.org/abs/1711.07697} {arXiv:1711.07697
  [astro-ph.HE]}\BibitemShut {NoStop}%
\bibitem [{\citenamefont {Hessels}\ \emph {et~al.}(2015)\citenamefont
  {Hessels}, \citenamefont {Possenti}, \citenamefont {Bailes}, \citenamefont
  {Bassa}, \citenamefont {Freire}, \citenamefont {Lorimer}, \citenamefont
  {Lynch}, \citenamefont {Ransom},\ and\ \citenamefont
  {Stairs}}]{Hessels:2014yja}%
  \BibitemOpen
  \bibfield  {author} {\bibinfo {author} {\bibfnamefont {J.~W.~T.}\
  \bibnamefont {Hessels}}, \bibinfo {author} {\bibfnamefont {A.}~\bibnamefont
  {Possenti}}, \bibinfo {author} {\bibfnamefont {M.}~\bibnamefont {Bailes}},
  \bibinfo {author} {\bibfnamefont {C.~G.}\ \bibnamefont {Bassa}}, \bibinfo
  {author} {\bibfnamefont {P.~C.~C.}\ \bibnamefont {Freire}}, \bibinfo {author}
  {\bibfnamefont {D.~R.}\ \bibnamefont {Lorimer}}, \bibinfo {author}
  {\bibfnamefont {R.}~\bibnamefont {Lynch}}, \bibinfo {author} {\bibfnamefont
  {S.~M.}\ \bibnamefont {Ransom}}, \ and\ \bibinfo {author} {\bibfnamefont
  {I.~H.}\ \bibnamefont {Stairs}},\ }\bibfield  {booktitle} {\emph {\bibinfo
  {booktitle} {{Proceedings, Advancing Astrophysics with the Square Kilometre
  Array (AASKA14): Giardini Naxos, Italy, June 9-13, 2014}}},\ }\href
  {\doibase10.22323/1.215.0047} {\bibfield  {journal} {\bibinfo  {journal}
  {PoS}\ }\textbf {\bibinfo {volume} {AASKA14}},\ \bibinfo {pages} {047}
  (\bibinfo {year} {2015})},\ \Eprint {http://arxiv.org/abs/1501.00086}
  {arXiv:1501.00086 [astro-ph.HE]}\BibitemShut {NoStop}%
\bibitem [{\citenamefont {Eatough}\ \emph {et~al.}(2015)\citenamefont {Eatough}
  \emph {et~al.}}]{Eatough:2015jka}%
  \BibitemOpen
  \bibfield  {author} {\bibinfo {author} {\bibfnamefont {R.~P.}\ \bibnamefont
  {Eatough}} \emph {et~al.},\ }\bibfield  {booktitle} {\emph {\bibinfo
  {booktitle} {{Proceedings, Advancing Astrophysics with the Square Kilometre
  Array (AASKA14): Giardini Naxos, Italy, June 9-13, 2014}}},\ }\href
  {\doibase10.22323/1.215.0045} {\bibfield  {journal} {\bibinfo  {journal}
  {PoS}\ }\textbf {\bibinfo {volume} {AASKA14}},\ \bibinfo {pages} {045}
  (\bibinfo {year} {2015})},\ \Eprint {http://arxiv.org/abs/1501.00281}
  {arXiv:1501.00281 [astro-ph.IM]}\BibitemShut {NoStop}%
\bibitem [{\citenamefont {Eatough}\ \emph {et~al.}(2013)\citenamefont {Eatough}
  \emph {et~al.}}]{efk+13}%
  \BibitemOpen
  \bibfield  {author} {\bibinfo {author} {\bibfnamefont {R.~P.}\ \bibnamefont
  {Eatough}} \emph {et~al.},\ }\href {\doibase10.1038/nature12499} {\bibfield
  {journal} {\bibinfo  {journal} {Nature}\ }\textbf {\bibinfo {volume} {501}},\
  \bibinfo {pages} {391--394} (\bibinfo {year} {2013})},\ \Eprint
  {http://arxiv.org/abs/1308.3147} {arXiv:1308.3147 [astro-ph.GA]}\BibitemShut
  {NoStop}%
\bibitem [{\citenamefont {{Dexter}}\ \emph {et~al.}(2017)\citenamefont
  {{Dexter}}, \citenamefont {{Deller}}, \citenamefont {{Bower}}, \citenamefont
  {{Demorest}}, \citenamefont {{Kramer}}, \citenamefont {{Stappers}},
  \citenamefont {{Lyne}}, \citenamefont {{Kerr}}, \citenamefont {{Spitler}},
  \citenamefont {{Psaltis}}, \citenamefont {{Johnson}},\ and\ \citenamefont
  {{Narayan}}}]{ddb+17}%
  \BibitemOpen
  \bibfield  {author} {\bibinfo {author} {\bibfnamefont {J.}~\bibnamefont
  {{Dexter}}}, \bibinfo {author} {\bibfnamefont {A.}~\bibnamefont {{Deller}}},
  \bibinfo {author} {\bibfnamefont {G.~C.}\ \bibnamefont {{Bower}}}, \bibinfo
  {author} {\bibfnamefont {P.}~\bibnamefont {{Demorest}}}, \bibinfo {author}
  {\bibfnamefont {M.}~\bibnamefont {{Kramer}}}, \bibinfo {author}
  {\bibfnamefont {B.~W.}\ \bibnamefont {{Stappers}}}, \bibinfo {author}
  {\bibfnamefont {A.~G.}\ \bibnamefont {{Lyne}}}, \bibinfo {author}
  {\bibfnamefont {M.}~\bibnamefont {{Kerr}}}, \bibinfo {author} {\bibfnamefont
  {L.~G.}\ \bibnamefont {{Spitler}}}, \bibinfo {author} {\bibfnamefont
  {D.}~\bibnamefont {{Psaltis}}}, \bibinfo {author} {\bibfnamefont
  {M.}~\bibnamefont {{Johnson}}}, \ and\ \bibinfo {author} {\bibfnamefont
  {R.}~\bibnamefont {{Narayan}}},\ }\href {\doibase10.1093/mnras/stx1777}
  {\bibfield  {journal} {\bibinfo  {journal} {Mon. Not. Roy. Astron. Soc.}\
  }\textbf {\bibinfo {volume} {471}},\ \bibinfo {pages} {3563--3576} (\bibinfo
  {year} {2017})},\ \Eprint {http://arxiv.org/abs/1707.03842}
  {arXiv:1707.03842}\BibitemShut {NoStop}%
\bibitem [{\citenamefont {Goddi}\ \emph {et~al.}(2016)\citenamefont {Goddi}
  \emph {et~al.}}]{Goddi:2017pfy}%
  \BibitemOpen
  \bibfield  {author} {\bibinfo {author} {\bibfnamefont {C.}~\bibnamefont
  {Goddi}} \emph {et~al.},\ }\bibfield  {booktitle} {\emph {\bibinfo
  {booktitle} {{Proceedings, 14th Marcel Grossmann Meeting on Recent
  Developments in Theoretical and Experimental General Relativity,
  Astrophysics, and Relativistic Field Theories (MG14) (In 4 Volumes): Rome,
  Italy, July 12-18, 2015}}},\ }\href {\doibase10.1142/S0218271817300014}
  {\bibfield  {journal} {\bibinfo  {journal} {Int. J. Mod. Phys.}\ }\textbf
  {\bibinfo {volume} {D26}},\ \bibinfo {pages} {1730001} (\bibinfo {year}
  {2016})},\ \Eprint {http://arxiv.org/abs/1606.08879} {arXiv:1606.08879
  [astro-ph.HE]}\BibitemShut {NoStop}%
\bibitem [{\citenamefont {{K{\i}z{\i}ltan}}\ \emph {et~al.}(2017)\citenamefont
  {{K{\i}z{\i}ltan}}, \citenamefont {{Baumgardt}},\ and\ \citenamefont
  {{Loeb}}}]{kbl17}%
  \BibitemOpen
  \bibfield  {author} {\bibinfo {author} {\bibfnamefont {B.}~\bibnamefont
  {{K{\i}z{\i}ltan}}}, \bibinfo {author} {\bibfnamefont {H.}~\bibnamefont
  {{Baumgardt}}}, \ and\ \bibinfo {author} {\bibfnamefont {A.}~\bibnamefont
  {{Loeb}}},\ }\href {\doibase10.1038/nature21361} {\bibfield  {journal}
  {\bibinfo  {journal} {Nature}\ }\textbf {\bibinfo {volume} {542}},\ \bibinfo
  {pages} {203--205} (\bibinfo {year} {2017})},\ \Eprint
  {http://arxiv.org/abs/1702.02149} {arXiv:1702.02149}\BibitemShut {NoStop}%
\bibitem [{\citenamefont {Freire}\ \emph {et~al.}(2017)\citenamefont {Freire}
  \emph {et~al.}}]{frk+17}%
  \BibitemOpen
  \bibfield  {author} {\bibinfo {author} {\bibfnamefont {P.~C.~C.}\
  \bibnamefont {Freire}} \emph {et~al.},\ }\href
  {\doibase10.1093/mnras/stx1533} {\bibfield  {journal} {\bibinfo  {journal}
  {Mon. Not. Roy. Astron. Soc.}\ }\textbf {\bibinfo {volume} {471}},\ \bibinfo
  {pages} {857--876} (\bibinfo {year} {2017})},\ \Eprint
  {http://arxiv.org/abs/1706.04908} {arXiv:1706.04908
  [astro-ph.HE]}\BibitemShut {NoStop}%
\bibitem [{\citenamefont {Perera}\ \emph {et~al.}(2017)\citenamefont {Perera},
  \citenamefont {Stappers}, \citenamefont {Lyne}, \citenamefont {Bassa},
  \citenamefont {Cognard}, \citenamefont {Guillemot}, \citenamefont {Kramer},
  \citenamefont {Theureau},\ and\ \citenamefont {Desvignes}}]{psl+17a}%
  \BibitemOpen
  \bibfield  {author} {\bibinfo {author} {\bibfnamefont {B.~B.~P.}\
  \bibnamefont {Perera}}, \bibinfo {author} {\bibfnamefont {B.~W.}\
  \bibnamefont {Stappers}}, \bibinfo {author} {\bibfnamefont {A.~G.}\
  \bibnamefont {Lyne}}, \bibinfo {author} {\bibfnamefont {C.~G.}\ \bibnamefont
  {Bassa}}, \bibinfo {author} {\bibfnamefont {I.}~\bibnamefont {Cognard}},
  \bibinfo {author} {\bibfnamefont {L.}~\bibnamefont {Guillemot}}, \bibinfo
  {author} {\bibfnamefont {M.}~\bibnamefont {Kramer}}, \bibinfo {author}
  {\bibfnamefont {G.}~\bibnamefont {Theureau}}, \ and\ \bibinfo {author}
  {\bibfnamefont {G.}~\bibnamefont {Desvignes}},\ }\href
  {\doibase10.1093/mnras/stx501} {\bibfield  {journal} {\bibinfo  {journal}
  {Mon. Not. Roy. Astron. Soc.}\ }\textbf {\bibinfo {volume} {468}},\ \bibinfo
  {pages} {2114--2127} (\bibinfo {year} {2017})},\ \Eprint
  {http://arxiv.org/abs/1705.01612} {arXiv:1705.01612
  [astro-ph.HE]}\BibitemShut {NoStop}%
\bibitem [{\citenamefont {{Perera}}\ \emph {et~al.}(2017)\citenamefont
  {{Perera}}, \citenamefont {{Stappers}}, \citenamefont {{Lyne}}, \citenamefont
  {{Bassa}}, \citenamefont {{Cognard}}, \citenamefont {{Guillemot}},
  \citenamefont {{Kramer}}, \citenamefont {{Theureau}},\ and\ \citenamefont
  {{Desvignes}}}]{psl+17b}%
  \BibitemOpen
  \bibfield  {author} {\bibinfo {author} {\bibfnamefont {B.~B.~P.}\
  \bibnamefont {{Perera}}}, \bibinfo {author} {\bibfnamefont {B.~W.}\
  \bibnamefont {{Stappers}}}, \bibinfo {author} {\bibfnamefont {A.~G.}\
  \bibnamefont {{Lyne}}}, \bibinfo {author} {\bibfnamefont {C.~G.}\
  \bibnamefont {{Bassa}}}, \bibinfo {author} {\bibfnamefont {I.}~\bibnamefont
  {{Cognard}}}, \bibinfo {author} {\bibfnamefont {L.}~\bibnamefont
  {{Guillemot}}}, \bibinfo {author} {\bibfnamefont {M.}~\bibnamefont
  {{Kramer}}}, \bibinfo {author} {\bibfnamefont {G.}~\bibnamefont
  {{Theureau}}}, \ and\ \bibinfo {author} {\bibfnamefont {G.}~\bibnamefont
  {{Desvignes}}},\ }\href {\doibase10.1093/mnras/stx1236} {\bibfield  {journal}
  {\bibinfo  {journal} {Mon. Not. Roy. Astron. Soc.}\ }\textbf {\bibinfo
  {volume} {471}},\ \bibinfo {pages} {1258--1258} (\bibinfo {year}
  {2017})}\BibitemShut {NoStop}%
\bibitem [{\citenamefont {Prager}\ \emph {et~al.}(2017)\citenamefont {Prager},
  \citenamefont {Ransom}, \citenamefont {Freire}, \citenamefont {Hessels},
  \citenamefont {Stairs}, \citenamefont {Arras},\ and\ \citenamefont
  {Cadelano}}]{brf+17}%
  \BibitemOpen
  \bibfield  {author} {\bibinfo {author} {\bibfnamefont {Brian~J.}\
  \bibnamefont {Prager}}, \bibinfo {author} {\bibfnamefont {Scott~M.}\
  \bibnamefont {Ransom}}, \bibinfo {author} {\bibfnamefont {Paulo C.~C.}\
  \bibnamefont {Freire}}, \bibinfo {author} {\bibfnamefont {Jason W.~T.}\
  \bibnamefont {Hessels}}, \bibinfo {author} {\bibfnamefont {Ingrid~H.}\
  \bibnamefont {Stairs}}, \bibinfo {author} {\bibfnamefont {Phil}\ \bibnamefont
  {Arras}}, \ and\ \bibinfo {author} {\bibfnamefont {Mario}\ \bibnamefont
  {Cadelano}},\ }\href {\doibase10.3847/1538-4357/aa7ed7} {\bibfield  {journal}
  {\bibinfo  {journal} {Astrophys. J.}\ }\textbf {\bibinfo {volume} {845}},\
  \bibinfo {pages} {148} (\bibinfo {year} {2017})},\ \Eprint
  {http://arxiv.org/abs/1612.04395} {arXiv:1612.04395
  [astro-ph.SR]}\BibitemShut {NoStop}%
\bibitem [{\citenamefont {{Chandrasekhar}}\ and\ \citenamefont
  {{Ferrari}}(1991)}]{1991RSPSA.434..449C}%
  \BibitemOpen
  \bibfield  {author} {\bibinfo {author} {\bibfnamefont {S.}~\bibnamefont
  {{Chandrasekhar}}}\ and\ \bibinfo {author} {\bibfnamefont {V.}~\bibnamefont
  {{Ferrari}}},\ }\href {\doibase10.1098/rspa.1991.0104} {\bibfield  {journal}
  {\bibinfo  {journal} {Proc. Roy. Soc. Lond.}\ }\textbf {\bibinfo {volume}
  {A434}},\ \bibinfo {pages} {449--457} (\bibinfo {year} {1991})}\BibitemShut
  {NoStop}%
\bibitem [{\citenamefont {Chandrasekhar}\ and\ \citenamefont
  {Ferrari}(1992)}]{Chandrasekhar:1992ey}%
  \BibitemOpen
  \bibfield  {author} {\bibinfo {author} {\bibfnamefont {Subrahmanyan}\
  \bibnamefont {Chandrasekhar}}\ and\ \bibinfo {author} {\bibfnamefont
  {V.}~\bibnamefont {Ferrari}},\ }\href {\doibase10.1098/rspa.1992.0051}
  {\bibfield  {journal} {\bibinfo  {journal} {Proc. Roy. Soc. Lond.}\ }\textbf
  {\bibinfo {volume} {A437}},\ \bibinfo {pages} {133--149} (\bibinfo {year}
  {1992})}\BibitemShut {NoStop}%
\bibitem [{\citenamefont {Correia}\ and\ \citenamefont
  {Cardoso}(2018)}]{Correia:2018apm}%
  \BibitemOpen
  \bibfield  {author} {\bibinfo {author} {\bibfnamefont {Miguel~R.}\
  \bibnamefont {Correia}}\ and\ \bibinfo {author} {\bibfnamefont {Vitor}\
  \bibnamefont {Cardoso}},\ }\href {\doibase10.1103/PhysRevD.97.084030}
  {\bibfield  {journal} {\bibinfo  {journal} {Phys. Rev.}\ }\textbf {\bibinfo
  {volume} {D97}},\ \bibinfo {pages} {084030} (\bibinfo {year} {2018})},\
  \Eprint {http://arxiv.org/abs/1802.07735} {arXiv:1802.07735
  [gr-qc]}\BibitemShut {NoStop}%
\bibitem [{\citenamefont {Vicente}\ \emph {et~al.}(2018)\citenamefont
  {Vicente}, \citenamefont {Cardoso},\ and\ \citenamefont
  {Lopes}}]{Vicente:2018mxl}%
  \BibitemOpen
  \bibfield  {author} {\bibinfo {author} {\bibfnamefont {Rodrigo}\ \bibnamefont
  {Vicente}}, \bibinfo {author} {\bibfnamefont {Vitor}\ \bibnamefont
  {Cardoso}}, \ and\ \bibinfo {author} {\bibfnamefont {Jorge~C.}\ \bibnamefont
  {Lopes}},\ }\href {\doibase10.1103/PhysRevD.97.084032} {\bibfield  {journal}
  {\bibinfo  {journal} {Phys. Rev.}\ }\textbf {\bibinfo {volume} {D97}},\
  \bibinfo {pages} {084032} (\bibinfo {year} {2018})},\ \Eprint
  {http://arxiv.org/abs/1803.08060} {arXiv:1803.08060 [gr-qc]}\BibitemShut
  {NoStop}%
\bibitem [{\citenamefont {Abedi}\ \emph
  {et~al.}(2017{\natexlab{a}})\citenamefont {Abedi}, \citenamefont {Dykaar},\
  and\ \citenamefont {Afshordi}}]{Abedi:2016hgu}%
  \BibitemOpen
  \bibfield  {author} {\bibinfo {author} {\bibfnamefont {Jahed}\ \bibnamefont
  {Abedi}}, \bibinfo {author} {\bibfnamefont {Hannah}\ \bibnamefont {Dykaar}},
  \ and\ \bibinfo {author} {\bibfnamefont {Niayesh}\ \bibnamefont {Afshordi}},\
  }\href {\doibase10.1103/PhysRevD.96.082004} {\bibfield  {journal} {\bibinfo
  {journal} {Phys. Rev.}\ }\textbf {\bibinfo {volume} {D96}},\ \bibinfo {pages}
  {082004} (\bibinfo {year} {2017}{\natexlab{a}})},\ \Eprint
  {http://arxiv.org/abs/1612.00266} {arXiv:1612.00266 [gr-qc]}\BibitemShut
  {NoStop}%
\bibitem [{\citenamefont {Pani}\ and\ \citenamefont
  {Ferrari}(2018)}]{Pani:2018flj}%
  \BibitemOpen
  \bibfield  {author} {\bibinfo {author} {\bibfnamefont {Paolo}\ \bibnamefont
  {Pani}}\ and\ \bibinfo {author} {\bibfnamefont {Valeria}\ \bibnamefont
  {Ferrari}},\ }\href@noop {} {\  (\bibinfo {year} {2018})},\ \Eprint
  {http://arxiv.org/abs/1804.01444} {arXiv:1804.01444 [gr-qc]}\BibitemShut
  {NoStop}%
\bibitem [{\citenamefont {Pani}(2015{\natexlab{a}})}]{Pani:2015tga}%
  \BibitemOpen
  \bibfield  {author} {\bibinfo {author} {\bibfnamefont {Paolo}\ \bibnamefont
  {Pani}},\ }\href {\doibase10.1103/PhysRevD.92.124030} {\bibfield  {journal}
  {\bibinfo  {journal} {Phys. Rev.}\ }\textbf {\bibinfo {volume} {D92}},\
  \bibinfo {pages} {124030} (\bibinfo {year} {2015}{\natexlab{a}})},\ \Eprint
  {http://arxiv.org/abs/1506.06050} {arXiv:1506.06050 [gr-qc]}\BibitemShut
  {NoStop}%
\bibitem [{\citenamefont {Uchikata}\ \emph {et~al.}(2016)\citenamefont
  {Uchikata}, \citenamefont {Yoshida},\ and\ \citenamefont
  {Pani}}]{Uchikata:2016qku}%
  \BibitemOpen
  \bibfield  {author} {\bibinfo {author} {\bibfnamefont {Nami}\ \bibnamefont
  {Uchikata}}, \bibinfo {author} {\bibfnamefont {Shijun}\ \bibnamefont
  {Yoshida}}, \ and\ \bibinfo {author} {\bibfnamefont {Paolo}\ \bibnamefont
  {Pani}},\ }\href {\doibase10.1103/PhysRevD.94.064015} {\bibfield  {journal}
  {\bibinfo  {journal} {Phys. Rev.}\ }\textbf {\bibinfo {volume} {D94}},\
  \bibinfo {pages} {064015} (\bibinfo {year} {2016})},\ \Eprint
  {http://arxiv.org/abs/1607.03593} {arXiv:1607.03593 [gr-qc]}\BibitemShut
  {NoStop}%
\bibitem [{\citenamefont {Yagi}\ and\ \citenamefont
  {Yunes}(2015)}]{Yagi:2015hda}%
  \BibitemOpen
  \bibfield  {author} {\bibinfo {author} {\bibfnamefont {Kent}\ \bibnamefont
  {Yagi}}\ and\ \bibinfo {author} {\bibfnamefont {Nicolás}\ \bibnamefont
  {Yunes}},\ }\href {\doibase10.1103/PhysRevD.91.123008} {\bibfield  {journal}
  {\bibinfo  {journal} {Phys. Rev.}\ }\textbf {\bibinfo {volume} {D91}},\
  \bibinfo {pages} {123008} (\bibinfo {year} {2015})},\ \Eprint
  {http://arxiv.org/abs/1503.02726} {arXiv:1503.02726 [gr-qc]}\BibitemShut
  {NoStop}%
\bibitem [{\citenamefont {Yagi}\ and\ \citenamefont
  {Yunes}(2013)}]{Yagi:2013bca}%
  \BibitemOpen
  \bibfield  {author} {\bibinfo {author} {\bibfnamefont {Kent}\ \bibnamefont
  {Yagi}}\ and\ \bibinfo {author} {\bibfnamefont {Nicolas}\ \bibnamefont
  {Yunes}},\ }\href {\doibase10.1126/science.1236462} {\bibfield  {journal}
  {\bibinfo  {journal} {Science}\ }\textbf {\bibinfo {volume} {341}},\ \bibinfo
  {pages} {365--368} (\bibinfo {year} {2013})},\ \Eprint
  {http://arxiv.org/abs/1302.4499} {arXiv:1302.4499 [gr-qc]}\BibitemShut
  {NoStop}%
\bibitem [{\citenamefont {Yagi}\ and\ \citenamefont
  {Yunes}(2017)}]{Yagi:2016bkt}%
  \BibitemOpen
  \bibfield  {author} {\bibinfo {author} {\bibfnamefont {Kent}\ \bibnamefont
  {Yagi}}\ and\ \bibinfo {author} {\bibfnamefont {Nicolás}\ \bibnamefont
  {Yunes}},\ }\href {\doibase10.1016/j.physrep.2017.03.002} {\bibfield
  {journal} {\bibinfo  {journal} {Phys. Rept.}\ }\textbf {\bibinfo {volume}
  {681}},\ \bibinfo {pages} {1--72} (\bibinfo {year} {2017})},\ \Eprint
  {http://arxiv.org/abs/1608.02582} {arXiv:1608.02582 [gr-qc]}\BibitemShut
  {NoStop}%
\bibitem [{\citenamefont {Alvi}(2001)}]{Alvi:2001mx}%
  \BibitemOpen
  \bibfield  {author} {\bibinfo {author} {\bibfnamefont {Kashif}\ \bibnamefont
  {Alvi}},\ }\href {\doibase10.1103/PhysRevD.64.104020} {\bibfield  {journal}
  {\bibinfo  {journal} {Phys. Rev.}\ }\textbf {\bibinfo {volume} {D64}},\
  \bibinfo {pages} {104020} (\bibinfo {year} {2001})},\ \Eprint
  {http://arxiv.org/abs/gr-qc/0107080} {arXiv:gr-qc/0107080
  [gr-qc]}\BibitemShut {NoStop}%
\bibitem [{\citenamefont {Hughes}(2001)}]{Hughes:2001jr}%
  \BibitemOpen
  \bibfield  {author} {\bibinfo {author} {\bibfnamefont {Scott~A.}\
  \bibnamefont {Hughes}},\ }\href {\doibase10.1103/PhysRevD.64.064004}
  {\bibfield  {journal} {\bibinfo  {journal} {Phys. Rev.}\ }\textbf {\bibinfo
  {volume} {D64}},\ \bibinfo {pages} {064004} (\bibinfo {year} {2001})},\
  \bibinfo {note} {[Erratum: Phys. Rev.D88,no.10,109902(2013)]},\ \Eprint
  {http://arxiv.org/abs/gr-qc/0104041} {arXiv:gr-qc/0104041
  [gr-qc]}\BibitemShut {NoStop}%
\bibitem [{\citenamefont {Taylor}\ and\ \citenamefont
  {Poisson}(2008)}]{Taylor:2008xy}%
  \BibitemOpen
  \bibfield  {author} {\bibinfo {author} {\bibfnamefont {Stephanne}\
  \bibnamefont {Taylor}}\ and\ \bibinfo {author} {\bibfnamefont {Eric}\
  \bibnamefont {Poisson}},\ }\href {\doibase10.1103/PhysRevD.78.084016}
  {\bibfield  {journal} {\bibinfo  {journal} {Phys. Rev.}\ }\textbf {\bibinfo
  {volume} {D78}},\ \bibinfo {pages} {084016} (\bibinfo {year} {2008})},\
  \Eprint {http://arxiv.org/abs/0806.3052} {arXiv:0806.3052
  [gr-qc]}\BibitemShut {NoStop}%
\bibitem [{\citenamefont {Poisson}(2009)}]{Poisson:2009di}%
  \BibitemOpen
  \bibfield  {author} {\bibinfo {author} {\bibfnamefont {Eric}\ \bibnamefont
  {Poisson}},\ }\href {\doibase10.1103/PhysRevD.80.064029} {\bibfield
  {journal} {\bibinfo  {journal} {Phys. Rev.}\ }\textbf {\bibinfo {volume}
  {D80}},\ \bibinfo {pages} {064029} (\bibinfo {year} {2009})},\ \Eprint
  {http://arxiv.org/abs/0907.0874} {arXiv:0907.0874 [gr-qc]}\BibitemShut
  {NoStop}%
\bibitem [{\citenamefont {Chatziioannou}\ \emph {et~al.}(2013)\citenamefont
  {Chatziioannou}, \citenamefont {Poisson},\ and\ \citenamefont
  {Yunes}}]{Chatziioannou:2012gq}%
  \BibitemOpen
  \bibfield  {author} {\bibinfo {author} {\bibfnamefont {Katerina}\
  \bibnamefont {Chatziioannou}}, \bibinfo {author} {\bibfnamefont {Eric}\
  \bibnamefont {Poisson}}, \ and\ \bibinfo {author} {\bibfnamefont {Nicolas}\
  \bibnamefont {Yunes}},\ }\href {\doibase10.1103/PhysRevD.87.044022}
  {\bibfield  {journal} {\bibinfo  {journal} {Phys. Rev.}\ }\textbf {\bibinfo
  {volume} {D87}},\ \bibinfo {pages} {044022} (\bibinfo {year} {2013})},\
  \Eprint {http://arxiv.org/abs/1211.1686} {arXiv:1211.1686
  [gr-qc]}\BibitemShut {NoStop}%
\bibitem [{\citenamefont {Cardoso}\ and\ \citenamefont
  {Pani}(2013)}]{Cardoso:2012zn}%
  \BibitemOpen
  \bibfield  {author} {\bibinfo {author} {\bibfnamefont {Vitor}\ \bibnamefont
  {Cardoso}}\ and\ \bibinfo {author} {\bibfnamefont {Paolo}\ \bibnamefont
  {Pani}},\ }\href {\doibase10.1088/0264-9381/30/4/045011} {\bibfield
  {journal} {\bibinfo  {journal} {Class. Quant. Grav.}\ }\textbf {\bibinfo
  {volume} {30}},\ \bibinfo {pages} {045011} (\bibinfo {year} {2013})},\
  \Eprint {http://arxiv.org/abs/1205.3184} {arXiv:1205.3184
  [gr-qc]}\BibitemShut {NoStop}%
\bibitem [{\citenamefont {Maselli}\ \emph {et~al.}(2018)\citenamefont
  {Maselli}, \citenamefont {Pani}, \citenamefont {Cardoso}, \citenamefont
  {Abdelsalhin}, \citenamefont {Gualtieri},\ and\ \citenamefont
  {Ferrari}}]{Maselli:2017cmm}%
  \BibitemOpen
  \bibfield  {author} {\bibinfo {author} {\bibfnamefont {Andrea}\ \bibnamefont
  {Maselli}}, \bibinfo {author} {\bibfnamefont {Paolo}\ \bibnamefont {Pani}},
  \bibinfo {author} {\bibfnamefont {Vitor}\ \bibnamefont {Cardoso}}, \bibinfo
  {author} {\bibfnamefont {Tiziano}\ \bibnamefont {Abdelsalhin}}, \bibinfo
  {author} {\bibfnamefont {Leonardo}\ \bibnamefont {Gualtieri}}, \ and\
  \bibinfo {author} {\bibfnamefont {Valeria}\ \bibnamefont {Ferrari}},\ }\href
  {\doibase10.1103/PhysRevLett.120.081101} {\bibfield  {journal} {\bibinfo
  {journal} {Phys. Rev. Lett.}\ }\textbf {\bibinfo {volume} {120}},\ \bibinfo
  {pages} {081101} (\bibinfo {year} {2018})},\ \Eprint
  {http://arxiv.org/abs/1703.10612} {arXiv:1703.10612 [gr-qc]}\BibitemShut
  {NoStop}%
\bibitem [{\citenamefont {Binnington}\ and\ \citenamefont
  {Poisson}(2009)}]{Binnington:2009bb}%
  \BibitemOpen
  \bibfield  {author} {\bibinfo {author} {\bibfnamefont {Taylor}\ \bibnamefont
  {Binnington}}\ and\ \bibinfo {author} {\bibfnamefont {Eric}\ \bibnamefont
  {Poisson}},\ }\href {\doibase10.1103/PhysRevD.80.084018} {\bibfield
  {journal} {\bibinfo  {journal} {Phys. Rev.}\ }\textbf {\bibinfo {volume}
  {D80}},\ \bibinfo {pages} {084018} (\bibinfo {year} {2009})},\ \Eprint
  {http://arxiv.org/abs/0906.1366} {arXiv:0906.1366 [gr-qc]}\BibitemShut
  {NoStop}%
\bibitem [{\citenamefont {Damour}\ and\ \citenamefont
  {Nagar}(2009)}]{Damour:2009vw}%
  \BibitemOpen
  \bibfield  {author} {\bibinfo {author} {\bibfnamefont {Thibault}\
  \bibnamefont {Damour}}\ and\ \bibinfo {author} {\bibfnamefont {Alessandro}\
  \bibnamefont {Nagar}},\ }\href {\doibase10.1103/PhysRevD.80.084035}
  {\bibfield  {journal} {\bibinfo  {journal} {Phys. Rev.}\ }\textbf {\bibinfo
  {volume} {D80}},\ \bibinfo {pages} {084035} (\bibinfo {year} {2009})},\
  \Eprint {http://arxiv.org/abs/0906.0096} {arXiv:0906.0096
  [gr-qc]}\BibitemShut {NoStop}%
\bibitem [{\citenamefont {Fang}\ and\ \citenamefont
  {Lovelace}(2005)}]{Fang:2005qq}%
  \BibitemOpen
  \bibfield  {author} {\bibinfo {author} {\bibfnamefont {Hua}\ \bibnamefont
  {Fang}}\ and\ \bibinfo {author} {\bibfnamefont {Geoffrey}\ \bibnamefont
  {Lovelace}},\ }\href {\doibase10.1103/PhysRevD.72.124016} {\bibfield
  {journal} {\bibinfo  {journal} {Phys. Rev.}\ }\textbf {\bibinfo {volume}
  {D72}},\ \bibinfo {pages} {124016} (\bibinfo {year} {2005})},\ \Eprint
  {http://arxiv.org/abs/gr-qc/0505156} {arXiv:gr-qc/0505156
  [gr-qc]}\BibitemShut {NoStop}%
\bibitem [{\citenamefont {G{\"u}rlebeck}(2015)}]{Gurlebeck:2015xpa}%
  \BibitemOpen
  \bibfield  {author} {\bibinfo {author} {\bibfnamefont {Norman}\ \bibnamefont
  {G{\"u}rlebeck}},\ }\href {\doibase10.1103/PhysRevLett.114.151102} {\bibfield
   {journal} {\bibinfo  {journal} {Phys. Rev. Lett.}\ }\textbf {\bibinfo
  {volume} {114}},\ \bibinfo {pages} {151102} (\bibinfo {year} {2015})},\
  \Eprint {http://arxiv.org/abs/1503.03240} {arXiv:1503.03240
  [gr-qc]}\BibitemShut {NoStop}%
\bibitem [{\citenamefont {Poisson}(2015)}]{Poisson:2014gka}%
  \BibitemOpen
  \bibfield  {author} {\bibinfo {author} {\bibfnamefont {Eric}\ \bibnamefont
  {Poisson}},\ }\href {\doibase10.1103/PhysRevD.91.044004} {\bibfield
  {journal} {\bibinfo  {journal} {Phys. Rev.}\ }\textbf {\bibinfo {volume}
  {D91}},\ \bibinfo {pages} {044004} (\bibinfo {year} {2015})},\ \Eprint
  {http://arxiv.org/abs/1411.4711} {arXiv:1411.4711 [gr-qc]}\BibitemShut
  {NoStop}%
\bibitem [{\citenamefont {Pani}\ \emph
  {et~al.}(2015{\natexlab{a}})\citenamefont {Pani}, \citenamefont {Gualtieri},
  \citenamefont {Maselli},\ and\ \citenamefont {Ferrari}}]{Pani:2015hfa}%
  \BibitemOpen
  \bibfield  {author} {\bibinfo {author} {\bibfnamefont {Paolo}\ \bibnamefont
  {Pani}}, \bibinfo {author} {\bibfnamefont {Leonardo}\ \bibnamefont
  {Gualtieri}}, \bibinfo {author} {\bibfnamefont {Andrea}\ \bibnamefont
  {Maselli}}, \ and\ \bibinfo {author} {\bibfnamefont {Valeria}\ \bibnamefont
  {Ferrari}},\ }\href {\doibase10.1103/PhysRevD.92.024010} {\bibfield
  {journal} {\bibinfo  {journal} {Phys. Rev.}\ }\textbf {\bibinfo {volume}
  {D92}},\ \bibinfo {pages} {024010} (\bibinfo {year} {2015}{\natexlab{a}})},\
  \Eprint {http://arxiv.org/abs/1503.07365} {arXiv:1503.07365
  [gr-qc]}\BibitemShut {NoStop}%
\bibitem [{\citenamefont {Pani}\ \emph
  {et~al.}(2015{\natexlab{b}})\citenamefont {Pani}, \citenamefont {Gualtieri},\
  and\ \citenamefont {Ferrari}}]{Pani:2015nua}%
  \BibitemOpen
  \bibfield  {author} {\bibinfo {author} {\bibfnamefont {Paolo}\ \bibnamefont
  {Pani}}, \bibinfo {author} {\bibfnamefont {Leonardo}\ \bibnamefont
  {Gualtieri}}, \ and\ \bibinfo {author} {\bibfnamefont {Valeria}\ \bibnamefont
  {Ferrari}},\ }\href {\doibase10.1103/PhysRevD.92.124003} {\bibfield
  {journal} {\bibinfo  {journal} {Phys. Rev.}\ }\textbf {\bibinfo {volume}
  {D92}},\ \bibinfo {pages} {124003} (\bibinfo {year} {2015}{\natexlab{b}})},\
  \Eprint {http://arxiv.org/abs/1509.02171} {arXiv:1509.02171
  [gr-qc]}\BibitemShut {NoStop}%
\bibitem [{\citenamefont {Porto}(2016{\natexlab{b}})}]{Porto:2016zng}%
  \BibitemOpen
  \bibfield  {author} {\bibinfo {author} {\bibfnamefont {Rafael~A.}\
  \bibnamefont {Porto}},\ }\href {\doibase10.1002/prop.201600064} {\bibfield
  {journal} {\bibinfo  {journal} {Fortsch. Phys.}\ }\textbf {\bibinfo {volume}
  {64}},\ \bibinfo {pages} {723--729} (\bibinfo {year} {2016}{\natexlab{b}})},\
  \Eprint {http://arxiv.org/abs/1606.08895} {arXiv:1606.08895
  [gr-qc]}\BibitemShut {NoStop}%
\bibitem [{\citenamefont {Cardoso}\ \emph
  {et~al.}(2017{\natexlab{a}})\citenamefont {Cardoso}, \citenamefont {Franzin},
  \citenamefont {Maselli}, \citenamefont {Pani},\ and\ \citenamefont
  {Raposo}}]{Cardoso:2017cfl}%
  \BibitemOpen
  \bibfield  {author} {\bibinfo {author} {\bibfnamefont {Vitor}\ \bibnamefont
  {Cardoso}}, \bibinfo {author} {\bibfnamefont {Edgardo}\ \bibnamefont
  {Franzin}}, \bibinfo {author} {\bibfnamefont {Andrea}\ \bibnamefont
  {Maselli}}, \bibinfo {author} {\bibfnamefont {Paolo}\ \bibnamefont {Pani}}, \
  and\ \bibinfo {author} {\bibfnamefont {Guilherme}\ \bibnamefont {Raposo}},\
  }\href {\doibase10.1103/PhysRevD.95.084014} {\bibfield  {journal} {\bibinfo
  {journal} {Phys. Rev.}\ }\textbf {\bibinfo {volume} {D95}},\ \bibinfo {pages}
  {084014} (\bibinfo {year} {2017}{\natexlab{a}})},\ \Eprint
  {http://arxiv.org/abs/1701.01116} {arXiv:1701.01116 [gr-qc]}\BibitemShut
  {NoStop}%
\bibitem [{\citenamefont {Flanagan}\ and\ \citenamefont
  {Hinderer}(2008)}]{Flanagan:2007ix}%
  \BibitemOpen
  \bibfield  {author} {\bibinfo {author} {\bibfnamefont {Eanna~E.}\
  \bibnamefont {Flanagan}}\ and\ \bibinfo {author} {\bibfnamefont {Tanja}\
  \bibnamefont {Hinderer}},\ }\href {\doibase10.1103/PhysRevD.77.021502}
  {\bibfield  {journal} {\bibinfo  {journal} {Phys. Rev.}\ }\textbf {\bibinfo
  {volume} {D77}},\ \bibinfo {pages} {021502} (\bibinfo {year} {2008})},\
  \Eprint {http://arxiv.org/abs/0709.1915} {arXiv:0709.1915
  [astro-ph]}\BibitemShut {NoStop}%
\bibitem [{\citenamefont {Hinderer}(2008)}]{Hinderer:2007mb}%
  \BibitemOpen
  \bibfield  {author} {\bibinfo {author} {\bibfnamefont {Tanja}\ \bibnamefont
  {Hinderer}},\ }\href {\doibase10.1086/533487} {\bibfield  {journal} {\bibinfo
   {journal} {Astrophys. J.}\ }\textbf {\bibinfo {volume} {677}},\ \bibinfo
  {pages} {1216--1220} (\bibinfo {year} {2008})},\ \bibinfo {note} {{Erratum:
  {\it ibid.} \href{https://dx.doi.org/10.1088/0004-637X/697/1/964}{{\bf 697},
  964 (2009)}}},\ \Eprint {http://arxiv.org/abs/0711.2420} {arXiv:0711.2420
  [astro-ph]}\BibitemShut {NoStop}%
\bibitem [{\citenamefont {Johnson-McDaniel}\ \emph {et~al.}(2018)\citenamefont
  {Johnson-McDaniel}, \citenamefont {Mukherjee}, \citenamefont {Kashyap},
  \citenamefont {Ajith}, \citenamefont {Del~Pozzo},\ and\ \citenamefont
  {Vitale}}]{Johnson-McDaniel:2018uvs}%
  \BibitemOpen
  \bibfield  {author} {\bibinfo {author} {\bibfnamefont {Nathan~K.}\
  \bibnamefont {Johnson-McDaniel}}, \bibinfo {author} {\bibfnamefont {Arunava}\
  \bibnamefont {Mukherjee}}, \bibinfo {author} {\bibfnamefont {Rahul}\
  \bibnamefont {Kashyap}}, \bibinfo {author} {\bibfnamefont {Parameswaran}\
  \bibnamefont {Ajith}}, \bibinfo {author} {\bibfnamefont {Walter}\
  \bibnamefont {Del~Pozzo}}, \ and\ \bibinfo {author} {\bibfnamefont
  {Salvatore}\ \bibnamefont {Vitale}},\ }\href@noop {} {\  (\bibinfo {year}
  {2018})},\ \Eprint {http://arxiv.org/abs/1804.08026} {arXiv:1804.08026
  [gr-qc]}\BibitemShut {NoStop}%
\bibitem [{\citenamefont {Pani}\ \emph {et~al.}(2010)\citenamefont {Pani},
  \citenamefont {Berti}, \citenamefont {Cardoso}, \citenamefont {Chen},\ and\
  \citenamefont {Norte}}]{Pani:2010em}%
  \BibitemOpen
  \bibfield  {author} {\bibinfo {author} {\bibfnamefont {Paolo}\ \bibnamefont
  {Pani}}, \bibinfo {author} {\bibfnamefont {Emanuele}\ \bibnamefont {Berti}},
  \bibinfo {author} {\bibfnamefont {Vitor}\ \bibnamefont {Cardoso}}, \bibinfo
  {author} {\bibfnamefont {Yanbei}\ \bibnamefont {Chen}}, \ and\ \bibinfo
  {author} {\bibfnamefont {Richard}\ \bibnamefont {Norte}},\ }\href
  {\doibase10.1103/PhysRevD.81.084011} {\bibfield  {journal} {\bibinfo
  {journal} {Phys. Rev.}\ }\textbf {\bibinfo {volume} {D81}},\ \bibinfo {pages}
  {084011} (\bibinfo {year} {2010})},\ \Eprint {http://arxiv.org/abs/1001.3031}
  {arXiv:1001.3031 [gr-qc]}\BibitemShut {NoStop}%
\bibitem [{\citenamefont {Macedo}\ \emph
  {et~al.}(2013{\natexlab{b}})\citenamefont {Macedo}, \citenamefont {Pani},
  \citenamefont {Cardoso},\ and\ \citenamefont {Crispino}}]{Macedo:2013qea}%
  \BibitemOpen
  \bibfield  {author} {\bibinfo {author} {\bibfnamefont {Caio F.~B.}\
  \bibnamefont {Macedo}}, \bibinfo {author} {\bibfnamefont {Paolo}\
  \bibnamefont {Pani}}, \bibinfo {author} {\bibfnamefont {Vitor}\ \bibnamefont
  {Cardoso}}, \ and\ \bibinfo {author} {\bibfnamefont {Luís C.~B.}\
  \bibnamefont {Crispino}},\ }\href {\doibase10.1088/0004-637X/774/1/48}
  {\bibfield  {journal} {\bibinfo  {journal} {Astrophys. J.}\ }\textbf
  {\bibinfo {volume} {774}},\ \bibinfo {pages} {48} (\bibinfo {year}
  {2013}{\natexlab{b}})},\ \Eprint {http://arxiv.org/abs/1302.2646}
  {arXiv:1302.2646 [gr-qc]}\BibitemShut {NoStop}%
\bibitem [{\citenamefont {Nakano}\ \emph {et~al.}(2017)\citenamefont {Nakano},
  \citenamefont {Sago}, \citenamefont {Tagoshi},\ and\ \citenamefont
  {Tanaka}}]{Nakano:2017fvh}%
  \BibitemOpen
  \bibfield  {author} {\bibinfo {author} {\bibfnamefont {Hiroyuki}\
  \bibnamefont {Nakano}}, \bibinfo {author} {\bibfnamefont {Norichika}\
  \bibnamefont {Sago}}, \bibinfo {author} {\bibfnamefont {Hideyuki}\
  \bibnamefont {Tagoshi}}, \ and\ \bibinfo {author} {\bibfnamefont {Takahiro}\
  \bibnamefont {Tanaka}},\ }\href {\doibase10.1093/ptep/ptx093} {\bibfield
  {journal} {\bibinfo  {journal} {PTEP}\ }\textbf {\bibinfo {volume} {2017}},\
  \bibinfo {pages} {071E01} (\bibinfo {year} {2017})},\ \Eprint
  {http://arxiv.org/abs/1704.07175} {arXiv:1704.07175 [gr-qc]}\BibitemShut
  {NoStop}%
\bibitem [{\citenamefont {Mark}\ \emph {et~al.}(2017)\citenamefont {Mark},
  \citenamefont {Zimmerman}, \citenamefont {Du},\ and\ \citenamefont
  {Chen}}]{Mark:2017dnq}%
  \BibitemOpen
  \bibfield  {author} {\bibinfo {author} {\bibfnamefont {Zachary}\ \bibnamefont
  {Mark}}, \bibinfo {author} {\bibfnamefont {Aaron}\ \bibnamefont {Zimmerman}},
  \bibinfo {author} {\bibfnamefont {Song~Ming}\ \bibnamefont {Du}}, \ and\
  \bibinfo {author} {\bibfnamefont {Yanbei}\ \bibnamefont {Chen}},\ }\href
  {\doibase10.1103/PhysRevD.96.084002} {\bibfield  {journal} {\bibinfo
  {journal} {Phys. Rev.}\ }\textbf {\bibinfo {volume} {D96}},\ \bibinfo {pages}
  {084002} (\bibinfo {year} {2017})},\ \Eprint
  {http://arxiv.org/abs/1706.06155} {arXiv:1706.06155 [gr-qc]}\BibitemShut
  {NoStop}%
\bibitem [{\citenamefont {Voelkel}\ and\ \citenamefont
  {Kokkotas}(2017)}]{Volkel:2017kfj}%
  \BibitemOpen
  \bibfield  {author} {\bibinfo {author} {\bibfnamefont {Sebastian~H.}\
  \bibnamefont {Voelkel}}\ and\ \bibinfo {author} {\bibfnamefont {Kostas~D.}\
  \bibnamefont {Kokkotas}},\ }\href {\doibase10.1088/1361-6382/aa82de}
  {\bibfield  {journal} {\bibinfo  {journal} {Class. Quant. Grav.}\ }\textbf
  {\bibinfo {volume} {34}},\ \bibinfo {pages} {175015} (\bibinfo {year}
  {2017})},\ \Eprint {http://arxiv.org/abs/1704.07517} {arXiv:1704.07517
  [gr-qc]}\BibitemShut {NoStop}%
\bibitem [{\citenamefont {Bueno}\ \emph {et~al.}(2018)\citenamefont {Bueno},
  \citenamefont {Cano}, \citenamefont {Goelen}, \citenamefont {Hertog},\ and\
  \citenamefont {Vercnocke}}]{Bueno:2017hyj}%
  \BibitemOpen
  \bibfield  {author} {\bibinfo {author} {\bibfnamefont {Pablo}\ \bibnamefont
  {Bueno}}, \bibinfo {author} {\bibfnamefont {Pablo~A.}\ \bibnamefont {Cano}},
  \bibinfo {author} {\bibfnamefont {Frederik}\ \bibnamefont {Goelen}}, \bibinfo
  {author} {\bibfnamefont {Thomas}\ \bibnamefont {Hertog}}, \ and\ \bibinfo
  {author} {\bibfnamefont {Bert}\ \bibnamefont {Vercnocke}},\ }\href
  {\doibase10.1103/PhysRevD.97.024040} {\bibfield  {journal} {\bibinfo
  {journal} {Phys. Rev.}\ }\textbf {\bibinfo {volume} {D97}},\ \bibinfo {pages}
  {024040} (\bibinfo {year} {2018})},\ \Eprint
  {http://arxiv.org/abs/1711.00391} {arXiv:1711.00391 [gr-qc]}\BibitemShut
  {NoStop}%
\bibitem [{\citenamefont {Maselli}\ \emph
  {et~al.}(2017{\natexlab{a}})\citenamefont {Maselli}, \citenamefont
  {Voelkel},\ and\ \citenamefont {Kokkotas}}]{Maselli:2017tfq}%
  \BibitemOpen
  \bibfield  {author} {\bibinfo {author} {\bibfnamefont {Andrea}\ \bibnamefont
  {Maselli}}, \bibinfo {author} {\bibfnamefont {Sebastian~H.}\ \bibnamefont
  {Voelkel}}, \ and\ \bibinfo {author} {\bibfnamefont {Kostas~D.}\ \bibnamefont
  {Kokkotas}},\ }\href {\doibase10.1103/PhysRevD.96.064045} {\bibfield
  {journal} {\bibinfo  {journal} {Phys. Rev.}\ }\textbf {\bibinfo {volume}
  {D96}},\ \bibinfo {pages} {064045} (\bibinfo {year} {2017}{\natexlab{a}})},\
  \Eprint {http://arxiv.org/abs/1708.02217} {arXiv:1708.02217
  [gr-qc]}\BibitemShut {NoStop}%
\bibitem [{\citenamefont {Wang}\ \emph
  {et~al.}(2018{\natexlab{a}})\citenamefont {Wang}, \citenamefont {Li},
  \citenamefont {Zhang}, \citenamefont {Zhou},\ and\ \citenamefont
  {Piao}}]{Wang:2018mlp}%
  \BibitemOpen
  \bibfield  {author} {\bibinfo {author} {\bibfnamefont {Yu-Tong}\ \bibnamefont
  {Wang}}, \bibinfo {author} {\bibfnamefont {Zhi-Peng}\ \bibnamefont {Li}},
  \bibinfo {author} {\bibfnamefont {Jun}\ \bibnamefont {Zhang}}, \bibinfo
  {author} {\bibfnamefont {Shuang-Yong}\ \bibnamefont {Zhou}}, \ and\ \bibinfo
  {author} {\bibfnamefont {Yun-Song}\ \bibnamefont {Piao}},\ }\href
  {\doibase10.1140/epjc/s10052-018-5974-y} {\bibfield  {journal} {\bibinfo
  {journal} {Eur. Phys. J.}\ }\textbf {\bibinfo {volume} {C78}},\ \bibinfo
  {pages} {482} (\bibinfo {year} {2018}{\natexlab{a}})},\ \Eprint
  {http://arxiv.org/abs/1802.02003} {arXiv:1802.02003 [gr-qc]}\BibitemShut
  {NoStop}%
\bibitem [{\citenamefont {Wang}\ and\ \citenamefont
  {Afshordi}(2018)}]{Wang:2018gin}%
  \BibitemOpen
  \bibfield  {author} {\bibinfo {author} {\bibfnamefont {Qingwen}\ \bibnamefont
  {Wang}}\ and\ \bibinfo {author} {\bibfnamefont {Niayesh}\ \bibnamefont
  {Afshordi}},\ }\href@noop {} {\  (\bibinfo {year} {2018})},\ \Eprint
  {http://arxiv.org/abs/1803.02845} {arXiv:1803.02845 [gr-qc]}\BibitemShut
  {NoStop}%
\bibitem [{\citenamefont {Ashton}\ \emph {et~al.}(2016)\citenamefont {Ashton},
  \citenamefont {Birnholtz}, \citenamefont {Cabero}, \citenamefont {Capano},
  \citenamefont {Dent}, \citenamefont {Krishnan}, \citenamefont {Meadors},
  \citenamefont {Nielsen}, \citenamefont {Nitz},\ and\ \citenamefont
  {Westerweck}}]{Ashton:2016xff}%
  \BibitemOpen
  \bibfield  {author} {\bibinfo {author} {\bibfnamefont {Gregory}\ \bibnamefont
  {Ashton}}, \bibinfo {author} {\bibfnamefont {Ofek}\ \bibnamefont
  {Birnholtz}}, \bibinfo {author} {\bibfnamefont {Miriam}\ \bibnamefont
  {Cabero}}, \bibinfo {author} {\bibfnamefont {Collin}\ \bibnamefont {Capano}},
  \bibinfo {author} {\bibfnamefont {Thomas}\ \bibnamefont {Dent}}, \bibinfo
  {author} {\bibfnamefont {Badri}\ \bibnamefont {Krishnan}}, \bibinfo {author}
  {\bibfnamefont {Grant~David}\ \bibnamefont {Meadors}}, \bibinfo {author}
  {\bibfnamefont {Alex~B.}\ \bibnamefont {Nielsen}}, \bibinfo {author}
  {\bibfnamefont {Alex}\ \bibnamefont {Nitz}}, \ and\ \bibinfo {author}
  {\bibfnamefont {Julian}\ \bibnamefont {Westerweck}},\ }\href@noop {} {\
  (\bibinfo {year} {2016})},\ \Eprint {http://arxiv.org/abs/1612.05625}
  {arXiv:1612.05625 [gr-qc]}\BibitemShut {NoStop}%
\bibitem [{\citenamefont {Abedi}\ \emph
  {et~al.}(2017{\natexlab{b}})\citenamefont {Abedi}, \citenamefont {Dykaar},\
  and\ \citenamefont {Afshordi}}]{Abedi:2017isz}%
  \BibitemOpen
  \bibfield  {author} {\bibinfo {author} {\bibfnamefont {Jahed}\ \bibnamefont
  {Abedi}}, \bibinfo {author} {\bibfnamefont {Hannah}\ \bibnamefont {Dykaar}},
  \ and\ \bibinfo {author} {\bibfnamefont {Niayesh}\ \bibnamefont {Afshordi}},\
  }\href@noop {} {\  (\bibinfo {year} {2017}{\natexlab{b}})},\ \Eprint
  {http://arxiv.org/abs/1701.03485} {arXiv:1701.03485 [gr-qc]}\BibitemShut
  {NoStop}%
\bibitem [{\citenamefont {Conklin}\ \emph {et~al.}(2017)\citenamefont
  {Conklin}, \citenamefont {Holdom},\ and\ \citenamefont
  {Ren}}]{Conklin:2017lwb}%
  \BibitemOpen
  \bibfield  {author} {\bibinfo {author} {\bibfnamefont {Randy~S.}\
  \bibnamefont {Conklin}}, \bibinfo {author} {\bibfnamefont {Bob}\ \bibnamefont
  {Holdom}}, \ and\ \bibinfo {author} {\bibfnamefont {Jing}\ \bibnamefont
  {Ren}},\ }\href@noop {} {\  (\bibinfo {year} {2017})},\ \Eprint
  {http://arxiv.org/abs/1712.06517} {arXiv:1712.06517 [gr-qc]}\BibitemShut
  {NoStop}%
\bibitem [{\citenamefont {Westerweck}\ \emph {et~al.}(2018)\citenamefont
  {Westerweck}, \citenamefont {Nielsen}, \citenamefont {Fischer-Birnholtz},
  \citenamefont {Cabero}, \citenamefont {Capano}, \citenamefont {Dent},
  \citenamefont {Krishnan}, \citenamefont {Meadors},\ and\ \citenamefont
  {Nitz}}]{Westerweck:2017hus}%
  \BibitemOpen
  \bibfield  {author} {\bibinfo {author} {\bibfnamefont {Julian}\ \bibnamefont
  {Westerweck}}, \bibinfo {author} {\bibfnamefont {Alex}\ \bibnamefont
  {Nielsen}}, \bibinfo {author} {\bibfnamefont {Ofek}\ \bibnamefont
  {Fischer-Birnholtz}}, \bibinfo {author} {\bibfnamefont {Miriam}\ \bibnamefont
  {Cabero}}, \bibinfo {author} {\bibfnamefont {Collin}\ \bibnamefont {Capano}},
  \bibinfo {author} {\bibfnamefont {Thomas}\ \bibnamefont {Dent}}, \bibinfo
  {author} {\bibfnamefont {Badri}\ \bibnamefont {Krishnan}}, \bibinfo {author}
  {\bibfnamefont {Grant}\ \bibnamefont {Meadors}}, \ and\ \bibinfo {author}
  {\bibfnamefont {Alexander~H.}\ \bibnamefont {Nitz}},\ }\href
  {\doibase10.1103/PhysRevD.97.124037} {\bibfield  {journal} {\bibinfo
  {journal} {Phys. Rev.}\ }\textbf {\bibinfo {volume} {D97}},\ \bibinfo {pages}
  {124037} (\bibinfo {year} {2018})},\ \Eprint
  {http://arxiv.org/abs/1712.09966} {arXiv:1712.09966 [gr-qc]}\BibitemShut
  {NoStop}%
\bibitem [{\citenamefont {Abedi}\ \emph {et~al.}(2018)\citenamefont {Abedi},
  \citenamefont {Dykaar},\ and\ \citenamefont {Afshordi}}]{Abedi:2018pst}%
  \BibitemOpen
  \bibfield  {author} {\bibinfo {author} {\bibfnamefont {Jahed}\ \bibnamefont
  {Abedi}}, \bibinfo {author} {\bibfnamefont {Hannah}\ \bibnamefont {Dykaar}},
  \ and\ \bibinfo {author} {\bibfnamefont {Niayesh}\ \bibnamefont {Afshordi}},\
  }\href@noop {} {\  (\bibinfo {year} {2018})},\ \Eprint
  {http://arxiv.org/abs/1803.08565} {arXiv:1803.08565 [gr-qc]}\BibitemShut
  {NoStop}%
\bibitem [{\citenamefont {Nielsen}\ \emph {et~al.}(2018)\citenamefont
  {Nielsen}, \citenamefont {Capano}, \citenamefont {Birnholtz},\ and\
  \citenamefont {Westerweck}}]{Nielsen:2018lkf}%
  \BibitemOpen
  \bibfield  {author} {\bibinfo {author} {\bibfnamefont {Alex~B.}\ \bibnamefont
  {Nielsen}}, \bibinfo {author} {\bibfnamefont {Collin~D.}\ \bibnamefont
  {Capano}}, \bibinfo {author} {\bibfnamefont {Ofek}\ \bibnamefont
  {Birnholtz}}, \ and\ \bibinfo {author} {\bibfnamefont {Julian}\ \bibnamefont
  {Westerweck}},\ }\href@noop {} {\  (\bibinfo {year} {2018})},\ \Eprint
  {http://arxiv.org/abs/1811.04904} {arXiv:1811.04904 [gr-qc]}\BibitemShut
  {NoStop}%
\bibitem [{\citenamefont {Lo}\ \emph {et~al.}(2018)\citenamefont {Lo},
  \citenamefont {Li},\ and\ \citenamefont {Weinstein}}]{Lo:2018sep}%
  \BibitemOpen
  \bibfield  {author} {\bibinfo {author} {\bibfnamefont {R.~K.~L.}\
  \bibnamefont {Lo}}, \bibinfo {author} {\bibfnamefont {T.~G.~F.}\ \bibnamefont
  {Li}}, \ and\ \bibinfo {author} {\bibfnamefont {A.~J.}\ \bibnamefont
  {Weinstein}},\ }\href@noop {} {\  (\bibinfo {year} {2018})},\ \Eprint
  {http://arxiv.org/abs/1811.07431} {arXiv:1811.07431 [gr-qc]}\BibitemShut
  {NoStop}%
\bibitem [{\citenamefont {Tsang}\ \emph {et~al.}(2018)\citenamefont {Tsang},
  \citenamefont {Rollier}, \citenamefont {Ghosh}, \citenamefont {Samajdar},
  \citenamefont {Agathos}, \citenamefont {Chatziioannou}, \citenamefont
  {Cardoso}, \citenamefont {Khanna},\ and\ \citenamefont {Van
  Den~Broeck}}]{Tsang:2018uie}%
  \BibitemOpen
  \bibfield  {author} {\bibinfo {author} {\bibfnamefont {Ka~Wa}\ \bibnamefont
  {Tsang}}, \bibinfo {author} {\bibfnamefont {Michiel}\ \bibnamefont
  {Rollier}}, \bibinfo {author} {\bibfnamefont {Archisman}\ \bibnamefont
  {Ghosh}}, \bibinfo {author} {\bibfnamefont {Anuradha}\ \bibnamefont
  {Samajdar}}, \bibinfo {author} {\bibfnamefont {Michalis}\ \bibnamefont
  {Agathos}}, \bibinfo {author} {\bibfnamefont {Katerina}\ \bibnamefont
  {Chatziioannou}}, \bibinfo {author} {\bibfnamefont {Vitor}\ \bibnamefont
  {Cardoso}}, \bibinfo {author} {\bibfnamefont {Gaurav}\ \bibnamefont
  {Khanna}}, \ and\ \bibinfo {author} {\bibfnamefont {Chris}\ \bibnamefont {Van
  Den~Broeck}},\ }\href {\doibase10.1103/PhysRevD.98.024023} {\bibfield
  {journal} {\bibinfo  {journal} {Phys. Rev.}\ }\textbf {\bibinfo {volume}
  {D98}},\ \bibinfo {pages} {024023} (\bibinfo {year} {2018})},\ \Eprint
  {http://arxiv.org/abs/1804.04877} {arXiv:1804.04877 [gr-qc]}\BibitemShut
  {NoStop}%
\bibitem [{\citenamefont {Carroll}(2001)}]{Carroll:2000fy}%
  \BibitemOpen
  \bibfield  {author} {\bibinfo {author} {\bibfnamefont {Sean~M.}\ \bibnamefont
  {Carroll}},\ }\href {\doibase10.12942/lrr-2001-1} {\bibfield  {journal}
  {\bibinfo  {journal} {Living Rev. Rel.}\ }\textbf {\bibinfo {volume} {4}},\
  \bibinfo {pages} {1} (\bibinfo {year} {2001})},\ \Eprint
  {http://arxiv.org/abs/astro-ph/0004075} {arXiv:astro-ph/0004075
  [astro-ph]}\BibitemShut {NoStop}%
\bibitem [{\citenamefont {Copeland}\ \emph {et~al.}(2006)\citenamefont
  {Copeland}, \citenamefont {Sami},\ and\ \citenamefont
  {Tsujikawa}}]{Copeland:2006wr}%
  \BibitemOpen
  \bibfield  {author} {\bibinfo {author} {\bibfnamefont {Edmund~J.}\
  \bibnamefont {Copeland}}, \bibinfo {author} {\bibfnamefont {M.}~\bibnamefont
  {Sami}}, \ and\ \bibinfo {author} {\bibfnamefont {Shinji}\ \bibnamefont
  {Tsujikawa}},\ }\href {\doibase10.1142/S021827180600942X} {\bibfield
  {journal} {\bibinfo  {journal} {Int. J. Mod. Phys.}\ }\textbf {\bibinfo
  {volume} {D15}},\ \bibinfo {pages} {1753--1936} (\bibinfo {year} {2006})},\
  \Eprint {http://arxiv.org/abs/hep-th/0603057} {arXiv:hep-th/0603057
  [hep-th]}\BibitemShut {NoStop}%
\bibitem [{\citenamefont {Peter}(2012)}]{Peter:2012rz}%
  \BibitemOpen
  \bibfield  {author} {\bibinfo {author} {\bibfnamefont {Annika H.~G.}\
  \bibnamefont {Peter}},\ }\href@noop {} {\  (\bibinfo {year} {2012})},\
  \Eprint {http://arxiv.org/abs/1201.3942} {arXiv:1201.3942
  [astro-ph.CO]}\BibitemShut {NoStop}%
\bibitem [{\citenamefont {Kashlinsky}(2016)}]{Kashlinsky:2016sdv}%
  \BibitemOpen
  \bibfield  {author} {\bibinfo {author} {\bibfnamefont {A.}~\bibnamefont
  {Kashlinsky}},\ }\href {\doibase10.3847/2041-8205/823/2/L25} {\bibfield
  {journal} {\bibinfo  {journal} {Astrophys. J.}\ }\textbf {\bibinfo {volume}
  {823}},\ \bibinfo {pages} {L25} (\bibinfo {year} {2016})},\ \Eprint
  {http://arxiv.org/abs/1605.04023} {arXiv:1605.04023
  [astro-ph.CO]}\BibitemShut {NoStop}%
\bibitem [{\citenamefont {Wang}\ \emph
  {et~al.}(2018{\natexlab{b}})\citenamefont {Wang}, \citenamefont {Wang},
  \citenamefont {Huang},\ and\ \citenamefont {Li}}]{Wang:2016ana}%
  \BibitemOpen
  \bibfield  {author} {\bibinfo {author} {\bibfnamefont {Sai}\ \bibnamefont
  {Wang}}, \bibinfo {author} {\bibfnamefont {Yi-Fan}\ \bibnamefont {Wang}},
  \bibinfo {author} {\bibfnamefont {Qing-Guo}\ \bibnamefont {Huang}}, \ and\
  \bibinfo {author} {\bibfnamefont {Tjonnie G.~F.}\ \bibnamefont {Li}},\ }\href
  {\doibase10.1103/PhysRevLett.120.191102} {\bibfield  {journal} {\bibinfo
  {journal} {Phys. Rev. Lett.}\ }\textbf {\bibinfo {volume} {120}},\ \bibinfo
  {pages} {191102} (\bibinfo {year} {2018}{\natexlab{b}})},\ \Eprint
  {http://arxiv.org/abs/1610.08725} {arXiv:1610.08725
  [astro-ph.CO]}\BibitemShut {NoStop}%
\bibitem [{\citenamefont {Carr}\ and\ \citenamefont
  {Hawking}(1974)}]{Carr:1974nx}%
  \BibitemOpen
  \bibfield  {author} {\bibinfo {author} {\bibfnamefont {Bernard~J.}\
  \bibnamefont {Carr}}\ and\ \bibinfo {author} {\bibfnamefont {S.~W.}\
  \bibnamefont {Hawking}},\ }\href {\doibase10.1093/mnras/168.2.399} {\bibfield
   {journal} {\bibinfo  {journal} {Mon. Not. Roy. Astron. Soc.}\ }\textbf
  {\bibinfo {volume} {168}},\ \bibinfo {pages} {399--415} (\bibinfo {year}
  {1974})}\BibitemShut {NoStop}%
\bibitem [{\citenamefont {Carr}(1975)}]{Carr:1975qj}%
  \BibitemOpen
  \bibfield  {author} {\bibinfo {author} {\bibfnamefont {Bernard~J.}\
  \bibnamefont {Carr}},\ }\href {\doibase10.1086/153853} {\bibfield  {journal}
  {\bibinfo  {journal} {Astrophys. J.}\ }\textbf {\bibinfo {volume} {201}},\
  \bibinfo {pages} {1--19} (\bibinfo {year} {1975})}\BibitemShut {NoStop}%
\bibitem [{\citenamefont {Byrnes}\ \emph {et~al.}(2012)\citenamefont {Byrnes},
  \citenamefont {Copeland},\ and\ \citenamefont {Green}}]{Byrnes:2012yx}%
  \BibitemOpen
  \bibfield  {author} {\bibinfo {author} {\bibfnamefont {Christian~T.}\
  \bibnamefont {Byrnes}}, \bibinfo {author} {\bibfnamefont {Edmund~J.}\
  \bibnamefont {Copeland}}, \ and\ \bibinfo {author} {\bibfnamefont {Anne~M.}\
  \bibnamefont {Green}},\ }\href {\doibase10.1103/PhysRevD.86.043512}
  {\bibfield  {journal} {\bibinfo  {journal} {Phys. Rev.}\ }\textbf {\bibinfo
  {volume} {D86}},\ \bibinfo {pages} {043512} (\bibinfo {year} {2012})},\
  \Eprint {http://arxiv.org/abs/1206.4188} {arXiv:1206.4188
  [astro-ph.CO]}\BibitemShut {NoStop}%
\bibitem [{\citenamefont {Raidal}\ \emph {et~al.}(2018)\citenamefont {Raidal},
  \citenamefont {Solodukhin}, \citenamefont {Vaskonen},\ and\ \citenamefont
  {Veermäe}}]{Raidal:2018eoo}%
  \BibitemOpen
  \bibfield  {author} {\bibinfo {author} {\bibfnamefont {Martti}\ \bibnamefont
  {Raidal}}, \bibinfo {author} {\bibfnamefont {Sergey}\ \bibnamefont
  {Solodukhin}}, \bibinfo {author} {\bibfnamefont {Ville}\ \bibnamefont
  {Vaskonen}}, \ and\ \bibinfo {author} {\bibfnamefont {Hardi}\ \bibnamefont
  {Veermäe}},\ }\href {\doibase10.1103/PhysRevD.97.123520} {\bibfield
  {journal} {\bibinfo  {journal} {Phys. Rev.}\ }\textbf {\bibinfo {volume}
  {D97}},\ \bibinfo {pages} {123520} (\bibinfo {year} {2018})},\ \Eprint
  {http://arxiv.org/abs/1802.07728} {arXiv:1802.07728
  [astro-ph.CO]}\BibitemShut {NoStop}%
\bibitem [{\citenamefont {Paczynski}(1986)}]{Paczynski:1985jf}%
  \BibitemOpen
  \bibfield  {author} {\bibinfo {author} {\bibfnamefont {Bohdan}\ \bibnamefont
  {Paczynski}},\ }\href {\doibase10.1086/164140} {\bibfield  {journal}
  {\bibinfo  {journal} {Astrophys. J.}\ }\textbf {\bibinfo {volume} {304}},\
  \bibinfo {pages} {1--5} (\bibinfo {year} {1986})}\BibitemShut {NoStop}%
\bibitem [{\citenamefont {Griest}(1991)}]{Griest:1990vu}%
  \BibitemOpen
  \bibfield  {author} {\bibinfo {author} {\bibfnamefont {Kim}\ \bibnamefont
  {Griest}},\ }\href {\doibase10.1086/169575} {\bibfield  {journal} {\bibinfo
  {journal} {Astrophys. J.}\ }\textbf {\bibinfo {volume} {366}},\ \bibinfo
  {pages} {412--421} (\bibinfo {year} {1991})}\BibitemShut {NoStop}%
\bibitem [{\citenamefont {De~Rujula}\ \emph {et~al.}(1991)\citenamefont
  {De~Rujula}, \citenamefont {Jetzer},\ and\ \citenamefont
  {Masso}}]{DeRujula:1990wq}%
  \BibitemOpen
  \bibfield  {author} {\bibinfo {author} {\bibfnamefont {A.}~\bibnamefont
  {De~Rujula}}, \bibinfo {author} {\bibfnamefont {P.}~\bibnamefont {Jetzer}}, \
  and\ \bibinfo {author} {\bibfnamefont {E.}~\bibnamefont {Masso}},\ }\href
  {\doibase10.1093/mnras/250.2.348} {\bibfield  {journal} {\bibinfo  {journal}
  {Mon. Not. Roy. Astron. Soc.}\ }\textbf {\bibinfo {volume} {250}},\ \bibinfo
  {pages} {348} (\bibinfo {year} {1991})}\BibitemShut {NoStop}%
\bibitem [{\citenamefont {Alcock}\ \emph {et~al.}(1997)\citenamefont {Alcock}
  \emph {et~al.}}]{Alcock:1996yv}%
  \BibitemOpen
  \bibfield  {author} {\bibinfo {author} {\bibfnamefont {C.}~\bibnamefont
  {Alcock}} \emph {et~al.} (\bibinfo {collaboration} {MACHO}),\ }\href
  {\doibase10.1086/304535} {\bibfield  {journal} {\bibinfo  {journal}
  {Astrophys. J.}\ }\textbf {\bibinfo {volume} {486}},\ \bibinfo {pages}
  {697--726} (\bibinfo {year} {1997})},\ \Eprint
  {http://arxiv.org/abs/astro-ph/9606165} {arXiv:astro-ph/9606165
  [astro-ph]}\BibitemShut {NoStop}%
\bibitem [{\citenamefont {Green}(2017)}]{Green:2017qoa}%
  \BibitemOpen
  \bibfield  {author} {\bibinfo {author} {\bibfnamefont {Anne~M}\ \bibnamefont
  {Green}},\ }\href {\doibase10.1103/PhysRevD.96.043020} {\bibfield  {journal}
  {\bibinfo  {journal} {Phys. Rev.}\ }\textbf {\bibinfo {volume} {D96}},\
  \bibinfo {pages} {043020} (\bibinfo {year} {2017})},\ \Eprint
  {http://arxiv.org/abs/1705.10818} {arXiv:1705.10818
  [astro-ph.CO]}\BibitemShut {NoStop}%
\bibitem [{\citenamefont {Inoue}\ and\ \citenamefont
  {Kusenko}(2017)}]{Inoue:2017csr}%
  \BibitemOpen
  \bibfield  {author} {\bibinfo {author} {\bibfnamefont {Yoshiyuki}\
  \bibnamefont {Inoue}}\ and\ \bibinfo {author} {\bibfnamefont {Alexander}\
  \bibnamefont {Kusenko}},\ }\href {\doibase10.1088/1475-7516/2017/10/034}
  {\bibfield  {journal} {\bibinfo  {journal} {JCAP}\ }\textbf {\bibinfo
  {volume} {1710}},\ \bibinfo {pages} {034} (\bibinfo {year} {2017})},\ \Eprint
  {http://arxiv.org/abs/1705.00791} {arXiv:1705.00791
  [astro-ph.CO]}\BibitemShut {NoStop}%
\bibitem [{\citenamefont {Garny}\ \emph {et~al.}(2016)\citenamefont {Garny},
  \citenamefont {Sandora},\ and\ \citenamefont {Sloth}}]{Garny:2015sjg}%
  \BibitemOpen
  \bibfield  {author} {\bibinfo {author} {\bibfnamefont {Mathias}\ \bibnamefont
  {Garny}}, \bibinfo {author} {\bibfnamefont {McCullen}\ \bibnamefont
  {Sandora}}, \ and\ \bibinfo {author} {\bibfnamefont {Martin~S.}\ \bibnamefont
  {Sloth}},\ }\href {\doibase10.1103/PhysRevLett.116.101302} {\bibfield
  {journal} {\bibinfo  {journal} {Phys. Rev. Lett.}\ }\textbf {\bibinfo
  {volume} {116}},\ \bibinfo {pages} {101302} (\bibinfo {year} {2016})},\
  \Eprint {http://arxiv.org/abs/1511.03278} {arXiv:1511.03278
  [hep-ph]}\BibitemShut {NoStop}%
\bibitem [{\citenamefont {Tremaine}\ and\ \citenamefont
  {Gunn}(1979)}]{Tremaine:1979we}%
  \BibitemOpen
  \bibfield  {author} {\bibinfo {author} {\bibfnamefont {S.}~\bibnamefont
  {Tremaine}}\ and\ \bibinfo {author} {\bibfnamefont {J.~E.}\ \bibnamefont
  {Gunn}},\ }\href {\doibase10.1103/PhysRevLett.42.407} {\bibfield  {journal}
  {\bibinfo  {journal} {Phys. Rev. Lett.}\ }\textbf {\bibinfo {volume} {42}},\
  \bibinfo {pages} {407--410} (\bibinfo {year} {1979})}\BibitemShut {NoStop}%
\bibitem [{\citenamefont {Baur}\ \emph {et~al.}(2016)\citenamefont {Baur},
  \citenamefont {Palanque-Delabrouille}, \citenamefont {Yèche}, \citenamefont
  {Magneville},\ and\ \citenamefont {Viel}}]{Baur:2015jsy}%
  \BibitemOpen
  \bibfield  {author} {\bibinfo {author} {\bibfnamefont {Julien}\ \bibnamefont
  {Baur}}, \bibinfo {author} {\bibfnamefont {Nathalie}\ \bibnamefont
  {Palanque-Delabrouille}}, \bibinfo {author} {\bibfnamefont {Christophe}\
  \bibnamefont {Yèche}}, \bibinfo {author} {\bibfnamefont {Christophe}\
  \bibnamefont {Magneville}}, \ and\ \bibinfo {author} {\bibfnamefont {Matteo}\
  \bibnamefont {Viel}},\ }\bibfield  {booktitle} {\emph {\bibinfo {booktitle}
  {{SDSS-IV Collaboration Meeting, July 20-23, 2015}}},\ }\href
  {\doibase10.1088/1475-7516/2016/08/012} {\bibfield  {journal} {\bibinfo
  {journal} {JCAP}\ }\textbf {\bibinfo {volume} {1608}},\ \bibinfo {pages}
  {012} (\bibinfo {year} {2016})},\ \Eprint {http://arxiv.org/abs/1512.01981}
  {arXiv:1512.01981 [astro-ph.CO]}\BibitemShut {NoStop}%
\bibitem [{\citenamefont {D'Agnolo}\ \emph {et~al.}(2017)\citenamefont
  {D'Agnolo}, \citenamefont {Pappadopulo},\ and\ \citenamefont
  {Ruderman}}]{DAgnolo:2017dbv}%
  \BibitemOpen
  \bibfield  {author} {\bibinfo {author} {\bibfnamefont {Raffaele~Tito}\
  \bibnamefont {D'Agnolo}}, \bibinfo {author} {\bibfnamefont {Duccio}\
  \bibnamefont {Pappadopulo}}, \ and\ \bibinfo {author} {\bibfnamefont
  {Joshua~T.}\ \bibnamefont {Ruderman}},\ }\href
  {\doibase10.1103/PhysRevLett.119.061102} {\bibfield  {journal} {\bibinfo
  {journal} {Phys. Rev. Lett.}\ }\textbf {\bibinfo {volume} {119}},\ \bibinfo
  {pages} {061102} (\bibinfo {year} {2017})},\ \Eprint
  {http://arxiv.org/abs/1705.08450} {arXiv:1705.08450 [hep-ph]}\BibitemShut
  {NoStop}%
\bibitem [{\citenamefont {Marsh}(2016)}]{Marsh:2015xka}%
  \BibitemOpen
  \bibfield  {author} {\bibinfo {author} {\bibfnamefont {David J.~E.}\
  \bibnamefont {Marsh}},\ }\href {\doibase10.1016/j.physrep.2016.06.005}
  {\bibfield  {journal} {\bibinfo  {journal} {Phys. Rept.}\ }\textbf {\bibinfo
  {volume} {643}},\ \bibinfo {pages} {1--79} (\bibinfo {year} {2016})},\
  \Eprint {http://arxiv.org/abs/1510.07633} {arXiv:1510.07633
  [astro-ph.CO]}\BibitemShut {NoStop}%
\bibitem [{\citenamefont {Pato}\ \emph {et~al.}(2015)\citenamefont {Pato},
  \citenamefont {Iocco},\ and\ \citenamefont {Bertone}}]{Pato:2015dua}%
  \BibitemOpen
  \bibfield  {author} {\bibinfo {author} {\bibfnamefont {Miguel}\ \bibnamefont
  {Pato}}, \bibinfo {author} {\bibfnamefont {Fabio}\ \bibnamefont {Iocco}}, \
  and\ \bibinfo {author} {\bibfnamefont {Gianfranco}\ \bibnamefont {Bertone}},\
  }\href {\doibase10.1088/1475-7516/2015/12/001} {\bibfield  {journal}
  {\bibinfo  {journal} {JCAP}\ }\textbf {\bibinfo {volume} {1512}},\ \bibinfo
  {pages} {001} (\bibinfo {year} {2015})},\ \Eprint
  {http://arxiv.org/abs/1504.06324} {arXiv:1504.06324
  [astro-ph.GA]}\BibitemShut {NoStop}%
\bibitem [{\citenamefont {Schive}\ \emph
  {et~al.}(2014{\natexlab{a}})\citenamefont {Schive}, \citenamefont {Chiueh},\
  and\ \citenamefont {Broadhurst}}]{Schive:2014dra}%
  \BibitemOpen
  \bibfield  {author} {\bibinfo {author} {\bibfnamefont {Hsi-Yu}\ \bibnamefont
  {Schive}}, \bibinfo {author} {\bibfnamefont {Tzihong}\ \bibnamefont
  {Chiueh}}, \ and\ \bibinfo {author} {\bibfnamefont {Tom}\ \bibnamefont
  {Broadhurst}},\ }\href {\doibase10.1038/nphys2996} {\bibfield  {journal}
  {\bibinfo  {journal} {Nature Phys.}\ }\textbf {\bibinfo {volume} {10}},\
  \bibinfo {pages} {496--499} (\bibinfo {year} {2014}{\natexlab{a}})},\ \Eprint
  {http://arxiv.org/abs/1406.6586} {arXiv:1406.6586 [astro-ph.GA]}\BibitemShut
  {NoStop}%
\bibitem [{\citenamefont {Bar}\ \emph {et~al.}(2018)\citenamefont {Bar},
  \citenamefont {Blas}, \citenamefont {Blum},\ and\ \citenamefont
  {Sibiryakov}}]{Bar:2018acw}%
  \BibitemOpen
  \bibfield  {author} {\bibinfo {author} {\bibfnamefont {Nitsan}\ \bibnamefont
  {Bar}}, \bibinfo {author} {\bibfnamefont {Diego}\ \bibnamefont {Blas}},
  \bibinfo {author} {\bibfnamefont {Kfir}\ \bibnamefont {Blum}}, \ and\
  \bibinfo {author} {\bibfnamefont {Sergey}\ \bibnamefont {Sibiryakov}},\
  }\href@noop {} {\  (\bibinfo {year} {2018})},\ \Eprint
  {http://arxiv.org/abs/1805.00122} {arXiv:1805.00122
  [astro-ph.CO]}\BibitemShut {NoStop}%
\bibitem [{\citenamefont {Aprile}\ \emph {et~al.}(2017)\citenamefont {Aprile}
  \emph {et~al.}}]{Aprile:2017iyp}%
  \BibitemOpen
  \bibfield  {author} {\bibinfo {author} {\bibfnamefont {E.}~\bibnamefont
  {Aprile}} \emph {et~al.} (\bibinfo {collaboration} {XENON}),\ }\href
  {\doibase10.1103/PhysRevLett.119.181301} {\bibfield  {journal} {\bibinfo
  {journal} {Phys. Rev. Lett.}\ }\textbf {\bibinfo {volume} {119}},\ \bibinfo
  {pages} {181301} (\bibinfo {year} {2017})},\ \Eprint
  {http://arxiv.org/abs/1705.06655} {arXiv:1705.06655
  [astro-ph.CO]}\BibitemShut {NoStop}%
\bibitem [{\citenamefont {Cui}\ \emph {et~al.}(2017)\citenamefont {Cui} \emph
  {et~al.}}]{Cui:2017nnn}%
  \BibitemOpen
  \bibfield  {author} {\bibinfo {author} {\bibfnamefont {Xiangyi}\ \bibnamefont
  {Cui}} \emph {et~al.} (\bibinfo {collaboration} {PandaX-II}),\ }\href
  {\doibase10.1103/PhysRevLett.119.181302} {\bibfield  {journal} {\bibinfo
  {journal} {Phys. Rev. Lett.}\ }\textbf {\bibinfo {volume} {119}},\ \bibinfo
  {pages} {181302} (\bibinfo {year} {2017})},\ \Eprint
  {http://arxiv.org/abs/1708.06917} {arXiv:1708.06917
  [astro-ph.CO]}\BibitemShut {NoStop}%
\bibitem [{\citenamefont {Hoferichter}\ \emph {et~al.}(2017)\citenamefont
  {Hoferichter}, \citenamefont {Klos}, \citenamefont {Menéndez},\ and\
  \citenamefont {Schwenk}}]{Hoferichter:2017olk}%
  \BibitemOpen
  \bibfield  {author} {\bibinfo {author} {\bibfnamefont {Martin}\ \bibnamefont
  {Hoferichter}}, \bibinfo {author} {\bibfnamefont {Philipp}\ \bibnamefont
  {Klos}}, \bibinfo {author} {\bibfnamefont {Javier}\ \bibnamefont
  {Menéndez}}, \ and\ \bibinfo {author} {\bibfnamefont {Achim}\ \bibnamefont
  {Schwenk}},\ }\href {\doibase10.1103/PhysRevLett.119.181803} {\bibfield
  {journal} {\bibinfo  {journal} {Phys. Rev. Lett.}\ }\textbf {\bibinfo
  {volume} {119}},\ \bibinfo {pages} {181803} (\bibinfo {year} {2017})},\
  \Eprint {http://arxiv.org/abs/1708.02245} {arXiv:1708.02245
  [hep-ph]}\BibitemShut {NoStop}%
\bibitem [{\citenamefont {Queiroz}(2017)}]{Queiroz:2017kxt}%
  \BibitemOpen
  \bibfield  {author} {\bibinfo {author} {\bibfnamefont {Farinaldo~S.}\
  \bibnamefont {Queiroz}},\ }in\ \href
  {http://inspirehep.net/record/1634808/files/arXiv:1711.02463.pdf} {\emph
  {\bibinfo {booktitle} {{2017 European Physical Society Conference on High
  Energy Physics (EPS-HEP 2017) Venice, Italy, July 5-12, 2017}}}}\ (\bibinfo
  {year} {2017})\ \Eprint {http://arxiv.org/abs/1711.02463} {arXiv:1711.02463
  [hep-ph]}\BibitemShut {NoStop}%
\bibitem [{\citenamefont {Hooper}(2017)}]{Hooper:2017eje}%
  \BibitemOpen
  \bibfield  {author} {\bibinfo {author} {\bibfnamefont {Dan}\ \bibnamefont
  {Hooper}},\ }\href {\doibase10.1103/Physics.10.119} {\bibfield  {journal}
  {\bibinfo  {journal} {APS Physics}\ }\textbf {\bibinfo {volume} {10}},\
  \bibinfo {pages} {119} (\bibinfo {year} {2017})}\BibitemShut {NoStop}%
\bibitem [{\citenamefont {Joyce}\ \emph {et~al.}(2016)\citenamefont {Joyce},
  \citenamefont {Lombriser},\ and\ \citenamefont {Schmidt}}]{Joyce:2016vqv}%
  \BibitemOpen
  \bibfield  {author} {\bibinfo {author} {\bibfnamefont {Austin}\ \bibnamefont
  {Joyce}}, \bibinfo {author} {\bibfnamefont {Lucas}\ \bibnamefont
  {Lombriser}}, \ and\ \bibinfo {author} {\bibfnamefont {Fabian}\ \bibnamefont
  {Schmidt}},\ }\href {\doibase10.1146/annurev-nucl-102115-044553} {\bibfield
  {journal} {\bibinfo  {journal} {Ann. Rev. Nucl. Part. Sci.}\ }\textbf
  {\bibinfo {volume} {66}},\ \bibinfo {pages} {95--122} (\bibinfo {year}
  {2016})},\ \Eprint {http://arxiv.org/abs/1601.06133} {arXiv:1601.06133
  [astro-ph.CO]}\BibitemShut {NoStop}%
\bibitem [{\citenamefont {Matos}\ and\ \citenamefont
  {Urena-Lopez}(2001)}]{Matos:2000ss}%
  \BibitemOpen
  \bibfield  {author} {\bibinfo {author} {\bibfnamefont {Tonatiuh}\
  \bibnamefont {Matos}}\ and\ \bibinfo {author} {\bibfnamefont {L.~Arturo}\
  \bibnamefont {Urena-Lopez}},\ }\href {\doibase10.1103/PhysRevD.63.063506}
  {\bibfield  {journal} {\bibinfo  {journal} {Phys. Rev.}\ }\textbf {\bibinfo
  {volume} {D63}},\ \bibinfo {pages} {063506} (\bibinfo {year} {2001})},\
  \Eprint {http://arxiv.org/abs/astro-ph/0006024} {arXiv:astro-ph/0006024
  [astro-ph]}\BibitemShut {NoStop}%
\bibitem [{\citenamefont {Hu}\ \emph {et~al.}(2000)\citenamefont {Hu},
  \citenamefont {Barkana},\ and\ \citenamefont {Gruzinov}}]{Hu:2000ke}%
  \BibitemOpen
  \bibfield  {author} {\bibinfo {author} {\bibfnamefont {Wayne}\ \bibnamefont
  {Hu}}, \bibinfo {author} {\bibfnamefont {Rennan}\ \bibnamefont {Barkana}}, \
  and\ \bibinfo {author} {\bibfnamefont {Andrei}\ \bibnamefont {Gruzinov}},\
  }\href {\doibase10.1103/PhysRevLett.85.1158} {\bibfield  {journal} {\bibinfo
  {journal} {Phys. Rev. Lett.}\ }\textbf {\bibinfo {volume} {85}},\ \bibinfo
  {pages} {1158--1161} (\bibinfo {year} {2000})},\ \Eprint
  {http://arxiv.org/abs/astro-ph/0003365} {arXiv:astro-ph/0003365
  [astro-ph]}\BibitemShut {NoStop}%
\bibitem [{\citenamefont {Hui}\ \emph {et~al.}(2017)\citenamefont {Hui},
  \citenamefont {Ostriker}, \citenamefont {Tremaine},\ and\ \citenamefont
  {Witten}}]{Hui:2016ltb}%
  \BibitemOpen
  \bibfield  {author} {\bibinfo {author} {\bibfnamefont {Lam}\ \bibnamefont
  {Hui}}, \bibinfo {author} {\bibfnamefont {Jeremiah~P.}\ \bibnamefont
  {Ostriker}}, \bibinfo {author} {\bibfnamefont {Scott}\ \bibnamefont
  {Tremaine}}, \ and\ \bibinfo {author} {\bibfnamefont {Edward}\ \bibnamefont
  {Witten}},\ }\href {\doibase10.1103/PhysRevD.95.043541} {\bibfield  {journal}
  {\bibinfo  {journal} {Phys. Rev.}\ }\textbf {\bibinfo {volume} {D95}},\
  \bibinfo {pages} {043541} (\bibinfo {year} {2017})},\ \Eprint
  {http://arxiv.org/abs/1610.08297} {arXiv:1610.08297
  [astro-ph.CO]}\BibitemShut {NoStop}%
\bibitem [{\citenamefont {Suárez}\ \emph {et~al.}(2014)\citenamefont
  {Suárez}, \citenamefont {Robles},\ and\ \citenamefont
  {Matos}}]{Suarez:2013iw}%
  \BibitemOpen
  \bibfield  {author} {\bibinfo {author} {\bibfnamefont {Abril}\ \bibnamefont
  {Suárez}}, \bibinfo {author} {\bibfnamefont {Victor~H.}\ \bibnamefont
  {Robles}}, \ and\ \bibinfo {author} {\bibfnamefont {Tonatiuh}\ \bibnamefont
  {Matos}},\ }\bibfield  {booktitle} {\emph {\bibinfo {booktitle} {{Proceedings
  of 4th International Meeting on Gravitation and Cosmology (MGC 4): Santa
  Clara, Cuba, June 1-4, 2009}}},\ }\href {\doibase10.1007/978-3-319-02063-1_9}
  {\bibfield  {journal} {\bibinfo  {journal} {Astrophys. Space Sci. Proc.}\
  }\textbf {\bibinfo {volume} {38}},\ \bibinfo {pages} {107--142} (\bibinfo
  {year} {2014})},\ \Eprint {http://arxiv.org/abs/1302.0903} {arXiv:1302.0903
  [astro-ph.CO]}\BibitemShut {NoStop}%
\bibitem [{\citenamefont {Zhang}\ and\ \citenamefont
  {Chiueh}(2017)}]{Zhang:2017dpp}%
  \BibitemOpen
  \bibfield  {author} {\bibinfo {author} {\bibfnamefont {Ui-Han}\ \bibnamefont
  {Zhang}}\ and\ \bibinfo {author} {\bibfnamefont {Tzihong}\ \bibnamefont
  {Chiueh}},\ }\href {\doibase10.1103/PhysRevD.96.063522} {\bibfield  {journal}
  {\bibinfo  {journal} {Phys. Rev.}\ }\textbf {\bibinfo {volume} {D96}},\
  \bibinfo {pages} {063522} (\bibinfo {year} {2017})},\ \Eprint
  {http://arxiv.org/abs/1705.01439} {arXiv:1705.01439
  [astro-ph.CO]}\BibitemShut {NoStop}%
\bibitem [{\citenamefont {Cedeño}\ \emph {et~al.}(2017)\citenamefont
  {Cedeño}, \citenamefont {González-Morales},\ and\ \citenamefont
  {Ureña-López}}]{Cedeno:2017sou}%
  \BibitemOpen
  \bibfield  {author} {\bibinfo {author} {\bibfnamefont {Francisco X.~Linares}\
  \bibnamefont {Cedeño}}, \bibinfo {author} {\bibfnamefont {Alma~X.}\
  \bibnamefont {González-Morales}}, \ and\ \bibinfo {author} {\bibfnamefont
  {L.~Arturo}\ \bibnamefont {Ureña-López}},\ }\href
  {\doibase10.1103/PhysRevD.96.061301} {\bibfield  {journal} {\bibinfo
  {journal} {Phys. Rev.}\ }\textbf {\bibinfo {volume} {D96}},\ \bibinfo {pages}
  {061301} (\bibinfo {year} {2017})},\ \Eprint
  {http://arxiv.org/abs/1703.10180} {arXiv:1703.10180 [gr-qc]}\BibitemShut
  {NoStop}%
\bibitem [{\citenamefont {Urena-Lopez}\ and\ \citenamefont
  {Gonzalez-Morales}(2016)}]{Urena-Lopez:2015gur}%
  \BibitemOpen
  \bibfield  {author} {\bibinfo {author} {\bibfnamefont {L.~Arturo}\
  \bibnamefont {Urena-Lopez}}\ and\ \bibinfo {author} {\bibfnamefont {Alma~X.}\
  \bibnamefont {Gonzalez-Morales}},\ }\href
  {\doibase10.1088/1475-7516/2016/07/048} {\bibfield  {journal} {\bibinfo
  {journal} {JCAP}\ }\textbf {\bibinfo {volume} {1607}},\ \bibinfo {pages}
  {048} (\bibinfo {year} {2016})},\ \Eprint {http://arxiv.org/abs/1511.08195}
  {arXiv:1511.08195 [astro-ph.CO]}\BibitemShut {NoStop}%
\bibitem [{\citenamefont {Hlozek}\ \emph {et~al.}(2015)\citenamefont {Hlozek},
  \citenamefont {Grin}, \citenamefont {Marsh},\ and\ \citenamefont
  {Ferreira}}]{Hlozek:2014lca}%
  \BibitemOpen
  \bibfield  {author} {\bibinfo {author} {\bibfnamefont {Renée}\ \bibnamefont
  {Hlozek}}, \bibinfo {author} {\bibfnamefont {Daniel}\ \bibnamefont {Grin}},
  \bibinfo {author} {\bibfnamefont {David J.~E.}\ \bibnamefont {Marsh}}, \ and\
  \bibinfo {author} {\bibfnamefont {Pedro~G.}\ \bibnamefont {Ferreira}},\
  }\href {\doibase10.1103/PhysRevD.91.103512} {\bibfield  {journal} {\bibinfo
  {journal} {Phys. Rev.}\ }\textbf {\bibinfo {volume} {D91}},\ \bibinfo {pages}
  {103512} (\bibinfo {year} {2015})},\ \Eprint {http://arxiv.org/abs/1410.2896}
  {arXiv:1410.2896 [astro-ph.CO]}\BibitemShut {NoStop}%
\bibitem [{\citenamefont {Mocz}\ \emph {et~al.}(2017)\citenamefont {Mocz},
  \citenamefont {Vogelsberger}, \citenamefont {Robles}, \citenamefont {Zavala},
  \citenamefont {Boylan-Kolchin},\ and\ \citenamefont
  {Hernquist}}]{Mocz:2017wlg}%
  \BibitemOpen
  \bibfield  {author} {\bibinfo {author} {\bibfnamefont {Philip}\ \bibnamefont
  {Mocz}}, \bibinfo {author} {\bibfnamefont {Mark}\ \bibnamefont
  {Vogelsberger}}, \bibinfo {author} {\bibfnamefont {Victor}\ \bibnamefont
  {Robles}}, \bibinfo {author} {\bibfnamefont {Jesus}\ \bibnamefont {Zavala}},
  \bibinfo {author} {\bibfnamefont {Michael}\ \bibnamefont {Boylan-Kolchin}}, \
  and\ \bibinfo {author} {\bibfnamefont {Lars}\ \bibnamefont {Hernquist}},\
  }\href {\doibase10.1093/mnras/stx1887} {\bibfield  {journal} {\bibinfo
  {journal} {Mon. Not. Roy. Astron. Soc.}\ }\textbf {\bibinfo {volume} {471}},\
  \bibinfo {pages} {4559} (\bibinfo {year} {2017})},\ \Eprint
  {http://arxiv.org/abs/1705.05845} {arXiv:1705.05845
  [astro-ph.CO]}\BibitemShut {NoStop}%
\bibitem [{\citenamefont {Schwabe}\ \emph {et~al.}(2016)\citenamefont
  {Schwabe}, \citenamefont {Niemeyer},\ and\ \citenamefont
  {Engels}}]{Schwabe:2016rze}%
  \BibitemOpen
  \bibfield  {author} {\bibinfo {author} {\bibfnamefont {Bodo}\ \bibnamefont
  {Schwabe}}, \bibinfo {author} {\bibfnamefont {Jens~C.}\ \bibnamefont
  {Niemeyer}}, \ and\ \bibinfo {author} {\bibfnamefont {Jan~F.}\ \bibnamefont
  {Engels}},\ }\href {\doibase10.1103/PhysRevD.94.043513} {\bibfield  {journal}
  {\bibinfo  {journal} {Phys. Rev.}\ }\textbf {\bibinfo {volume} {D94}},\
  \bibinfo {pages} {043513} (\bibinfo {year} {2016})},\ \Eprint
  {http://arxiv.org/abs/1606.05151} {arXiv:1606.05151
  [astro-ph.CO]}\BibitemShut {NoStop}%
\bibitem [{\citenamefont {Schive}\ \emph
  {et~al.}(2014{\natexlab{b}})\citenamefont {Schive}, \citenamefont {Liao},
  \citenamefont {Woo}, \citenamefont {Wong}, \citenamefont {Chiueh},
  \citenamefont {Broadhurst},\ and\ \citenamefont {Hwang}}]{Schive:2014hza}%
  \BibitemOpen
  \bibfield  {author} {\bibinfo {author} {\bibfnamefont {Hsi-Yu}\ \bibnamefont
  {Schive}}, \bibinfo {author} {\bibfnamefont {Ming-Hsuan}\ \bibnamefont
  {Liao}}, \bibinfo {author} {\bibfnamefont {Tak-Pong}\ \bibnamefont {Woo}},
  \bibinfo {author} {\bibfnamefont {Shing-Kwong}\ \bibnamefont {Wong}},
  \bibinfo {author} {\bibfnamefont {Tzihong}\ \bibnamefont {Chiueh}}, \bibinfo
  {author} {\bibfnamefont {Tom}\ \bibnamefont {Broadhurst}}, \ and\ \bibinfo
  {author} {\bibfnamefont {W.~Y.~Pauchy}\ \bibnamefont {Hwang}},\ }\href
  {\doibase10.1103/PhysRevLett.113.261302} {\bibfield  {journal} {\bibinfo
  {journal} {Phys. Rev. Lett.}\ }\textbf {\bibinfo {volume} {113}},\ \bibinfo
  {pages} {261302} (\bibinfo {year} {2014}{\natexlab{b}})},\ \Eprint
  {http://arxiv.org/abs/1407.7762} {arXiv:1407.7762 [astro-ph.GA]}\BibitemShut
  {NoStop}%
\bibitem [{\citenamefont {Urena-Lopez}\ \emph {et~al.}(2017)\citenamefont
  {Urena-Lopez}, \citenamefont {Robles},\ and\ \citenamefont
  {Matos}}]{Urena-Lopez:2017tob}%
  \BibitemOpen
  \bibfield  {author} {\bibinfo {author} {\bibfnamefont {L.~Arturo}\
  \bibnamefont {Urena-Lopez}}, \bibinfo {author} {\bibfnamefont {Victor~H.}\
  \bibnamefont {Robles}}, \ and\ \bibinfo {author} {\bibfnamefont
  {T.}~\bibnamefont {Matos}},\ }\href {\doibase10.1103/PhysRevD.96.043005}
  {\bibfield  {journal} {\bibinfo  {journal} {Phys. Rev.}\ }\textbf {\bibinfo
  {volume} {D96}},\ \bibinfo {pages} {043005} (\bibinfo {year} {2017})},\
  \Eprint {http://arxiv.org/abs/1702.05103} {arXiv:1702.05103
  [astro-ph.CO]}\BibitemShut {NoStop}%
\bibitem [{\citenamefont {Deng}\ \emph {et~al.}(2018)\citenamefont {Deng},
  \citenamefont {Hertzberg}, \citenamefont {Namjoo},\ and\ \citenamefont
  {Masoumi}}]{Deng:2018jjz}%
  \BibitemOpen
  \bibfield  {author} {\bibinfo {author} {\bibfnamefont {Heling}\ \bibnamefont
  {Deng}}, \bibinfo {author} {\bibfnamefont {Mark~P.}\ \bibnamefont
  {Hertzberg}}, \bibinfo {author} {\bibfnamefont {Mohammad~Hossein}\
  \bibnamefont {Namjoo}}, \ and\ \bibinfo {author} {\bibfnamefont {Ali}\
  \bibnamefont {Masoumi}},\ }\href@noop {} {\  (\bibinfo {year} {2018})},\
  \Eprint {http://arxiv.org/abs/1804.05921} {arXiv:1804.05921
  [astro-ph.CO]}\BibitemShut {NoStop}%
\bibitem [{\citenamefont {Palanque-Delabrouille}\ \emph
  {et~al.}(2013)\citenamefont {Palanque-Delabrouille} \emph
  {et~al.}}]{Palanque-Delabrouille:2013gaa}%
  \BibitemOpen
  \bibfield  {author} {\bibinfo {author} {\bibfnamefont {Nathalie}\
  \bibnamefont {Palanque-Delabrouille}} \emph {et~al.},\ }\href
  {\doibase10.1051/0004-6361/201322130} {\bibfield  {journal} {\bibinfo
  {journal} {Astron. Astrophys.}\ }\textbf {\bibinfo {volume} {559}},\ \bibinfo
  {pages} {A85} (\bibinfo {year} {2013})},\ \Eprint
  {http://arxiv.org/abs/1306.5896} {arXiv:1306.5896 [astro-ph.CO]}\BibitemShut
  {NoStop}%
\bibitem [{\citenamefont {Kobayashi}\ \emph {et~al.}(2017)\citenamefont
  {Kobayashi}, \citenamefont {Murgia}, \citenamefont {De~Simone}, \citenamefont
  {Iršič},\ and\ \citenamefont {Viel}}]{Kobayashi:2017jcf}%
  \BibitemOpen
  \bibfield  {author} {\bibinfo {author} {\bibfnamefont {Takeshi}\ \bibnamefont
  {Kobayashi}}, \bibinfo {author} {\bibfnamefont {Riccardo}\ \bibnamefont
  {Murgia}}, \bibinfo {author} {\bibfnamefont {Andrea}\ \bibnamefont
  {De~Simone}}, \bibinfo {author} {\bibfnamefont {Vid}\ \bibnamefont
  {Iršič}}, \ and\ \bibinfo {author} {\bibfnamefont {Matteo}\ \bibnamefont
  {Viel}},\ }\href {\doibase10.1103/PhysRevD.96.123514} {\bibfield  {journal}
  {\bibinfo  {journal} {Phys. Rev.}\ }\textbf {\bibinfo {volume} {D96}},\
  \bibinfo {pages} {123514} (\bibinfo {year} {2017})},\ \Eprint
  {http://arxiv.org/abs/1708.00015} {arXiv:1708.00015
  [astro-ph.CO]}\BibitemShut {NoStop}%
\bibitem [{\citenamefont {Brito}\ \emph
  {et~al.}(2015{\natexlab{c}})\citenamefont {Brito}, \citenamefont {Cardoso},\
  and\ \citenamefont {Pani}}]{Brito:2014wla}%
  \BibitemOpen
  \bibfield  {author} {\bibinfo {author} {\bibfnamefont {Richard}\ \bibnamefont
  {Brito}}, \bibinfo {author} {\bibfnamefont {Vitor}\ \bibnamefont {Cardoso}},
  \ and\ \bibinfo {author} {\bibfnamefont {Paolo}\ \bibnamefont {Pani}},\
  }\href {\doibase10.1088/0264-9381/32/13/134001} {\bibfield  {journal}
  {\bibinfo  {journal} {Class. Quant. Grav.}\ }\textbf {\bibinfo {volume}
  {32}},\ \bibinfo {pages} {134001} (\bibinfo {year} {2015}{\natexlab{c}})},\
  \Eprint {http://arxiv.org/abs/1411.0686} {arXiv:1411.0686
  [gr-qc]}\BibitemShut {NoStop}%
\bibitem [{\citenamefont {Brito}\ \emph
  {et~al.}(2017{\natexlab{a}})\citenamefont {Brito}, \citenamefont {Ghosh},
  \citenamefont {Barausse}, \citenamefont {Berti}, \citenamefont {Cardoso},
  \citenamefont {Dvorkin}, \citenamefont {Klein},\ and\ \citenamefont
  {Pani}}]{Brito:2017wnc}%
  \BibitemOpen
  \bibfield  {author} {\bibinfo {author} {\bibfnamefont {Richard}\ \bibnamefont
  {Brito}}, \bibinfo {author} {\bibfnamefont {Shrobana}\ \bibnamefont {Ghosh}},
  \bibinfo {author} {\bibfnamefont {Enrico}\ \bibnamefont {Barausse}}, \bibinfo
  {author} {\bibfnamefont {Emanuele}\ \bibnamefont {Berti}}, \bibinfo {author}
  {\bibfnamefont {Vitor}\ \bibnamefont {Cardoso}}, \bibinfo {author}
  {\bibfnamefont {Irina}\ \bibnamefont {Dvorkin}}, \bibinfo {author}
  {\bibfnamefont {Antoine}\ \bibnamefont {Klein}}, \ and\ \bibinfo {author}
  {\bibfnamefont {Paolo}\ \bibnamefont {Pani}},\ }\href
  {\doibase10.1103/PhysRevLett.119.131101} {\bibfield  {journal} {\bibinfo
  {journal} {Phys. Rev. Lett.}\ }\textbf {\bibinfo {volume} {119}},\ \bibinfo
  {pages} {131101} (\bibinfo {year} {2017}{\natexlab{a}})},\ \Eprint
  {http://arxiv.org/abs/1706.05097} {arXiv:1706.05097 [gr-qc]}\BibitemShut
  {NoStop}%
\bibitem [{\citenamefont {Press}\ and\ \citenamefont
  {Spergel}(1985)}]{Press:1985ug}%
  \BibitemOpen
  \bibfield  {author} {\bibinfo {author} {\bibfnamefont {William~H.}\
  \bibnamefont {Press}}\ and\ \bibinfo {author} {\bibfnamefont {David~N.}\
  \bibnamefont {Spergel}},\ }\href {\doibase10.1086/163485} {\bibfield
  {journal} {\bibinfo  {journal} {Astrophys. J.}\ }\textbf {\bibinfo {volume}
  {296}},\ \bibinfo {pages} {679--684} (\bibinfo {year} {1985})}\BibitemShut
  {NoStop}%
\bibitem [{\citenamefont {Gould}\ \emph {et~al.}(1990)\citenamefont {Gould},
  \citenamefont {Draine}, \citenamefont {Romani},\ and\ \citenamefont
  {Nussinov}}]{Gould:1989gw}%
  \BibitemOpen
  \bibfield  {author} {\bibinfo {author} {\bibfnamefont {Andrew}\ \bibnamefont
  {Gould}}, \bibinfo {author} {\bibfnamefont {Bruce~T.}\ \bibnamefont
  {Draine}}, \bibinfo {author} {\bibfnamefont {Roger~W.}\ \bibnamefont
  {Romani}}, \ and\ \bibinfo {author} {\bibfnamefont {Shmuel}\ \bibnamefont
  {Nussinov}},\ }\href {\doibase10.1016/0370-2693(90)91745-W} {\bibfield
  {journal} {\bibinfo  {journal} {Phys. Lett.}\ }\textbf {\bibinfo {volume}
  {B238}},\ \bibinfo {pages} {337} (\bibinfo {year} {1990})}\BibitemShut
  {NoStop}%
\bibitem [{\citenamefont {Goldman}\ and\ \citenamefont
  {Nussinov}(1989)}]{Goldman:1989nd}%
  \BibitemOpen
  \bibfield  {author} {\bibinfo {author} {\bibfnamefont {I.}~\bibnamefont
  {Goldman}}\ and\ \bibinfo {author} {\bibfnamefont {S.}~\bibnamefont
  {Nussinov}},\ }\href {\doibase10.1103/PhysRevD.40.3221} {\bibfield  {journal}
  {\bibinfo  {journal} {Phys. Rev.}\ }\textbf {\bibinfo {volume} {D40}},\
  \bibinfo {pages} {3221--3230} (\bibinfo {year} {1989})}\BibitemShut {NoStop}%
\bibitem [{\citenamefont {Bertone}\ and\ \citenamefont
  {Fairbairn}(2008)}]{Bertone:2007ae}%
  \BibitemOpen
  \bibfield  {author} {\bibinfo {author} {\bibfnamefont {Gianfranco}\
  \bibnamefont {Bertone}}\ and\ \bibinfo {author} {\bibfnamefont {Malcolm}\
  \bibnamefont {Fairbairn}},\ }\href {\doibase10.1103/PhysRevD.77.043515}
  {\bibfield  {journal} {\bibinfo  {journal} {Phys. Rev.}\ }\textbf {\bibinfo
  {volume} {D77}},\ \bibinfo {pages} {043515} (\bibinfo {year} {2008})},\
  \Eprint {http://arxiv.org/abs/0709.1485} {arXiv:0709.1485
  [astro-ph]}\BibitemShut {NoStop}%
\bibitem [{\citenamefont {Brito}\ \emph
  {et~al.}(2016{\natexlab{b}})\citenamefont {Brito}, \citenamefont {Cardoso},
  \citenamefont {Macedo}, \citenamefont {Okawa},\ and\ \citenamefont
  {Palenzuela}}]{Brito:2015yfh}%
  \BibitemOpen
  \bibfield  {author} {\bibinfo {author} {\bibfnamefont {Richard}\ \bibnamefont
  {Brito}}, \bibinfo {author} {\bibfnamefont {Vitor}\ \bibnamefont {Cardoso}},
  \bibinfo {author} {\bibfnamefont {Caio F.~B.}\ \bibnamefont {Macedo}},
  \bibinfo {author} {\bibfnamefont {Hirotada}\ \bibnamefont {Okawa}}, \ and\
  \bibinfo {author} {\bibfnamefont {Carlos}\ \bibnamefont {Palenzuela}},\
  }\href {\doibase10.1103/PhysRevD.93.044045} {\bibfield  {journal} {\bibinfo
  {journal} {Phys. Rev.}\ }\textbf {\bibinfo {volume} {D93}},\ \bibinfo {pages}
  {044045} (\bibinfo {year} {2016}{\natexlab{b}})},\ \Eprint
  {http://arxiv.org/abs/1512.00466} {arXiv:1512.00466
  [astro-ph.SR]}\BibitemShut {NoStop}%
\bibitem [{\citenamefont {Flauger}\ and\ \citenamefont
  {Weinberg}(2018)}]{Flauger:2017ged}%
  \BibitemOpen
  \bibfield  {author} {\bibinfo {author} {\bibfnamefont {Raphael}\ \bibnamefont
  {Flauger}}\ and\ \bibinfo {author} {\bibfnamefont {Steven}\ \bibnamefont
  {Weinberg}},\ }\href {\doibase10.1103/PhysRevD.97.123506} {\bibfield
  {journal} {\bibinfo  {journal} {Phys. Rev.}\ }\textbf {\bibinfo {volume}
  {D97}},\ \bibinfo {pages} {123506} (\bibinfo {year} {2018})},\ \Eprint
  {http://arxiv.org/abs/1801.00386} {arXiv:1801.00386
  [astro-ph.CO]}\BibitemShut {NoStop}%
\bibitem [{\citenamefont {Thorne}(1989 (accessed May 2017))}]{Thorne_notes}%
  \BibitemOpen
  \bibfield  {author} {\bibinfo {author} {\bibfnamefont {K.~S.}\ \bibnamefont
  {Thorne}},\ }\href@noop {} {\bibfield  {journal} {\bibinfo  {journal}
  {unpublished}\ } (\bibinfo {year} {1989 (accessed May 2017)})},\ \bibinfo
  {note}
  {\href{https://web.archive.org/web/20170519144752/http://elmer.tapir.caltech.edu/ph237/week6/KipNewWindow5.pdf}{http://elmer.tapir.caltech.edu/ph237/week6/KipNewWindow5.pdf}}\BibitemShut
  {NoStop}%
\bibitem [{\citenamefont {Annulli}\ \emph {et~al.}(2018)\citenamefont
  {Annulli}, \citenamefont {Bernard}, \citenamefont {Blas},\ and\ \citenamefont
  {Cardoso}}]{Annulli:2018quj}%
  \BibitemOpen
  \bibfield  {author} {\bibinfo {author} {\bibfnamefont {Lorenzo}\ \bibnamefont
  {Annulli}}, \bibinfo {author} {\bibfnamefont {Laura}\ \bibnamefont
  {Bernard}}, \bibinfo {author} {\bibfnamefont {Diego}\ \bibnamefont {Blas}}, \
  and\ \bibinfo {author} {\bibfnamefont {Vitor}\ \bibnamefont {Cardoso}},\
  }\href {\doibase10.1103/PhysRevD.98.084001} {\bibfield  {journal} {\bibinfo
  {journal} {Phys. Rev.}\ }\textbf {\bibinfo {volume} {D98}},\ \bibinfo {pages}
  {084001} (\bibinfo {year} {2018})},\ \Eprint
  {http://arxiv.org/abs/1809.05108} {arXiv:1809.05108 [gr-qc]}\BibitemShut
  {NoStop}%
\bibitem [{\citenamefont {Gondolo}\ and\ \citenamefont
  {Silk}(1999)}]{Gondolo:1999ef}%
  \BibitemOpen
  \bibfield  {author} {\bibinfo {author} {\bibfnamefont {Paolo}\ \bibnamefont
  {Gondolo}}\ and\ \bibinfo {author} {\bibfnamefont {Joseph}\ \bibnamefont
  {Silk}},\ }\href {\doibase10.1103/PhysRevLett.83.1719} {\bibfield  {journal}
  {\bibinfo  {journal} {Phys. Rev. Lett.}\ }\textbf {\bibinfo {volume} {83}},\
  \bibinfo {pages} {1719--1722} (\bibinfo {year} {1999})},\ \Eprint
  {http://arxiv.org/abs/astro-ph/9906391} {arXiv:astro-ph/9906391
  [astro-ph]}\BibitemShut {NoStop}%
\bibitem [{\citenamefont {Bertone}\ \emph
  {et~al.}(2005{\natexlab{b}})\citenamefont {Bertone}, \citenamefont
  {Zentner},\ and\ \citenamefont {Silk}}]{Bertone:2005xz}%
  \BibitemOpen
  \bibfield  {author} {\bibinfo {author} {\bibfnamefont {Gianfranco}\
  \bibnamefont {Bertone}}, \bibinfo {author} {\bibfnamefont {Andrew~R.}\
  \bibnamefont {Zentner}}, \ and\ \bibinfo {author} {\bibfnamefont {Joseph}\
  \bibnamefont {Silk}},\ }\href {\doibase10.1103/PhysRevD.72.103517} {\bibfield
   {journal} {\bibinfo  {journal} {Phys. Rev.}\ }\textbf {\bibinfo {volume}
  {D72}},\ \bibinfo {pages} {103517} (\bibinfo {year} {2005}{\natexlab{b}})},\
  \Eprint {http://arxiv.org/abs/astro-ph/0509565} {arXiv:astro-ph/0509565
  [astro-ph]}\BibitemShut {NoStop}%
\bibitem [{\citenamefont {Zhao}\ and\ \citenamefont
  {Silk}(2005)}]{Zhao:2005zr}%
  \BibitemOpen
  \bibfield  {author} {\bibinfo {author} {\bibfnamefont {Hong-Sheng}\
  \bibnamefont {Zhao}}\ and\ \bibinfo {author} {\bibfnamefont {Joseph}\
  \bibnamefont {Silk}},\ }\href {\doibase10.1103/PhysRevLett.95.011301}
  {\bibfield  {journal} {\bibinfo  {journal} {Phys. Rev. Lett.}\ }\textbf
  {\bibinfo {volume} {95}},\ \bibinfo {pages} {011301} (\bibinfo {year}
  {2005})},\ \Eprint {http://arxiv.org/abs/astro-ph/0501625}
  {arXiv:astro-ph/0501625 [astro-ph]}\BibitemShut {NoStop}%
\bibitem [{\citenamefont {Eda}\ \emph {et~al.}(2015)\citenamefont {Eda},
  \citenamefont {Itoh}, \citenamefont {Kuroyanagi},\ and\ \citenamefont
  {Silk}}]{Eda:2014kra}%
  \BibitemOpen
  \bibfield  {author} {\bibinfo {author} {\bibfnamefont {Kazunari}\
  \bibnamefont {Eda}}, \bibinfo {author} {\bibfnamefont {Yousuke}\ \bibnamefont
  {Itoh}}, \bibinfo {author} {\bibfnamefont {Sachiko}\ \bibnamefont
  {Kuroyanagi}}, \ and\ \bibinfo {author} {\bibfnamefont {Joseph}\ \bibnamefont
  {Silk}},\ }\href {\doibase10.1103/PhysRevD.91.044045} {\bibfield  {journal}
  {\bibinfo  {journal} {Phys. Rev.}\ }\textbf {\bibinfo {volume} {D91}},\
  \bibinfo {pages} {044045} (\bibinfo {year} {2015})},\ \Eprint
  {http://arxiv.org/abs/1408.3534} {arXiv:1408.3534 [gr-qc]}\BibitemShut
  {NoStop}%
\bibitem [{\citenamefont {Yue}\ and\ \citenamefont {Han}(2018)}]{Yue:2017iwc}%
  \BibitemOpen
  \bibfield  {author} {\bibinfo {author} {\bibfnamefont {Xiao-Jun}\
  \bibnamefont {Yue}}\ and\ \bibinfo {author} {\bibfnamefont {Wen-Biao}\
  \bibnamefont {Han}},\ }\href {\doibase10.1103/PhysRevD.97.064003} {\bibfield
  {journal} {\bibinfo  {journal} {Phys. Rev.}\ }\textbf {\bibinfo {volume}
  {D97}},\ \bibinfo {pages} {064003} (\bibinfo {year} {2018})},\ \Eprint
  {http://arxiv.org/abs/1711.09706} {arXiv:1711.09706 [gr-qc]}\BibitemShut
  {NoStop}%
\bibitem [{\citenamefont {Hannuksela}\ \emph {et~al.}(2018)\citenamefont
  {Hannuksela}, \citenamefont {Brito}, \citenamefont {Berti},\ and\
  \citenamefont {Li}}]{Hannuksela:2018izj}%
  \BibitemOpen
  \bibfield  {author} {\bibinfo {author} {\bibfnamefont {Otto~A.}\ \bibnamefont
  {Hannuksela}}, \bibinfo {author} {\bibfnamefont {Richard}\ \bibnamefont
  {Brito}}, \bibinfo {author} {\bibfnamefont {Emanuele}\ \bibnamefont {Berti}},
  \ and\ \bibinfo {author} {\bibfnamefont {Tjonnie G.~F.}\ \bibnamefont {Li}},\
  }\href@noop {} {\  (\bibinfo {year} {2018})},\ \Eprint
  {http://arxiv.org/abs/1804.09659} {arXiv:1804.09659
  [astro-ph.HE]}\BibitemShut {NoStop}%
\bibitem [{\citenamefont {Maselli}\ \emph
  {et~al.}(2017{\natexlab{b}})\citenamefont {Maselli}, \citenamefont
  {Pnigouras}, \citenamefont {Nielsen}, \citenamefont {Kouvaris},\ and\
  \citenamefont {Kokkotas}}]{Maselli:2017vfi}%
  \BibitemOpen
  \bibfield  {author} {\bibinfo {author} {\bibfnamefont {Andrea}\ \bibnamefont
  {Maselli}}, \bibinfo {author} {\bibfnamefont {Pantelis}\ \bibnamefont
  {Pnigouras}}, \bibinfo {author} {\bibfnamefont {Niklas~Grønlund}\
  \bibnamefont {Nielsen}}, \bibinfo {author} {\bibfnamefont {Chris}\
  \bibnamefont {Kouvaris}}, \ and\ \bibinfo {author} {\bibfnamefont
  {Kostas~D.}\ \bibnamefont {Kokkotas}},\ }\href
  {\doibase10.1103/PhysRevD.96.023005} {\bibfield  {journal} {\bibinfo
  {journal} {Phys. Rev.}\ }\textbf {\bibinfo {volume} {D96}},\ \bibinfo {pages}
  {023005} (\bibinfo {year} {2017}{\natexlab{b}})},\ \Eprint
  {http://arxiv.org/abs/1704.07286} {arXiv:1704.07286
  [astro-ph.HE]}\BibitemShut {NoStop}%
\bibitem [{\citenamefont {Ellis}\ \emph {et~al.}(2018)\citenamefont {Ellis},
  \citenamefont {Hektor}, \citenamefont {Hütsi}, \citenamefont {Kannike},
  \citenamefont {Marzola}, \citenamefont {Raidal},\ and\ \citenamefont
  {Vaskonen}}]{Ellis:2017jgp}%
  \BibitemOpen
  \bibfield  {author} {\bibinfo {author} {\bibfnamefont {John}\ \bibnamefont
  {Ellis}}, \bibinfo {author} {\bibfnamefont {Andi}\ \bibnamefont {Hektor}},
  \bibinfo {author} {\bibfnamefont {Gert}\ \bibnamefont {Hütsi}}, \bibinfo
  {author} {\bibfnamefont {Kristjan}\ \bibnamefont {Kannike}}, \bibinfo
  {author} {\bibfnamefont {Luca}\ \bibnamefont {Marzola}}, \bibinfo {author}
  {\bibfnamefont {Martti}\ \bibnamefont {Raidal}}, \ and\ \bibinfo {author}
  {\bibfnamefont {Ville}\ \bibnamefont {Vaskonen}},\ }\href
  {\doibase10.1016/j.physletb.2018.04.048} {\bibfield  {journal} {\bibinfo
  {journal} {Phys. Lett.}\ }\textbf {\bibinfo {volume} {B781}},\ \bibinfo
  {pages} {607--610} (\bibinfo {year} {2018})},\ \Eprint
  {http://arxiv.org/abs/1710.05540} {arXiv:1710.05540
  [astro-ph.CO]}\BibitemShut {NoStop}%
\bibitem [{\citenamefont {Pani}\ \emph
  {et~al.}(2012{\natexlab{b}})\citenamefont {Pani}, \citenamefont {Cardoso},
  \citenamefont {Gualtieri}, \citenamefont {Berti},\ and\ \citenamefont
  {Ishibashi}}]{Pani:2012vp}%
  \BibitemOpen
  \bibfield  {author} {\bibinfo {author} {\bibfnamefont {Paolo}\ \bibnamefont
  {Pani}}, \bibinfo {author} {\bibfnamefont {Vitor}\ \bibnamefont {Cardoso}},
  \bibinfo {author} {\bibfnamefont {Leonardo}\ \bibnamefont {Gualtieri}},
  \bibinfo {author} {\bibfnamefont {Emanuele}\ \bibnamefont {Berti}}, \ and\
  \bibinfo {author} {\bibfnamefont {Akihiro}\ \bibnamefont {Ishibashi}},\
  }\href {\doibase10.1103/PhysRevLett.109.131102} {\bibfield  {journal}
  {\bibinfo  {journal} {Phys. Rev. Lett.}\ }\textbf {\bibinfo {volume} {109}},\
  \bibinfo {pages} {131102} (\bibinfo {year} {2012}{\natexlab{b}})},\ \Eprint
  {http://arxiv.org/abs/1209.0465} {arXiv:1209.0465 [gr-qc]}\BibitemShut
  {NoStop}%
\bibitem [{\citenamefont {Arvanitaki}\ \emph {et~al.}(2015)\citenamefont
  {Arvanitaki}, \citenamefont {Baryakhtar},\ and\ \citenamefont
  {Huang}}]{Arvanitaki:2014wva}%
  \BibitemOpen
  \bibfield  {author} {\bibinfo {author} {\bibfnamefont {Asimina}\ \bibnamefont
  {Arvanitaki}}, \bibinfo {author} {\bibfnamefont {Masha}\ \bibnamefont
  {Baryakhtar}}, \ and\ \bibinfo {author} {\bibfnamefont {Xinlu}\ \bibnamefont
  {Huang}},\ }\href {\doibase10.1103/PhysRevD.91.084011} {\bibfield  {journal}
  {\bibinfo  {journal} {Phys. Rev.}\ }\textbf {\bibinfo {volume} {D91}},\
  \bibinfo {pages} {084011} (\bibinfo {year} {2015})},\ \Eprint
  {http://arxiv.org/abs/1411.2263} {arXiv:1411.2263 [hep-ph]}\BibitemShut
  {NoStop}%
\bibitem [{\citenamefont {Arvanitaki}\ \emph {et~al.}(2017)\citenamefont
  {Arvanitaki}, \citenamefont {Baryakhtar}, \citenamefont {Dimopoulos},
  \citenamefont {Dubovsky},\ and\ \citenamefont
  {Lasenby}}]{Arvanitaki:2016qwi}%
  \BibitemOpen
  \bibfield  {author} {\bibinfo {author} {\bibfnamefont {Asimina}\ \bibnamefont
  {Arvanitaki}}, \bibinfo {author} {\bibfnamefont {Masha}\ \bibnamefont
  {Baryakhtar}}, \bibinfo {author} {\bibfnamefont {Savas}\ \bibnamefont
  {Dimopoulos}}, \bibinfo {author} {\bibfnamefont {Sergei}\ \bibnamefont
  {Dubovsky}}, \ and\ \bibinfo {author} {\bibfnamefont {Robert}\ \bibnamefont
  {Lasenby}},\ }\href {\doibase10.1103/PhysRevD.95.043001} {\bibfield
  {journal} {\bibinfo  {journal} {Phys. Rev.}\ }\textbf {\bibinfo {volume}
  {D95}},\ \bibinfo {pages} {043001} (\bibinfo {year} {2017})},\ \Eprint
  {http://arxiv.org/abs/1604.03958} {arXiv:1604.03958 [hep-ph]}\BibitemShut
  {NoStop}%
\bibitem [{\citenamefont {Yoshino}\ and\ \citenamefont
  {Kodama}(2014)}]{Yoshino:2014wwa}%
  \BibitemOpen
  \bibfield  {author} {\bibinfo {author} {\bibfnamefont {Hirotaka}\
  \bibnamefont {Yoshino}}\ and\ \bibinfo {author} {\bibfnamefont {Hideo}\
  \bibnamefont {Kodama}},\ }\href {\doibase10.1093/ptep/ptv067} {\bibfield
  {journal} {\bibinfo  {journal} {PTEP}\ }\textbf {\bibinfo {volume} {2015}},\
  \bibinfo {pages} {061E01} (\bibinfo {year} {2014})},\ \Eprint
  {http://arxiv.org/abs/1407.2030} {arXiv:1407.2030 [gr-qc]}\BibitemShut
  {NoStop}%
\bibitem [{\citenamefont {Baryakhtar}\ \emph {et~al.}(2017)\citenamefont
  {Baryakhtar}, \citenamefont {Lasenby},\ and\ \citenamefont
  {Teo}}]{Baryakhtar:2017ngi}%
  \BibitemOpen
  \bibfield  {author} {\bibinfo {author} {\bibfnamefont {Masha}\ \bibnamefont
  {Baryakhtar}}, \bibinfo {author} {\bibfnamefont {Robert}\ \bibnamefont
  {Lasenby}}, \ and\ \bibinfo {author} {\bibfnamefont {Mae}\ \bibnamefont
  {Teo}},\ }\href {\doibase10.1103/PhysRevD.96.035019} {\bibfield  {journal}
  {\bibinfo  {journal} {Phys. Rev.}\ }\textbf {\bibinfo {volume} {D96}},\
  \bibinfo {pages} {035019} (\bibinfo {year} {2017})},\ \Eprint
  {http://arxiv.org/abs/1704.05081} {arXiv:1704.05081 [hep-ph]}\BibitemShut
  {NoStop}%
\bibitem [{\citenamefont {Brito}\ \emph
  {et~al.}(2017{\natexlab{b}})\citenamefont {Brito}, \citenamefont {Ghosh},
  \citenamefont {Barausse}, \citenamefont {Berti}, \citenamefont {Cardoso},
  \citenamefont {Dvorkin}, \citenamefont {Klein},\ and\ \citenamefont
  {Pani}}]{Brito:2017zvb}%
  \BibitemOpen
  \bibfield  {author} {\bibinfo {author} {\bibfnamefont {Richard}\ \bibnamefont
  {Brito}}, \bibinfo {author} {\bibfnamefont {Shrobana}\ \bibnamefont {Ghosh}},
  \bibinfo {author} {\bibfnamefont {Enrico}\ \bibnamefont {Barausse}}, \bibinfo
  {author} {\bibfnamefont {Emanuele}\ \bibnamefont {Berti}}, \bibinfo {author}
  {\bibfnamefont {Vitor}\ \bibnamefont {Cardoso}}, \bibinfo {author}
  {\bibfnamefont {Irina}\ \bibnamefont {Dvorkin}}, \bibinfo {author}
  {\bibfnamefont {Antoine}\ \bibnamefont {Klein}}, \ and\ \bibinfo {author}
  {\bibfnamefont {Paolo}\ \bibnamefont {Pani}},\ }\href
  {\doibase10.1103/PhysRevD.96.064050} {\bibfield  {journal} {\bibinfo
  {journal} {Phys. Rev.}\ }\textbf {\bibinfo {volume} {D96}},\ \bibinfo {pages}
  {064050} (\bibinfo {year} {2017}{\natexlab{b}})},\ \Eprint
  {http://arxiv.org/abs/1706.06311} {arXiv:1706.06311 [gr-qc]}\BibitemShut
  {NoStop}%
\bibitem [{\citenamefont {Essig}\ \emph {et~al.}(2013)\citenamefont {Essig}
  \emph {et~al.}}]{Essig:2013lka}%
  \BibitemOpen
  \bibfield  {author} {\bibinfo {author} {\bibfnamefont {Rouven}\ \bibnamefont
  {Essig}} \emph {et~al.},\ }in\ \href
  {https://inspirehep.net/record/1263039/files/arXiv:1311.0029.pdf} {\emph
  {\bibinfo {booktitle} {{Proceedings, 2013 Community Summer Study on the
  Future of U.S. Particle Physics: Snowmass on the Mississippi (CSS2013):
  Minneapolis, MN, USA, July 29-August 6, 2013}}}}\ (\bibinfo {year} {2013})\
  \Eprint {http://arxiv.org/abs/1311.0029} {arXiv:1311.0029
  [hep-ph]}\BibitemShut {NoStop}%
\bibitem [{\citenamefont {Rosa}\ and\ \citenamefont
  {Kephart}(2018)}]{Rosa:2017ury}%
  \BibitemOpen
  \bibfield  {author} {\bibinfo {author} {\bibfnamefont {Joao~G.}\ \bibnamefont
  {Rosa}}\ and\ \bibinfo {author} {\bibfnamefont {Thomas~W.}\ \bibnamefont
  {Kephart}},\ }\href {\doibase10.1103/PhysRevLett.120.231102} {\bibfield
  {journal} {\bibinfo  {journal} {Phys. Rev. Lett.}\ }\textbf {\bibinfo
  {volume} {120}},\ \bibinfo {pages} {231102} (\bibinfo {year} {2018})},\
  \Eprint {http://arxiv.org/abs/1709.06581} {arXiv:1709.06581
  [gr-qc]}\BibitemShut {NoStop}%
\bibitem [{\citenamefont {Cardoso}\ \emph
  {et~al.}(2017{\natexlab{b}})\citenamefont {Cardoso}, \citenamefont {Pani},\
  and\ \citenamefont {Yu}}]{Cardoso:2017kgn}%
  \BibitemOpen
  \bibfield  {author} {\bibinfo {author} {\bibfnamefont {Vitor}\ \bibnamefont
  {Cardoso}}, \bibinfo {author} {\bibfnamefont {Paolo}\ \bibnamefont {Pani}}, \
  and\ \bibinfo {author} {\bibfnamefont {Tien-Tien}\ \bibnamefont {Yu}},\
  }\href {\doibase10.1103/PhysRevD.95.124056} {\bibfield  {journal} {\bibinfo
  {journal} {Phys. Rev.}\ }\textbf {\bibinfo {volume} {D95}},\ \bibinfo {pages}
  {124056} (\bibinfo {year} {2017}{\natexlab{b}})},\ \Eprint
  {http://arxiv.org/abs/1704.06151} {arXiv:1704.06151 [gr-qc]}\BibitemShut
  {NoStop}%
\bibitem [{\citenamefont {East}(2017)}]{East:2017mrj}%
  \BibitemOpen
  \bibfield  {author} {\bibinfo {author} {\bibfnamefont {William~E.}\
  \bibnamefont {East}},\ }\href {\doibase10.1103/PhysRevD.96.024004} {\bibfield
   {journal} {\bibinfo  {journal} {Phys. Rev.}\ }\textbf {\bibinfo {volume}
  {D96}},\ \bibinfo {pages} {024004} (\bibinfo {year} {2017})},\ \Eprint
  {http://arxiv.org/abs/1705.01544} {arXiv:1705.01544 [gr-qc]}\BibitemShut
  {NoStop}%
\bibitem [{\citenamefont {Endlich}\ and\ \citenamefont
  {Penco}(2017)}]{Endlich:2016jgc}%
  \BibitemOpen
  \bibfield  {author} {\bibinfo {author} {\bibfnamefont {Solomon}\ \bibnamefont
  {Endlich}}\ and\ \bibinfo {author} {\bibfnamefont {Riccardo}\ \bibnamefont
  {Penco}},\ }\href {\doibase10.1007/JHEP05(2017)052} {\bibfield  {journal}
  {\bibinfo  {journal} {JHEP}\ }\textbf {\bibinfo {volume} {05}},\ \bibinfo
  {pages} {052} (\bibinfo {year} {2017})},\ \Eprint
  {http://arxiv.org/abs/1609.06723} {arXiv:1609.06723 [hep-th]}\BibitemShut
  {NoStop}%
\bibitem [{\citenamefont {Cardoso}\ \emph {et~al.}(2018)\citenamefont
  {Cardoso}, \citenamefont {Dias}, \citenamefont {Hartnett}, \citenamefont
  {Middleton}, \citenamefont {Pani},\ and\ \citenamefont
  {Santos}}]{Cardoso:2018tly}%
  \BibitemOpen
  \bibfield  {author} {\bibinfo {author} {\bibfnamefont {Vitor}\ \bibnamefont
  {Cardoso}}, \bibinfo {author} {\bibfnamefont {Óscar J.~C.}\ \bibnamefont
  {Dias}}, \bibinfo {author} {\bibfnamefont {Gavin~S.}\ \bibnamefont
  {Hartnett}}, \bibinfo {author} {\bibfnamefont {Matthew}\ \bibnamefont
  {Middleton}}, \bibinfo {author} {\bibfnamefont {Paolo}\ \bibnamefont {Pani}},
  \ and\ \bibinfo {author} {\bibfnamefont {Jorge~E.}\ \bibnamefont {Santos}},\
  }\href {\doibase10.1088/1475-7516/2018/03/043} {\bibfield  {journal}
  {\bibinfo  {journal} {JCAP}\ }\textbf {\bibinfo {volume} {1803}},\ \bibinfo
  {pages} {043} (\bibinfo {year} {2018})},\ \Eprint
  {http://arxiv.org/abs/1801.01420} {arXiv:1801.01420 [gr-qc]}\BibitemShut
  {NoStop}%
\bibitem [{\citenamefont {Frolov}\ \emph {et~al.}(2018)\citenamefont {Frolov},
  \citenamefont {Krtouš}, \citenamefont {Kubizňák},\ and\ \citenamefont
  {Santos}}]{Frolov:2018ezx}%
  \BibitemOpen
  \bibfield  {author} {\bibinfo {author} {\bibfnamefont {Valeri~P.}\
  \bibnamefont {Frolov}}, \bibinfo {author} {\bibfnamefont {Pavel}\
  \bibnamefont {Krtouš}}, \bibinfo {author} {\bibfnamefont {David}\
  \bibnamefont {Kubizňák}}, \ and\ \bibinfo {author} {\bibfnamefont
  {Jorge~E.}\ \bibnamefont {Santos}},\ }\href
  {\doibase10.1103/PhysRevLett.120.231103} {\bibfield  {journal} {\bibinfo
  {journal} {Phys. Rev. Lett.}\ }\textbf {\bibinfo {volume} {120}},\ \bibinfo
  {pages} {231103} (\bibinfo {year} {2018})},\ \Eprint
  {http://arxiv.org/abs/1804.00030} {arXiv:1804.00030 [hep-th]}\BibitemShut
  {NoStop}%
\bibitem [{\citenamefont {Yoshino}\ and\ \citenamefont
  {Kodama}(2012)}]{Yoshino:2012kn}%
  \BibitemOpen
  \bibfield  {author} {\bibinfo {author} {\bibfnamefont {Hirotaka}\
  \bibnamefont {Yoshino}}\ and\ \bibinfo {author} {\bibfnamefont {Hideo}\
  \bibnamefont {Kodama}},\ }\href {\doibase10.1143/PTP.128.153} {\bibfield
  {journal} {\bibinfo  {journal} {Prog. Theor. Phys.}\ }\textbf {\bibinfo
  {volume} {128}},\ \bibinfo {pages} {153--190} (\bibinfo {year} {2012})},\
  \Eprint {http://arxiv.org/abs/1203.5070} {arXiv:1203.5070
  [gr-qc]}\BibitemShut {NoStop}%
\bibitem [{\citenamefont {Pani}(2015{\natexlab{b}})}]{Pani:2015qhr}%
  \BibitemOpen
  \bibfield  {author} {\bibinfo {author} {\bibfnamefont {Paolo}\ \bibnamefont
  {Pani}},\ }\href {\doibase10.1103/PhysRevD.92.123530} {\bibfield  {journal}
  {\bibinfo  {journal} {Phys. Rev.}\ }\textbf {\bibinfo {volume} {D92}},\
  \bibinfo {pages} {123530} (\bibinfo {year} {2015}{\natexlab{b}})},\ \Eprint
  {http://arxiv.org/abs/1512.01236} {arXiv:1512.01236
  [astro-ph.HE]}\BibitemShut {NoStop}%
\bibitem [{\citenamefont {Caputo}\ \emph {et~al.}(2018)\citenamefont {Caputo},
  \citenamefont {Zavala},\ and\ \citenamefont {Blas}}]{Caputo:2017zqh}%
  \BibitemOpen
  \bibfield  {author} {\bibinfo {author} {\bibfnamefont {Andrea}\ \bibnamefont
  {Caputo}}, \bibinfo {author} {\bibfnamefont {Jesús}\ \bibnamefont {Zavala}},
  \ and\ \bibinfo {author} {\bibfnamefont {Diego}\ \bibnamefont {Blas}},\
  }\href {\doibase10.1016/j.dark.2017.10.005} {\bibfield  {journal} {\bibinfo
  {journal} {Phys. Dark Univ.}\ }\textbf {\bibinfo {volume} {19}},\ \bibinfo
  {pages} {1--11} (\bibinfo {year} {2018})},\ \Eprint
  {http://arxiv.org/abs/1709.03991} {arXiv:1709.03991
  [astro-ph.HE]}\BibitemShut {NoStop}%
\bibitem [{\citenamefont {Khmelnitsky}\ and\ \citenamefont
  {Rubakov}(2014)}]{Khmelnitsky:2013lxt}%
  \BibitemOpen
  \bibfield  {author} {\bibinfo {author} {\bibfnamefont {Andrei}\ \bibnamefont
  {Khmelnitsky}}\ and\ \bibinfo {author} {\bibfnamefont {Valery}\ \bibnamefont
  {Rubakov}},\ }\href {\doibase10.1088/1475-7516/2014/02/019} {\bibfield
  {journal} {\bibinfo  {journal} {JCAP}\ }\textbf {\bibinfo {volume} {1402}},\
  \bibinfo {pages} {019} (\bibinfo {year} {2014})},\ \Eprint
  {http://arxiv.org/abs/1309.5888} {arXiv:1309.5888 [astro-ph.CO]}\BibitemShut
  {NoStop}%
\bibitem [{\citenamefont {Blas}\ \emph {et~al.}(2017)\citenamefont {Blas},
  \citenamefont {Nacir},\ and\ \citenamefont {Sibiryakov}}]{Blas:2016ddr}%
  \BibitemOpen
  \bibfield  {author} {\bibinfo {author} {\bibfnamefont {Diego}\ \bibnamefont
  {Blas}}, \bibinfo {author} {\bibfnamefont {Diana~Lopez}\ \bibnamefont
  {Nacir}}, \ and\ \bibinfo {author} {\bibfnamefont {Sergey}\ \bibnamefont
  {Sibiryakov}},\ }\href {\doibase10.1103/PhysRevLett.118.261102} {\bibfield
  {journal} {\bibinfo  {journal} {Phys. Rev. Lett.}\ }\textbf {\bibinfo
  {volume} {118}},\ \bibinfo {pages} {261102} (\bibinfo {year} {2017})},\
  \Eprint {http://arxiv.org/abs/1612.06789} {arXiv:1612.06789
  [hep-ph]}\BibitemShut {NoStop}%
\bibitem [{\citenamefont {Ferreira}\ \emph {et~al.}(2017)\citenamefont
  {Ferreira}, \citenamefont {Macedo},\ and\ \citenamefont
  {Cardoso}}]{Ferreira:2017pth}%
  \BibitemOpen
  \bibfield  {author} {\bibinfo {author} {\bibfnamefont {Miguel~C.}\
  \bibnamefont {Ferreira}}, \bibinfo {author} {\bibfnamefont {Caio F.~B.}\
  \bibnamefont {Macedo}}, \ and\ \bibinfo {author} {\bibfnamefont {Vitor}\
  \bibnamefont {Cardoso}},\ }\href {\doibase10.1103/PhysRevD.96.083017}
  {\bibfield  {journal} {\bibinfo  {journal} {Phys. Rev.}\ }\textbf {\bibinfo
  {volume} {D96}},\ \bibinfo {pages} {083017} (\bibinfo {year} {2017})},\
  \Eprint {http://arxiv.org/abs/1710.00830} {arXiv:1710.00830
  [gr-qc]}\BibitemShut {NoStop}%
\bibitem [{\citenamefont {Bekenstein}\ and\ \citenamefont
  {Magueijo}(2006)}]{Bekenstein:2006fi}%
  \BibitemOpen
  \bibfield  {author} {\bibinfo {author} {\bibfnamefont {Jacob}\ \bibnamefont
  {Bekenstein}}\ and\ \bibinfo {author} {\bibfnamefont {Joao}\ \bibnamefont
  {Magueijo}},\ }\href {\doibase10.1103/PhysRevD.73.103513} {\bibfield
  {journal} {\bibinfo  {journal} {Phys. Rev.}\ }\textbf {\bibinfo {volume}
  {D73}},\ \bibinfo {pages} {103513} (\bibinfo {year} {2006})},\ \Eprint
  {http://arxiv.org/abs/astro-ph/0602266} {arXiv:astro-ph/0602266
  [astro-ph]}\BibitemShut {NoStop}%
\bibitem [{\citenamefont {Hall}\ \emph {et~al.}(2016)\citenamefont {Hall},
  \citenamefont {Callister}, \citenamefont {Frolov}, \citenamefont {Müller},
  \citenamefont {Pospelov},\ and\ \citenamefont {Adhikari}}]{Hall:2016usm}%
  \BibitemOpen
  \bibfield  {author} {\bibinfo {author} {\bibfnamefont {Evan~D.}\ \bibnamefont
  {Hall}}, \bibinfo {author} {\bibfnamefont {Thomas}\ \bibnamefont
  {Callister}}, \bibinfo {author} {\bibfnamefont {Valery~V.}\ \bibnamefont
  {Frolov}}, \bibinfo {author} {\bibfnamefont {Holger}\ \bibnamefont
  {Müller}}, \bibinfo {author} {\bibfnamefont {Maxim}\ \bibnamefont
  {Pospelov}}, \ and\ \bibinfo {author} {\bibfnamefont {Rana~X}\ \bibnamefont
  {Adhikari}},\ }\href@noop {} {\  (\bibinfo {year} {2016})},\ \Eprint
  {http://arxiv.org/abs/1605.01103} {arXiv:1605.01103 [gr-qc]}\BibitemShut
  {NoStop}%
\bibitem [{\citenamefont {Englert}\ \emph {et~al.}(2017)\citenamefont
  {Englert}, \citenamefont {Hild},\ and\ \citenamefont
  {Spannowsky}}]{Englert:2017det}%
  \BibitemOpen
  \bibfield  {author} {\bibinfo {author} {\bibfnamefont {Christoph}\
  \bibnamefont {Englert}}, \bibinfo {author} {\bibfnamefont {Stefan}\
  \bibnamefont {Hild}}, \ and\ \bibinfo {author} {\bibfnamefont {Michael}\
  \bibnamefont {Spannowsky}},\ }\href@noop {} {\  (\bibinfo {year} {2017})},\
  \Eprint {http://arxiv.org/abs/1712.04481} {arXiv:1712.04481
  [hep-ph]}\BibitemShut {NoStop}%
\end{thebibliography}%

\end{document}